\documentclass[a4paper,11pt,notitlepage,twoside]{report}
\pdfoutput=1

\usepackage{graphicx,amssymb}
\usepackage{amsmath}	
\usepackage{amsfonts}
\usepackage{color}
\usepackage{eso-pic}
\usepackage{subfigure}
\usepackage{lscape}
\usepackage[english]{babel}
\usepackage[T1]{fontenc}
\usepackage[utf8]{inputenc}
\usepackage{microtype}
\usepackage{url}
\usepackage{rotating}
\usepackage{pdflscape}
\usepackage{cite}
\usepackage[utf8]{inputenc}
\usepackage{enumerate}

\usepackage{setspace}
\usepackage{epstopdf}
\usepackage{makeidx}
\usepackage{lmodern}
\usepackage{booktabs}
\usepackage{units}
\usepackage{esint}
\usepackage{braket}
\usepackage{caption}
\usepackage[export]{adjustbox}

\usepackage{color}
\usepackage{bm}
\usepackage{enumitem}
\usepackage{epsfig}
\usepackage{subfigure}
\usepackage{array}
\usepackage[english]{babel}
\usepackage{tensor}
\usepackage{comment}
\usepackage[normalem]{ulem}
\usepackage[dvipsnames]{xcolor}

\usepackage{latexsym}
\usepackage{verbatim}
\usepackage{enumerate,mdwlist}

\usepackage{bbm}
\usepackage{tabularx}
\usepackage{slashed}
 
 \usepackage{etoolbox}
 \makeatletter
\patchcmd{\chapter}{\if@openright\cleardoublepage\else\clearpage\fi}{}{}{}
\makeatother

\usepackage{tikz}

\usepackage[pagestyles]{titlesec}
\usepackage{lipsum}

\usepackage{cancel}
\usepackage{dcolumn}
\usepackage{hhline}
\usepackage[dvipsnames]{xcolor}
\usepackage{caption}

\begingroup\expandafter\ifx\csname pdfsuppresswarningpagegroup\endcsname\relax\else\global\pdfsuppresswarningpagegroup=1\relax\fi\endgroup

\usepackage[top=1.in,inner=1.in,outer=1.in]{geometry} 

\usepackage{hyperref}
\usepackage{accents}
\usepackage{amsthm}

 \usepackage[safe]{tipa}

\usepackage{csquotes}

\makeatletter
\DeclareRobustCommand{\rcite}[1]{%
  \rcite@aux#1,\@nil{#1}%
}
\def\rcite@aux#1,#2\@nil#3{%
  \if\relax#2\relax
    Ref.~\cite{#3}%
  \else
    Refs.~\cite{#3}%
  \fi
}
\makeatother

\def\be{\begin{equation}}
\def\ee{\end{equation}}
\def\ba{\begin{eqnarray}}
\def\ea{\end{eqnarray}}

\def\f{\frac}
\def\g{\gamma}
\def\mp{M_\mathrm{pl}}

\def\nn{\nonumber}

\def\hiclass{\texttt{hi\_class }}
\def\eftcamb{\texttt{EFTCAMB} }
\def\coop{\texttt{COOP} }
\def\class{\texttt{CLASS} }
\def\camb{\texttt{CAMB} }

\def\eg{{\frenchspacing\it e.g.}}

\def\fg{\mathfrak{f}}

\newcommand\define{\equiv}

\renewcommand\i{\ensuremath{\mathrm{i}}}

\newcommand\U[1]{\:\mathrm{#1}}

\newcommand{\grad}{\vec{\nabla}}

\renewcommand\lim[2]{\underset{ #1 \rightarrow #2 }{ \mathrm{lim} } \,}

\newcommand{\delimiters}[4][]{
\ifthenelse{ \equal{#1}{1} }{  #2 #3 #4  }
					{ \ifthenelse{\equal{#1}{2}}{ \big#2 #3 \big#4 }
						{ \ifthenelse{\equal{#1}{3}}{ \Big#2 #3 \Big#4 }
							{ \ifthenelse{\equal{#1}{4}}{ \bigg#2 #3 \bigg#4 }
								{ \ifthenelse{\equal{#1}{5}}{ \Bigg#2 #3 \Bigg#4 }
									{ \left#2 #3 \right#4 }
								}
							}
						}
					}
													}

\newcommand{\HH}{\mathcal{H}}

\newcommand\eea{\end{eqnarray}}
\newcommand\bea{\begin{eqnarray}}

\newcommand{\bn}{\mathbf{n}}
\newcommand{\bx}{\mathbf{x}}

\newcommand{\bk}{\mathbf{k}}

\newcommand{\B}{\textrm{B}}
\newcommand{\F}{\textrm{F}}

\newcolumntype{C}[1]{>{\centering\arraybackslash}p{#1}}
\newcolumntype{L}[1]{>{\raggedright\arraybackslash}p{#1}}
\newcolumntype{R}[1]{>{\raggedleft\arraybackslash}p{#1}}

\newlength{\boxtitlelength}
\newlength{\halfrulelength}
\newcommand{\boxtitle}[1]{\footnotesize\bf{\:#1\:}}

\definecolor{blue4}{RGB}{0,0,143}
\definecolor{red4}{RGB}{143,0,0}
\definecolor{orange}{RGB}{255,128,0}
\definecolor{darkcyan}{RGB}{0,128,128}
\definecolor{olive}{RGB}{0,128,0}
\definecolor{purple}{RGB}{128,0,128}
\definecolor{cyan2}{RGB}{0,255,255}
\definecolor{fushia}{RGB}{255,0,255}
\definecolor{mygray}{gray}{0.5}
\definecolor{lightgray}{gray}{0.85}

\makeatletter
\def\@fpheader{\relax}
\makeatother

\def\de{\delta}

\def\H0{H_{0}}

\makeatletter

\newcommand{\Rmnum}[1]{\expandafter\@slowromancap\romannumeral #1@}
\makeatother

\newcommand{\symmC}{symm\_C\ }

\newcommand{\lcdm}{$\Lambda$CDM }

\newcommand{\bq}{\begin{eqnarray}}
\newcommand{\eq}{\end{eqnarray}}

\newcommand{\vek}[1]{\mathbf{#1}}

\newcommand*\intd[2]{
  \mathrm{d}
  \ifx\relax#1\relax\else
  \rule{-0.05em}{1.9ex}^{#1}\!
  \fi
  #2\,
}

\newcommand{\Mpch}{\,{\rm Mpc}\,\ifmmode h^{-1}\else $h^{-1}$\fi}
\newcommand{\Msh}{\,\ifmmode M_\odot\,h^{-1}\else $M_\odot\,h^{-1}$\fi}

\def \beq  {\begin{equation}}
\def \eeq  {\end{equation}}
\def \ber  {\begin{eqnarray}}
\def \eer  {\end{eqnarray}}
\def \Geff {G_{\rm eff} }
\def \Geffz {$G_{\rm eff}(z)$ }

\def \Geff {G_{\rm eff}}

\newcommand{\newc}{\newcommand}
\newc{\D}{\partial}
\newc{\lcdmnospace}{$\Lambda$CDM}
\newc{\wcdm}{wCDM }
\newc{\omom}{$\Omega_{0m}$ }
\newc{\omomnospace}{$\Omega_{0m}$}
\newc{\plcdm}{Planck18/$\Lambda$CDM }
\newc{\plcdmnospace}{Planck18/$\Lambda$CDM}
\newc{\wlcdm}{WMAP7/$\Lambda$CDM }
 
\newc{\fsz}{{\rm{\it f\sigma}}_8(z)}

\newc{\ra}{\Rightarrow}

\providecommand{\U}[1]{\protect\rule{.1in}{.1in}}

\newcommand{\mincir}{\raise
-3.truept\hbox{\rlap{\hbox{$\sim$}}\raise4.truept\hbox{$<$}\ }}
\newcommand{\magcir}{\raise
-3.truept\hbox{\rlap{\hbox{$\sim$}}\raise4.truept\hbox{$>$}\ }}

\newcommand{\Lag}{{\cal L}}

\def\widebar{\overline}

\def\b{\beta}
\def\gCg{\gamma}
\def\laCg{\lambda}

\def\Om{\Omega}
\def\GCg{\Gamma}
\def\vpCg{\varphi}
\def\NCg{\nabla}
\def\BCg{\Box}
\def\cC{\mathcal{C}}
\def\cK{\mathcal{K}}
\def\cL{\mathcal{L}}
\def\cR{\mathcal{R}}
\def\cV{\mathcal{V}}
\def\lpCg{\ell_{ Pl}}
\def\rmeCg{\text{e}}

\def\rmeCg{e}
\def\rmiCg{i}
\def\HCg{{\rm H}}
\newcommand{\Eq}[1]{(\ref{#1})}

\newcommand{\al}{\ensuremath{\alpha}}

\newcommand{\udt}[3]{#1^{#2}_{\phantom{#2}#3}}

\newcommand{\dut}[3]{#1_{#2}^{\phantom{#2}#3}}

\newcommand{\tp}[1]{\accentset{\bullet}{#1}\vphantom{#1}}

\newtheorem{theorem}{Theorem}

\theoremstyle{definition}
\newtheorem{definition}{Definition}
\theoremstyle{remark}

\newcommand{\bT}{\mathbf{T}}

\newcommand{\bK}{\mathbf{K}}

\newcommand{\bR}{\mathbf{R}}

\newcommand{\bQ}{\mathbf{Q}}

\newcommand{\bL}{\boldsymbol{\mathrm{L}}}

\newcommand{\st}[1]{\accentset{\diamond}{#1}\vphantom{#1}}

\newcommand{\lnm}{{\Lambda_{\rm{Q}}}}
\newcommand{\lrbg}{{\Lambda_{\rm{RBG}}}}
\newcommand{\bpsi}{{\bar{\psi}}}

\titleformat{\chapter}%
  {\normalfont\bfseries\Huge}{\thechapter.}{10pt}{}
\newpagestyle{mystyle}{
  \sethead[][\thechapter.\enspace\chaptertitle][]{}{\thesection~\sectiontitle}{}
\setfoot{}{\thepage}{}}

\begin{document}

{\title{\LARGE{\bf{  MODIFIED GRAVITY AND COSMOLOGY:\\ An Update by the CANTATA 
Network}}}}


\date{}
 \maketitle

\author{{\bf 
Emmanuel N. Saridakis$^{1,2,3}$,
Ruth Lazkoz$^{4}$,
 Vincenzo Salzano$^{5}$,
Paulo Vargas Moniz$^{6,7}$,\
Salvatore Capozziello$^{8,9,10}$,  \
Jose Beltr\'an Jim\'enez$^{11}$, \
 Mariafelicia De \\Laurentis$^{12,8,10}$,
 Gonzalo J. Olmo$^{13,14}$
(Editors)\\
}}

\author{
\bf \noindent
Yashar Akrami$^{15,16}$,\ \
Sebastian Bahamonde$^{17,18}$,\ \
Jose Luis Bl\'azquez-Salcedo$^{19}$, \\
Christian G. B{\"o}hmer$^{18}$,
Camille Bonvin$^{20}$,
Mariam Bouhmadi-L\'opez$^{21,4}$,
Philippe Brax$^{22}$,
Gianluca Calcagni$^{23}$,
Roberto Casadio$^{24,25}$,
Jose A. R. Cembranos$^{19}$,\\
\'Alvaro de la Cruz-Dombriz$^{26}$,
Anne-Christine Davis $^{7}$,
  Adrià Delhom$^{13}$, 
  Eleonora Di Valentino$^{27}$, 
Konstantinos F.  Dialektopoulos$^{28}$,
Benjamin Elder$^{29,30}$,
Jose Mar\'ia Ezquiaga$^{31}$,
Noemi Frusciante$^{32}$,
Remo Garattini$^{33,34}$,  
L\'aszl\'o \'A. Gergely$^{35}$,
Andrea Giusti$^{36}$,
Lavinia Heisenberg$^{37}$,
Manuel Hohmann$^{17}$, 
Damianos Iosifidis$^{38}$,\\
Lavrentios Kazantzidis$^{39}$,
Burkhard Kleihaus$^{40}$,
Tomi S. Koivisto$^{17,41,42,43}$,
Jutta Kunz$^{40}$,
Francisco S. N. Lobo$^{32}$,
Matteo Martinelli$^{44,45}$,
Prado Mart\'in-Moruno$^{19}$,
Jos\'e Pedro Mimoso$^{32}$,
David F. Mota$^{46}$, \
Simone Peirone$^{44}$, \
Leandros Perivolaropoulos$^{39}$, \
Valeria Pettorino$^{47}$,  
Christian Pfeifer$^{17}$,
Lorenzo Pizzuti$^{48}$,
Diego Rubiera-Garcia$^{19}$,
Jackson Levi Said$^{49,50}$,
Mairi Sakellariadou$^{51}$, 
Ippocratis D. Saltas$^{52}$, 
 Alessio Spurio Mancini$^{53}$,
Nicoleta Voicu$^{54}$,
 Aneta Wojnar$^{17}$
 (Section Contributors)\\
}


\vskip 2cm

\begin{abstract} 
General Relativity and the  $\Lambda$CDM framework are currently the standard 
lore and constitute the concordance paradigm. Nevertheless, long-standing open 
theoretical issues, as well as possible new observational ones arising from the 
explosive development of cosmology the last two decades, offer the motivation 
and lead a large amount of research to be devoted in constructing various 
extensions and modifications.

All extended theories and scenarios are first examined under the light of 
theoretical consistency, and then are applied to various 
geometrical backgrounds, such as the cosmological and the spherical symmetric 
ones. Their predictions at both the background and perturbation levels, and 
concerning cosmology at early, intermediate and late times, are then confronted 
with the huge amount of observational data that astrophysics and cosmology are 
able to offer recently. Theories, scenarios and models that 
successfully and efficiently pass the above steps are classified as viable and 
are candidates for the description of Nature.

This work is a Review of the recent developments in the fields of 
gravity and cosmology, presenting the state of the art,  high-lighting the 
open problems, and outlining the directions of future research. Its realization 
was performed in the framework of the  COST European Action  ``Cosmology and 
Astrophysics Network for Theoretical Advances and Training Actions''. 

\end{abstract}

 \newpage
 
\tableofcontents

\newpage

\newpage

\section*{Preface}

The dawn of the 21st century came with very positive prospects for gravity, 
cosmology and astrophysics. Technological progress made it possible for 
cosmology to enter to its adulthood and become a precision science, both for 
its own shake as well as for being the laboratory of gravity, which can now be 
accurately tested and investigated in scales different than the earth ones. As a 
result, the opinion that cosmology is one of the main directions that will lead 
to progress in physics in the near future, is now well established.

 ``Cosmology and Astrophysics Network for Theoretical Advances and Training 
Actions'' (CANTATA) is a COST European Action established in 2015 in order to 
contribute to the front of research in the fields of gravity, cosmology and 
astrophysics. It involves Institutions from 26 European countries, as well as 
from 5 countries abroad. CANTATA Collaboration has a variety of interests, 
which include: i) the classification and definition of theoretical and 
phenomenological aspects of gravitational interaction that cannot be enclosed 
in the standard lore scheme but might be considered as signs of alternative 
theories of gravity, ii) the confrontation of the theoretical predictions with 
observations at both the background and the perturbation levels, iii) the 
production of numerical codes to simulate astrophysical and cosmological 
phenomena, iv) the construction of self-consistent models at various scales and 
the investigation of the features capable of confirming or ruling out an 
effective theory of gravity, v) the study of how extended and modified theories 
of gravity emerge from quantum field theory and how mechanisms produced by the 
latter may explain cosmological dynamics. This Review presents the 
recent developments in the above fields.\\

 \noindent Emmanuel N. Saridakis
 
\noindent Ruth Lazkoz

\noindent  Vincenzo Salzano

\noindent Paulo Vargas Moniz

\noindent Salvatore Capozziello

\noindent Jose Beltr\'an Jim\'enez

\noindent Mariafelicia De Laurentis

\noindent Gonzalo J. Olmo

\newpage

\section*{Conventions}

\begin{table}[ht]
\begin{center}
\begin{tabular}{|c|l|}
\hline
Greek small letters $\alpha, \mu, \nu,$... &  space-time coordinates 
indices\\
Latin small letters $i,j,k$... & space coordinates indices\\
Latin capital  indices $A,B, $... & tangent space indices\\
& (only in chapters \ref{gravextradim} and \ref{Calcagninonlocal}\\
&D-dimensional 
coordinate indices)  
\\
$g_{\mu\nu}$ & metric tensor\\
(-+++) & metric signature   \\
$\Gamma^{\mu}_{\phantom{\mu} \nu \rho}$ & Levi-Civita connection\\
$R^{\mu}_{\phantom{\mu} \nu \alpha \beta} = \partial_{\alpha}
\Gamma^{\mu}_{\phantom{\mu} \nu \beta}- \partial_{\beta}
\Gamma^{\mu}_{\phantom{\mu} \nu \alpha} + \Gamma^{\mu}_{\phantom{\mu}
  \sigma \alpha} \Gamma^{\sigma}_{\phantom{\sigma} \nu \beta} -
\Gamma^{\mu}_{\phantom{\mu} 
  \sigma \beta} \Gamma^{\sigma}_{\phantom{\sigma} \nu \alpha}$ &  Riemann   
  curvature tensor\\
   $R_{\mu \nu} = R^{\alpha}_{\phantom{\alpha} \mu \alpha \nu}$ & Ricci tensor\\
  $R=R^{\alpha}_{\phantom{\alpha} \alpha}$ & Ricci scalar\\  
   $G_{\mu \nu} = R_{\mu \nu} - \frac{1}{2} g_{\mu \nu} R$ & Einstein tensor\\
   $\nabla_{\mu}$ & covariant derivative \\
   $\Box \equiv g^{\mu \nu} \nabla_{\mu} \nabla_{\nu}$ & 
d'Alembertian operator  \\
$2X_{[\alpha\beta]}=X_{\alpha\beta}-X_{\beta\alpha}$ &anti-symmetricity \\  
$2X_{(\alpha\beta)}=X_{\alpha\beta}+X_{\beta\alpha}$ &symmetricity \\  
  $ds^2 = -dt^2 + a^2 (t) \left[ \frac{dr^2}{1-k r^2} + r^2  \left(  d \theta^2 
+ \sin^2 
\theta d \phi^2 \right)   \right]$ & 
4-dimensional Friedmann-Lema\^{i}tre-\\
 & Robertson-Walker (FLRW) 
line-element\\
$\tau = \int dt/a(t)$ & conformal time\\
$\dot{\;} \equiv \frac{d}{dt}$ & cosmic time derivative\\
$\;^{\prime} \equiv \frac{d}{d\tau}$  & conformal time derivative\\
$ds_{(3)}^2=\gamma_{ij} dx^i dx^j = \frac{dr^2}{1-k r^2} +r^2
d\theta^2+r^2 \sin^2 \theta d \phi^2$ &  maximally symmetric  3-dimensional\\
 &
space-like hyper-surfaces metric\\
$\grad_i$ &  grad operator on the  3-dimensional \\
 &
space-like hyper-surfaces\\
$\Delta \equiv \gamma^{ij} \grad_i \grad_j$  &   Laplacian operator\\
$ds^2 =   -(1+2 \Psi)dt^2+      a^2 (t)(1-2 \Phi) \gamma_{ij}
  dx^i dx^j $ & Newtonian gauge scalar metric perturbations\\
$T^{\mu \nu} = \frac{2}{\sqrt{-g}} \frac{\delta \mathcal{L}_m}{\delta
  g_{\mu \nu}}$  &   energy-momentum tensor \\
 & of the Lagrangian density 
$\mathcal{L}$\\
$\kappa^2\equiv 8\pi G_N\equiv M_{Pl}^{-2}$ & gravitational constant\\
$\hbar=c=k_B=1$&natural units\\
\hline
\end{tabular}
\label{tab:notation}
\caption*{List of notational conventions used in this manuscript, unless 
otherwise stated.}
\end{center}
\end{table}

\begin{table}[htbp]
\begin{center}
\begin{tabular}{|c|l|}
\hline 
$\hat{\Gamma}^\alpha_{\mu\nu}$ & general affine connection\\
$\breve{\Gamma}^\alpha_{\mu\nu}$ & Palatini connection\\
$\tp{\Gamma}^\alpha_{\mu\nu}$ &  teleparallel affine  (Weitzenb\"ock) 
connection\\
$\st{\Gamma}^\alpha_{\mu\nu}$ &  symmetric teleparallel 
connection\\
$\widebar{\Gamma}^\alpha_{\mu\nu}$ &  Chern-Rund linear
connection\\
 $\bar{\bar{\Gamma}}^{\mu}{}_{\nu}$
  & canonical nonlinear connection\\
 $\bigcirc$ & arbitrary object wrt the Levi-Civita connection\\ 
$\hat{\bigcirc}$ & arbitrary object wrt the metric affine connection\\ 
$\breve{\bigcirc}$ & arbitrary object wrt the Palatini connection\\ 
$\tp{\bigcirc}$ & arbitrary object wrt the Weitzenb\"ock connection\\ 
$\st{\bigcirc}$ & arbitrary object wrt the symmetric teleparallel 
connection\\ 
$\widebar{\bigcirc}$ & arbitrary object wrt the Chern-Rund linear
connection\\ 
$\bar{\bar{\bigcirc}}$ & arbitrary object wrt the  canonical nonlinear 
connection \\
$\omega^A{}_{B\mu}$ & spin connection\\
$D_\mu$ & Fock-Ivanenko derivative\\
$T^{\mu}{}_{\nu\rho}$ & torsion tensor\\ 
$Q_{\alpha\mu\nu}=\nabla_{\alpha}g_{\mu\nu}$ & non-metricity 
tensor\\
$T_{\mu} = T^{\nu}{}_{\nu\mu}$ & torsion vector\\
$\hat{\tilde{R}}_{\alpha\beta}:=\hat{R}^{\mu}_{\ \  \mu\alpha\beta}$ & 
homothetic curvature \\
$\hat{\tilde{\mathcal{R}}}{}^{\lambda}_{\ \kappa}:=
\hat{R}^{\lambda}_{\ \mu\nu\kappa}g^{\mu\nu}$  & co-Ricci tensor  \\
$\mathcal{R} \equiv  g^{\mu\nu}\mathcal{R}_{\mu\nu}\equiv$ & 
\\
$ g^{\mu\nu}\left(
\breve{\Gamma}^\alpha_{\mu\nu , \alpha}
       - \breve{\Gamma}^\alpha_{\mu\alpha , \nu} +
\breve{\Gamma}^\alpha_{\alpha\lambda}\breve{\Gamma}^\lambda_{\mu\nu} -
\breve{\Gamma}^\alpha_{\mu\lambda}\breve{\Gamma}^\lambda_{\alpha\nu}\right)$ & 
Palatini curvature
\\
$e^A{}_{\mu}$ & tetrad (vielbein, coframe)\\
$e_A{}^{\mu}$ & frame dual to $e^A{}_{\mu}$\\
$K^{\mu}{}_{\nu\rho}$ & contortion tensor\\ 
$L^{\mu}{}_{\nu\rho}$ & distortion tensor\\ 
$S_{\mu}{}^{\nu\rho}$ & superpotential\\ 
$\mathbb{T}$ & torsion scalar\\ 
$\mathbb{Q}$ & non-metricity scalar\\ 
$\epsilon_{\mu\alpha\beta\gamma}$ & 4-dimensional totally 
antisymmetric\\
& Levi-Civita tensor\\
\hline
\end{tabular}
\label{tab:notation3}
\caption*{ (cont.) List of notational conventions used in this manuscript, 
unless 
otherwise stated.}
\end{center}

\end{table}

\phantomsection
\addcontentsline{toc}{part}{\bf Introduction}
\begin{center}
 
 \newpage
 
{\Huge \bf Introduction}
\end{center}
\begin{center}
Editors:   Emmanuel N. Saridakis,   Paulo Vargas Moniz
\end{center}


\chapter[Cosmophysics of modified gravity]{Cosmophysics of modified gravity}

{\em Ruth Lazkoz}\\

\section*{Breaking scientific and geographic boundaries  through the physics of 
gravitation  during  a societal plight}

One of the biggest achievements of Physics has been 
tailoring a standard cosmological model to describe the Universe quite 
successfully  through a vast multitude of observations. In this architectural 
miracle the role of buildings bricks and joists is played by exquisite 
observational data, informing us about different epochs, regions and regimes. 
Evidence ranges from  cosmic microwave background fluctuations to supernovae 
luminosity distances, along with baryon acoustic oscillations, cluster mass 
measurements and several other probes. Note that our access to an exquisite 
profusion of data is a fortunate sign of our times, while that was not certainly 
the case when the architects of modern cosmology set the foundations of our 
discipline. They certainly did not work in such gilded stage, as Nobel Laureate 
James Peebles recalls about his graduate years:
\begin{displayquote}
{\it ``I was very uneasy about going into cosmology because the experimental 
observations were so modest.''}
\end{displayquote}
However, our standard cosmological paradigm is a little patchy 
picture: in general, transitions between different epochs and scales are 
 somewhat  poorly depicted.  
Traditionally, cosmologists have aspired to fitting as much physics as possible  
into a full-fledged single model, but this goal faces challenges which have 
recently become quite an issue. In this respect we can mention two that have 
gathered some attention for different reasons.  
 
 Firstly, we witness  the discrepancy between reputed teams of cosmologists as 
to 
what is the speed at which astronomical bodies are hurtling away from us, 
unprivileged observers.  This results in the so called  tension on the current 
value of the $H_0$ parameter (see   Chapter \ref{sec:DiValentino} by 
Di Valentino),
a concern that 
has been around us for some years now and that does not cease to be a lively 
source of controversy, quite the opposite.

Secondly, I wish to add to this very short editor's choice of baffling problems 
another bold departure from (cosmological) orthodoxy that has been put forward 
according to reliable  evidence coming from \textit{Planck} data: the spatial 
curvature of the Universe might be non zero \cite{DiValentino:2019qzk}. The 
debate is alive and kicking, although it might be closed by a thoughtful 
identification of degeneracy-breaking data set, cosmic chronometers, which, 
perhaps not so anecdotally rely heavily on our understanding on galactic 
evolution \cite{Vagnozzi:2020dfn}.
 
 We may be tempted to regard these two quandaries as astrophysical rather 
gravitational, but
gravity is by all means  the dominant interaction on the scales of astrophysical 
interest, which are revealing to us every day more exquisitely. For this 
reason, 
it is clear that these are times when  a fluid dialogue between theory and 
experiment is very much needed. As Albert Einstein had it: 
\begin{displayquote}
{\it ``A theory is something nobody believes, except the person who made it. An 
experiment is something everybody believes, except the person who made it.''}
\end{displayquote}
It may seem from the previous paragraphs that the routes to test the 
gravitational interaction are restricted to macroscopic realms where 
concentrations of mass and/or energy are very significant, but even though full 
characterizations of effects in some modified gravity scenarios are still 
lacking,  our improved understanding  (and in particular beyond Riemannian 
frameworks) tells us that microscopic experiments might bring surprises (see 
Chapter \ref{Delhom2chapter} by A. Delhom). Just keep in mind that:
\begin{displayquote}
{\it ``We often fail to notice things that we are not expecting.''} - Lisa 
Randall
\end{displayquote}
 
Another possible summary from the preceding paragraphs is that the  array of 
fundamental problems the community is involved is certainly quite vast, and 
hence  
I could have easily arranged a  different set of questions to tackle rather than 
the pair up above (even without venturing into the quantum gravity dominion). 
Yet, in my personal view, the current wide spectrum of enigmas is nothing but a 
manifestation of our need to improve and extend the standard body of our 
knowledge in gravity. Fortunately, even though history tells us that early 
motivations to go beyond General Relativity were more intellectually motivated 
by Mathematics, it is Physics itself where such explorations find their  roots 
at present
(see Chapter \ref{Bohmerchapter} by C. Boehmer   
and references therein). Needless to say that a portion of our readership may 
benefit from the review chapter \ref{Mimosochapter}
by J.P. Mimoso  
on basic aspects of General Relativity that acts as a kick-off to this volume.
 
 One of the most tempting routes of modification of General Relativity (GR) has 
been the possibility that our spacetime has more than four dimensions, with the 
extra ones inaccessible to low-energy exploration methods (as explained in a 
much broader context in Chapter \ref{gravextradim} J.A.R. Cembranos). But it 
is well known that if we want to describe our macroscopic experience of the 
Universe in those scenarios, we have to roll up these extra dimensions. Such 
operation typically results in specific  modifications of the gravitational 
Lagrangian, which can be found among the many covered in the huge assortment of 
possibilities discussed in the present manuscript (endorsed with their physical 
motivation, definitely).
 
 But extra-dimensional gravity has not been the  major operation ground of the 
European network CANTATA, whose work is celebrated by this and all following 
chapters. In general, our large team has rather set the focus on other enticing 
possibilities, those in which the gravitational Lagrangian is modified in a 
``why not'' intellectually legit spirit, thus allowing other degrees of freedom  
to play a role in the construction of our theoretical settings. Put in a bit 
rough way: most modifications to be found in the following pages rely on 
modifying the action by resorting to scalars that can be built with either the 
metric tensor or different pieces of the affine connection.
 
 Note that I have inconspicuously introduced in my dissertation the two main 
building blocks of our gravitational theory:  differential geometry as its 
mathematical formulation on    one hand, and 
 the Lagrangian as the encoder of its physical content on the other hand. The 
necessary association among them emerges when we eventually find out how a 
manifold, furnished with a metric and a connection, dictates how all the 
particles that live in it move, and conversely, how all the physical 
manifestations of those particles cipher the metric and the connection. For a 
less over-scrupulous redemption of this powerful idea the reader can  resort to 
the {\it ad nauseam} quoted pronouncement by John Archibald Wheeler:
 \begin{displayquote}
 {\it ``Space-time tells matter how to move; matter tells space-time how to 
curve.''}
\end{displayquote}
 Let me elaborate further on this by working for a moment on a setting in which 
Newton's mechanics (and theory of gravity) is valid, say, Special Relativity 
(SR). In a covariant fashion, the (vanishing) acceleration we require to  
describe the motion of massive particles in free fall takes the following form 
for any observer other than the local inertial one:
 \begin{equation}     
a^{\lambda}\equiv\frac{d^2x^{\lambda}}{d\tau^2}+\Gamma^{\lambda}_{\mu\nu}\frac{
dx^{\mu}}{d\tau}\frac{dx^{\nu}}{d\tau},\label{geodesic}
 \end{equation}
 where $\Gamma^{\lambda}_{\mu\nu}$, according to usual notation, represents  the 
Christoffel symbol of the second kind, also known as the Levi-Civita connection, 
which is
 defined through the usual operation of combining partial derivatives of the 
metric and contractions (with the metric itself). Note that all the rest of 
notation used is so standard that explanations can hopefully be waived. 
 
 Now, in order to preserve our ability to switch consistently between the 
physical conclusions of different observers, we need to pinpoint rigorously some 
key aspects within the broad concept of ``variation''. For instance, in this 
specific problem I have resorted to a definition of acceleration that works the 
same in all coordinate systems. It turns out that the (mere) partial derivative 
of the velocity with respect to the proper time does not render a covariant 
vector which can be interpreted as the acceleration.  
 
 This is where the covariant derivative enters the picture and lets us define 
the acceleration rigorously and (in principle) exactly as above. But this 
difficulty with derivatives is not just a particular problem of the velocity:  
partial derivatives in general do not play as good tensor operators, whereas the 
covariant derivative does indeed allow for parallel transport of tangent vectors 
along curves, i.e. it produces derivatives with tensorial nature. In this 
fashion, the acceleration is nothing but the covariant derivative of the 
velocity along the curves which the velocity field itself constructs, that is, 
 $a^{\mu}\equiv u^{\mu}{}_{;\nu}u^{\nu}.$
 Nevertheless, there is a subtle point here, as this geometric route brings us 
back formally to the geodesics equation, that is,  Eq. 
 \ref{geodesic}, but with the caveat that in this setting 
$\Gamma^{\lambda}_{\mu\nu}$ can represent any connection whatsoever, and not 
necessarily the Levi-Civita one, which is metric compatible and torsion free 
(and therefore symmetric). Roadmaps to geometric concepts such as the last two, 
which will be revisited often in this volume, can be found in Chapter  
\ref{ref:Iosifidis} by D. Iosifidis and E.N. Saridakis and in Chapter 
\ref{Koivistochapter}  by T. 
Koivisto). 
 
Before proceeding any further, an apology  is in order, though, because I am 
taking advantage of the many more details other contributions are to offer in 
this collection to simplify my dissertation, and therefore I am giving myself 
permission to follow the Nobel Laureate Roger Penrose's recommendation:
\begin{displayquote}
\textit{``Do not be afraid to skip equations (I do this frequently myself).''}
\end{displayquote}
The ambitious and broad next step would then to perform experiments to inform us 
of the motion of test particles, which would then reveal the specific form of 
the connection through their geometric properties. In this respect, this volume 
offers deeper insights into the boundless question of how curvature affects the 
motion of particles (see  Chapter \ref{Gergelychapter} by L.\'{A}. Gergely 
for an overview on gravitational lensing). But in order to paint a master work 
of art and not a mere sketch, us physicists have to associate the equations of 
motion governing  the pertinent trajectories with a physical framework, that is, 
we need to match particles and fields\footnote{Not that I pretend by any means 
to be an unpaired artist of gravitation myself. I rather mean the work of art, 
that the edifice of gravitational theories represent, will be or has been the 
result of collective efforts such as our network's}. 

The, perhaps, most elegant way to accomplish this task is the application of 
the 
variational principle to an action. The advantage of this method is how 
(relatively) undemanding it becomes to compare GR and its modifications with 
other classical field theories (such as Maxwell's theory). Nevertheless, for the 
sake of fluidity of this chapter, I feel compelled to write here the so very 
well-known action 
\begin{equation}
    S=\int d^4 x\sqrt{-g}\left[\frac{1}{2\kappa} R+{\cal L}_m\right],
\end{equation}
where, again,  standard textbook notation is being used, and therefore $ \kappa 
={{8\pi G}/{c^{4}}}$, $R$ denotes the Ricci scalar, and $g$ is the determinant 
of the $g_{\mu\nu}$ metric. In the usual manner, I also let ${\cal L}_m$  
account for any matter fields. Once  the degrees of freedom of the action 
have 
been established, 
variation upon them will yield the (relevant) field equations, which will, in 
turn, be the route to the equations of motion. In this way, the geometric and 
matter pieces in the above Lagrangian yield the left and right hand side of the 
Einstein equations:
\begin{equation}
R_{\mu\nu}-\frac{1}{2}g_{\mu\nu}R=\kappa T_{\mu\nu}\, .
\end{equation}
The mathematical identity represented by the vanishing covariant derivative of 
the left hand side, implies the vanishing of the right hand side, and four 
constraint equations follow. In the particular case that the energy-momentum 
tensor is that of incoherent matter,  a straightforward\footnote{
Needless to say that in this framework straightforward actually means 
``inexperienced lecturer mind blowing''.} calculation can be followed to get 
exactly the same geodesics equation as above.

In case the reader finds this justification for the use of the variational 
principle not strong, I can rough out a two-fold additional justification. On 
  one hand, the action provides a most fundamental starting point, in the 
sense that it offers a link with the underlying high energy physics. On the 
other hand, it provides precisely the classical equations of motion, and this 
takes root on Wheeler's First Moral Principle:
\begin{displayquote}
{\it ``Never make a calculation until you know the answer. Make an estimate 
before every calculation, try a simple physical argument (symmetry! invariance! 
conservation!) before every derivation, guess the answer to every paradox and 
puzzle.''}
\end{displayquote}
In fact, this can be used as a {\it mantra} from now on, as any modification of 
GR we may fancy, must be reducible to it under some restrictions, or in the 
correct limit, as meekly as GR becomes SR in the absence of gravity, or 
Newton's 
gravity in the weak-field and low-velocity limit. But this would be a 
reduction 
pursued along the route of physical properties, whereas we can follow  a more  
mathematical track (or as least one which is not so physically grounded at its 
onset, which however converges with Physics as it progresses). This route to 
modify Einstein's gravity is that offered by  considering a non-trivial 
connection, as it allows to produce  two extra fundamental objects  with 
valuable relevant geometric information  (see Chapter \ref{Koivistochapter}). 
The first one is  the non-metricity tensor:
\begin{equation}
  {Q}_{\alpha\mu\nu}\equiv\nabla_\alpha g_{\mu\nu}.
\end{equation}
The second one is the torsion, which stems from the antisymmetric part of the 
connection:
\begin{equation}
    T^{\alpha}_{\mu\nu}\equiv\Gamma^\alpha_{\mu\nu}-\Gamma^\alpha_{\nu\mu}. 
\end{equation}
As explained in more detail in Chapter \ref{ref:Iosifidis}, the respective 
effects of torsion  under parallel transport is to crack parallelograms into 
pentagons whereas what non-metricity produces are  changes in dot products and 
vector lengths.

If the torsion and non-metricity tensors vanish
(that is, if connection is symmetric and metricity holds), the Levi-Civita 
connection is recovered, and gravity ``à la Einstein'' is (in which the metric 
is the only degree of freedom). Conversely, the metric and the connection could 
be considered as independent objects, whose relations would be given by the 
field equations. This is the so-called Palatini formalism (see here below).

In addition, and in view of the previous discernment, a general connection 
$\Gamma^{\alpha}_{\mu\nu}$ can be decomposed as  \cite{Jarv:2018bgs, 
BeltranJimenez:2019tjy}:
\bea
\hat{\Gamma}^{\alpha}_{\mu\nu}=\Gamma^{\alpha}_{\mu\nu}+K^{\alpha}_{\;\;\mu\nu}
+L^{
\alpha}_{\;\;\mu\nu},
\eea
where
\bea
K^{\alpha}_{\;\;\mu\nu}&=&\frac{1}{2}T^{\alpha}_{\;\;\mu\nu}+T^{\;\;\;\alpha}_{
(\mu\;\;\;\nu)}
\eea
is the contortion, and
\bea
L^{\alpha}_{\;\;\mu\nu}&=&\frac{1}{2}Q^{\alpha}_{\;\;\mu\nu}-Q^{\;\;\;\alpha}_{
(\mu\;\;\;\nu)}
\eea
is the disformation.

The Riemann tensor for this general connection turns out to have a too 
phenomenal look for me to avoiding to portrait it here, and I will restrict 
myself to just mentioning that, by resorting to some clever tricks, it can be 
made vanish, thus leading to two equivalent formulations of gravity. These two 
rely upon the definition of the  invariants $\mathbb{T}$ or $\mathbb{Q}$, 
obviously related to the torsion and non-metricity. Specifically the 
corresponding two new settings  stem from the Einstein-Hilbert Lagrangian upon 
the replacements $R\to -\mathbb{T}$ or $R\to -\mathbb{Q}$, which lead 
respectively to the metric teleparallel or symmetric metric teleparallel 
versions of GR. Full fledged chapters of this volume, namely Chapter   
\ref{Koivistochapter} and Chapter \ref{ref:teleparallelch}
by S. Bahamonde, K.F. Dialektopoulos, M. Hohmann, 
J.L. Said, are devoted to these modifications of gravity and the reader 
is invited to dwell into them.

However, it is not convenient to lose  sight of the physical perspective. Let us 
go back to the need of celebrating the ability of Einstein's relativity to 
describe most of the physical behaviours we have access too, and acknowledge at 
the same time that many puzzles are still standing, and therefore modifications 
attempting at solving them must have a sensible GR limit. This is, of course, 
one of the key aspects of one of the most popular ways to modify Einstein's 
gospel: the so called scalar-tensor theories. Historically, they 
originated as 
an arguably better way to incorporate Mach's principle into gravitational 
theory, a trick to make it impossible to gauge away the gravitational field 
completely, leaving behind a remnant  which depends on an universal  quantity 
such as the mass of the whole  Universe. Under these rules, the Lagrangian will 
then typically become dependent on a new quantity, the scalar field $\phi$.

In the pioneering presentation of this seductive idea, Brans and Dicke 
considered
\begin{equation}
    S=\int d^4 x\sqrt{-g}\left[\frac{\phi}{16\pi} R+{\cal 
L}_{\phi}(\phi,g_{\mu\nu},\phi^{;\alpha})+{\cal L}_m\right],
\end{equation}
with a carefully tailored ${\cal L}_{\phi}$ so that the field equations do not 
display derivatives of  order higher than second. This is important so as to 
guarantee the theory remains ghost free (that is, no Ostrogradski instabilities 
appear). One can of course consider to replace the  $\phi R$ with $f(\phi) R$ by 
means of an arbitrary function, which again requires a wise  choice of the term 
${\cal L}_{\phi}$  to avoid divergences. 

More generally, by relaxing some requirements, it is possible to engineer the 
family of all models with a Lagrangian containing  second order derivatives of 
the field leading to second order equations of motion. These are the so called 
Horndeski scenarios, which were rediscovered some decades after their first 
appearance, and were then dubbed Galileon models. In a compact way we can 
rewrite them as 
\begin{equation}
    S=\int d^4 x\sqrt{-g}\left[{\cal 
L}_{H}(\phi,g_{\mu\nu},\phi^{;\alpha},\phi^{;\alpha}_{~~;\beta},R^{\eta\sigma}_{
~~\lambda\tau})+{\cal L}_m\right]\, .
\end{equation}

\begin{figure}[ht!]
\centering
\includegraphics[width = 0.9\textwidth]{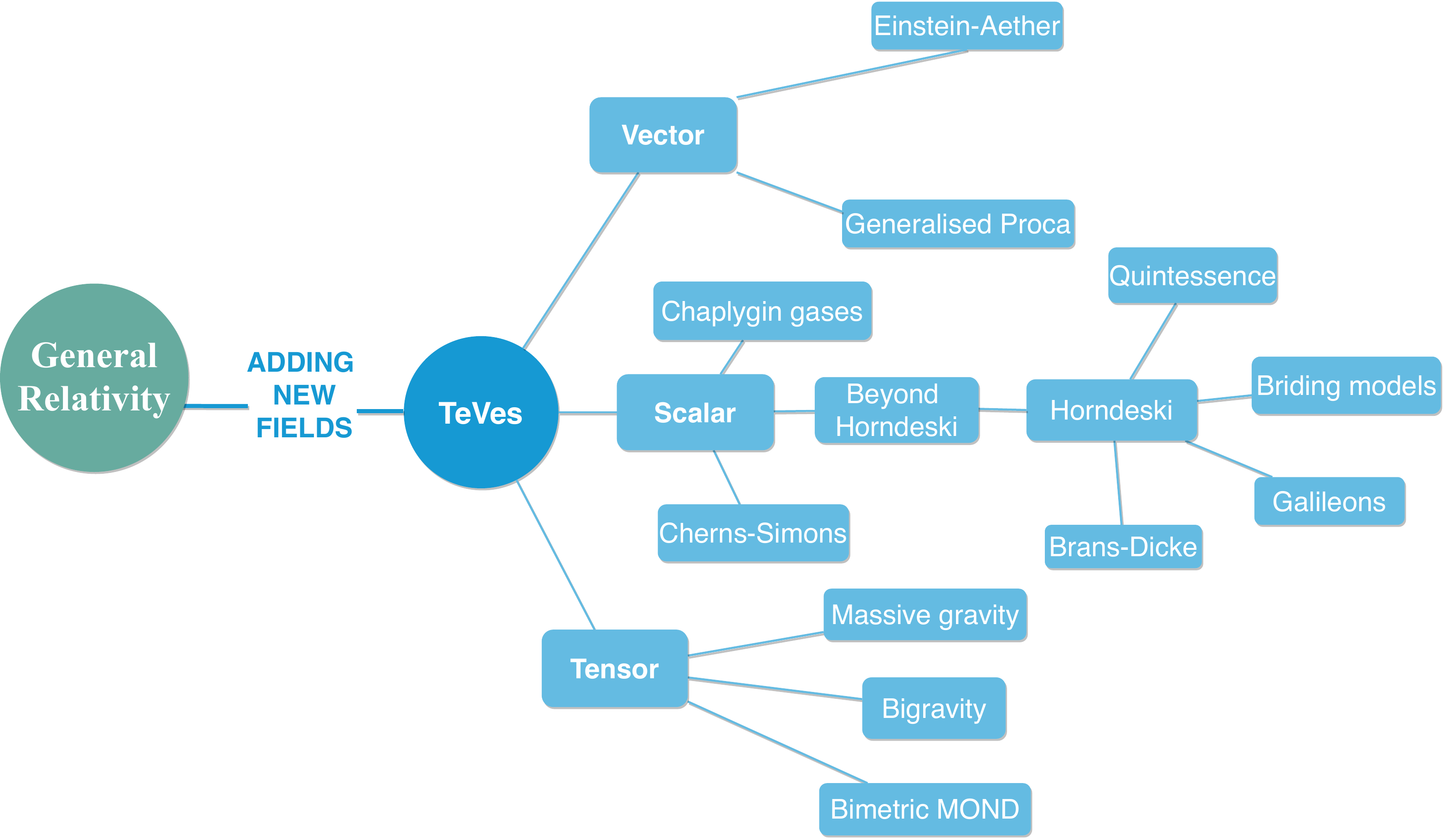}
\caption{{\it{Schematic 
categorisation of the   Tensor-Vector-Scalar  (TeVeS) class of theories, arising 
by 
adding 
new fields to General Relativity. }}}
\label{fig:TeVeS}
\end{figure}

This is the right place to invite the reader to 
have a look at the  contribution by P. Mart\'\i n Moruno in Chapter 
\ref{Morunochapter}, and references therein, where a more 
detailed account is offered on scalar-tensor theories such as those defined 
above and others beyond. In Fig. \ref{fig:TeVeS} we present a schematic 
categorisation of the   Tensor-Vector-Scalar class of theories, arising 
by 
adding 
new fields to General Relativity.

But unstoppable as curiosity and imagination are, the continuously growing set 
of generalisations of GR has one subset that stands above all others because of 
its popularity and (if we may) promises, the $f(R)$ proposal, where the Ricci 
scalar $R$ is replaced in the original Einstein-Hilbert (plus additions) action:
\begin{equation}
    S=\frac{1}{2\kappa}\int d^4 x\sqrt{-g}\left[ f(R)+{\cal L}_m\right]\, .
\end{equation}
With their generality, tractability and flexibility, these scenarios offer 
chances to reproduce a wide assortment of cosmological kinematics (see the 
staggering overview in Chapter \ref{ref:Cruz} by \'A. de la Cruz-Dombriz). 
They are constestably  able to provide mechanisms to explain either early or 
late-time acceleration (fitting {\it customer's} needs) and, in a sense, they 
represent the simplest general modification of Einstein's gravity than can be 
thought of. The understanding of the capability of this broad context to 
reproduce a succession of necessary stages (radiation-like, matter-like, and 
dark energy-like) has improved over the year along with the restrictions they 
are subject to. 
But then one should not be fooled by the potentialities offered by  $f(R)$ 
theories: hardly ever has there been in the history of Physics an one-size fit 
to 
all theoretical framework, which is actually a feature that has made our 
discipline precisely to play its unmatched role in science. From a strictly 
very 
personal perspective, it is somewhat frustrating how often we hear/read the 
claim that these scenarios can always be tweaked so they fit the data, which is 
quite a too loose statement if one does not provide an accompanying report of 
the quality of fit and other demanded statistical criteria. At the end of the 
day, if $f(R)$ theories (and other modified gravity routes) were the ``holy 
grail'' there would be no justification to write a volume such as ours. So, 
brushing aside this perhaps hypercritical tone, we should celebrate modified 
gravity models as laboratories to tests ideas, predict difficulties and venture 
solutions. Remember,  we get challenged and, as put in the words of Neil de 
Grasse Tyson,
\begin{displayquote}
{\it ``The Universe is under no obligation to make sense to you.''}
\end{displayquote}

Let me now, however, briefly sketch other relevant works shaping the section of 
the volume devoted to research in the theoretical front (that of Working 
Group (WG)1), so as to  
offer a tidbit of the high tea menu it represents. Clearly, a simple gaze at the 
table of contents reveals that the role played by the metric-affine or Palatini 
formalism is major in modified gravity, in general. It allows to relax the 
``standard'' convention between the metric and  the connection, so that the 
latter can be worked out from first principles \cite{Olmo:2012yv}. I will make 
no distinction between the Palatini and the metric-affine formalism, as it is 
customary in many works, and in some of our chapters, although formally a bit of 
extra rigour (and a higlight of differences) is in order if fermions are present 
in the matter Lagrangian \cite{Hehl:1994ue,Vagnozzi:2020dfn}, which is not 
really necessary for most cosmological applications.

\begin{figure}[ht!]
\centering
\includegraphics[width = 1.1\textwidth]{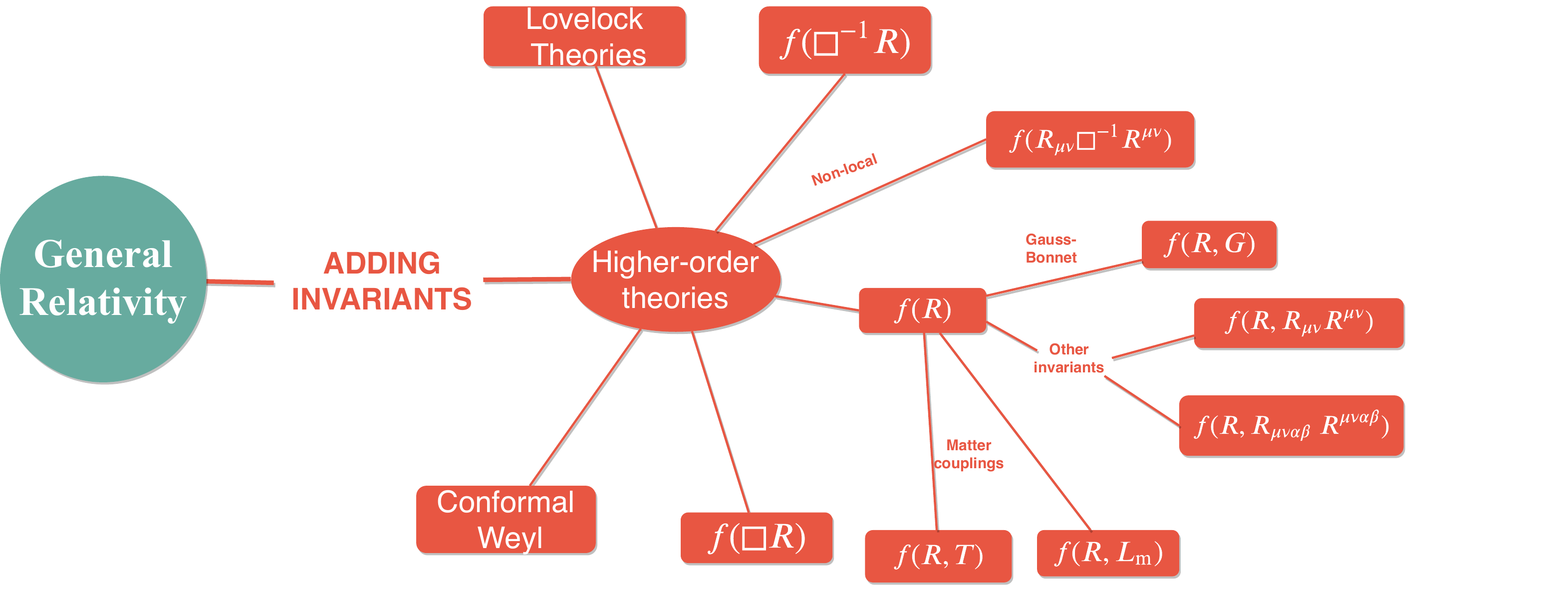}
\caption{{\it{Schematic 
categorisation of the  higher-order class of theories, arising by adding 
higher-order invariants in the Lagrangian of General Relativity.  }}}
\label{fig:Invariants}
\end{figure}

The metric-affine  scheme progresses initially from the construction of a 
curvature scalar using the connection and the metric tensor as independent 
quantities, but then it can be enlarged considerably by constructing 
Lagrangians 
with other scalar terms built from the symmetric part of the Ricci tensor and 
its contractions with the metric tensor. These settings offer many new 
attractive possibilities, such as the removal or smoothing out of cosmological 
singularities which are generic in GR scenarios, as discussed in  Chapter by 
\ref{Delhomchapter}  by A. Delhom and D. Rubiera-Garcia). Another 
remarkable point of the Palatini formulation is that,  unlike in the metric 
approach,  second-order field equation are always obtained, even for 
Lagrangians 
with terms which are non linear in the scalar curvature or include the above 
mentioned less standard scalars. A related approach is hybrid-metric Palatini 
gravity, where the standard $R$ term coming from the metric connection is added 
to extra terms depending on an alternative curvature scalar, $\cal R$, derived 
from an independent connection, such terms offering a richer phenomenology as 
to 
the evolution of matter inhomogeneities. These and many other aspects are 
covered in the contribution by F.S.N. Lobo in Chapter \ref{ref:Lobo}.
In     Fig. \ref{fig:Invariants} we present a schematic 
categorisation  of the theories, arising by adding higher-order invariants in 
the Lagrangian of General Relativity.

\begin{figure}[ht!]
\centering
\includegraphics[width = 0.98\textwidth]{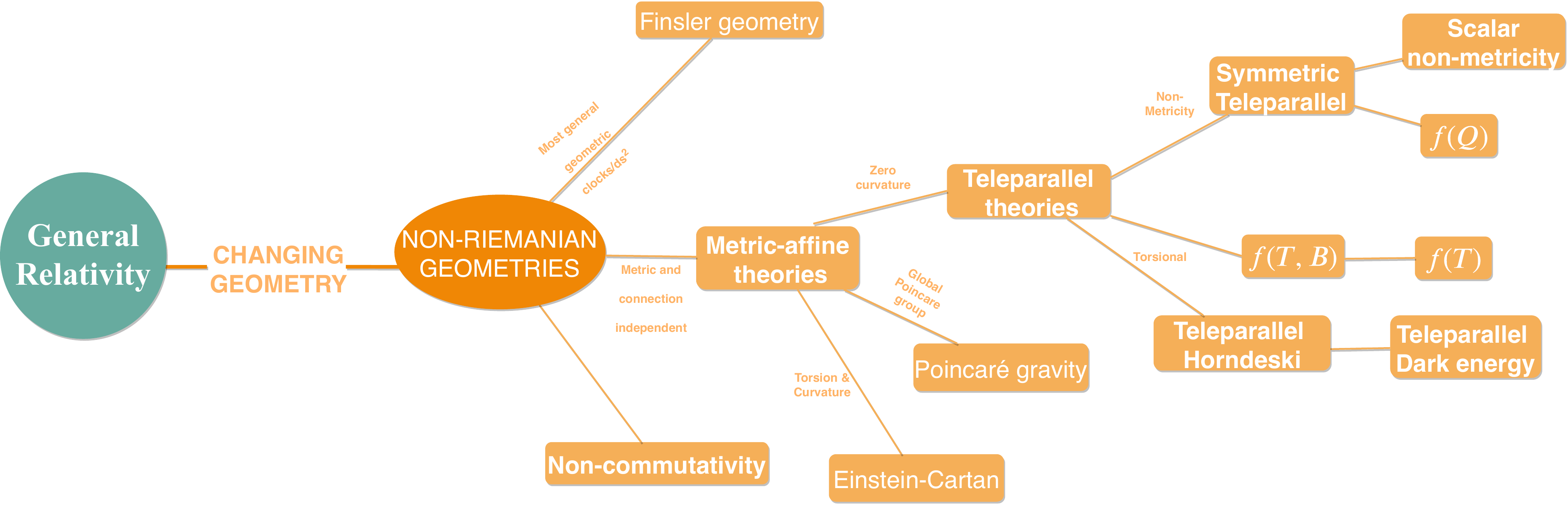}
\caption{{\it{Schematic 
categorisation of the theories arising by modifying the geometry of General 
Relativity. }}}
\label{fig:Geometry}
\end{figure}

Alternatively, a (wilder and) completely separate route to describe gravity as 
a 
manifestation of geometry can be pursued too: teleparallel gravity and its 
extensions.  These replace curvature with torsion and are build upon a 
curvature-free connection (therefore flat). Following the recipe I anticipated 
above, one possible Lagrangian starting point for teleparallel descriptions of 
gravity is:
\begin{equation}
    S=\frac{1}{2\kappa}\int d^4 x\sqrt{-g}\left[ f(\mathbb{T})+{\cal 
L}_m\right].
\end{equation}
Among the attractive features of teleparallel gravity we see again the 
second-order nature of the equations and the broadness of the whole setting, 
which allows to rewrite many theories built from the metric and the Levi-Civita 
connection. It is also certainly very appealing that the theory allows to waive 
the equivalence principle, thus re-framing gravitation as an interaction more 
similar to other fundamental ones. Many aspects of teleparallel gravity are 
covered into the sizeable contribution in Chapter   
\ref{ref:teleparallelch}, where the 
authors do not forget to acknowledge the absolute need to continue to examine 
these extensions in the light of observations. Actually, the collection of new 
features offered by these framework is tremendous, as the possibility of 
interpreting it as a gauge theory of translations contributes to filling the 
gap 
between gravity and other interactions in Nature, or as the intellectually not 
less challenging question of the definition and characterization of 
singularities in this alternative formulation of  gravitational physics.
  In Fig. \ref{fig:Geometry} we present a schematic 
categorisation  of the theories arising by modifying the geometry of General 
Relativity.

Recapitulating, there seem to be two main and disparate ways to manufacture 
modifications to GR: the addition of new fields on   one hand, and geometry 
based restylings on the other hand. This splitting is largely a matter of 
convention, as prescriptions to go from one to another can be typically found 
with manageable techniques (see again Chapters  \ref{ref:Cruz}  and 
\ref{ref:Lobo}). But this dual possibility has proved so far more enriching 
than 
a source of confusion, and it continues to provide motivation and inspiration.

 Nevertheless, this whole collection of possible alternatives to Einstein 
gravity may seem as if the community was meandering rather than following a 
straight course. But remembering Roger Penrose's words again, is just opportune:
 \begin{displayquote}
 {\it ``Sometimes it's the detours which turn out to be the fruitful ideas.'' } 
 \end{displayquote}
 In this sense, I also invite specialists to explore the contribution by 
N. Voicu and C. Pfeifer in Chapter \ref{Voicuchapter} and learn 
about the possibilites of relativistic extensions of  Finsler geometry, with 
their intrinsic unusual features such as multiple covariant derivatives and 
matter dynamics.

\begin{figure}[ht!]
\centering
\includegraphics[width = 0.9\textwidth]{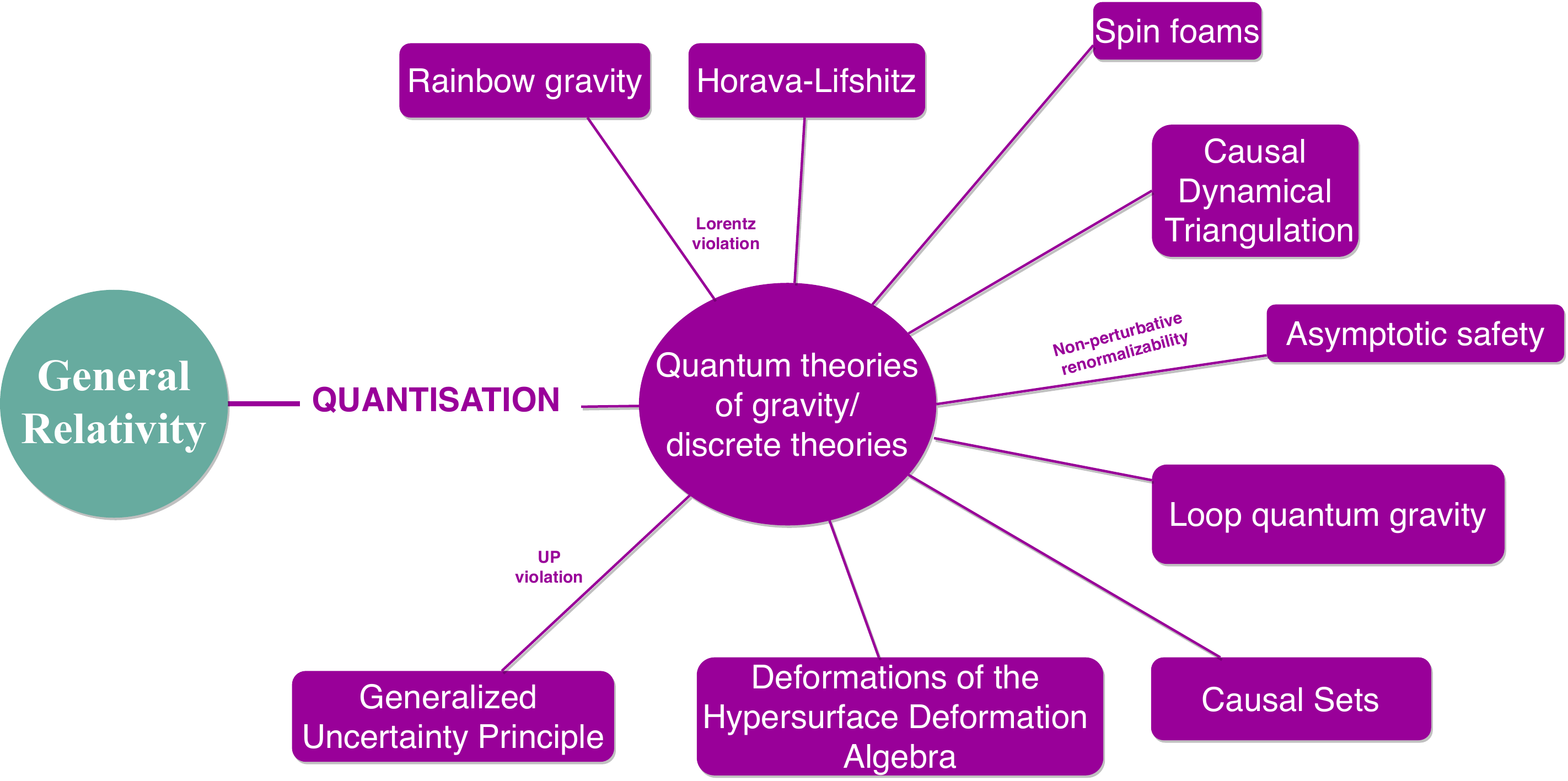}
\caption{{\it{Schematic 
categorisation of the theories arising by the use of quantum arguments. }}}
\label{fig:Quantisation}
\end{figure}

 Perhaps, some readers have noticed the absence of references to the colossal 
problem of our lack of substantial understanding of gravity at the quantum 
level. The reason is mainly that it falls beyond the major objectives of the 
Action. This choice is really not a gesture of contempt, but rather a need to 
concentrate efforts on an otherwise major attempt, which is to understand 
gravity at the mainly classical level. Our team however is happy to boast about 
intrepid representatives which explore the quantum side of gravity through 
(again) extensions of Einstein's framework. In this respect see Chapter 
\ref{Calcagninonlocal} by  G. Calcagni, Chapter \ref{Garattinichapter} by R. 
Garattini, Chapter \ref{Bouhmadichapter} by  M. Bouhmadi-L\'opez and P. 
Mart\'in-Moruno, and Chapter \ref{Casadiochapter} by R. Casadio and  A. 
Giusti.  In Fig. \ref{fig:Quantisation} we present a schematic 
categorisation  of the theories arising   by the use of quantum 
arguments.

Let me again stress that the two main routes to knowledge about modifications of 
our understanding of gravity are theory and data analysis, which must act 
synergistically, as it has been the case of the activities of our Action. This 
volume is a veracious reflection of that fact, with a carefully tailored choice 
of contributions at the forefront of the field in its broadest manifestation.

CANTATA researchers whose theoretical work makes strong contact with 
observations have contributed to WG2 (testing relativistic effects) an WG3 
(observational discriminators) mainly. Their daily research  encourages them to 
reflect pointedly about the questions  that these new observational findings 
pose. This allows these scientists to perform masterly forecasts of constraints 
to be placed on theoretical frameworks by the coming generation of cosmological 
observations. Remember in this context the wise words by Vera Rubin: 
\begin{displayquote}
\textit{``Science progresses best when observations force us to alter our 
preconceptions.''}
\end{displayquote}

One important area of operation where scientists in WG2 of our team have 
completed proficient works are tests of gravity at scales well below 1 Mpc, that 
is, non-cosmological ones (clusters are not covered in Part II, but rather in 
Part III), ranging from earth laboratory tests to orbits around compact objects. 
Within this domain, screening mechanisms in scalar-tensors theories with a 
scalar field coupled to matter have gathered a lot of interest. The reason is 
that the screening
treats differently traditional gravity tests  and laboratory tests in the sense 
that it allows effects to evade detection by the former, while revealing to the 
latter, as if some kind of reward for clever and inspired  novel experiments 
were offered. Several  screening mechanisms have been proposed so far 
\cite{Khoury:2010xi}, and among the most popular we find those with canonical 
kinetic terms on   one hand, such as the chameleon and the symmetron 
mechanisms, and then a third one on the other, the Vainshtein mechanism, which 
emerges from derivative non-linearities. In the contribution by Anne-Christine 
Davies and Benjamin Elder in Chapter \ref{Davischapter} the focus is mainly set 
on  tests on chameleon scalar fields through different approaches such as vacuum 
chambers, atom interferometry, torsion balances, Casimir force searchers and 
others. The symmetron field is covered as well, but with less extension, as 
research on this area is more recent. In a wide sense the dynamics of the scalar 
field is governed by an effective potential:
\begin{equation}
    V_{\rm eff}(\phi)=  V(\phi)+( A(\phi)-1)\rho_m.
\end{equation}
The (square of) the auxiliary function $A(\phi)$ encapsulates the coupling 
between matter and the scalar field in the form of a conformal rescaling between 
the metric used to express the matter sector of the action (Jordan metric/frame) 
and the metric used to render the Einstein equations in the ``orthodox'' fashion 
(Einstein metric/frame).  Further details can be found in  Chapter 
\ref{Braxchapter} by  P. Brax. 
But the zoo of screening mechanisms contains more species, and  when one lets 
curiosity (or imagination) run and non canonical kinetic terms enter the 
picture, we can find other mechanisms such as  K-mouflage (see Chapter 
\ref{Braxchapter}) described  in this volume, a screening possibility which has 
so far not been tightly constrained by gravitational wave observations.

Now, if Earth-based laboratories are at the bottom of the scale ladder of 
experiments to test modifications of gravity, the solar system is the obvious 
next: a weak-field limit realm which is typically treated under the parametrized 
post-Newtonian formalism, an in-between stratagem to wrap-up conjectures about 
both observations and theories being considered, which connects the former with 
the latter through a set of parameters. This is explained in detail in the 
contribution by M. Hohmann in Chapter \ref{Hohmannchapter}, where both the 
standard version and some extensions are discussed. In fact, a large class of 
extensions of GR can be accommodated into the customary formulation, but the 
community has not stopped there, and broader versions have been devised for 
cases demanding so. The (well-known) first two parameters of the expansion 
($\gamma$ and $\beta$)  estimate how mass creates curvature on   one hand and 
the degree of nonlinearity in the superposition law for the gravitational 
potentials on the other hand.  Stringent bounds have been obtained through 
refined experiments and can offer light on the viability of modifications of 
gravity with significant impact on those parameters and subsequent ones.

But the prevailing weak-field/strong-field battle in physics explorations 
demands we also turn our attention (as a team) to representatives of largest 
curvature and highest densities regimes: compact stars and black holes. 
Modelling stellar structure is a demanding task in the default GR setting, and 
modifications of gravity introduce additional difficulties, both from the 
mathematical and physical perspectives. Hardly any features can be studied from 
a general formulation and individual investigations are typically the way to go. 
New knowledge will, for instance, allow to spot (if any) changes in the 
mass-radius relation of neutron stars, as it will depend on new parameters 
characterizing the modifications (see   Chapter \ref{Olmochapter} by G. Olmo,  
D. Rubiera-Garcia,  and A. Wojnar). The analogous operates on black holes, in 
the 
sense that the famous absence of hair theorem may need a tweak in the presence 
of additional gravitational degrees of freedom.

 These 
are just some of the surprises that compact objects (whether extreme or not) may 
have in store. Another is spontaneous scalarization, which consists in the 
emergence of a non-trivial configuration of a scalar field without sources and 
which vanishes asymptotically. For this to occur  a scalar field must exist which 
is non-minimally coupled to gravity. This highly nonlinear effect may offer tight 
limits to the modified scenarios through observations of, for instance, binary 
systems as discussed in Chapter   \ref{Kunzchapter} 
by J.L. Bl\'azquez-Salcedo, B. Kleihaus, and J. Kunz.

When we move to (much) larger scales we must publicize that efforts of CANTATA 
researchers have served the  international effort of the community on the design 
of future accurate tests of gravity (modified or standard) and the improvement 
of current ones with their expertise on numerics and computation. Some of our 
colleagues have made distinguished contributions to large teams expected to lead 
the design and operation of terrestrial and space-based surveys which will be 
operating in the near future.
A glimpse of the breathtaking future is offered by observations of gravitational 
waves, which have brought strong implications for modified theories of gravity. 
At the cosmological background level \cite{Belgacem:2017ihm}, an additional 
friction that adds to the Hubble one appears, and this  modifies the amplitude 
of the waves, whereas an effective (anomalous) mass and consequent atypical
speed alter the phase, as discussed in Chapter 
\ref{sec:Ezquiaga} by J.M. Ezquiaga.
 Notoriously,  gravitational waves provides a new 
observational channel for testing gravity theories which predict a non-null mass 
for the graviton. In this context we can highlight the closely related massive 
gravity and bigravity settings, which belong to the field theory framework of 
gravity  and of which a glimpse is offered in Chapter \ref{Heisenbergchapter} 
by L. Heisenberg. 

But more sophisticated and challenging effects could also arise in the physics 
of gravitational waves when modified gravity scenarios are considered, 
as not only additional polarisations emerge (up to four extra ones), but also 
they could get mixed up  and frequency mutations might be produced too; see 
Chapter \ref{Sakellariadouchapter} by M. Sakellariadou.
Actually, the detection of these additional polarisation 
modes represents a significant technical challenge.

Into the bargain, the absolutely greatest promise of gravitational waves is the 
cosmological realm. Events that can be regarded as standard sirens 
\cite{Gray:2019ksv} (may be able to) probe the redshift evolution of the 
luminosity distance of the gravitational wave source, which is proportional to 
the inverse of the amplitude.
In principle, it can be different from the electromagnetic luminosity distance 
of a companion event, and therefore their ratio will offer a test for parameters 
associated with physics beyond GR and also with dark energy evolution (which is 
also referred to, sometimes, as ``new physics'', as differing from the 
$\Lambda$CDM setting).

Now, along those lines, the scheme which makes currently most sense is to 
combine those measurements with observations of the large scale structure coming 
from surveys which will probe the Universe in different scale regimes as 
compared to the size of the horizon. Further details on this classification are 
given in the contribution by Y. Akrami and M. Martinelli in Chapter 
\ref{sec:Akrami}, where readers can find a summary of one of the most popular 
formulations of perturbations valid for the linear regime (that is, for horizon 
size scales or intermediate scales). 

The starting point for this approach is to rewrite the 
Friedmann-Lema\^itre-Robertson-Walker metric in the Newtonian gauge as
\begin{equation}g_{\mu\nu}={\rm 
diag}\left[-(1+2\Phi),a^2(1-2\Psi)\delta_{ij}\right],
\end{equation}
where $a$ is the scale factor and $\Phi$ and $\Psi$ are the gauge-invariant 
Bardeen potentials. 

 One route to progress is making use of the quasi-static approximation, which is 
valid for the subhorizon linear regime and significant breadth of gravitational 
theories. This approach produces a sizeable simplification of the pertinent 
equations,
as the extra degrees of freedom associated with 
departures from GR can be jumbled up in the (generally) time and space dependent 
functions $\mu$ and $\eta$, respectively the effective Newton's constant and the 
gravitational slip. The second function (that is, $\eta$) makes the two Bardeen 
potentials differ from each other, unlike the GR case, and leads to observable 
weak lensing effects through the combination of $\mu$ and $\nu$; whereas $\nu$ 
(on its own) affects the growth of structure; see  Chapter 
\ref{sec:Perivolaropoulos} by   L. Kazantzidis and
L. Perivolaropoulos.
 
 Related to the previous discussion, one analytic method based on the 
perturbation theory framework I particularly like to highlight is the Effective 
Field Theory (EFT) approach,  which  can capture interesting effects  on scales 
smaller than the intermediate one, and in particular the onset of the transition 
from GR to a modified gravity regime, which is crucial for the research 
objectives of our Action. In particular, it depicts physical effects germane to 
macroscopic scales by integrating out short-distance features, so they appear on 
long-distance characteristics as extra/perturbative parameters. This chassis 
supports any dark energy or modified gravity model possessing one additional 
scalar degree of freedom in the way described in detail in Chapter 
\ref{sec:Frusciante} by N. 
Frusciante and S. Peirone.
 
 These theoretical developments are, of course, a concoction to make the 
community ready to take advantage of future large scale structure surveys when 
they become available. In this respect even though the current constraints are 
of  ${\cal O}(1)$, they are expected to get as low as ${\cal O}(10^{-1})$, and 
there is hope that 
 violations  of the current observational inference of  $\mu(z=0)=\nu(z=0)$ 
could be detected (see again Chapter \ref{sec:Akrami}).

 Among the abundance of outputs of future large scale structure surveys we must 
mention the sensitivity of the clustering of galaxies to the (specific) theory 
of gravity under play. Effects less tangible than density perturbations and 
redshift-space distorsions will place unprecedented constrains which are 
obtained by confronting   pertinent gauge-invariant quantities (the two Bardeen 
potentials being among them again) with the information provided by the power 
spectrum and its multipole expansion, exploring  effects  which are 
currently neglected in current surveys; see the contribution of C. Bonvin 
in Chapter \ref{sec:Bonvin}.
 Note that the importance of these studies is twofold. On   one hand they 
vindicate the role of galaxies as concentrations of baryonic (and therefore 
electromagnetically accessible) concentrations of matter for the study of the 
Universe,  and on the other they offer a fundamental channel to explore the 
vital role of dark matter.  In addition, it is relevant that the interplay 
between baryons and dark matter \cite{Chan:2019ukj}, which takes place in  
regions with high matter density, also allows to test the equivalence principle 
\cite{Bonvin:2020cxp}

The teamwork between dark and baryonic matter and its proportions in the 
Universe can be literarily cast into the dainty mould of these words by  one of 
the clearest minds in gravitation, Arthur S. Eddington:
\begin{displayquote}
\textit{``An ocean traveler has even more vividly the impression that the ocean 
is made of waves than that it is made of water.''}
\end{displayquote}
But having stated the paramount importance of galaxies to understand modified 
theories of gravity the Universe, we cannot forget that the Universe offers us 
even better laboratories, galaxy clusters. Their  ambivalence as both 
astrophysical and cosmological objects can help us discriminate between 
gravitational effects and cumbersome astrophysical phenomena as thoroughly 
covered in  Chapter \ref{sec:Saltas} by I.D. Saltas and L. Pizzuti, where 
prospect of kinematic, thermal and lensing explorations are reviewed yet again 
resorting to  perturbative quantities considered in other contributions, such as 
the gravitational slip $\eta$. Nevertheless, for a flavour on the sort of 
discriminating criteria clusters can offer  when time evolution of scalar fields 
is addressed, it is worth check the contribution by D. Mota in Chapter  
\ref{sec:Mota}, which closes those related to WG3.

But then again, going back to cosmological scales, I must stress that studies 
with perturbative ingredients will necessarily be flawed until some pending 
disquietudes regarding the cosmological background and its linear perturbations 
are relieved. The most talked about one being the tension between high and low 
redshift data estimates of the Hubble constant, $H_0$ (see Chapter 
\ref{sec:DiValentino} for a state-of-the-art review and a sketch of what lies 
road ahead). The second actor in this stage play of tensions is the apparent 
discrepancy in the amplitude of the matter power spectrum  as set by $\sigma_8$, 
the root-mean-square   fluctuations in the matter mass density in a comoving 
sphere of diameter 
$8$ Mpc. Large compilations of redhift-space distorsions and other dynamical 
probes show a statistically significant discrepancy with \textit{Planck} data 
and would favor an evolving and weaker effective Newton's constant (through 
$\mu$), and readers can head to the contribution of  Chapter 
\ref{sec:Perivolaropoulos} for a detailed discussion and review of the topic. 
Clearly, all complementary routes offered by tests of dynamical features are 
destined to play a most relevant role, for instance weak lensing. Its  
very-hard-to-spot effects are again encoded in the Bardeen potentials, and 
forecasts have been carried out for surveys such as \textit{Euclid} in the 
theoretical context of Horndeski theories; see Chapter \ref{sec:Pettorino} by 
V. Pettorino and A.S. Mancini.

It should be clear by now to the reader that has come so far, that the route to 
erudition in this limitless field is grievous, and will unavoidably require all 
the literature anybody can digest so as to get an exquisite insight on caveats 
of these scenarios. Such an exercise would undoubtedly lead the candidate to a 
multifaceted and more thorough understanding of (her/his favourite flavour of) 
gravity through questions such as whether the correct weak field limit is 
attainable, whether instabilities occur (a typical pathology of higher order 
theories), and whether the initial value problem is well posed. Nevertheless, 
and here comes the crux of the matter, no proficient understanding of a modified 
gravity framework can be reached without an analysis of the formation (and 
sustenance) of structures. In this context I cannot but highlight again the key 
question of  whether its cosmological perturbations  of the cosmological 
background are capable of leaving a blueprint agreeable with the currently 
observed patterns in the cosmic microwave background and the large-scale 
structure itself. The intimate connection between theory and observations is 
therefore an unbreakable bond, and the seed sown by our team's work will surely 
thrive and feed our knowledge hungry community, and, again, as a team, our 
feeling is that:
\begin{displayquote}
{\it ``This is the way''}  - The Mandalorian
\end{displayquote}

\chapter[General Relativity]{General Relativity}
\label{Mimosochapter}

{\em Jos\'e Pedro Mimoso}\\

Modern cosmology  stems from Einsteins's General Relativity
\label{GRref2}   as the 
fundamental description of gravitation. General Relativity mutates the  
gravitational force into 
the curvature of the four-dimensional space-time, which responds to the distribution 
of mass-energy. It was this  revolutionary viewpoint that rendered it possible for 
the first time to consider the whole Universe (Cosmos) as a single object of 
study~\cite{Einstein:1917ce}. 

A few fundamental building blocks underlie Einstein's theory, beginning with the  
Principle of Relativity, which prescribes  that the laws of physics should be the 
same in all reference frames~\cite{Rindler:2006km}. This means that there should be 
no preferred frame, and this is encapsulated in the Principle of General 
Covariance (also referred to as Diffeomorphism Invariance), by which the equations 
are tensorial and invariant under coordinate
transformations~\cite{BeltranJimenez:2019tjy}.
The Equivalence Principle  establishes that there are no local experiments that 
distinguish free-falling observers from inertial observers, and its weak 
formulation stands on the equivalence between inertial and gravitational masses. 
Finally, the theory is expected to recover Newtonian gravity in the weak field 
limit and slow motion of sources. This is denoted as the Correspondence Principle,
in analogy with the corresponding requirement met by quantum 
theory~\cite{Rindler:2006km}. 

In  General Relativity \cite{Einstein:1915ca}, the curvature of the spacetime 
is encoded into the 
Riemann tensor \label{Riemanntenref2} defined by the Ricci identities 
$\nabla_{[\gamma}\nabla_{\delta]} 
X^\alpha = {R^\alpha}_{\beta\gamma\delta}X^\beta$, where $X^\alpha$ is an 
arbitrary spacetime vector. It follows that
\begin{equation}
{R^\alpha}_{\beta\mu\nu} = \partial_\mu{\Gamma^\alpha}_{\beta\nu} - 
\partial_\nu{\Gamma^\alpha}_{\beta\mu} +{\Gamma^\delta}_{\beta\nu} 
{\Gamma^\alpha}_{\delta\mu}-{\Gamma^\delta}_{\beta\mu} 
{\Gamma^\alpha}_{\delta\nu}\; ,
\end{equation}
where ${\Gamma^\alpha}_{\mu\nu}$ is the affine connection that defines the 
concept of parallelism in the four-dimensional manifold.
However, the further prescription of a metric tensor $g_{\alpha\beta}$ to the 
space-time both generalises the special-relativistic  Minkowski line-element 
$ds^2 =\eta_{\alpha\beta}\,dx^\alpha\,dx^\beta=-{\rm 
d}t^2+dx^2+dy^2+dz^2$ with  
$
ds^2 =g_{\alpha\beta}\,dx^\alpha\,dx^\beta
$,
and specifies the metric connection
 \begin{equation}
{\Gamma^\alpha}_{\beta\gamma} = \frac{1}{2}g^{\alpha\delta}\left( 
\partial_\beta g_{\delta\gamma} + \partial_\gamma g_{\delta\beta} - 
\partial_\delta g_{\beta\gamma}\right).
 \end{equation}
It also means that the spacetime manifold has no torsion~\cite{Cai:2015emx}. 
Indeed, the connection is symmetric and this implies that
${\Gamma^\alpha}_{[\mu\nu]}=0$. The Riemann tensor can then 
be decomposed into  irreducible components
\begin{equation}
{R^\alpha}_{\beta\mu\nu} = {C^\alpha}_{\beta\mu\nu} + 
\frac{1}{2}\left(\delta^\alpha_{\mu}R_{\beta\nu}+g_{\beta\nu}R^\alpha_{\mu}
-\delta^\alpha_{\nu}R_{\beta\nu}-g_{\beta\mu}R^\alpha_{\nu}\right)+\frac{R}{6}
\left( \delta^\alpha_{\nu}g_{\beta\mu} - \delta^\alpha_{\mu}g_{\beta\nu}\right),
\end{equation}
where ${C^a}_{bcd}$ is the traceless conformally invariant Weyl tensor, and the 
traces $R_{ac}= g^{bd}{R}_{abcd}$, and $R=g^{ab}R_{ab}$ are, respectively, the 
Ricci tensor and the Ricci curvature scalar.

Einstein's field equations, which govern the interplay between the spacetime 
geometry and the non-gravitational fields, can be variationally derived from the 
fundamental Einstein-Hilbert  (EH) action~\cite{dInverno:1992gxs}  
\label{EHactionref2}
\begin{equation}
S_{EH}= \int dx^4  \sqrt{-g} \left[\frac{1}{2\kappa^2} 
(R-2\Lambda)+{\mathcal{L}}_m(g_{\mu\nu},\psi^\alpha)\right]
,\label{e12EHaction} 
\end{equation}
where $g_{\mu\nu}$ is the metric of space-time, $g$ the determinant of this 
metric and $R$ the corresponding Ricci curvature scalar. Additionally, 
$\Lambda\,$ is the 
cosmological constant,  ${\mathcal{L}} (g_{\mu\nu},\psi^\alpha)$ represents the 
Lagrangian of the matter fields generically denoted $\psi^a$, and $\kappa^2$ is 
the gravitational coupling constant $\kappa^2=8\pi G_N/c^4$, with $G_N$ the 
Newton's constant and $c$   the speed of light set to 1. The 
stationarity conditions  associated with the variational 
differentiation of the EH action with respect either to the metric or, 
alternatively, with respect  to the metric and the connections %
in the so-called Palatini formalism,
yield the celebrated equations~\cite{Einstein:1915by,Einstein:1915ca}
\begin{equation}
R_{\mu\nu} - \frac{1}{2}  g_{\mu\nu}  R = 
\kappa^2 T_{\mu\nu} \quad ,
\label{EFE_GR}
\end{equation}
where $T_{\mu\nu}$ is defined by
\begin{equation}
T^{\mu\nu} \equiv 
\frac{2}{\sqrt{-g}}
\frac{\delta{(\sqrt{-g}
{\cal{L}}_m)}}{\delta{g_{\mu\nu}}}
\label{e12defEMT}
\end{equation}
which is the energy-momentum tensor\label{Enmomtenref1}   of the matter fields. 
When we consider a time-like vector field $u^\alpha$ the energy-momentum tensor 
can be decomposed 
as 
\begin{equation}
    T^{\mu\nu} = \rho u^\mu u^\nu+p 
h^{\mu\nu}+\Pi^{\mu\nu}+2q^{(\mu}u^{\nu)}, \label{EMT_decomposed}
\end{equation}
where $h_{\mu\nu}=g_{\mu\nu}+u_\mu u_\nu$ is the metric induced on the spatial 
hypersurfaces orthogonal to $u^\alpha$,   $\Pi_{\mu\nu} = \left(h^\gamma_\mu 
h^\delta_\nu-\frac{1}{3}h_{\mu\nu}h^{\gamma\delta}\right)T_{\gamma\delta}$ is 
a transverse  traceless tensor, and $q^\mu 
=h^{\mu}_{\delta}T^{\delta}_{\nu}u^\nu$ is a spatial vector field.
The 
quantities $\rho$ and $p$ are defined as $\rho=T_{\mu\nu}u^\mu u^\nu$ and 
$3p=T_{\mu\nu}h^{\mu\nu}$, and  are 
 respectively   the energy-density and the pressure of the matter fluid
measured by an observer moving with 4-velocity $u^\alpha$, while the quantities 
$\Pi^{\mu\nu}$ and $q^\mu$ are the anisotropic stress tensor and the heat flow 
vector.

The diffeomorphism invariance of the action translates into the 
Bianchi-contracted identities  \label{Bianchiref1}
\begin{equation}
\nabla_\nu T^{\mu\nu} =0. 
\label{Bianchi_ids}
\end{equation}
Taking into consideration that $u^\mu\nabla_\mu ()=()\dot{}\,$,i.e., the dot  
represents differentiation with respect to the cosmic time $t$, and that the 
induced metric $h_{\mu\nu}$ is a projector onto the spatial hypersurfaces, we 
can make the following decomposition:
\begin{equation}
    \nabla_\mu u_\nu = -\dot u_\mu u_\nu 
+\frac{\theta}{3}\,h_{\mu\nu}+\sigma_{\mu\nu}+\omega_{\mu\nu},
\end{equation}
where $\dot u^\mu =u^\alpha \nabla_\alpha u^\mu$ is the acceleration, $\theta = 
h^{\mu\alpha}\nabla_\alpha u_\beta h^{\beta\mu}$ is the expansion scalar, 
$\sigma_{\mu\nu}=2h_{(\mu}^{\alpha}\nabla_\alpha u_\beta 
h^{\beta}_{\mu)}-\frac{1}{3}\theta h_{\mu\nu}$ is the shear, and 
$\omega_{\mu\nu} = 2h_{[\mu}^{\alpha}\nabla_\alpha u_\beta h^{\beta}_{\mu]}$ is 
the vorticity tensor. Thus, the Bianchi identities (\ref{Bianchi_ids}) can be 
cast as 
\begin{eqnarray}
   && \dot\rho = -(\rho+p)\theta- \Pi^{\mu\nu}\sigma_{\mu\nu} - q^\mu u_\mu 
-\nabla_\nu q^\nu\\ 
  && (\rho+p)\,\dot u^\mu
   = -h^{\mu\beta}\left(\nabla_\beta p +\nabla_\nu 
\Pi_\beta^{\nu}+ \dot q_\beta \right)+ 
\left(\omega^\mu_\nu+
\sigma^\mu_\nu-
\frac{4}{3}\theta\,h^\mu_\nu\right)q^\nu\; .
\end{eqnarray}

The field equations (\ref{EFE_GR}) are a system of non-linear 
partial differential equations on the components of the curvature tensor (see, 
for instance, J. Barrow's contribution in~\cite{Barrow:1991fj}). This implies 
that the gravitational field is itself also a source of curvature, and that a 
curved spacetime can thus  be a solution of the vacuum field equations.  The 
components of the metric  tensor can be perceived  as generalising the Newtonian 
concept of gravitational potential. This is particularly well illustrated  by 
the first exact solution discovered by K. Schwarzschild early in 
1916~\cite{1916AbhKP1916..189S} \label{Schwarzschild1}
\begin{equation}
ds^2 =  -\left(1-\frac{2G_NM}{r}\right)\,dt^2 + 
{dr^2}\left(1-\frac{2G_NM}{r}\right)^{-1}+r^2\left(d\theta^2 +\sin^2\theta 
d\phi^2 
\right),
\end{equation}
where one uses the standard spherical coordinates $r$, $\theta$ and $\phi$, and 
where $M$ represents the Misner-Sharp mass 
$ M=4\pi\,\int_0^r \rho(u)\,u^2\,du$~\cite{Misner:1964je}.
This metric represents the gravitational field of a central, spherical 
distribution of mass, $M$, on the exterior vacuum space.  Clearly, its 
coefficients $g_{00}$ and $g_{rr}$ relate to Newton's potential 
$\Phi(r)=-G_NM/r$, 
and the discovery of this solution allowed us to confirm both Einstein's  
explanation of  
Mercury's  perihelium shift, anomalous in the framework of Newtonian celestial 
mechanics, and the predicted value for the deflection of light rays passing 
close to the Sun in excess of the Newtonian value. The observational test of the 
latter forecast was made one century ago, on  
1919,  
and it was just  the first of many empirical successes  at  the Solar System  
\label{solarsystemref1}
level.  

The piling up of experimental  tests was  boosted in the early 1960's  by the 
challenge  placed by proposals of extended gravity theories such as the 
Brans-Dicke theory~\cite{Brans:1961sx}. This led to the development of a 
systematic scrutiny of the Post-Newtonian effects,
through the consideration of a consistent multipolar expansion of the 
gravitational potentials in the metric, dubbed Parametric 
Post-Newtonian~\cite{Will:1993ns}. Different gravitational theories yield  
diverse values or dependences on the parameters of the expansion for the 
competing theories. Experiments on the scale of the Solar System 
have led to very accurate  measurements of these coefficients and to  very tight 
 constraints on  gravitational theory~\cite{Will:1993ns,Will:2014kxa}. It 
must be emphasised at this point that the theories under scrutiny are mostly 
theories that share with General Relativity the paradigm set by its underlying 
principles. 

So far, General Relativity has passed the Solar System tests with flying colours. 
Moreover, its 
success has also been further  vindicated recently  by  the detection of 
gravitational waves by LIGO in 2015~\cite{Abbott:2016blz}, and by obtaining the 
first image of a black hole \label{BHref1} in 2019~\cite{Akiyama:2019eap}.  
These 
were 
staggering landmarks both in regard to the exquisite precision achieved in 
these observations, and  in regard to the confirmation of precocious 
predictions made by the theory decades ago.

Aiming at a completely different range of scales, modern cosmology began with 
the astonishing realisation that the Universe is 
expanding~\cite{Einstein:1917ce,Friedman:1922kd,Lemaitre:1931zz,Lemaitre:1931zzb
,Hubble:1929ig,Lemaitre:1933gd}  out  of an incredibly tiny 
size~\cite{Lemaitre:1931zzb}. It was necessary to repel the idea that the 
universe should be static, avoiding gravitational collapse onto itself, which 
led to  the introduction of the much-debated cosmological constant 
$\Lambda$~\cite{Einstein:1917ce,deSitter:1916zza,deSitter:1916zz,deSitter:1917zz}. 
The assumption of a Copernician or cosmological principle by which our location 
in the Universe is not special, and taking into account the observed uniformity 
of the distribution of matter  at large scales, is reflected in the adoption 
of the spatially homogeneous and isotropic metric 
Friedmann-Lema\^{\i}tre-Robertson-Walker (FLRW) metrics \label{FRWref1}
\begin{equation}
ds^2 =  a^2(\tau)\left[ -d\tau^2  + \frac{dr^2}{1-kr^2} + 
r^2\left(d\theta^2+\sin^2\theta d\phi^2\right)\right],
\label{Convension_setting_global}
\end{equation}
where $a(\tau)$ is termed as the scale factor, $k=0\pm 1$ accounts for the three 
possible curvatures of the spatial hypersurfaces, and $\tau$ is the so-called 
conformal time, which relates to the cosmological time $t$ through 
$dt=a(\tau)d\tau$.  This metric is characterised by the 
vanishing of the shear $\sigma_{\mu\nu}=0$, the vorticity $\omega_{\mu\nu}=0$ 
and  the acceleration $\dot u^\mu=0$. The expansion is the only non-vanishing 
kinematical quantity, and is $\theta=3\dot a/a$.

In addition, the admissible energy-momentum tensor (\ref{EMT_decomposed}) 
compatible with the assumed symmetries of the spacetime reduces to
\begin{equation}
T^{\mu\nu} = (\rho+p)u^\mu u^\nu + pg^{\mu\nu} \; , 
\end{equation}
usually interpreted as representing a perfect fluid with energy-density $\rho$, 
and pressure $p$ measured by the co-moving observer with 4-velocity $u^a$ (such 
that $u_\alpha u^\alpha=-1$, under our choice of 
signature)~\cite{Bernstein:1988bw,Kolb:1990vq}. Of course the latter 
energy-momentum tensor can be 
the sum of several components, i.e., $\rho =\sum_i \rho_{(i)}$, $p=\sum_i 
p_{(i)}$, with or without mutual 
interactions
\cite{Ellis:1973jva,Ellis:1998ct}. 

Applying the general Einstein field equations (\ref{EFE_GR}) in the case of the 
cosmological  FLRW metrics, one obtains the Friedmann equations, namely
\begin{eqnarray}
&&    H^2+\frac{k}{a^2} =\frac{\kappa^2}{3}\rho +\frac{\Lambda}{3}\\
  &&  \dot {H}+ H^2 = -\frac{\kappa^2}{6} (\rho+3p)+\frac{\Lambda}{3},
\end{eqnarray}
where we have introduced the so-called Hubble parameter
\begin{equation}
    H=\frac{\dot a}{a}\; .
\end{equation}
The cosmological model is fully prescribed when we assume some equation(s) of 
state. In the simplest scenario it is possible to model the expansion rate 
of the Universe, and its thermal history~\cite{Gamow:1948pob} as the interplay 
of two conventional matter sources: incoherent, baryonic matter ($p_b=0$) and 
radiation ($p_r=\rho_r/3$) -- summing up about $5\%$ of the overall budget of 
the Universe at the 
present time, plus two unconventional components dubbed ``dark matter'' (with 
$p_{dm}\simeq 0$) and ``dark energy'' (with $p_{DE}=p_{DE}(\rho_{DE})$).  The 
latter two are 
motivated, on the one hand, by the process of formation of the large-scale 
structure (LSS)  through gravitational instability, and on the other hand by 
the unexpected discovery that the Universe at low redshifts is inflating. Structure 
formation requires about $30\%$ of non-interacting matter, which is 
collectively denoted by $\rho_m(\equiv\rho_b+\rho_{dm})$. The accelerated 
expansion of the Universe, revealed by the observation of supernova type Ia 
(SNIa)~\cite{Perlmutter:1998np,Riess:1998cb}, demands a 
repulsive effect that can be simplest fitted by  the ellusive cosmological 
constant $\Lambda$ \cite{Copeland:2006wr,Clifton:2011jh}, representing about 
$70\%$ of the 
content of the Universe. In the latter case the cosmological constant is 
interpreted as a matter component of equation of state 
$p_\Lambda=-\rho_\Lambda$.

It must be remarked at this point that the adoption of the  spatially 
homogeneous  and isotropic FLRW models to describe the Universe was questioned 
in the 1970s~\cite{Dicke:1900mn}. It was pointed out that there are causality 
issues when one attempts  to explain the onset of such a highly  symmetrical 
Universe out of an initial Big-Bang singularity. Another problem would stem from 
the possibility that topological defects formed during the Early Universe would 
close it, in contrast with the observations that favour a Universe with flat 
spatial hypersurfaces of homogeneity.  The latter flatness was also a problem, 
as a Universe filled with matter and radiation required an extreme fine tuning 
of primordial conditions to fit the observations.  It was then devised  
~\cite{Guth:1980zm,Starobinsky:1980te} that  a brief stage of early inflation, 
i.e., of accelerated expansion close to the Planck epoch, would solve the 
problems under scrutiny (see, for instance, \cite{Brandenberger:1999sw}, and 
references therein). This inflationary epoch is  thought to be the outcome of 
primordial quantum conditions still not fully understood. The first scenarios 
involved  a scalar field, dubbed the inflaton, with the ability to produce a 
repulsive negative pressure, and hence inducing the accelerated stage of 
expansion. Moreover the quantum fluctuations of this field can be interpreted as 
the seeds for the density fluctuations that eventually grow into the observed 
large scale structure of the Universe. This would naturally happen at the 
transition between the epoch dominated by radiation and ultra-relativistic 
matter, and the subsequent epoch dominated by non-relativistic matter. 
Significantly this enables one to relate the power-spectrum of the CMB to the 
conditions fulfilled by inflation~\cite{Bertschinger:1998tv}, and thus it is a 
most relevant building block in regard to the possibility of testing the models  
of the Early Universe. There is a multitude of proposals and scenarios to 
promote the inflationary stage. Most of them involve the consideration of one or 
several scalar fields, but there are also scenarios in which other fields are 
envisaged~\cite{Brandenberger:1999sw}.

Among the notable successes of this cosmological model, we count the prediction 
of the abundances of the light elements~\cite{Alpher:1948ve} and of the number 
of particle species~\cite{Kolb:1990vq}, plus the explanation of the Cosmic Microwave  
Background (CMB) radiation found in 1965~\cite{Penzias:1965wn,Dicke:1965zz}. 
Moreover, the remarkable observational progress and the judicious combination of 
different observations, such as CMB~\cite{Ade:2013sjv}, Baryonic Acoustic 
Oscillations (BAO)~\cite{Percival:2009xn}, LSS power spectrum, and weak 
lensing~\cite{Bartelmann:1999yn} have led to a best fit that favours an 
$\Lambda$CDM model~\cite{Ferreira:2019xrr,Ade:2013sjv,Aubourg:2014yra}.

One description of the space-time metric which nicely adapts  both to the 
astrophysical and cosmological scales addressed by General Relativity, is the  
perturbed 
metric~\cite{Mukhanov:1990me,Efstathiou:1989yr}
\begin{equation}
ds^2 =  a^2(\tau)\,\left[ -\left(1+2\Psi\right)\,d\tau^2 + 2
\omega_i 
d\tau \,dx^i  +\left[\left(1-2\Phi \right)\, \gamma_{ij}+2h_{ij}\right]\,dx^i {\rm 
d}x^j\right]\; .
\end{equation}
In this expression $a^2(\tau)$ is a conformal factor 
that rescales the whole line element, and which for cosmological purposes is  
restricted to be only dependent on the so-called conformal time $\tau$. It 
accounts, of course, for the underlying  expansion of the Universe. $\Phi$ and 
$\Psi$ are scalar fields that relate to the Newtonian potential of the isotropic 
weak-field case.
The vector $\omega_i$  
represents the vorticity, and $\gamma_{ij}$ is the metric induced on the 
spatial surfaces, usually the co-moving part of the FLRW metric. Finally, 
$h_{ij}$ is a symmetric  traceless tensor that represents tensor perturbations, 
and hence relates to the gravitational waves.This form of the metric, denoted 
the Newtonian longitudinal 
gauge,  is particularly useful when investigating  the evolution of perturbations. 
In the weak-field approximation,  we assume $a(\tau)$ is effectively constant 
and the parameters $\Phi$, $\Psi$, $\omega^i$ and $h_{ij}$  respectively 
describe the scalar, vector and tensor modes of small perturbations away from 
Minkowski space-time  (in this case the induced 3-metric is 
$\gamma_{ij}=\delta_{ij}$). When we 
wish to investigate fluctuations in the cosmological context, we usually set 
$\omega^i=0$ due to the fast damping of rotational perturbations, the  
potentials $\Psi$ and $\Phi$  are equal due the background isotropy, and 
describe scalar perturbations. The equation for $h_{ij}$  yields 
the propagation of gravitational waves.  
The resulting gravitational wave 
equation corresponds to  a spin-2 particle, carrying the gravitational 
interaction -- the graviton. The investigation of the evolution of perturbations 
into their non-linear  regime is a crucial block to  contrast the cosmological 
models with observations~\cite{Bertschinger:1998tv}.

Despite the  impressive agreement of  General Relativity with most 
observational data, it has been recognised that it faces difficulties both at 
the small  scales (and high 
energies),  and  at the large scales 
(and small energies). These opposite limits have been dubbed the ultraviolet 
(UV) and infrared (IR) limits, respectively. In the former  case, one recognises 
the lack of success in building a consistent quantum gravity  and in the latter 
case, we are faced with  the tantalising absence of explanation for the 
existence of enigmatic  gravitational components in the large-scale Universe, 
called dark matter and dark energy problems~\cite{Debono:2016vkp}.

In order to address the latter caveats, the investigation of extended gravity 
theories sharing  the same fundamental principles of GR, but otherwise looking 
for improved field equations, has been a focal point of interest. 
This endeavour  can be traced back to the very first years after GR was put 
forward, when the efforts to  find a Grand Unified Theory (GUT) of all the 
fundamental forces started~\cite{Goenner:2004se}. At first the goal was 
circumscribed to  merging gravitation and the electromagnetic field, but the 
need to encompass the weak and strong nuclear forces further increased the 
proportions and difficulties of the task. The design of a Grand Unified Theory 
(GUT) has led to increasingly sophisticated theories of which the theory of 
Superstring is the present, and most promising  
paradigm~\cite{Green:1987sp,Green:1987mn}.
 
In regard to the restricted problem of extending GR without the specific 
consideration of  quantum gravity, it is  of great importance to be aware of 
the possible avenues for progress. On the one hand, we have three different 
alternative, variational derivations of the gravitational  equations of motion 
available: (i) the metric formalism; (ii) Palatini formalism; (iii) 
metric-affine formalism~\cite{Olmo:2004hj,Koivisto:2019jra}. In the metric 
formalism, the action is assumed to be only dependent on the metric tensor,  
whilst in the Palatini formalism the metric and the connection are taken to be 
independent degrees of freedom of the curvature. In the metric-affine approach 
the perspective is similar to the latter, but one no longer assumes the independence 
between the metric and the affine connection. We still vary the action with respect 
to both the metric and the affine connection, but the variation with respect to the 
affine connection yields a constraint to be articulated with the metric 
equation. 

On the other hand, Lovelock's theorem~\cite{Lovelock:1971yv} 
\label{Lovelocktheref1} clarifies the 
assumptions that must be met by the gravitational Lagrangian leading to 
second-order field equations, when the action is built from the metric tensor. 
Indeed, in four dimensions the most general Lagrangian density ${\mathcal{L}}$ 
built  just from
the metric and its derivatives that can yield Einstein's field equations reads
\begin{equation}
 {\mathcal{L}} = \eta \sqrt{-g}R - 2\lambda  \sqrt{-g} 
+\gamma\epsilon^{\alpha\beta\mu\nu} {R^{\gamma\delta}}_{\alpha\beta} 
R_{\gamma\delta\mu\nu} +\xi \sqrt{-g} \left(R^2-4R^\mu_\nu R^\nu_\mu + 
{R^{\alpha\beta}}_{\mu\nu}  \,{R^{\mu\nu}}_{\alpha\beta} \right)
\end{equation}
where $\gamma $ and $\lambda $ are constants and $\epsilon^{\alpha\beta\mu\nu} $ 
is the completely skew Levi-Civita tensor. Lovelock's 
theorem~\cite{Lovelock:1971yv} states that the third and fourth terms are 
boundary terms, and therefore do not contribute to the field equations. This 
reduces the contributing part of the previous Lagrangian only to the necessary 
terms to build the Einstein-Hilbert action plus a constant term.

By the same token, it becomes apparent which fields and/or couplings can evade 
the conditions of Lovelock's theorem. It tells us that if one wishes to 
introduce modifications in the gravitational 
sector~\cite{Avelino:2016lpj,Clifton:2011jh},  at 
least one of the following options must be taken:
\begin{itemize}
\item Include more fields beyond, or even instead of, the metric tensor; 
\item Allow for higher-order derivatives in the field equations;
\item Work in a space-time with more than four dimensions;
\item Break the diffeomorphism invariance.
\end{itemize}

In the next chapter, as well as in Part I of the present Review, these 
alternative extensions to General Relativity will be thoroughly addressed.

\chapter[Foundations of Gravity -- Modifications and Extensions]{Foundations 
of 
gravity
-- modifications and extensions}
\label{Bohmerchapter}
{\em Christian G. B{\"o}hmer }

\section{Preliminaries}

Shortly after the formulation of General Relativity 
\label{GRref}  was completed in 1915, it 
became clear that this theory could be extended in various different ways. The 
theory of General Relativity (GR) is formulated using the language of differential 
geometry, which - in itself - was a relatively new topic of research in 
mathematics 
at that time. The geometrical setting used by Einstein consists of a 
four-dimensional Lorentzian manifold, equipped with a metric structure $g$ and 
a 
covariant derivative $\nabla$, or equivalently, a connection $\hat{\Gamma}$. 
This derivative is assumed to be metric compatible and torsion free, which then
uniquely determines the connection coefficients to be the Christoffel symbol 
components $\Gamma$. 

Let us briefly dissect these assumptions to get an immediate idea of how one 
could modify GR. To begin with one does not have to restrict the geometry to 
four 
dimensions. Kaluza and Klein are credited with suggestions along those 
lines~\cite{Kaluza:1984ws,Klein:1926tv}. The use of four dimensions relies on 
our experience of three spatial dimensions and a sense of time, which acts as 
the 
fourth dimension, commonly denoted as the zeroth coordinate. One could now 
assume that there exist other spatial dimensions that have not yet been observed. 
Hypothetically, there could also be other time-like dimensions. These initial 
ideas were motivated by the idea of geometrically unifying the different 
physical theories known at that time; see \cite{Goenner2004,Goenner2014} for a 
comprehensive review on so-called unified field theories. It is probably fair to 
say that String Theory has followed that path, point particles (points are zero 
dimensional objects) being replaced by strings (strings or curves are 
one-dimensional objects). Bosonic string theory is formulated in a 
26-dimensional 
Lorentzian manifold, while superstring theory is formulated in 10 dimensions. 
These extra dimensions are dealt with by compactification, which means `rolling 
up' those dimensions in such a way that they are very small, hence, effectively 
leading to a four-dimensional space in which Special Relativity and General 
Relativity are formulated. 

The next generalisation concerns the connection $\hat{\Gamma}$ which   
neither has 
to be metric compatible nor torsion free. Both, non-metricity and torsion have 
neat geometrical interpretations~\cite{Hehl:1976kj,Hehl:1994ue}; one speaks of 
an affine connection. Let us consider an infinitesimal parallelogram that is 
constructed by parallelly transporting two given vectors along each other. There 
is no \emph{a priori} guarantee that this process does give a closed 
parallelogram. Indeed, torsion represents the failure of this infinitesimal 
parallelogram to close. In order to understand the effect of non-metricity on 
the manifold, let us consider a null vector $u^\mu$, which means it satisfies 
$g_{\mu\nu}u^\mu u^\nu = 0$. If the covariant derivative of the metric tensor 
does not vanish, then this vector may no longer be null when parallelly 
transported. In particular, the light cone structure would no longer be 
invariant under parallel transport. However, neither the lack of closed 
infinitesimal parallelograms nor the non-invariance of the light cone structure 
under parallel transport are reason enough to discard these geometrical concepts 
from a physical point of view. In the end, any theoretical model of the 
gravitational field will make certain predictions that an experiment can either 
verify or falsify.

The entire discussion up to now was independent of the Einstein field 
equations; it merely assumed that there exists a gravitational theory that can 
be formulated using differential geometry. Let us now start making some 
connections between the mathematical formulation and the physical content of our 
theories. It is a well-established everyday fact that light travels along 
straight lines, and so do massive particles in the absence of external forces. 
In 
classical physics one would refer to these as Fermat's principle and Newton's 
first law, respectively. In the context of differential geometry things start to 
get interesting now, as a manifold equipped with a metric structure and an 
affine 
connection gives rise to two distinct curves geodesics and autoparallels. 
Geodesics are the shortest possible curves between two fixed end points, 
autoparallels are the straightest possible curves between two points. Geodesics 
are generally introduced by studying curves $\mathcal{C}$ with tangent vectors 
$T^\mu = dX^\mu/d\lambda$ such that the quantity
\begin{align}
  s = \int\limits_{\lambda_1}^{\lambda_2} \sqrt{g_{\mu\nu}T^\mu T^\nu} d\lambda 
\,,
  \label{eqn1CGB}
\end{align}
is extremised. Here, $X^\mu(\lambda)$ are the local coordinates of the curve 
and 
$\lambda$ is the (affine) parameter of the curve. This yields the familiar 
geodesic equations  \label{geodesicsref1} \label{nonRiemannianref1}
\begin{align}
  \frac{d^2 X^\mu}{d\lambda^2} + {\Gamma}^{\mu}_{\sigma\tau}
  \frac{d X^\sigma}{d\lambda}\frac{d X^\tau}{d\lambda} = 0
  \quad \Leftrightarrow \quad
  \frac{d T^\mu}{d\lambda} + {\Gamma}^{\mu}_{\sigma\tau} T^\sigma 
T^\tau = 0 \,.
  \label{eqn2CGB}
\end{align}
It needs to be emphasised that the geodesic equation, defined via this 
variational approach, depends on the Christoffel symbol components 
${\Gamma}^{\mu}_{\sigma\tau}$ only. This follows from the fact that 
(\ref{eqn1CGB}) is independent of the affine connection\label{metaffconref1} - 
that is, it depends 
on the 
metric tensor and the curve. 

On the other hand, we can introduce the straightest possible curves or 
autoparallels  \label{autoparallelsref1}. Let us again consider a curve 
$\mathcal{C}$ with tangent vector 
$T^\mu$, then the vector $V^\sigma$ is parallelly transported along this curve 
if $T^\mu \nabla_\mu V^\sigma = 0$. The notion of parallel transport allows us 
to consider curves (defined indirectly) whose tangent vectors are parallelly 
transported along themselves, the tangent vector is kept as parallel as possible 
along the curve, hence autoparallel. Using the chain rule and the definition of 
covariant differentiation, the autoparallel equations are given by
\begin{align}
  T^\mu \nabla_\mu T^\sigma = 0
  \quad \Leftrightarrow \quad
  \frac{d T^\mu}{d\lambda} + \hat{\Gamma}^{\mu}_{\sigma\tau} T^\sigma T^\tau = 0 
\,.
  \label{eqn3CGB}
\end{align}
The key difference between (\ref{eqn2CGB}) and (\ref{eqn3CGB}) is that two 
different connections appear in these equations, while their form is identical. 
It is clear that (\ref{eqn3CGB}) depends on the symmetric part of the connection, 
since one can exchange $T^\sigma$ and $T^\tau$; however, it is important to 
state
\begin{align}
  \hat{\Gamma}^{\mu}_{(\sigma\tau)} \neq {\Gamma}^{\mu}_{\sigma\tau}
\,,
  \label{eqn4CGB}
\end{align}
which means that the symmetric part of the affine connection is not the 
Christoffel symbol. This symmetric part contains the Christoffel symbol, but it also 
depends on torsion and non-metricity, should these be present. 

General Relativity is special in the sense that the shortest possible lines 
coincide with the straightest possible lines.\footnote{If the affine connection 
differs from the Christoffel symbol components by a totally skew-symmetric 
piece, geodesics and autoparallels would also coincide.} These considerations 
have practical implications. By studying the geometric properties of 
trajectories of test particles one can, in principle, determine whether the 
connection contains contributions other than those from the Christoffel symbol; 
see 
the footnote again.

In its standard formulation, the dynamical variables of General Relativity are 
the 10 metric functions $g_{\mu\nu}$, which are the solutions of the ten 
Einstein field equations
\begin{align}
  G_{\mu\nu} := R_{\mu\nu} - \frac{1}{2}R\, g_{\mu\nu} = \kappa^2
T_{\mu\nu} \,.
  \label{eqn5CGB}
\end{align}
Here, $G_{\mu\nu}$ is the Einstein tensor, $R_{\mu\nu}$ is the Ricci tensor, 
$R$ is the Ricci scalar, and $T_{\mu\nu}$ stands for the metric 
energy-momentum-stress tensor. This is a true tensor equation in the sense that 
it is valid for all coordinate systems and hence diffeomorphism invariant. In 
four spacetime dimensions one has four coordinates, which can be arbitrarily 
changed, which implies that the Einstein field equations can be viewed as six 
independent equations. When a Hamiltonian analysis is performed on these 
equations, one finds four primary constraints, thereby reducing the number of 
propagating degrees of freedom of this theory to two (10 metric components minus 
four coordinate transformations minus four primary constraints). When General 
Relativity is introduced in a first course - see for 
instance \cite{Bohmer:2016ome} - the Einstein field equations are motivated by 
comparing Newton's equations with geometrical equations, which may describe the 
same physics.

A more elegant approach, which somewhat lacks physical motivation from first 
principles, is the variational approach. The field equations can also be derived 
from the so-called Einstein-Hilbert action \label{EHactionref1}
\begin{align}
  S_{\rm EH} &= \frac{1}{2\kappa^2}\int g^{\mu\nu} R_{\mu\nu} \sqrt{-g}\, d^4x 
= 
\frac{1}{2\kappa^2} \int R \sqrt{-g}\, d^4x\,,\\
  S_{\rm matter} &= \int \mathcal{L}_{\rm matter}(g,\phi,\nabla\psi) \, d^4x =
  \int L_{\rm matter}(g,\phi,\nabla\psi)\sqrt{-g} \, d^4x\,,\\
  S_{\rm total} &= S_{\rm EH} + S_{\rm matter}\,,
  \label{eqn6CGB}
\end{align}
where one varies with respect to the dynamical variable $g_{\mu\nu}$. Here, $g$ 
is the determinant of the metric tensor $g_{\mu\nu}$, so that $\sqrt{-g}\, 
d^4x$ 
is the appropriate volume element when integrating over the manifold. The matter 
fields are denoted by $\psi$ and the matter Lagrangian can depend on derivatives 
of the matter fields. The gravitational coupling constant $\kappa^2$ is given 
by 
$\kappa^2=8\pi G/c^4$. It is through this variational approach that one can 
introduce and motivate various gravitational theories, which can be seen as 
extensions or modifications of the original theory. In the following 
sections, some of 
these ideas will be discussed.

\section{Matter Couplings}
\label{sec_matter}

Before discussing other gravitational theories, let us briefly mention the issue 
of matter couplings \label{mattercouplingsref1}. This is, of course, of crucial 
importance as gravity is 
universal and is the dominant interaction in the macroscopic 
world~\cite{Blagojevic:2013xpa}. Lagrangians, which describe scalars (spin $0$ 
particles) or spinors (spin $1/2$ particles), typically depend on the fields 
and 
their first derivatives, thereby giving rise to equations of motion of at most 
second order. This also holds for Yang-Mills theories; however, we will focus 
our discussion on scalars and spinors for now. When the scalar or spinor field 
actions, formulated in Minkowski space, are formulated on an arbitrary manifold, 
one replaces the Minkowski metric $\eta$ with an arbitrary metric $g$. The 
partial derivatives are replaced with covariant derivatives. In the scalar field 
case one simply has $\nabla_\mu \phi = \partial_\mu \phi$, while for spinorial 
fields the covariant derivative also depends on the connection and we have 
$\nabla_\mu \psi \neq \partial_\mu \psi$, with $\psi$ standing for a spinor 
field. 
The immediate consequence of this is that theories in which variations with 
respect to the connection are considered will contain source terms when spinor 
fields are taken into account. Since protons, neutrons and electrons are all 
spin $1/2$ particles, this is an important issue to keep in mind. Finally, when 
considering Yang-Mills theories we recall that the currents which act as the 
source terms are conserved and couple to the gauge fields. General Relativity 
can also be formulated as gauge theories; however, it is not in the form of a 
typical Yang-Mills theory, 
see \cite{Obukhov:2006gea,Blagojevic:2013xpa,Aldrovandi:2013wha}. The above 
mentioned approach is often referred to as the principle of minimal coupling, 
however, many other coupling terms are in principle possible. There are 
Pauli-type terms and Jordan-Brans-Dicke-type terms where geometrical quantities 
like 
the Riemann tensor or the Ricci tensor couple to the matter fields; these 
mainly 
appear in phenomenological models or when models with symmetry breaking are 
concerned. Note that any coupling term which involves curvature will necessarily 
vanish in Special Relativity, therefore such terms require strong gravitational 
fields to affect the theory and to be, in principle, observable.

\section{The Einstein-Hilbert Action -- Linear Extensions}

On manifolds where the connection is metric compatible and torsion free, the 
Einstein-Hilbert action is the unique action that is linear in a curvature 
scalar, and the Ricci scalar is the unique linear curvature scalar. In more 
general spaces with torsion and non-metricity, one can also construct the scalar 
$\varepsilon^{\mu\nu\kappa\lambda}\hat{R}_{\mu\nu\kappa\lambda}$, which does not 
vanish 
in general. This term appears in the   so-called Palatini action of General 
Relativity or the Holst action \label{Palatiniformref1}. It becomes important 
in 
the context of Loop 
Quantum Gravity, where it appears in Ashtekar's choice of variables, which 
allows 
the formulation of GR as a Yang-Mills type theory 
 \cite{Ashtekar:1986yd,Ashtekar:1987gu,Barbero:1994ap,Holst:1995pc,
Immirzi:1996di,thiemann2007}. Let us return to the Einstein-Hilbert 
action~(\ref{eqn6CGB}) for now.

The Riemann curvature tensor and the Ricci tensor can be defined by the affine 
connection $\hat{\Gamma}$ alone, without requiring the metric tensor. To make 
this 
explicit, it is often written as $\hat{R}_{\mu\nu}$. Hence, in affine 
spacetimes  the Einstein-Hilbert action   
can be generalised   simply by writing\label{metaffgrref1}
\begin{align}
  S &= \frac{1}{2\kappa^2} \int g^{\mu\nu} \hat{R}_{\mu\nu} \sqrt{-g}\, 
d^4x\,,\\
  S_{\rm matter} &= \int L_{\rm matter}(g,\phi,\nabla\psi)\sqrt{-g} \, d^4x\,,\\
  S_{\rm total} &= S_{\rm EH} + S_{\rm matter}\,.
  \label{eqn7CGB}
\end{align}
One now considers the metric tensor $g$ and the connection $\hat{\Gamma}$ as 
\emph{a 
priori} independent dynamical variables. The matter action also depends on the 
connection through the covariant derivative; this is completely consistent with 
the principle of minimal coupling used in General Relativity. This principle 
states that first one writes all equations covariantly in a four-dimensional 
Lorentzian manifold, flat Minkowski space, then one replaces all partial 
derivatives with covariant derivatives and all Minkowski metric tensors with 
arbitrary metric tensors;  see Sec. \ref{sec_matter}.

If we assume  that the matter part of the action does not depend on the 
connection and we   make independent variations with respect to the metric 
and connection, we arrive  at Einstein's theory of General Relativity. This is 
often referred to as the Palatini variation; however, as we will soon discuss, 
things become  more subtle when geometries are more general.

In many ways the most natural generalisation of General Relativity is 
constructed when beginning with (\ref{eqn7CGB}) and allowing the matter part of 
the action to depend on the matter fields, the metric and an independent 
connection. When we now compute  the variations with respect to the metric and 
the independent connection, we arrive  at two sets of field equations. 
Variations with respect to the metric yield equations that resemble the 
Einstein field equations, while variations with respect to the connection give a 
new set of field equations which determine the connection. The source term that 
appears in the latter is often referred to as the hyper-momentum 
$\Delta^\lambda{}_{\mu\nu}$, following a commonly used 
notation~\cite{Hehl:1994ue}. As the affine connection has no symmetries, the 
hyper-momentum tensor has, in general, 64 independent components in four 
dimensions.

Let us now discuss how we can connect these different theories back to General 
Relativity, using a mathematically consistent approach. The perhaps most 
elegant 
way to do it is through the introduction of Lagrange multipliers in the total 
action (\ref{eqn7CGB}), so that this action is subsequently extremised subject 
to 
constraints. These constraints are introduced so that the geometrical properties 
of the manifold are controlled. More explicitly, let us, for the time being,  
extract General 
Relativity within the framework of metric affine theories. Recall that the two 
key geometrical assumptions are a metric compatible and torsion-free covariant 
derivative. In the language of constraints we would write\label{torsionref1}
\begin{align}
  S_{\rm GR} &= \frac{1}{2\kappa^2} \int \Bigl\{
  g^{\mu\nu} \hat{R}_{\mu\nu} + \lambda_{(1)}^{\mu\nu\lambda} 
T_{\mu\nu\lambda} +
  \lambda_{(2)}^{\mu\nu\lambda} Q_{\mu\nu\lambda}
  \Bigr\}\sqrt{-g}\, d^4x\,, \\
  S_{\rm total} &= S_{\rm GR} + S_{\rm matter}\,,
  \label{eqn8CGB}
\end{align}
where $T_{\mu\nu\lambda}$ is the torsion tensor and $Q_{\mu\nu\lambda}$ is the 
non-metricity tensor\label{nonmetricityref1}. Here, $\lambda_{(1)}$ and 
$\lambda_{(2)}$ are two 
Lagrange 
multipliers which ensure that the affine connection will become the usual 
Christoffel symbol. Clearly, variations with respect to $\lambda_{(1)}$ give 
$T_{\mu\nu\lambda} = 0$, while variation with respect to $\lambda_{(2)}$ yields 
$Q_{\mu\nu\lambda} = 0$.

A popular extension of General Relativity is the so-called Einstein-Cartan 
theory, which was proposed in the 1920s by Cartan; see~\cite{Hehl:1976kj}. 
Within 
the above framework, Einstein-Cartan theory is simply defined by
\begin{align}
  S_{\rm EC} &= \frac{1}{2\kappa^2} \int \Bigl\{
  g^{\mu\nu} \hat{R}_{\mu\nu} + \lambda_{(2)}^{\mu\nu\lambda} 
Q_{\mu\nu\lambda} +
  \Bigr\}\sqrt{-g}\, d^4x\,, \\
  S_{\rm total} &= S_{\rm EC} + S_{\rm matter}\,.
  \label{eqn9CGB}
\end{align}
The only difference with respect to General Relativity is the possible presence 
of torsion, 
which is no longer assumed to be zero. A natural source term for torsion would 
be fermions; their action depends on the connection, and hence variations with 
respect to the connection lead to source terms for torsion. A peculiar property 
of Einstein-Cartan theory is that the field equations for torsion are 
algebraic; 
they do not contain derivatives of the torsion tensor. This immediately implies 
that torsion cannot propagate, and consequently, only regions of spacetime that 
contain sources of torsion can contain torsion. This is in stark contrast to 
curvature, a fact well-known in GR. The Schwarzschild solution, for instance, 
is 
a vacuum (source-free) solution of the Einstein field equations, yet contains 
curvature. Likewise, gravitational waves can propagate through otherwise empty 
regions of space-time; torsional waves in this sense do not exist in 
Einstein-Cartan theory.

If we recall that Minkowski space is the setting of Special Relativity, we can 
of course also include this using the above approach, namely we consider the 
following theory
\begin{align}
  S_{\rm Mink} &= \frac{1}{2\kappa^2} \int \Bigl\{
  g^{\mu\nu} \hat{R}_{\mu\nu} + \lambda_{(0)}^{\mu\nu\lambda\kappa} 
\hat{R}_{\mu\nu\lambda\kappa} + \lambda_{(1)}^{\mu\nu\lambda} T_{\mu\nu\lambda} + 
\lambda_{(2)}^{\mu\nu\lambda} Q_{\mu\nu\lambda}
  \Bigr\}\sqrt{-g}\, d^4x\,.
  \label{eqn10CGB}
\end{align}
Minkowski space is the unique space that has vanishing torsion, vanishing 
non-metricity and is globally flat.

However, what makes this approach, using constraints, particularly useful is the 
ability to systematically study a variety of theories in a uniform setting; see 
also~\cite{BeltranJimenez:2019tjy}. Let us now discuss a theory, which is 
equivalent to General Relativity but is formulated rather differently. It was 
noted in the 1920s by Einstein and others that there exists a formulation of 
General Relativity based solely on the torsion tensor; this theory is now known 
as the Teleparallel Equivalent of General Relativity (TEGR). 

Start  as before, within the setting of metric-affine theories where the 
connection is assumed to be fully independent of the metric, we can define the 
Teleparallel Equivalent of General Relativity  by
\begin{align}
  S_{\rm TEGR} &= \frac{1}{2\kappa^2} \int \Bigl\{
  g^{\mu\nu} \hat{R}_{\mu\nu} + \lambda_{(0)}^{\mu\nu\lambda\kappa} 
\hat{R}_{\mu\nu\lambda\kappa} + \lambda_{(2)}^{\mu\nu\lambda} 
Q_{\mu\nu\lambda}
  \Bigr\}\sqrt{-g}\, d^4x\,.
  \label{eqn11CGB}
\end{align}
We note that the constraints force the connection to be metric compatible (no 
non-metricity, $Q_{\mu\nu\lambda}=0$) and make the manifold \emph{globally} 
flat, $\hat{R}_{\mu\nu\lambda\kappa} = 0$ everywhere. There is now a 
conceptual issue to understand: Is the theory so defined non-trivial? This is a 
natural question to ask, as we know that only Minkowski space satisfies 
$R_{\mu\nu\lambda\kappa} = 0$ everywhere in the usual GR setting. To answer 
this, let us begin by introducing the so-called contortion tensor $K$, defined 
by 
\begin{align}
  \hat{\Gamma}^{\mu}_{\sigma\tau} = {\Gamma}^{\mu}_{\sigma\tau} + 
K_{\sigma\tau}{}^\mu\,.
  \label{eqn12CGB}
\end{align}
The contortion tensor simply contains all the information of the connection 
that is not part of the Christoffel symbol. In other words, it contains the 
deviations from the standard GR framework. When we compute the Riemann curvature 
tensor for the full connection $\hat{\Gamma}^{\mu}_{\sigma\tau}$ and express 
the 
result using the Christoffel symbol and the contortion tensor, we arrive  at 
the neat result \label{Riemanntenref}
\begin{align}
  \hat{R}_{\nu\mu\lambda}{}^\kappa = R_{\nu\mu\lambda}{}^\kappa 
+
  \nabla^{\Gamma}_\nu K_{\mu\lambda}{}^\kappa - \nabla^{\Gamma}_\mu 
K_{\nu\lambda}{}^\kappa +
  K_{\nu\rho}{}^\kappa K_{\mu\lambda}{}^\rho - K_{\nu\rho}{}^\kappa 
K_{\mu\lambda}{}^\rho \,,
  \label{eqn13CGB}
\end{align}
which means that the curvature tensor splits into two parts: the Levi-Civita 
part $R_{\nu\mu\lambda}{}^\kappa$, which is constructed using the 
Christoffel symbols components only, and a second part that depends on the 
contortion tensor. The notation $\nabla^{\Gamma}$ stands for the covariant 
derivative involving the Christoffel symbol components. One normally defines the 
torsion tensor\label{torsionref1}\label{torsionref2}
 to be the skew symmetric part of the affine connection 
$T_{\mu\nu}{}^\kappa = (\hat{\Gamma}_{\mu\nu}^\kappa - 
\hat{\Gamma}_{\nu\mu}^\kappa)/2$ so 
that, in spaces where $Q_{\mu\nu\lambda}=0$, one has the simple relation
\begin{align}
  T_{\mu\nu}{}^\kappa = \frac{1}{2}\Bigl(K_{\mu\nu}{}^\kappa - 
K_{\nu\mu}{}^\kappa\Bigr) \,.
  \label{eqn14CGB}
\end{align}
Therefore, there is a linear relation between torsion and contortion. Going back 
to (\ref{eqn11CGB}), which imposes the constraint 
$\hat{R}_{\mu\nu\lambda\kappa} = 0$, we can now attempt to understand 
this 
condition in view of (\ref{eqn13CGB}). Is it always possible to choose a 
contortion or torsion tensor for a given curvature tensor 
$R_{\nu\mu\lambda}{}^\kappa$ such that the full metric-affine curvature 
vanishes?

The answer to this question is affirmative; this choice can be made but it 
requires 
a little bit more   mathematics. Let $e_\mu^A$ be four linearly independent 
orthonormal co-vector or co-frame fields, often called tetrads, which allow us 
to write the metric tensor at any point of the manifold as  
\begin{align}
  g_{\mu\nu} = e_\mu^A e_\nu^B \eta_{AB} \,.
  \label{eqn15CGB}
\end{align}
These fields can be used to define the frame components of any vector via $V^A = 
e_\mu^A V^\mu$. Moreover, one needs to introduce the spin connection 
$\hat{\omega}_\mu{}^A{}_B$, which is defined through the vanishing of the 
covariant 
derivative of the tetrad
\begin{align}
  \nabla_\mu e_\nu^A = 0
  \quad \Leftrightarrow \quad
  \partial_\mu e_\nu^A + \hat{\omega}^A{}_{B\mu} e_\nu^B - 
\hat{\Gamma}_{\mu\nu}^{\sigma} 
e_\sigma^A = 0 \,.
  \label{eqn16CGB}
\end{align}
We can now   express the complete (Levi-Civita plus torsional 
contributions) Riemann  curvature tensor using either the connection or the spin 
connection. In the latter case we have
\begin{align}
  \hat{R}^A{}_{B\mu\nu} = \partial_\mu \hat{\omega}^A{}_{B\nu} - 
\partial_\nu 
\hat{\omega}^A{}_{B\mu} +
  \hat{\omega}^A{}_{C\mu} \,\hat{\omega}^C{}_{B\nu} - \hat{\omega}^A{}_{C\nu} \,\hat{\omega}^C{}_{B\mu} 
\,,
  \label{eqn17CGB}
\end{align}
which now allows us to make the following useful observations. If we choose 
$\hat{\omega}^A{}_{B\mu} = 0$ everywhere, then $\hat{R}^A{}_{B\mu\nu}=0$ 
everywhere. 
This means that, applying the decomposition (\ref{eqn12CGB}) to the spin connection, 
we can write
\begin{align}
  \hat{\omega}^A{}_{B\mu} = \omega^A{}_{B\mu} + 
K_\mu{}^A{}_{B} \,,
  \label{eqn18CGB}
\end{align}
which implies that $\hat{\omega}^A{}_{B\mu} = 0$, or equivalently $\omega^A{}_{B\mu} = -K_\mu{}^A{}_B$. Therefore, for any Levi-Civita spin connection 
$\omega^A{}_{B\mu}$ there exists a contortion tensor 
$K_\mu{}^A{}_B$ such that $\hat{\omega}^A{}_{B\mu} = 0$. This result was found by 
Weitzenb\"ock, who noted that this connection can always be constructed, given 
a 
tetrad. Putting this result back into (\ref{eqn13CGB}) allows us to rewrite the 
complete Ricci scalar in terms of the Levi-Civita part and a torsion part; this 
gives
\begin{align}
  \hat{R} = 0
  \quad \Leftrightarrow \quad
  R +
  2 \nabla^{\Gamma}_\nu K^{\lambda}{}_{\lambda}{}^\nu +
  K_{\nu\rho}{}^\nu K^{\lambda}{}_{\lambda}{}^\rho - K_{\nu\rho}{}^\nu 
K^{\lambda}{}_{\lambda}{}^\rho = 0\,.
  \label{eqn19CGB}
\end{align}
Consequently, we have an alternative formulation of the Einstein-Hilbert action 
(\ref{eqn6CGB}) using the previous identity, namely
\begin{align}
  S_{\rm EH} = \frac{1}{2\kappa^2} \int R \sqrt{-g}\, d^4x
  \quad \Leftrightarrow \quad
  S_{\rm TEGR} = \frac{1}{2\kappa^2} \int  \Bigl\{K_{\nu\rho}{}^\nu 
K^{\lambda}{}_{\lambda}{}^\rho -  K_{\nu\rho}{}^\nu 
K^{\lambda}{}_{\lambda}{}^\rho\Bigr\} e\, d^4x \,,
  \label{eqn20CGB}
\end{align}
where we neglected the boundary term that does not contribute to the equations 
of motion. Here, $e$ denotes the determinant of the tetrad field, which 
satisfies 
$e=\sqrt{-g}$ due to (\ref{eqn15CGB}). This is the standard formulation of the 
Teleparallel Equivalent of General Relativity where the tetrad $e_\mu^a$ is 
the 
independent dynamical variable and the spin connection vanishes 
identically~\cite{Maluf:2013gaa}.

The issue of matter couplings was mentioned earlier, and the Teleparallel 
Equivalent of General Relativity is a good case study when it comes to matter 
couplings, especially for spin $1/2$ particles. The Lagrangian for a Dirac field 
contains the term $\nabla \psi$, where, as before, $\psi$ stands for the spinor 
field. Its covariant derivative depends explicitly on the connection, which is no 
longer a dynamical variable in the teleparallel formulation. This leads to 
issues regarding the coupling prescription of Dirac fields; see in particular 
the series of 
papers~\cite{Obukhov:2002tm,Maluf:2003fs,Mielke:2004gg,Obukhov:2004hv,
Leclerc:2004uu} and the references given therein. It was pointed out 
in~\cite{Leclerc:2004uu} that this problem is generic and affects all Poincare 
gauge theories that admit a teleparallel formulation.

\section{The Einstein-Hilbert Action -- Nonlinear Extensions}

All theories considered so far were based on an action linear in the curvature 
scalar as the key ingredient to formulate gravitational theories; however, we 
already noted in (\ref{eqn20CGB}) that such actions are quadratic in the 
contortion tensor. From a theoretical point of view it is well motivated to 
consider more general theories, which depend on other scalars constructed out 
of 
the Riemann curvature tensor or the Ricci tensor. There is no reason to exclude 
terms like $c_1 R_{\mu\nu} R^{\mu\nu}$, for example, in a gravitational action. 
Alternatively, one can consider  theories where an arbitrary function of the 
Ricci scalar is considered. In the following we will focus on the latter 
approach.

The key idea of this scheme is to consider the action 
\begin{align}
  S_{f(R)} &= \frac{1}{2\kappa^2} \int f(R) \sqrt{-g}\, d^4x\,,\\
  S_{\rm matter} &= \int L_{\rm matter}(g,\phi,\nabla\psi)\sqrt{-g} \, d^4x\,,\\
  S_{\rm total} &= S_{f(R)} + S_{\rm matter}\,,
  \label{eqn21CGB}
\end{align}
where $f(R)$ is a sufficiently regular function of the Ricci scalar. When 
choosing $f(R)=R$, one recovers General Relativity, while the choice 
$f(R)=R-2\Lambda$ introduces the cosmological constant into the field equations. 
Such a model was studied in the context of cosmology by~\cite{Barrow:1983rx}; 
however, it was only after the observation of the accelerated expansion of the 
Universe that models of this type became more mainstream and were subsequently 
thoroughly studied~\cite{Capozziello:2002rd,Capozziello:2003tk}; for reviews on 
$f(R)$ gravity the reader is referred 
to~\cite{Sotiriou:2008rp,DeFelice:2010aj,Nojiri:2010wj}. The basic idea 
underlying this approach is to view General Relativity as the lowest order 
theory. To see this, recall that Minkowski spacetime is the geometrical framework 
for Special Relativity that satisfies $R_{\mu\nu\lambda\kappa}=0$, the space 
being 
globally flat. Let us consider a series expansion of $f(R)$ around Minkowski 
spacetime, then $f(R) = f(0) + f'(0)R + f''(0)R^2/2 +\ldots$ so that a term linear 
in $R$ emerges quite naturally.

However, if one wishes to modify General Relativity for cosmological 
applications in particular, this expansion might not be ideal. Over very large 
scales the curvature becomes small, which motivates modifications that contain 
inverse powers of the Ricci scalar; clearly such terms pose problems when 
considering Minkowski space. Other models contain nonlinear functions of total 
derivative terms, like the Gauss-Bonnet term, for example. The Gauss-Bonnet 
term 
is related to a topological number, the Euler characteristic of the manifold. 
However, when any nonlinear function of any topological quantity is added to 
the action, it will yield some non-trivial field equations. Of course, one can 
also introduce new couplings between the geometry and the matter, different 
from 
the minimal coupling. Theories of this type have also received substantial 
attention; 
see in particular~\cite{Harko:2018ayt} for a comprehensive reference of such 
models. Let us add a small sceptic's remark: A function $f$ contains 
uncountably 
many degrees of freedom, so it is perhaps not too surprising that various 
models 
are able to fit a variety of observational data.

Going back to the Teleparallel Equivalent of General Relativity, one could 
apply 
the same ideas to the action (\ref{eqn20CGB}) and consider nonlinear models. 
The 
scalar that appears in the integrand of $S_{\rm TEGR}$ is often denoted by 
$\mathbb{T}$, so that (\ref{eqn19CGB}) can be written in the convenient form 
$R=-\mathbb{T}+B$, where $B$ stands for the boundary term. Consequently, one 
would 
consider the model\label{fTgraref1}
\begin{align}
  S_{f(\mathbb{T})} &= \frac{1}{2\kappa^2} \int f(\mathbb{T}) \sqrt{-g}\, 
d^4x\,,\\
  S_{\rm matter} &= \int L_{\rm matter}(g,\phi,\nabla\psi)\sqrt{-g} \, d^4x\,,\\
  S_{\rm total} &= S_{f(\mathbb{T})} + S_{\rm matter}\,,
  \label{eqn22CGB}
\end{align}
which was first suggested in~\cite{Ferraro:2006jd} and also led to a surge of 
interest; for reviews see~\cite{Capozziello:2011et,Cai:2015emx}. These models 
allow for cosmological solutions with accelerated expansion without the need to 
introduce dark energy. From a conceptual point of view, modified teleparallel 
theories of gravity are interesting, as these are no longer invariant under 
local 
Lorentz transformation in their standard formulation; this means the choice 
$\omega_\mu{}^A{}_B = 0$, where the spin connection vanishes identically. To 
see 
this, one only has to note that neither $\mathbb{T}$ not $B$ are Lorentz 
scalars; the combination $R=-\mathbb{T}+B$ is the unique Lorentz scalar that 
can be constructed, implying that General Relativity and $f(R)$ gravity are 
both locally Lorentz invariant, while any nonlinear theories based on 
combinations of $\mathbb{T}$ and $B$ are not; see~\cite{Bahamonde:2015zma}. We 
note that a fully invariant formulation of modified teleparallel gravity models 
has been proposed~\cite{Krssak:2018ywd}. 
It is very interesting to study the degrees of freedom in $f(\mathbb{T})$ 
gravity in four dimensions;  
see~\cite{Ferraro:2018tpu,Ferraro:2018axk,Blagojevic:2020dyq}. While in $f(R)$ 
gravity the extra degree of freedom is easily 
interpreted as a scalar due to the function $f$, an analogue interpretation in 
$f(\mathbb{T})$ cannot be made and the precise meaning of the extra degrees of 
freedom is an open question in the field.

Breaking local Lorentz invariance can be well  motivated by considering physics 
on very small scales. Quantum theory implies that positions and momenta 
cannot 
be measured simultaneously with unlimited accuracy. Consequently, one would 
expect a certain length scale at which local Lorentz transformations break 
down. 

Let us finish this section with another sceptic's  remark: Once one starts to 
consider nonlinear theories based on various scalar quantities, motivated 
either by the geometry or the matter content, one is able to create a plethora 
of theories. The entirety of such models is so large that it is (practically) 
impossible to study all of them. Clearly, many of these models can be built to 
pass a variety of observational tests due to their generality. What appears to 
be missing at the moment is an overarching guiding principle, which would allow 
us to restrict our attention to a small class of models based on some neat 
theoretical argument.



\phantomsection
\addcontentsline{toc}{part}{\bf Part I: Theories of Gravity}

\begin{center}
{\Huge \bf Part I: Theories of Gravity}
\end{center}
\begin{center}
Editors:  Salvatore Capozziello and 
Jose Beltr\'an Jim\'enez
\end{center}
 





\begingroup
\let\clearpage\relax 
\chapter[Introduction to Part I]{Introduction to Part I}
 
{\em  Salvatore Capozziello, Jose Beltr\'an Jim\'enez}\\

General Relativity   has not ceased its outstanding performance to explain 
gravitational phenomena with exquisite accuracy in an ever increasing range of 
scales. In this respect, local gravity experiments, Solar System tests and 
astrophysical objects have played a central role in establishing the 
fundamental 
properties of the gravitational interaction, showing no deviations with respect 
to General Relativity.  Cosmology, on the other hand, has traditionally been a 
driving force for 
speculations beyond General Relativity, whose only limitation was the 
imagination of the 
theoretical cosmologists. In the last two decades, however, the accumulation of 
precise cosmological measurements has substantially constrained the permitted 
theories and we have now the means to robustly rule out wide classes of 
theories. At the same time, these cosmological observations have triggered 
investigations seeking for theories beyond General Relativity, mainly motivated 
by the three 
fundamental missing ingredients of the standard cosmological model:  
dark 
matter, dark energy and inflation. 

A very distinctive feature of gravity that actually guided Einstein to its 
original formulation is its intimate relation with inertia, to the point that 
it 
is possible to interpret gravity (at least locally) as a purely inertial 
effect. 
This is rooted in the equivalence principle that dictates the universal 
character 
of 
gravity, which in turn lies at the very heart of the possibility to interpret 
gravity in geometrical terms. We thus arrive at the properties that could be 
used to {\it define} gravity from its geometrical side. From a field theory 
perspective, General Relativity is a theory that describes the interactions of 
a massless spin-2 
particle. It is profoundly remarkable that by starting with a massless spin-2 
particle 
and imposing some reasonable additional assumptions like Lorentz-invariance, it 
naturally follows that this particle {\it must} couple universally (at low 
energies) to matter fields, and its interactions are precisely those of General 
Relativity. Thus, 
from the field theory side, the fundamental defining property of gravity could 
be identified with its massless spin-2 nature. The structure of General 
Relativity then results 
as a particular consequence of the strict rules that govern the interactions of 
massless particles.

The approaches to modifications of General Relativity come in   several 
fashions, which can be 
broadly divided into those essentially based on adding new fields, and those 
that 
fully embrace its geometrical description and hence the modifications are 
based on 
modified geometrical scenarios. Definitely, this separation may be regarded as 
purely conventional and, as a matter of fact, it is not difficult to go from 
one 
to the other in some scenarios. This is clearly illustrated by, for example, 
gravity 
theories in a Weyl geometry that can equivalently be regarded as a theory with 
an extra vector field provided by the Weyl non-metricity trace. As it occurs 
many times, however, the starting point or interpretation of the same theory 
can 
serve as motivation and inspiration to explore different modified gravity 
scenarios. It is nevertheless important to keep in mind the basic properties 
that make General Relativity special among all gravity theories, so that the 
modifications can be 
clearly ascribed to the breaking of one of the fundamental assumptions for 
General Relativity. 
This is particularly important in helping us to discern truly modified theories 
from those that are simply General Relativity in disguise.

Modifying General Relativity is an arduous task, not only for its 
aforementioned exquisite 
performance to explain observations, but because its internal structure is 
tightly constrained by consistency conditions that are  ultimately imposed by 
the 
massless 
spin-2 nature of the graviton. This delicate structure causes   many 
(infrared) modifications of General Relativity to be doomed from their very 
conception, and this has 
fuelled an intense activity in recent years to find theoretically consistent 
modifications of General Relativity that in turn could play a role in 
describing the Universe's
dark sector or inflation. This Part will be devoted to disclosing some of the 
most popular and interesting modifications of gravity that have been explored, 
as 
well as their potential interest for cosmology.

\endgroup

\chapter[A Flavour on $f(R)$ Theories: Theory and Observations]{A Flavour on 
$f(R)$ Theories: Theory and Observations}
\label{ref:Cruz}

{\em \'Alvaro de la Cruz-Dombriz}



\section{Historia, Lux Veritatis}

Modifications to the General  Theory of Relativity (GR) emerged almost 
immediately upon its acceptance by the scientific community. By 1919, 
Weyl had already  toyed with the inclusion of higher-order curvature invariants 
in the usual 
Einstein-Hilbert action. 
Probably the first competitor to Einstein's GR  was Nordstr$\ddot{\text{o}}$m's 
1912 conformally flat scalar theory of gravity. In 1937, Dirac showed, by 
allowing the gravitational coupling $G$ to vary slowly over cosmological time 
scales, that there was a relationship that naturally emerged, between the 
cosmological constants and the fundamental physical constants. This idea was 
further developed by Jordan, a little over a decade later, using a scalar field 
to describe the gravitational coupling. In his theory, the gravitational scalar 
field behaved like a matter field, and satisfied a conservation law that was 
added to the theory \cite{Linden:1972up}.
 By 1961, these ideas, thanks to Brans and Dicke \cite{Brans:1961sx},  
culminated   a complete gravitational theory, containing one scalar field, 
which, together with the metric tensor, is responsible for the gravitational 
interaction. The so-called Brans-Dicke theory is indeed considered to be the 
prototype of alternative theories to GR \cite{Faraoni:2004pi}. 

While this and other attempts were driven solely  by curiosity, the 
physical  
motivation for such exercises was yet to come. In the following decades, the 
strong gravity regime of physics stimulated interest in higher-order theories 
of gravity, where it was shown that, unlike GR, higher-order theories are 
renormalisable \cite{Stelle:1976gc}. Moreover, taking  the consideration of 
quantum 
corrections into account requires the gravitational action to include 
higher-order curvature invariants \cite{Utiyama:1962sn, Birrell:1982ix, 
Buchbinder:1992rb, Vilkovisky:1992pb}. In such cases, the modification to GR 
would be  appreciable only at scales either of the order of the Planck length in 
the primordial Universe \cite{Starobinsky:1980te} or in the neighbourhood of 
extremely dense objects \cite{Brandenberger:1992sy, Resco:2016upv}. 
Complementary   late-time implications (in the low-energy regime) of    
modified, 
also dubbed extended, theories of gravity only became a field of interest later 
in the 20th century, when the revelation that over $75\%$ of the energy density 
of the Universe may be unknown  became evident from several experiments, 
including Supernovae type Ia \cite{Eisenstein:2005su},
large-scale structure power spectra measurements \cite{Riess:2004nr} and the  
study of CMB physics \cite{Spergel:2006hy}, among others.

 Thereafter, scalar field considerations towards viable alternatives to GR, 
which include the aforementioned Brans-Dicke theory as a subclass, developed 
into branches of research, including scalar fields coupled non-minimally to 
curvature and induced gravity. Further popularity of the addition of an extra 
gravitational scalar field stems from the fact that such a field is actually a 
vital part of theories such as supergravity, superstring and M-theories. 
Moreover, scalar fields have also been essential in the development of the 
inflationary paradigm following the increasing support of an inflationary epoch 
capable of curing several well-known shortcomings of the Cosmological 
Concordance ($\Lambda$CDM) Model  with an initial - Big Bang - singularity.

Thus, it became evident that the required dark energy component indicates the 
existence of either unknown forms of energy density or additional physics, or 
both. For instance, the Cosmological Concordance $\Lambda$CDM Model contains 
two 
{\it dark} forms of energy, namely dark matter and dark energy, as well as a
required, physically unexplained, period of rapid inflation. Although  
$\Lambda$CDM paradigm is able to match many observations with high precision, 
it suffers 
from several well-known drawbacks \cite{Martin:2012bt}.
Consequently, there is no   a priori epistemological reason -  not even a 
naive invocation to Ockham's razor! - to dismiss a 
theory of gravity 
for which the observed accelerated expansion of the Universe would emerge as  a 
low energy, large-scale geometrical consequence, rather than requiring the 
ad hoc introduction of exotic dark fluids with no observational ground.


Any extension of GR amounts to making one or more generalisations of the 
following form:  $(i)$ adding extra fields $(ii)$ including higher-order 
derivatives of the curvature or curvature-related invariants or other 
invariants and $(iii)$ adding dimensions to the spacetime. These modifications 
resulted in a plethora of extended gravity theories deserving study 
\cite{Clifton:2011jh, Capozziello:2010zz} and were  indeed far much richer than 
GR in 
complexity.
Without trying to be rigorous, a few paradigmatic examples would include 
Lovelock theories featuring field equations of second order in the metric 
\cite{Lovelock:1971yv}, Gauss-Bonnet theories \cite{Cognola:2006eg, 
Nojiri:2005jg, delaCruzDombriz:2011wn}, scalar-tensor theories 
\cite{Brans:1961sx, Brans:1962zz, GarciaBellido:1993wn, Cembranos:2009ds}, 
Tensor-Vector-Scalar (TeVeS)  \cite{Ford:1989me, Jimenez:2008au, 
Koivisto:2008xf}, theories with extra dimensions \cite{Alcaraz:2002iu}, and 
Dvali-Gabadadze-Porrati (DGP) \cite{Dvali:2000hr} theories for gravity. 

In this  chapter  we shall focus on a class of theories, dubbed $f(R)$,  that 
gained popularity after the discovery of the accelerating present epoch of the 
Universe, but which had been considered originally in the context of inflation 
\cite{Starobinsky:1980te}. Such theories involve introducing a generic function 
of the Ricci scalar into the gravitational action from which the field equations 
are subsequently derived. Obviously, once field equations are at hand, both 
theoretical and observational constraints must be carefully studied in order to 
assess - or dismiss - the validity of classes of models in the context of these 
theories.

This chapter is organised as follows: first, in Section  \ref{Sec:2-ST-theories}
 we shall provide a detailed review on the foundations of  the scalar-tensor 
theories of gravity, with some emphasis on the Brans-Dicke theory. 
The underlying idea would be then to present,   in Section  
\ref{Sec:3-fR-theories}, the equivalence between $f(R)$ theories and this 
subclass of scalar-tensor theories; namely, Brans-Dicke gravity. Therein, we 
shall first revise some of the most relevant formalisms to study such theories. 
Also, the viability requirements that $f(R)$ theories need to obey to be 
considered as a viable alternative to Einsteinian gravity will be displayed. 
Then, in Section \ref{Section:metric_f(R)_cosmology},  the $f(R)$ cosmological 
evolution technicalities within the metric formalism when applied to the usual 
four-dimensional Friedmann-Lem\^{a}itre-Robertson-Walker geometry will be 
provided so that the appearance of an effective curvature fluid becomes 
evident. 
Subsequently, as a natural step forward, in  Section 
\ref{Section:Scalar_Perturbations} we shall provide a thorough revision on how 
the use of 1+3 covariant gauge-invariant variables
can be applied to extended theories of gravity, $f(R)$ theories in particular.  
This route renders the analysis of cosmological scalar perturbations highly 
transparent, so that the latest  large-scale structure   data can be 
rigorously compared 
against theoretical predictions. A reader still unfamiliar with this technique 
will surely find this section very enlightening and accessible to follow.
The next two Sections, \ref{Sec:GDE} and  \ref{Section:Attractive_gravity}, are 
devoted to illustrate two paradigmatic gravitational features of $f(R)$ 
theories. 
On the one hand, in Section  \ref{Sec:GDE} the equivalent geodesic deviation  
equation for these theories is derived, and the 
consequences for the observer area distance extracted, a fact  with relevant 
cosmological implications in the measure of distances and comparison with 
eventual data. 
On the other hand, in Section \ref{Section:Attractive_gravity}  we shall briefly 
address the attractive or repulsive character that these theories exhibit in a 
cosmological context. This discussion is obviously related to the literature  
devoted to studying the so-called energy conditions in the context of extended 
theories of gravity and thus, it may help to shed some light on these 
controversial issues depending on the authors one follows.
Finally, we present our conclusions in Section \ref{Section:Conclusions}.


\section{Scalar-Tensor Theories}
\label{Sec:2-ST-theories}
In this section we review the action of general  scalar-tensor theories, 
\label{Scalartensoref1} their 
resulting field equations, and the equivalence between $f(R)$ gravity and 
scalar-tensor gravity.
Although the bulk of the considerations in subsequent  sections may be done from 
a purely metric $f(R)$ fourth-order gravity perspective, viewing such theories 
  from  a scalar-tensor framework may offer some insight into the physical 
dynamics of the resulting effective scalar field potential. Such a comparison 
proves useful when facing solutions containing singularities. 

\subsection{Field Equations of Scalar-Tensor Gravity}

Scalar-tensor theories of gravity include Brans-Dicke, Galileons, $f(R)$,  
quintessence and Horndeski theories as subclasses. The general Lagrangian for a 
scalar tensor theory has the form  \cite{Clifton:2011jh},
\begin{equation}
\mathcal{L} = \frac{1}{2\kappa^2} \sqrt{-g}\left[ f(\phi)R - 
g(\phi)\nabla_{\mu}\phi\nabla^{\mu}\phi - 2\lambda(\phi) \right] + 
\mathcal{L}_{m}(\psi,h(\phi) g_{\mu\nu}),
\end{equation}
 where $f$, $g$, $h$ and $\lambda$ are all arbitrary functions of the scalar 
field $\phi$ and $\mathcal{L}_{m}$ accounts for the Lagrangian of the matter 
fields 
$\psi$. Moreover, $\kappa^2=8\pi G$ holds for the usual gravitational 
  constant and $R$ is the Ricci scalar curvature. The function $h(\phi)$ 
may be absorbed into the metric following the conformal transformation
\begin{equation}
h(\phi)g_{\mu\nu}\rightarrow g_{\mu\nu}.
\end{equation}
  Should that transformation be performed, the resulting frame is known as the 
Jordan frame. Therein, the scalar field and matter are independent, the motion 
of 
test particles follows geodesics, and the weak equivalence principle is  
\label{equivprinref1}
satisfied for massless test particles. The effects of this transformation on the 
arbitrary functions $f$, $g$, $h$ and $\lambda$, may be absorbed by a suitable 
redefinition of these functions. For instance, by setting $f(\phi)\rightarrow 
\phi$, we can write the Lagrangian density as
\begin{equation}\label{eq:Lagrangian_scalarfield_jordan}
\mathcal{L}=\frac{1}{2\kappa^2} \sqrt{-g}\left[ \phi R - 
\frac{\omega(\phi)}{\phi}\nabla_{\mu}\phi\nabla^{\mu}\phi -2\Lambda(\phi)\right] 
+ \mathcal{L}_{m}(\psi,g_{\mu\nu}),
\end{equation} 
  where $\omega(\phi)$ represents  an arbitrary function, dubbed the coupling 
parameter, and the function $\Lambda$ provides 
a generalisation of the cosmological constant (in the 
case where $\omega(\phi)$ is a constant we shall denote it by 
$\omega_0$).  Variation of the 
corresponding 
action for (\ref{eq:Lagrangian_scalarfield_jordan}) with respect to the metric 
tensor leads to the following equations:
\begin{equation}
\phi\,G_{\mu\nu} + \left[ \square\phi +\frac{1}{2}\frac{\omega}{\phi} 
(\nabla\phi)^{2}+\Lambda \right]g_{\mu\nu} - \nabla_{\mu}\nabla_{\nu}\phi 
-\frac{\omega}{\phi}\nabla_{\mu}\phi\nabla_{\nu}\phi=\kappa^2
T^{(m)}_{\mu\nu}\,,
\end{equation}
where, as usual, $T^{(m)}_{\mu\nu}$ holds for the usual stress-energy matter 
tensor
\begin{equation}\label{T_munu}
T^{(m)}_{\mu\nu} = \frac{-2}{\sqrt{-g}}\frac{\delta}{\delta g^{\mu\nu}} 
\left(\sqrt{-g}\mathcal{L}_{m}\right)\, .
\end{equation}
As mentioned above, within the Jordan frame, the scalar field is also an  
independent dynamical quantity, so variation of the action with respect to 
$\phi$ yields the additional field equation \label{Jordanrref2}
\begin{equation}\label{eq:EOM_scalarfield}
(2\omega+3)\square\phi +\omega'(\nabla\phi)^{2}+4\Lambda-2\phi\Lambda'=\kappa^2 
T^{(m)}\,,
\end{equation}

\noindent where the prime indicates derivatives with respect to the scalar 
field $\phi$ and $T^{(m)}$ is just the trace of $T^{(m)}_{\mu\nu}$. 
Scalar-tensor theories are known to be conformally equivalent to GR. Indeed, 
under conformal transformations, we may find a metric tensor obeying a set of 
field equations analogous to the usual Einstein equations. In such equations, 
the scalar degree of freedom would source a resulting matter field. However, 
the 
latter would not couple to the geodesics like ordinary matter. 

\subsection{Brans-Dicke Theory}

Brans-Dicke (BD) theories were first studied in  an attempt to mathematically 
include Mach's principle in a theory consistent with GR. In BD theories, the 
gravitational coupling stops being constant and takes the form of a scalar 
field. Such a scalar field is in principle position-dependent and would feel 
the distance-scaled effects of all the matter contained in the Universe 
\cite{Brans:1961sx, Faraoni:2004pi}. A renewed interest in this theory emerged 
thanks to the discovery that string theories contain a low-energy limit given 
by a BD-type Lagrangian \cite{Blaschke:2004wa}. Moreover, further interest was 
fuelled by the fact that $f(R)$ theories, which, thanks to the chameleon 
mechanism \cite{Khoury:2003rn}, are once again contenders as viable candidates 
for weak field gravity, are dynamically equivalent to BD theories.

The action for BD theories in the Jordan frame, may  be obtained from the 
Lagrangian in (\ref{eq:Lagrangian_scalarfield_jordan}), just by considering 
$\omega\rightarrow\text{constant}$ provided $\Lambda\rightarrow 0$. Thus 
(\ref{eq:Lagrangian_scalarfield_jordan}) becomes\label{Bransref1}
\begin{equation}
S_{\rm BD} = \frac{1}{2\kappa^2} \int {d}^{4}x \sqrt{-g}\left[ \phi R - 
\frac{\omega}{\phi}g^{\mu\nu}\nabla_{\mu}\phi\nabla_{\nu}\phi-V(\phi) \right] + 
S_{m}.
\label{SB_action}
\end{equation}
\noindent Here, $S_{m}$ is the action of any matter fields present, and 
$\omega$ 
characterises the dimensionless BD parameter. As seen above, $\phi$ couples 
directly to the scalar curvature,  but not to the matter fields. The potential 
$V(\phi)$ above, although not constant in general, can be identified as a 
general form of the {\it cosmological constant}. Subsequently, the scalar field 
$\phi$ can be thought of as possessing an effective mass, defined by 
derivatives 
of $V$ with respect to $\phi$, as follows:
\begin{equation}
m^{2} = \frac{1}{2\omega + 3} \left( \phi\frac{{d}^{2}V}{{
d}\phi^{2}}-\frac{{d}V}{{d}\phi} \right).
\end{equation}
Variation of action (\ref{SB_action}) with respect to the metric yields the 
following  field equations: 
\begin{equation} \label{BD_EFE}
G_{\mu\nu}=\frac{\kappa^2}{\phi}T^{(m)}_{\mu\nu} + 
\frac{\omega}{\phi^{2}}\left( \nabla_{\mu}\phi\nabla_{\nu}\phi - 
\frac{1}{2}g_{\mu\nu}\nabla^{\alpha}\phi\nabla_{\alpha}\phi \right) + 
\frac{1}{\phi}\left( \nabla_{\mu}\nabla_{\nu}\phi - g_{\mu\nu}\square\phi 
\right) - \frac{V}{2\phi}g_{\mu\nu},
\end{equation}

As could be concluded from (\ref{BD_EFE}), in this theory, the gravitational 
coupling could be  identified as the inverse of the scalar field, $G_{eff} = 
\frac{G}{\phi}$, or equivalently  $\kappa^2_{eff} = 
\frac{\kappa^2}{\phi}$,
which obviously becomes a function of spacetime.  In order to obtain a positive 
gravitational coupling and to guarantee the attractive character of gravity, we 
require that $\phi>0$. If, then, variations of the action (\ref{BD_EFE}) with 
respect to $\phi$ are performed, the equation of motion for the scalar field,
\begin{equation}\label{BD_Eq_field}
\frac{2\omega}{\phi}\square\phi + R -  
\frac{\omega}{\phi^{2}}\nabla^{\alpha}\phi\nabla_{\alpha}\phi - \frac{{
d}V}{{d}\phi} = 0\,,
\end{equation}
is obtained. Using the trace of the field equations (\ref{BD_EFE}), we may 
eliminate $R$ in  (\ref{BD_Eq_field}), resulting in the dynamical equation for 
the BD scalar field, $\phi:$
\begin{equation}\label{trace-BD}
\square\phi = \frac{1}{2\omega+3}\left[\kappa^2 T^{(m)} + \phi\frac{{
d}V}{{d}\phi}-2V \right].
\end{equation}

With the pertinent redefinitions, this dynamical equation may also be  derived 
from (\ref{eq:EOM_scalarfield}). The BD parameter remains as a free parameter 
of the theory. It turns out that values of $\omega=1$ are consistent with 
results in string theory, but in order to satisfy local (Solar System) tests of 
 \label{solarsystemref2}
gravity, $\omega$ is preferred to be large. In fact, the larger $\omega$ is, 
the 
more closely BD theory resembles GR. The latest, most stringent constraints on 
   $\omega$ value come from the Cassini probe, and was found to satisfy 
$\omega>40000$, and it is this fine tuning of $\omega$ that renders BD 
theories unattractive. However, as shall be discussed later, this fine tuning 
may be avoided altogether, if the mass of the scalar field is large 
\cite{Perivolaropoulos:2009ak,Hohmann:2013rba}.


\section{Introduction to $f(R)$ Gravity}
\label{Sec:3-fR-theories}
Our choice to focus on $f(R)$ theories follows intensive research related  to 
cosmological applications of modified gravity theories, and the need for a 
wider 
understanding of the consequences and limitations of this class of theories. 
Such theories have been the subject of extraordinary attention over the last 
few decades, and this interest gave birth to a plethora of investigations into 
somewhat simple $f(R)$ models realisations, the cosmological scenarios that 
they govern, as well as some understanding into the theoretical caveats 
associated with these theories \cite{Faraoni:2004pi, Capozziello:2005mj, 
Sotiriou:2008rp, Sotiriou:2006hs, Sotiriou:2006sf}. 
As mentioned above, $f(R)$ models were originally considered as a means to 
facilitate  the early-time inflationary epoch required in the Big Bang Theory 
\cite{Starobinsky:1980te}. More recently, such models proved to be able to 
provide a mechanism to generating a late-time accelerated epoch in the 
expansion history of the Universe. 
This theory gives rise to a gravitational field propagated by a spin-2 massless 
graviton,  as well as a massive scalar degree of freedom that couples to 
density, resulting in an effective long-range \emph{fifth force}, which can  
\label{Fifthref1}
indeed facilitate the late-time acceleration observed.

Indeed, these theories result from a straightforward modification of the 
Einstein-Hilbert action,  which has been claimed to be its simplest general 
stable modification  that can be made \cite{Stelle:1976gc, 
Utiyama:1962sn}. In particular, the Ricci scalar $R$ is replaced with a general 
function of the Ricci scalar, namely:
\begin{equation}\label{eq:f(R)_action1}
S\,=\,\frac{1}{2\kappa^2}\int f(R)  \sqrt{-g} \,{d}^{4}x.
\end{equation}

The renewed interest also brought to light the scalar-tensor theory equivalence 
 with $f(R)$ gravity, and that such an action (\ref{eq:f(R)_action1}) was also 
able to produce the accelerated expansion, which was being sought at the time, 
{\it c.\,f.} \cite{Carroll:2003wy}. Early studies were pessimistic about $f(R)$ 
gravity, arguing that these theories are automatically ruled out since Solar 
System constraints required that the BD parameter be greater than 40000 
\cite{Bertotti:2003rm}, as mentioned above, while other results concluded that 
$f(R)\propto R^n$, i.e., power laws of $R$, could not produce standard 
cosmological evolution \cite{Amendola:2006kh, Amendola:2006eh},  for both large 
and small values of the Ricci scalar. Moreover, it was claimed that $f(R)$ 
theories did not contain a phase of matter-like expansion 
($t^{2/3}$, with $t$ cosmic time), when it was found that typically a 
radiation-like expansion  era ($t^{1/2}$) preceded a phase of dark energy 
expansion 
\cite{Amendola:2006we}. Since then, the understanding of the dynamics of these 
models and the various general restrictions that can be placed has developed 
significantly, and several models have been proposed that are able to sustain 
a period of matter-like expansion (see for example \cite{Hu:2007nk, 
Nojiri:2006gh, 
Nojiri:2006be, Evans:2007ch}) followed by an acceleration. In what follows, we 
shall provide the field equations, discuss the FLRW solutions corresponding to 
viable theories, as well as what exactly makes a theory viable in the first 
place.

Three remarks are pertinent at this stage. 
First, the generality $f(R)$ theories grant is fairly wide, and their resulting 
field equations  yield rich and interesting phenomenologies, providing ways to 
reproduce whatever dynamics one desires ({\it c.f.} 
\cite{delaCruzDombriz:2006fj,Dunsby:2010wg} and subsequent efforts), in 
particular that of $\Lambda$CDM in a two-fluid dominated scenario. 
Second, $f(R)$ theories are also arguably the most straightforward modification 
to GR  that can be made, meaning their solutions are tractable and their exact 
solutions can give insight into this class of theories, as well as GR.  Various 
viability considerations, which must be satisfied by viable $f(R)$ models, will 
be discussed in Section \ref{SectionViabilityConditions}.
Finally, the most attractive advantage is that   $f(R)$ theories avoid the 
Ostrogradsky  instability, from which many other higher-order modifications to 
the gravitational Lagrangian suffer.

The generalised form of the Lagrangian included in Eq. 
(\ref{eq:f(R)_action1}) supersedes  the addition of even higher-order quadratic 
curvature invariants such as 
$R^{\mu\nu}R_{\mu\nu}$, $R^{\mu\nu\alpha\beta}R_{\mu\nu\alpha\beta}$ or  
$\varepsilon^{\mu\nu\sigma\gamma}R_{\mu\nu\alpha\beta} 
R^{\alpha\beta}{}_{\sigma\gamma}
$, which in fact will reduce to a function of $R$, $f(R)$, when considering a 
maximally symmetric,  four-dimensional spacetime (a transparent proof of this 
argument may be found in \cite{DeWitt:1965jb}, p134).

\subsection{$f(R)$ Formalisms}
\label{fRref1}
The governing field equations for a general $f(R)$ theory may be derived by 
varying  the action (\ref{eq:f(R)_action1}) in three different ways, which 
result in three different formalisms within the $f(R)$ framework. In particular,

\begin{enumerate}

\item  The so-called \textit{metric formalism} considers the only independent  
variable to be the metric itself, and the action,
\begin{equation}\label{fmetric_action}
S_{met} = \frac{1}{2\kappa^2} \int {d}^{4}x \sqrt{-g} f(R) + 
S_{m}(g_{\mu\nu},\psi)
\end{equation}
\noindent is varied with respect to the metric alone, yielding the following 
field  equations:
\begin{equation}\label{fR_eqs}
f'(R)R_{\mu\nu} - \frac{1}{2}f(R)g_{\mu\nu} - \left( 
\nabla_{\mu}\nabla_{\nu}-g_{\mu\nu}\square\right)f'(R) = \kappa^2 
T^{(m)}_{\mu\nu}.
\end{equation}
\noindent These equations are fourth-order in the metric tensor, as per the 
usual definition of the Ricci scalar. Matter fluids are described by 
$T^{(m)}_{\mu\nu}$ as defined in (\ref{T_munu}).
Here, $\nabla_{\mu}$ is the covariant derivative corresponding to the usual 
Levi-Civita  connection of the metric, $\square=\nabla^{\mu}\nabla_{\mu}$, and 
primes denote differentiation with respect to  the Ricci scalar $R$.  \\

\item In what is known as the \textit{Palatini formalism}, the action from 
which we begin  (\ref{fmetric_action}) remains the same, however the 
connection 
is assumed as being an independent quantity from the metric. Thus the action 
must then be varied with respect to both quantities 
independently:\label{Palatiniformref2}
\begin{equation}\label{fpal_action}
S_{Pal} = \frac{1}{2\kappa^2} \int {d}^{4}x \sqrt{-g} f(\mathcal{R}) + 
S_{m}(g_{\mu\nu},\psi),
\end{equation}

\noindent where the Riemann tensor and the Ricci tensor are constructed with 
the independent  connection, and, in this section, we denote the difference 
between the metric Ricci scalar, $R$, and the Ricci scalar constructed with 
this independent connection as $\mathcal{R}=g^{\mu\nu}\mathcal{R}_{\mu\nu}$.  
Hence, varying (\ref{fpal_action}) with respect to both the metric and the 
connection yields \cite{Olmo:2011uz}
\begin{align}
& f'(\mathcal{R}) \mathcal{R}_{(\mu\nu)} - 
\frac{1}{2}f(\mathcal{R})g_{\mu\nu}=\kappa^2 T_{\mu\nu}^{(m)},\\
& \bar{\nabla}_{\gamma}\left(\sqrt{-g} f'(\mathcal{R}) g^{\mu\nu}\right) -  
\bar{\nabla}_{\sigma}\left( 
\sqrt{-g}f'(\mathcal{R})g^{\sigma(\mu}\right)\delta^{\nu)}_{\gamma}=0.
\end{align}
\noindent Here, $\bar{\nabla}_{\mu}$ represents the covariant derivative 
defined with  the independent connection used in the variation. 
The important point is that when $f(\mathcal{R})=\mathcal{R}$ in the Palatini  
formalism and when $f(R)=R$ in the metric formalism, both theories reduce to 
GR, and produce the same physics. However, they deviate significantly as soon 
as 
a more general function $f$ appears in the Lagrangian.

While, in the Palatini formalism, the connection is considered independent from 
 the metric, it does not appear explicitly in the Lagrangian. The Lagrangian 
density corresponding to  the matter fields is independent of the connection 
itself, and the covariant derivatives of the matter fields are defined as usual 
with the Levi-Civita connection. In fact it is possible to eliminate the 
independent connection entirely from the field equations of Palatini $f(R)$ 
gravity, since the independent connection is not actually related to the 
geometry. In this sense, the Palatini formalism is still a metric theory 
\cite{Sotiriou:2006hs}. The interested reader is referred to \cite{Olmo:2011uz} 
for a further insight.

\item A third and more general (and more complicated as well) approach to the  
variation of the $f(R)$ action is known as \textit{metric affine} $f(R)$. 
Herein, the geometric properties of the independent connection manifest in its 
coupling to the matter fields present in the theory. In this way, both of the 
aforementioned approaches can be generalised by taking both the metric and the 
connection to be independent variables, and also allowing the matter to depend 
explicitly on the connection. 
The interested reader is referred to \cite{Sotiriou:2006qn} for a further 
insight.

\end{enumerate}

In the following we shall choose the metric approach as the underlying 
formalism of our chapter. For the Palatini and metric-affine approaches, 
interested readers may consult the aforementioned references, 
Thus, retaking (\ref{fR_eqs}), these field equations can be rearranged to 
resemble  the Einstein field equations with an additional source term, 
comprising the higher order curvature contributions:
\begin{equation}
R_{\mu\nu}-\frac{1}{2}g_{\mu\nu}R \equiv 
G_{\mu\nu}=T^{(R)}_{\mu\nu}+\frac{1}{f'(R)}T^{(m)}_{\mu\nu},
\end{equation}
with 
\begin{equation}\label{TR}
T^{(R)}_{\mu\nu}\,=\,\frac{1}{f'(R)}\left[ \left(\nabla_{\mu}\nabla_{\nu}-  
g_{\mu\nu}\square \right)f'(R) + \frac{1}{2}g_{\mu\nu}\left(f(R) - 
f'(R)R\right) \right] 
\end{equation}
being   an effective stress-energy tensor, which behaves as a source for the 
resulting geometry.  This new source term is usually dubbed the \emph{curvature 
fluid},  $T^{(R)}_{\mu\nu}$. Definitely, this idea is not to be taken 
literally, 
and we should be aware that actually the resulting theory is completely 
different from GR. However, intuitively, it lends a useful perspective; the 
vacuum itself contains a source that is generated by the geometry, in an 
otherwise Einstein-like field. Obviously, when the stress energy tensor for the 
curvature fluid is exactly zero, we recover GR. Herein, an effective 
gravitational coupling can then be defined as $G_{eff}\equiv G/f'(R)$. The 
coupling is required to be positive, in analogy to requiring that the graviton 
is not a ghost in scalar-tensor gravity, and this amounts to the following 
condition on the form of $f(R)$: 
\begin{equation}\label{noghostcondition}
f'(R) > 0,
\end{equation}
\noindent which we touch on again later.
The trace of the field equations at 
(\ref{fR_eqs}) is given by 
\begin{equation}\label{ftrace}
f'(R)R+3\square f'(R)-2f(R)\,=\,\kappa^2 T^{(m)},
\end{equation}
\noindent $T^{(m)}=g^{\mu\nu}T^{(m)}_{\mu\nu}$. The trace equation 
(\ref{ftrace}) 
makes   the distinction between $f(R)$ gravity and GR  obvious, where for the 
latter $R=-\kappa^2 T^{(m)}$. Thus, in the $f(R)$ gravity case, a vanishing of 
ordinary matter sources does not imply a vanishing of Ricci curvature, hinting 
at the importance of the additional field, $f'(R)$, and the role it plays in 
the 
resultant physics of the given universe. In fact, we can interpret the trace 
equation as being an equation of motion for this emergent scalar degree of 
freedom, $f_{R}$, which is sometimes called the scalaron.
For maximally symmetric solutions of the above equation, where we have surfaces 
of  constant curvature ($R=const$), considering the vacuum scenario, i.e., 
$T^{(m)}_{\mu\nu} = 0$, Eq. (\ref{ftrace}) reduces to
\begin{equation}\label{MaxSymTrace}
f'(R)R - 2f(R) = 0,
\end{equation}
which is just an algebraic expression for $R$. When $R=0$ is a root of the 
above equation,  then the field equations at (\ref{fR_eqs}) become 
$R_{\mu\nu}=0$, corresponding to isotropic, homogeneous and flat Minkowski 
spacetime. When $R= const$ is taken to be a root of (\ref{MaxSymTrace}), the 
field equations (\ref{fR_eqs}) become $R_{\mu\nu} \propto g_{\mu\nu}$. This 
root gives an interesting and desirable result, as it represents exponential 
expansion of space-time, similar to the behaviour of a GR universe plus a 
cosmological constant.

\subsection{$f(R)$ Gravity From a Scalar-Tensor Perspective}

Performing coordinate transformations, normalisations or variable redefinitions 
 are standard  techniques in Classical Mechanics. Such techniques usually 
improve 
  the convenience of computations and the interpretation of results. Finding 
equivalence between theories is a tool that comes in handy, especially when a 
novel theory is being investigated. For the sake of definiteness, we can say 
that two theories are thought of as equivalent if after an appropriate 
redefinition, or transformation, or renormalisation, either their field 
equations or the action from which these are derived, are identical. In this 
case, the dynamics of the equivalent theories are indistinguishable, and 
previous progress in understanding   one theory will shed light on a theory 
that is found to be a dynamically equivalent and alternative representation to 
the former.

It is well known that it is possible to recast quadratic modifications to GR in 
the form of a BD theory, the  latter being a subclass of scalar-tensor 
theories. This equivalence is easily extended to the special case of $f(R)$ 
gravity \cite{Sotiriou:2006hs,Chiba:2003ir}. While metric $f(R)$ gravity 
explicitly seems not to include extra fields in the action, and the Palatini 
version, though containing an independent connection, is also fundamentally a 
metric theory, they both may be represented by various forms of a BD theory. 
Below, the equivalence is discussed  in terms of the equivalence of the 
actions.

Beginning with the action for metric $f(R)$ gravity, as given by 
(\ref{eq:f(R)_action1}), we introduce  an auxiliary field, $\chi$, to the 
action, which becomes,
\begin{equation}
S_{met} = \frac{1}{2\kappa^2}\int {d}^{4}x \sqrt{-g}\left[ f(\chi) + 
f'(\chi)(R-\chi) \right] + S_{m}(g_{\mu\nu},\psi),
\end{equation}
\noindent where $f(R)=f(\chi)+f'(\chi)(R-\chi)$ is an expansion around the 
Ricci scalar. Varying with  respect to $\chi$ reveals the following equation:
\begin{equation}
f''(\chi)(R-\chi)=0.
\end{equation}

\noindent Clearly, under the condition that $f''(\chi) \neq 0$, we obtain that 
$R=\chi$, which  of course reproduces the original action. If we identify 
$f'(\chi)$ as a scalar degree of freedom, and set $f'(\chi)=\phi$ to be a 
scalar field, we can then define its effective potential:
\begin{equation}
V(\phi)=\chi(\phi)\phi - f(\chi(\phi)).
\end{equation}

\noindent This allows us to rewrite the action (\ref{eq:f(R)_action1}) in 
terms of the scalar  field $\phi$ and its potential as follows:
\begin{equation}\label{f-BD-action}
S_{met}=\frac{1}{2\kappa^2}\int {d}^{4}x\sqrt{-g} \left[ \phi R - V(\phi) 
\right] + S_{m}(g_{\mu\nu},\psi),
\end{equation}
\noindent which is simply the action of a BD theory in the Jordan frame, with 
the BD parameter  $\omega_{0}=0$, known as {\it massive dilaton gravity} 
\cite{OHanlon:1972sdp}. Note that this scalar field $\phi$ is, unlike a matter 
field, able to violate all the energy conditions \cite{Faraoni:2004pi} (the 
energy conditions 
will be specified in subection (\ref{Sec_Fluid sources}) below). The field 
equations that result from metric variation 
of 
(\ref{f-BD-action}) are
\begin{align}
G_{\mu\nu} &= \frac{\kappa^2}{\phi}T_{\mu\nu}^{(m)} - 
\frac{V(\phi)}{2\phi}g_{\mu\nu} 
+ \frac{1}{\phi}(\nabla_{\mu}\nabla_{\nu}\phi - 
g_{\mu\nu}\square\phi)\,,\label{mfe_BD1}\\
R & = V'(\phi)\,. \label{mfe_BD2}
\end{align}
where in order to guarantee that an $f(R)$ theory is dynamically equivalent  to 
a scalar-tensor theory, the condition $f''\neq 0$ must be satisfied. Following 
a 
similar approach as before for the Palatini formalism of $f(R)$ gravity reveals 
that   the Palatini $f(R)$ theory is dynamically equivalent to 
the BD theory, for which $\omega_{0}=-\frac{3}{2}$  \cite{Sotiriou:2006hs, 
Teyssandier:1983zz,Wands:1993uu}, as introduced in 
(\ref{eq:Lagrangian_scalarfield_jordan}).

As mentioned above,  the trace of (\ref{mfe_BD1}) may be used to  eliminate 
$R$, resulting in Eq. (\ref{trace-BD}), which provides an equation of motion 
for 
the scalar field in terms of its potential in the presence of matter.
%
Such an equation highlights the difference between metric and  Palatini $f(R)$ 
theories, and their induced dynamics. It is clear to see that 
in the 
Palatini formalism the emergent scalar field, or scalaron, is not a dynamic 
variable, whereas in metric $f(R)$ theory the derivatives of the scalar field 
are non-zero, therefore resulting in an extra scalar degree of freedom 
\cite{Sotiriou:2006hs}.

\subsection{Viability}
\label{SectionViabilityConditions}

In this section we shall give a flavour of the required viability constraints 
on the  general $f(R)$ Lagrangians.
Given the apparent freedom on the form of the function $f$, it has become 
common practice to construct a theory that produces the desired results in the 
required energy level or time scale. It has been an interesting exercise for 
many years, given the $f(R)$ cosmological field equations, to find forms for 
$f$ that are consistent with data ({\it c.\,f.} \cite{Hu:2007nk, 
Lazkoz:2018aqk, Basilakos:2017rgc, Jaime:2012nj, Miranda:2009rs}). Even the 
reverse has been done, where given sets of data, a function, which best suits 
the data, has been reconstructed \cite{Dunsby:2010wg, delaCruzDombriz:2006fj}. 
Below we discuss the various considerations, which must be made prior to an 
$f(R)$ theory being accepted as a potentially viable candidate for the 
underlying theory of gravity.

\subsubsection{Cosmological dynamics}

It is not difficult to find a function, which, in principle, is consistent with 
the cosmological expansion history observations, in that it can produce   
late-time acceleration. The point of interest is making sure that other 
gravitational aspects and other predictions in the cosmological evolution are 
also respected, especially Big Bang Nucleosynthesis, the temperature 
anisotropies of the CMB and the expected growth of large-scale structures. 
Moreover, the theory must provide an inflationary period, which can solve the 
horizon, flatness and monopole problems.  After this a radiation domination 
phase of the Universe, leading to a matter domination phase, must be present.
Finally, the Universe should evolve towards a stable de Sitter type expansion, 
such that the 
theory agrees with observations. Note that the
transitions between the various phases must be smooth.

\subsubsection{Correct weak-field limit}

It took some time before consistent results regarding the weak-field limit in  
$f(R)$ theories of gravity were derived and understood. In 2003, Chiba 
\cite{Chiba:2006jp} concluded that, by virtue of the fact that observations 
required BD theories with $\omega_{0}\geq 40000$, $f(R)$ theories must be ruled 
out, having $\omega_{0}=0$. However, the Parametrised Post Newtonian slip 
parameter, $\gamma = - \Psi/\Phi$, defined by the ratio of the Newtonian 
potentials, is determined by both the mass of the scalar field in these 
theories, as well as the BD parameter, $\omega_{0}$. When the mass is small, 
constraints on $\omega_{0}$ are equivalent to constraints on $\gamma$. However, 
provided the mass of the scalar field is endowed with the so-called 
\textit{Chameleon mechanism},  \label{chameleonkiref2} and therefore is able to 
acquire a large mass 
locally depending on the environment, then in this regime the mass of the 
scalar field will dominate over $\omega_{0}$, allowing a select class of $f(R)$ 
theories to survive \cite{Khoury:2003rn, Khoury:2003aq}.  Once the scalar is 
massive and its range short, it will effectively be invisible to experiments 
performed within both the Newtonian and post Newtonian limits. 

Regarding the existence of a Newtonian limit, it was shown in 
\cite{Sokolowski:2008kf} that the existence of a stable Newtonian limit may be 
ascertained by the existence of a stable ground state of the theory, whether 
the ground state solution is Minkowski, de Sitter or anti-de Sitter. In 
order for a function to provide a theory with a stable Newtonian limit,  i.e. 
in the regime for $R$ where Newtonian gravity can be applied, compact objects, 
low velocities and relevant curvatures (curvatures much larger than the 
present background value in an FRLW universe,  but smaller than those interior 
to very compact bodies, e.g., neutrons stars or black holes), the following 
conditions on $f(R)$ must be met \cite{Appleby:2009uf}:
\begin{align}
& |f(R)-R|\ll R, \label{VC_Newt_1}\\ 
& |f_R(R)-1| \ll 1, \label{VC_Newt_2}\\
& Rf_{RR}(R)\ll 1 , \label{VC_Newt_3}
\end{align}
\noindent for $R \gg R(today)$, which guarantee that any deviations from GR to  
the metric are kept small. Additionally, the Compton wavelength of the scalar 
field is 
much smaller than the radius of curvature of the background 
\cite{Appleby:2009uf}. These are essential considerations regarding the form of 
$f$, and failing the weak-field limit renders a theory worthless.

 \subsubsection{Instabilities}\label{Section:Instabilities}
 
 An important issue that usually constitutes a problem for most of the
higher-order theories  of gravity is the appearance of ghost fields. These are 
massive states of negative norm and, in this context, result in the Hamiltonian 
for a given theory to be unbounded from below. 
 
  \begin{center}
      \textit{Classical and quantum stability}
  \end{center}

 If the modified Lagrangian includes  higher-order curvature invariants and 
derivatives, it has been shown that an additional spin-2 ghost field appears 
\cite{Stelle:1976gc, Utiyama:1962sn}, resulting in issues with the quantum 
stability of the theory \cite{Cline:2003gs}, as well as on a classical level.
To ensure the classical and quantum  stability of a theory in the physically 
relevant domain, i.e., $R>0$, the following requirements on the derivatives of 
$f$ are crucial: 
 \begin{align}
     &f_R(R) >0\,, \label{VC_firstDerivative}\\
     &f_{RR}(R)>0\,. \label{VC_secondDerivative}
 \end{align}
 \noindent Indeed, the condition  (\ref{VC_firstDerivative}) guarantees that 
gravity is attractive and that the effective gravitational constant is 
positive. It also ensures that the graviton is not a ghost 
\cite{Appleby:2009uf}, and a violation of this requirement has been shown to 
result in the loss of homogeneity and isotropy in regular FLRW models, and the 
formation of a strong space-like anisotropy curvature singularity 
\cite{Nariai:1973eg, Gurovich:1979xg}. 
 
The requirement (\ref{VC_secondDerivative})  conveys the avoidance of the 
so-called  Dolgov-Kawasaki instability \cite{Dolgov:2003px}, which may occur in 
very short time scales. Furthermore, related to the latter, a weak sudden 
singularity can also arise if $f_{RR}(R)=0$ for a finite value of the scalar 
curvature. Both problems are discussed below.\\

 \begin{center}
\textit{The Dolgov-Kawasaki Instability}\\
\end{center}
\noindent To summarise its importance, following \cite{Dolgov:2003px}, let us  
parametrise $f(R) = R + \epsilon g(R)$, such that $\epsilon$ is a small 
positive constant, containing the dimensions of mass squared, leaving the 
function $g(R)$ dimensionless. Substituting this into the $f(R)$ trace equation 
(\ref{ftrace}), we obtain
\begin{equation}
 \square R +\frac{g'''}{g''}\nabla^{\mu}R\nabla_{\mu}R + \left(\frac{\epsilon 
g' 
- 1}{3\epsilon g''}\right) R = \frac{\kappa^2 T^{(m)}}{3\epsilon g''} + \frac{2 
g}{3g''},
 \end{equation}   
 assuming $g''\neq0$ (to lighten the notation, here primes   denote  
derivatives with respect to $R$).
Now, we may approximate the local metric as an  expansion around the Minkowski  
metric $\eta$,
\begin{equation} 
g_{\mu\nu} = \eta_{\mu\nu} + h_{\mu\nu},
\end{equation}
\noindent in a weak-field region, where we can expand the scalar curvature $R$ 
as
\begin{equation}
R = -\kappa^2 T^{(m)} + R_{1},
\end{equation}
\noindent where $R_{1}$ is a small perturbation around the GR approximation.   
We may then consider the trace equation to study the dynamics of  $R_{1}$, 
which to a first order gives:

\begin{eqnarray}
&&
\!\!\!\!\!\!\!\!\!\!\!
\ddot{R_{1}} - \nabla^{2}R_{1} - \frac{2\kappa^2 
g'''}{g''}\dot{T}^{(m)}\dot{R}_{1} + 
\frac{2\kappa^2 g'''}{g''}\vec{\nabla}T^{(m)} \cdot \vec{\nabla}R_{1} + 
\frac{1}{3g''}\left( \frac{1}{\epsilon}-g' \right)R_{1}\nonumber\\
&&
\ \ \ \ \ \ \ \ \ \ 
= 
\kappa^2\ddot{T}^{(m)} - 
\kappa^2\nabla^{2}T^{(m)} -\frac{[\kappa^2 T^{(m)}g^{2}+2g]}{3g''}\,.
\end{eqnarray}
The coefficient of the last term on the left-hand side is effectively the 
square of  the mass of $R_{1}$ 
\begin{equation}
m^{2} \simeq \frac{1}{3\epsilon g''},
\end{equation}
and is dominated by the term $(\frac{1}{3\epsilon g''})$, since the value of 
$\epsilon$  is very small \cite{Dolgov:2003px, Faraoni:2006sy}.
Consequently, the introduced perturbation $R_1$ will remain under control  and 
the corresponding theory will be stable, provided $g''>0$ is satisfied, and 
unstable when the sign of the effective mass square is negative 
\cite{Faraoni:2006sy}. Thus, in order to protect against the Dolgov-Kawasaki 
instability, we require that $f_{RR}(R) > 0$. To include the GR limiting case, 
we require that $f''(R) \geq 0.$  Interestingly, because the scalar field in 
the Palatini formalism is non-dynamical, it does not suffer any Dolgov-Kawasaki 
instability. Of course, the above considerations only hold in a small 
neighbourhood confined to local expansion. However, we may derive a condition 
for the stability of the de Sitter space in $f(R)$ theories of gravity, by 
assuming a de Sitter background, and considering a general action that 
includes both $f(R)$ gravity and scalar-tensor gravity, and mixtures of the two 
\cite{Sotiriou:2006sf}. Considering, as we shall do in general in this work, an 
FLRW metric to describe the background, the following condition  for the 
existence of a stable de Sitter space solution is obtained:
\begin{equation} \label{cond_deSitter}
\frac{(f^0_{R})^{2}-2f^0f^0_{RR}}{f^0_{R}f^0_{RR}} \geq 0,
\end{equation}
where the zero super index denotes the functions to be evaluated at  the de 
Sitter solution $R_0$. Indeed, (\ref{cond_deSitter})
is consistent with the stability condition for homogeneous perturbations  
\cite{Sotiriou:2006hs, Faraoni:2007yn, Faraoni:2008mf}.

\begin{center}
   \textit{Sudden Singularities}\\
\end{center}

There exists another instability issue, which plagues $f(R)$ gravity even at a 
background level, due to non-linearities. Many, if not all, of the functions 
that are constructed to effectively produce GR dynamics in the high curvature 
regime result in scalar fields for which the potentials $V(\phi)$ contain an 
unprotected singularity \cite{Frolov:2008uf}.  

Some of the first considerations of such singularities, in $f(R)$ gravity,  
were those that occur as density increases inside compact objects. Such 
singularities were found to be curable by adding UV corrections to the action, 
specifically of the form $R^{2}$ \cite{Appleby:2009uf}, which are in fact the 
exact forms for $f$ that were considered as inflationary theories.

Furthermore, it was discovered that oscillations in the Ricci scalar in $f(R)$ 
theories evolving with  increasing redshift, or equivalently going back in 
time, are common in all viable theories of gravity \cite{Appleby:2009uf}. The 
oscillations, occurring about the GR limit, were found to increase in frequency 
and amplitude, and eventually result in a singularity. Interestingly, in the 
high-energy regime when considering an inflationary period driven by $f(R)$ 
theory, such oscillations are actually useful in driving gravitational particle 
production, enabling phases of reheating, the creation of ordinary matter and 
the transition to the radiation-dominated FLRW stage \cite{Starobinsky:1980te}. 
However, in the classical context these oscillations may become problematic 
since

\begin{enumerate}
    \item  The frequency\footnote{The frequency of oscillations of $R$ would 
correspond to the  rest mass of the scalarons.} grows rapidly with redshift to 
exceed the Planck value, rendering the classical description no longer valid 
\cite{Appleby:2007vb, Starobinsky:2007hu, Tsujikawa:2007xu}. 
    
    \item The amplitude of the linear oscillations also grows quickly with 
redshift, resulting in  an over-abundance of scalarons at early times, while we 
require its number density to be appropriately small during BBN 
\cite{Starobinsky:2007hu}.
    
\end{enumerate}

Perturbative approaches such that the Ricci scalar is defined by its value in 
GR plus a perturbation, $R=R_{GR}+\delta R$, were able to isolate an analytic 
expression for the oscillating part of the Ricci scalar 
\cite{Starobinsky:2007hu}. It was found that the oscillations become asymmetric 
and eventually evolve towards a singularity, and that an analysis neglecting 
nonlinear effects will be blind to pathological behaviour.  

These singularities have been widely considered in the scalar tensor framework, 
where  the scalar field $f_R(R)$ is identified and examined using the trace of 
the field equations (\ref{modtrace}), as a damped harmonic oscillator 
\cite{Frolov:2008uf}, 
 
\begin{equation}\label{eq_mod_trace_chap3}
\square f_{R} = \frac{1}{3}(2f-f_{R}R)+\frac{\kappa^2}{3}T^{(m)},
\end{equation}
\noindent where $f_{R}$ is the first derivative of $f(R)$ with respect to $R$, 
and thus, the  additional degree of freedom, the scalar field $\phi$, which 
herein for convenience we define as
\begin{equation}
\phi= f_{R}-1.
\end{equation}

\noindent Considering the above, Equation (\ref{eq_mod_trace_chap3}) in terms 
of  $\phi$ is
\begin{equation}
\square \phi = V'(\phi) - \mathcal{F},
\end{equation} 

\noindent where the effective scalar field potential can be determined by

\begin{equation} \label{eq_potential_phi}
V'(\phi) = \frac{{d}V}{{d}\phi}=\frac{1}{3}(2f - f_{R}R),
\end{equation}
\noindent and the force $\mathcal{F}$ driving the scalar field $\phi$ is the 
trace of the stress-energy tensor $T^{(m)}$. For a perfect fluid with 
energy density $\rho$ and pressure $p$, this is
\begin{equation}
\mathcal{F} = \frac{\kappa^2}{3}(\rho - 3p).
\end{equation}
The dynamics of the scalar field is determined by the field's potential, which 
is 
obtained by  integrating the following equation
\begin{equation}\label{eq:dVdR}
\frac{{d}V}{{d}R}=\frac{{d}V}{{d}\phi}\frac{{d}\phi}{{ d}R}  
=\frac{1}{3}(2f-f_{R}R)f_{RR}.
\end{equation}
The minimum of this potential, which we will identify as being located at 
$\phi_{min}$, is the point that corresponds to the  de Sitter solution. There 
also exists, in most, if not all \cite{Frolov:2008uf} viable cosmological 
$f(R)$ theories, a point $\phi_{sing}$, where the Ricci scalar diverges to 
infinity, resulting in a curvature singularity. It has been shown that these 
two 
points  are often easily separated by a finite value of the potential, thus it 
is possible (and usually highly likely) that the scalar field, in its 
oscillation about the potential minimum, may verge on the point leading to 
singularity. Theories that do not protect against this type of eventuality 
would be disqualified. 

This is a useful approach, and has led to an intuition in understanding the  
occurrence of these singularities, and insight into how to possibly avoid them, 
however it has also been argued that this approach is not reliable, and should 
be performed with caution \cite{Jaime:2012gc}.

\subsubsection{Cosmological Perturbations in $f(R)$ gravity}\label{LSSefs1}

Arranging a function to mimic a desirable cosmological expansion history  is 
straightforward. In principle there is an infinite number of such functions, 
and finding them is not a problem. In fact, the bigger problem is finding ways 
to discriminate between them. The study of perturbations around the chosen 
background cosmology is one of the most powerful methods to break such a 
degeneracy. Changing the underlying theory of gravity affects the way that
perturbations in the density of the matter components and curvature evolve, 
leaving an impression in the cosmic microwave background and its fluctuation 
spectrum, as well as in the large-scale structure of galaxies and galaxy 
clusters. Also, the tensor (gravitational-wave) spectrum may present 
differences with respect to the usual GR predictions. In this sense, recent 
investigations, although constraining the range of parameters of viable $f(R)$ 
models, have stressed the fact that such theories do not violate the 
gravitational waves signals as per their predicted propagation speed for the 
tensorial mode. For seminal references on this issue, refer to 
\cite{Ezquiaga:2017ekz} and \cite{Creminelli:2017sry}, as well as subsequent 
literature.   Once predictions of these first-order perturbations are at hand, 
it is required that the perturbation spectrum produced is consistent with both 
the cosmological perturbation observables and the gravitational-wave signals. 

 In $f(R)$ theory context, scalar perturbation modes\label{Scalarperfrref1} are 
affected by  the 
subsequent generalisation of the Einstein-Hilbert action. Consequences include 
a difference in the correlation between the CMB and the large-scale structure, 
as well as a decrease in the large angle anisotropy of the CMB 
\cite{Capozziello:2010zz}. The gravitational coupling in $f(R)$ gravity is 
stronger, resulting in less large-scale structure than in $\Lambda$CDM 
cosmology. Most 
studies of perturbations in the $f(R)$ gravity context are performed with 
respect to a quasi-static limit  \cite{Zhang:2005vt, Boisseau:2000pr, 
Zhang:2005vt, EspositoFarese:2000ij, Tsujikawa:2007gd, Bean:2006up}. 
While for specific functions this approximation may be valid in certain 
regimes,  it must be used with caution,
and the obtention of the full fourth-order perturbation  equations may be 
useful in order to study intermediate and Superhubble modes (see
\cite{delaCruzDombriz:2008cp} and references therein).

\subsubsection{The Initial Value Problem}

In order for any physical theory to be viable it must have the ability to  
predict the future of a system, given the details of an instance of its past 
(an 
initial vector specifying all the quantities present in the theory), including 
a description of all interactions at play in the system. Technically speaking, 
it must  have a \textit{well-formulated} and \textit{well-posed} initial value 
problem.  This issue has been widely studied in the scalar-tensor framework and 
in metric $f(R)$ gravity, and it has been discovered that for these type of 
theories it
is a well-formulated initial value problem in the presence of ``reasonable'' 
matter, and well-posed in   vacuum too \cite{Salgado:2005hx, Faraoni:2006sy}.

\section{Background Cosmology in the Metric Formulation}
\label{Section:metric_f(R)_cosmology}

In the late 1960's Ehlers, Geren and Sachs provided a compelling argument that  
if an observer measured the relic background radiation of the Universe as 
isotropic, assuming that the isotropy holds around every point in the Universe, 
then such a universe must be isotropic and homogeneous, and thus may be 
described completely as an FLRW spacetime \cite{Ehlers:1966ad}. While the 
measurements of the CMB from our observer perspective reveal striking isotropy, 
our modern ability to resolve tiny anisotropies  indicate that in fact the real 
Universe exhibits perturbations about what appears to be a ``nearly isotropic'' 
background radiation. The Ehlers-Geren-Sachs (EGS) theorem was shown to hold 
for a ``nearly'' isotropic measurement of the CMB as well 
\cite{Stoeger:1994qs}. The ``almost EGS'' theorem is also valid for $f(R)$ 
gravity, proven for both metric $f(R)$ and scalar tensor theory 
\cite{Clarkson:2001qc,Maartens:1994pb}, when the matter content is described by 
a barotropic equation of state. Thus, according to the above considerations, 
the 
usual FLRW space-time constitutes the ideal arena to study the expansion 
history 
of such theories.

Once this metric and a matter stress-energy tensor for a perfect fluid, with 
energy density $\rho$ and pressure $p$, are  
substituted in to the $f(R)$ field equations at (\ref{fR_eqs}), we obtain the 
following cosmological evolution equations, where the dot indicates derivatives 
with respect to cosmic time and $H$ is the usual Hubble parameter.
 The modified Friedmann equation is
\begin{equation}\label{modfried}
H^{2} = \frac{1}{3f_R}\left[  \rho + \frac{1}{2}(Rf_R - f)  - 3H\dot{R}f_{RR} 
\right]\,,
\end{equation}
and
the modified Raychaudhuri equation becomes
\begin{equation}\label{modraych}
2\dot{H}+3H^{2}=-\frac{1}{f_R}\left[ P + 2H\dot{R}f_{RR} +  \frac{1}{2}\left( 
f-Rf_R \right)+ \dot{R}^{2}f_{3R} + \ddot{R}f_{RR} \right]\,.
\end{equation}
  Additionally, we can also extract the trace of the field  
equations in this 
space-time, namely the modified trace equation:
\begin{equation}\label{modtrace}
3\ddot{R}f_{RR} \,=\, \rho(1-3w) + f_{R}R-2f-9Hf_{RR}\dot{R}-3f_{3R}\dot{R}^{2},
\end{equation}
where $w\equiv p/\rho$ is the equation-of-state parameter for the   perfect 
fluid under consideration (note that one may have the case where the latter can 
be the sum of 
several components, i.e. $\rho =\sum_i \rho_{(i)}$, $p=\sum_i p_{(i)}$, each 
one with its own equation-of-state parameter $w_i$).

The 
$f(R)$ cosmological field equations above are manifestly fourth-order in metric 
derivative, and since the spatial curvature has been settled to zero, we may in 
fact eliminate the scale factor altogether, in favour of the Hubble parameter 
being the only dynamical quantity, reducing the order of the equations by one.

In the spirit of the fact that we may identify all the extra terms in the field 
equations associated with the function $f(R)$ and its derivatives as an 
effective \textit{curvature fluid}, we may define its effective  equation of 
state, by using the modified Friedmann and Raychaudhuri equations to define its 
effective density and pressure as
\begin{align}
\rho^R_{\rm FLRW} &= \frac{1}{2f_R}\left( Rf_R-f - 6H\dot{R}f_{RR} \right),\\
p^R_{\rm FLRW} & = \frac{1}{f_R}\left[ \dot{R}^{2}f_{3R} + 2H\dot{R}f_{RR} +  
\ddot{R}f_{RR} + \frac{1}{2}(f-Rf_R) \right].
\end{align}

In this interpretation, in a vacuum, the curvature correction is viewed as an
effective fluid, which may be useful for gaining certain intuition, but 
certainly should not be taken too far. Even considering energy conditions for 
this effective geometric fluid would be meaningless, since it is well known 
that such fluids violate all energy conditions in general 
\cite{Faraoni:2004pi}.   The equation of state for this effective fluid may 
then 
be written as
\begin{equation}
w^R_{\rm FLRW} = \frac{p^R_{\rm FLRW}}{\rho^R_{\rm FLRW}}  = \frac{\left[ 
\dot{R}^{2}f_{3R} + 2H\dot{R}f_{RR} + \ddot{R}f_{RR} + \frac{1}{2}(f-Rf_R) 
\right]}{\frac{1}{2}\left( Rf_R-f - 6H\dot{R}f_{RR} \right)}.
\end{equation}
For our purposes, in the late-time regime, we require negative pressure to the  
end of generating accelerated expansion; and from this constraint, $w^R_{\rm 
FLRW}\approx-1$, we obtain the following relationship between the derivatives 
of the function $f$ and the Ricci scalar:
\begin{equation}
\frac{f_{3R}}{f_{RR}} \approx \frac{\dot{R}H-\ddot{R}}{\dot{R}^{2}}.
\end{equation}

\section{Scalar Perturbations: the 1+3 Formalism}
\label{Section:Scalar_Perturbations}

In Section \ref{Section:metric_f(R)_cosmology} \label{1p3formaref1}the metric 
approach to the 
derivation  of the cosmological field equations in $f(R)$ gravity was 
presented. As the field of modified theories matured, it became clear that 
their added complexity leads to practical trouble with analysis. Thus, other 
frameworks from which a space-time can be studied have been considered, such as 
decomposing the space-time into a set of 1+3 covariant variables, as has been 
briefly mentioned in Section \ref{Sec:2-ST-theories} \cite{Ellis:1998ct, 
Ehlers:1993gf, Maartens:1996hb}.  

The advantage of using the 1+3 covariant formalism to study FLRW universes is 
twofold;  the first advantage is that this formalism assists the process of 
conveniently extending the GR 1+3 formalism \cite{Ellis:1998ct} to modified 
theories, which must be a major consideration when the task of modifying 
gravity can come with significant complications, and the second is that in this 
formalism it is clear to track the physical meaning underlying calculations, 
which is another important consideration since it is easy to lose intuition at 
the expense of complexity in modified theories. The 1+3 formalism was indeed 
widely studied in the $f(R)$ gravity framework 
\cite{Carloni:2007br, Amendola:2006eh, Carloni:2007yv, Abebe:2013zua, 
Abebe:2011ry}. 

In the following we shall discuss the kinematical approach from the perspective 
of a  fundamental observer having four-velocity $u_{\alpha}$, and we present 
the 
constraint and propagation equations that may be derived using the Bianchi 
identities and conservation equations for momentum and energy.

The kinematic set-up is identical to that in GR; the extension to $f(R)$ 
gravity is  performed simply by adding the curvature fluid as an additional 
fluid component in the energy-momentum tensor. 
In this context, it is most natural and intuitive to choose what is known as 
the matter frame, $u^{m}_{\alpha}$, which is comoving with standard matter, 
representing motion of galaxies and galaxy clusters. The choice is also 
preferred for the obvious reason that this frame happens to coincide with the 
one we are in.

\subsection{Fluid Sources}\label{Sec_Fluid sources}

The first step is to write down the covariant decomposition of the stress 
energy  momentum tensor relative to the 4-velocity. The critical argument that 
allows this decomposition analysis in fourth-order theories of gravity is the 
ability to express the modification to GR in $f(R)$ gravity as an additional 
source term arising from the extra terms involving curvature and the correction 
$f$, where the purely matter part is influenced by a factor of 
$(\frac{1}{f'})$, such that the field equations resemble the Einstein field 
equations, including a curvature fluid, namely
\begin{equation}\label{mefe}
\left(R_{\alpha\beta}-\frac{1}{2}g_{\alpha\beta}R\right) = 
\tilde{T}^{m}_{\alpha\beta}+T^{(R)}_{\alpha\beta}=T_{\alpha\beta},
\end{equation}
\noindent where 
$\tilde{T}^{m}_{\alpha\beta}=\frac{1}{f_R}{T}^{m}_{\alpha\beta}$, and as given 
by Eq. (\ref{TR}),
give the expressions for the effective matter fluid  and the effective 
curvature  fluid respectively. 
The stress energy momentum tensor is given by (\ref{mefe}), where, in terms of 
the  individual contributing sources, the total effective  energy density for 
the combined matter and curvature fluid is
\begin{equation}\label{CF_rho}
\rho = \tilde{\rho}_{m}+\rho^{R} = T_{\alpha\beta}u^{\alpha}u^{\beta},  
\end{equation}

\noindent the total effective isotropic pressure is
\begin{equation}\label{CF_p}
p = \tilde{p}_{m}+p^{R} = \frac{1}{3}T_{\alpha\beta}h^{\alpha\beta},
\end{equation}
\noindent the total effective momentum density, or energy flux relative to 
$u^{\alpha}$,  is
\begin{equation}\label{CF_qa}
q_{\alpha} = \tilde{q}^{m}_{\alpha}+q^{R}_{\alpha} = 
-T_{\beta\mu}u^{\mu}h^{\beta}{}_{\alpha},
\end{equation}
\noindent and the total effective projected symmetric trace free anisotropic 
stress tensor  is
\begin{equation}\label{CF_piab}
\pi_{\alpha\beta}=\tilde{\pi}^{m}_{\alpha\beta}+\pi^{R}_{\alpha\beta} = 
T_{\mu\nu}^{(m)}h^{\mu}{}_{\left<\alpha\right.}h^{\nu}_{\left.\beta\right>}.
\end{equation}

\noindent Here, the tilde denotes the coupling to the $f_R$ field as follows,
\begin{equation}\label{CF_}
\tilde{\rho}_{m}=\frac{\rho_{m}}{f_R},~~\tilde{p}_{m}=\frac{p_{m}}{f_R}, 
~~\tilde{q}^{m}_{\alpha}=\frac{q^{m}_{\alpha}}{f_R}, 
~~\tilde{\pi}_{\alpha\beta}^{m}=\frac{\pi_{\alpha\beta}^{m}}{f_R}.
\end{equation}
\noindent Moreover, the following properties hold for $q_{\alpha}$ and  
$\pi_{\alpha\beta}$:
\begin{align}
&q_{\alpha} 
u^{\alpha}=0,~~~\pi^{\alpha}{}_{\alpha}=0,~~~\pi_{\alpha\beta}=\pi_{
(\alpha\beta)},\\
&\pi_{\alpha\beta}u^{\beta}=0, q_{\alpha}=q_{\left<\alpha\right>},  
\pi_{\alpha\beta}=\pi_{\left<\alpha\beta\right>}.
\end{align}

\noindent The physics will be contained in the specification of an  equation of 
state that relates the quantities above. The framework of a 
perfect fluid is widely used, and it is characterised by the following 
constraint:
\begin{equation}
q^{\alpha}=\pi_{\alpha\beta}=0 \implies T_{\alpha\beta}=\rho 
u_{\alpha}u_{\beta}+ph_{\alpha\beta}.
\end{equation}

Applying the twice-contracted Bianchi identities to the  \textit{total} stress 
energy tensor, $\nabla^{\beta}T_{\alpha\beta}^{(m)}=0$, reveals the 
conservation 
properties of the effective fluids. The effective matter fluid is not 
conserved,  since
\begin{equation}
\nabla^{\beta}\tilde{T}_{\alpha\beta}^{(m)}=\frac{\nabla^{\beta}T_{\alpha\beta}^
{(m)}}{f_R}-\frac{f_{RR}}{f_R^{2}}T^{(m)}_{\alpha\beta}\nabla^{\beta}R.
\end{equation}
\noindent Furthermore, the conservation of the \textit{total} energy-momentum 
implies
\begin{equation}
\nabla^{\beta}T^{(R)}_{\alpha\beta}=\frac{f_{RR}}{f_R^{2}}\tilde{T}^{m}_{
\alpha\beta}\nabla^{\beta}R.
\end{equation}

While the standard matter component is still subject to the energy conditions 
discussed previously, the effective matter and curvature fluids are free to (and 
in general do) violate the weak energy condition, leaving the natural 
choice of frame as the energy frame of the standard matter $u^{\alpha}_{m}$, 
since the thermodynamical properties of standard matter are always preserved.

The Bianchi identities, as applied to the total stress energy momentum tensor, 
show that as long as the stress energy momentum tensor for standard matter is 
conserved, $\nabla^{\beta}T^{(m)}_{\alpha\beta}=0$, then the total stress 
energy momentum tensor will satisfy conservation of energy.

\subsection{Geometry}

In the covariant formalism for GR, transpires that it is more useful to use the 
reverse  representation of the Einstein field equations
\begin{equation}\label{efe_rev}
R_{\alpha\beta}=T_{\alpha\beta}^{(m)}-\frac{1}{2}T^{(m)}g_{\alpha\beta}.
\end{equation} 
 
\subsubsection{The Curvature Fluid}

In order to discuss the ``thermodynamical'' properties of the curvature fluid, 
we consider 
the  right-hand side of Eq. (\ref{mefe}),  where the \textit{curvature 
fluid} in $f(R)$ gravity 
is given in Eq. (\ref{TR}). Should the terms in this equation be 
decomposed in derivative operators containing space and time parts, we have
\begin{align}\label{1+3Tcurv}
T_{\alpha\beta}^{R}=\frac{1}{f_R}\left[ \frac{1}{2}g_{\alpha\beta}(f-Rf_R)     
-\dot{f}_R\left( 
\frac{1}{3}h_{\alpha\beta}\Theta+\sigma_{\alpha\beta}+\omega_{\alpha\beta} 
\right) + \frac{1}{3}h_{\alpha\beta}\tilde{\nabla}^{2}f_R \right].
\end{align}

\noindent Now, using Equations (\ref{CF_rho}) -- (\ref{CF_}) and the above 
decomposition,  we may rewrite the thermodynamical quantities associated with 
the curvature fluid in terms of the $1+3$ variables:
\begin{align}
\rho^{R}~~ =~~& \frac{1}{f_R}\left[ \frac{1}{2}(Rf_R-f) + 
f_{3R}\tilde{\nabla}^{\alpha}R\tilde{\nabla}_{\alpha}R + 
f_{RR}\tilde{\nabla}^{2}R-\Theta f_{RR}\dot{R} \right], \label{eq_rho^R}\\
p^{R}~~ = ~~&\frac{1}{f_R}\left[ 
\frac{1}{2}(f-Rf_R)-\frac{2}{3}f_{RR}\tilde{\nabla}^{2}R - \frac{2}{3}f_{3R} 
\tilde{\nabla}^{\alpha}R\tilde{\nabla}_{\alpha}R +\frac{2}{3}\Theta 
f_{RR}\dot{R}+ f_{3R}\dot{R}^{2}  \right.\label{eq_p^R}\\
&\left.  + f_{RR}\ddot{R}-\dot{u}_{\mu}f_{RR}\tilde{\nabla}^{\mu}R\right], 
\nonumber\\ 
q^{R}_{\alpha} ~~=~~& -\frac{1}{f_R}\left[ 
f_{3R}\dot{R}\tilde{\nabla}_{\alpha}R + 
f_{RR}\tilde{\nabla}_{\alpha}\dot{R}-\frac{1}{3}\Theta 
f_{RR}\tilde{\nabla}_{\alpha}R -\sigma_{\alpha\mu}f_{RR}\tilde{\nabla}^{\mu}R - 
\omega_{\alpha\mu}f_{RR}\tilde{\nabla}^{\mu}R \right],\\
\pi_{\alpha\beta}^{R}~~ =~~& \frac{1}{f_R}\left[ 
f_{3R}\tilde{\nabla}_{\left<\alpha\right.}R\tilde{\nabla}_{\left.\beta\right>}
R+f_{RR} 
\tilde{\nabla}_{\left<\alpha\right.}\tilde{\nabla}_{\left.\beta\right>}R - 
\sigma_{\alpha\beta}f_{RR}\dot{R}\right].
\end{align}

The twice-contracted Bianchi identities 
are used to obtain evolution equations for $\rho^{m}, \rho^{R}$ and 
$q_{\alpha}^{R}$:
\begin{align}
&\dot{\rho}_{m} =- \Theta(\rho_{m}+p_{m}), \label{rhodotM_ev}\\
&\dot{\rho}^{R}+\tilde{\nabla}^{\alpha}q_{\alpha}^{R} = -\Theta(\rho^{R}+p^{R}) 
- 2\dot{u}^{\alpha}q_{\alpha}^{R}- 
\sigma^{\alpha\beta}\pi_{\beta\alpha}^{R}+\rho_{m}\frac{f_{RR}\dot{R}}{f_R^{2}},
\label{rhodotR_ev}\\
&\dot{q}^{R}_{\left< \alpha \right>} + 
\tilde{\nabla}_{\alpha}p^{R}+\tilde{\nabla}^{\beta}\pi_{\alpha\beta}^{R} =  
-\frac{4}{3}\Theta q_{\alpha}^{R} - 
\sigma_{\alpha}{}^{\beta}q_{\beta}^{R}-(\rho^{R}+p^{R})\dot{u}_{\alpha}  
-\dot{u}^{\beta}\pi^{R}_{\alpha\beta}\label{qdotR_ev}\\
&~~~~~~~~~~~~~~~~~~~~~~~~~~~~~~~~~ 
-\eta^{\beta\mu}_{\alpha}\omega_{\beta}q_{\mu}^{R} + \rho_{m}\frac{f_{RR} 
\tilde{\nabla}_{\alpha}R}{f_R^{2}}.\nonumber
\end{align}

We also have the following relation between the acceleration and the energy 
 density and pressure of the standard matter:
\begin{equation}
\tilde{\nabla}p_{m}=-(\rho_{m}+p_{m})\dot{u}^{\alpha},
\end{equation}
\noindent coming from the conservation of momentum for standard matter.

Substituting into (\ref{efe_rev}) for the total effective energy momentum 
tensor,  we may write the Ricci tensor and Ricci scalar in terms of the 
thermodynamic quantities of the total effective fluid as
\begin{align}
&R_{\alpha\beta}=\frac{1}{2}(\rho_{tot}+3p_{tot})u_{\alpha}u_{\beta}+\frac{1}{2}
(\rho_{tot}-p_{tot})h_{\alpha\beta}+2u_{(\alpha}q^{tot}_{\beta)}+\pi^{tot}_{
\alpha\beta}
,\label{RicciTensor1+3}\\
&R = \rho_{tot}-3p_{tot}. \label{RicciScalar1+3}
\end{align}

We can also construct the trace equation of the curvature fluid by considering  
$\tilde{T}^{m}$ and $T^{(R)}$, the traces of the effective matter and curvature 
fluids respectively:
\begin{align}
&\tilde{T}^{m}=\frac{1}{f_R}g^{\alpha\beta}T_{\alpha\beta}^{m}=\frac{1}{f_R}
\left( 3p_{m}-\rho_{m} \right), \label{trace_matter1+3}\\
&T^{(R)}=g^{\alpha\beta}T_{\alpha\beta}^{R}=\frac{1}{f_R}\left[ 2(f-Rf_R)  - 
3\left( f_{RR}\tilde{\nabla}^{2}R 
+f_{RR}'\tilde{\nabla}^{\alpha}R\tilde{\nabla}_{\alpha}R \right.\right. 
\nonumber\\
& \left.\left. 
-f_{3R}\dot{R}^{2}-f_{RR}\ddot{R}+\dot{u}_{\mu}f_{RR}\tilde{\nabla}^{\mu}R-f_{RR
}\theta\dot{R} \right) \right], \label{trace_curv1+3}
\end{align}
\noindent by taking the trace of (\ref{1+3Tcurv}). If we substitute  
(\ref{trace_matter1+3}) and (\ref{trace_curv1+3}) into the equation for the 
Ricci scalar, (\ref{RicciScalar1+3}), and considering only the curvature terms, 
we obtain the trace equation corresponding to the curvature fluid:
\begin{equation}
Rf_R - 2f = -3\left( f_{RR}\tilde{\nabla}^{2}R+f_{3R}\tilde{\nabla}^{\alpha}R 
\tilde{\nabla}_{\alpha}R-f_{3R}\dot{R}^{2}-f_{RR}\ddot{R}+\dot{u}_{\mu}f_{RR}
\tilde{\nabla}^{\mu}R-f_{RR}\theta\dot{R} \right).
\end{equation}

\subsection{Propagation and Constraint Equations}

The complete 1+3 decomposition for the Riemann tensor, in terms of both 
thermodynamic  and geometric quantities defined above, is given by
\begin{align}\label{eq_Riemann_decomp}
R^{\alpha\beta}{}_{\mu\nu} = R^{\alpha\beta}_{P}{}_{\mu\nu} + 
R^{\alpha\beta}_{I}{}_{\mu\nu}  + 
R^{\alpha\beta}_{E}{}_{\mu\nu}+R^{\alpha\beta}_{H}{}_{\mu\nu},
\end{align}

\noindent where
\begin{align}
& R^{\alpha\beta}_{P}{}_{\mu\nu} = \frac{2}{3} (\rho+3p-2\Lambda) u^{[\alpha} 
u_{[\mu}h^{\beta]}{}_{\nu]} + \frac{2}{3}(\rho+\Lambda) 
h^{[\alpha}{}_{[\mu}h^{\beta]}{}_{\nu]},\label{eq_Riemann_decomp_termsi}\\
& R^{\alpha\beta}_{I}{}_{\mu\nu} = -2u^{[\alpha}h^{\beta]}{}_{[\mu}q_{\nu]}  - 
2u_{[\mu}h^{[\alpha}{}_{\nu]}q^{\beta]} - 
2u^{[\alpha}u_{[\mu}\pi^{\beta]}{}_{\nu]} + 
2h^{[\alpha}{}_{[\mu}\pi^{\beta]}{}_{\nu]},\\
&R^{\alpha\beta}_{E}{}_{\mu\nu} = 4 u^{[a}u_{[c}E^{\beta]}{}_{\nu} + 
4h^{[\alpha}{}_{\mu}E^{\beta]}{}_{\nu]}, \\
&R^{\alpha\beta}_{H}{}_{\mu\nu} = 2\eta^{\alpha\beta\gamma} 
u_{[\mu}H_{\nu]\gamma}+2\eta_{\mu\nu\gamma}u^{[\alpha}H^{\beta]\gamma}.
\label{eq_Riemann_decomp_termsf}
\end{align}

Thus, using equations for the decomposition of the Riemann tensor  
(\ref{eq_Riemann_decomp}) - (\ref{eq_Riemann_decomp_termsf}), we can obtain 
three sets of propagation and constraint equations, coming from the 
Einstein equations, and their integrability conditions in the 1+3 covariant 
decomposition.

\subsubsection{1. Ricci identities}

The first set of propagation equations result from the Ricci identities for  
the velocity vector field $u^{\alpha}$, namely
\begin{equation}
2\nabla_{[\alpha}\nabla_{\beta]}u^{\mu}=R_{\alpha\beta}{}^{\mu}{}_{\nu}u^{\nu}.
\end{equation}

From the Ricci identities for the velocity vector field $u^{\alpha}$,\\
\noindent \textit{1. The Raychaudhuri propagation equation} is given by
\begin{equation}\label{P1}
\dot{\Theta} - \tilde{\nabla}_{\alpha}\dot{u}^{\alpha}  +\frac{1}{3}\Theta^{2} 
- (\dot{u}_{\alpha}\dot{u}^{\alpha}) + \sigma_{\alpha\beta}\sigma^{\alpha\beta} 
-2\omega_{\alpha}\omega^{\alpha}+\frac{1}{2}(\tilde{\rho}_{m}+3\tilde{p}_{m}
)=-\frac{1}{2}(\rho^{R}- 3p^{R}).
\end{equation}

\noindent {\textit{2. The vorticity propagation equation} is as before,
\begin{equation}
\dot{\omega}^{\left< \alpha \right>} - \frac{1}{2} 
\eta^{\alpha\beta\mu}\tilde{\nabla}_{\beta}\dot{u}_{\mu} = 
-\frac{2}{3}\Theta\omega^{\alpha}+\sigma^{\alpha}{}
_{\beta}\omega^{\beta}.
\end{equation}}
\noindent {\textit{3. The shear propagation equation } is
\begin{equation}
\dot{\sigma} ^{\left< \alpha\beta\right>}-  
\tilde{\nabla}^{\left<\alpha\right.}\dot{u}^{\left.\beta\right>}  
+\frac{2}{3}\Theta\sigma^{\alpha\beta} 
-\dot{u}^{\left<\alpha\right.}\dot{u}^{\left.\beta\right>} 
+\sigma^{\left<\alpha\right.}{}_{\mu}\sigma^{\left.\beta\right>\mu}+\omega^{
\left<\alpha\right.}w^{\left.\beta\right>}+E^{\alpha\beta} 
=\frac{1}{2}\pi^{\alpha\beta}_{R} .
\end{equation}
} \\
\noindent  {\textit{4. The $(0\alpha)$ shear divergence constraint}  is
\begin{equation}
0 = (C_{1})^{\alpha}=\tilde{\nabla}_{\beta}\sigma^{\alpha\beta}- 
\frac{2}{3}\tilde{\nabla}^{\alpha}\Theta+\eta^{\alpha\beta\nu}\left[ 
\tilde{\nabla}_{\beta}\omega_{\nu} + 2\dot{u}_{\beta}\omega_{\mu} \right] + 
q^{\alpha}_{R}.
\end{equation}
\\
\noindent \textit{5. The vorticity divergence constraint} is
\begin{equation}
0 = (C_{2}) = 
\tilde{\nabla}_{\alpha}\omega^{\alpha}-\dot{u}_{\alpha}\omega^{\alpha}.
\end{equation}

\noindent \textit{6. The gravito-magnetic $H_{\alpha\beta}$ constraint} is
\begin{equation}
0 = (C_{3})^{\alpha\beta}= 
H^{\alpha\beta}+2\dot{u}^{\left<\alpha\right.}\omega^{\left.\beta\right>} - 
\eta^{\mu 
\nu\left<\alpha\right.}\tilde{\nabla}_{\mu}\sigma^{\left.\beta\right>}{}_{\nu} 
+ 
\tilde{\nabla}^{\left<\alpha\right.}\omega^{\left.\beta\right>}.
\end{equation}

\subsubsection{2. Twice-contracted Bianchi identities}

The constraint obtained by projecting parallel to $u^{\alpha}$ yields 

\noindent 
\textit{7. The energy conservation equation}:

\begin{equation}
\dot{\rho}+\tilde{\nabla}_{\alpha}q^{\alpha}=- 
\Theta(\rho+p)-2\dot{u_{\alpha}}q^{\alpha}-\sigma_{\alpha\beta}\pi^{\alpha\beta}
,
\end{equation}
\noindent and by projecting orthogonally to $u^{\alpha}$ yields:
\noindent 
\textit{8. The conservation of momentum equation}:
\begin{equation}
\dot{q}^{\left<\alpha\right>} + \tilde{\nabla}^{\alpha}p+ 
\tilde{\nabla}_{\beta}\pi^{\alpha\beta}=-\frac{4}{3}\Theta 
q^{\alpha}-\sigma^{\alpha}{}_{\beta}q^{\beta}-(\rho+p)\dot{u}^{\alpha} - 
\dot{u}_{\beta}\pi^{\alpha\beta} - \eta^{\alpha\beta\mu}\omega_{\beta}q_{\nu}. 
\end{equation}

\noindent Additionally, for a perfect fluid we have
\begin{align}
\dot{\rho}_{m} &= -\Theta(\rho_{m}+p_{m}),\\
\tilde{\nabla}_{\alpha}p_{m} & = - (\rho_{m} + p_{m})\dot{u}_{\alpha}.
\end{align}}

\subsubsection{3. Once-contracted Bianchi identities}

\textit{9. The gravito-electric $\dot{E}$ propagation equation}:
\begin{align}
(\dot{E}^{\left<\alpha\beta\right>} +  
\frac{1}{2}\dot{\pi}^{\left<\alpha\beta\right>}) &- 
\eta^{\mu\nu\left<\alpha\right.}\tilde{\nabla}_{\mu}H^{\left.\beta\right>}{}_{
\nu} + \frac{1}{2}\tilde{\nabla}^{\left<\alpha\right.}q^{\left.\beta\right>} 
\nonumber \\
&= -\frac{1}{2}(\rho+p)\sigma^{\alpha\beta} - \Theta\left( E^{\alpha\beta} +  
\frac{1}{6}\pi^{\alpha\beta} \right) + 
3\sigma^{\left<\alpha\right.}{}_{\mu}\left( E^{\left.\beta\right>\mu} - 
\frac{1}{6}\pi^{\left.\beta\right>\mu} \right) \\
&- \dot{u}^{\left<\alpha\right.}q^{\left.\beta\right>} + 
\eta^{\mu\nu\left<\alpha\right.}\left[  2\dot{u}_{\mu} 
H^{\left.\beta\right>}{}_{\nu} + \omega_{\mu}\left( 
E^{\left.\beta\right>}{}_{\nu} + \frac{1}{2}\pi^{\left.\beta\right>}{}_{\nu}  
\right) \right] \nonumber.
\end{align}

\noindent\textit{10. The gravito-magnetic $\dot{H}$ propagation equation}:
\begin{align}\label{P}
\dot{H}^{\left<\alpha\beta\right>} + 
\eta^{\mu\nu\left<\alpha\right.}\tilde{\nabla}_{\mu}\left( E^{\left. \beta 
\right>}{}_{\nu}-\frac{1}{2}\pi^{\left.\beta\right>}{}_{\nu} \right) =& -\Theta 
H^{\alpha\beta}+3\sigma^{\left<\alpha\right.}{}_{\mu}H^{\left.\beta\right>\mu}
+\frac{3}{2}\omega^{\left<\alpha\right.}q^{\left.\beta\right>}\nonumber \\
& -\eta^{\mu\nu\left<\alpha\right.}\left[ 
2\dot{u}_{\mu}E^{\left.\beta\right>}{}_{\nu}-\frac{1}{2}\sigma^{
\left.\beta\right>}{}_{\mu}q_{\nu} - \omega_{\mu}H^{\left.\beta\right>}{}_{\nu} 
\right].
\end{align}
 
\noindent\textit{11. The gravito-electric divergence constraint:}

\begin{align}\label{C4}
0 = (C_{4})^{\alpha}=&\tilde{\nabla}_{\beta}\left( 
E^{\alpha\beta}+\frac{1}{2}\pi^{\alpha\beta} \right) - 
\frac{1}{3}\tilde{\nabla}^{\alpha}\rho + \frac{1}{3}\Theta q^{\alpha} - 
\frac{1}{2}\sigma^{\alpha}{}_{\beta}q^{\beta} - 3\omega_{\beta}H^{\alpha\beta} 
\\
&-\eta^{\alpha\beta\mu}\left[ \sigma_{\beta\nu}H^{\nu}{}_{\mu} - 
\frac{3}{2}\omega_{\beta}q_{\mu} \right].\nonumber
\end{align}

\noindent\textit{12. The gravito-magnetic divergence (div $H$) constraint}:

\begin{align}\label{C5}
0 = (C_{5})^{\alpha} = &\tilde{\nabla}_{\beta}H^{\alpha\beta} + 
(\rho+p)\omega^{\alpha}  + 3\omega_{\beta}\left( 
E^{\alpha\beta}-\frac{1}{6}\pi^{\alpha\beta} \right)\\
&+\eta^{\alpha\beta\mu}\left[ \frac{1}{2}\tilde{\nabla}_{\beta}q_{\mu} 
+\sigma_{\beta\nu} (E^{\nu}{}_{\mu} + \frac{1}{2}\pi^{\nu}{}_{\mu})\right].
\end{align}

We will recover the GR versions of these equations simply by setting $f(R)=R$,  
resulting in the matter parts being identical to standard matter, and all 
curvature fluid terms vanishing. In general, we can close the system by 
specifying the equation of state of the fluid sources, which amounts to 
choosing 
restrictions on the thermodynamic quantities. Very relevant for late-time 
considerations is that of pressureless non-relativistic matter, dust, with 
$p=q_\alpha=\pi_{\alpha\beta}=0 \Rightarrow \dot{u}_\alpha=0
$. 

Applying the decomposition to $f(R)$ modified gravity theories, for general 
spacetimes we  get

\begin{equation} \label{eq_1+3_f(R)}
\nabla_{\alpha}\nabla_{\beta}f_R = -\dot{f}_R\left( \frac{1}{3}h_{\alpha\beta} 
\Theta + \sigma_{\alpha\beta}+\omega_{\alpha\beta} \right) + 
u_{\beta}u_{\alpha}\ddot{f}_R+u_{\alpha}\dot{f}_R\dot{u}_{\beta}.
\end{equation}

Furthermore, we obtain
\begin{equation}
\square f_R = -\Theta\dot{f}_R - \ddot{f}_R,
\end{equation}
\noindent where any terms containing orthogonally projected derivatives have  
been neglected, since we are only considering isotropic and homogeneous space 
times. 

The kinematics and thermodynamics above, prescribed by Equations  
(\ref{P1})-(\ref{C5}),  along with (\ref{rhodotM_ev}) - (\ref{qdotR_ev}), 
formalise the physical interaction and evolution of the matter and 
gravitational 
fields in $f(R)$ gravity, and completely specify a cosmological model. This 
approach is invaluable in the investigations into alternative theories to GR, 
as 
it sets out a scheme that is both mathematically rigorous and intuitive. It 
has 
also been extremely useful in constructing and studying cosmological 
perturbations \cite{Carloni:2007yv, Ellis:1998ct, Kodama:1985bj}. We refer the 
reader to these three references
for a deeper insight about the foundations on  the 1+3 covariant gauge 
invariant 
 treatment of scalar perturbations in $f(R)$ gravity.

\section{Geodesic Deviation in $f(R)$ Gravity}
\label{Sec:GDE}

Historically, the geodesic deviation  \label{geodesicdevref1}
 equation (GDE) has played a key role in  
the realm of gravitational physics and,  consequently, in cosmology. Through 
this 
equation one could determine the relative geodesic deviation, dubbed $\eta$,  
between two neighbouring geodesics, which is pertinent to  deriving important 
results for cosmologies. This equation can be recast in terms of the set of 
expansion-normalised dynamical variables and thus allow us to  gain some 
insight into how for 
$f(R)$ theories it differs from GR predictions. We refer the reader to 
\cite{delaCruz-Dombriz:2013gfa} for further details.

\subsection{Formalism}

The general geodesic deviation equation is given by

\begin{equation}
\frac{\delta^{2}\eta^{\alpha}}{\delta v^{2}} =  
-R^{\alpha}{}_{\beta\sigma\gamma}V^{\beta}V^{\gamma}\eta^{\sigma},
\end{equation}

\noindent where $\eta$ is the deviation vector,  $V^{\alpha}$ is the normalised 
tangent vector field, and $v$ is an affine parameter. It is useful to be able 
to 
express this deviation in terms of the density  by substituting the expressions 
for the Riemann and Ricci tensors and the Ricci scalar. 

Using the Riemann tensor:
\begin{equation}\label{eq_Riemann_tensor}
     R_{\alpha\beta\sigma\gamma} = C_{\alpha\beta\sigma\gamma}   
+\frac{1}{2}\left(g_{\alpha\sigma}R_{\beta\gamma} - 
g_{\alpha\gamma}R_{\beta\sigma} + g_{\beta\gamma}R_{\alpha\sigma} - 
g_{\beta\sigma}R_{\alpha\gamma} \right) -\frac{R}{6}\left( 
g_{\alpha\sigma}g_{\beta\gamma}-g_{\alpha\gamma}g_{\beta\sigma}\right),
\end{equation}

\noindent and the fact that in FLRW space-times $C_{\alpha\beta\sigma\gamma}$,  
the Weyl tensor, vanishes, upon contracting (\ref{eq_Riemann_tensor}) with 
$V^{\beta}\eta^{\sigma}V^{\gamma}$, we obtain

\begin{align}\label{eq_Weyl_contracted2}
R^{\alpha}{}_{\beta\sigma\gamma}V^{\beta} \eta^{\sigma}V^{\gamma}  = 
&\frac{1}{2}\left( \eta^{\alpha}V^{\beta}V^{\gamma}R_{\beta\gamma} - 
V^{\alpha}V^{\beta}\eta^{\sigma}R_{\beta\sigma} + \epsilon 
R^{\alpha}{}_{\sigma}\eta^{\sigma}\right) - \frac{R}{6}\eta^{\alpha}\epsilon.
\end{align}

\noindent By writing  $E = -V_{\alpha}u^{\alpha}$,  
$\eta_{\alpha}u^{\alpha}=\eta_{\alpha}V^{\alpha}=0$, and 
$\epsilon=V_{\alpha}^{\alpha}$, we can simplify the terms of Equation 
(\ref{eq_Weyl_contracted2}) as follows:
\begin{align}
R^{\alpha}{}_{\beta\sigma\gamma}\eta^{\sigma} =& \frac{1}{f_R}\left[ 
\eta^{\alpha}\left( p_{m}+\frac{f}{2} - \square f_R \right) + 
(\nabla^{\alpha}\nabla_{\sigma}f_R)\eta^{\sigma}\right],\\
R_{\beta\sigma}V^{\alpha}V^{\beta}\eta^{\sigma}=&\frac{1}{f_R}\left[ 
(\nabla_{\beta}\nabla_{\sigma}f')V^{\alpha}V^{\beta}\eta^{\sigma} \right],\\
R_{\beta\gamma}V^{\beta}V^{\gamma}\eta^{\alpha} =& \frac{1}{f_R}\left[ 
(\rho_{m}+p_{m})E^{2} + \epsilon\left( p_{m}+\frac{f}{2}-\square f_R \right) + 
V^{\beta}V^{\gamma}\nabla_{\beta}\nabla_{\gamma} f_R \right]\eta^{\alpha}.
\end{align}
  Using Equation (\ref{eq_1+3_f(R)})  and that for FLRW spacetimes, 
$\omega_{\alpha\beta}=\sigma_{\alpha\beta}=0$, we can arrive at the following 
expressions:
\begin{align}
V^{\beta}V^{\gamma}\nabla_{\beta}\nabla_{\gamma}f_R &= 
-\frac{1}{3}\dot{f}_R\Theta(\epsilon + E^{2}) + E^{2}\ddot{f}_R,\\
(\nabla_{\beta}\nabla_{\sigma} f')V^{\alpha}V^{\beta}\eta^{\sigma} &=0 ,\\
(\nabla^{\alpha}\nabla_{\sigma}f')\eta^{\sigma}&=-\frac{1}{3}\dot{f}_R\Theta 
\eta^{\alpha}.
\end{align}
  Substituting the above results into Equation 
(\ref{eq_Weyl_contracted2}) gives
\begin{equation}\label{eq_Weyl_contracted3}
R^{\alpha}{}_{\beta\sigma\gamma}V^{\beta}V^{\gamma}\eta^{\sigma}=\frac{1}{2f_R}
\left[ \frac{f+\rho_{m}-2\dot{f}_R\Theta}{3} - \square f_R + p_{m} 
\right]\eta^{\alpha}\epsilon + \frac{1}{2f_R}\left[ 
\rho_{m}+p_{m}-\frac{1}{3}\dot{f}_R\Theta + \ddot{f}_R 
\right]\eta^{\alpha}E^{2}.
\end{equation}

We can then identify terms in Equation (\ref{eq_Weyl_contracted3})  to be 
combinations of the curvature fluid density and pressure, from
\begin{align}
\rho^{R}+p^{R}=&\frac{1}{f_R}\left[-\frac{1}{3}\dot{f}_R\Theta + \ddot{f}_R 
\right],\\ 
\rho^{R}+3p^{R}=& \frac{1}{f_R}\left[ f + \Theta\dot{f}_R + 3\ddot{f}_R\right] 
- 
R, 
\end{align}
 such that we may obtain the final result for the  geodesic deviation 
equation in metric $f(R)$ gravity \cite{delaCruz-Dombriz:2013gfa}:
\begin{equation} \label{eq:GDE_f(R)}
R^{\alpha}{}_{\beta\gamma\sigma} V^{\beta}V^{\gamma}\eta^{\sigma} = \frac{1}{2} 
\left(\rho_{tot} + p_{tot}\right)E^{2}\eta^{\alpha} + \left[ 
\frac{R}{6}+\frac{1}{6}\left(\rho_{tot} + 3p_{tot}\right) \right] \epsilon 
\eta^{\alpha}.
\end{equation}

The result is consistent with what is expected in a homogeneous and  isotropic 
geometry. The tidal force produced will only depend on $\eta^{\alpha}$, and thus
only the magnitude of the deviation vector $\eta$ will change along the  
geodesic, while its direction is preserved. 

Considering only the paths of photons, where $V^{\alpha}=k^{\alpha}$,  and 
$k_{\alpha}k^{\alpha}=0$, so $\epsilon=0$, Equation (\ref{eq:GDE_f(R)}) becomes
\begin{equation}
\label{eq:GDE_null_f(R)}
R^{\alpha}_{\beta\sigma\gamma}k^{\beta}k^{\gamma}\eta^{\sigma} =  
\frac{1}{2}(\rho_{tot}+p_{tot})E^{2}\eta^{\alpha},
\end{equation}
\noindent showing the focusing of all families of past directed null  geodesics 
as long as 
\begin{equation}
(\rho_{tot} +p_{tot})>0.
\end{equation}
Both Equations (\ref{eq:GDE_f(R)}) and (\ref{eq:GDE_null_f(R)}) will  reduce to 
the GR result when $f(R)=R$.

\subsection{Past-directed Null Geodesics and Area Distance in $f(R)$ Gravity}

We now consider how the case $V^{\alpha}=k^{\alpha}, k_{\alpha}k^{\alpha}=0,  
k^{0}<0$ affects Equation (\ref{eq:GDE_null_f(R)}). Let $\eta^{\alpha}=\eta 
e^{\alpha}$ and $e^{\alpha}e_{\alpha}=1$, and 
$e_{\alpha}u^{\alpha}=e_{\alpha}k^{\alpha}=0$, with a basis $e$ that is 
parallelly 
propagated and aligned, such that, $\delta e^{\alpha}/\delta v = 0 
=k^{\beta}\nabla_{\beta}e^{\alpha}$. Equation (\ref{eq:GDE_null_f(R)}) may thus 
be written as
\begin{equation}\label{eq:GDE_null_f(R)_2}
\frac{{d^{2}} \eta}{{{d}} v^{2}} = -\frac{1}{2}(\rho_{tot} + 
p_{tot})E^{2}\eta.
\end{equation}

Once again, all families of past directed null geodesics will be focused so  
long that $(\rho_{tot} + p_{tot})>0$. When the right-hand side of 
(\ref{eq:GDE_null_f(R)_2}) is zero (de Sitter universe in GR), the solution to 
this equations is the same as that in a flat Minkowski space-time: 
$\eta(v)=C_{1}(v)+C_{2}$. The chain rule gives
\begin{align}
\frac{{d}^{2}}{{{d}} v^{2}} &= \left( \frac{{{d}} z}{{{d}} v} 
\right)^{2} \left[\frac{{d}^{2}}{{{d}} z^{2}} - \frac{{{d}} 
{z}}{{{d}} {v}} \frac{{{d}}^{2}{v}}{{{d}} {z^{2}}} 
\frac{{{d}}}{{{d}}{z}} \right], \\ 
\frac{{{d}}z}{{{d}}v}& = E_{0}H(1+z).
\end{align}

Using this and the modified Friedmann and Raychaudhuri expressions, we obtain 
the  following evolution equation for $\eta$ with redshift, which depends only 
on the total equation of state:
\begin{equation} \label{eq:GDE_null_d2etadz2}
\frac{ {{d}}^{2}\eta }{ {{d}} z^{2}} + \frac{(7+3 w_{tot})}{2(1+z)} 
\frac{ {{d}}\eta}{{{d}}z} + \frac{3(1+w_{tot})}{2(1+z)^{2}}\eta = 0.
\end{equation}

\noindent We may then infer an expression for  the observer area distance 
$r_{0}(z)$ 
\begin{equation}
r_{0}(z)=\sqrt{\left|   \frac{ {{{d}} A_{0}(z)} }{{{d}}\Omega} \right|} = 
\left|  \frac{\eta(z')|_{z'=0}}{{{d}}\eta(z')/{{{d}} \ell }|_{z'=0}} 
\right|,
\end{equation}

\noindent where  $A_{0}$ is the area of the object, and $\Omega$ is the solid 
angle in the sky. Using the fact that ${{d}}/{{d}}\ell = 
E_{0}^{-1}(1+z)^{-1}{{d}}/{{d}}v=H(z+1){{d}}/{{d}}z$, we can express 
$r_{0}$ in terms of redshift derivatives as
\begin{equation}
r_{0}(z)=\left| \frac{\eta(z)}{H(0) {{d}}\eta(z')}/{{d}}z'|_{z'=0} \right|.
\end{equation}
In general, to find the observer  distance relation above, we need to resort to 
numerical integration. For instance, in \cite{delaCruz-Dombriz:2013gfa}, the 
GDE 
was expressed in terms of a set of dynamical systems 
variables.\label{Dynamicalref1}

\section{Gravitational Attractiveness in $f(R)$?}
\label{Section:Attractive_gravity}

In this section we study the positive contributions of the Raychaudhuri 
equation 
for time-like geodesics, which guarantee the attractive character of the 
gravitational interaction in $f(R)$ theories \cite{delaCruz-Dombriz:2015tye}.  
Following  \cite{Albareti:2012va}, we write the Raychaudhuri equation as 
\begin{equation}
    \frac{d\theta}{d\tau}=-\frac{1}{3}\theta^{2}-\sigma_{\mu\nu}\sigma^{\mu\nu} 
+ \omega_{\mu\nu}\omega^{\mu\nu}-R_{\mu\nu}\xi^{\mu}\xi^{\nu},
\end{equation}
  where $\theta$ is the expansion, $\sigma_{\mu\nu}$ is the shear, and 
$\omega_{\mu\nu}$ is the rotation of a congruence of time-like geodesics, 
generated by the tangent vector field $\xi^{\mu}$, and $\tau$ is an affine 
parameter. 

In GR, assuming the strong energy condition 
\begin{equation}\label{sec3}
    T_{\mu\nu}^{(m)}\xi^{\mu}\xi^{\nu}\geq -\frac{1}{2}T^{(m)},
\end{equation}
  implies that $R_{\mu\nu}\xi^{\mu}\xi^{\nu}\geq0$. This is an 
important 
 statement and results in the attractive nature of the gravitational 
interaction. It follows that the mean curvature \cite{Albareti:2012se}, defined 
by $\mathcal{M}_{\xi}=-R_{\mu\nu}\xi^{\mu}\xi^{\nu}$, must, in every time-like 
direction, be negative or zero in GR  for fluids for which (\ref{sec3}) 
holds. Following \cite{Albareti:2012se,Albareti:2012va}, the mean 
curvature in every time-like direction
\begin{equation}
    \mathcal{M}_{\xi}\equiv -R_{\mu\nu}\xi^{\mu}\xi^{\nu}
\end{equation}
is negative or zero in GR, provided that the strong energy condition holds.  If 
one chooses a congruence of time-like geodesics whose tangent vector field is 
locally hypersurface-orthogonal, then $\omega_{\mu\nu}=0$ for all the 
congruences. This result enables the use of the Raychaudhuri equation in the 
singularity theorems. Since the term $\sigma_{\mu\nu}\sigma^{\mu\nu}$ is 
non-negative, assuming $R_{\mu\nu}\xi^{\mu}\xi^{\nu}\geq 0$, then

\begin{equation} \label{eq:inequality_att_1}
\frac{ {{d}}\theta }{ {{d}} \tau} + \frac{1}{3}\theta^{2} \leq 0  \rightarrow 
\theta^{-1}(\tau) \geq \theta_{0}^{-1}+\frac{1}{3}\tau.
\end{equation}

Inequality (\ref{eq:inequality_att_1}) indicates that a congruence that is 
initially  converging $(\theta_{0}\leq 0)$ will converge to zero in a finite 
time. For this reasoning to be true, we require that 
$R_{\mu\nu}\xi^{\mu}\xi^{\nu}\geq 0$ for every non-space-like vector. In 
particular, for time-like geodesics, we consider this inequality in the 
late-time 
cosmological scenario, assuming a de Sitter phase of expansion, and negligible 
contributions from radiation and dust. In order to have accelerated expansion of 
time-like geodesics, the Ricci scalar, $R=R_{0}$, will be approximately 
constant. 

We followed the results \cite{Albareti:2012se}, from which  it can be proved 
that 
\begin{equation}\label{eq:inequality_2}
R_{\mu\nu}\xi^{\mu}\xi^{\nu} \geq \frac{f(R_{0}-R_{0}f'(R_{0}))}{2(1+f'(R_{0}))},
\end{equation}
  where we require that the field equations (\ref{fR_eqs}), with constant 
scalar  curvature and standard matter sources, satisfy  the strong energy 
condition. Thus, the 
right-hand side of (\ref{eq:inequality_2}) must be negative in order to have 
$R_{\mu\nu}\xi^{\mu}\xi^{\nu}<0$. Equivalently, $\mathcal{M}_{\xi}>0$, and this 
means we need $\mathcal{M}_{\xi}$ to be bounded from above. Thus the condition 
for time-like geodesics to diverge at late times becomes:
\begin{equation}
\frac{ f(R_{0}-R_{0}f'(R_{0})) }{2(1+f'(R_{0}))}<0.
\end{equation}
  If $1+f'(R_{0})>0$, we obtain
\begin{equation}\label{eq:inequality_3}
f(R_{0})-R_{0}f'(R_{0})<0.
\end{equation}
  If we take Equation (\ref{fR_eqs}) in a vacuum ($T=0$) for constant 
scalar  curvature solutions, the value of $R_{0}$ satisfies
\begin{equation}\label{eq:equation4_att}
R_{0}=\frac{-2f(R_{0})}{1-f'(R_{0})}.
\end{equation}
 Although  in general this cannot be solved analytically,  some 
$f(R)$ 
models exist, depending on their parameters, for which a closed solution can be 
found. Rearranging the terms in Equation (\ref{eq:inequality_3}), we can find 
for a given model whether (\ref{eq:equation4_att}) implies that $R_{0}>0$. A 
positive contribution to the Raychaudhuri equation from the space-time geometry 
$\mathcal{M}_{\xi}$ for every time like direction is obtained, when this is 
true. This is an important theoretical consideration, which may be used to 
constrain the parameter space of paradigmatic $f(R)$ models.

\section{Conclusions}
\label{Section:Conclusions}

Modifying   Einsteinian Relativity  is currently a popular field of 
research in many areas of physics, ranging from quantum theories of gravity to 
astrophysics and cosmology. The so-called $f(R)$ modification is derived by 
replacing the Ricci scalar, $R$, in the usual Einstein-Hilbert gravitational 
action with a general function $f$, which could in principle be $any$ function 
of $R$. As illustrated above, this modification can be equivalent to adding an 
extra scalar field to the theory, which turns out to be the first derivative of 
$f(R)$, provided the second derivative of $f$ is non-zero. 

In fact, at least 
pertaining to dark energy models, almost all modifications can be generalised 
under the umbrella of effective field theories. Over the years, it has become  
understood that only very special models for $f(R)$ are worth studying in a 
cosmological and astrophysical sense. Indeed, only specific forms for the 
function can produce satisfactory physical effects, such as reproducing the 
gravitational field in and around a compact object, accelerating the Universe at 
late times as well as generating an expansion history similar to the observable 
Universe and providing a growth of large-scale structures in agreement with 
observations.

In particular, a generally favoured approach  is to assume that the theory 
already contains General Relativity as a limit. Since one of the most relevant 
ways to determine the effects of this kind of modification is to examine 
dynamical changes in gravitating and large-scale systems, the behaviour of the 
gravitational field in $f(R)$ theories and the interaction between the emergent 
scalar field and matter have been widely considered in a range of different 
physical phenomena (see references in the bulk of the text). Degeneracy between 
$f(R)$ models has been  shown to be the biggest bottleneck for these theories, 
since 
in principle there are plenty of clever $f(R)$ proposals capable of satisfying 
the basic theoretical requirements and fitting the available observational - 
both cosmological and astrophysical - data. 

In this realm, one may have general classes of broken power-law models  
\cite{Miranda:2009rs, 
Hu:2007nk}, which are designed to reduce to General Relativity in low curvature 
regimes, and tend to   General Relativity plus a cosmological constant   
  when the curvature is large relatively to the local environment. These 
models also include, within their parameter space, previously considered $f(R)$ 
{\it viable} models, so in that sense, they parametrise a vast class of possible 
modification of the gravitational interaction with the sole addition of a scalar 
field. The redshift evolution of the Hubble rate and the deceleration parameter, 
in the class of models considered here, have very similar behaviour to that of 
the $\Lambda$CDM model. In general, deviations in the Hubble parameter begin to 
develop only at very low redshifts, $0<z<1$, while, the deceleration parameter, 
showing larger deviations at low redshifts, still converges towards a value of 
$\frac{1}{2}$ from around a value of $z=6$, consistent with the $\Lambda$CDM 
parametrisation. The exact behaviour will depend on the choices of model 
parameters, but this behaviour is typical of all the values that were tested. 

Additionally from a theoretical point of view, the sections above  devoted to 
first-order 
scalar perturbations, when expressed in the  $1+3$ formalism, show how this 
formalism allows us to perform a simple treatment of matter power spectra 
observables after solving second-order coupled equations, instead of 
fourth-order counterparts, which would be required in the usual metric 
approach.

Finally, Sections   \ref{Sec:GDE} and \ref{Section:Attractive_gravity} aimed at 
illustrating how simple geometric calculations, when having promoted the usual 
GR 
action  to $f(R)$ generalisations, may shed some light on unexpected 
consequences,
provided the Einsteinian paradigm is abandoned. First, in Section  \ref{Sec:GDE} 
we provided information on how the geodesic deviation equation 
can be obtained for $f(R)$ theories. We used a $1+3$ decomposition, as explained 
in previous sections, enabling us to render intermediate calculations 
manageable and determine that the new geometrical contributions contribute to 
the deviation for both null and time-like geodesics. Consequently,
we proved that the  additional terms introduced by $f(R)$ theories, together 
with the standard matter content, impact on the evolution of the geodesic 
deviation. The well-known fact that extended gravity theories do not need to
accomplish the standard energy conditions leads the geodesic deviation equation
to exhibit a model-dependent behaviour that may be useful to  constrain the 
viability of  classes of models in such theories. Thus, such a result can be 
used to study the evolution of the deviation for null geodesics in a 
cosmological background and the subsequent $f(R)$ model-dependent numerical 
results for the area distance formula to  be tested with observational data.
 
On the other hand, Section \ref{Section:Attractive_gravity}  showed how, should 
usual energy conditions be solely imposed upon the cosmological standard fluids, 
the capability of additional $f(R)$ terms to provide cosmological acceleration 
can be studied in a straightforward and systematic approach. With such a tool, 
this latter analysis can be extended to more involved cosmological scenarios and 
other alternative gravity theories beyond the concordance model.

With this brief and humble review on $f(R)$ theories we have  intended to 
provide the 
main milestones for a layman's understanding of the $f(R)$ framework and how 
the usual 
predictions, as provided within the concordance gravitational and cosmological 
scenario, need to be revisited should they be modified in the scalar-tensor 
frame of 
gravitational theories. Once this is done, the pieces of analysis above can be 
easily extended to other more convoluted theories, different cosmological 
parameters in models therein and physically well-motivated initial and boundary 
conditions for the problems under study. Obviously, viable models need to be 
tested against, amongst others, CMB and weak gravitational lensing data, the ISW 
effect and even gravitational waves in order to place a full set of constraints 
on the 
parameter space of broad classes of $f(R)$ models. Only with such a thorough 
analysis would   the class of  most likely theories, given observational 
cosmological 
data, be determined and duly constrained.

\chapter[Horndeski/Galileon theories]{Horndeski/Galileon theories}
\label{Morunochapter}
{\em Prado Mart\'in-Moruno}

\section{From Brans--Dicke to Horndeski} 

General Relativity describes gravity as a geometrical property of the spacetime 
that has its origin in the matter  contained therein.
This idea is inspired by what is known as Mach's principle, which argues that 
inertia should be the result of the  interaction between bodies.
Nevertheless, Brans and Dicke \cite{Brans:1961sx} considered that in order to 
suitably implement this principle,  the gravitational constant $\kappa^2=8\pi G$ 
has to depend on the mass distribution and, therefore, can vary with time and 
position.
They assumed that the scalar variable in which the gravitational ``constant'' 
depends is a new scalar field serving to determine the local value of 
$G$, so $G\propto 1/\phi$, and that
the weak equivalence principle  \label{equivprinref2} should be satisfied. 
Therefore, they proposed the 
following Lagrangian density to  describe gravity:
\begin{equation}\label{LBD}
\Lag=\phi R-\frac{\omega}{\phi} \phi_{;\nu}\,\phi^{;\nu}  +\Lag_{m},
\end{equation}
where $\omega$ is a positive constant, a semicolon denotes covariant 
derivative, 
and  $\Lag_{m}$ is the Lagrangian density of all (minimally coupled) 
non-gravitational fields. 
Whereas $G\propto 1/\phi$ is one of the simplest gravitational couplings one 
could imagine, it is possible  to consider more general functions. One can have 
generalized Brans--Dicke theories of the form \label{Bransref2}
\begin{equation}\label{LF}
\Lag=F(\phi) R-\phi_{;\nu}\,\phi^{;\nu} -V(\phi)+\Lag_{m},
\end{equation}
which lead to $G\propto1/F'(\phi)$.
This Lagrangian density is equivalent to allowing the constant
$\omega$ 
to be a function of the field, adding a potential term, and then performing a 
trivial field redefinition in the Brans–Dicke Lagrangian   (\ref{LBD}). With 
that field redefinition, Lagrangian  (\ref{LBD}) can also be written in the 
form 
of (\ref{LF}) with 
$F(\phi)\propto\phi^2$.

On the other hand, even within General Relativity, it is possible to go beyond 
the 
minimally coupled canonical scalar field. 
For example, regarding the applications of scalar fields to describe dark 
energy, quintessence models, which assume a canonical  kinetic term and a 
potential, have been generalized to K-essence \cite{ArmendarizPicon:2000ah}, 
whose Lagrangian is a more general function $K(\phi, \,\phi_{;\nu}\,\phi^{;\nu} 
)$.
Furthermore, in kinetic braiding models \cite{Deffayet:2010qz} the scalar field 
Lagrangian is further generalized:
\begin{equation}\label{LKB}
\Lag_\phi= K(\phi, \,X)+G(\phi, \,X)\Box \phi,
\end{equation}
where we have defined $X=-\phi_{;\nu}\,\phi^{;\nu} /2$, and $K(\phi, \,X)$ and  
$G(\phi, \,X)$ are arbitrary functions of the field and its kinetic term. 
Even if this Lagrangian density contains second derivatives of the field, the  
combination is such that the field equation is second order  
\cite{Deffayet:2010qz};
therefore, it is free of the Ostrogradski instability \cite{Woodard:2006nt}.

How far can we go in generalizing the scalar-field Lagrangian without 
introducing  
higher derivatives in the field equations?
Galileon models \cite{Nicolis:2008in} are shift-symmetric models in flat space, 
containing second-order derivatives in the Lagrangian and in the equation of 
motion.
When considering these Galileons in a dynamical spacetime,  one has to 
introduce 
a non-minimal coupling to curvature in order to avoid the introduction of 
higher 
order derivatives in the equations \cite{Deffayet:2009wt}.
This coupling can be more general than that considered by Brans and Dicke,  
Eq. (\ref{LF}), and entail $\phi^{;\mu}\, R_{\mu\nu}\,\phi^{;\nu}$ terms.
Dropping the requirement of shift symmetry, Generalized Galileons 
\cite{Deffayet:2011gz}  comprise all models with a Lagrangian containing 
second-order derivatives of the field and with second-order equations of 
motion.
The Lagrangian density is given by \cite{Kobayashi:2011nu}  
\label{Galileonsref1}
\begin{equation}\label{LGG}
\Lag_{\rm GG}= \Lag_2+\Lag_3+\Lag_4+\Lag_5,
\end{equation}
with
\begin{eqnarray}
\Lag_2&=&K(\phi,\,X)\\
\Lag_3&=&-G_3(\phi,\,X)\square\phi\\
\Lag_4&=&G_4(\phi,\,X)R+G_{4,X}(\phi,\,X)\left[(\square\phi)^2-\phi_{;\mu\nu}
\phi^{;\mu\nu}\right]\label{L4}\\
\Lag_5&=&G_5(\phi,\,X)G_{\mu\nu}\phi^{;\mu\nu}-\frac{1}{6}G_{5,X}\left[
(\square\phi)^3+2\phi_{;\mu}{}^\nu\phi_{;\nu}{}^\alpha\phi_{;\alpha}{}
^\mu-3\phi_{;\mu\nu}\phi^{;\mu\nu}\square\phi\right],
\label{L5}
\end{eqnarray}
with $K(\phi,\,X)$, $G_3(\phi,\,X)$, $G_4(\phi,\,X)$, and $G_5(\phi,\,X)$ being 
arbitrary functions and a comma representing a derivative.
Note that $\Lag_2$ is K-essence, $\Lag_2+\Lag_3$ is kinetic braiding 
(\ref{LKB}), and the Brans--Dicke theory, given by Lagrangian (\ref{LF}), is 
just a particular case of $\Lag_2+\Lag_4$.
 Generalized Galileons are equivalent \cite{Kobayashi:2011nu} to Horndeski 
theory 
\cite{Horndeski:1974wa}, which was constructed in 1974, but it was also 
practically 
forgotten for almost 40 years.
Horndeski Lagrangian density is \cite{Horndeski:1974wa} \label{Horndeskiref1}

\begin{eqnarray}\label{LH}
\mathcal{L}_{\rm H}&=&
\delta^{\alpha\beta\gamma}_{\mu\nu\sigma}\left[\kappa_1\left(\phi,\,
X\right)\phi^{;\mu}{}_{;\alpha} \,R_{\beta\gamma}{}^{\nu\sigma}
+\frac{2}{3}\kappa_{1,X}\left(\phi,\,X\right)\phi^{;\mu}{}_{;\alpha} 
\,\phi^{;\nu}{}_{;\beta}   \,   \phi^{;\sigma}{}_{;\gamma}   \right.\nonumber\\
&+&\left.\kappa_3\left(\phi,\,X\right)    \phi_{;\alpha} \,  \phi^{;\mu}    
R_{\beta\gamma}{}^{\nu\sigma}
+2\kappa_{3,X}\left(\phi,\,X\right) \phi_{;\alpha}\phi^{;\mu }    \,  
\phi^{;\nu}{}_{;\beta}  \,  \phi^{;\sigma}{}_{;\gamma}\right]\nonumber\\
&+&\delta_{\mu\nu}^{\alpha\beta}\left[F\left(\phi,\,X\right)\,R_{\alpha\beta}{}^
{\mu\nu}+2F_{,X}\left(\phi,\,X\right) \phi^{;\mu}{}_{;\alpha}\, 
\phi^{;\nu}{}_{;\beta}
+2\kappa_8\left(\phi,\,X\right)\phi_{;\alpha}\,   \phi^{;\mu}\, 
\phi^{;\nu}{}_{;\beta}\right]\nonumber\\
&-&6\left[F_{,\phi}\left(\phi,\,X\right)-X\,\kappa_8\left(\phi,\,X\right)\right]
\phi^{;\mu}{}_{;\mu}
+\kappa_9\left(\phi,\,X\right),
\end{eqnarray}
where $\kappa_i\left(\phi,\,X\right)$ are arbitrary functions (we have absorbed 
an additional $W(\phi)$ function in $F(\phi,\,X)$) \cite{Kobayashi:2011nu}, and
\begin{equation}\label{condF}
F_{,X}=2\left(\kappa_3+2X\kappa_{3,X}-\kappa_{1,\phi}\right).
\end{equation}
The antisymmetric character of the generalized delta functions in Lagrangian 
(\ref{LH}) is responsible for removing derivatives of an order higher than two
from 
the equations.
The dictionary between Lagrangians (\ref{LGG}) and (\ref{LH}) is 
\cite{Kobayashi:2011nu}
\begin{eqnarray}
K&=&\kappa_9+4X\int^X{d} X'\left(\kappa_{8,\phi}-2\kappa_{3,\phi\phi}\right),
\\
G_3&=&
6F_{,\phi}-2X\kappa_8-8X\kappa_{3,\phi}+2\int^X{d} 
X'(\kappa_8-2\kappa_{3,\phi}),
\\
G_4&=&2F-4X\kappa_3,
\\
G_5&=&-4\kappa_1\ .\label{dictionary}
\end{eqnarray}
Horndeski Lagrangian is the most general one leading to second-order equations 
of motion. 
However, it is worth noting that there are theories leading to higher-order 
equations of motion that, even so, avoid the Ostrogradski instability 
\cite{Zumalacarregui:2013pma,Gleyzes:2014dya,Gleyzes:2014qga}.
These scalar-tensor theories are known as beyond 
Horndeski.\label{Scalartensoref2}

\section{Background Cosmology}

Focusing our attention on cosmological models, we consider a spatially  flat 
Friedman-Lema\^itre-Robertson-Walker metric
\begin{equation}\label{FLRW}
ds^2=-dt^2+a(t)^2 d\vec{x}^2 ,
\end{equation}
where $a(t)$ is the scale factor. Deriving the field equations from the 
Horndeski  Lagrangian and then restricting to this highly symmetric background 
is equivalent to 
obtaining the point-like Lagrangian, which is defined in the minisuperspace 
formed by  the scale factor and the homogeneous field, $\{a,\,\phi\}$, and then 
deriving the equations of motion.
This is 
\begin{equation}\label{pL}
L_{\rm H}=\mathcal{V}_{(3)}^{-1}\int {d}^3 x\,\mathcal{L}_H,
\end{equation}
with $\mathcal{V}_{(3)}$ the spatial three-volume element.
Second derivatives in the point-like Lagrangian obtained from (\ref{LH}) can be 
removed by integrating  by parts \cite{Charmousis:2011ea}. That Lagrangian then 
takes the simple form \cite{Charmousis:2011ea}
\begin{equation}\label{Lsimple}
L_{\rm H}\left(\phi,\,\dot\phi,\,a,\,\dot 
a\right)=a^3\sum_{i=0..3}X_i\left(\phi,\,\dot\phi\right)\,H^i,
\end{equation}
where $H=\dot a/a$ is the Hubble parameter, an over-dot represents a derivative 
with respect to the  cosmic time $t$, and the functions $X_i$ are 
\cite{Charmousis:2011ea,Martin-Moruno:2015bda}
\begin{eqnarray}
X_0&=&-\widebar Q_{7,\phi}\dot\phi+\kappa_9,\label{X0}\\
X_1&=&-3\,\widebar Q_7+Q_7\dot\phi,\label{X1}\\
X_2&=&12\,F_{,X}X-12\,F,\label{X2}\\
X_3&=&-4\,\kappa_{1,X}\,\dot\phi^3,\label{X3}
\end{eqnarray}
where
\begin{eqnarray}
Q_7&=&\widebar Q_{7,\dot\phi}=6\,F_{,\phi}-3\,\dot\phi^2\kappa_8.\label{Q7}
\end{eqnarray}

As there is an Einstein--Hilbert term contained in the Horndeski Lagrangian  
(\ref{LH}), we can extract it from (\ref{Lsimple}) and express it explicitly:
\begin{equation}\label{Ltot}
L=L_{\rm EH}+ L_{\rm H}+ L_{m},
\end{equation}
with
\begin{equation}
L_{\rm EH}=- \frac{3}{\kappa^2}  a^3H^2
\quad {\rm and}\quad
L_{m}=-a^3\rho(a)=-a^3\sum_i\rho_i(a),
\end{equation}
where $\rho(a)$ is the  conserved total energy density, $\rho_i(a)$ corresponds 
to each cosmic fluid.
Therefore, the modified Friedmann equation and the field equation are
\begin{equation}\label{Friedmann}
\frac{3H^2}{\kappa^2}=
\sum_{i=0..3}\left[(i-1)X_i+X_{i,\dot\phi}\dot\phi\right]H^i+\rho(a),
\end{equation}
and
\begin{eqnarray}\label{field}
\sum_{i=0}^{3}   \left[X_{i,\phi}-3X_{i,\dot\phi}H- iX_{i,\dot\phi}\frac{\dot 
H}{H}-
X_{i,\dot\phi \phi}\dot\phi-X_{i,\dot\phi\dot\phi}\ddot\phi\right] H^i=0,
\end{eqnarray}
respectively. These equations can also be expressed using the Generalized 
Galileons  functions of Lagrangian density (\ref{LGG}). These are 
\cite{Kobayashi:2011nu}
\begin{equation}
\sum_i\mathcal{E}_i=0,
\end{equation}
with
\begin{eqnarray}
\mathcal{E}_2&=&2XK_{,x}-K\\
\mathcal{E}_3&=&6X\dot\phi H G_{3,X}-2XG_{3,\phi}\\
\mathcal{E}_4&=&-6H^2G_4+24H^2X(G_{4,X}+XG_{4,XX})-12H\dot\phi \,G_{4,\phi 
X}\nonumber\\
&&-6H\dot\phi\, G_{4,\phi}\\
\mathcal{E}_5&=&2H^3X\dot\phi \,(5G_{5,X}+2XG_{5,XX})-6H^2X 
(3G_{5,\phi}+2XG_{5,\phi X}),
\end{eqnarray}
and
\begin{eqnarray}
\frac{d}{dt}(a^3J)=a^3P_{,\phi},
\end{eqnarray}
with
\begin{eqnarray}
J&=&\dot\phi \,K_X+6HXG_{3,X}-2\dot\phi\,G_{3,\phi}+6H^2\dot\phi 
\,(G_{4,X}+2XG_{4,XX})-12HXG_{4,\phi X}\nonumber\\
&&+2H^3X\,(3G_{5,X}+2XG_{5,XX})-6H^2\dot\phi\,(G_{5,\phi}+XG_{5,\phi X})
\end{eqnarray}
and
\begin{eqnarray}
P_{,\phi}&=&K_{,\phi}-2X\,(G_{3,\phi \phi}+\ddot\phi\,G_{3,\phi 
X})+6G_{4,\phi}(2H^2+\dot H)+6HG_{4,\phi X}(\dot X+2HX)\nonumber\\
&&-6H^2XG_{5,\phi \phi}+2H^3X\dot\phi \,G_{5,\phi X}.
\end{eqnarray}
This field equation shows the special interest of shift-symmetric models 
\label{siftsymref1}. For 
these models, the  equations are invariant under a field redefinition 
$\phi\rightarrow\phi+c$. Therefore, $\mathcal{P}_{,\phi}=0$
and there is a conserved quantity, $\Sigma=Ja^3$; therefore, 
$J(\dot\phi,\,H)=\Sigma \, a^{-3} \rightarrow0$ when the scale factor goes to 
infinity \cite{Kobayashi:2011nu}.
Any trajectory in the phase space $\dot\phi(H,\,a)$, which can be obtained from 
condition  $J(\dot\phi,\,H)=\Sigma \, a^{-3}$, will tend to the \emph{tracker 
solution} $\dot\phi_{\rm tracker}(H)$, 
obtained from $\Sigma=0$ \cite{Germani:2017pwt}. Therefore, any attractor will 
be a point in the trajectory $J=0$, approached by any trajectory in the phase 
space. De Sitter attractors are of particular interest to describe the 
late-time 
cosmic acceleration \cite{Germani:2017pwt}.
Many interesting studies have investigated the potential of Horndeski theory to 
describe the late-time cosmic acceleration 
\cite{Silva:2009km,DeFelice:2010pv,DeFelice:2010nf,Appleby:2011aa,
Martin-Moruno:2015lha,Martin-Moruno:2015kaa},
with emphasis on the prediction of a de Sitter attractor,
and early-time cosmic acceleration 
\cite{Kobayashi:2011nu,Easson:2011zy,Qiu:2015nha}, pointing out the potential 
avoidance of the Big Bang singularity. It is impossible to include a complete 
list of references in the present paper.

\section{Cosmological Perturbations}

We have shown how one can study the background cosmic evolution described by 
Horndeski models. The next step should be, therefore, to investigate the 
stability of the obtained cosmic models.
In order to simplify the study of cosmic perturbations, Bellini and Sawicki 
introduced four useful functions that encapsulate all the necessary physics in 
reference~\cite{Bellini:2014fua}. Defining first the effective reduced Planck 
mass
\begin{equation}
M_* ^2=2\,(G_4-2XG_{4,X}+XG_{5,\phi}-\dot\phi \,HXG_{5,X}),
\end{equation}
where we remind the reader that $M_*^2$ is inversely proportional to the 
gravitational ``constant'',
these functions are:
\paragraph{Planck-mass run rate, $\alpha_{m}$}. The rate of evolution of the 
effective reduced Planck mass is
\begin{equation}
\alpha_{m}=H^{-1}\frac{d\,\ln\,M_*^2}{dt}.
\end{equation}

\paragraph{Kineticity, $\alpha_{\rm K}$}. The kinetic energy of scalar 
perturbations 
arising directly from the action
\begin{eqnarray}
H^2M_*^2\alpha_{\rm K}&=&2X\,(K_{,X}+2XK_{,XX}-2G_{3,\phi}-2XG_{3,\phi 
X})\nonumber\\
&&+12\dot\phi XH(G_{3,X}+XG_{3,XX}-3G_{4,\phi X}-2XG_{4,\phi XX})\nonumber\\
&&+12XH^2(G_{4,X}+8XG_{4,XX}+4X^2G_{4,XXX})\nonumber\\
&&-12XH^2(G_{5,\phi}+5XG_{5,\phi X}+2X^2G_{5,\phi XX})\nonumber\\
&&+4\dot\phi XH^3(3G_{5,X}+7XG_{5,xx}+2X^2G_{5,XXX}).
\end{eqnarray}

\paragraph{Braiding, $\alpha_{\rm B}$}. The mixing of the kinetic terms of the 
scalar 
and the metric is
\begin{eqnarray}
H M_*^2\alpha_{\rm B}&=&2\dot\phi (XG_{3,X}-G_{4,\phi}-2XG_{4,\phi 
X})\nonumber\\ 
&&+8XH(G_{4,X}+2XG_{4,XX}-G_{5,\phi}-XG_{5,\phi X})\nonumber\\
&&+2\dot\phi X H^2(3G_{5,X}+2XG_{5,XX}).
\end{eqnarray}

\paragraph{Tensor speed excess, $\alpha_{\rm T}$}. The deviation of the speed  
\label{gravitationalwavrefs4}
of 
gravitational waves from that of light is
\begin{equation}
M_*^2\alpha_{\rm T}=2X(2G_{4,X}-2G_{5,\phi}-(\ddot\phi-\dot\phi H)G_{5,X}).
\end{equation}
Then, the quadratic action for tensor and scalar cosmological perturbations for 
Horndeski theory can be expressed as \cite{Kobayashi:2011nu,Bellini:2014fua}
\begin{equation}
S_{(2)}=\int dt\,d^3x\,a^3\left\{Q_{\rm S}\left[\dot\zeta^2-\frac{c_{\rm 
S}^2}{a^2}(\partial_i\zeta)^2\right]+Q_{\rm T}\left[\dot h_{ij}^2-\frac{c_{\rm 
T}^2}{a^2}(\partial_k h_{ij})^2\right]\right\},
\end{equation}
where $h_{ij}$ and $\zeta$ denote the tensor and scalar perturbations, 
\label{Scalarperfrref2} \label{tensorpertref3}
respectively, and
\begin{eqnarray}
&&
\!\!\!\!\!\!\!\!\!\!\!\!\!\!\!
Q_{\rm S}=\frac{2M_*^2(\alpha_{\rm K}+3\,\alpha_{\rm B}^2/2)}{(2-\alpha_{\rm 
B})^2}>0,\\
&&
\!\!\!\!\!\!\!\!\!\!\!\!\!\!\!
c_{\rm S}^2=-\frac{(2-\alpha_B)[\dot H-H^2\alpha_{\rm 
B}(1+\alpha_{\rm T})/2-H^2(\alpha_{m}-\alpha_{\rm T})]-H\dot\alpha_{\rm 
B}+(\rho+p)/M_*^2}{H^2(\alpha_{\rm K}+3\,\alpha_{\rm B}^2/2)}>0,\\
&&
\!\!\!\!\!\!\!\!\!\!\!\!\!\!\!
Q_{\rm T}=\frac{M_*^2}{8}>0\\
&&
\!\!\!\!\!\!\!\!\!\!\!\!\!\!\!
c_{\rm T}= 1+\alpha_{\rm T}>0.
\end{eqnarray}
As stated, these four functions have to be positive to avoid ghost and gradient 
instabilities.

It can be noted that the four $\alpha$-functions, together with the function 
$H(t)/H_0$ and the constant $\Omega_{{m},0}$, encapsulate all the cosmic 
information of Horndeski models \cite{Bellini:2014fua}. Probably the most useful 
application of this formalism has been the development of \texttt{hi$\_$class}, 
which is an extension of the Boltzmann code CLASS to calculate predictions of 
Horndeski models \cite{Zumalacarregui:2016pph}. An alternative approach 
following the dark energy effective field theory (EFT) program 
\cite{Gubitosi:2012hu}, which includes Horndeski theory, is the extension of the 
code CAMB, leading to EFTCAMB \cite{Hu:2013twa}.

\section{Gravitational Waves Constraints}
\label{gravitationalwavrefs1}
As we have reviewed, we have all the tools needed to investigate the cosmic  
implications of complicated Horndeski models, but how complicated can those 
models 
be? The detection of a gravitational wave signal with an electromagnetic 
counterpart (GW170817) produced by a binary neutron star merger imposed strong 
constraints on some of the functions of Horndeski theory. Therefore, that event 
limited the application of Horndeski models to describe the late-time 
acceleration of the Universe 
\cite{Ezquiaga:2017ekz,Baker:2017hug,Creminelli:2017sry}. In particular, the 
detection of both signals led to \cite{Ezquiaga:2017ekz}
\begin{equation}
|c_{\rm T}/c-1|\leq5\cdot10^{-16},
\end{equation}
where we have explicitly written the speed of light $c=1$. Therefore, $\alpha_T$ 
has  to vanish during the recent cosmic evolution. (Potential caveats of this 
result have been pointed out in references~\cite{Baker:2017hug,deRham:2018red}.) 
This result can be accommodated in Horndeski theory, requiring
\begin{equation}\label{co}
G_{4,X}\approx0,\quad{\rm and}\quad G_5\approx{\rm constant},
\end{equation}
in Lagrangian (\ref{LGG}). Indeed, such kinds of constraints had already been 
suggested, taking into account indirect measures of gravitational waves 
\cite{Jimenez:2015bwa}.
Introducing the second expression of (\ref{co}) in (\ref{L5}) and using the 
Bianchi identity,  we can conclude that (\ref{L5}) is just equivalent to a 
total derivative; therefore, it can be dismissed. The first expression in 
(\ref{co}) implies that
\begin{equation}
G_{4}=F(\phi),
\end{equation}
and therefore we can consider non-minimal coupling to gravity only if it is 
of the 
generalized  Brans--Dicke form (\ref{LF}). Thus, for applications to the 
late-time 
cosmic evolution, we can just consider
\begin{equation}\label{LGG2}
\Lag_{\rm GG}=K(\phi,\,X)-G_3(\phi,\,X)\square\phi+F(\phi)R.
\end{equation}
Moreover, it is worth noting that the recent values of $F(\phi)$ are also 
constrained by  limits on the potential evolution of the gravitational 
``constant'' \cite{Damour:1992kf}. Those constraints have been obtained in the 
absence of screening mechanisms, which have been argued to be absent in 
self-accelerating surviving models \cite{Baker:2017hug}. Future observational 
data will reveal to us if the evolution of $F(\phi)$ is constrained in such a 
way to 
effectively lead only to GR plus a (potentially non-minimally coupled) scalar 
field, if generalized Brans--Dicke theory (with kinetic braiding) plays an 
important role in our Universe, or if we should go beyond the simple scalar 
field hypothesis by taking into account additional degrees of freedom or even 
revisiting our assumptions about the spacetime.

\chapter[Massive Gravity and Bimetric Gravity]{Massive Gravity and Bigravity}
\label{Heisenbergchapter}
 
{\em Lavinia Heisenberg} \\

The attempt at altering the underlying fundamental theory of gravity  relies 
strongly on abandoning one of the defining properties of General Relativity. 
Essentially, one either revises the assumed symmetries or the elemental 
geometries (see, for instance, \cite{Joyce:2014kja,Heisenberg:2018vsk}). 
General 
Relativity represents the fundamental theory of a Lorentz invariant local  
\label{loclinref3}
massless spin-2 field in the field-theoretical representation or a (pseudo-) 
Riemanian manifold with a fully determined connection in the geometrical 
formulation. Many efforts have been dovetailed into a unified endeavour of 
Lorentz-breaking theories \cite{Sotiriou:2010wn,Sotiriou:2009bx}. A similar 
undertaking was devoted to non-local theories. Apart from explicitly breaking 
Locality and Lorentz symmetry, one can alter the intrinsic geometrical 
properties. 

In the geometrical representation one can attribute gravity either 
to curvature, to torsion or to non-metricity \cite{BeltranJimenez:2019tjy}. In 
this context, models based on an arbitrary function of the curvature scalar 
$f(R)$ \cite{Sotiriou:2008rp,DeFelice:2010aj}, the torsion scalar $f(T)$ 
\cite{Li:2010cg,Cai:2015emx} and non-metricity scalar $f(Q)$ 
\cite{BeltranJimenez:2017tkd,Jimenez:2019ovq} received some attention. On the 
other hand, in the field-theoretical formulation, modifying gravity implies the 
introduction of additional degrees of freedom. These are typically   
additional 
scalar, vector or tensor fields. For instance, if one is willing to abandon the 
gauge symmetry of the massless spin-2 field, this introduces three additional 
propagating degrees of freedom in massive gravity. Since effective field 
theories for a 2-form naturally appear in string theory, one can also consider 
gravity theories in the presence of 2-forms. However, they will be either dual 
to a scalar field or to a vector field. 

The simplest modification of gravity relies on  the introduction of a scalar 
field. Without altering the fundamental symmetries of the spin-2 field, the most 
general scalar-tensor theories, the Horndeski theories, can successfully be 
constructed \cite{Horndeski:1974wa}. Even though they contain derivative 
self-interactions and non-minimal couplings, they yield second-order equations 
of motion. An analogue extension of gravity theories follows the idea of 
introducing an additional vector field into the gravity sector instead of a 
scalar field. In this way, one can establish the most general vector-tensor 
theories with derivative self-interactions but still second-order equations of 
motion. They constitute the generalised Proca theories with genuinely new vector 
interactions \cite{Heisenberg:2014rta,Jimenez:2016isa}, with tremendous 
implications for astrophysical and cosmological probes beyond the Horndeski 
case. The rich phenomenology of Horndeski and generalized Proca theories can be 
merged together in the form of scalar-vector-tensor theories. Furthermore, new 
effects emerge, due to the presence of genuinely new scalar-vector couplings.

\section{Massive and Bimetric Gravity}
\label{introsec}

Considering gravity in the field-theoretical framework,  a natural theoretical 
question arises to whether the graviton particle itself could be a massive 
particle. In the same way as the particles in the standard model acquire a mass, 
could the spin-2 particle be endowed with a mass term? The construction of a 
mass term for spin-1 particles is straightforward. One simply uses the inverse 
Minkowski metric in order to build a Lorentz scalar $A_\mu A_\nu \eta^{\mu\nu}$. 
The construction of a mass term for the spin-2 field requires the presence of an 
additional metric. This could be just a reference metric, as in massive gravity 
\cite{deRham:2010kj}, or a fully dynamical metric, as in bigravity 
\cite{Hassan:2011zd}. 

The action for ghost-free massive gravity is given by \cite{deRham:2010kj}
\begin{equation}
S = S_\mathrm{mG} + S_\mathrm{matter},
\end{equation}
with the gravity sector enforced to have the form\label{massivegravrefs1}
\begin{equation}
S_\mathrm{mG} =  \int\mathrm{d}^4x\left[\frac{1}{2\kappa^2}\sqrt{-g}R_g - 
\frac{m^2}{\kappa^2} 
\sqrt{-g}\displaystyle\sum_{n=0}^4\alpha_ne_n\left(\sqrt{g^{-1}f} 
\right)\right]\,,
\end{equation}
with $\kappa^2=1/M_{Pl}^2$ the gravitational constant and $M_{Pl}$ the reduced 
Planck 
mass.
One can generalize the action for ghost-free massive gravity into bigravity by 
including  an Einstein-Hilbert kinetic term $\frac{M_f^2}{2}\sqrt{-f}R_f$ for 
the reference metric. The two metrics, $g$ and $f$, are only allowed to interact 
through potential interactions encoded in the elementary symmetric polynomials 
$e_n(S)$ of the matrix square root $S^\mu_\nu = (\sqrt{g^{-1}f})^\mu_\nu$, which 
satisfies $S^\mu_\alpha S^\alpha_\nu \equiv g^{\mu\alpha}f_{\alpha\nu}$. The 
polynomials take
\begin{align}
e_0(S) &= 1, \nonumber\\
e_1(S) &= [S], \nonumber\\
e_2(S) &= \frac12\left(S]^2-[S^2]\right), \nonumber\\
e_3(S)  &= \frac16\left([S]^3-3[S][S^2]+2[S^3]\right), \nonumber\\
e_4(S)  &= \operatorname{det}(S).
\end{align}
One can rewrite the ghost-free interactions of massive gravity in terms of a 
deformed  determinant \cite{Hassan:2011vm}
\begin{eqnarray}
{\rm det}( \delta^\mu_\nu+S^\mu_\nu)=\sum_{i=0}^4\frac{-\alpha_i} 
{i!(4-i)!}\epsilon_{\mu_1\cdots \mu_i \alpha_{i+1}\cdots 
\alpha_4}\epsilon^{\nu_1\cdots \nu_i \alpha_{i+1}\cdots \alpha_4}
S^{\mu_1}_{\nu_1}\cdots S^{\mu_i}_{\nu_i},
\end{eqnarray}
where the antisymmetric structure of the elementary polynomials becomes transparent
\begin{align}\label{potentialsdRGT}
e_0[S] &= \epsilon^{\mu\nu\rho\sigma}   \epsilon_{\mu \nu\rho\sigma}, 
\nonumber\\
e_1[S] &= \epsilon^{\mu\nu\rho\sigma}   \epsilon^{\alpha}_{\quad \nu\rho\sigma} 
S_{\mu\alpha}, \nonumber\\
e_2[S] &= \epsilon^{\mu\nu\rho\sigma}   \epsilon^{\alpha\beta}_{\quad 
\rho\sigma} S_{\mu\alpha} S_{\nu\beta}, \nonumber\\
e_3[S] &=\epsilon^{\mu\nu\rho\sigma}   \epsilon^{\alpha\beta\kappa}_{\qquad 
\sigma} S_{\mu\alpha} S_{\nu\beta}  S_{\rho\kappa},  \nonumber\\
e_4[S] &=\epsilon^{\mu\nu\rho\sigma}  \epsilon^{\alpha\beta\kappa\gamma} 
S_{\mu\alpha} S_{\nu\beta}  S_{\rho\kappa}  S_{\sigma\gamma} \,.
\end{align}
The antisymmetric nature of the interactions together with the square root  
structure of the fundamental matrix render the Boulware-Deser ghost 
non-dynamical. In the case of only $g$ being dynamical, five physical degrees 
of 
freedom propagate, which can be decomposed into two helicity-2, two helicity-1, 
and one helicity-0 mode\label{helicityref1}. Promoting $f$ to be   dynamical 
too 
introduces two 
additional 
degrees of freedom, so that bigravity\label{bigravityrefs1} contains seven 
modes 
in its spectrum of 
one 
massless and one massive graviton. 

One can grasp some of the fundamental properties of massive gravity already in 
the  leading order interactions of the decoupling limit, i.e., the limit in 
which 
$m\to 0, M_{\rm Pl}\to \infty$, while keeping $\Lambda_3\equiv (M_{\rm Pl} 
m^2)^{1/3}$ fixed. Concentrating on the leading-order contributions of the 
helicity-2 and helicity-0 modes, the decoupling limit Lagrangian simplifies into 
\cite{deRham:2010ik}
\begin{equation}
\mathcal{L}=-\frac{1}{2} h^{\mu\nu}\mathcal{E}^{\alpha\beta}_{\mu\nu}  
h_{\alpha\beta}+ h^{\mu\nu}\sum_{n=1}^3 \frac{a_{n}}{\Lambda^{3(n-1)}_3} 
X^{(n)}_{\mu\nu} \left(\Pi\right),
\label{lagr1}
\end{equation}
where the helicity-2 mode $ h^{\mu\nu}$ couples to the helicity-0 mode through  
the matrices $X$'s
\begin{eqnarray}
X^{(1)}_{\mu\nu}\left(\Pi\right)&=&{\epsilon_{\mu}}^{\alpha\rho\sigma}
{{\epsilon_\nu}^{\beta}}_{\rho\sigma}\Pi_{\alpha\beta}, \quad  \nonumber \\
X^{(2)}_{\mu\nu}\left(\Pi\right)&=&{\epsilon_{\mu}}^{\alpha\rho\gamma}
{{\epsilon_\nu}^{\beta\sigma}}_{\gamma}\Pi_{\alpha\beta}
\Pi_{\rho\sigma}, \nonumber \\
X^{(3)}_{\mu\nu}\left(\Pi\right)&=&{\epsilon_{\mu}}^{\alpha\rho\gamma}
{{\epsilon_\nu}^{\beta\sigma\delta}}\Pi_{\alpha\beta}
\Pi_{\rho\sigma}\Pi_{\gamma\delta}\, ,
\label{Xs}
\end{eqnarray}
with the fundamental tensor $\Pi_{\mu\nu}=\partial_\mu\partial_\nu \pi$.
One immediate property of the   decoupling limit Lagrangian is the absence of 
any kinetic  term for the helicity-0 mode. One can diagonalise the first two 
interactions $hX^{(1)}$ and $hX^{(2)}$, which yields the appearance of Galileon 
interactions. They become strongly coupled at an energy scale $E\sim \Lambda_3$. 
These leading order interactions of the decoupling limit are characterised by 
the invariance under global field-space Galilean transformations 
$\pi\to\pi+b_\mu x^\mu+b$ of the helicity-0 mode and the linearised 
diffeomorphisms $h_{\mu\nu}\to h_{\mu\nu}+\partial_{(\mu}\xi_{\nu)}$ up to total 
derivatives of the helicity-2 mode.\\

The construction of the potential interactions in terms of the square root and 
the  relative tunings of $[S]^n$ is crucial for the absence of the 
Boulware-Deser ghost at all orders. In analogy to Galileon interactions, the 
leading-order interactions in the decoupling limit of massive gravity are not 
subject to large quantum corrections. Behind this is a non-renormalisation 
theorem: the classical operators of the decoupling limit Lagrangian are 
invariant under the aforementioned symmetries only up to total derivatives, 
whereas the Feynman diagrams generate operators that fulfill these symmetries 
exactly. Hence, the classical operators and their relative tunings in the 
high-energy limit remain unrenormalised. This indicates that the quantum 
corrections to the parameters of the theory (including the graviton mass) will 
receive small quantum corrections proportional to the graviton mass itself  
\cite{deRham:2012ew}. This property of technical naturalness is one of the most 
attractive characteristic of the theory. However, it still might be a matter of 
concern that    
the specific structure of the potential interactions might receive a detuning 
from quantum corrections beyond the decoupling limit. This has  led to the 
analysis 
of the radiative stability of the full theory 
\cite{Buchbinder:2012wb,deRham:2013qqa}. It transpires that the graviton loops 
do 
indeed destabilize the precise structure of the potential interactions, but in 
a 
way that the Boulware-Deser ghost remains harmless below the Planck scale, at 
least up to one loop \cite{deRham:2013qqa,Heisenberg:2014rka}. 

The linear theory of massive gravity suffers from the vDVZ discontinuity  
\cite{vanDam:1970vg,Zakharov:1970cc}, which is tightly related to the fact that 
the helicity-0 mode does not have its own kinetic term. It emerges only after 
diagonalising its mixing with the helicity-2 mode, which on the other hand 
results in a direct coupling of the helicity-0 mode to external matter fields. 
This coupling hinders the recovery of General Relativity in the vanishing mass 
limit. However, going beyond linear theory brings the rescue, where non-linear 
interactions for the massive graviton introduce Galileon-type derivative 
interactions for the helicity-0 mode. In this way, the same Vainshtein 
mechanism 
as for the Galileon scalar field cures this discontinuity: in the vicinity of 
matter, the non-linear interactions for the helicity-0 mode become large and 
ultimately suppress its coupling to matter. For a successful implementation of 
the Vainshtein mechanism in massive gravity, see, for instance, 
\cite{Babichev:2009jt,Babichev:2010jd}.

\section{Cosmological Applications}
\label{introsec}

The first attempt to apply massive gravity to cosmology  was pursued in 
\cite{deRham:2010tw}. There, self-accelerated solutions were successfully found 
in the decoupling limit of massive gravity, where the de Sitter metric was 
treated as a small perturbation on top of the Minkowski metric. An immediate 
observation was that the Hubble parameter is set by the graviton mass. As 
promising as these solutions were for the late-time acceleration, the 
helicity-1 
mode caused worries due to strong coupling issues. Cosmological solutions have 
then
been investigated beyond the decoupling limit in the full theory. Assuming 
a FLRW Ansatz for the metric $d_s^2=-N^2dt^2+a^2d\vec{x}^2$, the equations of 
motion immediately yield the Bianchi identity
\begin{equation}
m^2 \partial_0(a^3-a^2)=0,
\end{equation}
which is at the heart of the no-go result for flat FLRW solutions in massive 
gravity  \cite{DAmico:2011eto}. A similar no-go result is also there for 
spatially-closed FLRW solutions. On the other hand, open FLRW solutions do 
exist 
\cite{Gumrukcuoglu:2011ew}, in which $J=0$ (that depends on the parameters of 
massive gravity and the scale factor), even though their perturbations suffer 
either from strong coupling (the kinetic terms for the vector and scalar modes 
are proportional to $J$) or instabilities \cite{Gumrukcuoglu:2011zh}.

Promoting the reference metric to a dynamical metric, as in bigravity, brought  
some excitements in the cosmological applications of massive gravity 
\cite{Comelli:2011zm,vonStrauss:2011mq,Akrami:2012vf,Akrami:2013ffa,
Solomon:2014dua,Enander:2015vja}. A similar FLRW Ansatz can be assumed for the 
$f$ metric $d_{s{_f}}^2=-N_f^2dt^2+a_f^2d\vec{x}^2$. In this case, the above 
Bianchi identity changes into
\begin{equation}
J(H_g-\xi H_f)=0\,,
\end{equation}
where $H_g=\dot{a}_g/(N_ga_g)$, $H_f=\dot{a}_f/(N_fa_f)$ and $J$ depends on  the 
scale factors and the parameters of the theory. Hence, there are two branches of 
solutions. 

The first branch, $J=0$, is characterised by the fact that the ratio of 
the scale factors is constant, which enforces the potential interactions to 
contribute in the form of cosmological constants. Unfortunately, linear 
perturbations reveal that this branch suffers from strong coupling, due to 
vanishing kinetic terms for the vector and scalar modes. Therefore, this branch 
of solutions was disregarded quickly in the literature.

A more promising branch 
is $H_g=\xi H_f$, where the ratio of the scale factor $\xi$ can evolve in time. 
Two specific evolutions gained much attention in the literature 
\cite{Konnig:2014xva}. The infinite branch of solutions, with $\xi$ evolving 
from 
infinity to a finite value, suffers from ghost instabilities 
\cite{DeFelice:2014nja,Cusin:2014psa}. The finite branch of solutions, where 
the 
ratio evolves from zero to a finite value, is plagued by gradient instabilities 
\cite{Comelli:2011zm,Comelli:2012db,Comelli:2014bqa}. The gradient instabilities 
could be avoided by fine-tuning the mass $m \gg H $ 
\cite{DeFelice:2013nba,DeFelice:2014nja} or imposing $M_g \gg M_f$ 
\cite{Akrami:2015qga}.

More promising cosmological solutions can be found if the minimal coupling of 
the  matter fields to the $g$ metric is abandoned. Without introducing any ghost 
degrees of freedom within the decoupling limit,  the dark sector can be coupled 
to an effective metric 
\cite{deRham:2014naa,Enander:2014xga,Solomon:2014iwa,Akrami:2018yjz}
 \begin{eqnarray}\label{effective_metric}
g_{\mu\nu}^{\rm eff}=\alpha^2g_{\mu\nu}+2\alpha\beta  g_{\mu\rho}\left( 
\sqrt{g^{-1}f} \right)^\rho_\nu+\beta^2 f_{\mu\nu}.
\end{eqnarray}
In this way the constraint equation in massive gravity $J=0$ becomes
\begin{equation}
m^2J=\frac{\alpha\beta a_{\rm eff}^2\kappa^2}{a^2} p_m\,,
\end{equation}
where $p_m$ is pressure of the matter field that couples to  $g_{\mu\nu}^{\rm 
eff}$ and  $a_{\rm eff}=\alpha a+\beta$. Not only is the no-go theorem for flat 
FLRW solutions   circumvented, but  all the five physical degrees of freedom 
have 
non-vanishing kinetic terms and   strong coupling issues  are avoided. 
Similarly, 
bigravity with doubly coupled matter fields obtains a key change in the 
constraint equation
\begin{equation}
\left(m^2J- \frac{\alpha\beta a_{\rm eff}^2}{M_g^2a^2} p_m \right) (H_g-\xi 
H_f)=0\,,
\end{equation}
where the first branch of solutions $J=\frac{\alpha\beta a_{\rm 
eff}^2}{m^2M_g^2a^2} p_m$  is free of ghost and gradient instabilities. 


\chapter[Gravity in Extra Dimensions]{Gravity in Extra Dimensions}
\label{gravextradim}

{\em Jose A. R. Cembranos}\\



A large amount of present extensions of standard physics agree on
the existence of additional spatial dimensions. However,
the idea is quite old, since it dates back to  almost one hundred
years ago, when  Kaluza and Klein \cite{Kaluza:1984ws,Klein:1926tv} tried to unify
General Relativity proposed by Einstein
\cite{Einstein:1915ca} with classic electromagnetism completed by 
Maxwell half a century before \cite{Maxwell:1865zz}.

The existence of extra dimensions is interesting from different
theoretical points of view, but it also introduces new problems or questions.
The most basic one is why they have not been observed yet. The answer to 
this question may be very different, and depending on this answer,
the phenomenology of the extra dimensions varies in a broad range.
We will briefly summarise the main extra dimensional models following the
classification introduced in \cite{Cembranos:2003PhD}.

\section{Kaluza-Klein Model}
\label{extradimrefs1}
The first historical approach to this subject requires very small extra dimensions.
If the additional dimensions are compactified within a very small size,
of the order of the Planck length ($1/M_{Pl}$), their effects are negligible. 
This hypothesis was presented in the original models proposed by Kaluza 
(1917) and Klein (1926) \cite{Kaluza:1984ws,Klein:1926tv,Bailin:1987jd}. There, 
each field has 
a Kaluza-Klein (KK) tower of states associated with it. This tower is the 
result of the 
factorisation
of the wave functions associated with 
these fields that depend in one part on the ordinary 1+3 dimensions, and in 
another
one depending  on the coordinates of 
the 
extra dimensions. 

The effect of such factorisation is that every field propagating in these types
of compact extra dimensions with periodic boundary conditions has an infinite
numerable number of states of growing mass. In the most simple case, with a
flat extra space of toroidal topology: $T^{\delta}$
($\delta=D-4\geq 1$), the mass square of the KK towers depends quadratically 
on $\delta$ integer numbers weighted with the compactification radius of
each toroid:
\begin{eqnarray}
m^2_n=\sum_{i=1}^{\delta}\frac{n_i^2}{R_i^2}.\label{SKK}
\end{eqnarray}
Within these types of models, the scales are of the order of the standard 
four-dimensional Planck mass $M_{Pl}$,
and the standard model particles are interpreted as the zero KK modes,
whose mass needs to be explained by an other mechanism, such as the Higgs
mechanism within the standard model.

On the other hand, the metric associated with the bulk space is 
interpreted by a 1+3 observer as a KK tower of spin-2
(gravitons), $\delta-1\,$ KK towers associated with 
$U(1)$ gauge bosons (graviphotons) and $\delta(\delta-1)/2\,$ KK escalar
towers (graviscalars) \cite{Han:1998sg}. This division depends on the 
characteristics
of the extra space. 

The mode with spin-2 and zero mass has exactly the same coupling as 
the ordinary graviton, whereas the higher mass states can be interpreted
as massive gravitons with five degrees of freedom each. On the other hand,
the graviphotons have an  associated   algebra related to the isometries of the
extra space, but they do not couple (linearly) to the zero KK modes.
Finally, the graviscalars are  coupled to the zero KK modes through 
their trace of the energy-momentum tensor. Indeed, the zero modes of these
graviscalars can be identified with the Goldstone bosons (GBs) associated with
dilations. Sometimes, they are called dilatons, while at other times they are 
called radions. In any case,
in order to have a viable model, these radions need to be stabilized with
an important mass in order  to fix the size of the extra dimensions.

\section{Large Extra Dimensions}

In the above models, with small additional dimensions at the Planck scale, 
physics remains unchanged up to very high energies. However, in the late
1980s   several works started to study the possibilities of 
large extra
dimensions compactified with sizes much larger than  $\sim 1/M_{Pl}$, or even 
with non-compact extra dimensions.

\subsection{Brane Worlds}
\label{braneworldsrefs1}
A brane world is characterised by the fact that the standard model content
is confined to propagate within a manifold of three spatial dimensions.
This manifold is called brane. It has a fundamental origin in some
string-inspired models \cite{Antoniadis:1998ig}. On the contrary, it has also
studied the possibility that standard model particles can be confined
in a region of the extra space  through effective actions 
\cite{Andrianov:2003hx}.

\subsubsection{ADD Model}

The first proposal of large extra dimensions was introduced by 
Arkani-Hamed, Dimopoulos and Dvali \cite{ArkaniHamed:1998nn}. The ADD brane 
world 
model is
characterised by the fact that the brane is neglected as a source of gravity 
and the background geometry is assumed to be Minkowski.
The gravitational field is the only one that propagates in the 
$\delta$ extra dimensions that are typically supposed to be toroidal and
with the same radius, $R_B$.

One of the most distinctive features of this hypothesis is the following 
relation
between the fundamental mass scale of gravity in $D$ dimensions, $M_F$, 
and the   standard 
four-dimensional Planck mass $M_{Pl}=\frac{1}{\kappa}=\frac{1}{\sqrt{8\pi G}}$: 
\begin{equation}
M_{Pl}^2 = V_\delta \, M_F^{2+\delta}, \label{add1}
\end{equation}
where $V_\delta$ is the volume of the compactified extra space. For instance,
$V_\delta = (2\pi R_B)^\delta$ for the commented toroidal case. In this sense,
the Planck mass is not fundamental but just the effective scale of
gravity in 1+3 dimensions.

Eq. \eqref{add1} implies that it is possible to reduce $M_F$ up to the 
electroweak 
scale if the extra space is large enough. In particular, inverse radii of an  
order 
between 
$R_B^{-1}\sim$ $10^{-3}$ eV and $10$MeV provide this effect for a number of 
extra dimensions between  $\delta\sim 2$ and $7$ respectively.

On the other hand, the interaction among the fields confined within the brane
is not modified. The only modifications are introduced in the coupling with the
gravitational interaction. The metric can be linearized in the following way:
\begin{equation}
g_{_{AB}} = \eta_{_{AB}} + \frac{4\sqrt{\pi}}{M_F^{1+\delta/2}} \,
h_{_{AB}}, \label{add2}
\end{equation}
in order to expand $h_{_{AB}}$ in KK modes. They are labelled with $\delta$
integer numbers ({\it n}) in order to specify the KK spectrum introduced in Eq.
(\ref{SKK}).
%
In this model, the typical mass differences are so tiny, 
$\Delta m \sim R_B^{-1}$ $\sim 10^{-9}-10\hbox{\rm\,MeV}$, 
that the spectrum can be assumed continuous at very high energies 
($E \gg R_B^{-1}$). Therefore, the multiplicity of KK gravitons for a given 
energy scales as $N(E) \sim (ER_B)^\delta$.

The most important consequence of reducing the fundamental scale of gravity 
$M_F$ 
is that new gravitational effects can be observable at this new energy scale. 
This fact
can be understood trivially in the $D$-dimensional theory, or in an effective  
1+3-dimensional theory\label{effectivetherrefs1}, since the production 
cross-section for the  real 
production 
of
KK gravitons ($m_n \leq E$) is proportional to
\begin{equation}
\sigma_{KK} \sim \frac{1}{M_{Pl}^2} (ER_B)^\delta \sim
\frac{E^\delta}{M_F^{\delta+2}}, \label{add14}
\end{equation}
where we have taken into account that every KK graviton couples in the 
same way to the brane content:
\begin{equation}
{\mathcal L} = -\frac{2\sqrt{\pi}}{M_{Pl}} h_{\mu \nu}^{(n)} T^{\mu
\nu}_{\text{SM}}. \label{add4}
\end{equation}
Here, $T^{\mu\nu}_{\text{SM}}$ is the conserved energy-momentum tensor of the 
fields contained within the brane.
These KK gravitons are unstable with a lifetime given approximately by 
$\tau_n \simeq{M_{Pl}^2}/{m_n^3}$ \cite{Han:1998sg}. It means that they can be 
considered stable for collider phenomenology, giving a typical missing energy 
signal.

An infinite KK tower leads to divergent virtual effects even at tree level.
Indeed, KK graviton radiative effects generate new four-body interactions 
among standard model particles that need to be regularised (except for $\delta 
=1$)
  \cite{Han:1998sg,Giudice:1998ck,Hewett:1998sn}.
Finally, higher KK radiative corrections introduce  modifications
to different observables, such as anomalous moments, or electroweak
precision parameters that also demand regularisation \cite{Giudice:2003tu}.

\subsubsection{RS Model}

Immediately after the ADD model became popular, Randall and Sundrum
(RS) started the study of the phenomenology of similar scenarios when
the gravity of the brane is taken into account 
\cite{Randall:1999ee,Randall:1999vf}. The RS models
of gravity are built in a 1+4-dimensional anti-de Sitter. The extra dimension
can be compactified (RS1) or can be infinite (RS2). In the first case, the
additional dimension is typically assumed to have a $S^1/Z_2$ orbifold 
topology. 
In both models the metric can be written as:
\begin{eqnarray}
G_{MN}&=& \left(
\begin{array}{cccc}
\tilde g_{\mu\nu}(x,y)&0\\
0&-1
\end{array}\right)=
\left(
\begin{array}{cccc} e^{-2k|y|}\eta_{\mu\nu}&0\\ 0&1
\end{array}\right),
\end{eqnarray}
%
where $y \in [-R_B\pi, R_B\pi) $,   $R_B$ being  the size of the 
extra dimension in the compact case, whereas  it is $R_B\rightarrow \infty$ 
for the RS2 model. The $k$ parameter is related to the non-singular
part of the curvature scalar associated with the bulk space: 
$R_{(5)}(y\neq 0,y\neq R_B\pi)=-20k^2$.

From the (1+4)-dimensional action, the relation: $M_{Pl}=[(M_F^3/k)(1-e^{-2k\pi 
R_B})]^{1/2}$, 
can be obtained on the positive tension brane. It implies that all the 
physically relevant scales 
of the model are of the same order: $k\sim M_F \sim {M}_{P}$.

For the compactified model (RS1), in addition to the 3-brane with
positive tension placed at $y=0$, it is necessary to introduce another
3-brane at $y = \pi R_B$ with opposite tension in order to have
a consistent solution to the Einstein Equations in 1+4 dimensions.

In principle, it was proposed that the standard particle content was
placed in the 3-brane with negative tension. The reason is that the 
above metric generates a natural hierarchy between typical physical scales
on the 3-brane and the Planck scale. In this case: $M_{Pl}=[(M_F^3/k)(e^{2k\pi 
R_B}-1)]^{1/2}$,
which means that the Planck mass is exponentially related with the typical
scale of the model. The original idea was to fix this scale at 
the order of the electroweak scale  ($M_F\sim 1$ TeV), so the radius of the
extra dimension needs to be stabilized to be $k R_B \sim {\cal O} (10)$. This 
is 
the most studied scenario from a phenomenological point of view.

As it happens  in the ADD model, the most important phenomenological consequence
is the potential extreme reduction of the scale at which one can be sensitive
to new gravitational effects, typically at the order of $M_F$. By 
linearly expanding the metric over the RS background:
\begin{equation}
g_{\mu \nu} = e^{-2ky}\left( \eta_{\mu \nu} +
\frac{4\sqrt{\pi}}{M_F^{3/2}} h_{\mu \nu} \right)\,, \label{rs6}
\end{equation}
and by using a proper decomposition of the gravitational field in KK modes,
it is possible to obtain the following interaction term for KK gravitons, with
the energy-momentum tensor of the fields placed on the negative tension 3-brane:
\begin{equation}
{\mathcal L} = -\frac{2\sqrt{\pi}}{M_{Pl}} T^{\mu \nu}_{\text{SM}}\,\left(
h_{\mu \nu}^{(0)}(x) - \,e^{2k\pi R_B}\sum_{n\neq 0} h_{\mu
\nu}^{(n)}(x)\right)\,. \label{rs8}
\end{equation}
In this case, the masses of the KK gravitons read:
\begin{equation}
m_n = k x_n e^{-k R_B \pi}, \label{rs10}
\end{equation}
where $x_n$ are the roots of the Bessel function $J_1(x)$. 
Therefore, KK gravitons may have quite light masses,
which can be proved in colliders by the observation of resonances.
In such a case, the cleanest signal at hadron colliders, such as
the LHC, could be an excess in Drell-Yan processes ($q\bar q,gg \rightarrow
h^{(1)} \rightarrow l^+l^-$) \cite{Davoudiasl:1999jd,Allanach:2000nr}.
These light gravitons can also be produced directly, together with
a single photon or jet. It is interesting to note that the RS1 model predicts
a mono-energetic photon, whereas the ADD model does not.

The commented phenomenology is associated with the RS1 model. The RS2 model is
characterised by an infinite extra dimension and it contains a continuum of
KK gravitons~\cite{Randall:1999vf}. A summary of the RS model phenomenology can 
be found, for 
example, in \cite{Davoudiasl:2000wi}.

\subsubsection{Brane Dynamics}

In addition to KK states, brane worlds introduce another set of very distinctive 
states.
The dynamics of the brane itself can be parameterised by $\delta$ functions or 
fields, depending on the 
1+3 coordinates. A 1+3 observer identifies these fields as new scalar (or 
pseudo-scalar)
modes.  These scalars are called branons: $\pi^\alpha$, and their leading 
coupling with the
brane content is suppressed by the brane tension $\tau\equiv f^4$, which 
quantifies the
flexibility of the brane 
\cite{Dobado:2000gr,Cembranos:2001rp,Alcaraz:2002iu,Cembranos:2004eb,
Cembranos:2016jun}:
\begin{eqnarray}
&&
\!\!\!\!\!\!\!\!\!\!\!\!\!\!\!\!\!\!\!\!\!
S_B
=\int_{M_4}d^4x\sqrt{g}\left[-f^4+ {\mathcal L}_{\text{SM}}
 +
\frac{1}{2}g^{\mu\nu}\delta_{\alpha\beta}\partial_{\mu}
\pi^\alpha
\partial_{\nu}\pi^\beta-\frac{1}{2}M^2_{\alpha\beta}
\pi^\alpha\pi^\beta\right.
\nonumber\\
&& \
\ \ \ \ \ \ \ \ \ \ \ \ \,
+
\left.\frac{1}{8f^4}(4\delta_{\alpha\beta}\partial_{\mu}\pi^\alpha
\partial_{\nu}\pi^\beta-M^2_{\alpha\beta}\pi^\alpha\pi^\beta
\tilde g_{\mu\nu})
T^{\mu\nu}_{\text{SM}} \right]\,,
%
\label{lag}
\end{eqnarray}
where the branon mass matrix $M^2_{\alpha\beta}$ is determined by 
the local geometry of the bulk space where the brane is located
(see   \cite{Andrianov:2003hx} for a particular example).
Branons have zero mass only in highly symmetric bulk spaces.

A parity transformation in the extra space introduces a 
change of sign for the branon fields. The above Lagrangian (\ref{lag})
preserves this extra dimensional parity or brane parity. If this
symmetry is exact, branons interact by pairs with the particles 
confined into the brane and become stable. As they are generically massive
and weakly coupled, they are natural dark matter candidates.

Independent of their cosmological impact, branons can be searched for at 
colliders
\cite{Alcaraz:2002iu,Brax:2014vva,Achard:2004uu,Cembranos:2004jp,
Cembranos:2011cm,Landsberg:2015pka,Khachatryan:2014rwa,Cembranos:2005sr,
Cembranos:2005jc}.
Like many other dark matter candidates, their typical signatures include 
missing 
energy or missing transverse momenta 
signals 
\cite{Alcaraz:2002iu,Achard:2004uu,Landsberg:2015pka,Khachatryan:2014rwa,
Cembranos:2004jp,Cembranos:2011cm}.
On the other hand, as   happens with KK gravitons, branon radiative corrections
generate new couplings among SM particles. The leading terms are given by 
four-particle  interactions.
In addition, two-loop effects contribute to precision observables, such as 
anomalous magnetic moments \cite{Cembranos:2005sr,Cembranos:2005jc}.

With respect to the cosmological role of branon fields, they can achieve the 
required abundance of 
cold dark matter with the standard freeze-out mechanism 
\cite{Cembranos:2003mr,Cembranos:2003fu} or non-thermally 
\cite{Maroto:2003gm}, in  very much   the same way as axions 
\cite{Preskill:1982cy,Abbott:1982af,Dine:1982ah} or 
other bosonic degrees of freedom 
\cite{Frieman:1991qv,Hu:2000ke,Cembranos:2012kk,Cembranos:2016ugq}. 
This fact opens up the possibility for searching for branons through CMB and 
large-scale 
structure data 
\cite{Cembranos:2015oya,Cembranos:2018ulm,Brax:2019fzb,Brax:2019npi}, but also 
with direct dark matter experiments 
\cite{Cembranos:2011cm}, or indirectly through the analysis of cosmic rays. 
Indeed, two branons can annihilate 
contributing to the astrophysical flux of different particles. A series of 
studies have shown the 
possibilities for detecting branons through photons, neutrinos, positrons and 
antiprotons
\cite{Cembranos:2008bw,Cembranos:2011hi,Cembranos:2010dm,Cembranos:2012nj,
Cembranos:2013fya,Cembranos:2014yqa,Cembranos:2014wza,Bull:2018lat,
Cembranos:2019noa,Cembranos:2019amc}.

\subsection{Universal Extra Dimensions}

Motivated by the success of the brane world models discussed above, other 
extra dimension models without the existence of branes have also attracted   
important attention. In the Universal Extra Dimensional case there are no
branes, and every field can propagate in the entire bulk space.

In such a case, due to the translation invariance in the extra space, the
associated momenta, i.e., the KK number $n$, is conserved, at least at tree 
level. This implies that a vertex with $N$ particles with KK numbers: 
$n_1, n_2,
\ldots n_N$, can be non-zero only if:
\begin{equation}
n=n_1 + n_2 \ldots + n_{N-1} + n_N = 0. \label{ued1}
\end{equation}

Although the total KK number is not conserved due to radiative corrections,
the $K$-parity: $(-1)^{|n|}$ is   \cite{Rizzo:2001sd}. This implies that at 
least,
one of the lightest KK states is stable. This state is very interesting from
a cosmological or astrophysical approach, since this mode is a natural candidate
for dark matter \cite{Cheng:2002ej,Servant:2002aq,Kong:2005hn}. On the 
contrary, 
for collider physics, the sensitivity
of the different experiments is reduced, since the lightest KK excitations need 
to
be produced by pairs. In any case, within this framework, the expected 
degenerate 
spectrum of the first KK modes provides a distinctive phenomenology both for
collider and astrophysics searches \cite{Cembranos:2006gt,Cembranos:2007fj}.
\begin{figure}[bt]
\begin{center}
\resizebox{14cm}{!}
{\includegraphics{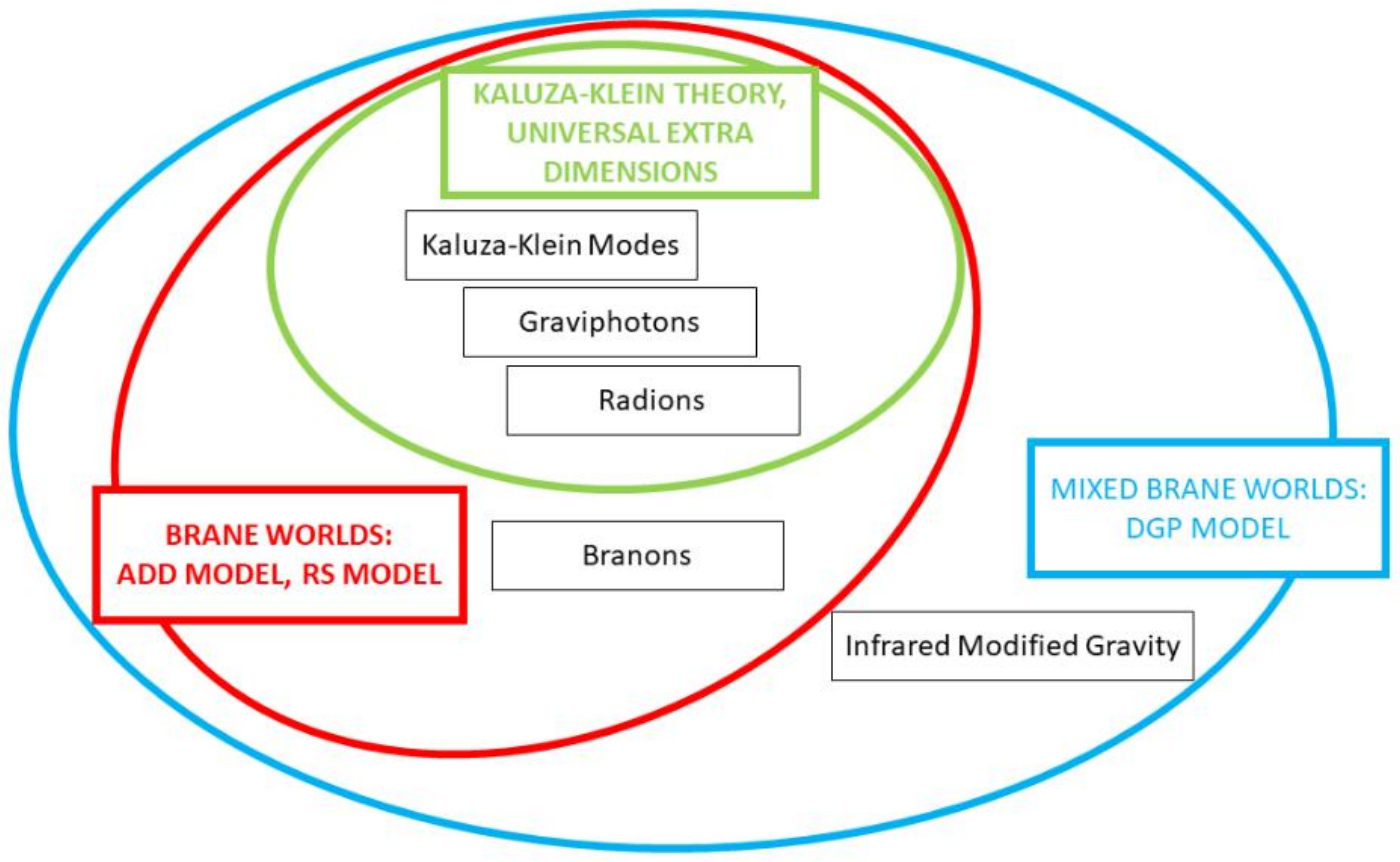}}
\caption {{\it{ Schematic representation of the basic phenomenology
associated with the different extra dimensional models discussed within this
chapter.}}}
\end{center}
\end{figure}

\subsection{Mixed Models}

Finally, we cannot finish this brief summary about extra dimensional models 
without
commenting about what we will call mixed models. All of them are built with the
presence of 3-branes, but they do not restrict the standard model particles to
propagate within such a brane. On the contrary, part of the standard model 
content
may have access to the bulk space. In other models, the situation is even more 
involved,
and different manifolds are available for different fields.

For example, standard model gauge bosons ($W^\pm$ and $Z$) can propagate in the 
entire bulk space and the zero modes can mix with their excited KK states.
These effects also depend on the Higgs field that can propagate in the bulk
space, on the branes, or on a combination of both 
\cite{Masip:1999mk,Rizzo:1999br}.

On the other hand, the fields can have kinetic terms in different domains.
For example, a field that has access to the bulk space can also have a 
kinetic term defined on the brane. An important detail  of such terms is
that they  are naturally produced by radiative corrections, due to translational 
invariance breaking in the extra space. This effect has been proved
when the fields that propagate in the extra dimensions couple to the fields 
that
are confined to the brane \cite{delAguila:2003bh}.

An interesting model motivated by the above discussion is the one proposed
by Dvali, Gabadadze and Porrati (DGP) \cite{Dvali:2000hr}. The action of this 
model
does not only introduce the scalar curvature associated with the bulk space,
but also the scalar curvature of the metric induced on the brane:
\begin{equation}
S_G=\frac{M_F^3}{16\pi} \int_{M_5}d^5Y\sqrt{G} R_D +
\frac{M_{Pl}^2}{16\pi}\int_{M_4}d^4x\sqrt{g} R.
\label{Dvaliag}
\end{equation}

This model is able to produce an accelerated expansion without introducing a 
cosmological constant or any other type of dark energy. Newtonian gravity can be 
approximately maintained with a non-compact extra dimension. Indeed, the Newton 
force does not suffer modifications at low distances but at large ones.

It has been argued that the branch of the model associated with this accelerated phase 
is unstable \cite{Gorbunov:2005zk} and in conflict with present observations
related to baryon acoustic oscillations, the anisotropies of the cosmic
microwave background and type Ia supernovae \cite{Fang:2008kc}. In any case,
the DGP model has attracted a lot of attention due to its relation to 
massive gravity and the avoidance of the vDVZ discontinuity due to 
non-perturbative effects.

\chapter[Non-local Gravity]{Non-local Gravity}
\label{Calcagninonlocal}
{\em Gianluca Calcagni}
\\

 \label{nonloccurnref1}

Quantum gravity is a collective name labelling all those theoretical  
frameworks 
combining quantum mechanics and the gravitational force in a consistent way. On 
one hand, there are approaches that quantise gravity as a fundamental force, or 
they embed it in a unified theory of elementary interactions. In this case, new 
physics is expected to emerge at short, ultraviolet (UV) scales of the  order 
of 
the 
Planck length $\lpCg$. On the other hand, quantum gravity may also indicate an 
effective field theory with infrared (IR) corrections manifesting themselves at 
large scales. This does not mean that the first group of theories is 
unobservable; on the contrary, Planckian scales may show up in the sky if 
quantum corrections were not negligible when cosmological inflation took place. 
Several examples of this amplification mechanism in quantum gravity exist 
\cite{Calcagni:2017sdq}.

Here we review a specific candidate of both of  the above groups  (theories or 
models with UV or IR corrections): nonlocal gravity. The latter is a field 
theory endowed with nonlocal form factors. A covariant nonlocal form factor 
$\gCg(\BCg)$ can be alternatively described either as an operator with 
infinitely many derivatives or as the convolution with an integral kernel 
\cite{Pais:1950za}. For a generic field $\vpCg$,
\be
\gCg(\BCg)\,\vpCg(x) = \int{d}^D y\,F(y-x)\,\vpCg(y)\,, \qquad 
F(z)=\int\frac{{d}^Dk}{(2\pi)^D}\,\rmeCg^{\rmiCg k\cdot 
z}\gCg(-k^2)\,,
\label{Calcker}
\ee
where $\BCg=\NCg_M\NCg^M$ is the Laplace--Beltrami operator. Since the 
derivative description can be misleading when the counting of degrees of freedom 
and the Cauchy problem of the theory are considered 
\cite{Calcagni:2018lyd,Calcagni:2018gke}, we prefer the name \emph{nonlocal}, 
without a  hyphen, in order to stress the uniqueness and originality of the 
dynamical properties of these theories with respect  to local theories.

As is well known, quantum field theory (QFT) is nonlocal at the quantum level  
because loop corrections to the bare propagator typically contain non-polynomial 
functions of momentum. In general, the quantum effective action contains 
nonlocal modifications that, in the case of gravity, may leave an imprint on 
the 
IR. This IR nonlocality is an effective description of a certain regime of a 
fundamentally local theory. In contrast, UV nonlocality refers to theories that 
are fundamentally nonlocal already at the classical level. Historically, UV 
nonlocality is as old as IR nonlocality. QFT pioneers such as Wataghin 
\cite{Wataghin:1934ann} and Yukawa \cite{Yuk49,Yukawa:1950eq} considered 
nonlocality as a means to give particles a finite radius and, thus, to cure the 
infinities of their self-energy \cite{Pau33}. Nowadays, fundamental nonlocality 
is invoked for roughly the same reason (to remove the classical singularities 
of 
General Relativity and to improve its renormalizability at the quantum level), 
as well as others such as preservation of unitarity (absence of ghosts and 
conservation 
of probability). In the following, we first review UV nonlocal theories and 
their phenomenology, followed by IR nonlocal models. Note that, strictly 
speaking, IR nonlocal models are not models of quantum gravity because, despite 
their loose motivation from nonlocal corrections found in quantum theories, 
they 
do not arise from the latter.


\section{UV Nonlocal Gravity}

UV nonlocal gravity, or nonlocal quantum gravity, is a perturbative quantum 
field  theory of the gravitational force, such that the bare action and its 
classical dynamics are endowed with nonlocal operators of a certain form. The 
foundations of fundamentally nonlocal quantum field theory have been set by 
Efimov and collaborators in the case of a scalar field 
\cite{Efimov:1967dpd,Efimov:1967pjn,Alebastrov:1973vw,Alebastrov:1973np,
Efimov:1974cmh,Efimov:1977bpa}, while gauge and gravitational theories were 
considered about a decade later by Krasnikov and Kuz'min 
\cite{Krasnikov:1987yj,Kuzmin:1989sp}. Since then, a surge of interest in 
classical and quantum nonlocal scalar field theory, gauge theory and gravity, 
as 
well as in singularity resolution and cosmology in nonlocal gravity, has been 
taking place and is gradually increasing 
\cite{Tseytlin:1995uq,Tomboulis:1997gg,Siegel:2003vt,Biswas:2005qr,Khoury:2006fg
,Calcagni:2010ab,Biswas:2010zk,Moffat:2010bh,Modesto:2010uh,Modesto:2011kw,
Biswas:2011ar,Alexander:2012aw,Modesto:2012ys,Biswas:2012bp,Briscese:2012ys,
Modesto:2013ioa,Calcagni:2013vra,Modesto:2013jea,Briscese:2013lna,Biswas:2013cha
,Modesto:2014xta,Calcagni:2014vxa,Biswas:2014yia,Modesto:2014lga,Conroy:2014dja,
Talaganis:2014ida,Modesto:2015lna,Dona:2015tra,Modesto:2015foa,Li:2015bqa,
Tomboulis:2015esa,Talaganis:2016ovm,Edholm:2016hbt,Giaccari:2016kzy,
Modesto:2016max,Biswas:2016egy,Koshelev:2017ebj,Calcagni:2017sov,Cornell:2017irh
,Edholm:2018wjh,Buoninfante:2018xiw,Koshelev:2018hpt,Koshelev:2018rau,
Calcagni:2018lyd,Calcagni:2018gke,Calcagni:2018pro,Giaccari:2018nzr,
Briscese:2018oyx,Buoninfante:2018rlq,Buoninfante:2018stt,Buoninfante:2018mre,
Briscese:2018bny,Buoninfante:2018lnh,Briscese:2019rii}. The bare (classical) 
action of the theory in $D$ dimensions is
\ba\label{Calcnlqg}
S &=&\frac{1}{2\kappa_D^2}
\int{d}^Dx\,\sqrt{|g|}\,[R-2\Lambda+ R\gCg_0(\BCg)\,R + 
R_{MN}\gCg_2(\BCg)\,R^{MN}\nonumber\\
&&\qquad\qquad+ R_{MNKL}\gCg_4(\BCg)
\,R^{MNKL}+\cV(\cR)]\,,
\ea
where the form factors $\gCg_{0,2,4}(\BCg)$ are certain functions of  the 
Laplace--Beltrami operator $\BCg$ and $\cV(\cR)$ is a ``potential'' term made 
of 
local curvature operators. When $\gCg_{0,2,4}=0$, one recovers the 
Einstein--Hilbert action, while when $\gCg_{0,2,4}={\rm const.}$, one gets 
fourth-order Stelle gravity and its generalisations 
\cite{Stelle:1976gc,Stelle:1977ry,Buchbinder:1992rb,Asorey:1996hz,Salles:2014rua
}. To study non-trivial form factors, one can use the generic parametrisation
\be\label{Calcgamma}
\gCg(\Box) = \frac{\rmeCg^{\HCg(\Box)}-1}{\Box}\simeq 
\frac{\HCg(\BCg)}{\BCg}=c_0+c_1\BCg+\dots\,,
\ee
where the function $\HCg(\BCg)$ depends on the dimensionless combination  
$\ell_*^2\BCg$ and $\ell_*$ is a fixed length scale, presumably $O(\lpCg)$. 
Assuming that $\HCg(0)=0$, then by construction the $1/\BCg$ operator does not 
carry any of the field-redefinition issues that  we will discuss in Section 
\ref{Calcirnlg} because it is absorbed by the $O(\BCg)$ term of 
$\HCg(\BCg)=c_0\BCg+c_1\BCg^2+\dots$.

The choice of form factors $\gCg_{0,2,4}$ is not unique but the UV properties 
of 
 the theory greatly reduce the possibilities. The absence of extra degrees of 
freedom at the linearized tree level around Minkowski spacetime requires
\be
\gCg_0(\BCg) = - \frac{(D-2) ( \rmeCg^{\HCg_0} -1 ) + D ( \rmeCg^{\HCg_2} -1 
)}{4 (D-1) \BCg} + \gCg_4(\BCg)\,,\qquad
\gCg_2(\BCg) = \frac{\rmeCg^{\HCg_2} -1 }{\BCg} - 4 \gCg_4(\BCg) 
\,,\label{Calcgamma2}
\ee
which reduce to the relation $2\gCg_0(\BCg)+\gCg_2(\BCg)+2\gCg_4(\BCg)=0$ when 
$\HCg_1=\HCg_2$.  Renormalisability of the theory constrains the form factor 
$\gCg_4(\Box)$ to have the same (or lower in the number of derivatives) 
asymptotic UV behaviour as the other two form factors $\gamma_{0,2}(\Box)$. 
Since $\gamma_4(\Box)$ does not appear in the graviton propagator, the minimal 
choice compatible with unitarity corresponds to $\gCg_4(\Box)=0$, but a 
non-vanishing form factor affects the admissible metric solutions of the 
theory. 
In particular, Ricci-flat spacetimes  with $R_{MNKL}\neq 0$ are not 
classical solutions if $\gCg_4\neq 0$.

Two classes of form factors have been widely studied in the literature:
\begin{itemize}
\item \emph{Exponential-monomial form factors}, which include the \emph{minimal 
form  factor} \cite{Biswas:2005qr,Calcagni:2014vxa}
\be\label{Calcmini}
\gCg(\BCg)=\frac{\rmeCg^{-\ell_*^2\BCg}-1}{\BCg}\,,
\ee
and the \emph{Krasnikov form factor} \cite{Krasnikov:1987yj}
\be\label{Calckras}
\gCg(\BCg)=\frac{\rmeCg^{\ell_*^4\BCg^2}-1}{\BCg}\,.
\ee
When linearizing in the metric $g_{MN}=g^{(0)}_{MN}+h_{MN}$ and 
expanding around  the background $g^{(0)}_{MN}$, the minimal form factor 
produces a kinetic term $\sim h_{MN}\BCg\exp(-\ell_*^2\BCg)h^{MN}$, 
where the kinetic operator is called Wataghin form factor 
\cite{Wataghin:1934ann} and is typically found in string field theory.
\item \emph{Asymptotically polynomial form factors}.
Here, $\exp[\HCg(z)]$ is such that: (i) it is an entire function; (ii) it has 
no 
zeros in the whole complex plane, hence   $\HCg(z)$ is also   entire; (iii) it 
is 
real and positive on the real axis; (iv) $\HCg(0) = 0$ ($\gamma(z)$ is 
non-singular); (v) in a suitable conical region $\mathcal{C}$, 
$\exp[\HCg_\infty(z)]:=\lim_{|z| \rightarrow +\infty}\exp[\HCg(z)]= |z|^{n}$, 
where 
$n$ is a natural number that depends on the order of local derivative operators 
in the action, on the spacetime dimension $D$ and on the renormalisability of 
the theory; (vi) in the conical region $\cC$, $\lim_{|z| \rightarrow +\infty} 
\{\exp[\HCg(z)]/\exp[\HCg_\infty(z)]-1\}\, z^n = 0$ for all $n \in \mathbb{N}$, 
hence 
the name is  asymptotically polynomial. Two representatives of this class are  
the
\emph{Kuz'min form factor} \cite{Kuzmin:1989sp}
\be
\HCg(\BCg)=\HCg_{\rm Kuz}(\BCg):= 
\ln(-\ell_*^2\BCg)+\Gamma(0,-\ell_*^2\BCg)+\gamma_{\rm E}\,,\label{Calckuz}
\ee
and  the \emph{Tomboulis form factor} \cite{Tomboulis:1997gg,Modesto:2011kw}
\be
\HCg(\BCg)=\HCg_{\rm Tom}(\BCg):=\frac{1}{2} \left\{\ln 
p^2(\BCg)+\Gamma[0,p^2(\BCg)]+\gamma_{\rm E}\right\},\label{Calctom}
\ee
where $\Gamma(0,x)$ is the upper incomplete gamma function with  its first 
argument vanishing, $\gamma_{\rm E}\approx 0.577$ is the Euler--Mascheroni 
constant and $p(z)$ is a real polynomial of degree $n$ with $p(0) = 0$.
\end{itemize}

Exponential-monomial form factors are simpler but asymptotically  polynomial 
form factors that  are especially well behaved at the quantum level. Within the 
above 
choices, and depending on the spacetime dimension $D$ and the local potential 
$\cV(\cR)$, the theory is super-renormalisable (only one-loop divergences 
survive and one must add a finite number of operators to absorb them) or finite 
(no divergences, all beta functions vanish), tree-level unitary (no classical 
instabilities, no ghost modes) and perturbative unitary (conservation of 
probability). The job of the nonlocal form factors in gravity is, on one hand, 
to modify the renormalisation properties of the theory and, on the other hand, 
to guarantee unitarity. In contrast, higher-order local gravity is 
renormalizable but non-unitary.

Particle physics places a lower bound on the energy scale of the  theory, 
$E_*=1/\ell_*>1-3\,{\rm TeV}$ \cite{Biswas:2014yia}. Above that scale,   
interesting physical effects  are obtained, which we describe in the following.

\subsubsection{Singularity Problem}
 \label{singularitiesrefs2}

The infinities of General Relativity are not supposed to survive in a  
renormalisable theory of gravitation. However, this statement depends on the 
background and on the specific form of the Lagrangian. Consider first the 
question about the fate of black-hole singularities. There are two 
diametrically 
opposed claims in the literature of UV nonlocal gravity: singularities do not 
disappear just because the dynamics is nonlocal 
\cite{Calcagni:2010ab,Calcagni:2017sov,Calcagni:2018pro,Briscese:2019rii} or 
they actually do 
\cite{Edholm:2016hbt,Cornell:2017irh,Buoninfante:2018xiw,Koshelev:2018hpt,
Buoninfante:2018rlq,Buoninfante:2018stt}. There is no contradiction between 
these views, since the former holds in theories with $\gCg_4(\BCg)=0$, while 
the 
latter holds when $\gCg_4(\BCg)\neq 0$.

In the $\gCg_4= 0$ case, Ricci-flat spacetimes are not only exact vacuum  
solutions of the theory \cite{Li:2015bqa}, but they are also stable against 
perturbations of linear order \cite{Calcagni:2017sov,Calcagni:2018pro} or even 
arbitrary order \cite{Briscese:2019rii}, if they are stable in General 
Relativity. Any Ricci-flat stable vacuum solution of Einstein gravity is also a 
solution of nonlocal gravity and, consequently,   the Schwarzschild metric 
is also 
a stable solution. This implies that   singular solutions exist  in the 
theory. 
However, the issue of whether such black holes \label{BHref2} are physical or 
not 
is still 
open 
and it depends on the matter distribution (calculated in the sense of 
generalised functionals) associated with them. A matter distribution 
incompatible 
with the singularity at $r=0$ could indicate that the vacuum solution is purely 
mathematical.

In the $\gCg_4\neq 0$ case, Ricci-flat metrics are not solutions to the  theory 
because the Riemann tensor does not vanish therein. Finding alternative exact 
solutions is difficult, both because the equations of motion are complicated 
and 
because in theories with higher-order curvature terms there is no Birkhoff 
theorem guaranteeing the uniqueness of spherically symmetric solutions. Still, 
one can look at linearised perturbations of the Minkowski background and, in 
particular, at the Newton potential. The latter is the static potential 
$\Phi(\bm{x})$ in $(D-1)$-dimensional flat space defined by the Green equation
\be\label{Calcnewt1}
\gCg(\NCg^2)\,\Phi(\bm{x})=\de^{D-1}(\bm{x})\,.
\ee
When $\gCg(\NCg^2)=\NCg^2\exp(-\ell_*^2\NCg^2)$, $\Phi$ is regular near  the 
origin $r:=|\bm{x}|=0$ \cite{Tseytlin:1995uq,Siegel:2003vt} (see also 
\cite{Biswas:2005qr,Modesto:2010uh,Edholm:2016hbt}; all these papers work in 
$D=4$ dimensions):
\ba
\Phi(r) &=& 
-\frac{1}{4\pi^{\frac{D-1}{2}}r^{D-3}}\left[\GCg\left(\frac{D-3}{2}
\right)-\GCg\left(\frac{D-3}{2},\frac{r^2}{4\ell_*^2}\right)\right]\\
&\stackrel{D=4}{=}&-\frac{{\rm erf}\left(\frac{r}{2\ell_*}\right)}{4\pi r}\,,
\ea
where $\GCg(\nu,z)$ is the upper incomplete gamma function of order $\nu$ and 
erf is the error function. Near the origin,
\be
\Phi(r)= -\frac{1}{(D-3)2^{D-2}\pi^{\frac{D-1}{2}}\ell_*^{D-3}}+O(r^2)\ 
\stackrel{D=4}{=} \ 
-\frac{1}{4\pi^{\frac{3}{2}}\ell_*}+O(r^2)\,,\label{Calccosteq}
\ee
i.e., the Newton potential tends to a non-zero constant.  Consistently, in the 
limit $\ell_*
\to 0$ one recovers the local potential singular at $r=0$. A more 
general no-go theorem is that the theory with the Riemann--Riemann term does 
not 
allow any metric potential of the form $1/r^\alpha$ with $\alpha>0$ 
\cite{Koshelev:2018hpt}.

The nonlocal generalization of the Schwarzschild black-hole $D=4$ linearised 
line  element features this potential in its 00 and $rr$ components 
\cite{Buoninfante:2018xiw}:
\be
{d} s^2 = -[1+2\Phi(r)]\,{d} t^2+[1-2\Phi(r)]\,({d} r^2+r^2{d}\Om^2)\,,
\ee
where ${d}\Om^2$ is the angular line element. Although this line element 
represents a linear perturbation of the Minkowski background, one can argue 
that 
it is sufficient to assess the microscopic properties of spherically symmetric 
solutions, since in the UV, nonlocal gravity is weak, due to asymptotic freedom 
\cite{Buoninfante:2018rlq}. The linear size $r_{\rm nonlocal}\simeq 2\ell_*$ of 
the nonlocal region is larger than the Schwarzschild radius of black holes in 
General Relativity, while the interior is regular and described by an effective 
Euclidean field theory \cite{Buoninfante:2018xiw,Buoninfante:2018rlq}. 

Therefore, black holes are non-singular in the theory with $\gCg_4\neq 0$  
\cite{Buoninfante:2018xiw,Koshelev:2018hpt}. These results were generalised to 
non-singular rotating black holes (the counter part of the Kerr metric   
in 
General Relativity) \cite{Cornell:2017irh} and charged black holes (the 
counter part of the Reissner--Nordstr\"om metric in General Relativity) 
\cite{Buoninfante:2018stt}.

The Big Bang is another singularity that nonlocality could smear  out. 
Friedmann--Lema\^itre--Robertson--Walker (FLRW) classical solutions to the 
equations of motion typically have a bouncing scale factor $a(t)$, as shown in 
a 
toy model with pure Ricci-scalar Lagrangian ($\gCg_2=0=\gCg_4$) 
\cite{Biswas:2005qr,Biswas:2010zk,Biswas:2012bp,Conroy:2014dja}. Also, the 
Kasner metric with scale factor $a(t)=t^{p_i}$ ($i=1,2,3$) is not a solution of 
the theory, neither with $\gCg_4=0$ \cite{Calcagni:2013vra} nor with 
$\gCg_4\neq 
0$ \cite{Koshelev:2018rau}. In the first case, it was shown that the 
Belinsky--Khalatnikov--Lifshitz singularity is replaced by an anisotropic 
bounce 
\cite{Calcagni:2013vra}.\label{Bouncerefs1}

In the theory with $\gCg_2\neq 0$ and $\gCg_4=0$, a bouncing profile $a(t)$ is  
an approximate solution of the full equations of motion valid both at very 
early 
times (when, roughly, the cosmological horizon scale is near the UV 
asymptotically free fixed point) and at late times. This profile can be 
described by the effective Friedmann equation \cite{Calcagni:2013vra}
\be\label{Calcfreq}
H^2= \frac{\kappa^2}{3}\rho_{\rm eff}:=\frac{\kappa^2}{3} \rho\left[1-  
\left(\frac{\rho}{\rho_*}\right)^\b\right],
\ee
where $H:=\dot a/a$ is the Hubble parameter  (not to be confused with the form 
factor ${\rm H}$), $\kappa^2=8\pi G_N$ with $G_N$ the four-dimensional Newton's 
constant, $\rho$ is the total
energy density of the Universe content, $\rho_*\leq\rho_{\rm Pl}$ is the 
critical 
energy density at which the bounce occurs, $\rho_{\rm Pl}$ is the Planck 
energy density, and $\b>0$ is a real parameter. The 
exponent $\b$ is determined by plugging the profile $a(t)$ found under the 
provision of asymptotic freedom (weak coupling in the UV) into \Eq{Calcfreq} 
for 
a given energy density profile $\rho(a)$. Since this fit is generally rather 
good, one can conclude that the asymptotic bouncing solution is also  
reasonably valid 
  at intermediate times, and that the bouncing accelerating scenario of the 
theory is well described by the effective Friedmann equation \Eq{Calcfreq}.

The presence of non-singular solutions does not guarantee the removal of  
infinities at the classical level unless such solutions are typical. As an 
example, where nonlocal operators are not enough to make the dynamics 
singularity-free, we cite a scalar-tensor nonlocal theory with both bouncing 
and 
singular solutions \cite{Calcagni:2010ab}. In that case, conformal invariance 
seems to play a more pivotal role than nonlocality in the resolution of the 
Big Bang.

\subsubsection{Inflation}

The realization of early-universe acceleration in UV nonlocal gravity  is 
anticipated by the observation that de Sitter cosmology (exponential scale 
factor, constant $H$) is a solution of the theory. Its stability and the 
absence 
of extra propagating degrees of freedom therein was checked both at the linear 
order \cite{Biswas:2016egy,Koshelev:2017ebj} and at all perturbative orders 
\cite{Briscese:2019rii}.

On a less formal side, acceleration is automatically implemented by the 
non-singular  scale factors $a(t)$ discussed above. All these solutions realize 
a geodesically complete inflationary scenario because they accelerate near the 
bounce. Geometry can drive an early acceleration era both in a toy model with 
pure Ricci-scalar Lagrangian ($\gCg_2=0=\gCg_4$) 
\cite{Biswas:2010zk,Biswas:2012bp,Conroy:2014dja} and the minimal theory with 
$\gCg_2\neq 0=\gCg_4$ in the asymptotically free regime 
\cite{Khoury:2006fg,Calcagni:2013vra}. The model is compatible with 
observations, 
since it yields the local Starobinsky action in the limit of small $\ell_*$ 
\cite{Briscese:2013lna}.

\subsubsection{Dark Energy}

The cosmological constant problem has not been addressed yet in this class of  
theories beyond some preliminary considerations \cite{Moffat:2010bh}.

\subsubsection{Gravitational Waves}
\label{gravitationalwavrefs5}

The golden age of gravitational-wave (GW) observations, from those emitted  by 
small-redshift astrophysical compact objects 
\cite{Abbott:2016blz,Monitor:2017mdv} to the inflationary tensor spectrum 
generated in the early Universe \cite{Akrami:2018vks}, has  been among the most 
notable achievements of General Relativity. To date, the 
predictions of Einstein gravity related to GWs, small ripples of spacetime 
propagating through cosmological scales, have been confirmed, and no evidence 
of 
new physics has been found. While with the available LIGO data we are closing 
in 
on many models beyond General Relativity, new astrophysical and cosmological 
constraints have been devised to be tested in near-future experiments such as 
KAGRA,  \label{LISAref1}
LISA or DECIGO. One of the observables we can use to discriminate among 
different cosmological models is the luminosity distance of standard sirens, 
which are sources of both photons (light) and gravitons (GWs) that will become 
gradually available when populating the catalogue of GWs 
\cite{Belgacem:2019pkk}.

Quantum gravity is among the things that could happen beyond General Relativity 
and, although most theories of quantum gravity predict too small effects to be 
detected, some could generate a non-negligible signal 
\cite{Calcagni:2019kzo,Calcagni:2019ngc}. Unfortunately, nonlocal quantum 
gravity is among the candidates failing to produce such a signal 
\cite{Briscese:2019rii}. The intuitive reason, common to many other (albeit not 
all) quantum gravities, is that nonlocal corrections are confined to Planckian 
scales. Still, it is instructive to see the details of the result, because on 
one hand they introduce the basics of luminosity distance, and on the other 
hand they are an example of how to extract predictions from the theory and 
pitch 
them against observations. The smoking gun of nonlocal
quantum gravity cannot be found in other GW observables either, such as
the amplitude of the primordial stochastic GW background
\cite{Calcagni:2020tvw}, which we will not discuss here.

We first
consider  General Relativity in four dimensions. The linearised 
propagation 
 equation of GWs on a FLRW background is $\BCg h=0$, where 
$h(t,\bm{x})=h_{+,\times}(t,\bm{x})$ is the amplitude of either tensor 
polarisation mode \cite{Mukhanov:1990me}. From the solution of this equation, 
or 
from a simple scaling argument, one can recast the GW amplitude in terms of the 
redshift $z:=1/a-1$ and a physical observable, the luminosity distance of an 
optical source
\be
d_L^\textsc{em}=(1+z)\int_0^z\frac{{d} z}{H}\,.
\ee
It turns out that, up to coefficients \cite{Maggiore:1900zz},
\be
\frac{1}{d_L^\textsc{gw}}:=h= \frac{1}{d_L^\textsc{em}}\,,
\ee
where we defined the GW luminosity distance as the inverse of the amplitude.  
Therefore, for a standard siren in General Relativity 
$d_L^\textsc{gw}/d_L^\textsc{em}=1$.

In UV nonlocal gravity, the linearized perturbation equation is  
\cite{Modesto:2011kw,Kostelecky:1990mi}
\be
\BCg\tilde h=0\,,\qquad \tilde h:=\rmeCg^{\HCg} h\,.
\ee
Using the same scaling argument as in General Relativity, with $h$  replaced by 
$\tilde h$, for entire form factors we have
\be
\tilde h= \frac{1}{d_L^\textsc{gw}}\qquad\Longrightarrow\qquad h=  
\rmeCg^{-\HCg} \frac{1}{d_L^\textsc{gw}}\,.
\ee
We can estimate the nonlocal correction in the right-hand side for the  minimal 
form factor \Eq{Calcmini}, 
$\HCg(\BCg)=-\ell_*^2\BCg=\ell_*^2(\partial_t^2+3H\partial_t)$ 
in 
the homogeneous approximation and, crudely, an approximately constant Hubble 
parameter $H\simeq H_0$, so that $z\simeq\rmeCg^{-H_0 (t-t_0)}-1$ and 
$d_L\simeq 
(z+1)z/H_0\simeq [\rmeCg^{-2H_0(t-t_0)}-\rmeCg^{-H_0(t-t_0)}]/H_0$. Since 
$\rmeCg^{-\HCg(\BCg)}\rmeCg^{n H_0t}=\rmeCg^{-n(n+3)(\ell_*H_0)^2}\rmeCg^{n 
H_0t}$, at large redshift $h\simeq 
H_0\rmeCg^{-10(\ell_*H_0)^2}\rmeCg^{2H_0(t-t_0)}$, while at small redshift 
$z\simeq -H_0 (t-t_0)\ll 1$ one has $h\simeq 
H_0\rmeCg^{-3(\ell_*H_0)^2}\rmeCg^{2H_0(t-t_0)}$. Overall,
\be
h\simeq \frac{\rmeCg^{-c (\ell_*H_0)^2}}{d_L^\textsc{gw}}\,,\qquad 
c=O(1)\!-\!O(10)\,.
\ee

The binary neutron star merger GW170817 is the first known example  of the  
standard 
siren \cite{Monitor:2017mdv}. If propagation of electromagnetic waves was 
affected in the same way by the form factor, then the ratio between the 
luminosity distance $d_L^\textsc{gw}$ measured by an interferometer and the 
luminosity distance $d_L^\textsc{em}$ measured for the optical counter-part 
would be equal to 1, as in General Relativity. However, even if light was not 
affected by nonlocality, we would have
\be\label{Calcdd}
\frac{d_L^\textsc{gw}}{d_L^\textsc{em}}\simeq 1+c (\ell_*H_0)^2\,,
\ee
and, for $\ell_*=\lpCg$, the right-hand side would be of the order of 
$1+10^{-120}$,  an effect completely unobservable compared with the estimated 
error $\Delta d_L/d_L\sim 0.001\!-\!0.1$ of present and future interferometers 
\cite{Dalal:2006qt,Nissanke:2009kt,Camera:2013xfa,Tamanini:2016zlh}. For a 
power-law expansion $a=(t/t_0)^p$, $d_L\propto (t_0/t)^{2p}(t_0-t)$, and one 
can 
show that, again, the correction in the ratio \Eq{Calcdd} is of the order of 
$(\ell_*/t_0)^2\sim 10^{-120}$. Increasing $\ell_*$ to particle-physics scales 
does not magnify this correction enough, since it is governed by the 
cosmological scale $H_0^{-1}\sim t_0\sim 10^{17}\,{\rm s}$.


\section{IR Nonlocal Gravity}\label{Calcirnlg}

At the level of the quantum effective action in perturbative gravity,  one can 
add all admissible higher-order curvature terms featuring four derivatives. 
These terms, of the form $\cR^L \BCg^{-n} \cR^M$, where $\cR$ is a generic 
curvature tensor, are such that $n=M+L-2\geq 0$, so that the effective action 
is 
decorated with inverse powers of the Laplace--Beltrami operator $\BCg$ 
\cite{Frolov:1979tu}. Since in momentum space $\BCg^{-n}\to (-k^2)^{-n}$, these 
terms can dominate in the IR (small momenta). However, one cannot expand 
perturbative one-loop corrections from light particles, which have been 
calculated and take the well-known logarithmic form $\ln(-\BCg)$ 
\cite{Barvinsky:1985an,Shapiro:2008sf,Barvinsky:1987uw,Barvinsky:1990up,
Barvinsky:1994hw,Barvinsky:1994ic,Gorbar:2002pw,Gorbar:2003yt}, to get 
$\BCg^{-n}$ operators with a cosmological impact, since loop corrections are 
valid in the UV, while the latter are an IR effect (deviations from General 
Relativity at large scales) \cite{Maggiore:2016fbn}. Therefore, the main 
motivation to consider Lagrangians of the form $f(\BCg^{-n}R)$ is 
phenomenological. Although there may still be high-energy scenarios justifying 
them \cite{Maggiore:2016fbn,Belgacem:2017cqo,Barvinsky:2003kg}, they are not 
strictly necessary, since here the focus is on IR divergences that could 
generate 
an effective mass dynamically.

The operator $1/\BCg$ admits an integral representation via the solution of the 
 Green equation
\be\label{CalcbcK}
\BCg_x\cK(x-x')=\frac{\de^{D}(x-x')}{\sqrt{|g|}}\,.
\ee
The kernel $\cK$ is nothing but the representation of the form factor $1/\BCg$. 
 
Nonlocal field redefinitions $\vpCg(x)\to\bar\vpCg(x)=\gCg(\BCg)\vpCg(x)$ do 
not 
change the spectrum of the theory in the case of form factors 
$\gCg(\BCg)=c_0+c_1\BCg+O(\BCg^2)$ with $c_0\neq 0$, i.e., those with trivial 
kernel. However, the operator $1/\BCg$ does not enjoy this property and 
nonlocal 
field redefinitions are defined up to a harmonic function $\laCg$, 
$\bar\vpCg(x)=\BCg^{-1}\vpCg(x)+\laCg(x)$, where $\BCg\laCg(x)=0$. This 
delicate 
point must be dealt with carefully, since dynamical solutions do not admit 
$\laCg=0$ in general. 

In \Eq{CalcbcK}, one must specify a contour prescription for the Green function 
$\cK$.  However, even if the causal (retarded) propagator $\cK_{\rm ret}(x-y)$ 
is used to define the $\BCg^{-1}$ operator, the variation of an action with 
$\BCg^{-1}$ operators acting on the fields always gives rise to the even 
combination $\cK_{\rm ret}(x-y)+\cK_{\rm ret}(y-x)=:\cK_{\rm ret}(x-y)+ 
\cK_{\rm adv}(x-y)$. The retarded Green function is not even, and changing the 
 sign 
to its argument gives the advanced Green function $\cK_{\rm ret}(y-x) = 
\cK_{\rm adv}(x-y)$, which is anti-causal. Therefore, the equations of motion 
for the field expectation values on in-out states (obtained from the variation 
of the quantum effective action) in theories with nonlocalities of the type 
$\BCg^{-1}$ are acausal \cite{Foffa:2013sma,Zhang:2016ykx,Belgacem:2017cqo}. 
However, in-in expectation values obey equations of motion obtained in the 
Schwinger--Keldysh formalism, which automatically give the retarded propagator. 
Considered in this way, IR nonlocal theories are causal 
\cite{Foffa:2013sma,Belgacem:2017cqo}.

A class of models with such peculiarities is
\be\label{Calcirnlac}
\cL=\frac{1}{2\kappa_D^2}\left[R+\cR\,f(\BCg^{-n}\cR)\right]\,,
\ee 
with different choices of parameters and operators. The best studied  
candidates in $D=4$ dimensions
are  the  Deser--Woodard model \cite{Deser:2007jk}, nonlocal massive gravity 
\cite{Maggiore:2013mea} and the Maggiore--Mancarella model 
\cite{Maggiore:2014sia}:
\begin{itemize}
\item \emph{Deser--Woodard} (or $f(\BCg^{-1} R)$) \emph{model}  ($\cR=R$, 
$n=1$, 
$f$ arbitrary) 
\cite{Deser:2007jk,Nojiri:2007uq,Koivisto:2008xfa,Koivisto:2008dh,
Koshelev:2008ie,Deffayet:2009ca,Elizalde:2012ja,Elizalde:2013dlt,Deser:2013uya,
Dodelson:2013sma,Zhang:2016ykx,Park:2016jym,Nersisyan:2017mgj,Park:2017zls,
Belgacem:2018lbp,Park:2019btx,Belgacem:2018wtb,Amendola:2019fhc,Chen:2019wlu}:
\be\label{Calctolag}
\cL=\frac{1}{2\kappa^2}\, R\,[1-f(\BCg^{-1}R)]\,.
\ee 
When $f=\mathbbm{1}$ and 
$\cL\propto R-R\BCg^{-1}R$,  one obtains an 
approximation of the first IR nonlocal model of this type, due to Wetterich 
\cite{Wetterich:1997bz}.
\item \emph{Nonlocal massive gravity} (or \emph{RT model}), where an $\cR=R$, 
$n=1$ correction arises at the level of the equations of motion 
\cite{Maggiore:2013mea,Foffa:2013sma,Foffa:2013vma,Nesseris:2014mea,
Dirian:2014bma,Dirian:2016puz,Belgacem:2018wtb,Belgacem:2019lwx}:
\be
G_{\mu\nu}-\frac{m^2}{3}(g_{\mu\nu}\BCg^{-1}R)^{\rm T}=\kappa^2 T_{\mu\nu}\,,
\ee
where T indicates the transverse part of the tensor. This version  of the 
theory 
improves a previous one \cite{Jaccard:2013gla} with unstable cosmological 
evolution \cite{Foffa:2013vma}.
\item \emph{Maggiore--Mancarella} (or 
$R\BCg^{-2} R$, or \emph{RR}) 
\emph{model} ($\cR=R$,  $n=2$, $f=\mathbbm{1}$) 
\cite{Maggiore:2014sia,Dirian:2014ara,Barreira:2014kra,Dirian:2014bma,
Dirian:2016puz, 
Nersisyan:2016hjh,Maggiore:2016fbn,Belgacem:2017cqo, 
Belgacem:2018wtb}:
\be\label{Calctolag2}
\cL=\frac{1}{2\kappa^2}\, R\,\left(1-\frac{m^2}{6}\frac{1}{\BCg^2}\,R\right).
\ee
\item \emph{Other models}: with the Ricci tensor ($\cR=R_{\mu\nu}$,  
$f=\mathbbm{1}$), $R_{\mu\nu}\BCg^{-1}R^{\mu\nu}$ model ($n=1$) 
\cite{Ferreira:2013tqn,Nersisyan:2016jta}, $R_{\mu\nu}\BCg^{-1}G^{\mu\nu}$ 
model ($n=1$) 
\cite{Barvinsky:2003kg,Barvinsky:2005db,Barvinsky:2011hd,Barvinsky:2011rk} and 
$R_{\mu\nu}\BCg^{-2}R^{\mu\nu}$ model ($n=2$) 
\cite{Cusin:2015rex,Zhang:2016ykx}; conformal extension with 
$\cR=R$, $n=1$ 
\cite{Cusin:2016nzi}, similar to Wetterich's early proposal 
\cite{Wetterich:1997bz}; new Deser--Woodard model 
\cite{Deser:2019lmm,Ding:2019rlp}.
\end{itemize}
The Deser--Woodard model \cite{Deser:2007jk}  does not contain any extra scale, 
while the massive-gravity model 
\cite{Jaccard:2013gla,Maggiore:2013mea,Foffa:2013vma} and the 
Maggiore--Mancarella model \cite{Maggiore:2014sia} feature a mass scale $m$. 
Furthermore, while the Deser--Woodard and the Maggiore--Mancarella models have 
simple actions and complicated equations of motion, massive gravity has simple 
equations of motion and a possibly complicated non-linear action, which is 
still 
unknown. Finally, the massive-gravity model and the Maggiore--Mancarella model 
give the same linearized equations on Minkowski background but they differ on 
other backgrounds or at the non-linear level, the second model deviating more 
strongly from $\Lambda$CDM \cite{Dirian:2014ara}.

Auxiliary fields  can be introduced so that, for instance in the $\cR=R$ case, 
the Lagrangian $\tilde\cL\propto R\,[1+f(\phi)]+\psi(\BCg\phi-R)$ replicates on 
shell the dynamics of the original system \Eq{Calcirnlac} \cite{Nojiri:2007uq}. 
However, this version is not completely equivalent to the former because the 
equation of motion $\BCg\phi-R=0$ of the Lagrange multiplier $\psi$ is used to 
obtain $\phi$ as a nonlocal function of $R$. The problem is that the solution 
of this relation is of the form
\be\label{CalcPhifr}
\phi=\BCg^{-1}R+\laCg\,,
\ee
where $\laCg$ is a scalar mode obeying  the homogeneous equation $\BCg\laCg=0$ 
\cite{Koshelev:2008ie}. This extra mode is responsible for extending the space 
of solutions to dynamics not admitted by the original nonlocal system 
\cite{Koshelev:2008ie}. Also, it makes an otherwise immaterial ghost degree of 
freedom dynamical. Integrating by parts,
\be\nonumber
\psi\BCg\phi\to-\partial_\mu\psi\partial^\mu\phi=-\frac14\partial_\mu(\psi+\phi)
\partial^\mu (\psi+ \phi) 
+\frac14\partial_\mu(\psi-\phi)\partial^\mu(\psi-\phi)\,.
\ee
Suitable conditions on $f$, found along the same lines of ghost constraints in  
$f(R)$ or higher-order theories, remove this ghost \cite{Deser:2013uya}.

All these models can be recovered in a unified way as the integer-order
limit of the action \cite{Calcagni:2021ljs}
\begin{equation}
S=\frac{1}{2\kappa^2}\int
{\text{d}}^4x\,\sqrt{|g|}\,
\left[R+\ell_*^2
G_{\mu\nu}(-\ell_*^2\Box)^{\gamma-2}\,R^{\mu\nu}\right],
\label{Calcact4fin}
\end{equation}
where $0<\gamma<1$. This theory has been built as a perturbative quantum
field theory of gravity realizing a change of spacetime spectral
dimension across scales. While renormalizability and unitarity are
difficult to achieve simultaneously, unitarity is generally preserved in
the above range of $\gamma$. Therefore, it may be possible to recover
the phenomenology of IR nonlocal gravity while at the same time avoiding
stability problems.

\subsubsection{Singularity Problem}

IR nonlocal models have been considered in the early Universe only    recently. 
The Deser--Woodard mode admits bouncing solutions, thus removing the 
Big-Bang singularity \cite{Chen:2019wlu}.

\subsubsection{Inflation}

Given that the generalisation $\cL\propto F(R)+\cR\, f(\BCg^{-n}\cR)$ of the  
total Lagrangian \Eq{Calctolag} contains both local and nonlocal higher-order 
curvature terms, there is the possibility to unify early-time (inflation) and 
late-time (dark energy) acceleration. For instance, the local term $F(R)$ and 
the nonlocal term $R\,f(\BCg^{-n}R)$ can drive, respectively, inflation and 
dark 
energy. In both cases, acceleration is sustained by the scalar fields hidden in 
the dynamics and made explicit by, respectively, a local ($\phi=F'(R)$) or 
nonlocal (\ref{CalcPhifr}) field redefinition. However, unification 
scenarios require a certain degree of engineering of the functions $F$ and $f$ 
\cite{Nojiri:2007uq}.

Both the massive-gravity and the Maggiore--Mancarella models are compatible  
with observations of the cosmic microwave background (CMB) multipole spectrum 
\cite{Barreira:2014kra,Dirian:2014bma,Belgacem:2017cqo}.

\subsubsection{Dark Energy}

All the IR nonlocal models listed above  are endowed with self-accelerating 
late-time solutions without a  cosmological constant, where the role of dark 
energy 
is played by a nonlocal curvature component with an effective phantom equation 
of state (effective barotropic index $w<-1$).

Since the field redefinition (\ref{CalcPhifr}) entails a non-trivial 
homogeneous 
solution, the Deser--Woodard model in its nonlocal form \Eq{Calctolag} differs 
from its ``localized'' form obtained via such a  field transformation. 
Therefore, 
predictions are frame-dependent. The Deser--Woodard model was mainly studied in 
its localised form. Since $R=0$ during the radiation epoch, $O(\BCg^{-1} R)$ 
terms start to grow only after the onset of matter domination and do not spoil 
early-universe constraints, which makes this model a viable candidate for dark 
energy.  Functions $f$ fitting late-time observations can be extracted by data 
\cite{Deffayet:2009ca,Elizalde:2012ja}. For instance, the $\Lambda$CDM model is 
reproduced when
\be
f(\phi)= c\left[\tanh\left(\sum_{l=0}^3c_l\phi^l\right)-1\right],
\ee
for certain numerical coefficients $c$ and $c_l$ \cite{Deffayet:2009ca}.  The 
form of $f$ is completely \emph{ad hoc} and not especially attractive, but it 
does not involve fine tuning. Depending on the data set of structure-evolution 
observations and on the type of analysis, the Deser--Woodard model is favoured 
\cite{Nersisyan:2017mgj,Park:2017zls} or comparable \cite{Amendola:2019fhc} 
with 
respect to standard General Relativity. However, the model does not pass 
Solar-System tests, due to failure of its screening mechanism 
\cite{Belgacem:2018wtb}. The new Deser--Woodard model seems able to avoid this 
issue \cite{Deser:2019lmm,Ding:2019rlp}.

The localised version of the Maggiore--Mancarella model gives a viable  
dark-energy evolution with $m=O(H_0)$ 
\cite{Maggiore:2014sia,Dirian:2014ara,Dirian:2014bma,Dirian:2016puz,
Belgacem:2017cqo} and is compatible with structure formation 
\cite{Barreira:2014kra,Dirian:2014bma,Dirian:2016puz,Belgacem:2017cqo}. 
Unfortunately, the lack of a screening mechanism leads to non-negligible 
deviations in Solar-System tests, which rule the model out 
\cite{Barreira:2014kra,Belgacem:2018wtb}. 

The massive-gravity model passes a battery of observational tests 
\cite{Nesseris:2014mea} and,  in fact, it is indistinguishable from $\Lambda$CDM 
with present data \cite{Dirian:2016puz}. Since it is compatible also with 
Solar-System tests \cite{Belgacem:2018wtb}, it is a stronger candidate than the 
Deser--Woodard and the Maggiore--Mancarella models and, perhaps, the only 
survivor of the three. 

Finally, the $n=2$, $\cR=R_{\mu\nu}$ model is unviable, due to instabilities in 
tensor perturbations \cite{Cusin:2015rex,Zhang:2016ykx}.

\subsubsection{Gravitational Waves}

IR nonlocal gravity could leave an imprint in GW propagation  
\cite{Belgacem:2017cqo,Belgacem:2018lbp,Belgacem:2019pkk}. In particular, the 
ratio of the gravitational and electromagnetic luminosity distance for a 
standard siren fits the parametrisation
\be
\frac{d_L^\textsc{gw}}{d_L^\textsc{em}} \simeq \Xi_0+\frac{1-\Xi_0}{(1+z)^n}\,,
\ee
where $\Xi_0$ and $n$ are constants. The Mancarella--Maggiore model (RR model, 
under  strong pressure by Solar-System tests) predicts $\Xi_0\approx 0.970$ and 
$n\approx 2.5$ \cite{Belgacem:2018lbp}, while for nonlocal massive gravity (RT 
model) $\Xi_0\approx 0.934$ and $n\approx 2.6$ when setting the initial 
conditions in the radiation-domination era \cite{Belgacem:2019pkk}. These values 
are within the sensitivity of LISA, for which an $O(1\%)$ relative error on 
$\Xi_0$ has been estimated \cite{Belgacem:2019pkk}. Setting instead the initial 
conditions during inflation, one can get even larger deviations form General 
Relativity (around $\Xi_0\sim 1.5$) while respecting other constraints 
\cite{Belgacem:2019lwx}.




%
%
%
%
%
%

%
%

\chapter[Metric-Affine Gravity]{Metric-Affine Gravity}
\label{ref:Iosifidis}

{\em Damianos Iosifidis, Emmanuel N. Saridakis }\\



Probably one of the most beautiful characteristics of General Relativity  (GR)
is its clear geometric interpretation. Having been acquainted with this idea of
geometrization of gravity, theories that incorporate  geometrical notions are of
particular interest and well motivated. Let us recall that in Einstein's
Gravity, the fundamental field is the metric $g_{\mu\nu}$ and the connection is
in some sense a secondary object, since it can be obtained by the former and its
derivatives(the Levi-Civita connection). One is  then dealing with the familiar
Riemannian geometry \cite{eisenhart1997riemannian}, where the metric of the
manifold completely determines its structure. In this case the connection is by
definition symmetric (torsionless) and metric compatible (the metric is
covariantly constant), and is the unique Levi-Civita connection.

From the above discussion it becomes clear what would be the most natural way
to  extend the geometry, and this is of course the relaxation of the Riemannian
constraints. In other words, by starting with a general affine connection that
admits both torsion and non-metricity (see definitions in what follows), one
enriches the geometric arena and is in the realm of the so-called
non-Riemannian geometry \cite{eisenhart2012non}. The underlying gravity theory
formulated in such generalised geometry is the Metric-Affine
Gravity (MAG) \cite{Hehl:1994ue}.
As we noted above in the non-Riemannian arena of MAG the manifold apart from
curvature  is also endowed with torsion and non-metricity. In this space,
vectors rotate (torsion) and experience a length change (non-metricity) when
moved from one point of the manifold to another. Both of those degrees of
freedom (i.e. torsion and non-metricity) are beautifully encoded  into the
general affine connection $\hat{\nabla}$. By restricting the general connection in
certain ways, one obtains the teleparallel
equivalents \cite{Aldrovandi:2013wha,Nester:1998mp,
BeltranJimenez:2018vdo,Jimenez:2019ghw} that are elaborated in great detail
in subsequent chapters.

 Going back to the MAG framework, the fundamental variables in this case are
both the metric and the affine  connection (note that in the
equivalent differential form formalism of MAG the independent fields are the
frame $e^{A}$ and the linear connection $\udt{\hat{\omega}}{A}{B}$ one forms
\cite{Hehl:1994ue}). The field equations are obtained by varying
independently with respect to both. The relation between the metric and the
connection may be found only after solving the associated field equation. In
what follows we briefly discuss the geometric and physical setup of
Metric-Affine Theories. It is also worth mentioning that MAG is a gauge theory
of gavity. More precisely, when expressed in the language of differential forms
it is clear that MAG is the gauge theory of the ${\rm GL}(4,\mathbb{R})$ group
\cite{Hehl:1994ue}. It also offers interesting possibilities for the quantisation
of Gravity \cite{Percacci:2020bzf}.  For further details on the matter of MAG,
see, for instance,
\cite{Hehl:1994ue,Hehl:1999sb,Iosifidis:2019jgi,Iosifidis:2018jwu,
Obukhov:1996ka,Vitagliano:2010sr}.

\section{Geometrical Objects: Torsion, Curvature and non-Metricity}
\label{gaugetheref1}
The structure of a given manifold is completely determined once a metric
tensor $g$ and an affine-connection\label{metaffconref2} with coefficients 
$\hat{\Gamma}$ is given. The manifold is then
denoted as ($\mathcal{M},g,\hat{\Gamma}$). Geometrically speaking, the above two
notions of metric and connection serve different purposes  and are, in general,
unrelated. The metric is needed in order to define distances, dot products
between vector fields, and also define mappings among covariant and
contravariant tensor fields. On the other hand, the connection defines parallel
transport of tensor fields (through covariant differentiation) and allows
to compare vectors  living at different vector spaces. In a general setting, the
affine-connection possesses both torsion and non-metricity, namely it is
neither symmetric nor metric compatible. This generalized geometrical setup is
known as non-Riemannian Geometry \cite{eisenhart2012non}. As we have mentioned,
this non-Riemannian geometry offers a playground for the Metric-Affine
Theories to be constructed. It is therefore appropriate to discuss some basic
concepts of non-Riemannian geometries, that will help delve into the
Metric-Affine framework.


Let us start with some basic concepts of non-Riemannian 
geometry.\label{nonRiemannianref2}
 Firstly, in our conventions, the covariant derivative of, say, a $(1,1)$ type
tensor reads
\beq
\hat{\nabla}_{\alpha}A^{\mu}_{\thickspace\thickspace\nu}=\partial_{\alpha}
A^{\mu}_{\thickspace\thickspace\nu}-\hat{\Gamma}^{\lambda}_{
\thickspace\thickspace\thickspace\nu\alpha}A^{\mu}_{
\thickspace\thickspace\lambda}+\hat{\Gamma}^{\mu}_{
\thickspace\thickspace\thickspace\lambda\alpha}A^{\lambda}_{
\thickspace\thickspace\nu}.
\eeq
Considering the commutator of two covariant derivatives and acting it on a
scalar, it  follows that
\beq
2
\hat{\nabla}_{[\mu}\hat{\nabla}_{\nu]}\phi=-\hat{T}^{\lambda}_{\thickspace\thickspace\mu\nu}
\hat{\nabla}_{\lambda}\phi,
\eeq
where
\beq
\hat{T}^{\lambda}_{\thickspace\thickspace\mu\nu}:=
-2\hat{\Gamma}^{\lambda}_{\thickspace\thickspace\thickspace[\mu\nu]}
\eeq
is the torsion tensor\label{torsionref3}, which as seen from the above equation 
is defined as the
antisymmetric part of the affine-connection.   Out of the torsion tensor we can
construct the torsion vector  according to
\beq
\hat{T}_{\mu}:=\hat{T}^{\lambda}_{\thickspace\thickspace\lambda\mu},
\eeq
which exists for arbitrary space dimension-$n$. For $n=4$ in particular we can
also define the  torsion pseudo-vector
\beq
\hat{A}_{\mu}:=\epsilon_{\mu\alpha\beta\gamma}\hat{T}^{\alpha\beta\gamma},
\eeq
where  $\epsilon_{\mu\alpha\beta\gamma}$ is the 4-dimensional totally
antisymmetric Levi-Civita tensor.

Acting now, with the antisymmetrized covariant derivative on a vector field
$u^{\mu}$ we obtain
\begin{equation}
[\hat{\nabla}_{\alpha} ,\hat{\nabla}_{\beta}]u^{\mu}=2\hat{\nabla}_{[\alpha}
\hat{\nabla}_{\beta]}u^{\mu}=\hat{R}^{\mu}_{\thickspace\thickspace\thickspace\nu\alpha\beta}
 u^{\nu}- \hat{T}^{\nu}_{\thickspace\thickspace\alpha\beta}\hat{\nabla}_{\nu}u^{\mu},
\end{equation}
where
\begin{equation}
\hat{R}^{\mu}_{\thickspace\thickspace\thickspace\nu\alpha\beta}:=
2\partial_{[\alpha}\hat{\Gamma}^{\mu}_{\thickspace\thickspace\thickspace|\nu|\beta]}
+2\hat{\Gamma}^{\mu}_{\thickspace\thickspace\thickspace\rho[\alpha}\hat{\Gamma}^{\rho}_{
\thickspace\thickspace\thickspace|\nu|\beta]} \label{Riem}
\end{equation}
is the usual Riemann or Curvature tensor and the horizontal bars around an
index denote that this index is  left out of the (anti)-symmetrization. It is
worth mentioning that in general (non-Riemannian Geometries), the only symmetry
of the Riemann tensor is antisymmetry in its last two indices, as seen from
(\ref{Riem}).  Without the use of any metric, we can construct the  two
independent contractions
\begin{eqnarray}
\label{Damosdef11}
&&\hat{R}_{\nu\beta}:=
\hat{R}^{\mu}_{\thickspace\thickspace\thickspace\nu\mu\beta}\thickspace,
\thickspace\\
&&\hat{\tilde{R}}_{\alpha\beta}:=\hat{R}^{\mu}_{
\thickspace\thickspace\thickspace\mu\alpha\beta}.
\label{Damosdef22}
\end{eqnarray}
Expression (\ref{Damosdef11})
  defines as usual the Ricci tensor,  while expression (\ref{Damosdef22}) 
defines the 
homothetic curvature (the field strength of the Weyl vector) it is of purely 
non-Riemannian origin and it is anti-symmetric. Note that in
our discussion so far, no metric was involved. If we enrich the geometrical
structure with a metric   we can form yet another  contraction,
\beq
\hat{\tilde{\mathcal{R}}}{}^{\lambda}_{\thickspace\thickspace\kappa}:=
\hat{R}^{\lambda}_{\thickspace\thickspace\mu\nu\kappa}g^{\mu\nu},
\eeq
which is often called, ``co-Ricci'' tensor, and in general is different from the 
Ricci tensor (\ref{Damosdef11}) (in General Relativity we have that $
\hat{\tilde{\mathcal{R}}}{}^{\lambda}_{\thickspace\thickspace\kappa}=-\hat{R}{}^
{\lambda}_{\thickspace\thickspace\kappa}$).
However, the Ricci scalar is still uniquely defined since
\begin{eqnarray}
&&\hat{R}:=\hat{R}_{\mu\nu}g^{\mu\nu}=-\hat{\tilde{\mathcal{R}}}_{\mu\nu}g^{
\mu\nu }\\
&&\hat{\tilde{R}}_{\mu\nu}g^{\mu\nu}=0.
\end{eqnarray}
Having introduced a metric tensor on a manifold the latter is not, in general,
covariantly constant.
A metric is said to be compatible with a given
connection, if and only if its covariant derivative with respect to this
connection
vanishes, i.e. $\hat{\nabla}_{\alpha}g_{\mu\nu}\equiv0$. Hence, it is
exactly this deviation from
the compatibility condition, defines the non-metricity
tensor\label{nonmetricityref2}
\beq
\hat{Q}_{\alpha\mu\nu}=\hat{\nabla}_{\alpha}g_{\mu\nu}.
\eeq
We can then contract the above in two independent ways, to get the two
non-metricity vectors
\beq
\hat{Q}_{\alpha}:=\hat{Q}_{\alpha\mu\nu}g^{\mu\nu}\thickspace,\thickspace\thickspace
\hat{\tilde{Q}}_{\nu}=\hat{Q}_{\alpha\mu\nu}g^{\alpha\mu}.
\eeq
The former goes by the name Weyl vector, and the latter is the second
independent vector that can  be extracted from non-mnetricity. The general
non-Riemannian space that has all of its geometrical objects unconstrained is
often denoted as $L_{n}$, and is the playground of general Metric-Affine
Theories. As a last note, let us mention that with the above definitions it is
trivial to show that the affine connection can be decomposed according
to \cite{schouten2013ricci}(see also
\cite{Hehl:1994ue,Iosifidis:2018jwu}),
\begin{equation}
\hat{\Gamma}^{\lambda}_{\thickspace\thickspace\thickspace\mu\nu}=
\Gamma ^{\lambda}_{\thickspace\thickspace\thickspace\mu\nu}+\frac{1}{2}g^
{\alpha\lambda}(-\hat{Q}_{\mu\nu\alpha}-\hat{Q}_{\nu\alpha\mu}+\hat{Q}_{\alpha\mu\nu})
+\frac{1}{2}g^{\alpha\lambda}(\hat{T}_{\nu\alpha\mu}+\hat{T}_{\mu\alpha\nu}-\hat{T}_{\alpha
\mu\nu}) \label{affconection},
\end{equation}
where $$\Gamma ^{\lambda}_{\thickspace\thickspace\thickspace\mu\nu} = \frac{1}{2}g^{\lambda\kappa} \left(g_{\kappa\nu,\mu} + g_{\mu\kappa,\nu}-g_{\mu\nu,\kappa}\right)\,,$$
represents the usual Levi-Civita  connection, and the combination
\beq
\hat{L}_{\alpha\mu\nu}+\hat{K}_{\alpha\mu\nu}=\frac{1}{2}(-\hat{Q}_{\mu\nu\alpha}-\hat{Q}_{\nu\alpha\mu}+\hat{Q}_{\alpha\mu\nu}
)  +\frac{1}{2}(\hat{T}_{\nu\alpha\mu}+\hat{T}_{\mu\alpha\nu}-\hat{T}_{\alpha\mu\nu}) \label{l}
\eeq
is known as the distortion tensor, measuring how much the affine connection
deviates  from the Riemannian Levi-Civita connection. This decomposition allows
one to split any object into its Riemannian part, plus non-Riemannian
contributions.

 Let us now touch upon the geometrical interpretation of the geometrical
objects.  Since the role of curvature is widely known, we shall concentrate on
the notion of torsion and non-metricity.

\section{Geometrical Meaning of Torsion and Non-metricity}

\subsection{Geometrical Meaning of Torsion}\label{torsionref4}
\begin{center}\begin{figure}
    \centering
    \includegraphics[width=0.45\textwidth]{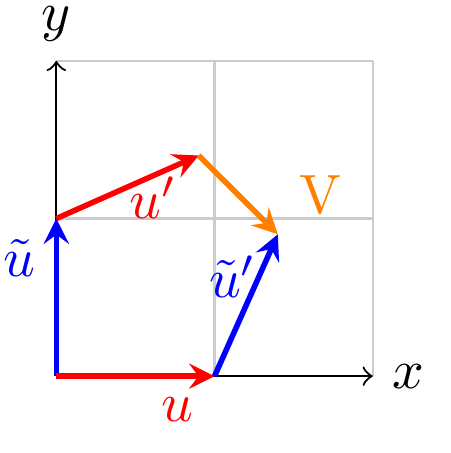}
	\caption{{\it{Representation of torsion in 2 dimensions.}}}
    \label{fig:torsion1}
\end{figure}
\end{center}

 Let us examine the geometric effect of torsion with the following simple
example.  Consider two curves $\mathcal{C}:x^{\mu}=x^{\mu}(\lambda)$ \textrm{and}
$\mathcal{\tilde{C}}:\tilde{x}^{\mu}=\tilde{x}^{\mu}(\lambda)$ with associated
tangent vectors
\begin{equation}
u^{\mu}=\frac{{\rm d}x^{\mu}}{{\rm d}\lambda}
\thickspace\thickspace\thickspace\thickspace
and\thickspace\thickspace\thickspace\thickspace
\tilde{u}^{\mu}=\frac{{\rm d}\tilde{x}^{\mu}}{{\rm d}\lambda}
\end{equation}
respectively. Now, let  ${\rm d}\tilde{x}^{\mu}$ represent a displacement of
$u^{\alpha}$  along $\mathcal{\tilde{C}}$ and obtain $u'^{\alpha}$ which in
first order is given by
\begin{equation}
u'{}^{\alpha}=u^{\alpha}+(\partial_{\mu}u^{\alpha}){\rm d}\tilde{x}^{\mu} \label{toru},
\end{equation}
but recalling that  $u^{\alpha}$ is parallel transported along
$\mathcal{\tilde{C}}$,  it follows that
\begin{equation}
\frac{{\rm d}\tilde{x}^{\mu}}{{\rm d}\lambda}\hat{\nabla}_{\mu}u^{\alpha}=0=
\frac{{\rm d}\tilde{x}^{\mu}}{{\rm d}\lambda}\partial_{\mu}u^{\alpha}+\hat{\Gamma}^{\alpha}_{
\thickspace\thickspace\thickspace\nu\mu}\frac{{\rm d}\tilde{x}^{\mu}}{{\rm d}\lambda}u^{\nu}
, \nonumber
\end{equation}
or
\begin{equation}
(\partial_{\mu}u^{\alpha}){\rm d}\tilde{x}^{\mu}=
-\hat{\Gamma}^{\alpha}_{\thickspace\thickspace\thickspace\nu\mu}u^{\nu}\tilde{u}^{\mu}
{\rm d}\lambda,
\end{equation}
which when placed against $(\ref{toru})$ gives
\beq
u'^{\alpha}=u^{\alpha}
-\hat{\Gamma}^{\alpha}_{\thickspace\thickspace\thickspace\nu\mu}u^{\nu}\tilde{u}^{\mu}
{\rm d}\lambda.
\eeq
Following the exact same procedure,  but now for a $dx^{\mu}$-displacement  of
$\tilde{u}^{\alpha}$  along $\mathcal{C}$, we acquire
\beq
\tilde{u}'^{\alpha}=\tilde{u}^{\alpha}-
\hat{\Gamma}^{\alpha}_{\thickspace\thickspace\thickspace\nu\mu}\tilde{u}^{\nu}u^{\mu}
{\rm d}\lambda=
\tilde{u}^{\alpha}-\hat{\Gamma}^{\alpha}_{\thickspace\thickspace\thickspace\mu\nu}
\tilde{u}^{\mu}u^{\nu}{\rm d}\lambda.
\eeq
From the last two we obtain
\beq
(\tilde{u}^{\alpha}+u'^{\alpha})-(u^{\alpha}+\tilde{u}'^{\alpha})=
-\hat{T}^{\alpha}_{\thickspace\thickspace\mu\nu}\tilde{u}^{\mu}u^{\nu}{\rm d}\lambda.
\eeq
Notice now that for the infinitesimal parallelogram to exist, the vectors
$(\tilde{u}^{\alpha}+u^{'\alpha})$ and $(u^{\alpha}+\tilde{u}^{'\alpha})$
should be equal, and as is clear from the above, this is not true in the
presence of torsion. Defining the vector that shows this deviation as
$\hat{V}^{\alpha}{\rm d}\lambda=(\tilde{u}^{\alpha}+u^{'\alpha})-(u^{\alpha}+\tilde{u}^{
'\alpha})$, the latter can also be written as
\beq
\hat{V}^{\alpha}=-\hat{T}^{\alpha}_{\thickspace\thickspace\mu\nu}\tilde{u}^{\mu}u^{\nu},
\eeq
which represents how much the parallelogram has been cracked. This
is definitely true
for small displacements in the directions of $\tilde{u}^{\mu}$ and $u^{\nu}$
which themselves are computed at the starting point of the path. Hence, simply
put, the presence  of torsion cracks parallelograms into pentagons, as illustrated in Fig.(\ref{fig:torsion1}) (in a $2$-dim flat space).

\subsection{Geometrical Meaning of Non-Metricity}

\begin{center}\begin{figure}
    \centering
    \includegraphics[width=0.4\textwidth]{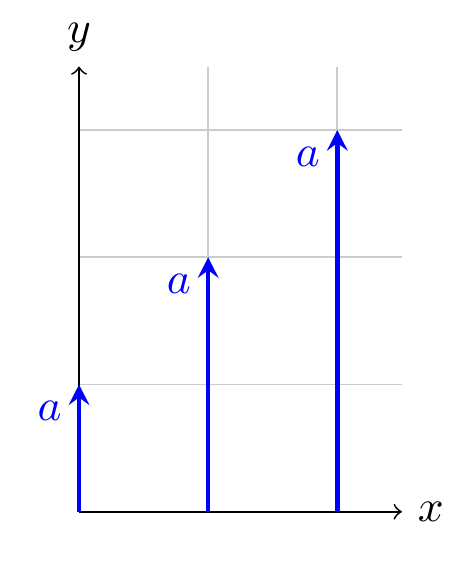}
	\caption{{\it{Representation of non-metricity in 2 dimensions.}}}
    \label{fig:torsion}
\end{figure}
\end{center}

Let us explore the geometric property of non-metricity. We consider a
differential manifold, and we assume that a metric and a connection
are also given for this manifold. Moreover, we consider two vectors
$a^{\mu}$ and $b^{\mu}$ on this manifold. Defining their inner product $a\cdot
b=a^{\mu}b^{\nu}g_{\mu\nu}$  let us parallel transport both vectors along a
given curve $\mathcal{C}: x^{\mu}=x^{\mu}(\lambda)$. We have
\beq
\frac{\hat{D}}{{\rm d}\lambda}(a\cdot b)=\frac{{\rm d}x^{\alpha}}{{\rm d}\lambda}(\hat{\nabla}_{\alpha}
a^{\mu})b_{\mu}+\frac{{\rm d}x^{\alpha}}{{\rm d}\lambda}(\hat{\nabla}_{\alpha}
b^{\nu})a_{\nu}+\frac{{\rm d}x^{\alpha}}{{\rm d}\lambda}(\hat{\nabla}_{\alpha}g_{\mu\nu})a^{\mu}b^
{\nu}.
\eeq
Now, since $a^{\mu}$ and $b^{\mu}$ are parallel transported along the curve, it
 follows that
\beq
\frac{{\rm d}x^{\alpha}}{{\rm d}\lambda}(\hat{\nabla}_{\alpha}
a^{\mu})=0\thickspace,\thickspace\thickspace
\frac{{\rm d}x^{\alpha}}{{\rm d}\lambda}(\hat{\nabla}_{\alpha} b^{\nu})=0,
\eeq
and thus
\beq
\frac{\hat{D}}{{\rm d}\lambda}(a\cdot
b)=\hat{Q}_{\alpha\mu\nu}\frac{{\rm d}x^{\alpha}}{{\rm d}\lambda}a^{\mu}b^{\nu},
\eeq
which tells us that  the inner product of two vectors changes when we parallel
transport them  along a curve. Note that, for $b^{\mu}=a^{\mu}$ the above becomes
\beq
\frac{\hat{D}}{{\rm d}\lambda}(\vert a \vert^{2}
)=\hat{Q}_{\alpha\mu\nu}\frac{{\rm d}x^{\alpha}}{{\rm d}\lambda}a^{\mu}a^{\nu} \label{fixedlvq},
\eeq
which signals a length change of the vectors magnitude when we parallel
transport  it along a given curve. Therefore, in a space with non-metricity, dot
products and lengths of vectors change under parallel transport (for
more concrete examples on the geometrical meaning of both, as well as
non-metricity, see  \cite{Iosifidis:2018jwu}). Fig.(\ref{fig:torsion}) illustrates
a two-dimensional example of a vector dragged along the $x-$axis in the presence
of non-metricity.

\section{Identities of non-Riemannian Geometry}

	Many (if not all) of the identities that are familiar from  GR
generalise and receive further contributions from torsion and non-metricity
when passing to a non-Riemannian background.
Let us start with the generalised Bianchi identities for a torsionful,
non-metric connection. These identities are also known as Weitzenb\"ock
identities. The first one we obtain by taking the covariant derivative of the
Riemann tensor and antisymmetrise in three indices, which results in
\cite{schouten2013ricci,Iosifidis:2019jgi}
\beq
\hat{\nabla}_{[\rho}\hat{R}^{\alpha}_{\thickspace\thickspace\thickspace|\beta|\mu\nu]}=-
\hat{R}^{\alpha}_{\thickspace\thickspace\thickspace\beta\lambda[\rho}\hat{T}^{\lambda}_{
\thickspace\thickspace\mu\nu]} \label{bian},
\eeq
where the vertical bars around an index indicate that this index is left out of the (anti-)symmetrisation. The above constitute the generalised differential
Bianchi identities in a non-Riemannian setup. Another identity comes about by
applying the commutator of two covariant derivatives on the metric tensor.
Then, using the definitions of curvature, torsion and non-metricity, one can
easily show \cite{schouten2013ricci,Iosifidis:2019jgi}
\beq
\hat{R}_{(\alpha\mu)\beta\nu}=-\hat{\nabla}_{[\beta}\hat{Q}_{\nu]\alpha\mu}+
\hat{T}^{\lambda}_{\thickspace\thickspace\beta\nu}\hat{Q}_{\lambda\alpha\mu},
\eeq
which shows that in the presence of non-metricity, the curvature tensor  is no
longer antisymmetric in its first two indices. If we contract the latter
 with $g^{\alpha\beta}$ and employ the Leibniz rule we get the relation
\begin{gather}
\hat{R}_{\mu\nu}+\hat{\tilde{\cal R}}_{\mu\nu}=
-\hat{\nabla}_{\lambda}\hat{Q}_{\nu\mu}^{\thickspace\thickspace\thickspace\thickspace\lambda
}+\hat{Q}_{\nu\alpha\beta}\hat{Q}^{\alpha\beta}_{
\thickspace\thickspace\thickspace\thickspace\mu}+\hat{\nabla}_{\nu}\hat{\tilde{Q}}_{\mu}
-\hat{Q}_{\nu\mu\alpha}\hat{\tilde{Q}}^{\alpha}+2
\hat{T}_{\alpha\beta\nu}\hat{Q}^{\beta\alpha}_{
\thickspace\thickspace\thickspace\thickspace\mu} \label{B2}.
\end{gather}
showing that the third contraction $\hat{\tilde{R}}_{\mu\nu}$ of the  curvature is
not completely independent from the Ricci tensor. In addition, using the
definition of the Riemann tensor and antisymmetrizing in its last three indices,
we easily obtain \cite{schouten2013ricci,Iosifidis:2019jgi}
	\beq
\hat{R}^{\alpha}_{\thickspace\thickspace\thickspace[\beta\mu\nu]}=
\hat{\nabla}_{[\beta}\hat{T}^{\lambda}_{\thickspace\thickspace\mu\nu]}-
\hat{T}^{\lambda}_{\thickspace\thickspace\thickspace[\beta\mu}\hat{T}^{\alpha}_{
\thickspace\thickspace\nu]\lambda} \label{B1},
\eeq
which is the analogue of the algebraic Bianchi identity in a non-Riemannian
geometry.  Notice that the Riemannian geometry identity
$\hat{R}^{\alpha}_{\thickspace\thickspace\thickspace[\beta\mu\nu]}=0$ is spoiled now
by the presence of torsion only. It should be clear by now that apart from the
antisymmetry in the last two indices, the Riemann tensor possesses no other
symmetry in a generalized geometry. In particular, in $\hat{R}_{\alpha\beta\mu\nu}$
one cannot interchange the set of indices $\alpha\beta$ with $\mu\nu$ that is
$\hat{R}_{\alpha\beta\mu\nu}\neq \hat{R}_{\mu\nu\alpha\beta}$. However, a thing that is
never mentioned in the literature is that there is a certain identity that
measures the difference between the aforementioned two Riemann tensors. Indeed,
starting with a general tensor identity holding for arbitrary type $(0,4)$
tensors and applying it  to the Riemann tensor, after using the identities we
derived above, it follows that \cite{Iosifidis:2019jgi}
\begin{align}
\hat{R}_{\mu\nu\kappa\lambda}-\hat{R}_{\kappa\lambda\mu\nu}=
&\hat{\nabla}_{[\lambda}\hat{Q}_{\kappa]\nu\mu}+\nabla_{[\nu}\hat{Q}_{\lambda]\kappa\mu}+\hat{\nabla}_{[\kappa}\hat{Q}_{\mu]\lambda\nu}+\hat{\nabla}_{[\mu}\hat{Q}_{\nu]\lambda\kappa}+\hat{\nabla}_{[\mu}\hat{Q}_{\lambda]\nu\kappa}+\hat{\nabla}_{[\nu}\hat{Q}_{\kappa]\lambda\mu} \nonumber \\
&-\frac{3}{2}\Big(
g_{\mu\alpha}\hat{T}^{\beta}_{\,\,\,\,[\nu\lambda}\hat{T}^{\alpha}_{\,\,\,\,\kappa]\beta}+g_{\nu\alpha}\hat{T}^{\beta}_{\,\,\,\,[\mu\lambda}\hat{T}^{\alpha}_{\,\,\,\,\kappa]\beta}
+g_{\kappa\alpha}\hat{T}^{\beta}_{\,\,\,\,[\lambda\mu}\hat{T}^{\alpha}_{\,\,\,\,\nu]\beta}+g_{\lambda\alpha}\hat{T}^{\beta}_{\,\,\,\,[\kappa\mu}\hat{T}^{\alpha}_{\,\,\,\,\nu]\beta} \Big) \nonumber \\
&+\frac{1}{2}\Big(\hat{T}^{\alpha}_{\,\,\,\,\lambda\kappa}\hat{Q}_{\alpha\nu\mu}+
\hat{T}^{\alpha}_{\,\,\,\,\nu\lambda}\hat{Q}_
{\alpha\kappa\mu}+ \hat{T}^{\alpha}_{\,\,\,\,\kappa\mu}\hat{Q}_{\alpha\lambda\nu} + \hat{T}^{\alpha}_{\,\,\,\,\mu\nu}\hat{Q}_{\alpha\lambda\kappa}+ \hat{T}^{\alpha}_{\,\,\,\,\mu\lambda}\hat{Q}_{\alpha\nu\kappa}+ \hat{T}^{\alpha}_{\,\,\,\,\nu\kappa}\hat{Q}_{\alpha\lambda\mu}\Big) \nonumber \\
&- \frac{3}{2}\Big( g_{\mu\alpha}\hat{\nabla}_{[\nu}\hat{T}^{\alpha}_{\,\,\,\,\lambda\kappa]} + g_{\nu\alpha}\hat{\nabla}_{[\mu}\hat{T}^{\alpha}_{\,\,\,\,\lambda\kappa]}+
g_{\kappa\alpha}\hat{\nabla}_{[\lambda}\hat{T}^{\alpha}_{\,\,\,\,\mu\nu]}+g_{\lambda\alpha}\hat{\nabla}_{[\kappa}\hat{T}^{\alpha}_{\,\,\,\,\mu\nu]}\Big)  \label{xamoulhs},
\end{align}
and we see that the symmetry $\hat{R}_{\mu\nu\kappa\lambda}=\hat{R}_{\kappa\lambda\mu\nu}$
only holds when both torsion and non-metricity vanish. This last identity
clearly shows that when one is working in the general Metric-Affine framework,
with a generic affine connection, things get quite serious computationally!

\subsection{The  Sources of Metric-Affine Gravity}
\label{metaffgrref2}

As we have noted, in the general Metric-Affine Gravity  (MAG) framework, the
action is a functional of the metric, the frame and the independent affine
connection and the matter fields. More precisely, the matter sector of general
MAG Theories reads\footnote{The independent geometrical variables are truly two,
since it can be shown that the metric field equations are redundant and are
related to the ones coming from the connection and the
frame \cite{Hehl:1994ue}. In other words, the set of field equations for
($e,\hat{\Gamma}$) and ($g,\hat{\Gamma}$) are fully equivalent, and it is a matter of
personal preference,  which one to solve.} 
\beq
{S}_{\rm matter}[g,e,\hat{\Gamma},\phi]=\int {\rm d}^{4}x \sqrt{-g} 
\mathcal{L}_{\rm matter}(g,\hat{\Gamma},\phi)
\eeq \label{Enmomtenref2}
where $\phi$ collectively denotes the matter fields. One then defines as usual
the metrical (symmetric) energy-momentum tensor by the metric variation of the
matter part
\beq
T_{\alpha\beta}:= -\frac{2}{\sqrt{-g}}\frac{\delta {S}_{\rm matter}}{\delta
g^{\alpha\beta}} =-\frac{2}{\sqrt{-g}}\frac{\delta(\sqrt{-g}
\mathcal{L}_{\rm matter})}{\delta g^{\alpha\beta}}.
\eeq
In MAG, the connection naturally couples to matter, so we now also have the
variation
\beq
\hat{\Delta}_{\lambda}^{\thickspace\thickspace\thickspace\mu\nu}:=
-\frac{2}{\sqrt{-g}}\frac{\delta {S}_{\rm matter}}{\delta
\hat{\Gamma}^{\lambda}_{\thickspace\thickspace\thickspace\mu\nu}}=-\frac{2}{
\sqrt{-g}}
\frac{\delta ( \sqrt{-g} \mathcal{L}_{\rm matter})}{\delta
\hat{\Gamma}^{\lambda}_{\thickspace\thickspace\thickspace\mu\nu}},
\eeq
which is the hypermomentum tensor \cite{hehl1976hypermomentum}  and encompasses
the microscopic characteristics of matter, such as spin, dilation and
shear \cite{Hehl:1994ue}. Additionally,  there is also the canonical
energy-momentum tensor given by
\beq
t^{\mu}_{\thickspace\thickspace C}=\frac{1}{\sqrt{-g}}\frac{\delta
{S}_{\rm matter}}{\delta e^{C}{}_{\mu}},
\eeq
which is not symmetric in general. The above three currents
($T_{\alpha\beta}$,\thickspace$\hat{\Delta}_{\lambda}^{
\thickspace\thickspace\thickspace\mu\nu}$,\thickspace$t^{\mu}_{
\thickspace\thickspace C}$) represent the sources of
MAG \cite{Hehl:1994ue,Obukhov:2014nja}, which produce spacetime
curvature, torsion and non-metricity through the field equations of MAG.

\section{Field Equations of Metric-Affine
Gravity}

Let us know proceed with the field equations of MAG theories.  A fairly general
class of theories consists of a gravitational Lagrangian constructed out of
curvature, torsion and non-metricity\footnote{And on the metric $g_{\mu\nu}$
obviously.}, i.e., $\mathcal{L}=\mathcal{L}(g_{\mu\nu},\hat{R}^{\alpha}_{
\thickspace\thickspace\beta\gamma\rho},
\hat{T}^{\lambda}_{\thickspace\thickspace\mu\nu},\hat{Q}_{\alpha\mu\nu})$.
After all, these are the three covariant geometrical objects building up  a
non-Riemannian Geometry. Then, adding some matter fields, the general action
reads
\beq
{S}[g,\hat{\Gamma},\phi]=\frac{1}{2 \kappa ^2}\int {\rm d}^{4}x
\sqrt{-g}\mathcal{L}(g_{\mu\nu},\hat{R}^{\alpha}_{
\thickspace\thickspace\beta\gamma\rho},
\hat{T}^{\lambda}_{\thickspace\thickspace\mu\nu},\hat{Q}_{\alpha\mu\nu})+{S}_{m}
.
\eeq
Defining now the field excitations
\beq
\hat{\Omega}_{\lambda}^{\thickspace\thickspace\thickspace\mu\alpha\nu} :=
\frac{\partial \mathcal{L}}{\partial
\hat{R}^{\lambda}_{\thickspace\thickspace\mu\alpha\nu}}\thickspace,
\thickspace\thickspace
\hat{V}^{\mu\nu}_{\thickspace\thickspace\thickspace\thickspace\lambda} := -2
\frac{\partial \mathcal{L}}{\partial
\hat{T}^{\lambda}_{\thickspace\thickspace\mu\nu}}\thickspace,\thickspace
\hat{W}^{\alpha\mu\nu}  :=- \frac{\partial \mathcal{L}}{\partial
\hat{Q}_{\alpha\mu\nu}},
\eeq
we vary independently with respect to $g$,\thickspace$\Gamma$ and $\phi$ to
obtain
\begin{equation}
-2\frac{\hat{\nabla}_{\alpha}
(\sqrt{-g}\hat{\Omega}_{\lambda}^{\thickspace\thickspace\thickspace\mu\alpha\nu}
)}{
\sqrt{-g}}+2
\hat{\Omega}_{\lambda}^{\thickspace\thickspace\thickspace\mu\alpha\nu}\hat{T}_{
\alpha}+\frac
{1}{2}\hat{\Omega}_{\lambda}^{\thickspace\thickspace\thickspace\mu\gamma\delta}
\hat{T}^{\nu}
_{\thickspace\thickspace\gamma\delta} +2 
\hat{W}^{\mu\nu}_{\thickspace\thickspace\thickspace\thickspace\lambda} +
\hat{V}^{\mu\nu}_{\thickspace\thickspace\thickspace\thickspace\lambda}= \kappa^2
\hat{\Delta}_{\lambda}^{\thickspace\thickspace\thickspace\mu\nu},
\end{equation}
\beq
-\frac{1}{2}g_{\mu\nu}\mathcal{L}+\frac{\partial \mathcal{L} }{\partial
g^{\mu\nu}}
+\frac{1}{\sqrt{-g}}( \hat{T}_{\alpha}-\hat{\nabla}_{\alpha})\sqrt{-g}  
\frac{\partial
\mathcal{L}}{\partial
\hat{Q}_{\alpha}^{\thickspace\thickspace\thickspace\mu\nu}}=\kappa^2 T_{\mu\nu},
\eeq
\beq
\frac{\delta {S}_{m} }{\delta \phi}=0.
\eeq
The above field equations  cover a fairly wide spectrum of Metric-Affine
Theories of Gravity. Notice that we chose to present them in a coordinate-based
formalism (our basic variables were $g_{\mu\nu}$ and
$\hat{\Gamma}^{\lambda}_{\thickspace\thickspace\thickspace\alpha\beta}$), but one 
can
just as well work with the equivalent differential forms
formalism \cite{Hehl:1994ue} and regard as fundamental gravitational fields
the coframe $e^{A}$ and linear connection
$\hat{\omega}^{A}_{\thickspace\thickspace B}$  $1-$ forms, as we briefly discuss 
below.

\section{The Differential Form Formulation of Metric-Affine
Gravity}

The true gauge character of MAG is most transparent when  using its exterior
differential  form  formulation. In such a formalism, the independent gauge
potentials are the orthonormal coframe $e^{A}$, the linear connection
$\hat{\omega}^{A}_{\thickspace\thickspace B}$  and the metric $\eta_{AB}$ and 
the
associated gauge field strengths are the torsion, curvature and non-metricity
\beq
\hat{T}^{A}:=\hat{D} e^{A} \label{To}
\eeq
\beq
\hat{R}^{A}_{\thickspace\thickspace B}:={\rm 
d}\hat{\omega}^{A}_{\thickspace\thickspace 
B}+\hat{\omega}^{A}_{\thickspace\thickspace C} \wedge
\hat{\omega}^{C}_{\thickspace\thickspace B} \label{Ri}
\eeq
\beq
\hat{Q}_{AB}:=\hat{D}\eta_{AB}
\eeq
respectively. Here, $\hat{D}$ represents the gauge exterior covariant derivative.
Equations $(\ref{To})$ and $(\ref{Ri})$ are often referred to as Cartan's
first and second structure equations. Note that torsion and curvature are
$2$-forms while non-metricity  is an $1$-form. For instance, in components,
torsion is expanded as
\beq
\hat{T}^{A}=\frac{1}{2}\hat{T}^{A}_{\thickspace\thickspace\mu\nu}dx^{\mu}\wedge 
dx^{\nu}.
\eeq
Acting on the above three field strengths with $\hat{D}$, we easily obtain the 
Bianchi
identities (see,  for instance,  \cite{Hehl:1994ue})
\beq
\hat{D}\hat{T}^{A}=\hat{R}^{A}_{\thickspace\thickspace B}\wedge e^{B}
\eeq
\beq
\hat{D}\hat{R}^{A}_{\thickspace\thickspace B}=0 
\thickspace\thickspace\thickspace {\rm or}
\thickspace\thickspace\thickspace\thickspace
\hat{D}\hat{R}^{AB}=\hat{R}^{A}_{\thickspace\thickspace C} \wedge \hat{Q}^{CB}
\eeq
\beq
\hat{D} \hat{Q}_{AB}=2 \hat{R}_{(AB)}	,
\eeq
which are just the identities $(\ref{B1})$, $(\ref{bian})$ and  $(\ref{B2})$,
expressed in the language of differential forms. In this formulation, the field
equations are obtained by varying with respect to the three gauge fields
$e^{A}$, $\hat{\omega}^{A}_{\thickspace\thickspace B}$, $\eta_{AB}$. However, it
can be shown \cite{Hehl:1994ue} that the field equations obtained by varying
with respect to the metric are not independent from the field equations coming
from the coframe and the linear connection and therefore, as independent
fields can be regarded only the latter two.

\section{Conservation Laws and Hyperfluid Models}

As one may have guessed, in a general MAG framework the familiar conservation
law for the energy momentum tensor of matter no longer applies. In this case,
the diffeomorphism  along with the GL (local general linear group) invariance
of the matter action,
yield \cite{Hehl:1994ue,Obukhov:2013ona,Babourova:2004xx,
Obukhov:2014nja,Iosifidis:2020gth}
\begin{eqnarray}    
\frac{1}{\sqrt{-g}}\left(\hat{T}_{\mu}-\hat{\nabla}_{\mu}\right)(\sqrt{-g}t^{\mu
}_{\thickspace\thickspace\alpha})&=&\frac{1}{2}\hat{\Delta}^{\lambda\mu\nu}\hat{
R}_{\lambda\mu\nu\alpha}-\frac{1}{2}\hat{Q}_{\alpha\mu\nu}T^{\mu\nu}- 
\hat{T}_{\nu\alpha\mu}t^{\mu\nu}\,,\\
    t^{\mu}_{\thickspace\thickspace\lambda} &=& 
T^{\mu}_{\thickspace\thickspace\lambda}- \frac{1}{2 
\sqrt{-g}}\left(\hat{T}_{\nu}-\hat{\nabla}_{\nu}\right)(\sqrt{-g}\hat{\Delta}_{
\lambda}^{\thickspace\thickspace\mu\nu}) \,,
\end{eqnarray}
provided that the matter fields satisfy their field equations (i.e. the
on-shell condition  $\frac{\delta {S}_{m}}{\delta \phi}=0$ holds). The above
constitute the conservation laws for energy-momentum and hyper-momentum currents
in MAG. Notice that, combining the above one can get the divergence of the
energy momentum tensor \cite{Iosifidis:2020gth}
\begin{eqnarray}
&&
\!\!\!\!\!\!\!\!\!\!\!\!\!\!\!\!\!\!\!\!\!\!\!\!\!\!\!\!\!\!\!\!\!\!\!\!\!\!
\sqrt{-g}(2 
\hat{\nabla}_{\mu}T^{\mu}_{\thickspace\thickspace\alpha}-\hat{\Delta}^{
\lambda\mu\nu
}
\hat{R}_{\lambda\mu\nu\alpha})
+\left(\hat{T}_{\mu}-\hat{\nabla}_{\mu}\right)
\left(\hat{T}_{\nu}-\hat{\nabla}_
{\nu}\right)(\sqrt{-g}\hat{\Delta}_{
\alpha}^{\thickspace\thickspace\mu\nu})  
\nonumber\\
&& 
\ \ \ \ \ \ \ \ \ \ \ \ \ \ \ \ \ \ \ 
-\hat{T}^{\lambda}_{
\thickspace\thickspace\mu\alpha}\left(\hat{T}_{\nu}-\hat{\nabla}_{\nu}\right)(\sqrt{-g}\hat{\Delta}_{\lambda}^{
\thickspace\thickspace\thickspace\mu\nu})=0 \label{ccc}\,.
\end{eqnarray}
The above conservation laws combined with the field equations of MAG along with
making a proper ansatz for the sources,   constitute a complete setup for the
study of the effects of MAG theories.

\subsection{Hyperfluids in Cosmology}
\label{hyperfluidsref1}

The question then arises what is the current status of non-Riemannian
Cosmology?  The answer is that it is not quite so clear at the moment.
There are many interesting works, see, for instance,
 \cite{Puetzfeld:2001ur}, but almost all of them depend on specific
ansatz that restricts the underlying structure. For a historical review
regarding the chronological ordering of such investigations, see
\cite{Puetzfeld:2004yg} and references therein (some inflationary
scenarios in MAG were recently discussed in  \cite{Shimada:2018lnm}).
However, as we pointed out, most of these works make some restrictive
assumptions about the form of torsion and non-metricity (like the assumption
that the geometry is of Weyl-Cartan type) and thus cannot give an adequate
description of the effects of torsion and non-metricity. Perhaps the reason why
the role of non-Riemannian degrees of freedom in cosmology has been obscure so
far is because a complete cosmological hyperfluid model was lacking. 

A
development in this direction was the  model of the unconstrained hyperfluid
\cite{Obukhov:1996mg}, which is described by the energy tensors
	\beq
\hat{\Delta}_{\alpha\mu\nu}=\hat{J}_{\alpha\mu}u_{\nu},
\eeq
where $T^{\mu\nu}$ is the usual  energy-momentum tensor of perfect fluid and
$J_{\mu\nu}$  is the hyper-momentum density of the
hyperfluid \cite{Obukhov:1996mg} (another interesting model was
developed in  \cite{Babourova:2004xx}). Even though the above model is
certainly interesting, it has some serious limitations, since when applied to
cosmology it cannot produce any torsional degrees of freedom and only allows
for a restricted form of non-metricity. As a result it does not allow confident conclusions, about the effects of torsion and non-metricity.

A
model taking into account all the non-Riemannian degrees of freedom in
cosmology (two for torsion and three for non-metricity) was only recently
developed \cite{Iosifidis:2020gth}. This is the model of the Perfect
Cosmological Hyperfluid, which is the most general non-Riemannian fluid that is
compatible with the Cosmological Principle. In this model the energy tensors
read
 \beq
t_{\mu\nu}=T_{\mu\nu}=\rho u_{\mu}u_{\nu}+p h_{\mu\nu}
\eeq
\beq
\hat{\Delta}_{\alpha\mu\nu}=\phi h_{\mu\alpha}u_{\nu}+\chi
h_{\nu\alpha}u_{\mu}+ \psi u_{\alpha}h_{\mu\nu}+\omega u_{\alpha}u_{\mu}
u_{\nu}+\epsilon_{\alpha\mu\nu\kappa}u^{\kappa}\zeta ,
\eeq
where $T^{\mu\nu}$ again has the usual  expression (corresponding to a perfect 
fluid with energy density $\rho$ and pressure $p$) and the above
form  of the hyper-momentum is the most general one that is allowed in an FLRW
universe (i.e. the most general form compatible with the Cosmological
Principle). Finally, $h_{\mu\nu}=g_{\mu\nu}+u_{\mu}u_{\nu}$ is the projection 
operator. The functions $\phi,\chi,\psi,\omega,\zeta$ can only depend on time 
and are
associated with the microscopic characteristics of matter (have a purely
non-Riemannian origin). These five fields are the sources for the two
torsional \cite{1979PhLA...75...27T} and three
non-metricity \cite{Minkevich:1998cv} degrees of freedom that span a
homogeneous cosmological background \cite{Iosifidis:2020gth}.

The most
straightforward application of the Perfect Hyperfluid Model is to consider the
Einstein-Hilbert action and the presence of such a fluid in an FLRW background.
Then, the modified Friedmann equations in the presence of hyperfluid induced
torsion and non-metricity, read \cite{Iosifidis:2020gth}
\begin{gather}
    H^{2}=-H\left[ \frac{3}{2}X-\frac{1}{2}Y+Z+V \right] 
-\frac{1}{2}(\dot{X}+\dot{Y}) -\frac{1}{2}(X-Y)(Z+V)+XY+W^{2}+\frac{ 
\kappa^2}{3}\rho_m\,,    \\
    \dot{H}+H^2=-\frac{\kappa^2}{6}\Big[\rho_m +3p_m 
\Big]+\dot{Y}+H(Y+Z+V)-Y(V+Z)\,,
\end{gather}
with $\rho_m$ and $p_m$ the  energy density and pressure of usual matter, and
where $X,Y,Z,V,W$ are given by the decomposition of \eqref{l}
\beq
\hat{L}_{\alpha\mu\nu}+\hat{K}_{\alpha\mu\nu} = X(t) h_{\mu\alpha}u_{\nu} + Y(t)
h_{\nu\alpha}u_{\mu} + Z(t) u_{\alpha}h_{\mu\nu} + V(t) u_{\alpha}u_{\mu}
u_{\nu} + \epsilon_{\alpha\mu\nu\kappa}u^{\kappa}W(t) ,
\eeq
and in this case, they are linearly related to the sources $\phi,\chi,\psi,\omega,\zeta$, see  \cite{Iosifidis:2020gth}. The above modified Friedmann equations combined with the conservation laws of the perfect hyperfluid
\begin{gather}
    \dot{\rho}_m + 3H(\rho_m+p_m) = -\frac{1}{2}u^{\mu}u^{\nu}(\chi 
\hat{R}_{\mu\nu}+\psi
\hat{\tilde{\mathcal{R}}}_{\mu\nu})\label{rhop}\,,\\
-\delta^{\mu}_{\lambda}    \frac{\partial_{\nu}(\sqrt{-g}\phi
u^{\nu})}{\sqrt{-g}}-u^{\mu}u_{\lambda}
\frac{\partial_{\nu}\Big(\sqrt{-g}(\phi+\chi +\psi +\omega) u^{\nu}\Big)}{\sqrt{-g}} + \left[
\Big(\hat{T}_{\lambda}-\frac{\hat{Q}_{\lambda}}{2}\Big)u^{\mu}-\hat{\nabla}_{\lambda}u^{\mu}
\right]\chi+
\nonumber \\
 +\left[
\Big(\hat{T}^{\mu}-\frac{\hat{Q}^{\mu}}{2}+\hat{\tilde{Q}}^{\mu}\Big)u_{\lambda}-g^{\mu\nu}\hat{\nabla}_{\nu}u_{
\lambda}\right]\psi + u^{\mu}u_{\lambda}(\dot{\chi}+\dot{\psi})
-(\phi+\chi+\psi+\omega)(\dot{u}^{\mu}u_{\lambda}+ u^{\mu}\dot{u}_{\lambda})
=0 \label{conl22}\,,
\end{gather}
constitute a cosmological model  where all the non-Riemannian degrees of
freedom are taken into account in a minimal way (with this terminology
we mean that we do not go beyond the Einstein-Hilbert action as far as the
gravitational part of the action is concerned). In a sense, these represent
the most natural generalisation of the Friedmann equations in MAG. The
cosmological implications of the latter are currently under investigation. For
an analytic review on the various other non-Riemannian models that have been
considered in the literature, see  \cite{Puetzfeld:2004yg}.

%
%
%
%
%
%
%
%
%
%
%
%

\chapter[Geometric Foundations of Gravity]{Geometric Foundations of Gravity}
\label{Koivistochapter}

{\em Tomi S. Koivisto}

\section{Metric-affine geometry}
\label{metaffgrref3}

Gravity theories are naturally considered in a geometrical setting. In the original formulation of General Relativity, in terms of coordinates $x^\mu$ of spacetime and a metric tensor $g_{\mu\nu}$ defined on them, the setting is a (pseudo-)Riemannian space. Cartan's  work 
provided a more suitable, coordinate-independent, framework for the geometrical 
approach to theories of gravity.
There, local frames are described by a set of vectors 
${\boldsymbol{\mathrm{e}}}_A$ whose  components 
$e_A{}^\mu$ are known as the vierbein 
${\boldsymbol{\mathrm{e}}}_A = 
e_A{}^\mu\partial_\mu$. 
Unless the frame field
is degenerate, it has an inverse, the coframe one-form ${\boldsymbol{\mathrm{e}}}^A$, 
which is  dual 
to the frame with respect to the inner product ${\boldsymbol{\mathrm{e}}}^A\cdot {\boldsymbol{\mathrm{e}}}_B 
= \delta^A_B$. The relationship of the 
frames at different points in spacetime can be specified by referring to the 
concept of parallel transport.  This is determined by the connection  
\label{metaffconref3}
$\hat{\boldsymbol{\omega}}_A{}^B$ which is a one-form,
 $\hat{{\boldsymbol{\omega}}}_A{}^B=\hat{\omega}_A{}^B{}_\mu {\rm d} x^\mu$. We denote by 
${\hat{\rm D}}$ the 
covariant derivative associated with this  connection. 
 
To translate tensorial objects from the frame formulation into the coordinate 
formulation, the projections by  the frame and the coframe are used. It should, however, be taken into account that the
 connection $\hat{{\boldsymbol{\gamma}}}_A{}^B$ is not a tensor but has an inhomogeneous 
transformation rule. If we consider a general  linear transformation 
$\Lambda^A{}_B$ with the inverse $\Lambda_A{}^B$,
 we have ${\boldsymbol{\mathrm{e}}}^A \rightarrow \Lambda^A{}_B {\boldsymbol{\mathrm{e}}}^B$ 
and ${\boldsymbol{\mathrm{e}}}_A \rightarrow 
\Lambda_A{}^B {\boldsymbol{\mathrm{e}}}_B$ but $\hat{{\boldsymbol{\gamma}}}_A{}^B  \rightarrow 
\Lambda_A{}^C\hat{{\boldsymbol{\gamma}}}_C{}^D\Lambda^B{}_D - {\rm d}
\Lambda_A{}^B$. The connection 
describing the
 corresponding parallel transport in the tensor formalism is called 
$\hat{\Gamma}^\alpha{}_{\mu\nu}$, and is defined by  $\hat{\Gamma}^\alpha{}_{\mu\nu} = 
e_A{}^\alpha{\hat{\rm D}}_\mu e^A{}_\nu = 
-e^A{}_\nu{\hat{\rm D}}_\mu  e_A{}^\alpha$. 
 In the Riemannian formulation of General Relativity the connection was not an 
independent field, but was rather given  as
 \be \label{levi}
 {\Gamma}{}^\alpha{}_{\mu\nu} \overset{\text{}}{=} 
g^{\alpha\lambda}\left( g_{\lambda(\mu,\nu)}  
-\frac{1}{2}g_{\mu\nu,\lambda}\right)\,.
 \ee 
However, in hindsight, ``[...] it seems of secondary importance, in some sense, 
that some particular $\hat{\Gamma}$  field can be deduced from a Riemannian metric'' 
\cite{einstein2015meaning}.

There are two gauge-invariant properties of a generic connection. These are the 
curvature two-form and the  torsion two-form,
defined as
\ba
    \hat{\bR}_A{}^B & = & {\hat{\rm D}}{\hat{\boldsymbol{\omega}}}_A{}^B = {\rm d}{\hat{\boldsymbol{\omega}}}_A{}^B 
+ 
\hat{{\boldsymbol{\gamma}}}_A{}^C\wedge{\hat{\boldsymbol{\omega}}}_C{}^B\,,  \label{curvature} \\
\hat{\bT}^A & = & {\hat{\rm D}}{\boldsymbol{\mathrm{e}}}^A = {\rm d}{\boldsymbol{\mathrm{e}}}^A + 
\hat{\boldsymbol{\omega}}_A{}^B\wedge{\boldsymbol{\mathrm{e}}}_B\,, 
\ea 
respectively. By construction, these satisfy the Bianchi identities 
${\hat{\rm D}}\hat{\bR}_A{}^B=0$ and  ${\hat{\rm D}}\hat{\bT}^A = \hat{\bR}_{B}{}^{A}\wedge{\boldsymbol{\mathrm{e}}}^B$. 
From the 
definition of the 
affine connection $\hat{\Gamma}^\alpha{}_{\mu\nu}$, it follows that the corresponding 
objects in  the tensor formulation are
\ba
\hat{R}^{\alpha}{}_{\beta\mu\nu} & = &  2\partial_{[\mu}\hat{\Gamma}^\alpha{}_{\nu]\beta} 
-2\hat{\Gamma}^\alpha{}_{[\mu\mid\lambda\mid}\hat{\Gamma}^\lambda{}_{\nu]\beta} = 
-e^A{}_\beta \hat{R}_A{}^B{}_{\mu\nu} e_B{}^\alpha \,, \\
\hat{T}^\alpha{}_{\mu\nu} & = & 2\hat{\Gamma}^\alpha{}_{[\mu\nu]} = 
\hat{T}^A{}_{\mu\nu} e_A{}^\alpha\,.
\ea
These objects have a clear geometrical interpretation. By using the identity
\be
[\hat{\nabla}_\mu,\hat{\nabla}_\nu]V^\alpha = \hat{R}^{\alpha}{}_{\beta\mu\nu}V^\beta  - 
\hat{T}^\lambda{}_{\mu\nu}\hat{\nabla}_\lambda V^\alpha\,,
\ee
one may see that when a vector $V^\mu$
is parallel transported around a closed loop, 
torsion describes the displacement of the vector, and curvature describes  the 
change of the vector. These concepts do not require a metric, but are the 
properties solely of the connection.  

\begin{figure}[ht]
  \begin{center}
    \includegraphics[width = 0.95\textwidth]{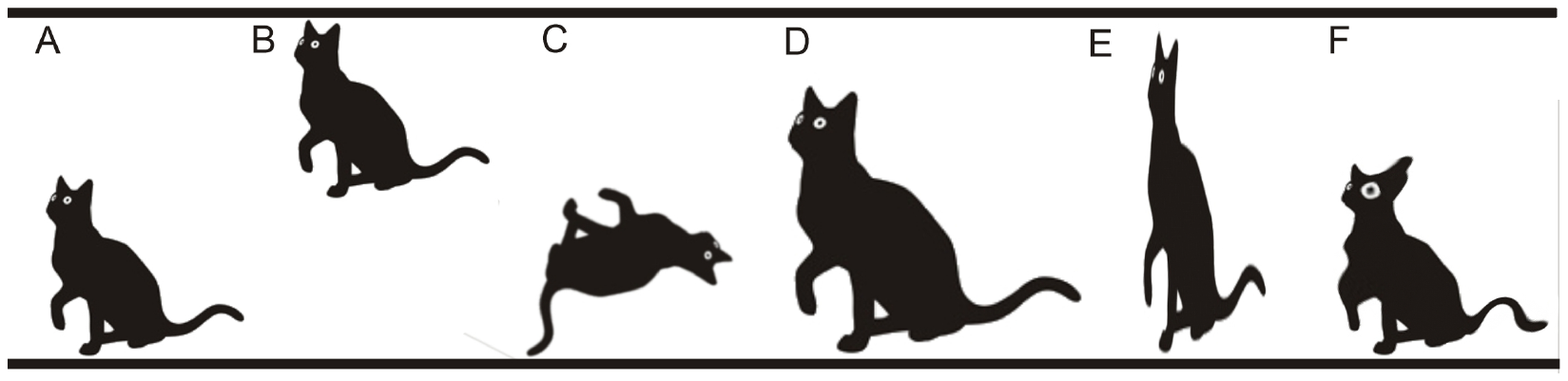}
    \vspace{-3.5cm}
    \caption{{\it{Exploring the metric-affine space wherein (A) 
${\hat{\boldsymbol{R}}}_A{}^B={\hat{\boldsymbol{T}}}{}^A=0\,$, (B) 
${\hat{\boldsymbol{T}}}{}^A \neq 0\,$, (C) ${{\boldsymbol{R}}}_A{}^B \neq 0\,$, 
(D) ${\hat{\boldsymbol{R}}}_A{}^A \neq 0\,$,
     (E) ${\hat{\boldsymbol{R}}}_A{}^B \neq 0$ for $A=B\,$, and (F) 
${\hat{\boldsymbol{R}}}_A{}^B - \delta^A_B\hat{\boldsymbol{R}}_C{}^C \neq 0$\,. 
\label{Tomiinterpretation}}}}
    \end{center}
  \end{figure}

If a metric structure $g_{\mu\nu}$ is also available, it is  possible to refer 
to concepts such as lengths and angles.
Then we may define the contraction $\hat{R}_{\alpha\beta\mu\nu}  = 
g_{\alpha\lambda}\hat{R}^{\lambda}{}_{\beta\mu\nu}$ and see that the
metric curvature $\hat{R}_{[\alpha\beta]\mu\nu}$ describes the  rotation of the 
vector 
that is parallel-transported around a closed loop,
and  the non-metric curvature $\hat{R}_{(\alpha\beta)\mu\nu}$  describes the change 
of 
the magnitude of the components of the vector;
in particular, the piece $g^{\alpha\beta}\hat{R}_{\alpha\beta\mu\nu}$ describes an 
overall rescaling, and the rest of the symmetric components describe the shape distortions. The presence of non-metric 
curvature implies that that the inner products are not conserved  by the connection. The 
incompatibility of the two structures can be \label{nonmetricityref3}
quantified by the non-metricity defined as $\hat{Q}_{\alpha\mu\nu}=\hat{\nabla}_\alpha 
g_{\mu\nu}$.  Note that even if the affine geometry 
is integrable, $\hat{R}^{\alpha}{}_{\beta\mu\nu}=0$, $\hat{T}^\alpha{}_{\mu\nu}=0$, it can 
be incompatible  with the metric geometry,
$\hat{Q}_{\alpha\mu\nu} \neq 0$. The geometric interpretation of the invariants 
is illustrated in Fig. \ref{Tomiinterpretation}.
\vspace{-3.5cm}

The generic connection can be decomposed, with respect to a given metric, as
\be \label{ggamma}
{\hat{\Gamma}}{}^\alpha{}_{\mu\nu} = {\Gamma}{}^\alpha{}_{\mu\nu}  + 
\hat{K}^\alpha{}_{\mu\nu} + \hat{L}^\alpha{}_{\mu\nu}\,,
\ee \label{torsionref7}
where the contributions of the torsion and the non-metricity,
\ba
\hat{K}^\alpha{}_{\mu\nu} & = & \frac{1}{2}\hat{T}^\alpha{}_{\mu\nu} - 
\hat{T}_{(\mu\nu)}{}^\alpha\,, \\
\hat{L}^\alpha{}_{\mu\nu} & = & \frac{1}{2}\hat{Q}^\alpha{}_{\mu\nu} - 
\hat{Q}_{(\mu\nu)}{}^\alpha\,, 
\ea
are called the contortion and the disformation, respectively. These two tensors 
can be related to the corresponding
one-forms with latin indices. The contortion one-form $\hat{\bK}_A{}^B$ could be 
defined  via $\hat{\bK}_B{}^A\wedge{\boldsymbol{\mathrm{e}}}^B = \hat{\bT}^A$. 
The disformation one-form is given by $\hat{\bL}_{AB}=\frac{1}{2}{\hat{\rm D}}\eta_{AB}  
- 
{\boldsymbol{\mathrm{e}}}_{[A}\cdot{\hat{\rm D}}\eta_{B]C}{\boldsymbol{\mathrm{e}}}^C$, where 
$\eta_{AB}$
is the tangent space metric, whose covariant derivative is related to the  
non-metricity one-form $\hat{\bQ}_{AB}$ as 
${\hat{\rm D}}\eta_{AB}=2(\hat{\bQ}_{AB}+\hat{{\boldsymbol{\gamma}}}_{(AB)})$.
Finally, we may note that defining the Levi-Civita one-form 
$\boldsymbol{{\omega}}_A{}^B$  via 
$\boldsymbol{{\omega}}_B{}^A\wedge{\boldsymbol{\mathrm{e}}}^B=-{\rm d}\boldsymbol{\mathrm{e}}^A$, we
can display the decomposition
\be \label{ggamma2}
\hat{{\boldsymbol{\gamma}}}_A{}^B = \boldsymbol{{\omega}}_A{}^B + \hat{\bK}_A{}^B + \hat{\bL}_A{}^B\,,
\ee
which is just the rewriting of (\ref{ggamma}).

\section{The Geometrical Trinity}
\label{trinityref1}

In the standard description of General Relativity, the tangent space is  
endowed 
with the Minkowski metric $\eta_{AB}$, and the
spacetime metric is then its projection 
$g_{\mu\nu}=e^A{}_\nu e^B{}_\nu\eta_{AB}$.  The 
Einstein-Hilbert action can 
then be cast in the 
well-known equivalent forms
\ba \label{gr}
S_{\text{GR}} & = & \frac{1}{2\kappa^2}\int  
\epsilon_{ABCD}{\bR}{}^{AB}\wedge\boldsymbol{\mathrm{e}}^C\wedge{\boldsymbol{\mathrm{e}}}^D = \frac{1}{2\kappa^2}\int {\rm d}^4 x \sqrt{-g}g^{\beta\nu}{R}{}^\alpha{}_{\beta\alpha\nu}\,.
\label{e-h}
\ea
It is also possible to consider the Einstein-Palatini action that does not 
impose $\hat{\Gamma}{}^\alpha{}_{\mu\nu}  = 
{\Gamma}{}^\alpha{}_{\mu\nu}$
but obtains this (up to an undetermined but irrelevant piece of torsion) as a 
dynamical condition.  It is very interesting to consider further alternatives, 
from the 
viewpoint of the decomposition of the geometric elements we found above in 
\eqref{affconection}. We split the connection $\hat{\boldsymbol{\omega}}_A{}^B= 
{{\boldsymbol{\omega}}}_A{}^B + \hat{\boldsymbol{x}}_A{}^B$
into the metrical part ${{\boldsymbol{\omega}}}_A{}^B$ and the rest that 
we'll call the distortion  $\hat{\boldsymbol{x}}_A{}^B$, and we can now split the curvature
\ba \label{x}
    \hat{\bR}_A{}^B & = & {\hat{\rm D}}{\hat{\boldsymbol{\omega}}}_A{}^B = {\rm d}{\hat{\boldsymbol{\omega}}}_A{}^B 
+ {\hat{\boldsymbol{\omega}}}_A{}^C\wedge{\hat{\boldsymbol{\omega}}}_C{}^B = {\bR}{}_A{}^B + {{\hat{\rm D}}} 
\hat{\boldsymbol{x}}_A{}^B 
+  \hat{\boldsymbol{x}}_A{}^C\wedge\hat{\boldsymbol{x}}_C{}^B\,.
\ea
Therefore, if one imposes teleparallelism, i.e. $\hat{\bR}_A{}^B=0$, it appears that the action 
\be \label{e-h2}
S_{\text{GR}_\parallel}  =   -\frac{1}{2\kappa^2}\int \epsilon_{ABCD}
\hat{\boldsymbol{x}}^{AE}\wedge\hat{\boldsymbol{x}}_E{}^B\wedge\boldsymbol{\mathrm{e}}^C\wedge{\boldsymbol{\mathrm{e}}}^D\,,
\ee
is equivalent to (\ref{e-h})  up to a total derivative \cite{Jimenez:2019ghw}. This change of 
geometrical variables opens new perspectives on Einstein's theory. 
In particular, the equivalence of the formulations has been established in  two 
extensively studied cases, which we shall now review.

\begin{figure}[ht]
\begin{center}
\hspace{-0.5cm}\mbox{\includegraphics[width=0.73\textwidth,angle 
=-90]
{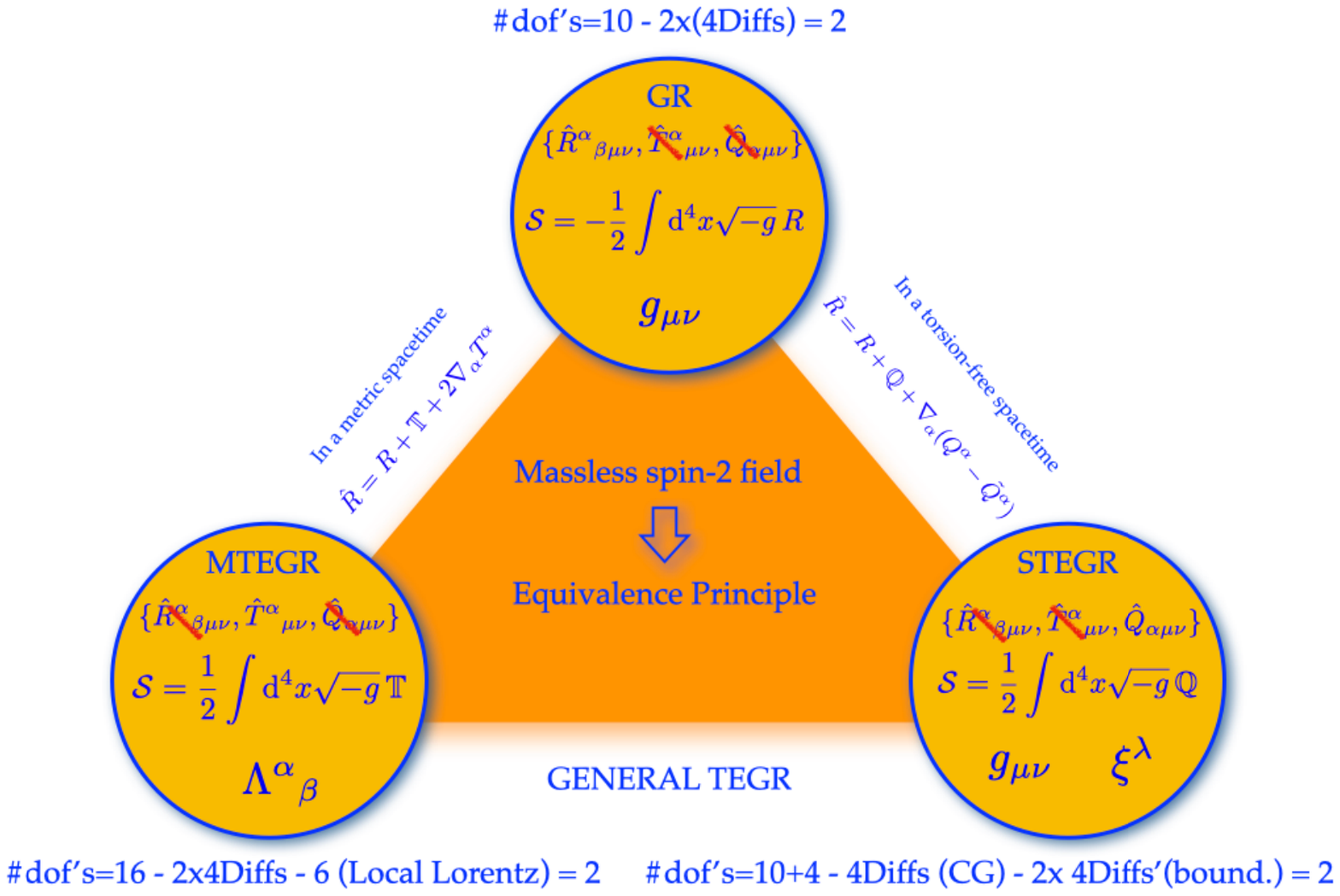}}
\end{center}
\caption{{\it{A schematic illustration of the geometrical trinity.}}}
\label{GravitationalTrinity2}
\end{figure}

In the case of metric teleparallelism, one imposes metric-compatibility.  Then, (\ref{e-h2}) becomes the action of TEGR (read:
the metric Teleparallel Equivalent of General Relativity\footnote{It is a convention to denote quantities in this geometry (with some exceptions like that of the frame and the coframe) with the bullets on top of them; and another convention is to drop the epithet ``metric'' when referring to the special case with metric-compatible connection.}  
\cite{Aldrovandi:2013wha,Maluf:2013gaa};),
\ba \label{tegr}
S_{\text{TEGR}}:=S_{\tp{\rm GR}}   =   -\frac{1}{2\kappa^2}\int  \epsilon_{ABCD}
\tp{\bK}{}^{AE}\wedge\tp{\bK}_E{}^B\wedge{\boldsymbol{\mathrm{e}}}^C\wedge{\boldsymbol{\mathrm{e}}}^D =  -\frac{1}{2\kappa^2}\int{\rm d}^4 x \sqrt{-g}\mathbb{T}\,. 
\ea
We have introduced the torsion scalar
\be
\mathbb{T} = -\tp{T}_{\alpha\mu\nu}\left(\frac{1}{4}\tp{T}^{\alpha\mu\nu} 
+\frac{1}{2}\tp{T}^{\mu\alpha\nu}-g^{\alpha\nu}\tp{T}^{\lambda\mu}{}_\lambda\right)\,.
\ee
In the case of symmetric teleparallelism, one precludes torsion.  Then, 
(\ref{e-h2}) becomes the action of $\st{\rm GR}$ (read:
the symmetric Teleparallel Equivalent of General Relativity 
\cite{Nester:1998mp,Adak:2005cd}), 
\ba \label{stgr}
S_{\text{STEGR}}:=S_{\st{\rm GR}}   =   -\frac{1}{2\kappa^2}\int \epsilon_{ABCD}
\st{\bL}^{AE}\wedge\st{\bL}_E{}^B\wedge{\boldsymbol{\mathrm{e}}}^C\wedge{\boldsymbol{\mathrm{e}}}^D  =   -\frac{1}{2\kappa^2}\int{\rm d}^4 x \sqrt{-g}\mathbb{Q}\,,
\ea
where we have introduced the non-metricity scalar  \cite{BeltranJimenez:2017tkd}
\be
\mathbb{Q}:= \frac{1}{2}\st{Q}_\alpha{}^{\mu\nu}\left( 
\st{L}^\alpha{}_{\mu\nu}-\st{\bar{L}}^\alpha{}_{\mu\nu}\right)\,, 
\ee
which may be defined in terms of the Weyl connection 
\be \label{weyl}
{\bar{L}}^\alpha{}_{\mu\nu} = 
\frac{1}{2}{Q}^\alpha{}g_{\mu\nu}-\delta^\alpha_{(\mu}{Q}_{\nu)}\,,  \quad 
{Q}_\alpha=g_{\mu\nu}{Q}_\alpha{}^{\mu\nu}\,. 
\ee
A technical remark is that the connection in each of the three cases 
(\ref{e-h},\ref{tegr},\ref{stgr}) can  be either restricted to the desired form 
a priori
or by the use of Lagrange multipliers, with the same result in each case. We 
also emphasise  that in each of the three cases either of the two given 
forms of the action can equally well be taken as the definition of the theory, 
since teleparallel theories admit a consistent Palatini formulation  
\cite{BeltranJimenez:2017tkd,BeltranJimenez:2018vdo}. 
The  geometrical trinity is schematically presented in Fig. 
\ref{GravitationalTrinity2}. 

At the conceptual level, (\ref{e-h}) most naturally describes gravity as 
geometry,  (\ref{tegr}) as force, and (\ref{stgr}) as inertia. 
 Consider, for example, the equation of motion for the trajectory 
$x^\alpha(\tau)$ 
of a test particle, 
\be \label{rg}
\ddot{x}^\alpha + 
{\Gamma}{}^\alpha{}_{\mu\nu}\dot{x}^\mu\dot{x}^\nu =  0\,,
\ee
which in the standard Riemannian context is seen as the statement that the 
particle moves  along the geodesics, coinciding with the 
shortest path as measured by the metric. The coordinate system wherein 
${\Gamma}{}^\alpha{}_{\mu\nu}=0$  is taken
to be freely falling. The very same equation (\ref{rg}) can be rewritten in 
terms of the metric teleparallel connection  and contortion,
\be \label{mtgeod}
\ddot{x}^\alpha + \tp{\Gamma}{}^\alpha{}_{\mu\nu}\dot{x}^\mu\dot{x}^\nu  = 
\tp{K}{}^\alpha{}_{\mu\nu}\dot{x}^\mu\dot{x}^\nu = F^\alpha\,,
\ee
where in the Weitzenb\"ock frame the left hand side describes the gravitational 
force $F^\alpha$ that tends to displace the particle,
and is orthogonal to the four-velocity, $F^\alpha\dot{x}_\alpha=0$. Now we 
cannot,  in general, find a frame wherein $\tp{\Gamma}{}^\alpha{}_{\mu\nu}$ 
would
vanish, but there is a frame $\ddot{x}^\mu=0$ wherein the inertial and 
gravitational  forces exactly cancel each other out. In symmetric teleparallel 
geometry, the equation reads 
\be \label{stgeod}
\ddot{x}^\alpha + \st{\Gamma}^\alpha{}_{\mu\nu}\dot{x}^\mu\dot{x}^\nu 
=  \st{L}^\alpha{}_{\mu\nu}\dot{x}^\mu\dot{x}^\nu = I^\alpha\,.
\ee
Now we can choose the coincident gauge 
$\st{\Gamma}^\alpha{}_{\mu\nu}=0$, where  the deviation of   
\label{geodesymmtegTref1} \label{geodesicsref2}
the 
trajectory from
straight lines is seen to be solely due to the inertial effects described by 
the 
left hand side.  The latter are not appropriately interpreted as
forces, since $I_\alpha\dot{x}^\alpha= 
-\frac{1}{2}\st{Q}_{\alpha\mu\nu}\dot{x}^\alpha \dot{x}^\mu \dot{x}^\nu \neq 0$.  
Instead, the disformation
describes the apparent change of magnitudes, i.e. rescalings of units of 
measurements, due  to the non-preservation of the metric. The frame
$\ddot{x}^\alpha=0$ is simply the inertial frame of the particle, coinciding 
with what we call below  the canonical frame.

Though these are merely matters of interpretation at the level of classical 
dynamics, the different  formulations of the action 
principle can lead to very different results in calculations of the quantum 
theory, energetics, or  thermodynamics of gravity.
 In particular, the framework of (\ref{stgr}) has recently lead to some 
interesting new results,  which we would like to
review in the following. 

\section{Purified Gravity}

Let us first have a closer look at the action principle (\ref{stgr}). 
As a direct consequence of (\ref{x}), we have, in the absence of curvature and  
torsion,
\be
g^{\mu\nu}{\st{R}}{}^\alpha{}_{\mu\alpha\nu} = \mathbb{Q} +  
{\nabla}_\mu\left( \st{Q}^\mu-\st{Q}_{\alpha}{}^{\mu\alpha}\right)\,.
\ee
Thus, the action (\ref{stgr}) differs from (\ref{gr}) by a total derivative. In 
fact, it can be  regarded as the minimal covariantisation of Einstein's
original action principle \cite{1916SPAW......1111E}. Noting that a symmetric
teleparallel 
connection  $\st{\Gamma}^\alpha{}_{\mu\nu}$ can always be expressed in 
terms
of a 4$\times$4 matrix $\Lambda^\mu{}_{\nu}$ that has an inverse  
$\Lambda_\nu{}^{\mu}$ as
\be \label{lambda}
\st{\Gamma}^\alpha{}_{\mu\nu} =  \Lambda_\lambda{}^{\alpha} 
\partial_\mu\Lambda^\lambda{}_{\nu}\,,
\ee  
since the most general flat linear connection must be obtained by a general 
linear transformation  of the vanishing connection, we can
choose the gauge in which this connection vanishes, 
$\st{\Gamma}^\alpha{}_{\mu\nu}=0$. 
It is straightforward to see that in this gauge we have 
\be
\mathbb{Q} = 2 
g^{\mu\nu}\Gamma{}^\alpha{}_{\beta[\mu}
\Gamma{}^\beta{}_{\alpha]\nu}\,,
\ee
where the left-hand side is the Einstein Lagrangian \cite{1916SPAW......1111E}. 
The action  (\ref{stgr}) thus combines the desired properties of both Hilbert's 
and Einstein's formulations,
since it is a scalar and quadratic in first derivatives. Next, let us note that 
since we  also require the connection to be free of torsion, the matrix 
$\Lambda^\alpha{}_\mu$
in (\ref{lambda}) has to reduce to a Jacobian of a coordinate transformation,  
i.e. $\Lambda^\alpha{}_\mu = \xi^\alpha{}_{,\mu}$ for some $\xi^\alpha$. Thus 
the connection
is always of the form \cite{BeltranJimenez:2017tkd,BeltranJimenez:2018vdo}
\be \label{lambda}
\st{\Gamma}^\alpha{}_{\mu\nu} = \frac{\partial 
x^\alpha}{\partial\xi^\lambda} \frac{\partial^2 \xi^\lambda}{\partial 
x^\mu\partial x^\nu}\,,
\ee  
that is, a (passive) translation of the vanishing connection. This proves that  
purified gravity is the unique framework that provides the minimal covariantisation of the 
Einstein Lagrangian and the canonical gauge theory of translation. 

\subsection{Field Equations}

Let us now consider the field equations. For that it is convenient to introduce 
 the non-metricity conjugate, defined as 
\be 
{P}^\alpha{}_{\mu\nu} = -\frac{1}{2}{L}^{\alpha}{}_{\mu\nu} + \frac{1}{4}\left( 
{Q}^\alpha  - {Q}_{\lambda}{}^{\alpha\lambda}\right) g_{\mu\nu} 
- \frac{1}{4}\delta^\alpha_{(\mu}{Q}_{\nu)}\,.
\ee
The field equations derived by the variation with respect to the metric are 
\be \label{field}
\st{\tau}^\mu{}_\nu = T^\mu{}_\nu + \st{t}^\mu{}_\nu\,,
\ee
where $T^\mu{}_\nu$ is the energy-momentum tensor of matter, defined as usual by
\be
T_{\mu\nu} = \frac{-2}{\sqrt{-g}} \frac{\delta {S}_{\text{matter}}}{\delta 
g^{\mu\nu}}\,, 
\ee
where ${S}_{\text{matter}}$ is the action for the matter fields, and
\ba \label{tau}
\st{\tau}^\mu{}_\nu & = &   
\frac{2\kappa^2}{\sqrt{-g}}\st{\nabla}_\alpha\left(\sqrt{-g}\st{P}^{\alpha\mu}{}_\nu\right)\,, \\
\st{t}^\mu{}_\nu & = & \frac{\kappa^2}{2}\mathbb{Q}\delta^\mu_\nu - 
\kappa^2 \st{P}^\mu{}_{\alpha\beta}\st{Q}_\nu{}^{\alpha\beta}\,.  \label{t}
\ea
It is interesting to note that $\st{t}^\mu{}_\nu$ is the canonical 
current 
according  to Noether's theorem \cite{Noether1918}, and indeed Einstein had
proposed this pseudotensor to describe the energy-momentum of the gravitational 
field  \cite{1916SPAW......1111E}.  Likewise, one finds that $2\st{\nabla}_\alpha 
(\sqrt{-g}\st{P}^{\alpha\mu}{}_\nu) = \st{\nabla}_\alpha \st{H}^{\alpha\mu}{}_\nu$, and for the 
object $\st{H}^{\alpha\mu}{}_\nu=\st{H}^{[\alpha\mu]}{}_\nu$,
\be
 \st{H}^{\mu\nu}{}_\alpha = \kappa^2 \sqrt{-g}\left( \st{Q}^{[\mu\nu]}{}_\alpha - 
\st{Q}^{[\mu}\delta^{\nu]}_\alpha  + 
\st{Q}^{\beta[\mu}{}_\beta\delta^{\nu]}_\alpha\right)\,,
\ee
for which $\st{H}^{\alpha\mu}{}_\nu$ is what is known as the Einstein 
energy-momentum  complex. Thus, we
have arrived at the covariant version of the canonical energy-momentum split.  
In addition to the usual metric-covariant conservation law
\be
\nabla_\mu \st{\tau}^\mu{}_\nu = \nabla_\mu 
\left( T^\mu{}_\nu + \st{t}^\mu{}_\nu\right) = 0\,, 
\ee
we now also have the covariant form of the usual conservation law
\be
\st{\nabla}_\mu \left(\sqrt{-g}\st{\tau}^\mu{}_\nu\right) = \st{\nabla}_\mu \left(\sqrt{-g} 
T^\mu{}_\nu + 
 \sqrt{-g}\st{t}^\mu{}_\nu\right) =0\,, 
\ee
from which it is directly seen that $2\st{\nabla}_\alpha (\sqrt{-g}\st{P}^{\alpha\mu}{}_\nu) 
=  \st{\nabla}_\alpha \st{H}^{\alpha\mu}{}_\nu$, 
but it can also be derived as the equation of motion for the connection which, we  
recall, is now an independent variable. 

\subsection{Energy and Entropy}
\label{eande}

We should note that the field equation (\ref{field}) now possesses two gauge 
 invariances: one may transform only the metric 
or both the metric and the connection. In fact, this is related to the unique 
property of  the quadratic action (\ref{stgr}), which is that
the translational gauge connection decouples from it at the linear order  
\cite{BeltranJimenez:2017tkd}. Now, however, the split into $\st{t}^\mu{}_\nu$ and 
$\st{\tau}^\mu{}_\nu$
is not invariant under independent transformations of the connection or of the 
metric. This introduces an ambiguity into the definition of the
gravitational energy-momentum and related quantities. This situation is the  
same as in GR,  where the results depend
arbitrarily on the coordinate frame, and in TEGR, where the results 
depend  arbitrarily on the local Lorentz frame. \label{loclinref2}
Recently, a possible resolution was proposed by suggesting that the canonical 
frame of  purified gravity should be defined by the vanishing of 
the energy-momentum associated with the metric field  \cite{Jimenez:2019yyx}. 
This is  the natural consequence of the interpretation of gravitation as 
fundamentally an inertial phenomenon, 
despite its alternative descriptions as force or geometry. Let us emphasise 
that 
the proposed  prescription, which is nothing but the minimal covariant version 
of the canonical prescription, uniquely determines any physical quantity of  
interest.

The energy-momentum of any gravity-matter system according to Noether's  
theorem
can in our case be determined from the integral, which after applying the Gauss 
 \label{Enmomtenref3}
theorem becomes
\be \label{e}
E_\mu = \int {\rm d}^2 x H^{i0}{}_\mu n_i\,,
\ee
where $n_i$ is the outward normal vector of the surface of the volume under consideration. The energy of the system can be also 
determined in the Euclidean path integral approach wherein, in the saddle point 
approximation,  the partition function $Z(\beta)$ is given by the exponential 
of 
the 
Euclidean action. At the leading (on-shell) order we have
\ba
-\log{Z(\beta)}  \approx  \frac{\kappa^2}{2}\int {\rm d}^3 
x\sqrt{-g}\left( \st{Q}^\alpha 
-  
\st{Q}_\mu{}^{\alpha\mu}\right) n_\alpha + \int{\rm d}^4 x\sqrt{-g}\left( 
L_{\rm matter}-\frac{1}{2}T\right)\,, \label{s}
\ea
where for the gravitational sector we have again used the Gauss theorem, and 
written the contribution  from the matter sector in terms of 
the Lagrangian $L_{\rm matter}$ and the trace of the matter energy-momentum tensor $T$. 
The 
partition function is considered a  function of the inverse temperature $\beta$, while the energy $E_0$ and entropy $S$ of the system are given by
\be \label{thermo}
\langle E_0 \rangle = -\frac{\partial}{\partial\beta}\log{Z}\,, \quad S =  
\beta\langle E_0 \rangle + \log{Z}\,,
\ee
respectively.
The formulae (\ref{e}) and (\ref{s}) are to be evaluated in the canonical frame 
specified by $\st{t}^\mu{}_\nu=0$,  but they are valid in any coordinate system. 
Notice that both the universal expressions have been reduced to integrals over 
the boundaries of spatial and  spacetime volumes, respectively, and that the 
path integral
needs the introduction of neither a surface nor a counter term.

It is highly nontrivial that the above expressions (\ref{e}) and (\ref{s}) 
yield 
results that are consistent  with each other, with the first law of 
thermodynamics, 
and with the area law of black hole  \label{BHref3}thermodynamics. We omit the 
details of the 
calculations but we will report the  results in the two most important cases with 
horizons: the black 
hole spacetime and the cosmological spacetime \cite{Jimenez:2019yyx}. Consider 
first a black hole described by  the metric
\be \label{ss}
{\rm d} s^2 = -f(r) {\rm d} t^2  + 
f^{-1}(r) {\rm d} r^2 + 
r^2 {\rm d} \Omega_2^2\,, 
\ee
where $\Omega_2$ is the metric on the unit 2-sphere. Recall that the 
temperature 
is  $\beta^{-1}= -f'(r)/4\pi$ and the horizon radius $r_+$ is the largest $r$ 
for which $f(r) =0$. 
Let us consider a charged black hole, described by $f(r)=1-2m/r + q/r^2$. When  
the charge is $q \neq 0$ we need to also take into account the electromagnetic 
field that supports
the solution. The formula
(\ref{e}) then gives $E_\mu = (m-q^2/(2r_+))\delta^0_\mu$. The canonical energy 
 thus reduces to the expected $E_0=m$ in the case of the Schwarzschild black
hole, whilst 
a charged black hole with the same mass, together with the surrounding  
electromagnetic field, has less total energy, in the extremal case $E_0 
\rightarrow m/2$. From
(\ref{s}) we obtain, by direct calculation, that $-\log{Z} \approx \pi r_+^2 
0+\beta q^2/(2r_+)$.  This is precisely the consistent result for which 
(\ref{thermo}) implies
that $E_0= (m-q^2/(2r_+))$, $S=\pi r_+^2$. In the case of de Sitter  space, we 
may consider the line element (\ref{ss}) where now $f(r) = 1-2r^2/r_+^2$. Then, we find vanishing energy-momentum from (\ref{e}), $E_\mu=0$. Using (\ref{s}) and  
(\ref{thermo}), we obtain consistently that $E_0=0$, $S=\pi r_+^2$.

It is important to note that the canonical frame wherein $\st{t}^\mu{}_\nu=0$ is reached  
only with a non-trivial connection for the form of the metric (\ref{ss}). The 
results computed in any other
frame would describe the measurement of a non-inertial observer, and their 
physical  interpretation would thus be very difficult. The correct results that 
were quoted above can also
be computed with a vanishing connection by simply transforming into the 
coincident  gauge. After such a transformation one finds the metric in the form 
\be
g_{\mu\nu} = \eta_{\mu\nu} + \left[ f(r)-1\right] \ell_\mu\ell_\nu\,,
\ee
where 
\be
\ell_\mu{\rm d} x^\mu = {\rm d} t + 
\delta_{ij}\frac{x^i}{r} {\rm d} x^j\,,  \quad r^2 
= \delta_{ij}x^i x^j\,.
\ee
It is clear to see that now $\st{t}^\mu{}_\nu=0$ and the 
formulae  (\ref{e}) and (\ref{s}) yield the energy-momentum and the entropy as 
quoted above for
the two physical systems. In contrast, if we consider the coordinate system 
(\ref{ss})  in the coincident gauge, we are not in the canonical frame where 
the 
energy-momentum of 
the metric field would vanish, but instead we obtain from (\ref{t}) that, for example
$\st{t}^0{}_0 = -(1/\kappa r)^2$, meaning that there appears to be nonzero (in fact, negative) energy
due to the gravitational
field. Now we have $\mathbb{Q}=-1/r^2$, and therefore the action is positive, but 
divergent. In particular, there is a logarithmic divergence at $r \rightarrow 
\infty$, which is fully consistent 
with the interpretation that this frame is non-inertial. In precisely such a case we might expect to find spurious effects that do not vanish even infinitely 
far 
away from the sources. 
Note that these effects are now completely independent of the actual  mass of 
the black hole, but vanish when the gravitational coupling is set to vanish. 

\subsection{On Quantum Theory}
\label{quantumgrefs3}

It may be useful to consider the analogy of purified gravity and massive 
electromagnetism.  The action (\ref{stgr}) is manifestly a mass term for the 
connection, and the metric
could thus be considered as a field restoring the broken symmetry, in 
analogy 
to Stueckelberg's  version of Proca's theory. Perhaps the analogy is complete 
in 
that there
is also a (canonical) kinetic term for the connection, though at the classical 
level we may neglect the  field strength of the connection. The reason is that 
since the connection is massive, it 
interacts only at finite distances. To wit, the  range of the force is of the 
order of the Planck length, $\sim$ 10$^{-35}$ meters. It is interesting to 
consider that though at the macroscopic
level the gauge field of purified gravity does not propagate,  gravity should 
become impure as microscopic distances approaching the Planck scale are probed. One might even speculate that such ripples of spacetime at its  tiniest scales 
could eventually vindicate the idea of Clifford's visionary geometric theory of 
gravity according to which 
matter is nothing but a disturbance in the spatial curvature, so that matter  
in 
motion can be understood as a simple variation in space of these wave-like 
disturbances.

To elaborate on the analogy of $\st{\rm GR}$ with massive vector  field 
theory, let us consider the strength tensor $F_{\mu\nu}$ and the excitation 
tensor density $H^{\mu\nu}$ sourced by matter current density $T^\mu$ (usually denoted $J^\mu$). The 
premetric  form of the equations is then
\be \label{premetric}
\st{\nabla}_\mu H^{\mu\nu} = T^\mu + \st{t}^\mu\,, \quad \st{\nabla}_{[\alpha}F_{\mu\nu]} =0\,.
\ee       
The difference with massless electromagnetism is that the field also sources 
itself  through the term $\st{t}^\mu$. For the theory to be predictive, one has to 
specify the constitutive 
relations that determine the forms of the extensive $H^{\mu\nu}$ and $\st{t}^\mu$ in 
 terms of the intensive quantities. In the case of Proca, the constitutive 
relations involve a metric, $H^{\mu\nu}=\sqrt{-g}F^{\mu\nu}$ and $\st{t}^\mu= m^2 
A^\mu$, 
where $m$ is the mass of the gauge potential $A_\mu$, whose existence one 
deduces 
 from the second premetric equation (\ref{premetric}). As we have seen, the 
equations
of motion in purified gravity are 
\be \label{premetric2}
\st{\nabla}_\mu \st{H}^{\mu\nu}{}_\alpha = T^\mu{}_\alpha + \st{t}^\mu{}_\alpha\,,  \quad 
F^{\alpha\beta}{}_{\mu\nu}  =0\,.
\ee    
The different number of indices in the first equations (\ref{premetric2})  with 
respect to that of (\ref{premetric}) only reflects that the latter is deduced 
from the conservation of the electric charge,
whereas the former is deduced from the conservation of the four translational  
charges, but the form of the equations is exactly the same. The question arises as to whether the analogy of theories 
is complete, the second equation in (\ref{premetric2}) then being a valid 
approximation  only at super-Planckian length scales, the full theory only being subject to
$\st{\nabla}_{[\rho}F^{\alpha\beta}{}_{\mu\nu]}=0$. Though massive Abelian 
gauge theories are renormalisable even
without the Higgs mechanism \cite{Ruegg:2003ps}, from the perspective of 
purified gravity  it is natural to consider a spontaneous emergence of the 
Planck scale, since one wants to
recover scale invariance at the most fundamental level of physics. The idea of 
the metric  as a Goldstone boson of spontaneous symmetry breaking goes back 
to at least Isham, Salam and Sthrahdee \cite{Isham:1971dv}, and has been 
previously discussed in various different contexts 
\cite{Percacci:1990wy,Percacci:2009ij,Pagani:2015ema,Isham:1971dv,
Tresguerres:2000qn,Leclerc:2005qc,Tiemblo:2005js,Ali:2007hu,Westman:2014yca,
Zlosnik:2018qvg,Koivisto:2019ejt}. However, the analogy with massive gauge theory and the Planck mass 
as the rationale of teleparallelism are very recent theoretical advances \cite{Koivisto:2019jra}. 

Even regardless of those, we should recall the results  of 
\ref{eande}, which we believe could be highly relevant in view of the eventual 
reconciliation of gravity and quantum physics. In the canonical approach to 
quantum gravity, the notorious problem of time might be taken into 
reconsideration from the perspective wherein besides the conventional ADM 
Hamiltonian formalism, we also have available the unique consistent definition of 
localisable energy in a gravitational system. 
 The other main approach to quantisation, the path integral formalism, can  
obviously also be reconsidered from a more promising starting point, since the 
action (\ref{stgr}) in the canonical frame
 is well-defined without the addition of boundary terms or counter terms. We 
already  took advantage of this in the zeroth order approximation to Euclidean 
quantum gravity by determining the entropy of charged black holes
 and the de Sitter Universe by a considerably simplified calculation compared to 
the 
standard ones. It could also be mentioned that the origin of the otherwise 
mysterious simplifications that have been found to occur in the perturbative 
quantum gravity calculations via the famous technique dubbed the double copy 
\cite{Cheung:2017pzi} can presumably be traced back to the realisation of a 
translation gauge theory in the framework of purified gravity 
\cite{BeltranJimenez:2018vdo}.

\subsection{Matter Coupling}
\label{mattercouplingsref2}

Having seen that the geometrical trinity encloses an infinite theory space of 
equivalent  formulations, the question may have arisen whether the geometry of 
spacetime can be decided by experiments, or   
whether it is merely a matter of convention.  Considering solely a theory of a 
vacuum, the latter may well be the case, but by taking into account matter 
besides the gravitational field(s), criteria could be found that
distinguish the ``physical'' geometry. Matter particles in the Standard Model are  
described by spin-half fields, whose gravitational coupling is intriguing and, 
especially in metric-affine geometries, often an issue of some controversy 
\cite{Maluf:2003fs,Obukhov:2004hv,Mielke:2004gg}. We present in careful detail 
what  we find to be the most reasonable approach. 

Consider the Hermitean Dirac action for a (for simplicity, massless)  spinor 
$\psi$ 
\be \label{dirac1}
S_D = -\frac{1}{2}\int{\rm d}^4 x   \sqrt{-g}\left[ \left( 
i\bar{\psi}\gamma^\mu 
\hat{\nabla}_\mu\psi\right) + \left( i\bar{\psi}\gamma^\mu 
\hat{\nabla}_\mu\psi\right)^\dagger\right]\,,
\ee
where $\gamma^\mu=\gamma^A e_A{}^\mu$ are the  Dirac matrices 
satisfying the 
Clifford algebra $\gamma^{(A}\gamma^{B)}=-\eta^{AB}$, and 
$\bar{\psi}=\psi^\dagger\gamma^0$ is the conjugate spinor. Let  us denote the 
spinor covariant derivative as 
$\hat{\nabla}_\mu\psi=\partial_\mu\psi+\hat{\Gamma}_\mu\psi$. Without making any assumptions about the geometry or the connection  
$\hat{\Gamma}_\mu$, we obtain the equation of motion for the spinor
\be \label{dirac2}
\left[ 2\gamma^\mu\partial_\mu+\frac{1}{\sqrt{-g}}\partial_\mu\left(\sqrt{-g} 
\gamma^\mu\right) +\gamma^\mu 
\hat{\Gamma}_\mu-\gamma^0\hat{\Gamma}^\dagger_\mu\gamma^0\gamma^\mu\right]\psi=0\,.
\ee
The equation of motion for the spinor conjugate is the same, with 
$\hat{\Gamma}_\mu\rightarrow -\hat{\Gamma}_\mu$, $\psi\rightarrow\bar{\psi}$. Now we should 
deduce how to present the connection \eqref{affconection} for fermions.

To extend the Lorentz algebra of the rotation generators $r_{AB}$
\begin{subequations}
\label{cr}
\be \label{cr1}
[r_{AB},r_{CD}]    =    2\left( \eta_{D[A}r_{B]C} -  
\eta_{C[A}r_{B]D}\right)\,, 
\ee
to the general linear algebra, we need to also introduce the shear generators  
$q_{AB}$, entailing the additional commutation relations
\ba
\left[ q_{AB},q_{CD}\right]   & = &  2\left( \eta_{C(A}r_{B)D} + 
\eta_{D(A}r_{B)C} \right)\,,  
\label{crq}  \\
\left[ q_{AB},r_{CD}\right]   & = &  2\left( 
\eta_{C(A}q_{B)D}-\eta_{D(A}q_{B)C} \right)\,.  
\label{crm} 
\ea
\end{subequations}
It is clear to see that the generators can be represented as
\be \label{basis}
r_{AB} = 2x_{[A}\partial_{B]} + \Delta_{[AB]}\,, \quad q_{AB} = 
2x_{(A}\partial_{B)} + \Delta_{(AB)}\,,
\ee
where the vectors are called the orbital parts, while the $\Delta_{ab}$, the form 
of which depends  on the fields we are acting upon, are called the matrix parts 
of the generators. 
The infinitesimal transformations of the fields are nothing but their Lie 
derivatives along the  Killing vectors of the group, in the case at hand given 
by (\ref{basis}). For example,
 the Lie Derivative $\mathsterling$ of a vector $V$ along $X$ is
\be
\left(\mathsterling_X V\right)^\mu  = X^\alpha\partial_\alpha V^\mu -   
V^\alpha\partial_\alpha X^\mu\,.
\ee
Quite intuitively, the first term gives the orbital action and the second term 
gives the matrix part.  From the 16 Killing vectors $X$ we now have, it is easy to obtain the matrix components 
$\Delta^{1}_{AB\mu}{}^\nu=- 2\eta_{\mu A}\delta^{\nu}_{B}$ and check that (\ref{basis}) then generates the correct rotations and shear transformations of any 
vector $V$, taking into account its argument.
Now, for a spinor $\psi$, the Lie derivative is given as
\be
\mathsterling_X\psi = X^A{\nabla}_A\psi -  
\frac{1}{4}{\nabla}_A X_B\gamma^A\gamma^B\psi.
\ee
In the case that one antisymmetrises the above,  $\gamma^A\gamma^B \rightarrow 
\gamma^{[A}\gamma^{B]}$, it is called the Kosmann lift. 
We obtain  
\be
\Delta^\frac{1}{2}_{AB}= 
-{\frac{1}{2}}\gamma_A\gamma_B=\eta_{AB}-{\frac{1}{2}}\gamma_{[A}\gamma_{B]}\,. 
\ee
We can now plug the decomposition \eqref{affconection} into this spinor basis. We get
\be \label{spinc}
\hat{\boldsymbol{\omega}} = -\frac{1}{4}\eta^{AB}{\hat{\rm D}}\eta_{AB} + 
\frac{1}{4}\left(\boldsymbol{{\omega}}_{AB} 
+\hat{\bK}_{AB} -  \boldsymbol{e}_{[A}\cdot{\hat{\rm D}}\eta_{B]C}\boldsymbol{\mathrm{e}}^C\right)\gamma^{[A}\gamma^{B]}\,.
\ee   
In particular, it should be noted that non-metricity enters not only into the Weyl 
part,  but also into the Lorentz part, since though $\hat{\bQ}_{[AB]}=0$, $\hat{\bL}_{[AB]} 
\neq 0$. If one adopts the Kosmann lift,
only the trace part is dropped.

Now we can return to the equation of motion (\ref{dirac2}). First let us remark 
that although now, the Clifford algebra is dictating that 
$Q_\alpha{}^{\mu\nu}=\hat{\nabla}_\alpha\gamma^{(\mu}\gamma^{\nu)}$,
the parallel transport does not preserve the matrices $\hat{\nabla}_\mu \gamma^\alpha 
= \hat{Q}_{\mu\beta}{}^{\alpha}\gamma^\beta$;  the matrices nevertheless are 
invariant 
with respect to the metric-covariant derivative, as usual. In the following, it 
is more transparent to write (\ref{spinc}) in a coordinate system,
\be
\hat{\Gamma}_\mu = -\frac{1}{4}\hat{Q}_\mu + 
\frac{1}{4}\left(\omega_{\alpha\beta\mu}+ \hat{K}_{\alpha\beta\mu}-\hat{Q}_{[\alpha\beta] 
\mu}\right)\gamma^{[\alpha}\gamma^{\beta]}\,,
\ee
noting, though, that 
$\omega_{\alpha\beta\mu}=e^A{}_\alpha e^B{}_\beta\omega_{AB\mu}
$ is  not a 
tensor. 
We immediately see that the trace part decouples  from the action 
(\ref{dirac1}). This reflects the 
scale invariance of massless spinors. However, the possible imaginary part of 
the trace would remain, leaving us with a U(1) gauge  field to contemplate the 
geometric unification of gravitation and electromagnetism 
 \cite{Koivisto:2018aip,Janssen:2018exh} (a gravitoelectroweak extension was 
considered in  \cite{Koivisto:2019ejt}).

The property of Lorentz generators that 
$\gamma^0(\gamma^{[\mu}\gamma^{\nu]})^\dagger\gamma^0=-\gamma^{[\mu}\gamma^{\nu]
}$ is readily checked. Only a few more lines of Clifford algebra is required to
show that the combination appearing in (\ref{dirac2}) reduces to
\be
\{\gamma^\mu, \gamma^{[\alpha}\gamma^{\beta]}\}  = 
-2i\epsilon^{\alpha\beta\mu\nu}\gamma_\nu\gamma^5\,.
\ee
Wrapping up the final result then we have
\ba
\gamma^\mu\partial_\mu\psi +
\frac{1}{2\sqrt{-g}}\partial_\mu\left(\sqrt{-g}\gamma^\mu\right)\psi -  
\frac{1}{4}\left(  \omega_{\alpha\beta\mu}+\hat{K}_{\alpha\beta\mu}\right) 
i\epsilon^{\alpha\beta\mu\nu}\gamma_\nu\gamma^5\psi = 0\,.
\ea
All the non-metricity has disappeared. We mention that the Dirac equation has also been considered written directly in terms of the connection (\ref{spinc}) 
\cite{Adak:2002pq,Adak:2003qg,Adak:2008gd}. However, that equation (which 
features non-metricity) does not follow from the Hermitean action 
(\ref{dirac1}) 
of a unitary theory. A derivation in an index-free formalism 
\cite{Formiga:2012ns} seems to agree with our conclusion.

This is one of the most important arguments supporting purified gravity.  In a 
generic Teleparallel Equivalent of General Relativity (\ref{e-h2}), there is both 
torsion and non-metricity.
Whereas the latter does not appear in the spinor connection 
$\hat{\boldsymbol{\omega}}$,   
 the presence of the former
 should be stringently constrained. The case of TEGR is obviously ruled out,  since there, matter would couple only to the Weitzenb\"ock 
connection - so actually the right hand side in (\ref{mtgeod}) is 
 an ad hoc non-minimal coupling. In contrast, for the case of TEGR, 
 Eq.(\ref{stgeod}) is exactly what the theory predicts. Even when the spacetime 
connection vanishes, $\hat{\Gamma}{}^\alpha{}_{\mu\nu}=0$,
 matter moves along the metric geodesics, i.e. following the connection 
${\Gamma}{}^\alpha{}_{\mu\nu}$,  due to the peculiar property 
of fermions. In this way, causal structure is filtered into the reality of 
material objects ``from nothing''. 
 
Finally we comment on the case of gauge fields, where the same conclusion 
follows much  more directly: promoting the kinetic term of (for simplicity, the 
Abelian) gauge field $A_\mu$ into manifest general linear covariance,
 $F_{\mu\nu}=2\partial_{[\mu}A_{\nu]} \rightarrow  2\hat{\nabla}_{[\mu}A_{\nu]} = 
F_{\mu\nu}  - \hat{T}^\alpha{}_{\mu\nu}A_\alpha$, makes no difference unless the 
$\hat{\nabla}_\mu$ has torsion that would spoil the gauge invariance.

\section{Modified Gravity}

The results above suggest that the complementary perspectives to the theory of  
General Relativity could be very useful in unifying gravity with
quantum mechanics and the other fundamental interactions of nature. The 
geometrical trinity  also provides fresh starting points for building new 
extended theories of gravity.
A plethora of such theories have been proposed, motivated by the cosmological 
observations that  cannot be explained by General Relativity without invoking 
exotic matter and energy components. Many of the extended gravity theories are 
also based on introducing additional fields into the gravitational sector 
\cite{Heisenberg:2018vsk}. Indeed, in the Riemannian corner of the trinity
the only viable purely gravitational non-trivial modification is the well-known 
$f({R})$ theory, which is in fact is a scalar-tensor theory   discussed in 
Chapter \ref{ref:Cruz}.
Some of the limitations in generalising the action are due to the 
higher-derivative  property of the curvature of the connection $\udt{\Gamma}{\alpha}{\mu\nu} = g^{\alpha\beta}(g_{\beta(\mu,\nu)}-g_{\mu\nu,\beta}/2)$. 
While theories with an independent connection
 avoid this, the degrees of freedom in the connection can introduce ghosts 
unless  they are suitably eliminated by symmetries.
In any case, in the framework of metric-affine theories, where the connection is 
 independent of the metric,
there are at least in principle many new possibilities for extensions, as we 
learned in Chapter \ref{ref:Iosifidis}.

\begin{figure}[ht]
\begin{center}
\mbox{\includegraphics[width=.6\textwidth]
{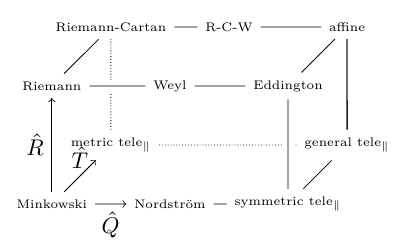}}
\end{center}
\caption{{\it{Special cases of metric-affine geometries are obtained by 
switching on/off the curvature $\hat{\bR}^A{}_B$ and/or the torsion 
$\hat{\bT}^A$ and/or the non-metricity $\hat{\bQ}_{AB}$. The labels in the box 
refer to spacetime geometries in the context of gravity theories. }}}
\label{Tomifig1}
\end{figure}

An obvious generalisation of Einstein's theory in this framework is to consider 
more  general curvature invariants in the action than the scalar $R$. Along the 
lines of $f({R})$
models, we can consider actions defined by a function $f(\hat{R})$. We recall 
from Chapter \ref{ref:Iosifidis} that these theories 
have a very interesting structure, where the scalar $\hat{R}$ does not introduce 
additional dynamics but can
be determined algebraically from the trace of the field equations as a function 
of the matter  fields. While such models could account for the accelerating 
expansion of the background Universe, they fail to reproduce the observed matter 
power spectrum \cite{Koivisto:2006ie} (see \cite{Koivisto:2007sq}). In 
Chapter \ref{ref:Lobo}   the hybrid metric-Palatini theories were introduced, 
which interpolate between the $f(\hat{R})$ and
$f({R})$ models  
\cite{Harko:2011nh,Leanizbarrutia:2017xyd,VargasdosSantos:2017ggl,Danila:2018xya
,Wojnar:2018hal,Bronnikov:2019ugl,Rosa:2019ejh}, and which (barring 
\cite{Amendola:2010bk}) can also be recast as scalar-tensor theories. In the 
metric-affine context it is also possible to consider, without introducing higher 
derivatives, more generic curvature invariants that the Ricci scalar of the 
connection, such as in the so called Ricci-based models \cite{Afonso:2018bpv}. 
Such models do not reduce to simple scalar-tensor theories, but there is some 
evidence that they usually contain pathological degrees of freedom, at least 
unless one restricts to symmetric connections \cite{BeltranJimenez:2019acz}. The 
latter case is labeled ``Eddington'' in Fig. \ref{Tomifig1} since 
symmetric 
affine connections had been utilised in, e.g., Born-Infeld type constructions of 
gravity \cite{BeltranJimenez:2017doy}. Delhom and Rubiera-Garcia have reviewed some of the many
cosmological applications of these models.

As special cases of generic metric-affine formulations, the metric and the 
symmetric  teleparallel geometries provide novel starting points for 
generalisation. As an example,
it is natural to consider the nonlinear generalisation of (\ref{stgr}),  
\label{fQref1}
\be \label{fq}
S_{f(\mathbb{Q})}   =  -\frac{1}{2\kappa^2}\int{\rm d}^4 x 
\sqrt{-g}f(\mathbb{Q})\,,
\ee
which clearly reduces to (\ref{stgr}) when $f=\mathbb{Q}$. The cosmological  
background equations of the $f(\mathbb{Q})$ models 
\cite{Lu:2019hra,Lazkoz:2019sjl}  contain the same solutions as the extensively
studied $f(\mathbb{T})$ models  \cite{BeltranJimenez:2017tkd}, but the evolution 
of cosmological perturbations and, consequently, the predicted large-scale 
structure of the Universe, can be considerably different \cite{Jimenez:2019ovq}. 
 The dark energy problem was also considered in the generalisation of the model 
with non-minimal matter couplings at cosmological scales 
\cite{Harko:2018gxr,Lobo:2019xwp,Xu:2019sbp}, and the models might be relevant 
for dark matter at galactic scales as well \cite{Milgrom:2019rtd,DAmbrosio:2020nev}. However, an 
action of the form (\ref{fq}) was shown to propagate two extra scalar degrees in the general cosmological background, but one of those decouples at the
limit of de Sitter space \cite{Jimenez:2019ovq}. This degree of freedom thus  
appears to be strongly coupled, which could be a problem for the viability of 
the model. Dialektopoulos classified the cosmological Noether symmetries as an action that can be an arbitrary function of the six invariants 
\cite{Dialektopoulos:2019mtr}.  The special 5-parameter quadratic case had been 
dubbed Newer General Relativity \cite{BeltranJimenez:2017tkd}. Gravitational 
waves were recently considered in this quadratic theory 
\cite{Hohmann:2018wxu,Soudi:2018dhv}, although Conroy had analysed in detail the 
full spectrum of symmetric teleparallel gravity, including the ghost-free and 
singularity-free ultra-violet completions inspired by string field theory 
\cite{Conroy:2017yln}. Some of the interesting classes of extensions in symmetric 
teleparallel gravity are models of conformal gravity 
\cite{Iosifidis:2018zwo,Gakis:2019rdd}. 

Modified gravity in the metric teleparallel corner of Fig.~\ref{Tomifig1} 
has attracted
considerable attention and was described in Chapter  \ref{ref:teleparallelch}. 
Additionally, whilst investigations most often focus on astrophysical and 
cosmological 
phenomenology,  some studies considering the theoretical consistency of 
teleparallel modified gravity have pointed out generic problems related to 
strong coupling \cite{Jimenez:2019ovq,Jimenez:2019tkx}, formulation of the 
Cauchy problem \cite{Cheng:1988zg,Ferraro:2018tpu} and ghosts 
\cite{BeltranJimenez:2017tkd,Koivisto:2018loq}. In fact, in our opinion there 
still remains the open question that we
could now pose as a theoretical challenge: {\it are there any  consistent 
teleparallel modified gravity models that do not have a metric equivalent?}

\chapter[Palatini 
Theories of Gravity and Cosmology]{Palatini 
Theories of Gravity and Cosmology}
\label{Delhomchapter}

{\em Adrià Delhom and Diego Rubiera-Garcia}\\


%
%
%
%
%
In the Palatini formulation\label{Palatiniformref3}, the metricity (or 
Riemannian) postulate,  which 
renders the affine connection as a secondary object completely determined by the 
metric, is abandoned to view metric and connection as equally fundamental fields 
in describing the gravitational interaction. Indeed, whether the affine 
structure is determined by the metric one or otherwise is a fundamental question 
in the understanding of the gravitational interaction as a manifestation of a 
curved spacetime, whose associated phenomenology has begun to be unravelled 
very recently.  A large family of Palatini theories of gravity can be nicely 
accommodated by considering those Lagrangian densities built out of the 
(symmetric part of the) Ricci tensor and its contractions with the metric. More 
specifically, such theories are constructed as traces of powers of the object 
${M^\mu}_{\nu}\equiv g^{\mu\alpha}R_{\alpha \nu}$ and   include, as 
particular examples, GR itself, $f(R)$ theories, Ricci-squared theories 
$f(R,R_{\mu\nu}R^{\mu\nu})$, or Born-Infeld-inspired theories of gravity 
\cite{Afonso:2017bxr,BeltranJimenez:2017doy}. 

An Einstein-frame representation  \label{Einsteinfrref4}
for this family of theories can be conveniently introduced as
\begin{equation} \label{eq:Gmunuq1}
{G^\mu}_{\nu}(q)=\frac{\kappa^2 }{|\Omega|^{1/2}}\left[ T^{\mu}{}_\nu- 
\left(\mathcal{L}_G+\frac{T}{2}\right)\delta^\mu{}_\nu\right],
\end{equation}
where the independent connection $\Gamma$ is Levi-Civita of the auxiliary 
metric 
$q_{\mu\nu}$  (that is, $\nabla_{\alpha}(\sqrt{-q}q^{\mu\nu})=0)$,  
${T^\mu}_{\nu}$ represents the energy-momentum tensor of the matter fields 
(with 
$T$ its trace) and $\vert \Omega \vert$ is the determinant of the 
\emph{deformation} matrix $\Omega$, which relates the auxiliary and spacetime 
metrics as  $q_{\mu\nu}=g_{\mu\alpha}\Omega^\alpha{}_\nu$ and can always be 
written 
on-shell as a function of the matter fields and (possibly) the spacetime metric 
itself (and so does the gravitational Lagrangian $\mathcal{L}_G$). This 
(projectively-invariant) class of theories, dubbed as \emph{Ricci-based 
gravities} (RBGs) features a number of relevant properties: second-order field 
equations and absence of ghost-like instabilities (with the projective symmetry 
requirement playing a key role in incorporating this feature 
\cite{BeltranJimenez:2019acz,Jimenez:2020dpn}), and reduction to GR equations 
in 
vacuum\footnote{Although the equations for the deformation matrix 
$\Omega^{\mu}{}_\nu$ are non-linear, and there might be several branches of 
solutions, there always exists a branch which recovers GR in vacuum 
\cite{Jimenez:2020iok}.}, which  \label{gravitationalwavrefs6}
ensures their compatibility with Solar System observations and the propagation 
of gravitational waves at the speed of light in vacuum. In high-energy density 
environments, however, the dynamics encoded in the new couplings engendered by 
the matter fields yield relevant deviations from GR results  that can be 
exploited, in particular, in cosmological scenarios, which we review in this 
section for different kinds of RBG theories.

For completeness, we will write down the modified Friemdan equations for 
isotropic cosmologies and for the isotropic branch of solutions of 
$\Omega^{\mu}{}_\nu$ of the form 
$\Omega^\mu{}_\nu=diag(\Omega_0,\Omega_x,\Omega_x,\Omega_x)$. Assuming a FLRW 
metric of the form $ds_g=-d t^2+a^2(t)d\vec{x}^2$ we will have 
$ds_q^2=-\Omega_{0} d t^2+\tilde{a}^2(t)d\vec{x}^2$ with $\tilde{a}^2=\Omega_x 
a^2$. 
By using (\ref{eq:Gmunuq1}) and the fact that $M^{\mu}{}_{\nu}$ is also 
diagonal $M^\mu{}_\nu=diag(M,M_x,M_x,M_x)$, we arrive at the following modified 
Friedman equation
\begin{equation}
6 H^{2}=\frac{3\frac{\Omega_{0}}{\Omega_{x}} 
M_{x}-M_{0}}{\left[1-3(\rho+p)\left(\partial_{\rho} \log 
\sqrt{\Omega_x}+c_{s}^{2} \partial_{p} \log \sqrt{\Omega_x}\right)\right]^{2}}
,
\end{equation}
where $M_i$ and $\Omega_i$ can be written in terms of $\rho$ and 
$p$ by using the field equations of the theory \cite{Jimenez:2020iok}.

\section{Smoothing out Cosmological Singularities}
A great deal of activity within this context is based upon the removal or  
smoothing out of cosmological singularities\label{singularitiesrefs1} 
encountered 
in GR. While it is 
known 
that within GR the singularities cannot be avoided for matter fields that 
satisfy the energy conditions, this is not the case in Palatini RBGs. This can 
be traced back to Einstein-frame representation (\ref{eq:Gmunuq1}) of RBGs, 
where the matter fields become non-linearly interacting. Presumably, if one 
couples matter that satisfies the energy conditions to an RBG, it is plausible 
that in the passage to the Einstein frame the modified matter sector violates 
some of the energy conditions. Thus, the removal of cosmological singularities 
can be understood in light of the equivalence between RBGs and GR with 
non-linearly modified matter sectors that have recently been introduced in a 
series of works 
\cite{BeltranJimenez:2017doy,Afonso:2017bxr,Afonso:2018mxn,Afonso:2018hyj,
Afonso:2018bpv,Delhom:2019zrb}.
 
The main focus of research on smoothing cosmological singularities has been  
the 
finding of bounces replacing the Big Bang singularity. One of the first works  
builds  an $f(R)$ action such that the corresponding cosmological evolution 
matches that predicted by Loop Quantum Cosmology with a scalar field, i.e., a 
singularity-free bouncing cosmology \cite{Olmo:2008nf}. The existence of such 
an 
action explicitly demonstrated the covariance of LQC at the background 
evolution 
level. The next obvious step was to perform an analysis allowing  
conditions to be found   that a generic Palatini $f(R)$ has to satisfy in order 
to describe 
such non-singular bouncing cosmologies\label{fRref2}. This was done in 
\cite{Barragan:2009sq}, 
where it was found that for flat models either the condition $f_R=0$ or 
$\kappa^2\rho+(Rf_R-f)/2=0$ (with $w>-1$) have to be satisfied at some point in 
the evolution  in order  to find a bounce (here $\rho$ is the energy density of 
the total 
or the dominant sector at the bounce phase).\label{Bouncerefs2}

In the non-flat universe case the discussion is 
more involved, although bouncing solutions can be found under quite generic 
conditions. The particular case of $R+R^2/R_P$ (with $R_P$ being Planck's 
curvature) is also analysed, finding bouncing solutions for flat as well as 
non-flat universes. Also adding torsion, a Nieh-Yan term and a dynamical 
Immirzi 
field to quadratic $f(R)$ gravity yields bouncing solutions in some branch of 
the theory \cite{Bombacigno:2018tyw}. After finding such bouncing solutions one 
must guarantee their stability against perturbations. A general formalism for 
computing the evolution of perturbations in $f(R)$ models was devised in 
\cite{Koivisto:2010jj}, with the result that solutions with $f_R=0$ at the 
bounce have an unstable behaviour in a flat universe with dust, although the 
behaviour can improve in the non-flat case. Nonetheless, it is noticed that the 
relation between Jordan and Einstein frames becomes ill-defined at $f_R=0$, and 
that the perturbation expansion is not valid at the bounce, which implies that 
the unstable behaviour could be due to the breakdown of the perturbative 
expansion. A non-perturbative analysis is therefore necessary. 
 
Moving on from $f(R)$, Eddington-inspired Born-Infeld (EiBI) 
\label{EddingtonBIref1}gravity has been  
the pivotal point of many discussions of bouncing solutions and their 
perturbations. A cosmological bounce was indeed shown to arise within this 
theory for a radiation fluid in the early work of \cite{Banados:2010ix}. In 
several subsequent works 
\cite{EscamillaRivera:2012vz,Avelino:2012ue,Yang:2013hsa}, linear perturbations 
on a cosmological background were studied. It was found that that linear tensor 
perturbations are singular for the EiBI parameter $\epsilon<0$  
\cite{EscamillaRivera:2012vz}, but that they can be stable for $\epsilon>0$ and 
a time-varying equation of state parameter $w$ \cite{Avelino:2012ue}.

The 
evolution of a universe driven by a  perfect fluid within this model was 
investigated in \cite{Cho:2012vg}, finding that for $w>0$ there is a 
non-singular initial state of finite size at which $H=0$. For $w=0$, the 
Universe approaches a de Sitter evolution at high energy densities (early 
times). Other bouncing cosmologies can be found in power-law extensions of the 
fundamental object defining EiBI gravity \cite{Odintsov:2014yaa} or in 
elementary polinomial extensions \cite{Jimenez:2014fla}, where for 
$-2/3<w\leq0$ 
a quasi de Sitter inflationary phase can also be achieved.

As shown in 
\cite{Odintsov:2014yaa,BeltranJimenez:2017uwv}, in all Palatini-type models 
tensorial perturbations\label{tensorpertref1} propagate upon the auxiliary 
metric 
$q_{\mu\nu}$ (and 
not $g_{\mu\nu}$), which entails that for all these power-law type extensions 
of 
EiBI gravity, a severe instability  is found for bouncing solutions 
($\epsilon<0$) 
due to a strong coupling problem that yields a diverging speed of propagation. 
A 
milder instability is found for loitering solutions ($\epsilon>0$, which 
interpolate with an asymptotic Minkowski past) due to the vanishing of the 
propagation speed.

It is, moreover, seen that such instabilities might be cured by 
adding higher-derivative terms to the pure EiBI theory and its power-law 
extensions, and that for elementary polynomial extensions the propagation speed 
remains non-vanishing, thus solving the problem \cite{BeltranJimenez:2017uwv}. 
A combination of EiBI+$f(R)$ has also been investigated in 
\cite{Makarenko:2014lxa}. This allows selection of the parameters in  
$R^n$-type 
terms (but not other terms such as those of $Tr[(R^\mu{}_\nu)^n]$) of the 
curvature expansion of the EiBI Lagrangian. It is seen that bouncing solutions 
are robust against such modifications of the coefficients for $\epsilon<0$, 
while 
for $\epsilon>0$ the robustness is lost. It is also shown that these 
modifications 
can account for an inflationary stage without the need of extra degrees of 
freedom. Similar results are found for an $f(|\Omega|)$ extension in 
\cite{Makarenko:2014cca}. The stability of an Einstein static universe filled 
with a perfect fluid within EiBI has also been studied in  \cite{Li:2017ttl}. 
There is no spatially flat Einstein universe and it is seen that scalar 
perturbations are unstable in Einstein closed universe, while Einstein open 
universes could be stable if $\epsilon<0$ and $\omega<-1$. 
 
The possibility of resolving other cosmological singularities than the Big Bang 
 within Palatini theories such as future cosmological  singularities has also 
been investigated.
It has been shown in  \cite{Bouhmadi-Lopez:2014jfa} that that for the spacetime 
metric  the Big Rip and Little Rip singularities driven by phantom dark energy 
are not cured, but some other phantom dark energy-related singularities 
occurring in GR are smoothed out. For the auxiliary metric, only a past type IV 
singularity persists on a tiny region of the parameter space, while all the 
other singularities disappear. Furthermore, bound structures are found to be 
destroyed 
near Big Rip and Little Rip singularities, while they can remain bounded at 
sudden, Big Freeze and type  IV singularities. Moreover, by exploiting the 
cosmographic approach, it is concluded that EiBI is a good alternative that 
smoothes curvature singularities occurring in GR \cite{Bouhmadi-Lopez:2014jfa}. 

Recently, a modified Wheeler-DeWitt \label{DeWittref2} equation was derived 
from the EiBI 
Hamiltonian, and it is seen that quantum effects could avoid the Big Rip 
singularity when it is sourced either by phantom dark energy or by a phantom 
scalar field \cite{Bouhmadi-Lopez:2018tel,Albarran:2018mpg}. Additionally, 
future 
singularities within the quadratic $f(R)$ model have been investigated by 
 finding a 
new type of sewn singularities, where the first derivative of the potential (in 
scalar-tensor representation) has a discontinuity. However, the model appears 
to 
be disfavoured as compared to $\Lambda$CDM regarding $\Omega_{\gamma}$  
\cite{Stachowski:2016zio}.
\section{Inflationary Models}
Inflation in the quadratic $f(R)$ model has been considered in several works.  
In \cite{Meng:2004yf} it was shown that the new corrections to GR modify the 
kinetic-energy dominated epoch of the inflationary 
regime\label{inflationrefs1}, 
constraining the 
kinetic energy density of the scalar field to be sub-Planckian, while the 
potential-energy dominated epoch is not affected by these corrections. Since 
Palatini $f(R)$ models have no extra propagating degrees of freedom, one way to 
trigger inflation is to consider non-minimal couplings. 

With this ingredient, 
several potentials such as quadratic, quasi-scale invariant Coleman-Weinberg, 
induced gravity model, or Higgs-type, were considered within the slow roll 
inflationary paradigm, all of which turn out to be viable models when compared 
to Planck's data \cite{Antoniadis:2018ywb}. The predictions of these models 
deviate from those of their metric counterparts in the preheating era in terms 
of resonance bands and transfer of energy from the scalar field to fluctuations 
\cite{Fu:2017iqg}, as well as in the number of e-folds \cite{Almeida:2018oid, 
Racioppi:2017spw}. The quadratic $f(R)$ model also yields a suppression of the 
tensor-scalar-ratio and the tensor spectral index \cite{Enckell:2018hmo}, which 
are compatible with Planck's data \cite{Antoniadis:2018yfq}. Further analysis 
of 
this question \cite{Rubio:2019ypq} shows that, for Higgs inflation, the 
reduction of the number of e-folds and smaller spectral index is channelled  
through tachyonic production of Higgs excitation, while preheating occurs 
almost  instantaneously in such a way that almost all of the background energy 
density at the end of inflation is turned into radiation. Though in both the 
metric and the Palatini  formulations the Universe evolves into a slow-roll 
inflationary era, discriminating them could be possible via a  stochastic 
background of primordial gravitational waves in future high-frequency 
detectors. 
These conclusions also hold when the inflation potential is considered a 
periodic function  of the inflaton field \cite{Jinno:2018jei}. On the other 
hand, 
$f(R)$ models with $\alpha R^{m}-\beta/R^{n}$ terms are able to generate an 
inflationary epoch, although this cannot be followed by a radiation era 
\cite{Fay:2007gg}.
 
Regarding EiBI gravity, in \cite{Cho:2013pea} it was coupled to a scalar  field 
with a quadratic potential, where it was shown to provide a natural precursor 
of 
inflation. This is due to the fact that the maximal pressure state (MPS) of the 
Universe in this theory, though being non-singular, is unstable and evolves to 
an inflationary period. The tensor-to-scalar ratio within this theory is 
smaller 
than in GR \cite{Cho:2014jta,Cho:2014xaa}, while tensor perturbations give 
similar results to standard chaotic inflation \cite{Cho:2014ija}; however, 
if the attractor of the model does not account for 60 e-folds, then the 
perturbations near the MPS have to be taken into account in explaining the 
behaviour of low-angular modes. High-$k$ modes show a similar behaviour to those 
in 
the attractor regime, but low-$k$ modes are enhanced with respect to standard 
inflation, which could serve as a discriminator. 

Furthermore, potentials with asymptotic 
form $\phi^{2+2n}$ are seen to allow for a non-singular initial state for 
$n\leq0$, leading to an inflationary epoch for $-1\leq n\leq 2$ 
\cite{Kim:2013noa}. In \cite{Cho:2015yza} a large (though constrained) value of 
$\epsilon$ is considered to analyse the behaviour of scalar and tensor 
perturbations within the strong gravity regime of the $\phi^2$ model. The 
scalar 
power spectrum receives little corrections, while the tensor power spectrum, as 
well as the tensor-to-scalar ratio, can be suppressed significantly with 
respect 
to GR+$\phi^2$. The correction to the spectral index also fits well with 
observations, and therefore, although the $\phi^2$ model is ruled out within 
GR, 
it is not within EiBI gravity.  An inflationary model for EiBI coupled to a 
Born-Infeld-like scalar field has also been considered in \cite{Jana:2016uvq}, 
for which a particular form of the kinetic term is chosen, and it is seen to 
satisfactorily fit 
  late-time acceleration for any sign of $\epsilon$.  \label{latetimerefss1}

Different behaviours for 
the early Universe may arise for the different sign of $\epsilon$, but only a 
particular example is treated. As shown in \cite{Jimenez:2015jqa}, inflation in 
EiBI gravity can also be induced by gravitating dust, with a graceful exit 
naturally provided by the dilution of the matter density. Reheating can be 
implemented if it is assumed that the dust components conform an unstable chain 
that 
ends by  decaying in radiation, with two coexisting dust components being the 
minimum amount of components that give predictions compatible with 
observations. 
This scenario is predictive thanks to the dependence of background evolution in 
the sound speed and, in particular, it predicts no primordial gravitational 
waves, therefore it could be ruled out if B-modes are found within the CMB. 
\label{CMBrefs1}By 
contrast to Palatini $\alpha R^m-\beta/R^n$ models, EiBI+$f(R)$ gravity for an 
FLRW 
background can generate an inflationary epoch followed by a radiation era, as 
shown in \cite{Banik:2018nyz}, using a dynamical systems approach. There it is 
seen how an initial radiation-dominated epoch can be followed by inflation and 
evolve into radiation-dominated, matter-dominated and accelerated expansion 
eras.
Similar conclusions apply for a Bianchi I cosmology within this theory.

 In 
\cite{Makarenko:2014lxa} it is shown that the fact that an inflationary stage 
without extra degrees of freedom can be achieved in EiBI+$f(R)$ can be traced 
back to the possibility introduced by the $f(R)$ term to modify the $R^n$ 
coefficients of the curvature expansion of the EiBI Lagrangian. Other models 
such as $R+(R_{\mu\nu}R^{\mu\nu})^n$ corrections have also been shown to 
account 
for an inflationary epoch if $n>3/4$ \cite{Allemandi:2004wn}, although they 
have 
not received much attention in the literature regarding the inflationary 
scenario. It has also been argued in \cite{Avelino:2012ue} that bouncing models 
in EiBI gravity can by themselves account for the solution of the typical 
problems solved by inflation (without actually having an inflationary phase). 
Particularly, if the contracting branch lasts long enough, this could solve the 
horizon problem, as well as the size age and entropy problems. Given that the 
Universe meets the maximum density after the contracting branch, the flatness 
problem can also be alleviated. If the bounce occurs at a scale below Grand 
Unification, the monopole problem can also be solved. Furthermore, a 
matter-bounce scenario can generate a nearly scale-invariant primordial 
spectrum.

\section{Background Evolution, Late-time Acceleration, and Observational  
Constraints}

The first relevant cosmological application of Palatini theories of gravity was 
a possible  explanation of   late-time accelerated expansion without the need 
of a cosmological constant, driven by the simple inverse model $R-\alpha^2/R$ 
\cite{Vollick:2003aw}. It was shown that the Palatini formulation of this 
theory 
is free from ghosts, and when coupled to dust it approaches  
a 
de Sitter Universe exponentially fast  at late times.  Shortly after its 
birth, however, this 
model 
was ruled out as an explanation for cosmic speed-up due to the appearance of 
new 
matter interactions that were not in agreement with observations 
\cite{Flanagan:2003rb}. Moreover, the approach employed in 
\cite{Flanagan:2003rb} was criticized later in \cite{Vollick:2004ws} on the 
basis that field re-definitions may not be allowed in curved spacetime. 

Cosmologically accelerated solutions at late-times without dark energy have 
been 
implemented in other Palatini models. The FLRW cosmology of a generic Palatini 
$f(R)$ and the possibility of using generic $f(R)$ infrared corrections to GR 
(beyond the $\mu^4/R$ one) as an explanation of the cosmic speed-up problem was 
devised in \cite{Meng:2003uv,Nojiri:2003wx}. In \cite{Allemandi:2004wn}, 
explicit 
solutions for the models $R^n$, $\ln(R)$, and $\sinh^{-1}(R)$ were found, and 
it 
is explicitly seen that $f(R)$ corrections that grow at low curvatures in the 
Palatini formalism lead to cosmic speed-up. Moreover, corrections to GR of the 
form 
$f(R_{\mu\nu}R^{\mu\nu})$ were analysed in \cite{Allemandi:2004wn} and, in 
particular, $(R_{\mu\nu}R^{\mu\nu})^n$ corrections were shown to account for 
cosmic speed-up for $n<0$. Indeed, after much effort it was shown that 
in 
(Palatini-type) infrared modifications to GR that account for late-time 
acceleration, the  behaviour of low-curvature $f(R)$ corrections to GR aimed at 
explaining the cosmic speed-up (with $\mu^4/R$ as a particular case) inevitably 
leads 
to instabilities in the hydrogen atom, forcing its fundamental state to decay 
after a short time and thus being incompatible with experiments 
\cite{Olmo:2008ye}.  In the same work it was also shown that ultraviolet $f(R)$ 
corrections to GR do not introduce these instabilities, and thus are still not 
excluded.
 
Palatini $f(R)$ and $f(R^{\mu\nu}R_{\mu\nu})$ cosmological models have also 
been 
confronted  with different sets of observational data.  For the model 
$R+\lambda_1 \exp(R/\lambda_2)$, the CMB, the matter power spectrum  
\cite{Li:2006vi} and linear perturbations \cite{Li:2006ag} have been 
considered. For the CMB it is shown that the integrated Sachs-Wolfe effect 
receives significant corrections for intermediate multipole moments 
$l\sim(10,500)$, recovering the $\Lambda$CDM behaviour for both larger and 
smaller $l$'s. The matter power spectrum picks up oscillatory corrections for 
high momentums, leading to effective pressure fluctuations and a diminished 
growth of small-scale density perturbations \cite{Li:2006vi}. In 
\cite{Li:2006ag}, WMAP, SNLS and SDSS data are used to constrain Palatini 
$\alpha(-R)^\beta$ corrections to GR, being able to substantially reduce the 
parameter space by demanding small departures from $\Lambda$CDM. The matter 
power spectrum was also computed for this model in  \cite{Koivisto:2006ie}, 
where the observational constraints were found to favour values of the 
parameter 
close to $\Lambda$CDM. For the model $\alpha(-R)^\beta$, constraints to 
$(\alpha,\beta)$ parameters are obtained from CMB data, with a best fit of 
$\alpha=-3.6$ and $\beta=0.09$, and the predictions of this model  lying within 
1$\sigma$ from $\Lambda$CDM model predictions \cite{Amarzguioui:2005zq}.

$R^n$ 
models were analysed in \cite{Lee:2008ek}, finding that super-horizon density 
perturbations evolve as in GR, but sub-horizon evolutions differ from it. Later 
on, by studying the evolution of the deceleration parameter, it was seen that 
positive values of $n$ representing ultraviolet corrections to GR are very 
suppressed by cosmographic constraints \cite{Pires:2010fv}. The model 
$\sqrt{R^2-R_0^2}$ has been confronted with data from type Ia supernovae, CMB, 
BAO and large-scale structure formation \label{LSSefs2} \cite{Movahed:2007cs}. 
Using both 
separate and combined analysis, similar results to the $\Lambda$CDM for the 
best 
fit of $R_0\sim (6.3\pm0.2)H_0^2$ are obtained, although inconsistencies with 
the 
age of the quasar APM 08279+5255 are found. A generalisation of the form 
$(R^n-R_0^n)^{1/n}$ was later analysed in \cite{Baghram:2009fr}, with the same 
data plus data from gas mass fraction, in galaxy clusters. The best fits are 
$n=0.98\pm 0.08$ and $R_0\sim (4.4\pm0.4)H_0^2$, being also close to 
$\Lambda$CDM (note that for $n=1$ this model is exactly GR+$\Lambda$ with 
$R_0=2\Lambda$), but the same problem for the quasar APM 08279+5255 arises.

The 
quadratic + inverse $f(R)$ model was compared to  SNIa data using the SNLS 
catalogue of 115 SNIA in \cite{Astier:2005qq} and the Union2.1 catalogue of 580 
SNIa  \cite{Amanullah:2010vv}, whose best fit has large confidence intervals, 
with the conclusion that there is too much degeneracy in the space of 
parameters 
to fit them all by just considering  SNIa data \cite{Pinto:2018rfg}. We point 
out that the analysis of the Hubble drift (the relation between the Hubble 
parameter and the redshift) of this model gives a different prediction for 
large 
$z$ as compared to $\Lambda$CDM paradigm \cite{DelVecchio:2018abv}. 

Corrections of the 
form  $\beta/R^n$, $\alpha \,ln(R)-\beta$, which are able to generate phase 
transitions corresponding to the radiation-matter-dark energy dominated eras, 
were constrained with SNIa and BAO data in \cite{Fay:2007gg}. For the model 
including Ricci-squared corrections $R+F(R_{\mu\nu}R^{\mu\nu})$, the background 
evolution is obtained by interpreting the corrections to GR as an effective 
stress-energy tensor \cite{Li:2007xw}.  A particular power-law model $F=\alpha 
(R_{\mu\nu}R^{\mu\nu})^\beta$ is then confronted to SNIa luminosity distance 
and 
CMB shift parameter data, leading to a preferred small positive value for 
$\beta$ at $95\%$ confidence level, but with the standard $\Lambda$CDM lying in 
the $68\%$ confidence level. Late-time matter density perturbations growth is 
also analysed within the model, finding similar results to $f(R)$, having also 
an effective scale-dependent sound-speed-like term in the equations.

Cosmographic approaches have  also been used to constraint these models. This 
is 
the case with $f(R)$ models using power-law forms \cite{Capozziello:2018aba}, 
while an analysis of more general models can be found at 
\cite{Capozziello:2019cav}. Via the cosmographic approach, for $R$ + power-law 
and $R$ + logarithmic models, their parameters are fitted using CMB, SNIa and 
BAO data, together with Hubble parameter estimations, finding that these models 
can be made compatible with the right sequence of cosmological eras but with 
large confidence intervals \cite{TeppaPannia:2018ale}.
 
Linear scalar perturbations in the matter-dominated era and large-scale 
structure  formation within EiBI gravity are studied in \cite{Du:2014jka}. The 
growth rate deviates (it is suppressed for $\epsilon>0$ and enhanced for 
$\epsilon<0$) from $\Lambda$CDM at early times (high energy-densities) and the 
deviation increases with $k$, but as the Universe expands, it quickly 
approaches 
the $\Lambda$CDM prediction. 

The early-times deviations are insignificant if one 
takes into account the tightest constraints on $\epsilon$. The influence of the 
integrated SW effect on the CMB power spectrum is seen to have no deviations 
from $\Lambda$CDM. In \cite{Avelino:2012ge}, BBN data was used to constrain the 
free parameter of EiBI, finding the bound $\epsilon\leq6\times10^8 m^5 kg^{-1} 
s^{-2}$ (where $\epsilon>0$ was assumed), although it is much weaker than other 
bounds from astrophysical scenarios, nuclear physics or high-energy experiments 
\cite{Avelino:2012qe,Avelino:2019esh,Latorre:2017uve,Delhom:2019wir}. FRWL 
solutions in a model combining an EiBI + $f(R)$ were found in  
\cite{Banik:2018nyz} by means of a dynamical systems analysis with several 
choices for $f(R)$ (quadratic, inverse, power-law), finding a plethora of 
accelerated/decelerated solutions. 

Additionally, a hybrid version of EiBI gravity was 
introduced in \cite{Banados:2008rm}, where it was found to reproduce both 
late-time acceleration and flat rotation curves. In \cite{Skordis:2009zza} it 
was also shown that in this theory dark matter effects are relevant at the 
linearized cosmological level, which agrees with observations on large scales.

Cosmological effects were also analysed in \cite{Banados:2008fj}, finding that 
although the growth of structure occurs as in $\Lambda$CDM, if the theory is 
demanded to produce cosmic accelerated expansion then the gravitational 
potentials grow (and some combinations diverge), leading to an enhancement of 
the 
integrated Sachs-Wolfe effect that does not match the data. However, if the 
theory is not demanded to reproduce dark energy effects but dark matter effects 
only, then the entire evolution is indistinguishable from $\Lambda$CDM 
cosmology. In 
\cite{DeFelice:2012hq}, WMAP7, BAO and SNIa data are used to conclude that the 
model cannot be used as an explanation for both dark energy and dark matter 
either. In \cite{Rodrigues:2008kv} an analysis of Bianchi I cosmology is made 
within this theory, finding that the isotropic solution is an attractor, 
according with the data, but the decay of the shears is 
damped-oscillations-like. 
 
Cosmological models corresponding to gravity theories with non-minimal 
matter-curvature  couplings have also been analysed. For instance, in  $f(R,T)$ 
gravity, using quadratic and inverse forms for the $f(R)$ sector, de Sitter 
type 
solutions at late times  are found \cite{Wu:2018idg}. Other non-minimal 
couplings of the form  $f(R)+F(R)L_{\phi}$, as considered in 
\cite{Allemandi:2005qs}, were used in \cite{Koivisto:2005yc} in the more 
general 
class of models 
$f(R,\phi)$ to find that the amplitude of vector and tensor perturbations gets 
modulated  by a factor depending on $f_R$, and also that a damping term appears 
for gravitational waves. Scalar perturbations get a more complex modification, 
and it is also seen that an effective pressure gradient appears when the 
expansion of the Universe is driven by non-linear curvature terms, which is 
problematic when compared to structure formation data. 

Considering  models including 
torsion in $f(R)$ theories, their incorporation within this framework yields 
hard equations to crack \cite{Minkevich:1998cv}, although some preliminary 
ideas were 
considered in  \cite{Hehl:1999sb,Capozziello:2009mq}. For instance, the 
torsional degrees of freedom can be treated as a scalar field, in such a way 
that a cosmological toy model can be constructed, where torsion dynamically 
determines a relation between the amounts of dark matter and dark energy at 
present times \cite{Capozziello:2007tj}. 

Finally, accelerated cosmologies can also be 
found in a two-parameter scalar-tensor theory with derivative coupling of the 
scalar field to the curvature scalar and to the Ricci tensor 
\cite{Galtsov:2018xuc}. More general models can be considered: for instance, 
assuming an anisotropic deformation matrix in general RBG theories yields the 
result that when the spacetime metric describes a standard FLRW metric, the 
auxiliary one describes a Bianchi type-I \cite{AndresBletranDelhom}.

\chapter[Hybrid Metric-Palatini Gravity  and 
Cosmology]{Hybrid Metric-Palatini Gravity  and 
Cosmology}
\label{ref:Lobo}

{\em Francisco S. N. Lobo}\\

 \label{Palatiniformref4}
 \label{metricPalatress1}


Given the success of General Relativity (GR) at relatively short scales (such as 
the Solar System, stellar models, or  compact binary systems), the idea that 
modified dynamics could arise at larger scales has been investigated in much 
detail over recent years. Theories in which the gravitational action consists
of more general combinations of curvature invariants than the pure
Einstein-Hilbert term have been investigated with special emphasis 
\cite{Harko:2018ayt}. From these investigations it was soon noticed that the 
usual metric formulation of alternative theories of gravity is generically 
different from its Palatini  (or metric-affine) counterpart (see 
\cite{Olmo:2011uz} for a recent \label{metaffgrref4}
review on the Palatini approach). Whereas the metric approach typically leads to 
higher-order derivative equations, in the Palatini formulation the resulting 
field equations are always second-order. The appealing character of the 
second-order equations of the Palatini formalism, however, is accompanied by 
certain algebraic relations between the matter fields and the affine connection, 
 \label{metaffconref4}
which is now determined by a set of equations coupled  to the matter fields and 
the metric. 

The case of $f(R)$ theories is particularly useful to illustrate the differences 
between these two approaches.  In the metric formulation, the object $\phi\equiv 
{d}f/{d}R$ behaves as a dynamical scalar field, which satisfies a 
second-order equation with self-interactions that depend on the form of the 
Lagrangian $f(R)$. In order to have an impact at large astrophysical and 
cosmological scales, the scalar field $\phi$ should have a very low mass, 
implying a long interaction range. It is well known, however, that light 
scalars 
do have an impact at shorter scales, where their presence is strongly 
constrained by laboratory and Solar System observations unless some kind of 
screening mechanism is invoked. In the Palatini case, a scalar-tensor 
representation is also possible, but with the scalar field satisfying an 
algebraic rather than a differential equation. It is then found that the 
scalar 
$\phi$ turns out to be an algebraic function of the trace of the stress-energy 
tensor of the matter, $\phi=\phi(T)$, which may lead, in models of late-time 
cosmic speed-up, to undesired gradient instabilities at various contexts, as 
has 
been shown by studies of cosmological perturbations 
\cite{Koivisto:2006ie,Koivisto:2005yc} and atomic physics 
\cite{Olmo:2006zu,Olmo:2008ye}.

In this chapter we will review the {\it hybrid} variation of $f(R)$ gravity, 
in 
which the (purely metric)  Einstein-Hilbert action is supplemented with 
(metric-affine) correction terms constructed \`a la Palatini 
\cite{Capozziello:2013uya,Harko:2011nh}.
Given that the metric and Palatini $f(R)$ theories allow the construction of 
simple extensions of GR with interesting properties and, at the same time, 
suffer from different types of drawbacks, it proves interesting to establish 
bridges between these two seemingly disparate approaches, hoping to find ways 
to 
cure or improve their individual deficiencies. For that purpose, in a number of 
works  
\cite{Harko:2011nh,Capozziello:2012ny,Boehmer:2013oxa,Borowiec:2014wva,
Carloni:2015bua,Harko:2018ayt,Capozziello:2013uya} a hybrid combination of 
metric and Palatini elements to construct the gravity Lagrangian was 
considered, 
and it was found that viable models sharing properties of both formalisms are 
possible. An interesting aspect of these theories is the possibility to 
generate 
long-range forces without entering into conflict with local tests of gravity 
and 
without invoking any kind of screening mechanism (which would, however, require 
that at the present time the cosmological evolution reduces to GR). 

The possibility of expressing these hybrid $f(R)$ metric-Palatini theories 
using 
 a scalar-tensor representation simplifies the analysis of the field equations 
and the construction of solutions.
In some sense, considering a theory like $R+f({\cal R})$ means that one retains 
all  the positive results of GR, represented by the Einstein-Hilbert part of 
the 
action $R$, while the further ``gravitational budget'' is endowed in the 
metric-affine $f({\cal R})$ component, where ${\cal R}$ is the Palatini 
curvature scalar constructed in terms of an independent connection. In fact, it 
is well known that metric-affine and purely metric formalisms coincide in GR, 
 i.e., considering the action $R$.  On the  contrary, the two formalisms 
lead to different results considering more generic functions $f({\cal R})$.

\section{Hybrid Metric-Palatini Gravity: The General Formalism}

We start our considerations by providing the general features of the theory.  
More specifically, we present the action and the field equations both in the 
so-called hybrid metric-Palatini gravity and its equivalent scalar-tensor 
representations in both the Jordan and the Einstein frames. 
\label{Einsteinfrref1}\label{Jordanrref1}

\subsection{Action and Gravitational Field Equations}
\label{fRref4} \label{Riemanntenref3}

The action of the hybrid metric-Palatini gravity is specified as 
\cite{Harko:2011nh,Capozziello:2012ny}
\begin{equation} \label{eq:S_hybrid}
S= \frac{1}{2\kappa^2}\int {d}^4 x \sqrt{-g}  
\left[ R + f(\mathcal{R})\right] +S_m ,
\end{equation}
where $S_m$ is the matter action, $\kappa^2\equiv 8\pi G$, $R$ is
the Einstein-Hilbert term, $\mathcal{R}  \equiv g^{\mu\nu}\mathcal{R}_{\mu\nu} 
$ 
is
the Palatini curvature, defined in terms of
an independent connection $\breve{\Gamma}^\alpha_{\mu\nu}$  as
\be \mathcal{R}
\equiv  g^{\mu\nu}\mathcal{R}_{\mu\nu} \equiv g^{\mu\nu}\left(
\breve{\Gamma}^\alpha_{\mu\nu , \alpha}
       - \breve{\Gamma}^\alpha_{\mu\alpha , \nu} +
\breve{\Gamma}^\alpha_{\alpha\lambda}\breve{\Gamma}^\lambda_{\mu\nu} -
\breve{\Gamma}^\alpha_{\mu\lambda}\breve{\Gamma}^\lambda_{\alpha\nu}\right)\,,
\label
{ r_def}
\ee 
which generates the Ricci curvature tensor $\mathcal{R}_{\mu\nu}$ as
\begin{equation}
\mathcal{R}_{\mu\nu} \equiv \breve{\Gamma}^\alpha_{\mu\nu ,\alpha} -
\breve{\Gamma}^\alpha_{\mu\alpha , \nu} +
\breve{\Gamma}^\alpha_{\alpha\lambda}\breve{\Gamma}^\lambda_{\mu\nu}
-\breve{\Gamma}^\alpha_{\mu\lambda}\breve{\Gamma}^\lambda_{\alpha\nu}\,.
\end{equation}

Now, varying the action (\ref{eq:S_hybrid}) with respect to the metric, one  
obtains the following gravitational field equation  
\be
\label{efe} G_{\mu\nu} +
F(\mathcal{R})\mathcal{R}_{\mu\nu}-\frac{1}{2}f(\mathcal{R})g_{\mu\nu} = 
\kappa^2 
T_{\mu\nu}\,,
\ee
where the matter stress-energy tensor is defined, as usual, through
 \be \label{memt}
 T_{\mu\nu} \equiv -\frac{2}{\sqrt{-g}} \frac{\delta
 (\sqrt{-g}\mathcal{L}_m)}{\delta(g^{\mu\nu})}.
 \ee
Varying the action with respect to the independent connection 
$\breve{\Gamma}^\alpha_{\mu\nu}$,  it is then found as the solution to the 
resulting 
equation of motion that $\breve{\Gamma}^\alpha_{\mu\nu}$ is compatible with the 
metric
$F(\mathcal{R})g_{\mu\nu}$, conformally related to the physical metric 
$g_{\mu\nu}$, with 
the conformal  factor given by $F(\mathcal{R}) \equiv 
df(\mathcal{R})/d\mathcal{R}$. This implies the 
following relation:
\ba
\!
\label{ricci} \mathcal{R}_{\mu\nu}  =  R_{\mu\nu} +
\frac{3}{2}\frac{1}{F^2(\mathcal{R})}F(\mathcal{R})_{,\mu}F(\mathcal{R})_{,\nu}
  - \frac{1}{F(\mathcal{R})}\nabla_\mu F(\mathcal{R})_{,\nu} -
\frac{1}{2}\frac{1}{F(\mathcal{R})}g_{\mu\nu}\nabla_\alpha \nabla^\alpha 
F(\mathcal{R})\,. \ea 
%

\subsection{Scalar-tensor Representation}
\label{sts}
\label{Scalarperfrref4}

In a similar manner to the pure metric and Palatini approaches 
\cite{Olmo:2005zr,Olmo:2005hc},  the action (\ref{eq:S_hybrid}) for the hybrid 
metric-Palatini theory can be represented as that of a
scalar-tensor theory by introducing an auxiliary field $A$, such that
\begin{equation} \label{eq:S_scalar0}
S= \frac{1}{2\kappa^2}\int {d}^4 x \sqrt{-g} \left[\Omega_A R + 
f(A)+f_A(\mathcal{R}-A)\right] +S_m \ ,
\end{equation}
where $f_A\equiv df/dA$ and a coupling constant $\Omega_A$ has been included 
for 
generality, where $\Omega_A=1$ reduces  to the original hybrid metric-Palatini 
theory 
\cite{Harko:2011nh}. Now, rearranging the terms and defining $\phi\equiv f_A$, 
$V(\phi)=A f_A-f(A)$,
Eq. (\ref{eq:S_scalar0}) becomes
\begin{equation} \label{eq:S_scalar1}
S= \frac{1}{2\kappa^2}\int {d}^4 x \sqrt{-g} \left[\Omega_A R + 
\phi\mathcal{R}-V(\phi)\right] +S_m \,,
\end{equation}
which is equivalent to our original starting point (\ref{eq:S_hybrid}).

In order obtain the equations of motion, one varies  the action 
(\ref{eq:S_scalar1}) 
with respect  to the metric, the scalar $\phi$ and the connection, which leads 
to the following field equations:
\begin{eqnarray}
\Omega_A R_{\mu\nu}+\phi \mathcal{R}_{\mu\nu}-\frac{1}{2}\left(\Omega_A 
R+\phi\mathcal{R}-V\right)g_{\mu\nu}&=&\kappa^2 T_{\mu\nu}\,,
\label{eq:var-gab}\\
\mathcal{R}-V_\phi&=&0 \label{eq:var-phi} \,, \\
\breve{\nabla}_\alpha\left(\sqrt{-g}\phi g^{\mu\nu}\right)&=&0 \,, 
\label{eq:connection}\
\end{eqnarray}
respectively. 

The solution of Eq.~(\ref{eq:connection}) implies that the independent 
connection is  the Levi-Civita connection of
a metric $h_{\mu\nu}=\phi g_{\mu\nu}$. This implies that the relation 
(\ref{ricci}) between the  tensors $\mathcal{R}_{\mu\nu}$ and $R_{\mu\nu}$ 
reduces to
\begin{equation} \label{eq:conformal_Rmn}
\mathcal{R}_{\mu\nu}=R_{\mu\nu}+\frac{3}{2\phi^2}\partial_\mu \phi \partial_\nu 
\phi-\frac{1}{\phi}\left(\nabla_\mu
\nabla_\nu \phi+\frac{1}{2}g_{\mu\nu}\nabla_\alpha \nabla^\alpha \phi\right) \ ,
\end{equation}
which can be inserted in the action (\ref{eq:S_scalar1}) to eliminate the 
independent connection and to  obtain the following scalar-tensor 
representation,    ultimately arriving at:
\begin{equation} \label{eq:S_scalar2}
S= \frac{1}{2\kappa^2}\int {d}^4 x \sqrt{-g} \left[ (\Omega_A+\phi)R 
+\frac{3}{2\phi}\partial_\mu \phi \partial^\mu \phi
-V(\phi)\right] +S_m \ .
\end{equation}

In the limit $\Omega_A\rightarrow 0$, the theory (\ref{eq:S_scalar2}) presents 
the 
Palatini-$f(\mathcal{R})$ gravity,  and in the limit $\Omega_A\rightarrow 
\infty$ the metric 
$f(R)$ gravity \cite{Koivisto:2009jn}. Apart from these singular cases, the 
more 
generic theories with a finite $\Omega_A$ thus lie in the ``hybrid'' regime, 
which 
from this perspective provides a unique interpolation between the two a priori 
completely distinct classes of gravity theories. In fact, we have arrived at 
Brans-Dicke type of theories specified by the non-trivial coupling function 
\label{Bransref3}
\be
\omega_{BD}=\frac{3\phi}{2\phi-2\Omega_A}\,,
\ee
which generalises the $\omega_{BD}=0$ and $\omega_{BD}=-3/2$ cases, which in 
turn 
correspond to the  scalar-tensor representations of the metric $f(R)$ and the 
Palatini-$f(\mathcal{R})$ gravities \cite{fRgravity6}, respectively. 

Using Eq.~(\ref{eq:conformal_Rmn}) and Eq.~(\ref{eq:var-phi}) in 
Eq.~(\ref{eq:var-gab}), the metric field equation can be written as
\begin{equation}
(\Omega_A+\phi) R_{\mu\nu}=\kappa^2\left(T_{\mu\nu}-\frac{1}{2}g_{\mu\nu} 
T\right)+\frac{1}
{2}g_{\mu\nu}\left(V+\nabla_\alpha \nabla^\alpha 
\phi\right)+\nabla_\mu\nabla_\nu\phi-\frac{3}{2\phi}\partial_\mu \phi
\partial_\nu \phi \ \label{eq:evol-gab} ,
\end{equation}
or equivalently as
\begin{equation}\label{einstein_phi}
(\Omega_A+\phi)G_{\mu\nu}=\kappa^2T_{\mu\nu} + \nabla_\mu\nabla_\nu\phi -
\nabla_\alpha \nabla^\alpha \phi\/g_{\mu\nu}
 -\frac{3}{2\phi}\nabla_\mu\phi \nabla_\nu\phi +
\frac{3}{4\phi}\nabla_\lambda\phi\nabla^\lambda\phi g_{\mu \nu}-
\frac{1}{2}Vg_{\mu\nu},
\end{equation}
from which it is seen that the spacetime curvature is generated by both the 
matter and the scalar field.
The scalar field equation can be manipulated in two different ways that 
illustrate further how the hybrid models  combine physical features of the 
$\omega_{BD}=0$ and $\omega_{BD}=-3/2$ scalar-tensor models. 

First, tracing Eq.~(\ref{eq:var-gab}) with $g^{\mu\nu}$,  we find $-\Omega_A 
R-\phi\mathcal{R}+2V=\kappa^2T$, and using Eq.~(\ref{eq:var-phi}), it takes the 
following form:
\begin{equation}\label{eq:phi(X)}
2V-\phi V_\phi=\kappa^2T+\Omega_A R \ .
\end{equation}
Similar to the Palatini ($\omega_{BD}=-3/2$) case, this equation tells
us that the field $\phi$ can be expressed as an algebraic function
of the scalar $X\equiv \kappa^2T+\Omega_A R$, i.e., $\phi=\phi(X)$. In the
pure Palatini case, however, $\phi$ is just a function of $T$. The
right-hand side of Eq.~(\ref{eq:evol-gab}), therefore, besides
containing new matter terms associated with the trace $T$ and its
derivatives, also contains the curvature $R$ and its derivatives.
Thus, this theory can be seen as a higher-derivative theory in
both  matter and  metric fields. However, such an
interpretation can be avoided if $R$ is replaced in
Eq. (\ref{eq:phi(X)}) with the relation 
\begin{equation}
R=\mathcal{R}+\frac{3}{\phi}\nabla_\mu \nabla^\mu 
\phi-\frac{3}{2\phi^2}\partial_\mu \phi \partial^\mu \phi
\end{equation}
together with $\mathcal{R}=V_\phi$. It is then found that the scalar field is
governed by the second-order evolution equation that becomes, when $\Omega_A=1$,
\begin{equation}\label{eq:evol-phi}
-\nabla_\mu \nabla^\mu \phi+\frac{1}{2\phi}\partial_\mu \phi \partial^\mu
\phi+\frac{\phi[2V-(1+\phi)V_\phi]} {3}=\frac{\phi\kappa^2}{3}T\,,
\end{equation}
which is an effective Klein-Gordon equation.
This last expression shows that, unlike in the Palatini ($\omega_{BD}=-3/2$)
case, the scalar field is dynamical. The theory is therefore not
affected by the microscopic instabilities that arise in Palatini
models with infrared corrections \cite{Olmo:2011uz}.

Finally, we can make a conformal transformation into the Einstein frame of 
these 
theories. Thus,  considering the following conformal rescaling 
$\tilde{g}_{\mu\nu} 
\equiv \left( \phi + \Omega_A\right) g_{\mu\nu}$,
 the Einstein frame Lagrangian becomes
\be
\label{einsteinframe}
\tilde{\mathcal{L}} = \tilde{R} + 
\frac{3\Omega_A}{2\phi}
\frac{\tilde{g}^{\alpha\beta}\phi_{,\alpha}\phi_{,\beta}}{\left( \phi 
+ \Omega_A\right)^2} - \left( \phi + \Omega_A\right)^2 V(\phi).
\ee
This can be further put into its canonical form by introducing the rescaled 
field $\psi$ as $\phi = \Omega_A \tan^2\left[ \psi/(2\sqrt{3})\right]$.
Thus, the vacuum theory then becomes a canonical scalar theory with a very 
specific potential (stemming, of course,  from the original function 
$f(\mathcal{R})$) in 
the Einstein frame.


\section{Hybrid-gravity Cosmology}
 
In order to explore the cosmology of the metric-Palatini gravitational 
theories, 
we employ the scalar-tensor  formulation derived above (\ref{eq:S_scalar2}):
\be \label{st2}
S=\frac{1}{2\kappa^2}\int {d}^4 x \sqrt{-g}\left[ (\Omega_A+\phi) R + 
\frac{3}{2\phi}\left( \partial\phi\right)^2  - 2\kappa^2V(\phi)\right] + S_m\,,
\ee
where
\be
\kappa^2V(\phi)=\frac{1}{2}\left[ r(\phi)\phi-f(r(\phi))\right], \quad r(\phi) 
\equiv {f'}^{-1}(\phi) \,.
\ee
In this section, we first write down the cosmological equations in the 
formulation (\ref{st2}) and then  have a brief look at the phase space of exact 
solutions for these equations (we refer the reader to the recent phase space 
analysis for the most complete global analysis of the cosmological dynamics of 
these theories \cite{Carloni:2015bua}). Then we will analyse the formation of 
cosmological large-scale structure in these models.

\subsection{Background Expansion}

The flat Friedmann-Lema\^{i}tre-Robertson-Walker metric is defined as
\be
{d}s^2 = -{d}t^2 + a^2(t)\left( {d} x^2 +  {d} y^2 +  {d} 
z^2\right)\,,  
\ee
where the rate of time-evolution of the scale factor $a(t)$ is parameterised by 
the Hubble parameter $H=\dot{a}/a$, and  the overdot denotes a derivative with 
respect to cosmic time $t$. In the following we will mainly be interested in 
accelerating dark energy-like dynamics; for a study of Einstein static spaces, 
see Ref. \cite{Boehmer:2013oxa}.

\subsubsection{The Friedmann Equations}

The Friedmann equations that govern the evolution of $H$ can be written in 
terms 
of an effective energy  density and pressure, respectively, in the following 
manner
\ba
3H^2 & = & \kappa^2\rho_{\rm eff}\,, \label{fr1} \\
\dot{H} & = & -\frac{\kappa^2}{2}\left(\rho_{\rm eff}+p_{\rm 
eff}\right)\label{fr2}\,,
\ea
where for the theory defined by the action (\ref{st2}) we obtain the following 
effective source terms
\ba
\left( \Omega_A+\phi \right) \kappa^2\rho_{\rm eff} & = & 
-\frac{3}{4\phi}\dot{\phi}^2 + 
\kappa^2V(\phi)  - 3H\dot{\phi} + \kappa^2\rho_m\,, \\
\left( \Omega_A+\phi \right) \kappa^2 p_{\rm eff} & = & 
-\frac{3}{4\phi}\dot{\phi}^2 - \kappa^2V(\phi) + \ddot{\phi} + 2H\dot{\phi} + 
\kappa^2 p_m\,,
\ea
respectively.
The conservation equations for the matter component and the scalar field are
\ba
\dot{\rho}_m + 3H(\rho_m + p_m) & = & 0\,, \\
\ddot{\phi} + 3H\dot{\phi} - \frac{\dot{\phi}^2}{2\phi} + \frac{1}{3}\phi R - 
\frac{2}{3}\kappa^2  \phi V'(\phi) & = & 0 \label{kg}\,.
\ea
Recalling that $R=6(2H^2+\dot{H})$ and using Eqs. (\ref{fr1}) and (\ref{fr2}), 
we can rewrite the Klein-Gordon equation as
\be
\ddot{\phi}+3H\dot{\phi} - \frac{\dot{\phi}^2}{2\phi} + U'(\phi) + 
\frac{\kappa^2 \phi}{3\Omega_A}\left( \rho_m-3p_m\right) =0\,, \label{kg2}
\ee
where for notational simplicity, $U'(\phi)$ is defined by
\be
U'(\phi) \equiv \frac{2\kappa^2\phi}{3\Omega_A}\left[ 2V(\phi) - \left( 
\Omega_A + \phi \right) V'(\phi)\right]\,. \label{veff}
\ee
Note that as a consistency check we can verify that the Klein-Gordon equation 
together with the matter conservation, allows one to derive Eq. (\ref{fr2}) 
from 
(\ref{fr1}).
By combining Eqs. (\ref{kg}) and (\ref{kg2}), we find that
\be
2V(\phi)-V'(\phi)\phi =\frac{1}{2}\left( \Omega_A R + \kappa^2 T \right) 
\equiv \frac{1}{2}X\,. \label{alg}
\ee
The solution for $\phi=\phi(X=0)$ gives us the natural initial condition for 
the field in the early Universe. The asymptotic value of the field in the far 
future may then be deduced by studying the minima of the function $U(\phi)$ 
defined by Eq. (\ref{veff}).

\subsubsection{Dynamical System Analysis}

The cosmological dynamics can be analysed  by taking into account a suitable  
dynamical system. Consider the following dimensionless variables
\be
\Omega_m  \equiv  \frac{\kappa^2\rho_m}{3H^2}\,, \quad
x  \equiv  \phi\,, \quad y = x_{,N}\,, \quad z = \frac{\kappa^2V}{3H^2}\,,
\ee
where $N=\log{a}$ is the $e$-folding time. The Friedmann equation (\ref{fr1}) 
can then be rewritten as
\be
\Omega_A + x + y -z +\frac{y^2}{4x} = \Omega_m\,.
\ee
Due to this constraint, the number of independent degrees of freedom is three 
instead of four. We choose to span our phase space by the triplet $\{x,y,z\}$, 
so that the autonomous system of equations  reads as
\ba
x_{,N} & = & y\,, \\
y_{,N} & = & \frac{2x+y}{8\Omega_A x}\Big\{ \left( 3w_m-1 \right) y^2 + 
4x\left[ \left( 3w_m-1 \right) y-3\left( 1+w_m\right) z\right] \nonumber \\
&& - 4x^2\left( 1-3w_m-2u(x)z\right) +4 \Omega_A\left[ 3x\left( w_m-1\right) 
y+y^2-x^2\left( 2-6w_m-4u(x)z\right)\right]\Big\}\,, \\
z_{,N} & = & \frac{z}{4\Omega_A x}\Big\{\left( 3w_m-1\right) y^2  +  
4x\left[\left( 3w_m-1\right) y - 3\left( 1+w_m\right) z\right]
\nonumber \\ 
&& + 4\Omega_A x \left( 3+3w_m+u(x)y\right) + 4x^2\left( 
3w_m-1+2u(x)z\right)\Big\}\,,
\ea
respectively, with $w_m\equiv p_m/\rho_m$ the matter equation-of-state 
parameter. Additionally, 
we have defined $u(x) \equiv V'(\phi)/V(\phi)$. The relevant fixed points 
appear 
in this system.  In particular, we have the matter-dominated fixed point where 
$x=y=z=0$ and $w_{\rm eff}=w_m$, and the de Sitter fixed point described by 
$w_{\rm eff}=-1$ and
\be
x_*=(2-\Omega_A u_*)/u*\,, \quad y_*=0\,, \quad z_* = 2/u_*\,. \label{dS}
\ee
We denote the asymptotic values corresponding to this fixed point by a 
subscript 
star. In particular,  the asymptotic value of the field $x_*$ is solved from 
the 
first equation in (\ref{dS}) once the form of the potential is given. As 
expected, this value corresponds to the  minimum of the effective potential 
(\ref{veff}), $U'(x_*)=0$. To construct a viable model, the potential should be 
such that we meet the two requirements:
\begin{itemize}
\item The matter-dominated fixed point should be a saddle point and the de 
Sitter fixed point an attractor.  Then we naturally obtain a transition to 
acceleration following standard cosmological evolution.
\item At the present epoch the field value should be sufficiently close to 
zero, 
in order to avoid  conflict with the Solar System tests of gravity 
\cite{Harko:2011nh,Harko:2018ayt}. \label{solarsystemref3}
\end{itemize}
Note that the simplest metric $f(R)$ theories that provide acceleration fail in 
both predicting a viable structure formation era and the Solar System as we 
observe it. On the other hand, it can be argued that the 
Palatini-$f(\mathcal{R})$ models can be ruled out as a  dark energy alternative 
by 
considering their structure formation or implications to microphysics, if such 
a theory is regarded as  consistent in the first place. As shown here and 
explored 
further below, hybrid metric-Palatini gravity models exist that are free of 
these problems.
\newline
\newline
To summarise: the field goes from $\phi_i$ to $\phi_*$, where the former is 
given by $2V(\phi_i)=V'(\phi_i)\phi_i$ and the latter by 
$2V(\phi_*)=(\Omega_A+\phi_*)V'(\phi_*)$. We just need a suitable function 
$V(\phi)$, i.e., a form of $f(\mathcal{R})$ in such a way that the slope will 
be downwards and $\phi_*$ near the origin. 
We refer the reader to \cite{Carloni:2015bua} for a more complete and detailed 
phase space analysis  of the cosmological background dynamics.

\subsubsection{On Cosmological Solutions}

As a specific simple example, let us consider in more detail the specific case 
of the  de Sitter solution in vacuum when $\Omega_A=1$. For this case, the 
modified 
Friedmann equations take the form 
\begin{eqnarray}
3H^2&=& \frac{1}{1+\phi }\left[\kappa^2\rho_m
+\frac{V}{2}-3\dot{\phi}\left(H+\frac{\dot{\phi}}
{4\phi}\right)\right] \,,\label{field1d} \\
2\dot{H}&=&\frac{1}{1+\phi }\left[
-\kappa^2(\rho_m+p_m)+H\dot{\phi}+\frac{3}
{2}\frac{\dot{\phi}^2}{\phi}-\ddot{\phi}\right] \,,\label{field2d}
\end{eqnarray}
and the scalar field equation (\ref{eq:evol-phi}) becomes
\begin{equation}
\ddot{\phi}+3H\dot{\phi}-\frac{\dot{\phi}^2}{2\phi}+\frac{\phi}{3}
[2V-(1+\phi)V_\phi]=-\frac{\phi\kappa^2}{3}(\rho_m-3p_m) .  \label{3d}
\end{equation}

Furthermore, consider a model that arises by demanding
that matter and curvature satisfy the same relation as in GR.
Taking
\begin{equation} \label{pot1d}
V(\phi)=V_0+V_1\phi^2\,,
\end{equation}
the trace equation automatically implies $R=-\kappa^2T+2V_0$
\cite{Harko:2011nh,Capozziello:2012ny}. As $T\to 0$ with the
cosmic expansion, this model naturally evolves into a de Sitter
phase,  which requires $V_0\sim \Lambda$ for consistency with
observations. If $V_1$ is positive, the de Sitter regime
represents the minimum of the potential \cite{Harko:2011nh,Harko:2018ayt}.   
The 
effective mass for
local experiments, $m_\varphi^2=2(V_0-2 V_1 \phi)/3$, is then
positive and small as long as $\phi<V_0/V_1$. For sufficiently
large $V_1$ one can make the field amplitude small enough to be in
agreement with Solar System tests. It is interesting that the
exact de Sitter solution is compatible with dynamics of the scalar
field in this model  \cite{Harko:2011nh,Harko:2018ayt,Capozziello:2012ny}.

The accelerating dynamics that drive the hybrid metric-Palatini gravitational 
theory  towards its general relativistic limits today have indeed been realised 
in several specific models 
\cite{Harko:2011nh,Capozziello:2012ny,Lima:2014aza,Lima:2015nma}. Our 
preliminary phase space analysis confirmed the existence of de Sitter attractor 
solutions, and the recent study of cosmology in terms of dynamical system 
analysis extends this result to more general models \cite{Carloni:2015bua}. 
Analytic solutions were also presented  in Ref. \cite{Capozziello:2012ny} as 
well 
as in Ref. \cite{Borowiec:2014wva},   using a N\"oether symmetry technique. 
A designer approach was formulated by Lima \cite{Lima:2014aza} to reconstruct 
precisely the standard $\Lambda$CDM expansion history by a nontrivial hybrid 
metric-Palatini model, and finally, two families of models were constrained by 
confronting their predictions with a combination of cosmic microwave 
background, 
supernovae Ia and baryonic acoustic oscillations background data 
\cite{Lima:2015nma}.

\subsection{Cosmological Perturbations}\label{sec:pert}

To understand the implications of these models on the cosmological structure  
formation, we derive the perturbation equations and analyse them in some 
specific cases of interest. This paves the way for a detailed comparison of the 
predictions with the cosmological data on large-scale structure and the cosmic 
microwave background. For generality, we will keep the parameter $\Omega_A$ in 
the formulae in this section.

\subsubsection{Field Equations and Conservation Laws}

Consider the Newtonian gauge \cite{Ma:1995ey}, which can be parameterised by 
the 
 two gravitational potentials $\Phi$ and $\Psi$,
\be \label{newtongauge}
{d}s^2= -\left( 1+2\Psi \right) {d}t^2+a^2(t)\left( 1-2\Phi\right) {\rm 
d}\vec{x}^2\,. 
\ee
As a matter source we consider a perfect fluid, with the background equation of 
state $w$ and with density perturbation $\delta = \delta\rho_m/\rho_m$, 
pressure 
perturbation $\delta p_m=c_s^2 \delta\rho_m$ and velocity perturbation $v$.

The $0$-$0$ part of the field equations is given by
\ba
-\frac{k^2}{a^2}\Phi 
- 3\left( H-\frac{\dot{\phi}}{2\left(\Omega_A+\phi\right)}\right) \dot{\Phi} 
- 3\left( H^2 +\frac{H\dot{\phi}}{\F}-\frac{\dot{\phi}^2}{4\phi 
\left(\Omega_A+\phi\right)}\right)\Psi      \nonumber \\ 
=\frac{1}{2\F}\left[ \kappa^2\delta\rho_m + \left( 
\frac{3}{4\phi^2}\dot{\phi}^2 
+V'(\phi)-3H^2-\frac{k^2}{a^2}\right)\varphi - 3\left( H + 
\frac{\dot{\phi}}{2\phi}\right) 
\dot{\varphi}\right]\,,
\ea
where we have denoted $\varphi=\delta\phi$.
The Raychaudhuri equation for the perturbations reads
\begin{eqnarray}
 &&
 \!\!\!\!\!\!\!\!\!\!\!  \!\!\!\!\!\!\!\!\!\!\!\!\!\!
 \left[ 6\left( H^2+2\dot{H}\right) - 2\frac{k^2}{a^2} + \frac{6}{\F}\left( 
\ddot{\phi}-\frac{\dot{\phi}^2}{\phi^2}+H\dot{\phi}\right) \right] \Psi 
- 3\left( 2H - \frac{\dot{\phi}}{\F}\right) \left( \dot{\Phi} + 
\dot{\Psi}\right) 
+6\ddot{\Phi}=   
 \nonumber \\
 &&
  \!\!\!\!\!\!\!\!\!\!\!  \!\!\!\!\!\!\!\!\!\!\!\!\!\!
 \frac{1}{\F}  \left\{\kappa^2 \left( \delta\rho_m+3\delta p_m\right) 
+\left[ 6H^2+6\dot{H} + 
3\frac{\ddot{\phi}}{\phi^2}-2V'(\phi)+\frac{k^2}{a^2}\right]\varphi
+3\left( H-\frac{2\dot{\phi}}{\phi}\right) \dot{\varphi}
+ 3\ddot{\varphi}
\right\}.
\end{eqnarray}
The $0$-$i$ equation is
\be
  \left( H+\frac{\dot{\phi}}{2\left(\Omega_A+\phi\right)}\right)\Phi - 
\dot{\Phi}
= \frac{1}{2\left(\Omega_A+\phi\right)}\left[ \kappa^2\left(\rho_m+p_m\right) a 
v_m + \left( H+\frac{3\dot{\phi}}{2\phi}\right)\varphi+\dot{\varphi}\right]\,.
\ee
Note that the set of perturbed field equations is completed by the off-diagonal 
spatial piece:
\be
\Psi-\Phi=-\frac{\varphi}{\F}\,.
\ee
Assuming a perfect fluid, the continuity and Euler equations for the matter 
component are 
\ba
\dot{\delta} + 3H\left( c_s^2-w\right)\delta & = &  \left( 1+w \right) \left( 
3\dot{\Phi}+\frac{k^2}{a}v\right)\,, \\
\ddot{v}+\left( 1-3c_a^2\right) Hv & = &  \frac{1}{a}\left( \Psi + 
\frac{c_s^2}{1+w}\delta\right)\,, 
\ea
respectively. The linear part of the Klein-Gordon equation is then compatible 
with the above system, which  for completeness, is given by 
\begin{eqnarray}
&&
\!\!\!\!\!\!\!\!\!\!\!\!\!\!\!\!\!\!\!\!\!\!\!\!\!\!\!\!\!\!\!\!\!\!\!\!\!\!\!\!
\!\!
\ddot{\varphi}+\left( 3H+\frac{1}{\phi}\right) \dot{\varphi} + \left( 
\frac{k^2}{a^2}+\frac{\dot{\phi}^2}{2\phi^2}- 
\frac{2}{3}V''(\phi)\right)\varphi =  \nonumber
\left( 
2\ddot{\phi}+6H\dot{\phi}-\frac{3}{2\phi}\dot{\phi}^2\right)\Psi\nonumber\\
&& \ \ \ \ \ \ \ \ \ \ \ \ \ \ \ \ \ \ \  \ \ \ \ \ \ \ \ \ \ \ \ \ \ \ \ \ \ \  
\ \ \ \ \, + 
\dot{\phi}\left( \dot{\Psi}+3\dot{\Phi}\right) - \frac{\phi}{3}\delta R\,.
\end{eqnarray}
This completes the presentation of the field equations and the conservation 
laws.

\subsubsection{Matter-dominated Cosmology}

In this subsection, we consider the formation of structure in the 
matter-dominated  Universe, where $w=c_s^2=0$, and assume scales deep inside 
the 
Hubble radius. This so-called quasi-static approximation is well-known in the 
literature, having been applied to scalar-tensor theories since early studies 
\cite{Boisseau:2000pr},  and which
  more recently   has been generalised to a wide variety of coupled dark 
sector models \cite{coupled}.
In this limit the   spatial gradients are more important than the time 
derivatives and,  consequently, the 
matter density perturbations are much stronger than the gravitational 
potentials.  Combining the continuity and the Euler equation at this 
quasi-static subhorizon limit, one obtains
\be
\ddot{\delta}=-2H\dot{\delta}-\frac{k^2}{a^2}\Psi\,.
\ee  

We then need  to solve the gravitational potential. Let us define $\Pi = 
a^2\rho_m\delta/k^2$  and write the field equations and the Klein-Gordon 
equation at this limit in a very simple way as
\ba
\left(\Omega_A+\phi\right)\Phi &=& \varphi-\Pi\,, \\
\left(\Omega_A+\phi\right)\left(\Psi-\Phi\right) & = & -\varphi\,, \\
-2\left(\Omega_A+\phi\right)\Psi & = & \Pi + \varphi\,, \\
3\varphi & = & -2\phi\left( \Psi - 2\Phi\right)\,.
\ea
We immediately see that one of the equations is (as expected) redundant, and 
that the $\Psi$ is  (as usual) proportional to $\Pi$, where now the 
proportionality is given as a function 
of the field $\phi$. Our result is \label{Geffref2}
\be \label{delta_evol}
\ddot{\delta} + 2H\dot{\delta} = 4\pi G_{\rm eff}\rho_m\delta\,,
\ee
with
\be \label{delta_evol2}
 G_{\rm eff} \equiv 
\frac{\Omega_A-\frac{1}{3}\phi}{\Omega_A\left(\Omega_A+\phi\right)}G_N\,. 
\ee

This shows that instabilities can be avoided in the evolution of the matter 
inhomogeneities,  in contrast to the Palatini-$f(\mathcal{R})$ models and some 
matter-coupled scalar field models (recall that  our theory 
can be mapped into such in the Einstein frame). Equation (\ref{delta_evol}) 
provides a very  simple approximation to track the growth of structure 
accurately within the linear regime during matter-dominated cosmology.

\subsubsection{Vacuum Fluctuations}

The propagation of the scalar degree of freedom in vacuum is also a crucial 
consistency check on the theory. Setting $\rho_m=0$, consider the curvature 
perturbation in the uniform-field 
gauge $\zeta$, which in terms of the Newtonian gauge perturbations is given by 
$\zeta= -\Phi- H \varphi/\dot{\phi}$.
After somewhat more tedious algebra than in the previous case, we obtain the 
exact (linear)  evolution equation  \cite{Harko:2018ayt,Capozziello:2012ny}.
\be \label{eta_evol}
\ddot{\zeta}+\left\{ 3H
-2\left[
\frac{\ddot{\phi}+2\dot{H}\left(\Omega_A+\phi\right) 
-\frac{\dot{\phi}^2}{\F}}{\dot{\phi}+2H\left(\Omega_A+\phi\right)} \right]
+ 
\frac{\phi }{\dot{\phi}^2}\left[\frac{2\ddot{\phi}\dot{
\phi}}{\phi }
 + 
\frac{\dot{\phi}^3\left(\Omega_A+\phi\right)^3\phi}{
1-\phi^3\left(\Omega_A+\phi\right)^3}\right]\right\}\dot{\zeta} = - 
\frac{k^2}{a^2}\zeta\,.
\ee

The friction term depends on the perturbation variable we consider, but the 
perturbations at  small scales still propagate with the speed of light, as in 
canonical scalar field 
theory. This also excludes  gradient and tachyon instabilities in the 
graviscalar 
sector. Now, Eq. (\ref{eta_evol}) can be used to study the generation of 
fluctuations in hybrid metric-Palatini-inflation. Construction of specific 
models and their observational tests are left for forthcoming studies; let us 
only note in passing that the Einstein-frame formulation (\ref{einsteinframe}) 
might present a convenient starting point for this: in fact, we note that the 
quadratic $\mathcal{R}^2$ curvature correction results in an interesting 
generalisation 
of the Starobinsky inflation in the hybrid  metric-Palatini context, where the 
parameter $\Omega_A$ controls the flow along the so-called alpha attractor.

\section{Discussions and Final Remarks}\label{sect6}

In this work we have presented a hybrid metric-Palatini framework for modified 
theories of gravity, and we have tested the new theories it entails, using a 
number 
of theoretical consistency checks and observational constraints. Having 
established the theoretical consistency and interest on the hybrid 
metric-Palatini family of theories, we considered applications in which these 
theories provide gravitational alternatives to dark energy  
\cite{Harko:2018ayt,Capozziello:2012ny}. As shown by the post-Newtonian analysis 
 \cite{Harko:2011nh,Harko:2018ayt},   hybrid theories are promising  in this 
respect, as they can avoid the local gravity constraints but modify the 
cosmological dynamics at large scales. 
Cosmological perturbations have also  been   analysed in these models up to 
the 
linear order  \cite{Capozziello:2012ny,Harko:2011nh,Lima:2014aza}, and the 
results imply that the formation of large-scale structure in the aforementioned 
accelerating cosmologies is viable although it exhibits subtle features that 
might be 
detectable in future experiments.

At an effective level, the hybrid theory modifications involve both (the trace 
of) the matter stress energy and (the Ricci scalar of) the metric curvature, and 
from this point of view it appears appealing to speculate on the possible 
relevance of these theories to both the problems of dark energy and dark matter, 
in a unified theoretical framework and without distinguishing {\it a priori} 
matter and geometric sources.  Various aspects of dark-matter phenomenology, 
from astronomical to galactic and extragalatic scales, have also been discussed 
\cite{Capozziello:2013uya,Capozziello:2012qt,Capozziello:2013yha}. The 
generalised virial theoreom can acquire, in addition to the contribution from 
the baryonic masses, effective contributions of geometrical origin to the total 
gravitational potential energy \cite{Capozziello:2013uya,Capozziello:2012qt}, 
which may account for the well-known virial theorem mass discrepancy in clusters 
of galaxies.
In the context of galactic rotation curves, the scalar-field modified relations 
between the various  physical quantities such as tangential velocities of test 
particles around galaxies, Doppler frequency shifts and stellar dispersion 
velocities were derived \cite{Capozziello:2013uya,Capozziello:2013yha}. More 
recently, observational data of stellar motion near the Galactic centre were 
compared with simulations of the hybrid gravity theory, which turned out to be 
particularly suitable to model star dynamics. 

To conclude, whilst the physics of the metric and the Palatini versions of 
$f(R)$ gravity have  been uncovered in exquisite detail in a great variety of 
different contexts 
\cite{Capozziello:2002rd,Capozziello:2011et,DeFelice:2010aj,Lobo:2008sg, 
Nojiri:2010wj}, those studies largely wait to be extended for the hybrid version 
of the theory. We believe the results this far, as reported in this work, 
provide compelling motivation for the further exploration of these particular 
theories.

\chapter[Teleparallel   Gravity: Foundations and 
Cosmology]{Teleparallel   Gravity: 
Foundations and Cosmology}
\label{ref:teleparallelch}

{\em Sebastian Bahamonde, Konstantinos F. Dialektopoulos, Manuel Hohmann, 
Jackson Levi Said}

\section{Foundations of Teleparallel Gravity}\label{sec:foundations}
\label{nonRiemannianref3}\label{torsionref5}
Teleparallel Gravity theories have received growing attention during the last 
decade.  Their most distinguishing feature is the use of a different geometric 
setting compared to General Relativity, which features a flat (curvature-free) 
connection, and torsion instead of curvature. This section gives a brief 
introduction to the mathematical background and foundations of Teleparallel 
Gravity. Section~\ref{ssec:telegeom} gives an overview of the geometric 
setting and dynamical field content. Its relation to a gauge theory of 
translations is explained in Section~\ref{ssec:transgauge}. Local Lorentz 
invariance is discussed in Section~\ref{ssec:lorlorinv}. 
Section~\ref{ssec:matter} briefly elucidates how matter couples to the 
teleparallel geometry. Finally, Section~\ref{ssec:tegr} gives an account of the 
most simple Teleparallel Gravity theory, which is equivalent to General 
Relativity at the level of its field equations. The notation used here and in 
the following sections on Teleparallel Gravity is summarized in the Convention 
Table in the beginning of the book.

\subsection{Teleparallel 
Geometry}\label{ssec:telegeom}\label{telegraref1}\label{torsionref6}

The most important, distinguishing feature of Teleparallel 
Gravity~\cite{Aldrovandi:2013wha}  is its use of an affine connection, whose 
connection coefficients will be denoted by $\tp{\Gamma}^{\mu}{}_{\nu\rho}$, and 
which is different from the Levi-Civita connection of the metric \(g_{\mu\nu}\). 
This connection is chosen to be flat, in the sense that has vanishing 
curvature
\begin{equation}
\tp{R}^{\mu}{}_{\nu\alpha\beta} = 
\partial_{\alpha}\tp{\Gamma}^{\mu}{}_{\nu\beta} -  
\partial_{\beta}\tp{\Gamma}^{\mu}{}_{\nu\alpha} + 
\tp{\Gamma}^{\mu}{}_{\rho\alpha}\tp{\Gamma}^{\rho}{}_{\nu\beta} - 
\tp{\Gamma}^{\mu}{}_{\rho\beta}\tp{\Gamma}^{\rho}{}_{\nu\alpha} \equiv 0\,,
\end{equation}
where we denote quantities related to the teleparallel  connection with a 
bullet \(\tp{ }\)  to distinguish them from their Levi-Civita counterparts. For 
instance, \(\tp{\nabla}_{\mu}\) denotes the teleparallel covariant derivative, 
while \(\nabla_{\mu}\) denotes the Levi-Civita covariant derivative. The 
flatness of the connection allows a path-independent parallel transport, hence 
maintaining a notion of being parallel at a distance, which is the reason for 
calling it ``teleparallel''~\cite{Einstein1928}. Further, the teleparallel 
connection is metric compatible, so that its non-metricity tensor vanishes,
\begin{equation}
\tp{Q}_{\rho\mu\nu} = \tp{\nabla}_{\rho}g_{\mu\nu} = \partial_{\rho}g_{\mu\nu}  
- \tp{\Gamma}^{\sigma}{}_{\mu\rho}g_{\sigma\nu} - 
\tp{\Gamma}^{\sigma}{}_{\nu\rho}g_{\mu\sigma} \equiv 0\,,
\end{equation}
while its torsion
\begin{equation}\label{eq:torsion}
\tp{T}^{\rho}{}_{\mu\nu} = \tp{\Gamma}^{\rho}{}_{\nu\mu} - \tp{\Gamma}^{\rho}{}_{\mu\nu}
\end{equation}
is allowed to be non-vanishing. In Teleparallel Gravity, the torsion takes the 
role of the  gravitational field strength, in contrast to General Relativity, 
where this role is attributed to the Levi-Civita curvature.

There are different, equivalent possibilities to implement the teleparallel 
connection as a  dynamical field. In the original formulation by Einstein~\cite{Einstein1928}, the only fundamental dynamical field is a tetrad 
(or vielbein) field \(e^A = e^A{}_{\mu}{\rm d}x^{\mu}\). Here and in the remainder of 
this section, capital Latin letters \(A, B = 0, \ldots, 3\) denote Lorentz 
indices. The tetrad defines both the metric
\begin{equation}\label{eq:metric}
g_{\mu\nu} = \eta_{AB}e^A{}_{\mu}e^B{}_{\nu}
\end{equation}
and the coefficients
\begin{equation}\label{eq:weitz}
\tp{\Gamma}^{\mu}{}_{\nu\rho} = e_A{}^{\mu}\partial_{\rho}e^A{}_{\nu}
\end{equation}
of the teleparallel affine connection, where \(\eta_{AB} = \mathrm{diag}(-1, 1, 
1, 1)\)  is the Minkowski metric and \(e_A = e_A{}^{\mu}\partial_{\mu}\) is the 
inverse tetrad satisfying \(e^A{}_{\mu}e_A{}^{\nu} = \delta_{\mu}^{\nu}\) and 
\(e^A{}_{\mu}e_B{}^{\mu} = \delta^A_B\). This particular choice of the 
connection is known as the Weitzenb\"ock connection, and it belongs to a family 
of flat, metric compatible connections. The constituents of this family can be 
expressed in terms of the connection one-forms \(\tp{\omega}^A{}_B = 
\tp{\omega}^A{}_{B\mu}{\rm d}x^{\mu}\) of a flat Lorentz spin connection via the 
relation
\begin{equation}\label{eq:teleconn}
\tp{\Gamma}^{\mu}{}_{\nu\rho} = e_A{}^{\mu}(\partial_{\rho}e^A{}_{\nu}  + 
\tp{\omega}^A{}_{B\rho}e^B{}_{\nu})\,.
\end{equation}
Here, flatness corresponds to vanishing curvature of the spin connection,
\begin{equation}\label{eq:telespincurv}
\tp{R}^A{}_{B\alpha\beta} = \partial_{\alpha}\tp{\omega}^A{}_{B\beta} -  
\partial_{\beta}\tp{\omega}^A{}_{B\alpha} + 
\tp{\omega}^A{}_{C\alpha}\tp{\omega}^C{}_{B\beta} - 
\tp{\omega}^A{}_{C\beta}\tp{\omega}^C{}_{B\alpha} \equiv 0\,,
\end{equation}
while metric compatibility follows from the antisymmetry 
\(\tp{\omega}^{(AB)}{}_{\mu} \equiv 0\).  In the covariant formulation of 
teleparallel gravity theories~\cite{Krssak:2015oua}, the spin connection is 
promoted to a dynamical field, and its flatness must be imposed either through 
Lagrange multipliers in the gravitational action or by explicitly allowing only 
for flat connections and accordingly restricting the variation with respect to 
the spin connection in the derivation of the Euler-Lagrange 
equations~\cite{Golovnev:2017dox}. Yet another possibility to ensure the 
flatness of the spin connection is to consider it as a dependent quantity 
\(\tp{\omega}^A{}_{B\mu} = \Lambda^A{}_C\partial_{\mu}\Lambda_B{}^C\) derived 
from a local Lorentz transformation \(\Lambda^A{}_B\), and to promote the latter 
to a fundamental dynamical field next to the 
tetrad~\cite{Blixt:2018znp,Blixt:2019mkt}.

Another implementation of teleparallel geometry is the Palatini approach, which 
considers as  fundamental fields the metric \(g_{\mu\nu}\) and the affine 
connection coefficients \(\tp{\Gamma}^{\rho}{}_{\mu\nu}\), and enforces metric 
compatibility and flatness of the connection via Lagrange 
multipliers~\cite{BeltranJimenez:2018vdo}.

\subsection{Translation Gauge Theory}\label{ssec:transgauge}\label{gaugetheref2}

One argument that is commonly mentioned in favour of Teleparallel Gravity is its 
 possible interpretation as a gauge theory of translations. Various approaches 
and realisations of this gauge theory have been 
studied~\cite{Blagojevic:2013xpa}. It was first found in the non-covariant 
formulation of Teleparallel Gravity that the tetrad allows for a gauge symmetry, 
which can be related to infinitesimal translations, and that imposing this 
symmetry leads to an action equivalent to the Einstein-Hilbert 
action~\cite{Hayashi:1967se,Cho:1975dh}. Relaxing the condition of local Lorentz 
invariance yields a more general class of theories~\cite{Hayashi:1979qx}. A more 
sophisticated approach relates Teleparallel Gravity to higher gauge theory, and 
generalisations of Cartan geometry~\cite{Baez:2012bn}. Cartan geometry has also 
been suggested as a possible interpretation of the non-standard nature of 
translational gauge transformations, which act not only on an internal space, is 
usual in gauge theory, but also on the underlying spacetime 
manifold~\cite{Fontanini:2018krt}. This interpretation has been contrasted with 
a formulation making use of a principal bundle of 
translations~\cite{Pereira:2019woq}. Despite their differences, the 
aforementioned approaches have in common that the tetrad field is related to a 
gauge potential of translations, while the spin connection is related to an 
external Lorentz gauge symmetry. However, for a more fundamental understanding of 
how the tetrad and spin connection in the covariant formulation of Teleparallel 
Gravity arise, it is helpful to take a step backwards and view Teleparallel 
Gravity in the more general context of Poincar\'e gauge theory, and even more 
generally in the context of metric-affine 
gravity~\cite{Blagojevic:2002du,Obukhov:2002tm}. We therefore briefly review 
their description in terms of gauge connections.

We denote by \(FM\) the general linear frame bundle of the spacetime manifold  
\(M\). The fiber \(F_xM\) at a point \(x \in M\) is the set of all frames, i.e., 
ordered bases of the tangent space \(T_xM\). Any frame can be expressed by a 
bijective linear map \(f: \mathbb{R}^4 \to T_xM\). Given coordinates 
\((x^{\mu})\) on \(M\), we can introduce coordinates \((x^{\mu}, f_A{}^{\mu})\) 
on \(FM\), where the frame \(f\) maps an element \(v \in \mathbb{R}^4\) with 
components \(v^A\) to \(v^Af_A{}^{\mu}\partial_{\mu} \in T_xM\). Note that 
\(FM\) is a principal bundle: the group \(H_0 = \mathrm{GL}(4, \mathbb{R})\) 
acts from the right on the fibres of \(FM\). Writing the matrix components of an 
element \(\Lambda \in H_0\) as \(\Lambda^A{}_B\), the action reads
\begin{equation}
(x^{\mu}, f_A{}^{\mu}) \mapsto (x^{\mu}, f_A{}^{\mu}) \cdot \Lambda = (x^{\mu},  
f_B{}^{\mu}\Lambda^B{}_A)
\end{equation}
in the coordinates we used.

A frame allows expressing tensor fields in a basis that is different from the  
coordinate basis. Changing the frame changes the basis, while keeping the point 
fixed at which the tensor field is evaluated. In order to incorporate  
translations as well, one considers a larger bundle, which may be constructed as 
follows. Let \(G_0 = \mathrm{GA}(4, \mathbb{R}) = \mathbb{R}^4 \rtimes 
\mathrm{GL}(4, \mathbb{R})\) be the general affine group. The group \(H_0\), 
being a subgroup of \(G_0\), acts on \(G_0\) by left multiplication: given 
\(\tilde{\Lambda} \in H_0\) and \((\Lambda, v) \in G_0\), one has
\begin{equation}
\tilde{\Lambda} \cdot (\Lambda, v) = (\tilde{\Lambda}\Lambda, \tilde{\Lambda}v)\,.
\end{equation}
This action gives rise to an associated bundle \(AM = FM \times_{H_0} G_0\), 
which we  call the affine frame bundle. To understand the geometry of \(AM\), 
recall that for each \(x \in M\) the elements of the fiber \(A_xM\) are given by 
equivalence classes
\begin{equation}
[\tilde{f}, (\Lambda, v)] = [\tilde{f} \cdot \tilde{\Lambda}^{-1}, 
\tilde{\Lambda}  \cdot (\Lambda, v)]\,,
\end{equation}
where \(\tilde{f} \in F_xM\) is a frame at \(x\) and \((\Lambda, v) \in G_0\),  
and equivalence is given by the simultaneous action of \(\tilde{\Lambda} \in 
H_0\) on both of these objects, following the equation above. One can see that 
every such equivalence class is defined by a frame \(f = \tilde{f} \cdot 
\Lambda\), as well as a tangent vector \(y = \tilde{f}v\), or in components
\begin{equation}
f_A{}^{\mu} = \tilde{f}_B{}^{\mu}\Lambda^B{}_A \quad \text{and} \quad y^{\mu}  = 
v^A\tilde{f}_A{}^{\mu}\,,
\end{equation}
since these combinations are invariant under the application of 
\(\tilde{\Lambda}\).  Hence, the affine frame bundle has the structure
\begin{equation}
AM = TM \underset{M}{\times} FM\,,
\end{equation}
or in other words, every element of \(AM\) consists of a base point \(x \in M\), 
a  vector \(y \in T_xM\) and a frame \(f \in F_xM\). Coordinates on \(AM\) thus 
take the form \((x^{\mu}, y^{\mu}, f_A{}^{\mu})\). One finds that \(AM\) is a 
principal $G_0$-bundle, where the right action is given by
\begin{equation}
(x^{\mu}, y^{\mu}, f_A{}^{\mu}) \mapsto (x^{\mu}, y^{\mu}, f_A{}^{\mu}) \cdot 
(\Lambda, v) =  (x^{\mu}, y^{\mu} + f_A{}^{\mu}v^A, f_B{}^{\mu}\Lambda^B{}_A)
\end{equation}
for \((\Lambda, v) \in G_0\).

The bundle \(AM\) is the arena for the gauge theory we model. The gauge 
connection is  given by a one-form \(\hat{\mathbf{A}
}\in \Omega^1(AM, 
\mathfrak{g}_0)\) on \(AM\), taking values in the Lie algebra \(\mathfrak{g}_0\) 
of \(G_0\). Since this Lie algebra splits as \(\mathfrak{g}_0 = \mathfrak{h}_0 
\oplus \mathfrak{z}\) into the homogeneous part \(\mathfrak{h}_0\) and 
translations \(\mathfrak{z} \cong \mathbb{R}^4\), the same holds for the 
gauge connection \(\hat{\mathbf{A}} = \hat{\boldsymbol{\omega}} + \mathbf{e}\). Using the 
canonical matrix and vector representations of these Lie algebras, it turns out 
that the most general one-form connection can be written in the form
\begin{subequations}
\begin{align}
\hat{\boldsymbol{\omega}}^A{}_B &= 
f^{-1\,A}{}_{\mu}(f_B{}^{\nu}\hat{\Omega}^{\mu}{}_{\nu\rho}{\rm d}x^{\rho}  + 
{\rm d}f_B{}^{\mu})\,,\label{eq:gaugespicon}\\
\mathbf{e}^A &= f^{-1\,A}{}_{\mu}(E^{\mu}{}_{\nu}{\rm d}x^{\nu} + 
\hat{\Omega}^{\mu}{}_{\nu\rho}y^{\nu}{\rm d}x^{\rho}  + {\rm d}y^{\mu})\label{eq:gaugetet}
\end{align}
\end{subequations}
in our chosen coordinates, where \(E^{\mu}{}_{\nu}\) and 
\(\hat{\Omega}^{\mu}{}_{\nu\rho}\) are functions of  \(x\) only. This form is a 
consequence of the demand that \(\hat{\mathbf{A}}\) is a principal connection.

Finally, to obtain the gauge fields on the spacetime manifold \(M\), we need to 
choose a \emph{gauge};  this corresponds to choosing a section \(\sigma: M \to 
AM\). In our coordinates this section can be expressed as
\begin{equation}
\sigma: x \mapsto (x^{\mu}, \sigma^{\mu}, \sigma_A{}^{\mu})\,.
\end{equation}
The section allows us to take the pullbacks
\begin{subequations}
\begin{align}
\sigma^*\hat{\boldsymbol{\omega}}^A{}_B &= \hat{\omega}^A{}_{B\mu}{\rm d}x^{\mu} =  
\sigma^{-1\,A}{}_{\mu}(\sigma_B{}^{\nu}\hat{\Omega}^{\mu}{}_{\nu\rho} + 
\partial_{\rho}\sigma_B{}^{\mu}){\rm d}x^{\rho}\,,\label{eq:pullspicon}\\
\sigma^*\mathbf{e}^A &= e^A{}_{\mu}{\rm d}x^{\mu} = 
\sigma^{-1\,A}{}_{\mu}(E^{\mu}{}_{\rho} +  \hat{\Omega}^{\mu}{}_{\nu\rho}\sigma^{\nu} 
+ \partial_{\rho}\sigma^{\mu}){\rm d}x^{\rho}\,.\label{eq:pulltet}
\end{align}
\end{subequations}
Here we have already suggestively identified these pullbacks with the tetrad 
\(e^A{}_{\mu}\)  and teleparallel spin connection \(\hat{\omega}^A{}_{B\mu}\). To 
justify this identification and to resort to more familiar notation, we 
introduce new coordinates
\begin{equation}
(x^{\mu}, y^{\mu}, f_A{}^{\mu}) \mapsto (x^{\mu}, \xi^A, f_A{}^{\mu}) = 
(x^{\mu},  f^{-1\,A}{}_{\mu}y^{\mu}, f_A{}^{\mu})\,,
\end{equation}
and accordingly replace \(\sigma^{\mu} = \sigma^A\sigma_A{}^{\mu}\). Then  the 
tetrad~\eqref{eq:pulltet} becomes
\begin{equation}
\sigma^*\mathbf{e}^A = e^A{}_{\mu}{\rm d}x^{\mu} = (B^A{}_{\rho}  + 
\hat{\omega}^A{}_{B\rho}\sigma^B + \partial_{\rho}\sigma^A){\rm d}x^{\rho}\,,
\end{equation}
where \(B^A{}_{\rho} = \sigma^{-1\,A}{}_{\mu}E^{\mu}{}_{\rho}\).  This is the 
form most commonly encountered in the literature~\cite{Aldrovandi:2013wha}. Note 
that changing the section \(\sigma^A{}_{\mu}\) simply corresponds to a change of 
the Lorentz frame, while a change of \(\sigma^A\) can be interpreted as a 
translation gauge transformation. Finally, calculating the torsion yields
\begin{equation}
\hat{T}^A{}_{\mu\nu}{\rm d}x^{\mu} \wedge {\rm d}x^{\nu} = {\rm d}\mathbf{e}^A + \hat{\omega}^A{}_B \wedge \mathbf{e}^B =  
(\partial_{\mu}B^A{}_{\nu} + \hat{\omega}^A{}_{B\mu}B^B{}_{\nu} + 
\partial_{\mu}\hat{\omega}^A{}_{B\nu}\sigma^B + 
\hat{\omega}^A{}_{C\mu}\hat{\omega}^C{}_{B\nu}\sigma^B){\rm d}x^{\mu} \wedge {\rm d}x^{\nu}\,.
\end{equation}
In the teleparallel geometry the spin connection has vanishing 
curvature~\eqref{eq:telespincurv},  and so the two terms involving \(\sigma^B\) 
vanish. The torsion then becomes the field strength of the translation gauge 
potential \(B^A{}_{\mu}\).

\subsection{Local Lorentz Invariance}\label{ssec:lorlorinv}
\label{telecovarref1}
\label{loclinref1}

When the concept of teleparallelism was introduced, the only dynamical field was 
the  tetrad \(e^A{}_{\mu}\), which has 16 components. It was believed that 
gravity is described by the 10 components of the metric, while the additional six
components could be attributed to the electromagnetic field 
strength~\cite{Einstein1928}. However, this turned out not to be the case, and 
it was realised that these additional components are related to local Lorentz 
transformations \(e^A{}_{\mu} \mapsto \Lambda^A{}_Be^B{}_{\mu}\) instead. 
Further, it was found that in general Teleparallel gravity theories are not 
invariant under such local Lorentz transformations, due to the presence of the 
Weitzenb\"ock connection \(\tp{\Gamma}^{\rho}{}_{\mu\nu} = 
e_A{}^{\rho}\partial_{\nu}e^A{}_{\mu}\): in order to solve the field equations 
of such theories, one cannot choose an arbitrary tetrad corresponding to a 
particular metric, but only specific tetrads are 
allowed~\cite{Li:2010cg,Sotiriou:2010mv,Ferraro:2011us,Tamanini:2012hg,
Tamanini:2013xya}. This means that the extra components present in the tetrad 
cannot be regarded as pure gauge degrees of freedom. This fact raised a debate 
regarding the nature of the additional degrees of freedom, whether they might be 
acausal or even superluminal, or whether some of them can be absorbed by a 
remnant gauge symmetry that is still present despite the otherwise broken 
Lorentz 
symmetry~\cite{Kopczynski_1982,Cheng:1988zg,Nester_1988,Chen:1998ad,Ong:2013qja,
Ferraro:2014owa,Chen:2014qtl,Bejarano:2019fii,Ferraro:2020tqk}.

In order to resolve the aforementioned issues, a covariant formulation of 
Teleparallel  Gravity theories was developed, which features the flat Lorentz 
spin connection \(\tp{\omega}^A{}_{B\mu}\) as an additional dynamical 
field~\cite{Krssak:2015oua,Golovnev:2017dox,Hohmann:2018rwf,Krssak:2018ywd}. In 
the covariant formulation, local Lorentz transformations take the form
\begin{equation}\label{eq:loclortrans}
e^A{}_{\mu} \mapsto \Lambda^A{}_Be^B{}_{\mu}\,, \quad
\tp{\omega}^A{}_{B\mu} \mapsto 
\Lambda^A{}_C(\Lambda^{-1})^D{}_B\tp{\omega}^C{}_{D\mu} +  
\Lambda^A{}_C\partial_{\mu}(\Lambda^{-1})^C{}_B\,,
\end{equation}
and thus act on both the tetrad and the spin connection. The teleparallel 
connection~\eqref{eq:teleconn},  and hence also its torsion~\eqref{eq:torsion}, 
are invariant under this combined transformation. It thus follows that any 
action constructed from the (in any case invariant) metric, the torsion and 
their derivatives are locally Lorentz invariant.

Note that the spin connection~\eqref{eq:pullspicon} also naturally arises in the 
gauge  theory approach shown in section~\ref{ssec:transgauge}. In this picture, 
local Lorentz transformations are simply transformations of the Lorentz part 
\(\sigma^A{}_{\mu}\) of the section defining the gauge.

\subsection{Matter Coupling}\label{ssec:matter}
An important issue in Teleparallel Gravity is the question about how to couple matter 
to the teleparallel  geometry. The most commonly considered procedure is given 
by the minimal coupling prescription, according to which the metric 
\(g_{\mu\nu}\) in the matter action is chosen to be the metric~\eqref{eq:metric} 
obtained from the tetrad. This prescription is sufficient for bosonic fields, 
which couple to the metric only, without direct coupling to a spin connection. 
This differs from the case of fermions, where a spin connection must be 
specified. The question about a consistent coupling of fermions and the proper 
choice of the spin connection is a highly 
topic~\cite{Obukhov:2002tm,Maluf:2003fs,Mielke:2004gg,Obukhov:2004hv,
Formiga:2013dja,BeltranJimenez:2020sih}. We will not enter this debate here, and assume that fermions 
couple to the metric geometry through the spin connection associated with the 
Levi-Civita connection only. Following this assumption, the matter action does 
not depend on the teleparallel spin connection, but only on the tetrad. For the 
variation with respect to the tetrad we may write
\begin{equation}\label{eqn:enmomtens}
\delta_e{S}_{\text{matter}} = -\int {\rm d}^4x \,  eT_A{}^{\mu}\delta 
e^A{}_{\mu}\,.
\end{equation}
The quantity \(T_A{}^{\mu}\) introduced here is the matter energy-momentum 
tensor, which we also  write in the form \(T_{\mu\nu} = 
e^A{}_{\mu}g_{\mu\nu}T_A{}^{\rho}\)
\footnote{While $T_{\mu\nu}$ refers to the 
energy-momentum tensor, $\udt{\tp{T}}{\rho}{\mu\nu}$ refers to the torsion 
tensor which is not related.}.
Demanding that the matter action is invariant 
under local Lorentz transformations, which is equivalent to demanding that it 
depends only on the metric obtained from the tetrad, then implies that the 
energy-momentum tensor is symmetric, \(T_{[\mu\nu]} = 0\)~\cite{Obukhov:2006ge}.

\subsection{Teleparallel Equivalent of General Relativity 
(TEGR)}\label{ssec:tegr}
\label{TEGRref1}
An interesting feature of Teleparallel Gravity is that it allows for an 
alternative  formulation of General Relativity, in which gravity is mediated by 
torsion instead of curvature~\cite{Maluf:2013gaa}. One possibility to derive the 
action for the Teleparallel Equivalent of General Relativity (TEGR) is by 
starting from the Einstein-Hilbert action
\begin{equation}
{S}_{\text{GR}} = \frac{1}{2\kappa^2}\int {\rm d}^4x\,\sqrt{-g}\,R\,,
\end{equation}
where $\kappa^2=8\pi G$, and by rewriting the Ricci scalar in terms of the 
torsion of  the teleparallel connection. For this purpose, we introduce the 
contortion tensor
\begin{equation}\label{eq:contortion}
\tp{K}^{\rho}{}_{\mu\nu} = \tp{\Gamma}^{\rho}{}_{\mu\nu} - 
\Gamma^{\rho}{}_{\mu\nu} =  \frac{1}{2}\left(\tp{T}_{\mu}{}^{\rho}{}_{\nu} + 
\tp{T}_{\nu}{}^{\rho}{}_{\mu} - \tp{T}^{\rho}{}_{\mu\nu}\right)
\end{equation}
as the difference between the teleparallel and Levi-Civita connection 
coefficients. By  making use of this relation, one can write the Riemann 
curvature tensor of the Levi-Civita connection in the form
\begin{equation}\label{eq:lccurv}
R^{\mu}{}_{\nu\rho\sigma} = \tp{R}^{\mu}{}_{\nu\rho\sigma} -  
\nabla_{\rho}\tp{K}^{\mu}{}_{\nu\sigma} + 
\nabla_{\sigma}\tp{K}^{\mu}{}_{\nu\rho} - 
\tp{K}^{\mu}{}_{\tau\rho}\tp{K}^{\tau}{}_{\nu\sigma} + 
\tp{K}^{\mu}{}_{\tau\sigma}\tp{K}^{\tau}{}_{\nu\rho}\,,
\end{equation}
keeping in mind that the curvature of the teleparallel connection vanishes,  
\(\tp{R}^{\mu}{}_{\nu\rho\sigma} \equiv 0\), while the curvature of the 
Levi-Civita connection does not vanish in general. Taking the appropriate 
contractions, and using the antisymmetry \(\tp{K}^{(\mu\nu)\rho} = 0\), the 
Ricci scalar reads
\begin{equation}\label{eq:ricstor}
R = -2\nabla_{\mu}\tp{K}^{\mu\nu}{}_{\nu} + 
\tp{K}^{\rho\mu}{}_{\mu}\tp{K}_{\rho\nu}{}^{\nu} -  
\tp{K}^{\rho\mu\nu}\tp{K}_{\rho\nu\mu} = -\mathbb{T} + 
2\nabla_{\mu}\tp{T}_{\nu}{}^{\nu\mu}  = - \mathbb{T} + B\,.
\end{equation}
Here we introduced the torsion scalar \(\mathbb{T}\), which can be written in the form
\begin{equation}\label{eq:torsscal}
\mathbb{T} = \frac{1}{2}\tp{T}^{\rho}{}_{\mu\nu}\tp{S}_{\rho}{}^{\mu\nu} =  
\frac{1}{4}\tp{T}^{\mu\nu\rho}\tp{T}_{\mu\nu\rho} + 
\frac{1}{2}\tp{T}^{\mu\nu\rho}\tp{T}_{\rho\nu\mu} - 
\tp{T}^{\mu}{}_{\mu\rho}\tp{T}_{\nu}{}^{\nu\rho}\,,
\end{equation}
where we used the superpotential
\begin{equation}\label{eq:suppot}
\tp{S}_{\rho}{}^{\mu\nu} = \tp{K}^{\mu\nu}{}_{\rho} -  
\delta_{\rho}^{\mu}\tp{T}_{\sigma}{}^{\sigma\nu} + 
\delta_{\rho}^{\nu}\tp{T}_{\sigma}{}^{\sigma\mu}\,.
\end{equation}
Finally, the last term in the expression~\eqref{eq:ricstor}  
is a total divergence, and thus appears as a boundary term in the action, which 
does not contribute to the field equations. Omitting this term and using the 
fact that the determinants of the metric and the tetrad are related by
\begin{equation}
e = \sqrt{-g}\,,
\end{equation}
the TEGR action finally reads
\begin{equation}\label{eq:tegraction}
{S}_{\text{TEGR}} = -\frac{1}{2\kappa ^2}\int {\rm d}^4x\, e\,\mathbb{T}\,.
\end{equation}
Variation of this action, together with a matter action, with respect to the 
tetrad,  and transforming the resulting Lorentz index into a spacetime index, 
yields the field equations
\begin{equation}
\nabla_{\rho}\tp{S}_{(\mu\nu)}{}^{\rho} - 
\frac{1}{2}\tp{S}_{(\mu}{}^{\rho\sigma}\tp{T}_{\nu)\rho\sigma}  + 
\frac{1}{2}\mathbb{T}g_{\mu\nu} = \kappa^2T_{\mu\nu}\,.
\end{equation}
The right-hand side is given by the energy-momentum 
tensor~\eqref{eqn:enmomtens}.  The left-hand side of these field equations is 
most easily understood by realising the geometric identity
\begin{equation}\label{eqn:stor}
\nabla_{\rho}\tp{S}_{(\mu\nu)}{}^{\rho} - 
\frac{1}{2}\tp{S}_{(\mu}{}^{\rho\sigma}\tp{T}_{\nu)\rho\sigma}  + 
\frac{1}{2}\mathbb{T}g_{\mu\nu} = R_{\mu\nu} - 
\frac{1}{2}Rg_{\mu\nu}=G_{\mu\nu}\,,
\end{equation}
and so it turns out that the field equations indeed agree with those  of General 
Relativity, as one would expect. An interesting property of the 
action~\eqref{eq:tegraction} is the fact that the teleparallel spin connection 
\(\tp{\omega}^A{}_{B\mu}\) enters only in the form of a boundary 
term~\cite{Krssak:2015lba}. This can be seen from the 
relation~\eqref{eq:ricstor} between the Ricci scalar of the Levi-Civita 
connection and the torsion scalar. The left-hand side is independent of the 
teleparallel spin connection, and so its variation \(\delta_{\omega}R\) 
vanishes. This implies
\begin{equation}
\delta_{\omega}\mathbb{T} = 2\delta_{\omega}\nabla_{\mu}\tp{T}_{\nu}{}^{\nu\mu}  
= 2\nabla_{\mu}\delta_{\omega}\tp{T}_{\nu}{}^{\nu\mu}\,,
\end{equation}
so that variation of the TEGR action with respect to the teleparallel spin 
connection yields a  boundary term only. One consequence is that the 
teleparallel spin connection drops out of the field equations, as can be seen 
from the identity~\eqref{eq:ricstor}. From this, further follows that the theory 
is invariant not only under the local Lorentz 
transformations~\eqref{eq:loclortrans}, but also under the pure tetrad 
transformations \(e^A{}_{\mu} \mapsto \Lambda^A{}_Be^B{}_{\mu}\). This 
invariance, which does not hold for general teleparallel theories, as discussed 
in the following section, is one reason why Teleparallel Gravity was originally 
developed without appealing to a non-trivial spin connection, and the viability 
of the covariant formulation of TEGR has been challenged~\cite{Maluf:2018coz}.

\section{Teleparallel Gravity Extensions}\label{sec:TeleExt}
\label{fTgraref2}

Even if $\Lambda$CDM has passed most of the observational tests with flying 
colours,  General Relativity (GR), and thus the Teleparallel Equivalent of GR 
(TEGR) as well, contain some shortcomings. Motivated mostly by the accelerated 
expansion of the Universe, many people initiated the hunt for a modification of 
gravity. In the same way as in the curvature case, the literature abounds 
different theories of Teleparallel Gravity, that introduce new degrees of 
freedom to describe several phenomena.

Either by introducing scalar fields to the TEGR Lagrangian, or by generalising  
it to an arbitrary function of the torsion scalar, i.e. $f(\mathbb{T})$, by 
introducing non-localities, non-linear boundary terms and other topological 
invariants, there has been a great plethora of models studied. In this section we 
will give a brief review on the most well-known modifications.

\subsection{\texorpdfstring{$f(\mathbb{T})$}{f(T)} Gravity}\label{f_T_sec}

In the curvature case, the most straightforward modification, and maybe the 
simplest one,  is the so-called $f(R)$ gravity. As the name witnesses, it is a 
generalisation of the Einstein-Hilbert action to an arbitrary function of the 
Ricci scalar that offers richer phenomenology 
\cite{Sotiriou:2008rp,Nojiri:2006ri,DeFelice:2010aj,Nojiri:2010wj} by 
introducing a new scalar degree of freedom 
\cite{Myung:2016zdl,Berry:2011pb,Capozziello:2008rq}. In the same spirit, 
$f(\mathbb{T})$ theory was proposed almost a decade ago 
\cite{Bengochea:2008gz,Ferraro:2008ey,Ferraro:2006jd,Ferraro:2011us} and its 
action reads
\begin{equation}\label{f(T)action}
\mathcal{S}_{f(\mathbb{T})} = \frac{1}{2\kappa^2} \int {\rm d}^4x\, e f(\mathbb{T})  + 
\mathcal{S}_{\text{matter}}\,.
\end{equation}
It is worth mentioning that even though at the level of field equations GR and 
TEGR are totally equivalent theories,  the same does not happen for $f(R)$ and 
$f(\mathbb{T})$ theories. As already discussed in the previous section, $R = - 
\mathbb{T} + B$ \eqref{eq:ricstor}, where $B = -2 \partial _{\alpha} (e 
\tp{T}^{\mu\alpha}{}_{\mu})/e$ is a boundary term, that in TEGR does not contribute 
to the field equations. However, the arbitrary function $f(\mathbb{T})$ is 
non-linear and thus the two theories are no longer equivalent.

Varying the action \eqref{f(T)action} with respect to the tetrad $e^A{}_{\mu}$ 
we  get its field equations that read
\begin{equation}\label{f(T)field_equations}
f_\mathbb{T} \partial _{\nu}\left(e \tp{S}_A{}^{\mu\nu}\right) + e 
\left(f_{\mathbb{T}\mathbb{T}} \tp{S}_A{}^{\mu\nu}\partial _{\nu} \mathbb{T} - 
f_\mathbb{T} \tp{T}^B{}_{\nu A}\tp{S}_B{}^{\nu\mu} + f_\mathbb{T} \tp{\omega} ^B{}_{A\nu} 
\tp{S}_B{}^{\nu\mu} + \frac{1}{2}f e_A {}^{\mu} \right) =  \kappa ^2 e T _A 
{}^{\mu}\,,
\end{equation}
with $f_\mathbb{T}$ and $f_{\mathbb{T}\mathbb{T}}$ being respectively the first, 
and second-order derivatives of $f$ with respect to $\mathbb{T}$ and 
$T_{\mu}{}^{\nu} = e^A{}_{\mu}T_A{}^{\nu}$ is the energy-momentum tensor of the 
matter fields. As one can immediately notice, the equations for the tetrad are of 
second order, in contrast with those in $f(R)$ gravity.
For the unimodular formulation of $f(\mathbb{T})$ gravity, where the determinant  
of the tetrad is kept constant, one can check \cite{Nassur:2016yhc}.

Variation of \eqref{f(T)action} with respect to the spin connection  
\cite{Krssak:2017nlv,Golovnev:2017dox,Hohmann:2017duq} gives the antisymmetric 
part of the tetrad equations \eqref{f(T)field_equations}, meaning
\begin{equation}
    f_{\mathbb{T}\mathbb{T}}\tp{S}_{[AB]}{}^{\nu}\tp{T}_{\nu} = 0
\end{equation}

There was a period when it was believed that $f(\mathbb{T})$ gravity violates local Lorentz  invariance \cite{Sotiriou:2010mv,Li:2010cg}. Indeed, if 
one considers the theory with the tetrad being the only variable, the discussion 
is constrained on a very specific class of frames where the spin connection 
vanishes. That is why the existence of \textit{good} and \textit{bad tetrads} 
was proposed \cite{Tamanini:2012hg}, referring to tetrads in the same 
equivalence class that solve and do not solve respectively the field equations. 
An illuminating example is that the diagonal tetrad in spherical symmetry
\begin{equation}
    e^A{}_{\mu} = {\rm diag}\left(A,B,r,r\sin  \theta \right)\,,
\end{equation}
corresponding to the metric $g_{\mu\nu} = {\rm 
diag}\left[A^2(r).B^2(r),r^2,r^2\sin^2\theta  \right]$, is a bad tetrad, not 
satisfying the field equations \eqref{f(T)field_equations} with a vanishing spin 
connection for $f_{\mathbb{T}\mathbb{T}}\neq0$, while the non-diagonal tetrad 
associated to the same metric
\begin{equation}
    e^A{}_{\mu} = \begin{pmatrix}
    A & 0 & 0 & 0\\
    0 & B \cos \phi\sin\theta & r \cos \phi \cos \theta &   -r \sin \phi \sin \theta \\
    0 & -B \cos \theta & r \sin \theta & 0 \\
    0 & B\sin \phi \sin \theta & r \sin \phi \cos \theta & r \cos \phi \sin \theta
    \end{pmatrix}\,,
\end{equation}
is a good tetrad. The problem was resolved when the covariant formulation  of 
$f(\mathbb{T})$ gravity was proposed \cite{Krssak:2015oua}, where both the tetrad 
$e^A{}_{\mu}$ and the spin connection $\tp{\omega} ^A{}_{B\mu}$ are determined by the 
field equations \eqref{f(T)field_equations}. For a more detailed study about tetrads in spherical symmetry in Teleparallel theories, see~\cite{Hohmann:2019nat}.

\subsection{New General Relativity and Extensions}

This modification of TEGR is the first one, and it was proposed by Hayashi  and 
Shirafuji in \cite{Hayashi:1979qx}.
The torsion tensor can be decomposed into its three irreducible parts as
\begin{equation}
\tp{T}_{\lambda\mu\nu} = \frac{2}{3}\left(\tp{t}_{\lambda\mu\nu} - 
\tp{t}_{\lambda\nu\mu}\right) +  
\frac{1}{3}\left(g_{\lambda\mu}\tp{v}_{\nu}-g_{\lambda\nu}\tp{v}_{\mu} \right) + \epsilon 
_{\lambda\mu\nu\rho}\tp{a}^{\rho}\,,
\end{equation}
where $\tp{v}_{\mu} = \tp{T}^{\lambda}{}_{\lambda\mu},$ $\tp{a}_{\mu} = \frac{1}{6}\epsilon 
_{\mu\nu\sigma\rho}\tp{T}^{\nu\sigma\rho},$  and $\tp{t}_{\lambda\mu\nu} = 
\frac{1}{2}(\tp{T}_{\lambda\mu\nu}+\tp{T}_{\mu\lambda\nu})+\frac{1}{6}(g_{\nu\lambda}\tp{v}_{
\mu}+g_{\nu\mu}\tp{v}_{\lambda})-\frac{1}{3}g_{\lambda\mu}\tp{v}_{\nu}.$

Contracting these components, one can construct the following scalars up to  
quadratic order
\begin{equation}\label{irreducible_scalars}
T_{\rm ten} = \tp{t}_{\lambda\mu\nu}\tp{t}^{\lambda\mu\nu} \,,\,\,T_{\rm ax}  = 
\tp{a}_{\mu}\tp{a}^{\mu} \,,\,\, T_{\rm vec} =\tp{ v}_{\mu}\tp{v}^{\mu}\,.
\end{equation}
The torsion scalar $\mathbb{T}$ is equal to
\begin{equation}
    \mathbb{T} = \frac{3}{2}T_{\rm ax} + \frac{2}{3}T_{\rm ten} -  \frac{2}{3} 
T_{\rm vec}\,,
\end{equation}
and an immediate generalisation of this, with arbitrary coefficients,  is the 
action of the New General Relavitiy, i.e.
\begin{equation}\label{NGRaction}
    \mathcal{S}_{\rm NGR} = \frac{1}{2 \kappa ^2} \int {\rm d}^4x e \left(c_1 T_{\rm 
ax}  + c_2 T_{\rm vec} + c_3 T_{\rm ten} \right)\,.
\end{equation}
For completeness we should mention that there are two more quadratic scalars  
that one can construct from the torsion tensor,
\begin{equation}
    P_1 = \tp{v}_{\mu}\tp{a}^{\mu} \quad \text{and} \quad P_2 = \epsilon _{\mu\nu\rho 
\sigma}  \tp{t}^{\lambda\mu\nu}\tp{t}_{\lambda}{}^{\rho\sigma}\,.
\end{equation}
However, both of them are parity violating and we do not consider them here. For 
more  details one can check Ref.~\cite{MuellerHoissen:1983vc}, where the authors 
argue that the parity violating sector could play a crucial role for the 
well-posedness of the Cauchy problem. It is not clear whether these parity  
\label{parityef2}
violating scalars can play a fully consistent role in gravitational theories.

The theory \eqref{NGRaction} has been studied and constrained since its proposal 
 in Refs~\cite{1982JPhA...15..493K,1988CQGra...5.1003N,Cheng:1988zg}. In 
greater detail, Solar System tests were studied \cite{Hayashi:1979qx}, 
singularities of Schwarzschild-like spacetimes in \cite{Kawai:1989qt} , axially 
symmetric solutions in \cite{Fukui:1981si} and its weak-field limit in 
\cite{Fukui:1984rj}. More recently, in \cite{Hohmann:2018jso} the propagation of 
gravitational waves was studied, as well as its Hamiltonian analysis was 
considered in \cite{Okolow:2011np,Nester:2017wau,Blixt:2018znp}. Finally, the 
linearized theory was studied in \cite{ortin2004gravity}, were the author shows 
that in order for the theory to be viable, the 2-form field has to feature a 
gauge symmetry so that it describes a massless Kalb-Ramond field. However, cubic 
order interactions show~\cite{Jimenez:2019tkx} that the above gauge symmetry of 
the 2-form is not preserved at higher orders.

Beyond the NGR theory, extensions of different kinds were considered as well.  In~\cite{Maluf:2011kf} it was found that is it possible to construct a conformally invariant theory considering a quadratic Lagrangian based on the tensorial and axial part of torsion, which has the following form, 
\begin{equation}\label{Conformalaction}
    \mathcal{S}_{\rm Conformal-TG} = \frac{1}{2 \kappa ^2} \int {\rm d}^4x e \left( \frac{3}{2}T_{\rm ax}+\frac{2}{3}T_{\rm ten} \right)^2\,.
\end{equation}
Note that the vectorial part of the torsion tensor does not appear in this theory.
Specifically, in \cite{Koivisto:2018loq} a generalisation of NGR by nine 
functions of the d'Alembertian operator was considered and the authors show that 
it can accommodate the ghost- and singularity-free structure that was realised 
in the metric theories \cite{Conroy:2017yln,Heisenberg:2018vsk,Biswas:2011ar}. 
In addition, in the same spirit with $f(\mathbb{T})$, a straightforward 
generalisation of NGR is the $f(T_{\rm ax},T_{\rm vec},T_{\rm ten})$ theory 
proposed in \cite{Bahamonde:2017wwk}. Apart from richer phenomenology compared 
to $f(\mathbb{T})$ gravity, this theory provides us with the ability to study 
conformal transformations of teleparallel theories in a simple way. In 
particular, the quadratic scalars constructed by the irreducible parts of the 
torsion tensor \eqref{irreducible_scalars} transform under the conformal 
transformation of the tetrad $\tilde{e}^A{}_{\mu} = \Omega e^A{}_{\mu}$ (or of 
the metric $\tilde{g}_{\mu\nu} = \Omega ^2 g_{\mu\nu}$), with $\Omega$ being the 
conformal factor, as
\begin{equation}
    T_{\rm ax} = \Omega ^2 \tilde{T}_{\rm ax} \,,\quad T_{\rm ten} =  \Omega ^2 
\tilde{T}_{\rm ten} \,,\quad T_{\rm vec} = \Omega ^2 \tilde{T}_{\rm vec} + 6 
\Omega v^{\mu}\tilde{\partial} _{\mu} \Omega +   9 
\tilde{g}^{\mu\nu}\tilde{\partial} _{\mu}\Omega\tilde{\partial} _{\nu}\Omega\,.
\end{equation}
Obviously, unlike the $f(R)$ case \cite{Sotiriou:2008rp}, in modified 
teleparallel theories  there cannot be an Einstein frame because of the way 
$T_{\rm vec}$ transforms. More studies on conformal transformations in the 
teleparallel framework can be found in Refs.~\cite{Bamba:2013jqa,Yang:2010ji}.

\subsection{Higher-order Derivatives,   
\texorpdfstring{$f(\mathbb{T},B,T_{\mathcal{G}},B_{\mathcal{G}})$}{f(T,B,T,G,B)}
}
\label{higherordtorref1}
In the $f(\mathbb{T})$ theory only the torsion scalar takes part in the action, 
and  since it contains only first derivatives of the tetrads, the resulting 
equations are of second order. This is not a necessity though (if you can ignore 
or screen out ghosts), and thus the field equations can be of higher order as 
well; as is in $f(R)$.

One of the most well-studied theories include, in addition to the torsion 
scalar, the  boundary term $B = -2 \partial_{\alpha}(e\tp{T}^{\mu\alpha}{}_{\mu})/e$. 
The action of this theory 
\cite{Bahamonde:2015zma,Bahamonde:2016cul,Bahamonde:2016grb} reads
\begin{equation}
\mathcal{S}_{f(\mathbb{T},B)}= \frac{1}{2\kappa ^2}\int {\rm d}^4x e f(\mathbb{T},B) + 
 \mathcal{S}_{\text{matter}}
\end{equation}
and varying this with respect to the tetrad we take the field equations
\begin{align}
2 e e_A{}^{\nu} \square f_B - 2 e e_A {}^{\mu} \nabla ^{\nu} \nabla_{\mu} f_B &+ 
 e B f_B e_A{}^{\nu} + 4 e \left(\partial _{\mu}f_B + \partial 
_{\mu}f_{\mathbb{T}} \right)\tp{S}_A{}^{\mu\nu} + \nonumber \\
&+4 \partial _{\mu} (e\tp{S}_A{}^{\mu\nu}) f_{\mathbb{T}} - 4 e f_{\mathbb{T}}  
\tp{T}^{\lambda}{}_{\mu A}\tp{S}_{\lambda}{}^{\nu\mu} - e f e_{A}{}^{\lambda} = 2 \kappa e 
T _A{}^{\nu}\,.
\end{align}
It is interesting to notice that such theories are much more general than the 
$f(R)$ theories,  since the last ones are just a subclass when $f(\mathbb{T},B) 
= f(-\mathbb{T}+B) = f(R)\,.$ In addition to these, theories with higher 
derivative terms of the torsion scalar, e.g. $\nabla ^2 \mathbb{T}, \square 
\mathbb{T},$ etc, where also proposed \cite{Otalora:2016dxe}.

Last but not least, theories with higher-order invariants were also considered 
in the  literature. An interesting example is the inclusion of the Gauss-Bonnet 
invariant,
\begin{equation}
    \mathcal{G} = R^2 - 4 R_{\mu\nu}R^{\mu\nu}  + 
R^{\alpha\beta\mu\nu}R_{\alpha\beta\mu\nu}\,.
\end{equation}
Its teleparallel version reads, as with the Ricci scalar,
\begin{equation}
    \mathcal{G} = -T_{\mathcal{G}} + B_{\mathcal{G}}\,,
\end{equation}
where $T_{\mathcal{G}}$ is the teleparallel Gauss-Bonnet term and 
$B_{\mathcal{G}}$  its boundary term. Such theories have an action of the form
\begin{equation}
    \mathcal{S}_{f(\mathbb{T},T_{\mathcal{G}})} = \frac{1}{2  \kappa ^2} \int 
{\rm d}^4x \,e f(\mathbb{T},T_\mathcal{G})\,,
\end{equation}
or even more complicated functions like 
$f(\mathbb{T},B,T_{\mathcal{G}}.B_\mathcal{G})$.  Such theories present some 
interesting features in cosmology 
\cite{Capozziello:2016eaz,Bahamonde:2016kba,Kofinas:2014owa,Kofinas:2014aka}. 
Other theories considering non-minimal couplings between matter and gravity have 
also been considered in Teleparallel Gravity. One example for this is the so-called 
$f(\mathbb{T},T)$ gravity, where $T$ is the trace of the energy-momentum 
tensor~\cite{Harko:2014aja}. This theory is analogous to the famous $f(R,T)$ 
gravity considered in the GR framework~\cite{Harko:2011kv}. Other theories based 
on Lagrangian like $f_1(\mathbb{T})+f_2(\mathbb{T})(1+\lambda \mathcal{L}_{\rm 
m})$, where $\mathcal{L}_{\rm m}$ is the matter Lagrangian density and $\lambda$ 
is a constant, have been studied in~\cite{Harko:2014sja}. Furthermore, a more 
general theory concerning $f(\mathbb{T},B,\mathcal{L}_{\rm m})$ gravity has also 
been studied in order to connect and generalise these kind of 
theories~\cite{Bahamonde:2017ifa}.

\subsection{Teleparallel Non-local Theories}
\label{nonloctorref1}  \label{equivprinref3}
Apart from adding new degrees of freedom, modifications can be done in  the 
foundations of a theory, like violating general covariance, locality, abandoning 
the equivalence principle, etc. In the Riemaniann geometry, the first non-local 
proposal came a bit more than a decade ago, with the introduction of a 
\textit{distortion function} $f(\square ^{-1}R)$ of the inverse d'Alembertian 
operator acting on the Ricci scalar. The action of that theory reads
\begin{equation}
    \mathcal{S}_{\rm non-local} = \frac{1}{2 \kappa ^2} \int {\rm d}^4x \sqrt{-g}  
R(1+f(\square ^{-1}R))\,.
\end{equation}
The argument of $f$ can be expressed using the retarded Green's function $G(x,x')$ as
\begin{equation}
\square ^{-1}F(x) = \int {\rm d}^4x' e(x') F(x')G(x,x')\,.
\end{equation}
The motivation came clearly from high energies, since such terms arise in 
quantum loop  corrections and they are also considered as possible solution to 
the black hole information paradox \cite{Donoghue:1994dn,Giddings:2006sj}. It 
was seen however, that even at larger scales, such non-local terms can unify the 
inflation with the late-time acceleration era and they have been proven 
ghost-free and stable.

Based on that, it was natural to study what the effect of such terms would be in 
the teleparallel  framework. That is why the teleparallel non-local (TNL) theory
\begin{align}
\mathcal{S}_{\rm TNL} &= - \frac{1}{2\kappa ^2}\int {\rm d}^4x e \mathbb{T} + 
\frac{1}{2\kappa ^2}  \int {\rm d}^4x e \mathbb{T}f\left(\square 
^{-1}\mathbb{T}\right) \,. \label{nonlocaltele}
\end{align}
was proposed in \cite{Bahamonde:2017bps}. The authors show that the theory is 
consistent with  cosmological data from SNe Ia + BAO + CC + $H_0$ observations. 
A generalisation of \eqref{nonlocaltele} was proposed in  
\cite{Bahamonde:2017sdo}, where they also introduced the effect of the 
d'Alembertian operator on the boundary term $B$. The action of that theory is
\begin{align}
\mathcal{S}_{\rm TNL} &= - \frac{1}{2\kappa ^2}\int {\rm d}^4x e \mathbb{T} + 
\frac{1}{2\kappa ^2}  \int {\rm d}^4x e \left(\xi \mathbb{T} + \chi B \right) 
f\left(\square ^{-1}\mathbb{T},\square ^{-1}B\right) \,. 
\label{generalizednonlocaltele}
\end{align}
A localised version of this has been studied, introducing scalar fields. In 
addition, using  symmetries, a classification of the distortion function has 
been done.

\subsection{Horndeski Analog and Subclasses}\label{sec:BDLS}\label{teleHornref1}

Horndeski theory is the most general scalar tensor theory  (with a single scalar 
field) that leads to second-order field equations in four dimensions~\cite{Horndeski:1974wa,Kobayashi:2011nu}. Most of 
the modified theories of gravity can be mapped onto its action; from 
Brans-Dicke~\cite{Brans:1961sx}, extended quintessence~\cite{Perrotta:1999am,Copeland:2006wr}, kinetic gravity braiding~\cite{Deffayet:2010qz} to $f(R)$~\cite{Sotiriou:2008rp}. However, there is no physical requirement that forces us 
to use the Levi-Civita connection as the fundamental connection to formulate the 
theory, and thus we have the freedom to build the theory in any affine geometry. 
For this reason, in \cite{Bahamonde:2019shr} S.~Bahamonde, K.~F.~Dialektopoulos 
and J.~Levi Said formulated the BDLS theory, that is the teleparallel analog of 
the Horndeski gravity.

In greater detail, they wanted to build a theory which has the properties: i. 
leads  to second order field equations; ii. the scalar invariants are not parity 
violating; and iii. only quadratic contractions of the torsion tensor are 
included. In that way, BDLS action reads
\begin{equation}\label{BDLS_Action}
    \mathcal{S}_{\text{BDLS}} = \frac{1}{2\kappa^2}\int  {\rm d}^4x\,  e 
\mathcal{L}_{\rm Tele}   +\frac{1}{2\kappa^2} \sum_{i=2}^{5}\int {\rm d}^4x\, 
e\mathcal{L}_{i}\,,
\end{equation}
where
\begin{equation}\label{eq:LTele}
    \mathcal{L}_{\rm Tele} = G_{\rm Tele}\left(\phi, X,  \mathbb{T},T_{\rm 
ax},T_{\rm vec},I_2,J_1,J_3,J_5,J_6,J_8,J_{10}\right)\,,
\end{equation}
and
\begin{align}
\mathcal{L}_2 &:= G_2(\phi,X)\,,\\
\mathcal{L}_3 &:= G_3(\phi,X)\Box\phi\,,\\
\mathcal{L}_4 &:= G_4(\phi,X)\left(-\mathbb{T}+B\right) + G_{4,X}(\phi,X) 
\left[\left(\Box\phi\right)^2 - \phi_{;\mu\nu}\phi^{;\mu\nu}\right]\,,\\
\mathcal{L}_5 &:= G_5(\phi,X)\mathcal{G}_{\mu\nu}\phi^{;\mu\nu} - 
\frac{1}{6}G_{5,X}(\phi,X) \Big[\left(\Box\phi\right)^3 + 
2\dut{\phi}{;\mu}{\nu}\dut{\phi}{;\nu}{\alpha}\dut{\phi}{;\alpha}{\mu} - 
3\phi_{;\mu\nu}\phi^{\mu\nu}\left(\Box\phi\right)\Big]\,.
\end{align}
where comma denotes the partial derivative, semi-colon the covariant derivative 
with respect  to the Levi-Civita connection, and $\square$ the associated 
d'Alembertian operator. The scalars that appear in the \eqref{eq:LTele} are
\begin{gather}
    I_2 = \tp{v}^{\mu}\partial _{\mu}\phi\,,\quad J_1 = \tp{a}^{\mu}\tp{a}^{\nu}\partial 
_{\mu}\phi \partial _{\nu} \phi\,,  \quad J_3 = \tp{v}_{\mu}\tp{t}^{\mu\lambda\nu}\partial 
_{\lambda}\phi \partial _{\nu}\phi\,,\quad J_5 = 
\tp{t}^{\sigma\mu\nu}\tp{t}^{\alpha\beta\gamma}\partial _{\sigma} \phi \partial _{\mu} 
\phi \partial _{\nu} \phi \partial _{\alpha} \phi \partial _{\beta} \phi 
\partial _{\gamma} \phi \,, \\
    J_6= \tp{t}^{\sigma\mu\nu}\tp{t}_{\sigma}{}^{\kappa\lambda}\partial _{\mu} \phi 
\partial _{\nu}  \phi \partial _{\kappa} \phi \partial _{\lambda} \phi \,,\quad 
J_8 = \tp{t}^{\sigma\mu\nu}\tp{t}_{\sigma\mu}{}^{\alpha}\partial _{\nu} \phi\partial 
_{\alpha} \phi\,, \quad J_{10} = \epsilon ^{\mu}{}_{\nu\rho\sigma} 
\tp{a}^{\nu}\tp{t}^{\alpha\sigma\rho}\partial _{\mu} \phi\partial _{\alpha} \phi \,,
\end{gather}
all the rest they are either parity violating or can be obtained combining 
these~\cite{Bahamonde:2019shr,Bahamonde:2019ipm,Hohmann:2019gmt}.

\begin{figure} 
    \centering
    \includegraphics[width=1.02\textwidth]{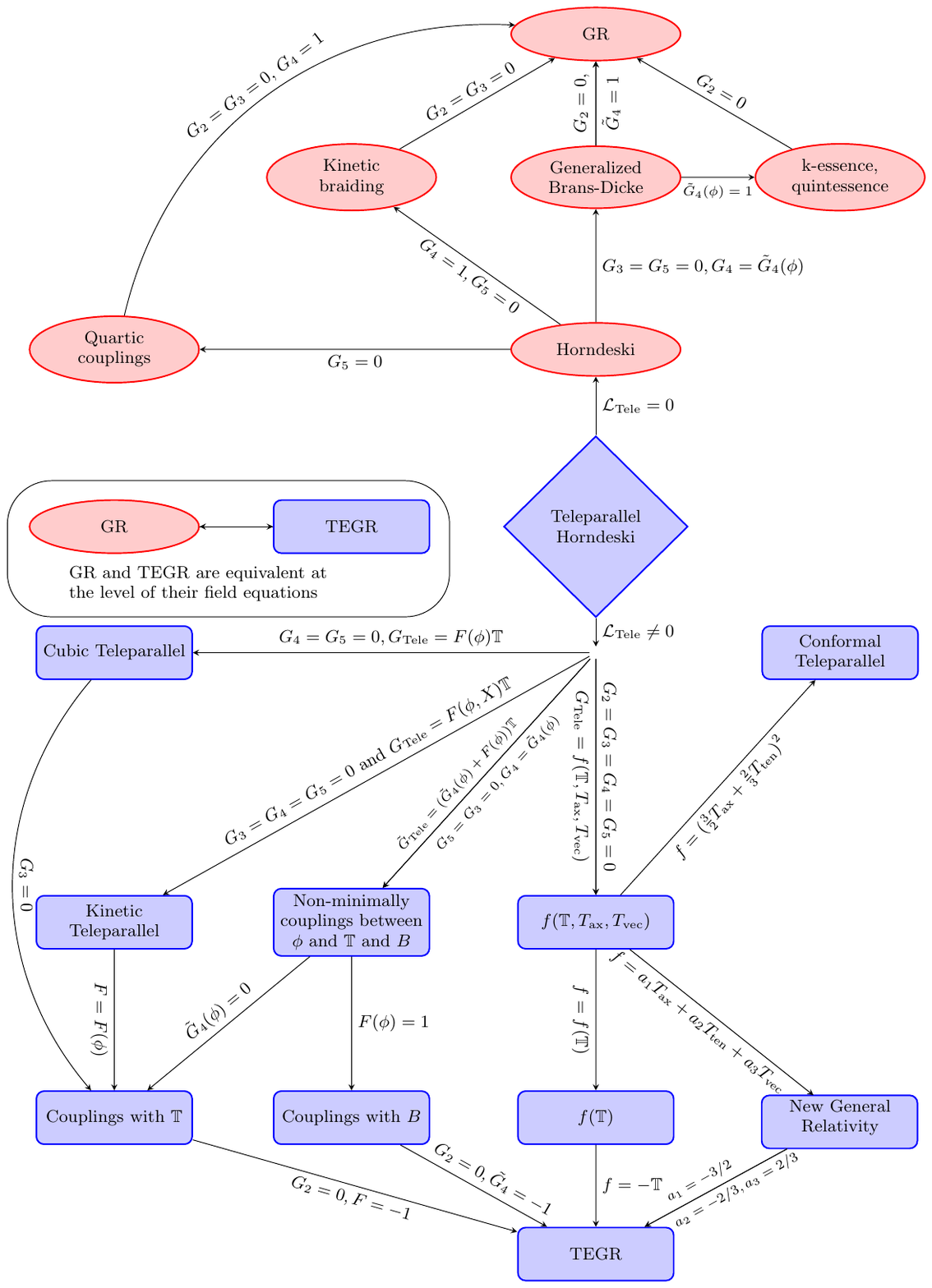}
	\caption{ Relationship between BDLS and various theories.}
    \label{fig:BDLS}
\end{figure}

Obviously, because of the existence of $G_{\rm Tele}$ in the action, this theory 
contains  much more phenomenology compared to its curvature analog. 
Specifically, it contains the curvature analog as a subclass when $G_{\rm Tele} 
= 0$\,. Its relation with various known theories is depicted in 
Fig.(\ref{fig:BDLS}).

The curvature case of Horndeski was severely constrained 
\cite{Kreisch:2017uet,Gong:2017kim}  after the observation of GW170817 and its 
companion GW170817A \cite{TheLIGOScientific:2017qsa}, since the speed of the 
gravitational waves \label{gravitationalwavrefs7}is constrained to the value of 
the speed of light
\begin{equation}
\Big|\frac{c_{g}}{c}-1\Big|\gtrsim10^{-15}\,.\label{con1}
\end{equation}
That is another reason the BDLS theory is worth studying. In 
\cite{Bahamonde:2019ipm}  the authors study the tensor mode perturbations on a 
flat FLRW background and they show that the models that survive the above 
constraint \eqref{con1} are given by
\begin{align}
  \mathcal{L}&=\tilde{G}_{\rm tele}(\phi ,X,\mathbb{T}, T_{\rm vec},T_{\rm 
ax},I_2,J_1,J_3,J_6,J_8-4J_5,J_{10})+G_2(\phi,X)+G_3(\phi,X)\Box 
\phi\,,\nonumber\\
   &+G_4(\phi,X)+G_{4,X}\left[\left(\Box\phi\right)^2 - \phi_{;\mu\nu} 
\phi^{;\mu\nu}+4J_5\right]+G_5(\phi)\mathcal{G}_{\mu\nu}\phi^{;\mu\nu}-4J_5G_{5,
\phi}\,,
   \label{revivedLagrangian}
\end{align}
meaning that many models that were eliminated in the curvature case,  like 
quartic and quintic Galileons, the Fab-Four and more, will survive in the 
teleparallel framework because of the appearance of the $J_5$ scalar.

\subsection{Teleparallel Dark Energy Models}\label{teleDEref1}
All the scalar torsion theories in four dimensions with a  single scalar field 
are subclasses of the BDLS theory 
\cite{Geng:2011aj,Kofinas:2015hla,Geng:2011ka,Zubair:2016uhx,Bahamonde:2018miw,
Jarv:2015odu,Kofinas:2015zaa,Horvat:2014xwa,Jamil:2012vb}. The first one with a 
non-minimal coupling between the scalar field and the torsion scalar was 
proposed in \cite{Geng:2011aj} under the name teleparallel dark energy, and its 
action reads
\begin{equation}
    \mathcal{S}_{\rm TDE} = \int {\rm d}^4x \,e \left[\frac{\mathbb{T}}{2\kappa ^2}  + 
\frac{1}{2}\xi \mathbb{T}\phi ^2 +\frac{1}{2}\nabla_{\mu}\phi\nabla^{\mu}\phi - 
V(\phi)\right]\,.
\end{equation}
Such models present interesting cosmological behaviour, not necessarily the  
same compared to $f(\mathbb{T})$ models \cite{Wu:2011kh,Wei:2011yr}. There are 
similar to the $\xi R \phi ^2 $ models in the Riemannian  geometry and such 
models will be discussed in detail in Sec.~\S.\ref{sec:phenomenology}. Moreover, 
in~\cite{Bahamonde:2015hza}, a generalisation of the above theory with an extra 
term $\chi B\phi^2$ was considered. This theory contains the theory with the 
coupling $\xi R \phi ^2 $ as a special case for the case $\chi=-\xi$. Apart from 
these ones, in a recent series of 
papers~\cite{Hohmann:2018rwf,Hohmann:2018vle,Hohmann:2018dqh,Hohmann:2018ijr} 
the most general scalar-torsion theories have been presented in their covariant 
formulation, as well as theories involving kinetic and derivative couplings of 
the scalar field with torsion.

\section{Phenomenology of Teleparallel Gravity}
\label{sec:phenomenology}
\label{telecosmoref1}

After formulating a possible well-motivated gravitational theory, it is then  
important to see its viability in terms of its confrontations with observations. It is well-known that 
GR works very well at Solar System scales, therefore, any meaningful theory must 
not deviate too much from these predictions at this scale of phenomenology. The 
easiest way to verify if a modified theory passes these constraints is by 
computing the so-called \textit{post-Newtonian parameters} (PPN). Basically, 
these parameters measure the weak-field approximation of a theory with the 
corresponding post-Newtonian terms. For a detailed review about this method, see 
Ref.~\cite{Will:2014kxa}. Usually, this method is constructed for theories 
concerning the metric but since Teleparallel Gravity uses tetrads, a 
different approach is needed as discussed in~\cite{Hayward:1981bk}. These parameters 
already put some constraints on different modified gravity models. In the 
context of Teleparallel Gravity, it was found that the PPN parameters of both 
$f(\mathbb{T})$ gravity and teleparallel dark energy (see 
Sec.~\S.\ref{sec:TeleExt}) are exactly the same as GR, therefore, these theories 
automatically pass all the Solar System \label{solarsystemref4}
observations~\cite{Chen:2014qsa,Li:2013oef,Ualikhanova:2019ygl}, as confirmed in \cite{Iorio:2015rla,Iorio:2012cm,Farrugia:2016xcw}.
If one further generalises $f(\mathbb{T})$ to $f(T_{\rm ax},T_{\rm vec},T_{\rm 
ten})$ or  teleparallel dark energy with an extra coupling between the boundary 
term and the scalar field as $\chi B\phi^2$, the PPN parameters $\gamma$ and 
$\beta$ deviate from GR, constraining these 
models~\cite{Ualikhanova:2019ygl,Sadjadi:2016kwj,Emtsova:2019qsl}. These studies were further generalised to more general teleparallel scalar tensor theories such as $f(\mathbb{T},X,Y,\phi)$ and Teleparallel Horndeski in~\cite{Flathmann:2019khc,Bahamonde:2020cfv}, finding again that only $\alpha$ and $\beta$ can differ from GR. For the general $f(\mathbb{T},B)$ gravity scenario, the precise form of this deviation is investigated in \cite{Farrugia:2020fcu,Capozziello:2019msc}, where the solar system tests and gravitomagnetic effects are probed against current observations. Moreover, in \cite{Pourbagher:2019zhq} the stability of theory and its thermodynamics are probed. One of the strongest bounds 
coming from these PPN parameters are the observations from Cassini, which is 
$|\gamma-1|\lesssim 2\cdot 10^{-5}$ and also, using the lunar laser ranging 
experiments, $|\beta-1|\lesssim2\cdot 10^{-4}$~\cite{Will:2014kxa}.

There are other new observational bounds that one needs to take into account for 
constructing  a coherent theory. For example, in Ref.\cite{Monitor:2017mdv}, it 
was found that the speed of the propagation of gravitational waves $c_g$ is very close 
to the speed of light $|c/c_g-1|\lesssim 3\cdot 10^{-15}$. The theories 
$f(\mathbb{T}), f(\mathbb{T},T_{\mathcal{G}})$ and $f(\mathbb{T},B)$ predict 
that $c_g=c$~\cite{Cai:2018rzd,Farrugia:2018gyz}, so that they are not 
observationally constrained through this test. Furthermore, there are only two propagating modes 
 in both $f(\mathbb{T})$ and $ f(\mathbb{T},T_{\mathcal{G}})$ 
gravity~\cite{Bamba:2013ooa,Farrugia:2018gyz}, exactly like GR. On the other hand, as in $f(R)$ 
gravity, $f(\mathbb{T},B)$ exhibits an extra polarisation mode (longitudinal or 
breathing mode)~\cite{Farrugia:2018gyz}. Further generalisations such as 
Teleparallel Horndeski predicts a different speed of the gravitational waves, 
and hence, the theory needs to be constrained for a certain type of them. 
However, one notices that the number of theories respecting $c_g=c$ is larger 
than in standard Horndeski gravity. Furthermore, the coupling functions $G_4(\phi,X)$ 
and $G_5(\phi)$ that were highly constrained in the standard version of 
Horndeski, can now be restored in its teleparallel 
analogue~\cite{Bahamonde:2019ipm}.

The majority of the work produced in modified Teleparallel Gravity has been done in the context of 
cosmology.  There are several works regarding this for different kinds of 
theories. Some important results in these theories are:
\begin{itemize}
    \item The possibility of explaining the acceleration of the Universe without evoking a cosmological constant (see, for example, Secs.~\S.\ref{sec:DSTele} and ~\S.\ref{sec:NoetherTele}).
    \item The possibility of roughly describing the evolution of the observed Universe eras (see, for example, Sec.~\S.\ref{sec:DSTele}) and crossing of the phantom divide line~\cite{Bahamonde:2015hza}.
    \item The possibility of reducing the tension for the value of the $H_0$ parameter and the growing $f\sigma_8$ tension (see Sec.~\S.\ref{sec:H0}).
    \item The existence of cosmological bouncing solutions (see Sec.~\S.\ref{sec:BounceTele}).
    \item The possibility of avoiding dark matter for describing the galactic rotation curves~\cite{Finch:2018gkh}.
\end{itemize}
In the following sections, we will briefly describe some of these results, and also some techniques used for studying cosmology in these theories.

\subsection{\texorpdfstring{$f(\mathbb{T})$}{f(T)} Cosmology and the Power-law 
Model}

The framework that Teleparallel Gravity offers in the form of its $f(\mathbb{T})$ gravity 
formulation can be investigated  against cosmological observations, where 
tensions appear to be growing with $\Lambda$CDM 
\cite{Aghanim:2018eyx,Riess:2019cxk,Wong:2019kwg}. The analysis takes the good 
tetrad $\udt{e}{a}{\mu} = \left(1,a(t),a(t),a(t)\right)$ for a homogeneous and 
isotropic universe (this reproduces the standard FLRW metric in its Cartesian coordinates form). By 
choosing the Lagrangian density to take the form $-\mathbb{T}+F(\mathbb{T})$, 
the resulting field equations turn out to be 
\cite{Cai:2015emx,Bengochea:2008gz,Krssak:2015oua}
\begin{align}
    H^2 &= \frac{\kappa^2}{3}\left(\rho + \rho_{\text{DE}}\right), 
\label{Friedmann1}\\
    2\dot{H} &= -\kappa^2\left(\rho + p + \rho_{\text{DE}} + 
p_{\text{DE}}\right)\,, \label{Friedmann2}
\end{align}
where $\kappa^2:=8\pi G$, and the effect of the extension to the TEGR Lagrangian 
 is to act as an exotic fluid with components
\begin{align}
    \rho_{\text{DE}} &= \frac{1}{2\kappa^2}\left(2\mathbb{T}F_{\mathbb{T}} - F\right)\,,\\
    p_{\text{DE}} &= \frac{1}{2\kappa^2}\left[\frac{F - \mathbb{T}F_{\mathbb{T}} 
 + 2\mathbb{T}^2F_{\mathbb{T}\mathbb{T}}}{1 + F_{\mathbb{T}} + 
2\mathbb{T}F_{\mathbb{T}\mathbb{T}}}\right]\,.
\end{align}
Together, these fluid properties satisfy the continuity equation
\begin{equation}
    \dot{\rho}_{\text{DE}} + 3H\left(\rho_{\text{DE}} + p_{\text{DE}}\right) = 0\,,
\end{equation}
as well as make up the effective EoS
\begin{equation}
    w_{\text{DE}} = -\frac{F/\mathbb{T} - F_{\mathbb{T}} + 
2\mathbb{T}F_{\mathbb{T} 
\mathbb{T}}}{\left(1+F_{\mathbb{T}}+2\mathbb{T}F_{\mathbb{T}\mathbb{T}}
\right)\left(F/\mathbb{T}-2F_{\mathbb{T}}\right)}\,,\label{weff}
\end{equation}
which returns a constant $w_{\text{DE}} = -1$ for the appearance of a  
cosmological constant through the condition $F_{\mathbb{T}}=0$. Here, we assume 
that $F(\mathbb{T}) \neq 0$, so that gravity is indeed modified by $f(\mathbb{T})$ gravity.

Finally, using the EoS in Eq.~(\ref{weff}), it follows that $F/\mathbb{T}  - 
F_\mathbb{T} + 2\mathbb{T}f_{\mathbb{T}\mathbb{T}}=0$ has solutions 
$F(\mathbb{T})=c_1\sqrt{\mathbb{T}} + c_2$, where the first part plays no role in 
the cosmic dynamics in four dimensions, while the second part plays the role of 
the cosmological constant \cite{Zheng:2010am,Linder:2010py,Farrugia:2016qqe}. 
Thus, this case is neglected since it reduces to $\Lambda$CDM.

\subsection{Cosmography in \texorpdfstring{$f(\mathbb{T})$}{f(T)} Gravity \label{f_T_Cosmography}}

Cosmography offers a model independent way in which to determine viable models  
of gravity through standard candle data (such as SNeIa) \cite{de2018towards}. 
That is, by using the Hubble diagram, gravitational Lagrangians can be 
constrained in their parameter space by using expansion data. In 
Refs.~\cite{Aviles:2013nga,Capozziello:2011hj,Capozziello:2019cav}, this is 
considered for the following parameters: Hubble ($H=\dot{a}/a$), deceleration 
($q=-\frac{1}{a}\frac{{\rm d}^2a}{{\rm d}t^2}H^{-2}$), jerk 
($j=\frac{1}{a}\frac{{\rm d}^3a}{{\rm d}t^3}H^{-3}$), and snap 
($s=\frac{1}{a}\frac{{\rm d}^4a}{{\rm d}t^4}H^{-4}$). These cosmographic parameters can then 
be correlated with the Hubble diagram by considering the Taylor expansion of the 
scale factor about present time, $a_0=1$, together with the luminosity distance 
relation
\begin{equation}
    H(z) = \left[\frac{{\rm d}}{{\rm d}z}\left(\frac{d_L(z)}{1+z}\right)\right]^{-1}\,,
\end{equation}
and redshift relation $a=\left(1+z\right)^{-1}$, which result in the Hubble 
cosmographic  relations
\begin{equation}
    H(z) \simeq H_0 \left[1 + H^{(1)} z + \frac{H^{(2)}}{2}z^2 + \frac{H^{(3)}}{6} z^3\right]\,,
\end{equation}
where
\begin{equation}
    H^{(1)} = 1+q_0\,, \quad H^{(2)} = j_0 - q_0^2\,, \quad H^{(3)} = 3q_0^2 + 3q_0^3 - j_0 \left(3+4q_0\right) s_0\,,
\end{equation}
with $(H,q,j,s)=(H_0,q_0,j_0,s_0)$ are all determined at current times. 
This expansion  is considered up to fourth-order derivatives, due to the lack of 
accuracy of the cosmological data beyond that point
\cite{Suzuki:2011hu,Riess:2009pu}. Fitting each of these parameters, using the 
Hubble diagram, it 
is then straightforward to infer an $F(\mathbb{T})$ model by using the modified 
Friedmann equations in Eqs.~(\ref{Friedmann1})-(\ref{Friedmann2}). To do this, 
consider $f(\mathbb{T})=-\mathbb{T}+F(\mathbb{T})$, so that we can impose the 
following constraints: (i) the effective gravitational constant must be equal to 
Newton's constant at present times~\cite{Capozziello:2019cav}
\begin{equation}
    \frac{{\rm d}f}{{\rm d}z}\bigg\vert_{z=0} = 1\,,
\end{equation}
which emerges by considering again the Friedmann equation in  
Eq.~(\ref{Friedmann1}) as $H^2 = \kappa^2 (\rho - 
2F/(3\kappa^2))/(6F_{\mathbb{T}})$ and recognising the effective coupling 
parameter, $G_{\text{eff}}=G/f_{\mathbb{T}}$ ($f_{\mathbb{T}}={\rm d}f/{\rm d}\mathbb{T}$). 
The requirement can then be written as $G_{\text{eff}}\vert_{z=0} = G 
\Rightarrow f_{\mathbb{T}}\vert_{z=0} = 1$. This means that at current time we 
recover TEGR. (ii) The second constraint is an evaluation of the Friedmann 
equation in Eq.~(\ref{Friedmann1}) at current times, such that the present value 
of the Lagrange density must be  \label{Geffref3}
\begin{equation}
    f(\mathbb{T}(z))\big\vert_{z=0} = 6H_0^2\left(\Omega_{M0} - 2\right)\,,
\end{equation}
where $\Omega_{M0}$ is the present value of the matter density parameter.

In Refs.~\cite{Aviles:2013nga,Capozziello:2011hj,Capozziello:2019cav}, various  
$f(\mathbb{T})$ models were considered, but we focus on the power-law model here. 
This can be represented by \cite{Myrzakulov2011}
\begin{equation}
    f(\mathbb{T}) = \alpha \mathbb{T} + \beta \mathbb{T}^n\,,
\end{equation}
where $\alpha,\beta,n$ are arbitrary constants, and the constraints on the 
Lagrangian  density result in the relations
\begin{align}
    \alpha = \frac{\left(2-\Omega_{M0}\right)n - 1}{n-1}\,,& &\beta =  
\frac{\left(\Omega_{M0} - 1\right)T_0^{1-n}}{1-n}\,.
\end{align}
While the parameter $\beta$ is used, it can be made dimensionless by taking the  
transformation $\beta\rightarrow \beta_0/\mathbb{T}_0^n$. To preserve TEGR for 
Solar System scale physics and the astrophysics regime, the $\alpha$ parameter 
can be set to $-1$ so that this is recovered as a first approximation. The 
best-fit cosmographic parameters then give $n=-0.011$~\cite{Capozziello:2011hj}. 
The result is an expansion rate very close to $\Lambda$CDM but not exactly 
equal. Along a similar vein, in \cite{Chakrabarti:2019bed} cosmography was used 
to reconstruct various $f(\mathbb{T})$ gravity Lagrangians by imposing 
conditions in the jerk parameter.

This analysis relies heavily on the assumption that higher-order contributions,  
which have much less accurate observational data, are sub-dominant, which may 
not always hold. Also, this analysis has only been applied to the popular 
$F(\mathbb{T})$ extension to TEGR. It would be interesting to explore other 
avenues of Teleparallel Gravity such as, for example, the ones presented in 
Sec.~\S.~\ref{sec:TeleExt}.

Along a similar rationale, in \cite{ElHanafy:2019zhr,ElHanafy:2020pek} the 
$f(\mathbb{T})$ Lagrangian is reconstructed against phenomenological data, 
achieving very interesting constraints on viable models.

\subsection{The Growth Factor}
\label{growthref1}

The inflationary epoch rendered an early Universe that was nearly uniform. It 
was small  quantum fluctuations that then resulted in the seeds of structure 
formation. Over the cosmic timescale, these seeds then grew into the structure 
we can observe today. This effect was amplified during the early 
matter-dominated phase of the Universe, where density perturbations were 
intensified by gravity. The growth factor can propagate how this growth changes between different theories of gravity.

To explore this aspect of cosmology, the evolution of linear scalar 
perturbations must be  considered, which at the level of the metric appear as 
\label{Scalarperfrref3}
 \begin{equation}\label{conformal_newt_gauge_scalar_pert}
    ds^2 = a^2(\tau)\left[ -\left(1+2\Psi\right){\rm d}\tau^2 + 
\left(1-2\Phi\right)\gamma_{ij}{\rm d}x^i {\rm d}x^j\right],
\end{equation}
where the spatial metric 
is chosen 
to be the Cartesian coordinate system. The problem then becomes, what perturbed 
tetrad to consider? One approach is given in Refs.~\cite{Chen:2010va}, but the resulting field equations turn out to restrict the $f(\mathbb{T})$ 
Lagrangian to its TEGR value which is not allowable with a good tetrad. It was 
later in 
Refs.~\cite{Zheng:2010am,Izumi:2012qj,Farrugia:2016pjh,Golovnev:2018wbh,
Wu:2012hs} that the correct good tetrad was studied. In perturbation theory, a 
good tetrad must adhere to the already discussed conditions for being a good 
tetrad up to perturbative order (see Sec.~\ref{f_T_sec} for further details). 
This is ultimately represented by
\begin{equation}\label{TG_scalar_pert}
\udt{e}{a}{\mu}=\left(\delta^a_b + \udt{\chi}{a}{b}\right)\udt{\bar{e}}{b}{\mu}\,,
\end{equation}
where $\udt{\bar{e}}{b}{0} = \delta^a_0$ and $\udt{\bar{e}}{b}{i} = 
a\delta^a_i$,  and the scalar perturbations are given by
\begin{equation}
    \chi_{ab}=
    \begin{pmatrix}
    \phi & \partial_i w\\
    \partial_i \tilde{w} & \delta_{ij} \psi + \partial_i \partial_j h +  
\epsilon_{ijk} \partial^{k} \tilde{h}\,,
    \end{pmatrix}
\end{equation}
where $w$ and $\tilde{w}$ are 2 scalar degrees of freedom (DoFs) of mass  
dimension and $h$ and $\tilde{h}$ are parity-violating terms. To obtain the 
correct scalar perturbations in the Newtonian gauge, the setting $w=-\tilde{w}$ 
and $h=0$~\cite{Zheng:2010am} needs to be taken, while $\tilde{h}$ vanishes 
naturally at the level of the metric. Using this tetrad setting, the correct 
scalar perturbations at the level of the metric are obtained, as in 
Eq.~(\ref{conformal_newt_gauge_scalar_pert}) (in cosmic time rather than 
conformal time). We consider only the scalar perturbations in this work, but 
interesting results have also been obtained for tensor perturbations in 
Refs.~\cite{Nunes:2016qyp,Nunes:2018xbm,Nunes:2018evm,Wu:2012hs,Nunes:2019bjq,
Cai:2018rzd,Golovnev:2018wbh,Izumi:2012qj}. The end result indeed appears in the 
linear perturbations of the torsion scalar as
\begin{equation}
    \mathbb{T}=6H^2 - 12H\left(\dot{\psi} + H\phi\right),
\end{equation}
which means that the scalar perturbations have an effect in the $F(\mathbb{T})$ 
gravity section, while, the matter perturbations are taken by considering only 
dust ($p=0=\delta p$). The density perturbations, $\delta \rho$, can then be 
encapsulated in the so-called gauge invariant fractional matter perturbation 
given by
\begin{equation}
    \delta_m = \frac{\delta \rho_m}{\rho_m} - 3Hv\,,
\end{equation}
where $v$ is the magnitude of the velocity of the fluid, $v^i=u^i/u^0$ 
\cite{dodelson2003modern}.  This regime is best studied by going into the 
Fourier domain and considering subhorizon modes where $\phi\sim\psi$. By 
combining the $f(\mathbb{T})$ field equations, the following linear matter 
evolution equation is obtained in Ref.~\cite{Zheng:2010am}
\begin{equation}\label{f_T_matter_evolution}
    \ddot{\delta}_m + 2H \dot{\delta}_m - 4\pi G_{\text{eff}}\rho_m\delta_m = 0\,,
\end{equation}
where $G_{\text{eff}}= G/(-1+F_{\mathbb{T}})$ is the effective gravitational 
constant.  This is generalised in Ref.~\cite{Farrugia:2016pjh} to include the 
Lagrangians formed with arbitrary combinations of the contraction of the stress-energy tensor, 
$T=\udt{T}{\mu}{\mu}$.

\noindent To probe the growth of matter density perturbations, 
Ref.~\cite{Zheng:2010am}  defines the variable
\begin{equation}
    g(a) := \frac{D(a)}{a}\,,
\end{equation}
which is defined this way to avoid scale factor dependence during  matter-dominated eras, and where the reasonable initial conditions 
$g(a_i)=1,\,\left(dg/d\ln a\right)\vert_{a=a_i} = 0$ are chosen. Here, $D(a) := 
\delta_m(a) / \delta_m(a_i)$ (for some reference scale factor $a_i$) is the 
growth factor.

\begin{figure} [ht]
    \centering
    \includegraphics[scale=1.1]{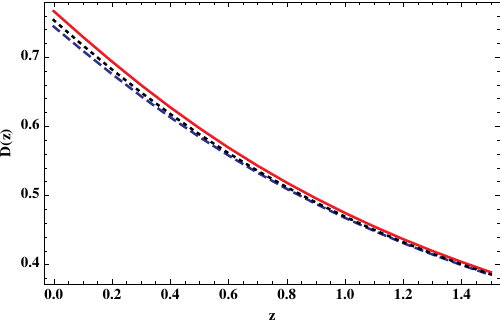}
    \caption{{\it{Matter perturbation evolution for $\Lambda$CDM (solid red 
line), 
power-law $F(\mathbb{T})$  (dashed blue line), and constant dark energy EoS in 
GR (dotted black line) \cite{Zheng:2010am}.}}}\label{fig:D_Power_law}
\end{figure}

Considering again the power-law model with $F(\mathbb{T}) = \beta \mathbb{T}^n$, 
the growth factor as a function of redshift can be solved numerically and turns 
out to give the evolution depicted in Fig.~(\ref{fig:D_Power_law}). Given that 
$F_{\mathbb{T}}>0$ for the power-law, it follows that the growth factor will be 
dampened due to the effective gravitational constant relation. The main result 
of this is that over-dense perturbations grow slower when compared with GR. In 
Ref.~\cite{Wu:2012hs}, it was also found that the vector perturbations are 
well-behaved for sub-horizon modes. These results are confirmed in 
Refs.~\cite{Izumi:2012qj,Golovnev:2018wbh}, where it is also noted that the cause 
of the lack of extra propagating DoFs could be the symmetric nature of the 
background cosmology.

\subsection{The \texorpdfstring{$H_0$}{H0} Tension Problem}\label{sec:H0}
\label{H0tensionjref1}\label{CMBrefs2} \label{madelindepeefs3} 
\label{Hubbleefs2}
The discrepancy between model-independent measurements of the current value of 
the  Hubble parameter~\cite{Riess:2019cxk,Wong:2019kwg} and those inferred from 
the CMB using flat $\Lambda$CDM \cite{Aghanim:2018eyx,Alam:2016hwk} is now 
corroborated by an overwhelming wealth of evidence. The $H_0$ tension problem 
then points to the necessity of new physics beyond flat $\Lambda$CDM, such as 
modified gravity within the Teleparallel Gravity regime \cite{Wang:2020zfv}.

\begin{figure}[ht]
\begin{center}
\includegraphics[scale=0.69]{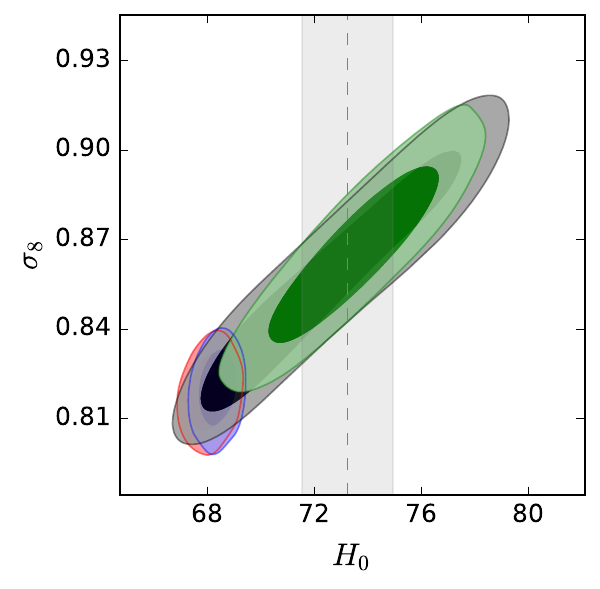}~~~
\includegraphics[scale=0.89]{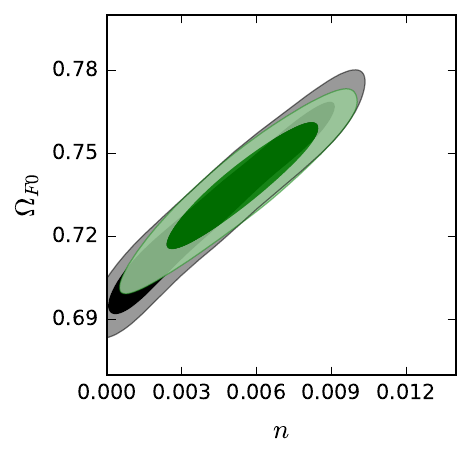}
\caption{{\it{ The comparison of $\Lambda$CDM with the $f(\mathbb{T})$ 
gravity 
power-law model as  reported in Ref.~\cite{Nunes:2018evm}. Left panel: Plot of 
the $\sigma_8 - H_0$ parameter space comparing the 
$\Lambda$CDM model in red (blue) for  CMB+BAO (CMB+BAO+$H_0$) data 
respectively. Additionally, we present the results for the    power-law 
$f(\mathbb{T})$ model in black (green) for the 
same data, where the vertical gray band corresponds to 
$H_0=73.24\pm1.74\,\text{km}\,\text{s}^{-1}$ \cite{2018ApJ...855..136R}. The 
extended model has best fit index parameter $n=0.0043^{+0.0033}_{0.0039}$ 
($0.0054^{0.0020}_{0.0020}$) \cite{Nunes:2018xbm} (more details here about the 
data used). Right panel: 68\% and 95\% confidence levels   for CMB+BAO 
(CMB+BAO+$H_0$) data
in black  (green) \cite{Nunes:2018xbm}.}}}
\label{fig:f_sigma_8-H0_plot}
\end{center}
\end{figure}

To explore this possibility in Teleparallel Gravity, we need to consider again the perturbations 
in Eq.~(\ref{TG_scalar_pert}).  However, in this scenario the contributions of 
matter and radiative pressures are not neglected even at perturbative 
level~\cite{Nesseris:2013jea,El-Zant:2018bsc}. Also, the gravitational 
potentials $\phi$ and $\psi$ are no longer equal. As already discussed in 
Refs.~\cite{Golovnev:2018wbh,Nunes:2018evm}, the following modified Poisson equation is derived
\begin{equation}
    k^2 \psi = 4\pi G_{\text{eff}}a^2\delta\rho\,,
\end{equation}
where we have transformed to Fourier space. By keeping to the  power-law model, Ref.~\cite{Nunes:2018evm} explores 
this possibility in terms of the $H_0$ problem as well as the growing 
$f\sigma_8$ tension. In this work, the authors show that the $H_0$ tension can 
be reduced in conjunction with reducing the $f\sigma_8$ tension. As they show in
Fig.~(\ref{fig:f_sigma_8-H0_plot}), a consistent cosmological setup can be 
constructed for a small value of index $n$. 

One of the principal motivations for exploring modified gravity in cosmology is 
to better explain the  appearance of dark energy without modifying the matter 
content of the Universe. This entails reinterpreting the Friedmann equation as 
an effective equation in which the modified gravity component acts as a separate 
contribution to cosmic evolution beyond GR. This can easily be done by writing
\cite{Nesseris:2013jea}
\begin{equation}
    \frac{H^2(z,\mathbf{r})}{H_0^2} = \Omega_{m0}\left(1+z\right)^3 + 
\Omega_{r0} \left(1+z\right)^4 +  \Omega_{F0} y(z,\mathbf{r})\,,
\end{equation}
where $\Omega_{F0} = 1 - \Omega_{m0} - \Omega_{r0}$ is the $F(\mathbb{T})$ 
density parameter at current times,  and
\begin{equation}
    y(z,\mathbf{r}) = \frac{1}{\mathbb{T}_0\Omega_{F0}}\left(F  - 
2\mathbb{T}F_{\mathbb{T}}\right)\,,
\end{equation}
represent the background evolution of the $F(\mathbb{T})$ model. In the right 
panel of
Fig.~(\ref{fig:f_sigma_8-H0_plot}),  the $n-\Omega_{F0}$ plane is shown, where 
Ref.~\cite{Nunes:2018evm} reports promising results for small values of $n$.
\label{sigmaref2}

\begin{figure} [ht]
    \centering
    \includegraphics[scale=0.505]{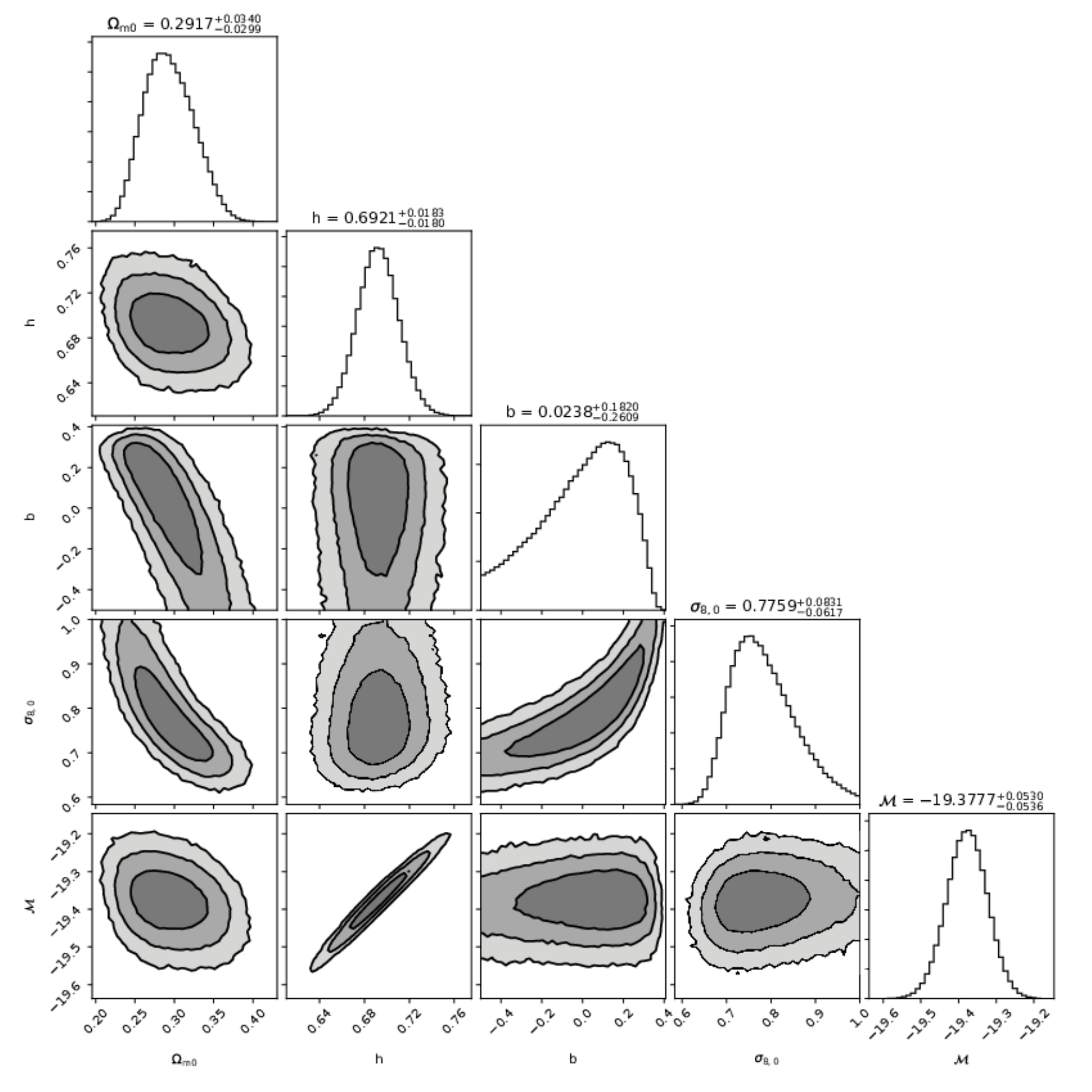}
	\caption{{\it{ The  $1\sigma$, $2\sigma$ and $3\sigma$ 
likelihood 
contours  for the $F(\mathbb{T})$ power-law model (with $F(\mathbb{T})=\alpha 
(\mathbb{T})^b$)\cite{Anagnostopoulos:2019miu}.}}}\label{fig:power_law_f_T}
\end{figure}

Along a similar vein, the authors of Ref.~\cite{Anagnostopoulos:2019miu}  
confront the growth of structure in the Universe by using several cosmological 
probes. The study involves three $F(\mathbb{T})$ models, but we highlight the 
results for the power-law model here. In their analysis, they use growth rate 
data for $f\sigma_8$ that has been verified for internal robustness 
\cite{Sagredo:2018ahx}, which differs from some other approaches where 
inconsistencies can arise due to overlaps between separate studies. The second 
data set used in this study is the updated Hubble expansion through the cosmic 
chronometric method~\cite{Yu:2017iju}, while the third is the latest standard 
candle data in Ref.~\cite{Scolnic:2017caz}. These three data sets are used in a 
joint analysis for the power-law $F(\mathbb{T})$ model resulting in the 
likelihood plots shown in Fig.~(\ref{fig:power_law_f_T}).

In this work, the authors use the Akaike Information Criterion (AIC) 
\cite{1100705}, Bayesian  Information Criterion (BIC) \cite{schwarz1978} and 
Deviance Information Criterion (DIC) \cite{burnham2004multimodel} to compare the 
different models. The power-law is favoured using the AIC and BIC comparisons 
when compared to $\Lambda$CDM. While not performing best using the DIC to 
compare the models, it still fares relatively well compared to other 
prominent models in the literature.

\begin{figure}[ht]
    \centering
\includegraphics[scale=0.85]
{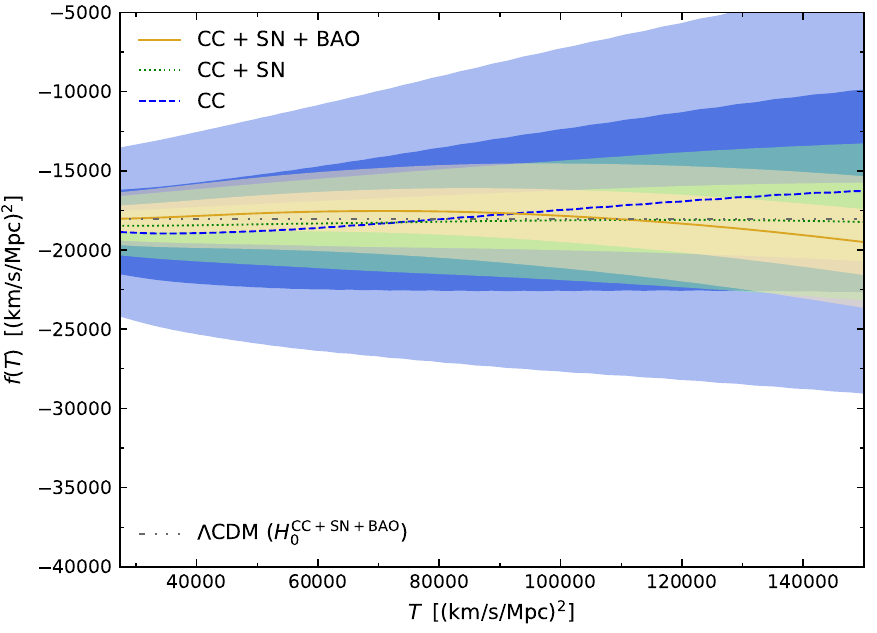}
	\caption{{\it{The regions are produced by a model-independent reconstruction from Hubble data with 1 and 2 $\sigma$ error regions coming from the Gaussian process method (the present matter density parameter is imposed to be $\Omega_{m0}=0.302$). The blues regions represent cosmic chronometer data, green includes also supernova data, while orange represents the inclusion of baryonic acoustic data  \cite{Briffa:2020qli}.}}}\label{Gaussian_Process}
\end{figure}

Another important contribution to the reduction of the $H_0$ tension are 
Refs.\cite{Cai:2019bdh,Yan:2019gbw,Briffa:2020qli}, where Hubble data is 
interpreted as a stochastic process such that the various model ansatz choices 
must 
reproduce. By taking this approach, the authors determine a region 
for acceptable Lagrangian forms for the $F(\mathbb{T})$ model. While this region 
is model-independent and goes up to $z=2.4$, various models can be constrained 
against cosmic data. In Ref.\cite{Briffa:2020qli}, Gaussian processes are used on Hubble data which is interpreted as being sourced by stochastic processes. Here, the authors reconstruct values of the arbitrary Lagrangian through the Friedmann equation and then use Gaussian processes to determine the best fit for this function. The result is a model independent reconstruction of the Lagrangian $\mathbb{F}(T)$ (except for the assumption that $\Lambda$CDM dominates at current times, which resonates with the results from cosmography in Sec.~\ref{f_T_Cosmography}). These values, together with their 1 and 2 $\sigma$ errors are shown in Fig.~(\ref{Gaussian_Process}), which depicts the allowable regions in which all cosmological models must predict values for the arbitrary Lagrangian.

There has also been a growing body of work of confronting observations within other extensions of Teleparallel Gravity, such as \cite{Escamilla-Rivera:2019ulu}, where Pantheon data is used in $f(\mathbb{T},B)$ gravity. In this work, the $H_0$ tension problem is also confronted with new constraints on literature models within this framework of gravity.

\subsection{Inflation in Teleparallel Theories of 
Gravity}\label{sec:Teleinflation}
\label{inflationrefs2}
The first modification of Teleparallel Gravity was introduced in~\cite{Ferraro:2006jd}, with the 
aim of  studying inflation. In this paper, the authors found that for a 
Born-Infield $f(\mathbb{T})$ gravity model, it is possible to cure the horizon 
problem without an inflation field, and describe a de-Sitter expansion. This 
paper was the crucial starting point for investigating cosmology in the context 
of Teleparallel Gravity. So far, the majority of the inflationary models in modified teleparallel 
theories have been carried out in $f(\mathbb{T})$ gravity. By performing a 3+1 
decomposition for the tetrad, the cosmic inflationary perturbations for 
$f(\mathbb{T})$ were computed in~\cite{Wu:2011kh}. Similar to the Starobinsky 
inflationary model $R+\alpha R^2$, in~\cite{Bamba:2014zra} it was found that the 
case $-\mathbb{T}+\alpha \mathbb{T}^2$ behaves differently since in this case, a 
de-Sitter solution occurs (not a quasi de-Sitter). This means that in the 
teleparallel version of the Starobinsky model, only eternal de Sitter inflation is 
possible. In~\cite{Nashed:2014lva} it was further found that these models can 
have good agreements with a hot big bang nucleosynthesis, and also that the 
system evolves towards a flat FLRW universe naturally, even if we start with a 
non-flat one. Later in~\cite{Rezazadeh:2015dza}, it was found that power-law
and intermediate inflationary models in $f(\mathbb{T})$ gravity are compatible 
with  Planck measurements, and that a self-interacting quartic
potential $V(\phi)\propto \phi^4$, which has an interesting reheating process is 
viable in $f(\mathbb{T})$ inflation. Further, the standard inflationary model is 
not observationally compatible with this kind of potential. After introducing a 
suitable scale factor, it was found that depending on a parameter, 
$f(\mathbb{T})$ gravity can have a graceful exit inflation or can have a 
bounce~\cite{Bamba:2016gbu}.  In that study it was obtained that the problem of 
a large tensor-to-scalar ratio is not present in the bouncing models. The 
standard logamediate inflation is not compatible with observations but 
in~\cite{Rezazadeh:2017edd} it was concluded that in $f(\mathbb{T})$ gravity, 
this model can be compatible with Planck observations. Finally, 
in~\cite{Keskin:2018gev} it was found that a super inflation scenario can be 
achieved in $f(\mathbb{T})$ gravity. Concerning extensions to $f(\mathbb{T})$ 
gravity, in~\cite{Bamba:2012mi}, the authors introduced a coupling between a 
Maxwell field and the torsion scalar, $I(\mathbb{T})F_{\mu\nu}F^{\mu\nu}$, 
finding the possibility of generating large-scale magnetic fields from 
inflation. In $f(\mathbb{T},T_{\mathcal{G}})$ gravity, there was obtained a 
model that unifies inflation with dark energy with a super inflation 
mechanism~\cite{Keskin:2017yzy}. A unimodular $f(\mathbb{T})$ gravity model was 
also analysed in the context of inflation, finding another alternative 
inflationary models with, graceful inflationary exit~\cite{Bamba:2016wjm}.

The first study related to inflation in the context of teleparallel scalar 
tensor  theories was done in~\cite{Jamil:2013nca}, where an extended 
$f(\mathbb{T})$ plus an inflation field with a kinetic, a potential term and an 
interaction term was introduced. In this work, it was found that a possible 
warm-inflation model is compatible with the Planck data. Later 
in~\cite{Goodarzi:2018feh}, a similar model was studied, finding a reheating 
process, with the scalar field being responsible to reheat the Universe after 
the inflationary era. In~\cite{Abedi:2017ijd}, another scalar tensor model 
concerning two non-minimally couplings with the scalar field, one with the 
torsion scalar $F(\phi)\mathbb{T}$ and the other with a vector field 
$G(\phi)F_{\mu\nu}F^{\mu\nu}$, was considered in the context of anisotropic 
inflation, obtaining that in the strong coupling regime, the anisotropy shear to 
expansion ratio has a different value than its standard form. The constant-roll 
inflation in $f(\mathbb{T})$, minimally coupled with a scalar field, was analysed 
in~\cite{Awad:2017ign}, allowing the theory to have a wide range of viability in 
terms of observations. In~\cite{Gonzalez-Espinoza:2019ajd} the authors studied slow-roll inflation in a more general teleparallel scalar-tensor theory, with a canonical scalar field non-minimally coupled to torsion with a Galileon-type field $G(\phi,X)\Box \phi$ and a monomial scalar field potential $V\propto \phi^n/n$. This theory is a particular case of the BDLS theory (see Sec.~\S.\ref{sec:BDLS}). Based on Planck 2018 data for both the spectral index $n_s$ and the tensor-to-scalar ratio $r$, standard inflation with monomial scalar field potential with $n\geq 2$ is ruled out. However, in this teleparallel version, $n=2$ (chaotic quadratic inflation) is in agreement with data and $n=1$ and $n=2/3$ are even more favoured than previous models. In a model with non-minimally couplings between a scalar field and both the boundary term and the torsion scalar, it was shown that the scalar field does not source linear scalar perturbations, unless the coupling functions satisfy certain conditions~\cite{Raatikainen:2019qey}. Only in these situations, can one have successful Higgs inflationary models. Finally, considering more exotic models, a tachyonic 
teleparallel one explaining inflation and agreeing with the current
observational Planck limits was studied in~\cite{Akbarieh:2018oie}. Further, it was also found in~\cite{Bahamonde:2019gjk} that by extending the study with a boundary term non-minimally coupled with the scalar field, accelerated expansion and scaling solutions are attained.

\subsection{Dynamical System in Cosmology for Teleparallel Theories of  
Gravity}\label{sec:DSTele}

 When one starts modifying or extending the Einstein field equations, the 
cosmological  equations become more involved to solve. The FLRW equations can be 
written as an autonomous system of differential equations, and for this, they can be 
recast as a dynamical system. Mathematicians have studied these systems for a 
long time, giving an easy way to understand the dynamics of a model and how the 
stability of their critical points behave. Hence, this is a powerful 
mathematical technique, which is very useful for studying cosmology in modified 
theories of gravity. For a detailed review about dynamical systems, both 
mathematically and in terms of cosmology, 
see~\cite{Bahamonde:2017ize,arrowsmith,Coley:2003mj,WainwrightEllis}.
\label{Dynamicalref2}

There are several works in modified Teleparallel Gravity that use dynamical systems in the 
context of cosmology.  The first work for $f(\mathbb{T})$ cosmology was 
presented in~\cite{Wu:2010xk}, where the authors  studied a power-law 
$f(\mathbb{T})$ finding one critical point behaving as a late-time attractor and 
another two describing matter and radiation eras as saddle points. Later, in 
\cite{Zhang:2011qp}, a logarithmic $f(\mathbb{T})$ cosmology was studied, finding 
a de-Sitter late-time attractor. In 
\cite{Jamil:2012nma,Jamil:2012yz,Biswas:2015cva}, it was also found that by 
introducing an interaction between the dark fluids and considering a power law 
$f(\mathbb{T})$, it is possible to get tracker cosmological solutions. Other 
dynamical system studies concerning more general approaches for $f(\mathbb{T})$ 
cosmology have been done. For example, in~\cite{Feng:2014fsa} the authors used 
the nullcline method to study the bifurcation phenomenon to study the global 
dynamical properties of the dynamical system. In \cite{Mirza:2017vrk}, it was 
found that there are three conditions which ensure that $f(\mathbb{T})$ 
cosmology could roughly describe the cosmological history of the evolution of 
the Universe. Using non-standard dimensionless variables, Hohmann et. 
al~\cite{Hohmann:2017jao} found de-Sitter fixed points, accelerated expansion, 
crossing the phantom divide, and finite time singularities in $f(\mathbb{T})$ 
cosmology. They also found some bounce solutions in this model. Finally, in 
\cite{Awad:2017yod}, the authors were able to rewrite the $f(\mathbb{T})$ 
dynamical system as a one dimensional one by using the fact that $\mathbb{T}$ 
depends only on the Hubble parameter as $\mathbb{T}=6H^2$ in flat FLRW. Doing 
this, they found that it is possible to reconstruct the whole history of the 
Universe starting from a big bang singularity and finalising in an accelerating 
expansion. In addition, they also found some other exotic solutions, such as 
cosmological bounce and turnaround, the phantom-divide crossing, the Big Brake 
and the Big Crunch, and also they found that it may exhibit various 
singularities.

\begin{figure} 
    \centering
    \includegraphics[scale=0.75]{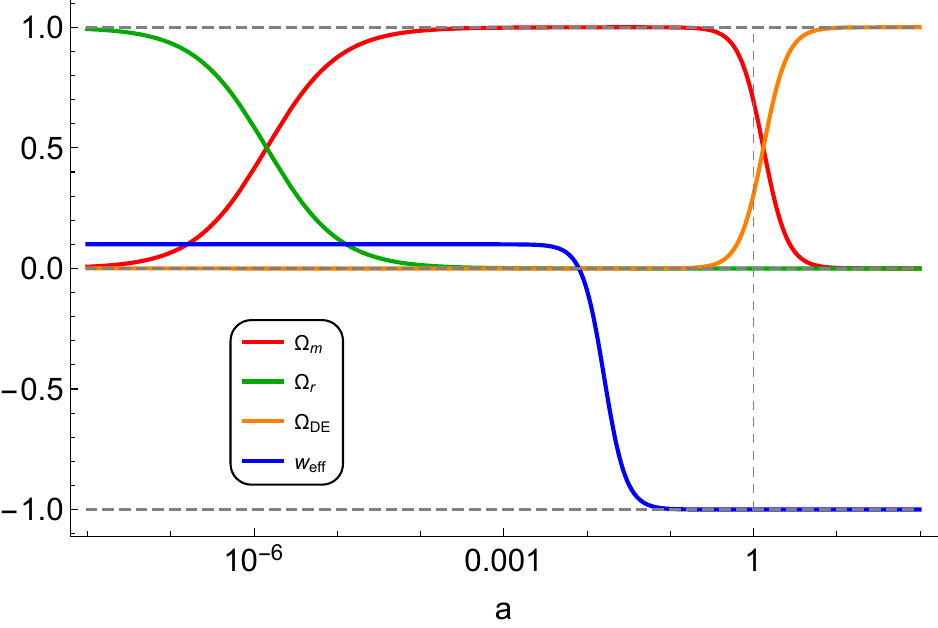}
    \caption{{\it{ Evolution of energy density matter $\Omega_{\rm m}$,  
radiation 
$\Omega_{\rm rad}$ and dark energy $\Omega_{\rm DE}$ for a 
$f(\mathbb{T})=-\mathbb{T}+\frac{C_1 }{m+1}\mathbb{T}^{m+1}$ model with 
$m=0.1$.}}}
    \label{fig:DSf(T)}
\end{figure}

Let us now briefly review the dynamical system of $-\mathbb{T}+F(\mathbb{T})$  
(TEGR plus $F(\mathbb{T})$) cosmology described by the modified flat FLRW 
equations~\eqref{Friedmann1} and \eqref{Friedmann2}. For a universe composed of 
two fluids with effective energy density described by $\rho=\rho_{\rm 
rad}+\rho_{\rm m}$, where the first fluid represents a radiation fluid and the 
second one a pressureless fluid, one can introduce the following dimensionless 
variables
\begin{equation}
    x=-\frac{F(\mathbb{T})}{6H^2}\,,\quad 
y=\frac{\mathbb{T}F_\mathbb{T}}{3H^2}\,,\quad  \Omega_{\rm rad}=z=\frac{\kappa^2 
\rho_{\rm rad}}{3H^2}\,,\quad \Omega_{\rm m}=\frac{\kappa^2  \rho_{\rm 
m}}{3H^2}\,,
\end{equation}
to then rewrite the first FLRW equation \eqref{Friedmann1} as follows
\begin{equation}
    \Omega_{\rm m}=1-x-y-z\,,
\end{equation}
which gives a constraint and reduces the dynamical system to be a 3 dimensional 
one.  By introducing $N=\log(a)$, the dynamical system for 
this model becomes
\begin{align}
     \frac{{\rm d}x}{{\rm d}N}&=-(2x+y)\frac{z+3-3x-3y}{2my-2+y}\,,\label{DS1}\\
      \frac{{\rm d}y}{{\rm d}N}&=2my\frac{z+3-3x-3y}{2my-2+y}\,,\label{DS2}\\
       \frac{{\rm d}z}{{\rm d}N}&=-4z-2z\frac{z+3-3x-3y}{2my-2+y}\,,\label{DS3}
\end{align}
where we have introduced the quantity
\begin{equation}
    m=\frac{\mathbb{T}F_{\mathbb{T}\mathbb{T}}}{F_\mathbb{T}}\,.
\end{equation}
In order to close the dynamical system, one needs to assume a form for 
$F(\mathbb{T})$.  The easiest case is to assume that $m=\textrm{constant}$, which 
closes the dynamical system and includes two kind of $F(\mathbb{T})$, one 
behaving as a power-law $F(\mathbb{T})=C_1 \mathbb{T}^{m+1}/(m+1)+C_2$ when 
$m\neq-1$ and also including a logarithmic case $F(\mathbb{T})=C_1 \log 
(\mathbb{T})+C_2$ for $m=-1$. We will assume this case for simplicity. For other 
kinds of $F(\mathbb{T})$, see the papers mentioned before. The dark energy state 
parameter~\eqref{weff} in the dimensionless variables reads
\begin{equation}
    w_{\rm DE}=-\frac{x+y/2-my}{(1-y/2-my)(x+y)}\,.
\end{equation}
For $m=\textrm{constant}$, one finds that the dynamical  system 
\eqref{DS1}-\eqref{DS3} has three critical points. The first critical point is 
$P_1=(x,y,z)=(0,0,0)$ which is the origin of the phase space and represents a 
matter-dominated era, since $\Omega_{\rm m}=1$. This critical point has three 
eigenvalues with different signs, $\{3,-1,-3m\}$, so that, this point is always 
a saddle point. This behaviour is expected for describing the standard matter-dominated era, since it is known that this point needs to be represented by a 
point which attracts trajectories in some directions but repels them along 
others. The second critical point is $P_2=(x,y,z)=(0,0,1)$, representing a 
universe dominated by radiation, $\Omega_{\rm rad}=1$. This point has also three 
eigenvalues $\{4,1,-4 m\}$ but now depending if $m>0$, the point is saddle, and if 
$m<0$ the point is unstable. This again has the correct cosmological behaviour, 
since it is known that there was a radiated-dominated era at some point of the 
history of the Universe but after this era, this era changed to be a matter-dominated era. This is then achieved by having either an unstable or a saddle 
point for $P_2$ (neglecting inflation). The final point is represented by the 
critical line  $P_3=(x,1-x,0)$ whose cosmological behaviour is representing an 
accelerating universe with an effective state parameter of $-1$, which denotes a 
de-Sitter accelerating expansion. This critical line has three eigenvalues 
$\{0,-4,-3\}$. Since one of them is zero, one cannot study its stability 
property with the standard linear stability theory since it is a non-hyperbolic 
point, and this method fails for analysing such points. Other 
stability methods can then be used, such as Lyapunov functions or centre manifold theory. 
See~\cite{Bahamonde:2017ize} for a detailed description about these methods. If 
one uses the second mentioned method, one needs to first shift the critical 
point to the origin, then introduce new variables in such a way that one can 
diagonalise the Jacobian matrix associated with the dynamical system. After 
doing this, the centre manifold can then be constructed according to the theorem 
described in Sec.~\S.2.4 in \cite{Bahamonde:2017ize}. By doing this for our 
dynamical system, we find that the leading term in the dynamical system is 
reduced to the centre manifold $\dot{z}=-4 z+\mathcal{O}(z^2)$, which tells us 
that the point $P_3$ is always stable. Thus, the critical line $P_3$ represents 
a late-time accelerating attractor behaving as a de-Sitter. Then, 
$-\mathbb{T}+F(\mathbb{T})$ cosmology can describe a transition from radiation- 
to matter-dominated eras, finalising in a dark energy era with a 
de-Sitter accelerating expansion of the Universe. This analysis would be valid 
for both logarithmic and power-law kinds of $F(\mathbb{T})$. As an example, 
Fig.~(\ref{fig:DSf(T)}) shows the evolution of the relative energy density of 
matter $\Omega_{\rm m}$, radiation $\Omega_{\rm rad}$ and dark energy 
$\Omega_{\rm DE}$ related to the modifications coming from $F(\mathbb{T})$ for 
the case where $m=0.1$, which represents a power-law type of $F(\mathbb{T})$. 
One can notice that the evolution of the Universe is roughly described as one 
expected from our Universe, starting from a radiation-dominated era, then 
passing to a matter dominated era and finalising in a dark energy dominated era.

Other different modified Teleparallel Gravity cosmological models extending $f(\mathbb{T})$  
gravity have also been analysed using dynamical systems techniques. In the 
papers~\cite{Karpathopoulos:2017arc,Paliathanasis:2017flf,Paliathanasis:2017efk}
, a $f(\mathbb{T},B)$ gravity model including the boundary term $B$,  which 
connects the Ricci scalar computed with the Levi-Civita connection and the 
torsion scalar, was analysed, finding scaling solutions, a matter epoch of the 
Universe, and that two accelerated phases can be
recovered describing de-Sitter universes. Later, using non-minimally 
torsion-matter  theories like $f_1(\mathbb{T})+f_2(\mathbb{T})\mathcal{L}_{\rm 
m}$, with $\mathcal{L}_{\rm m}$ being the matter density Lagrangian, some scaling
decelerated solutions, dark-matter dominated, or dark-energy dominated 
accelerated solutions were found.  Further generalisations such as 
$f(\mathbb{T},B,\mathcal{L}_{\rm m})$ also found similar 
results~\cite{Bahamonde:2017ifa}. Other models with higher-order torsion 
invariants such as $f(\mathbb{T},T_{\mathcal{G}})$ found scaling solutions and 
found past, present and future singularities, depending on the parameters of the 
theory~\cite{Kofinas:2014aka}. In two recent papers~\cite{Gonzalez:2019tky,Bahamonde:2020vfj}, the 
authors studied the dynamical system of a teleparallel Lovelock gravity theory and string-inspired theories 
finding a rich phenomenology that can also describe a late-time acceleration of 
the Universe. Further higher-order derivative torsion theories such as 
$f(\mathbb{T},(\nabla \mathbb{T})^2,\Box \mathbb{T})$ or non-local theories 
$\mathbb{T}f(\Box^{-1}\mathbb{T})$ found vacuum de-Sitter solutions and phantom 
divide line crossing in agreement with 
observations~\cite{Otalora:2016dxe,Bamba:2017ufh}. Regarding higher dimensional models, in~\cite{Bohmer:2019qfi} the authors used dynamical systems to find that $f(\mathbb{T})$ gravity in eleven dimensions can give rise to an early inflationary epoch driven by the presence of extra dimensions without other matter
sources.

In~\cite{Wei:2011yr}, the authors analysed the first teleparallel scalar-tensor 
theory using  dynamical system techniques. He analysed a model called 
teleparallel dark energy (see Sec.~\S.\ref{sec:TeleExt}), which is constructed by 
a non-minimally coupling between the torsion scalar $\mathbb{T}$ and a scalar 
field $\phi$ as $(1+\xi\phi^2)\mathbb{T}$, with $\xi$ being a constant, finding 
that these models contain certain similarities to Elko spinor dark energy 
models. In a non-minimally coupled model between the Ricci scalar and a scalar 
field, it is possible to find scaling solutions but in teleparallel dark energy, 
this cannot be achieved~\cite{Wei:2011yr}. However, if the coupling is changed to $F(\phi)\mathbb{T}$ scaling solutions can be obtained \cite{Otalora:2013tba}. 
Further, in ~\cite{Xu:2012jf} the authors analysed the phase space of this model, 
finding similar results as standard quintessence models but having an additional 
late-time solution behaving as de-Sitter without any fine-tuning. They also 
found that the crossing of the phantom divide line is possible for this model. 
In a series of two papers, Skugoreva et. al. \cite{Skugoreva:2014ena,Skugoreva:2014gka} also studied this model, giving a 
detailed comparison between it with the standard non-minimally coupled case constructed 
from the Ricci scalar, finding that in teleparallel dark energy the presence
of oscillatory behaviors is more frequent \cite{Skugoreva:2014ena,Skugoreva:2014gka}. If one uses 
the  teleparallel dark energy model and adds an additional coupling between dark 
matter and dark energy, is a deceleration to acceleration 
phase transition via a $Z_2$ symmetry breaking results \cite{Sadjadi:2015fca}. Another 
proposed teleparallel model was introduced in~\cite{Bahamonde:2015hza} by adding 
a new coupling $\chi B \phi^2$, where $\chi$ is a constant and $B$ is the 
boundary term. Clearly, the standard Ricci case non-minimally coupled with the 
scalar field is recovered by setting $\chi=-\xi$. The authors studied the case 
where only a boundary term coupling exists, and they found that the evolution of 
the model evolves towards a late-time accelerating attractor without any 
fine-tuning. They also found the possibility of the crossing of the phantom 
divide line in this model. Later, in \cite{Marciu:2017sji}, it was found that 
using this coupling, one can also get scaling solutions. Other more exotic 
models have been also proposed, such as tachyonic teleparallel 
models~\cite{Bahamonde:2019gjk,Otalora:2013dsa,Banijamali:2016ozr,
Otalora:2013tba,Fazlpour:2014qaa,Fazlpour:2014qla} finding scaling solutions, 
alleviating the coincidence problem without fine-tuning, obtaining a late-time 
accelerating attractor and also finding a field-matter-dominated era. Moreover, quintom models concerning two scalar fields (one phantom and the other 
canonical), non-minimally coupled two both the torsion scalar and the boundary 
term, have also been  studied using the dynamical system, finding similar results to 
the other models obtaining the correct picture of the history of the 
Universe~\cite{Bahamonde:2018miw}.

As discussed in this section, the correct use of recasting the FLRW equations  
into a dynamical system form helps in understanding the main behaviour of the 
system without the need for analytically solving the equations directly. In the 
next section, a different approach that is related to obtain analytical 
cosmological solutions will be presented.

\subsection{Noether Symmetry Approach in Teleparallel Theories of 
Gravity}\label{sec:NoetherTele}
\label{NoetherTref1}

The Einstein field equations are a system of ten partial differential equations  
and certain symmetries need to be assumed (such as spherical or cylindrical 
symmetries) in order to find analytical solutions. When they are modified, usually, the 
equations become even more involved, and for this reason, it is not so easy to find 
analytical solutions in those models (even for the maximally symmetric cases). 
One useful tool to obtain analytical solutions for a certain Lagrangian is
Noether's symmetry approach, which allows us to reduce dynamics for a certain 
model by using their symmetries and conserved quantities from the Noether theorem. 
This allows us to get an exact integration of a system  because their symmetries 
are first integrals.

The procedure to get analytical solutions using Noether's theorem is quite 
simple, but in practice it sometimes becomes a hard task. First, the point-like canonical Lagrangian associated with the studied action in 
the examined geometry needs to be written down. After this, Noether's theorem is used which can be 
stated in two parts~\cite{Noether:1918zz,Dialektopoulos:2018qoe,Paliathanasis:2015mxa}:
\begin{theorem}[\textit{Noether's theorem part one}]\label{noethertheorem1}
Let $q^i$ be some generalised coordinates in the configured space 
$\mathcal{Q}=\{q^{i}\}$ of  a non-higher derivative order Lagrangian 
$L=L(t,q^i,\dot{q}^i)$ whose tangent space is 
$\mathcal{T}\mathcal{Q}=\{q^{i},\dot{q}^i\}$ with dots representing 
differentiation with respect to the time coordinate.
The existence of a Noether symmetry that leaves the Euler Lagrange equations 
$E_i(L)=0$ associated  with the Lagrangian invariant under the transformations 
$\bar{t}=t+\epsilon \xi(t,q^i)$ and $\bar{q}^i=q^i+\epsilon \eta^i(t,q^k)$, with 
$\epsilon$ being a parameter, implies the existence of a function $g(t,q^k)$ 
which satisfies the condition
\begin{equation}
{\bf X}^{[1]} L + L \frac{{\rm d}\xi}{{\rm d}t} = \frac{{\rm d}g}{{\rm d}t}\,, \label{ngs-eq}
\end{equation}
where ${\bf X}^{[1]}$ is the first prolongation of the generator vector field 
given  by
\begin{equation}
{\bf X}^{[1]}= \xi(t,q^i) \frac{\partial}{\partial t} + \eta^i(t,q^i) 
\frac{\partial}{\partial  q^i}+\dot{\eta}^i(t,q^i) \frac{\partial}{\partial 
\dot{q}^i} ={\bf X}+\dot{\eta}^i(t,q^i) \frac{\partial}{\partial \dot{q}^i}  \,,
\end{equation}
and ${\bf X}$ is the Noether symmetry vector.
\end{theorem}
\begin{theorem}[\textit{Noether's theorem part two}]\label{noethertheorem2}
For any Noether symmetry vector $\bf{X}$ associated with the Lagrangian 
$L=L(t,q^i,\dot{q}^i)$,  there corresponds a function called the \textbf{Noether 
integral} of the Euler Lagrange equations, which is given by
\begin{equation}
    I=\Big(\dot{q}^i\frac{\partial L}{\partial \dot{q}^i}-L\Big)- 
\eta^i\frac{\partial L}{\partial\dot{q}^i}+g
\end{equation}
and is the first integral of the equations of motion ${\rm d}I/{\rm d}t=0$.
\end{theorem}
Using this theorem, one can get the conserved quantities related to the 
symmetries of  a certain theory. This method has been used in TG in the context 
of cosmology and also in the case of spherically symmetric spacetimes. It is 
important to mention that different authors have used an incomplete version of Noether's theorem just by considering the uncompleted Noether's condition 
${\bf X} L=0$ instead of the complete Noether's condition~\eqref{ngs-eq}. This 
method is also correct but it does not give all the possible symmetries of a 
model. The first work related to this method in teleparallel theories was done 
in~\cite{Wei:2011aa} for vacuum $f(\mathbb{T})$ cosmology, finding power-law 
$f(\mathbb{T})=c_1 \mathbb{T}^n$ solutions using the incomplete Noether's 
condition and then finding power-law analytical cosmological solutions 
$a(t)\propto t^{2n-3}$. Some days later, independently, other authors found the 
same results in~\cite{Atazadeh:2011aa}. Using the complete Noether's condition, 
a more detailed analysis was carried out in~\cite{Sadjadi:2012xa} and 
\cite{Basilakos:2013rua} in $f(\mathbb{T})$ cosmology, obtaining similar 
cosmological solutions to the first work. In~\cite{Jamil:2012fs} the authors 
added a minimally coupled scalar field, obtaining $f(\mathbb{T})\propto 
\mathbb{T}^{3/4}$ with a potential $V(\phi)\propto \phi^2$. One of the first 
non-trivial spherically symmetric solutions in $f(\mathbb{T})$ theories was 
found using Noether's symmetry approach in~\cite{Paliathanasis:2014iva}. 
Among these solutions, one of them behaves similarly to the Schwarzschild 
solution. For $f(\mathbb{T},B)$ gravity, in~\cite{Bahamonde:2016grb}, several 
types of power-law cosmological solutions were obtained, along with a new 
logarithmic boundary term solution $f(\mathbb{T},B)=-\mathbb{T}+(1/3) B\log B$, 
which admits a new cosmological solution $a(t)=(c_2 e^{C_1t}+3 C_{3}(t 
+c_4))^{1/3}$. The authors also made a comparison between the symmetries found 
in $f(\mathbb{T})$, $f(R)$ and $-\mathbb{T}+f(B)$ gravity. Further, in~\cite{Bahamonde:2019jkf}, new spherically symmetric exact solutions were found in $f(\mathbb{T},B)$ gravity. The modified 
teleparallel Gauss-Bonnet theory 
$f(\mathbb{T},T_{\mathcal{G}})$~\cite{Capozziello:2016eaz} and then the extended 
case with the boundary terms 
$f(\mathbb{T},B,T_{\mathcal{G}},B_{\mathcal{G}})$~\cite{Bahamonde:2018ibz} also 
found different kinds of analytical power-law types of gravity solutions for $f$, 
with some non-trivial scale factors behaving as a combination of exponential with 
power-laws or hyperbolic functions. 

Later, using Noether's symmetry approach 
for a generalised non-local teleparallel theory $(\xi \mathbb{T}+\chi 
B)f(\Box^{-1}\mathbb{T},\Box^{-1}B)$ and without assuming any condition, the 
authors found that the non-local coupling functions are constrained from the 
symmetries to be either linear combination or 
exponential~\cite{Bahamonde:2017sdo}. It is interesting to remark that some 
non-local theories were using exponential non-local coupling functions by hand 
to get renormalisable theories, but in~\cite{Bahamonde:2017sdo}, this result 
appears directly from the symmetries of the theory. Regarding teleparallel 
scalar tensor theories, there have been a large amount of papers on different 
theories, such as teleparallel dark energy or its extended version to 
non-minimally coupling 
$F(\phi)\mathbb{T}$~\cite{Sharif:2014fla,Kucukakca:2013mya}, adding non-minimal 
couplings with the boundary term and the scalar 
field~\cite{Gecim:2017hmn,Bahamonde:2016jqq}, or adding vector 
fields~\cite{Motavalli:2018ien,Tajahmad:2016bjs}, or with a fermionic 
field~\cite{Kucukakca:2014vja}, or even adding an unusual coupling 
$F(\phi,\partial_\mu \phi \partial^\mu \phi)\mathbb{T}$~\cite{Tajahmad:2017ywa}. 
In these works, new wormhole spherically symmetric solutions and also 
non-trivial cosmological solutions were found in flat FLRW and in Bianchi models 
concerning different non-trivial scale factors, along with different types of 
coupling functions $F(\phi)$ between both the torsion scalar and the boundary 
term $B$. Therefore, the teleparallel community has used Noether's symmetry 
approach in many different works in order to get cosmological and spherically 
symmetric solutions.

As an example, following~\cite{Basilakos:2013rua}, let us here briefly review  
the simplest non-trivial modified TG case, which is vacuum $f(\mathbb{T})$ 
gravity in flat FLRW cosmology. The canonical point-like Lagrangian for 
$f(\mathbb{T})$ gravity in the minisuperspace of flat FLRW can be found by 
considering the canonical variables $a,\mathbb{T}$ in the 
action~\eqref{f(T)action} and then using $\mathbb{T}=6H^2$ in flat FLRW 
cosmology. This lets us rewrite the action using Lagrange multipliers to then 
finally obtain the following point-like Lagrangian
\begin{equation}
    L=a^3(f(\mathbb{\mathbb{T}})-\mathbb{T}f(\mathbb{\mathbb{T}}))+ 
6a\dot{a}^2f_\mathbb{T}\,.
\end{equation}
Now, we replace the above point-like Lagrangian into Noether's 
condition~\eqref{ngs-eq}, finding a  set of partial differential equations for Noether's vector, and also depending on the function $f(\mathbb{T})$. For 
this model, there are seven differential equations
\begin{align}
    \quad a^{3}\left( f_{\mathbb{T}}\mathbb{T}-f\right) \xi 
_{,\mathbb{T}}=g_{,\mathbb{T}}\,,\quad    3a^{2}\eta _{1}\left( 
f_{\mathbb{T}}\mathbb{T}-f\right) +a^{3}f_{\mathbb{T}\mathbb{T}}\mathbb{T}\eta 
_{2}+a^{3}\left(
f_{\mathbb{T}}\mathbb{T}-f\right) \xi _{,t}=g_{,t}\,,\label{noether1}\\
\xi _{,a}=0\,,\quad \xi _{,\mathbb{T}}=0\,,\quad \eta _{1,\mathbb{T}}=0\,, \quad 
12f_{\mathbb{T}}a\eta _{1,t}+a^{3}\left( f_{\mathbb{T}}\mathbb{T}-f\right) \xi 
_{,a}=g_{,a}\,,\label{noether2}\\
f_{\mathbb{T}}\eta _{1}+f_{\mathbb{T}\mathbb{T}}\mathbb{T}a\eta 
_{2}+2f_{\mathbb{T}}a \eta _{1,a}-f_{\mathbb{T}}a\xi _{,t}=0\,,\label{noether3}
\end{align}
where $\eta^i \partial 
q_i=\eta_1(t,a,\mathbb{T})\partial_a+\eta_2(t,a,\mathbb{T})\partial_\mathbb{T}$. 
 We now need to solve the above system of partial differential equations to get 
the symmetries of the model. It is easy to see from \eqref{noether2} that Noether's vector is constrained to $\xi=\xi(t), \eta_1=\eta_1(a)$ and $g=g(t)$. 
If one uses these conditions into \eqref{noether3}, one gets 
$\eta_2=f_\mathbb{T} S(a,\mathbb{T})/f_{\mathbb{T}\mathbb{T}}$ with 
$S(a,\mathbb{T})$ being an arbitrary function that must be of the form 
$S(a,\mathbb{T})=M(a)+N(\mathbb{T})$ due to \eqref{noether1}. Considering all of 
these equations, one finds that $g=\textrm{constant}$, and then the above system for 
$f(\mathbb{T})\neq e^{C_1\mathbb{T}}$ is reduced to be
\begin{align}
    \frac{f_{\mathbb{T}}\mathbb{T}}{f_{\mathbb{T}}\mathbb{T}-f}=\frac{n}{n-1}\,,\\
   N=c+\xi _{,t}\,,\quad 2\eta _{1,a}+\frac{\eta _{1}}{a}+M=c\,,\quad  
3\frac{\eta _{1}}{a}+\frac{n}{n-1}M=m\,,\quad \frac{n}{1-n}N-\xi _{,t}=m\,,
\end{align}
where $m,n$ and $c$ are constants. This yields the power-law solution  $f\left( 
\mathbb{T}\right)=f_{0}\mathbb{T}^{n}$ with Noether's vector and Noether's integral for $n\neq3/2$ and $n\neq1/2$ being equal to
\begin{eqnarray}
X &=&\left( \frac{3C}{2n-1}t\right) \partial _{t}+\left( Ca+c_{3}a^{1-%
\frac{3}{2n}}\right) \partial _{a} \nonumber \\
&&+\left[ \frac{1}{n}\left( \left( c-m\right) n+3c_{3}a^{-\frac{3}{2n}%
}\right) +\frac{3C}{2n-1}+c\right] \mathbb{T}\partial _{\mathbb{T}}\,,\\
I&=&\left( \frac{3C}{2n-1}t\right) {\cal H}-12f_{0}n\left( Ca^{2}+c_{3}a^{2-\frac{%
3}{2n}}\right) \mathbb{T}^{n-1}\dot{a}\,,
\end{eqnarray}
with $\cal H$ being the Hamiltonian and $c_i$ some integration constants. The 
other cases  $n=3/2$ and $n=1/2$ have different Noether's symmetries. Finally, 
if one assumes the power-law $f(\mathbb{T})=f_{0}\mathbb{T}^{n}$ obtained by the 
symmetries, it is easy to find from the modified FLRW 
equations~\eqref{Friedmann1}-\eqref{Friedmann2} with $\rho=p=0$, that 
$a(t)\propto t^{2n-3}$ is an analytical solution of the system. Thus, even in 
vacuum, $f(\mathbb{T})$ power-law admits power-law $a(t)$ solutions. In this 
way, one is not putting the function $f(\mathbb{T})$ by hand, instead, the 
symmetries of the model based on Noether's theorem are choosing the form of 
the function. One can follow the other branch of Noether's equation and find 
that exponential $f(\mathbb{T})=f_0 e^{C_1\mathbb{T}}$ are also part of the 
symmetries of the model. For this case, one can obtain de-Sitter cosmological 
solutions from the modified FLRW equations.

\subsection{Bounce Solutions in Modified Teleparallel 
Cosmology}\label{sec:BounceTele}
\label{Bouncerefs3}

One key problem in General Relativity is the existence of both cosmological and 
black hole singularities. \label{BHref4} The $\Lambda$CDM model, which is based 
on GR, states that 
the Universe started from an unnatural initial big bang cosmological singularity 
and this cannot be alleviated without evoking new physics. This means that the 
cosmological equations break down at $t=0$ and some divergences in the curvature 
appears. In other words, there exists a singularity due to the incompleteness of 
the geodesic deviation equation \label{geodesicdevfTref1}. Possible solutions 
for 
this in cosmology are the so-called \textit{bouncing cosmological solutions}, 
which essentially are solutions describing a contraction of the Universe until a 
minimum non-zero radius to then describing an expanded Universe passing through 
a bounce. This means that generically, there needs to be a model starting with a 
contracting universe whose Hubble parameter is $H<0$, then passing to a 
continuous bounce with $H=0$, to then finalising in an expansion with $H>0$. 
There are several types of bounce solutions. Some of them have a different 
evolution, for example, exhibiting a discontinuity in the Hubble parameter. 
These solutions do not appear in GR, but it is possible to achieve them if either exotic matter or modified gravity are introduced. The simplest way to achieve 
this in GR is by introducing a quintom scalar field model with two scalar 
fields. The first scalar field behaves as a standard canonical scalar field and 
the other one has an incorrect sign in the kinetic term,  and then behaves as a 
ghost scalar field. It is unclear how physical it is to introduce these kind of 
exotic scalar fields, due to the possible existence of instabilities. If one 
modifies GR, on the other hand, it is possible to obtain bouncing solutions 
without introducing such exotic elements. Basically, the method is very simple, a form of the scale factor (or the Hubble parameter) is assumed, which can describe bounces. Then, a reconstruction technique can be used to get a 
model (or a theory) that has these kind of solutions. For a more detailed 
description about bouncing cosmological solutions, 
see~\cite{Brandenberger:2016vhg}.

\begin{figure} 
    \centering
    \includegraphics[scale=0.8]{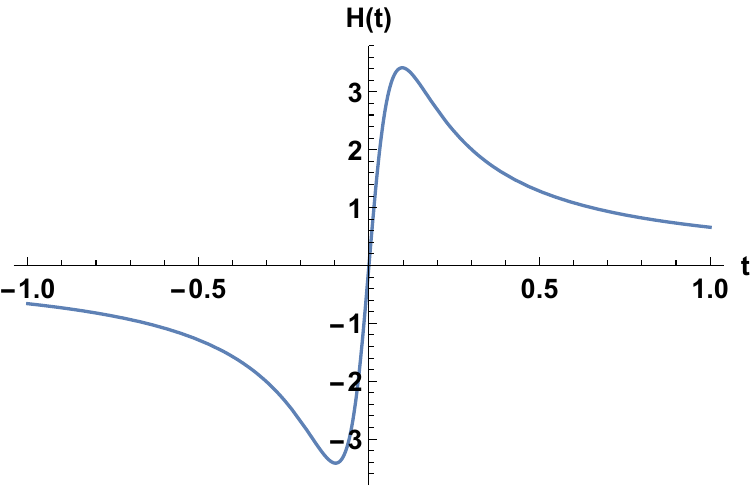}
	\caption{{\it{ Hubble parameter versus time for the model~\eqref{HHis} with 
$\sigma=70$, in units where $\kappa^2=1$,  describing a bounce behaviour.}}}
    \label{fig:bounce}
\end{figure}

In the context of TG, the first study that found bouncing cosmological solutions 
was conducted in  $f(\mathbb{T})$ gravity~\cite{Cai:2011tc}. To achieve this 
scenario, it is necessary to first check the existence of bounce solutions at the 
background level and then study the evolution of perturbations through the 
bounce. Let us here briefly review what happens at the background level. For the 
perturbation study, see more details in~\cite{Cai:2011tc}. One type of bouncing 
cosmological solution can be described by having the following scale factor
\begin{equation}
    a(t)=a_0\Big(1+\frac{3}{2}\sigma t^2\Big)^{1/3}\,,\label{scale1}
\end{equation}
where $\sigma>0$ and $a_0$ are constants. The first parameter determines how 
fast the bounce is, whereas $a_0$ is the scale factor evaluated at the bouncing 
point. The Hubble parameter then behaves as \begin{equation}
H(t)=\frac{\sigma t}{1+\frac{3}{2}\sigma t^2}\,, \label{HHis}
\end{equation}
with the time varying from $-\infty$ to $+\infty$. Fig.~(\ref{fig:bounce}) shows 
the behaviour of the  above Hubble parameter achieving the expected bounce 
scenario, starting from a matter contraction, then following with a bounce (static) 
and finalising in an expansion.  If we assume a pressureless matter, from the 
conservation equation the energy density becomes $\rho=C_1/(3 
\sigma  t^2+2)$. Since $\mathbb{T}=6H^2$, we can then find that 
$t=t(\mathbb{T})$, and we can then use this expression for the first flat FLRW 
equation~\eqref{Friedmann1}, to then solve this differential equation for 
$f(\mathbb{T})$, yielding
\begin{equation}
    f(\mathbb{T})=\mathbb{T}+\frac{1}{2}C_1 
\kappa^2\sqrt{\frac{\mathbb{T}}{\sigma}}   
\arcsin\left(\sqrt{\frac{\mathbb{T}}{\sigma}} \right)+\frac{C_1 \kappa ^2 }{2 
}\left(1+\sqrt{1 -\frac{\mathbb{T}}{\sigma}}\right)\,.
\end{equation}
The scale factor~\eqref{scale1} that gives the Hubble parameter~\eqref{HHis} 
describing a  bouncing cosmological solution is a solution of the above form of 
$f(\mathbb{T})$ gravity, which is a GR term plus some correction terms, depending 
on the scalar torsion.
There are other works finding other kinds of bounce solutions in the context of  
modified TG. In~\cite{Astashenok:2013kka}, different types of non-singular 
bounce solutions were found in $f(\mathbb{T})$, such as $\Lambda$CDM with a 
bounce in the past, with the scale factor behaving as $a(t)\propto 
\sinh[\sqrt{t^2+\tau^2}/t_0]$ or future bounces with Quasi-rip and Little-rip 
behaviours. In~\cite{Odintsov:2015uca} some superbounce 
solutions in $f(\mathbb{T})$ gravity that have a ekpyrotic contracting phase 
preventing large anisotropies are found. The authors also made some comparisons between 
other modified models starting from GR, such as $f(R)$ and $f(\mathcal{G})$ 
gravity. Later, it was found that in $f(\mathbb{T})$ gravity it is possible to have 
a bounce inflation model with a graceful decelerated exit~\cite{Bamba:2016gbu}. 
Using dynamical systems techniques, in \cite{Hohmann:2017jao}, the authors also 
found the possibility of bounces and turnaround solutions in $f(\mathbb{T})$ 
gravity but the impossibility of cyclic and oscillating universes. 
In~\cite{Qiu:2018nle}, the authors studied a model with $f(\mathbb{T})$ and a 
scalar field to then perform perturbations for the bounce inflationary models, 
finding that it is difficult to obtain a stable bounce inflation solution, since 
there are many conditions that the models need to satisfy in order to have this property, though, they were able to show that a combination of power-law types of 
$f(\mathbb{T})$ can achieve these stable bounce inflation solutions. If one 
understands the cosmological singularities as a break down of GR at very high 
energies, then some studies have argued that Loop quantum cosmology could 
alleviate them. Some works 
like~\cite{Haro:2014wha,Amoros:2013nxa,deHaro:2017yll} have also found bounces 
in a toy model in $f(\mathbb{T})$ gravity, assuming Loop quantum cosmological 
considerations. Further generalisations to $f(\mathbb{T})$ gravity, such as 
considering the teleparallel Gauss-Bonnet invariant $T_{\mathcal{G}}$ in the 
so-called $f(\mathbb{T},T_{\mathcal{G}})$ gravity, have also been used to 
analyse possible bounce solutions, finding five types of 
them~\cite{delaCruz-Dombriz:2018nvt,delaCruz-Dombriz:2017lvj}. In~\cite{Fiorini:2013kba} the case of 
Born-Infeld gravity was studied. These kind of theories are 
achieved from $f(\mathbb{T})$ gravity with 
$f(\mathbb{T})=\frac{\lambda}{2\kappa^2}((1+\frac{\mathbb{T}}{2\lambda})^{1/2}
-1)$, with $\lambda$ being a parameter of the theory that becomes important at 
high energies. In this paper, the author found that at high energy regimes, the 
big bang singularity is absent, and either a de-Sitter inflationary stage of 
geometrical character or a bounce is present. Later, 
in~\cite{Bouhmadi-Lopez:2014tna}, they showed that this behaviour is only valid 
for a certain region of the parameter space, and other singularities such as Big 
Rip, Big Bang, Big Freeze, and Sudden singularities can emerge. It is 
interesting to mention that recently, in~\cite{Boehmer:2019uxv}, a regular 
Schwarzschild black hole solution was found in this theory. In this context, the 
central Schwarzschild curvature singularity is replaced by an infinitely long 
cosmic string related to the parameter $\lambda$.

As was seen in the previous example and also in the papers mentioned here, it is not so complicated to get bouncing cosmological solutions in modified TG. It 
is important to emphasise that in standard GR, this cannot be achieved without 
introducing matter that violates the energy conditions.

\section{What Can Teleparallel Theories Have to Offer? What Are the Open 
Problems in Teleparallel Theories?}

Teleparallel theories give an alternative starting point for understanding  
gravity compared with the standard curvature-based approach to gravity used 
in GR. Some interesting features about them compared to GR are:
\begin{itemize}
    \item \textit{Modified teleparallel theories are broader than standard 
modified  theories starting from GR:}

	We have seen that theories equivalent to GR, scalar-tensor gravity,  
Horndeski theory, $f(R)$ and Gauss-Bonnet gravity based on the teleparallel 
geometry exist are subclasses of more general families of Teleparallel 
Gravity theories. In this spirit, Teleparallel Gravity may be used to formulate 
broader classes of theories than could be constructed by using only the 
metric and its Levi-Civita connection. This should not be too surprising, since 
any theory of the latter type can be rewritten into teleparallel language by 
using the relations~\eqref{eq:metric}, \eqref{eq:contortion} 
and~\eqref{eq:lccurv} to replace the metric, its Levi-Civita connection and the 
corresponding curvature by the tetrad and the contortion, and hence the torsion. 
This teleparallel formulation then gives rise to further extensions and 
generalisations by splitting off boundary terms or including additional terms 
that have no equivalent expression in terms of the metric geometry only. An 
easy example would be to take the case of $f(R)$ and $f(\mathbb{T})$ gravity. If 
we start by formulating a theory from GR, then torsion is zero and it is not 
possible to derive a similar theory to $f(\mathbb{T})$. On the other hand, if 
we start from TG, we can generalise $f(\mathbb{T})$ to $f(\mathbb{T},B)$ and 
then we can formulate a teleparallel equivalent version of $f(R)$ by 
considering $f(\mathbb{T},B)=f(-\mathbb{T}+B)=f(R)$. 

	\item \textit{It is easy to get second-order field equations, since torsion 
contains  only first-order derivatives:}

	Since most physical phenomena are described by up to second-order field 
equations,  it is expected that gravity will not have a behaviour different to 
this. Moreover, higher-order field equations often lead to ghost instabilities. 
In Teleparallel Gravity it is easy to get second order field equations because 
the torsion tensor contains only first-order derivatives of the tetrad, in 
contrast with the Riemann tensor that contains second-order derivatives in the 
metric. This means that in modifications like $f(\mathbb{T})$ or even more 
complicated ones like $f(T_{\rm ax}, T_{\rm vec}, T_{\rm ten})$, BDLS, etc, the 
field equations will be of second order, unlike in, e.g., $f(R)$ theory where they 
are fourth-order in all cases beyond GR. In principle, one could construct a 
theory with infinite contractions of the torsion tensor and still get second-order field equations.

	\item \textit{Similarity to Yang-Mills theory - possible connections to gauge 
 theory and particle physics:}

	The TEGR action in Eq.\eqref{eq:tegraction} has an obvious similarity with a 
Yang-Mills  type action, since it is quadratic in the first-order derivatives of 
the tetrad, the latter being the fundamental dynamical field of the theory. This 
means that has more in common with actions encountered in particle physics 
than it is the case for the Einstein-Hilbert action. Furthermore, the potential 
interpretation as a gauge theory of translations makes it more similar to 
theories describing the other fundamental interactions in Nature, hence paving a 
potential road map towards a unification of gravity and other forces. While 
still being non-renormalisable, it may provide an alternative and possibly more 
promising starting point for constructing a UV complete theory.

    \item \textit{The possibility of formulating the theory without the 
equivalence  principle:}  \label{equivprinref4}
   In GR, gravitation is characterised by curvature 
through the Ricci scalar, which then acts on particles through the geodesic 
equation. TG is wholly different in that the geometric torsion that is produced 
to expressed gravity, acts through a Lorentz force-type equation. In this way, 
we recover the original concept of gravitation as a force similar to the other 
fundamental forces of Nature. While the equations turn out to be equivalent for 
TEGR, they differ for modifications to this model. Coupled with the potential of 
being described as a gauge theory of translations, TG does not rely on the weak 
equivalence principle in that it would survive a violation 
\cite{2011SHPMP..42..264K,Aldrovandi:2013wha}.  However, in these settings, the 
Newtonian limits remain intact resulting in a more natural weak-field limit 
compared with GR.

    \item \textit{Defining a gravitational energy-momentum tensor:}
    
    TG offers the possibility of describing gravitation as a gauge current  
\cite{Aldrovandi:2013wha,Blixt:2018znp,Blixt:2019mkt}. Then, by separating 
inertial and gravitation, it may become possible to define an energy-momentum tensor 
for gravitation. However, this feature of the theory may require selection of a 
Lorentz frame, which would limit the applicability of this result 
\cite{Jimenez:2019yyx,Koivisto:2019ejt}.
    
    \item \textit{The regularity of the TEGR action. One does not need to 
introduce  a Gibbons-Hawking-York boundary term:}

	One may argue that the TEGR action in~Eq.\eqref{eq:tegraction} is a  more 
natural formulation of the dynamics of GR than the Einstein-Hilbert action, 
since the latter also contains second order derivatives of the metric. The 
latter mandates the inclusion of a Gibbons-Hawking-York boundary term in order 
to possess a well-defined variational problem and Hamiltonian formulation, which 
is not the case for TEGR. While this does not affect the classical equations of 
motion, it may affect the quantum behavior and thermodynamics of black holes. Further, it was found in~\cite{Hammad:2019oyb} that the black hole entropy is more naturally expressed as a volume integral in TEGR than in GR.

    \item \textit{Torsional or Curvature-based Gravity}:
    
TEGR is dynamically equivalent to GR in that the identical field equations 
emerge  due to the respective Lagrangian densities being equal up to a boundary 
term. However, the action may contain further information, such as the 
renormalisability of the theory, as well as the symmetries of the theory. 
Moreover, modifications to TEGR offer a much richer avenue to construct new 
theories, due to the second-order nature of the torsion scalar 
\cite{Ferraro:2012wp}.
\end{itemize}

\noindent Although teleparallel theories have attained a lot of attention in the 
 past ten years, the community effort is still not comparable to that in GR. 
For this reason, there are several open problems/questions that have not been 
addressed properly in the literature. Some of these open problems are:
\begin{itemize}
    \item \textit{Confrontation with cosmological data}: Teleparallel  theories 
have shown great promise for consistently describing observations at cosmic 
scales while also satisfying Solar System and astrophysical-scale physics 
\cite{Nunes:2018xbm,Nunes:2016plz,Nunes:2019bjq,Nunes:2018evm,Nesseris:2013jea,
Oikonomou:2016jjh,Capozziello:2017bxm,Nunes:2016qyp,Cai:2018rzd,
Anagnostopoulos:2019miu,Geng:2011ka,Basilakos:2018arq}. Further work needs to be 
done in this  \label{solarsystemref5}
direction to better understand the explaining power of the theory in its various 
manifestations.
    \item \textit{Gravitational Waves:} \label{gravitationalwavrefs8} A strong 
effort has been made in  
regards to understanding the cosmological consequences of gravitational waves in 
$f(\mathbb{T})$ gravity 
\cite{Cai:2018rzd,Li:2018ixg,Nunes:2019bjq,Nunes:2018evm}. However, there 
remains many other contexts in which TG has shown promise, and more than this, 
further work is needed in the astrophysical context. In 
Refs.~\cite{Abedi:2017jqx,Farrugia:2018gyz,Hohmann:2018jso,Bamba:2013ooa}, it is 
shown that $f(\mathbb{T})$ gravity contains only two propagating modes, while 
other variants of TG can provide more. It would be interesting to investigate 
further the waveform for astrophysical events, which would entail obtaining 
further information about the $3+1$ formalism of the theory and performing 
simulations of astrophysical events. Ultimately, it is crucial to understand how 
current and future observatories can be exploited to test TG.
    \item \textit{The Galactic Rotation Curve Problem}: $f(\mathbb{T})$ gravity  
has shown positive results for describing the rotation curves without the 
requirement of adding dark matter \cite{Finch:2018gkh}. Weak lensing tests have 
also contributed to constraints on one of the potential models of the theory 
\cite{Chen:2019ftv}. However, further analysis needs to be done on considering 
more cosmologically inspired models, as well as increasing the breadth of the 
analysis. One possibility is to use the SPARC (Spitzer Photometry \& Accurate 
Rotation Curves) obtained in Ref.~\cite{2017ApJ...836..152L}.
    \item \textit{Teleparallel quantum gravity}: The majority of quantum gravity 
 approaches have been attempted with GR as the starting point, but not so many 
have been initiated from TG. For example, the loop quantum corrections of the 
TEGR action~\eqref{eq:tegraction} are not so well-known. One of the first attempts was done in~\cite{Dupuis:2019unm}. In GR, this has been 
studied with a lot of effort, even obtaining corrections of an action with 
$\sum_{i=1}^{70}R^i$~\cite{Falls:2018ylp}. This has been studied within the 
asymptotic safety approach to quantum gravity~\cite{Niedermaier:2006wt} that 
 has not also been applied to Teleparallel Gravity. Another interesting route 
would be to formulate a teleparallel version of loop quantum gravity, since both 
formulations use a similar mathematical language.
     \item \textit{What are singularities in Teleparallel Gravity?}: Hawking  
and Penrose proposed a way to define singularities in GR, which is based on the 
incompleteness of the geodesic equation. Then, a simple way to quantify these 
singularities is by defining some invariants such as the Kretschmann scalar 
$K\equiv R^{\mu\nu\alpha\sigma}R_{\mu\nu\alpha\sigma}$. In TG, it is not clear 
whether the Hawking-Penrose theorem would be the same, or if one would need to 
define different invariants to check if there are singularities. Then, it is 
still open as to  how a singularity in the context of TG  can be defined. 
\end{itemize}


 \chapter[Finsler Gravity]{Finsler Gravity}
\label{Voicuchapter}

{\em Nicoleta Voicu, Christian Pfeifer}\\

This chapter summarises some recent developments in the application of
Finsler geometry as the extended geometry of spacetime. As original sources, we
refer to the articles \cite{Hohmann:2018rpp,Pfeifer:2019wus, Hohmann:2019sni}.

\section{Physical Motivations}

Finsler geometry appears in the description of physical systems, as a
suitable mathematical tool at various stages. We will briefly discuss the
appearance of Finsler geometry in physics and motivate the formulation of a
consistent extension of General Relativity based on Finsler geometry.

\subsection{Finsler Geometry in Physics}
\label{sec:FGIP} 
\label{Finslergrref1}

In physics, Finsler geometry naturally appears in the study
of dispersion relations, as the most general geometric realization of the
clock postulate of General Relativity and as a minimal modification of the
Ehlers-Pirani-Schild (EPS) axiomatic approach to spacetime geometry.

A {\it{dispersion relation}} is a constraint that  the position $x$ and the
four-momentum $p$ of a particle have to satisfy, so that the particle is
physically viable. Technically, the dispersion relation is described by a
Hamilton function $H$, depending on $x$ and $p$ (i.e.,
defined on   the point particle phase space, or the cotangent bundle of 
spacetime). The level
sets $H(x,p)=const$ represent the dispersion relation and the Hamilton
equations of motion $\dot{x}^{\mu}=\partial _{p_{\mu}}H$ and $\dot{p}%
_{\mu}=-\partial _{x^{\mu}}H$ determine the particle trajectory. Dispersion
relations lead to a Finslerian spacetime geometry, when the Hamilton
function is mapped to a Finsler function via the Helmholtz action. This
procedure has been pointed out in \cite{Raetzel:2010je} for general
dispersion relations and has been applied to the $\kappa $-Poincar\'{e}
dispersion relation used in quantum gravity 
\cite{Amelino-Camelia:2014rga,Lobo:2020qoa},
as well as to weak premetric-, or minimal standard model extension
electrodynamics \cite{Gurlebeck:2018nme}. The resulting Finsler geometries
are candidates to extend the geometry of spacetime beyond pseudo-Riemannian
Lorentzian one. Non-quadratic dispersion relations, resulting in a 
non-pseudo-Riemannian Finsler geometry, are used in the study of physical fields
and waves propagating through media 
\cite{Gibbons:2011ib,Yajima2009,MARKVORSEN2016208}  they emerge in the context 
of quantum gravity
phenomenology from quantum deformations of the Poincar\'{e} algebra 
\cite{Barcaroli:2017gvg,Amelino-Camelia:2014rga} and in the study of non-local 
Lorentz invariant physics, such as in the standard model extension or
the very special relativity framework \cite{Kostelecky:2011qz}, \cite%
{Cohen:2006ky}.

One of the fundamental pillars of Special and General Relativity is the 
{\it{clock postulate}}: "The time which passes for an observer between two
events is given by the length of its worldline connecting these two events".
To realize this physical axiom, mathematically the spacetime manifold must
be equipped with a length measure for curves. The most general geometric
length measure for curves is given by a generic Finsler length measure.
Among all possible Finsler length measures, the one induced by a Lorentzian
metric is singled out by demanding invariance under local Lorentz
transformations. Relaxing the symmetry demand, more general length measures can 
be obtained, for example,   the one found by Bogoslovsky in the study 
of
transformations that leave the massless wave equation invariant 
\cite{Bogoslovsky1977}. Later, field theories built upon this symmetry were
summarised under the name Very Special/Very General Relativity 
\cite{Cohen:2006ky,Gibbons:2007iu,Fuster:2018djw}. Further interesting length
measures can be identified as they appear in the description of point
particle trajectories in certain physical situations, such as, for example:
the Randers length measure \cite{Randers}, describing a particle in an
electromagnetic field as well as the Zermelo navigation problem and the 
influence of wind on a physical system in general 
\cite{Gibbons:2011ib,MARKVORSEN2016208,Javaloyes:2018lex}, or $m$-th root length
measures, which describe the point particle limit of quantisable
bi-hyperbolic field equations \cite{Raetzel:2010je} - and, in particular for 
$m=4$, the propagation of light on the basis of premetric or area metric
electrodynamics~\cite{Punzi:2007di}.

In their {\it{axiomatic approach to spacetime geometry}}, Ehlers, Pirani
and Schild deduce a pseudo-Riemannian (more precisely, Lorentzian) spacetime
geometry from a set of physical demands \cite{Ehlers2012}. One of the axioms
is, however, of a  rather   technical mathematical nature than  is physically
necessary. It states that for a sufficiently small neighborhood $V$ for
every $p\in V,$ the map $g_{p}:p\mapsto t_{e}t_{a}$, which associates the
product between the emission time $t_{e}$ and the return time $t_{a}$ of
radar echo between any particle worldline passing through $V$ (but not
through $p)$ and $p$, must be at least twice differentiable. Tavakol and van
den Bergh demonstrated that  if one relaxes this demand and only requires $%
g_{p}$ to be   differentiable once, the EPS spacetime axiomatic leads to
a Finslerian spacetime geometry \cite{Tavakol2009}. The EPS axiomatic has been 
reviewed from a  Finsler spacetime perspective in \cite{Bernal:2020bul}.

\subsection{Finsler Gravity}
\label{sec:FGrav}

The natural appearance in physics and its close relation
to Special and General Relativity make Finsler geometry one of the
candidates for a description of the gravitational interaction, beyond
General Relativity. It has the potential to shed light on the shortcomings
of Einstein's theory of gravity, such as dark energy and dark matter \cite%
{Clowe:2006eq,Corbelli:1999af,Peebles:2002gy,Riess:1998cb}, from a
geometric point of view.

The idea to use Finsler geometry as geometry of spacetime has been   long
known in the literature - and is ongoing \cite%
{Beem,Rutz,Pfeifer:2011tk,Javaloyes:2018lex,Hohmann:2018rpp,Pfeifer:2011xi}. The
difficulty in the construction of a consistent theory of Finsler gravity is,
on   one hand, to find a precise definition of Finsler spacetimes, and  on
the other hand, to prove that Finsler spacetime geometry realises the
threefold role of the geometry of spacetime: encoding a causal structure,
via the identification of causal (time-like and lightlike) directions,
encoding the description of observers and their measurements, and encoding
the gravitational interaction and its dynamics. In particular, the coupling
between Finsler geometry and physical matter is a point of debate in the
literature. The question is whether a  scalar, or a tensor Finsler
gravitational field equation on the tangent bundle,\label{Finslcobundlelref1} 
should be employed in order 
to
determine the (scalar) Finsler function - or, accordingly, the Finsler metric 
tensor,
and how the matter source side term for the respective equation can be
constructed. 
Another possibility is  that instead of giving dynamics to the Finsler function
as a whole, to give dynamics to various tensor fields on the base
manifold, which then compose a Finsler length element \cite{Basilakos:2013hua}. 
Recently, strong arguments in favour of a scalar
field equation -- sourced by a kinetic gas -- have been 
given~\cite{Hohmann:2019sni,Hohmann:2018rpp}. \label{Finslskineticrref1}

Indeed, it is possible to realise this threefold role of the geometry of
spacetime in physics with Finsler geometry \cite{Pfeifer:2019wus}. The
existence of a causal structure is ensured by the definition of Finsler
spacetime, which we will discuss in Section~ \ref{sec:FSTDef}; observers and
their time and length measurements can be constructed with the help of a
radar experiment~\cite{Gurlebeck:2018nme,Pfeifer:2014yua}, and we will
present the description of the gravitational field in term of Finsler
geometry, directly sourced by a kinetic gas, in Section~\ref%
{sec:kingasgravity}.

\subsection{Finsler Cosmology}

Applying the cosmological principle to Finsler spacetime geometry restricts our 
study to Finsler spacetime geometries that are spatially homogeneous and 
isotropic \cite{Hohmann:2020mgs}. On these spacetime geometries, the dark 
energy phenomenology becomes particularly visible. Finsler geometry can address 
the dark energy issue on two levels: in the description of light propagation 
and through modified gravitational dynamics.

The accelerated expansion of the Universe reveals itself by extracting the
so-called deceleration parameter from the magnitude-redshift relation of
distant supernovae. A negative deceleration parameter implies an accelerated
expansion of the Universe. The precise prediction of the deceleration
parameter depends on the redshift, induced by the background geometry, and
on the magnitude of the light signal observed at a distance from its
emission. Using a Finslerian spacetime geometry, the redshift differs from
the one obtained on the basis of General Relativity, due to Finslerian
deformation of the light cone structure. The general Finslerian redshift
formula has been obtained in \cite{Hasse:2019zqi}.  Explicit expressions for 
the deceleration parameter have been derived
for several Finsler geometry models, such as Randers geometry, $4th$-root
geometry and general Finsler perturbations of
Friedmann-Lemaitre-Robertson-Walker geometry \cite{Hohmann:2016pyt}. A
future perspective is to determine the coefficients in the deceleration
parameter, obtained in the perturbative approach, from the observed value of
this parameter.

To fully address dark energy with Finsler geometry, the kinematic analysis
of the magnitude-redshift relation does not suffice. It is furthermore
necessary to derive the Finslerian version of the Friedmann equations  in
order to predict the evolution of the Universe. For the earlier mentioned
scalar field equation coupled to a kinetic gas, this task is work in
progress -- only preliminary results in the context of the so-called Very
General Relativity exist \cite{Fuster:2018djw}. For other strategies to
determine a Finslerian length element dynamically, more progress has already
been made. For example, in Randers geometries, which are Finsler geometries
built from a Lorentzian metric $g$ and a $1$-form $B$ on spacetime,
resembling dark energy can be obtained in a geometric manner \cite%
{Basilakos:2013hua}; this is achieved by determining the Lorentzian metric $%
g $ via the usual Einstein equations and coupling the additional $1$-form
via the Finslerian geodesic deviation equation. More generally, in the
context of non-Lorentz invariant cosmological models, it has been shown that
there exist Finsler models which predict a cosmological bounce \cite%
{Minas:2019urp} and that Finsler-inspired generalized geometries give a
geometric picture of scalar tensor cosmologies \cite{Ikeda:2019ckp}.

The aforementioned results merely represent   an incomplete excerpt of the
literature on the explicit applications of Finsler geometry in cosmology.

\section{Definition of Finsler Spacetimes}
\label{sec:FSTDef}

Before defining Finsler spacetimes, we briefly recall the
definition of classical, positive definite Finsler spaces, which are
straightforward generalisations of Riemannian ones. The idea of equipping a
manifold with a non-quadratic length measure for vectors -- accordingly, of
curves -- goes back to Riemann himself \cite{Riemann}, but only Finsler
analysed such spaces systematically \cite{Finsler}. Since then, Finsler
geometry became an established field in mathematics \cite{Bao}.

Roughly speaking, a Finslerian manifold, respectively a Finslerian spacetime, 
is a
smooth manifold $M$ equipped with a length measure for curves $L(x,\dot{x}),$
where the function $L$ is just 2-homogeneous in the tangent vectors $\dot{x}$
(i.e., not necessarily quadratic, as in the case of pseudo-Riemannian
manifolds).

We introduce the following notations on the tangent bundle $TM$ of a
connected, orientable smooth manifold $M$. Any local coordinate chart $%
(U,x^{\mu})$ on $M$ induces a local coordinate chart $(TU,$ $x^{\mu},\dot{x}%
^{\mu}) $ on $TM$ as follows. For an element $(x,\dot{x})\in TU,$ 
$(x^{\mu},\dot{%
x}^{\mu})$ are given by the coordinates $(x^{\mu})$ of the point $x\in U$ and
the decomposition $\dot{x}=\dot{x}^{\mu}\partial _{\mu}|_{x}$ of the vector $%
\dot{x}\in T_{x}M$ in the natural local basis. If there is no risk of
confusion, we will sometimes omit the indices on the coordinate
representation, i.e., we write briefly $(x,\dot{x})$ for 
$(x^{\mu},\dot{x}^{\mu})
$. The natural coordinate bases of the tangent and cotangent spaces of $M$
are denoted by $(\partial _{\mu}=\frac{\partial }{\partial 
x^{\mu}},\dot{\partial%
}_{\mu}=\frac{\partial }{\partial \dot{x}^{\mu}})$ and 
$(dx^{\mu},d\dot{x}^{\mu}),$
respectively.

\subsection{Positive Definite Finsler Manifolds}

A Finslerian manifold $(M,F)$ is a smooth $n$-dimensional manifold $M$
equipped with a continuous \textit{Finsler norm} $F:TM\rightarrow (0,\infty)%
;(x,\dot{x})\mapsto F(x,\dot{x})$, which satisfies:

1)$\ F$ is positively $1$-homogeneous in $\dot{x}:$ $F(x,\ell \dot{x}%
)=\ell F(x,\dot{x}),~\ \forall \ell \in \mathbb{R}^{+}$.

2)$\ F$ is smooth on $TM\backslash \{0\}$.

3)\ The Finsler metric, defined by the Hessian of $L:=F^{2}$ with respect to 
$\dot{x}^{\mu}$, 
\begin{equation*}
g_{\mu\nu}^{F}=\frac{1}{2}\dot{\partial}_{\mu}\dot{\partial}_{\nu}L\,,
\end{equation*}%
is positive definite.

As a consequence of the above axioms, $F$ obeys (pointwise) the triangle
inequality: $F(x,\dot{x}+\dot{\bar{x}})\leq F(x,\dot{x})+F(x,\dot{\bar{x}})$.
Moreover, the length measure for curves $\gamma :[a,b]\rightarrow M,$ $\tau
\mapsto \gamma (\tau )$ on $M$: 
\begin{equation}\label{eq:FinAct}
S[\gamma ]=\overset{b}{\underset{a}{\int }}F(\gamma (\tau ),\dot{\gamma}%
(\tau ))d\tau
\end{equation}%
is independent of the choice of the parameter $\tau $ along $\gamma .$

Riemannian geometry is obtained as a particular case, for $F(x,\dot{x})=%
\sqrt{g_{\mu\nu}(x)\dot{x}^{\mu}\dot{x}^{\nu}}$ i.e., for quadratic functions 
$L(x,%
\dot{x})=g_{\mu\nu}(x)\dot{x}^{\mu}\dot{x}^{\nu}$ in $\dot{x},$ where 
$g_{\mu\nu}$
define a Riemannian metric. In this case, $g_{\mu\nu}=g_{\mu\nu}^{F}$ only 
depend
on the point $x$ (which is not true for general Finsler metrics).

The geometry of a Finsler manifold can be derived from the Finsler function $%
F$ and the Finsler metric $g^{F}$, in very close analogy to the derivation
of the geometry of a Riemannian manifold from its metric $g$.

\subsection{Finsler Spacetime}

Changing from positive definite Finsler manifolds to Lorentzian Finsler
spaces, there arises the problem that, typically, the Finsler function can not 
be
smooth everywhere on $TM\backslash \left\{ {0}\right\} $, due to the
existence of non-trivial zero length vectors, i.e., null vectors. We have to
carefully impose further restrictions upon $L$ in order to ensure the
existence of a well-defined causal structure. 
To formulate the definition in a precise way, we need the notion of conic
subbundle.

Let $\overset{\circ }{TM}:=TM\backslash \{0\}$ be the tangent bundle without 
the zero section $x\mapsto (x,0)$. A conic subbundle of $TM$ is 
 a non-empty open submanifold $\mathcal{Q}%
\subset \overset{\circ }{TM}$, with $\pi _{TM}(\mathcal{Q})=M,$ possessing
the so-called \textit{conic property:} if $(x,\dot{x})\in \mathcal{Q}$,
then, for any $\ell >0:$ $(x,\ell \dot{x})\in \mathcal{Q}$ 
\cite{Javaloyes:2018lex}.

\begin{definition}
\label{def:FST}By a Finsler spacetime we will understand in the following a pair 
$(M,L)$, where $M$ is a smooth $n$-dimensional manifold and the Finsler 
Lagrangian $L:\mathcal{A} \to \mathbb{R}$ is a smooth function on a conic 
subbundle $\mathcal{A}\subset TM$, such that:
\begin{itemize}
	\item $L$ is positively homogeneous of degree two with respect to $\dot x$: 
$L(x,\lambda \dot x) = \lambda^2 L(x,\dot x)$ for all $\lambda \in 
\mathbb{R}^+$;
	\item on $\mathcal{A}$, the vertical Hessian of $L$, called $L$-metric, is 
non-degenerate,
	\begin{align}\label{g_def}
		g^L_{\mu\nu}=\dfrac{1}{2}\dfrac{\partial ^{2}L}{\partial 
\dot{x}^{\mu}\partial \dot{x}^{\nu}};
	\end{align}
	\item there exists a conic subbundle $\mathcal{T}\subset \mathcal{A}$ such 
that on $\mathcal{T}$: $L>0$, has Lorentzian signature $(+,-,-,-)~$ and, on the 
boundary $\partial \mathcal{T}$, $L$ can be continuously extended as 
$L|_{\partial \mathcal{T}} = 0$.\footnote{It is possible to equivalently 
formulate this property with the  opposite sign of $L$ and the  metric $g^L$ of 
signature $(-,+,+,+)$. We fixed the signature and sign of $L$ here to simplify 
the discussion.}
\end{itemize}
This is a refined version of the definition of Finsler spacetimes in 
\cite{Hohmann:2018rpp} and basically covers, if one chooses $\mathcal{A} = 
\mathcal{T}$, the improper Finsler spacetimes defined in \cite{Bernal:2020bul}.
\end{definition}

The non-degeneracy, respectively the signature condition on $g^{L}$, are well
defined, i.e., they do not depend on the choice of the local chart on $TM.$
Moreover, the 2-homogeneity of $L$ implies: 
$L(x,\dot{x})=g_{\mu\nu}^{L}(x,\dot{x%
})\dot{x}^{\mu}\dot{x}^{\nu}.$

We notice the presence of four conic subbundles of $TM$:

\begin{itemize}
\item $\mathcal{A}$, called the set of \textit{admissible vectors,}\emph{\ }%
is the subbundle where $L$ is defined, smooth and $g^{L}$ is nondegenerate.

\item $\mathcal{N} = L^{-1}(0)$ is the set of \textit{null or lightlike 
}vectors.

\item $\mathcal{A}_{0}=\mathcal{A}\backslash \mathcal{N}$ is the set of
admissible, non-lightlike vectors; it is the subset of $TM$ where $L$ can be
used for homogenizing geometric objects.

\item $\mathcal{T}$ is a maximally connected conic subbundle where $L>0$,
the $L$ metric exists and has Lorentzian signature $(+,-,-,-)$. This will be
interpreted as the set of \textit{future pointing time-like vectors, }which
are allowed as tangent vectors to the trajectories of physical observers.
The fibres $\mathcal{T}_{x}:=\mathcal{T}\cap T_{x}M$ are convex cones.
\end{itemize}

We note that $\mathcal{T\subset A}_{0}\subset \mathcal{A},$ but $\mathcal{N}$
is not necessarily contained in $\mathcal{A}.$ 

The set 
\begin{equation*}
\mathcal{O}:=\left\{ \left( x,\dot{x}\right) \in \mathcal{T}~|~L(x,\dot{x}%
)=1\right\} ,
\end{equation*}%
of normalized future-directed time-like vectors, is called the \textit{%
observer space }and\textit{\ }is interpreted as the set of admissible
4-velocities of massive particles.

\bigskip

Our definition includes large classes of Finsler spacetimes according to the
older definitions. It allows, for example, Finsler spacetime geometries of:
\label{Finslspacetibeccigrref1}

\begin{itemize}
\item Randers type: $L=\epsilon 
(\sqrt{|g_{\mu\nu}\dot{x}^{\mu}\dot{x}^{\nu}|}+A_{\sigma}%
\dot{x}^{\sigma})^{2}$, where $\epsilon 
=sign(\sqrt{|g_{\mu\nu}\dot{x}^{\mu}\dot{x}^{\nu}|%
}+A_{\sigma}\dot{x}^{\sigma});$

\item Bogoslovsky/Kropina type $L=\epsilon |g_{\mu\nu}\dot{x}^{\mu}\dot{x}%
^{\nu}|^{1-q}(A_{\sigma}(x)\dot{x}^{\sigma})^{2q},$ $\epsilon 
=sign(g_{\mu\nu}\dot{x}^{\mu}%
\dot{x}^{\nu});$

\item polynomial $m$-th root type $L=\epsilon |P|^{\frac{2}{m}},$ where $%
P=G_{\mu_{1}\cdots \mu_{m}}(x)\dot{x}^{\mu_{1}}\ldots \dot{x}^{\mu_{m}},$ 
$\epsilon
=sign(P).$
\end{itemize}

\begin{figure}[h!]
	\centering
	\includegraphics[width=0.79\textwidth]{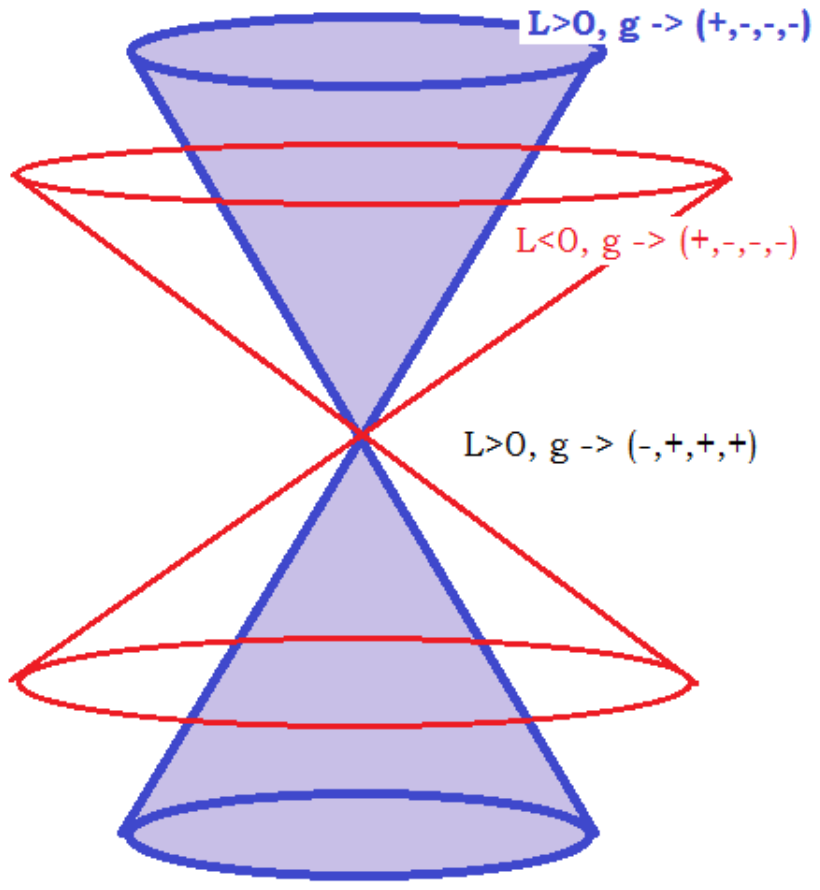}
	\caption{{\it{The local causal structure of a bi-metric Finsler 
spacetime.}}}
	\label{fig:bimetricnull}
\end{figure}

To illustrate the interesting causal structures, which can be described
geometrically in terms of Finsler geometry, Fig. \ref{fig:bimetricnull} 
presents the
situation for a polynomial 4-th root metric $L(x,\dot{x})=\epsilon \sqrt{%
\left\vert 
(g_{\mu\nu}\dot{x}^{\mu}\dot{x}^{\nu})(h_{\sigma\rho}\dot{x}^{\sigma}\dot{x}%
^{\rho})\right\vert }$ (where $g$ and $h$ are Lorentzian metrics). Such
bimetric geometries naturally appear in the study of birefringent crystals.
The null set $\mathcal{N}_{x}$ of $L$ at $x\in M$ is the union of two cones;
the set $\mathcal{T}_{x}$ is the interior of the sharper cone (in blue). We
notice that $g^{L}$ is allowed to change signature on $TM;$ but, on $%
\mathcal{T},$ the sign of $L$ and the signature of $g^{L}$ must agree. $L$
is, obviously, not smooth on $\mathcal{N}$; yet, it can be proven, \cite%
{Pfeifer:2011xi,Pfeifer:2011tk}, that the geodesic equation coefficients admit a 
$%
\mathcal{C}^{\infty }$-smooth prolongation to $\mathcal{N}$.

A particularly important class of Finsler spacetime metrics is represented
by the ones with cosmological symmetry, i.e., spatial homogeneity and
isotropy. Denoting by $(t,r,\theta ,\varphi ,\dot{t},\dot{r},\dot{\theta},%
\dot{\varphi})$ the coordinates on $TM$ induced by the
local spherical coordinates $(t,r,\theta ,\varphi )$ on the spacetime manifold 
$M$, the most
general cosmologically symmetric Finsler-Lagrange function $L:TM\rightarrow 
\mathbb{R}$ is locally described as,  \cite{Pfeifer:2011xi,Hohmann:2016pyt, 
Hohmann:2020mgs},
as:%
\begin{equation*}
L=L(t,\dot{t},w), \textrm{ where } 
w^2:=\frac{\dot{r}^{2}}{1-kr^{2}}+r^{2}\dot{\theta}^{2}+r^{2} 
\dot{\varphi}^{2}\sin ^{2}\theta. 
\end{equation*}%
This implies that the cosmological models for the most studied Finsler 
geometries in the
literature take the form:

\begin{itemize}
\item Randers type: $L=\epsilon \sqrt{|-n(t)\dot{t}^{2}+a(t)w^{2}|}+A_{t}(t)%
\dot{t};$

\item Bogoslovsky/Kropina type: $L=\epsilon |-n(t)\dot{t}%
^{2}+a(t)w^{2}|^{1-q}(A_{t}(t)\dot{t})^{2q};$

\item polynomial $4$-th root type: $L=\epsilon |a(t)\dot{t}^{4}+b(t)\dot{t}%
^{3}w+c(t)\dot{t}^{2}w^{2}+d(t)\dot{t}w^{3}+e(t)w^{4}|^{\frac{1}{4}}$.
\end{itemize}

Generically they have more degrees of freedom than just the scale factor of
the Friedmann-Lemaitre-Robertson-Walker metric.

\subsection{Geodesics, Geodesic Deviation and Curvature Scalar}
\label{geodeFinslerfTref1}

Arc length parametrised geodesics $\gamma :s\mapsto \gamma (s)$ of $(M,L)$
are characterized by the Euler-Lagrange equations of~(\ref{eq:FinAct}). On 
$\mathcal{A},$ these will take the form:%
\begin{equation*}
\ddot{\gamma}^{\mu}+2{\bar{\bar{\Gamma}}}^{\mu}(\gamma ,\dot{\gamma})=0,
\end{equation*}%
where the geodesic coefficients 
${\bar{\bar{\Gamma}}}^{\mu}={\bar{\bar{\Gamma}}}^{\mu}(x,\dot{x})$ are given 
by: 
$%
2{\bar{\bar{\Gamma}}}^{\mu}=\frac{1}{2}g^{L\mu\nu}(\dot{x}^{\sigma}L_{,
\sigma\cdot \nu}-L_{,\nu}).$ In turn, $%
{\bar{\bar{\Gamma}}}^{\mu}{}$ define the coefficients 
${\bar{\bar{\Gamma}}}^{\mu}{}_{\nu}:=\dot{\partial}_{\nu}{\bar{\bar{\Gamma}}}^{
\mu}$ of
the canonical \emph{nonlinear connection}. The latter defines an adapted
basis $\{\delta _{\mu},\dot{\partial}_{\mu}\}$ of the tangent spaces 
$T_{(x,\dot{%
x})}TM,$ with 
\begin{equation*}
\delta _{\mu}=\partial 
_{\mu}-{\bar{\bar{\Gamma}}}^{\nu}{}_{\mu}\dot{\partial}_{\nu}.
\end{equation*}%
The elements of the adapted basis transform under manifold induced
coordinate transformations as if they were vectors on the base manifold $M$
(more technically, they define a splitting of the tangent spaces of $TM$
into horizontal and vertical parts: $T_{(x,\dot{x})}TM=H_{(x,\dot{x}%
)}TM\oplus V_{(x,\dot{x})}TM$).

The nonlinear connection and the $L$-metric $g$ are the only necessary
ingredients for our gravitational field Lagrangian. Yet, in order to handle
tensors on $\mathcal{A}$, we will need a notion of covariant differentiation
- given by an affine connection on $\mathcal{A}\subset TM.$ In the
literature, there are multiple such examples, all of which project onto the
Levi-Civita connection in the pseudo-Riemannian case. In the following, we
will use the so-called Chern-Rund linear connection, locally defined by 
\begin{equation*}
\mathrm{D}_{\delta _{\sigma}}\delta 
_{\nu}=\widebar{\Gamma}^{\mu}{}_{\nu\sigma}\delta _{\mu},\quad 
\mathrm{D}_{\delta 
_{\sigma}}\dot{\partial}_{\nu}=\widebar{\Gamma}^{\mu}{}_{\nu\sigma}\dot{\partial
}%
_{\mu},\quad \mathrm{D}_{\dot{\partial}_{\sigma}}\delta 
_{\nu}=\mathrm{D}_{\dot{%
\partial}_{\sigma}}\dot{\partial}_{\nu}=0\,,
\end{equation*}%
where $\widebar{\Gamma}^{\mu}{}_{\nu\sigma}:=\frac{1}{2}g^{L\mu\rho}(\delta 
_{\sigma}g_{\nu\rho}^{L}+\delta
_{\nu}g_{\sigma\rho}^{L}-\delta _{\rho}g_{\nu\sigma}^{L}).$ The difference 
\begin{equation*}
P^{\mu}{}_{\sigma\nu}=\dot{\partial}_{\sigma}{\bar{\bar{\Gamma}}}^{\mu}{}_{\nu}
-\widebar{\Gamma}^{\mu}{}_{\nu\sigma}
\end{equation*}%
defines the so-called \emph{Landsberg tensor }on $\mathcal{A};$ we will
denote by $P_{\mu}=P^{\nu}{}_{\mu\nu}$ its trace.

It is worth noticing that the derivative operator $\nabla :=\dot{x}^{\mu}%
\mathrm{D}_{\delta _{\mu}}$ of vector fields on $TM,$ called the dynamical
covariant derivative, actually only depends on 
${\bar{\bar{\Gamma}}}^{\mu}{}_{\nu}$ (i.e., on the
nonlinear connection only).

Let $c=(\gamma ,\dot{\gamma})$ be the lift of a Finsler geodesic $\gamma $
to $TM$. Its tangent vector $\dot{c}$ can be expressed in the adapted basis
as $\dot{c}=\dot{\gamma}^{\mu}\delta _{\mu}$, i.e., it is horizontal. If $V$ is
a deviation vector field on spacetime, then the geodesic deviation equation
is: 
\begin{equation*}
(\nabla \nabla \hat{V})_{|(\gamma ,\dot{\gamma})}=\mathbf{R}(\dot{\gamma},%
\hat{V}),
\end{equation*}%
where $\hat{V}=V^{\mu}\delta _{\mu}$ is the canonical horizontal lift of $V.$
The geodesic deviation operator $\mathbf{R}=R^{\mu}{}_{\nu}dx^{\nu}\otimes 
\delta
_{\mu}$ is derived from the curvature of the nonlinear connection as 
\begin{equation*}
R^{\mu}{}_{\nu}=R^{\mu}{}_{\nu\sigma}\dot{x}^{\sigma},\quad 
R^{\mu}{}_{\nu\sigma}\dot{\partial}%
_{\mu}=[\delta _{\nu},\delta _{\sigma}]=(\delta 
_{\sigma}{\bar{\bar{\Gamma}}}^{\mu}{}_{\nu}-\delta
_{\nu}{\bar{\bar{\Gamma}}}^{\mu}{}_{\sigma})\dot{\partial}_{\mu}\,.
\end{equation*}%
The non-homogenized Finsler Ricci scalar $R$ is defined as its trace 
\label{Finslerriccigrref1}
\begin{equation}
R=R^{\mu}{}_{\mu}=R^{\mu}{}_{\nu\sigma}\dot{x}^{\sigma}\,.  
\label{eq:FinslerRicciS}
\end{equation}%
The curvature tensors appearing here are defined solely in terms of the
canonical nonlinear connection. The Finsler linear connections, which one
may define, are not entering here.

\bigskip

In the particular case when $L=g_{\mu\nu}(x)\dot{x}^{\mu}\dot{x}^{\nu}$, where 
$%
g_{\mu\nu}(x)$ are the components of a Lorentzian metric, the geometry of 
$(M,L)$
becomes essentially the geometry of the pseudo-Riemannian spacetime manifold 
$(M,g)$. 

The $L$-metric becomes the Lorentzian metric $g$, the nonlinear connection 
coefficients and the nonlinear curvature tensor become the Christoffel symbols 
$\Gamma^\mu{}_{\nu\rho}(x)$ and the Riemann curvature tensor 
$r^\mu{}_{\nu\rho\sigma}$ of the Levi-Civita connection of g, up to a 
contraction with a velocity $\dot x$.  Yet, the
Finslerian Ricci scalar becomes 
$R(x,\dot{x})=-r_{\nu\sigma}(x)\dot{x}^{\nu}\dot{x}%
^{\sigma}$ and is not equal to the Riemannian Ricci scalar 
$r=r_{\mu\nu}g^{\mu\nu}$ in
this case.

In the following, when constructing an action for Finsler gravity, we will
need to work with $0$-homogeneous objects in $\dot{x}$. On $\mathcal{A}_{0}$
we can introduce the $0$-homogenized Ricci scalar%
\begin{equation}
R_{0}=\frac{1}{L}R\,.  \label{R0}
\end{equation}

\section{Finslerian Scalars as Physical Fields}

By a \textit{Finslerian scalar, }we will understand in the following a
function $\Phi :TM\rightarrow \mathbb{R},$ $\Phi =\Phi (x,\dot{x}),$
homogeneous of some degree $k$ in $\dot{x}.$ As physical fields, we have in
mind two such examples:

1) The Finsler-Lagrange function $L:TM\rightarrow \mathbb{R}$ of a Finsler
spacetime manifold, obeying $L(x,\ell \dot{x})=\ell^{2}L(x,\dot{x}),$ $%
\ell >0$.

2) The 1-particle distribution function $\phi $ of a kinetic 
gas.\label{Finslsdistributionref1} Typically, 
$\phi =\phi (x,\dot{x})$ is understood as a function on the observer 
space\label{Finslsobserverrref1} $%
\mathcal{O}$. Equivalently, we can understand it as a function on the entire 
$TM$, setting $\phi (x,\ell \dot{x}):=\phi (x,\dot{x})$ for $\ell >0$
(and $\varphi (x,0)=0$).

\bigskip

As we want to apply the standard tools of variational calculus, \cite%
{Krupka-book}, we need to understand homogeneous functions $\Phi
:TM\rightarrow \mathbb{R}$ as \textit{sections} of some fibred manifold.
Naively, $\Phi $ can be understood as a section of $TM\times \mathbb{R}.$
Yet, this description is problematic for the calculus of variations
machinery, since, due to the $\dot{x}$-homogeneity requirement, variations $%
\delta \Phi $ cannot be compactly supported on $TM$ (since, imposing that $%
\delta \Phi (x,\dot{x})=0$ for some $(x,\dot{x})$ will force $\delta \Phi
(x,\ell \dot{x})=0$ for all $\ell >0$).

The solution comes from considering the so-called positive (or oriented)
projective tangent bundle\textit{\ }$PTM^{+}$\textit{, }obtained by treating
each half-line $\left\{ (x,\ell \dot{x})~|~\ell >0\right\} $ as a single
point. Intuitively, $PTM^{+}$ can be interpreted as the \textit{space of
directions }over $M;$ more precisely, $PTM^{+}$ is a set of equivalence
classes\textbf{\ }$PTM^{+}:=\{[(x,\dot{x})]_{\sim }~|~(x,\dot{x})\in \overset%
{\circ }{TM}\}$, where: 
\begin{equation*}
(x,\dot{x})~\sim ~\left( x,u\right) \Leftrightarrow ~u=\ell \dot{x}~~\text{%
for~~some~\ }\ell >0.
\end{equation*}%
The set $PTM^{+}$ is an  orientable seven-dimensional manifold.
Moreover, the slit tangent bundle $\overset{\circ }{TM}=TM\backslash \{0\}$
is a principal bundle over $PTM^{+},$ with fibre $\mathbb{R}_{+}^{\ast }$
and projection:%
\begin{equation*}
\pi ^{+}:\overset{\circ }{TM}\rightarrow PTM^{+},~\ \ \left( x,\dot{x}%
\right) \mapsto \left[ (x,\dot{x})\right] .
\end{equation*}

\bigskip

{\it{Local coordinates on}} $PTM^{+}.$ Coordinate charts on $PTM^{+}$ can
be taken of the form $U_{\mu}=\left\{ [(x^{0},\ldots ,x^{3},\dot{x}^{0},\ldots
,\dot{x}^{\mu},\ldots ,\dot{x}^{3})]~|~\dot{x}^{\mu}\text{ has constant 
sign}\right\} $%
. An element $[(x,\dot{x})]\in U_{\mu}$ will have the coordinates $%
(x^{\mu},u^{i})$, where $u^{i}=\dfrac{\dot{x}^{i}}{\dot{x}%
^{\mu}},$ for all $i \in \{0,..,3\},$ $i \not=\mu.$

But, working in coordinates $(x^{\mu},u^{i })$ is usually inconvenient.
Fortunately, for $[(x,\dot{x})]\in U_{\mu},$ one can introduce \textit{%
homogeneous coordinates, }which are nothing but the coordinates $(x^{\mu},\dot{%
x}^{\mu})$ of the representative $(x,\dot{x})\in TM.$ In these coordinates, the 
calculations (including differentiation and integration) are identical to
those on $TM$. It simply has to be ensured that the involved geometric objects
are well-defined objects on $PTM^{+}$. This can be seen as follows:

- Functions $f^{+}=f^{+}(x^{\mu},\dot{x}^{\mu})$ and vector fields $%
X^{+}=X^{+}(x^{\mu},\dot{x}^{\mu})$ are well-defined on $PTM^{+}$ if and only if
they are 0-homogeneous in $\dot{x}.$

- Differential forms $\rho ^{+}=\rho ^{+}(x^{\mu},\dot{x}^{\mu})$ are 
well-defined on $PTM^{+}$ if and only if they are 0-homogeneous in $\dot{x}$ 
and $%
\pi ^{+}$-horizontal, i.e., $\mathbf{i}_{\mathbb{C}}\rho ^{+}=0,$ where $%
\mathbb{C}=\dot{x}^{\mu}\dot{\partial}_{\mu}$ is the generator of the fibres of
the principal bundle $(\overset{\circ }{TM},\pi ^{+},PTM^{+},\mathbb{R}%
_{+}^{\ast }).$

We will denote by a superscript $^{+}$ objects on $PTM^{+},$ to distinguish
them from their $TM$-correspondents (which have the same coordinate
expression), e.g: $f=f^{+}\circ \pi ^{+},$ $\rho =\left( \pi ^{+}\right)
^{\ast }\rho^{+} $.

\bigskip

The space of directions $PTM^{+}$ has one more property, which is
particularly important when dealing with kinetic gases. Integrals on \textit{%
compact} subsets $V$ of the observer space $\mathcal{O}\subset TM$ can be
understood as integrals on compact subsets $V^{+}:=\pi ^{+}(V)$ of $PTM^{+}$%
: 
\begin{equation*}
\underset{V}{\int }\rho =\underset{V^{+}}{\int }\rho ^{+},
\end{equation*}%
where $\rho :=(\pi ^{+})^{\ast }\rho ^{+}$; this is possible since the
restriction $\pi ^{+}:V\rightarrow V^{+}$ is a diffeomorphism. This allows
us to express action integrals on $PTM^{+}$ as action integrals on $\mathcal{%
O}$ and vice-versa.

\bigskip

{\it{Construction of the configuration bundle}}. Now, having described the
space of directions $PTM^{+},$ it remains to describe the homogeneous
functions $L,\phi :\overset{\circ }{TM}\rightarrow \mathbb{R}$ as sections
of some bundle $Y$ sitting over $PTM^{+}.$

\begin{itemize}
\item The 0-homogeneous function $\phi $ poses no problems, since $\phi (x,%
\dot{x})$ can be identified as a function $\phi ^{+}([x,\dot{x}])$ on $%
PTM^{+}$. Hence, in this case, $Y=PTM^{+}\times \mathbb{R}.$
Briefly: we can equivalently understand $\phi $\ either as a function on $%
\mathcal{O},$\ or as a function on (a subset of) $PTM^{+}.$

\item The situation is more complicated in the case of $L$ - which, because
of its non-zero homogeneity degree, cannot be understood as a function on $%
PTM^{+}.$ Still, we will be able to describe $L$ as a section of a fibred
manifold sitting over $PTM^{+},$ as follows. We start from the remark that $(%
\mathbb{R}_{+}^{\ast },\cdot )$ is a Lie group, acting on both $\overset{%
\circ }{TM}$ and $\mathbb{R}$: 
\begin{equation*}
\cdot :\overset{\circ }{TM}\times \mathbb{R}_{+}^{\ast }\rightarrow \overset{%
\circ }{TM},~\ (x,\dot{x})\cdot \ell =(x,\ell \dot{x}),~\ \ \ast :%
\mathbb{R}_{+}^{\ast }\times \mathbb{R}\rightarrow \mathbb{R},\ell \ast
z=\ell ^{2}z.
\end{equation*}%
The 2-homogeneity of $L$ can be understood as \textit{equivariance} under
these actions. This points out\textbf{\ }to the associated bundle to the
principal bundle $\overset{\circ }{TM},$ with fibre $\mathbb{R}$ (and base $%
PTM^{+}$) 
\begin{equation*}
Y:=(\overset{\circ }{TM}\times \mathbb{R)}_{/\sim }
\end{equation*}%
where the equivalence relation $\sim $ is given by: $(x,\dot{x},L)~\sim
~(x,\ell \dot{x},\ell ^{2}L),~\ell \in \mathbb{R}_{+}^{\ast }.$ The
set $Y$ is a fibred manifold over $PTM^{+},$ with projection: $\ \pi
:Y\rightarrow PTM^{+},[(x,\dot{x},L)]\rightarrow \lbrack (x,\dot{x})].$
Using homogeneous local coordinates $(x^{\mu},\dot{x}^{\mu})$ on $PTM^{+},$ the
local coordinates on $Y$ will be simply $(x^{\mu},\dot{x}^{\mu},L)$. Then,
Finsler-Lagrange functions $L=L(x,\dot{x})$ are in one-to-one correspondence
with sections $\gamma $ of $Y:$%
\begin{equation}
L\mapsto \gamma :PTM^{+}\rightarrow Y,~\gamma ([x,\dot{x}])=[x,\dot{x},L(x,%
\dot{x})].  \label{corresp_L_sections}
\end{equation}%
or, in local coordinates, $\gamma :(x^{\mu},\dot{x}^{\mu})\mapsto 
(x^{\mu},\dot{x}%
^{\mu},L(x^{\mu},\dot{x}^{\mu})).$
\end{itemize}

Having constructed the configuration bundle $Y,$ the gravitational field
Lagrangian will be understood as a 7-form $\Lambda =\mathcal{L}(x,\dot{x}%
,L,DL,...,D^{k}L)d(Vol)$ on a jet bundle $J^{k}Y$. The corresponding action
is constructed by substituting $L=L(x,\dot{x})$ into $\Lambda $ (i.e.,
pulling back this 7-form to $PTM^{+}$ by sections $J^{k}\gamma $) and
integrating over arbitrary compact domains $V^{+}\subset PTM^{+}.$ A similar
construction holds for $\phi .$

\section{Gravitational Dynamics}
\label{sec:kingasgravity}

The mathematical construction presented in the previous section lays the
foundation to construct consistent Finslerian theories of gravity and matter
fields. Since the geometry of Finsler spacetimes is described on the tangent
bundle, it is necessary to also describe physically viable matter on the same
level, so that the matter can actually determine the geometry dynamically.
The construction of such a matter-geometry coupling has for long been an
open issue. A natural candidate for such coupling is the kinetic description
of gases. By using Finsler geometry to describe kinetic gases, we
constructed for the first time a coupling between a Finsler geometric
description of gravity and physical matter, which naturally lives on the
tangent bundle \cite{Hohmann:2019sni,Hohmann:2020yia}.

\subsection{The kinetic Gas Action on the Tangent Bundle}

The properties of a large system of P (interacting or non-interacting),
particles can be encoded into the so-called 1-particle distribution function
(1PDF) $\phi =\phi (x,\dot{x})$  \cite{Ehlers2011,Rezzolla2013}, which
is a $0$-homogeneous real scalar function defined on $TM$ - hence
equivalently on $PTM^{+},$ or on the observer space~$\mathcal{O}$. It is
defined by the particle number counting integral 
\begin{equation*}
N(\sigma ):=\underset{\sigma }{\int }\phi \Omega \,,
\end{equation*}%
which gives the number of particles intersecting an arbitrary oriented
six-dimensional hypersurface $\sigma \subset O$, i.e., it counts the
intersection of tangent bundle particle trajectories $c = (\gamma, \dot
\gamma)$ with $\sigma $. The volume form $\Omega $ is canonically determined
from the Finsler-Lagrange function $L,$ see \cite{Hohmann:2019sni}.

Since in all physical realistic situations, there will be particles of
maximal velocity, we can assume that the restriction of the 1PDF to the
observer\textsl{\ }space $\phi _{x}(\cdot )=\phi (x,\cdot )$ is compactly
supported.

For collissionless gases, the 1PDF satisfies the \emph{Liouville equation} 
\begin{equation}
\frac{\dot{x}^{\mu}}{\sqrt{L}}\delta _{\mu}\phi =0,  \label{Liouville equation}
\end{equation}%
which can be derived from the demand that the number of particles is
constant between two different hypersurfaces, that are transversal to
particle trajectories. In the following, we will see that the Liouville
equation can be understood as an energy-momentum conservation equation.

The action of the kinetic gas is 
\begin{equation}
S_{\text{gas}}=\int_{V}\phi \Sigma =\int_{D}\left( \int_{\mathcal{O}%
_{x}}\phi \Sigma _{x}\right) d^{4}x\,,  \label{eq:gasact}
\end{equation}%
where the compact subset $V\subset \mathcal{O}$ is generated by the flow of
particle worldlines from a given hypersurface $\sigma _{0}$ for a finite
flow parameter. The second equal sign above is justified as follows. Taking
into account that $\phi (x,\cdot )$ is compactly supported, the integral of $%
\phi \Sigma $ on $V$ can be understood as an integral on $\bigcup_{x\in D}%
\mathcal{O}_{x}$, where $D=\pi (V)$ \textsl{and }$\mathcal{O}_{x}=\mathcal{O}%
\cap T_{x}M$ is the set of unit time-like directions at $x$; in  this way, the
latter expression of $S_{\text{gas}}$ is well defined and equal to an
integral on a corresponding domain $V^{+}\subset PTM^{+}.$ The volume forms $%
\Sigma $ on $V$ and $\Sigma _{x}$ on $\mathcal{O}_{x}$ are, again,
canonically determined by the Finsler Lagrangian $L$ and can be found in 
\cite{Hohmann:2019sni}.

The action \eqref{eq:gasact} is invariant under coordinate transformations
on $TM$ induced from coordinate transformations on $M$. As a consequence of
this invariance, we can identify the \emph{energy-momentum distribution 
\label{Enmomtenref4}
tensor} $\Theta $ of the gas on $\mathcal{O}$ as 
\begin{equation*}
\Theta _{~\nu}^{\mu}=\phi \dfrac{\dot{x}^{\mu}\dot{x}_{\nu}}{L}\,.
\end{equation*}%
It satisfies an averaged conservation equation 
\begin{equation}
\int_{\mathcal{O}_{x}}D_{\delta _{\mu}}\Theta ^{\mu}{}_{\nu}\ \Sigma _{x}=0\,.
\label{eq:avconlaw}
\end{equation}%
For a collisionless gas, we see that even a pointwise conservation equation
holds, since 
\begin{equation*}
D_{\delta _{\mu}}\Theta 
^{\mu}{}_{\nu}=\frac{\dot{x}^{\mu}\dot{x}_{\nu}}{L}D_{\delta
_{\mu}}\phi =\frac{\dot{x}^{\mu}\dot{x}_{\nu}}{L}\delta _{\mu}\phi =0\,
\end{equation*}%
is precisely the Liouville equation. The other way around, demanding that %
\eqref{eq:avconlaw} holds pointwise, the Liouville equation is implied.

The usual fluid energy-momentum tensor $T$ of the kinetic gas on the base
manifold $M$ is obtained by averaging $\Theta$ over the unit time-like
directions: 
\begin{equation}
T^{\mu}{}_{\nu}(x):=\underset{\mathcal{O}_{x}}{\int }\Theta^{\mu}{}_{\nu}(x,\dot{x}%
)\Sigma _{x}=\underset{\mathcal{O}_{x}}{\int }\phi \dfrac{\dot{x}^{\mu}\dot{ x}%
_{\nu}}{L}\Sigma _{x}.  \label{relation Theta-T}
\end{equation}%
It is worth noting that in the particular case where $L$ is
pseudo-Riemannian, the operator $D _{\delta _{\mu}}$ becomes the usual
Levi-Civita covariant derivative $_{; \mu}$ and it can be pulled out of the
integral. Hence, in this case, averaged conservation law 
\eqref{eq:avconlaw}\label{Finslconservationnref1}
becomes equivalent to the usual covariant conservation law $T^{\mu}{}_{\nu;\mu}=0.$

\subsection{The Finsler Gravity Action}

One of the first Finsler gravitational field equations proposed as a
Generalization of the Einstein vacuum equations was suggested by Rutz \cite%
{Rutz} as the vanishing of the Finsler Ricci scalar 
\begin{equation*}
R_{0}=0\,.
\end{equation*}%
This equation represents the vanishing of the trace of the geodesic
deviation operator, a method suggested by Pirani to find the Einstein vacuum
equations.

We proved that  the Rutz equation has one drawback, namely that it cannot be
derived from an action. The variational completion algorithm \cite{Voicu-Krupka} revealed that the closest Finsler gravity equation, which can
be derived from an action, must be obtained from the action on $PTM^{+}$ 
\begin{equation*}
S_{\text{grav}}=\int_{V^{+}}R_{0}^{+}\Sigma ^{+}=\int_{V}R_{0}\Sigma \,,
\end{equation*}%
where actually, since we assume $V^{+}\subset PTM^{+}$ to be compact, it can
be equivalently expressed on $V\subset \mathcal{O}$; see \cite{Hohmann:2018rpp}, 
where   the 1PDF of the kinetic gas action is defined, too.
After this preparation it is straightforward to derive the coupled Finsler
spacetime gravitational dynamics.

\subsection{Kinetic Gases as Physical Sources for Finsler Gravity}

The combined action for a kinetic gas coupled to a Finslerian spacetime
geometry, which encodes gravity, is 
\begin{equation*}
S=\frac{2}{\kappa ^{2}}\int_{V}R_{0}\Sigma +\int_{V}\phi \Sigma \,.
\end{equation*}%
Variation with respect to the Finsler-Lagrange function $L$ yields 
\cite{Hohmann:2019sni} 
\begin{equation}
3R_{0}-\dfrac{1}{2}g^{\mu\nu}R_{\cdot \mu \nu}+g^{\mu\nu}[(\nabla P_{\mu})_{\cdot
\nu}+D_{\delta _{\nu}}P_{\mu}-P_{\mu}P_{\nu}]=\kappa ^{2}\phi \,.  
\label{eq:fgravgas}
\end{equation}

The corresponding vacuum equation, setting $\phi =0$, has already been
discussed in \cite{Pfeifer:2011xi}. Its completion through the addition of a viable
consistent physical matter coupling has only been achieved recently.

Equation \eqref{eq:fgravgas} determines the geometry of spacetime, i.e., the
gravitational field, directly from the 1PDF of a kinetic gas. It takes into
account the influence of the generally nontrivial velocity distribution
over the different gas particles. In contrast, in the general relativistic
coupling between a system of $P$ particles and gravity in terms of the
energy-momentum tensor, the velocity distribution of the particles is
averaged out. Hence, we expect that the Finslerian description of the
gravitational field of kinetic gases gives a more accurate result than its
general relativistic description.

The future application of the gravitational field equation is expected to
highly improve the understanding of systems that are described by
gravitating fluids, such as the Universe as a whole in cosmology, ordinary
and neutron stars, as well as accretion disks of black holes, by replacing
the averaged gravitating fluid by the more accurate and finer notion of a
kinetic gas.

\chapter[Gravity’s Rainbow]{Gravity’s Rainbow}
\label{Garattinichapter}
{\em Remo Garattini}\\
\label{rainbowref1}
 

According to Quantum Mechanics, Zero Point Energy (ZPE) is the lowest possible
energy \label{zeropointref1}that a quantum mechanical system may have. Unlike in 
classical
mechanics, quantum systems constantly fluctuate in their lowest energy state
as described by the Heisenberg uncertainty principle. The same behaviour
appears also in Quantum Field Theory and, if one wishes to also  apply the 
quantum
world   to General Relativity, one should obtain a quantum theory of the
gravitational field, better known as \textit{Quantum Gravity}. Unfortunately,
such a theory is still lacking. However, there is no barrier to     searching
for
a possible ZPE candidate, even if it is well known that every ZPE calculation
is plagued by divergences. Usually, the divergences are kept under control with
the help of a regularisation and renormalisation procedure. In ordinary
gravity the computation of ZPE for quantum fluctuations of the \textit{pure
gravitational field} can be extracted by rewriting the Wheeler-DeWitt equation 
\label{DeWittref1}
(WDW)\cite{DeWitt:1967yk} in a form that looks like an expectation value
computation\cite{Garattini:2004zu,AmelinoCamelia:2002tc}. Its derivation is a 
consequence of the
Arnowitt-Deser-Misner ($\mathcal{ADM}$) decomposition\cite{Arnowitt:1962hi} of 
spacetime based on the following line element%
\begin{equation}
ds^{2}=g_{\mu\nu}\left(  x\right)  dx^{\mu}dx^{\nu}=\left(  -N^{2}+N_{i}%
N^{i}\right)  dt^{2}+2N_{j}dtdx^{j}+\gamma_{ij}dx^{i}dx^{j},
\end{equation}
where $N$ is the \textit{lapse }function and $N_{i}$ the \textit{shift
}function. In terms of the $\mathcal{ADM}$ variables, the four-dimensional
scalar curvature $R$ can be decomposed in the following way%
\begin{equation}
R=\text{ }^{(3)}R+K_{ij}K^{ij}-\left(  K\right)  
^{2}-2\nabla_{\mu}\left(
Ku^{\mu}+a^{\mu}\right)  ,\label{RemoR}%
\end{equation}
where%
\begin{equation}
K_{ij}=-\frac{1}{2N}\left[  \partial_{t}\gamma_{ij}-N_{i|j}-N_{j|i}\right]
\end{equation}
is the second fundamental form, $K=$ $\gamma^{ij}K_{ij}$ is its trace, $ 
^{(3)}R$ is 
the
three-dimensional scalar curvature and $\sqrt{\gamma}$ is the three-dimensional
determinant of the metric. The last term in $\left(  \ref{RemoR}\right)  $
represents the boundary terms contribution where the four-velocity $u^{\mu}$
is the timelike unit vector normal to the spacelike hypersurfaces
($t$=constant) denoted by $\Sigma_{t}$ and $a^{\mu}=u^{\alpha}\nabla_{\alpha
}u^{\mu}$ is the acceleration of the timelike normal $u^{\mu}$. Thus%
\begin{equation}
\mathcal{L}\left[  N,N_{i},\gamma_{ij}\right]  =\sqrt{- g}\left(  
R-2\Lambda\right)
=\frac{N}{2\kappa^2}\sqrt{\gamma}\text{ }\left[  K_{ij}K^{ij}-K^{2}+\,\text{ 
}^{(3)}R-2\Lambda
-2\nabla_{\mu}\left(  Ku^{\mu}+a^{\mu}\right)  \right]  \label{RemoLag}%
\end{equation}
represents the gravitational Lagrangian density where $\kappa^2=8\pi G$ with $G$
the Newton's constant, and for the sake of generality we have also included a
cosmological constant $\Lambda$. After a Legendre transformation, the WDW
equation simply becomes%
\begin{equation}
\mathcal{H}\Psi=\left[  \left(  2\kappa^2\right)  G_{ijkl}\pi^{ij}\pi^{kl}%
-\frac{\sqrt{\gamma}}{2\kappa^2}{}\left(  \,\text{ }^{(3)}R-2\Lambda\right)  
\right]
\Psi=0,\label{RemoWDWO}%
\end{equation}
where $G_{ijkl}$ is the super-metric and where the conjugate super-momentum
$\pi^{ij}$ is defined as%
\begin{equation}
\pi^{ij}=\frac{\delta\mathcal{L}}{\delta\left(  \partial_{t}\gamma_{ij}\right)
}=\left(  \gamma^{ij}K-K^{ij}\text{ }\right)  \frac{\sqrt{\gamma}}{2\kappa^2
}.\label{Remomom}%
\end{equation}
Note that $\mathcal{H}=0$ represents the classical constraint that guarantees
the invariance under time reparametrisation. The other classical constraint
represents the invariance by spatial diffeomorphism and it is described by
$\pi_{|j}^{ij}=0$, where the vertical stroke \textquotedblleft%
$\vert$%
\textquotedblright\ denotes the covariant derivative with respect to the $3D$
metric $\gamma_{ij}$. It is interesting to note that formally, the WDW equation 
can
be transformed into an eigenvalue equation. Let us see how, with a concrete
example.

\section{The Cosmological Constant as a Sturm-Liouville Eigenvalue Problem}
\label{Liouvilleref1}

The Friedmann-Lema\^{\i}tre-Robertson-Walker (FLRW) line element is defined as%
\begin{equation}
ds^{2}=-N^{2}dt^{2}+a^{2}\left(  t\right)  d\Omega_{3}^{2},\label{RemoFRW}%
\end{equation}
describing a homogeneous, isotropic and closed universe, where%
\begin{equation}
d\Omega_{3}^{2}=\gamma_{ij}dx^{i}dx^{j}\label{Remodomega}%
\end{equation}
is the line element on the three-sphere, $N$ is the lapse function and $a(t)$
denotes the scale factor. Let us consider a very simple mini-superspace model
described by the metric of Eq.$\left(  \ref{RemoFRW}\right)  $. In this
background the Einstein-Hilbert action in $(3+1)$-dim becomes%
\begin{equation}
S=-\frac{6\pi^2}{\kappa^2}\int_{I}\left[  \dot{a}^{2}a-a+\frac{\Lambda}{3}%
a^{3}\right]  dt~,\label{Remoaction}%
\end{equation}
where $\Lambda$ is the cosmological constant.
In Eq.~$\left(  \ref{Remoaction}\right)  $, we have integrated every degree of
freedom except the scale factor. In addition we have computed the volume
associated with the three-sphere, namely $V_{3}=2\pi^{2}$, and we have set 
$N=1$.
The canonical momentum reads%
\begin{equation}
\pi_{a}=\frac{\delta S}{\delta\dot{a}}=-\frac{12\pi^2}{\kappa^2}\dot{a}a~,
\end{equation}
and the resulting Hamiltonian density is%
\begin{align}
\mathcal{H} &  =\pi_{a}\dot{a}-\mathcal{L}   =-\frac{\kappa^2}{24\pi^2 
a}\pi_{a}^{2}-\frac{6\pi^2}{\kappa^2}a+\frac{6\pi^2}{\kappa^2}\frac
{\Lambda}{3}a^{3}~.\label{RemoH0}%
\end{align}
Following the canonical quantization prescription, we promote $\pi_{a}$ to a
momentum operator, setting%
\begin{equation}
\pi_{a}^{2}\rightarrow-a^{-q}\left[  \frac{\partial}{\partial a}a^{q}%
\frac{\partial}{\partial a}\right]  ,\label{Remoordering}%
\end{equation}
where we have introduced a factor order ambiguity $q$. The generalisation to
$k=0,-1$ is straightforward. The WDW equation for such a metric is%
\begin{eqnarray}
&&
\!\!\!\!\!\!
H\Psi\left(  a\right)  =\left[  -a^{-q}\left(  \frac{\partial}{\partial
a}a^{q}\frac{\partial}{\partial a}\right)  +\frac{144\pi^{4}}{\kappa^{2}}\left(
a^{2}-\frac{\Lambda}{3}a^{4}\right)  \right]  \Psi\left(  a\right)
\,,\nonumber\\
&&
\ \ \ \ \ \ \ 
=
\left[  -\frac{\partial^{2}}{\partial a^{2}}-\frac{q}{a}\frac{\partial
}{\partial a}+\frac{9\pi^{2}}{4G^{2}}\left(  a^{2}-\frac{\Lambda}{3}%
a^{4}\right)  \right]  \Psi\left(  a\right)  =0.\label{RemoWDW_0}%
\end{eqnarray}
It represents the quantum version of the invariance with respect to time
reparametrisation. If we define the following reference length $a_{0}%
=\sqrt{3/\Lambda}$, then Eq.$\left(  \ref{RemoWDW_0}\right)  $ assumes the
familiar form of a one-dimensional Schr\"{o}dinger equation for a particle
moving in the potential%
\begin{equation}
U\left(  a\right)  =\frac{144\pi^{4}a_{0}^{2}}{\kappa^4}\left[  \left(  \frac
{a}{a_{0}}\right)  ^{2}-\left(  \frac{a}{a_{0}}\right)  ^{4}\right]
\,,\label{RemoU(a)}%
\end{equation}
with zero total energy. The potential $U\left(  a\right)  $ resembles a
potential well, which is unbounded from below.\textbf{\ }When $0<a<a_{0}$,
Eq.~$\left(  \ref{RemoU(a)}\right)  $ implies $U\left(  a\right)  >0$, which
is the classically forbidden region, while for $a>a_{0}$, one gets $U\left(
a\right)  <0$, which is the classically allowed region. It is interesting to
note that for for the special case of the operator ordering $q=-1$,   exact 
solutions can be determined  \cite{Vilenkin:1987kf}. Even if the WDW equation
$\left(  \ref{RemoWDW_0}\right)  $ has a zero-energy eigenvalue, it also has a
hidden structure. Indeed, Eq.$\left(  \ref{RemoWDW_0}\right)  $ has the
structure of a Sturm-Liouville eigenvalue problem with the cosmological
constant as the  eigenvalue. We ask the reader to recall  that a Sturm-Liouville
differential equation is defined by%
\begin{equation}
\frac{d}{dx}\left(  p\left(  x\right)  \frac{dy\left(  x\right)  }{dx}\right)
+q\left(  x\right)  y\left(  x\right)  +\lambda w\left(  x\right)  y\left(
x\right)  =0\label{RemoSL},
\end{equation}
while the normalisation is defined by%
\begin{equation}
\int_{a}^{b}dxw\left(  x\right)  y^{\ast}\left(  x\right)  y\left(  x\right)
.
\end{equation}
In the case of the FLRW model we have the following correspondence%
\begin{align}
p\left(  x\right)   &  \rightarrow a^{q}\left(  t\right)  \,,\nonumber\\
q\left(  x\right)   &  \rightarrow\left(  \frac{12\pi^2}{\kappa^2}\right)  ^{2}%
a^{q+2}\left(  t\right)  \,,\nonumber\\
w\left(  x\right)   &  \rightarrow a^{q+4}\left(  t\right)  \,,\nonumber\\
y\left(  x\right)   &  \rightarrow\Psi\left(  a\right)  \,,\nonumber\\
\lambda &  \rightarrow\frac{\Lambda}{3}\left(  \frac{12\pi^2}{\kappa^2}\right)  
^{2}\,,
\end{align}
and the normalisation becomes%
\begin{equation}
\int_{0}^{\infty}daa^{q+4}\Psi^{\ast}\left(  a\right)  \Psi\left(  a\right)
.\label{RemoNorm1}%
\end{equation}
It is a standard procedure  to convert the Sturm-Liouville problem $\left(
\ref{RemoSL}\right)  $ into a variational problem of the
form\footnote{Actually the standard variational procedure prefers the
following form%
\begin{equation}
F\left[  y\left(  x\right)  \right]  =\frac{-\left[  y^{\ast}\left(  x\right)
p\left(  x\right)  \frac{d}{dx}y\left(  x\right)  \right]  _{a}^{b}+\int
_{a}^{b}dxp\left(  x\right)  \left[  \frac{d}{dx}y\left(  x\right)  \right]
^{2}-q\left(  x\right)  y\left(  x\right)  }{\int_{a}^{b}dxw\left(  x\right)
y^{\ast}\left(  x\right)  y\left(  x\right)  }\,,
\end{equation}
with appropriate boundary conditions.}%
\begin{equation}
F\left[  y\left(  x\right)  \right]  =\frac{-\int_{a}^{b}dxy^{\ast}\left(
x\right)  \left\{  \frac{d}{dx}\left[  p\left(  x\right)  \frac{d}{dx}\right]
+q\left(  x\right)  \right\}  y\left(  x\right)  }{\int_{a}^{b}dxw\left(
x\right)  y^{\ast}\left(  x\right)  y\left(  x\right)  }\,,\label{RemoFunct}%
\end{equation}
with a boundary condition to be specified. If $y\left(  x\right)  $ is an
eigenfunction of $\left(  \ref{RemoSL}\right)  $, then%
\begin{equation}
\lambda=\frac{-\int_{a}^{b}dxy^{\ast}\left(  x\right)  \left\{  \frac{d}%
{dx}\left[  p\left(  x\right)  \frac{d}{dx}\right]  +q\left(  x\right)
\right\}  y\left(  x\right)  }{\int_{a}^{b}dxw\left(  x\right)  y^{\ast}\left(
x\right)  y\left(  x\right)  }\,,
\end{equation}
is the eigenvalue, otherwise%
\begin{equation}
\lambda_{1}=\min_{y\left(  x\right)  }\frac{-\int_{a}^{b}dxy^{\ast}\left(
x\right)  \left\{  \frac{d}{dx}\left[  p\left(  x\right)  \frac{d}{dx}\right]
+q\left(  x\right)  \right\}  y\left(  x\right)  }{\int_{a}^{b}dxw\left(
x\right)  y^{\ast}\left(  x\right)  y\left(  x\right)  }\,.
\end{equation}
\textbf{\ }The minimum of the functional $F\left[  y\left(  x\right)  \right]
$ corresponds to a solution of the Sturm-Liouville problem $\left(
\ref{RemoSL}\right)  $ with the eigenvalue $\lambda.$ In the mini-superspace
approach with a FLRW background\footnote{For apllications of this procedure to
Varying Speed of Light Cosmology (VSL), see \cite{Capozziello:2007gm}. For 
applications
to the  Generalized Uncertainty Principle (GUP), see \cite{Garattini:2015aca} 
and 
finally, for
applications to Inflation, see \cite{Garattini:2012ca}.}, one finds%
\begin{equation}
\frac{\int\mathcal{D}aa^{q}\Psi^{\ast}\left(  a\right)  \left[  -\frac
{\partial^{2}}{\partial a^{2}}-\frac{q}{a}\frac{\partial}{\partial a}%
+\frac{9\pi^{2}}{4G^{2}}a^{2}\right]  \Psi\left(  a\right)  }{\int
\mathcal{D}aa^{q+4}\Psi^{\ast}\left(  a\right)  \Psi\left(  a\right)  }%
=\frac{48\Lambda\pi^{4}}{\kappa^4}.\label{RemoWDW_1}%
\end{equation}
Note that the original WDW Eq.$\left(  \ref{RemoWDW_0}\right)  $ is always
preserved. This means that the cosmological constant, in this approach,
represents the degree of degeneracy of the original eigenvalue $E=0$. It is
immediate to generalise the quantum mechanical problem into a problem of
quantum field theory with the help of Eq.$\left(  \ref{RemoWDWO}\right)  $.

\section{ From Quantum Mechanics to Quantum Field Theory}

The Sturm-Liouville problem represented by the functional $\left(
\ref{RemoFunct}\right)  $ can be generalised to describe a problem of quantum
field theory. Indeed to this purpose, one can write%
\begin{equation}
\frac{1}{V}\frac{\int\mathcal{D}\left[  \gamma_{ij}\right]  \Psi^{\ast}\left[
\gamma_{ij}\right]  \int_{\Sigma}d^{3}x\hat{\Lambda}_{\Sigma}\Psi\left[
\gamma_{ij}\right]  }{\int\mathcal{D}\left[  \gamma_{ij}\right]  
\Psi^{\ast}\left[
\gamma_{ij}\right]  \Psi\left[  \gamma_{ij}\right]  
}=\frac{1}{V}\frac{\left\langle
\Psi\left\vert \int_{\Sigma}d^{3}x\hat{\Lambda}_{\Sigma}\right\vert
\Psi\right\rangle }{\left\langle \Psi|\Psi\right\rangle }=-\frac{\Lambda
}{\kappa^2},\label{RemoVEVO}%
\end{equation}
where we have integrated over the hypersurface $\Sigma$ and we have defined%
\begin{equation}
V=\int_{\Sigma}d^{3}x\sqrt{\gamma}\label{RemoVol}%
\end{equation}
as the volume of the hypersurface $\Sigma$ with%
\begin{equation}
\hat{\Lambda}_{\Sigma}=\left(  2\kappa^2\right)  G_{ijkl}\pi^{ij}\pi^{kl}%
-\sqrt{\gamma}\text{ }^{(3)}R/\left(  2\kappa^2\right)  
.\label{RemoLambdaSigma}%
\end{equation}
In this form, Eq.$\left(  \ref{RemoVEVO}\right)  $ can be used to compute Zero
Point Energy (ZPE) provided that $\Lambda/\kappa^2$ is considered as an
eigenvalue of $\hat{\Lambda}_{\Sigma}$. Nevertheless, solving Eq.$\left(
\ref{RemoVEVO}\right)  $ is a quite impossible task, therefore we are oriented
to use a variational approach with trial wave functionals. The related
boundary conditions are dictated by the choice of the trial wave functionals,
which, in our case, are of the Gaussian type; this choice is
justified by the fact that ZPE should be described as a good candidate of the
\textquotedblleft\textit{vacuum state}\textquotedblright. However, if we change
the form of the wave functionals we also  change   the corresponding boundary
conditions and therefore the description of the vacuum state. It is better to
observe that the obtained eigenvalue $\Lambda/\kappa^2$,  is far to be a
constant, so it will be dependent on some parameters and therefore it will
be considered more like a \textquotedblleft\textit{dynamical cosmological
constant}\textquotedblright. Usually, when we compute Eq.$\left(
\ref{RemoVEVO}\right)  $ to one loop or higher loops, ultra-violet (UV)
divergences appear. In ordinary gravity, to take under control such
divergencies we need a regularisation/renormalisation 
scheme\cite{Garattini:2004zu,AmelinoCamelia:2002tc}. Basically, we find 
that the one loop evaluation on a spherically symmetric
background is represented by the following expression%
\begin{equation}
\rho_{i}=-\frac{1}{4\pi^{2}}\int_{\sqrt{m_{i}^{2}\left(  r\right)  }}%
^{+\infty}d\lambda_{i}\lambda_{i}^{2}\sqrt{\lambda_{i}^{2}-m_{i}^{2}\left(
r\right)  };\qquad i=1,2,\label{Remorho}%
\end{equation}
where $i=1,2$ represents the two modes of the TT tensor. Of course, the
expression $\left(  \ref{Remorho}\right)  $ is divergent. The divergence can
be kept under control, adopting the zeta function regularization scheme so that
the integral in $\left(  \ref{Remorho}\right)  $ becomes%
\begin{equation}
\rho_{i}\left(  \varepsilon,\mu\right)  =-\frac{1}{4\pi^{2}}\mu^{2\varepsilon
}\int_{\sqrt{m_{i}^{2}\left(  r\right)  }}^{+\infty}d\lambda_{i}\frac
{\lambda_{i}^{2}}{\left(  \lambda_{i}^{2}-m_{i}^{2}\left(  r\right)  \right)
^{\varepsilon-\frac{1}{2}}};\qquad i=1,2.\label{Remozeta}%
\end{equation}
The integration has to be meant in the range where $\lambda_{i}^{2}-m_{i}%
^{2}\left(  r\right)  \geq0$ and the additional mass parameter $\mu$ has been
introduced in order to restore the correct dimension for the regularised
quantities. Following the same steps as in 
  \cite{Garattini:2004zu,AmelinoCamelia:2002tc}, we obtain
\begin{equation}
\rho_{i}\left(  \varepsilon,\mu\right)  =\frac{m_{i}^{4}\left(  r\right)
}{64\pi^{2}}\left[  \frac{1}{\varepsilon}+\ln\left(  \frac{4\mu^{2}}{m_{i}%
^{2}\left(  r\right)  \sqrt{e}}\right)  \right]  ,\qquad i=1,2.
\end{equation}
In order to renormalise the divergent ZPE, we write%
\begin{equation}
\frac{\Lambda}{\kappa^2}\rightarrow\frac{\Lambda_{0}}{\kappa^2}+\frac
{\Lambda^{div}}{\kappa^2}=\frac{\Lambda_{0}}{\kappa^2}+\frac{m_{1}^{4}\left(
r\right)  +m_{i}^{4}\left(  r\right)  }{64\pi^{2}\varepsilon}.
\end{equation}
Thus, the renormalisation is performed via the absorption of the divergent
part into the re-definition of the bare classical constant $\Lambda$. The
remaining finite value for the cosmological constant reads%
\begin{equation}
\frac{\Lambda_{0}\left(  \mu\right)  }{\kappa^2}=\sum_{i=1}^{2}\rho_{i}\left(
\mu\right)  =\frac{1}{64\pi^{2}}\sum_{i=1}^{2}m_{i}^{4}\left(  r\right)
\ln\left(  \frac{4\mu^{2}}{m_{i}^{2}\left(  r\right)  \sqrt{e}}\right)
=\rho_{eff}^{TT}\left(  \mu,r\right)  .\label{Remolambda0}%
\end{equation}
To avoid the dependence on the arbitrary mass scale $\mu$ in Eq.$\left(
\ref{Remolambda0}\right)  $, we adopt the renormalisation group equation and
we impose that%
\begin{equation}
\frac{1}{\kappa^2}\mu\frac{\partial\Lambda_{0}\left(  \mu\right)  }{\partial\mu
}=\mu\frac{d}{d\mu}\rho_{eff}^{TT}\left(  \mu,r\right)  .
\end{equation}
Solving it, we find that the renormalised constant $\Lambda_{0}$ should be
treated as a running one in the sense that it varies, provided that the scale
$\mu$ is changing\footnote{Note that the same procedure can be applied also to
a $f\left( R\right)$ theory\cite{Capozziello:2007gm}.}
\begin{equation}
\frac{\Lambda_{0}\left(  \mu,r\right)  }{\kappa^2}=\frac{\Lambda_{0}\left(
\mu_{0},r\right)  }{\kappa^2}+\frac{m_{1}^{4}\left(  r\right)  +m_{2}^{4}\left(
r\right)  }{32\pi^{2}}\ln\frac{\mu}{\mu_{0}}.\label{RemoRGsol}%
\end{equation}
It is interesting to note that the whole regularisation/renormalisation
procedure can be avoided  if we consider appropriate distortions of General
Relativity; one good candidate is represented by Gravity's Rainbow.

\subsection{The Wheeler-DeWitt Equation Distorted by Gravity's Rainbow}

We refer the reader to Ref. \cite{Garattini:2011kp} for details, even if a brief
outline will be presented. As a first step, one has to introduce two arbitrary
functions $g_{1}\left(  E/E_{P}\right)  $ and $g_{2}\left(  E/E_{P}\right)  $,
which have the following property\cite{Magueijo:2002xx}%
\begin{equation}
{\underset{ E/E_{P} \rightarrow 0 }{ \mathrm{lim} } \,}
 g_{1}
\left(  E/E_{P}\right)  =1\qquad\text{and}%
\qquad
{\underset{ E/E_{P} \rightarrow 0 }{ \mathrm{lim} } \,}
g_{2}\left(  E/E_{P}\right)
=1.\label{Remolim}%
\end{equation}
For a spherically symmetric line element, we can write the \textit{rainbow}
version as%
\begin{equation}
ds^{2}=-\frac{N^{2}\left(  r\right)  }{g_{1}^{2}\left(  E/E_{P}\right)
}dt^{2}+\frac{dr^{2}}{\left[ 1-\frac{b\left(  r\right)  }{r}\right]
g_{2}^{2}\left(  E/E_{P}\right)  }+\frac{r^{2}}{g_{2}^{2}\left(
E/E_{P}\right)  }\left(  d\theta^{2}+\sin^{2}\theta d\phi^{2}\right)
,\label{RemodS}%
\end{equation}
where $N$ is the lapse function and $b\left(  r\right)  $ is subject to the
only condition $b\left(  r_{t}\right)  =r_{t}$. Following 
Ref.\cite{Garattini:2011kp},
one can write the distorted classical constraint in the following way%
\begin{equation}
\mathcal{H}=\left(  2\kappa^2\right)  \frac{g_{1}^{2}\left(  E/E_{P}\right)
}{g_{2}^{3}\left(  E/E_{P}\right)  }\tilde{\gamma}_{ijkl}\tilde{\pi}^{ij}\tilde
{\pi}^{kl}\mathcal{-}\frac{\sqrt{\tilde{\gamma}}}{2\kappa^2 g_{2}\left(
E/E_{P}\right)  }{}\left( ^{(3)}\tilde{R}-\frac{2\Lambda_{c}}{g_{2}^{2}\left(
E/E_{P}\right)  }\right)  =0,
\end{equation}
where we have used the following property on $\text{ }^{(3)}R$%
\begin{equation}
\text{ }^{(3)}R=\gamma^{ij}\,\text{ }^{(3)}R_{ij}=g_{2}^{2}\left(  
E/E_{P}\right)  
\text{ }^{(3)}\tilde{R},
\end{equation}
and where
\begin{equation}
G_{ijkl}=\frac{1}{2\sqrt{\gamma}}\left(  
\gamma_{ik}\gamma_{jl}+\gamma_{il}\gamma_{jk}-\gamma_{ij}%
\gamma_{kl}\right)  =\frac{\tilde{\gamma}_{ijkl}}{g_{2}\left(  E/E_{P}\right)  
}.
\end{equation}
The symbol \textquotedblleft$\sim$\textquotedblright\ indicates the quantity
computed in absence of the  rainbow's functions $g_{1}\left(  E/E_{P}\right)  $ 
and
$g_{2}\left(  E/E_{P}\right)  $. The corresponding vacuum expectation value
$\left(  \ref{RemoVEVO}\right)  $ becomes%
\begin{equation}
\frac{g_{2}^{3}\left(  E/E_{P}\right)  }{\tilde{V}}\frac{\left\langle
\Psi\left\vert \int_{\Sigma}d^{3}x\tilde{\Lambda}_{\Sigma}\right\vert
\Psi\right\rangle }{\left\langle \Psi|\Psi\right\rangle }=-\frac{\Lambda
}{\kappa^2},\label{RemoWDW1}%
\end{equation}
with%
\begin{equation}
\tilde{\Lambda}_{\Sigma}=\left(  2\kappa^2\right)  \frac{g_{1}^{2}\left(
E/E_{P}\right)  }{g_{2}^{3}\left(  E/E_{P}\right)  }\tilde{\gamma}_{ijkl}\tilde
{\pi}^{ij}\tilde{\pi}^{kl}\mathcal{-}\frac{\sqrt{\tilde{\gamma}}\ \text{ 
}^{(3)}\tilde{R}}{\left(
2\kappa^2\right)  g_{2}\left(  E/E_{P}\right)  }{}.\label{RemoLambdaR}%
\end{equation}
Extracting the TT tensor contribution from Eq.$\left(  \ref{RemoWDW1}\right)
$, we find%
\begin{eqnarray}
&&
\!\!\!\!\!\!\!\!\!\!\!\!\!\!\!\!\!\!\!\!\!\!\!\!\!\!\!\!\!\!\!\!\!\!\!\!\!\!\!\!
\!\! 
\hat{\Lambda}_{\Sigma}^{\bot}=\frac{g_{2}^{3}\left(  E/E_{P}\right)  }%
{4\tilde{V}}\int_{\Sigma}d^{3}x\sqrt{\overset{\sim}{\bar{g}}}\tilde{\gamma}%
^{ijkl}
\left\{  \left(  2\kappa^2\right)  \frac{g_{1}^{2}\left(  E/E_{P}\right)
}{g_{2}^{3}\left(  E/E_{P}\right)  }\tilde{K}^{-1\bot}\left(  x,x\right)
_{ijkl}
\right.\nonumber\\
&&\left.
\ \ \ \ \ \ \ \ \ \ \ \ \ \ \ \ \ \ \ \ \ \ \ \ 
+\frac{1}{\left(  2\kappa^2\right)  g_{2}\left(  E/E_{P}\right)  }%
{}\left[  \tilde{\bigtriangleup}_{L}^{m}\tilde{K}^{\bot}\left(
x,x\right)  \right]  _{ijkl}\right\}  ,\label{Remop22}%
\end{eqnarray}
with the prescription that the corresponding eigenvalue equation transforms
as
\begin{equation}
\left(  \hat{\bigtriangleup}_{L}^{m}{}h^{\bot}\right)  _{ij}=E^{2}%
h_{ij}^{\bot}\qquad\rightarrow\qquad\left(  \tilde{\bigtriangleup}_{L}%
^{m}{}\tilde{h}^{\bot}\right)  _{ij}{}=\frac{E^{2}}{g_{2}^{2}\left(
E/E_{P}\right)  }\tilde{h}_{ij}^{\bot}\label{RemoEE},
\end{equation}
in order to re-establish the correct way of transformation of the perturbation.
$\tilde{\bigtriangleup}_{L}^{m}$is the modified Lichnerowicz
operator \cite{Garattini:2011kp}, defined as%
\begin{equation}
\left(  \tilde{\bigtriangleup}_{L}^{m}{}h^{\bot}\right)  _{ij}=\left(
\tilde{\bigtriangleup}_{L}{}h^{\bot}\right)  _{ij}-4R{}_{i}^{k}{}%
h_{kj}^{\bot}+\text{ }^{(3)}R{}{}h_{ij}^{\bot},\label{RemoM Lichn}%
\end{equation}
where $\tilde{\bigtriangleup}_{L}$is the Lichnerowicz operator defined by%
\begin{equation}
\left(  \tilde{\bigtriangleup}_{L}h\right)  _{ij}=\bigtriangleup
h_{ij}-2\text{ 
}^{(3)}R_{ikjl}h^{kl}+\text{ 
}^{(3)}R_{ik}h_{j}^{k}+\text{ }^{(3)}R_{jk}h_{i}^{k}\qquad\bigtriangleup
=-\nabla^{a}\nabla_{a}.\label{RemoDeltaL}%
\end{equation}
Finally, the propagator $K^{\bot}\left(  x,x\right)  _{iakl}$ will transform
as
\begin{equation}
K^{\bot}\left(  \overrightarrow{x},\overrightarrow{y}\right)  _{iakl}%
\rightarrow\frac{1}{g_{2}^{4}\left(  E/E_{P}\right)  }\tilde{K}^{\bot}\left(
\overrightarrow{x},\overrightarrow{y}\right)  _{iakl}.\label{Remoproptt}%
\end{equation}
Thus the total one-loop energy density for the graviton for the distorted GR
becomes%
\begin{equation}
\frac{\Lambda}{\kappa^2}=-\frac{1}{2\tilde{V}}\sum_{\tau}g_{1}\left(
E/E_{P}\right)  g_{2}\left(  E/E_{P}\right)  \left[  \sqrt{E_{1}^{2}\left(
\tau\right)  }+\sqrt{E_{2}^{2}\left(  \tau\right)  }\right]  \label{RemoVEVR},
\end{equation}
and can be rearranged to give%
\begin{equation}
\frac{\Lambda}{\kappa^2}=-\frac{1}{3\pi^{2}}\sum_{i=1}^{2}\int_{E^{\ast}%
}^{+\infty}E_{i}g_{1}\left(  E/E_{P}\right)  g_{2}\left(  E/E_{P}\right)
\frac{d}{dE_{i}}\sqrt{\left(  \frac{E_{i}^{2}}{g_{2}^{2}\left(  E/E_{P}%
\right)  }-m_{i}^{2}\left(  r\right)  \right)  ^{3}}dE_{i},\label{RemoLoverG}%
\end{equation}
where $E^{\ast}$ is the value that annihilates the argument of the root. It
is easy to see that if $g_{1}\left(  E\right)  =g_{2}\left(  E\right)  =1$,
one recovers the expression $\left(  \ref{Remorho}\right)  $ with a different
normalisation factor. It is clear that not every choice of the rainbow's
functions will produce a finite integral. This is the case for the following
choice on the range $\left[  E^{\ast},+\infty\right)  $%
\begin{equation}
g_{1}\left(  E/E_{P}\right)  =1-\eta\left(  E/E_{P}\right)  ^{n}%
\qquad\text{and}\qquad g_{2}\left(  E/E_{P}\right)  =1,
\end{equation}
where $\eta$ is a dimensionless parameter and $n$ is an
integer\cite{Ling:2006az,Ling:2005bp,Ling:2005bp}. An interesting and promising
choice seems to be%
\begin{equation}
g_{1}\left(  E/E_{P}\right)  =\sum_{i=0}^{n}\beta_{i}\frac{E^{i}}{E_{P}^{i}%
}\,
\exp\left(-\alpha\frac{E^{2}}{E_{P}^{2}}\right),\qquad g_{2}\left(  
E/E_{P}\right)
=1;\qquad\alpha>0,\beta_{i}\in\mathbb{R}.\label{Remog1g2}%
\end{equation}
The use of a \textquotedblleft Gaussian\textquotedblright\ form is dictated by
the possibility of doing a comparison with NCG models\cite{Garattini:2010dn}. 
That one
could obtain a finite result from a one-loop calculation modifying gravity in
the high energy sector is not a surprise. Indeed, as shown by Ho\v{r}ava, a
modification of Einstein gravity motivated by the Lifshitz theory in solid
state physics\cite{Horava:2008ih,Garattini:2010dn}  allows the theory to be
power-counting UV renormalisable with the prescription to recover General
Relativity in the infrared (IR) limit.

\section{Correspondence of Gravity's Rainbow With Ho\v{r}ava-Lifshitz Gravity}
\label{Lifshitzref1}

  Following
\cite{Garattini:2010dn,Garattini:2014rwa} one finds that the Hamiltonian 
constraint  for a 
FRLW
background in Ho\v{r}ava-Lifshitz gravity  becomes%
\begin{equation}
H=\pi_{a}^{2}+\frac{\left(  3\lambda-1\right)  }{\kappa^{4}}24\pi^{4}%
a^{4}\left(  t\right)  \left[  \frac{6}{a^{2}\left(  t\right)  }%
-\frac{12\kappa^2 b}{a^{4}\left(  t\right)  }-\frac{24\kappa^{4}c}{a^{6}\left(
t\right)  }-2\Lambda\right]  =0,
\end{equation}
where%
\begin{gather}
3g_{2}+g_{3}=b,\nonumber\\
9g_{4}+3g_{5}+g_{6}=c\label{Remobcgs}%
\end{gather}
and%
\begin{gather}
g_{0}\kappa^{-2}\equiv2\Lambda,\nonumber\\
g_{1}\equiv-1.\label{Remocoupling}%
\end{gather}
General Relativity is recovered when $b=c=0$, which does not necessarily mean
that all the couplings are vanishing. The potential part is obtained by
imposing the \textquotedblleft projectability\textquotedblright\ condition,
which is a weak version of the invariance with respect to time
reparametrisations, namely that the lapse function is just a function of time,
i.e., $N=N(t)$ \cite{Horava:2009uw}. Such a condition also allows  for a 
significant
reduction of terms in the potential, since it eliminates the spatial
derivatives of $N$. In this case, and neglecting parity-violating terms, the
potential part of the action becomes \cite{Garattini:2005ky}
\begin{eqnarray}
&& 
\!\!\!\!\!\!\!\!\!\!\!\!\!\!\!\!\!\!\!\!\!
\mathcal{L}_{P}=N\sqrt{\gamma}\left\{  g_{0}\kappa^{-2}+g_{1}\text{ 
}^{(3)}R+\kappa^2\left[
g_{2}\text{ }^{(3)}R^{2}+g_{3}\text{ }^{(3)}R^{ij}\text{ }^{(3)}R_{ij}\right] 
\right.  \nonumber\\
&&  \ \ \ \ \ \ \,  +\kappa^{4}\left[  g_{4}\text{ 
}^{(3)}R^{3}+g_{5}\text{ }^{(3)}R\text{ }^{(3)}R^{ij}%
\text{ }^{(3)}R_{ij}+g_{6}\text{ }^{(3)}R_{j}^{i}\text{ 
}^{(3)}R_{k}^{j}\text{ }^{(3)}R_{i}^{k}\right.\nonumber\\
&&
\left. \left.
\ \ \ \ \ \ \, 
\ \ \ \ \ \ \, 
+g_{7}\text{ }^{(3)}R\nabla^{2}\text{ }^{(3)}R+g_{8}\nabla
_{i}\text{ }^{(3)}R_{jk}\nabla^{i}R^{jk}\right] \right\}  ,
\label{Remolpnodb}
\end{eqnarray}
where the couplings $g_{a}\left(  a=0\ldots8\right)  $ are all dimensionless
and running, and moreover we can set $g_{1}=-1$. Let us mention here that the
scenario described by the distorted potential Lagrangian $\left(
\ref{Remolpnodb}\right)  $, in the specific case of FLRW geometry, could be
considered to arise equivalently in the framework of 
$f\left( ^{(3)}R\right)  $
gravity, with $\text{ }^{(3)}R$ the three-dimensional scalar curvature 
\cite{Garattini:2012ec}.
Indeed, if one starts from the Lagrangian
\begin{equation}
\mathcal{L}_{f\left(  \text{ }^{(3)}R\right)  
}=N\sqrt{\gamma}f\left(^{(3)}R\right)
\label{RemoLpf(R)}%
\end{equation}
with
\begin{align}
f\left(^{(3)}R\right)   &  =g_{0}\kappa^{-2}+g_{1}\text{ 
}^{(3)}R-\frac{\kappa^2 b}{3}\text{ }^{(3)}R^{2}%
-\frac{\kappa^{4}c}{9}\text{ }^{(3)}R^{3},\nonumber\\
&  =2\Lambda+\text{ }^{(3)}R\left(  1-2\pi 
b\frac{\text{ }^{(3)}R}{R_{0}}-4\pi^{2}c\frac{\text{ }^{(3)}R^{2}}{R_{0}^{2}%
}\right)  ,
\end{align}
and $b$ and $c$ given by $\left(  \ref{Remobcgs}\right)  $, one will obtain the
same equations as those extracted from $\mathcal{L}_{P}$ in $\left(
\ref{Remolpnodb}\right)  $. Lastly, note that we have used the definitions
$\left(  \ref{Remocoupling}\right)  $, while we have furthermore set%
\begin{equation}
R_{0}\equiv\frac{48\pi}{\kappa^2}=\frac{6}{l_{p}^{2}}.\label{RemoR0}%
\end{equation}
Note that a direct correspondence arises if we allow the energy $E$ to evolve
depending on $t$. In this case, we find  that the classical Hamiltonian
constraint reduces to%
\begin{equation}
\mathcal{H}=\tilde{\pi}_{a}^{2}+\frac{12\left(  3\lambda-1\right)  \pi
^{4}a^{4}\left(  t\right)  }{\kappa^{4}}{}\left[  g_{2}^{2}\left(
E\left(  a\left(  t\right)  \right)  /E_{P}\right)  \frac{6}{a^{2}\left(
t\right)  }-2\Lambda\right]  =0,\label{RemoHGR}%
\end{equation}
and if we assume that and we use the definition $\left(  \ref{RemoR0}\right)
$, we find 
\begin{align}
g_{2}^{2}\left(  E\left(  a\left(  t\right)  \right)  /E_{P}\right)  \frac
{6}{a^{2}\left(  t\right)  } &  =\frac{6}{a^{2}\left(  t\right)  }\left[
1-\frac{2\kappa^2 b}{a^{2}\left(  t\right)  }-\frac{4\kappa^{4}c}{a^{4}\left(
t\right)  }\right]  \nonumber\\
&  =1-\frac{16b \text{ 
}^{(3)}R}{R_{0}}-\frac{256c \text{ }^{(3)}R^{2}}{R_{0}^{2}}.\label{Remoide}%
\end{align}
Although at first site identification $\left(  \ref{Remoide}\right)  $ seems
to be imposed \textit{ad hoc}, it can be supported by invoking the dispersion
relation of a massless graviton, which (see the appendix of
Ref.\cite{Garattini:2014rwa} for details) for a FLRW background, acquires the 
form
\begin{equation}
E^{2}=\frac{k^{2}}{a^{2}\left(  t\right)  },
\end{equation}
with $k$ the constant dimensionless radial wavenumber. Thus, when Gravity's
Rainbow comes into play, we find
\begin{equation}
\frac{E^{2}}{g_{2}^{2}\left(  E\left(  a\left(  t\right)  \right)
/E_{P}\right)  }=\frac{k^{2}}{a^{2}\left(  t\right)  }.\label{Remograv}%
\end{equation}
Since the dispersion relation $\left(  \ref{Remograv}\right)  $ is valid at
the Planck scale too, we can write
\begin{equation}
\frac{E^{2}}{g_{2}^{2}\left(  E\left(  a\left(  t\right)  \right)
/E_{P}\right)  }\rightarrow\frac{E_{P}^{2}}{g_{2}^{2}\left(  E_{P}%
/E_{P}\right)  }=E_{P}^{2}=\frac{k^{2}}{a_{P}^{2}}.
\end{equation}
Hence, Eq.$\left(  \ref{Remoide}\right)  $ becomes
\begin{align}
g_{2}^{2}\left(  E\left(  a\left(  t\right)  \right)  /E_{P}\right)   &
=1-\frac{16b\pi \text{ 
}^{(3)}R}{R_{0}}-\frac{256c\pi^{2}\text{ }^{(3)}R^{2}}{R_{0}^{2}}\nonumber\\
&  =1-c_{1}\frac{E^{2}\left(  a\left(  t\right)  \right)  }{E_{P}^{2}}%
-c_{2}\frac{E^{4}\left(  a\left(  t\right)  \right)  }{E_{P}^{4}%
}.\label{Remog2E}%
\end{align}
Therefore, we deduce that
\begin{equation}
E^{2}=\text{ }^{(3)}R/6k^{2}%
\end{equation}
with
\begin{equation}
E_{P}^{2}=\frac{8\pi}{\kappa^2},\qquad 
c_{1}=16b\pi\text{\qquad\textrm{and\qquad}}%
c_{2}=256c\pi^{2}.
\end{equation}
To summarise, the connection between Gravity's Rainbow and Ho\v{r}ava-Lifshitz
gravity offers an interesting opportunity to investigate the possible quantum
nature of General Relativity. This opportunity is also corroborated by the
fact that, in this analysis, it is the pure gravitational field and its
fluctuations that come into play. Since the quantum fluctuations are
represented by the graviton, and it is the only particle present in absence
of matter fields, we can argue that it is the graviton itself that is able to
distort the gravitational field. From a certain point of view this is not
surprising, since gravity is nonlinear. As a consequence of such a distortion,
some of the usual divergencies that naturally appear in quantum field theory
can be kept under control. This property is also present in
Ho\v{r}ava-Lifshitz gravity. Note that the correspondence between the two
theories is established through the examination of their Wheeler-De Witt
equations. However, although we have explicitly shown this in the case of two
physically interesting spacetimes, namely the FLRW and the spherically
symmetric ones (see  \cite{Garattini:2014rwa} for further details), 
and thus we
have a strong indication that this correspondence is not an artifact of the
spacetime symmetries but rather it arises from the features of the two
theories, a general proof (or disproof) in the case of arbitrary metrics is
still needed. It is interesting to mention that Gravity's Rainbow, in the FLRW
background, generates Ho\v{r}ava-Lifshitz gravity under a specific form of
$f\left(^{(3)}R\right)  $ theory. A
similar result was pointed out in \cite{Olmo:2011sw}, where a connection between
the rainbow's functions and a specific $f\left(^{(3)}R\right)  $ form 
seems to be
evident. These issues reveal that the  bridge between Gravity's Rainbow and
Ho\v{r}ava-Lifshitz gravity could be much richer, and deserves further investigation.




\chapter[Quantum Cosmology in Modified Theories of Gravity]{Quantum Cosmology in 
Modified Theories of Gravity}
\label{Bouhmadichapter}

{\em Mariam Bouhmadi-L\'opez,  Prado Mart\'in-Moruno}\\


Singularities are commonplace in gravitational theories. In particular,
singularities tend to form at the beginning of the Universe through the 
Big Bang or during gravitational collapse, like in black holes. 
Curvature  becomes increasingly large as one approaches those singularities, 
which indicates that the  new physics
could come into play near them. Therefore, quantum gravitational effects are 
usually expected to cure spacetime singularities and this effect is intrinsic to 
any gravitational theory, in particular to any modified or extended theory of 
gravity. In the next few pages  we will give a brief account of quantum 
cosmology in modified theories of gravity within the metric and the Palatini 
approach. Please bear in mind that so far there is no consensus on how a 
fundamental theory encapsulating gravity and quantum
physics should be constructed, and it is still currently a very active area of 
research. However, it is expected that a fundamental quantum theory of gravity 
is necessary, such
that some pathologies within gravitational theories at high energy scales can be 
resolved, such as the non-renormalisability of the theory  or the appearance of 
singularities (in case that those pathologies are also present in the new 
theory).\label{quantumgrefs1}

One of the promising approaches to tackle this issue is based on the quantum
geometrodynamics in which the Wheeler-DeWitt (WDW) equation 
\label{DeWittref3} describes the 
quantum state of the Universe as a
whole through its wave function \cite{kiefer2007quantum}. Alternative approaches 
to quantum gravity include  path integrals, loop quantum gravity and string 
theory, but they will not be covered here. The WDW equation in General 
Relativity 
(GR) is deduced through the Hamiltonian constraint defined by the  Einstein 
equation. 
Therefore, as it is natural whenever the gravitational action, or equivalently 
the equation of motions are modified, we expect to get a new modified WDW 
equation.
In addition, if the solution to the WDW equation satisfies the DeWitt (DW) 
boundary condition \cite{kiefer2007quantum}, that is, it vanishes at the
classical singularity, we may claim that the singularity is expected to be
avoidable through quantum effects. We expect,  this condition, in 
principle,  to be
independent of the WDW equation. In the coming subsections, we will briefly 
summarise how this approach  can be applied to some modified theories of 
gravity.

\section{Quantum Cosmology in a Metric Theory}
\label{quantumetricf1}

As this extensive review shows, there are plenty of metric theories. The 
simplest  - 
and in many aspect best-suited for observations of the early- and the 
late-universe - is the 
$f(R)$ metric approach, which, for example, describes perfectly the early 
inflationary era through the  Starobinsky model  \cite{Starobinsky:1980te}.
The Starobinsky inflationary model was quantised
about thirty years ago. To this end, it was necessary to obtain  the correct 
modified WDW equation for a
homogeneous and isotropic universe after introducing at the classical level a
proper Lagrange multiplier that takes into account the relation between the
size of the Universe, i.e., the scalar factor,  and the curvature as measured 
through the scalar curvature \cite{Vilenkin:1985md}. We remind readers at this 
point
that an $f(R)$ metric theory (different from GR) has a further degree of 
freedom, named the scaleron. It is therefore not surprising that for
a homogeneous and isotropic universe, the WDW equation in this
case has two degrees of freedom, even for an empty  universe. A quantum
approach for a more general class of higher-derivative theories was analysed 
in the references \cite{Hawking:1984ph,Horowitz:1984wv}.

In addition, dark energy singularities in this type of theories have been 
recently analysed: (i) the Big Rip singularity in the framework of $f(R)$ 
quantum geometrodynamics and invoking the DeWitt criterion has been analysed in 
\cite{Alonso-Serrano:2018zpi}, 
where the existence was shown  of solutions to the WDW equation that vanishes 
when approaching the singularity, i.e.,
fulfilling this condition; similarly, (ii) the Little Sibling of the Big Rip 
in 
the framework of $f(R)$ quantum geometrodynamic was analysed in 
\cite{Vasilev:2019iuh}.
For a recent review on dark energy singularities, please see 
\cite{Bouhmadi-Lopez:2019zvz}. It is very important to highlight that this 
equation is always hyperbolic for any $f(R)$-cosmology, even if the classical 
model mimics a phantom
expansion.

\section{Quantum Cosmology in a Palatini Theory}
 \label{quantumPalef1}

Likewise, there is a plethora of extended theories of gravity within the 
Palatini approach; here we will focus on the Eddington-inspired-Born-Infeld 
(EiBI) scenario proposed in \cite{Banados:2010ix} which is  appealing for 
several theoretical aspects: (i) it reduces to GR in vacuum, unlike $f(R)$ 
metric theories,
and deviates from it when matter fields are included. (ii) Due to the structure 
of the gravitational
action (there is a square root in front of the  linear combination that 
involves 
the 
curvature of spacetime), 
the curvature scale and the energy scale seem to be bounded from above (at 
least when the strong energy condition is fullfilled) and the Big Bang
singularity is
naturally avoided in the EiBI gravity \cite{Banados:2010ix}. (iii) The theory 
is 
simple in the sense that it only contains one free additional
parameter, the Born-Infeld constant, as compared with GR. (iv) This scenario is 
free of ghost instabilities because the theory
is constructed through a Palatini variational principle; i.e., the connection 
is 
different from the Levi-Civita connection. The interested reader can see 
\cite{BeltranJimenez:2019acz} to check the conditions that Palatini 
higher-order theories have to satisfy to avoid those instabilities. In fact, the 
idea 
of including the
Born-Infeld structure into the gravitational theory was proposed in 
\cite{Deser:1998rj} but  within  a
metric variational principle inducing, in principle,  ghost degrees of freedom, 
due to the 
higher-order derivative terms in the field equations. The
EiBI theory, however, is formulated via the Palatini variational principle. The 
field equations only contain up to
second-order derivatives, and consequently no ghost is present in the theory. 
The applications and several properties of
the EiBI gravity have been studied widely in the literature; see 
\cite{BeltranJimenez:2017doy} for a nice review on the topic.
Some approaches to quantise EiBI gravity have been proposed in 
\cite{Bouhmadi-Lopez:2016dcf,Arroja:2016ffm,Albarran:2017swy,
Bouhmadi-Lopez:2018tel, Bouhmadi-Lopez:2018sto,Albarran:2018mpg}. 

In fact, in \cite{Arroja:2016ffm,Bouhmadi-Lopez:2018sto} instantons were 
analysed showing, for example, that $O(4)$-symmetric regular instanton 
solutions 
can deviates from those of GR, and in particular the singular Vilenkin 
instanton 
and the Hawking-Turok instanton (in GR) can be regular under certain 
conditions. Moreover, in  
\cite{Bouhmadi-Lopez:2016dcf,Albarran:2017swy,Bouhmadi-Lopez:2018tel,
Albarran:2018mpg} the modified WDW equation in the EiBI scenario was analysed. 
In fact, this involves a very careful analysis of the classical Hamiltonian, 
and 
in particular of the primary and secondary constraints. It can be shown that 
the total Hamiltonian is a first class constraint. After identifying the 
independent constraints of the theory and using the Poisson brackets of all 
those constraints, one can identify any gauge degree of freedom of the theory 
and fix it. Finally, the quantisation of the system requires a proper use of 
Dirac brackets and the promotion of the first
class constraint of the total Hamiltonian as a restriction on the Hilbert 
space, 
where the wave function of the Universe is defined. This approach has been used 
in several EiBI models with several kinds of matter contents: perfect fluids, 
standard scalar field as well as phantom scalar fields, and it has been shown 
successfully that cosmological singularities are removed at the quantum level.

In summary, modified theories of gravity in many cases need to be quantised 
like GR,  and at the same time they offer a splendid new arena   to further 
explore the most suitable or even the correct path to get a consistent quantum 
gravity theory. In this sense, there is still a lot of interesting and 
important work to be carried on.


 \newpage

 \phantomsection
\addcontentsline{toc}{part}{\bf Part II: Testing Relativistic Effects}
\begin{center}
{\Huge \bf Part II: Testing Relativistic Effects}
\end{center}
\begin{center}
Editors:  Mariafelicia De Laurentis  and    Gonzalo J. Olmo  
\end{center}






\begingroup
\let\clearpage\relax 
\chapter[Introduction to Part II]{Introduction to Part II}

{\em  Mariafelicia De Laurentis,    Gonzalo J. Olmo }\\

The confrontation of gravitation theories with experimental and observational 
data is a fundamental step in the scientific process. The analysis of 
relativistic effects is not only necessary but essential for this purpose. 
Laboratory tests typically search for fifth force effects in the form of short 
range interactions, which can introduce departures from Newton’s law via 
Yukawa-type corrections mediated by some kind of massive degree of freedom. In 
some cases this may require going beyond the linearized approximation due to the 
existence of screening mechanisms that may hide these (chameleon) interactions, 
which poses severe experimental challenges for current technologies such as 
atomic interferometry, torsion balance experiments, Casimir force, dipole moment 
tests, \ldots Screening mechanisms can also be constrained via Lunar ranging, by 
measuring cosmic filaments, and by probing the nonlinear regime of cosmological 
perturbations.  

The effects of modified gravitational dynamics may also arise via nonlinearities 
induced by the stress-energy densities rather than by new propagating degrees of 
freedom, thus leading to new phenomena, which do not involve fifth force 
interactions. This can have nontrivial effects even in scenarios involving 
elementary particles if one focuses on aspects not related to curvature but to 
non-metricity and/or torsion. However, from an effective field theory 
perspective, 
such new interactions could fit naturally in an extended matter framework; the 
universality of certain couplings could reveal an underlying geometric 
structure, thus showing that elementary particle experiments could complement 
astrophysical tests to unveil modified gravity effects. The possibility of 
having new gravitational physics induced by non-linearities in the matter sector 
may also have an impact on the structural properties of self-gravitating 
systems. In very low mass stars, where the equation of state of the gas is well 
understood, modifications in the Newtonian dynamics can change the threshold for 
sustained hydrogen burning reactions, offering new observables in the search for 
departures from the predictions of GR. In compact objects such as neutron stars, 
on the other hand, these interactions could lead to new degeneracies with the 
matter sector, complicating even more the quest for the properties of the 
nuclear matter equation of state. In order to break such degeneracies, it is 
important to identify observables that may lead to universal relations able to  
tell different gravity theories apart. Some of these relations involve the 
moment of inertia, asteroseismology, quasi-normal modes, etc,
and it has been shown that massive scalar degrees of freedom could manifest 
themselves clearly in sufficiently separated binary systems and in quasi-normal 
modes spectra.

Orbital motions and lensing are also key probes for modifications of gravity. 
The parametrised post-Newtonian formalism developed in the 1970s allows us to 
confront very different types of theories with observations by just computing 
certain key coefficients in the appropriate limit and gauge 
choice. This formalism must be extended in order to accommodate   new theories, 
which do not quite fit within this original framework.  

Beyond stellar objects and the slow motion limit, strong gravity effects, such 
as gravitational waves and strong lensing, typically involving black holes, 
also offer a glimpse of potentially new gravitational phenomena, including the 
quantum regime.

\endgroup

\chapter[Laboratory Constraints]{Laboratory Constraints}
\label{Davischapter}
\label{laboratoryref1}

{\em Anne-Christine Davis,
Benjamin Elder}\\


 
 







%
%
%
Dark energy is a cosmological phenomenon per se \cite{Brax:2017idh}. In this 
chapter  we will describe attempts to detect effects of the physics
of modified gravity, motivated by dark energy and the cosmological constant 
problem, in the laboratory.
Classical effects of modified gravity  can be tested by fifth force searches,
where new classical interactions could influence the motion of test masses 
\cite{Kapner:2006si}. The quantum nature of the modifications can also be probed 
using the Casimir effect \cite{Lamoreaux:1996wh}, atom \cite{Hamilton:2015zga, 
Sabulsky:2018jma} and neutron \cite{Lemmel:2015kwa} interferometry, the neutron 
energy levels in vacuum \cite{Nesvizhevsky:2003ww}, atomic spectra 
\cite{Brax:2010gp, Wong:2017jer} and the electron magnetic moment 
\cite{Brax:2018zfb}. For a recent review, see \cite{Brax:2018iyo}.   We will 
restrict our attention to scalar-tensor theories with a coupling between a 
scalar field and matter.

Laboratory tests of gravity have a long history, and the need to make 
cosmological theories of dark energy and modified gravity compatible with 
laboratory and Solar System tests was a key motivation for the introduction of 
screening mechanisms.
Screening allows for modified gravity theories
to evade traditional tests of gravity, whilst still modifying gravity 
cosmologically.  However,  
carefully designed laboratory experiments can allow the effects of the scalar 
field to be  unscreened.  The additional level of precision and control that we 
have in the laboratory then means that these measurements  tend to be extremely 
constraining for cosmological modified gravity theories. Laboratory tests probe 
the theory in the nonlinear regime, 
unlike cosmological tests where the linear regime is easiest to test.
Probing the nonlinear regime necessitates a case-by-case analysis, since a 
model independent parameterised description  is not yet available for 
laboratory 
tests.

\label{chameleonkiref3} \label{Fifthref2}
Laboratory experiments are most effective at constraining theories, which 
screen  through a chameleon-like mechanism, i.e., {\it thin-shell} theories.   
Thin-shell theories have the advantage that the scalar field responds rapidly 
to changes in 
density, meaning that even if the effects of the scalar are screened in the 
Solar System they can be unscreened by a laboratory vacuum chamber.  Other 
theories, notably Galileons~\cite{Nicolis:2008in}, do not have this property 
and 
are therefore not as tightly constrained by laboratory tests, so in this 
chapter 
we restrict our attention to thin-shell theories.

In this section we will first describe how a thin-shell scalar behaves in a  
laboratory vacuum, and then go on to detail the laboratory experiments, which 
currently are the most constraining for chameleon models. 
For convenience's sake, we focus on the prototypical  chameleon model 
\cite{Khoury:2003aq,Brax:2004qh}
\be
V_\mathrm{eff}(\phi)= \frac{\Lambda^{4+n}}{\phi^n} +  
\frac{\beta}{M_{Pl}} \phi \rho_\mathrm{m}+ \dots,
\ee
where $\rho_\mathrm{m}$ is the ambient matter density, $M_{Pl} \equiv (8 
\pi G)^{-1/2}$, and the $\dots$ includes a cosmological constant piece, as  
chameleons do not lead to self-acceleration.   The coupling of the chameleon 
field $\phi$ to matter may be understood as originating from a coupling 
function 
$A(\phi)= e^{\beta \phi/m_{\rm Pl}}$,
which rescales the metric between the Jordan and Einstein frames. 
\label{Einsteinfrref5}\label{Jordanrref6}

\section{Chameleons in Laboratory Vacuums}

The chameleon scalar field changes its mass $m \equiv 
\frac{\mathrm{d}^2}{\mathrm{d}  \phi^2}V_\mathrm{eff}(\phi)$ as a function of 
the local matter density $\rho_\mathrm{m}$. We consider here an idealised 
vacuum chamber, which is spherical with internal radius $L$, internal density 
$\rho_{\rm vac}$ and walls of density $\rho_{\rm wall}$ and thickness $T$. If 
$T > 1/ m(\rho_{\rm wall})$, then the field has sufficient room to minimise its 
effective potential within the vacuum chamber walls. This greatly simplifies 
our calculations, as it guarantees that the interior of the vacuum chamber is 
effectively shielded from external objects.
Whilst the condition $T> 1/ m(\rho_{\rm wall})$  needs to be checked experiment 
by experiment, and  model by model, in general we find that this is satisfied 
for chameleon models of interest if $T\gtrsim 1 \mbox{ mm}$.

In the interior of the vacuum chamber the density is much smaller than within 
the walls,  therefore inside the vacuum chamber the chameleon field will try to 
reach the minimum of its effective potential in this lower density 
$\rho_\mathrm{vac}$. This is only possible if there is sufficient room for the 
scalar field to do so, that is, if $L>1/m(\rho_{\rm vac})$.
If this condition is not met, then the chameleon takes a value such that its 
Compton wavelength is of the order of the size of the vacuum chamber $L \approx 
1/m(\rho_{\rm vac})$.
In general, one needs to compute the chameleon profile inside the vacuum 
chamber 
 numerically~\cite{Elder:2016yxm,Schlogel:2015uea}.


Whether or not an object is screened depends on the value of the scalar field 
in the interior of the object $\phi_\mathrm{obj}$, and the background value 
that the scalar would take if the source were absent, $\phi_\mathrm{vac}$. The 
advantage of performing experiments in laboratory vacua is that the difference 
$|\phi_\mathrm{vac} - \phi_\mathrm{obj}|$ can be very large.  Furthermore, the 
range of the chameleon force inside the vacuum chamber is approximately 
$1/m_\mathrm{vac}$, which can stretch to $\sim$ cm scales, making it far easier 
to detect the fifth force in vacuum than in the atmosphere.

\begin{figure}[!]
\begin{center}
\includegraphics[width=7.5cm]{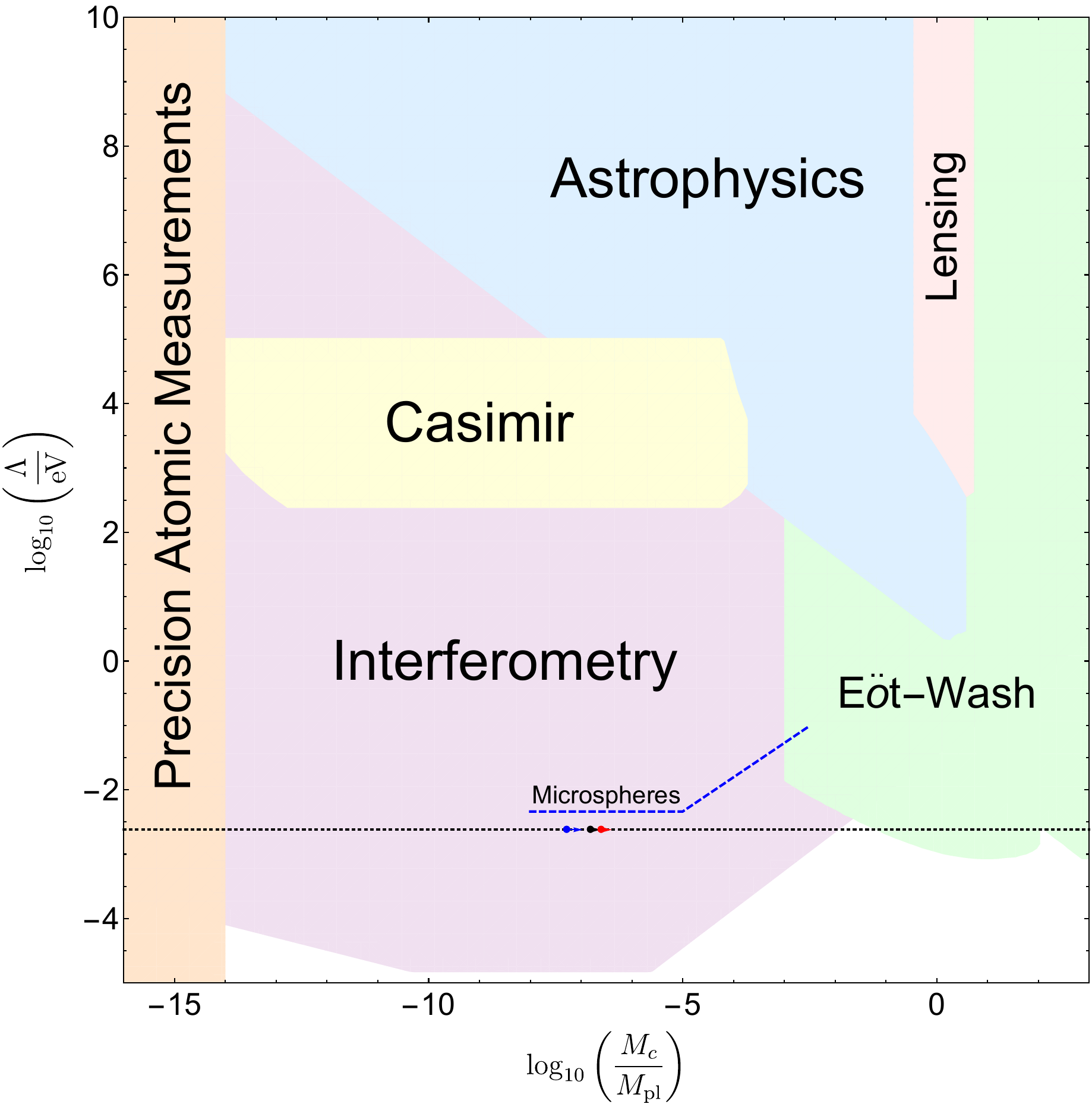}
\hspace{40pt}
\includegraphics[width=11.5cm,rotate=-90]{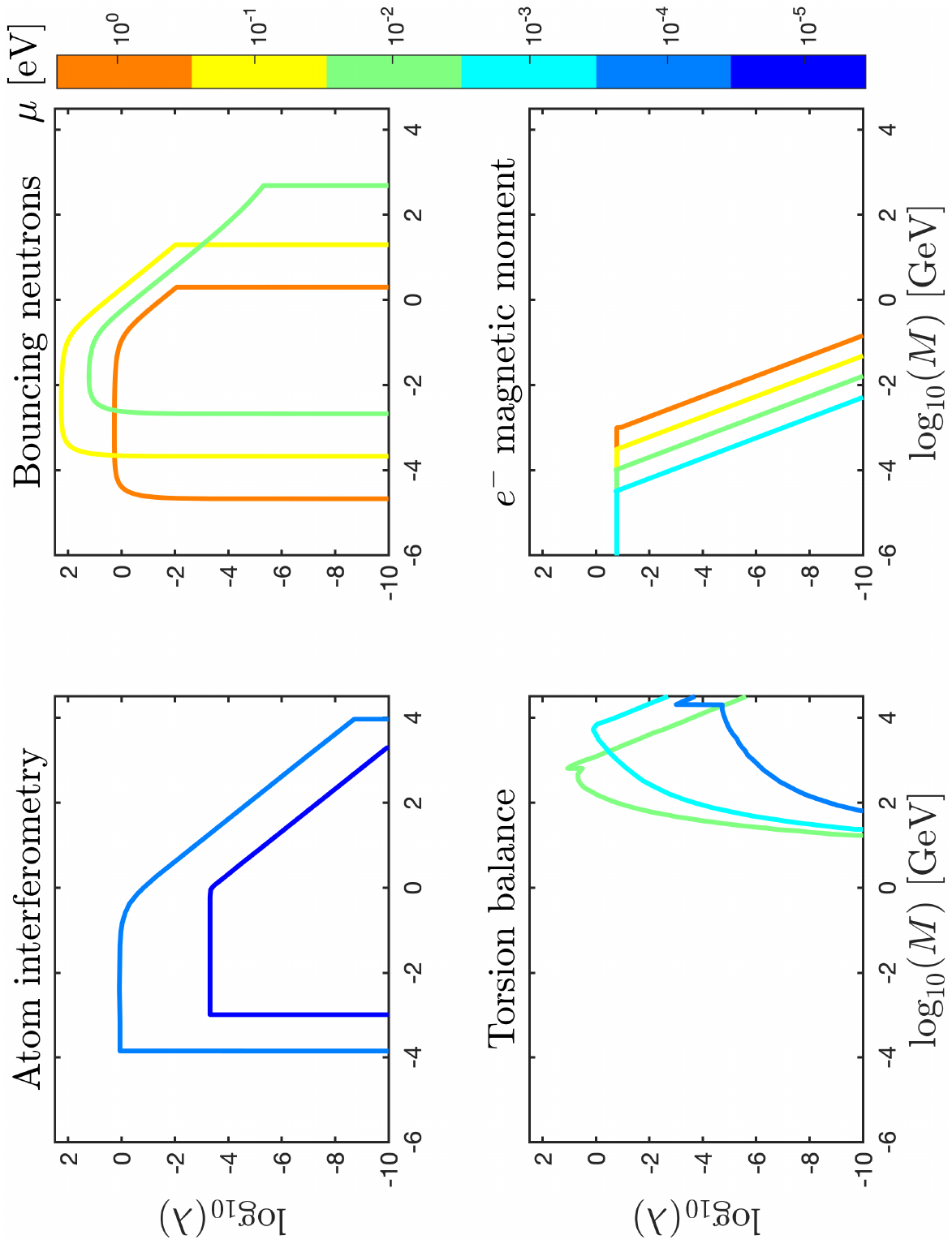}
\vspace*{8pt}
\caption{{\it{ {\bf Upper:} Experimental constraints on the chameleon 
self-coupling  $\Lambda$ and coupling to matter $M_c= \frac{M_{\rm 
Pl}}{\beta}$, for the chameleon potential $V(\phi)= \Lambda^5/\phi$ 
\cite{Burrage:2016bwy}.  Shaded regions are ruled out. 
{\bf Lower:} Constraints on the symmetron 
parameters $\mu, \lambda, M$.  The area below each curve is ruled out.  The 
symmetron has three parameters (as opposed to two parameters for the simplest 
chameleon model), hence bounds are plotted separately for improved 
readability.  The constraints deriving from measurements of the electron's 
magnetic moment extend into other regions of parameter space as well; see 
\cite{Brax:2018zfb} for details.  Some caution is warranted in interpreting 
constraints on exceedingly large or small values of $\lambda$; again,  see 
\cite{Brax:2018zfb} for a discussion.}}}
\label{fig:constraints}
\end{center}
\end{figure}

In large parts of the chameleon parameter space, very small (but dense) objects 
like neutrons, atomic nuclei and {silica}  microspheres can be unscreened in 
vacua with $L \sim 10 \mbox{~cm}$ and $\rho_{\rm vac}\sim  10^{-17} \mbox{ 
g/cm}^3$, which makes them excellent test masses.
In the following sub-sections we review the most constraining experiments for  
chameleon models. These constraints are summarised in 
Fig.~\ref{fig:constraints}.

\section{Atom Interferometry}
\label{atominteref1}

Atom interferometry is a well-established technique to measure the  external 
forces acting on a single atom~\cite{ChuAI}.  It functions in a manner somewhat 
analogous to the classic double slit experiment.  The atom's wave function is 
split into two parts, which are sent along different trajectories and then 
overlapped again at a later point in time.  Any difference in the quantum 
mechanical phase that is accumulated along the paths results in interference at 
the end point.

This process is accomplished in the following manner.  First, a small cloud of  
cold atoms are placed in free-fall inside a vacuum chamber.  The atoms are then 
subjected to a pulse of laser light, the frequency and duration of which are 
chosen such that each atom has a $50\%$ probability of absorbing a photon and 
its momentum and being kicked into an excited state.  This splits the 
wavefunction of each individual atom into two wavepackets, travelling along 
slightly different trajectories.  A short time $T$ later, a second laser pulse 
is timed such that atoms in the ground state have a $100\%$ probability of 
absorbing a photon, and the atoms in the excited state have a $100\%$ 
probability of undergoing stimulated emission.  This reverses the relative 
motion of the wavepackets, so that they overlap after another $T$ seconds.  At 
that time, a third $50\%$ pulse recombines and interferes the wavepackets, and 
the relative phase accumulated along the two paths may be measured by   
counting the numbers of atoms in the ground and excited states.

Relative phase differences may be due to a difference in the accumulated action 
along each path, or from differences in the phase inherited from the photon at 
each photon absorption and stimulated emission event.  If the atoms experience 
a constant acceleration along the direction of the laser pulses, then the total 
phase difference is particularly simple, and the probability to find the atoms 
in the initial state is
\begin{equation}
P \propto \cos^2\left(a k T^2\right),
\end{equation}
where $k$ is the photon momentum, $T$ is  the time between laser pulses (so 
that $2T$ is the duration of the experiment), and $a$ is the constant 
acceleration.

The acceleration $a$ includes all external forces acting on the atom, including 
 Newtonian gravity and any new fifth forces.  To constrain the latter, the 
experiment is performed with an {\it in-vacuum} source mass near the atoms.  
The experiment is performed twice, with the source mass nearby and again with 
the source mass far away, such that the force due to the source mass may be 
isolated.  After subtracting out the ordinary Newtonian gravitational force, 
any leftover signal would be due to new interactions, such as chameleons.
 
The source mass is typically sufficiently large that it will be screened, but  
the atomic nuclei are small enough that, as discussed in the previous 
subsection, they are unscreened over a large range of the parameter space
 \cite{Burrage:2014oza,Burrage:2015lya,Elder:2016yxm},  making them sensitive 
probes of the chameleon field.
Experiments searching for chameleon accelerations with atom interferometry have 
 reached a sensitivity of ${\approx 
10^{-8}g}$~\cite{Hamilton:2015zga,Burrage:2016rkv,Jaffe:2016fsh,Sabulsky:2018jma
}, where $g \equiv G M_\mathrm{Earth} / R_\mathrm{Earth}$ is the gravitational 
acceleration at the surface of the  Earth.

\section{E\"{o}t-Wash}
 \label{Washref1}

Torsion balance experiments have a long history of searching for fifth  forces 
and modifications of gravity.  The underlying principle is to have one or more 
test masses suspended, and to look for deflections of the test masses towards 
source masses by measuring the torsion in the suspension of the test masses.
Commonly, the source and test masses are arranged so that the inverse-square  
contribution to the total force is cancelled, and the experiment is only 
sensitive to deviations from standard gravity.

The current best constraints come from the  E\"{o}t-Wash  experiment 
\cite{Adelberger:2003zx}, which uses circular disks for the masses. The disks 
have holes bored into them and are arranged one above the other, so that if 
there are no modifications to gravity then there is expected to be no net 
torque between 
the two plates. Recent constraints on fifth forces are discussed in 
\cite{Upadhye:2012qu, Upadhye:2012rc}.

\section{Casimir}
 \label{Casimirref1}

The Casimir force is an effect predicted by quantum electrodynamics,  which is 
absent in classical physics. It is the force that arises between two parallel 
plates, placed in vacuum, due to the quantum fluctuations of the 
electromagnetic field in the space between the plates. This force scales as 
$d^{-4}$, where $d$ is the distance between the plates, and therefore is most 
easily detected when the plates are placed close together. Current experiments 
probe sub-mm and sub-micron distance scales \cite{Lamoreaux:2005zza}.
If fifth forces exist they could also be detected by an experiment searching 
for  Casimir effects.  These experiments are particularly sensitive to 
screening through the thin-shell effect, as close to the surface of a source 
the 
field is changing rapidly, giving rise to potentially detectable forces. The 
chameleon force (per unit area) between  two plates  scales as 
\cite{Brax:2007vm}
\begin{equation}
\frac{F_{\rm cham}}{A}\propto d^{-\frac{2n}{n+2}}~.
\end{equation}

The experimental challenge for such a search is to make the two plates 
perfectly smooth and to keep the plates perfectly parallel. In practice it may 
be easier to search for the Casimir effect between a plate and a sphere.
Current searches for the {Casimir} force are most constraining for chameleon 
models with  $n=-4$ and $n=-6$,  when $\Lambda_c$ is fixed to the dark energy 
scale.
Further sensitivity to the chameleon could be obtained by varying the  density 
of the gas between the two plates \cite{Brax:2010xx}, or by using a sphere and 
plate experiment, which enables different distances to be investigated, as in 
\cite{Chen:2014oda}.

\section{Neutron, Atomic and Electron Dipole Moment Tests}
\label{dipolemomref1}

New fifth forces would also change the quantum energy spectrum of ultra-cold 
neutrons in the  gravitational potential of the Earth. A spectroscopic 
technique based on the Rabi resonance method has been used to constrain the 
parameters of popular dark energy models \cite{Brax:2013cfa}.

As previously discussed, atomic nuclei can be unscreened in a laboratory 
vacuum.  The nuclei can be considered as the source of a chameleon field that 
is 
probed by the orbiting electrons. If the unscreened chameleon force is very 
strong then this could cause measurable perturbations to atomic energy levels.
The most precise measurements currently are of the structure of hydrogenic 
atoms. Shifts to the lowest energy levels due to a chameleon force can occur 
and are discussed in \cite{Brax:2010gp, Wong:2017jer}. In \cite{Wong:2017jer},
constraints were imposed on the coupling of the chameleon to matter and also to 
photons.
Similarly, fifth forces induce additional quantum corrections to the  electron 
dipole moment, which also leads to constraints \cite{Brax:2018zfb}.

\section{The Symmetron}

The symmetron is another quintessential example of a screened modification to 
gravity \cite{Hinterbichler:2010es}.   
The symmetron is similar to the chameleon in that it has a  canonical kinetic 
term, and its screening is through terms that are nonlinear in the field.  
However, the difference between the models is that while the chameleon screens 
because mass varies depending on its environment, the symmetron screens 
primarily because the strength of its coupling to matter is allowed to vary 
dynamically.
This is possible thanks to the symmetron's spontaneous symmetry  breaking 
potential, which contains a coupling to matter such that regions of high 
density restore the symmetry, effectively shutting off the coupling to matter.

To see this concretely, the symmetron effective potential is
\begin{equation}
V_{\rm eff} (\phi) = \frac{1}{2}\left(-\mu^2 +\frac{\rho}{M^2}\right)\phi^2  
+\frac{\lambda}{4 !} \phi^4,
\label{eq:symmpot}
\end{equation}
where $\mu$ is the bare mass of the symmetron, $M$ the energy scale controlling 
strength of the coupling to matter, and $\lambda$ the dimensionless constant 
controlling the self-interactions of the field.  The form of the coupling to 
matter also means that the symmetron fifth force experienced by a test particle 
is $\vec{F}=\phi \vec{\nabla}\phi /M^2$.  Note that the strength of the force 
is scaled by the local field value  consequently, in dense regions where $\rho 
> \mu^2 M^2$, the field is driven to $\phi \to 0$ and the fifth force shuts 
off.    \label{Fifthref2}

As can be seen from Eq.~\eqref{eq:symmpot} the mass of the symmetron in the  
symmetry broken phase is approximately $\mu$. Unlike the chameleon, therefore, 
the symmetron does not have the ability to adjust its mass in the low density 
environment of a laboratory vacuum chamber.  If the Compton wavelength of the 
symmetron is larger than the size of the vacuum chamber, $\mu L \ll 1$, the 
field is not able to vary within the chamber and so no fifth force can be 
present.  Conversely, if the Compton wavelength of the symmetron is smaller 
than the distances probed in the experiment (for example, the distance between 
test and source masses), then the fifth force will be exponentially suppressed 
by the Yukawa term $e^{-m d}$, where $m$ is the symmetron mass and $d$ the 
distance 
between two objects.  This means that any experiment is only sensitive to 
symmetron models whose masses fall between these two limits.

Constraints on the symmetron model have recently been computed for all of  the 
experiments described above, and the bounds are summarised in 
Fig.~\ref{fig:constraints}.  Casimir constraints are very 
recent~\cite{Elder:2019yyp} and use a sphere and plate, with the experiment 
probing 
two different distances by a clever set-up similar to that described in 
\cite{Chen:2014oda}.  As the full details of that experiment are not yet 
available, those constraints are not presently included.

\section{Conclusions}

In summary, laboratory tests of gravity contain a wealth of  information about 
dark energy and modifications to gravity.  The rich phenomenology of screened 
theories in particular leads us to look for new physics in novel and sometimes 
surprising ways.  As these experiments probe the nonlinear regime, there is no 
model-independent way to connect these constraints to the parametrised linear 
and quasi-linear theories used to obtain cosmological constraints.  However, 
once a model is specified, that connection can be made clear.



\chapter[Screening Mechanisms]{Screening Mechanisms}
\label{Braxchapter}

{\em Philippe Brax}


\section{Screening}
 \label{screeningref1}

Quintessence models   are meant to reproduce the late-time acceleration of the 
Universe, using a scalar field sufficient for this one particular problem. 
Requiring 
that the equation of state of the scalar field model is close to $-1$ implies that 
the mass of the scalar must be of the order of the Hubble rate. This would not 
be a problem if the scalar field were decoupled from matter. However, the 
scalar 
couples to gravity and standard model particles too. This induces at one loop a 
coupling between the scalar and matter, which is logarithmically divergent. In 
the absence of symmetry reasoning, this coupling is not naturally vanishing and 
should be considered as a new parameter of quintessence models.

The coupling between matter and the scalar can be embodied using a conformal 
rescaling between  the Jordan metric and the Einstein one. The Jordan metric 
\label{Jordanrref4}
$\tilde g_{\mu\nu}$ is the one that is used to write down the matter action. 
The Einstein metric $g_{\mu\nu}$ is the one where the Enstein equation takes a 
natural form. The simplest relation between them is
\be
\tilde g_{\mu\nu}= A^2(\phi) g_{\mu\nu},
\ee
and the coupling is then identified with
\be
\beta= m_{\rm Pl} \frac{\partial \ln A(\phi)}{\partial \phi}.
\ee \label{solarsystemref6} \label{Fifthref3}
In the Solar System, the Cassini bound on fifth forces implies that $\beta^2  
\le 
10^{-5}$  which would exclude metric $f(R)$ theories as their 
$\beta=1/\sqrt{6}$ 
is fixed. \label{chameleonkiref4}
This is not the case when the chameleon mechanism is at play. Indeed, in the 
presence of matter,  the dynamics of the scalar is governed by the effective 
potential
\be
V_{\rm eff}(\phi)= V(\phi)+ (A(\phi)-1)\rho_m,
\ee
where $\rho_m$ is the conserved matter density and $V(\phi)$ is the potential 
of 
the scalar model.  When the effective potential admits a density-dependent 
minimum for which the mass of the scalar at the minimum, i.e., the second 
derivative of the effective potential at the minimum, is larger than the Hubble 
rate, then the minimum is a cosmological tracker. In this case the model is a 
chameleon model. For large curvature $f(R)$ theories, the minimum is a tracker 
and they are chameleon models. In this case  and for static configurations, 
screening occurs when the mass of the scalar field inside a dense body is so 
large that its Yukawa suppression in $e^{-m r}$ from each point inside the body 
prevents any radiation of the field outside the body. Only a thin shell of mass 
$\Delta M$ can in fact generate a scalar field outside the body and hence 
contribute to a fifth force.
The mass of the thin shell is given by
\be
\frac{\Delta M}{M}= \frac{\vert \phi_\infty-  \phi_c\vert}{2 \beta_\infty 
\Phi_N 
m_{\rm Pl}},
\ee
where $\phi_c$ is the value of the field at the minimum associated with the 
density $\rho_c$ of the body.  Outside, the density is $\rho_\infty$ and the 
associated minimum is $\phi_\infty$. The coupling in the vacuum outside is 
$\beta_\infty$ and $\Phi_N$ is the Newton potential at the surface of the body, 
e.g., $10^{-6}$ for the Sun. Screening occurs when $\Delta M\lesssim M$. In 
this 
situation, the scalar interaction between two bodies contributes an extra 
factor 
$2\beta_A \beta_B$ between bodies $A$ and $B$. If they are not screened, then 
$\beta_{A,B}=\beta_\infty$. If the bodies are screened then $\beta_A= 
\beta_\infty \frac{\Delta M_A}{M_A}$.
In particular, the Sun, the Earth and the Moon must be screened to pass Solar 
System tests of gravitation. The Lunar  Ranging experiment has constrained the 
difference between the accelerations of the Moon and the Earth towards the Sun 
at the $10^{-13}$ level. This implies that
\be
\beta_\oplus \le 10^{-6},\ \beta_\odot \le 10^{-9}.
\label{charge}
\ee
and for an unscreened satellite the Cassini bound only requires $\beta_\infty 
\beta_\odot  \le 10^{-5}$, which is always satisfied when $\beta_\infty={\cal 
O}(1)$.
This formalism has been successfully applied to the case of Hybrid-Metric 
Palatini theories  \cite{VargasdosSantos:2017ggl} where the first derivative 
$f_{R_g}$ of the function $f(\hat R)$ of the Palatini curvature that 
parameterises the deviation from General Relativity must be less than 
$10^{-4}$. 
This result is similar to the bound on $f_{R_g}$ in large curvature metric 
models.

Screening operates differently when the density of the compact objects 
\label{compactobrefs1} is not 
constant. For instance, in the case  of stars on the Red Giant Branch,  
checks must be made as to whether the field can really stay at the minimum of 
the effective 
potential when the density evolves. For this,  one must compare the chameleon 
time $t_\phi=1/m_\phi$ where $m_\phi$ is the mass at the putative minimum and 
the collapse time $t_{\rm ast}=\frac{1}{(G_N \rho)^{1/2}}$ for a star of 
average 
density $\rho$. When $t_\phi \ll t_{\rm ast}$, the field tracks the evolution 
of 
the density all the way to the Red Giant formation. Different chameleon models 
with different couplings to matter have been considered in 
\cite{Najafi:2018apy},
where the effects of the chameleon are found to be  small.

For compact objects such as neutron stars,\label{neutronstarsref1} the coupling 
between the scalar 
field 
and gravity cannot be neglected, resulting in modified 
Tolman-Oppenheimer-Volkov 
equations. In this case an instability occurs due to the unboundedness from 
below of the effective potential when the density falls below the pressure deep 
inside the star, i.e.,
\be
\rho<3 p.
\ee
When this is not the case, the effects of modified gravity are almost 
degenerate 
with a change of the equation  of state of matter in the neutron star. This has 
been observed in \cite{Brax:2017wcj}, where it has been suggested that the 
degeneracy can be lifted using the relationship between the reduced moment of 
inertia $\bar I= I/G_N^2 M^3$ and the compactness $C= G_N M/R$ of the star,  
which does not depend on the equation of state \footnote{Here, $R$ is the 
radius 
where the pressure vanishes and $M$ is the mass of the star within this 
radius.}.

Another astrophysical probe of screening can be found on astrophysical scales: 
cosmic filaments. Indeed,  the constraints from the Solar System such as the 
Lunar Ranging experiment, lead to (\ref{charge}), which in turn implies for 
typical models such as the Hu-Sawicki $f(R)$ model that the mass of the scalar 
field 
on cosmological scales is typically larger than $10^2 H_0$, where $H_0$ is the 
Hubble rate now. This implies that the range of the scalar interaction is 
shorter than a few Mpcs. Filaments of sizes between 1 and 20 Mpcs are 
particularly suited to see effects of screened modified gravity. Indeed, it has 
been shown in \cite{Ho:2018byw}, using the N-body simulations generated by the 
ISIS and RAMSES codes, that filaments are shorter and denser than in the 
$\Lambda$CDM model. Overall they are good candidates to observe sizeable 
differences with the standard model of cosmology on astrophysical scales. A 
similar type of observables probing the nonlinear to the  quasi-linear regime 
of cosmological perturbations on scales between 0.1 and 10 Mpcs is the power 
spectrum of the Lyman-$\alpha$ flux decrement. Lyman-$\alpha$ clouds at high
redshifts absorb the light emitted by distant quasars. For the 
Hu-Sawicki $f(R)$ model the deviations from the standard model are small. On 
the 
other hand, for models subject to another type of screening mechanism, i.e.,
K-mouflage, this can be more significant.
K-mouflage operates when the gravitational acceleration $\vert \nabla 
\Phi_N\vert$ is large enough.  Typically this happens in scalar models of the 
K-essence type where the kinetic terms are a function $K(X)$, which is 
nonlinear 
and where $X= -\frac{(\partial \phi)^2}{2M^4}$ with $M$ taken to be the dark 
energy scale $M\simeq 10^{-3}$ eV. The coupling to matter $\beta$ is typically 
required to be less than 0.1 to satisfy the Solar System tests.
In this context, a precise comparison of K-mouflage screened models and the 
Lyman-$\alpha$ effects   will require dedicated numerical simulations of the 
intergalactic medium in the presence of modified gravity \cite{Brax:2018zbd}.  
On larger scales, the effects of screening necessitate the study of  
cosmological 
perturbations beyond leading order. This has been performed in 
\cite{Aviles:2018qot}, where effects on the higher-order statistics of the 
density contrast of matter have been investigated. These tools will be of 
relevance with the advent of   large-scale surveys in the next decade.

\section{Laboratory Experiments and Quantum Effects}
 \label{laboratoryref2} \label{quanteffectsfs1}

Screened models of the chameleon type could induce large effects in laboratory 
experiments. The prime  example is the Casimir effect, whereby two metallic 
plates are attracted due to the quantum fluctuations of the photon field. 
Scalar 
fields can induce a classical force between screened plates, as they are very 
dense, due to the particular profile of the scalar field in the vacuum between 
the plates. A particularly relevant model in this context is the symmetron,
whose 
potential and coupling functions read
\be
V(\phi)= -\frac{\mu^2}{2} \phi^2 + \frac{\lambda}{4} \phi^4, \  \ A(\phi)=1 + 
\frac{\phi^2}{2\Lambda^2}.
\ee
This Higgs-like models has a $Z_2$ breaking transition in the  presence of 
matter. Between plates the field vanishes when the distance between the plates 
is small enough, $d\lesssim \frac{\pi}{\mu}$. When the distance is large 
enough, 
a soliton-like solution can be found involving elliptic Jacobi functions. These 
solutions allow for an exact calculation of the classical Casimir force between 
two plates for symmetrons \cite{Brax:2017hna}. Another consequence of the 
symmetron phase transition is that starting in the symmetric phase, 
corresponding 
to a vanishing field and lowering the density in the presence of boundaries, 
e.g., plates, can result in the spontaneous appearance of domain walls 
interpolating between the two vacua at $\phi= \pm \frac{\mu}{\sqrt \lambda}$. 
These domain walls could be triggered in dedicated experiments by removing the 
gas inside a chamber.
A probe of the existence of domain walls in an experimental cavity could be 
obtained by  monitoring the trajectories of neutral particles across the domain 
walls. Their crossing time of the cavity would be affected by the scalar force 
induced by the domain wall. This has been investigated in 
\cite{Llinares:2018mzl}.

So far we have only considered the classical interaction mediated by a scalar 
field between matter  particles or objects. In fact, coupled scalar fields also
mediate  a quantum interaction, which manifests itself when the coupling 
function $A(\phi)$ has a non-vanishing second derivative. The force acting on a 
body separated from another one by a distance $L$ is given by
\be
F= \int d^3 x \partial_L J <A>_J,
\ee
where $J(x)$ is the matter density of all the objects and the derivative is 
taken  by moving one object with respect to the other one. The quantum average 
$<A>_J$ is the average of the quantum operator in the presence of the source 
$J$. This formal results can be loop-expanded. The tree level force is the 
classical interaction that we have already presented. The first correction 
occurs at one loop and reads
\be
F_{\rm 1-loop}= \frac{1}{2}\int d^3 x \partial_L J A'' \Delta_J (x,x),
\ee
involving the second derivative of the coupling function $A(\phi)$ and the 
Feynman  propagator in the presence of the source $J$ evaluated at coinciding 
points. This formulation allows one to calculate the quantum contribution to 
the 
Casimir effect from light scalars. In particular, one retrieves that for a 
nearly 
massless scalar between two plates where the coupling is nearly infinite, as 
they are very dense, the quantum force for a scalar  is one half of the photon 
case. These results have been exploited in \cite{Brax:2018grq} - in particular 
for 
symmetron models. One-loop effects can also induce a scalar correction to the 
magnetic moment of the electron. This happens when the usual photon propagator 
between two electron lines coupled to a photon is replaced by the scalar field 
propagator. The constraints thus obtained are particularly stringent for the 
symmetron \cite{Brax:2018zfb}.

\section{Other Screening Effects}

We have dealt with the chameleon screening mechanisms and touched upon the 
K-mouflage screening.  There is a good reason for selecting these effects: they 
have not been tightly constrained by  the speed of gravitons as deduced from 
the \label{eLIGOfref2}
LIGO/VIRGO experiments. Indeed, in all these models the speed of gravitons is 
equal to the speed of light. This is not the case for models such as the 
quartic 
and quintic Galileons, where the speed of the gravitons deviate significantly 
from unity when the scalar field is responsible for the acceleration of the 
Universe. The cubic Galileon, whose leading nonlinear term in its Lagrangian is 
$ \Box \phi (\partial \phi)^2$, is not affected by these constraints.
The induced modification of gravity is locally reduced by the Vainshtein 
mechanism, whereby the nonlinear  terms dominate inside the Vainshtein radius 
and Newtonian gravity is retrieved. In this context, it is highly relevant that 
inside clusters of galaxies, where the density profile is not constant, the 
Vainshtein screening could be lifted, resulting in a discrepancy between the 
two 
Newtonian potentials. Although no significant deviation from General Relativity 
can be inferred from data using this effect yet, it has to be noted that for 
certain relaxed clusters the violation of the Vainshtein mechanism inside the 
cluster could help in  resolving the tension between lensing and X-ray cluster 
data 
\cite{Salzano:2017qac}.

Finally, new screening mechanisms have been proposed in \cite{Brax:2019koq},
where 
in a 5d scenario with a  warped bulk, the dark sector lives on the IR brane and 
the standard model on the UV brane. At low energy the two scalars are coupled 
via a bulk scalar. At high energy the bulk scalar can decay into gravitons, 
leading to a finite width for its propagator. This imaginary part operates at 
high energy, larger than the energy of the IR brane,  and induces a complete 
decoupling of the two branes \cite{Brax:2019koq}.

\chapter[Small-scale effects associated to \\ non-metricity and 
torsion]{Small-scale effects associated to \\ non-metricity and torsion}
\label{Delhom2chapter}
{\em Adrià Delhom}\\




%
%
%
%
%

\section{Small-scale Effects and Gravity}
 \label{smallscaleeffetsfs1}

The fact that the gravitational coupling constant is so weak compared to the 
other fundamental couplings may suggests that gravity plays a negligible role 
in 
physical scenarios where other fundamental forces are present, unless Planck 
scale processes are involved. This is actually not true, due to the fact that, 
unlike the other forces, the \textit{charge} that sources the gravitational 
field is additive, and there are scenarios where the presence of a huge amount 
of mass/energy makes gravity the dominant force at macroscopic scales, as in 
astrophysical or cosmological processes. Nonetheless, one could expect that in 
microscopic experiments, where the amounts of mass/energy are much smaller, 
gravity indeed plays a negligible role.

However, the fact that the charges 
corresponding to the other interactions are not additive allows looking for 
special set-ups, where the strength of the other interactions can be tuned to 
be 
smaller than possible gravitational effects occurring in some microscopic 
systems.  Given that gravity is still unknown at small scales (in the UV), it 
is then interesting to characterise what kind of effects due to gravity could 
appear in microscopic systems.

Keeping with metric theories, the only 
gravitational effects that can appear are those due to the spacetime curvature 
(i.e., the coupling to the graviton), thus pushing those effects all the way up 
to the Planck scale. However, when metric-affine gravity theories are explored, 
the phenomenology becomes much richer due to the fact that they allow for 
non-Riemannian geometries, which brings new effects that could be relevant 
even in 
processes where other interactions play a central role. We mention 
here that  in the present chapter, and similarly to \ref{ref:Iosifidis},
``non-Riemannian'' may refer to both departure from   the Levi-Civita conection 
as well as from  
\textit{metricity}. \label{nonRiemannianref4}

From the geometrical 
perspective, this fact is related to the non-Riemannian geometric objects that 
arise within the metric-affine formalism, namely the \textit{torsion} and  
\textit{non-metricity} tensors, defined respectively by 
$\mathcal{T}^\al{}_{\mu\nu}=2{\hat{\Gamma}}^{\al}{}_{[\mu\nu]}$ and 
$Q_{\al\mu\nu}=\nabla_\al g_{\mu\nu}$. These objects can generate new 
effective 
interactions within the matter sector due to the coupling between matter and 
geometry, and such interactions will be suppressed by the length scale at which 
the geometry deviates from a metric one. Below that scale, there could in 
principle be propagation of the non-Riemannian geometrical features in the whole
spacetime, and non-metricity and torsion would play a central role in the 
description of gravity. 

From the field theory perspective, there are two 
different cases: when these new fields $\mathcal{T}^\al{}_{\mu\nu}$ and 
$Q_{\al\mu\nu}$ do not propagate any extra degrees of freedom, they can be 
algebraically solved in terms of the matter fields and (possibly) their 
derivatives and the metric tensor, which allows   to integrate them out of 
the 
action and obtain these new effective interactions that will be suppressed by a 
universal energy, related to the geometrical length scale mentioned above. In 
the case that they propagate new degrees of freedom, they are usually assumed 
to 
be massive, so that they can also be integrated out by  producing similar 
effects 
that will be now suppressed by the mass scale of the corresponding new 
particles 
that have been integrated out. This mass scale would characterise the  scale at 
which the particles associated with torsion and/or non-metricity propagate and 
their effects become non-perturbative. While the theories of propagating 
torsion 
and non-metricity are not well developed   to date, we will see below that the 
new geomtry-related effective interactions that arise in metric-affine theories 
of gravity can be constrained by observations, which allow us to set a bound to 
the scale at which non-Riemannian features can play a nontrivial role in the 
description of gravity.

\section{Small-scale Effects Associated to Non-metricity} 
\label{nonmetricityref4}

In general, metric-affine theories of gravity are not well understood, due to  
the difficulty in solving their dynamics, and the role played by non-metricity 
in the microscopic regime is not yet fully characterised. However, there is a 
broad class of metric-affine theories dubbed as Ricci-Based Gravities (RBGs) 
where small-scale perturbative effects associated to non-metricity are currently 
understood. The class comprises all metric-affine gravity theories with 
diffeomorphism and projective symmetries, where the action is an arbitrary 
scalar function of the Ricci tensor and the metric (see    e.g. 
\cite{Jimenez:2020dpn}).
A generic RBG action   reads:
\begin{equation}\label{RBGaction}
{S}_{\rm RBG}=\frac{1}{2\kappa^2} \int 
d^4x\sqrt{-g}F(g^{ab},\cR_{(ab)},\lrbg),
\end{equation}
where  $\kappa^2=8\pi G_N=M^{-2}_{Pl}$, 
$\cR_{\rm ab}$ is the Ricci tensor of the independent affine connection 
${\hat{\Gamma}}^a{}_{bc}$, and $\lrbg$ is a high-energy scale parametrising 
departures from it. We stress that the energy scale $\lrbg$ need not to be 
related to
the cosmological constant.
Examples of widely explored gravity theories that fall within the RBG are, for 
instance, GR, as well as the extensions  $f(\mathcal{R})$,  Ricci-squared,
$f(\cR,\cR^{(ab)}\cR_{(ab)})$,
 the Eddington-inspired Born-Infeld gravity, and the majority of the 
metric-affine curvature-based models of the literature. For consistency 
with 
experimental data, F is also required to recover GR at low energy scales, i.e.,
in the limit $E/\lrbg\rightarrow 0$. Note also that projective symmetry 
requires that only the symmetric part of the Ricci tensor appears in the 
action, 
thus guaranteeing the absence of ghostly degrees of freedom 
\cite{BeltranJimenez:2019acz,Jimenez:2020dpn}. 
The field equations derived from (\ref{RBGaction}) imply that the connection is 
non-dynamical and that it is the Levi-Civita connection of some symmetric 
2-tensor $q_{ab}$, usually called \textit{auxiliary metric}. By performing a 
field redefinition and integrating out metric and connection, it can be shown 
that these theories admit an Einstein-frame representation in terms $q_{ab}$. 
In the process, one can see that there exists an on-shell relation between the 
spacetime metric and this auxiliary metric given by 
$g_{ab}=q_{ac}(\Omega^{-1})^c{}_{b}$ and where the \textit{deformation matrix} 
is defined as
\beq
\sqrt{\Omega}(\Omega^{-1})^a{}_b=\frac{\partial F}{\partial \cR_{(ac)}}g_{cb},
\eeq
being $\Omega=det(\Omega^a{}_b)$.
It is in general possible to write the deformation matrix as a combination of 
one of the metrics, the matter fields and their derivatives, and as shown in 
\cite{Jimenez:2020iok}, due to the the non-linearities of the equations for the 
deformation matrix, they usually admit several solutions. Noteworthy, there is 
always one solution that connects with GR at low energies and for which the 
deformation matrix satisfies the same symmetries than the metric $g_{ab}$. In 
this branch, the deformation matrix will satisfy (at least) the same internal 
symmetries as the stress energy tensor, and the metric $q_{ab}$ will also share 
the symmetries of  satisfied by the metric $g_{ab}$. This is strictly true if 
the 
matter sector does not couple to the connection. For matter fields that do 
couple to the connection,   even for the branch that recovers GR at low 
energies, the dependence of the deformation matrix on the matter fields can be 
more general 
\cite{Afonso:2017bxr,BeltranJimenez:2017doy,Jimenez:2020dpn}. However, this will 
only imply 
that 
the spacetime, as a function of the auxiliary metric and matter fields, will 
have 
more general combinations of matter fields that may in general have other 
symmetries than those of the the stress-energy tensor.

In vacuum, the 
deformation matrix is trivial and all RBG theories exactly reproduce  the 
dynamics of vacuum GR. Nonetheless, in the presence of matter the deformation 
matrix becomes nontrivial and introduces nonlinear modifications in the 
matter 
sector of the Einstein-frame, which couples to $q_{ab}$. In fact, in this 
frame, 
the field equations can be written as
\beq
\mathcal{G}^{a}{}_b(q)=\kappa^2 \tilde T^a{}_b
\eeq
where $\tilde T^a{}_b$ is the stress-energy tensor of the  Einstein-frame 
matter 
sector. 

 The fact that the equations for the auxiliary metric are formally  identical 
to 
those for the spacetime metric in GR puts forward that the role played by 
$q_{ab}$ in RBG theories is analogous to the role of $g_{ab}$ in GR. The 
auxiliary metric is the object that accounts for the usual effects associated 
with gravity: a long-range force mediated by a massless spin-2 field. Thus 
$q_{ab}$ 
accounts for the gravitational force due to \textit{total} amounts of 
matter-energy and propagates gravitational waves. However, in the presence of 
matter fields, there arise new effects that depend on \textit{local} amounts 
of 
energy density, and which have their origin in the gravitational sector. 
Indeed, 
due to the nontrivial form of the deformation matrix, the spacetime metric 
picks up corrections that depend on local amounts of energy-density. This can 
be explicitly seen by  writing the deformation perturbatively as
\beq\label{omegastressenergy}
{(\Omega^{-1})^a}_b={\delta^\al}_\nu+\frac{1}{\lnm^4}\bigg(\alpha T 
\delta^a{}_b+ \beta T^a{}_b\bigg)+\mathcal{O}(\lnm^{-8}),
\eeq
where $\alpha,\;\beta$ are model-dependent dimensionless parameters that are  
completely specified once we choose a particular RBG Lagrangian, and 
$\lnm=\sqrt{M_{Pl}\lrbg}$ is a high-energy scale whose meaning will be 
explained 
later in detail. Notice that the requirement that GR must be recovered at low 
energies requires the zeroth order term in the expansion to be $\delta^a{}_b$ 
(actually it suffices to be proportional to $\delta^a{}_b$, in which case it 
would only affect the normalisation of the 
fields). 
As a consequence, in vacuum, all RBG theories are 
exactly GR.

The above expansion for the deformation matrix allows us to write
\begin{eqnarray}
g_{ab}=q_{ab}+\frac{1}{\lnm^4}\bigg(\alpha T  q_{ab}+\beta T_{ab}\bigg) 
+\mathcal{O}(\lnm^{-8})\label{metricstressenergy}\\
Q_{abc}=\frac{1}{\lnm^4}\bigg[\alpha (\nabla_a T)  q_{ab}+\beta  
(\nabla_a 
T_{bc})\bigg]+\mathcal{O}(\lnm^{-8})\label{nonmetricitystressenergy},
\end{eqnarray}
where $Q_{abc}\equiv\nabla_a g_{bc}$ is the non-metricity tensor,  and 
notice that 
we defined the auxiliary metric such that its covariant derivative vanishes 
on-shell \cite{Afonso:2017bxr,BeltranJimenez:2017doy}. Although the above 
expansion is only valid below the scale $\lnm$, the general form of the 
deformation matrix suggests that these effects will also generally have a 
non-perturbative counterpart. Indeed, here we can see how the scale $\lnm$ is 
the scale at which non-perturbative corrections to GR related to the 
non-Riemannian nature of the underlying spacetime arise within the RBG 
framework. Below that scale the non-metricity tensor has perturbative effects 
and becomes suppressed by $(E/\lnm)^4$, although it plays a fundamental role in 
the non-perturbative behaviour of these theories, as is apparent from 
(\ref{nonmetricitystressenergy}). From here, we can see that the spacetime 
metric, besides taking into account the typical effects of gravity,   is also 
point-wise sensitive to the distribution of energy density, which represents a 
distinctive feature of RBG theories among other modified gravities. 

Thus, in the RBG frame, we see that the spacetime metric accounts for two 
different effects 
of gravitational origin: a long-range force mediated by a spin-2 field 
associated with gravitational waves, and new effects controlled by the energy 
scale $\lnm$ that depend upon the local distribution of energy-density, and 
which can be associated with a non-dynamical non-metricity tensor sourced by 
matter-energy density which occurs within RBGs, as seen in 
(\ref{nonmetricitystressenergy}). Given that all matter fields couple to the 
metric, an expansion of the spacetime metric in powers of $1/\lnm$ also allows 
us to interpret these non-metricity related effects as perturbative 
interactions 
within the RBG frame. 

From the point of view of the Einstein frame, however, 
these new effects related to non-metricity are translated into nonlinear 
interactions within the matter sector whose fields propagate in a General 
Relativistic spacetime. The fact that these new effects become more relevant 
the higher the energy-density scale of the process or, as viewed in the 
Einstein 
frame, that they introduce higher-dimensional operators (we are 
referring to the mass-dimension of the new field operators that describe these 
nonlinear interactions) in the matter Lagrangian, already points out that they 
can be relevant at high-energy microscopic processes. Let us illustrate this 
with some results in this direction available in the literature.

\subsection{Small-scale Effects in $f(\cR)$ Theories}
\label{fRref5}
\label{Palatiniformref5}

The first context in which the physical implications of these  effective 
interactions were found was that of Palatini $f(\cR)$ theories (see, e.g.,  
\cite{Olmo:2011uz}). Let us review these results within the general framework 
of RBG theories developed above. To that end, we must first look at the form of 
the deformation matrix within metric-affine $f(\cR)$ theories, which is given by
\beq
\Omega^a{}_b= f_{\cR} \delta^a{}_b,
\eeq
being $\cR$ (and thus $f_\cR$), an on-shell function of the stress-energy 
tensor 
trace obtained from the trace of the metric field equations $\cR 
f_\cR-2f=\kappa^2 T$. Given a particular model of $f(\cR)$, we will be able 
to 
write the corresponding deformation matrix only in terms of the trace of the 
stress-energy tensor, which implies that the $\beta$ coefficient in 
(\ref{omegastressenergy}) will vanish for all $f(\cR)$ models, having an 
on-shell parametrisation for an arbitrary $f(\cR)$ of the form
\beq\label{omegastressenergy}
{(\Omega^{-1})^a}_b\Big|_{f(\cR)}={\delta^\al}_\nu+\frac{\alpha }{\lnm^4} T  
\delta^a{}_b+\mathcal{O}(\lnm^{-8}).
\eeq

 Once a particular $f(\cR)$ is chosen,  then the algebraic equation  $\cR 
f_\cR-2f=\kappa^2 T$ can be solved for $\cR(T)$, leading to the function 
$f_\cR(T)$ that fixes the dimensionless $\alpha$ coefficient. This deformation 
matrix introduces new interactions in the matter sector due to the coupling 
between the metric and all the matter fields. For instance, a spin 1/2 field is 
coupled to an $f(\cR)$ theory that  will be described by the Lagrangian  
\begin{eqnarray}\label{FermionLag}
&&
\!\!\!\!\!\!\!\!\!\!\!\!\!\!\!\!\!\!\!\!\!
\mathcal{L}_{1/2}=\sqrt{-q}\left[\frac{i}{2}
\left(\bpsi\gamma_q^\mu\nabla_\mu\psi- 
(\nabla_\mu\bpsi)\gamma_q^\mu\psi\right)
-\bpsi m 
\psi\right]\nonumber
\\&&
+\sqrt{-q}\frac{3\al}{2\lnm^4}T\left[\frac{i}{2}\left(
\bpsi\gamma_q^\mu\nabla_\mu\psi-(\nabla_\mu\bpsi)\gamma_q^\mu\psi\right)
-2\bpsi 
m 
\psi\right]
+\mathcal{O}(\lnm^{-8}),
\end{eqnarray}
where $\{\gamma^\mu_q,\gamma^\nu_q\}=2q^{\mu\nu}$. This Lagrangian describes  a 
fermion field with contact interactions with  all matter fields through the 
trace of the stress-energy tensor. Notice that the perturbative expansion 
breaks 
down at the scale $\lnm$, where the interactions described by the above 
Lagrangian become non-unitary, and non-perturbative effects are expected to 
dominate.

The fact that matter fields in metric-affine $f(\cR)$ theories develop 
new contact interactions below $\lnm$ was first noticed in the context of 
$\cR-\lrbg^4/\cR$ by Flannagan in \cite{Flanagan:2003rb} (where originally 
$\mu$ 
was written instead of $\lrbg$). For that particular model we have the relation
$$f_\cR=1+\frac{4\lrbg^4}{-\kappa^2 T\pm\sqrt{\kappa^4 T^2+12\lrbg^4}},$$
which after the constant re-scaling of the auxiliary metric 
$q_{\mu\nu}\rightarrow  3/4 q_{\mu\nu}$ leads to the dimensionless coefficient 
$\alpha=\mp 1/4\sqrt{3}$. We can then derive with the form of the stress-energy 
tensor for a spin 1/2 from (\ref{FermionLag}) and use the above value of $\al$ 
to obtain
\begin{eqnarray}
&&
\!\!\!\!\!\!\!\!\!\!\!\!\!\!\!\!\!\!\!\!
\mathcal{L}_{1/2}=\sqrt{-q}\left[
\frac{i}{2} 
\left(
\bpsi\gamma_q^\mu\nabla_\mu\psi-(\nabla_\mu\bpsi)\gamma_q^\mu\psi\right)
-\bpsi m 
\psi\right]
\nonumber\\
&&+\sqrt{-q}\Bigg\{ 
\pm\frac{\sqrt{3}}{16\lnm^4}\left[
\bpsi\gamma_q^\mu\nabla_\mu\psi-
(\nabla_\mu\bpsi)\gamma_q^\mu\psi\right]^2\mp
\frac { \sqrt{3}
m^2}{2\lnm^4}(\bpsi\psi)^2\nonumber
\\
&&
\ \ \ \ \ \ \ \ \ \ \ \,
\pm i\frac{5\sqrt{3} m}{16\lnm^4}\left[\bpsi\gamma_q^\mu\nabla_\mu\psi- 
(\nabla_\mu\bpsi)\gamma_q^\mu\psi\right]
(\bpsi\psi)\Bigg\}
+\mathcal{O}(\lnm^{-8}
).
\end{eqnarray}

As Flanagan pointed out in \cite{Flanagan:2003rb}, these effective 
interactions 
 would have physical implications for the spin 1/2 sector of the Standard 
Model. 
Indeed, the value that the energy scale $\lnm$ should take for 
$\cR-\lrbg^4/\cR$ 
to account for   late-time acceleration (i.e.,
$\lnm=\sqrt{M_{Pl}\lrbg}\sim 10^{-3}eV)$ was seen to be incompatible with 
experimental data on, for instance, electron-electron scattering experiments, 
which rules out the model as an alternative explanation for dark energy 
\cite{Flanagan:2003rb}. Shortly afterwards, two remarks to the work by Flanagan 
were 
made by Vollick.

Firstly, in \cite{Vollick:2003ic} it was argued that the two frames 
admitted by $f(\cR)$ theories   were not physically equivalent, due to the fact 
that the 
mapping between frames may become singular in some spacetime regions (here we 
call these frames   the RBG  and 
Einstein frames, since we discuss $f(\cR)$ as a subset of theories of the RBG 
class, although traditionally, within $f(\cR)$ theories they were called Jordan 
and Einstein ones respectively).  \label{Einsteinfrref6}
However, 
since a field re-definition is known to leave physics invariant, both frames 
should be equivalent at least where the field re-definition is well defined. 
Vollick also argued \cite{Vollick:2004ws} that there is no proof that field 
re-definitions leave the S-matrix invariant in curved spacetimes. 

Secondly,  Vollick in
\cite{Vollick:2004ws} also discussed how Flanagan did not consider the 
potential role played by torsion by assuming that the spin connection is given 
by the Levi-Civita connection, which is not consistent with the metric-affine 
formalism. Indeed, the corresponding modifications to include torsion were 
introduced, and as expected, the coupling between the fermion fields and the 
connection led to a non-propagating torsion field that is algebraic in the 
femrion fields and can be integrated out to produce further four-fermion 
effective 
interactions. However, given that the coupling constant for the interactions 
was 
the Planck scale, rather than $\lnm$, they are negligible compared to the ones 
in Flanagan's analysis. Even if they were characterised by the same 
scale, since the non-metricity related interaction terms contain powers of the 
derivatives of the fields, the dependence on the momenta of the different 
interaction terms when calculating decay rates or cross-sections will be 
different than that of the torsion-related interactions. Thus, cancellation 
between the different terms cannot take place and Vollick's result would 
presumably still apply. Hence,  in this case the interactions do not contain 
derivatives of the fields, which implies that the contribution to the 
cross-sections will grow with a different power, although from the field theory 
perspective it appears that the result can be generalised in a rather 
straightforward manner, because  as far as we know there is still no rigorous 
proof. 

  These effective interactions were also used by 
Olmo in \cite{Olmo:2008ye} to derive corrections to the hydrogen atom behaviour 
in general metric-affine $f(\cR)$ theories. In order to do so, the 
non-relativistic limit of the modified Dirac equation  that resulted from 
taking 
into account these new interactions in a general Palatini $f(\cR)$ theory was 
computed. Then it was shown that while UV  $f(\cR)$ corrections to GR were 
perfectly compatible with the observed behaviour of the hydrogen atom (provided 
that the UV scale is high enough), IR $f(\cR)$ corrections characterised by an 
IR scale that accounted for late-time acceleration generically have dramatic 
consequences for the stability of atomic systems \cite{Olmo:2008ye}. The 
particular cases of $\cR+\cR^2/R_P$ and $\cR-\lrbg^4/\cR$ were analysed, and 
while in the first case the hydrogen atom was perfectly stable, $1/\cR$ 
modifications where seen to introduce a potential well in the outskirts of the 
atom to which the electron in the ground state would tunnel in a short time. 
Indeed, the stability of the ground state of the hydrogen atom in a universe 
described by  $\cR-\lrbg^4/\cR$ was found to be of around 12 minutes if $\mu$ 
is 
chosen to account for late-time acceleration, in clear contradiction with 
observations. These previous results from particle physics reasoning ruling out 
Palatini $\cR-\lrbg^4/\cR$ as a candidate for explaining late-time 
acceleration, were 
confirmed in a different physical system.

Other sources of incompatibility with 
data of $f(\cR)$ theories with corrections that grow at low curvature are also 
potential violations of the Einstein Equivalence Principle (EEP, see 
\cite{Thorne:1970wv,Will:2014kxa,will_2018} for a rigorous definition). In an 
earlier work \cite{Olmo:2006zu}, Olmo pointed out that the corrections induced 
by the non-minimal interactions introduce effects in atomic-like systems that 
cannot in general be removed by a suitable choice of freely falling 
coordinates, 
thus not recovering (locally) the Minkowski metric and introducing potential 
violations of the EEP. While these violations would be negligible for UV 
$f(\cR)$ corrections (i.e., those that become relevant at high curvatures), 
they 
would introduce observable violations of the EEP in $f(\cR)$ theories with IR 
corrections (i.e., corrections that become relevant at low curvatures), such as 
the $\cR-\lrbg^4/\cR$ model.

Finally, another interesting consequence of the effective 
interactions within the $f(\cR)$ framework is that due to these corrections, 
the 
gravitational potential in Newtonian limit picks up energy-density dependent 
corrections \cite{Allemandi:2005tg,Olmo:2005zr,Olmo:2005hc,Sotiriou:2005xe}. As 
also pointed out in  \cite{Olmo:2006zu}, this dependence allows for the 
possibility of having different extenal gravitational fields sourced by 
extended 
objects with the same total mass and symmetries but a different mass-density 
distribution. 

\subsection{Small-scale Effects in Generic RBGs}
\label{EddingtonBIref2}

The characteristic microscopic effects of RBG theories have  also been used to 
constrain other gravity theories whithin the RBG class that lie beyond 
$f(\cR)$. 
Particularly, Eddington-inspired Born-Infeld (EiBI) theories have been 
constrained by 
means of these effective interactions in the contexts of astrophysics, nuclear 
physics and high-energy particle physics experiments. Furthermore, some effects 
related to
the behaviour of collapsing matter have been studied. Particularly in 
\cite{Pani:2012qb,Avelino:2012ge}, Pani \textit{et. al.} and Avelino first 
showed 
how these non-minimal couplings can have effects in astrophysical scenarios in 
both the relativistic and non-relativistic regimes. 

As shown previously by 
Ba\~nados and Ferreira \cite{Banados:2010ix}, in the non-relativistic regime of 
the theory, the Poisson equation picks up a correction due to these new 
effective interactions, leading to the result that non-interacting collapsing 
particles produce a pressure-less star that is seen to be stable under linear 
radial perturbations, instead of black holes \cite{Pani:2012qb}. In the 
relativistic regime, though black holes \label{BHref5} (or wormholes) may form 
after collapse, it is found that the modifications introduced by the effective 
matter couplings give birth to a plethora of compact objects that do not exist 
within GR \cite{Pani:2012qb}. Indeed, the non-perturbative effects of these new 
interactions are also responsible for the existence of wormhole solutions in 
EiBI when coupled to a free Maxwell field.

The first 
constraints to EiBI were derived by Avelino in \cite{Avelino:2012ge}, also by 
exploiting the modified Poisson equation in the non-relativistic limit, which 
implies a constraint in the energy scale of EiBI due to the mere existence of 
compact objects of reaius $R$. Avelino also used the modified Poisson equation 
in a nuclear physics context to constrain the EiBI energy scale by assuming 
that 
the corrections due to these effective interactions were smaller than 
electromagnetic effects inside atomic nuclei \cite{Avelino:2012qe}. Later, he 
also showed how measurements of the pressure distribution inside the proton, 
which get modifications due to these new effective interactions, could further 
constrain the EiBI energy scale by using nuclear physics data 
\cite{Avelino:2019esh}.

A systematic analysis of the effect of these effective 
interactions without the use of the non-relativistic approximation has been 
carried out in \cite{Latorre:2017uve,Delhom:2019wir}. Concretely, in 
\cite{Latorre:2017uve} we developed the framework presented above to describe 
these new effective interactions for a generic RBG theory when coupled to a 
free 
spin 1/2 field. There, the $\mathcal{O}(\lnm^{-4})$ terms in 
(\ref{omegastressenergy}) were used to set general bounds to the RBG parameter 
space, using data from $e^+e^-\rightarrow e^+e^-$ collisions at LEP. Notice 
that 
while in the $f(\cR)$ case only the $\alpha$ parameter is non-vanishing, in a 
general RBG model there will also feature the $\beta$ parameter. However, in 
high 
energy processes, since the $\alpha$ terms are on-shell-proportional to some 
power of the particle mass, they are negligible in front of the $\beta$ terms 
that are  typically proportional to some power of the momentum of the 
particles. Thus, the $\beta$ terms will be the dominant corrections in any 
tree-level process at a scale above the mass of all the particles involved. 

In  \cite{Latorre:2017uve} the authors  used data of  $e^+e^-$ collisions at 
$\sqrt{s}=207$ GeV in LEP, in order to extract a bound, by making the following 
assumptions: i) neglect the influence of post-Newtonian corrections in 
experiments carried out in LEP (i.e., assuming 
$q_{\mu\nu}\approx\eta_{\mu\nu}$)  
ii)  neglect   the 
$\alpha$-dependent terms, i.e., the $\alpha$ term is suppressed  
by a factor $m_e/\sqrt{s}\sim10^{-5}$ with respect to $\beta$-dependent terms, 
where $s$ is the characteristic energy scale of LEP $e^+e^-$, and iii) neglect 
the torsion corrections. The last approximation was based on arguments 
in \cite{Boos:2016cey}, where it was explained that torsion effects would only 
be relevant in scenarios with strong magnetic fields, such as neutron stars,  
\label{neutronstarsref2}
due 
to the need of a high spin-density for torsion effects to be observable. 
Indeed, within RBGs, although non-metricity related effects are suppressed by 
$\lnm$, torsion sourced by fermions is suppressed by the Planck mass, which 
further justifies the approximation. Under these three assumptions, 
the bound $\lnm\geq \beta^{1/4}0.6$ TeV was 
obtained  \cite{Latorre:2017uve}.

This bound is valid for generic RBG models, and it 
was also particularised for the EiBI case, for which $\beta=1$, obtaining the 
most stringent constraints for this theory   to date.  Recently, the method has 
been extended to arbitrary spin fields in \cite{Delhom:2019wir}, and using data 
from  $\ensuremath{\gamma} e^-\rightarrow \ensuremath{\gamma} e^-$ collisions, 
a 
similar bound for $\beta/\lnm^4$ 
was obtained. Moreover, X-Ray 
$\ensuremath{\gamma}\ensuremath{\gamma}\rightarrow\ensuremath{\gamma}\ensuremath
{\gamma}$ collisions were analysed, 
obtaining a much milder bound due to the fact that photon-photon scattering 
experiments that are  available are performed in the keV range.


\section{Small-scale Effects Associated with Torsion}
\label{torsionref8}

The torsion tensor was first introduced by Cartan in 
\cite{Cartan1,Cartan2,Cartan3,Cartan4},  and shortly after that, Einstein tried 
to unify - without success - gravity and electromagnetism by using this brand 
new 
geometrical object. Already, Cartan had the intuition that torsion had to be 
related with matter spin.  Inspired by the work of Utiyama 
\cite{Utiyama:1956sy},  Kibble \cite{Kibble:1961ba} and Sciama 
\cite{Sciama_1962,Sciama:1964wt} realized that the gauge theory of the 
Poincar\'e group naturally features a tetrad and an affine connection as the 
gauge fields of translations and Lorentz rotations respectively.  
\label{loclinref4}

The simplest 
Lagrangian for these gauge fields was found to be $\sqrt{-g}\cR$ 
(note that we have 
here traded the natural gauge variables used in the original work, which were 
vierbein and spin connection, for the metric and affine connection in order to 
be consistent with the rest of the text), i.e., the 
metric-affine/Palatini Lagrangian for General Relativity.   We mention that  we
prefer to use 
the name Einstein-Cartan-Sciamma-Kibble (ECKS), standing for   this minimal 
version of Poincar\'e gauge theory, since   the name  
metric-affine/Palatini GR  could 
cause confusion (the different names have historical roots, 
particularly in the different approaches, namely  gauge vs. metric-affine 
approaches). 
Although the above Lagrangian leads to 
a 
vanishing non-metricity, it was seen to lead to a nontrivial torsion tensor if 
matter fields couple to the affine connection, as is the case for spin 1/2 
fields. Indeed, this coupling leads to a non-vanishing hyper-momentum (or 
spin-density), which sources torsion, and which for spin 
1/2 fields was seen to be
\begin{equation}\label{Hypmom}
\Delta_{\mu\al\ensuremath{\beta}}=\sqrt{-g} 
\frac{i}{2}\epsilon_{\mu\al\ensuremath{\beta}\sigma}\bpsi\ensuremath{\gamma}
^\sigma\ensuremath{ \gamma}_5\psi.
\end{equation}
Actually, spin 1/2 fields couple 
minimally only to the pseudo-vector part of the torsion tensor, defined by 
\begin{equation}
 S^\mu=\epsilon^{\mu\al\ensuremath{\beta}\ensuremath{\gamma}}\mathcal{T}_{
\al\ensuremath{\beta}\ensuremath{ \gamma}}
\label{MinimalTorsionCoupling},
\end{equation}
  since the standard 
kinetic term of Dirac fields has a hidden interaction term of the form $-1/8 
S^\mu\bpsi\ensuremath{\gamma}_\mu\ensuremath{\gamma}_5\psi$.
Then, by separating the torsion contribution from the metric ones in the  
gravitational Lagrangian, and integrating out the torsion tensor generated by 
(\ref{Hypmom}), they arrived at the well known effective operator 
\begin{equation}\label{TorsionInt}
\bpsi\ensuremath{\gamma}_\mu\ensuremath{\gamma}_5\psi\bpsi\ensuremath{\gamma}
^\mu\ensuremath{\gamma}_5\psi
\end{equation}
that describes torsion-induced four-fermion interactions within minimal 
Poincar\'e 
 gauge theory. 
 
 As we see, similar to the case of non-metricity in RBG theories, 
torsion in metric-affine GR does not propagate and is algebraic in the matter 
fields, although instead of being sourced by energy density, it is sourced by 
spin density (or hyper-momentum). As in the case for the non-metricity related 
effective interactions in RBGs, the effective interactions resulting from 
integrating out the torsion tensor also have implications in the behaviour of 
matter fields. 

In the review by Hehl \textit{et. al.} \cite{Hehl:1976kj} it is 
emphasized that within minimal Poincar\'e gauge theory, torsion only exists in 
the bulk of matter fields with non-vanishing spin density, and it was estimated 
that the effects of torsion within these theories could become relevant right 
before singularity formation in gravitational collapse, close to the Big Bang, 
or in some quantum gravity processes; situations where the number of spin 1/2 
particles per unit volume volume is expected to reach the critical value 
$n=m/\kappa\hbar^2$, where $m$ is the particle mass. These effective 
interactions 
have been suggested as mechanisms for singularity avoidance in early Universe 
cosmology \cite{Poplawski:2010kb}. 

The idea of treating torsion as an external classical field, or as a new  
propagating degree of freedom that can be quantised, has also been treated in 
the 
literature. To do that, one needs to go beyond the metric-affine GR action, 
which 
can be done in several ways. Since it is a very extensive topic, and the main 
results have already been presented in different works, we will only give here 
a brief summary of such results. The interested reader is referred to the 
review 
by Saphiro  \cite{Shapiro:2001rz} and references therein. 

The possible 
parity-preserving couplings of the torsion field to a scalar, and also to  spin 
1/2 
fields, are explored. There are four different couplings for a real scalar 
field, an 
extra one for complex scalar field, five couplings for scalar-pseudoscalar 
couplings, and two couplings for fermion fields. With the minimal coupling 
prescription, only one of these couplings appear, which is the coupling between 
torsion and fermions through the spin connection given in  
(\ref{MinimalTorsionCoupling}). Starting from the classical theory given by 
minimally-coupled scalar and spin 1/2 fields in a curved background with 
torsion, it is seen that two non-minimal couplings must be added: the well 
known 
$\xi_1 R\phi^2$, and a general version of the minimal coupling between fermions 
and torsion given by $\eta_1 \bpsi 
S^\mu\ensuremath{\gamma}_\mu\ensuremath{\gamma}_4\psi$ (notice the arbitrary 
coefficient). If an extra Yukawa interaction between the fermion and scalar is 
added, renormalisability requires the addition of an interaction term between 
the torsion pseudo-vector and the scalar of the form $\xi_4 S^\mu S_\mu\phi^2$. 

The renormalisation group (RG) equations that give the running of the masses 
and 
matter couplings are identical to their flat spacetime version. Moreover, the 
running of the couplings to curvature and of the vacuum parameters not related 
to torsion are not influenced by torsion couplings. However, in general, the 
running of torsion couplings might be influenced by the other couplings present 
in the theory. For the $\eta_1$ parameter, it is seen that its running is 
always 
positive, thus making the interaction between spinors and torsion stronger in 
the UV for different choices of matter sector, but the running is seen to be 
not 
relevant physically. For a non-minimally coupled massless scalar field, it is 
seen that the non-minimal couplings to torsion modify the potential in such a 
way that spontaneous symmetry breaking can always  be achieved   either at the 
classical or at the quantum level. Additionally, first-order phase transitions 
might be 
induced by such non-minimal couplings. Also, anomalies were studied in the 
presence of torsion, finding that the anomaly cancellation that takes place 
within the Standard Model is respected by the non-minimal couplings for weak 
external torsion fields. However, the leptonic current picks up a correction 
due 
to curvature and torsion. This correction could induce anisotropies in the 
polarisation of light that comes from distant galaxies, although the current 
constraints in the coupling parameter make this effect un-observable. 
 
  The possibility of propagating torsion has been mainly studied for its 
pseudo-vector part  from an effective field theory approach. As shown in 
\cite{Carroll:1994dq}, this pseudo-vector carries four degrees of freedom 
encoded 
in a scalar and a spin-1 field. However, if both fields propagate, the 
Hamiltonian 
would not be bounded below, thus restricting the parameters of the effective 
theory. The case of the pseudo-scalar field is treated within 
\cite{Carroll:1994dq}, showing that it can induce interaction between torsion 
and minimally-coupled gauge fields due to the chiral anomaly. Several bounds on 
the torsion mass and the coupling to gauge fields are obtained in that 
reference. For the case of the spin-1 field, we refer the reader again to the 
review \cite{Shapiro:2001rz}. 

Due to   an extra gauge symmetry for the torsion 
pseudo-vector that is broken at low energies by the fermionic mass terms, the 
pseudo-vector is found to be massive, with a mass that is required to be higher 
than fermion masses due to non-renormalisability of the theory. Notice 
that, given its non-renormalisability, it only makes sense as an effective 
field 
theory below the torsion mass. If the torsion mass was of the order of the most 
massive fermion or lower, this would have important corrrections to the 
Standard Model that 
have not been observed, thus ruling out the possibility. However, one has to 
keep in mind that these results only apply to a theory of the torsion 
pseudo-vector, and general theories including the full tensor have not yet been 
well understood in the quantum regime. Hence, below the torsion mass, it 
introduces new contact interactions between fermion fields described again by 
(\ref{TorsionInt}). If $\eta^2/M_S^2$ is taken as the coupling constant of such 
an 
effective operator, the bound $M_s>1.4\eta\,$TeV was found from axial-axial 
$eeqq$ interactions. 

The influence of the torsion pseudo-vector in 
forward-backward asymmetries has also been discussed in \cite{Belyaev:1997zv}. 
Following a different approach, and concerned with the fact that the 
four-fermion 
interactions  are non-renormalisable and become non-unitary in the UV, Boos and 
Hehl recently suggested in \cite{Boos:2016cey} that by adding suitable kinetic 
terms to the gauge fields of the Poincar\'e group, which are given by the 
square 
of torsion and curvature tensors, new propagating gravitational degrees of 
freedom related to the affine sector will appear that mediate these 
interactions. These new degrees of freedom could be understood as gauge bosons 
of the Poincar\'e group in close analogy to the W and Z bosons of the 
electroweak model that gave a more fundamental explanation of Fermi's effective 
theory for $\beta$-decay.

The non-relativistic counterpart of these effective interactions has also been 
widely  treated in the literature. First,  Hehl suggested that in the geometric 
optics approximation,   torsion could 
modify the trajectories of particles with non-vanishing spin-density in 
\cite{HEHL1971225} (note that in \cite{HEHL1971225} the approximation was 
called  semi-classical   but we prefer to call it
``geometric optics'' to avoid confusion with semi-classical gravity). 
Later in \cite{Rumpf:1979vh,Audretsch:1981xn}, Audretsch 
confirmed Hehl's claim by two different methods, showing how in the geometric 
optics approximation a free spin 1/2 particle, as well as the spin vector, do 
not propagate along metric geodesics, but instead they follow 
geodesics of the connection 
$\Gamma^{*}{}^\al{}_{\mu\nu}=\Gamma(g)^\al{}_{\mu\nu}+3/2 
g^{\al\sigma}\mathcal{T}_{[\sigma\mu\nu]}$, which can be written in terms of 
the 
metric connection and the contortion tensor. From a field theory perspective, 
since the contortion tensor is sourced by the spin 1/2 field hyper-momentum, it 
is seen how the effect of the effective interaction of (\ref{TorsionInt}) in 
the 
macroscopic regime is that of a self-force that pushes objects with 
spin-density 
away from metric geodesics. Audretsch also showed that the spin vector would 
precess around the trajectory. After some years, Nomura \textit{et. al.} 
\cite{Nomura:1991yx} confirmed previous results using the Fock-Papapetrou 
method. 

More recently, Cembranos \textit{et. al.} made a revision of all the 
previous work about spin 1/2 particle trajectories. In this revision, they 
found 
that the WKB method used by Audretsch is cleaner than the others regarding the 
derivation of the trajectories and spin precession. Also, by using the 
Raychaudhuri equations for spin 1/2 particles, they also showed the existence 
of 
a new parameter that characterises the difference between the accelerations 
felt 
between   spin 1/2 particle trajectories and minimal-length trajectories (i.e., 
metric geodesics), as measured by local observers, which could be employed to 
detect torsion effects. However, they also concluded that in order to observe 
the effects, huge magnetic fields that align the spins of a large amount of 
spin 
1/2 particles should be present. Concretely, they propose that a difference in 
the angles of incidence of photons and neutrinos coming from the same neutron 
star could be an interesting situation, in which the deviation of the neutrino 
geodesics with respect to the photon ones due to these torsion effects could 
become observable.

Recently, Obukhov \textit{et.al.} \cite{Obukhov:2014fta} 
derived the Foldy-Whitehouse Hamiltonian for a minimally-coupled spin 1/2 field 
without taking any weak-field limit. They showed how the gravitational moments 
played a major role in building a covariant extension of the Dirac Lagrangian 
that features intrinsic dipole moments (induced by Noether charges) in a 
systematic way. They also show that, contrary to electromagnetism, anomalous 
gravito-electric and gravito-magnetic moments cannot be introduced for a Dirac 
particle in a covariant way (see, e.g., \cite{Mashhoon:2003ax}  an analogy 
between 
gravity and Maxwell's theory in the weak-field limit in the context of    
gravitoelectromagnetism).
There have also been many attempts to observe torsion in the lab due to these  
high-energy/small-scale effects. Two main classes of experiments involving 
torsion can be distinguished: those which regard torsion as a background field, 
and those which understand it as an extra field described by an effective 
Lagrangian. We will here give a brief taste of the current status of the field. 
The interested reader is referred to the broad reviews on the topic by Ni 
\cite{Ni:2009fg,Ni:2015poa}.
The first observational tests that tried to observe 
a background torsion field were proposed by Audretsch and Lammerzahl 
\cite{AudretschLammerzahl_1983}, by means of neutron interferometry,  and 
Hughes-Drever type experiments also proposed by Lammerzahl 
\cite{Lammerzahl:1997wk}. Assuming minimal coupling to fermions, the bound 
$|S_i|<10^{-31}$ GeV was found for the norm of the spatial component of the 
torsion pseudo-vector. Later, Mohanty and Sarkar \cite{Mohanty:1998vs} used 
measurements on CPT violation parameters in Kaon systems also  allowed to 
constrain 
  the temporal component to be $|S_0|<10^{-26}$ GeV. 

Singh and Ryder  
\cite{Singh_1997} have shown how the pseudo-vectorial torsion could also have 
observable effects in spectroscopy experiments due to a correction of the 
energy 
levels by a spin-rotation coupling. Indeed, in experiments with nuclear spins 
freely precessing in the bulk of a constant magnetic field, the measurements on 
the Zeeman frequencies for different nuclei allowed  setting bounds to the 
 spatial 
part of the torsion pseudo-vector of the order of $|S_i|\,|\cos{\theta}|\leq 
8.1 
\times 10^{-31}$ GeV, with $\theta$ being the angle between $S_i$ and the 
magnetic field, as shown by Obukhov \textit{et.al}.  in \cite{Obukhov:2014fta}.

The possibility of constraining torsion 
\cite{Kostelecky:2007kx} and background non-metricity \cite{Foster:2016uui} (see 
also \cite{Delhom:2019gxg,Delhom:2020gfv} for Lorentz Violating effects linked 
to non-metricity in metric-affine bumblebee gravity)  
by 
constraints on Lorentz violation parameters has also been discussed. Besides the 
observables named 
above, torsion effects were seen to potentially induce Weak Equivalence 
Principle violation by Ni \cite{Ni:2009fg}, and universality of free-fall 
experiments, was also used to constrain torsion gradients by Duan 
\textit{et.al.} \cite{Duan:2015zmf}. From the effective field theory 
perspective, the possibility of discovering torsion at CERN was discussed by 
deAlmeida \textit{et.al.} in \cite{deAlmeida:2008dk}. Indeed, using data on 
Drell-Yang and $t\bar t$ processes at LHC, a possible non-universal coupling 
between torsion and fermions was severely constrained by Saphiro 
\textit{et.al.} 
\cite{Belyaev:2007fn}, and also lower bounds on the torsion mass 
$M_S\geq\mathcal{O}($TeV$)$ for a coupling constant of the order of 
$\eta_1\sim10^{-1}$ (remember that minimal coupling is given by $\eta_1=1/8$) . 
Constraints to in-matter torsion (such as that of ECKS theory) were first 
placed 
in neutron spin rotation experiments using liquid $^4$He by Lehnert 
\textit{et.al.} \cite{Lehnert:2013jsa}. 

\section{Outlook}

As a final remark we would like to mention that although much work has  been 
done, especially in characterising the effects of torsion in metric-affine 
gravities, the implications of the possible non-minimal couplings to matter are 
not yet fully explored. Regarding non-metricity, only the consequences through 
the nontrivial modifications that it induces in the metric tensor in a limited 
class of metric-affine theories are fairly understood, and an understanding of 
the effects of the possible non-minimal couplings to matter is lacking. 
Additionally, 
even though we have an effective theory for the torsion pseudo-vector, such a 
construction for the full torsion is still lacking, and for the non-metricity 
tensor we do not even  have an effective theory   for any of its irreducible 
components. 

There are also questions on which we should reflect further that 
stem from the fact that although non-metricity and torsion are geometrical, in 
most cases (if not all) they can be treated just as extra fields added to the 
Lagrangian with no relation to geometry at all, and especially in the cases where 
they do not propagate, they are apparently not distinguishable from just a bunch 
of irrelevant operators. Do geometrical fields have a different physical 
behaviour than non-geometrical ones? And if so, is there any canonical way in 
which the demarcation can be made? The only remainder that we have (for instance 
in non-metricity related couplings in RBGs) is that these interactions may be 
interpreted as having geometrical origin comes from their {\it{universality}} 
in the 
sense that all the effective operators are characterised by the same coupling 
$1/\lnm$. Thus it would be important to understand whether geometrical fields 
have further implications that are not apparent from the pure field theory 
perspective, or if the geometrical interpretation is just a convenient tool that 
allows us to gain insight about the behavior of gravitational theories. 
In this 
sense, it has been recently argued that although classical GR can be described 
in three different geometrical ways involving only torsion, only non-metricity 
or only curvature; the field theory description is the unique theory of a 
unitary and Lorentz invariant massless spin-2 field, thus implying that the 
field theory point of view is in some sense a more fundamental one 
\cite{BeltranJimenez:2019tjy}. On the other hand, it is also acknowledgeable 
that many central results in GR (such as, for instance, the singularity 
theorems, 
or the positivity of mass) are only fully understood in geometrical terms, and 
as far as the author knows, there is no straightforward translation to the field 
theory 
language for these results. We thus encourage the interested readers to reflect 
further on this issues.

\chapter[Stars as Tests of Modified Gravity]{Stars as Tests of 
Modified Gravity}
\label{Olmochapter}

{\em Gonzalo J. Olmo,
Diego Rubiera-Garcia,
Aneta Wojnar}
\\



%
%
%
%

Compact stars, both individual and in binary mergers, represent suitable 
scenarios to test General Relativity (GR) in its strong-field regime and to 
eventually find any 
deviations from its predictions. This is so because compacts stars are the 
objects (excluding black holes) where the largest curvatures and higher 
densities can be reached in Nature \cite{Lattimer:2004pg}.  To model stellar 
structure within modified theories of gravity in order to be able to explore 
possible new physics, one has to address several additional difficulties 
associated with the mathematical structure of these theories or to the new 
dynamics introduced by them within this context \cite{Olmo:2019flu}. In this 
sense, in addition to the well-known troubles with the  higher-order equations 
of motion and propagation of ghost-like instabilities introduced by many such 
theories, finding theoretically consistent models of stellar 
structure\label{stellarstrref1} beyond GR 
requires additional improvements on a case-by-case basis. This is well 
illustrated in theories such as $f(R)$ and, more generally, scalar-tensor 
theories. In such cases, the fact that the scalar field contributes to the mass 
of the star beyond its surface \cite{Olmo:2011uz} introduces important technical 
difficulties, in particular  regarding the consistency of their weak-field and 
slow-motion limit with Solar System experiments and also with respect to how to 
perform the right matching to an external Schwarzschild solution, which has 
been 
the source of a heated debate in the literature 
\cite{Chiba:2003ir,Erickcek:2006vf,Kainulainen:2007bt,Faulkner:2006ub,
Multamaki:2006ym,Henttunen:2007bz}. As another example, those theories inducing 
additional dependences on the local energy density of the matter fields are 
known to be potentially problematic in the description of stellar surfaces 
\cite{Pani:2011mg,Barausse:2007pn,Barausse:2007ys}, requiring a more careful 
analysis of the physical description at the matching surface with the external 
solution \cite{Olmo:2008pv}.

\section{Modified Tolman-Oppenheimer-Volkoff Equations}

Gravitational theories extending GR typically modify the 
Tolman-Oppenheimer-Volkoff (TOV)  equations\label{TOVtrref1} of hydrostatic 
equilibrium
\begin{eqnarray} \label{eq:TOVGR1}
p'&=&-(\rho+p) \frac{m(r)+4\pi p r^3}{r\left[r-2m(r)\right]} \\
m(r)&=&4\pi \int_0^r d \tilde{r} \tilde{r}^2 \rho(\tilde{r}) \label{eq:TOVGR2}
\end{eqnarray}
where $\rho$ and $p$ are the energy density and pressure of a  perfect fluid as 
the matter source, respectively, while $m(r)$ is the mass function.
In a typical case, such modifications will consist of additional 
terms/contributions to these equations.

One of the most characteristic and observationally accessible predictions of 
models of compact stars within these theories corresponds to the mass-radius 
relations of neutron stars \label{neutronstarsref3} as compared to those of 
GR. Such relations are 
obtained once a given equation of state (EOS) relating energy density and 
pressure $p=p(\rho)$ inside the different layers of a neutron star is given, 
which allows for the resolution of Eqs.(\ref{eq:TOVGR1}) and (\ref{eq:TOVGR2}) 
and their extensions within modified gravity, typically via numerical methods. 
Since the EOS of dense matter at the supernuclear densities reached in the 
innermost region of neutron stars is unknown, the mass-radius relations in GR 
largely depend on the assumptions on the EOS derived from extrapolations of 
laboratory nuclear physics. Therefore, the EOS of dense matter can be 
constrained via observations of neutron stars masses and radii 
\cite{Lattimer:2006xb}, among other quantities, or via observations derived 
from 
gravitational wave astronomy \cite{Llanes-Estrada:2019wmz}.

Furthermore, for 
every modified theory of gravity chosen, such mass-radius relations 
\label{massradiausref1}will also 
depend on the extra gravitational parameter(s) that appear in the gravitational 
sector \cite{Shao:2019gjj}. Since tracking the full mass-radius curve with 
measurements of enough neutron stars masses and radii is a hard task, the focus 
is placed instead on the predictions of any such gravitational models on the 
maximum available mass for a given EOS, and their comparison with the reported 
measurements of several neutron stars with masses around $2M_{\odot}$ 
\cite{Antoniadis:2013pzd,Crawford:2006xb,Linares:2018ppq,Cromartie:2019kug}. In 
this sense, since the TOV equations of $f(R)$ gravity are known in closed form 
\cite{Olmo:2011uz}, several $f(R)$ models have been thoroughly studied in the 
literature, including the quadratic one, $f(R)=R+\alpha R^2$ 
\cite{Resco:2016upv,Cooney:2009rr,Orellana:2013gn,Yazadjiev:2014cza,
Astashenok:2017dpo,Astashenok:2018iav} (also considering the effects of 
anisotropies \cite{Folomeev:2018ioy}), cubic \cite{Capozziello:2015yza}, 
exponential \cite{Cognola:2007zu}, logarithmic 
\cite{Astashenok:2013vza,Alavirad:2013paa}, power-law 
\cite{Capozziello:2015yza,DeLaurentis:2018odx}, Hu-Sawicki 
\cite{Resco:2016upv}, 
and others \cite{Kase:2019dqc}, using both perturbative and non-perturbative 
analyses, and dealing with the definition of the measurable mass in these 
theories. The hyperonic puzzle, namely, the fact that the presence of hyperons 
in realistic EOS at the neutron star's core tends to soften it, and decrease 
the 
maximum mass perhaps even below the $2M_{\odot}$ has also been addressed within 
these theories \cite{Astashenok:2014pua}.

Beyond $f(R)$ gravity, the mass-radius relations as well as other aspects of  
the phenomenology of compact stars have been studied within many other theories 
of gravity, such as in the more general scalar-tensor theories 
\cite{Horbatsch:2010hj,Cisterna:2015yla,Wojnar:2016bzk,Sotani:2017pfj} where, 
in particular, 
evidence for scalarised solutions has been found 
\cite{Silva:2014ora,Horbatsch:2015bua,Novak:1997hw}. Further theories with 
additional scalar fields include the Horndeski and beyond Horndeski families,  
\label{beyondHorndeskiref2}
and their degenerate higher-order scalar-tensor (DHOST) extensions 
\cite{Crisostomi:2017pjs,Babichev:2016jom,Chagoya:2018lmv}. The list of 
modified 
theories with other assumptions/dynamics beyond GR investigated within this 
context is varied, and includes Proca theories \cite{Kase:2017egk}, 
Einstein-dilaton-Gauss-Bonnet gravity 
\cite{Pani:2011xm,Panotopoulos:2019zxv,Silva:2017uqg,Doneva:2017duq,
Blazquez-Salcedo:2015ets}, massive gravity 
\cite{Katsuragawa:2015lbl,Kareeso:2018xum,Hendi:2017ibm} and bigravity 
\cite{Enander:2015kda,Aoki:2016eov}, Rainbow's gravity \cite{Hendi:2015vta}, 
teleparallel gravity \cite{Ilijic:2018ulf,DeBenedictis:2016aze,Deb:2018voz}, 
mimetic gravity \cite{Astashenok:2015qzw,Fabris:2019qvy}, theories breaking the 
Lorentz 
invariance \cite{Xu:2019gua,Kim:2018dbs,Eling:2007xh,Barausse:2019yuk} or the 
conservation of the energy-momentum tensor 
\cite{Moraes:2015uxq,Das:2016mxq,Deb:2018gzt,Maurya:2019hds,Maurya:2019hds,
Maurya:2019sfm,Carvalho:2019gzs,Oliveira:2015lka,Hansraj:2018reh}, 
$f(\mathcal{R})$ gravity in metric-affine spaces 
\cite{Kainulainen:2006wz,Pannia:2016qbj,Panotopoulos:2017jdc,Wojnar:2017tmy}, 
or 
Eddington-inspired Born-Infeld gravity \cite{Pani:2011mg}, among many others. 
The parameter(s) of any such theories are typically 
experimentally/observationally constrained by Solar System experiments   
\label{solarsystemref7}
\cite{Will:2014kxa}, cosmological observations \cite{Bull:2015stt}, fundamental 
physics, etc, thus restricting the space of parameters one can play with.

The main observation regarding the phenomenology of such modified gravity models 
 for neutron stars is a modification of the mass-radius relations in the sense 
of producing larger masses and smaller radii, or the other way round, depending 
on the specific combinations of parameters and EOS, though some specific models 
have more involved behaviours. However, regarding the maximum allowed mass for a 
given EOS, and within experimental/observational constraints of every theory 
from different sources (when available), most such models yield   meagre
increases that may be typically larger than the  observational errors of 
current capabilities to measure neutron stars' masses and radii. This yields a 
bottleneck for testing these theories via these observations, unless significant 
improvements in the precision of these probes are obtained in the near future 
\cite{Berti:2015itd}. 

There are some exceptions to this general rule, with 
several models yielding significant increases of the maximum mass. In turn, this 
might be able to bring back to life some cases of soft EOS unable to reach the 
$2M_{\odot}$ observational threshold within GR. On the other hand, this offers 
the possibility of further constraining these theories and testing their 
observational viability as compared to GR, should heavier neutron stars be 
detected in the future.

While these are examples of static stars studied within modified  theories of 
gravity, the addition of rotation has several interesting effects and 
modifications to the above picture. It is indeed known that in GR the presence 
of rotation may raise the maximum available mass for a given EOS by a factor up 
to $\sim 15\%-20\%$ as compared to the static case \cite{Stergioulas:2003yp} 
and, overally, a similar conclusion
can be reached for modified gravity, as has been investigated for several 
specific models. 

However, an even more interesting aspect of the corresponding
rotating structure models arises from the consideration of the moment of 
inertia 
and other  associated quantities, which, being also observationally accessible, 
typically show
much larger increases in their values when moving from GR to modified gravity 
than in the case  of maximum masses. This has been explored and confirmed for
several theories, including $f(R)$ \cite{Staykov:2014mwa,Staykov:2015kwa,Staykov:2015cfa},
scalar-tensor theories \cite{Silva:2018yxz,Staykov:2015mma,Popchev:2018fwu, 
Staykov:2018hhc,Silva:2014ora},
the Horndeski family \cite{Silva:2016smx,Cisterna:2016vdx,Maselli:2016gxk},
and so on \cite{Sullivan:2017kwo}, also including  anisotropic contributions 
\cite{Silva:2014fca}.  Though this analysis is more readily performed for the 
slowly-rotating case, where the effects of rotation can be simulated via a small 
perturbation upon a static, spherically symmetric star \cite{Hartle:1967he}, the 
case of
rapidly and differentially rotating stars has a particular interest for  the 
analysis of tidal deformability and universal (I-love-Q)
relations \cite{Doneva:2013qva, Yazadjiev:2015zia, Yazadjiev:2017vpg,  
Kleihaus:2016dui,Doneva:2018ouu},
as well as for the generation 
of gravitational waves out of binary neutron star
mergers \cite{Pani:2014jra, Barausse:2012da, Yagi:2013mbt, Okounkova:2019dfo,  
Carson:2019fxr,Zhang:2019iim,Zhao:2019kif} in order to place constraints on 
these theories.

\section{Modified Lane-Emden  Equation}
\label{maEmdenusref1}

In addition to neutron stars, the case of non-relativistic stars (such as white, 
brown and red dwarfs)  is also of interest. In this case, $p \ll \rho$, the TOV
equations can be reduced down to their Newtonian counterpart, namely, the 
Poisson equation, which is called Lane-Emden equation on its dimensionless form. 
Moreover, different types of non-relativistic stars can be well modelled using 
polytropic EOS, $p=K\rho^{\frac{n+1}{n}}$, according to specific values of the 
polytropic constant $K$ and index $n$. Despite central densities and curvatures 
being much weaker than in their relativistic counterparts, this scenario has the 
advantage of removing the large uncertainty in the EOS ascribed to dense matter 
inside neutron stars and, therefore, the degeneracy in the predictions of the 
corresponding models. 

The main prediction of many modified gravity models for non-relativistic stars 
is the weakening/strengthening of the gravitational
force inside astrophysical bodies, as given by the presence of extra terms  in 
the Lane-Emden equation
\cite{Capozziello:2011nr,Farinelli:2013pza,Andre:2017usv,Koyama:2015oma, 
Saito:2015fza,  Wibisono:2017dkt,Wojnar:2018hal,Sergyeyev:2019aul}. This has 
several physical consequences
for the structure of non-relativistic stars, with some observational 
discriminators  attached to them. For instance, for white 
dwarfs,\label{dwarfsref1} many such 
models
modify the Chandrashekar $1.40M_{\odot}$ limit of GR \label{masslimitingusref1}
\cite{Chandrasekhar:1935zz}, including scalar-tensor theories and their  
Horndeski extensions and beyond
\cite{Wibisono:2017dkt,Jain:2015edg,Chowdhury:2018qrf,Carvalho:2017pgk, 
Crisostomi:2019yfo}, allowing us to place constraints on some of these theories. 
In particular, this feature may allow us to generate the so-called 
Super-Chandrashekar white dwarfs, with masses as high as $2.8M_{\odot}$, to meet 
some recent controversial observations \cite{Kalita:2019yaj, Banerjee:2017uwz}.

Another aspect of non-relativistic stars affected by modified theories  of 
gravity is their luminosity, as demonstrated in 
\cite{Koyama:2015oma,Olmo:2019qsj}. Moreover, the modified Lane-Emden equation 
in this case also has  a direct impact in the minimum mass required  for a star 
to burn sufficiently stable fuel to compensate for  
photospheric losses, that is, to belong to the main sequence. Using a crude but  
surprisingly accurate analytic model, GR yields an estimate of roughly $\sim 
0.09M_{\odot}$ \cite{Kumarnum,Burrows:1992fg} for this minimum main
sequence mass (MMSM), which is just less than $\sim 10\%$ {smaller} than  the 
one obtained via numerical simulations. For three theories of modified gravity, 
namely,
scalar-tensor \cite{Sakstein:2015zoa}, DHOST \cite{Crisostomi:2019yfo},  and 
metric-affine $f(\mathcal{R})=\mathcal{R}+\alpha \mathcal{R}^2$ 
\cite{Olmo:2019qsj}, this same calculation has been carried out, yielding 
constraints
upon the parameters of each theory resulting from the comparison with the least  
massive star ever observed, Gl 866 C, with $M_{MMSM} \approx 0.0930\pm 0.0008 
M_{\odot}$ \cite{Segransan:2000jq}.
Observations of the radii of low-mass brown dwarfs are another possibility  to 
constrain theories of gravity, as provided by the future mission GAIA 
\cite{Brown:2018dum}, since for such objects the theoretical models given by the 
(modified) Lane-Emden equation are mass-independent 
\cite{Sakstein:2015zoa,Saumon:1995bn,Sergyeyev:2019aul}. Moreover, some modified 
theories of gravity shift
the stability bound of GR (with a critical value of the polytropic index 
$n=3$), 
allowing for a wider range of the polytropic index $n$, and at
the same time providing an additional dependence on the theory's parameter  
\cite{Pani:2011mg, Pani:2012qb, aneta4}.

It turns out that modified gravity has also impact on evolution of stellar 
objects: the main sequence and red giant stars in modified gravity framework 
were studied in \cite{chang2011stellar,davis2012modified,Chowdhury:2020wfr} 
while low-mass stellar objects in \cite{Wojnar:2020txr}, where it was 
demonstrated that Hayashi tracks and radiative core development are gravity 
model dependent phenomena. It was also argued that the upper mass limit of 
fully convective stars on the main sequence might be different than the one 
provided by numerical simulations whose codes are based on Newtonian equations. 
Moreover, due to the fact that one does not observe any stars in the 
so-called Hayashi forbidden zone, studying Hayashi tracks may provide a tool for 
constraining gravitational theories \cite{Wojnar:2020txr}.

	Another important problem that needs to be deeper examined is the age of 
the stellar objects. One of the methods to determine the clusters’ age, as well 
as the age of their individual stars and white dwarfs, uses lithium abundance 
in low-mass stars. The lithium depletion method has been believed so far to be 
the most reliable technique for young globular clusters’ age determination, and 
this is why this procedure is utilized to calibrate other techniques used for 
the age estimation.

An additional use of lithium abundance is to distinguish brown dwarfs from Main 
Sequence stars,
because the true low-mass stars burn out lithium before they reach the Main 
Sequence, while in the
brown dwarf’s atmosphere this light element is observed. It was however 
demonstrated \cite{Wojnar:2020frr} that the lithium abundance in low-mass 
stellar objects 
depends on the gravitational model, introducing an
additional uncertainty to age determination techniques, if they rely on the 
light element depletion
method. This causes a significant shortening of the early stellar evolution 
phases, having the following consequences: it will contribute to the explanation 
of ``too old'' white dwarfs, as for example the one in the binary system KIC 
8145411 \cite{Masuda_2019,Laughlin_1997}. Furthermore, one also deals with the 
changes in the number of stars in the pre- and main sequence phases, giving 
different impact to the
distant galaxies brightness in comparison to the GR prediction 
\cite{davis2012modified}. Moreover, 
the lithium abundance
may also be a tool to test theories of gravity: theories which prominently 
prolong the low-mass
stars’ lifetimes, in comparison to the current accepted models, would rise 
doubts on those that
introduce such an effect.

In summary,   stars represent yet another suitable scenario to test the  
predictions of modified theories of gravity. Using both relativistic and 
non-relativistic stars, the combination of the predictions of any theory for 
both such types of stars may allow us to narrow down the range of observationally 
viable values of the corresponding parameters, as has been proven in the case of 
some scalar-tensor theories \cite{Babichev:2016jom}. In addition, such 
constraints can be connected with others coming for gravitational wave 
astronomy and cosmological tests (such as inflationary and dark energy models). 
It thus represents a promising avenue in which  to explore the strong field 
regime of our 
gravitational theories.










\chapter[Compact Objects  
in General Relativity and Beyond]{Compact Objects  
in General Relativity and Beyond}
\label{Kunzchapter}

{\em Jose Luis Bl\'azquez-Salcedo, Burkhard Kleihaus, Jutta Kunz}\\





Recent years have seen tremendous progress in gravitation
and the astrophysics of compact objects. 
The new window of gravitational wave (GW) multimessenger astronomy
opened by the LIGO/VIRGO observations \label{eLIGOfref3}
has provided us with the first glimpses of the cataclysmic events
of stellar type black hole (BH) mergers \label{BHref6}
\cite{Abbott:2016blz,Abbott:2016nmj,Abbott:2017oio,Abbott:2017vtc,LIGOScientific:2018mvr}
as well as the merger of neutron stars (NSs)
\cite{Abbott:2017xzg}.
In particular, the latter event has been associated with numerous observations
in the electromagnetic spectrum 
\cite{Coulter:2017wya, TheLIGOScientific:2017qsa, GBM:2017lvd,Abbott:2018wiz},
allowing far-reaching conclusions
concerning our understanding of NSs on the one hand
and gravity on the other. Moreover, the EHT collaboration has recently presented 
the first observation of a BH shadow,
giving us an amazing image of the central supermassive BH of M87
\cite{Akiyama:2019cqa,Akiyama:2019fyp,Akiyama:2019eap}.


Depending on the mass of the progenitor star,
NSs or BHs arise after stellar core collapse. 
From the observational side, exploiting 
a large variety of techniques
from radio and X-ray observations to GWs, 
much progress has been
made in recent years in extracting basic properties of
neutron stars, such as their masses and radii
\cite{Lattimer:2012nd,Ozel:2016oaf,Baym:2017whm,Abbott:2018wiz}.
From the theoretical side, however, the still unknown
equation of state (EOS) of the nuclear matter
under the extreme conditions present in NSs
allows for a multitude of NS models,
all satisfying the observational constraints.
Here, substantial progress will be possible,
when in the analysis of future observations
the so-called \textit{universal relations}
will be exploited
\cite{Yagi:2016bkt,Doneva:2017jop}.
These represent rather accurate relations between
various appropriately scaled NS properties, 
including their ringdown frequencies, that are 
to a large extent EOS-independent.

Due to their extreme compactness, BHs and NSs
represent ideal laboratories to test 
GR and alternative theories of gravity
\cite{Will:2005va,Capozziello:2010zz,Berti:2015itd}.
Alternative gravitational theories introduce additional
degrees of freedom, and most prominently a 
gravitational scalar field.
Depending on the theory,
the resulting compact objects \label{compactobrefs2} may then differ
significantly from their GR counterparts
\cite{Will:2005va,Capozziello:2010zz,Berti:2015itd}.
NSs and BHs may be scalarized, as discussed below,
and the GW spectrum will get much richer,
allowing, for instance, for scalar radiation.
The presence of these new degrees of freedom
must be consistent with present and future observations,
and thus strong constraints on the parameters of the 
theories can arise.

\section{Neutron Stars}\label{neutronstarsref4}

\subsection{Neutron Stars in General Relativity}

\begin{figure}[ht]
\begin{center}
\mbox{\includegraphics[width=.65\textwidth, angle 
=270]{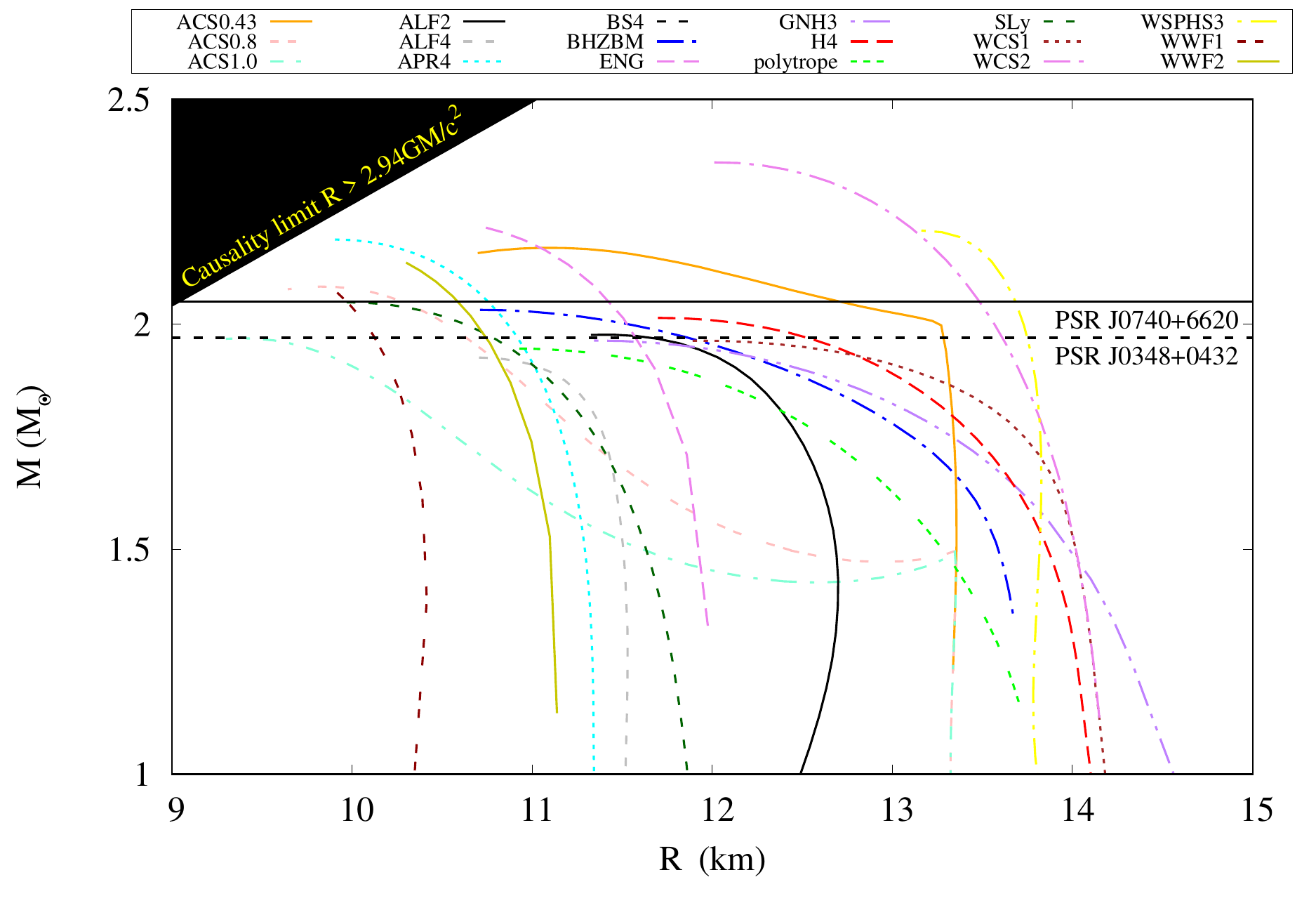}}
\end{center}
\vspace{-0.5cm}
\caption{{\it{NS mass $M$ (in solar masses $M_\odot$) vs radius $R$ (in km)
for a variety of EOSs 
(see, e.g.,\cite{BlazquezSalcedo:2012pd,Blazquez-Salcedo:2013jka,Motahar:2017blm,AltahaMotahar:2019ekm}). 
Additionally, we show the high mass pulsars
\cite{Demorest:2010bx,Antoniadis:2013pzd,Cromartie:2019kug}
and the causality limit for NSs.}}}
\label{Kunzfig1}
\end{figure}

To obtain NSs, the Tolman-Oppenheimer-Volkoff (TOV) equations of GR
need to be supplemented by an equation of state (EOS)
describing the dense strongly interacting NS matter 
(see, e.g., the reviews \cite{Lattimer:2012nd,Ozel:2016oaf,Baym:2017whm}).
In particular, in the NS core, the EOS is largely unknown,
and depending on the assumptions on its decomposition
and the hadronic interactions, very diverse EOSs result,
leading to very diverse NS properties.
While observations of high mass NSs with
$M \sim 2$ solar masses ($M_\odot$)
\cite{Demorest:2010bx,Antoniadis:2013pzd,Cromartie:2019kug}
have been able to rule out various EOSs,
there still remains a plethora of viable EOSs,
and thus the challenge to find the appropriate theoretical models 
to describe highly dense nuclear matter.
The mass-radius relation for static NSs is demonstrated 
in Fig.~\ref{Kunzfig1} for various EOSs.

If we are to exploit NSs as astrophysical labs 
to not only learn about nuclear and particle physics,
but also about gravity, we need to eliminate,
or at least strongly reduce, the EOS dependence
of the observables. Interestingly,
this can indeed be done by considering appropriately
scaled (typically) dimensionless quantities,
that allow us to extract so-called \textit{universal relations} 
\label{cuniversalefs1}
for the NSs (see, e.g., the reviews \cite{Yagi:2016bkt,Doneva:2017jop}).
These \textit{universal relations} not only concern
NS properties like the moment of inertia $I$, the
rotational quadrupole moment $Q$ and the tidal quadrupole moment $\lambda$
or Love number, which form the famous $I$-Love-$Q$ relations
\cite{Yagi:2013bca,Yagi:2013awa,Yagi:2014qua},
but also concern asteroseismology and thus
the quasi-normal modes (QNMs) \label{tuasnormiref1} describing the ringdown of 
NSs
\cite{Andersson:1997rn,Kokkotas:1999bd,Berti:2015itd}.
The key point is then, that gravity theories beyond GR
may also yield \textit{universal relations} which, however,
could be distinctly different from those of GR,
thus allowing us to put bounds on the parameters of the respective theories.

Let us begin by briefly recalling the \textit{universal relations} in GR.
The mass $M$ and the radius $R$ of the NSs yield the compactness
$C=M/R$, which is dimensionless in geometric units,
and thus a good candidate for considering \textit{universal relations}.
For a Schwarzschild BH the compactness is maximal, $C=0.5$, 
while for a NS it mostly resides in the range $0.1<C<0.3$.
The moment of inertia $I=J/\Omega$ is obtained from 
the angular momentum $J$ and the rotational velocity $\Omega$ of the fluid.
Its dimensionless counterpart for the \textit{universal relations}
is $\bar I=I/M^3$.

In general the star will carry 
also higher multipole moments, the mass moments $M_{2l}$
and the current moments $S_{2l+1}$.
The mass and the angular momentum may then be viewed as its lowest
multipole moments, $M=M_0$ and $J=S_1$. 
The higher moments can be obtained following the procedures of
Geroch and Hansen \cite{Geroch:1970cd,Hansen:1974zz}
(see also \cite{Hoenselaers:1992bm,Sotiriou:2004ud})
or Thorne \cite{Thorne:1980ru}.
The rotational quadrupole moment then corresponds
to the second mass moment $Q=M_2$, 
with its dimensionless counterpart 
for the \textit{universal relations} $\bar Q=QM/J^2$.

The tidally induced quadrupole moment
corresponds to the tidal deformability or Love number $\lambda_2$
\cite{Flanagan:2007ix,Hinderer:2007mb}.
Its dimensionless counterpart
for the \textit{universal relations} is given by $\bar \lambda_2=
\lambda_2/M^5$.
However, the response of NSs to external tidal fields
may also yield the corresponding higher tidal mass and current moments
$\lambda_n$ \cite{Damour:2009vw},
i.e., the higher Love numbers,
which need to be appropriately scaled for 
their use in the \textit{universal relations}.

When evaluating the $I$-Love and $Q$-Love relations,
Yagi and Yunes \cite{Yagi:2013bca,Yagi:2016bkt}
found remarkably little dependence of the corresponding
scaled quantities $\bar I$, $\bar \lambda_2$, $\bar Q$
on the EOS, exhibiting deviations below 1\% from a best fit of the type
$$
\ln y_i 
= a_i + b_i \ln x_i + c_i (\ln x_i)^2 + d_i (\ln x_i)^3+ e_i (\ln x_i)^4 ,
$$
with coefficients shown in Table I \cite{Yagi:2013bca,Yagi:2016bkt}:

\begin{center}
\begin{tabular}{cccccccc}
\hline
\hline
\noalign{\smallskip}
$y_i$ & $x_i$ &&  \multicolumn{1}{c}{$a_i$} &  \multicolumn{1}{c}{$b_i$}
&  \multicolumn{1}{c}{$c_i$} &  \multicolumn{1}{c}{$d_i$} &  \multicolumn{1}{c}{$e_i$}  \\
\hline
\noalign{\smallskip}
$\bar{I}$ & $\bar{\lambda}_2$ && 1.496 & 0.05951  & 0.02238 & $-6.953\times 10^{-4}$ & $8.345\times 10^{-6}$\\
$\bar{I}$ & $\bar Q$ && 1.393  & 0.5471 & 0.03028  & 0.01926 & $4.434 \times 10^{-4}$\\
$\bar Q$ & $\bar{\lambda}_2$ && 0.1940  & 0.09163 & 0.04812  & $-4.283 \times 10^{-3}$ & $1.245\times 10^{-4}$\\
\noalign{\smallskip}
\hline
\hline
\end{tabular}
\end{center}
\noindent
Interestingly, these relations are much better than the previously
considered relations of these scaled quantities versus the compactness
\cite{Yagi:2016bkt}.
When going to higher multipole moments, the so-called  \label{threehairef1}
\textit{three hair relations} emerge \cite{Stein:2013ofa,Yagi:2016bkt}.
These are no more as well satisfied as the $I$-Love-$Q$ relations, 
but still exhibit considerable EOS independence. As examples we
give the universal $\bar S_3$-$\bar Q$ and $\bar M_4$-$\bar Q$ relations
in Table II \cite{Stein:2013ofa,Yagi:2016bkt}:
\begin{center}
\begin{tabular}{cccccccc}
\hline
\hline
\noalign{\smallskip}
$y_i$ & $x_i$ &&  \multicolumn{1}{c}{$a_i$} &  \multicolumn{1}{c}{$b_i$}
&  \multicolumn{1}{c}{$c_i$} &  \multicolumn{1}{c}{$d_i$} &  \multicolumn{1}{c}{$e_i$}  \\
\hline
\noalign{\smallskip}
$\bar S_3$ & $\bar Q$ && $3.131 \times 10^{-3}$  & 2.071 & $-0.7152$  & 0.2458 & $-0.03309$\\
$\bar M_4$ & $\bar Q$ && $-0.02287$  & 3.849 & $-1.540$  & 0.5863 & $-8.337\times 10^{-2}$\\
\noalign{\smallskip}
\hline
\hline
\end{tabular}
\end{center}
\noindent
where universality holds to 4\% and 10\%, respectively.
While the accuracy of the universal relations deteriorates with
increasing $l$, they would still be highly useful
to determine the lower moments within good accuracy
\cite{Stein:2013ofa,Yagi:2016bkt}.

The name \textit{three hair relations}\label{hairtheref1} \label{nohairtheref2} 
results from a generalization
of the \textit{no-hair} or \textit{two hair relation} of black holes in GR,
since uncharged rotating black holes are fully described by 
only two numbers, their mass and their spin. For neutron stars it turns out,
that besides their mass and their spin, a third number is needed, 
their quadrupole moment, to approximately describe them 
\cite{Stein:2013ofa,Yagi:2016bkt}.
In a Newtonian setting, the three hair relations can be understood
by making the elliptical isodensity approximation,
assuming that stellar isodensity surfaces are self-similar
ellipsoids with a fixed stellar eccentricity,
and the density as a function of the isodensity radius for a rotating
configuration matches the one of a non-rotating configuration
with the same volume \cite{Stein:2013ofa,Yagi:2016bkt}.
The \textit{three hair relations} also remain valid in the fully 
relativistic regime.
As discussed in \cite{Yagi:2014qua,Yagi:2016bkt}, an emergent approximate
symmetry -- isodensity self-similarity -- seems to be the (most probable) cause
of the these universal relations for compact stars.

Whereas most known NSs are well described in the perturbative limit
of slow rotation, rapidly rotating NSs require the full solution
of the rotating generalization of the TOV equations 
(see, e.g., \cite{Stergioulas:2003yp}).
The domain of existence of rotating NSs is
delimited by the NSs rotating at the Kepler limit, 
the NSs along the secular instability line,
and the static NSs. The domain thus forms a two-dimensional surface
and not a single line, as in the static/slow rotation limit.
Accordingly, when rapid rotation is considered,
the universal relations generalize from lines to
surfaces in a three-dimensional space, 
where the third axis is, e.g., represented by 
the scaled dimensionless angular momentum $\bar J=J/M^2$
\cite{Doneva:2013rha,Pappas:2013naa,Chakrabarti:2013tca,Yagi:2014bxa,Yagi:2014qua}.

Asteroseismology is the second area where universal relations have
received much attention in recent years, since they arise in NS
QNMs, and QNMs emitted during the ringdown phase of merger events 
should be quite well discernible by future GW detectors.
QNMs of NSs come in various types
\cite{Andersson:1997rn,Kokkotas:1999bd,Berti:2015itd}.
The general pulsations are usually described in terms of 
radial and non-radial pulsations, where the non-radial ones
are decomposed into parity-even (polar) and parity-odd (axial) QNMs.
The axial modes decouple from the nuclear matter, and thus represent
pure space-time modes of oscillation, called $w$-modes.
The polar modes couple to the nuclear matter, and thus to the fluid
oscillations. All Newtonian fluid modes find their analogues
in GR, yielding in particular the fundamental $f$-mode, and the
excited $p$-modes. However, there are also polar $w$-modes,
that are without Newtonian counterparts \cite{Kokkotas:2003mh}.
Since in GR no gravitational monopolar or dipolar radiation exists, 
the ringdown spectrum of neutron stars is expected to be dominated by the 
quadrupolar $f$-, $p$- and $w$-modes.

To obtain the QNM spectrum one employs perturbation theory,
involving metric and fluid perturbations
\cite{Andersson:1997rn,Kokkotas:1999bd,Berti:2015itd}.
Decomposing these in terms of tensor spherical harmonics,
and performing a Laplace transformation, one is left with
an eigenvalue problem for
a complex spectral parameter $\omega=\omega_R + i \omega_I$.
The real part $\omega_R$ corresponds to the ringdown frequency
of the mode, while the imaginary part $\omega_I$ determines
whether the mode is stable ($\omega_I>0$) or unstable ($\omega_I<0$).
For stable modes, $\omega_I$ is the damping rate,
and its inverse is the decay time $\tau = 1/\omega_I$.

\begin{figure}[ht]
\begin{center}
\mbox{
(a) \hspace{-0.5cm}
\includegraphics[width=.37\textwidth, angle =270]{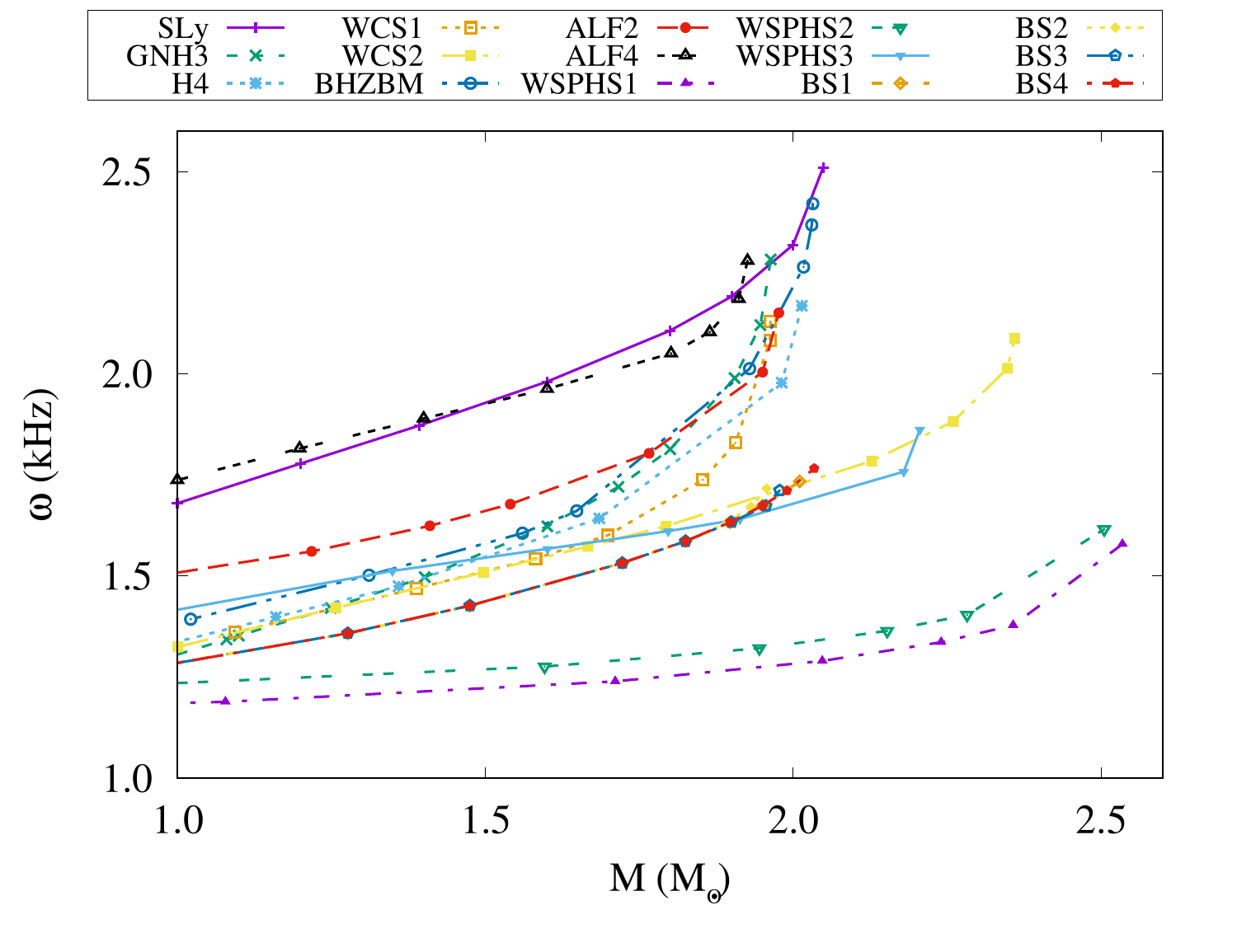}
(b) \hspace{-0.5cm}
\includegraphics[width=.37\textwidth, angle =270]{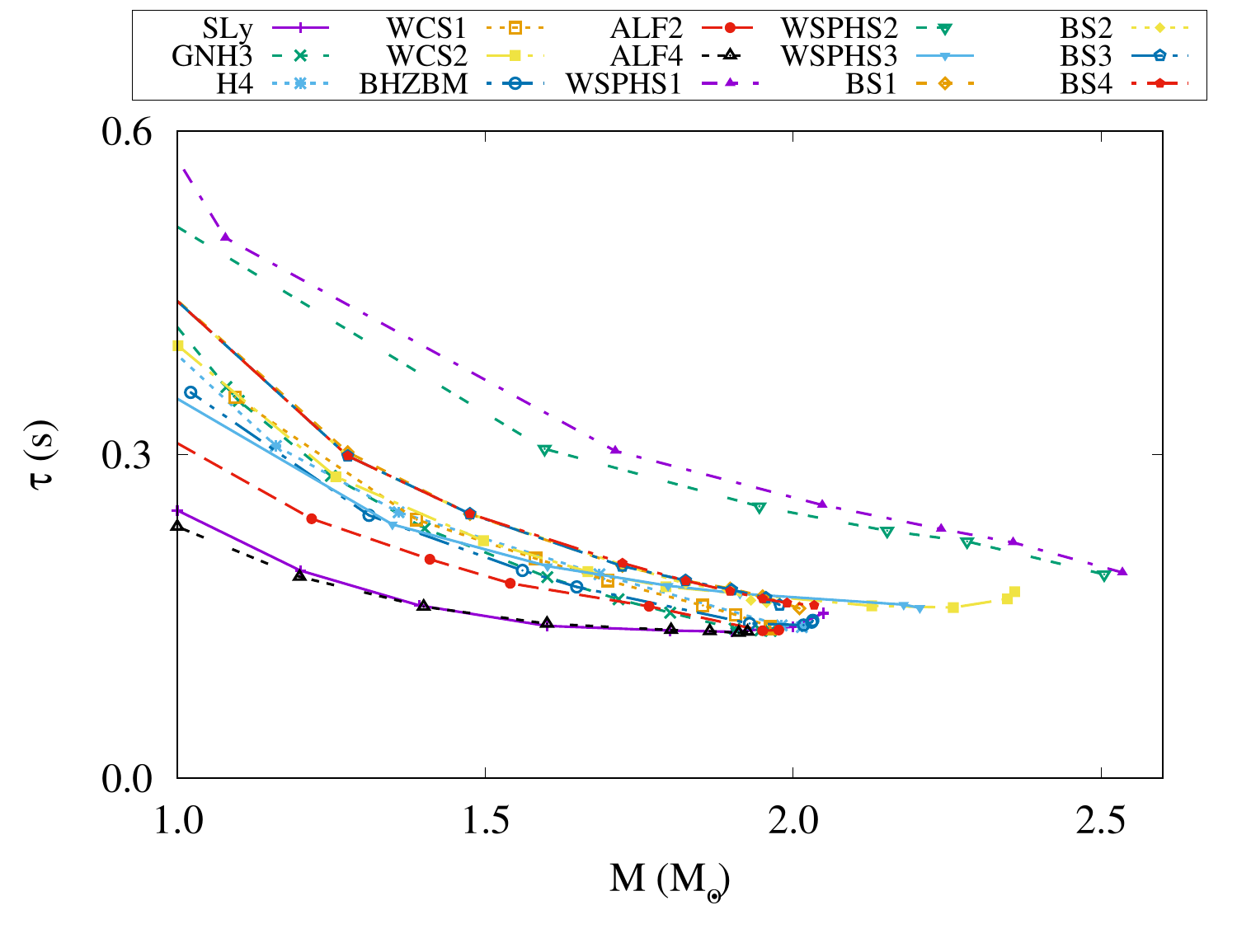}
}
\mbox{
(a) \hspace{-0.5cm}
\includegraphics[width=.37\textwidth, angle =270]{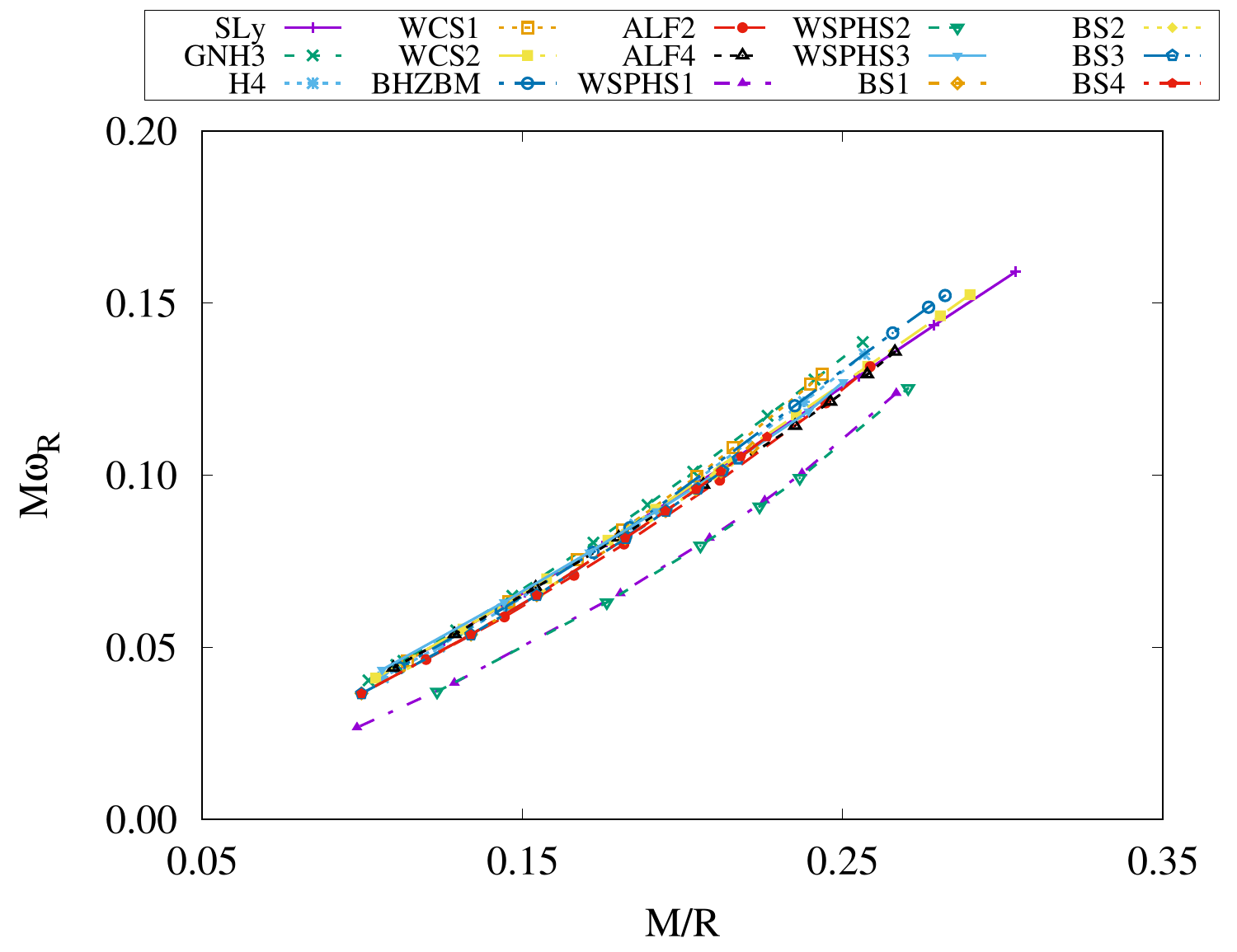}
(b) \hspace{-0.5cm}
\includegraphics[width=.37\textwidth, angle =270]{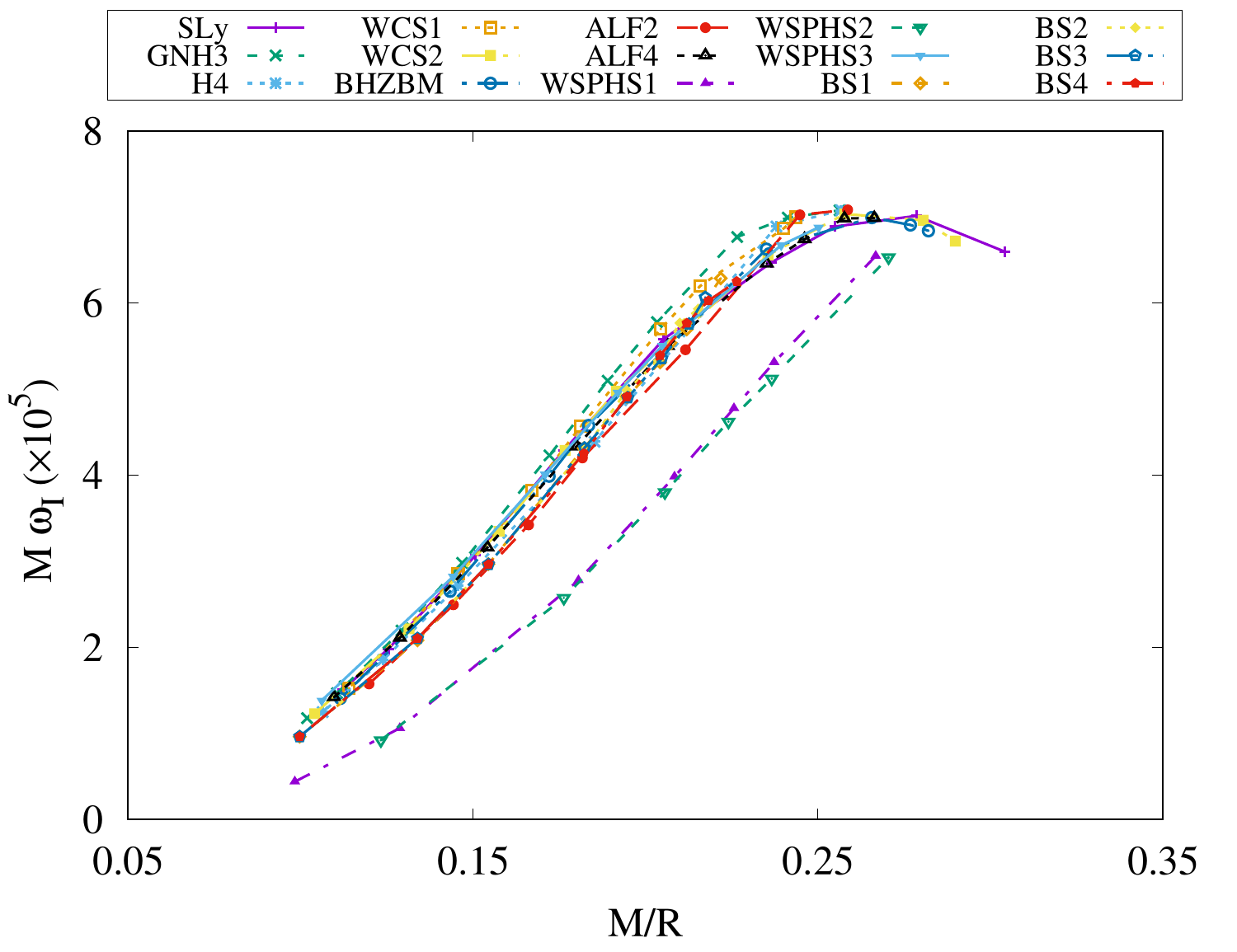}
}
\end{center}
\vspace{-0.5cm}
\caption{{\it{NS $f$-mode
for a variety of EOSs
(see, e.g.,\cite{BlazquezSalcedo:2012pd,Blazquez-Salcedo:2013jka,Motahar:2017blm,AltahaMotahar:2019ekm}):
(a) frequency $\omega_R$ (in kHz) vs mass $M$ (in $M_\odot$);
(b) decay time $\tau$ (in s) vs mass $M$ (in $M_\odot$);
a set of universal relations for the $f$-mode:
(c) scaled frequency $M\omega_R$ vs compactness $M/R$;
(d) scaled decay rate $M\omega_I$ vs compactness $M/R$.
Note the different universal relations for quark matter (lower curves).
}}}
\label{Kunzfig2}
\end{figure}

Generically, the QNM spectrum is quite EOS-dependent,
but as noticed by Anderson and Kokkotas 
\cite{Andersson:1996pn,Andersson:1997rn} early on,
the oscillation frequency $\omega_R$, 
the damping rate $\omega_I$ of the $f$-mode 
and the first $w$-mode exhibit 
some EOS independence and thus universal behavior.
This observation has been followed up by a number of
further QNM studies in GR
\cite{Kokkotas:1999mn,Benhar:1998au,Benhar:2004xg,Tsui:2004qd,Lau:2009bu,BlazquezSalcedo:2012pd,Blazquez-Salcedo:2013jka,Chirenti:2015dda}.
Figs.~\ref{Kunzfig2}(a) and (b) illustrate the strong EOS dependence for the
$f$-mode frequency and decay time
for a varity of EOSs, while Figs.~\ref{Kunzfig2}(c) and (d)
demonstrate approximate EOS independence for
the scaled frequency and scaled decay rate versus compactness (we note that for quark stars a slightly different set of universal relations
holds for the $f$-mode).
As in the case of the $I$-Love-$Q$ relations, compactness $C$ turned out
to be not the best parametrization for the universal relations.
Instead, a parametrization in terms of the scaled moment of inertia
$\bar I=I/M^3$ proved to work rather well (on the 1-2\% level), yielding 
for instance for the $f$-mode the relations \cite{Lau:2009bu}
\begin{equation}
M \omega_R = -0.0047 + 0.133 \eta + 0.575 \eta^2
, \ \ \  M \omega_I  = 0.00694 \eta^4 - 0.0256 \eta^6,
\nonumber
\end{equation}
where $\eta=\bar I^{-1/2}$ has been called the effective compactness.
Other promising universal relations obtained in
\cite{BlazquezSalcedo:2012pd,Blazquez-Salcedo:2013jka}
represent relations between the scaled frequency of a mode and the scaled 
damping rate. 

In the above discussion of the QNMs, only non-rotating NSs have been 
addressed. The presence of rotation in a star complicates the
mode analysis tremendously.
Most studies have therefore employed the so-called Cowling approximation,
where the space-time is frozen \cite{Gaertig:2010kc}.
The presence of rotation leads to a splitting
of co- and counter-rotating perturbations,
which for slow rotation will simply be a shift
proportional to the rotation frequency $\Omega$ of the star.
Only recently has a full calculation of the $f$-mode of rapidly
rotating NSs been reported \cite{Kruger:2019zuz},
providing also a universal relation for the $f$-mode
frequency 
\begin{equation}
M \omega_R= (c_1 +c_2 \bar \Omega + c_3 \bar \Omega^2)
   +(d_1 + d_3 \bar \Omega^2) \eta,
\nonumber
\end{equation}
where $\bar \Omega = M \Omega$, and the coefficients
$c_i$ and $d_i$ differ for co- and counter-rotation
\cite{Kruger:2019zuz}.

Addressing briefly the usefulness of the \textit{universal relations}
in terms of a few examples within GR, 
we note that the relations can be
employed to gain additional information on NS properties.
For instance, a measurement of any of the three quantities constituting 
the $I$-Love-$Q$ relations
will automatically yield the remaining two quantities
\cite{Yagi:2016bkt}.
Invoking the three-hair-relations,
a measurement of the mass, rotation period and moment of inertia
would allow to obtain the seven lowest moments
within an accuracy of 10\%
\cite{Stein:2013ofa}.
If an axial or polar mode would be measured,
we could use the relation between the frequency
and the effective compactness ($M\omega_R$-$\eta$),
together with the relation between 
the damping rate and the effective compactness ($M\omega_I$-$\eta$)
to determine the mass $M$ and the moment of inertia $I$ of the star
\cite{Lau:2009bu}. Moreover, we might then extract
the radius $R$ and obtain valuable
information on the EOS (obtaining information about the EOS from macroscopic data is sometimes called the inverse stellar structure problem \cite{Kokkotas:1999mn}).

\subsection{Neutron Stars in Generalized Theories of Gravity}
\label{neutronstarsref}

\begin{figure}[ht]
\begin{center}
\mbox{\includegraphics[width=.5\textwidth, angle =270]{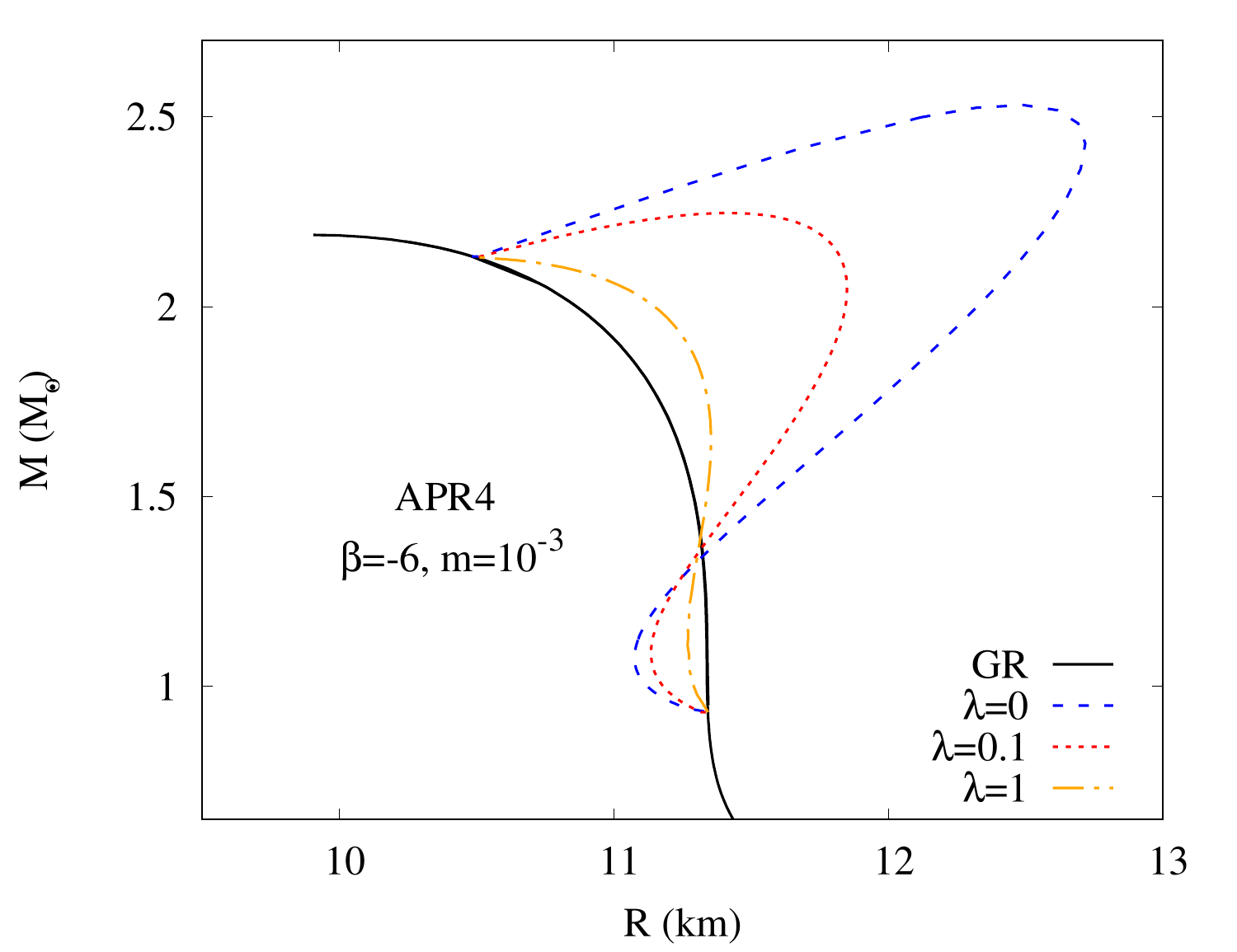}}
\end{center}
\vspace{-0.5cm}
\caption{{\it{NS mass $M$ (in solar masses $M_\odot$) vs radius $R$ (in km)
for a viable nuclear matter EOS 
(see, e.g.,\cite{AltahaMotahar:2019ekm})
for GR and for STT with ``Gaussian'' coupling function
parameter $\beta$, scalar mass $m$ and several values of
the quartic self-interaction coupling constant $\lambda$.}}
}
\label{Kunzfig3}
\end{figure}

Generalized theories of gravity can affect the properties of
neutron stars considerably, the most well-known and well-studied effect
being their scalarization \label{scalarizationiref1} in the presence of 
gravitational scalar fields,
discovered by Damour and Esposito-Farese \cite{Damour:1993hw}.
In scalar-tensor theories (STTs) \label{Scalartensoref3} 
\cite{Brans:1961sx,Damour:1992we,Fujii:2003pa}
the physical action is given in the Jordan frame, \label{Jordanrref3}
where the scalar field does not couple directly to the matter fields. 
The transformation to the Einstein frame then involves  \label{Einsteinfrref7}
a coupling function $A(\varphi)$ relating the metric in both frames,
$\tilde g_{\mu\nu}= A^2(\varphi)  g_{\mu\nu}$ (the tilde referring
to the Jordan frame),
with $\varphi$ representing the scalar field in the Einstein frame.
The Einstein equations in the Einstein frame then contain 
a stress-energy tensor with appropriate powers of the coupling function
$A(\varphi)$. 
At the same time, an equation for the scalar field is obtained
\begin{equation}
 \nabla^{\mu}\nabla_{\mu}\varphi = - 4\pi k(\varphi)T,
\nonumber
\end{equation}
where $T$ is the trace of the stress-energy tensor
and $k(\varphi)= \frac{d\ln({A}(\varphi))} {d\varphi}$
is the logarithmic derivative of the coupling function $A(\varphi)$.
Examples for the mass-radius relation of scalarized NSs
are shown in Fig.~\ref{Kunzfig3} and compared with the respective GR relation.

Let us consider choices of the coupling function $A(\varphi)$, which allow
for GR NS solutions, i.e., solutions with $\varphi=0$ reproducing the
GR solutions.
The phenomenon of matter-induced spontaneous scalarization then arises, 
when it becomes energetically favorable for a NS to acquire
a non-zero value for the scalar field inside the star \cite{Damour:1993hw}.
A branch of spontaneously scalarized NSs then bifurcates from 
the GR branch, when the coupling and the compactness are sufficiently
strong 
\cite{Damour:1993hw,Damour:1996ke,Harada:1998ge,Harada:1997mr,Salgado:1998sg, 
Sotani:2012eb,Pani:2014jra,Silva:2014fca,Sotani:2017pfj,Motahar:2017blm,
Doneva:2013qva,Doneva:2014faa,Staykov:2016mbt,Yazadjiev:2016pcb,Doneva:2016xmf}.
In particular, in \cite{Motahar:2017blm} it could be shown, that
the onset of scalarization depends in a universal, i.e., 
almost EOS-independent, way on the gravitational potential
at the center of the star.

Depending on the chosen coupling function,
the NS properties, like their mass $M$, radius $R$, 
moment of inertia $I$, quadrupole moment $Q$, etc.,
can change distinctively, in principle,
when the NSs get scalarized.
This concerns, in particular, the rapidly rotating case,
where for a given coupling function the effect
of scalarization is significantly enhanced
\cite{Doneva:2013qva,Doneva:2014faa}.
Clearly, an investigation of the universal $I$-Love-$Q$ relations 
of NSs in STT is called for \cite{Pani:2014jra,Doneva:2014faa}. 
However, the spontaneously scalarized NSs possess 
universal $I$-Love-$Q$ relations, that differ from those in GR
only a little (i.e., within less than a few percent) \cite{Pani:2014jra},
when the present observational constraints on
the coupling functions are taken into account
\cite{Demorest:2010bx,Freire:2012mg,Archibald:2018oxs}.
Thus these $I$-Love-$Q$ relations cannot serve to discriminate
between GR and the respective STTs.
This also holds for the three hair relations involving the higher
multipole moments, holding in GR as well as for scalarized NSs
\cite{Pappas:2018csu}.

While the observational constraints on the coupling functions 
have been getting quite strong in recent years
\cite{Demorest:2010bx,Freire:2012mg,Archibald:2018oxs},
the presence of a mass term can evade these constraints.
In fact, the scalar field becomes short-ranged,
i.e., the effect of the mass is to suppress the scalar field 
exponentially at distances on the order of its Compton wavelength
(which should be smaller than the separation of NSs in a binary system
precisely to evade the constraints) \cite{Doneva:2016xmf,Danchev:2020zwn}.
Thus a much broader range of parameters is allowed than in the massless case.
From the universal $I$-Love-$Q$ relations, mainly the $I$-$Q$ relation
has been considered \cite{Doneva:2016xmf}, showing that
relatively large deviations from the GR relation can occur.
Thus massive STT could be tested with these $I$-Love-$Q$ relations.

\begin{figure}[ht]
\begin{center}
\mbox{
(a) \hspace{-0.5cm}
\includegraphics[width=.505\textwidth, angle =270]{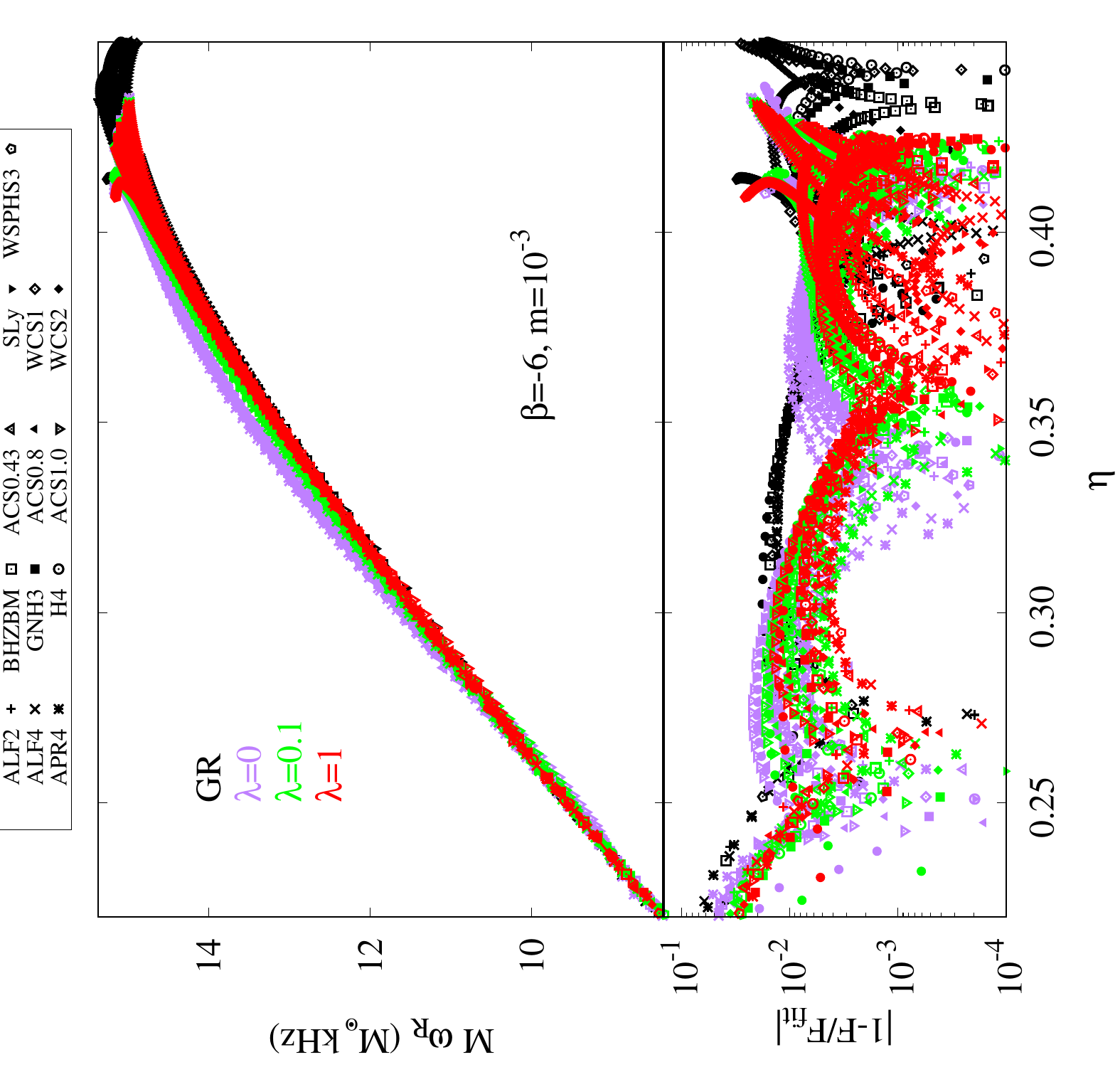}
(b) \hspace{-0.5cm}
\includegraphics[width=.505\textwidth, angle =270]{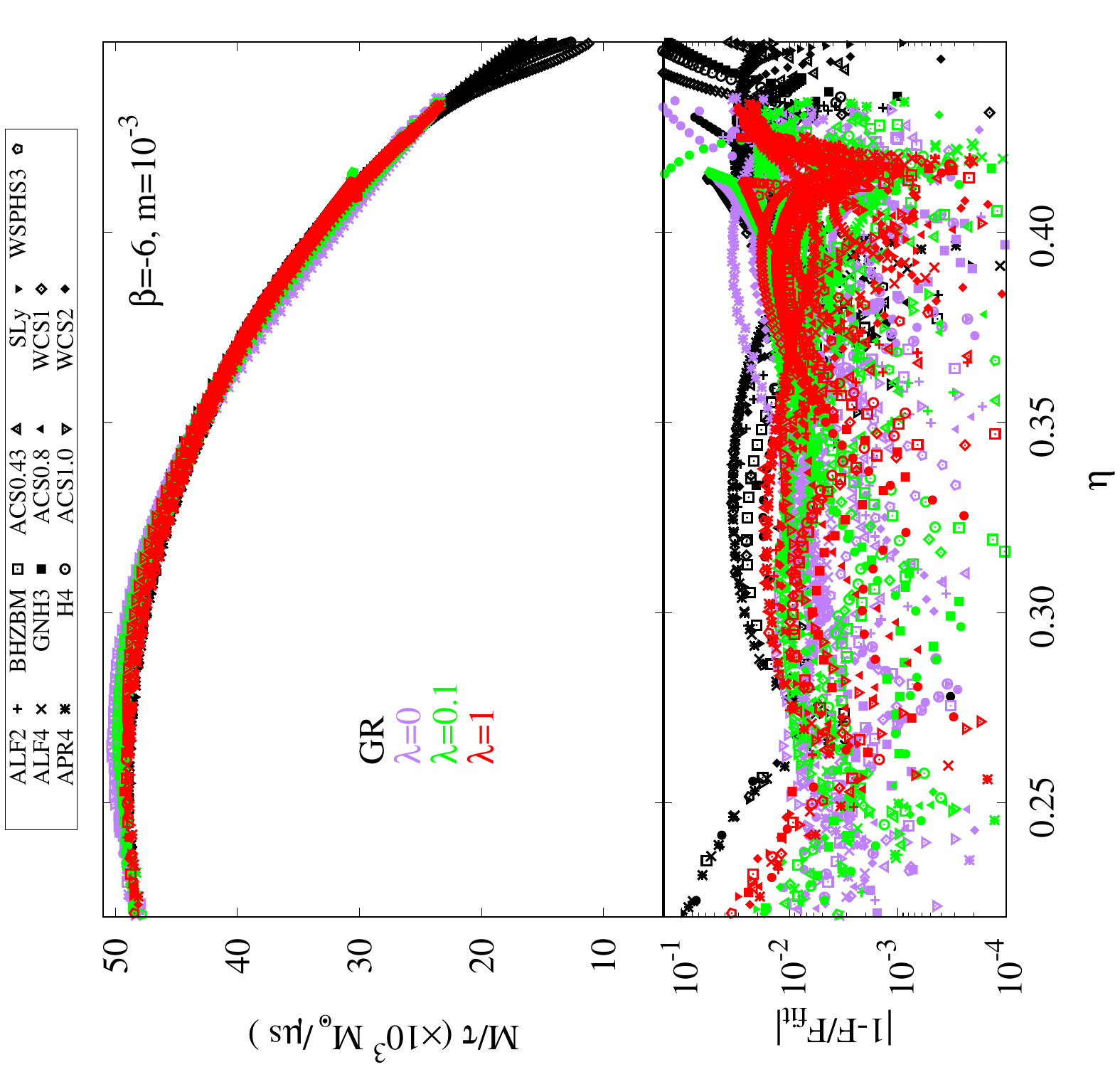}
}
\end{center}
\vspace{-0.5cm}
\caption{{\it{Universal relations for NS fundamental axial mode
for GR and for STT with ``Gaussian'' coupling function
parameter $\beta$, scalar mass $m$ and several values of
the quartic self-interaction coupling constant $\lambda$
(see, e.g.,\cite{AltahaMotahar:2019ekm}):
(a) scaled frequency $M\omega_R$ vs generalized compactness $\eta$;
(b) scaled decay rate $M\omega_I$ vs generalized compactness $\eta$.
Note the same universal relations for quark matter.
}}}
\label{Kunzfig4}
\end{figure}

Turning to the QNMs of NSs in STTs, 
we note that QNMs of nonrotating NSs were first considered in 
\cite{Sotani:2004rq},
where the polar $f$ and $p$ modes were obtained in the Cowling
approximation, i.e., the perturbations of the space-time 
and the scalar field were frozen.
Subsequently, also the gravitational 
axial $w$ modes were studied \cite{Sotani:2005qx}. 
In the axial case, there is, however,
no coupling to the fluid and the scalar field to begin with.
The exact results for the QNMs are therefore obtained in a 
considerably simpler way. Consequently, also the
universal relations have been investigated for axial QNMs
\cite{AltahaMotahar:2018djk}.
Whereas, in general, the frequency and the damping time
of the modes are higher in GR than in STT,
the STT universal relations do not deviate significantly 
from the respective GR relations, either for a massless scalar field
\cite{AltahaMotahar:2018djk}
or for a massive scalar field with self-interaction
\cite{AltahaMotahar:2019ekm}, 
as illustrated in Fig.~\ref{Kunzfig4}.
Exploratory studies of the polar modes for the rapidly rotating case
have again been performed in the Cowling approximation 
\cite{Yazadjiev:2017vpg}.
Going beyond the Cowling approximation would allow
for qualitatively new types of polar modes, 
since scalar radiation would be produced.
This would represent an important probe, if detectable.
So far, their presence has only been studied
in the sector of radial NS oscillations 
\cite{Mendes:2018qwo}.

We next consider $f(R)$ theories
\cite{Sotiriou:2008rp,DeFelice:2010aj,Capozziello:2011et},
which can be reformulated in terms of STTs.
Here, in particular, $R^2$ gravity has received much attention,
which is based on the Lagrangian $f(R) = R + a R^2$,
and has been reformulated and studied in the Einstein frame
\cite{Yazadjiev:2014cza, Staykov:2014mwa, Yazadjiev:2015zia, 
Staykov:2016mbt,Doneva:2015hsa,Staykov:2015kwa,Yazadjiev:2018xxk,
Astashenok:2017dpo}.
The resulting scalar field potential leads to a mass of the scalar field,
$m_\varphi \sim a^{-1/2}$.
For $a\to 0$ the GR limit is recovered, whereas for
$a \to \infty$ a certain Brans-Dicke theory is approached.
Various properties of NSs have been studied in detail, both for
slowly and rapidly rotating NSs
\cite{Yazadjiev:2014cza, Staykov:2014mwa, Yazadjiev:2015zia, 
Staykov:2016mbt,Doneva:2015hsa,Staykov:2015kwa,Yazadjiev:2018xxk,
Astashenok:2017dpo}.
As in STTs, also in $R^2$ gravity, faster rotation implies larger deviations from GR.
The universal $I$-$Q$ relations are retained in $R^2$ gravity,
but deviations from the GR relations can become large ($>$20\%)
and will thus possibly allow for observational constraints on
the theory \cite{Doneva:2015hsa}.

Axial QNMs for NSs in $R^2$ gravity have been considered in
\cite{Blazquez-Salcedo:2018qyy,Blazquez-Salcedo:2018pxo}.
As compared to GR, the frequencies and the damping times decrease 
in $R^2$ gravity, when the parameter $a$ is increased. 
But whereas the frequencies typically deviate significantly from the GR values,
the damping times differ significantly only for large scalar mass.
The universal relations in $R^2$ gravity exhibit some qualitative
differences with respect to the GR ones,
which increase with increasing parameter $a$ 
\cite{Blazquez-Salcedo:2018qyy,Blazquez-Salcedo:2018pxo}.

Recently, the full polar QNMs of NSs in $R^2$ gravity have been studied without 
use of the Cowling approximation \cite{Blazquez-Salcedo:2020ibb}. 
It was shown that the spectrum of these stars includes ultra-long lived modes 
in the radial sector, and that it is enriched by a family of scalar modes. 
The analysis was also done for a particular type of massive STT with very similar results. 
The analysis of the universal relations in this context is currently under way.

NSs have also been considered in more general theories involving
scalar fields.
These theories are particularly attractive, 
when they give rise to second order equations and do not contain ghosts,
as discussed long ago by Horndeski 
\cite{Horndeski:1974wa,Nicolis:2008in,Kobayashi:2011nu,Kobayashi:2019hrl}.
Horndeski theories in fact comprise a wide variety of theories.
A subset of Horndeski theories is known as the \textit{Fab Four} 
\cite{Charmousis:2011bf,Charmousis:2011ea}.
They are motivated largely by cosmology, since they allow for a
\textit{dynamical self-tuning mechanism}.
The \textit{Fab Four} comprise:
(i) General Relativity (``George''),
(ii) Einstein-dilaton-Gauss-Bonnet gravity (``Ringo''),
(iii) theories with a non-minimal coupling with the Einstein tensor (``John''),
(iv) and theories involving the double-dual of the Riemann tensor (``Paul'').
Whereas all combinations were considered in \cite{Maselli:2016gxk},
let us first focus on the combination of (i) and (iii),
yielding spherically symmetric NSs with a time-dependent scalar field
$\varphi(r,t)=Qt + F(r)$, where $Q$ is a constant. 
Such a time-dependence is allowed because of the shift symmetry
of the action.
\begin{figure}[ht]
\begin{center}
\mbox{
(a) \hspace{-1.0cm}
\includegraphics[width=.53\textwidth, angle =270]{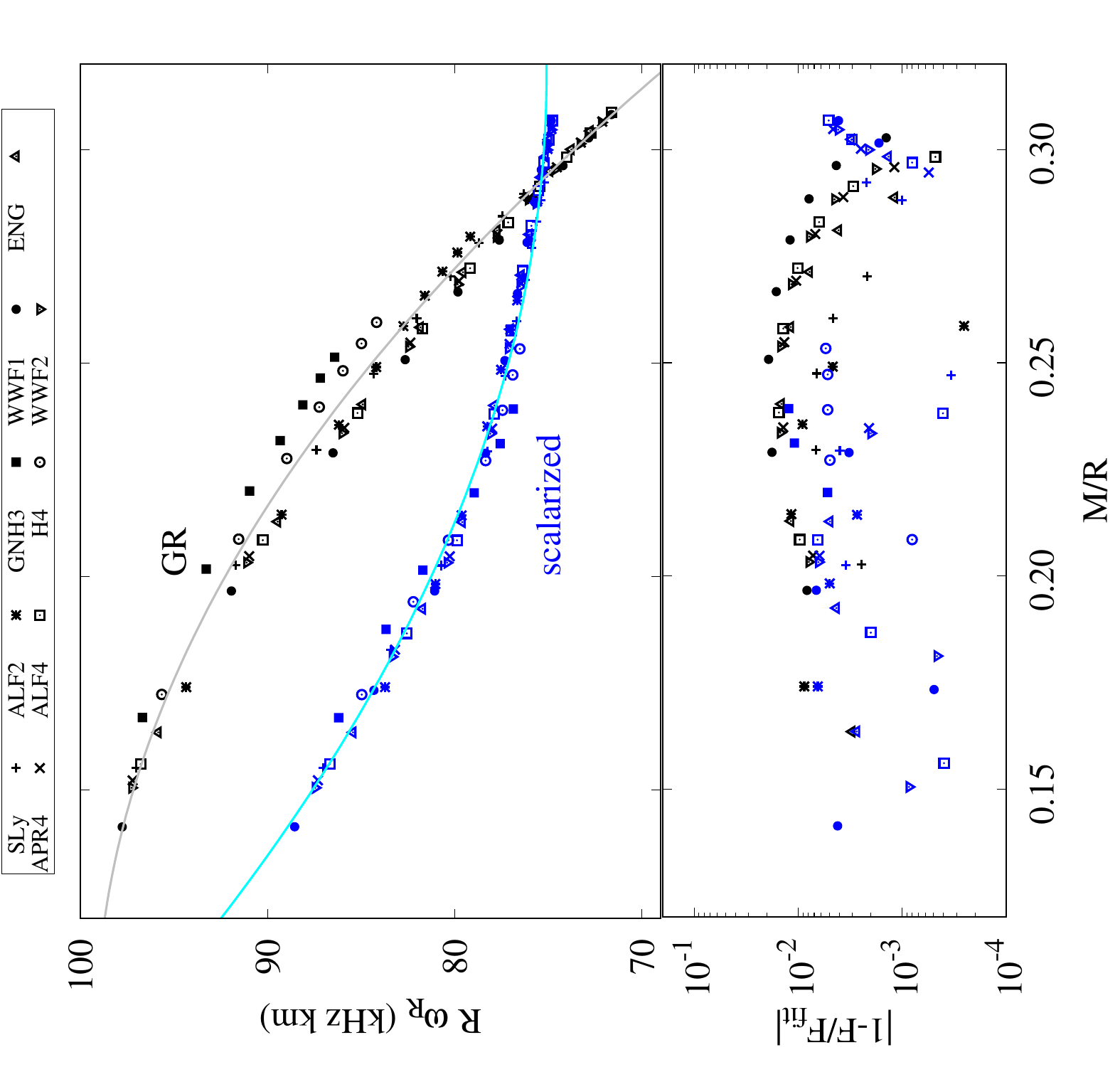}
(b) \hspace{-0.5cm}
\includegraphics[width=.53\textwidth, angle =270]{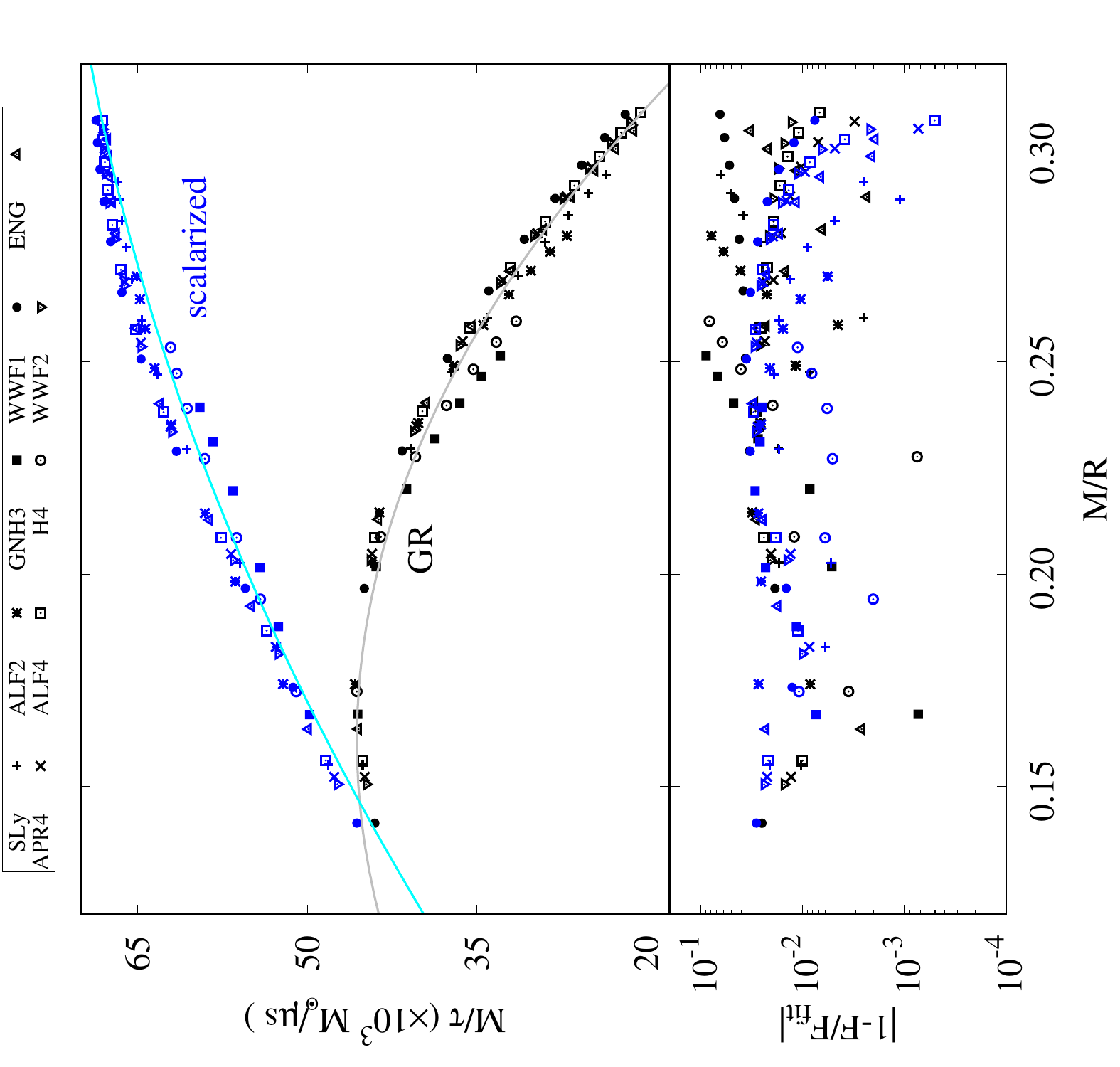}
}
\end{center}
\vspace{-0.5cm}
\caption{{\it{Universal relations for NS fundamental axial mode
for GR and for Horndeski theory (i) and (iii) \cite{Blazquez-Salcedo:2018tyn}
with parameter $Q=0$: 
(a) scaled frequency $M\omega_R$ vs compactness $M/R$;
(b) scaled decay rate $M\omega_I$ vs compactness $M/R$.
}}}
\label{Kunzfig5}
\end{figure}

For the resulting NSs 
\cite{Maselli:2016gxk,Cisterna:2015yla,Cisterna:2016vdx,Babichev:2016jom,
Blazquez-Salcedo:2018tyn} 
the external metric is simply the Schwarzschild metric,
thus the stars pass all Solar System tests,
but the structure of the NSs is considerably different from GR.
The axial QNMs possess different frequencies and damping times 
compared to GR. The QNMs exhibit still the same type of universal relations 
as present in GR, which do, however, differ strongly from those of GR,
as seen in Fig.~\ref{Kunzfig5}
\cite{Blazquez-Salcedo:2018tyn}.

Let us focus next on Einstein-dilaton-Gauss-Bonnet (EdGB) theory,
i.e., considering (i) and (ii),
which is motivated by string theory
\cite{Gross:1986mw,Metsaev:1987zx}.
In this case the scalar dilaton field couples
with coupling function $A(\varphi)= \exp(-\gamma \varphi)$
to the higher-order curvature Gauss-Bonnet (GB) term.
Without a coupling function, the GB term would not
contribute to the field equations.
For this specific coupling function the NSs are always
scalarized, but the scalarization is not spontaneous.
In fact, the GR solutions are not solutions of the EdGB field equations.

The theory then involves two coupling constants, the GB coupling constant
$\alpha$
in front of the GB term and the coupling constant $\gamma$ in the exponential
coupling function. 
As noted in \cite{Pani:2011xm}, in the small scalar field limit,
the solutions depend only on the combination $\alpha \gamma$.
Requiring that the scalar field be real leads to constraints on the parameters,
since NSs then exist only for a certain range of values 
\cite{Pani:2011xm,Kleihaus:2014lba,Kleihaus:2016dui}.
Static and slowly rotating NSs have been studied in \cite{Pani:2011xm},
whereas the full domain of existence of rotating NSs has been explored
in \cite{Kleihaus:2014lba,Kleihaus:2016dui}.
Here the universal $I$-$Q$ relations were also considered,
showing that the GR relations remain basically valid for EdGB NSs.

Axial QNMs have also been investigated in EdGB theory
\cite{Blazquez-Salcedo:2015ets}.
The frequency of the modes increases distinctively
in the presence of the GB coupling, 
whereas the damping time varies only slightly.
The respective universal relations of GR still hold in EdGB theory
\cite{Blazquez-Salcedo:2015ets}.

EdGB theory can be considered a particular case of
Einstein-scalar-Gauss-Bonnet (EsGB) theory
\cite{Doneva:2017bvd,Silva:2017uqg,Antoniou:2017acq},
where the dilatonic coupling function is generalized to some arbitrary
coupling function.
In this case the coupling function can allow for spontaneous scalarization,
when the first derivative of the coupling function vanishes
for vanishing scalar field, and at the same time, the second derivative 
is positive. The GR solutions then remain solutions,
while the GB term serves as a source for the scalar field,
yielding curvature-induced scalarized NSs \cite{Doneva:2017bvd,Doneva:2017duq}.
As in the EdGB case, neutron star solutions cannot exist for
arbitrary values of the parameters of the coupling function,
since the scalar field should be real.
Also, NSs have been found where the scalar field has one or more zeros,
representing radially excited NSs  \cite{Doneva:2017duq}.

As our final example let us consider dynamical Chern-Simons (dCS) gravity,
which involves third-order equations
and introduces parity-violation into gravity \label{parityef3}
\cite{Alexander:2009tp}.
The static NS solutions in dCS gravity are the same as in GR.
Therefore, one has to consider rotation, to see effects of the Chern-Simons (CS) 
term.
In the slow rotation limit, the moment of inertia and the quadrupole moment
were first calculated in \cite{Yunes:2009ch,AliHaimoud:2011fw,Yagi:2013mbt}. 
While these differ from GR, the electric-type tidal deformability 
and thus the Love number agrees with GR \cite{Yagi:2011xp}.
The $I$-Love-$Q$ relations have been studied in dCS gravity 
in \cite{Yagi:2013bca,Yagi:2013awa,Yagi:2016bkt,Gupta:2017vsl}.
Since the dCS relations can deviate considerably from the GR ones,
this will allow to place strong constraints on dCS gravity
with future observations,

Concluding this section, we note that,
whereas there is a large degeneracy with respect to the EOSs
and the generalized models of gravity, the universal relations
represent excellent (largely) EOS-independent tests, that may be used
to constrain gravity models, whenever they possess distinctly different
universal relations from GR.

\section{Black Holes}

\subsection{Black Holes in General Relativity}

The existence of BHs is a genuine prediction of GR; however, 
their properties are highly constrained.
Leaving aside electric (and magnetic) charge, since 
astrophysical BHs are expected to be basically uncharged,
black holes in GR are described by the Kerr family
of rotating BHs, and in the static limit,
by the Schwarzschild BHs.
These asymptotically flat BH solutions of the vacuum Einstein equations
are uniquely characterized by their global charges, the mass $M$ and the
angular momentum $J$ (see, e.g. \cite{Chrusciel:2012jk}),
thus they carry \textit{no hair}. \label{nohairtheref2} 
Their multipole moments are completely determined by these two quantities
\cite{Geroch:1970cd,Hansen:1974zz}
$$M_l + i S_l = M \left( i \frac{J}{M} \right)^l , $$
where $M_0=M$, $S_1=J$, and the odd mass moments $M_l$ and 
the even current moments $S_l$ are identically zero. 
The Kerr BHs are subject to the angular momentum bound,
$j = \frac{J}{M^2} \le  1$, which is saturated by extremal BHs.
Beyond this bound only naked singularities reside.
The \textit{no-hair} hypothesis will be testable 
in various future experiments \cite{ Cardoso:2016ryw}.

The QNMs of Schwarzschild and Kerr BHs are well studied
\cite{Kokkotas:1999bd,Nollert:1999ji,Rezzolla:2003ua,Berti:2009kk,
Konoplya:2011qq}.
As in the case of NSs, the QNMs of BHs are of utmost importance 
in the ringdown phase after merger events,
and thus represent crucial observables.
All Schwarzschild QNMs possess positive imaginary parts, and therefore
represent damped modes, showing that Schwarzschild BHs are linearly stable.
The Kerr BHs are linearly stable as well \cite{Whiting:1988vc}.
The Schwarzschild and Kerr QNMs are called \textit{isospectral}, since the
polar and axial perturbations possess identical 
complex eigenvalues $\omega$.
Of course,
the Kerr QNMs again feature splitting of the QNM frequencies due to rotation.
Intuitive understanding of the modes is known in the high-overtone limit, 
where the frequencies can be related to the surface gravity of the horizon
\cite{Nollert:1993zz},
and in the eikonal limit, where they can be described in terms of the Keplerian
frequency of the circular photon orbit and the Lyapunov exponent of the orbit
\cite{Ferrari:1984zz,Yang:2012he}.

Also of tremendous observational significance is the shadow of a BH.
It represents the region around a BH, where any light from 
background stars will be captured, making this region appear dark.
For Schwarzschild black holes this region is spherical and determined
by the photon ring, i.e., the unstable photon orbit at a radius of
$3M$. Calculated first by Synge \cite{Synge:1966okc},
the angle $\alpha$ determining the size of the
shadow of a Schwarzschild black hole is
$$
\sin^2 \alpha = \frac{27}{4} \,
\frac{(\rho_{\rm O} -1)}{\rho_{\rm O}^{3}} ,
$$
where $\rho_{\rm O} = r/(2m)$ represents the observer coordinate.
For Kerr BHs one has to distinguish co-rotating and counter-rotating
orbits, and a photon region results. The shadow is then no longer spherical,
unless viewed from the rotation axis,
and its deviation from a circle is a measure for the BH spin
\cite{Bardeen:1973,Perlick:2004tq,Grenzebach:2014fha}.

The \textit{no-hair} theorem of GR can be evaded in a number of ways. For instance, when including non-Abelian gauge and scalar fields
(see, e.g., \cite{Volkov:1998cc,Kleihaus:2016rgf}),
or rotating complex scalar and Proca fields
\cite{Herdeiro:2014goa,Herdeiro:2015waa,Brito:2015pxa}.
In particular, the latter case has drawn considerable interest,
and various physical properties of these rotating hairy BHs have
been addressed and compared to their pure Kerr counterparts
\cite{Herdeiro:2014goa,Herdeiro:2014jaa,Herdeiro:2015gia,Kleihaus:2015iea,
Cunha:2015yba,Shen:2016acv}.
For instance, they can exceed the Kerr bound, $J/M^2>1$, 
they can possess a quadrupole moment larger than the Kerr value $J^2/M$, 
or even multiple disconnected shadows may arise.

\subsection{Black Holes in Generalized Theories of Gravity}

\begin{figure}[ht]
\begin{center}
\mbox{
(a) \hspace{-1.0cm}
\includegraphics[width=.495\textwidth, angle =0]{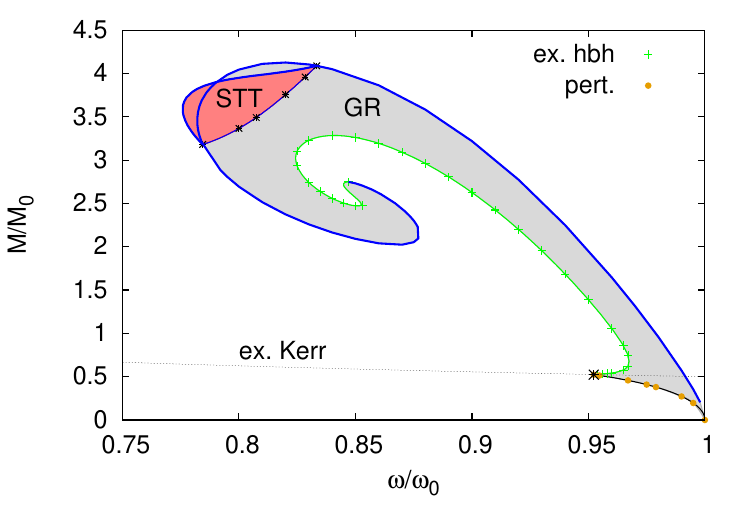}
(b) \hspace{-0.5cm}
\includegraphics[width=.495\textwidth, angle =0]{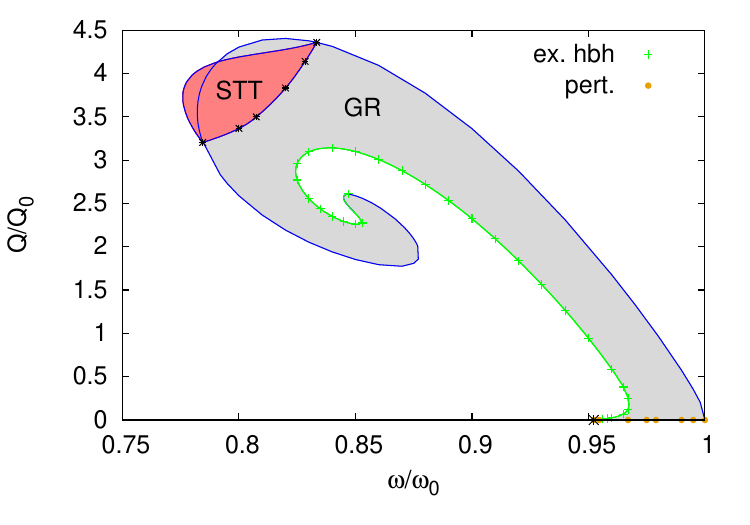}
}
\end{center}
\vspace{-0.5cm}
\caption{{\it{Domain of existence of 
fundamental rotating BHs with complex scalar hair
for GR (grey) and with additional scalarization for STT (red):
(a) normalized mass $M$ and (b) normalized scalar charge $Q$ 
vs normalized parameter $\omega$ (of the harmonic time dependence)
\cite{Kleihaus:2015iea}.
Moreover, in (a) we indicate the domain of Kerr BHs.
}}}
\label{Kunzfig6}
\end{figure}

Addressing BHs in the same set of gravity theories discussed above
in the context of NSs, we first note that the STTs 
which give rise to matter-induced spontaneous scalarization of
NSs, possess only the Schwarzschild and Kerr BH solutions,
but no additional scalarized branches of BHs. 
In fact, the \textit{no-hair} theorems do not allow for scalar hair
in these STTs, when only a single real scalar field is present.
For massive complex scalars, on the other hand, 
besides the Kerr BHs the rotating hairy BHs
\cite{Herdeiro:2014goa,Herdeiro:2015waa} are also recovered.
Also, rotating hairy BHs with additional scalarization
are possible \cite{Kleihaus:2015iea},
as demonstrated in Fig.~\ref{Kunzfig6}, where their domain
of existence is shown for the fundamental scalarized BHs.
Since $f(R)$ theories may be considered special cases of STTs,
the Schwarzschild and Kerr BHs are also recovered there.

Turning to Horndeski theories, we note that, 
depending on the chosen action, hairy BHs may arise
(see, e.g., \cite{Babichev:2017guv} for a review).
The \textit{no-hair} theorem can, for instance, be evaded
by allowing for a time-dependent scalar field
in the combination (i) and (iii) of the \textit{Fab Four}
\cite{Babichev:2013cya,Kobayashi:2014eva}.
But even when within the Horndeski theories
the Schwarzschild BHs is recovered,
the presence of additional degrees of freedom 
can lead to non-GR signatures in the QNM spectrum
\cite{Tattersall:2017erk,Tattersall:2018nve,Tattersall:2019pvx,Tattersall:2019nmh}.

Considering next the case of
curvature-induced scalarization, we note that
the \textit{no-hair} theorem is evaded in  the presence of the GB term.
The field equations now possess an \textit{effective} energy momentum tensor
containing the higher curvature terms.
EdGB BHs possess distinctive features as compared to Schwarzschild and Kerr BHs.
Since there are no known closed form solutions,
static BHs have first been constructed perturbatively
\cite{Mignemi:1992nt,Mignemi:1993ce}
and numerically \cite{Kanti:1995vq,Torii:1996yi,Guo:2008hf}.
As in the case of NSs discussed above, 
there arises a condition for the EdBG BHs from the requirement
that the dilaton field should be real \cite{Kanti:1995vq}.
For given coupling constants $\alpha$ and $\gamma$ of the theory, 
this translates into a lower bound on the mass of these dilatonic BHs.
The minimum mass solution is often referred to as the critical  solution.
In fact, for larger values of $\gamma$ (including $\gamma=1$)
there are two branches of BH solutions: the fundamental branch,
which extends from the minimum mass to infinity, 
and a short secondary branch, which extends from the minimum mass
to a naked singularity 
\cite{Kanti:1995vq,Torii:1996yi,Guo:2008hf}.
For a given mass, the EdGB BHs have a smaller size than
the Schwarzschild BHs.
 For large masses, the fundamental branch of EdGB BHs approaches
the branch of Schwarzschild BHs from below. But mass and radius
are not proportional. Therefore, the scaled horizon area
$A_H/M^2$ is not constant as for Schwarzschild.

\begin{figure}[ht]
\begin{center}
\mbox{
(a) \hspace{-1.0cm}
\includegraphics[width=.49\textwidth, angle =0]{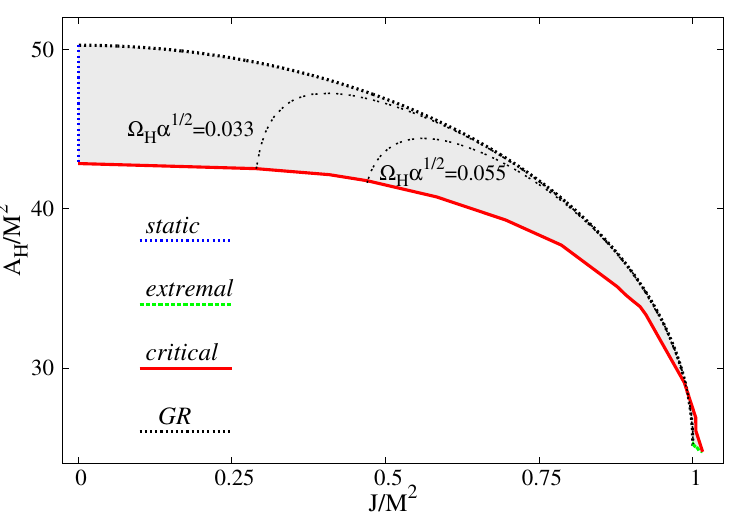}
(b) \hspace{-0.5cm}
\includegraphics[width=.5\textwidth, angle =0]{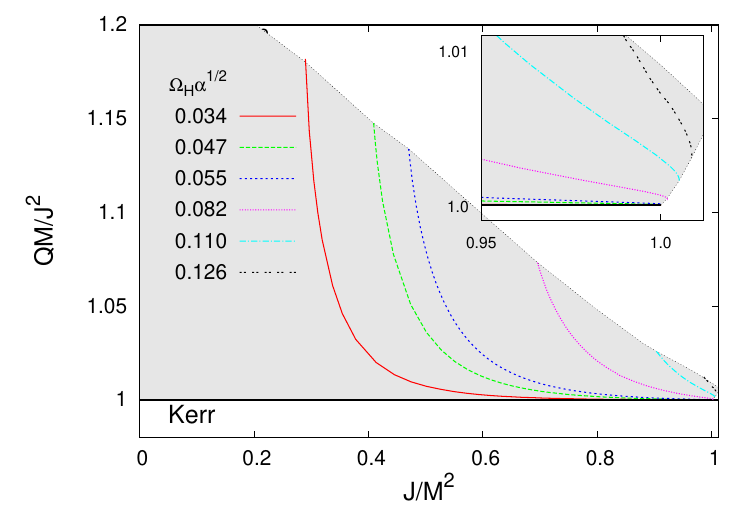}
}
\end{center}
\vspace{-0.5cm}
\caption{{\it{Domain of existence of
rotating BHs with dilaton hair in EdGB theory:
(a) scaled horizon area $A_{\rm H}/M^2$ and (b) scaled quadrupole
moment $QM/J^2$ vs scaled angular momentum $J/M^2$.
Furthermore, we present families of solutions with fixed scaled
horizon angular velocities \cite{Kleihaus:2015aje}.
The Kerr BHs form part of the boundary (black).
}}}
\label{Kunzfig7}
\end{figure}

For fixed coupling constants,
the slowly \cite{Pani:2009wy,Pani:2011gy,Ayzenberg:2014aka,Maselli:2015tta}
and fast \cite{Kleihaus:2011tg,Kleihaus:2014lba,Kleihaus:2015aje,Chen:2018jed}
rotating EdGB BHs possess a domain of existence,
that is limited by the static EdGB BHs, the critical EdGB BHs, 
the extremal EdGB BHs, and the Kerr BHs,
as shown in Fig.~\ref{Kunzfig7}(a), where the scaled area
is shown versus the scaled angular momentum.
Interestingly, the Kerr bound can be slightly exceeded by fast rotating 
EdGB BHs  (see Fig.~\ref{Kunzfig7}(a)).
The EdGB quadrupole moment can differ considerably from the Kerr case
(see Fig.~\ref{Kunzfig7}(b)),
where $QM/J^2=1$  \cite{Kleihaus:2014lba,Kleihaus:2015aje}.
The shadow of EdGB BHs is always smaller than the shadow
of the comparable Kerr BHs; however, the deviations 
are always smaller than a few percent
\cite{Cunha:2016wzk}.
Let us note,  that area and entropy
differ in the presence of the GB coupling,
since the entropy receives a contribution from the GB term
\cite{Wald:1993nt}. Since the entropy of EdBH BHs 
is  larger than for Kerr BHs, this indicates stability
of these solutions.

\begin{figure}[ht]
\begin{center}
\mbox{
(a) \hspace{-1.0cm}
\includegraphics[width=.385\textwidth, angle =270]{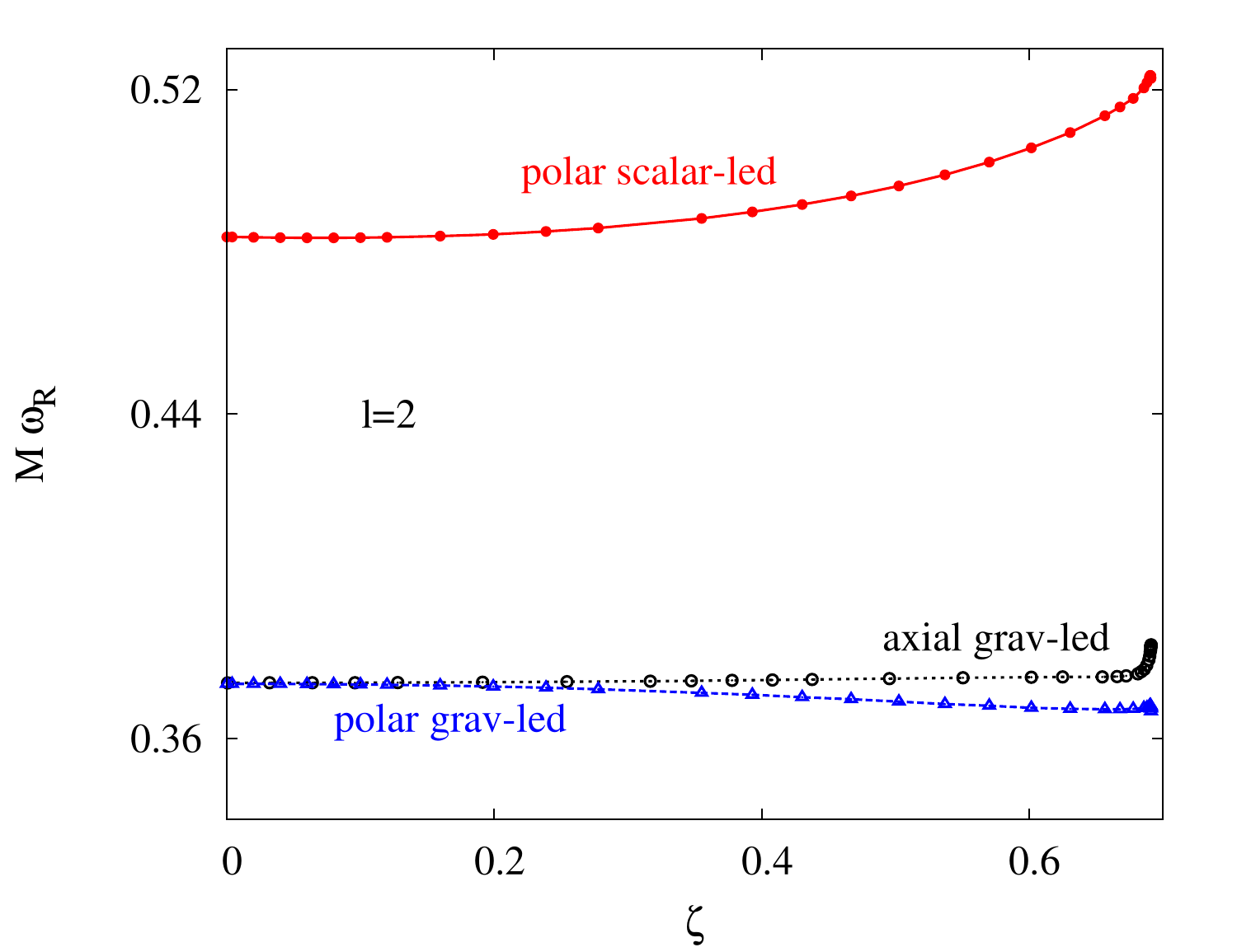}
(b) \hspace{-0.5cm}
\includegraphics[width=.385\textwidth, angle =270]{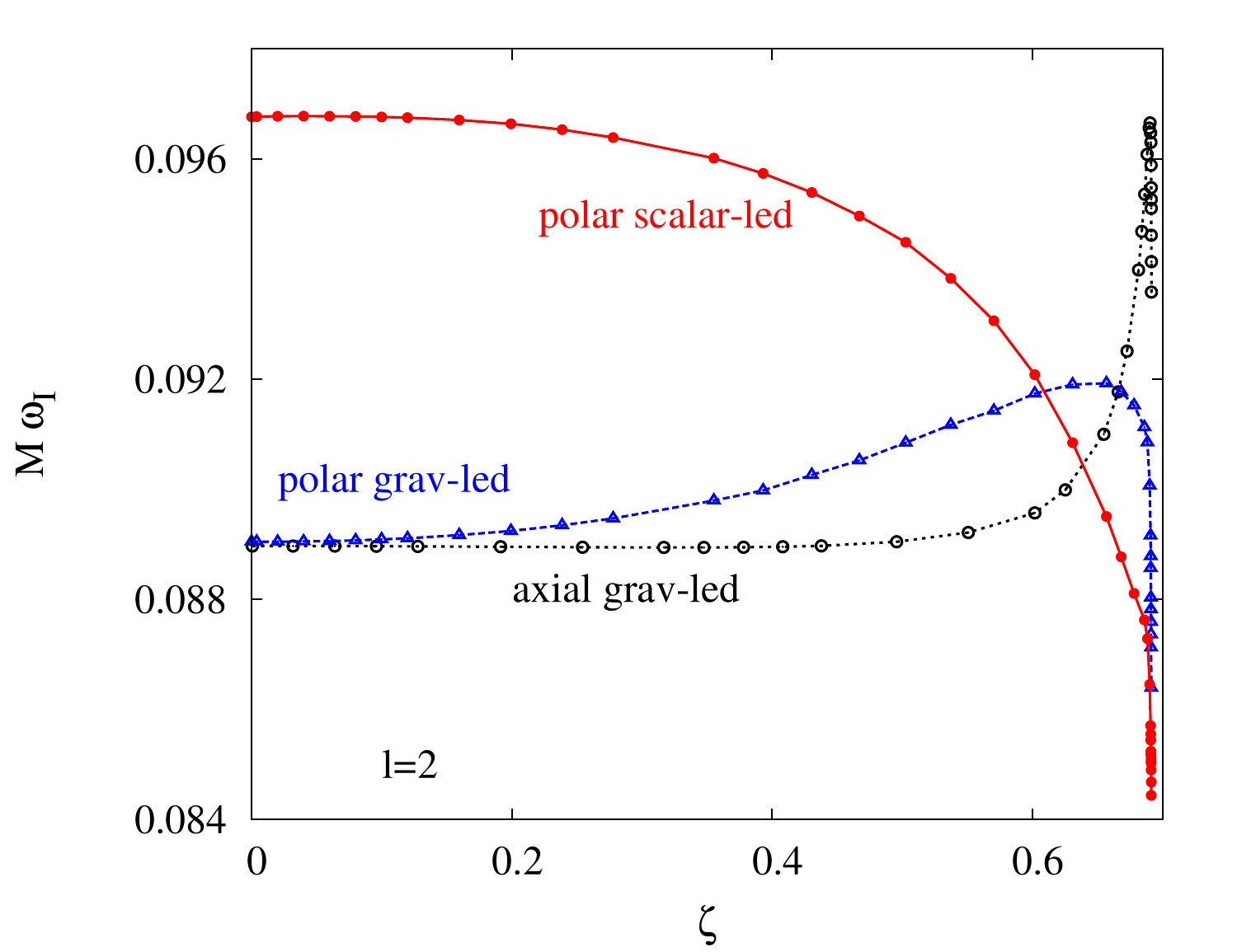}
}
\end{center}
\vspace{-0.5cm}
\caption{{\it{QNMs of EdGB BHs:
(a) scaled frequency $M \omega_R$ and (b) scaled decay rate $M \omega_I$
vs scaled GB coupling constant $\zeta=\alpha/M^2$ (for $\gamma=1$)
for the fundamental axial, polar gravitational-led, and polar scalar-led
modes \cite{Blazquez-Salcedo:2016enn}.}}
}
\label{Kunzfig8}
\end{figure}

Mode stability of EdGB BHs has been investigated in the static case,
where the QNMs have been obtained and analyzed
\cite{Kanti:1997br,Pani:2009wy,Ayzenberg:2013wua,Blazquez-Salcedo:2016enn,Blazquez-Salcedo:2017txk,Konoplya:2019hml,Zinhailo:2019rwd}.
All modes except for the $l=0$ mode on the short secondary 
branch are stable.
Because of the scalar field, there is monopolar and dipolar
radiation, not present  in GR.
The axial modes are pure space-time modes. Here the geodesic correspondence
of the eikonal approximation works quite well, except for the
parameter region, where the critical solution is approached.
The polar modes are coupled to the dilaton.
Here  we need to distinguish scalar-led and gravitational-led modes,
which in the GR limit reduce to the QNMs of Schwarzschild BHs.
The deviations of the polar modes from GR modes are larger 
than for axial modes. Not surprisingly, the presence of the
dilaton breaks the isospectrality of the GR BHs.
This is illustrated in Fig.~\ref{Kunzfig8}, where the
fundamental axial modes and polar gravitational-led and scalar-led 
modes are shown.
While there would be scalar radiation from EdGB BHs,
its detection would depend on the coupling of the dilaton to matter.
\begin{figure}[ht]
\begin{center}
\mbox{
\includegraphics[width=.6\textwidth, angle =270]{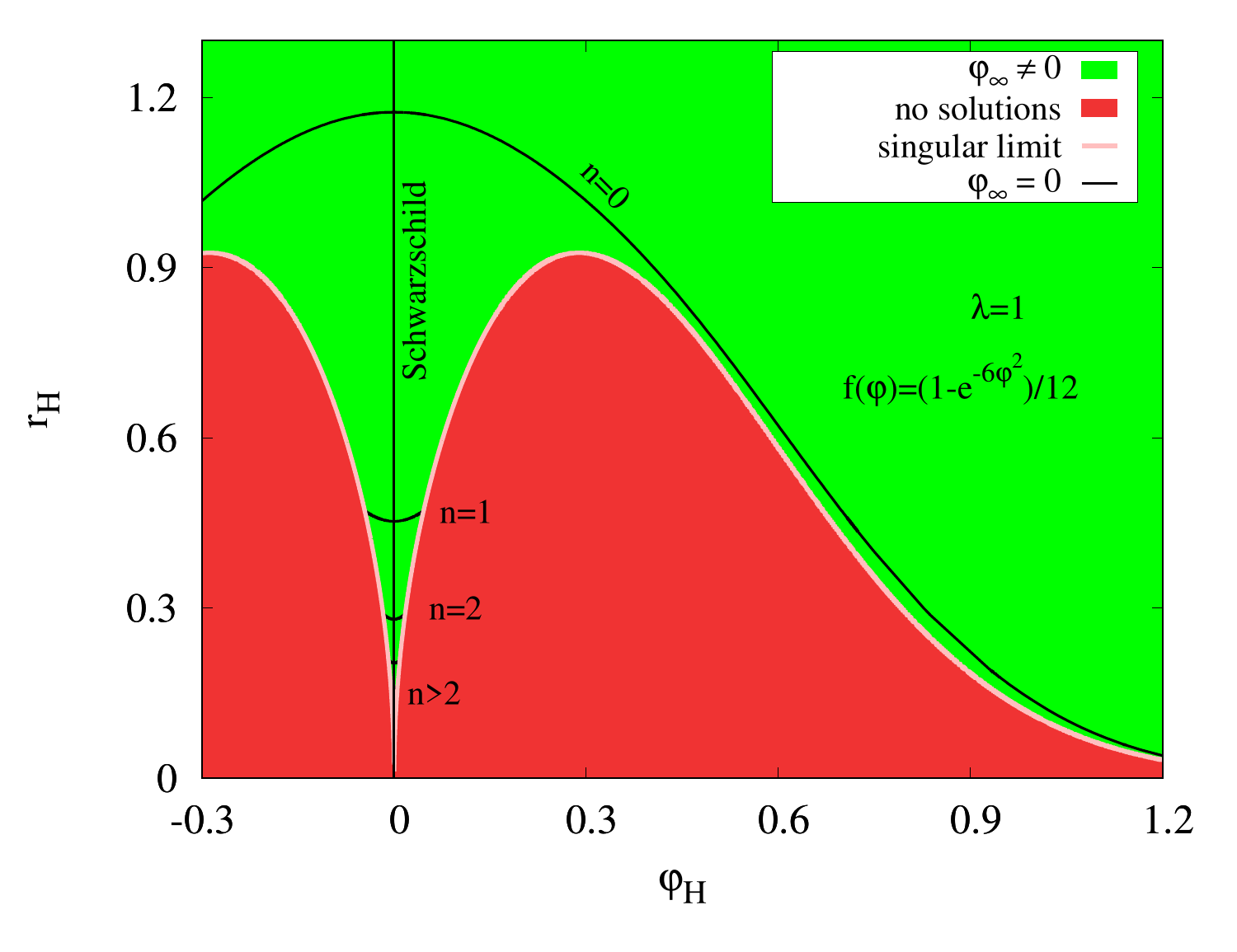}
}
\end{center}
\vspace{-0.5cm}
\caption{{\it{Domain of existence of static scalarized EsGB BHs:
horizon radius $r_{\rm H}$ vs the horizon scalar field $\phi_{\rm H}$.
\cite{Blazquez-Salcedo:2018jnn}.
The fundamental spontaneously scalarized BHs are labeled by $n=0$,
$n\ge 1$ labels radially excited BHs.
Additionally, we indicate the   Schwarzschild BHs.
}}}
\label{Kunzfig9}
\end{figure}

We next turn to the BHs of EsGB theory with a coupling function
chosen such, as to allow for curvature-induced spontaneous scalarization
\cite{Doneva:2017bvd,Silva:2017uqg,Antoniou:2017acq,Antoniou:2017hxj,
Blazquez-Salcedo:2018jnn,Doneva:2018rou,Silva:2018qhn,Cunha:2019dwb,
Macedo:2019sem,Collodel:2019kkx}.
Then, scalarized BHs arise for a certain range of the coupling constant,
i.e., at certain values, branches of scalarized BHs
bifurcate from the Schwarzschild solution.
These branches represent the fundamental solutions and their
excitations (see Fig.~\ref{Kunzfig9}).
QNMs of static EsGB black holes have received much interest
\cite{Blazquez-Salcedo:2018jnn,Silva:2018qhn,Macedo:2020tbm,Blazquez-Salcedo:2020rhf,Blazquez-Salcedo:2020caw}.
The fundamental scalarized solutions can be stable (in part of their
domain of existence), depending on the choice of coupling function.
This is, for instance, the case for a ``Gaussian'' coupling function
\cite{Doneva:2017bvd,Blazquez-Salcedo:2018jnn}.
Stability under spherical perturbations can also be achieved, when a mass term and a potential
for the scalar are included
\cite{Macedo:2019sem}.
The analysis of the non-radial perturbations of the fundamental scalarized solutions 
of the case with a ``Gaussian'' coupling function was recently performed 
for the axial channel \cite{Blazquez-Salcedo:2020rhf} and polar channel \cite{Blazquez-Salcedo:2020caw}, showing stability of this branch 
under these higher multipole perturbations (except for the region where 
hyperbolicity is lost).

The fundamental as well as excited rapidly rotating scalarized BHs 
have been investigated recently 
for ``Gaussian'' and quadratic coupling functions \cite{Cunha:2019dwb,Collodel:2019kkx}.
In the Gaussian case the domain of existence of the fundamental BHs
is quite broad for small rotation  rates, but becomes narrower
for faster rotation.
Interestingly, this can be exploited to obtain bounds on the
coupling constant of the theory, by studying the shadow
of these rotating scalarized BHs.
For slowly rotating BHs, the deviations  in the shadow size
between EsGB and Kerr BHs can be large,
e.g., in the (stable) static limit above 40\%
\cite{Cunha:2019dwb}.
Considering the shadow of M87, observations suggest that
the deviation from  Kerr should be below 10\%,
allowing to obtain a (weak) bound on the coupling constant
\cite{Cunha:2019dwb}.
Moreover, the phenomenon of spin-induced spontaneous scalarization
of   sufficiently fast rotating BHs has been observed
\cite{Dima:2020yac,Hod:2020jjy,Doneva:2020nbb,Herdeiro:2020wei,Berti:2020kgk}.

Finally, in dCS theory, static BHs are given by the Schwarzschild  metric.
They have been shown to be mode stable 
\cite{Cardoso:2009pk,Molina:2010fb,Kimura:2018nxk}.
However, rotating dCS BHs carry scalar hair. 
Slowly rotating BHs have been obtained in 
\cite{Yunes:2009hc,Konno:2009kg,Yagi:2012ya},
where the dCS correction to the quadrupole moment was calculated,
and rapidly rotating BHs were obtained in
\cite{Delsate:2018ome}, employing a linear coupling function.
Appropriately chosen coupling functions
may allow for spontaneously scalarized rotating dCS BHs.

\section{Conclusions}

Here we have addressed the properties of NSs and BHs 
in GR and generalized theories of gravity.
For NSs we have highlighted the power of the EOS-independent,
\textit{universal relations}, which may differ 
significantly from their GR counterparts \cite{Yagi:2016bkt,Doneva:2017jop}.
Non-rotating and slowly rotating compact objects
and their QNMS have been studied already for numerous
gravity theories. However, much less is known about
rapidly rotating NSs and BHs and their QNMs
in generalized theories of gravity.
Here, theoretical progress will be valuable and timely,
to provide predictions and templates for future
highly precise observations.
These will then allow us
to put stronger bounds on these generalized theories
\cite{Barausse:2012da,Shibata:2013pra,Taniguchi:2014fqa,Sennett:2016rwa,Ponce:2014hha,Sennett:2017lcx,Sagunski:2017nzb,Berti:2018cxi,Berti:2018vdi,Vivanco:2019qnt}. 

Bounds must, however, be employed with caution.
Consider, for example, the joint observation of GW170817 
in gravitational waves and in the electromagnetic spectrum,
allowing to constrain the speed of gravitational waves.
This observation has been interpreted to rule out a large number 
of alternative theories of gravity 
\cite{Baker:2017hug,Ezquiaga:2017ekz,Sakstein:2017xjx}. 
Closer inspection revealed that when viewing these
generalized theories as effective field theories,
subtleties arise, that challenge these conclusions
\cite{deRham:2018red}.
Moreover, generalized theories of gravity may remain viable in the context of
compact objects, when they are not required to resolve the cosmological issues at the same time \cite{Kobayashi:2012kh,Ayzenberg:2013wua,Kobayashi:2014wsa,Antoniou:2019awm}.








\chapter[Parametrized Post-Newtonian Formalism]{Parametrized Post-Newtonian 
Formalism}
\label{Hohmannchapter}
\label{PPNref1}

{\em Manuel Hohmann}

\section{Historical remarks}
In order to assess the viability of different gravity theories and compare their 
predictions to a large number of observations, several frameworks have been 
developed. The common idea behind these frameworks is to characterise each 
gravity theory by a number of certain parameters, which can then be compared to 
a corresponding set of observations. The parameters introduced as an 
intermediate step in these formalisms serve as an abstraction both to the 
observations and to the theories under consideration, and thus allow us to 
divide testing gravity theories into two parts: deriving a fixed set of 
predicted values of the parameters from any given gravity theory, and comparing 
these to their experimentally determined values.

An early example of such a formalism is the Eddington-Robertson-Schiff formalism 
for tests of post-Newtonian gravity in the Solar   \label{solarsystemref7}
System~\cite{Eddington:1922,Robertson:1962,Schiff:1967}. It is based on the 
assumption that the Sun is a point-like, non-rotating mass generating a 
spherically symmetric gravitational field, which is described by a metric 
tensor, while the planets are test bodies moving along the geodesics of this 
metric. The Eddington-Robertson-Schiff metric contains two free parameters, 
\(\beta\) and \(\gamma\), which are determined by solving the field equations of 
a given (metric) gravity theory for a point-like source mass. This theoretical 
prediction of the two parameters is complemented by their experimental 
measurement, as they describe, among other observables, the light deflection 
angle and perihelion precession of mercury. The advantage in using the 
Eddington-Robertson-Schiff formalism lies in the fact that instead of 
calculating the trajectories of light and mercury (and potentially other 
observables) for any given gravity theory, it suffices to calculate only two 
parameters, whose values can be compared to different observations. This 
advantage comes with the drawback that the formalism may be applied only to 
theories satisfying a particular set of assumptions - in this case metric 
gravity theories, whose metric takes a particular form for a point-like source.

Further developments led to a generalisation of the Eddington-Robertson-Schiff 
formalism, first by relaxing the assumption of a static point mass as the 
gravitational source. These generalisations are allowed to include rotation to 
describe the Lense-Thirring effect~\cite{Schiff:1967}, as well as a simple 
extension to fluids~\cite{Baierlein:1967zz}. A system of several gravitating 
point masses was assumed in~\cite{Nordtvedt:1968qr,Nordtvedt:1968qs}, and the 
formalism was subsequently extended to a full perfect fluid 
description~\cite{Thorne:1970wv,Will:1971zzb,Will:1971wt}. The modern form of 
the parametrized post-Newtonian (PPN) formalism was then developed from 
combining these preceding approaches into a single 
formalism~\cite{Will:1972zz,Nordtvedt:1972zz}, and complementing it with another 
extension in order to also describe preferred location effects occurring in a 
particular class of gravity theories~\cite{Will:1973zz}. This is the formalism 
we will discuss in the following section.

\section{Parametrized post-Newtonian Formalism}\label{ssec:ppnformalism}

The central ingredient of the parametrized post-Newtonian 
formalism~\cite{Will:1993ns} is a perturbative expansion of the metric tensor 
around a flat Minkowski background,
\begin{equation}\label{eq:metperturb}
g_{\mu\nu} = \eta_{\mu\nu} + h_{\mu\nu}\,,
\end{equation}
where the perturbation \(h_{\mu\nu}\) is due to a localised source matter 
distribution, which vanishes asymptotically far away from the source. This 
matter source is described by the energy-momentum tensor of a perfect fluid,
\begin{equation}\label{eq:tmunu}
T^{\mu\nu} = (\rho + \rho\Pi + p)u^{\mu}u^{\nu} + pg^{\mu\nu}\,,
\end{equation}
with density \(\rho\), pressure \(p\), specific internal energy \(\Pi\) and 
four-velocity \(u^{\mu}\), normalised by the metric to 
\(g_{\mu\nu}u^{\mu}u^{\nu} = -1\). It is further assumed that the velocity 
\(v^i = u^i/u^0\) of the source matter in a given frame of reference, in which 
the expansion~\eqref{eq:metperturb} is valid, is small compared to the speed of 
light, \(|\vec{v}| \ll c \equiv 1\). Based on this assumption, we promote  the 
velocity to a perturbation parameter, and assigns so-called velocity orders 
\(\mathcal{O}(n) \propto |\vec{v}|^n\) to all quantities present in the theory 
under investigation, based on their orders of magnitude in the Solar System. For 
the matter variables,   velocity orders of \(\mathcal{O}(2)\) are used for 
\(\rho\) and \(\Pi\) and \(\mathcal{O}(4)\) for \(p\). The metric perturbation 
is expanded in the form 
\begin{equation}\label{eq:metvelord}
h_{\mu\nu} = \overset{\scriptscriptstyle 1}{h}_{\mu\nu} 
+ \overset{\scriptscriptstyle 2}{h}_{\mu\nu} + \overset{\scriptscriptstyle 
3}{h}_{\mu\nu} + \overset{\scriptscriptstyle 4}{h}_{\mu\nu} + \mathcal{O}(5)\,,
\end{equation}
where we used overscripts to denote the velocity orders 
\(\overset{\scriptscriptstyle n}{h}_{\mu\nu} \sim \mathcal{O}(n)\). Finally, 
it is assumed that the gravitational field is quasi-static, i.e., its time 
evolution is governed entirely by the motion of the localised sources. This 
assumption justifies assigning an additional velocity order to time 
derivatives, \(\partial_0 \sim \mathcal{O}(1)\).

In order to determine which terms in the expansion~\eqref{eq:metvelord} are 
relevant, one considers the action
\begin{equation}\label{eq:partaction}
S[\gamma] = -m\int\sqrt{-g_{\mu\nu}\dot{\gamma}^{\mu}\dot{\gamma}^{\nu}}dt = 
-m\int\sqrt{-g_{00} - 2g_{0i}\dot{\gamma}^i - 
g_{ij}\dot{\gamma}^i\dot{\gamma}^j}dt
\end{equation}
of a test particle of mass \(m\) moving along a trajectory \(\gamma^{\mu} = 
(t,\vec{x})\), which we chose to parametrise by the coordinate time \(t\). Here 
one 
assumes that the velocity of the test particle is of the same order 
\(|\dot{\vec{\gamma}}| \sim \mathcal{O}(1)\) as that of the source matter. 
Expanding the action~\eqref{eq:partaction} into velocity orders, we find that 
the second velocity order corresponds to the Newtonian limit, with 
\(\overset{\scriptscriptstyle 2}{h}_{00} = 2U\) given by the Newtonian 
potential, while the post-Newtonian limit appears as the fourth velocity order, 
thus containing the metric components \(\overset{\scriptscriptstyle 2}{h}_{ij}, 
\overset{\scriptscriptstyle 3}{h}_{0i}, \overset{\scriptscriptstyle 
4}{h}_{00}\). Odd velocity orders do not appear, since they are antisymmetric 
under time reversal, and thus correspond to dissipative processes. These are 
prohibited by energy-momentum conservation: conservation of rest mass prohibits 
terms of the first velocity order, while the third velocity order is prohibited 
by Newtonian energy conservation.

In order to determine the aforementioned components of the metric perturbations, 
it is assumed that they are expressed in a generic form \label{Scalarperfrref5}
\begin{subequations}\label{eq:ppnmetric} 
\begin{eqnarray}
&&\overset{\scriptscriptstyle 2}{h}_{00} = 2U\,,\\
&&\overset{\scriptscriptstyle 2}{h}_{ij} = 2\gamma U\delta_{ij}\,,\\
&&\overset{\scriptscriptstyle 3}{h}_{0i} = -\frac{1}{2}(3 + 4\gamma + \alpha_1 
- 
\alpha_2 + \zeta_1 - 2 \xi )V_i - \frac{1}{2}(1 + \alpha_2 - \zeta_1 + 
2\xi)W_i\,,\\
&&\overset{\scriptscriptstyle 4}{h}_{00} = -2\beta U^2  + (2 + 
2\gamma + \alpha_3 + \zeta_1  -2\xi)\Phi_1 + 2(1 + 3\gamma - 2\beta + \zeta_2 + 
\xi)\Phi_2\nonumber\\
&&\ \ \ \ + 2(1 + \zeta_3)\Phi_3 + 2(3\gamma + 3\zeta_4 - 2\xi)\Phi_4 - 
(\zeta_1 - 2\xi)
 {A}- 2\xi \Phi_W\,,
\end{eqnarray}
\end{subequations}
where     a number of post-Newtonian potentials  are introduced, which are 
defined as 
Poisson-like integrals over the source matter distribution. In particular, they 
are given by the Newtonian potentials at the second velocity order
\begin{equation}
U(t,\vec{x}) = \int\frac{\rho(t,\vec{x}')}{|\vec{x} - \vec{x}'|}d^3x'\,, \quad
U_{ij}(t,\vec{x}) = \int\frac{\rho(t,\vec{x}')(x_i - x_i')(x_j - 
x_j')}{|\vec{x} 
- \vec{x}'|^3}d^3x'\,,
\end{equation}
the third-order vector potentials
\begin{equation}
V_i(t,\vec{x}) = \int\frac{\rho(t,\vec{x}')v_i(t,\vec{x}')}{|\vec{x} - 
\vec{x}'|}d^3x'\,,\quad
W_i(t,\vec{x}) = \int\frac{\rho(t,\vec{x}')v_j(t,\vec{x}')(x_i - x_i')(x_j - 
x_j')}{|\vec{x} - \vec{x}'|^3}d^3x'\,,
\end{equation}
as well as the fourth-order scalar potentials
\begin{gather}
\Phi_1(t,\vec{x}) = \int\frac{\rho(t,\vec{x}')v^2(t,\vec{x}')}{|\vec{x} - 
\vec{x}'|}d^3x'\,,\quad
\Phi_2(t,\vec{x}) = \int\frac{\rho(t,\vec{x}')U(t,\vec{x}')}{|\vec{x} - 
\vec{x}'|}d^3x'\,,\nonumber\\
\Phi_3(t,\vec{x}) = \int\frac{\rho(t,\vec{x}')\Pi(t,\vec{x}')}{|\vec{x} - 
\vec{x}'|}d^3x'\,,\quad
\Phi_4(t,\vec{x}) = \int\frac{p(t,\vec{x}')}{|\vec{x} - 
\vec{x}'|}d^3x'\,,\nonumber\\
 {A}(t,\vec{x}) = \int\frac{\rho(t,\vec{x}')\left[v_i(t,\vec{x}')(x_i - 
x_i')\right]^2}{|\vec{x} - \vec{x}'|^3}d^3x'\,,\quad
 {B}(t,\vec{x}) = \int\frac{\rho(t,\vec{x}')}{|\vec{x} - \vec{x}'|}(x_i - 
x_i')\frac{dv_i(t,\vec{x}')}{dt}d^3x'\,.\nonumber\\
\Phi_W(t,\vec{x}) = \int\rho(t,\vec{x}')\rho(t,\vec{x}'')\frac{x_i - 
x_i'}{|\vec{x} - \vec{x}'|^3}\left(\frac{x_i' - x_i''}{|\vec{x} - \vec{x}''|} - 
\frac{x_i - x_i''}{|\vec{x}' - \vec{x}''|}\right)d^3x'd^3x''\,.
\end{gather}
Here we have also included the PPN potentials \(U_{ij}\) and \( {B}\), which do 
not appear in the standard PPN metric~\eqref{eq:ppnmetric}, but must be 
included 
to describe theories with broken diffeomorphism invariance, as mentioned in 
Section~\ref{sssec:nodiffeo}. The coefficients \(\gamma, \beta, \alpha_1, 
\alpha_2, \alpha_3, \zeta_1, \zeta_2, \zeta_3, \zeta_4, \xi\) are the PPN 
parameters, which characterise the particular gravity theory under 
consideration, and which are determined by solving the gravitational field 
equations using the generic metric~\eqref{eq:ppnmetric}. They are usually 
assumed to be constant, but this assumption may be relaxed, as discussed in 
Section~\ref{ssec:extmod}.

\section{Comparison to Observations}

The specific form of the PPN parameters in the metric~\eqref{eq:ppnmetric} is 
chosen such that they can be given a physical interpretation, which is closely 
linked to the observational properties of the gravity theory under 
consideration:
\begin{enumerate}
\item
\(\gamma\) measures the amount of spatial curvature produced by unit rest mass.
\item
\(\beta\) measures the non-linearity in the gravitational superposition law.
\item
\(\alpha_1, \alpha_2, \alpha_3\) measure the violation of local Lorentz  
\label{loclinref5}
invariance, i.e., the presence of preferred-frame effects.
\item
\(\alpha_3, \zeta_1, \zeta_2, \zeta_3, \zeta_4\) measure the violation of total 
energy-momentum conservation.
\item
\(\xi\) measures the violating of local position invariance, i.e., the presence 
of preferred-location effects.
\end{enumerate}
Further, they are normalised such that for General Relativity they take the 
values \(\beta = \gamma = 1\), while all other parameters vanish due to the 
absence of the corresponding effects. Finally, they are chosen such that they 
are linked to observations either in the Solar System or in other compact 
gravitating systems, such as stellar or neutron star binaries. Numerous bounds 
on the parameters have been obtained from such observations; see 
Table~\ref{tab:bounds} for an overview.

\begin{table}[ht]
\centering
\begin{tabular}{|c|c|c|c|}
\hline
Par. & Bound & Effects & Experiment\\
\hline
${2}^{*}\gamma - 1$ 
& \(2.3 \cdot 10^{-5}\) & time delay & Cassini 
tracking~\cite{Bertotti:2003rm}\\
& \(2 \cdot 10^{-4}\) & light deflection & VLBI~\cite{Fomalont:2009zg}\\
\(\beta - 1\) & \(8 \cdot 10^{-5}\) & perihelion shift & Solar System 
ephemeris~\cite{Verma:2013ata,Viswanathan:2017inp,Viswanathan:2017vob}\\
\(\xi\) & \(4 \cdot 10^{-9}\) & spin precession & millisecond pulsars\\
$2^* \alpha_1$ & \(10^{-4}\) & $ 2^{*}${orbital 
polarisation} & lunar laser ranging\\
& \(4 \cdot 10^{-5}\) & & PSR J1738+0333\\
\(\alpha_2\) & \(2 \cdot 10^{-9}\) & spin precession & millisecond pulsars\\
\(\alpha_3\) & \(4 \cdot 10^{-20}\) & self-acceleration & pulsar spin-down 
statistics\\
\(\eta_N\) &  \(9 \cdot 10^{-4}\) & Nordtvedt effect & lunar laser ranging\\
\(\zeta_1\) & \(0.02\) & combined PPN bounds & —\\
\(\zeta_2\) & \(4 \cdot 10^{-5}\) & binary pulsar acceleration & PSR 1913+16\\
\(\zeta_3\) & \(10^{-8}\) & Newton's 3rd law & lunar acceleration\\
\(\zeta_4\) & \(0.006\) & — & Kreuzer 
experiment~\cite{Kreuzer:1968zz}\footnote{Assuming \(6\zeta_4 = 3\alpha_3 + 
2\zeta_1 - 3\zeta_3\)~\cite{Will:1976zza}.}\\
\hline
\end{tabular}
\caption{Current bounds on the PPN parameters according to~\cite{Will:2014kxa}, 
unless otherwise indicated.}
\label{tab:bounds}
\end{table}

In the table we have  also included the Nordtvedt 
parameter~\cite{Nordtvedt:1968qr,Nordtvedt:1968qs}
\begin{equation}\label{eq:nvparam}
\eta_N = 4\beta - \gamma - 3 - \frac{10}{3}\xi - \alpha_1 + \frac{2}{3}\alpha_2 
- \frac{2}{3}\zeta_1 - \frac{1}{3}\zeta_2\,,
\end{equation}
since this particular combination is closely related to the Nordtvedt effect, 
which can be tested separately by  using lunar laser ranging.

\section{Extensions and Modifications}\label{ssec:extmod}

Although the standard PPN formalism as described in 
section~\ref{ssec:ppnformalism} is suitable to describe the post-Newtonian 
limits of a large class of gravity theories, it still allows for numerous 
extensions to include further physical effects or adapt it to gravity theories 
which are not covered by the standard PPN formalism. In the following, we list 
a few such extensions and modifications.

\subsection{Invariant density formulation}

In a more recent adaptation of the PPN formalism~\cite{Will:2018bme}, the 
density variable \(\rho\) appearing in the definition of the post-Newtonian 
potentials is replaced by a conserved quantity \(\rho^* = \rho\sqrt{-g}u^0\), 
which satisfies the continuity equation
\begin{equation}\label{eq:continuity}
\partial_0\rho^* + \partial_i(\rho^*v_i) = 0\,.
\end{equation}
The advantage of this modification is that it allows for simpler transformation 
rules between potentials, since the continuity equation~\eqref{eq:continuity} 
together with integration by parts yields
\begin{equation}
\partial_0\int\rho^*(t, \vec{x})f(t, \vec{x})d^3x = \int\rho^*(t, 
\vec{x})\frac{df}{dt}(t, \vec{x})d^3x
\end{equation}
for any function \(f\). Note, however, that by using this modified definition, 
several of the formulas mentioned in Section~\ref{ssec:ppnformalism} 
change~\cite{Will:2018bme}.

\subsection{Broken Diffeomorphism Invariance}\label{sssec:nodiffeo}

In order to achieve the standard form~\eqref{eq:ppnmetric} of the 
post-Newtonian 
metric, certain PPN potentials, which would otherwise appear in the most 
general 
post-Newtonian metric, have been eliminated by applying a specific coordinate 
transformation. This is possible only if the theory under investigation is 
invariant under the full diffeomorphism group, and does not exhibit any 
preferred coordinate system. If diffeomorphism invariance is (partially) 
broken, 
additional potentials, together with corresponding PPN parameters, remain in 
the 
PPN metric. This is the case, for example, in Ho\v{r}ava-Lifshitz  
\label{Lifshitzref2}
gravity~\cite{Horava:2009uw}, where the broken invariance under changes of the 
time coordinate leads to the presence of the PPN potential \( {B}\) in 
the fourth order metric component \(\overset{\scriptscriptstyle 4}{h}_{00}\), 
together with a PPN parameter 
\(\zeta_B\)~\cite{Lin:2012bs,Lin:2012ea,Lin:2013tua}. Similarly, breaking of 
spatial diffeomorphism invariance would require   introducing a term 
proportional 
to the potential \(U_{ij}\) in the second-order metric component 
\(\overset{\scriptscriptstyle 2}{h}_{ij}\).

\subsection{Yukawa-type Couplings}\label{sssec:yukawa}

For the standard PPN formalism it is assumed that the metric perturbations, as 
well as perturbations of other relevant fields, appear in the gravitational 
field equations in terms of second total derivative order, i.e., either as 
second-order derivatives of the fields or as products of two first-order 
derivatives. In this case the field equations are of Poisson type and can be 
solved by the PPN potentials introduced in Section~\ref{ssec:ppnformalism}. 
This corresponds to the physical interpretation that these fields are massless. 
However, if mass terms are present in the field equations, which carry no 
derivatives, one must introduce Yukawa-type potentials in order to solve the 
field equations. Such an extension of the PPN formalism has been developed in 
the context of scalar-tensor gravity in~\cite{Zaglauer:1990yh,Helbig:1991pk}.

\subsection{Higher Derivative Orders}

The opposite situation, compared to the previous case, is the appearance of 
terms in the perturbative field equations with higher than second derivative 
order, i.e., either higher than second-order derivatives acting on the 
perturbations, or products of terms such that their total number of derivatives 
exceeds two. Such terms require the introduction of additional PPN potentials 
that contain derivatives of the matter variables. PPN potentials of this form 
have been introduced in the context of Poincaré gauge theory 
in~\cite{Gladchenko:1990nw}.

\subsection{Parity-violating Terms}
\label{parityef1}

As discussed in Section~\ref{ssec:ppnformalism}, the PPN potentials are defined 
as solutions to a set of Poisson-like equations, and as such are of even 
parity. 
In order to solve the field equations on theories involving parity-violating 
terms, additional, parity-odd potentials must be introduced. An example is 
given 
by Chern-Simons gravity, which features the parity-violating term \(\ast RR = 
\frac{1}{2}\epsilon^{\mu\nu\rho\sigma}R^{\alpha}{}_{\beta\mu\nu}R^{\beta}{}_{
\alpha\rho\sigma}\), which requires the introduction of a new potential 
\(\epsilon^{ijk}\partial_jV_k\) in the metric component 
\(\overset{\scriptscriptstyle 3}{h}_{0i}\), together with a corresponding PPN 
parameter~\cite{Alexander:2007zg,Alexander:2007vt,Alexander:2009tp}.

\subsection{Screening Mechanisms}
\label{screeningref2}

An important assumption of the PPN formalism is the validity of the linear and 
quadratic order perturbation theory around the asymptotically flat Minkowski 
background. While this assumption holds for most theories, there are also 
counter-examples, such as Horndeski or bimetric gravity, in which strong 
coupling effects are present and lead to a screening on Solar System 
scales~\cite{Vainshtein:1972sx,Babichev:2013usa}. Such effects may be included 
into the PPN formalism by introducing the characteristic scale of the screening 
effect, known as the Vainshtein radius, as another perturbation 
parameter~\cite{Avilez-Lopez:2015dja}.

A different kind of screening is present in the so-called chameleon theory. 
\label{chameleonkiref5}  In 
this case the scalar field is suppressed inside matter, and so only a thin 
shell, forming the outer layer of the source mass, contributes to the 
post-Newtonian scalar 
field~\cite{Khoury:2003aq,Hees:2011mu,Scharer:2014kya,Burrage:2016bwy,
Burrage:2017qrf}. General screening effects are discussed 
in~\cite{McManus:2017itv}.

\subsection{Cosmological Background Evolution}

In the standard PPN formalism it is assumed that the background around which 
the perturbation is performed is given by a Minkowski spacetime, and thus in 
particular stationary, i.e., invariant under time translation. As a 
consequence, the values of the post-Newtonian parameters are constant in time. 
In~\cite{Sanghai:2016tbi} this assumption has been dropped and an evolving 
cosmological background assumed, which leads to a possible time dependence of 
the PPN parameters.

\subsection{Multiple Metrics}

Another assumption of the standard PPN formalism, whose footing is the validity 
of the Einstein equivalence principle,  \label{equivprinref5} is the existence 
of a single, dynamical 
metric, which governs the behaviour of physical systems within a gravitational 
field. Note that this assumption does not exclude theories with an additional, 
non-dynamical metric, such as Rosen's theory~\cite{Rosen:1974ua}, whose PPN 
parameters agree with General Relativity except for 
\(\alpha_2\)~\cite{Lee:1975kc}, or massive 
gravity~\cite{Hinterbichler:2011tt,deRham:2014zqa}, provided that one allows 
for Yukawa-type terms.

Including another, dynamical metric, however, requires an extension of the 
standard PPN formalism, which features additional PPN potentials and in which 
both metrics are treated 
perturbatively~\cite{Clifton:2010hz,Hohmann:2010ni,Hohmann:2013oca}.

\subsection{Tetrad Formulation}

The aforementioned approaches have a common feature that the fundamental field 
which carries the gravitational interaction, and which is thus expanded in the 
post-Newtonian perturbation orders, is the metric (or one of several metrics). 
However, there are also gravity theories in which one considers a tetrad as the 
fundamental field variable, so that the metric becomes a dependent quantity. 
For such theories one must therefore perform a perturbative expansion of the 
tetrad. Such a post-Newtonian expansion has been developed in the context of 
Poincaré gauge theory in~\cite{Smalley:1980em}, and in the case of translation 
gauge theory in~\cite{Nitsch:1979qn}. An adaptation for a scalar-tetrad theory 
of gravity has been developed in~\cite{Hayward:1981bk}. For the covariant 
formulation of teleparallel gravity, it has been developed 
in~\cite{Ualikhanova:2019ygl}.

\subsection{Gauge-invariant Approach}

An important assumption of the PPN formalism is the existence of a particular 
gauge, or choice of coordinates, in which the post-Newtonian metric takes the 
standard form~\eqref{eq:ppnmetric}. Determining the post-Newtonian limit of a 
given theory of gravity conventionally involves solving the perturbative field 
equations either in this particular gauge, or choosing a gauge in which the 
equations become possibly simpler, and then transforming the resulting solution 
into the standard PPN gauge. Either method may turn out to be cumbersome, since 
in general there is no canonical gauge choice that simplifies the field 
equations, and even if such a gauge is found, the equations may still exhibit a 
highly non-trivial coupling between the tensor components of the metric and 
other fields. In cosmology, such difficulties are overcome by making use of a 
gauge-invariant 
approach~\cite{Bardeen:1980kt,Kodama:1985bj,Mukhanov:1990me,Malik:2008im}. 
Using the same mathematical footing, which requires the use of second-order 
perturbations~\cite{Nakamura:2004rm,Nakamura:2006rk}, a similar approach to the 
PPN formalism has been developed~\cite{Hohmann:2019qgo}.

\section{Post-Newtonian Limit of Particular Theories}

The PPN formalism has been applied to a vast number of gravitational theories, 
some of which we mentioned in Section~\ref{ssec:extmod} in the context of 
extensions of the formalism. Here we list further results on the post-Newtonian 
limit for selected classes of gravity theories, which belong to the most 
relevant and most actively studied modified gravity theories.

\subsection{Scalar-tensor and $f(R)$ Theories}
\label{sssec:stg}
\label{fRref6}

An important class of gravity theories is described by the general 
scalar-tensor action~\cite{Flanagan:2004bz}
\begin{equation}\label{eq:stgactiong}
S = \frac{1}{2\kappa^2}\int\left[\mathcal{A}(\phi)R - 
\mathcal{B}(\phi)g^{\mu\nu}\partial_{\mu}\phi\partial_{\nu}\phi - 
2\kappa^2\mathcal{V}(\phi)\right]\sqrt{-g}\,d^4x + S_m[\chi, 
e^{2\alpha(\phi)}g_{\mu\nu}]\,.
\end{equation}
Here,
\(\chi\) collectively denotes any set of matter fields, while 
\(\mathcal{A}, \mathcal{B}, \mathcal{V}, \alpha\) are free functions of the 
scalar field. This class of theories has the property that by a redefinition of 
the scalar field and a conformal transformation of the metric, equivalent 
theories can be related to each other, which are described by different choices 
of the free functions in the action. This allows us to represent such an 
equivalence class of theories in different conformal frames and 
representations. Making use of this freedom, the PPN parameters for different 
subclasses of these theories have been calculated, mostly choosing to work in 
the Jordan frame  \pageref{Jordanrref4} \(\alpha(\phi) \equiv 0\), using the 
scalar field 
parameterisation \(\mathcal{A}(\phi) \equiv \phi\) and redefining the kinetic 
term as \(\mathcal{B}(\phi) \equiv \omega(\phi)/\phi\). The only non-trivial 
parameters for this case are \(\gamma\) and \(\beta\):
\begin{enumerate}
\item
For an arbitrary function \(\omega\), but vanishing potential \(\mathcal{V}\), 
the PPN parameters are given by
\begin{equation}
\gamma = \frac{\omega_0 + 1}{\omega_0 + 2}\,, \quad
\beta = 1 + \frac{\omega_1\Phi}{(2\omega_0 + 3)(2\omega_0 + 4)^2}\,,
\end{equation}
where \(\omega_0 = \omega(\Phi)\), \(\omega_1 = \omega'(\Phi)\) and \(\Phi\) is 
the cosmological background value of the scalar field~\cite{Nordtvedt:1970uv}. 
Light propagation in the Solar System at the post-Newtonian level has been 
studied for this class of theories in~\cite{Minazzoli:2010pr,Deng:2012rx}.
\item
In~\cite{Olmo:2005zr,Olmo:2005hc,Perivolaropoulos:2009ak} the PPN parameter 
\(\gamma\) for constant \(\omega\) and non-vanishing potential \(\mathcal{V}\) 
has been derived.
\item
For general free functions \(\omega\) and \(\mathcal{V}\), the post-Newtonian 
field equations have been derived in~\cite{Xie:2007gq}. The PPN parameters 
\(\beta\) and \(\gamma\) for this class of theories have been calculated under 
the assumption of a static point mass source in the Jordan frame \(\alpha(\phi) 
\equiv 0\)~\cite{Hohmann:2013rba} as well as the Einstein frame  
\label{Einsteinfrref8}
\(\mathcal{A}(\phi) \equiv 1\)~\cite{Scharer:2014kya}. A frame-independent 
calculation in terms of conformal invariants is shown in~\cite{Jarv:2014hma}. 
The result has been extended to a homogeneous sphere replacing the point mass 
source~\cite{Hohmann:2017qje}.
\item
Non-standard kinetic terms and matter couplings of the scalar field have been 
considered in~\cite{Moffat:2010ek,Saaidi:2011zza,Minazzoli:2012ym,Devi:2011zz} 
for the calculation of \(\gamma\).
\item
Screening mechanisms in the context of the post-Newtonian approximation of 
scalar-tensor theories have been discussed 
in~\cite{Roshan:2011kz,Minazzoli:2013ara}.
\end{enumerate}
The obtained results have also been applied to obtain the post-Newtonian limit 
of \(f(R)\) type gravity theories, through their dynamical equivalence with 
scalar-tensor 
gravity~\cite{Capozziello:2005bu,Capozziello:2006jj,Capozziello:2007ms,
Capozziello:2010wt,Clifton:2008jq,Capone:2009xk,Capozziello:2010gu,Harko:2011nh}
.

Care must be taken when the resulting PPN parameters for the case of a massive 
scalar field and a point mass source are to be interpreted physically, since the 
former leads to non-constant values of the PPN parameters, which contrasts the 
standard assumption of constant PPN 
parameters~\cite{Hohmann:2013rba,Scharer:2014kya}, while the latter is only a 
very coarse approximation of the Sun, neglecting its gravitational 
self-energy~\cite{Hohmann:2017qje}. This influences, for example, the derivation 
of the light deflection by the S2un, so that in general a full calculation of 
light trajectories in the post-Newtonian solution of the spacetime geometry must 
be performed~\cite{Deng:2016moh,Zhang:2016njn}.

\subsection{Multi-scalar-tensor Theories}\label{sssec:mstg}
\label{Scalartensoref5}

A straightforward generalisation of the scalar-tensor theories discussed above 
is to include multiple scalar fields~\cite{Damour:1992we,Berkin:1993bt}. For 
the case of a vanishing potential, the PPN parameters have been obtained in the 
Einstein frame in~\cite{Damour:1992we} and in the Jordan frame for a diagonal 
kinetic term in~\cite{Berkin:1993bt}, and a general kinetic term 
in~\cite{Randla:2014saa}. For an arbitrary frame, they have been expressed in 
terms of invariant quantities in~\cite{Kuusk:2015dda}. For the general case 
including a potential, the PPN parameter \(\gamma\) has been calculated 
in~\cite{Hohmann:2016yfd}.

Various gravity theories can be reformulated as particular multi-scalar-tensor 
gravity theories, and also for such classes of theories the post-Newtonian limit 
has been considered: for C-theory with a vanishing 
potential~\cite{Koivisto:2011tp}, as well as for non-local gravity both in the 
biscalar representation~\cite{Koivisto:2008dh} and independent 
thereof~\cite{Conroy:2014eja}.

\subsection{Horndeski Gravity}\label{ssec:horndeski}
\label{Horndeskiref4}

Another generalisation of the scalar-tensor action~\eqref{eq:stgactiong} is 
given by the Horndeski class of gravity 
theories~\cite{Horndeski:1974wa,Deffayet:2011gz,Kobayashi:2011nu}. Their 
post-Newtonian limit has been derived in~\cite{Hohmann:2015kra} for the case of 
negligible Vainshtein screening. An alternative approach using an effective 
energy-momentum tensor including gravitational waves is presented 
in~\cite{Hou:2017cjy}. In general, however, also screening effects must also  
be taken into account~\cite{Kase:2013uja}.

\subsection{Bimetric and Multimetric Gravity}
\label{bigravityrefs2}

Since their development about a decade ago, theories of massive 
gravity~\cite{deRham:2010kj} and bimetric 
gravity~\cite{Hassan:2011zd,Hinterbichler:2012cn}, being the unique ghost-free 
theories of thes types, have received growing attention; 
see~\cite{Hinterbichler:2011tt,deRham:2014zqa,Schmidt-May:2015vnx} for a number 
of reviews. One interesting aspect of these theories is the possibility to 
include a dark matter sector that is only coupled gravitationally to 
ordinary matter. The post-Newtonian parameter \(\gamma\) is calculated under 
this assumption, and the deflection of light by visible and dark matter sources 
is obtained in~\cite{Hohmann:2017uxe}.

\subsection{Teleparallel Gravity}
\label{telegraref2}\label{fTgraref3}

Another large class of theories, which has gained growing interest during the 
last decade, is based on the framework of teleparallel 
gravity~\cite{Einstein:1928,Aldrovandi:2013wha}. The most simple contender 
theory in this class is the Teleparallel Equivalent of General 
Relativity~\cite{Maluf:2013gaa}, which constitutes one of the three physically 
equivalent, but geometrically distinct formulations of General 
Relativity~\cite{BeltranJimenez:2019tjy}. Due to this equivalence, its PPN 
parameters agree with that of General Relativity.

Modifications of TEGR, however, differ also in their phenomenology from 
analogous modifications of the standard Einstein-Hilbert formulation of General 
Relativity in terms of curvature. A prominent example is given by 
$f(\mathbb{T})$
theories~\cite{Bengochea:2008gz,Linder:2010py}, whose PPN parameters are 
identical to those of General Relativity~\cite{Ualikhanova:2019ygl}, while 
\(f(R)\) type modifications lead to a deviation of the PPN parameters, as 
discussed in Section~\ref{sssec:stg}.

Another class of modified teleparallel gravity theories is based on New General 
Relativity~\cite{Hayashi:1979qx}. Using the Eddington-Robertson-Schiff 
formalism, it has been found that their PPN parameters \(\beta\) and \(\gamma\) 
are of the form \(\beta = 1 - \epsilon/2\) and \(\gamma = 1 - 2\epsilon\), with 
a theory-dependent constant parameter \(\epsilon\). This result has later been 
generalised to a larger class of theories~\cite{Bahamonde:2017wwk}, which is 
based on a covariant formulation of teleparallel 
gravity~\cite{Krssak:2015oua,Krssak:2018ywd}, and using a tetrad-based framework 
to derive the full set of PPN parameters~\cite{Ualikhanova:2019ygl}. The result 
has shown that also in this larger class of theories the PPN parameters 
\(\beta\) and \(\gamma\) are of the same form as in New General Relativity, 
while all other PPN parameters vanish. An interesting consequence of this result 
is the fact that the Nordvedt parameter~\eqref{eq:nvparam} vanishes identically, 
even for those theories whose individual PPN parameters differ from that of 
General Relativity. This means that such theories lead to different predictions, 
e.g., for the deflection of light or the precession of the Mercury orbit, but 
are indistinguishable by lunar laser ranging experiments.

In analogy to scalar-tensor theories of gravity (which could also more precisely 
be called scalar-curvature theories), as described in Section~\ref{sssec:stg}, 
also a large class of scalar extensions to teleparallel gravity has also been 
studied, which are summarised under the term scalar-torsion 
gravity~\cite{Hohmann:2018rwf,Hohmann:2018vle,Hohmann:2018dqh,Hohmann:2018ijr}. 
In the most simple cases, in which the scalar field is minimally coupled to 
torsion, it transpired that the PPN parameters are identical to those of 
General Relativity~\cite{Li:2013oef,Chen:2014qsa}. However, this degeneracy is 
broken if   a coupling between the scalar field and the teleparallel 
boundary term  is considered \cite{Sadjadi:2016kwj}. Equivalently, one could 
study a kinetic coupling of the scalar field to the vector part of the torsion, 
and also in this case an effect on the PPN parameters \(\beta\) and \(\gamma\) 
is found for both massive and massless scalar 
fields~\cite{Emtsova:2019qsl,Flathmann:2019khc}.






%








\chapter[Gravitational Waves]{Gravitational Waves}
\label{Sakellariadouchapter}

{\em Mairi Sakellariadou}\\




The direct detection of gravitational-waves  \label{gravitationalwavrefs2} from 
the LIGO and Virgo  \label{eLIGOfref4}
collaborations 
\cite{Abbott:2016blz,TheLIGOScientific:2017qsa,LIGOScientific:2018mvr} opened a 
new window into the Universe, offering a novel and powerful way to test not only 
astrophysical models about compact objects \label{compactobrefs3} 
\cite{TheLIGOScientific:2016htt}, but 
also cosmological models 
\cite{Caldwell:2019vru,Sathyaprakash:2019nnu,Auclair:2019wcv}, particle physics 
beyond the Standard Model 
\cite{Ringeval:2017eww,Abbott:2017mem,Jenkins:2018lvb,Brdar:2018num,
Auclair:2019wcv,Caprini:2019egz}, General Relativity 
\cite{TheLIGOScientific:2016src,Callister:2017ocg,Abbott:2018lct,
LIGOScientific:2019fpa}, modified gravity \cite{Berti:2019xgr,Belgacem:2019pkk}, 
and even quantum gravity proposals 
\cite{Abbott:2018lct,Calcagni:2019kzo,Calcagni:2019ngc}.

\section{Tests of General Relativity}

Einstein's theory of General Relativity (GR) allows for the existence of only 
two gravitational-wave polarisations (the tensor plus and cross modes), and 
predicts a massless graviton propagating at the speed of light. On the contrary, 
any general metric theory of gravity may allow for up to four additional 
polarisations: two vector modes (helicity $\pm 1$), and two scalar (helicity 
$0$) --  the breathing and longitudinal -- modes. The detection of any such 
additional polarisation modes would imply violation of General Relativity, while 
a non-detection may constrain any extended theory of gravity.
\newline
Interferometers with different orientations will respond differently to GW 
signals emitted from a given sky location as a function of their polarisation. 
When allowing for all six polarisation modes, a network of at least five 
detectors is required in order to uniquely determine the polarisation of 
transient gravitational-wave signals, like the ones detected from compact binary 
coalescences (CBC). This is because our detectors are only sensitive to the 
traceless scalar mode, and hence we expect to be able to distinguish only five 
polarisations.

In addition to the  gravitational-waves from CBC, one expects an astrophysical 
stochastic gravitational-wave background (SGWB) formed by the superposition of 
many weak or distant, independent and unresolved sources. 
 The isotropic SGWB can be described in terms of the energy density per 
logarithmic frequency interval $f$ as
\begin{eqnarray}
\Omega_{\rm GW}={f\over \rho_{\rm c}} {{\rm d}\rho_{\rm GW}\over {\rm d}f}~,
\end{eqnarray}
where ${\rm d}\rho_{\rm GW}$ stands for the energy density in GW in the 
frequency interval between $f$ and $f+{\rm d}f$, and $\rho_{\rm c}$ denotes the 
critical energy density.
The SGWB from compact binaries is well approximated by the power law
\begin{eqnarray}
\Omega_{\rm GW}=\Omega_{\rm ref}\left( {f\over f_{\rm ref} }   \right)^{2/3}~,
\end{eqnarray}
where $\Omega_{\rm ref}$ denotes the background's amplitude at a reference 
frequency $f_{\rm ref}$, usually chosen to be equal to 25 Hz.
 
Considering a SGWB of astrophysical origin, in \cite{Callister:2017ocg} a 
Bayesian approach allowing for full parameter estimation on the SGWB, assuming a 
power-law model for the GW energy density in each polarisation mode, was 
proposed.
This method can be applied not only within the context of simple cases of purely 
tensor, vector or scalar-polarised backgrounds, but also in the situations 
appearing in alternative theories of gravity that predict a mixed background of 
multiple polarisation modes.

Utilising this Bayesian method, it was demonstrated  \cite{Callister:2017ocg} 
that one may detect and identify, with the existing detectors, an SGWB 
containing any combination of GW polarisation modes. In particular, it was shown 
that after three years of observation at design sensitivity, Advanced LIGO will 
be able to impose upper bounds to the amplitudes of tensor, vector, and scalar 
polarisation to respectively, $\Omega_{\rm ref}^{\rm T} < 1.6 \times 10^{-10}, 
\Omega_{\rm ref}^{\rm V} < 2.0 \times 10^{-10}$ and $\Omega_{\rm ref}^{\rm S} < 
5.0 \times 10^{-10}$, at $95\%$ credibility \cite{Callister:2017ocg}.

Let us use the currently available compact binary signals observed by Advanced 
LIGO and Advanced Virgo during the first two observing runs of the advanced 
detector era (O1 and O2), assuming 
either a log uniform or a uniform prior on the reference GW amplitude 
$\Omega_{\rm ref}$ for each polarisation mode, and considering the presence of 
  tensor, vector and scalar backgrounds. Applying the Bayesian method of 
\cite{Callister:2017ocg} and marginalising over the spectral indices and the 
amplitudes for the three different polarisation modes, we obtain 
\cite{LIGOScientific:2019vic} the following upper limits on tensor, vector and 
scalar polarisations, respectively:
\begin{eqnarray} 
\Omega_{\rm ref}^{\rm T} < 8.2 \times 10^{-8} ,  \Omega_{\rm ref}^{\rm V} < 1.2 
\times 10^{-7} , \Omega_{\rm ref}^{\rm S} < 4.2 \times 10^{-7},
\nonumber
\end{eqnarray}  
for uniform prior, and 
\begin{eqnarray}
\Omega_{\rm ref}^{\rm T} < 3.2 \times 10^{-8}, \Omega_{\rm ref}^{\rm V} < 2.9 
\times 10^{-8}, \Omega_{\rm ref}^{\rm S} < 6.1 \times 10^{-8},
\nonumber
\end{eqnarray}
for log-uniform.

For compact binary coalescences we can also investigate deviations from GR by 
introducing separate modifications to the emitted gravitational-wave waveform 
and its propagation. The former corresponds to GR modifications in the 
strong-field region close to the compact binary, and the latter to the 
weak-field region away from the emitting source. For instance, deviations from 
GR may introduce modifications of the dynamics of compact binaries, hence 
leading to modifications of the orbital phase which consequently imply shifts in 
the GW phase coefficients as a function of the intrinsic parameters of the 
compact binaries.

Introducing a phenomenological (but well-motivated) parametrised deviation in 
the waveform model for binary black holes, \label{BHref7} including the 
post-Newtonian 
coefficients, it was shown using
data from the catalogue \cite{LIGOScientific:2018mvr}, as detected by the 
Advanced LIGO and Advanced Virgo during the first two runs of these detectors,
that these deviations are consistent with their GR value of zero 
\cite{LIGOScientific:2019fpa}.

Introducing a phenomenological modification of the GW dispersion relation
\begin{eqnarray}
E^2=p^2c^2+A_\alpha p^\alpha c^\alpha,
\end{eqnarray}
in terms of the phenomenological parameters $A_\alpha, \alpha$, where $E, p, c$ 
are the energy and momentum of GWs and the speed of light, respectively,
one can use GWs data in order to  test GR,\label{testsGRefs1} where the 
additional power-law term 
in momentum is absent. Considering $\alpha$ values from 0 to 4 in steps of 0.5, 
and utilising data from the catalogue \cite{LIGOScientific:2018mvr}, 90$\%$ 
credible upper bound on the absolute value of the parameter $A_\alpha$ and the 
mass of the graviton $m_{\rm g}$ were imposed \cite{LIGOScientific:2019fpa}. In 
particular, it was found that
\begin{eqnarray}
m_{\rm g}\leq 5.0\times 10^{-23} {\rm eV}/c^2,
\nonumber
\end{eqnarray}
at $90\%$ credible limit.

The first detection of a GW signal from a neutron star merger, 
\label{neutronstarsref5} the GW170817 
event, accompanied  by a short-duration gamma-ray burst (SGRB), the GRB 
170817 A, with an observed time delay of $(+1.74 \pm 0.05)$s, offered a powerful 
tool to test General Relativity \cite{Monitor:2017mdv} and constrain alternative 
theories of gravity.

Coupling the standard electromagnetic theory with General Relativity, 
gravitational and electromagnetic waves propagate at the same speed. 
Consequently,
the temporal offset of $(+1.74 \pm 0.05)$s can be used to constrain theories 
that predict a deviation of the speed of gravity from the speed of light, or on 
theories leading to Lorentz violation. Also, it can be used to  test the 
equivalence principle \cite{Monitor:2017mdv}.

Assuming that the SGRB signal was emitted 10s after the GW signal, and knowing 
the travel distance  the temporal offset can be used  to set a conservative 
bound on the fractional speed difference $\Delta v/v_{\rm EM}$, where $\Delta 
v=v_{\rm GW}-v_{\rm EM}$,  with $v_{\rm GW}$ the speed of gravitational  and 
$v_{\rm EM}$ the speed of electromagnetic waves. This conservative constraint on 
the fractional speed difference is \cite{Monitor:2017mdv}
\begin{eqnarray}
-3\times 10^{-15}\leq {\Delta v\over v_{\rm EM}}\leq  7\times 10^{-16}.
\end{eqnarray}

Lorentz symmetry is a cornerstone of both the standard model (SM) of particle 
physics and Einstein's theory of General Relativity. However, particle physics 
beyond the SM or modified gravity may lead to Lorentz symmetry violation. 
\label{loclinref6}
Consequently, an observation of Lorentz violation would be an indicator of new 
physics. 

Let us consider the gravitational Standard-Model Extension (SME) 
\cite{Colladay:1998fq}, an effective field-theory based framework for studying 
Lorentz symmetry. The  Lagrange density of the SME contains the SM and GR,  
along with all possible Lorentz-violating terms constructed by coupling observer 
vector or tensor coefficients for Lorentz violation to SM operators. Each 
Lorentz-violating term is the coordinate-independent product of a coefficient 
for Lorentz violation with a Lorentz-violating operator. Hence, the 
Lorentz-violating process associated with any operator is dictated by the 
corresponding coefficient, implying that any experimental signal for Lorentz 
violation can be expressed in terms of one or more of these coefficients.
The full SME includes an infinite number of SM operators of  increasing mass 
dimension. A limiting case is the minimal SME, a restriction of the SME to 
include only Lorentz-violating operators of mass dimensions of four or less. 

The difference $\Delta v$ in group velocities between EM and GWs, in the minimal 
SME case, reads
\begin{eqnarray}
\Delta v = -\sum_{\ell m, \ell\leq 2} Y_{\ell m}(\hat{n})
\left[{1\over 2}(-1)^{1+\ell}\bar{s}^{(4)}_{\ell m} -c^{(4)}_{(I)\ell m} 
\right],
\end{eqnarray}
where $Y_{\ell m}$ are spherical harmonics, $\hat{n}$ denotes the sky position 
and $\bar{s}^{(4)}_{\ell m}, c^{(4)}_{(I)\ell m}$ are spherical-basis 
coefficients 
for Lorentz violation in the gravitational and electromagnetic sectors, 
respectively.

Using the GW170817 binary neutron star merger, accompanied by gamma-ray burst 
(GRB), namely the GRB 170817A event, constraints have been imposed 
\cite{Monitor:2017mdv} on the gravity sector coefficients $\bar{s}^{(4)}_{\ell 
m}$, by considering such coefficients one at a time, while setting all other 
coefficients, including those from the electromagnetic sector, equal to zero. It 
is worth noting that the isotropic ($\ell =0$) upper bound on 
$\bar{s}^{(4)}_{00}$was improved  by more than 10 orders of magnitude, with the 
current upper bound reaching $5\times 10^{-15}$.

An attempt can also be made  to use GWs followed by EM radiation in order to 
test the equivalence principle through the Shapiro effect. 
\label{equivprinref6}  The Shapiro effect 
predicts the difference between the propagation time of massless particles in 
curved spacetime, with respect to that in a flat background; the former being 
slightly higher. However, as   has been recently pointed out 
\cite{Minazzoli:2019ugi},
the Shapiro delay is not an observable in General Relativity, and is computed by 
comparing it  with a fiducial Euclidean distance. Moreover, the Shapiro delay 
calculated in the usual away, namely by taking the Newtonian potential to vanish 
at infinity, diverges as one incorporates many far-away sources. Thus,  a 
conservative lower bound to the Shapiro delay cannot be obtained  by just 
considering a subset of the sources of the gravitational field.

\section{Modified Gravity}

The Universe is at present in a phase of accelerated expansion, and several 
theoretical, and often phenomenological, proposals have been made in order to 
account for such a late-time evolution. 

Considering the propagation of tensor perturbations in an FLRW universe, in the 
context of GR, we have
\begin{eqnarray}
h''_A  +2 {\cal H}[1-\delta(\eta)] h'_A+k^2h_A=\,\Pi_A,
\end{eqnarray}
where $A$ denotes the two (plus and cross) polarisations and prime stands for conformal time; $\Pi_A$ is is the source term related to the anisotropic stress tensor.

In a modified gravity model the above equation becomes
\begin{eqnarray} \label{gen-ev-eq1}
 h''_A+ 2\left[1-\delta(\eta) \right]{\cal H}\, h'_A+[ c_T^2(\eta)\,{k^2}+m_T^2(\eta) ] h_A\,=\,\Pi_A,
\end{eqnarray} 
where $\delta(\eta), c_T, m_T$ are functions that can in principle be constrained from GW data.

The function $\delta(\eta)$ modifies the friction term in the propagation 
equation and can therefore affect the amplitude of GWs propagating across 
cosmological distances, introducing the so-called {\sl GW luminosity distance}, 
which can be probed by LISA standard sirens. \label{earlyDEefs2} 
\label{LISAref2}

Following a phenomenological approach, let us write \cite{Belgacem:2018lbp} 
\begin{eqnarray}
\Xi(z)\equiv{d_{\rm L}^{\rm GW}(z)\over d_{\rm L}^{\rm EM}(z)}=\Xi_0+{1-\Xi_0\over (1+z)^n},
\end{eqnarray}
where $\Xi_0, n$ denote the two parameters of the phenomenological model.

The value $\Xi_0=1$ corresponds to GR, while for large redshifts $\Xi(z\gg 1)=\Xi_0$, constant. This simple expression smoothly interpolates between these asymptotic values with a power-law determined by the index $n$. 
Following this two-parameter parametrisation approach,  we can first construct 
simulated catalogues of LISA massive black hole binaries with electromagnetic 
counterparts, and consequently use them in order to constrain modified gravity 
models \cite{Belgacem:2019pkk}.

The important ingredients in order to build these mock catalogues are the initial mass function of the massive black hole seeds, and the delays between galaxy and massive black hole mergers. In Ref.\cite{Belgacem:2019pkk} two distinct seeding models were used: a light seed scenario in which the massive black holes evolve from the remnants of population III stars (with remnant masses $\sim 10^2 M_{\rm Sun}$) at high redshifts $z\geq 15$ ; and a heavy seed scenario in which the black hole seeds form from bar instabilities of protogalactic disks, with seed masses $\sim 10^5 M_{\rm Sun}$,  again at $z\geq 15$. 

Regarding delay times, the model used in Ref.\cite{Belgacem:2019pkk} followed the evolution of dark matter halos driven by dynamical friction, from the moment when they first touch to the final halo/galaxy merger, including any  environmental effects.

Investigating LISA forecasts, it was consequently shown that LISA will be able 
to measure the parameter $\Xi_0$ to an accuracy that reaches $1.1\%$, and in the 
worst scenario will still be $4.4\%$ \cite{Belgacem:2019pkk}. 
Since the parameters $\Xi_0, n$ are related to parameters of different modified 
gravity models, by  using data from LISA we will be able to constrain modified 
gravity models at a high level of  accuracy.

\section{Quantum Gravity}

Quantum Gravity (QG) can affect the generation of gravitational waves and modify their propagation. In what follows, we will briefly describe modifications of the propagation of GWs in different QG theories. 

Within the context of brane theories with extra dimensions, we may wonder 
whether there is damping of the waveform due to gravitational leaking into the 
extra dimensions. In such theories there is a high dimensional bulk within which 
there are embedded branes of any dimensionality (not only three) allowed from 
the particular string theory under consideration. Damping of the waveform may be 
expected, since gravitons can escape into the bulk, whereas matter fields are 
trapped on the branes. Therefore, sources of gravitational waves with 
electromagnetic counterparts may provide a test of spacetime dimensionality. 

In the context of GR the luminosity distance $d_{\rm L}$ is related to the amplitude of gravitational waves through
\begin{eqnarray}
 h_{\rm GR}\propto d_{\rm L}^{-1},
 \end{eqnarray} 
 where the electromagnetic luminosity distance reads
\begin{eqnarray}
d_{\rm L}^{\rm EM} \simeq {z(1+z)\over H_0}
\simeq \
{z\over H_0}\ \mbox{for} \ z\ll1.
\end{eqnarray}
Any deviation from the above GR estimation depends on the number of spacetime 
dimensions D; in particular, we may expect a systematic overestimation of the 
electromagnetic luminosity distance inferred from GW data. 

Assuming that light and matter propagate in a four-dimensional spacetime, we 
have
\begin{eqnarray}
h\propto {1\over d_{\rm L}^{\rm GR}}={1\over d_{\rm L}^{\rm EM}}\left [1+\left({d_L^{\rm EM}\over R_c}\right)^n\right]^{-(D-4)/(2n)},
\end{eqnarray}
where $R_c$ and $n$ stand respectively for the distance scale of the screening and the transition steepness, and have different values for different string theory models.

In the above equation $d_{\rm L}^{\rm GR}$ is obtained from the strain measured in a GW interferometer, whereas $d_{\rm L}^{\rm EM}$ is the luminosity distance measured for the optical counterpart of the standard siren. Performing a Bayesian analysis from the
joint posterior probability for $D, d^{\rm GW}_{\rm L}, d^{\rm EM}_{\rm L}$, given the
two statistically independent measurements of electromagnetic and gravitational-wave data from GW170817, constraints were imposed on the number of spacetime dimensions \cite{Abbott:2018lct}. The results of the analysis have shown consistency with the GR prediction of $D = 4$.
 
We may then wonder whether sources of gravitational waves with an 
electromagnetic counterpart could also be used to constrain non-perturbative 
approaches to quantum gravity. Most quantum gravity 
\label{quantumgrefs2}theories 
exhibit a 
long-range non-perturbative mechanism, known as dimensional flow, with 
important consequences for the propagation of GWs over cosmological distances. 
Quantisation of spacetime geometry leads to an anomalous spacetime measure 
$\delta\rho(x)$ and a kinetic operator ${\cal K}(\vartheta)$; the former 
represents how volumes scale and the latter encodes modified dispersion 
relations.

Consider a spin-2 perturbation \label{tensorpertref2} $h_{\mu\nu}$, described 
by 
two polarisations, over a background metric 
$g_{\mu\nu}^{(0)}=g_{\mu\nu}-h_{\mu\nu}$. The perturbed action reads 
\cite{Calcagni:2019kzo,Calcagni:2019ngc}
\begin{eqnarray}
S=\frac{1}{2\ell_*^{2\Gamma}}\!\!\int\!
d
\varrho\sqrt{-g^{(0)}}
\left[h_{\mu\nu}{\cal K} h^{\mu\nu}\!+{\cal O}(h_{\mu\nu}^2)+{\cal J}^{\mu\nu}h_{\mu\nu}\right],
\end{eqnarray}
where ${\cal J} $ denotes a source and $\ell_\star, \Gamma$ define respectively a characteristic scale of geometry and a scaling parameter.
Note that there are two limits of $\Gamma$; the $\Gamma_{\rm UV}$ corresponding 
for QG corrections, which are important, and the $\Gamma_{\rm meso}$ 
corresponding to contributions to GR, which are small but cannot be neglected.

Specifying a spacetime measure $\rho$, a kinetic operator ${\cal K}$, and a source ${\cal J}$, the strain of the emitted gravitational waves is determined by the convolution $h \propto \int d\rho {\cal J} G$ of the source with the retarded Green's function. 

In radial coordinates and in the local wave zone, the amplitude of GWs becomes 
the product of a dimensionless function $f_h$ that depends on the source, and a 
power-law distance behaviour  \cite{Calcagni:2019kzo,Calcagni:2019ngc}
\begin{eqnarray}
h(t,r)\sim f_h(t,r)\,\left({\ell_*}/{r}\right)^{\Gamma},\qquad [f_h]=0\,. 
\end{eqnarray}

For GWs propagating over cosmological distances, the relation between the 
luminosity distance, as measured in a GW interferometer through the strain and 
the luminosity distance measured for the counterpart of the standard siren, is 
\cite{Calcagni:2019kzo,Calcagni:2019ngc}
\begin{eqnarray}
h \propto \frac{1}{d_L^\textsc{gw}}\,,\qquad 
\frac{d_L^\textsc{gw}}
{d_L^\textsc{em}}=1+\varepsilon
\left(\frac{d_L^\textsc{em}}{\ell_*}\right)^{\g-1}, 
\end{eqnarray} 
with $\epsilon=\pm(\gamma-1)$, and $\gamma\neq 0$. 

The above equation is exact in the presence of only one fundamental length scale
$\ell_\star={\cal O}(l_{\rm Pl})$ and then $\g=\Gamma_{\rm UV}$, with a specific 
value for a given quantum gravity theory. Conversely, this equation is valid 
only near the IR regime if $\ell_\star$ is a mesoscopic scale, and in this case 
$\g=\Gamma_{\rm meso}\approx 1$.
  \label{madelindepeefs4}
  
Constraints can then be placed on the parameters $\ell_\star$ and $\gamma$ in a 
model-independent way, by constraining the ratio $d_{\rm L}^{\rm GW}(z) /d_{\rm 
L}^{\rm EM}(z)$ as a function of the redshift $z$ of the source.

Using the neutron star merger GW170817 and a  simulated $z=2$ supermassive 
black hole merger within LISA detectability, it was shown 
\cite{Calcagni:2019kzo,Calcagni:2019ngc} that the only quantum gravity theories 
that can be constrained in this way are those with $\Gamma_{\rm 
meso}>1>\gamma_{\rm UV}$, and such theories are only group field theory, spin 
foams and loop quantum gravity.

The constraint reads \cite{Calcagni:2019kzo,Calcagni:2019ngc}
\begin{eqnarray}
0<\Gamma_{\rm meso}-1<0.02.
\end{eqnarray}

Complementary constraints   can be certainly found, which can be  tighter 
than the one above, using for instance solar system tests, nevertheless such 
constraints require additional assumptions and therefore are not 
model-independent.

 
 





\chapter[Gravitational Lensing]{Gravitational Lensing}
\label{Gergelychapter}

{\em L\'{a}szl\'{o} \'{A}. Gergely}


\section{Deflection of Light in Schwarzschild Geometry}
\label{deflectionlirefs1}

One of the first verified predictions of General Relativity was the
deflection of light when it passes close to massive objects. Its value is
twice the one estimated through Newtonian reasoning by Soldner in 1801 \cite%
{Soldner1804} (equivalent to applying the Equivalence Principle, 
\label{equivprinref7} as Einstein
proved in 2011). This half of the correct result can be understood as
arising from the weak-field and slow-motion limit of General Relativity,
giving the metric component $g_{tt}=-1-2\phi \left( r\right) /c^{2}$ in
terms of the Newtonian gravitational potential $\phi \left( r\right) 
=-G_NM/r$, 
with $G_N$ the Newton's constant (note that here, obviously, for the moment we 
keep 
the 
light speed in our calculations, setting it to 1 in the next subsection 
and onward). 
This modification of the Minkowski metric induces a gravitational blueshift,
as verified by the Pound-Rebka experiment \cite{PoundRebka1959}.

Due to the
Equivalence Principle, gravitational effects can be replaced locally by
acceleration. Hence, there is a system where they locally cancel, with the 
proper
time $\tau $ (defined as $ds^{2}=-c^{2}d\tau ^{2}$), which is related to
the coordinate time $t$ (this being the proper time the of another observer
at infinity, where the metric is flat) as $d\tau =\left( -g_{00}\right)
^{1/2}dt$. The distant observer would not only see a severely redshifted
spectrum of a source, but also an apparent light velocity 
\begin{equation}
c\left( r\right) =\frac{dr}{dt}=c\left( 1+\frac{2\phi \left( r\right) }{c^{2}%
}\right) ^{1/2}~,
\end{equation}%
where $c=dr/d\tau $ is the invariant speed of light. Hence an effective
index of refraction emerges as \cite{TPChengRel2010}:%
\begin{equation}
n(r)=\frac{c}{c\left( r\right) }=\left( 1+\frac{2\phi \left( r\right) }{c^{2}%
}\right) ^{-1/2}\approx 1-\frac{\phi \left( r\right) }{c^{2}}~.
\end{equation}%
Light appears to be both  redshifted and slowed down through the
gravitational pull of a mass, as seen by a distant observer.

A radius-dependent index of refraction is not unencountered in nature; in
this respect it bears similarities to gradient-index (GRIN) lenses, which
model both the lens of the eye and the varying temperature and density
layers above a hot road, causing the mirage. The analogy, however, stops here:
the GRIN-lens eye ideally has almost no aberration at both short and low
distances. The relaxed eye has the focal point at infinity, but flexing the
eye muscles can accommodate for close focal points. The gravitational lens,
however, has no focal point at all. Instead it has a focal line.

The other half of the deflection angle comes from the fact that not only
time  but also  space is   affected by gravity. Indeed, from the 
Schwarzschild
solution%
\begin{equation}
ds^{2}=-c^{2}\left( 1+\frac{2\phi \left( r\right) }{c^{2}}\right)
dt^{2}+\left( 1+\frac{2\phi \left( r\right) }{c^{2}}\right)
^{-1}dr^{2}+r^{2}d\Omega ^{2},
\end{equation}%
as seen by an observer at infinity, the radially moving ($d\Omega ^{2}=0$)
photon ($ds^{2}=0$) has the apparent velocity 
\begin{equation}
c\left( r\right) =\frac{dr}{dt}=\pm c\left( 1+\frac{2\phi \left( r\right) }{%
c^{2}}\right) ~,
\end{equation}%
the plus (minus) sign referring to outward (inward) movement. The
corresponding effective index of refraction becomes 
\begin{equation}
n\left( r\right) =\frac{c}{\left\vert c\left( r\right) \right\vert }=\left(
1+\frac{2\phi \left( r\right) }{c^{2}}\right) ^{-1}\approx 1-\frac{2\phi
\left( r\right) }{c^{2}}~.
\end{equation}%
The factor of two appeared by taking into account the contribution of $%
g_{rr} $, which is mandatory as long as the motion is not slow, in
particular for light.

\begin{figure}[ht]
\begin{center}
\includegraphics[height=8.5cm]{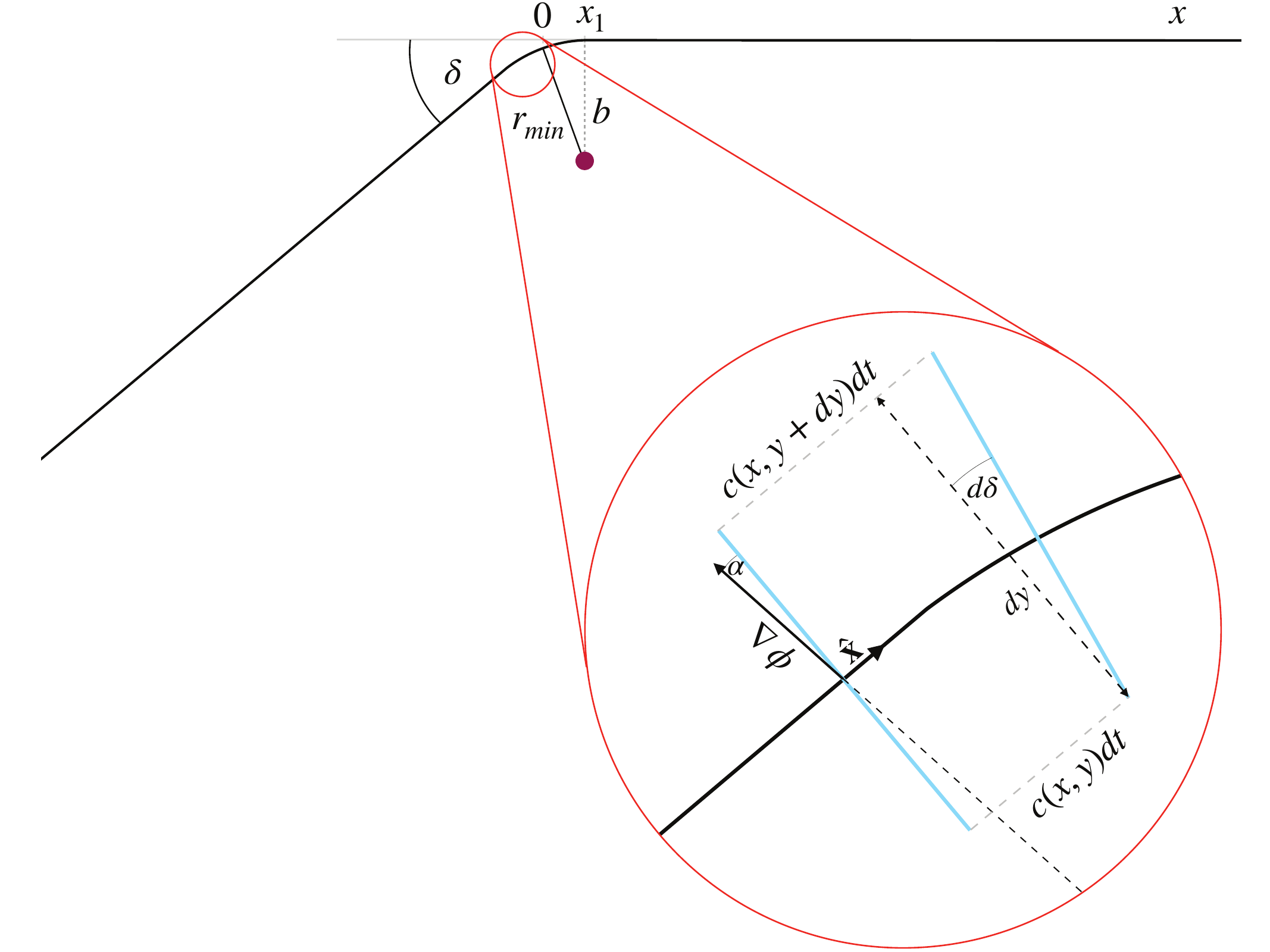}
\caption{{\it{Light deflection through the Huygens construction. }}}
\label{bendingfig}
\end{center}
\end{figure}

This varying index of refraction yields the elementary bending $d\delta $ of
a planar wavefront travelling in the $\mathbf{\hat{x}}$ direction (which
subtends an angle $\pi /2+\alpha $ with the gradient of the gravitational
field, the $\mathbf{\hat{r}}$ direction) through the Huygens construction.
Let $y$ be the coordinate perpendicular to $x$, such that $r^{2}=x^{2}+y^{2}$
(see the magnified region of Fig. \ref{bendingfig}), then 
\begin{eqnarray*}
d\delta &\approx &\tan d\delta =\frac{c\left( x,y+dy\right) dt-c\left(
x,y\right) dt}{dy} \\
&=&\frac{\partial c\left( r\right) }{\partial r}\frac{\partial \left(
r\left( x,y\right) \right) }{\partial y}\left( \frac{dx}{c}\right) \\
&=&\frac{2}{c^{2}}\frac{\partial \phi \left( r\right) }{\partial r}\frac{y}{r%
}dx=\frac{2G_NM}{c^{2}}\frac{y}{r^{3}}dx~.
\end{eqnarray*}%
The full deflection angle is found by integration over the trajectory (Fig. %
\ref{bendingfig}): 
\begin{equation*}
\delta \approx \frac{2G_NM}{c^{2}}\int_{-\infty }^{\infty }\frac{ydx}{r^{3}}=%
\frac{4G_NM}{c^{2}}\int_{0}^{\infty }\frac{ydx}{\left( y^{2}+x^{2}\right)
^{3/2}}~.
\end{equation*}%
For the second equality we explored that the orbit is symmetric on the two
sides of the point of closest approach. The coordinate line $x$ bends
together with the trajectory; however, at large distances the potential goes
to zero, hence the bending becomes negligible for large $x$ (say $x>x_{1}$)
and $y$ can be chosen as the impact parameter. To zeroth order that can be
approximated by the distance of the closest approach, $r_{\min }$. When $x$
is small, in order to have $x$ running along the orbit, at each point one
can rotate the system of reference by an angle $\gamma \left( r\right) $ of
the order $\mathcal{O}\left( \gamma \right) =\mathcal{O}\left( \delta
\right) $. As the bending angle is small, to first order $y=y^{\prime
}+\gamma x^{\prime }$ and $x=-\gamma y^{\prime }+x^{\prime }$, also $%
dx=-\gamma dy^{\prime }+dx^{\prime }$. To leading order one can also replace 
$y^{\prime }=r_{\min }$, getting the same integrand as for large $x$. Hence 
\begin{equation}
\delta \approx \frac{4G_NM}{c^{2}}\int_{0}^{\infty }\frac{r_{\min }dx}{\left(
r_{\min }^{2}+x^{2}\right) ^{3/2}}=\frac{4G_NM}{c^{2}r_{\min }}~.
\label{bendingangle}
\end{equation}%
This gives $\delta =1.75$ arcseconds for the deflection of light rays
grazing the Sun, the prediction being first verified by Eddington and Dyson
in 1919 \cite{Dyson:1920cwa} and confirmed several times since.

The effect increases with gravity, thus with curvature. While in the Solar
System the deflection angles are small, for compact objects the values can
be significantly larger. In the strong lensing \label{stronglensefs1} regime 
the 
deflection angle can 
be extreme.

Deflection of light, however, is not the only effect pertinent to the realm of
gravitational lensing. The multiplication
of images of a bright source, the light of which is
overpassing a gravitational lens is given by a lens equation.  
The number of its 
solutions depends on the geometry of the
problem. The number of images and their angular separation are observables. 
Gravitational
lensing, however, also magnifies the source image. The ratio of the
magnifications of two images is another variable, depending on the particular 
geometry. We
will discuss these in a specific example in what follows.

\section{Deflection of Light by Spherically Symmetric, Static Tidal Charged
Brane Black Holes}

Considering non-vacuum solutions or a modified gravity theory, the generic, 
spherically symmetric, static line element reads 
\begin{equation}
ds^{2}=-f\left( r\right) dt^{2}+h\left( r\right) dr^{2}+r^{2}\left( d\theta
^{2}+\sin ^{2}\theta d\varphi ^{2}\right) \, ,  \label{ds2sphe}
\end{equation}
where from now on we  follow the usual convention of this manuscript and we set 
the light speed to $c=1$.
The more special case%
\begin{equation}
f\left( r\right) =h^{-1}\left( r\right) =1-\frac{2m}{r}+\frac{q}{r^{2}}%
-\Lambda r^{2}\,,  \label{RNdS}
\end{equation}%
beyond the Schwarzschild vacuum solution (\thinspace $q=0=\Lambda $), also
incorporates other metrics. For \thinspace $%
\Lambda =0$ and $q=Q^{2}$ (with $Q$ the central electric charge) it
represents the Reissner-Nordstr\"{o}m black hole, \label{BHref8} which for 
$Q\leq 
m$ has
horizons at $r_{h}=m\pm \sqrt{\left( m^{2}-Q^{2}\right) }$, while for $Q>0$
it encompasses a naked singularity of the Einstein-Maxwell system. A generic 
$q$ leads to the interpretation of a tidal charged brane black hole \cite%
{Dadhich:2000am} embedded in a higher dimensional spacetime, with its Weyl 
curvature generating   the tidal charge $q$. This solution of the 
effective
Einstein equation \cite{Shiromizu:1999wj,Gergely:2003pn,Gergely:2008fw}\ on 
the brane has a
similar horizon structure, a black hole for positive $q\leq m^{2}$ and a naked
singularity above, while for negative $q$ it has only one horizon at $%
r_{h}=m+\sqrt{\left( m^{2}+\left\vert q\right\vert \right) }$, gravity being
strengthened by the tidal charge in this case. For $m=0=q$ and a positive
(negative) cosmological constant $\Lambda $ the metric (\ref{ds2sphe})-(\ref%
{RNdS}) describes the de Sitter (anti-de Sitter) spacetime. Rindler and
Ishak proved that contrary to previous claims, although by a small
amount, the cosmological constant affects the deflection angle generated by
a concentrated spherically symmetric mass in a Schwarzschild-de Sitter
geometry \cite{Rindler:2007zz}.

In what follows, we illustrate how to derive the deflection of light when
more than one black hole parameter is present, and to higher order in these
parameters on the example of the tidal charged black hole metric.

The deflection angle of light rays caused by brane black holes with tidal
charge has been studied in Refs. \cite{Kar:2003gh,Majumdar:2004mz}, to 
first order
in the lensing parameters. Second order gravielectric contributions were
derived in Ref. \cite{Gergely:2006qt} in both $m$ and $q$ from the Lagrangian 
given by 
$2\mathcal{L}=\left( ds^{2}/d\lambda ^{2}\right) $, where $\lambda $ is a
parameter of the null geodesic curve (a dot will denote the derivative with
respect to $\lambda $). Hence, 
\begin{equation}
0=2\mathcal{L}=-f\left( r\right) \dot{t}^{2}+f^{-1}\left( r\right) \dot{r}%
^{2}+r^{2}\dot{\varphi}^{2}\, , \label{Lag}
\end{equation}%
and the cyclic variables $t$ and $\varphi $ lead to the constants of motion 
\begin{equation}
E\equiv -p_{t}=f\dot{t}\,,\qquad L\equiv p_{\varphi }=r^{2}\dot{\varphi}\,.
\label{first_integrals}
\end{equation}%
Then, Eq. (\ref{Lag}) can be rewritten (in terms of the reciprocal radial
variable $u=1/r$ and the dependent variable $\varphi$) as 
\begin{equation}
\left( u^{\prime }\right) ^{2}=\frac{E^{2}}{L^{2}}-u^{2}f\,\left( u\right)
\,,  \label{radial}
\end{equation}%
where a prime denotes differentiation with respect to $\varphi $.
\allowbreak Differentiating this leads to 
\begin{equation}
u^{\prime \prime }=-uf\,-\frac{u^{2}}{2}\frac{df}{du}\,.  \label{radial2}
\end{equation}%
For the flat metric $f=1$, this simplifies to $u^{\prime \prime }+u=0$, with
solution $u=u_{0}=b^{-1}\cos \varphi $, with $b=L/E$ the impact parameter
and the polar angle $\varphi $ measured from the line indicating the point
of closest approach. For the tidal charged black hole, Eq. (\ref{radial2})
gives 
\begin{equation}
u^{\prime \prime }+u=3mu^{2}-2qu^{3}\,\,.  \label{lensing}
\end{equation}%
A perturbative solution in the small parameters 
\begin{equation}
\varepsilon =mb^{-1}\,,\qquad \eta =qb^{-2}  \label{params}
\end{equation}%
was sought for in the second-order accuracy form 
\begin{eqnarray}
u &=&b^{-1}\cos \varphi +\varepsilon u_{1}+\eta v_{1}+\varepsilon
^{2}u_{2}+\eta ^{2}v_{2}+\varepsilon \eta w_{2}  \notag \\
&&+\mathcal{O}\left( \varepsilon ^{3},\eta ^{3},\varepsilon \eta
^{2},\varepsilon ^{2}\eta \right) \,,  \label{ansatz}
\end{eqnarray}%
generating differential equations for the unknown functions $%
u_{1},\,u_{2},\,v_{1},\,v_{2}$ and $w_{2}$, solved for%
\begin{eqnarray}
bu &=&\cos \varphi +\frac{\varepsilon }{2}\left( 3-\cos 2\varphi \right) 
\notag \\
&&-\frac{\eta }{16}\left( 9\cos \varphi -\cos 3\varphi +12\varphi \sin
\varphi \right)  \notag \\
&&+\frac{\varepsilon \eta }{16}\left( -87+40\cos 2\varphi -\cos 4\varphi
+12\varphi \sin 2\varphi \right)  \notag \\
&&+\frac{\varepsilon ^{2}}{16}\left( 37\cos \varphi +3\cos 3\varphi
+60\varphi \sin \varphi \right)  \notag \\
&&+\frac{\eta ^{2}}{256}\bigl(271\cos \varphi -48\cos 3\varphi +\cos 5\varphi
\notag \\
&&+384\varphi \sin \varphi -36\varphi \sin 3\varphi -72\varphi ^{2}\cos
\varphi \bigr)~.  \label{u}
\end{eqnarray}%
At infinities $u=0$ and $\,\varphi =\pm \pi /2\pm \delta /2$, with $\delta $
representing the light deflection by the lensing object with mass $m$ and
tidal charge $q$. To second-order:%
\begin{equation}
\delta =4\varepsilon -\frac{3\pi }{4}\eta +\frac{15\pi }{4}\varepsilon ^{2}+%
\frac{105\pi }{64}\eta ^{2}-16\varepsilon \eta \,.  \label{phivariation}
\end{equation}%
This can also be expressed in terms of the distance of minimal approach
(thus for $u=1/r_{\min }$ and $\varphi =0$):%
\begin{equation}
r_{\min }=b\left( 1\allowbreak -\varepsilon +\allowbreak \frac{1}{2}\eta -%
\frac{3}{2}\varepsilon ^{2}-\frac{5}{8}\eta ^{2}+2\varepsilon \eta \right) ~,
\end{equation}%
as 
\begin{eqnarray}
\delta &=&\allowbreak \allowbreak \frac{4m}{r_{\min }}-\frac{3\pi q}{%
4r_{\min }^{2}}\allowbreak \allowbreak +\frac{\left( 15\pi -16\right) m^{2}}{%
4r_{\min }^{2}}  \notag \\
&&+\frac{57\pi q^{2}}{64r_{\min }^{4}}+\frac{\left( 3\pi -28\right) mq}{%
2r_{\min }^{3}}\,.  \label{deflection2}
\end{eqnarray}

While the first-order solutions were in agreement with the results of Ref. 
\cite{Briet:2008mz} (when $q=Q^{2}$), the second-order contributions did not 
agree
with the result of Ref. \cite{Boehmer:2008zh}, obtained by a
Hamilton-Jacobi approach. By carefully employing the eikonal method, the
Lagrangian result was reproduced in Ref. \cite{Gergely:2009xg}, re-establishing
the unicity of the expression for light deflection. This result was used to
impose upper limits on the tidal charge from Solar System measurements.
\label{solarsystemref8}

\section{The Lens Equation}

Due to light deflection and the common perception of assuming that images  lie 
in
the direction from which light arrives to the observer ($O$), the light
source ($S$) may appear in one or more different locations. As rays passing
both above and below the lensing object ($L$)  arrive at $O$, in general, two
images ($I$) appear. We denote the angular positions of the source and image
as $\beta =\widehat{SOL}$ (chosen positive, e.g., $S$ is above the optical
axis, defined by $L$ and $O$) and $\theta =\widehat{IOL}$ (with $s=$sgn $%
\theta $), respectively.

A lens equation for generic curved spacetimes was presented in Ref. \cite%
{Frittelli:1998hr}. However, we employ the exact lens equation for a
spherically symmetric lens, derived in Ref. \cite{Horvath:2010xq} by
trigonometric considerations only (no approximations applied) \label{lensrefs1}
\begin{eqnarray}
0 &=&\frac{2D_{l}}{D_{s}}\cos \left( \frac{\delta }{2}-\left\vert \theta
\right\vert \right) \cos \left\vert \theta \right\vert \sin \frac{\delta }{2}
\notag \\
&&+\cos \left( \delta -\left\vert \theta \right\vert \right) \left( \sin
\left\vert \theta \right\vert -s\cos \left\vert \theta \right\vert \tan
\beta \right) -\sin \delta ~.  \label{LE}
\end{eqnarray}%
Here, $\delta $ is the deflection angle, with the convention that $\delta >0$
whenever the light is bent towards the optical axis and $\delta <0$
otherwise, cf. Ref. \cite{Virbhadra:2008ws}, while $D_{l}$ and $D_{s}$ are the
distances of
$O$ from $L$ and from the projection of $S$ to the optical axis, respectively.

Whenever $S$ lies on the optical axis (such that $\beta =0$), Eq. (\ref{LE})
simplifies as 
\begin{equation}
0=D_{l}\sin \left\vert \theta \right\vert +\left( D_{s}-D_{l}\right) \sin
(\left\vert \theta \right\vert -\delta )~,  \label{ringeq}
\end{equation}%
and the two images degenerate into the Einstein ring
\begin{equation}
\left\vert \theta \right\vert =\arctan \frac{\left( D_{s}-D_{l}\right) \sin
\delta }{D_{s}\cos \delta }~.
\end{equation}

As shown in Ref. \cite{Horvath:2010xq}, a first-order expansion of Eq. 
(\ref%
{LE}) in the parameter $\epsilon :=\delta -2\left\vert \theta \right\vert $,
representing the deviation from perfect alignment of the system, gives the
Virbhadra-Ellis lens equation \cite{Virbhadra:1999nm}%
\begin{equation}
0=\tan \left\vert \theta \right\vert -\tan \left( s\beta \right) -\frac{%
D_{s}-D_{l}}{D_{s}}\left[ \tan \left\vert \theta \right\vert +\tan \left(
\delta -\left\vert \theta \right\vert \right) \right] ~.  \label{lens_Ellis}
\end{equation}%
A similar approximate lens equation has been derived by Dabrowski and
Schunck \cite{Dabrowski:1998ac}. The zeroth order approximation,
explored in Ref. \cite{Capozziello:2006dp}, yields to an even simpler equation 
\begin{equation}
\left\vert \theta \right\vert -s\beta -\frac{D_{s}-D_{l}}{D_{s}}\delta =0~.
\label{lens0}
\end{equation}%
Differences among the predictions of these lensing equations are in general
of the order of microarcseconds, hence a proper selection should be made once
instruments capable of resolving angular differences on the order of
microarcseconds will be available. 

\section{Image Positions}

In terms of the more convenient set of small parameters 
\begin{equation}
\bar{\varepsilon}=\frac{m}{L}~,\quad \bar{\eta}=\frac{q}{L^{2}}~,
\end{equation}%
with $L=D_{s}D_{l}/\left( D_{s}-D_{l}\right) $, to second-order Eq. (\ref{LE}%
) formally decomposes as \cite{Horvath:2010xq}:%
\begin{equation}
0=L_{0}+\bar{\varepsilon}L_{10}+\bar{\eta}L_{01}+\bar{\varepsilon}^{2}L_{20}+%
\bar{\varepsilon}\bar{\eta}L_{11}+\bar{\eta}^{2}L_{02}\,.  \label{LE1}
\end{equation}

If the mass term dominates, standard Schwarzschild lensing arises as a solution 
of the equation 
\begin{eqnarray}
0 &=&\cos \left\vert \theta \right\vert \left[ \cos \left\vert \theta
\right\vert ~\left( \tan \left\vert \theta \right\vert -s\tan \beta \right)
\right.  \notag \\
&&\left. -4\bar{\varepsilon}\left( \cot \left\vert \theta \right\vert +s%
\frac{L}{D_{l}}\tan \beta \right) \right] ~,
\end{eqnarray}%
which, to leading order yields%
\begin{equation}
0=~\theta ^{2}-\beta \theta -4\bar{\varepsilon}\equiv \mathcal{S}~,
\label{LE_Schw}
\end{equation}%
with the image positions 
\begin{equation}
\theta _{1,2}=\frac{\beta \pm \sqrt{\beta ^{2}+16\bar{\varepsilon}}}{2}~.
\label{sg}
\end{equation}%
With perfect alignment of $S$, $L$ and $O$ the Einstein ring with radius $%
\theta _{E}=2\bar{\varepsilon}^{1/2}$ emerges.

If, by contrast, the tidal charge term dominates, then a similar analysis
will give different results depending on its sign. For negative tidal charge one
image appears above, the other below the optical axis, similar to  
Schwarzschild lensing, although the location of the images is distinctly
different, as summarised in Fig. 4 of Ref. \cite{Horvath:2010xq}.

A positive tidal charge, however, generates a negative deflection angle - in
other words, a scattering behaviour - similar to  a negative mass \cite%
{Cramer:1994qj}. There is also an upper limit $\bar{\eta}_{\max }$, above which 
no such scattered images could emerge.

Other cases and second-order corrections were discussed in Ref. \cite%
{Horvath:2010xq}. For mass-dominated weak lensing, \label{weaklensefs1}the 
position of the images
is similar to the Reissner-Nordstr\"{o}m black hole lensing 
\cite{Sereno:2003nd},
however a sensible tidal charge qualitatively modifies the image properties.

The lensing effect of boson stars has also been studied. Neutral boson stars
are transparent, charged ones opaque, but they deflect light in the same way
as the Schwarzschild or Reissner-Nordstr\"{o}m black holes 
\cite{Dabrowski:1998ac}. In the neutral case, however, beside the two images a
third, much dimmer one may appear, due to the transparency of the lens.

\section{Magnification Ratios}

The magnification factor is defined as the ratio of the solid angle
subtended by the image, divided by the solid angle subtended by the source.
For a lens with axial symmetry it is given as \cite{Schneider1984AA}, \cite%
{BlandfordNarayan1986}%
\begin{equation}
\mu =\left\vert \frac{\theta }{\beta }\frac{d\theta }{d\beta }\right\vert ~,
\label{amplification}
\end{equation}%
with both $\theta $ and $\beta $ small. For the images (\ref{sg}) obtained
in the case of Schwarzschild lensing the magnification factors are 
\begin{equation}
\mu _{1,2}=\frac{1}{4}\left( \frac{1}{\beta }\sqrt{\beta ^{2}+4\theta
_{E}^{2}}+\frac{\beta }{\sqrt{\beta ^{2}+4\theta _{E}^{2}}}\pm 2\right) ~.
\label{amplificationS}
\end{equation}%
For a collinear configuration both magnifications diverge as $\mu
_{1,2}=\allowbreak \theta _{E}/2\beta $,\ however their ratio (the flux
ratio) goes to unity.

The magnifications in the tidal charge dominated regime are given by more
cumbersome expressions. Comparing the ratio of the magnifications for the
various subcases transpired to be  instructive. This ratio $\mu _{1}/\mu _{2}$ 
shows
significant deviations from the Schwarzschild case even when the tidal
charge is a second-order correction \cite{Horvath:2010xq}.

Moreover, by plotting the magnification ratio 
\label{magnificationratrefs1}(flux 
ratio, brightness ratio)
as a function of the image separation \label{limagesepfs1}in the case of tidal 
charged black
holes yields a power-law relation, which is significantly different from that
for Schwarzschild black holes. \label{BHref9} This might be interpreted as a 
smoking gun for
detecting the underlying black hole structure with precision observations,
provided they are not blurred by environmental noise.

\section{Strong Lensing by Spherically Symmetric, Static Tidal Charged Black
Holes}

At a closer approach to the gravitational centre, with lower values of the
impact parameter, the deflection angle ceases to be small, and increases to
finite values. If it passes over $2\pi $, for all numbers of exact windings,
relativistic images are formed on both sides of the lens, in an infinite
sequence. Analytic methods for studying strong gravitational lensing in
spherically symmetric geometries have been developed in Refs. 
\cite{Bozza:2001xd,Bozza:2002zj} for the case of almost perfect collinearity of 
the 
source,
lens and observer, a configuration allowing for series expansion. It has
been found that the deflection angle diverges logarithmically at a minimal
impact parameter. 

The above formalism has been applied to Schwarzschild,
Reissner-Nordstr\"{o}m, and Janis-Newman-Winicour black holes, for which the
position and the magnification of the relativistic images could be
distinguished. If the mass and the distance of the lens is known by other
methods and observations, then the position of the relativistic images could
distinguish among these geometries. The optical resolution needed to
investigate the strong field behaviour of light in the vicinity of the black
hole at the centre of our galaxy through its relativistic images has been
estimated.

The investigation of the lensed images of stars orbiting close to Sgr A*
with a possible tidal charge was advanced in Ref. \cite{BinNun:2010se}, in the
regime between the weak and strong field lensing. It has been conjectured,
that with the next generation of instruments, the detection of secondary
images will place constraints on the size of the tidal charge. Strong
gravitational lensing has also been discussed in Ref. \cite{Horvath:2012ru}, 
where an upper bound for the tidal charge of the supermassive black hole
in the centre of our galaxy is derived, based on the margin of error of the 
detecting
instrument in the measurements of the radius of the first (also of the
second) relativistic Einstein ring, due to strong lensing. With constraints 
from the designed
sensitivity of GRAVITY \cite{Gillessen:2010ei}, a four-telescope beam combiner
instrument for NIR interferometry with the Very Large Telescope
Interferometer (VLTI), the supermassive black hole could be allowed to 
have a much larger tidal charge than derived for the Sun or neutron stars.

\section{Gravitational Lensing by Other Spherically Symmetric, Static Brane
Black Holes}

The gravitational lensing of spherically symmetric, static brane black holes
has been extensively studied. Such black holes include the tidal charged
brane black hole \cite{Dadhich:2000am}, although the five-dimensional spacetime
encompassing it has not been found.

Another possibility is the Garriga-Tanaka black hole \cite{Garriga:1999yh},
embedded in a five-dimensional AdS spacetime with curvature radius\textit{\ }$l$
and five-dimensional cosmological constant $\tilde{\Lambda}=-6l^{-2}$. The 
weak-field limit of the Schwarzschild metric on the brane receives corrections
from the Kaluza-Klein modes of the five-dimensional AdS spacetime: 
\begin{eqnarray}
f\left( r\right) &=&1-\frac{2m}{r}\left( 1-\frac{4}{\tilde{\Lambda}r^{2}}%
\right) ~,  \notag \\
h\left( r\right) &=&1+\frac{2m}{r}\left( 1-\frac{2}{\tilde{\Lambda}r^{2}}%
\right) ~.  \label{GT}
\end{eqnarray}%
The third-order corrections fade away with $r^{-3}$. Ref. \cite{Kar:2003gh}
calculated the bending of null geodesics by both the tidal charged and
Garriga-Tanaka spherically symmetric, static brane black holes.

Brane-world black hole (both tidal charged and Garriga-Tanaka) lensing has also 
been studied in Ref. \cite{BinNun:2009jr} by numerical techniques. Image
magnifications of such braneworld primordial black holes were found to be 
significantly  increased   comparing to their four-dimensional analogues.

Ref. \cite{Majumdar:2004mz} has calculated light deflection in the weak-field 
limit of
yet another braneworld black hole \cite{Guedens:2002km}, representing the near 
horizon
approximation of the five-dimensional Schwarzschild geometry:%
\begin{equation}
f\left( r\right) =h^{-1}\left( r\right) =1-\frac{r_{0}^{2}}{r^{2}}~,
\label{nhS5}
\end{equation}%
where $r_{0}$ is the radius of the five-dimensional black hole horizon, assumed
much smaller than the size of the AdS radius. The bending angle of a light
ray emerges as the result given in Eq. (\ref{bendingangle}) is  multiplied by $%
2l/r_{\min }$, indicating a more prominent effect for small impact
parameters. Strong field images were analysed in Ref. \cite{Eiroa:2004gh}.

Ref. \cite{Majumdar:2004mz} also addressed the modifications of the Einstein 
ring radius
and magnification for a point light source. It was shown that the radius of
the Einstein ring decreases by a factor of $\left( l/2D_{l}\right) ^{1/2}$,
where $D_{l}$ is the distance of the lens from the observer. The
magnification of the Schwarzschild black hole is corrected by several
factors, as given by their Eq. (27). Brane black holes of type (\ref{nhS5})
of our galactic halo, could be dimmed by the extra dimensional extension of
gravity by  as much as six orders of magnitude, which would make them extremely
difficult to detect by gravitational lensing. That is an interesting
scenario in the quest to explain dark matter.

Strong lensing brane black holes could have significantly different
observational signatures compared to the Schwarzschild black holes, as 
discussed 
in
Ref. \cite{Whisker:2004gq}. This work investigated the tidal charged black hole 
and
another black hole that has been originally proposed in Ref. \cite%
{Casadio:2001jg} in the area gauge as%
\begin{eqnarray}
f\left( r\right) &=&\left[ \left( 1+\epsilon \right) \left( 1-\frac{2m}{r}%
\right) ^{1/2}-\epsilon \right] ^{2}~,  \notag \\
h\left( r\right) &=&\left( 1-\frac{2m}{r}\right) ^{-1}~.  \label{CasFM}
\end{eqnarray}%
The parameter $\epsilon >0$ scales with the difference of the horizon radius
and a turning point in the area function. It is also related through $1+\epsilon
=\left( 1+\eta \right) ^{-1}$ to the parameter $\eta $ of Ref. \cite%
{Casadio:2001jg}, measuring the difference between the gravitational and 
inertial
masses. The gravitational mass is given by $f\left( r\right) $ with $m\left(
1+\epsilon \right) $, where $m$ is the ADM\ mass, the asymptotic limit of the
Misner-Sharp mass, appearing in $h\left( r\right) $. This metric is singular
at $r_{0}=2m$ and $r_{h}=2m\left( 1+\epsilon \right) ^{2}/\left( 1+2\epsilon
\right) \,$, which is larger than $r_{0}$, but only by a second-order
contribution in $\epsilon .$ The event horizon at $r_{h}$ is singular. Most
notably, the radius of the photon sphere in this geometry is modified, being
pulled inwards or pushed outwards together with the inwards / outwards
shifted horizon relative to the Schwarzschild case. With black hole shadow
imaging already a reality through the Event Horizon Telescope
very-long-baseline interferometry (VLBI) measurements, this may turn out to be 
another test of General Relativity in the future.

Lensing properties of brane black holes are also discussed in the review 
\cite{Majumdar:2005ba}. It is argued that Myers--Perry black holes 
\cite{Myers:1986un} (the projection onto the brane of the five-dimensional
Schwarzschild solution given by the Myers--Perry metric) residing in the
galactic halo would be difficult to detect by their weak lensing features, as
their magnification is reduced in comparison to the Schwarzschild case.

Based on a formalism with two gravitational potentials, Ref. \cite{Pal:2007ap}
investigated gravitational lensing, obtaining general expressions for the
time delay, deflection angle, Einstein ring, image positions, magnification
and critical curves (the loci of image points in the lens plane with
infinite magnification).

\section{Gravitational Lensing in Ho\v{r}ava-Lifshitz Gravity}
\label{Lifshitzref3}

A decade ago, Ho\v{r}ava proposed a renormalisable field theoretical model that
can be interpreted as a complete theory of gravity \cite%
{Horava:2008ih,Horava:2009uw} (see also the review \cite{Visser:2011mf}). At
low energies it reduces to Einstein gravity with a non-vanishing cosmological
constant. At the ultraviolet energy scales, however, it exhibits an
anisotropic Lifshitz scaling between time and space given by $%
x^{i}\rightarrow lx^{i}$ and $t\rightarrow l^{z}t$, with $z$ the scaling
exponent. 

Ho\v{r}ava-Lifshitz gravity admits the Kehagias-Sfetsos asymptotically flat
solution \cite{Kehagias:2009is}, given by 
\begin{equation}
f\left( r\right) =h^{-1}\left( r\right) =1+\omega r^{2}\left[ 1-\left( 1+%
\frac{4G_Nm}{\omega r^{3}}\right) ^{1/2}\right] \,,  \label{ks}
\end{equation}%
where $m$ is the total mass of the black hole and $\omega $ is the Ho\v{r}%
ava-Lifshitz parameter. The origin represents a curvature singularity, while 
\begin{equation}
r_{\pm }=G_Nm\left( 1\pm \sqrt{1-\frac{1}{2G_N^{2}m^{2}\omega }}\right) ~
\label{hors}
\end{equation}%
are coordinate singularities, disappearing in Eddington-Finkelstein  
coordinates. The outer apparent horizon $r_{+}$ is an event horizon that
goes to the Schwarzschild radius in the limit $\omega \rightarrow \infty $
(in the same limit, the inner horizon degenerates into the central
singularity).

The deflection angle of light \label{deflectionlirefs2} can be calculated based 
on the result that
light follows null geodesics (although massive particles do not). Image
locations and the radius of the Einstein ring emerge by a similar procedure
as described before.

For every $\omega $ there is a maximal deflection angle $\delta _{\text{max}}
$, corresponding to certain $r_{\text{crit}}$, and light rays passing both
above and below $r_{\text{crit}}$ will experience less deflection than $%
\delta _{\text{max }}$\cite{Horvath:2011xr}, a feature inexistent in 
Schwarzschild weak lensing. The critical distance $r_{\text{crit}}$
decreases with increasing $\omega $, so when approaching the Schwarzschild
limit, $r_{\text{crit}}$ is confined below the horizon (goes to $0$ when $%
\omega \rightarrow \infty $), generating the expected decreasing $\delta (r_{%
\text{min}})$ function out of the horizon, also producing two images, as in
the Schwarzschild lensing. 

Far from the Schwarzschild regime, however, the Kehagias-Sfetsos spacetime
approaches flatness, therefore there is only one  undeflected and
unmagnified image. In the intermediate range only the upper focused image is
produced, due to the existence of the maximal deflection angle $\delta _{%
\text{max}}$. 

Concerning the Einstein rings, it has been found that for a given mass and
lensing geometry, the Einstein angles in the Kehagias-Sfetsos spacetime are
always smaller than their Schwarzschild counterpart. In the strong lensing 
regime the detection of the first two relativistic Einstein rings could
constrain the parameter range of the the Kehagias-Sfetsos spacetime \cite%
{Horvath:2011xr}. Their observation with the $10^{-5}$ arcsec expected accuracy
of future instruments \cite{Trippe:2009hf} would set the strongest observational
constraint on the Ho\v{r}ava-Lifshitz parameter.

\section{Gravitational Lensing in $f\left( R\right) $ Gravity}
\label{Bransref4}

In the quest to explain dark matter and dark energy, higher-order theories
of gravity were advanced, both in metric and Palatini (see the review \cite%
{DeFelice:2010aj}) approaches. In the action describing such theories in the 
metric
formulation, the curvature scalar $R$ of the Einstein-Hilbert action is
replaced by an arbitrary function $f\left(R\right) $. It is known that this
can be transformed to a particular form of Brans-Dicke scalar-tensor theory,
with the Brans-Dicke parameter $\omega _{0}$ vanishing (a massive dilaton
gravity model). 

It has been proven in Ref. \cite{Sotiriou:2011dz} that the the Schwarzschild
solution is the unique spherically symmetric, asymptotically flat vacuum
black hole solution for a scalar field of a generalised Brans-Dicke class
with linear kinetic term, with the energy-momentum tensor of the scalar
field obeying the weak energy condition in the Einstein frame. 
\label{Einsteinfrref9}

Nevertheless, for $f=R^{n}$ a weak-field solution has been found in Ref. \cite%
{Capozziello:2006ph} as 
\begin{equation}
f\left( r\right) =h^{-1}\left( r\right) =1-\frac{G_Nm}{r}\left[ 1+\left( \frac{%
r}{r_{c}}\right) ^{\sigma }\right] ~,
\end{equation}%
with a characteristic distance $r_{c}$ (much larger than the Schwarzschild
radius) and a strength parameter 
\begin{equation*}
\sigma =\frac{12n^{2}\!-\!7n\!-\!1\!-\!\sqrt{36n^{4}+12n^{3}-83n^{2}+50n+1}}{%
6n^{2}-4n+2}<1~
\end{equation*}%
monotonically increasing with $n$. The Schwarzschild solution is recovered
for $\sigma =0$ ($n=1$). This solution evades the conditions of the unicity
theorem \cite{Sotiriou:2011dz}, as the deviation from General Relativity can
be formulated in terms of an effective energy-momentum tensor, which
violates all energy conditions \cite{Horvath:2012kf}. This solution can be
considered as an approximation of a compact object, whose parameters fall
outside the conditions of the above theorem.

Weak lensing characteristics in this geometry are as follows. Above (below) $
r_{c}$, gravity is strengthened (weakened), and as a consequence, weak lensing
features are modified compared to the Schwarzschild case. There is a
critical impact parameter (depending upon $r_{c}$) for which the behaviour of
the deflection angle changes. In Ref. \cite{Capozziello:2006dp} the effects on 
the
amplification of the images and the Paczynski curve in microlensing
experiments were estimated. 

Image positions, Einstein ring radii, magnification factors and the
magnification ratio have been computed \cite{Horvath:2012kf}. Most importantly 
it
was found that the magnification ratio as a function of image separation obeys
a power-law depending on the parameter $\sigma $, with a double degeneracy.
No $\sigma \neq 0$ value gives the same power as the one characterizing
Schwarzschild black holes. As the magnification ratio and the image
separation are the lensing quantities most conveniently determined by direct
measurements, future lensing surveys will be able to constrain the parameter 
$\sigma $ based on this prediction.

\section{Gravitational Lensing in Scalar-Tensor Theories}\label{stronglensefs2}

The Janis-Newman-Winicour black hole \cite{Janis:1968zz} represents a
spherically symmetric solution of the Einsten -- massless scalar field
equations, with a point singularity replacing the Schwarzschild horizon. It
has been discussed in a more general context by Gautreau 
\cite{GautreauNeuvoC1969},
rediscovered by Wyman \cite{Wyman:1981bd}, as shown in Ref. 
\cite{Virbhadra:1997ie}%
. The strong lensing features were analysed by Bozza \cite{Bozza:2002zj} and the
effects of the scalar charge $q<m$ of the black hole (expressed through a
parameter $\gamma =m/\sqrt{m^{2}+q^{2}}$) on image positions and
magnifications quantified. Such effects could in principle be measurable by
VLBI\ techniques, provided the relativistic images survive environmental noise.

Gravitational lensing has been studied for the case of a static and
circularly symmetric lens, characterised by both mass and scalar charge
parameters \cite{Virbhadra:1998dy}. With increasing scalar charge parameter,
the deviations from Schwarzschild lensing become important, manifested in
either the disposition or the emergence of two Einstein rings. The number
and position of images may also vary.

Recently, the strong gravitational lensing by a Kiselev black hole \cite%
{Younas:2015sva} has been investigated. This black hole \cite{Kiselev:2002dx}\ 
arises
under the assumption of spherical symmetry and it has quintessence ($%
p=w_{q}\rho ,$ $-1<w_{q}</1/3$) hair. \label{hairtheref1}For these parameters 
it 
has the
asymptotics of de Sitter spacetime. The position and total magnification of
relativistic images were computed and compared with the case of the
Schwarzschild black hole.\label{BHref10} A possible observational signature was 
identified
in the bending angle, being larger than for the Schwarzschild black hole with
the same mass parameter.

\section{Gravitational Lensing in Teleparallel Gravity}

In the Teleparallel Equivalent of General Relativity (TEGR) the torsion $T$
or in its symmetric counterpart (STEGR) the non-metricity $Q$ generate forces
representing gravity, causing the deflection of trajectories from the
straight lines of the flat metric. In the equivalent general relativistic
picture there is a curved metric, which exhibits these trajectories as its
geodesics. Hence the description of weak and strong lensing, developed in
terms of this metric, can be unequivocally rewritten in the language of torsion
or non-metricity, with no new physical effects to be expected.

However, in the generalised $f\left( T\right) $ and $f\left( Q\right) $
theories the lensing properties could be different and are worth studying in
order to derive effects and predictions that could either support or
falsify these theories when confronted with high-sensitivity observations.

\section{Gravitational Lensing, Galaxies and Cosmology}

The baryonic (luminous) mass of the galaxies, although plagued by the
uncertainties in the mass-to-light ratio, is assessed by observations on their
luminosity. Realistic mass-to-light models, however, do not in general  
provide
enough baryonic mass   to explain the behaviour of galactic rotation
curves, observed spectroscopically. The flatness of these curves is one
indication for the existence of galactic dark matter halos. Many dark matter
models compatible with galactic rotation curves have been proposed. They may
differ, especially in the outer regions of the galaxies, where no
spectroscopic information is available.

Gravitational lensing
observations, however, are able to assess the combined baryonic and dark mass
of the galaxy at distances far exceeding the validity of the galactic
rotation curves. Hence, lensing surveys of galaxies and galactic clusters may
distinguish from among the dark matter models. These surveys measure the 
convergence
(magnification) and shear deformation of the objects, and are equally
important for cosmological considerations. Notable examples are the
first-year shear catalogue of the Subaru Hyper Suprime-Cam Subaru Strategic
Program Survey \cite{Mandelbaum:2017dvy} and the Kilo-Degree Survey (KiDS) 
\cite{deJong:2012zb},
carried out with the VLT Survey Telescope (VST) and OmegaCAM camera, which is  
able to
map the large-scale matter distribution in the Universe and constrain the
equation-of-state of Dark Energy.

In a cosmological setting the perturbed Friedmann metric reads 
\begin{equation}
ds^{2}=-\left( 1+2\Psi \right) dt^{2}+a^{2}\left( 1-2\Phi \right) d\mathbf{x}%
^{2}~,
\end{equation}%
with \thinspace $a$ being the scale factor. Lensing properties are sensitive to 
$%
\nabla ^{2}\left( \Phi +\Psi \right) $ along the line of sight. The Poisson
equation, on the other hand, contains $\nabla ^{2}\Phi $, sourced by the
fractional overdensity $\delta _{o}$. In General Relativity $\Psi =\Phi $
(provided there are no anisotropic stresses), thus lensing is intimately
related to  $\delta _{o}$\ along the line of sight. By contrast, in
alternative theories of gravity this relation is modified. Ref. 
\cite{Zhang:2007nk}
proposed to determine the matter overdensity at a given redshift by
measuring the velocity field, rather than from galaxy overdensities, and
discussed the differences among the cold dark matter model with a
cosmological constant, the Dvali-Gabadadze-Porrati, the TeVeS, and $f(R)$
gravity theories.

As an example of another lensing feature that could be tested by these
methods, we mention again brane-world models. The higher-dimensional Weyl
curvature induces a new source of gravity on the brane. In the spherically
symmetric, static configuration it reduces to dark radiation and dark
pressure components. These are able to modify spacetime geometry around
galaxies consistently, with the flatness of galactic rotation curves \cite%
{Mak:2004hv,Harko:2005fq,Harko:2005yk,Boehmer:2007xh,Gergely:2011df}. 
Gravitational lensing 
could provide a test to
discern between this geometrical model and other dark matter models. Indeed,
in the asymptotic regions, light deflection is enhanced, as compared to dark
matter halo predictions. For a linear equation of state of the Weyl fluid,
there is a critical radius below which braneworld effects reduce, while
above it they amplify the deflection of light \cite{Wong:2012sd}. This is in
contrast to other dark matter models, in which the deflection angle
increases due to dark matter at any radius.

\section{Concluding Remarks\label{conclusec}}
\label{geometricaloprfs1}

All results discussed above refer to the geometrical optics or high
frequency approximation of light. In a gravitational theory this means that
the wavelength of the light is much smaller than the gravitational curvature
radius. This is an approximation, which always holds in flat spacetime and
most of the time in weak gravity sectors as well. Information about the
polarisation of the electromagnetic waves is not contained in the
geometrical optics approximation. It has been argued that high-frequency
wave propagation is insensitive to compositions of certain conformal,
Kerr-Schild, and related transformations of the background metric, while the
inclusion of the polarisation information breaks some of the degeneracy
\cite{Harte:2019tid}.

In the geometrical optics approximation we have seen that the occurrence of
light deflection depends on the underlying geometry. While the general
relativistic mass term of the Schwarzschild solution is expected to dominate
gravitational lensing in most of the spherically symmetric cases, other
charges of a hairy black hole could modify the deflection value in an
essential way, leading to modified image positions. Moreover, the
magnification or brightness of the images will also be affected.

For weak gravitational lensing, testing the power-law relation obeyed by the
brightness ratio as a function of the image separation (both being
observable quantities) could unequivocally distinguish between Schwarzschild
lensing or hairy characteristics. For strong lensing, the detection of the
positions of the relativistic images would provide information about the
possible types of hair. Future precision measurements, able to penetrate the
environmental noise, are expected to shed more light on these features,
confirming or falsifying hairy features and thus the underlying
gravitational theories.

\chapter[Classicalizing Gravity]{Classicalizing Gravity}
\label{Casadiochapter}
{\em Roberto Casadio, Andrea Giusti}


%
%
%
%
%
%
%
%
\section{Semiclassical Gravity and Localized Quantum States}

 \label{classicalizingprfs1}
  \label{csemiclassicalrfs1}

One of the defining results of the twentieth century was the unravelling of the 
key  feature of curvature singularities
and what leads to their formation in General Relativity.
This was achieved in the 1960s, mostly through the Penrose-Hawking singularity 
 theorems~\cite{Hawking:1969sw},
which roughly state that if a trapped surface appears during the gravitational 
collapse,  the spacetime will either be
geodesically incomplete or form a region containing closed timelike curves.
In other words, when a trapped region occurs, the sourcing matter will either  
be forced to collapse into a curvature
singularity or the very notion of causality gets lost.
However, the emergence of singularities is not just puzzling from a classical 
perspective, if the   truth be told it turns out
to be unacceptable if we take into account the basic principles of quantum 
physics.
Indeed, from Heisenberg's uncertainty principle
\begin{eqnarray}
\Delta x \, \Delta p \geq \hbar / 2
\ , 
\end{eqnarray}
it follows that if we try to localise a quantum particle within a region of  
size $\Delta x \lesssim \ell_{Pl}$,
the particle will become hidden behind its own gravitational horizon,  
i.e., $R_{\rm H} \gtrsim \Delta x$.
This suggests the existence of a minimal length, of the order of the  {\rm 
Planck scale\/}, beyond which the quantum
nature of gravity cannot be neglected (for a review,  see 
Ref.~\cite{Hossenfelder:2012jw}).
\par
Hawking figured out that, in order to describe the innermost regions of  a 
black hole, we cannot really forget
about quantum effects.
We could then wonder whether this quantum region should be treated as a source
localised deep inside the horizon.
To put this idea to the test, we could employ, for instance, the {\em horizon  
quantum mechanics\/} (HQM) formalism
(see, e.g., \cite{Casadio:2013tma,Casadio:2013aua, 
Casadio:2015qaq,Casadio:2016fev,  Casadio:2013uga, Casadio:2015sda, 
Casadio:2015rwa, Casadio:2017nfg, Giugno:2017xtl, Casadio:2018vae, 
Casadio:2018rlc}).
Let us consider a quantum particle of mass $m$ described by a Gaussian 
wave-packet of  typical spatial extension
$\sigma = \lambda _m \simeq \hbar / m$, namely
\begin{eqnarray} 
\Psi _{\rm S} (r) = \left(\frac{1}{\pi \lambda _m^2}\right)^{3/4} 
\exp \left( - \frac{r^2}{2 \, \lambda _{m}^2} \right)
\ ,
\end{eqnarray}
with $r = |\bm{x}|$. Taking the Fourier transform of $\Psi _{\rm S} (r)$,we 
obtain
\begin{eqnarray}
\label{eq:momentum}
\widetilde{\Psi} _{\rm S} (\bm{p}) = \left(\frac{1}{\pi m^2}\right)^{3/4} 
\exp \left( - \frac{|\bm{p}|^2}{2 \, {m}^2} \right)
\ .
\end{eqnarray}
If we now assume the validity  of the flat mass--shell relation, $E^2 = 
|\bm{p}|^2 + m^2$,
the momentum--space representation of $\Psi _{\rm S}$ yields the spectral 
decomposition of the system,
\begin{eqnarray}
\Psi _{\rm S} (\bm{x}) \sim \int {d} ^3 p \, e^{i \bm{p} \cdot \bm{x}} 
\widetilde{\Psi} _{\rm S} (\bm{p})
\sim \int {d} E \, \varphi _{E} (\bm{x}) \, C_{\rm S} (E)
\ ,
\end{eqnarray}
with $C_{\rm S} (E) \sim \widetilde{\Psi} _{\rm S} (|\bm{p}|^2 = E^2 - m^2)$  
and $\varphi _{E} (\bm{x})$
the eigenfunctions of an appropriate Hamiltonian operator.
Interpreting the Schwarzschild expression $R_{\rm H} = 2\, G_{N}\, M$, with 
$G_{N}$ Newton's constant, as a 
kind of  Gupta--Bleuler
condition for the admissible physical states, from  \eqref{eq:momentum} we can 
compute the
{\em horizon wave--function}
\begin{eqnarray}
\label{eq:HWF}
\Psi _{\rm H} (r _{\rm H}) = \mathcal{N} _{\rm H} \, \Theta (r_{\rm H} - R_{\rm 
H})
\, 
\exp \left( - \frac{M_{Pl} ^2 \, r_{\rm H} ^2}{8 \, m^2 \, \ell _{Pl} ^2} 
\right)
\ , 
\end{eqnarray}
which determines the size of the gravitational radius ($r_{\rm H} = 2 \, G_{\rm 
N} \, E$)
at the quantum level (see~\cite{Casadio:2015qaq} for details).
From here, we can compute the probability for a quantum system to lie inside 
its own event horizon as
\begin{eqnarray}
P_{\rm BH} = 
\int _0 ^\infty \mathcal{P} _{\rm S} (r < r_{\rm H}) \, \mathcal{P} _{\rm H} 
(r_{\rm H}) \, {d} r_{\rm H},
\end{eqnarray}
with 
$\mathcal{P} _{\rm H} (r_{\rm H}) = 4 \pi r _{\rm H}^2 |\Psi _{\rm H} (r _{\rm 
H})|^2$
and
\begin{eqnarray}
\mathcal{P} _{\rm S} (r < r_{\rm H}) = 4 \pi \int_0 ^{r_{\rm H}} \bar{r}^2|\Psi 
_{\rm H} (\bar{r})|^2 \, {d} \bar{r}
\ .
\end{eqnarray}
Computing $P_{\rm BH}$ for the case~\eqref{eq:HWF}, we find that $P_{\rm BH} 
(m) \ll 1$
for $m \ll M_{Pl}$, while it quickly approaches unity as $m \simeq M_{Pl}$, 
with 
$M_{Pl}$
denoting the Planck mass.
This result is consistent with the standard semi-classical expectation  
however, 
if we compute the corresponding
quantum fluctuations $\Delta R_{\rm H}$ for the size of the horizon, we find
that
\begin{eqnarray}
\Delta R _{\rm H} ^2 = 
\langle{R^2_{\rm H}}\rangle_{\Psi _{\rm H}} - \langle{R _{\rm H}}\rangle_{\Psi 
_{\rm H}} ^2 
\sim (2 \, G_{N} \, m)^2 \equiv R_{\rm H} ^2 \, .
\end{eqnarray}
In other words, for such a localised object the quantum fluctuations of the 
horizon scale
as the size of the horizon itself,  i.e., $\Delta R_{\rm H} \sim R_{\rm H}$.
This is, definitely, rather unreasonable, since one expects that the bigger a 
black 
hole, the more classical it will appear.
This conclusion thus seems to suggest that modelling the interior of a black 
hole in terms of very localised quantum
objects might not be a viable option.
\section{Corpuscular Gravity}
\label{corpuscularrfs1}
Corpuscular gravity (see, e.g., ~\cite{Dvali:2011aa, Dvali:2012rt, 
Dvali:2012en, Casadio:2015bna, Casadio:2015lis, Casadio:2018ukt}
and~\cite{Giusti:2019wdx} for a review of the topic) offers a way to 
effectively 
describe the interior of black holes
as extended objects.
Specifically, in this picture, black holes are treated as 
self-sustained~\cite{Casadio:2014vja} marginally bound states
constituted of a very large number of soft and off-shell gravitons with typical 
wavelength $\lambda _{\rm G} \simeq R_{\rm H}$.
It is worth noting that this picture of gravity in the strong coupling regime 
represents a possible implementation
of a much more general scheme, known as {\em classicalisation\/}, whose scope 
is 
to provide an alternative
to the standard Wilsonian UV-completion of non-renormalisable 
theories~\cite{Dvali:2010jz}.
Denoting $\epsilon _{\rm G} = \hbar / \lambda _{\rm G}$, the typical energy 
of these constituent gravitons,
and introducing the effective (dimensionless) gravitational coupling 
$\alpha_{\rm G} = \epsilon _{\rm G} ^2 / M_{Pl} ^2$,
we find  that these gravitons satisfy an energy balance that reads
\begin{eqnarray}
K + U \simeq 0
\ ,
\end{eqnarray}
with $K \simeq \epsilon_{\rm G}$ denoting the ``kinetic energy'' of each 
constituent graviton
and $U \simeq - \alpha_{\rm G} N \hbar / \lambda _{\rm G}$ representing a first 
rough approximation
of the binding potential felt by each of the $N \gg 1$ constituents. This 
condition then implies
\begin{eqnarray}
\label{criticality}
\alpha_{\rm G} \, N \simeq 1
\ .
\end{eqnarray}
As a consequence, the softness of the constituents will be measured by the 
multiplicity of the collective state, since
\begin{eqnarray} \label{eq:scaling-e}
\epsilon _{\rm G} = M_{Pl} \sqrt{\alpha _{\rm G}} \simeq 
\frac{M_{Pl}}{\sqrt{N}} 
\, .
\end{eqnarray} 
Furthermore, since $\lambda _{\rm G} \simeq R_{\rm H}$, it is easy to see that
\begin{eqnarray}
R_{\rm H} \simeq \lambda _{\rm G} \simeq \frac{\hbar}{\epsilon _{\rm G}} \simeq 
\sqrt{N} \, \ell _{Pl}
\ ,
\end{eqnarray}
which also suggests that the mass of the black hole should scale as
\begin{eqnarray}
\label{eq:scaling-M}
M \simeq \sqrt{N} \, M_{Pl}
\ ,
\end{eqnarray}
or equivalently $M \simeq N \, \epsilon _{\rm G}$.  
\par
From the perspective of condensed matter physics, the quantity $g = \alpha_{\rm 
G} \, N$ represents
the collective coupling of the system.
Hence, the condition~\eqref{criticality} tells us that, whereas the 
constituents 
of the bound state interact
very weakly with each other through the coupling $\alpha_{\rm G} \simeq 1/N \ll 
1$, the collective interaction
$g \simeq 1$ is indeed strong.
The system is therefore at its quantum {\em critical point\/}, meaning that its 
constituents will tend to leak
out because of (however small) quantum fluctuations.
The peculiar feature of this leaking of gravitons out of the black hole is that 
the system still remains
at the critical point throughout the whole depletion process (at least until 
the 
gravitational interaction is no longer
able to overcome the internal quantum pressure generated by the confined 
matter~\cite{Casadio:2019tfz}).
This description therefore offers a way to produce the Hawking 
process~\cite{Hawking:1974sw}
as the result of the quantum depletion~\cite{Dvali:2012en, Casadio:2015bna, 
Casadio:2015lis, Muck:2016stv}
rather than a pure vacuum effect.
In detail, let us focus on a tree--level $2 \to 2$ scattering between two 
gravitons within the marginally bound state.
Assuming that one of them gains enough energy to escape the ground state, we 
can then naively estimate
the decay rate of the system as
\begin{eqnarray}
\label{eq:decayrate}
\Gamma \sim \alpha_{\rm G} ^2 \, N (N-1) \, \frac{\epsilon _{\rm G}}{\hbar}
\ ,
\end{eqnarray}  
where $\alpha_{\rm G} ^2$ comes from the two vertices of the scattering process,
$N (N-1)$ is a combinatorial factor that accounts for the fact that each 
constituent interacts
with the remaining $N-1$ gravitons in the bound state, and $\epsilon _{\rm G}$ 
is the scale of the
the momentum transfer.
From the criticality condition~\eqref{criticality} and the scaling of $\epsilon 
_{\rm G}$
in~\eqref{eq:scaling-e}, we find 
\begin{eqnarray}
\Gamma \simeq \frac{1}{\sqrt{N} \, \ell _{Pl}} + \mathcal{O} \left( 
\frac{1}{N^{3/2} \, \ell _{Pl}} \right)
\ ,
\end{eqnarray}
which leads to the master equation for the quantum depletion
\begin{eqnarray}
\label{eq:depletion}
\dot{N} \simeq - \Gamma \simeq 
- \frac{1}{\sqrt{N} \, \ell _{Pl}} + \mathcal{O} \left( \frac{1}{N^{3/2} \, 
\ell _{Pl}} \right)
\ .
\end{eqnarray}
As discussed above, the quantum criticality of the system also relates the 
total 
mass of a black hole\label{BHref11}
to the number of constituents of the bound state.
Hence, from~\eqref{eq:scaling-M}, we can rephrase the depletion 
law~\eqref{eq:depletion} in terms
of the mass $M$, which yields
\begin{eqnarray}
\dot{M} \sim - \frac{M_{Pl} ^3}{\ell _{Pl} \, M^2} \sim - \frac{T_{\rm 
H}^2}{\hbar}
\ , 
\end{eqnarray}
with $T_{\rm H} \simeq \hbar/G_{N} M \simeq \epsilon_{\rm G}$.
In other words, the quantum depletion acts as a corpuscular precursor of the 
Hawking process
and it reduces to the latter in the semi-classical limit (i.e., for $N \to 
\infty$, $\ell _{Pl} \to 0$,
and $\sqrt{N} \ell _{Pl} = \mbox{constant}$).
 
Applying the HQM to this model, we finds that the expectation value of the 
gravitational radius
is very close to the classical one and that the fluctuations are suppressed 
according
to $\Delta R_{\rm H} / R_{\rm H} \sim 1/N$.
Indeed, let us consider, for the sake of simplicity, a collection of $N$ 
marginally bound scalar
particles acting as ``toy gravitons''.
The marginally bound condition can be implemented at the level of the 
single-particle Hilbert space
by requiring that  each particle's spectrum contain:
i) a ground state $\ket{\epsilon}$ of energy $\epsilon = M_{Pl} / \sqrt{N}$;
ii) a continuous part of quasi-gapless excitations $\ket{\omega _i}$,
meant to model the depleted states.
Then, since the collective state sources the deformation of the spacetime,
the single-particle wave-function for each constituent toy graviton reads
\begin{eqnarray}
\ket{\Psi _{\rm S} ^{(i)}} = \frac{\ket{\epsilon} + \gamma \ket{\psi _{\rm c} 
^{(i)}}}{\sqrt{1 + \gamma^2}}
\ ,
\end{eqnarray}
with
\begin{eqnarray}
\ket{\psi _{\rm c} ^{(i)}} = \int _{\epsilon} ^\infty {d} \omega _i \, f_i 
(\omega _i  | \epsilon) \ket{\omega _i}
\ ,
\end{eqnarray}
where $f_i (\omega _i  | \epsilon)$ represents the energy distribution function 
for the states
in the continuous part of the spectrum and $\gamma \geq 0$ denotes the 
likelihood of a constituent
to be in the continuum.
The $N$-particle wave-function of the system is given by the symmetrised tensor 
product
\begin{eqnarray}
\ket{\Psi _{\rm S}} =  
\bigotimes _{i = 1, \ldots, N} ^{\mathcal{S}} \ket{\Psi _{\rm S} ^{(i)}}
\ .
\end{eqnarray}
The total Hamiltonian $\widehat{H}$ will correspondingly have a spectrum that 
can be split into a discrete
part $\ket{M} = \otimes^{\mathcal{S}} \ket{\epsilon}$, such that $\widehat{H} 
\ket{M} = M \ket{M} = N \epsilon \ket{M}$,
and a continuous part $\ket{E}$, with $\widehat{H} \ket{E} = E \ket{E}$ and $E 
> 
M$.
If we then model $f_i (\omega _i  | \epsilon)$ with a thermal spectrum at a 
temperature $T \simeq \epsilon$,
we find that
\begin{eqnarray}
\frac{\Delta R_{\rm H}}{R_{\rm H}} \sim \frac{1}{N}
\ ,
\end{eqnarray}  
which means that for a black hole of large mass $M\simeq \sqrt{N}\,M_{Pl} 
\gg M_{Pl}$ the quantum corrections
to the classical results are suppressed by a very small factor of order $1/N$.
\section{Gravitational Collapse}
\label{citationalcollapss1}
A more precise derivation of~\eqref{criticality}, based one the Hamiltonian 
formulation
of gravity, was provided in~\cite{Casadio:2016zpl}, where the Newtonian 
potential was explicitly described as
a coherent state of many soft virtual gravitons~\cite{Mueck:2013mha}.
More precisely, let us consider a spherically symmetric distribution of matter 
of radius $R$ and total mass $M$.
In General Relativity we know that the Hamiltonian constraint for an 
asymptotically flat spacetime reads
\begin{eqnarray}
H = H_{\rm G} + H_{m} = M
\ ,
\end{eqnarray}
with $H_{\rm G}$ and $H_{m}$ representing the (super-)Hamiltonian of 
gravity 
and matter, respectively.
Since we are interested in the gravitational collapse, let us prepare the 
distribution of matter in such a way that,
in its initial configuration, all the components are infinitely far apart from 
each other.
In other words, in the initial configuration the gravitational interaction is 
negligible and $H_{\rm in} \simeq M$.
As the system shrinks down to the size $R$, assuming we can neglect the 
contribution of any emission of radiation,
the total energy in the final configuration reads
\begin{eqnarray}
H (R) \simeq M + K_{m} (R) + U_{\rm Grav} (R) + U_{\rm P} (R)
\ ,
\end{eqnarray}
with $K_{m} (R)$ the total kinetic energy of the matter content, $U_{\rm 
Grav} (R) \leq 0$ accounts for
the gravitational interaction, and $U_{\rm P} (R) \geq 0$ describes any 
non-gravitational repulsive effects
( e.g., Pauli exclusion principle).
At the corpuscular level we can split  $U_{\rm Grav} (R)$ into two 
contributions
\begin{eqnarray}
U_{\rm Grav} (R) = U_{\rm G} (R) + U_{\rm GG} (R)
\ ,
\end{eqnarray}
with $U_{\rm G} (R)$ keeping track of the gravitational interaction among the 
matter constituents
mediated by gravitons and $U_{\rm GG} (R)$ representing the energy content due 
to the gravitons
self-interactions.
At leading order, we can approximate $U_{\rm G} (R)$ with the Newtonian 
potential energy
\begin{eqnarray}
U_{\rm G} (R) \simeq M V_{\rm N} (R) \simeq - \frac{G_{N} \, M^2}{R}
\ .
\end{eqnarray} 
From a field-theoretic perspective, the Newtonian potential can be understood 
as 
a coherent state
of $N \simeq M^2 / M_{Pl} ^2$ soft virtual gravitons with typical size $\lambda 
_{\rm G} \simeq R$.
Hence, in this picture  $U_{\rm G} (R) \simeq - N \, \epsilon _{\rm 
G}$.
Since gravitons self-interact, we can then estimate $U_{\rm GG} (R)$ as the 
energy due to the interaction
between each constituent graviton with the collective state,
\begin{eqnarray}
U_{\rm GG} (R) \simeq 
N (- \epsilon _{\rm G} (R)) V_{\rm N} (R) \simeq 
\frac{G_{N}^2 \, M^3}{R^2}
\ ,
\end{eqnarray}
which clearly resembles a {\em post-Newtonian correction}.
Taking the limit $R \to R_{\rm H}$, we obtain
\begin{eqnarray}
\label{eq:criticality-2}
U_{\rm GG} (R_{\rm H}) \simeq - U_{\rm G} (R_{\rm H}) \simeq M
\ ,
\end{eqnarray}
which implies the quantum criticality condition~\eqref{criticality} and the 
scaling of the typical momenta
of the constituent gravitons~\eqref{eq:scaling-e}.
\section{Bootstrapping Newton}
A refinement of this study was then presented in~\cite{Casadio:2017cdv}, where 
an effective theory for the gravitational
potential of a static spherically symmetric distribution of matter was 
investigated with the addition of a term accounting
for the interaction of each graviton with the collective state.
Specifically, we can start from the Einstein--Hilbert action for some minimally 
coupled matter fields
\begin{eqnarray}
S = \int {d}^4 x \, \sqrt{- g} \left[- \frac{R}{16 \pi G_{N}} + 
\mathcal{L}_{m} \right]
\ .
\end{eqnarray}
Linearising the action and considering static, non-relativistic, and 
spherically 
symmetric matter profiles
characterised by an energy density $\rho = \rho (r)$, we find a Lagrangian of 
the form
\begin{eqnarray}
L_{\rm N} [V_{\rm N}] \simeq - 4 \pi \int _0 ^\infty r^2 \, {d} r \left[ 
\frac{(\partial _r V_{\rm N})^2}{8 \pi G_{N}}
+ \rho \, V_{\rm N}
 \right]
 \ ,
\end{eqnarray} 
which clearly leads to the Poisson equation and 
\begin{eqnarray}
\left. H_{\rm N} [V_{\rm N}] \right| _{\rm on-shell} = 
- \left. L_{\rm N}[V_{\rm N}] \right| _{\rm on-shell} \sim
\int _0 ^R r^2 \, {d} r \, \rho (r) \, V_{\rm N} (r)
\ ,
\end{eqnarray}
as expected.
This allows us to make the identification $U_{\rm G} (R) \sim \left. H_{\rm 
N}[V_{\rm N}] \right| _{\rm on-shell}$
at the leading order of the approximation.
At the {\rm next-to-leading order}, we acquire
\begin{eqnarray}
L_{\rm NLO} [V]
&\simeq&
- 4 \pi \int _0 ^\infty r^2 \, {d} r \left[ \frac{\left(\partial _r V 
\right)^2}{8 \pi G_{N}} 
\left( 1 - 4 V \right) + \rho V (1 - 2 V) \right]
\nonumber
\\
&\simeq&
L_{\rm N} [V] - 4 \pi \int _0 ^\infty r^2 \, {d} r \left[ J_V \, V + J_\rho 
\, \rho \right]
\ ,
\label{eq:NLO-lagrangian}
\end{eqnarray}
with $J_V = - (\partial _r V)^2 / (2 \pi G_{N})$ denoting the gravitational 
current and $J_\rho = - 2 V^2$
representing the higher-order correction to the matter part.
Note that the self-sourcing feature, at the heart of the corpuscular model, 
manifests itself through
$J_V \sim (\partial _r V)^2$.
The corresponding field equation then reads
\begin{eqnarray}
\label{eq:NLO-eq}
\triangle V = 4 \pi G_{N} \rho + \frac{2 (\partial _r V)^2}{1 - 4 V}
\ ,
\end{eqnarray}
and the corresponding next-to-leading-order expansion of the Hamiltonian 
functional, computed on-shell,
is therefore given by
\begin{eqnarray}
\left. H_{\rm NLO} [V] \right| _{\rm on-shell} \sim \int _0 ^\infty r^2 \, { 
d} r 
\left[
\rho V (1 - 4 V) - \frac{3\,(\partial _r V)^2 \, V}{2 \pi G_{N}}
\right]
\ .
\end{eqnarray}
The first term in the square brackets contains information about $\rho$ and can 
be understood as
the gravitational energy of the matter constituents mediated by gravitons,
\begin{eqnarray}
U _{\rm G} (R) \sim \int _0 ^R r^2 \, {d} r \rho V (1 - 4 V)
\ ,
\end{eqnarray}
whereas the second term keeps track of the self-sourcing and can be interpreted 
as the energy
associated with the gravitons self-interaction, namely
\begin{eqnarray}
U _{\rm GG} (R) \sim - \frac{3}{2 \pi G_{N}} \int _0 ^R r^2 \, {d} r 
(\partial _r V)^2 \, V
\ .
\end{eqnarray}
The classical analogue of the criticality condition~\eqref{eq:criticality-2} is 
recovered for
$\left. H_{\rm NLO} [V] \right| _{\rm on-shell} \simeq 0$, that for some 
well-behaved matter
distributions leads to $R \sim R_{\rm H}$.
\par
From a quantum-mechanical perspective, \label{quanteffectsfs3} denoting by $\Phi 
= V / \sqrt{G_{N}}$ 
and 
$J_{m} = 4 \pi \sqrt{G_{N}} \rho $,  \eqref{eq:NLO-lagrangian} leads
to an effective scalar field theory defined by the Lagrangian
\begin{eqnarray}
L[\Phi] = 4 \pi \int _{0} ^{\infty} r^2 \, {d} r 
\left[\frac{1}{2} \Phi \Box \Phi - J_{m} \Phi (1 - 2 \sqrt{G_{N}} \, 
\Phi) 
+ 4 \pi \sqrt{G_{N}} \, (\partial_\mu \Phi)^2 \Phi \right] \, .
\end{eqnarray}
Considering the linear order and recalling that $\Phi = \Phi (r)$, we then 
obtain  the classical field equation
\begin{eqnarray}
\triangle \Phi = J_{m} \, ,
\end{eqnarray}
as mentioned above.
Then, we can construct a {\em coherent state} $\ket{g}$ such that
\begin{eqnarray}
\langle g | \widehat{\Phi} | g\rangle = \Phi (r) \, ,
\end{eqnarray}
namely a state that reproduces the classical behaviour, and is defined by
\begin{eqnarray}
\widehat{a} _{\bm k} \ket{g} = e^{i \gamma _{\bm k} (t)} \, g _{\bm k} \ket{g}
\ ,
\qquad 
g _{\bm k} = - \frac{\widetilde{J} _{m} (k)}{\sqrt{2 \hbar k^3}}
\ .
\end{eqnarray}
We can easily infer the total occupation number for such a state, which reads
\begin{eqnarray}
N
\simeq
\int _{0} ^\infty \frac{k^2 \, {d} k}{2 \pi^2} |g_{\bm k}|^2 \sim 
\frac{M^2}{M_{Pl} ^2} \, \log \left( \frac{R_{\infty}}{R} \right)
\ ,
\end{eqnarray} 
where we considered the case of a source of mass $M$ and spatial extension $R$,
while $R_\infty$ denotes an infrared cut-off related to the size of the 
Universe.
In other words, we have shown that the Newtonian potential can be recovered 
from 
a coherent state of scalar toy gravitons
and that the occupation number of this state scales holographically with the 
size of the horizon.
This justifies the naive arguments in the previous section. 
\par
It is then interesting to consider what sort of modification of this coherent 
state would allow us to also reproduce the post-Newtonian correction, that is
\begin{eqnarray}
\sqrt{G_{N}} \langle g' | \widehat{\Phi} | g'\rangle = V_{\rm N} (r) + 
V_{\rm PN} (r)
\ ,
\end{eqnarray}
with $\ket{g'} \simeq \mathcal{N} \left(\ket{g} + \ket{\delta g} \right)$ such 
that
$\widehat{a} _{\bm k} \ket{g} = g_{\bm k} \ket{g} + \delta g_{\bm k} 
\ket{\delta 
g}$.
It turns out that, assuming that most of the toy gravitons belong to one mode 
$\bar\lambda \simeq R$, we find
\begin{eqnarray}
\delta g_{\bm k} \sim \delta g_{\bar{\bm k}} \sim - \ell _{Pl} \bar{k}^{5/2} 
|g_{\bar{\bm k}}|^2
\  ,
\end{eqnarray}
with $\bar{k} \sim 1/R$.
\par
A further improvement of this model was then carried out 
in~\cite{Casadio:2018qeh, Casadio:2019cux},
with the aim of understanding the effects of the matter pressure in this 
simplified setting. 
This last approach was termed {\em bootstrapped Newtonian gravity\/} and can be 
understood as an explicit
effective implementation for investigating the classicalisation of the 
gravitational interaction.
In more detail, if we are interested in describing general compact objects 
\label{compactobrefs4}, 
the 
sole energy density $\rho$
of a non-relativistic matter distribution is not enough to give a substantial 
insight on the fundamental
physics governing the gravitational collapse.
The simplest way to go beyond this restriction consist of introducing a 
pressure 
term in the model.
To this aim, we can replace the Lagrangian \eqref{eq:NLO-lagrangian} with
\begin{eqnarray}
L_{\rm BN} [V]
\simeq
L_{\rm N} [V] - 4 \pi \int _0 ^\infty r^2 \, {d} r \left[ J_V \, V + J_\rho 
\left( \rho + p \right) \right]
\ ,
\end{eqnarray}
where we replaced $\rho \to \rho + p$ in order to add a pressure $p$ that 
satisfies $\partial _r p = - (\partial _r V) (\rho +p)$.
Most notably, for this model it was found that a finite value of the pressure 
can support sources
of arbitrarily large compactness $\mathcal{C} \equiv G_{N} M / R$, thus 
concluding that there is
no equivalent of the Buchdahl limit in this effective model.
Note that, even though the strong energy condition is preserved in the limit 
$\mathcal{C} \gg 1$,
the dominant energy condition is violated in this regime.
This is rather consistent with the idea that a high-compactness configuration 
should be sourced
by a system in a fully quantum regime.  
\section{Quantum Compositeness of Gravity at Cosmological Scales}
 \label{quanteffectsfs2}
Similar arguments to the one presented for black holes  can be applied to 
cosmological spaces~\cite{Giusti:2019wdx, Dvali:2013eja}.
Indeed, as argued in~\cite{Casadio:2017twg}, this can be achieved by recalling 
that the Friedmann equation
provides the Hamiltonian constraint for cosmology, allowing us to extend the 
arguments in~\cite{Casadio:2016zpl}
to such spaces.
More precisely, the Friedmann equation
\begin{eqnarray}
H^2 = \left( \frac{\dot a}{a} \right)^2 = \frac{8 \pi G_{N}}{3} \rho_m
\end{eqnarray}
implements the Hamiltonian constraint for Robertson-Walker geometries, namely 
it 
corresponds to
\begin{eqnarray}
\mathcal{H} = \mathcal{H}_{\rm G} + \mathcal{H}_{m} = 0 \, ,
\end{eqnarray} 
where the total Hamiltonian $H_{\rm Tot} \sim \mathcal{N}(t) \, \mathcal{H}$, 
with $\mathcal{N}(t)$ the lapse function. 
\par
Now, considering the de~Sitter solution of the Einstein field equations with a 
cosmological constant $\Lambda$,
we find 
\begin{eqnarray}
\label{eq:friedmann-ds}
H^2 = \left( \frac{\dot a}{a} \right)^2 = \frac{8 \pi G_{N}}{3} \rho 
_\Lambda
\ ,
\qquad 
\rho _\Lambda = \frac{\Lambda}{8 \pi G_{N}}
\ , 
\end{eqnarray}
which corresponds to an exponentially growing scale factor leading to a 
spacetime with an horizon at
\begin{eqnarray}
L_{\Lambda} = \sqrt{\frac{3}{\Lambda}} \, .
\end{eqnarray}
Upon integrating \eqref{eq:friedmann-ds} over the volume inside $L_{\Lambda}$, 
we obtain
\begin{eqnarray}
L_{\Lambda} \simeq G_{N} L_{\Lambda} ^3 \rho _\Lambda \simeq G_{N} 
M_\Lambda
\ ,
\end{eqnarray}
with $M_\Lambda \simeq L_{\Lambda} ^3 \rho _\Lambda$ denoting the fraction of 
dark energy
contained within this volume.
This last expression is akin to the one for the gravitational radius for a 
Schwarzschild black hole.
Thus, we are tempted to try and extend the same Hamiltonian analysis carried 
out 
for black holes
and compact objects to the case of cosmological spaces.
To this aim, let us consider a universe filled only with dark energy 
(considered 
as a purely gravitational effect),
 i.e., $\mathcal{H}_{m} = 0$. 
Then, we are left with the constraint
\begin{eqnarray}
\mathcal{H} = \mathcal{H}_{\rm G} \simeq 0
\ .
\end{eqnarray}    
In the corpuscular picture we can split $\mathcal{H}_{\rm G}$ into two 
contributions:
\begin{itemize}
\item[(i)]
the energy associated with the coherent state of $N$ marginally bound gravitons,
\begin{eqnarray}
U_{\rm G} \simeq - N \epsilon _{\rm G} \simeq
- N \, \frac{\hbar}{L_\Lambda}
\ ;
\end{eqnarray}
\item[(ii)]
the energy due to the gravitational interaction of each constituent with the 
collective (Newtonian)
potential generated by the remaining $N-1$ quanta,
\begin{eqnarray}
U_{\rm GG} \simeq - N \epsilon _{\rm G} V_{\rm N}(L_\Lambda) \simeq
N^{3/2} \, \frac{\hbar \, \ell _{Pl}}{L_\Lambda ^2}
\ .
\end{eqnarray}
\end{itemize}
Hence, the Hamiltonian constraint reads
\begin{eqnarray}
\mathcal{H} = \mathcal{H}_{\rm G} \simeq U_{\rm G} + U_{\rm GG} \simeq 0,
\end{eqnarray}
and yields
\begin{eqnarray}
L_{\Lambda} \simeq \sqrt{N} \, \ell _{Pl}
 .
\end{eqnarray}
This also implies the criticality condition~\eqref{criticality}, since 
\begin{eqnarray}
\epsilon _{\Lambda} \simeq \frac{\hbar}{L_\Lambda} \simeq 
\frac{M_{Pl}}{\sqrt{N}}
\ ,
\end{eqnarray}
hence $\alpha _{\rm G} = \epsilon ^2 _{\rm G} / M_{Pl} ^2 \simeq 1 / N$, as in 
the case of black holes.
This allows interpretation of the de~Sitter space and, in general, any 
cosmological spacetime as the result
of a bound state of a large number of soft virtual gravitons.
The peculiarity of the (quasi) de~Sitter case is that gravitons in it should be 
at a critical point.
\par
Along this line of reasoning, if we assume that the quasi-de~Sitter state, which 
is usually employed to model the inflationary
phase of the Universe, is the result of a marginally bound configuration of a 
large number of soft off-shell gravitons,
it is possible to show that the Starobinsky 
model~\cite{Starobinsky:1980te,Starobinsky:1979ty} naturally emerges from the 
corpuscular
picture without the need of an inflaton field~\cite{Casadio:2017twg, 
Giugno:2018zty}.
Besides, if one also takes into account the effect of the quantum depletion, it 
turns out that the pure de~Sitter
space is excluded at the quantum mechanical level~\cite{Dvali:2018fqu, 
Casadio:2017twg,Casadio:2019rrc}.
On a different note, if one applies the corpuscular theory to model the 
late-time large scale structure of the
Universe, some of the phenomena, usually explained by assuming the existence of 
{\em dark matter\/},
can be simply understood as the response of a cosmological reservoir of 
gravitons,
responsible for the accelerated expansion, to the presence of local impurities,
namely baryonic matter~\cite{Cadoni:2017evg, Cadoni:2018dnd,Tuveri:2019zor}.
\section{Outlook}
There are three main directions of development for the above project, 
respectively, regarding:
\begin{description}
\item[a)]
compact astrophysical sources and black holes;\label{BHref12}
\item[b)]
cosmological models of the early and present Universe, and
\item[c)]
the fundamental classicalisation of the gravitational 
interaction.\label{gphenomenologicalprfs1}
\end{description}
 
First of all, the bootstrapped Newtonian model of compact sources appears 
amenable to further
investigate the quantum corpuscular nature of very compact astrophysical sources 
and their transition
to black hole states.
From the more theoretical point of view, we expect to deepen our understanding 
of the nature of
real black holes by continuing to refine the quantum description of the 
bootstrapped potential generated by
static sources in the form of coherent states of gravitons.
On the more phenomenological side, we expect that the implementation of specific 
equations of state to describe
the matter source will lead to differences with respect to the predictions of 
General Relativity.
 
From the cosmological point of view, the effective dark matter phenomenology obtained as a reaction
of the cosmological (quasi) de~Sitter condensate to the presence of localised matter sources will 
benefit from a more detailed description of the latter.
{\em Vice versa}, the unified description of the bootstrapped potential for local sources in terms of a coherent state
built out of gravitons in the cosmological condensate might shed more light on the nature of both dark matter
and black holes.

Finally, pushing the model of compact sources to their quantum limit, roughly defined as states
with a small number of gravitons in the bootstrapped gravitational potential, should provide an explicit construction
of the classicalisation process and the final stages of black hole evaporation, as well as a way to clarify
the role played by the matter.
%
%
%
%

 \newpage

\phantomsection
\addcontentsline{toc}{part}{\bf Part III: Cosmology and Observational 
Discriminators}
\begin{center}
{\Huge \bf Part III:  Cosmology and Observational 
Discriminators}
\end{center}
\begin{center}
Editors: 
Ruth Lazkoz and Vincenzo Salzano
\end{center}






\begingroup
\let\clearpage\relax 
\chapter[Introduction to Part III]{Introduction to Part III}

{\em  Ruth Lazkoz, Vincenzo Salzano }\\

The contributions that can be found in this section are aimed to characterise 
what has been the main nature of the Working Group 3 (WG3) within the CANTATA 
Cost Action: a ``cross-group'', a link between members and participants from all 
working groups. Although the main footprint of the WG3 has always been strictly 
connected to the \textit{direct} use of observations, the interplay and the 
close synergy with WG1 and WG2 has characterised the hybrid nature of its 
scientific outcomes, where theory and data analysis have never gone on separate 
roads, but have been always interconnected.

The contributions from this chapter are no exception to this rule. We have 
decided to select those scientists whose research trajectory might somehow 
summarise part of the scientific investigation that has been performed under 
CANTATA coverage, who have explored new interesting ways to test gravity, and 
who could outline a sort of road map to follow in the very near future, which 
looks very promising for observational cosmology. In   forthcoming years we 
expect 
many terrestrial and space-based advanced surveys to be launched and/or become 
fully operative (among them, \textit{Euclid} and SKA), throwing us directly into 
a new highly-upgraded era of \textit{precision cosmology}. All of them will 
provide us with data of unprecedented precision about the large-scale 
structure, 
giving us the possibility for very accurate tests of gravity (and modified 
gravity theories, specifically) on scales spanning many orders of magnitude. A 
\textit{phenomenological} summary of all the possible insights we can gain from 
using these data will be provided in Sec.~\ref{sec:Akrami}. One of the main 
outputs of the above mentioned surveys will be, among other things, data related 
to the clustering of galaxies. In Sec.~\ref{sec:Bonvin}, we have left our 
contributor to explore some more subtle effects, which might influence those 
data, i.e., \textit{relativistic effects} on the number counts, which emerge as
powerful complementary tool, to be used in addition to more standard and 
well-established ones.

A further distinctive element that has characterised the activity of this WG is 
the research performed on the numerical and computational side. In 
Sec.~\ref{sec:Frusciante} we have a clear example of the entanglement between 
theory and numerics: the contributors introduce a theoretical framework, the 
Effective Field Theory, which has become   very fruitful in recent
years by providing interesting and stringent constraints on both (standard) dark 
energy models and modified gravity theories, thanks mainly to some numerical 
codes (now widely used in the cosmological community), which have been 
developed 
and improved by the same members of CANTATA, and which have helped to optimise 
the calculation of the most important quantities that are needed to apply such 
framework to real data  in particular, cosmological perturbations.

Sticking more closely to the observational arena, we have to point out that the 
cosmological debate nowadays is particularly heated up by the so-called 
\textit{observational tensions}, namely, statistically-grounded conflicts 
between different, complementary probes. The most (in)famous is the 
\textit{$H_0$ tension}, based on a discrepancy between the measurement of the 
expansion rate of our Universe performed locally (by means of Cepheids and Type 
Ia Supernovae) and the same quantity inferred from cosmological analysis using 
high redshift/early times data (the Cosmic Microwave Background radiation as 
measured by the most update project in that field, the \textit{Planck} 
telescope). This topic is dealt with in Sec.~\ref{sec:DiValentino}. The other 
less frequently reported tension, but equally controversial and decisive, is the 
so-called \textit{$\sigma_8$ tension}: again, a statistically significant 
discordance is found when comparing the estimation of this parameter (related to 
matter density fluctuations, thus, to the evolution of cosmological 
perturbations) from \textit{Planck} with low redshift data from galaxy 
clustering. This problem is discussed more closely in 
Sec.~\ref{sec:Perivolaropoulos}.

A special role in this series of contributions has been necessarily devoted to a 
field which, although old in its theoretical background and in its practical 
design, has reached full maturity only in the most recent years, eventually 
accomplishing extraordinary results (the Physics
Nobel Prize in 2017): \textit{gravitational wave astronomy}. A full section, 
Sec.~\ref{sec:Ezquiaga}, is devoted to the enormous implications that this 
field has on constraining alternative theories of gravity and on the future 
role 
played by gravitational waves astronomy in our understanding of how gravity 
works. Another kind of observational data that seems to be very promising in 
this perspective is   \textit{cosmological weak lensing}. Although not as new 
as the previous one, this is a very sensitive and hard-to-retrieve probe, which 
will become crucial in the near future, thanks to some of the surveys we have 
introduced above, which will be   accurate  enough, and will observe such a 
huge 
amount of galaxies as  to make it feasible at unique levels. A discussion about 
weak 
lensing will be found on Sec.~\ref{sec:Pettorino}.

Finally, our two last contributions will focus not on cosmological scales, nor 
on the statistical properties of the gravitational structures, but will describe 
which constraints or information about modifications of gravity can be derived 
from analysing clusters of galaxies and galaxies as single objects. In 
Sec.~\ref{sec:Saltas}, the focus of the attention is galaxy clusters, the 
largest clearly-observable self-gravitating structures for which we can retrieve 
multi-messenger astronomical data covering a wide range of wavelengths and 
complementary information. Instead, in Sec.~\ref{sec:Mota}, we will mainly use 
numerical simulations to study possible characteristic features that should be 
imprinted in galaxies and clusters of galaxies by modified gravity theories, 
and 
which could help to eventually discriminate among General Relativity and 
alternative gravities.

\endgroup






\chapter[Phenomenological Tests of Gravity on Cosmological Scales]
{Phenomenological Tests of \\ Gravity on Cosmological Scales}
\label{sec:Akrami}
\label{gphenomenologicalprfs2}

{\em Yashar Akrami, Matteo Martinelli}







\section{Cosmological Tests of Gravity}
\label{sec:tests}\label{cosmosurveysefs1}
The late-time accelerated expansion of the Universe \cite{Riess:1998cb,Perlmutter:1998np} (see Refs.~\cite{Caldwell:2009ix,Weinberg:2012es,Joyce:2014kja,Bull:2015stt} for recent comprehensive reviews on the subject) is still a mystery and needs to be addressed by theoretical physics. This cosmic acceleration could be driven by the cosmological constant or some dynamical dark energy, or could be a consequence of deviations from General Relativity (GR); see, e.g., Refs.~\cite{2010deto.book.....A,Clifton:2011jh,Koyama:2015vza,Ishak:2018his,Ferreira:2019xrr} for reviews.

There are reasons to believe that GR may not be the ultimate theory of gravity on large scales, and despite its striking and continued observational successes, there is a persistent interest in extending GR at cosmological scales. The need for unknown ingredients of dark matter and dark energy to explain the observed cosmic evolution, and the unsolved and related \emph{problem of the cosmological constant}~\cite{Weinberg:1988cp,Martin:2012bt}, all suggest that gravity might be different in the infrared (IR), as it should be in the ultraviolet (UV) in the quest for a theory of quantum gravity. More importantly, GR has been verified only over a surprisingly small range of scales, and there is no reason to presume its validity on all scales; the theory {\it has} to be tested in all regimes in order to avoid unjustified extrapolation. Finally, if GR is ``the'' theory of gravity on large scales, we want to know what makes it so special, why gravity is described by only massless spin-2 particles, and why other degrees of freedom are forbidden in the gravitational sector.

Several alternative (or modified) theories of gravity on the cosmological scales have been proposed, some of which are able to explain cosmic acceleration~\cite{2010deto.book.....A,Clifton:2011jh,Koyama:2015vza,Ishak:2018his,Ferreira:2019xrr}. While the efforts for building new models continue and the existing cosmological surveys have already provided us with important constraints on modified theories of gravity, several upcoming surveys are expected to eventually pin down the nature of gravity. There have been remarkable efforts in performing systematic studies of gravity beyond GR on large scales, and various techniques have been developed for testing the models in different regimes in connection with the wealth of information that is expected to be provided by the upcoming surveys, especially those observing the large-scale structure (LSS) of the Universe.

We are all excited by the recent milestones that observations of gravitational waves (GWs)~\cite{TheLIGOScientific:2017qsa,Goldstein:2017mmi} and black-hole images have marked for physics, especially their strong implications for theories of gravity, many of which are now significantly disfavoured or constrained by GWs~\cite{Creminelli:2017sry,Sakstein:2017xjx,Ezquiaga:2017ekz,Baker:2017hug,Nojiri:2017hai,Boran:2017rdn,Amendola:2017orw,Crisostomi:2017lbg,Langlois:2017dyl,Gumrukcuoglu:2017ijh,Heisenberg:2017qka,Kreisch:2017uet,Dima:2017pwp,Peirone:2017ywi,Crisostomi:2017pjs,Linder:2018jil,Kase:2018iwp,Battye:2018ssx,Akrami:2018yjz} (see also Refs.~\cite{Lombriser:2015sxa,Brax:2015dma,Lombriser:2016yzn,Pogosian:2016pwr,Bettoni:2016mij}) -- this is perhaps a call to go back to the basics, and rethink the justifications behind the models of gravity that we have constructed over the past two decades. The incredible knowledge that GWs have provided and will continue to provide will, however, need to be complemented by the precision measurements of the LSS, as one undoubtedly leading cosmological probe in coming years, in order to achieve a full understanding of gravity.

Cosmological surveys probe gravitational interactions in a large range of scales and in different regimes; the larger the scale of interest, the more linear the evolution and growth of the cosmological structure. We can therefore define four regimes of cosmological interest that can be probed by LSS surveys: {\it ultra-large scales}, i.e., scales that are comparable to the size of the horizon with the structure that can be treated linearly, {\it intermediate scales} that are well inside the horizon but are still linear, subhorizon scales that are {\it mildly nonlinear}, and very small scales where the structure is {\it fully nonlinear}, i.e., the regime in which linear perturbation theory breaks down completely. In the next subsections, we focus on the first two regimes, where the cosmological structure is studied through the theory of linear perturbations. We then briefly discuss the mildly and fully nonlinear regimes in Section~\ref{sec:nonlin}.

\subsection{Large Scales and the Linear Regime: Phenomenological Departures from GR}
\label{sec:phenom}
\label{LSSefs}

On sufficiently large scales, we can apply the theory of linear cosmological  
perturbations to the large-scale structure of the Universe. This can be done by 
perturbing the Friedman-Lema\^{i}tre-Robertson-Walker (FLRW) metric as (in the 
Newtonian gauge)
\begin{equation}
    g_{\mu\nu}=\text{diag}[-(1+2\Psi), a^2(1-2\Phi)\delta{ij}],
\end{equation}
where $\Phi$ and $\Psi$ are the so-called gravitational (or Bardeen) potentials and $a$ is 
the  scale factor of the Universe. The energy densities and pressures of 
different components $X$ are also perturbed,
\begin{equation}
    \rho_X=\bar\rho_X(1+\delta_X) \quad \mathrm{and} \quad p_X=\bar p_X(1+\delta^p_X),
\end{equation}
where $\bar\rho_X$ and $\bar p_X$ are, respectively, the background density  
and 
pressure for the component $X$. In GR, we can combine the $00$ and $0i$ 
components of the linearised  Einstein equations to obtain the so-called 
(cosmological) {\it Poisson equation} for a mode $k$ in Fourier space,
\begin{equation}\label{eq:PoissonGR}
    -k^2\Psi=\frac{\kappa^2}{2}\delta^\ast\rho,
\end{equation}
where
\begin{equation}
\delta^\ast\rho=\sum_X\bar\rho_X[\delta_X+3(1+w_X)\frac{H}{a}\frac{\theta_X}{k^2}],
\end{equation}
with $\kappa^2=8\pi G$ the gravitational constant, $H$ the Hubble parameter, and $\theta_X=\vec\nabla.\vec v_X$, where 
$\vec v_X$ is the  peculiar velocity for the component $X$ with the equation of 
state parameter $w_X$. The traceless, transverse component of the Einstein 
equations yields the so-called {\it slip relation},
\begin{equation}\label{eq:slipGR}
    \Phi-\Psi=0.
\end{equation}
Eqs.~(\ref{eq:PoissonGR}) and~(\ref{eq:slipGR}), together with the perturbed 
conservation equations  for the components $X$, fully describe the cosmological 
evolution at the linear level in standard cosmology.

In theories of gravity beyond GR, the forms of Eqs.~(\ref{eq:PoissonGR}) 
and~(\ref{eq:slipGR}) are modified  in general, as new degrees of freedom enter 
the perturbation equations. In order to study the evolution of structure in 
modified gravity, one can derive similar linearised equations for any models of 
interest and use those equations to constrain the models by observations. One 
can also work with theoretically motivated parameterisations of deviations from 
GR, e.g., through the effective field theory (EFT) of dark energy and modified 
gravity~\cite{Creminelli:2008wc,Gubitosi:2012hu,Bloomfield:2012ff,
Gleyzes:2013ooa,Gleyzes:2014dya} or covariant parameterisations such as in 
Horndeski theory and beyond~\cite{Bellini:2014fua}. Here, however, we focus on a 
popular phenomenological parameterisation directly at the level of the 
linearised Einstein equations, i.e., the Poisson equation and the slip relation, 
Eqs.~(\ref{eq:PoissonGR}) and~(\ref{eq:slipGR}). We later discuss the advantages 
and disadvantages of this approach.

Let us start with subhorizon and linear perturbations, where the so-called {\it 
quasi-static  approximation} is valid. This regime covers a large fraction of 
the scales probed by current and future LSS surveys. In this subhorizon, 
quasi-static regime, for any perturbative quantity $X$ in the equations we 
assume
\begin{equation}
    \ddot X \sim H\dot X \sim H^2X \ll k^2X,
\end{equation}
where $k$ is the wavenumber of the mode under consideration and an overdot 
denotes a derivative  with respect to time. This approximation dramatically 
simplifies the equations and has been shown to hold for almost all popular 
theories of gravity. One can then remove the extra degrees of freedom (which 
usually appear in extensions to GR) from the linearised Einstein equations and 
replace their effects by two functions of space and time, $\mu(a,k)$ and 
$\eta(a,k)$, in the Poisson equation~(\ref{eq:PoissonGR}) and the slip 
relation~(\ref{eq:slipGR}), respectively, to obtain
\begin{align}
        -k^2\Psi&=\frac{\kappa^2}{2}\mu(a,k)\delta^\ast\rho, 
\label{eq:PoissonMG} \\ \Phi&=\eta(a,k)\Psi.\label{eq:slipMG}
\end{align}
Here, $\mu$ quantifies the modification to the effective Newton's constant and 
$\eta$ is called  the gravitational 
slip~\cite{Hu:2007pj,Bertschinger:2008zb,Amin:2007wi,Pogosian:2016pwr}. The two 
quasi-static parameters are generally functions of both time and scale. The 
scale dependence, however, becomes important typically on very small scales, 
i.e., deeply in the nonlinear regime, or on scales close to the horizon. 
Therefore, for the intermediate scales, as we defined above, $\mu$ and $\eta$ 
can be considered as functions of time only. While the parameter $\mu$ sources 
the growth rate of structure, it is the combination of $\mu$ and $\eta$,
\begin{equation}\label{eq:SigmaDef}
    \Sigma\equiv\frac{1}{2}\mu(1+\eta),
\end{equation}
that sources the weak gravitational lensing through the combination of the Poisson 
equations for $\Phi$ and $\Psi$, leading to
\begin{equation}\label{eq:SigmaEq}
    -k^2(\Phi+\Psi)=\kappa^2\Sigma(a,k)\delta^\ast\rho,
\end{equation}
where the combination $\Phi+\Psi$ is known as the {\it Weyl potential}.

The parameterisation of departures from GR in terms of the quantities $\mu$, 
$\eta$  and $\Sigma$ has the advantage of providing us with a simple, 
model-independent way of testing gravity on cosmological scales. On the one 
hand, any deviations from $\mu=1$ and $\eta=1$ at any time or scale are 
signatures of violations of GR, and this therefore gives a direct way of testing 
the standard theory. On the other hand, any model of modified gravity provides 
particular forms for these functions in terms of $k$ and $a$, which can then be 
used in constraining the model with observational data. The forms of $\mu$ and 
$\eta$ can be obtained for any modified gravity models of interest.

There are, however, issues with this phenomenological approach. First of all, the 
functional  forms of $\mu$ and $\eta$, especially in terms of $k$, are typically 
the same for most interesting theories of gravity~\cite{Baker:2014zva}, and there 
are therefore degeneracies between different theories. For example, the 
functions $\mu$ and $\eta$ obtained for Horndeski models of scalar-tensor 
gravity have the general forms\cite{DeFelice:2011hq}
\begin{align}
    \mu=\frac{a_1+a_2k^2}{b_1+b_2k^2},\nonumber\\
    \eta=\frac{b_1+b_2k^2}{c_1+c_2k^2},
\end{align}
where $a_i$, $b_i$ and $c_i$ are functions of background quantities and 
therefore of time.  A similar structure has been obtained for bimetric theories 
of gravity~\cite{Solomon:2014dua,Konnig:2014xva}.
This means that if we find a favoured non-GR form for any of these functions 
from observations,  it is not easy to uniquely identify the underlying theory. 
In addition, one needs to parameterise the forms of $\mu$ and $\eta$ when 
constraining the functions observationally, and these parameterisations add 
extra levels of arbitrariness to the analysis. The improved constraining power 
of future data has, however, the potential to solve this issue, allowing us to 
constrain these free functions through reconstruction techniques without the 
need to assume any specific parameterisation (see, e.g., 
Ref.~\cite{Casas:2017eob}). Finally, as we discussed before, the $\{\mu,\eta\}$ 
parameterisation of departures from GR is valid only for the quasi-static 
regime, which holds only for subhorizon modes.

Since most alternative theories of gravity in the IR aim at explaining cosmic 
acceleration,  which corresponds to horizon-scale physics, they directly affect 
observables on ultra-large (horizon-size) scales. The fact that physics on such 
scales is linear makes them of particular interest. Even though the starting 
point for our phenomenological parameterisation of modified gravity was based on 
subhorizon scales where the quasi-static approximation is believed to be valid, 
there are arguments in the literature also justifying the use of a similar 
parameterisation for ultra-large-scale modes~\cite{Baker:2015bva}. These 
modes, however, suffer from one particular complication, and that is the problem 
of {\it cosmic variance}, which generates intrinsic and relatively large 
uncertainties on the measured quantities because of the low number of available 
modes on such large scales. Combining the data from different large-angle 
surveys taken at different redshifts will, however, help us tackle cosmic variance 
and better constrain theories of gravity. In addition, {\it relativistic 
effects} are important on horizon-size scales and cannot be neglected. This adds 
an extra level of complication when working with those modes. Finally, various 
other approximations that we usually make in cosmological data 
analysis\label{cosmoldatanalysefs1} to 
reduce the computation time may introduce additional uncertainties to the 
results that are comparable with departures from GR and signatures of modified 
gravity on ultra-large scales \cite{dalal:inprep}.

In the next section, we review the cosmological data and various probes that are 
 used to test GR on cosmological (and linear) scales,\label{testsGRefs2} and to 
place constraints on 
alternative theories of gravity.

\subsection[Cosmological Observables and Phenomenological  
Constraints]{Cosmological Observables and Phenomenological \\ Constraints}
\label{sec:observables}
Deviations from the Theory of General Relativity have imprints  on several 
cosmological observables,
which may therefore be used to constrain both phenomenological departures from 
standard gravity and possible alternative theories. In this section, we focus on 
the signatures that modified theories of gravity would have on  cosmological 
observables, especially those of interest to current and upcoming surveys. In 
particular, we discuss the impacts of modifying gravity on the observations of 
large-scale structure, as well as their effects on the propagation of the cosmic 
microwave background photons.

\subsubsection{Cosmic Microwave Background}
\label{sec:CMB}\label{CMBrefs3}
The cosmic microwave background (CMB) is a relic radiation coming from  the 
epoch of recombination, when the free electrons in the Universe reached a sufficiently low abundance to allow photons to travel freely; this point in time, when the 
photons decoupled from free electrons, is known as the {\it last scattering 
surface}. Although extremely homogeneous, the tiny differences in the 
temperature of CMB photons ($\Delta T/T_{\rm CMB}\approx 10^{-5}$ K for an 
average temperature of $T_{\rm CMB}\approx2.726$ K) carry a significant amount 
of information. Before recombination, photons were coupled via Thomson 
scattering to the primordial plasma, and deviations from the average temperature 
$T_{\rm CMB}$ can therefore be connected to primordial perturbations of the 
homogeneous matter distribution. As the CMB is the observable that allows us to 
look furthest in the past, such information is crucial to reconstruct the 
initial conditions of our Universe.

The information brought by CMB anisotropies is encoded in the angular 
power spectrum
\begin{equation}
C^{TT}(\ell) \equiv \langle |a_{\ell m}|^2\rangle,\label{eq:ClSH}
\end{equation}
where the $a_{\ell m}$ are the coefficients of the spherical harmonics expansion 
 of temperature perturbations
\begin{equation}
\frac{\Delta T}{T}({\mathbf x},\tau)=\sum^\infty_{\ell=1} 
\sum^\ell_{m=-\ell}a_{\ell m}({\mathbf x},\tau)Y_{\ell m}(\hat{n}),
\end{equation}
with $Y_{\ell m}$ the spherical harmonics, $\tau$ the conformal time, 
and $\ell$  and $m$ conjugates of the direction of incoming photons 
$\hat{n}$. $\langle...\rangle$ in Eq.~(\ref{eq:ClSH}) is the average over different values 
of $m$.

Theories alternative to GR are usually only expected to give deviations from the 
standard behaviour at low redshifts, as they are introduced, in most 
cases, to explain the late-time accelerated expansion of the Universe. For that 
reason, modified theories of gravity are not expected in general to impact the 
physics of recombination, and therefore to affect the primary anisotropies of the 
CMB described above.

Nevertheless, CMB photons are influenced by a number of secondary effects in their 
propagation from the  last scattering surface to the observer; in particular, 
two effects are relevant for investigation of alternative theories of gravity; 
the {\it integrated Sachs-Wolfe} effect (ISW) and gravitational lensing.  
\label{ISWref1}

The ISW effect is the cumulative variation of photons' energy due to temporal 
changes of gravitational  potentials along their path, which can add or 
substract energy,
\begin{equation}
 \frac{\Delta T}{T}=\int_{\tau_0}^\tau d\tau (\Phi'+\Psi')\, ,
\end{equation}
where primes denote derivatives with respect to $\tau$. As this is  an 
integrated effect, late-time deviations from the GR evolution of the Bardeen 
potentials $\Phi$ and $\Psi$ are imprinted on the temperature anisotropies of 
the CMB at large scales, and such deviations can therefore be observed to 
constrain departures from GR.

The second effect of interest is the weak gravitational lensing of CMB photons;  
along their path, these photons are affected by the gravitational potentials 
generated by the distribution of matter in the Universe, which leads to deflections 
of their path. Such an effect impacts the baryon acoustic oscillations imprinted 
on the temperature power spectrum $C^{TT}(\ell)$, decreasing its amplitude. By 
observing the effects on CMB spectra induced by gravitational lensing, it is 
possible to reconstruct $C^{\phi\phi}(\ell)$~\cite{Okamoto:2003zw}, the power spectrum of the lensing potential $\phi$, which can then be used to constrain 
theories of gravity.

\subsubsection{Galaxy Clustering and Redshift-space Distortions}
\label{sec:GCRSD}

The evolution of matter distribution, from the primordial perturbations in the 
matter  density until the current distribution, is driven by the laws of 
gravity, determining the collapse of overdensities into the structure observed 
in the Universe. The matter perturbation, indicated in Fourier space as 
$\delta_\mathrm{m}(k,z)\equiv(\rho_\mathrm{m}(k,z)-\bar{\rho}_\mathrm{m})/\bar{\rho}_\mathrm{m}$, evolves in time in GR as

\begin{equation}\label{eq:deltaGR}
    \ddot{\delta}_\mathrm{m}(k,z)+2H(z)\dot{\delta}_\mathrm{m} (k,z)-\frac{3H_0^2\Omega_\mathrm{m}^{0}(1+z)^3}{2}\delta_\mathrm{m}(k,z)=0,
\end{equation}
where the label `$\mathrm{m}$' denotes non-relativistic matter with the present density parameter $\Omega_\mathrm{m}^{0}$, $H_0$ is  the 
present value of the Hubble parameter and $z$ is redshift. As the evolution of 
$\delta_\mathrm{m}$ is connected with the Poisson equation described in the previous 
section, by employing the parameterisation of deviations from GR introduced in 
Eq.~(\ref{eq:PoissonMG}), we can obtain a generalized form
\begin{equation}\label{eq:deltaMG}
    \ddot{\delta}_\mathrm{m}(k,z)+2H(z)\dot{\delta}_\mathrm{m} (k,z)-\frac{3H_0^2\Omega^{0}_\mathrm{m}(1+z)^3}{2}\mu(k,z)\delta_\mathrm{m}(k,z)=0.
\end{equation}
Modifications of gravity, therefore, enter this evolution through the  
phenomenological function $\mu(k,z)$, and additionally, through the 
modification of the background expansion rate $H(z)$.

Given a specific cosmological model, we can therefore obtain the matter  power 
spectrum $P_\mathrm{m}(k,z)$ defined as
\begin{equation}
    P_\mathrm{m}(k,z)\propto\langle\delta_\mathrm{m}(k,z)\delta_\mathrm{m}(k,z)\rangle.
\end{equation}
From Eq.~(\ref{eq:deltaMG}) it is clear how departing from GR affects  the 
evolution of $\delta_\mathrm{m}$ and, consequently, how it impacts $P_\mathrm{m}(k,z)$. This 
quantity, however, refers to the evolution of the total matter density, 
including dark matter, and therefore cannot be directly observed. Nevertheless, 
galaxy surveys provide observations of the distribution of galaxies, and can 
therefore obtain the observed power spectrum $P_{\rm obs}(k,z)$, which can be 
used as a proxy to the total matter distribution.
The relation between these two spectra contains several physical effects, which 
also carry cosmological information~\cite{Blanchard:2019oqi}:
\begin{itemize}
    \item The galaxy bias $b(z)$, which relates the linear galaxy and matter  
power spectra as
    \begin{equation}
        P_\text{g}(k,z)=b^2(z)P_\mathrm{m}(k,z).
    \end{equation}
    This term comes from the fact that the galaxy distribution is a biased  
tracer of the underlying matter distribution; this bias term can in principle 
be  \label{Darkmatterharef1}
obtained from models of galaxy evolution and dark matter halo collapse (see, 
e.g., Ref.~\cite{Desjacques:2016bnm} for an extensive review), thus it can in 
principle be affected by modifications of the standard theory of gravity.
    \item The non-cosmological contribution to galaxy redshifts due to their  
peculiar velocities. Assuming redshift to be due only to the cosmological 
expansion introduces distortions in the density field that imprint a specific 
pattern in the power spectrum, known as {\it redshift space distortions 
(RSD)}. However, peculiar velocities are sourced by the density fields and can 
therefore themselves provide cosmological information. These velocities depend 
on the growth rate $f(z)$, defined as
    \begin{equation}
        f(z)\equiv-\frac{d\ln D(z)}{d\ln(1+z)},
    \end{equation}
    with $D(z)$ the growth factor defined through $\delta_\mathrm{m}(k,z)/ 
\delta_\mathrm{m}(k,z_i)=D(z)/D(z_i)$, where the label $i$ denotes `initial values'.

    In the linear regime, this allows us to define a redshift-space power 
spectrum,  $P_{\rm rs}(k,z)$ \cite{Kaiser:1987qv}, \label{sigmaref4}
    \begin{equation}
    P_{\rm rs}(k,\gamma;z) =
    \left[b(z)\sigma_8(z) +f(z)\sigma_8(z)\gamma^2\right]^2 
\frac{P_\mathrm{m}(k,z)}{\sigma_8^2(z)}\, ,
    \end{equation}
    where $\sigma_8(z)$ is the root mean  square of linearly evolved density 
fluctuations in spheres of $8h^{-1}$ Mpc at redshift $z$ with $h\equiv 
H_0/100$, 
while $\gamma=\cos\theta$, with $\theta$ the angle between the wave vector 
${\mathbf k}$ and the line-of-sight direction.
    \item The Alcock-Paczynski effect, arising from the fact that a measurement 
of the galaxy power spectrum requires the assumption of a reference cosmology 
in 
order to transform redshifts into distances. If this assumption  does not 
correspond to the true underlying cosmology, it is necessary to rescale the 
wave 
vector components, $k_\parallel$ and $k_\perp$, as
    \begin{equation}
    k_{\perp} = \frac{k_{\perp,{\rm ref}}}{q_\perp}
    \qquad \text{and} \qquad
    k_{\parallel} = \frac{k_{\parallel,{\rm ref}}}{q_\parallel} \, ,
    \end{equation}
    where the coefficients $q_\perp$ and $q_\parallel$ are given, respectively, by the ratios of the angular diameter distance $D_\mathrm{A}$ and expansion rate $H$ to the corresponding 
quantities 
in the reference cosmology,
    \begin{equation}
    q_{\perp} = \frac{D_{\rm A}(z)}{D_{\rm A,\, ref}(z)}
    \qquad \text{and} \qquad
    q_{\parallel} = \frac{H_\text{ref}(z)}{H(z)} \, .
    \end{equation}
    It can be shown that overall, the Alcock-Paczynski effect rescales the 
power 
 spectrum by a factor of $\left(q_\perp^2 q_\parallel\right)^{-1}$ 
\cite{Ballinger:1996cd}.
\end{itemize}
At the linear level, the combination of these effects allows us to relate the 
observed galaxy power spectrum to the total matter one as~\cite{Blanchard:2019oqi}
\begin{equation}
P_{\rm obs}(k,z)=\frac{1}{q_\perp^2 q_\parallel} 
P_\text{rs,lin}(k(k_\text{ref},\gamma_\text{ref}),\gamma(\gamma_\text{ref});z).
\end{equation}
Therefore, by solving Eq.~(\ref{eq:deltaMG}) using the phenomenological 
departures from GR described in Section \ref{sec:phenom}, one can obtain 
theoretical predictions for the galaxy power spectrum and use observations to 
constrain deviations from standard gravity.

\subsubsection{Weak Gravitational Lensing}
\label{sec:WL}
Weak gravitational lensing does not affect only the CMB photons travelling from 
the  last scatteing surface, but also the light emitted by any cosmological 
sources due to the distribution of matter anisotropies along the line of sight. 
Using galaxy surveys, it is possible to measure the distortions in the shape of 
distant light sources, e.g., distant galaxies, and to infer through this the 
amplitude of the gravitational lensing that these have endured. Weak lensing is a 
particularly powerful probe for cosmology, since it measures simultaneously the 
growth of structure through the matter power spectrum and the geometry of the 
Universe. The reconstruction of the matter density field can be conducted by 
looking 
at the correlations of the image distortions. The observable we need to work 
with is, therefore, the shear angular power spectrum $C^{\gamma\gamma}(\ell)$, which can be 
obtained as (see, e.g., Ref.~\cite{Blanchard:2019oqi})
\begin{equation}\label{eq:cijdef}
C_{ij}^{\gamma\gamma}(\ell) = \frac{c}{H_0}
\int{\frac{{\hat{W}}_i^\gamma(z) {\hat{W}}_{j}^\gamma(z)}{E(z) r^2(z)}
P_{\Phi+\Psi}\left ( k_{\ell}, z \right ) dz},
\end{equation}
where $E(z)\equiv H(z)/H_0$, $r(z)$ is the comoving distance, $P_{\Phi+\Psi}$ 
is \label{multipolesref2}
 the power spectrum of the Weyl potential $\Phi+\Psi$, and we have made use of 
the 
Limber approximation~\cite{Kaiser:1991qi,LoVerde:2008re,Giannantonio:2011ya,Kitching:2016zkn,Kilbinger:2017lvu,Lemos:2017arq}, which relates scales $k$ to 
multipoles $\ell$ as $k_\ell=(\ell+1/2)/r(z)$; see, e.g., Ref.~\cite{Taylor:2018qda} for the full computation. Notice that here we assume a 
tomographic weak lensing survey, with the indices $i$ and $j$ running on the 
redshift bins of the survey; such a tomographic reconstruction that is used by 
weak lensing surveys allows us to obtain information on the evolution of the 
Weyl potential in time, thus carrying significant information for investigating 
the time evolution of the potential.

By defining the lensing kernel $\widetilde{W}^{\gamma}_i(z)$, a purely geometrical quantity expressed 
as
\begin{equation}\label{eq:lenskernel}
\widetilde{W}_{i}^\gamma(z) =
\int_{z}^{z_\text{max}}{n_{i}(z^{\prime}) \frac{r(z^{\prime}-z)}{r(z^{\prime})} dz^{\prime}} \ ,
\end{equation}
with $n_i(z)$ the normalised observed galaxy number density in the $i$th 
redshift bin, the quantity ${\hat{W}}_i^\gamma(z)$ in Eq.~(\ref{eq:cijdef}) is
\begin{equation}\label{eq:hatw}
\hat{W}_{i}^\gamma(z) = \frac{r(z)}{2}\widetilde{W}_{i}^\gamma(z)\,.
\end{equation}

In $\Lambda$CDM cosmology, which relies on the assumption of GR, it is easy to 
relate the weak lensing  \label{weaklensefs2}power spectrum $P_{\Phi+\Psi}$ to 
the matter power spectrum $P_\mathrm{m}$ using the 
relation
\begin{equation}\label{eq:weylLAM}
    P^\text{GR}_{\Phi+\Psi} =  \left[3 \left(\frac{H_0}{c}\right)^2\Omega^{0}_\mathrm{m} (1 + z)\right]^{2} P^\text{GR}_\mathrm{m},
\end{equation}
which leads to
\begin{equation}\label{eq:cijmatter}
C_{ij}^{\gamma\gamma,\mathrm{GR}}(\ell) = \frac{c}{H_0}
\int{\frac{W_i^{\gamma,\mathrm{GR}}(z) W_{j}^{\gamma,\mathrm{GR}}(z)}{E(z) r^2(z)}
P^\mathrm{GR}_\mathrm{m}\left ( k_{\ell}, z \right ) dz},
\end{equation}
with the window function $W_i^{\gamma,\mathrm{GR}}(z)$ given by
\begin{equation}
    W_i^{\gamma,\mathrm{GR}}(z)= \left [ 3 \left(\frac{H_0}{c}\right)^2\Omega^{0}_\mathrm{m} (1 + z)\right ]\frac{r(z)}{2}\widetilde{W}_{i}^{\rm \gamma}(z)\,.
\end{equation}
Eq~(\ref{eq:weylLAM}), however, does not hold if we depart from GR, and one 
would therefore have to use the full expression~(\ref{eq:cijdef}) to obtain 
theoretical predictions on this observable. Using Eqs.~(\ref{eq:PoissonMG}), 
(\ref{eq:slipMG}) and (\ref{eq:SigmaEq}), it is possible to rewrite the window 
function as \cite{SpurioMancini:2019rxy}
\begin{equation}
W_{i}^\gamma(z) =
\left [ 3 \left(\frac{H_0}{c}\right)^2\Omega^{0}_\mathrm{m} (1 + z) \right ]
\left \{ \frac{\mu(k, z) [1 + \eta(k, z)]}{2} \right \} \frac{r(z)}{2}\widetilde{W}_{i}^{\rm \gamma}(z),
\end{equation}
for modified theories of gravity, where we used the fact that in this parameterisation
\begin{equation}\label{eq:weylMG}
P_{\Phi+\Psi} =  \left [ 3 \left(\frac{H_0}{c}\right)^2\Omega^{0}_\mathrm{m} (1 + z) \Sigma(k,z)\right]^{2} P_\mathrm{m},
\end{equation}
with $\Sigma$ given by Eq.~(\ref{eq:SigmaDef}).

Weak lensing is therefore sensitive to the modifications that an extended 
theory 
of gravity would bring to the evolution of  perturbations through its effect on 
the Weyl potential, which is related to the matter power spectrum through 
Eq.~(\ref{eq:weylMG}). Notice that since weak lensing probes the dark matter 
power spectrum directly, this observable is not limited by any assumptions 
about 
the galaxy bias, which represents one of the main limitations of galaxy surveys.

\subsection{Einstein-Boltzmann Codes: from Theoretical Predictions to Data 
Analysis}
\label{sec:codes}
Having obtained the expressions for theoretical predictions of  cosmological 
observables in Section \ref{sec:observables}, we can in principle compare 
these 
with observations and constrain the parameters that measure deviations from GR. 
In order to compute these predictions, however, one has to solve the equations 
for cosmological perturbations in the presence of the modified Poisson 
equations 
of Section \ref{sec:phenom}. For this purpose, we can rely on the publicly 
available Einstein-Boltzmann solvers \texttt{CAMB} \cite{Lewis:1999bs} and 
\texttt{CLASS} 
\cite{Blas:2011rf}, which are able to compute the evolution of the cosmological 
background and perturbations from the early Universe to the present time.

These codes work in the standard $\Lambda$CDM/GR model and its simple 
extensions,  but their modifications based on phenomenological 
parameterisations of modified gravity are also publicly available. One example 
of these codes is \texttt{MGCAMB}~\cite{Zhao:2008bn,Hojjati:2011ix,Zucca:2019xhg}, which implements deviations 
from GR parameterised by the $\mu$, $\eta$ and $\Sigma$ functions, as well as 
some specific models of modified gravity (e.g., the Hu-Sawicki $f(R)$ 
model~\cite{Hu:2007nk}).

\begin{figure}[ht]
\begin{center}
\includegraphics[width=.48\columnwidth]{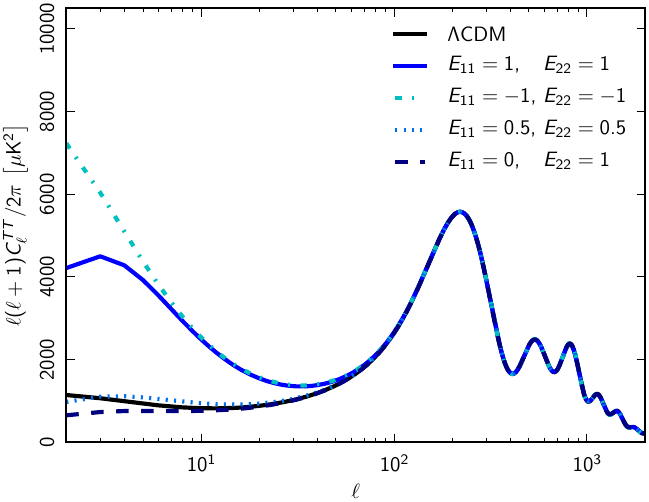}~~~
\includegraphics[width=.48\columnwidth]{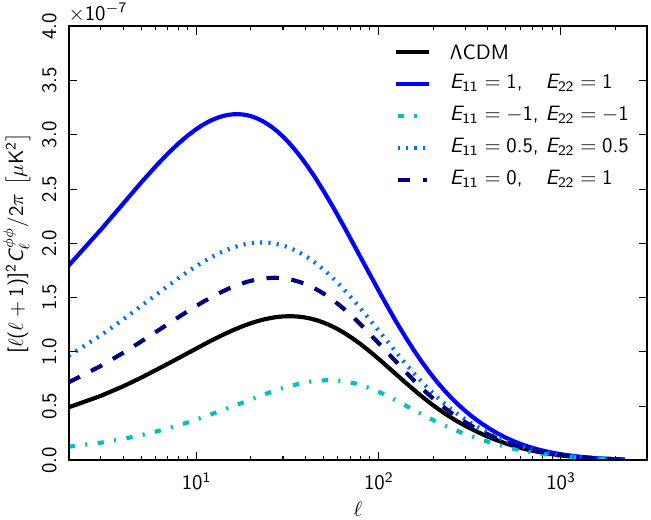}
\caption{{\it{Effects  of modifications of gravity on CMB power spectra, 
demonstrated 
through the parameterisations given in Eqs.~(\ref{eq:muetapk}) for deviations 
from the standard Poisson equation and slip relation, taken from the results of 
the {\it Planck} collaboration \cite{Ade:2015rim}.}}}
\label{fig:cmbMG}
\end{center}
\end{figure}

As an example, Fig.~\ref{fig:cmbMG} shows predictions for the CMB 
temperature  and lensing-potential spectra, using the parameterisation of 
Eqs.~(\ref{eq:PoissonMG}) and~(\ref{eq:slipMG}), obtained by the {\it Planck} 
collaboration using the \texttt{MGCAMB} code \cite{Ade:2015rim}, with $\mu$ and 
$\eta$ parameterized as
\begin{align}\label{eq:muetapk}
    \mu(z) = 1+E_{11}\Omega_{\rm DE}(z), \nonumber \\
    \eta(z) = 1+E_{22}\Omega_{\rm DE}(z),
\end{align}
where $\Omega_\mathrm{DE}(z)$ is the dark energy density parameter, and $E_{11}$ and $E_{22}$ are free parameters.

Having Einstein-Boltzmann codes able to produce the quantities  highlighted in Section 
\ref{sec:observables}, they can then be interfaced with codes for sampling 
cosmological parameter spaces (e.g., \texttt{Cobaya},\footnote{\url{https://github.com/CobayaSampler/cobaya}} \texttt{CosmoMC} 
\cite{Lewis:2002ah}, \texttt{CosmoSIS} \cite{Zuntz:2014csq} and 
\texttt{MontePython} \cite{Audren:2012wb}), enabling us to compare theoretical 
predictions to observational data. Such a pipeline, therefore, allows us to 
reconstruct the posterior probabilities of cosmological parameters and to 
obtain 
observational constraints on models and parameters.

\subsection{Small Scales and Nonlinearities}
\label{sec:nonlin}

Nonlinear structure formation is of extreme interest for testing gravity, as 
nonlinear observables are highly  sensitive to the properties of gravitational 
interactions. In addition, the smaller the scales, the larger the statistical 
weight in the data, as more modes contribute to cosmological observables. 
Analysing the small scales is, however, much more complicated and computationally 
expensive, because on the one hand, the evolution of quantities is nonlinear, 
which on its own makes the numerical analysis complicated and slow, and on the 
other hand, the physics on small scales is considerably more complex than the 
linear and large scales because of the nonlinear gravitational collapse and the 
important roles that non-gravitational physics plays on such scales; this 
includes the Baryonic effects of gas, stars and highly energetic astrophysical 
processes such as supernova explosions and active galactic nuclei, which affect 
the evolution of the structure on sub-Megaparsec scales~\cite{Chisari:2018prw}. 
For these reasons, the computation of nonlinear observables usually relies on 
numerical simulations of the large-scale structure, which are computationally expensive and time-consuming, 
especially because of the variety of modified gravity models with several free 
parameters each. Recently, there has been considerable progress in developing 
numerical ($N$-body and hydrodynamic) simulations, as well as employing 
cutting-edge artificial intelligence (e.g., machine learning and deep learning) and inference 
techniques to tackle the complexity of the small and nonlinear scales for 
testing the standard model and beyond; see, e.g., 
Refs.~\cite{Alsing:2019xrx,Ntampaka:2019udw,He:2018ggn,Chartier:2020pmu} and references therein.

In addition to $N$-body simulations and numerical techniques, various analytical 
methods have also been developed, based on the frameworks of standard perturbation theory (SPT)~\cite{Jain:1993jh,1986ApJ...311....6G}, Lagrangian perturbation theory (LPT)~\cite{Bouchet:1994xp,Matsubara:2007wj}, renormalised perturbation theory (RPT)~\cite{Crocce:2005xy}, effective field theory (EFT)~\cite{Carrasco:2012cv,Vlah:2015sea,2016arXiv161009321P,Cusin:2017mzw,Cusin:2017wjg} and kinetic field theory (KFT)~\cite{Bartelmann:2019unp} (see also Refs.~\cite{Bernardeau:2001qr,Desjacques:2016bnm} for reviews) for modelling the large-scale structure in the mildly nonlinear 
regime, where the theory of cosmological perturbations can still be applied by 
going beyond linear orders. In particular, in the framework of the effective 
field theory of large-scale structure (EFTofLSS), one can capture interesting 
effects on scales smaller than the intermediate ones that we discussed in 
Section~\ref{sec:phenom}, while the perturbative regime is valid, computations 
are analytical, and physics is well understood and under control. In an EFT 
description of a system, all the physics relevant at a macroscopic scale of 
interest is captured by integrating out the short-distance physics, affecting 
long-distance modes only through a few parameters in a perturbative expansion. 
Since in our Universe, matter perturbations are strongly (weakly) coupled in the 
UV (IR), a perturbative EFT can be consistently applied to the formation of 
structure on a relatively wide range of cosmologically interesting scales, 
making precise analytical predictions possible. The mildly nonlinear regime 
studied by analytic, perturbative techniques such as EFTofLSS are particularly 
interesting from the point of view of cosmological modifications to gravity, as 
they can capture the onset of ``GR to modified gravity'' transition, which is 
central to modified gravity models.

Even though there has been considerable progress in developing numerical  and 
analytic techniques to study the nonlinear regime and small scales, the level 
of 
accuracy needed for analysing the wealth of data provided by the upcoming and 
future cosmological surveys requires significantly more work in this direction 
in the coming years.

\section{Existing Constraints and Tensions}
\label{sec:current}
In recent years, observational collaborations performing galaxy or CMB surveys  
have started to constrain possible alternative theories of gravity. A first 
example is the {\it Planck} 2015 release, which dedicated a paper to dark 
energy 
and modified gravity models, including phenomenological deviations from GR 
\cite{Ade:2015rim}. The analysis was further improved in the 2018 release of 
the 
{\it Planck} collaboration \cite{Aghanim:2018eyx}. In this latest release, {\it Planck} constrained deviations from GR following the approach of 
subsection~\ref{sec:phenom}, considering $\mu(z,k)$ and $\eta(z,k)$ as free 
functions and computing $\Sigma(z,k)$ as $\Sigma(z,k)=[1+\eta(z,k)]\mu(z,k)/2$, 
while keeping the background expansion history to mimic that of a $\Lambda$CDM 
cosmology. Note that the {\it Planck} collaboration assumed no scale 
dependence for these functions.

\begin{figure}[t!]
\begin{center}
\begin{tabular}{cc}
\includegraphics[width=.7\columnwidth]{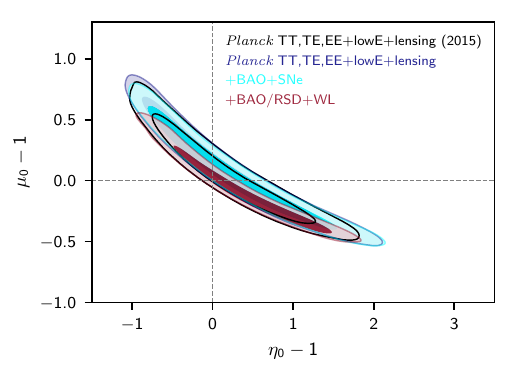}\\
\includegraphics[width=.7\columnwidth]{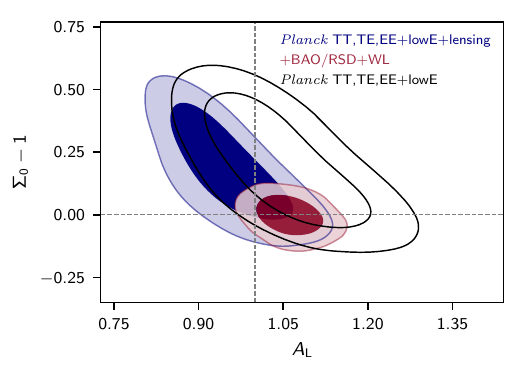}
\end{tabular}
\caption{
{\it{
Constraints on parameterised deviations from General Relativity  
obtained using {\it Planck} data~\cite{Aghanim:2018eyx}. The upper panel shows the contours for $\mu$ and 
$\eta$ values at the present time, while the lower panel shows the degeneracy 
between the modified gravity lensing function $\Sigma$ and the phenomenological 
lensing amplitude $A_\text{L}$. The figures are taken from 
Ref.~\cite{Aghanim:2018eyx}.}}}
\label{fig:planckres}
\end{center}
\end{figure}

In the upper panel of Fig.~\ref{fig:planckres}, the {\it Planck} results on the 
 present values of the $\mu$ and $\eta$ functions are shown, using only the CMB 
observables (both temperature and polarisation spectra, and CMB lensing 
reconstruction), as well as in combination with external data sets, i.e., 
measurements of the baryon acoustic oscillations (BAO) and redshift space 
distortions from the Sloan Digital Sky Survey (SDSS) and the Baryon Oscillation Spectroscopic Survey (BOSS) 
\cite{Beutler:2011hx,Ross:2014qpa,Alam:2016hwk}, the Pantheon supernovae catalogue 
\cite{Scolnic:2017caz} and the weak lensing data from the Dark Energy Survey (DES) 
\cite{Abbott:2017wau}. 
These results show no significant preference towards deviations from GR, as the 
limit $\mu(z=0)=\eta(z=0)=1$ is within $\sim1\sigma$ confidence region. As 
discussed in Section~\ref{sec:CMB}, the CMB constrains modified gravity 
theories 
mainly through their effects on the ISW and CMB lensing; given that the 
constraining power of the former is severely limited by cosmic variance, CMB 
lensing contributes the most to these constraints. This, interestingly, 
connects 
the constraints on modified gravity to one of the open ``curious'' implications 
of the {\it Planck} results: assuming a $\Lambda$CDM cosmology but allowing for 
a phenomenological departure from its predicted lensing amplitude through the 
parameter $A_\text{L}$ defined through
\begin{equation}
    C^{\phi\phi}(\ell)=A_\text{L}C^{\phi\phi}_{\Lambda \text{CDM}}(\ell),
\end{equation}
{\it Planck} finds a preference for $A_\text{L}>1$ when analysing temperature 
and  polarization spectra, while the value of $A_\text{L}$ is shifted back 
towards the $\Lambda$CDM limit when the CMB lensing reconstruction is included  
\cite{Aghanim:2018eyx}. In the lower panel of Fig.~\ref{fig:planckres}, we show 
the {\it Planck} results on the degeneracy between $A_\text{L}$ and the 
modified 
gravity quantity $\Sigma_0\equiv\Sigma(z=0)$, which encodes the effect of 
modified gravity on the Weyl potential $\Phi+\Psi$ affecting the lensing 
amplitude: the results show how, in order to recover $A_\text{L}=1$, an approximately
$1\sigma$ shift away from the GR limit $\Sigma_0=1$ is required when the CMB 
lensing reconstruction is not included; by adding this further observable the 
results are, however, consistent with GR within $1\sigma$. This highlights how 
improving the lensing reconstruction, from both CMB and LSS surveys, is crucial 
in order to improve the constraints on modified gravity models. 
\label{sigmaref3}
\begin{figure}[h!]
\begin{center}
\begin{tabular}{cc}
\includegraphics[width=.6\columnwidth]{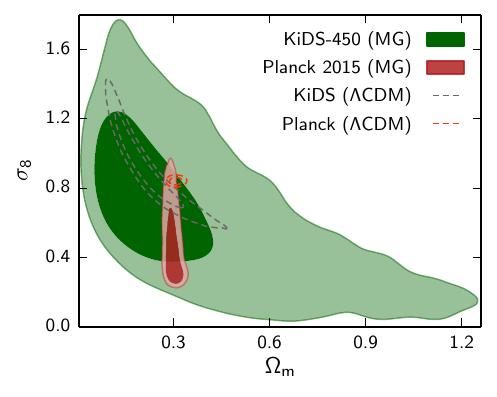}\\
\includegraphics[width=.6\columnwidth]{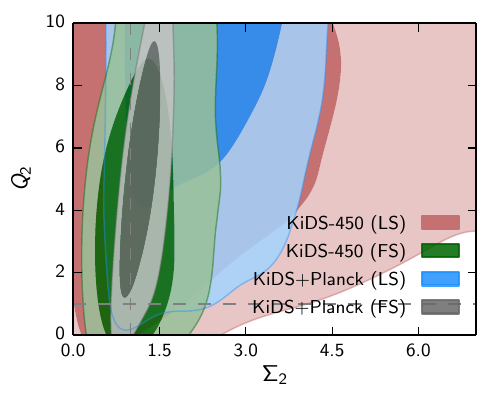}
\end{tabular}
\caption{{\it{Upper panel: Effect of allowing gravity to be 
modified on the contours of $\sigma_8$ and $\Omega_\mathrm{m}$ (note that here $\Omega_\mathrm{m}$ is the present value of the matter density parameter). These results are 
provided by the KiDS collaboration~\cite{Joudaki:2016kym}. Lower panel: Constraints on the binned 
modified gravity functions $Q(=\mu)$ and $\Sigma$ obtained using KiDS weak 
lensing data alone and combined with {\it Planck} CMB observations. The figures are taken from 
Ref.~\cite{Joudaki:2016kym}}}
}\label{fig:kidsres}
\end{center}
\end{figure}

Other phenomenological investigations of departures  from GR have been 
performed 
by recent galaxy surveys, mainly the Kilo-Degree Survey (KiDS)~\cite{Joudaki:2016kym} and DES 
\cite{Abbott:2018xao}. The KiDS collaboration decided not to rely on a 
particular parameterisation of the $\mu$ and $\eta$ functions (labelled $Q$ and 
$R$, respectively, in their paper), but rather to constrain their values in 4 
redshift and 4 scale bins. Note that given that no adequate 
prescriptions for computing the matter power spectrum on nonlinear scales are currently
available for modified gravity theories, the KiDS collaboration provided 
results 
using both their standard range of scales (results labelled as FS) and limiting 
the analysis only to linear scales (results labelled as LS).

In Fig.~\ref{fig:kidsres}, we show the results obtained  by KiDS on the 
modified 
gravity functions, using their weak lensing data and comparing with CMB 
constraints. The upper panel compares the results to those obtained in 
$\Lambda$CDM in an attempt to ease the tension on the $\sigma_8$ parameter 
between {\it Planck} and KiDS measurements (at the level of $\sim2\sigma$ in 
$\Lambda$CDM; see, e.g., Ref.~\cite{Asgari:2020wuj} for updates on the tension), highlighting how introducing eight modified gravity degrees of 
freedom, together with the conservative assumption about the scales used in the 
analysis, significantly washes out the constraining power of this survey. The 
lower panel shows instead the constraints on the modified gravity parameters, 
once again showing the impact of the assumptions on the analysis of nonlinear 
scales and, at the same time, showing how the KiDS survey is significantly more 
sensitive to $\Sigma$ (obtained as a combination of $Q=\mu$ and $R=\eta$), 
which 
impacts the Weyl potential, rather than to $Q$, which is instead related to the 
clustering of matter.

\begin{figure}[ht!]
\begin{center}
\begin{tabular}{cc}
\includegraphics[width=.485\columnwidth]{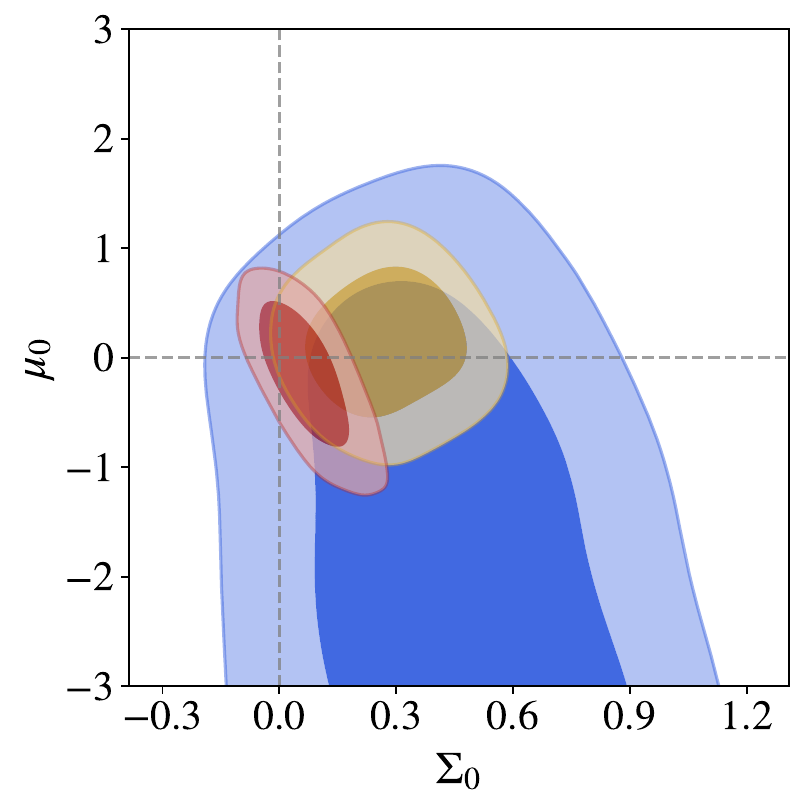}
\includegraphics[width=.485\columnwidth]{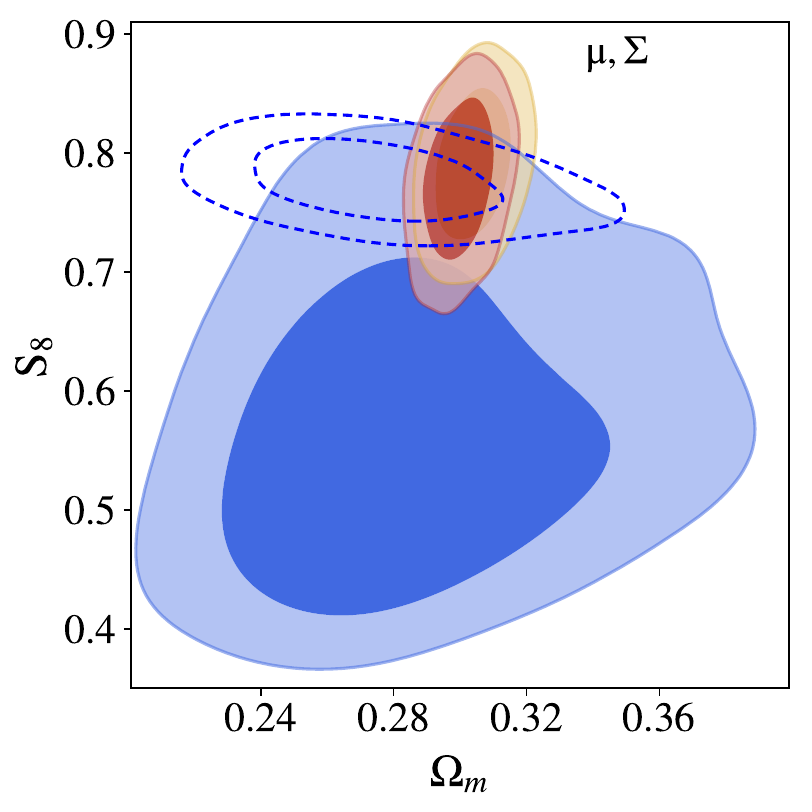}
\\
\end{tabular}
\caption{{\it{The left panel depicts the constraints on the parameters $\mu_0$ and 
$\Sigma_0$ 
for  deviations from GR obtained by DES (blue contour), {\it Planck} and 
BAO/RSD/SNIa (yellow contours), as well as their combination (red contour). For 
the same data sets, the right panel shows the effect of allowing for departures 
from GR on the quantities $S_8\equiv\sigma_8\sqrt{\Omega^{0}_\mathrm{m}/0.3}$ and 
$\Omega^{0}_\mathrm{m}$ (denoted as $\Omega_\mathrm{m}$ in the figure). The figures are taken from 
Ref.~\cite{Abbott:2018xao}}.}}
\label{fig:desres}
\end{center}
\end{figure}
The DES collaboration decided instead to constrain parameterised modified 
gravity functions with a parameterisation similar to the one adopted by {\it 
Planck} but using $\mu(z)$ and $\Sigma(z)$ as primary functions and 
constraining $\eta(z)$ as a derived function. In Fig.~\ref{fig:desres}, they 
have compared, similarly to the KiDS approach, their results (blue contours) 
with those obtained from {\it Planck} together with BAO/RSD and supernovae data 
(yellow contours), as well as the combination of the two data sets (red 
contours). The left panel shows how, as for KiDS, the DES measurements are 
more 
sensitive to changes in the lensing equations (through $\Sigma$) rather than in 
the Poisson equation for non-relativistic particles (through $\mu$), with no 
significant evidence for non-GR cosmologies. The right panel depicts instead 
how 
allowing for departures from GR significantly worsens and shifts the constraint 
on $S_8$ with respect to the $\Lambda$CDM case (dashed line), highlighting once 
again how future LSS surveys have the potential to improve the  constraints on 
these theories.

Overall, the results from these current surveys do not show significant 
evidence 
for  modified gravity. All the results are compatible with the $\Lambda$CDM 
limit, with small deviations ($\sim1\sigma$) in the $\Sigma$ function affecting 
the evolution of the lensing potential. Moreover, the results of these 
collaborations demonstrate how the constraints are significantly degraded with 
respect to the analyses involving other extensions of the $\Lambda$CDM model, 
showing how improvements in the near future in both the sensitivity of the 
surveys and the theoretical treatment of small scales are necessary to further 
investigate these non-standard cosmologies.

Nevertheless, allowing for modifications of gravity provides a promising way to ease the tension on the measured value of the 
$S_8\equiv\sigma_8\sqrt{\Omega^{0}_\mathrm{m}/0.3}$ parameter between CMB and LSS 
observations (estimated to be $\sim2\sigma$ for both KiDS and LSS data). 
However, as the results shown in this section highlight, allowing for deviations 
from GR and the necessity to cut out the measurements at small scales 
significantly lower the constraining power of the data; therefore it will be 
necessary to wait for more sensitive observations in the future to properly address the 
possibility that the $S_8$ tension might be a hint of modifications to the 
standard theory of gravity.

Finally, it is important to stress that here we did not discuss the 
(statistically)  most significant tension between currently available data, i.e., 
the tension in the measured values of the Hubble constant $H_0$ 
\label{Hubbleefs3} (see, e.g., Ref.~\cite{Verde:2019ivm} for a review). For 
what 
concerns this parameter, the values inferred from high-redshift measurements through the CMB (and LSS), which assume a 
$\Lambda$CDM expansion of the Universe, and those obtained through local, 
model-independent methods~\cite{Riess:2019cxk,Wong:2019kwg} are in tension with 
 \label{madelindepeefs5}
$\sim4.5\sigma$. Such a significant tension can be seen as a strong motivation 
to go beyond the $\Lambda$CDM model, and modifications of gravity might suggest 
ways to achieve this. However, the model-independent results presented in this 
section rely on the assumption of a background expansion which perfectly mimics 
that of $\Lambda$CDM and therefore do not significantly affect the constraints on 
$H_0$. Such an assumption is necessary for two main reasons: on the one hand, 
the sensitivity of the current data is not sufficient for simultaneously 
constraining modifications of the background expansion and of the evolution of 
perturbations, while on the other hand, in order to test realistic theories of 
gravity, modifications of the two sectors must be connected to each other, which 
is unfeasible with the simple parameterisations presented here. While the first 
issue will be eased with the upcoming plethora of LSS data from future surveys 
(see Section \ref{sec:futres}), the second point is also currently under 
investigation and promising results have been achieved, although for more 
restricted classes of theoretical models rather than for fully general 
parameterisations~\cite{Espejo:2018hxa}.

\section{Upcoming Surveys and the Road Ahead}
\label{sec:futres}
\label{tensionsrefs1}\label{cosmosurveysefs2}

As we discussed in the previous section, the current cosmological constraints  
on deviations from GR and on modified theories of gravity are relatively weak, 
and no significant deviations from $\Lambda$CDM can be inferred from the 
existing data, despite small tensions that are, to some extent, not even 
consistent. The situation may, however, change dramatically in the future, when the 
upcoming cosmological surveys with much higher levels of precision start to 
deliver data. Currently, the constraints on the modified gravity parameters 
$\mu$, $\eta$ and $\Sigma$ are of $\mathcal{O}(1)$, which we expect to reduce to 
$\mathcal{O}(10^{-1})-\mathcal{O}(10^{-2})$ in the next few years.

We expect an exciting time to come when the Stage IV LSS surveys are fully 
operational,  providing a large amount of highly precise cosmological 
data. These surveys include several ground-based experiments such as the Dark 
Energy Spectroscopic Instrument 
(DESI)~\cite{Aghamousa:2016zmz,Aghamousa:2016sne}, the Rubin Observatory Legacy Survey of Space and Time (LSST)~\cite{Ivezic:2008fe,Abell:2009aa,Mandelbaum:2018ouv}, the Square Kilometre 
Array 
(SKA)~\cite{Bull:2014rha,Jarvis:2015tqa,Bacon:2015dqe,Kitching:2015fra,
Yahya:2014yva,Santos:2015gra,Bacon:2018dui}, and the Subaru Hyper Suprime-Cam (HSC) and Prime Focus Spectrograph (PFS) surveys~\cite{Aihara:2017paw,Tamura:2016wsg}, as well as the space-based 
experiments {\it Euclid}~\cite{Laureijs:2011gra,Amendola:2016saw,Blanchard:2019oqi}, the Wide 
Field InfraRed Survey Telescope 
(WFIRST)~\cite{Spergel:2015sza,Hounsell:2017ejq}, and the Spectro-Photometer for the History of the Universe, Epoch of Reionization, and Ices Explorer (SPHEREx)~\cite{Dore:2014cca,Dore:2018kgp}. In addition to these LSS 
surveys, future ground-based CMB experiments, such as 
CMB-S4~\cite{Abazajian:2016yjj,Abazajian:2020dmr}, the Simons Observatory (SO)~\cite{Ade:2018sbj} and 
CMB-HD~\cite{Sehgal:2019ewc,Sehgal:2020yja}, will provide extremely high-resolution CMB maps 
covering a large fraction of the sky, which can also help us test gravity and 
constrain modified gravity quantities.
\begin{figure}[ht!]
\begin{center}
\begin{tabular}{cc}
\includegraphics[width=.485
\columnwidth]{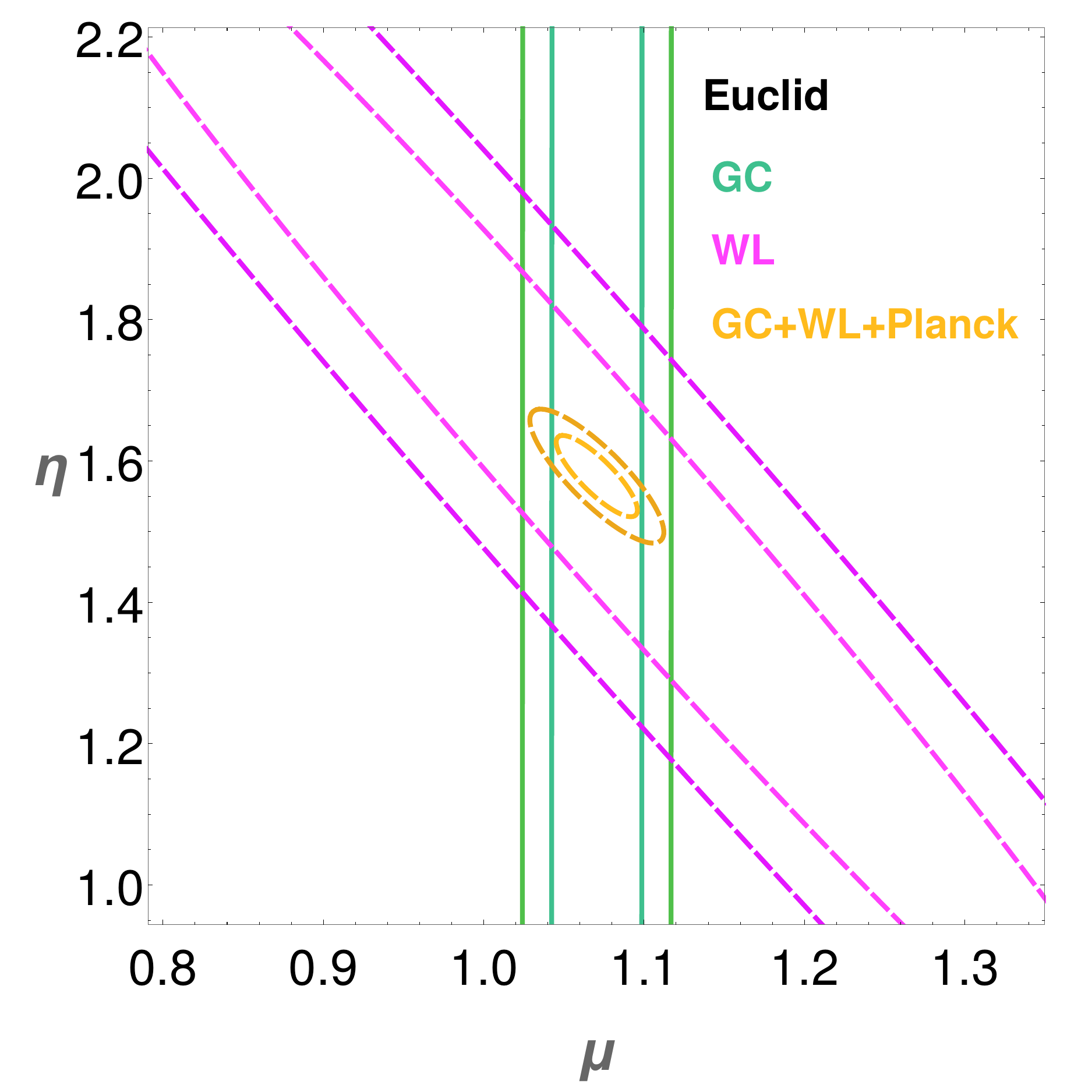}~~~
\includegraphics[width=.485\columnwidth]{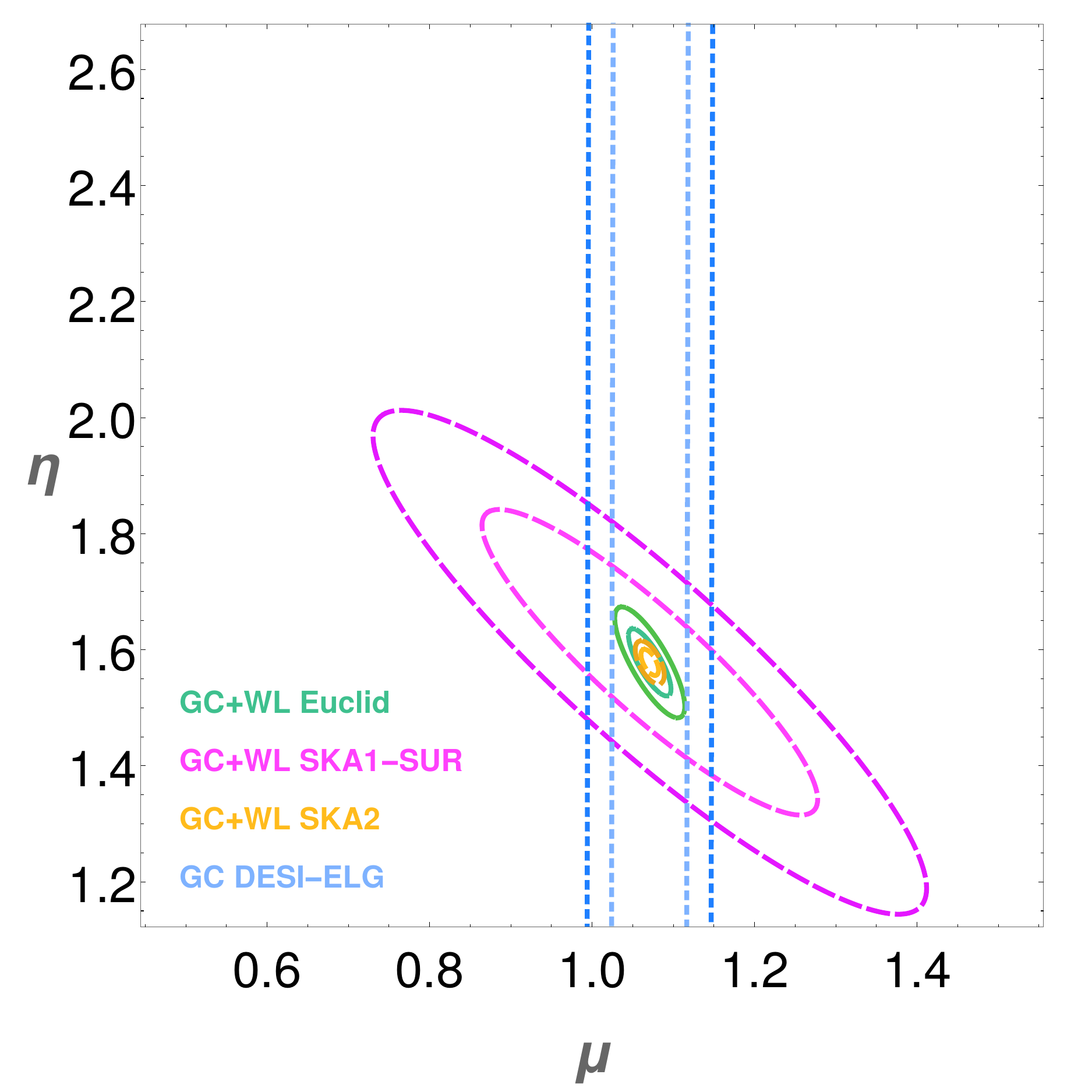}\\
\end{tabular}
\caption{{\it{Left panel: Results of a Fisher matrix forecast analysis   
\label{cforecastsefs1}
for 
{\it Euclid} and for a late-time parameterisation of modified gravity functions $\mu$ and $\eta$ where the amplitude of deviations from GR is assumed to be
proportional to the dark energy density parameter $\Omega_\mathrm{DE}(z)$. Linear and mildly nonlinear scales have been included and the 
contours show the predicted $1\sigma$ and $2\sigma$ confidence regions for 
galaxy clustering (GC) \label{galaxyclurefs1}  and weak lensing (WL) surveys, 
as well as their 
combination with the {\it Planck} constraints. Right panel: Fisher matrix 
forecast confidence regions for $\mu$ and $\eta$ and for {\it Euclid} and two 
phases of the SKA, SKA1 and SKA2, when combinations of GC and WL have been 
considered. Constraints from DESI are also shown with only GC observables 
included. The figures are taken from 
Ref.~\cite{Casas:2017eob}.}}}
\label{fig:forecast1}
\end{center}
\end{figure}

\begin{figure}[ht!]
\begin{center}
\begin{tabular}{cc}
\includegraphics[width=.89\columnwidth]{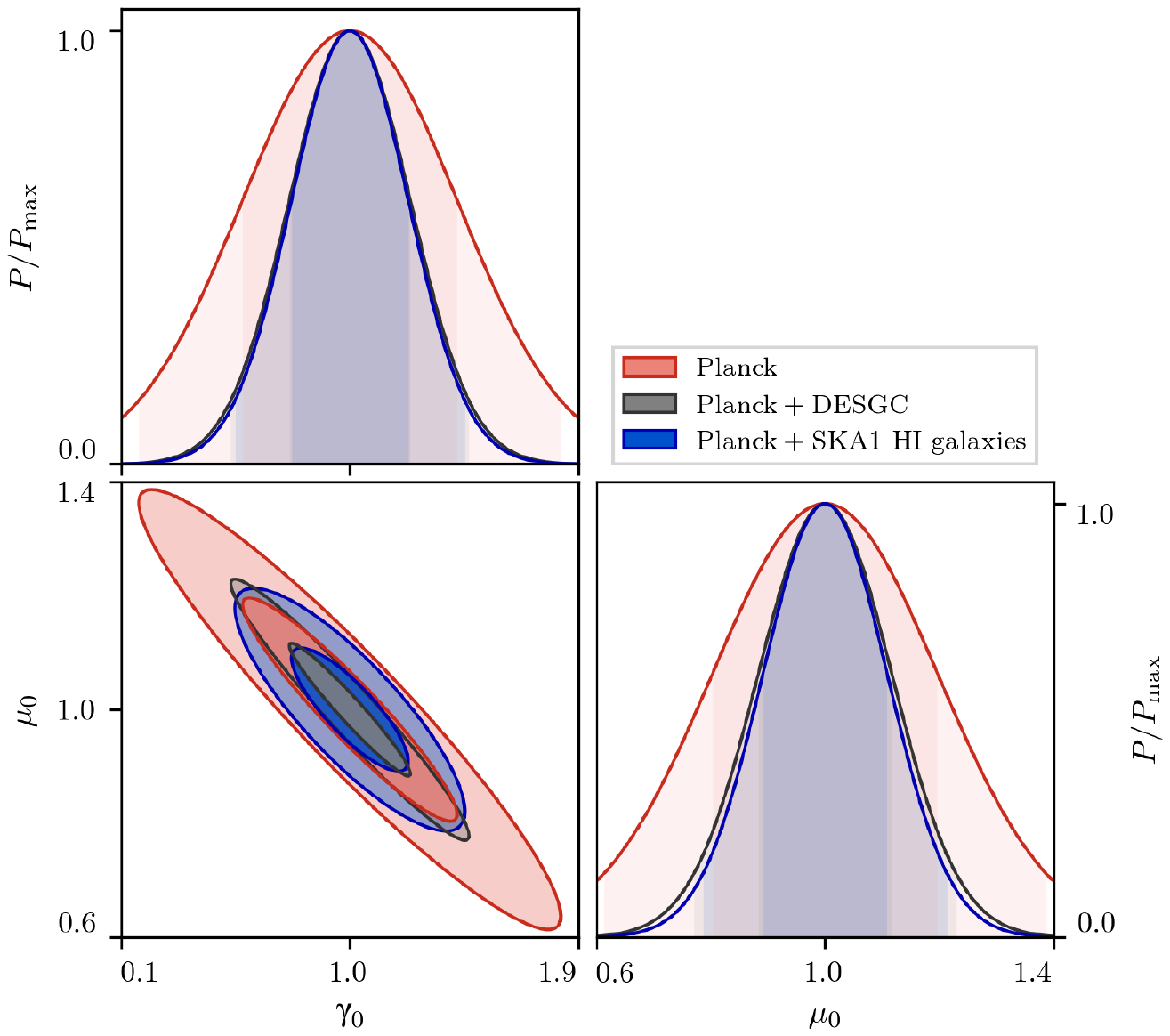} 
\end{tabular}
\caption{{\it{Fisher matrix forecast constraints on the phenomenological  
modified 
gravity parameters $\mu_0\equiv\mu(z=0)$ and $\gamma_0=\eta_0\equiv\eta(z=0)$ 
when the broadband shape of the power spectrum detected by the HI galaxy sample 
of the {\it Medium-Deep Band 2 Survey} is used for Phase 1 of the SKA (SKA1). 
Constraints from {\it Planck} and DES (galaxy clustering only) are also 
presented for comparison. The figure is taken from 
Ref.~\cite{Bacon:2018dui}.}}}\label{fig:forecast2}
\end{center}
\end{figure}

Several forecast analyses for these experiments and surveys are currently  being 
performed, and here we only present a few representative examples of the 
constraints we expect to achieve by using some of these surveys. 
Fig.~\ref{fig:forecast1} shows the results of a Fisher matrix forecast analysis 
for {\it Euclid} and the SKA for the two phenomenological modified gravity 
functions $\mu$ and $\eta$, when the late-time parameterisation of 
Eq.~(\ref{eq:muetapk}) has been used and linear and mildly nonlinear scales have 
been considered~\cite{Casas:2017eob}. The constraints are based on galaxy 
clustering (GC) and weak lensing (WL) observables and their combinations. The 
left panel of the figure indicates that combining the {\it Euclid} GC and WL 
measurements with {\it Planck} breaks many degeneracies in the parameter space 
and provides extremely tight constraints on $\mu$ and $\eta$. On the other hand, 
the right panel of the figure presents the constraining power for the 
combination of GC and WL surveys for {\it Euclid} and phases 1 and 2 of the SKA, 
as well as GC only for DESI. Although all the constraints in the figure are 
parameterisation-dependent (i.e., they depend on how $\mu$ and $\eta$ are 
parameterised in terms of time and scale), the contours prove a dramatic 
reduction in the uncertainties compared to the current constraints. In order to 
see this more clearly, we have provided in Fig.~\ref{fig:forecast2} the forecast 
constraints on the phenomenological modified gravity parameters 
$\mu_0\equiv\mu(z=0)$ and $\gamma_0=\eta_0\equiv\eta(z=0)$, to be provided by the 
HI galaxy sample of the {\it Medium-Deep Band 2 Survey} for phase 1 of the SKA 
(SKA1), as defined in the SKA Red Book 2018~\cite{Bacon:2018dui}. The figure includes 
also current constraints from {\it Planck} and DES (GC only) for comparison. The 
improvement from adding SKA1 is comparable to DES.

With the level of precision we expect to reach in the next few years, we will 
be 
 able to strongly constrain departures from GR and test a large class of its 
interesting and theoretically well-motivated alternatives. In case the current 
tensions in cosmological observations that seem to indicate preference for 
departures from standard gravity hold against the tide of various future 
high-quality
data, they will then provide undeniable evidence for new physics in the 
gravitational sector.







\chapter[Relativistic Effects]{Relativistic Effects}
\label{sec:Bonvin}
 \label{relativistiffectcefs1}
 
{\em Camille Bonvin}\\







The clustering of galaxies is highly sensitive to the theory of gravity and 
provides, therefore, a powerful way to test for deviations from General 
Relativity. The two dominant contributions to the galaxy number counts are 
density perturbations and redshift-space distortions (RSD). These contributions 
have been measured in surveys like BOSS~\cite{BOSSWeb} and 
WiggleZ~\cite{WigglezWeb} and used to test the consistency of General Relativity 
and place constraints on alternative models~\cite{Alam:2016hwk,Parkinson_2012}. 
Here we will show that other more subtle effects, called {\it relativistic 
effects}, contribute to the galaxy number 
counts~\cite{Yoo:2009au,Bonvin:2011bg,Challinor:2011bk}. We will see that these 
effects contain additional information with respect to density and RSD, and 
that they can therefore be used to place new constraints on the theory of 
gravity. In particular we will show how one of these effects, called 
gravitational redshift, can be used to test the equivalence principle for dark 
matter~\cite{Bonvin:2018ckp}.

\section{Number Counts}
\label{sec:cb counts}
\label{numbercountsefs1}

We start by deriving the general expression describing the clustering of 
galaxies. Maps of galaxies can be pixelised, i.e., separated in bins of solid 
angle and redshift. An observer can then count how many galaxies he detects in 
each pixel, $N(z,\bn)$, where $z$ denotes the redshift of the pixel and $\bn$ 
its direction in the sky, and construct the galaxy fractional number overdensity
\be
\Delta(z,\bn)=\frac{N(z,\bn)-\bar N(z)}{\bar N(z)}\, .\label{cb:N}
\ee
Here, $\bar N(z)$ is the mean number of galaxies per pixel at redshift $z$. The 
number of galaxies can be expressed in terms of the galaxy number density, 
$\rho$, and the volume of the pixel, $\mathcal{V}$, as 
$N(z,\bn)=\rho(z,\bn)\mathcal{V}(z,\bn)$. Inserting this   in~\eqref{cb:N}, 
and keeping only terms at linear order in the perturbations we 
find~\cite{Bonvin:2011bg}
\be
\Delta(z,\bn)=\delta_g(z,\bn)-\frac{3\,\delta z}{1+\bar z}+\frac{\delta\mathcal{V}(z,\bn)}{\bar{\mathcal{V}}(\bar z)}\, ,
\ee
where $\delta_g(z,\bn)$ denotes the departure from the background number density of galaxies at the background redshift $\bar z$
\be
\delta_g(z,\bn)=\frac{\rho(z,\bn)-\bar\rho(\bar z)}{\bar\rho(\bar z)}\, .
\ee
The perturbations in the redshift $\delta z$ and the perturbations in the volume $\delta\mathcal{V}(z,\bn)$ can be calculated by solving the propagation of null geodesics in a perturbed Friedmann-Lema\^itre-Robertson-Walker Universe. At linear order in perturbation theory we find the relativistic expression for $\Delta$~\cite{Yoo:2009au,Bonvin:2011bg,Challinor:2011bk}
\begin{eqnarray}
&&
\!\!\!\!\!\!\!\!\!\!\!\!\!\!\!\!\!\!\!\!\!\!\!\!\!\!
\Delta(z,\bn)=b\cdot 
D-\frac{1}{\mathcal{H}}\partial_r(\mathbf{V}\cdot\mathbf{n})\label{cb:Delta}
\nonumber\\
&&
+\left(5s-2\right)\int_0^{r}d r' 
\frac{r-r'}{2rr'}\Delta_\Omega(\Phi+\Psi)\nonumber\\
&&
+\left(1-5s-\frac{\dot{\mathcal{H}}}{\mathcal{H}^2}+\frac{5s-2}{r\mathcal{H}}
+f_{\rm evo} 
\right)\mathbf{V}\cdot\mathbf{n}+\frac{1}{\mathcal{H}}\dot{\mathbf{V}}
\cdot\mathbf{n}+\frac{1}{\mathcal{H}}\partial_r\Psi\nonumber\\
&&
+\frac{2-5s}{r}\int_0^{r}dr'(\Phi+\Psi)+3\mathcal{H}\nabla^{-2}(\nabla\mathbf{
V})+\Psi+(5s-2)\Phi\nonumber\\
&&
+\frac{1}{\mathcal{H}}\dot{\Phi}+\left(\frac{\dot{\mathcal{H}}}{\mathcal{H}^2}
+\frac{2-5s}{r\mathcal{H}}+5s-f_{\rm evo} 
\right)\left[\Psi+\int_0^{r}dr'(\dot{\Phi}+\dot{\Psi})\right],
\end{eqnarray}
where $\partial_r$ denotes a derivative along the line-of-sight, a  dot is a 
derivative with respect to conformal time $\eta$, $\mathcal{H}$ is the Hubble 
parameter in conformal time, $\mathcal H=\dot a/a$, $b$ is the bias, $s$ is the 
magnification bias, and $f_{\rm evo}$ denotes the evolution bias. The variable 
$D$ is the gauge-invariant density perturbation in the comoving gauge, $V$ 
denotes the gauge-invariant velocity potential in the Newtonian gauge, and 
$\Phi$ and $\Psi$ are the two gauge-invariant Bardeen potentials.

The first line in~\eqref{cb:Delta} contains the density and redshift-space 
distortion (RSD)  contributions, which we call hereafter the {\it standard 
terms}. These terms have been measured with great precision in past 
surveys~\cite{BOSSWeb,WigglezWeb} and used to place constraints on modified 
theories of gravity~\cite{Alam:2016hwk,Parkinson_2012}. The second line contains 
the effect of {\it lensing magnification} (also called magnification bias). This 
effect is a combination of the fact that gravitational lensing increases the 
volume of observation, consequently diluting the number of galaxies that we 
observe, whereas at the same time it increases the luminosity of galaxies, 
allowing us to see galaxies that would otherwise be fainter than the flux limit 
of our detector~\cite{Scranton:2005ci,Duncan:2013haa}. Lensing magnification has 
been measured by cross-correlating background galaxies (or quasars) at high 
redshift, with foreground galaxies at low 
redshift~\cite{Scranton:2005ci,Garcia-Fernandez:2016oud}. The last three lines 
contain the so-called {\it relativistic 
effects}~\cite{Yoo:2009au,Bonvin:2011bg,Challinor:2011bk}. These effects are 
neglected in current surveys, since they are suppressed by powers of $\mathcal 
H/k$ with respect to the standard terms. Hereafter, we will see, however,   
that it 
is possible to construct an estimator to isolate some of the relativistic 
effects, which makes them detectable with the coming generation of surveys.

\section{Correlation Function}
\label{sec:cb corr}

\subsection{Estimators}

The standard strategy to extract information from the galaxy number counts is  
to measure the two-point correlation function of $\Delta$. Various two-point 
estimators have been used in galaxy surveys. The estimator that is the closest 
to observations is the angular power spectrum, 
$C_\ell(z,z')$~\cite{Bonvin:2011bg,Challinor:2011bk}. This is the estimator used 
in Cosmic Microwave Background analyses, since it is perfectly adapted to 
observations covering the whole sky. It has, furthermore, the advantage of 
being 
directly constructed from observable quantities, namely the angular separation 
between galaxies (which is directly translated into the multipole $\ell$) and 
the redshifts $z$ and $z'$. However, for spectroscopic surveys, the angular 
power spectrum has the disadvantage of requiring a very large number of redshift 
bins to recover all the available information~\cite{Asorey:2012rd} and of not 
providing a simple separation between density and redshift-space distortions, 
which is needed to test modified gravity in a model-independent way. 
\label{madelindepeefs5} As such, 
the 
angular power spectrum is most useful for photometric surveys and we will not 
use it here.

The second well-known estimator is the power spectrum, $P(k,\mu)$ (here $\mu$  
represents the angle between the direction of observation $\bn$ and the vector 
$\bk$), and more precisely its first three even multipoles: the monopole 
$P_0(k)$, the quadrupole $P_2(k)$ and the hexadecapole $P_4(k)$. The multipoles 
expansion has the advantage of allowing separate measurements of the bias $b$ 
and  
the growth rate of structure $f$ (more precisely, one measures the two 
combinations $b\sigma_8$ and $f\sigma_8$). Contrary to the angular power 
spectrum, the power spectrum requires the use of a fiducial cosmology to 
translate the measurements of angle and redshifts into cartesian coordinates, 
used to build the Fourier mode $\bk$. However, this can  be accounted for by 
introducing scaling parameters, which allow the true cosmology to differ from 
the fiducial one~\cite{Xu_2013}. The true problem of the power spectrum is the 
fact that it is built on the flat-sky approximation, i.e.,  the approximation 
that the directions to the two pixels in the correlation are parallel. As such 
it is not well adapted to surveys with large sky coverage, like the coming 
generation of surveys. \label{multipolesref1}

In particular, since relativistic effects are expected to 
be of the same order of magnitude as wide-angle effects~\cite{Bonvin:2013ogt}, 
the power spectrum cannot be consistently used to study relativistic effects. 
Methods have been proposed to overcome this problem, but they complicate 
significantly the measurement of the power spectrum~\cite{Yamamoto:2005dz}. The 
second more fundamental problem of the power spectrum is that it does not allow 
for a consistent calculation of the integrated terms in Eq.~\eqref{cb:Delta}, 
in 
particular of the lensing magnification, which is important at high 
redshift~\cite{Tansella:2017rpi}. The power spectrum indeed requires  the 
knowledge of $\Delta$ on three-dimensional hypersurfaces of constant time, 
whereas 
lensing can only be calculated consistently on the past-light cone of the 
observer. The power spectrum is therefore badly adapted to the coming generation 
of surveys that will probe large scales and high redshifts.

The last two-point estimator that is widely used is the two-point correlation  
function in redshift space, $\xi(d,\sigma)$, where $d$ represents the 
separation 
between pairs of galaxies, and $\sigma$ denotes the angle between this 
separation and the direction of observation $\bn$. As for the power spectrum, 
the first three even multipoles of the correlation function, $\xi_0(d), 
\xi_2(d)$ and $\xi_4(d)$ are routinely measured in galaxy surveys. These 
multipoles provide a separate measurement of $b$ and $f$. The correlation 
function also requires  a fiducial cosmology to translate the measurement of 
angles and redshift into a measurement of $d$ and $\sigma$, but as for the power 
spectrum, this can consistently be accounted for. The correlation function has, 
furthermore, the advantage over the power spectrum that it can be calculated in 
the full sky~\cite{Bonvin:2013ogt,Reimberg:2015jma}. One can then consistently 
expand around the flat-sky approximation and control the regime in which this 
approximation is valid. Moreover, lensing magnification can be included in the 
modelling of the correlation function and of its 
multipoles~\cite{Tansella:2017rpi}. The correlation function is therefore 
ideally adapted to the coming generation of surveys. Its only disadvantage is 
that its covariance matrix is non-diagonal, which complicates the construction 
of the likelihood. In the following we will use the correlation function to 
extract relativistic effects.

\subsection{Even and Odd Multipoles}
\label{multipoles}

Let us start by calculating the correlation function in General Relativity,  
i.e.,  using the continuity equation to relate the peculiar velocity $V$ to the 
matter density $D$ (in the standard terms), using the  Euler equation to relate 
the 
gravitational potential $\Psi$ to the peculiar velocity $V$ (in the relativistic 
terms), and assuming no anisotropic stress and Poisson equation to relate the 
two metric potentials to the density $D$ (in the lensing magnification term).

The full-sky expression for the correlation function from the standard terms 
(density and RSD)  can be found in~\cite{Bonvin:2013ogt,Gaztanaga:2015jrs}. This 
expression is valid at all scales. In the flat-sky approximation it reduces to 
the well-known expression
\begin{align}
\label{cb:xist}
\xi^{\rm st}(z, d, \sigma)=&\Bigg[b^2+\frac{2bf}{3}+ \frac{f^2}{5}\Bigg]\mu_0(d,z)
-\Bigg[\frac{4bf}{3} +\frac{4f^2}{7}  \Bigg]\mu_2(d,z) \cdot P_2(\cos\sigma)\\
&+\frac{8f^2}{35}\mu_4(d,z)\cdot P_4(\cos\sigma)\, , \nonumber
\end{align}
where $P_\ell$ is the Legendre polynomial of degree $\ell$, $f$ is the growth 
rate of structure defined  through $f=d\ln D_1/d\ln a$, with $D_1$ the growth 
function and
\be
\mu_\ell(d,z)=\frac{1}{2\pi^2}\int dk k^2 P(k,z)j_\ell(kd)\, .
\ee
Here, $P$ denotes the density matter power spectrum at present defined 
as
\be
\langle D(\bk,z)D(\bk',z)\rangle=(2\pi)^3P(k,z)\delta_D(\bk+\bk')\, ,
\ee
where we have used the Fourier transform convention $f(\bx, 
z)=\frac{1}{(2\pi)^3}\int d^3\bk\,  e^{-i\bk\cdot\bx} f(\bk,z)$. The redshift 
$z$ denotes the mean redshift of the pair, or the mean redshift of the bin in 
which we average the correlation function. In the flat-sky approximation we 
automatically neglect the evolution between the two galaxies in the pair.

The lensing magnification adds contributions to the correlation function. The 
full-sky  and flat-sky expressions for the lensing can be found 
in~\cite{Tansella:2017rpi}. The multipoles can then be numerically extracted 
using
\be
\xi_\ell^{\rm lens}(z, d)=\frac{2\ell+1}{2}\int_{-1}^1d\mu\, \xi^{\rm lens}(z, d,\mu)P_\ell(\mu)\, .
\ee
In this chapter, we will neglect the lensing magnification, since we will work 
at relatively 
small  redshift ($z\leq 1.2$), where the lensing is strongly subdominant.

Finally, the relativistic effects also contribute to the correlation function. 
The dominant relativistic effects are due to the third line in 
Eq.~\eqref{cb:Delta}, which contains one gradient of the potentials. As such 
they are suppressed by one power of $\mathcal{H}/k$ with respect to the standard 
terms. However, these terms are anti-symmetric, which implies that  their 
correlation with the density and RSD exactly vanishes. Hence their contribution 
to the correlation function comes from their auto-correlation, which is 
suppressed by  $\left(\mathcal{H}/k\right)^2$ with  respect to the standard 
terms. Due to this suppression, these terms become relevant only at very large 
separations, where cosmic variance is important. It is therefore very difficult 
to detect these terms in the monopole, quadrupole and hexadecapole of the 
correlation function. To overcome this difficulty, the following method has been 
proposed: if one splits the population of galaxies into two (or more) 
populations, then the correlation function can contain an anti-symmetric part. 
Consequently, the correlation between the standard terms and the relativistic 
effects does not vanish anymore, and it generates odd multipoles in the 
correlation function. By measuring these odd multipoles, one can therefore {\it 
isolate} the contribution from relativistic effects. The full-sky expression for 
the relativistic correlation function can be found in~\cite{Bonvin:2013ogt}.
\label{crpssprobesref2}

In the flat-sky, cross-correlating two populations of galaxies, one bright and one faint population, we obtain
\begin{align}
\xi^{\rm rel}(z, d, \sigma)&=\frac{\mathcal{H}}{\mathcal{H}_0}\Bigg\{\Bigg[(b_{\rm B}-b_{\rm F})\left(\frac{2}{r\mathcal{H}}+\frac{\dot{\mathcal{H}}}{\mathcal{H}^2} \right)+3(s_{\rm F}-s_{\rm B})f^2\left(1-\frac{1}{r\mathcal{H}} \right) \label{cb:rel}\\
&+5(b_{\rm B}s_{\rm F}-b_{\rm F}s_{\rm B})f\!\left(1-\frac{1}{r\mathcal{H}} \right)\!\!\Bigg]\nu_1(d,z)P_1(\cos\sigma) \nonumber \\
&+2\left(1-\frac{1}{r\mathcal{H}} \right)\!(s_{\rm B}-s_{\rm F})f\nu_3(d,z)P_3(\cos\sigma)\Bigg\}\, ,\nonumber
\end{align}
where $b_\B$ and $s_\B$ denotes the bias and magnification bias of the bright galaxies, $b_\F$ and $s_\F$ those of the faint galaxies, and
\be
\nu_\ell(d,z)=\frac{1}{2\pi^2}\int dk k\HH_0 P(k,z)j_\ell(kd)\, .
\ee

In the flat-sky the standard terms do not contribute to the odd multipoles. 
However,  because the standard terms are larger than the relativistic terms by 
one factor $k/\mathcal{H}$, their full-sky correction is of the same order of 
magnitude as the flat-sky relativistic effects. For consistency we therefore 
need to include this correction, which reads
\begin{align}
\xi^{\rm st\,corr}(z, d, \sigma)&=
\Bigg[-(b_{\rm B}-b_{\rm F})\frac{2f}{5}\,\mu_2(d,z)
+(b_{\rm B}-b_{\rm F})\frac{rf'}{6}\Big(\mu_0(d,z)\label{cb:wide}\\
&-\frac{4}{5}\mu_2(d,z)\Big) -\frac{rf}{6}\big(b'_{\rm B}-b'_{\rm F}\big)\Big(\mu_0(d,z)-\frac{4}{5}\mu_2(d,z)\Big)
\nonumber \\
&+\frac{r}{2}\big(b_{\rm B} b'_{\rm F}-b'_{\rm B} b_{\rm 
F}\big)\mu_0(d,z)\Bigg]\,\frac{d}{r}\,  P_1(\cos\sigma)+\Bigg[(b_{\rm B}-b_{\rm 
F})\frac{2f}{5} \nonumber\\
&-(b_{\rm B}-b_{\rm F})\frac{rf'}{5}
+\big(b'_{\rm B}-b'_{\rm F}\big)\frac{rf}{5}\Bigg]\mu_2(d,z)\,\frac{d}{r}\, P_3(\cos\sigma)\, ,\nonumber
\end{align}
where a prime denotes a derivative with respect to conformal distance $r$. The 
first term is  a wide-angle correction, due to the fact that the line-of-sights 
to the two galaxies are not parallel. The other terms are evolution corrections, 
coming from the fact that the bias and the growth rate $f$ are evolving between 
the two galaxies. As shown in~\cite{Bonvin:2013ogt}, the evolution corrections 
are always much smaller than the relativistic contributions and we can safely 
neglect them. The wide-angle correction is, however, of the same order as the 
relativistic effects and we have to keep it.

\section{Test of the Equivalence Principle}
\label{sec:cb equiv} \label{equivprinref8}

Let us now explore how the monopole, quadrupole, hexadecapole and dipole can be 
used to test  gravity. Note that the octupole is also non-zero, but its 
signal-to-noise is significantly lower that the one of the dipole, thus we 
prefer not to consider it here (see, e.g., \cite{Kodwani:2019onz} for a 
discussion about the octupole in theories with a screening mechanism). As 
explained above, the even multipoles are sensitive to density and RSD. A 
combined measurement of these quantities allows, therefore, to probe the 
density 
$D$ and velocity $V$ separately. They have been used to constrain modifications 
of gravity through their impact on the growth rate $f$. The dipole is sensitive 
to the time component of the metric $\Psi$, through the effect of gravitational 
redshift (the  last term in the third line of~\eqref{cb:Delta}) and through the 
velocity via Doppler effects (the first two terms in the third line 
of~\eqref{cb:Delta}). The dipole  therefore adds new information, via the 
measurement of $\Psi$, and can be used to test the  Euler equation which 
relates 
$\Psi$ to $V$. This provides a direct test of the equivalence principle for dark 
matter~\cite{Bonvin:2018ckp}. $V$ is indeed the peculiar velocity of galaxies, 
which are mainly made of dark matter, and it tells us therefore how halos of 
dark 
matter fall inside a gravitational potential. $\Psi$, on the other hand, is 
measured from gravitational redshift, and tells us how light escapes from that 
same gravitational potential. The dipole therefore provides  a way of testing 
the 
equivalence between the fall of light and the fall of dark matter. Any violation 
of this principle would leave an imprint on the dipole.

Note that such a test of the equivalence principle cannot be performed without 
the dipole. Lensing measurements indeed provide  a measurement of the sum of 
the 
two metric potentials $\Phi+\Psi$. If one does not know the value of the 
anisotropic stress, this cannot be used to test the equivalence principle. 
Standard tests of gravity usually assume that the equivalence principle is 
valid, to translate measurements of $V$ from RSD into measurements of $\Psi$, 
using the Euler equation. They then compare the $\Psi$ obtained in this way 
with 
lensing measurements of $\Phi+\Psi$ to test for the presence of anisotropic 
stress. They also compare $\Phi$ with a measurement of the density $D$ to test 
the validity of the  Poisson equation. With this method, we see that if a 
non-zero 
anisotropic stress is detected, we cannot know if this anisotropic stress is 
real, or if it is a consequence of the fact that the Euler equation is not 
valid, 
and that the inferred $\Psi$ is wrong. Similarly, any observed deviation in the
Poisson equation could either be real, or due to the fact that $\Psi$ is 
incorrect (leading to an incorrect $\Phi$). Observing the dipole is therefore 
crucial for these tests, since it will break the degeneracies by providing a 
direct measurement of $\Psi$. Note that this method is highly complementary to 
the tests of the equivalence principle that have been proposed using consistency 
relations between the two-point and three-point correlation 
functions \cite{Kehagias:2013rpa,Creminelli:2013nua}. These consistency 
relations indeed
provide  a test of the  equivalence principle between baryons and dark 
matter, but they do not provide a measurement of the gravitational potential 
$\Psi$.

As discussed above, the dipole is a combination of relativistic 
effects~\eqref{cb:rel} and wide-angle effects (the first term 
in~\eqref{cb:wide}). 
To measure $\Psi$ with the dipole, it is therefore necessary  to remove the 
wide-angle 
effects. As shown in~\cite{Bonvin:2013ogt}, the wide-angle effects are 
proportional to the difference between the quadrupole of the bright and of the 
faint populations. We can therefore remove the wide-angle effects  by 
constructing the following estimator~\footnote{Note that the sign 
in~\cite{Hall:2016bmm} and~\cite{Bonvin:2018ckp} is incorrect, however this is 
a typo and the results are correct.}~\cite{Hall:2016bmm}
\begin{align}
\hat{\xi}_{1}&=\frac{3}{8\pi}\left(\frac{\ell_{Pl}}{d}\right)^2\frac{\ell_{Pl}^3
} {\mathcal{V}_z}\sum_{ij}
\big[\Delta_{\B}(\bx_i)\Delta_{\F}(\bx_j)-\Delta_{\F}(\bx_i)\Delta_{\B}
(\bx_j)\big]P_1(\cos\sigma_{ij}) \delta_{K}(d_{ij}-d)\label{cb:estimator} 
\nonumber \\
&-\frac{3}{10}\frac{d}{r}\frac{5}{4\pi}\left(\frac{\ell_{Pl}}{d}\right)^2\frac{
\ell_{Pl}^3}{\mathcal{V}_z}\sum_{ij}
\big[\Delta_{\B}(\bx_i)\Delta_{\B}(\bx_j)-\Delta_{\F}(\bx_i)\Delta_{\F}
(\bx_j)\big]  P_2(\cos\sigma_{ij})\delta_{K}(d_{ij}-d)\, ,
\end{align}
where $\ell_{Pl}$ denotes the size of the cubic pixels, and $\mathcal{V}_z$ is 
the volume of the redshift bin in which we average the dipole.

\begin{figure}[!hb]
\centering
\includegraphics[width=0.6\columnwidth]{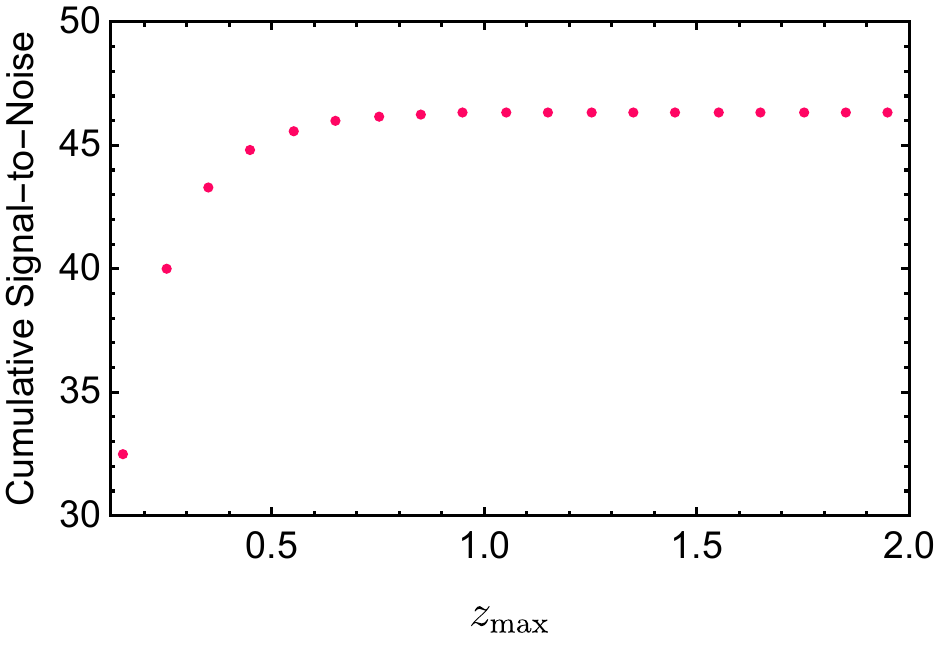}
\caption{\label{fig:cb_sn} {\it{Signal-to-noise for the dipole for a survey 
like SKA 
phase 2, cumulative over separations from $10\leq d\leq 200$\,Mpc/$h$ and up to 
redshift $z_{\rm max}$. We use pixels of size $\ell_{Pl}=2$\,Mpc/$h$.}}}
\end{figure}

To test the Euler equation, we modify it in the following 
way~\cite{Bonvin:2018ckp}
\be
\dot{\mathbf{V}}\cdot\mathbf{n}+\mathcal{H}\big[1+\Theta(z)  
\big]\mathbf{V}\cdot\mathbf{n}+\big[1+\Gamma(z) \big]\partial_r\Psi=0\, ,
\ee
where $\Theta(z)$ and $\Gamma(z)$ are two free parameters, which vanish  if the 
equivalence principle is valid. For simplicity we assume that these two 
parameters depend only on redshift and not on scale. This is the case in 
particular models, like scalar-tensor and vector-tensor theories, that are 
non-minimally coupled to dark matter (in the quasi-static approximation). The 
dipole is then modified in the following way by $\Theta(z)$ and $\Gamma(z)$
\begin{align}
\xi_1=\langle \hat \xi_1\rangle&= \frac{\HH}{\HH_0}\Bigg\{ (b_\B-b_\F)  
\left[\left(\frac{2}{r\HH} + \frac{\dot\HH}{\HH^2}\right)f + \Upsilon(z)\right]
+3 (s_\F-s_\B)f^2\left( 1-\frac{1}{r\HH}\right) \nonumber\\
&+ 5  (b_\B s_\F-b_\F s_\B) f  
\left(1-\frac{1}{r\HH}\right)\Bigg\}\nu_1(d,z)\, , \label{cb:dipmod}
\end{align}
where
\begin{equation}\label{cb:Ups}
\Upsilon(z) \define \frac{\Theta-\Gamma}{1+\Gamma} \,  f - 
\frac{\Gamma}{1+\Gamma} \left(\frac{\dot\HH}{\HH^2}f+f^2+\frac{\dot f}{\HH} 
\right)\, .
\end{equation}
We now forecast the precision with which future experiments,  like 
DESI~\cite{Aghamousa:2016zmz} and the SKA~\cite{SKAWeb}, will be able to 
constrain $\Theta(z)$ and $\Gamma(z)$. Details of the forecasts can be found 
in~\cite{Bonvin:2018ckp}. Here we summarise the results.

In Fig.~\ref{fig:cb_sn} we show the cumulative  signal-to-noise of the dipole  
for a survey like SKA phase 2, cumulative over separations (from $10\leq d\leq 
200$\,Mpc/$h$), and up to redshift $z_{\rm max}$. We use the specifications 
of~\cite{Bull:2015lja} for the number density, volume and bias. We split the 
populations of galaxies into two populations with the same number of galaxies, 
with bias $b_\B=b+\Delta b/2$ and $b_F=b-\Delta b/2$, and we choose $\Delta 
b=0.5$.
The signal-to-noise is directly proportional  to $\Delta b$. As an example, a 
bias difference of $\Delta b\simeq1$ between the bright and faint populations of 
luminous red galaxies has been measured in BOSS~\cite{Gaztanaga:2015jrs}. For 
the HI galaxies targeted by the SKA, the expected bias difference is less well 
known, therefore we choose a conservative bias difference of $\Delta b=0.5$. For 
simplicity we set here $s_{\B}=s_{\F}=0$.
From Fig.~\ref{fig:cb_sn}, we see that above $z\simeq 1-1.2$,  
 there is no improvement of the 
signal-to-noise. This is due to the fact that the dipole decreases 
with redshift and is therefore more easily detectable at low redshift. We can 
therefore safely neglect the contribution from lensing magnification to the 
dipole, and also to the other even multipoles, since it is strongly subdominant 
below $z=1.2$. The cumulative signal-to-noise over the whole range of redshift 
reaches 46.4, showing that the dipole will be robustly measured with the SKA.

\begin{figure}[ht!]
\centering
\includegraphics[width=0.66\columnwidth]{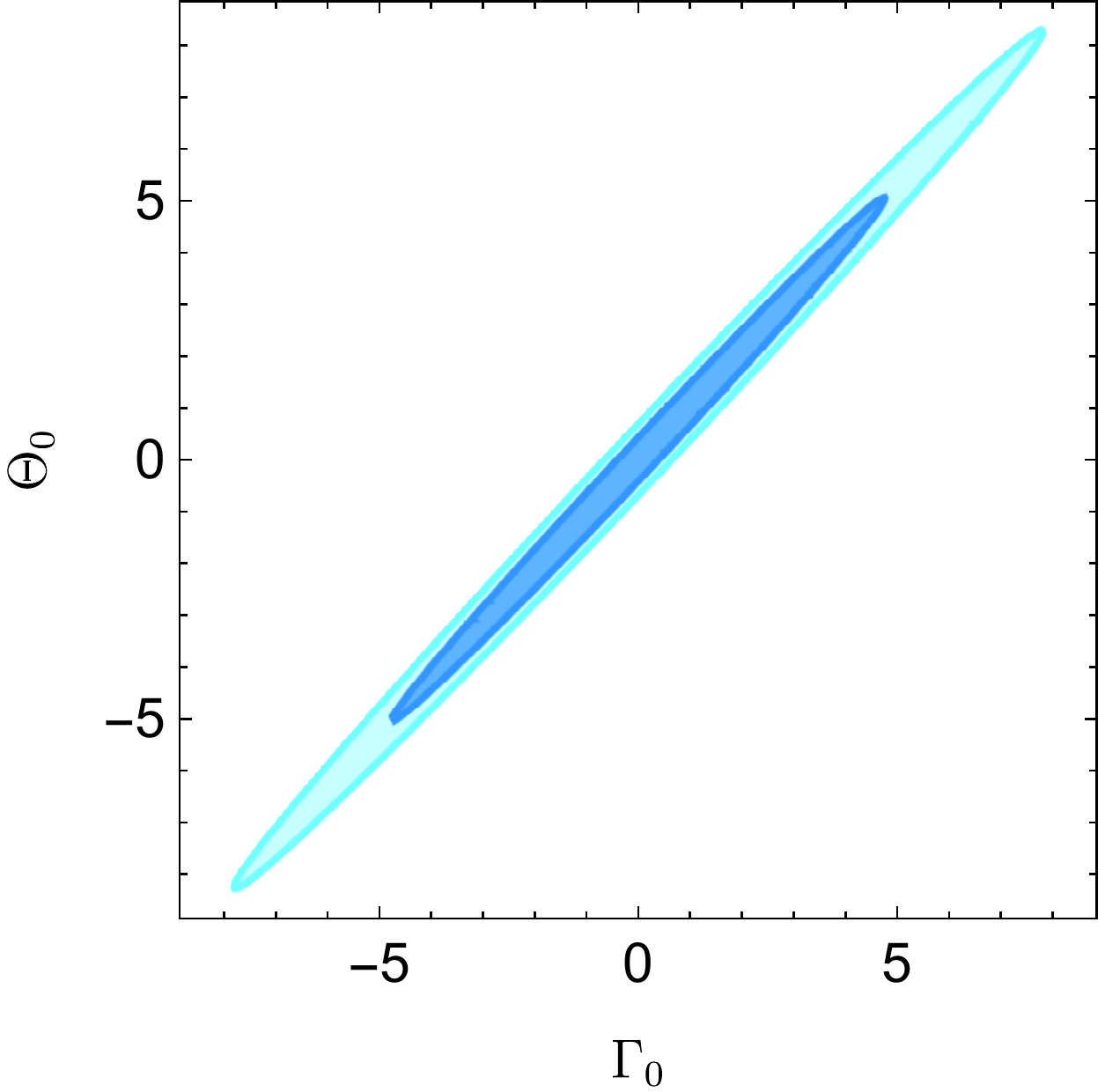}
\caption{\label{fig:cb_dip}{\it{Joint $1\sigma$ and $2\sigma$ constraints on 
$\Theta_0$ and $\Gamma_0$ obtained from the dipole for a survey like SKA phase 
2. All other cosmological parameters and the biases are fixed to their fiducial 
values. Published in  \cite{Bonvin:2018ckp} (© OP Publishing Ltd and SISSA 
Medialab Srl.  Reproduced by permission of IOP Publishing.  All rights 
reserved).}}}
\end{figure}

Following~\cite{Gleyzes:2014rba}, we assume that the functions  $\Theta(z)$ and 
$\Gamma(z)$ evolve as
\be
\Theta(z)=\Theta_0\frac{\Omega_\Lambda(z)}{\Omega_\Lambda(0)}\, , \label{cb:evol}
\ee
and similarly for $\Gamma(z)$. This means that the breaking of the equivalence 
principle happens  at late time, when dark energy or modified gravity are 
relevant. We forecast the constraints on $\Theta_0$ and $\Gamma_0$, using a 
Fisher matrix analysis. In Fig.~\ref{fig:cb_dip} we show the $1\sigma$ and 
$2\sigma$ contours obtained by fixing all other cosmological parameters and the 
biases to their fiducial values, and assuming no other deviation from General 
Relativity. We see that the constraints are extremely degenerated. This can be 
understood from Eq.~\eqref{cb:Ups}, which shows that the first term in 
$\Upsilon$ is sensitive to $\Theta-\Gamma$ (at lowest order in $\Theta$ and 
$\Gamma$ the denominator reduces to 1). This first term dominates in $\Upsilon$, 
which is therefore mainly sensitive to this combination. What we are therefore 
really interested in is rather the sensitivity of the dipole to $\Upsilon$, or 
similarly to one of the parameters $\Theta$ or $\Gamma$, since our goal is to 
detect any generic violation of the equivalence principle.

\begin{figure}[ht!]
\centering
\includegraphics[width=0.66\columnwidth]{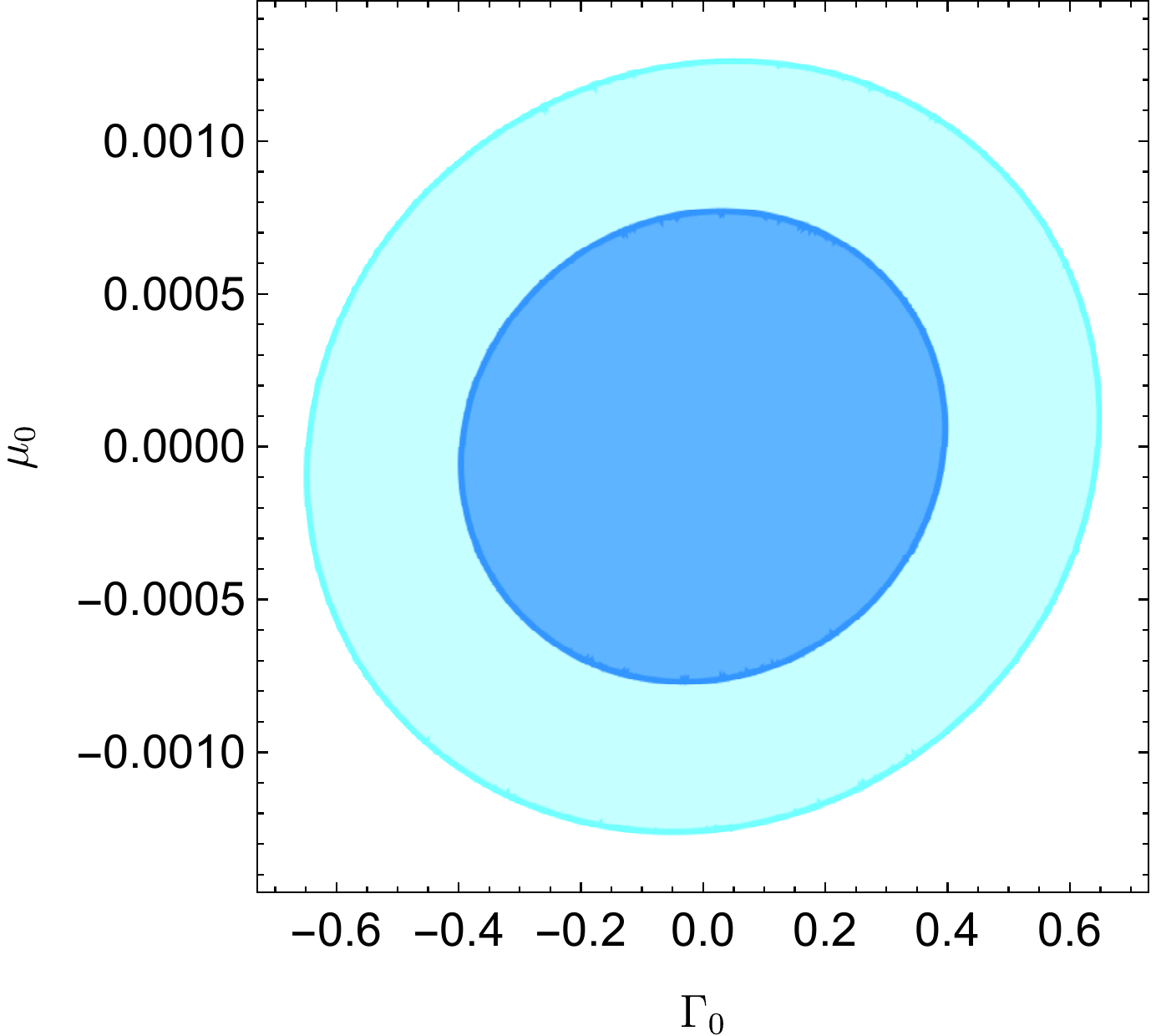}
\caption{\label{fig:cb_all}
{\it{
Joint $1\sigma$ and $2\sigma$ constraints on $\mu_0$ 
and $\Gamma_0$ obtained from  a combination of the dipole, monopole, quadrupole 
and hexadecapole for a survey like SKA phase 2. We marginalise over the biases 
of the bright and faint populations. $\Theta_0$ is fixed to zero and all other 
cosmological parameters are fixed to their fiducial values.  Published in  
\cite{Bonvin:2018ckp} (© OP Publishing Ltd and SISSA 
Medialab Srl.  Reproduced by permission of IOP Publishing.  All rights 
reserved).}}}
\end{figure}

In the following we therefore  fix   $\Theta=0$, and we forecast the 
constraints on 
$\Gamma_0$. We also  include this time    the monopole, quadrupole and 
hexadecapole 
in the forecasts. These terms are not sensitive to $\Gamma_0$ but they are 
sensitive to the biases, which we model in the following way~\cite{Bull:2015lja}
\bea
b_\B(z)&=&b_{1} e^{b_{2} z}+\frac{\Delta b}{2}\, ,\\
b_\F(z)&=&b_3 e^{b_4 z}-\frac{\Delta b}{2}\, ,
\eea
where as before we fix the bias difference $\Delta b=b_\B-b_\F=0.5$. We  
therefore have four additional free parameters  $b_1, b_2, b_3$ and $b_4$ that 
we 
want to constrain. We choose their fiducial value as in~\cite{Bull:2015lja} (see 
table 6): $b_1=b_3=0.554$ and $b_2=b_4=0.783$. In this way the bias of the 
bright and of the faint population are uncorrelated, but their difference is 
always   0.5.
The even multipoles are also sensitive to any modification in the growth of 
structure. We model this by a function $\mu(z)$, which modifies the growth 
function
\be
\label{cb:D1mod}
D_1(z)=\bar D_1(z)\big[1+\mu(z) \big]\, ,
\ee
where $\bar D_1$ denotes the growth function in a $\Lambda$CDM universe, and we 
let $\mu(z)$ evolve as in Eq.~\eqref{cb:evol}.  The growth rate can then be 
written as a function of $\mu_0$
\be
\label{cb:fmod}
f(z)=\bar f(z)+3\Omega_{{m}0}(1+z)^3\left[\frac{1-\Omega_{m}(z)}{1-\Omega_{
m0}}\right]^2\mu_0\, ,
\ee where $\bar f$ denotes the growth rate in a $\Lambda$CDM universe. The 
parameter $\mu_0$ and the four  bias parameters are constrained by the even 
multipoles and by the dipole, whereas $\Gamma_0$ is only constrained by the 
dipole. In Fig.~\ref{fig:cb_all} we show the $1\sigma$ and $2\sigma$ contours on 
$\Gamma_0$ and $\mu_0$, marginalised over the bias parameters. We see that the 
constraints on $\Gamma_0$ are much tighter, now that $\Theta_0=0$. This really 
shows the sensitivity of the dipole to a generic breaking of the equivalence 
principle. The constraints on $\mu_0$ are tighter than the constraints on 
$\Gamma_0$ by two orders of magnitude. This simply reflects the fact that the 
even multipoles, which are sensitive to $\mu_0$, have a significantly larger 
signal-to-noise than the dipole, which is sensitive to $\Gamma_0$. A 
constraint  
of $\sim$ 20-30 percent on the validity of the equivalence principle is, 
however, 
a very significant constraints at low redshift, since standard RSD and lensing 
measurement are completely insensitive to this.

\section{Conclusions}
\label{sec:cb conclusion}

Galaxy clustering is a powerful probe of the theory of gravity. Here we have  
demonstrated that the relativistic effects in the galaxy number counts do carry 
additional information, which is complementary to the standard density and RSD 
contributions. In particular, one extremely interesting relativistic effect is 
gravitational redshift, which is sensitive to the time component of the metric, 
$\Psi$. We have shown that by splitting the population of galaxies into two 
populations, and by fitting for a dipole in the cross-correlation between these 
two populations, we can isolate the effect of gravitational redshift. We can 
then use this effect in conjunction with RSD to test the equivalence principle 
for dark matter. We have forecasted the precision with which a violation of the 
equivalence principle can be detected with a survey like SKA phase two and we 
have 
seen that it reaches 20-30 percent. This shows the interest of relativistic 
effects in galaxy clustering.







\chapter[Cosmological Constraints From the Effective Field Theory of Dark 
Energy]
{Cosmological Constraints From the Effective Field Theory of Dark 
\\ Energy}
\label{sec:Frusciante}

{\em Noemi Frusciante,
Simone Peirone}



\section{The Effective Field Theory for Dark Energy in a 
Nutshell}\label{NFSPsec:intro}
\label{eftrefs1}

The Effective Field Theory (EFT) framework for dark energy (DE)  provides   an 
effective description for cosmological perturbations of genuine departures from 
General Relativity  
(GR)~\cite{Gubitosi:2012hu,Bloomfield:2012ff,Gleyzes:2013ooa,Piazza:2013pua,
Hu:2014oga,Tsujikawa:2014mba,Li:2018ixg}.  It  encompasses any dark energy  and 
modified 
gravity (MG) models with one additional scalar degree of freedom (DoF)  through 
a variety of geometrical operators compatible with the residual symmetries of 
unbroken spatial diffeomorphisms. The operators are organised in powers of the 
number of perturbations and spatial derivatives, and each of them multiplies a 
time-dependent function, called EFT function. The EFT action is constructed in  
the unitary gauge around a Friedmann-Lema\^{i}tre-Robertson-Walker (FLRW) 
background, and up to second order in perturbations it 
reads~\cite{Gubitosi:2012hu}
\begin{eqnarray}
&&
\!\!\!\!\!\!\!\!\!\!\!\!\!\!\!\!\!\!\!\!\!\!\!\!\!\!\!\!\!\!\!
S= \frac12\int d^4x \sqrt{-g} \left[\mp^2  \fg(t) R \,  - 2 \Lambda(t)     - 
2c(t)  g^{00} +  M_2^4(t) (\delta g^{00})^2 \right.
\nonumber\\
&&
\ \ \ \ \ \ \ \ \ \ 
- \bar m_1^3(t)\,  \delta g^{00} \delta K - \bar M_2^2(t)\,  \delta K^2 - \bar 
M_3^2(t)\,  \delta K_{\mu}^{\ \nu} \delta K_{\ \nu}^\mu \nn\\
&&
\ \ \ \ \ \ \ \ \ \ \left. + \mu_1^2(t) \delta g^{00} \delta R+ m_2^2(t) h^{\mu 
\nu} \partial_\mu g^{00} \partial_\nu g^{00} + \dots \,\right]\nn\\
&&
\ \ \ \ \ \ \ \ \ \ + S_m[g_{\mu\nu},\chi_m]\,,	\label{NFSP:eftaction}
\end{eqnarray}
where $\mp^2$ is the Planck mass, $g^{00}= -1+\delta g^{00}$ is the time-time 
component of the metric and $\delta g^{00}$ its perturbation, $g$ is the  
determinant of the metric, $g^{\mu\nu}$, $h_{\mu\nu}=g_{\mu\nu}+n_\mu n_\nu$ is 
the induced metric, with $n_\mu$ being the unit vector  perpendicular to the 
time 
slicing, $\delta R$ and $\delta R_{\mu \nu} $ are the perturbations of the Ricci 
scalar and tensor, and $\delta K$ and $\delta K^\mu_\nu$ are the perturbations 
of 
the extrinsic curvature scalar and tensor.   $\fg$, $\Lambda$, $c$, $M_i$, 
$m_i$, $\bar M_i$, $\bar m_i$ and $\mu_i$ are  the EFT  functions. We note that 
$\fg,\Lambda,c$ are the sole  EFT functions entering in both the background 
equations and linear perturbations, while the others affect only the 
perturbations.  $S_m$ is the matter action, with the metric $g_{\mu\nu}$ 
universally coupled to the matter fields $\chi_m$.  The ellipsis stand for  
possible extensions of the action in several directions to include:
 additional second order operators, e.g., $(\delta R)^2$  
\cite{Gleyzes:2013ooa}, or  higher order terms in  
derivatives~\cite{Kase:2014cwa,Frusciante:2015maa,Frusciante:2016xoj}, direct 
gravitational interaction between the additional scalar DoF and the matter 
fields~\cite{Gleyzes:2015pma,Gleyzes:2014qga,Tsujikawa:2015upa,DAmico:2016ntq}, 
 theories with second-order derivative equations of motion with a vector or 
 tensor additional field \cite{Lagos:2016wyv,Lagos:2017hdr},  non-linear 
perturbative effects 
\cite{Bellini:2015wfa,Frusciante:2017nfr,Yamauchi:2017ibz,Cusin:2017mzw,
Cusin:2017wjg,Kennedy:2019nie}.
  
 The EFT framework preserves a direct link between the model-independent   
\label{madelindepeefs1}
approach of the EFT basis and specific DE/MG models, such as $f(R)$, Horndeski  
\label{Horndeskiref2}
and others. It is indeed possible to obtain a mapping recipe, which allows us 
to 
write any EFT function in term of the free functions  characterising a specific 
theory~\cite{Gubitosi:2012hu,Bloomfield:2012ff,Bloomfield:2013efa,
Gleyzes:2013ooa,Gleyzes:2014rba,Frusciante:2015maa,Frusciante:2016xoj}.  It 
follows that considering sub-sets of EFT functions, we can model specific 
classes of DE/MG models. For example, the sub-set of background EFT functions  
$\{\fg, \Lambda, c\}$ describes the Generalised Brans-Dicke class of theories  
\label{Bransref5}
(GBD), while imposing $\bar M_2^2=-\bar M_3^2=2\mu_1^2$ (and $m_2^2=0$)    
restricts us to Horndeski theory~\cite{Horndeski:1974wa,Deffayet:2009mn} and 
relaxing the latter condition to have $\bar M_2^2=-\bar M_3^2$ (and $m_2^2=0$) 
 we can construct scalar-tensor theories beyond Horndeski, the so called 
Gleyzes-Langlois-Piazza-Vernizzi (GLPV) theories~\cite{Gleyzes:2014dya}.  
\label{GLPV1}
Finally,  Lorentz violating theories are characterised by $m^2_2\neq 0$ 
\cite{Gubitosi:2012hu,Frusciante:2015maa}.

An alternative basis of the EFT action \eqref{NFSP:eftaction}, called  
$\alpha$-basis, \label{alphabasisrefs}was  introduced in~\cite{Bellini:2014fua}. 
This basis encodes 
specific physical properties of the Horndeski theory in four time-dependent 
phenomenological functions, namely: the {\it running Planck mass} 
$\alpha_M(t)$, 
which defines the time variation of the  \textit{effective Planck mass} 
$M^2(t)$, the  {\it braiding} $\alpha_B(t)$ accounting for the interaction 
between the extra DoF and the metric, the {\it kineticity} $\alpha_K(t)$, which 
is purely a kinetic function, and the {\it speed of tensor excess} 
$\alpha_T(t)$, 
describing any modification  in the speed of propagation of tensor modes. The 
inclusion of  GLPV models requires an extra function, $\alpha_H(t)$, defining 
the departure from Horndeski models~\cite{Gleyzes:2014qga,Gleyzes:2014rba}, and 
two 
additional   functions are necessary to include  Lorentz violation effects,  
\label{loclinref7}
$\alpha_{K_2}(t),\alpha_B^{GLPV}(t)$~\cite{Frusciante:2016xoj}, respectively an 
extension of the {\it kineticity} and deviation from GLPV models.
These functions can be identified in terms of the EFT basis as follows
\ba\label{NFSP:alphabasis}
&&\alpha_M=\frac{1}{H}\frac{d\ln M^2}{d\ln t} \,,\qquad 
\alpha_B(t)=-\frac{\mp^2\dot{\fg}+\bar{m}^3_1}{H M^2}\,,  \nn \\ 
&&\alpha_T(t)=\frac{\bar{M}^2_3}{M^2}\equiv c_t^2-1\,,\qquad 
\alpha_K(t)=\frac{2c+4M_2^4}{H^2 M^2}\,, \nn\\
&&\alpha_{K_2}(t)=\frac{8m_2^2}{M^2H^2}\,,\qquad \alpha_H(t)=\f{2\mu^2_1+\bar{M}^2_3}{M^2}\,, \nn \\
&&\alpha^{GLPV}_B(t)=\f{\bar{M}^2_3+\bar{M}^2_2}{M^2}\,,
\ea
where $M^2(t)=\mp^2\fg-\bar{M}^2_3$, $c_t$ is the  speed of propagation  of 
gravitational waves (GWs) \label{gravitationalwavrefs9} (tensor modes) and 
$H=\f{1}{a}\f{da}{dt}$ is the 
Hubble function.   While all the $\alpha$ functions enter into the linear 
scalar 
perturbation equations, only two of them impact the propagation of GWs, i.e., 
$\alpha_M$, which modifies the friction term and $\alpha_T$, which as previously 
mentioned,
is the deviation in the speed of propagation of GWs.

The advantage of having at our disposal a framework  to perform 
model-independent explorations of gravity models    allowed to identifying clear 
patterns~\cite{Bloomfield:2012ff,Piazza:2013pua,DeFelice:2016ucp,Gleyzes:2015pma
,Tsujikawa:2015upa,DeFelice:2017mwa,Gleyzes:2014qga,Tsujikawa:2015upa,
DAmico:2016ntq,Pogosian:2016pwr,Perenon:2015sla,Perenon:2016blf,Peirone:2017ywi,
Lombriser:2015cla,Espejo:2018hxa,Frusciante:2018vht,Kennedy:2017sof,
Kennedy:2018gtx,Lombriser:2018olq,Tsujikawa:2015mga,Traykova:2019oyx,
DAmico:2016ntq,Linder:2019bqp,Amendola:2014wma,Bellini:2015wfa,
Salvatelli:2016mgy,Renk:2016olm,Zumalacarregui:2016pph,Yamauchi:2017ibz,
Brush:2018dhg,Garcia-Garcia:2018hlc,Frusciante:2018jzw,Hirano:2018uar,
Traykova:2019oyx,Duniya:2019mpr,Pace:2019uow} to be tested  against cosmological 
data.  In the following we will overview the state of the art about cosmological 
constraints obtained in the EFT context using present day measurements and 
forecasts from future missions.

\section{Einstein Boltzmann Codes}\label{NFSPsec:ebcodes}

\label{EinBoltcodesref1}

The large number of DE/MG proposals introduces the challenge  of being able to 
constrain them against observational data, such as the cosmic microwave 
background (CMB) radiation, galaxy clustering (GC), \label{galaxyclurefs2}weak 
lensing (WL),  \label{weaklensefs3}
redshift-space distortions (RSD), Supernovae Ia (SNIa)  and baryon acoustic 
oscillations (BAO). To this purpose one needs to accurately compute a range of 
theoretical predictions for the considered cosmologies. This is possible thanks 
to  Einstein-Boltzmann (EB) codes that numerically solve  the linear evolution 
of relevant perturbed quantities (e.g., gravitational potentials, matter 
density 
fluctuations) on an expanding FLRW background. \label{CMBrefs4}

For the standard $\Lambda$CDM scenario, many different EB codes 
exist~\cite{Peebles:1970ag,Wilson:1981yi,Ma:1995ey,Seljak:1996is,Seljak:2003th,
Kaplinghat:2002mh,Doran:2003sy}. Among these, the most widely used in the 
cosmological analyses  are: \camb~\cite{Lewis:1999bs} and 
\class~\cite{Blas:2011rf}.
EB codes based on the EFT framework have also been developed in order  to test a 
broad class of DE/MG models. In such codes the linear perturbation equations are 
written in terms of the EFT functions preserving the model-independent approach 
of the EFT framework. The EFT functions can be then fixed according to two 
procedures:
\begin{itemize}
\item The \textit{pure} EFT approach: each EFT function (or a sub-set) is 
directly parametrised,  for example as a function of the scale factor, $a$, or 
the DE density parameter $\Omega_{\rm DE}$. In this case the background 
expansion has to be chosen as well, by fixing the DE equation of state $w_{\rm 
DE}$;
\item The  {\it mapping} approach: the EFT functions evolve according to a 
specific DE/MG model  through the mapping 
recipe~\cite{Gubitosi:2012hu,Bloomfield:2012ff,Bloomfield:2013efa,
Gleyzes:2013ooa,Gleyzes:2014rba,Frusciante:2015maa,Frusciante:2016xoj}, and 
their 
behaviour is fixed after solving the background Friedmann equations for the 
chosen model.
\end{itemize}

 As a first example of such codes, \eftcamb~\cite{Hu:2013twa,Raveri:2014cka} 
(\url{www.eftcamb.org}) is an implementation of the EFT framework into  \camb.
The code can compute cosmological  observables for specific models  ({\it 
designer} $f(R)$-gravity~\cite{Raveri:2014cka}, $f(R)$ Hu-Sawicki  
\label{fRref3}
model~\cite{Hu:2016zrh}, minimally coupled quintessence~\cite{Hu:2014oga},  
low-energy Ho\v rava gravity~\cite{Frusciante:2015maa}, covariant 
Galileon~\cite{Peirone:2017vcq}, K-mouflage~\cite{Benevento:2018xcu}, Galileon 
Ghost Condensate \cite{Peirone:2019aua}, beyond Horndeski model 
\cite{Peirone:2019yjs}, generalised cubic covariant Galileon 
\cite{Frusciante:2019puu}) as well as for phenomenological 
parameterisations\label{gphenomenologicalprfs3} of 
the time dependence of the EFT functions and $w_{\rm DE}$.
The code is interfaced with a Markov Chain Monte-Carlo (MCMC)  code, 
\texttt{EFTCosmoMC}~\cite{Raveri:2014cka}, which allows  constraints to be 
placed  on 
specific MG models, see e.g. 
refs.~\cite{Hu:2016zrh,Frusciante:2015maa,Raveri:2014cka,Peirone:2017vcq,
Benevento:2018xcu,Peirone:2019aua,Peirone:2019yjs}, to explore the interplay 
between massive neutrinos and DE~\cite{Hu:2014sea},
the tension between primary and weak lensing signal in  CMB 
data~\cite{Hu:2015rva}, as well as the behaviour and impact of theoretical 
priors~\cite{Raveri:2017qvt,Peirone:2017lgi,Frusciante:2018vht}.

Among the same category, \hiclass~\cite{Zumalacarregui:2016pph}  
(\url{www.hiclass-code.net}) implements in  \class the evolution equations in 
terms
of the $\alpha$-basis defining Horndeski~\cite{Zumalacarregui:2016pph} and GLPV 
models~\cite{Traykova:2019oyx}.
\hiclass  is interfaced with \texttt{MontePython} 
\cite{Audren:2012wb,Brinckmann:2018cvx}  to perform cosmological constraints. It 
has been used to place constraints on parameterisations of the $\alpha_i 
(\tau)$ 
with current CMB data~\cite{Bellini:2015xja},  study relativistic effects on 
ultra-large scales~\cite{Renk:2016olm}, forecast constraints  forecastwith Stage 
4 
clustering, lensing and CMB data~\cite{Alonso:2016suf}, and constraints on 
Brans-Dicke theory~\cite{Alonso:2016suf} and Covariant Galileon 
models~\cite{Renk:2017rzu}.

Another example implementing the $\alpha$-basis is \coop~\cite{Huang:2012mt} 
(\url{www.cita.utoronto.ca/~zqhuang}).
The code then outputs CMB power spectra by using the line of sight
integral approach~\cite{Seljak:1996is, Hu:1997hp}, while the matter power 
spectra  are computed via a gauge transformation from
 the Newtonian to the cold dark matter (CDM) rest frame in synchronous gauge.   
\coop  includes the dynamics of the GLPV operator, and it has been used in 
order 
to study the signatures of a non-vanishing $\alpha_\text{H}$ on the matter 
power 
spectrum  and on the primary and lensing CMB signals~\cite{DAmico:2016ntq}.

 Finally, there is \texttt{EoS\_class}  \cite{Pace:2019uow} 
(\url{https://github.com/fpace}),  which  implements  the equation of state  
approach \cite{Battye:2013aaa} in \class, using  the $\alpha$-basis description 
of the Horndeski theory. As specific theory the code includes the {\it designer} 
$f(R)$-model (\texttt{CLASS\_EOS\_FR})~\cite{Battye:2015hza}.

The aforementioned codes have been validated~\cite{Bellini:2017avd,Pace:2019uow} 
by comparing  the shapes of the CMB  angular  power spectra and of   dark 
matter power spectrum, showing an excellent sub-percent agreement on the scales 
of interest for present and upcoming surveys.

A common feature of the above codes is that they  have built-in modules to 
enforce stability conditions.  Such conditions are derived in the EFT framework  
encompassing general classes of DE/MG models 
\cite{Bloomfield:2012ff,Piazza:2013pua,Kase:2014cwa,DeFelice:2016ucp} to ensure 
that the model under consideration is stable against ghosts, negative speeds of 
propagation and tachyonic instabilities.  All of the above EB codes check for 
the no-ghost and positive speeds of propagation. \eftcamb is the only one that 
includes the no-tachyonic condition \cite{Frusciante:2018vht}. Stability 
requirements, when used as priors in MCMC codes, reduce the viability space, 
these 
codes have to explore, and  they can even dominate over the constraining power 
of 
observational 
data~\cite{Piazza:2013pua,Salvatelli:2016mgy,Peirone:2017lgi,Raveri:2014cka,
Frusciante:2015maa,Peirone:2017lgi,Frusciante:2018vht}.

Another advantage in using such EB codes is that the set of  linear perturbative 
 equations   are evolved without assuming any Quasi Static (QS) approximation. 
The latter  considers the time derivatives of linear perturbations 
as sub-leading contributions compared to the spatial derivatives inside the 
sound horizon of the DE mode, i.e. 
$k/aH>c_s$~\cite{Sawicki:2015zya,Frusciante:2018jzw}, hence they are neglected. 
However, such approximation in some cases is not sufficient to exploit the full 
dynamics of the extra DoF. For example, in Horndeski theory, time derivatives 
revealed to modify the evolution of the linear gravitational potentials even  
within the sound horizon of the scalar field ($k <0.001$ 
h/Mpc)~\cite{Peirone:2017ywi} and a  semi-dynamical treatment of the linear 
perturbations at a pivot scale showed that the QS approximation is not 
sufficiently accurate for GLPV models \cite{Lombriser:2015cla}.

\section[Cosmological Constraints on Horndeski and GLPV Models]{Cosmological 
Constraints on Horndeski  and GLPV \\ Models}\label{NFSPsec:constraints}

Modifications in the gravitational interaction can largely affect the shapes of 
the cosmological power spectra  
\cite{Sachs:1967er,Kofman:1985fp,Hu:1996vq,Acquaviva:2005xz,Amendola:2011ie}. 
The EFT approach allows inspection of the DE/MG effects on these  large scale 
observables in a model-independent fashion 
\cite{Amendola:2014wma,Bellini:2015wfa,Salvatelli:2016mgy,Renk:2016olm,
Zumalacarregui:2016pph,DAmico:2016ntq,Yamauchi:2017ibz,Brush:2018dhg,
Garcia-Garcia:2018hlc,Frusciante:2018jzw,Hirano:2018uar,Traykova:2019oyx,
Duniya:2019mpr,Pace:2019uow}, i.e.,  no covariant theory is considered, while a 
pure EFT approach is adopted. In this case the effects of a specific operator in 
the EFT action can be easily investigated. Let us notice that although switching 
on/off a specific operator can introduce a modification in the observables, the 
magnitude of such modification strictly depends on the adopted parameterisation 
for the corresponding EFT function.  For example,  a positive braiding term, 
$\alpha_B$, impacts  both the lensing and matter power spectra that are 
enhanced compared to $\Lambda$CDM scenario, and depending on its magnitude  can 
generate 
either a suppressed or an enhanced   ISW (integrated Sachs-Wolfe)  tail at 
low-$\ell$ in  the temperature-temperature power spectrum \cite{Pace:2019uow}. 
A suppression in the lensing power spectrum and an enhanced  TT power spectrum 
at low-$\ell$ with respect to $\Lambda$CDM scenario can be obtained for values 
of the 
effective Planck mass larger than $M_{Pl}$ \cite{Salvatelli:2016mgy}. 
Furthermore, $\alpha_M$ also alters   the amplitude of the GWs; in particular, 
 if 
$\alpha_M < 0$, the GW amplitude is smaller than that predicted by $\Lambda$CDM 
scenario, and 
the opposite holds for positive values \cite{Nunes:2018zot}. In this case,   in 
the GWs and  electromagnetic luminosity distance are different and can be  then 
 tested  by GWs experiments and standard sirens  \label{earlyDEefs3}
\cite{Abbott:2017xzu,Nissanke:2013fka,Lagos:2019kds,Ezquiaga:2018btd,
Belgacem:2019pkk}. 

Concerning the propagation of GWs, $\alpha_T$ changes the 
location of the inflationary peak of the BB spectrum \cite{Amendola:2014wma}.  
$\alpha_H$ modulates the matter power spectrum, the lensing potential and the 
CMB TT power spectrum at low-$\ell$ 
\cite{Traykova:2019oyx,Duniya:2019mpr,DAmico:2016ntq}.
Finally, the kineticity coupling, $\alpha_K$, modifies the low-$\ell$ TT power  
spectrum due to the late-time ISW effect \cite{Frusciante:2018jzw}.  This 
coupling does not  affect the constraints on the other model parameters 
regardless of the chosen parametrisation \label{ISWref2}
\cite{Bellini:2015xja,Kreisch:2017uet,Frusciante:2018jzw,Perenon:2019dpc}, and 
it 
is hard to be constrained   because its effect has been proven to be below the 
cosmic variance \cite{Frusciante:2018jzw}. Nevertheless, it gives a 
non-negligible contribution to the viable parameter space of the theory and thus 
it 
cannot be neglected in cosmological 
analysis~\cite{Kreisch:2017uet,Frusciante:2018jzw}.

In the following we discuss the cosmological constraints obtained in the 
\textit{pure} EFT approach  (we refer the reader to \cite{Frusciante:2019xia} 
for a review). Then one has to make a choice about the parametrisation to use 
for the  EFT functions.
Usually, such parameterisations are chosen in terms of  the scale factor, $a$ 
(e.g., 
linear, exponential, e-fold),   the DE density parameter $\Omega_{\rm DE}(a)$, 
or 
they are constants. As expected, the results of an analysis depend dramatically 
on the parametrisation adopted as well as on the choice for the background 
evolution: $w_{\rm DE}=-1$ ($\Lambda$CDM), $w_{\rm DE}=w_0$, where $w_0$ is a 
constant (wCDM) or $w_{\rm DE}=w_0+w_a(1-a)$, where $w_a$ is also a constant 
(CPL) \cite{Chevallier:2000qy,Linder:2002et}. Nevertheless, one can identify 
some clear trends.

In the literature, the set of EFT functions largely considered is the one  
describing the
 Horndeski  class of models or sub-classes, i.e.   
$\{\fg,\bar{m}^3_1,M^4_2,\bar{M}^2_2\}$ in the EFT basis and 
$\{\alpha_M,\alpha_K,\alpha_B,\alpha_T\}$ in the $\alpha$-basis. The 
modifications induced by the running Planck mass function, $\alpha_M$, have 
been 
constrained using different datasets. Planck CMB data (2015) and  the $H_0$ 
prior by Riess et al. \cite{Riess:2011yx}  constrain the free parameter in 
$\alpha_M=\alpha_{M,0}\Omega_{\rm DE}(a)$ to be positive at $2\sigma$ 
\cite{Huang:2015srv}. Let us note that this model, if representative of a 
covariant theory, would lead to a ghost instability. Indeed the kinetic term in 
this case is vanishing, since both $\alpha_B$ and $\alpha_K$ are zero. 
Considering instead  the relation $\alpha_M=-\alpha_B$ will  avoid the presence 
of scalar ghosts, and additionally it mimics the dynamics of 
$f(R)$~\cite{Sotiriou:2008rp,DeFelice:2010aj}  and 
Brans-Dicke~\cite{Brans:1961sx,Boisseau:2000pr} theories. In this  case the 
amplitude of a power low parametrisation for $\alpha_M$, is found to be 
$<0.097$ 
at 95\% C.L., using the combination of CMB+WL+BAO+RSD data~\cite{Ade:2015rim} 
in 
agreement with the upper bound  obtained using Planck13+WP+BAO+lensing 
($<0.061$ 
at 95\% C.L.) \cite{Raveri:2014cka}. The recent analysis by Planck 2018 shows 
that CMB data alone favour this parameter to be negative 
\cite{Aghanim:2018eyx}. 
On the other hand, assuming $\alpha_M=-\alpha_B/2$, which is the case of the No 
Slip Gravity model  \cite{Linder:2018jil}, stability conditions force 
$\alpha_M$ 
to be positive if the background is $\Lambda$CDM, while a larger viability 
space, allowed by a CPL background, leads the combination of CMB+BAO+RSD+SNIa 
measurements to favour a negative amplitude for $\alpha_M$ (parametrised with 
the 
e-fold form) \cite{Brando:2019xbv}.

\begin{figure}[ht!]
\vspace{-4.1cm}
\includegraphics[width = 
1.0\textwidth]{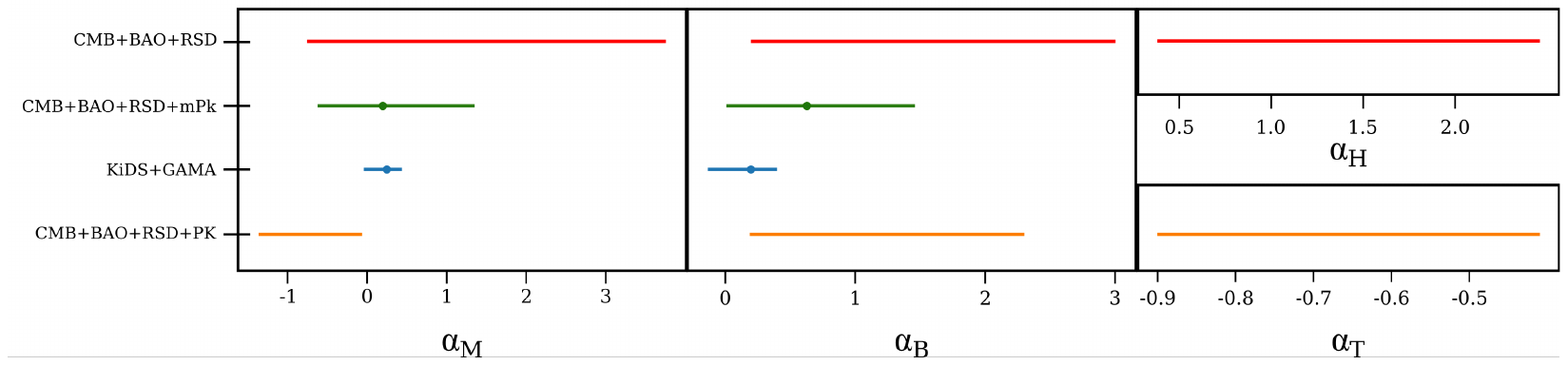}
\vspace{-4.3cm}
\caption{
{\it{
Collection of cosmological constraints obtained on the  linear 
DE-parametrisation, i.e. $\alpha_i(t)=\alpha_i \Omega_{DE}(a)$ on a 
$\Lambda$CDM background.  This is the most commonly used parametrisation for 
which 
we can show  more results and which covers different combinations of data sets. 
The $\alpha$ functions sharing the same colour are those switched on in the 
analysis. The constraints showed are at 95\% C.L. The blue error bars are from 
Ref. \cite{SpurioMancini:2019rxy}, the green ones from ref. 
\cite{Noller:2018wyv}, the orange ones are from Ref.  \cite{Bellini:2015xja}, 
and 
finally the red ones are  from Ref. \cite{Traykova:2019oyx}. 
\label{fig:NFSPalphaconstarints}
}}}
\end{figure}

The complete class of Horndeski theory is investigated assuming a 
parameterisation for  the four $\alpha$-functions  proportional to $\Omega_{\rm 
DE}$  on a $\Lambda$CDM background~\cite{Bellini:2015xja}. Also in this case, a 
combination of  CMB+BAO+RSD data and the shape of the power spectrum of 
galaxies 
from the WiggleZ survey favour a negative value for the amplitude of $\alpha_M$ 
at more than $2 \sigma$.  Instead, when assuming luminal propagation of tensor 
modes ($\alpha_T=0$), the MCMC analysis   shows  a preference for a positive 
value at 2$\sigma$ using KiDS+GAMA datasets \cite{SpurioMancini:2019rxy}, while 
CMB, RSD, matter power spectrum from SDSS and BAO measurements do not 
particularly favour any sign for $\alpha_{M,0}$ ($=0.20^{+1.15}_{-0.82}$ 95\% 
C.L.) \cite{Noller:2018wyv}. The latter combination of datasets, when used on a 
linear parametrisation in the scale factor, instead prefers a positive 
amplitude 
($0.27^{+0.54}_{-0.26}$ 95\% C.L.) \cite{Noller:2018wyv}. Some of these results 
are summarised in Fig. \ref{fig:NFSPalphaconstarints}.

Additionally, it has been shown that the constraints on  $\alpha_{M}$ will 
improve considerably  for Stage IV CMB experiments, even when considering a CPL 
parametrisation for the DE equation of state parameter~\cite{Reischke:2018ooh}.
Considering the propagation of tensor modes, $\alpha_M$ is degenerate with  the 
tensor-to-scalar ratio $r$,  as they both affect the amplitude of the 
primordial 
peak.  Demanding for $r$ to be close to zero requires $\alpha_M$ to assume 
negative values according to BICEP2 data \cite{Pettorino:2014bka}.   The sign 
of the amplitude of the braiding function, $\alpha_B$, is favoured to be 
positive 
regardless of the parametrisation considered when tested against the 
combinations  KiDS+GAMA ($\alpha_{B,0}=0.20^{+0.20}_{-0.33}$ 95\% 
C.L.)~\cite{SpurioMancini:2019rxy} and CMB+BAO+RSD with both the power spectrum 
of galaxies from the WiggleZ survey ($0.19 < \alpha_{B,0} < 2.30$ 95\% C.L.) 
\cite{Bellini:2015xja} and matter power spectrum from SDSS 
($\alpha_{B,0}=0.48^{+0.83}_{-0.46}$ 95\% C.L.) \cite{Noller:2018wyv}. 
However, when it shares a relation with $\alpha_M$, as previously discussed, it 
might show a negative amplitude in the case  where $\alpha_M$ is found to be 
positive.

Cosmological constraints on the amplitude of the parameter encoding the 
variation in the speed of GWs, $\alpha_{T,0}$, regardless of the 
parametrisation 
adopted, show that data (CMB, RSD, BAO) prefer sub-luminal propagation of 
GWs~\cite{Salvatelli:2016mgy,Noller:2018wyv}. Super-luminal propagation of GWs 
is instead possible when considering the B-modes data of the CMB  in 
polarisation (Planck and BICEP2 datasets)  \cite{Raveri:2014eea}. The speed of 
propagation of GWs is further constrained by the detection of the 
GW170817/GRB170817A  event that sets $\alpha_{T,0} \sim 
\mathcal{O}(10^{-15})$~\cite{Monitor:2017mdv} (see Section~\ref{NFSP:Astro}). 
In that case setting $\alpha_T=0$ has the impact to improve the constraints on 
the parameters of the remaining $\alpha$-functions and some investigations 
using present-day data and forecasts for next generation surveys have been 
performed with this prior 
\cite{Kreisch:2017uet,Frusciante:2018jzw,Noller:2018wyv,Perenon:2019qmd,
Bellini:2015wfa}.

Constraints on {\it pure} GLPV models with luminal propagation of tensor modes  
are derived using a combination of CMB, BAO and RSD datasets 
\cite{Traykova:2019oyx}. The beyond Horndeski parameter, $\alpha_H$, is found 
to   \label{beyondHorndeskiref1}
be degenerate with  $\alpha_B$  and   $\alpha_M$.  The data favour generally 
positive values of $\alpha_H$ and of order $\mathcal{O}(1)$.  We note that a 
stringent constraint on the present-day value of $\alpha_H$ of the  order  
$\mathcal{O}(10^{-6})$ is found when testing a specific model that extends the 
Galileon ghost condensate  to the domain of beyond Horndeski  theories 
\cite{Peirone:2019yjs}, and the phenomenon of GWs decay into DE sets a bound of 
$\mathcal{O}(10^{-10})$ \cite{Creminelli:2018xsv}.

The original EFT formalism given by the action (\ref{NFSP:eftaction}) assumes 
the validity  of the weak equivalence principle \label{equivprinref9} (WEP) and  
the matter fields, 
$\chi_m$, are hence minimally coupled to gravity through a unique metric 
$g_{\mu\nu}$.   However, direct gravitational interactions between the 
additional scalar DoF and the matter fields can also be  considered, for 
example 
considering  couplings between dark matter and neutrinos. In the latter case, 
  observational constraints are indeed  less severe than the case of 
baryons and 
photons~\cite{Hui:2009kc,Creminelli:2013nua}. In the context of the EFT, 
framework  couplings with matter fields have been investigated for Horndeski 
theory~\cite{Gleyzes:2015pma} and later generalised to GLPV 
theory~\cite{Gleyzes:2014qga,Tsujikawa:2015upa,DAmico:2016ntq}. In particular,  
a 
direct coupling between DE and dark matter has been investigated in the  
Horndeski case \cite{Gleyzes:2015rua}, with the coupling function of the form:
\be
\gamma_c(t)=\f{\beta_\gamma}{2\sqrt{2}}c_s(t)\sqrt{\alpha(t)}\,,
\ee
where $\beta_\gamma$ is a constant and $c_s(t)$ and $\alpha(t)$ correspond 
respectively to  the speed of propagation of the DE mode and kinetic function 
given by the following combination $\alpha_K +3/2 \alpha_B^2$. The forecast 
analysis based on  spectroscopic and photometric surveys with specifications 
similar to  Euclid  gives  the error on $\beta_\gamma$ of the order $\sim 
10^{-4}$ 
\cite{Gleyzes:2015pma}.

Improvements in the cosmological constraints of the EFT function parameters  
are 
obtained by considering specifications similar to those of future generation 
surveys. The constraints on the amplitude of the $\alpha$-functions  increased 
by a factor 5 \cite{Alonso:2016suf} forecasting LSST \cite{Abell:2009aa}, SKA 
\cite{Bacon:2018dui} and CMB-S4 \cite{Abazajian:2016yjj} experiments.    
\label{shearref2}
Considering  future CMB primaries, CMB lensing, GC and  cosmic shear data, a 
variation of the effective Newtonian gravitational coupling larger than 10\%  
is 
excluded  and  setting $c_t=1$ constrains the remaining parameters  at the 10\% 
level \cite{Reischke:2018ooh}. Furthermore, for  a Euclid-like experiment, the 
results of two cosmic shear methods have been compared using  the 
parametrisation $\alpha_i=\alpha_{i,0}\Omega_{\rm DE}$: the 3D analysis better 
constrains the model's parameters   by about 20\% compared to the tomographic 
approach \cite{Mancini:2018qtb}.

Observational measurements of RSD, GC and WL show a tension with Planck  CMB 
data in the estimation of $\sigma_8$, by predicting a lower growth of 
structures~\cite{Hildebrandt:2016iqg,deJong:2015wca,Kuijken:2015vca,
Conti:2016gav}. Horndeski models have been investigated in the EFT framework by 
using   large samples of models obtained with the Monte-Carlo approach. The 
results show that the models satisfying $M^2>M_{Pl}^2$ and with sub-luminal 
propagation of tensor modes can produce with respect to $\Lambda$CDM a lower 
growth of structure  at low-$z$ and a larger one at high-$z$  
\cite{Perenon:2015sla}.  An investigation on the growth index $\gamma(a)$ 
\cite{Linder:2005in} shows that  the majority of Horndeski models with the same 
expansion history as the standard cosmological  statistically generate a  growth 
rate of cosmic structures that is $12\%$ smaller than $\Lambda$CDM one, and no 
viable theory has  a larger  leading order growth index $\gamma_0$  than that of 
$\Lambda$CDM.\cite{Piazza:2013pua}. A suppression of the matter power spectrum 
is also possible in GLPV models, thanks to the extra free function $\alpha_H$ 
\cite{Traykova:2019oyx,Duniya:2019mpr,DAmico:2016ntq}, as discussed earlier.  
These aspects can be used to construct viable models able to ease the $\sigma_8$ 
tension. \label{sigmaref5}

 The above constraints rely on a specific choice of the time evolution  of EFT 
functions further specified through a certain number of free parameters.
Such a choice should try to model realistically cosmological observables, with  
the 
aim of revealing the nature of cosmic acceleration.  At the same time, it 
should 
not include too  many free parameters, to avoid loosening the constraining 
power 
of data \cite{Salvatelli:2016mgy}. Smooth parameterizations that 
are acceptable to describe the theory space as cosmological observables are only 
slightly sensitive to short time-scale variations \cite{Gleyzes:2017kpi}. 
Furthermore, one should be careful, because  sometimes the selected forms of 
the 
EFT functions generate simplified shapes of relevant physical quantities such as 
$\mu(t,k)$ or $\Sigma(t,k)$,  if compared to the complex behavior they show when 
specific covariant models  are considered \cite{Linder:2015rcz,Linder:2016wqw}. 
The risk in this case might be to  underestimate the modifications of the 
underlying  gravity  force or even miss its signatures.  An alternative approach 
relies on data-driven analysis to reconstruct the EFT functions across cosmic 
times using cosmological data, from which one can then derive specific model 
properties \cite{Espejo:2018hxa,Raveri:2019mxg}.  The reconstruction of 
\textit{pure} Horndeski models from CMB+WL+BAO+SNIa+$H_0$ shows that between 
$0.1<a<1$: the variation of the Planck mass and $\alpha_T$ are positive and 
$\alpha_B$ is mostly negative \cite{Raveri:2019mxg}.

Despite the uncertainties in fixing the EFT functions,  the EFT framework has   
already  boosted our knowledge about the real nature of  gravity force and 
helped in deriving novel predictions.

\section{Astrophysical Constraints}\label{NFSP:Astro}

Astrophysical constraints can be used to complement those obtained from 
measurements  at cosmological scales. In particular, the parameter space 
identified by the EFT functions can be constrained using  the bounds from the  
joint detection of the GWs event GW170817 and its electromagnetic counterpart 
GRB170817A and those from  massive astrophysical bodies. In details:

\begin{itemize}

\item \textit{GW170817 and GRB170817A:} the joint detection of the GWs event  
GW170817 from a binary neutron stars \label{neutronstarsref6} 
merger~\cite{TheLIGOScientific:2017qsa} and 
its gamma-ray burst GRB170817A~\cite{Monitor:2017mdv} placed a strong constraint 
on the deviation of  the speed of GW from the speed of light, $c$. It is 
estimated  to be $-3\times 10^{-15}\leq c_t-c\leq 7\times 
10^{-16}$~\cite{Monitor:2017mdv}. In the EFT formalism it implies 
$|\alpha_T|<10^{-15}$, with a very tiny amount of   room left for this kind of 
deviation in 
DE/MG models \cite{Creminelli:2017sry,Baker:2017hug,Ezquiaga:2017ekz}.  For 
example, assuming that this condition holds at any time and setting exactly 
$\alpha_T=0$, we obtains
\begin{eqnarray} 
\label{NFSP:GWsconstraints}
 &&\bar{M}^2_3=-\bar{M}^2_2=0\quad \mbox{(in GLPV)} \,;\\
 &&\bar M_3^2=-\bar M_2^2=-2\mu_1^2=0\quad  \mbox{(in Hordenski})\,. \nonumber
\end{eqnarray}
For the Horndeski case the above relations imply that the quintic Horndeski  
Lagrangian is  ruled out and the quartic Lagrangian reduces to $f(\phi)R$ where 
$f(\phi)$ is a function of the scalar 
field~\cite{Creminelli:2017sry,Baker:2017hug}. Going beyond the Horndeski case 
and considering GLPV models  culminates in the quintic GLPV  vanish  as 
well~\cite{Creminelli:2017sry,Baker:2017hug}.  As  a consequence,  some 
well-known 
MG models were ruled out \cite{Ezquiaga:2017ekz}, and we refer the reader to 
\cite{Kase:2018aps} for a review about viable models after GW170817.
Additionally, the phenomenon of the decay of GWs into DE fluctuations is found 
to be  driven by  a coupling proportional to $\alpha_H$ 
\cite{Creminelli:2018xsv}.  A large  decay rate of the GWs,  which implies that  
no 
wave would reach the detector, is associated with $\alpha_H\neq 0$, hence 
$\alpha_H$ 
is constrained  to be of the  order $10^{-10}$.  According to this result GLPV  
would 
be completely ruled out.
However, one has also to consider that  the bound on $\alpha_T$  should be 
imposed only at $z < 10^{-2}$ \cite{Kennedy:2018gtx,Kase:2018aps}, since   the 
source of GWs is at redshift $z \simeq 0.009$.

Another source of debate about  \label{eLIGOfref5}
the applicability of the LIGO/Virgo bound on models  that modify gravity at 
large scales concerns the LIGO/Virgo measured frequencies.  Indeed, those are 
close to the cut-off scale of DE/MG models, and if  ultra-violet effects come 
into play  to recover the GR propagation of GWs around the frequency
$f \sim 100$~Hz, the previous bounds on $c_t$ and $\alpha_{H}$ cannot be applied 
 \cite{deRham:2018red}.  In this regard, measurements at lower frequencies, 
e.g., 
those that will be provided by the future space-based mission LISA 
\cite{Audley:2017drz},  would offer a proper testing ground. Further theoretical 
bounds \cite{Creminelli:2019kjy}, which follow the requirement $c_t^2=1$, were 
set on the braiding function $\alpha_B$ and $\alpha_H$ respectively  within the 
Horndeski and GLPV models. It has been shown that  GWs of  large amplitude 
generated by binary systems can lead to ghost and gradient instabilities in the 
dark energy perturbations. This contingency is avoided for Horndeski models 
when 
$|\alpha_B|< 10^{-2}$ and for GLPV models when $|\alpha_H|<10^{-20}$.

\item \textit{Massive astrophysical bodies:} MG theories are characterised by 
screening  mechanisms that work to hide the fifth force on small scales or high 
density environments in order to recover GR \cite{Joyce:2014kja}.  Solar-System 
and astrophysical tests indeed constrain gravity in these cases to be that 
described by  GR with high accuracy \cite{Uzan2011,Will2014}.   The Vainshtein  
screening mechanism 
\cite{Vainshtein:1972sx,Nicolis:2008in,Koyama:2013paa,Kimura:2011dc}, which is 
generally characteristic of Horndeski and GLPV theories, exhibits a very 
peculiar feature in GLPV:
 a ``partial breaking'' inside astrophysical bodies where the fifth force is 
not  \label{Fifthref3}
completely screened while outside the object GR is restored 
\cite{Kobayashi:2014ida}. In case the screening mechanism is fully operating,  
the  Minkowski potentials  $\phi(r)$,  $\psi(r)$ inside an overdensity  are 
equal. In GLPV theories  the equations for these potentials can be written as 
\cite{Kobayashi:2014ida,Sakstein:2016ggl}
\ba\label{NFSP:upsilon}
&&\f{d\phi}{dr}=\f{G_N\tilde{M}(r)}{r^2}+\f{\Upsilon_1G_N}{4}\f{d^2\tilde{M}(r)}{dr^2}\,,\\
&&\f{d\psi}{dr}=\f{G_N\tilde{M}(r)}{r^2}-\f{5\Upsilon_2G_N}{4r}\f{d\tilde{M}(r)}{dr}\,,
\ea
where  $\tilde{M}(r)$ is the mass inside the object and $\Upsilon_i$ are 
dimensionless constants,  which in the EFT formulation read  
\cite{Koyama:2015oma,Saito:2015fza}
\ba
\label{NFSP:upsilon1EFT}
&&\Upsilon_1 = \frac{4 \alpha_H^2}{c_t^2(1+\alpha_B)-\alpha_H-1}\,,\\
\label{NFSP:upsilon2EFT}
&&\Upsilon_2 = \frac{4\alpha_H(\alpha_H-\alpha_B)}{5(c_t^2(1+\alpha_B)-\alpha_H-1)}\,.
\ea
Thus in models with $\alpha_H \neq 0$ the potentials are no longer equal because 
the screening is  not fully efficient. Constraints on these parameters have been 
obtained using   dwarf, neutron, hyperon and quark stars and galaxy clusters 
\cite{Sakstein:2015zoa,Saito:2015fza,Koyama:2015oma,Sakstein:2016ggl,
Sakstein:2016oel}.  The stringent constraint on $\Upsilon_1$ and the first bound 
on $\Upsilon_2$ are obtained at $0.1<z<1.2$, using X-ray and lensing profiles 
of 
galaxy clusters from XMM Cluster Survey \cite{Romer:1999qt} and CFHTLenS 
\cite{Heymans:2012gg}. They are
 $\Upsilon_1 =  -0.11^{+0.93}_{-0.67}$ and  $\Upsilon_2 = -0.22^{+1.22}_{-1.19}$ 
at $2\sigma$ \cite{Sakstein:2016ggl}.
These bounds can be used to constrain the EFT functions from Eqs. 
\eqref{NFSP:upsilon1EFT}-\eqref{NFSP:upsilon2EFT}. If  on top of these bounds 
one also considers the GWs constraint on $c_t$~\cite{Monitor:2017mdv}, the EFT 
parameter space further reduces  \cite{Sakstein:2017xjx}.

Another peculiar characteristic of the Vainshtein mechanism  is the 
\textit{piercing effect} \cite{Jimenez:2015bwa}, i.e., within screened 
environments,   the gradient of the scalar field is not bound to  vanish for  
shift 
symmetric models. It is possible to further bound $\alpha_T$ and $\alpha_H$ by
combining the Hulse-Taylor pulsar  constraint on the local value of $c_t$, 
which 
is  of the  order $10^{-2}$ \cite{Jimenez:2015bwa}  with the parameterised 
post-Newtonian constraint from the Cassini  experiment on the screened  
gravitational 
slip parameter $\eta_{sc} -1 = (2.3 \pm 2.1) \times 
10^{-5}$\cite{Bertotti:2003rm}. Finally, note that on super-Compton scales, 
$\eta$ reduces to  $\eta_{\rm sc}  = 
\frac{1}{1+\alpha_T}$ \cite{Pogosian:2016pwr}. This is connected to  the 
modification of gravity that remains in a screened environment.  

\end{itemize}

\chapter[The \texorpdfstring{$H_0$}{H0} Tensions to Discriminate Among 
Concurring Models] {The \texorpdfstring{$H_0$}{H0} Tensions to Discriminate \\ 
Among Concurring Models}
 \label{sec:DiValentino}
\label{H0tensionjref2}

\label{Hubbleefs1}

{\em Eleonora Di Valentino}\\
 

\label{tensionsrefs2}
The latest 2018 legacy release from the Planck satellite \cite{Akrami:2018vks}  
of the Cosmic Microwave Background (CMB) temperature and polarization 
anisotropies power spectra, has provided a fantastic confirmation of the 
standard $\Lambda$ Cold Dark Matter ($\Lambda$CDM) cosmological model. However, 
some anomalies and tensions between Planck and external cosmological probes, 
present above the three standard deviations, are becoming even more stronger, 
justifying possible extensions of the standard cosmological scenario. The most 
famous one is the so-called {\it Hubble constant $H_0$ tension} between the CMB 
estimation of $H_0$, under the assumption of the $\Lambda$CDM model, and the 
direct local distance ladder measurements.
\label{CMBrefs5}

In particular, the Planck 
collaboration found $H_0=67.27\pm0.60$ km/s/Mpc at 68\% CL for Planck TTTEEE + 
lowE 2018~\cite{Aghanim:2018eyx} ($H_0=67.36\pm0.54$ km/s/Mpc at 68\% CL for 
Planck TTTEEE + lowE + CMB lensing 2018), but this bound is model-dependent and 
affected by the degeneracy with the other parameters of the model. This 
constraint is in tension at about $3.3\sigma$, with the 2.4\% determination of 
the Hubble constant by Riess et al. in 2016 (R16)~\cite{Riess:2016jrr}, 
$H_0=73.24 \pm 1.74$ km/s/Mpc at 68\% CL, based on the analysis of the Hubble 
Space Telescope (HST) observations using four geometric distance calibrations of 
Cepheids. Moreover, the tension increases to $3.6\sigma$ if we consider the 
2.3\% 2018 Riess et al. (R18)~\cite{Riess:2018uxu} estimate of the Hubble 
constant $H_0=73.48 \pm 1.66$ km/s/Mpc at 68\% CL, including parallax 
measurements of seven long-period Milky Way Cepheids. Finally, the tension arises to 
$4.4\sigma$ with the latest 2019 SH0ES collaboration (R19) 
constraint~\cite{Riess:2019cxk} on the Hubble constant $H_0=74.03 \pm 1.42$ 
km/s/Mpc at 68\% CL, obtained using the HST observations of 70 long-period 
Cepheids in the Large Magellanic Cloud (LMC).

\begin{figure}[ht!]
\vspace{-0.4cm}
\hspace{-2cm}
\includegraphics[width=1.16\textwidth]{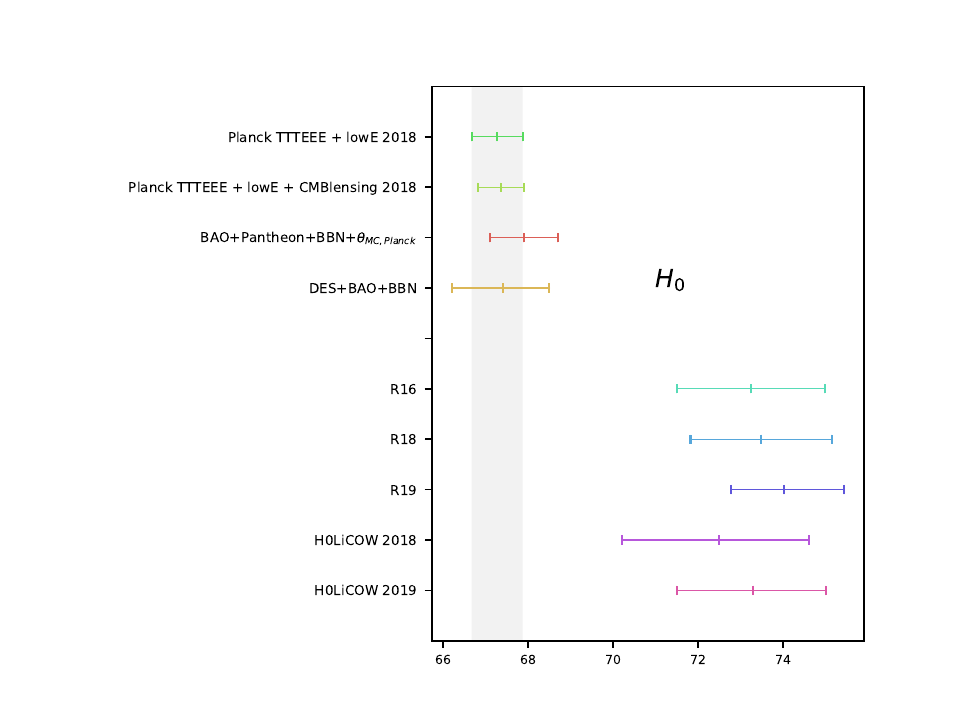}
\caption{{\it{Whisker plots showing the 68\% CL constraint on the Hubble 
constant  $H_0$ from different cosmological probes. The grey band is the 2018 
Planck TTTEEE + lowE 68\% CL bound on $H_0$. }}}
\label{Valentwhisker}
\end{figure}

Actually, the Hubble constant tension is referring to two independent blocks,  
as we can see in Fig~\ref{Valentwhisker}. On the left side, preferring smaller 
values, we have estimates of $H_0$ as obtained by Planck 2018, and some 
combinations of data. In particular, we are showing two different combinations of 
Baryon Acoustic Oscillation (BAO) 
measurements~\cite{Beutler:2011hx,Ross:2014qpa,Alam:2016hwk}, the first-year 
measurements of the Dark Energy Survey (DES) experiment~\cite{Troxel:2017xyo, 
Abbott:2017wau, Krause:2017ekm}, supernovae from the recent Pantheon 
catalogue~\cite{Scolnic:2017caz} (Pantheon), a prior on the baryon density of 
$\Omega_bh^2=0.0222 \pm 0.005$ derived from measurements of primordial 
$D$~\cite{Cooke:2017cwo} assuming Big Bang Nucleosynthesis (BBN), and a prior on 
$\theta_{MC}$ as obtained by Planck in a $\Lambda$CDM scenario. Namely, we have 
that BAO+Pantheon+BBN+$\theta_{MC,Planck}$ gives $H_0=67.9 \pm 0.8$ km/s/Mpc at 
68\% CL, and DES+BAO+BBN find $H_0=67.4_{-1.2}^{+1.1} $ km/s/Mpc at 68\% CL. On 
the right side, instead, preferring larger values, we find the late Universe 
measurements, as obtained by the SH0ES collaboration and the H0LiCOW experiment, 
which is performing a cosmographic analysis of multiply-imaged quasars systems 
through the time-delay strong lensing.\label{stronglensefs3} Therefore, we have 
the H0LiCOW 
2018~\cite{Birrer:2018vtm} Hubble constant estimate $H_0=72.5_{-2.3}^{+2.1} $ 
km/s/Mpc at 68\% CL and the 2.4\% H0LiCOW 2019~\cite{Wong:2019kwg} constraint 
$H_0=73.3_{-1.8}^{+1.7} $ km/s/Mpc at 68\% CL.
Therefore, the late Universe estimates (R19 and H0LiCOW 2019) combined together 
give  $H_0=73.8\pm1.1 $ km/s/Mpc at 68\% CL, showing a $5.3\sigma$ tension with 
the early Universe values obtained in a $\Lambda$CDM model, and resulting in a 
Universe expanding faster at late times than one driven by the cosmological 
constant.

In addition, we have the re-analysis of the R16 Cepheid data  by using Bayesian 
hyper-parameters (HPs)~\cite{Cardona:2016ems}, which gives $H_0=73.75\pm2.11 $ 
km/s/Mpc at 68\% CL. Then, we have to mention the local determination of the 
Hubble constant~\cite{Camarena:2019moy} as obtained by considering the 
cosmographic expansion of the luminosity distance, i.e., $H_0=75.35\pm1.68 $ 
km/s/Mpc at 68\% CL, in agreement with R19, but in tension at about $4.5\sigma$ 
with Planck 2018.
Moreover, there is the debated independent determination of the Hubble constant based on 
the Tip of  the Red Giant Branch (TRGB). In fact, a first measurement made 
by~\cite{Freedman:2019jwv} (F19), which gives $H_0=69.8 \pm 0.8 (stat) \pm 1.7 
(sys) $ km/s/Mpc at 68\% CL, has been followed by the re-analysis 
of~\cite{Yuan:2019npk} (Y19), which gives $H_0=72.4\pm2.0 $ km/s/Mpc at 68\% CL, 
contesting the F19 overestimate of the LMC extinction, and concluded by the 
revised measurement made by~\cite{Freedman:2020dne} (F20), which gives $H_0=69.6 
\pm 0.8 (stat) \pm 1.7 (sys) $ km/s/Mpc at 68\% CL, discussing the Y19 
neglection of the correction for reddening. However, these $H_0$ measurements are not useful to discriminate between Planck and R19 because they are in agreement with both within 95\% CL.

Finally, many others independent late Universe measurements point towards a 
larger value for  the Hubble constant, like MIRAS (variable red giant 
stars)~\cite{Huang:2019yhh}, STRIDES (STRong-lensing Insights into Dark Energy 
Survey)~\cite{Shajib:2019toy}, water masers and the use of the Surface 
Brightness Fluctuations method~\cite{Verde:2019ivm}. If the late Universe 
measurements are averaged by excluding each time a different method or geometric 
calibration or team, these $H_0$ estimates will be in tension between 
4.5$\sigma$ and 6.3$\sigma$ with the Planck one (see the recent 
review~\cite{Riess:2020sih}).

However, since the CMB constraints are model-dependent, by changing the underlying 
cosmological  scenario we can have completely different cosmological parameter 
constraints. Even if possible systematic effects can always be present in some 
(or all) of the cosmological experiments, there is a possibility of restoring 
the concordance between these measurements taken at a face value, considering the
extended model beyond the $\Lambda$CDM one.
For a list of recent works (from 2015 onwards) trying to solve or  
alleviate the $H_0$ tension, see, for 
example,~\cite{DiValentino:2015ola,DiValentino:2016hlg,Bernal:2016gxb,
Kumar:2016zpg,Kumar:2017dnp,DiValentino:2017iww,DiValentino:2017oaw,
DiValentino:2017rcr,Binder:2017lkj,Khosravi:2017hfi,DiValentino:2017zyq,
Renk:2017rzu,DiValentino:2017gzb,Belgacem:2017cqo,
Sola:2017znb,Nunes:2018xbm,Yang:2018euj,Colgain:2018wgk,DEramo:2018vss,
Yang:2018uae,Guo:2018ans,Lin:2018nxe,Yang:2018qmz,Poulin:2018cxd,
Banihashemi:2018oxo,Banihashemi:2018has,Mortsell:2018mfj,
Kreisch:2019yzn,Martinelli:2019dau,Vattis:2019efj,Kumar:2019wfs,Agrawal:2019lmo,
Yang:2019jwn,Yang:2019qza,Yang:2019uzo,DiValentino:2019exe,Desmond:2019ygn,
Yang:2019nhz,Pan:2019gop,Visinelli:2019qqu,Martinelli:2019krf,Cai:2019bdh,
Pan:2019hac,DiValentino:2019dzu,DiValentino:2019ffd, 
DiValentino:2019jae,Schoneberg:2019wmt,Shafieloo:2016bpk,Li:2019san,
Cuceu:2019for,Colgain:2019joh,Pan:2019jqh,Berghaus:2019cls,Knox:2019rjx,
Pandey:2019plg,Vagnozzi:2019ezj,Adhikari:2019fvb,Lancaster:2017ksf,
Niedermann:2019olb,Raveri:2019mxg,Hart:2019dxi,Yan:2019gbw,Yadav:2019jio,
Kasai:2019yqn,Amirhashchi:2020qep,Perez:2020cwa,Pan:2020bur,DAgostino:2020dhv,Benevento:2020fev}.~\footnote{While the book was under review, many new $H_0$ measurements and new proposed solutions to the tension appear. For an updated perspective, see the letter submitted for SNOWMASS 2021~\cite{DiValentino:2020zio} about $H_0$ (and references therein). In particular, here a list of updated references about the solutions~\cite{Tram:2016rcw,DiValentino:2016ziq,Gomez-Valent:2020mqn,Lucca:2020zjb,vandeBruck:2017idm,DiValentino:2020naf,
Keeley:2019esp,Yang:2018prh,Karwal:2016vyq,Smith:2019ihp,Lucca:2020fgp,Lin:2019qug,Kumar:2018yhh,
Yang:2020zuk,Pan:2020zza,Wu:2020nxz,Blinov:2020hmc,Wang:2020zfv,Alestas:2020mvb,
Clark:2020miy,Keeley:2020rmo,Hazra:2018opk,
Niedermann:2020dwg,Archidiacono:2020yey,DiValentino:2020kha,Capozziello:2020nyq,Anchordoqui:2019yzc,
Ivanov:2020mfr,Gonzalez:2020fdy,Hryczuk:2020jhi,Carneiro:2018xwq,Paul:2018njm,Gelmini:2019deq,
Anchordoqui:2020znj,Sakstein:2019fmf,Das:2020wfe,Ye:2020btb,Hart:2017ndk,Chiang:2018xpn,
Jedamzik:2020krr,Yang:2020myd,Chudaykin:2020igl,Sekiguchi:2020teg,Bose:2020cjb,Agrawal:2019dlm,
Anchordoqui:2019amx,Frusciante:2019puu,Braglia:2020iik,Ballardini:2020iws,Li:2019yem,Rezaei:2020mrj,
Li:2020ybr,Choi:2019jck,Choi:2020tqp,Berezhiani:2015yta,Anchordoqui:2015lqa,Desai:2019pvs,
Alcaniz:2019kah,Chudaykin:2016yfk,Chudaykin:2017ptd,Hill:2020osr,Ivanov:2020ril,Rezaei:2020lfy,Wang:2020dsc,
Leonhardt:2020qam,Ballesteros:2020sik,Blinov:2019gcj,Hernandez-Almada:2020uyr,Feeney:2017sgx,
Banerjee:2020xcn,Adler:2019fnp,Gu:2020ozv,Akarsu:2019pwn,Nygaard:2020sow,
Benaoum:2020qsi,Krishnan:2020vaf,DiValentino:2020vnx,Kitazawa:2020qdx,Anchordoqui:2020djl,Yao:2020hkw,
Artymowski:2020zwy,Jedamzik:2020zmd,Smith:2020rxx,Murgia:2020ryi,LinaresCedeno:2020uxx,Lin:2020jcb,
DeFelice:2020cpt,Alvarez:2020xmk,Arias-Aragon:2020qip,Kameli:2020kao,Ortiz:2020noa,
Mandal:2020buf,Hashim:2020sez,Braglia:2020bym,Briffa:2020qli,Chudaykin:2020acu,Odintsov:2020qzd,
Yao:2020pji,Cruz:2020cje,daSilva:2020mvk,daSilva:2020bdc} and here about the measurements~\cite{Dhawan:2020xmp,Hinshaw:2012aka,Henning:2017nuy,Aiola:2020azj,Ivanov:2019pdj,
DAmico:2019fhj,Alam:2020sor,Zhang:2018air,Pogosian:2020ded,Philcox:2020xbv,Reid:2019tiq,Bonvin:2016crt,
Birrer:2020tax,Birrer:2020jyr,Dhawan:2017ywl,Freedman_2012,Kim:2020gai,Yu:2017iju,Gomez-Valent:2019lny,
Haridasu:2018gqm,Dutta:2019pio,Nunes:2020uex,Liao:2019qoc,Liao:2020zko,Kourkchi:2020iyz,Schombert:2020pxm,
deJaeger:2020zpb,Fernandez-Arenas:2017isq,Pesce:2020xfe,Qi:2020rmm,Baxter:2020qlr,Freedman:2000cf,
Gayathri:2020fbl,Mukherjee:2020kki,Ashton:2020kyr,Bonilla:2020wbn,Harvey:2020lwf,Renzi:2020fnx,Wang:2020hqq}.}

In the following sections of the paper we will investigate some interesting departures 
from the $\Lambda$CDM model, starting from the most famous extensions, namely a 
neutrino effective number different from $3.046$ in Sec.~\ref{nnu} or a dark 
energy equation of state different from the cosmological constant in 
Sec.~\ref{EoS}, until the most recent proposed solutions. In particular, we will 
revise in Sec.\ref{cosmoscope} a multi-parameters approach, in Sec.~\ref{EDE} an 
Early Dark Energy scenario, in Sec~\ref{IDE} the Interacting Dark Energy model, 
 \label{IDErefs1} \label{earlyDEefs001}
in Sec.~\ref{MG} the Modified Gravity solution, and in Sec.~\ref{many} a list of 
other specific models solving the $H_0$ tension. In Sec.~\ref{guide} we will see 
which characteristic a model should have for alleviating the Hubble constant 
tension, and finally in Sec.~\ref{future} we will discuss which will be the 
constraints on $H_0$ expected in the near future.

\section{The Effective Number of Relativistic Degrees of Freedom} \label{nnu}
One of the most famous extensions considered in the literature for solving the 
$H_0$  tension is the possibility of having extra relativistic degrees of 
freedom at the recombination. We can parametrise the radiation density $\rho_r$ 
as a function of the photon density $\rho_\gamma$ in the following way:
\begin{equation}
    \rho_r = \Bigg[1+\frac{7}{8}\Bigg(\frac{4}{11}\Bigg)^{4/3} N_{\rm 
eff}\Bigg]\rho_\gamma,
\end{equation}
where $(4/11)^{1/3}=T_\nu / T_\gamma$ is the ratio between the background 
temperatures of neutrinos and photons derived under the approximation of an 
instantaneous neutrino decoupling. $N_{\rm eff}$ is expected to be equal to 
$3.046$ for three active massless neutrino 
families~\cite{Mangano:2005cc,deSalas:2016ztq}. If we have more relativistic 
degrees of freedom, we then have more radiation and we can measure the effect on 
the CMB temperature power spectrum from the smearing and shifting of the 
acoustic peaks in the damping tail. More radiation also means a delay in the 
matter to radiation  equivalence, producing an enhancement of the early 
Integrated Sachs Wolfe (eISW)  \label{ISWref3} effect, which peaks around the 
multipole $\ell\sim 
200$~\cite{Archidiacono:2013fha}.

Letting the $N_{\rm eff}$ parameter free to vary, thanks to the well known 
degeneracy between $H_0$ and  $N_{\rm eff}$, could increase the $H_0$ value at 
the price of additional radiation at recombination, due to additional 
relativistic species, like, for example, sterile neutrinos or thermal axions 
\cite{Carneiro:2018xwq,Paul:2018njm,DiValentino:2015sam,Green:2019glg,
Ferreira:2018vjj,DiValentino:2015wba}. With Planck 2015 TT+lowP we had in fact 
$N_{\rm eff}=3.13\pm0.32$ at 68\% CL, corresponding to $H_0=68.0\pm2.8$ km/s/Mpc 
at 68\% CL, solving the Hubble constant tension at $2.1\sigma$. Now, with the 
inclusion of the Planck 2018 polarization data, which is insensitive to the eISW, 
we have help in breaking the degeneracy between the cosmological parameters. 
The improved constraint on $N_{\rm eff}=2.92\pm0.19$ at 68\% CL from Planck 2018 
TTTEEE + lowE corresponds to $H_0=64.4\pm 1.4$ km/s/Mpc at 68\% CL, in tension 
at about $3.8\sigma$ with R19. With the addition of the Planck polarization data 
and the new $\tau$ estimate, that shifts the mean value of $N_{\rm eff}$ towards 
smaller values because of their correlation, an extra dark radiation, producing a 
neutrino effective number different from the standard value $3.046$, seems no 
more a suitable solution for solving the Hubble constant tension between Planck 
and R19.

\section{Dark Energy Equation of State}\label{EoS}
The second most important extension considered in the $H_0$ tension discussion, 
is a dark energy component  with an equation of state $w$ different from -1, 
i.e., different from a cosmological constant. Varying $w$ means to change the 
expansion rate of the Universe through the first Friedmann equation. From Planck 
2018 TTTEEE + lowE we have $w=-1.58^{+0.16}_{-0.35}$ at 68\% CL, to which 
corresponds $H_0>69.9$ km/s/Mpc at 95\% CL, which solves the $H_0$ tension within 
two standard deviations. The Hubble constant is in fact almost unconstrained in 
this model, because of the geometrical degeneracy present with the DE equation 
of state $w$, so it can be perfectly in agreement with R19 at the price of a 
phantom-like dark energy.
Phantom dark energy has several theoretical problems and is expected to produce 
a Big Rip in the future  of the Universe. Moreover, it violates the energy 
condition $\rho \geq |p|$, which means that the matter could move faster than 
light and a comoving observer could measure a negative energy density, and the 
Hamiltonian could have vacuum instabilities due to negative kinetic energy. 
Nevertheless, there exist models expecting an effective energy density with a 
phantom 
equation of state without showing the problems mentioned before, as - for example -
the Parker Vacuum 
Metamorphosis~\cite{Parker:2000pr,Parker:2003as,Caldwell:2005xb}.

Another possibility is allowing the DE equation of state varying with redshift. 
For example, we can consider the well-known Chevallier - Polarski - Linder 
parametrisation (CPL)~\cite{Chevallier:2000qy,Linder:2002et} of the DE equation 
of state $w(a)$ with two parameters:
\begin{equation}
    w(a) = w_0+(1-a)w_a \,,
\end{equation}
where $a$ is the dimensionless cosmological scale factor normalized to unity 
today. Here,  $w_0$ gives an idea of the behaviour of the DE equation of state 
today, while $w_a$ of the evolution with time. If $w(a)$ is increasing with $a$ 
we will have $w_a < 0$, while if $w(a)$ is become more negative with $a$ we will 
have a positive $w_a$. In this scenario from Planck 2018 TTTEEE + lowE we have 
$w_0=-1.21^{+0.33}_{-0.60}$ and $w_a<-0.85$ at 68\% CL, and $H_0>63$ km/s/Mpc at 
95\% CL, solving the tension with R19 within $2\sigma$.

\section{Multi-parameters Extension}\label{cosmoscope}
In order to understand which is the best extension preferred by the data for 
solving the $H_0$  tension, we can try to vary the cosmological parameters all 
together. In practice, instead of solving the $H_0$ tension with a specific 
mechanism, we look for a combination of parameters that can ameliorate it. In 
fact, even if the $\Lambda$CDM model provides a good fit of the data, some of 
the assumptions and simplifications made in this six parameters approach are no 
more fully justified. This simplification can be excessive and can hide some 
physical aspects driving the evolution of the Universe. 

If we consider, for 
example, a multi-parameter space, we can understand where data wants us to go, 
avoiding biases due to the choice of the model. In~\cite{DiValentino:2019dzu} we 
see a multi-parameter approach varying 10, 11 or 12 parameters at the same time. 
In particular, in the 10 parameters space the extensions of the standard 
$\Lambda$CDM model under consideration are a running of the scalar spectral 
index $\alpha_s$, a total neutrino mass $\Sigma m_\nu$, the effective number of 
relativistic degrees of freedom $N_{\rm eff}$ and a dark energy equation of 
state $w$. In the 11 parameters space there is the addition of the $A_{\rm 
lens}$ parameter~\cite{Calabrese:2008rt}, parametrising the lensing potential in 
the temperature power spectrum of the CMB. This phenomenological parameter is 
just a consistency check that the Planck data fails, because it is different from 
the expected value at about $2.8\sigma$ in the last 2018 data release. Finally, 
in the 12 parameters approach the CPL parametrisation $w(a) = w_0 + (1-a)w_a$ of 
the dark energy equation of state with redshift is taken under consideration.

The bottom line of this multi-parameters approach is that $H_0$ is almost 
unconstrained in these very extended scenarios, mainly because of the strong 
geometrical degeneracy present with the dark energy equation of state $w(a)$. 
For Planck 2018 TTTEEE + lowE in a 10 parameters scenario we have $H_0>65.2$ 
km/s/Mpc at 95\% CL, in a 11 parameters space $H_0=73^{+10}_{-20}$ km/s/Mpc at 
68\% CL and in a 12 parameters approach $H_0=72\pm20$ km/s/Mpc at 68\% CL. For 
this reason, a safe combination of Planck 2018 and R19 can be performed, and the 
new "concordance" model has $N_{\rm eff}$ exactly in agreement with the standard 
expectation, while the cosmological constant is ruled out at more than three 
standard deviations, in favour of a phantom-like dark energy.
Unfortunately, this "concordance" model is strongly in tension with BAO and 
Pantheon data at more  than $3\sigma$, and the combination of Planck with these 
datasets restores the $H_0$ tension at high significance.

\section{Early Dark Energy}\label{EDE}
Very important models, effective in reducing the $H_0$ tension, are Early Dark 
Energy  (EDE) scenarios. In these models, the Dark Energy behaves like a 
cosmological constant at early times ($z \geq 3000$) and then decays away like 
radiation or faster at later times. For EDE scenarios, the sound horizon at 
decoupling is reduced, resulting in a larger value of the Hubble parameter $H_0$ 
inferred from the CMB.

In~\cite{Poulin:2018cxd} the authors consider two physical models for the EDE:
\begin{itemize}
    \item An oscillating scalar field with a potential $V(\phi) \propto 
(1-cos[\phi/f])^n$  \cite{Kamionkowski:2014zda}. At early times this field acts 
like a cosmological constant, but at a certain point it starts to oscillate, and 
behaves like a fluid with an equation of state $w_n = (n-1)/(n+1)$.
    \item A field that slowly rolls down a potential that is linear in $\phi$ at 
early times and  reaches asymptotically zero at late times.
\end{itemize}
Depending on $n$, the homogeneous EDE density dilutes like matter ($n=1$), 
radiation ($n=2$) or faster  than radiation ($n \geq 3$). For $n \sim \infty$ 
the energy density dilutes instead as $a^{-6}$.
For Planck 2015 TTTEEE~\cite{Aghanim:2015xee} + lowTEB + CMBlensing 2015 +R18 + 
BAO + Pantheon, the Hubble  constant is $H_0=70.3\pm1.2$ km/s/Mpc at 68\% CL for 
$n=2$, $H_0=70.6\pm1.3$ km/s/Mpc at 68\% CL for $n=3$ and $H_0=69.9\pm1.1$ 
km/s/Mpc at 68\% CL for $n=\infty$. Within these models, the $H_0$ tension with 
R19 is reduced at about $2\sigma$, $1.8\sigma$ and $2.2\sigma$ respectively, 
with an improvement of the $\chi^2$ of $9.5$, $14.5$ and $9.1$ in fitting the 
data, with respect to the standard $\Lambda$CDM model. However, it is difficult 
to assess this result as a possible solutions of the Hubble constant tension 
because the authors of~\cite{Poulin:2018cxd} are combining together R18 and BAO, 
which are potentially in tension with each other and can bias the final 
findings. A re-analysis with the Planck 2018 TTTEEE + lowE can be found in Ref.~\cite{Hill:2020osr},
that for $n=3$ gives $H_0=68.3\pm1.0$ km/s/Mpc at 68\% CL, in tension with R19 at $3.5\sigma$.

A possibility for the realization of an EDE is that an axion could interact with 
a dilaton,  as proposed by~\cite{Alexander:2019rsc}.

Moreover, EDE models suffer a fine-tuning problem (see, for 
example,~\cite{Pettorino:2013ia}),  because, in order to solve the Hubble 
tension, the early component of dark energy needs to become active around the 
time of matter-radiation equality, and these quantities are completely 
disconnected. However, in~\cite{Sakstein:2019fmf} the possibility that the EDE 
scalar couples to neutrinos, receiving a large injection of energy when 
neutrinos become non-relativistic, i.e., around the time of matter-radiation 
equality, is considered to solve the fine-tuning.

In addition, in~\cite{Niedermann:2019olb} the authors investigate a first-order 
phase transition  in a dark sector in the early Universe, before recombination. 
This will lead to a short phase of a New EDE component ameliorating the Hubble 
constant tension. For Planck 2015 TTTEEE + lowTEB + CMBlensing 2015 +R19 + BAO + 
Pantheon, the Hubble constant is $H_0=70.4^{+0.9}_{-1.0}$ km/s/Mpc at 68\% CL, 
reducing the tension with R19 at $2.1\sigma$. A re-analysis with the Planck 2018 TTTEEE + lowE + CMBlensing + BAO + Pantheon can be found in Ref.~\cite{Niedermann:2020dwg}, which gives $H_0=69.6^{+1.0}_{-1.3}$ km/s/Mpc at 68\% CL, reducing the Hubble constant tension at $2.5\sigma$.

Finally, in \cite{Ye:2020btb} the authors studied a phenomenological EDE model  
\label{gphenomenologicalprfs4}
with an Anti-de Sitter  ({\it AdS}) vacua around recombination, obtaining for 
Planck 2018 + BAO + Pantheon + R19 $H_0=72.64^{+0.57}_{-0.64}$ km/s/Mpc at 68\% 
CL, reducing the tension within one standard deviation.

\section{Interacting Dark Energy}\label{IDE}
Another possibility is to consider an Interacting Dark Energy (IDE) scenario, 
where Dark Matter (DM) and Dark Energy  (DE) share interactions other than 
gravitational. In this model the conservation equations for DM and DE can be 
modified with the introduction of an interaction rate $Q=\xi \mathcal{H} 
\rho_{DE}$, proportional to the comoving Hubble rate $\mathcal{H}$ and the dark 
energy density $\rho_{DE}$ via a negative dimensionless parameter $\xi$ 
quantifying the strength of the coupling.

In the IDE scenario, as recently analysed 
in~\cite{DiValentino:2019ffd,DiValentino:2019jae}, for Planck 2018 TTTEEE  + 
lowE the Hubble constant is $H_0=72.8^{+3.0}_{-1.5}$ km/s/Mpc at 68\% 
CL~\cite{DiValentino:2019ffd}, solving the tension with R19 within one standard 
deviation. The "concordance" model obtained combining Planck 2018 + R19 gives a 
non-zero coupling $\xi$ at more than five standard deviations, indicating a flux 
of energy from DM to DE. Planck 2018 TTTEEE + lowE + BAO data estimates instead 
$H_0=69.4^{+0.9}_{-1.5}$ km/s/Mpc at 68\% CL~\cite{DiValentino:2019jae}, 
increasing the tension with R19 again at $2.5\sigma$. However, the procedure 
that leads to the extraction of the BAO data is assuming a $\Lambda$CDM model 
(for a discussion about the BAO model dependence see, for 
example,~\cite{Heinesen:2019phg}), so it might need to be revised when applied to 
extended DE cosmologies.

It is important to mention the following  
works~\cite{Kumar:2016zpg,DiValentino:2017iww,Kumar:2017dnp,Yang:2018euj,
Yang:2018uae,Yang:2019uzo,Martinelli:2019dau,Gomez-Valent:2020mqn} pointing 
towards the same 
conclusions, i.e., the possibility that an IDE model could alleviate the Hubble 
constant tension with a coupling different from zero. However, these papers make 
use of the old Planck 2015 likelihood and need to be updated in light of the new 
released polarisation data.

\section{Modified Gravity} \label{MG}
An argument in favor of Modified Gravity models (see Ref.~\cite{Ade:2015rim}, 
and  references therein) is the possibility of relieving the Hubble constant 
tension. In fact, if gravity is weaker at intermediate scales than expected in 
General Relativity (GR), then the $H_0$ estimate from CMB can have larger 
values.

In~\cite{Raveri:2019mxg} is presented a data-driven reconstruction  of 
gravitational theories and Dark Energy models on cosmological scales, making use 
of the Effective Field Theory (EFT) approach, which compresses the freedom in 
defining such models into a finite set of functions that can be reconstructed 
across cosmic times using cosmological data. Using the EFT approach, some of the 
models, in particular Scalar Horndeski (SH) and Full Horndeski (FH), can 
alleviate the present discrepancy in the determination of the Hubble constant 
between Planck and R19, and are statistically preferred against the standard 
$\Lambda$CDM model.

Moreover, in~\cite{Yan:2019gbw} the EFT approach is applied  to the torsional 
gravity, finding the possibility of alleviating the $H_0$ tension.
The use of the EFT formalism allows us to study systematically the evolution 
equations  at the background and perturbation levels separately, in a  
\label{madelindepeefs2}
model-independent way, examining the role of the coefficients in relaxing the 
Hubble constant tension. In~\cite{Yan:2019gbw} the authors focus on a well-known 
class of torsional gravity, namely the $f(T)$ gravity, where $T$ is the torsion 
scalar. They find that, imposing initial conditions at the last
scattering reproducing the $\Lambda$CDM scenario, and imposing the late times values preferred by local measurements, $f(T)$ models described by this parameterization:
\begin{equation}
    f(T)=-T-2\Lambda /M_P^2+\alpha T^\beta,
\end{equation}
can alleviate the Hubble constant tension. Moreover, using this information, 
extended models  can be considered, like, for example, the $f(T,B)$ gravity, 
where B is the boundary term $B=-2\nabla_\mu T_\nu^{\nu\mu}$. However, an actual 
fit of the data with these promising models is missing, and for this reason it is 
difficult to quantify how much they are effectively able in alleviating the 
$H_0$ tension between Planck and R19.

Finally, a promising class of theories, within the context of the Modified 
Gravity proposals solving  the $H_0$ tension, are the Horndeski ones. On this 
topic we can find \cite{Frusciante:2019puu}, where the authors studied a {\it 
generalized cubic covariant Galileon} (GCCG) model, obtaining for Planck 2015 TT 
+ lowTEB $H_0=72^{+8}_{-5}$ km/s/Mpc at 95\% CL, alleviating the tension with 
R19. However, in the paper the Planck high-$\ell$ polarization is not included 
and the tension with R19 is restored when Planck data are combined with external 
datasets.

\section{More specific models} \label{many}
In the literature, we can find many more specific models that can relieve the Hubble constant tension, but at the moment most of them make use of the old Planck 2015 likelihood. Here we list just a few of them:

\begin{itemize}

    \item In \cite{DiValentino:2017rcr} the authors analyse the {\it Parker 
vacuum metamorphosis}  (PVM) model, physically motivated by quantum 
gravitational effects and with the same number of parameters as $\Lambda$CDM, 
showing that it can remove the $H_0$ tension, and give an improved fit of the 
data. For Planck 2015 TTTEEE + lowTEB $H_0=78.61\pm0.38$ km/s/Mpc at 68\% CL for 
the original PVM, reducing the tension with R19 at $2.2\sigma$, while 
$H_0=71.6^{+2.8}_{-5.1}$ km/s/Mpc at 68\% CL for the elaborated PVM, with one 
more degree of freedom, removing the $H_0$ tension within one standard 
deviation. An updated re-analysis for Planck 2018 TTTEEE + lowE can 
be found in Ref.~\cite{DiValentino:2020kha}, which gives $H_0=76.7^{+3.9}_{-2.6}$ km/s/Mpc at 68\% CL 
for the elaborated PVM.

    \item In \cite{Khosravi:2017hfi} the authors propose the $\ddot u 
\Lambda$CDM, based  on {\it $\ddot u$ber gravity}, to alleviate the $H_0$ 
tension. In this model, there is a sharp transition between $\Lambda$CDM to a 
phase in which the Ricci scalar is constant. For Planck 2015 TT + lowTEB + R16 + 
BAO $H_0=70.6^{+1.1}_{-1.3}$ km/s/Mpc at 68\% CL, improving the agreement with 
R19 at $1.9\sigma$. Again this bound is obtained by combining together the SH0ES 
collaboration R16 constraint on $H_0$ with BAO measurements, and considering 
just the temperature power spectrum for the CMB at higher multipoles. However, 
as we can see from the figures in the paper~\cite{Khosravi:2017hfi}, a Planck 
only constraint on $H_0$ is really relaxed within this model, so in agreement 
with R19 within one standard deviation, with only one additional free parameter.

    \item In \cite{Renk:2017rzu} the authors show how the {\it Galileon Gravity} 
is able to  alleviate the Hubble tension. Considering Planck 2015 TT + lowTEB + 
CMBlensing 2015 + BAO and varying the total neutrino mass too, for the Cubic 
Galileon $H_0=71.6\pm 2.1$ km/s/Mpc at 95\% CL, for the Quartic Galineon 
$H_0=72.4\pm 2.0$ km/s/Mpc at 95\% CL and for the Quintic Galileon $H_0=72.3\pm 
2.1$ km/s/Mpc at 95\% CL, reducing, respectively, the tension between the CMB 
and R19 at $1.4\sigma$, $0.9\sigma$ and $1\sigma$.

    \item In \cite{Belgacem:2017cqo} the authors analyse {\it Nonlocal 
Gravity},  and in particular the RR model, obtained by the requirement of 
providing a viable cosmological evolution severely and corresponding to a 
dynamical mass generation for the conformal mode. In this scenario, for Planck 
2015 TTTEEE + lowTEB + BAO + JLA, $H_0=69.49^{+0.79}_{-0.80}$ km/s/Mpc at 68\% 
CL, reducing the tension with R19 at $2.8\sigma$.

    \item In \cite{Sola:2017znb} the authors investigate the {\it running  
vacuum model} (RVM), finding that in the dynamical quasi-vacuum models ($w$DVMs) 
with a varying dark energy equation of state can alleviate the tension if Large 
Scale Structure (LSS)  \label{LSSefs3}data are not considered. In particular, 
for 
Planck 2015 
TTTEEE + lowTEB + CMBlensing 2015 + BAO + R16 $H_0=71.0\pm1.5$ km/s/Mpc at 68\% 
CL for wRVM, i.e., at $1.5\sigma$ tension with R19. However, this result is 
driven by the addition of the R16 prior on the Hubble constant and is mainly due 
to the phantom behaviour of the dark energy.

    \item In \cite{Nunes:2018xbm} the author proposes the {\it $f(T)$ parallel  
telegravity} that performs in a fantastic way in removing the $H_0$ tension. For 
Planck 2015 TTTEEE + lowTEB + BAO $H_0=72.4^{+3.3}_{-4.1}$ km/s/Mpc at 68\% CL, 
improving the agreement with R19 within $1\sigma$, even in presence of the BAO 
data, which are difficult to put in agreement with a larger $H_0$ value.  An updated re-analysis for 
Planck 2018 TTTEEE + lowE can be found in Ref.~\cite{Wang:2020zfv}, which gives $H_0=66.51\pm 3.65$ km/s/Mpc at 68\% CL.

    \item In~\cite{Yang:2018qmz} the authors studied several DE equations of state varying with redshift, but with just one free parameter. All the models considered in the paper favour a phantom DE equation of state at the present time, while leading the $H_0$ values in fantastic agreement with the Hubble constant direct measurement R19 within one standard deviation, for Planck 2015 TTTEEE + lowTEB.

    \item In~\cite{Banihashemi:2018oxo} the authors show that with a phase transition in the dark sector, based on a simplified model such that the cosmological constant has two values before a transition redshift and afterwards it becomes single-valued, the Hubble constant tension can be relieved. This model is called {\it double-$\Lambda$ Cold Dark Matter} and using BAO volume distances listed in \cite{Banihashemi:2018oxo}, R18 and a gaussian prior from the Planck 2015 value on $\Theta$, i.e., the distance of the last scattering surface to us, $H_0=72.5 ^{+2.5}_{-3.0}$ km/s/Mpc at 68\% CL with a $\chi^2$ analysis. The tension with R19 is therefore solved within one standard deviation.

    \item In \cite{Banihashemi:2018has} the authors analysed a DE model  based 
on {\it Ginzburg-Landau theory of phase transition} (GLTofDE). In the mean field 
approximation, a phase transition happens which causes a spontaneous symmetry 
breaking. Using a $\chi^2$ analysis with BAO, R18 and quasars $H(z)$ data points 
as listed in~\cite{Banihashemi:2018has}, CMB distance data point and a prior on 
$\Omega_mh^2$ based on the Planck 2015 value, $H_0=71.89 \pm 0.93$ km/s/Mpc at 
68\% CL, i.e., in agreement with R19 at $1.3\sigma$.

    \item In \cite{Kreisch:2019yzn} the authors claim that delaying  the onset 
of neutrino free-streaming until close to the matter-radiation equality epoch, 
introducing a neutrino self-interaction in presence of a total neutrino mass 
different from zero, $H_0$ can be larger than in the standard cosmological 
model. In the case of a strongly interacting neutrino cosmology for Planck 2015 
TTTEEE + lowTEB the Hubble constant is $H_0=66.2^{+2.3}_{-1.9}$ km/s/Mpc at 68\% 
CL, reducing the $H_0$ tension at $2.9\sigma$.

    \item In \cite{Vattis:2019efj} the authors analyse the effect of {\it  
two-body dark matter decays}, where the products of the decay include a massless 
and a massive particle. For Planck 2015 TTTEEE + lowTEB + CMBlensing 2015 + R18 
+ BAO $H_0=70^{+4}_{-3}$ km/s/Mpc at 68\% CL, improving the agreement with R19 
within $1\sigma$. However, again the presence of the R18 prior in the data 
combination makes it difficult to assess how good the model is in solving the 
Hubble tension.  An updated re-analysis for 
Planck 2018 TTTEEE + lowE + CMBlensing 2018 can be found in Ref.~\cite{Clark:2020miy}, which gives $H_0=67.31^{+0.53}_{-0.56}$ km/s/Mpc at 95\% CL, consistent with the $\Lambda$CDM value.

    \item In \cite{Agrawal:2019lmo} the authors consider a scenario with an 
evolving scalar field  $\phi^{2n}$, asymptotically {\it rocking and rolling}. 
For Planck 2015 TTTEEE + lowTEB + Pantheon + R18 + BAO $H_0=70.1^{+1.0}_{-1.2}$ 
km/s/Mpc at 68\% CL, reducing the tension with R19 at $2.3\sigma$, for the case 
in which $n=2$. However, even in this analysis for the presence of the R18 prior 
on $H_0$ it is difficult to assess how much the model can solve the Hubble tension 
between the data.

    \item In \cite{Pan:2019hac} the authors analysed a {\it Phenomenologically 
Emergent Dark  Energy} (PEDE) scenario, introduced in Ref.~\cite{Li:2019yem}, 
finding that the $H_0$ tension with R19 
is alleviated within one standard deviation without additional degrees of 
freedom with respect to the $\Lambda$CDM model. In fact, for Planck 2015 TTTEEE 
+ lowTEB $H_0=72.58^{+0.79}_{-0.80}$ km/s/Mpc at 68\% CL.  An updated re-analysis for 
Planck 2018 TTTEEE + lowE can be found in Ref.~\cite{Yang:2020myd}, which finds $H_0=72.35^{+0.78}_{-0.79}$ km/s/Mpc at 68\% CL, in agreement with R19.

    \item In \cite{Li:2019san} the authors investigate two {\it metastable dark 
energy models}:  (I) a DE decaying exponentially and (II) a DE decaying into DM. 
For Pantheon + BAO $H_0=75.0^{+4.7}_{-5.8}$ km/s/Mpc at 68\% CL for the model I, 
while $H_0=71.8^{+4.7}_{-4.6}$ km/s/Mpc at 68\% CL for the model II, 
respectively. Unfortunately, when CMB distance priors from Planck 2018 are added 
to the analysis, the tension with R19 is restored at more than $3\sigma$.

    \item In~\cite{Pandey:2019plg} the authors consider a {\it decaying dark matter} (DDM) model, where a fraction of dark matter density decays into dark radiation. For Planck 2015 TT + lowTEB + R18 $H_0=70.6^{+1.1}_{-1.3}$ km/s/Mpc at 68\% CL, improving the agreement with R19 at $1.9\sigma$. However this result is driven by the addition of the R18 prior on the Hubble constant, therefore it is difficult to quantify the ability of the model in solving the Hubble tension by itself.
An updated re-analysis for Planck 2018 TTTEEE + lowE can be found in Ref.~\cite{Nygaard:2020sow}, which finds $H_0=67.7\pm1.2$ km/s/Mpc at 68\% CL.

    \item In \cite{Adhikari:2019fvb} the authors show that assuming a 
non-Gaussian covariance  contribution from a primordial trispectrum ({\it 
super-$\Lambda$CDM}), $H_0$ is larger than in the classic $\Lambda$CDM model and 
the $\Delta \chi^2=-7.8$ with the addition of just two degrees of freedom. In 
fact, for Planck 2015 TT + $\tau$ prior $H_0=68.4^{+2.5}_{-2.3}$ km/s/Mpc at 
95\% CL, reducing the tension with R19 at $2.9\sigma$.

    \item In \cite{Lancaster:2017ksf} the authors investigated a {\it 
self-interacting neutrino mode}.  For Planck 2015 TT + lowTEB $H_0=70.4\pm1.3$ 
km/s/Mpc at 68\% CL, alleviating the $H_0$ tension within $1.9\sigma$. When 
the high-$\ell$ polarization is also considered, however, the tension increases 
again, but is alleviated at $2.8\sigma$, giving $H_0=69.59^{+0.74}_{-0.71}$ 
km/s/Mpc at 68\% CL.

    \item In \cite{Hart:2019dxi} the authors propose a modification of the effective electron rest mass $m_e$ during the cosmological recombination era, for ameliorating the Hubble constant tension. For Planck 2018 TTTEEE + lowE + BAO in this model $H_0=69.1\pm1.2$ km/s/Mpc at 68\% CL, reducing the Hubble tension with R19 at $2.7\sigma$, at the price of an indication for a larger electron rest mass $m_e = (1.0078 \pm 0.0067) m_{e,0}$ at 68\% CL.

\end{itemize}

\section{Requirements: Hubble Hunter's Guide} \label{guide}\label{CMBrefs6}
Guidance to model builders about where in redshift departures from $\Lambda$CDM 
are expected  for solving the Hubble constant tension can be found in 
\cite{Knox:2019rjx}. The estimation of $H_0$ from the CMB data can be done in 
three steps: determination of the sound horizon at the CMB last-scattering 
$r_s^*$ from the baryon density and the matter density, determination of the 
comoving angular diameter distance to last scattering $D_A^*=r_s^*/\theta_s^*$ 
from the position of the acoustic peaks, determination of H(z) for all the 
redshift $z$ from $D_A^*=\int_0^{z_*}dz/H(z)$ adjusting the last free density 
parameter.

The authors divide the models into post- and pre-recombination solutions.
Since BAO constrains the product $Hr_s$, the Hubble tension can be seen 
also as a tension between a high and a low sound 
horizon, as  preferred by $\Lambda$CDM and distance ladder, 
respectively. Therefore, to have an agreement between all the data, the authors conclude that the 
late-time solutions (for example a phantom DE) to the Hubble 
constant tension, which do not require a departure from the $CMB$ prior to 
recombination, are unlikely to be a viable path, because they can change the 
Hubble constant estimate, leaving unaltered the sound horizon (and the tension with BAO data).
The most likely solutions to the Hubble tension have instead to be found in the 
pre-recombination  possibilities  (for example an EDE). Moreover, within the pre-recombination 
solutions, those increasing $H(z)$ and decreasing $r_s$ are the best. These should lead to features 
in the CMB power spectra that we could already see in Planck, like, for example, 
the excess of lensing in the temperature power spectrum of Planck ($A_{lens}>1$ 
at $2.8\sigma$).

Their findings are in agreement with a previous work~\cite{Mortsell:2018mfj}, 
where the authors find that late-time dark energy explanations are slightly 
disfavoured, whereas a pre-CMB decoupling extra dark energy component is better 
in alleviating the Hubble constant tension.
An updated view can instead be found in~\cite{Arendse:2019hev}, where it is shown
that the early solutions are primising but cannot solve completely the $H_0$ tension.

\section{Standard Sirens} \label{future}
\label{earlyDEefs1}

The next decade of experiments will be decisive in confirming or ruling out the 
scenarios listed  in the previous sections. In fact, together with the next 
generation of CMB experiments, like the Simon Observatory or CMB-S4, and cosmic 
surveys, like Euclid and LSST, which are expected to reach an uncertainty of 
about a $1\%$ in the $H_0$ estimate, an important role will play complementary 
probes like Standard Sirens~\cite{DiValentino:2018jbh,Palmese:2019ehe}.

In fact, the observations of gravitational-wave and electromagnetic emission 
produced by  the merger of the binary neutron-star system GW170817 have opened 
the possibility of using Standard Sirens, the gravitational-wave analogue of 
astronomical standard candles, to constrain the value of the Hubble constant. 
Indeed, in~\cite{Abbott:2017xzu} a constraint of $H_0=70_{-\,8}^{+12}$ km/s/Mpc 
at 68\% CL has been reported. While this constraint is significantly weaker than 
those derived from observations of Cepheid variables, it does not require any 
form of cosmic ‘distance ladder’ and can be considered, in principle, as more 
conservative. The GW170817 determination of $H_0$ is relatively poor and 
strongly non-Gaussian, but it is possible to investigate what is the impact on 
the Planck constraints in an extended parameter space in which the Planck data 
alone is unable to strongly constrain the Hubble 
constant~\cite{DiValentino:2017clw}.

At least more than 25 additional observations of Standard Sirens are needed for 
reaching a  useful uncertainty on $H_0$ to discriminate between Planck and the 
SH0ES collaboration value. Actually, an uncertainty of about $2\%$ in the 
$H_0$ determination is expected in the early(mid)-2020s~\cite{Chen:2017rfc}, for 
the analysis of Gravitational Waves events with electromagnetic counterparts.
These H0 estimates from bright Standard Sirens will be model-independent,  
oppositely to  the measurements from CMB experiments.

\chapter[$\sigma_8$ Tension. Is Gravity Getting Weaker at Low z? Observational 
Evidence and 
Theoretical Implications]
{$\sigma_8$ Tension. Is Gravity Getting Weaker at Low z? Observational 
Evidence and 
Theoretical Implications}
\label{sec:Perivolaropoulos}
\label{sigmaref1}

{\em Lavrentios Kazantzidis,
Leandros 
Perivolaropoulos}\\

%

\label{tensionsrefs3}

The simplest model consistent with current cosmological observations
is the \lcdm model, which assumes the existence of a fine tuned cosmological  
constant that drives the accelerating expansion of the Universe 
\cite{Carroll:2000fy}.  A wide range of cosmological observations have imposed 
strong constraints on the six free parameters of the model. These observations 
include Type Ia supernovae (SnIa) used as distance indicators 
\cite{Riess:1998cb,Perlmutter:1998np,Betoule:2014frx,Scolnic:2017caz}, the 
Cosmic Microwave Background (CMB) angular power spectrum \label{CMBrefs7}
\cite{Hinshaw:2012aka,Ade:2015xua,Aghanim:2018eyx}, the Baryon Acoustic 
Oscillations (BAO) \cite{Aubourg:2014yra,Alam:2016hwk}, Cluster Counts (CC) 
\cite{Rozo:2009jj,Rapetti:2008rm,Ade:2015fva,Bocquet:2014lmj,Ruiz:2014hma}, 
Weak 
Lensing (WL) 
\cite{Hildebrandt:2016iqg,Joudaki:2017zdt,Troxel:2017xyo,Kohlinger:2017sxk,
Abbott:2017wau,Abbott:2018xao} and Redshift Space Distortions (RSD) 
\cite{Macaulay:2013swa,Johnson:2015aaa,Basilakos:2016nyg,Nesseris:2017vor,
Kazantzidis:2018rnb}.
The first three of the above (SnIa, CMB power spectrum peak locations and BAO), 
act as cosmological distance indicators and   directly probe  the cosmic metric 
independent of the underlying theory of gravity. These are known as ``geometric
probes'' \cite{Nesseris:2006er,Basilakos:2013nfa,Ruiz:2014hma}. The other three 
types of observations, probe  simultaneously the cosmic metric and the growth 
rate of cosmological perturbations. They are  sensitive to the 
dynamics of 
growth and thus to the type of the underlying theory of gravity. These are 
known 
as ``dynamical probes'' \cite{Nesseris:2006er,Basilakos:2013nfa,Ruiz:2014hma}.

\begin{figure}[ht!]
\centering
\includegraphics[width = 0.75\textwidth]{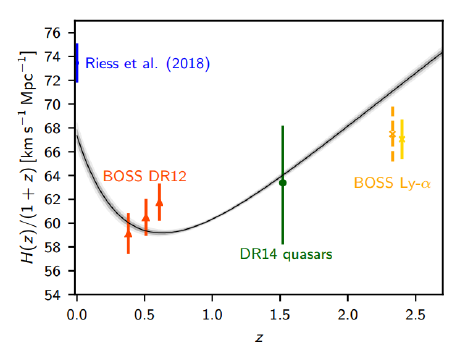}
\caption{
{\it{
The comoving Hubble parameter as a function of $z$ superimposed with 
BAO data from BOSS  DR12 \cite{Alam:2016hwk} survey (orange points), BOSS DR14 
quasar sample \cite{Zarrouk:2018vwy} (green point), SDSS DR12 Ly$\alpha$ sample 
\cite{Bautista:2017zgn}  (yellow points) and the Hubble Space Telescope survey 
\cite{Riess:2018uxu} (blue point). The black line corresponds to the best fit 
obtained from the Planck18 CMB data under the assumption of a \lcdm background, 
while the grey areas are the $1 \sigma$ regions (from Ref. 
\cite{Aghanim:2018eyx}).}}}
\label{fig:Hubblez}
\end{figure}

The consistency of the standard \lcdm model with cosmological observations 
requires that the model  passes two types of tests:
\begin{itemize}
\item
The quality of fit of the model is acceptable in the context of each one of the 
above observational probes.
\item
The best fit values of the six free parameters of the model obtained with each 
individual probe are consistent with each other at a level of about 
$1-2\sigma$. 
If this is not the case, then we have a ``tension'' of \lcdmnospace.
\end{itemize}
The \lcdm model appears to pass the first test in the context of practically 
all 
current observational probes.  However, there seem to be some issues for \lcdm 
in the context of the second test \cite{Bull:2015stt}. In particular, two 
classes of tension have appeared to persist and amplify during the past decade. 
The 
first is the $H_0$ tension, where $H_0$ is the Hubble parameter.  The Planck 
mission \cite{Ade:2015xua,Aghanim:2018eyx} reports that $H_0=67.4 \pm 0.5 km \, 
s^{-1} \, Mpc^{-1}$ at the $1 \sigma$ level, whereas local measurements mainly 
from  Cepheid \cite{Tammann:2008xf} and SnIa luminosity distance indicators 
\cite{Riess:2019cxk,Riess:2016jrr} report that $H_0=74.03 \pm 1.42 km \, s^{-1} 
\, Mpc^{-1}$ at the $1 \sigma$ level, a value approximately $4\sigma$ away from 
the Planck reported one. This tension indicates that the local measurement of 
the Hubble parameter (at scales up to $400 Mpc$) obtained mostly using SnIa, is 
higher than the global value obtained from the Hubble volume on scales of $10 
Gpc$ \cite{Margalef-Bentabol:2012kwa} through an extrapolation of $H(z)$ from 
the last scattering surface to the present time in the context of \lcdm 
scenario, as 
shown 
in Fig. \ref{fig:Hubblez}

Possible explanations of this tension include systematic errors of the CMB 
and/or the SnIa distance  indicators. Alternatively this tension could be an 
early hint for physical deviations from the \lcdm model (see, e.g., Ref. 
\cite{Huterer:2017buf} for a recent review). The later possibility is more 
likely in view of the fact that other local cosmological observations, 
including 
other SnIa data analysis methods 
\cite{Efstathiou:2013via,Cardona:2016ems,Zhang:2017aqn},  gravitational lensing 
\cite{Suyu2012} and Tully-Fisher type calibration of SnIa \cite{Sorce_2012}, 
appear to be consistent with the SnIa measurement 
\cite{Riess:2019cxk,Riess:2016jrr}. In contrast, measurements involving BAO 
\cite{Aubourg:2014yra} and SnIa calibration using the tip of the red-giant 
branch distances \cite{Tammann:2012ut}  are consistent with the extrapolated 
global CMB measurements of $H_0$.

A natural cause  for this tension could be cosmic variance. If it happens that 
we live in a locally  underdense region of the Universe we would locally 
measure 
a value of $H_0$ that would be higher than the mean value over the whole 
Universe. It has been shown, however \cite{Wu:2017fpr,Kazantzidis:2020tko}, that 
the required 
magnitude of such an underdensity  on the required scales of $150 Mpc$ is very 
unlikely in a \lcdm Universe. In such a Universe cosmic variance adds a 
$1\sigma$ error to the locally measured $H_0$ of only $\sigma_{H_0}=0.31 km 
sec^{-1}Mpc^{-1}$, which is negligible compared to the $6 km s^{-1}Mpc^{-1}$ 
needed to resolve the $H_0$ tension.

Non-gravitational physical mechanisms that can reduce the $H_0$ tension include 
the following:
\begin{itemize}
\item
Modifications of expansion rate at late times in the context of alternative 
dark 
energy models  
\cite{Yang:2018qmz,Yang:2019jwn,Belgacem:2017cqo,Alestas:2020mvb}, decaying 
dark 
matter models \cite{Pandey:2019plg,Vattis:2019efj}, or the presence of massive 
sterile neutrinos \cite{Zhao:2017urm} that tend to amplify the accelerating 
expansion at late times. Such modifications could drive upwards the low z part 
of 
the $H(z)$ curve shown in Fig. \ref{fig:Hubblez}, thus bringing the $z=0$ 
prediction of the CMB closer to the $H_0$ result of the local measurements of 
Ref. \cite{Riess:2016jrr}.
\item
Inhomogeneous cosmologies \cite{Lukovic:2018ljo} that would make a local deep 
underdensity more  likely than in the case of \lcdmnospace.
\item
A new component of dark radiation \cite{Bernal:2016gxb} that would tend to 
decrease the sound horizon $r_s$ at radiation drag, thus leading to a predicted 
increase of $H_0$ by shifting the whole curve of Fig. \ref{fig:Hubblez} 
upwards. 
This approach has  the advantage of shifting, at the same time, the BAO points 
shown in Fig. \ref{fig:Hubblez}.
\end{itemize}
The origin of these mechanisms is non-gravitational and the consensus is that 
since $H_0$  is a geometric parameter it can not be affected by modifications 
of 
GR. However, as discussed in more detail below, the physics of SnIa is heavily 
based on the assumption of validity of GR. For example, an evolving Newton's 
constant at low redshifts would directly affect the absolute magnitude of SnIa, 
leading to a requirement for a new interpretation of the SnIa distance moduli. 
Therefore, even though the SnIa absolute magnitude is usually assumed constant 
and is marginalised as being a nuisance parameter, its possible evolution may 
carry useful information about the robustness of the determination of $H_0$ 
using SnIa and about possible modifications of GR. Two interesting question, 
therefore arise:
\begin{itemize}
\item
Are there indications for evolution of the SnIa absolute magnitude at low $z$?
\item
What would be the implications of such evolution on the derived value of $H_0$ 
and on the  possible evolution of the effective Newton's constant?
\end{itemize}
These are among the questions discussed in what follows.

The second tension in the context of \lcdm is the $\sigma_8$ tension, where 
$\sigma_8$ is the density rms matter fluctuations within spheres of radius $8 
h^{-1} Mpc$ and is determined by the amplitude of the primordial fluctuations 
power spectrum and by the growth rate of cosmological fluctuations. 
\label{growthref2} In 
particular, dynamical probes, (mainly RSD 
\cite{Macaulay:2013swa,Nesseris:2017vor,Basilakos:2017rgc,Kazantzidis:2018rnb} 
  WL   
\cite{Hildebrandt:2016iqg,Kohlinger:2017sxk,Joudaki:2017zdt,Abbott:2017wau}), 
and  EG data  \cite{Skara:2019usd},
favour lower values of $\sigma_8$ and/or \omomnospace than the corresponding 
values reported by Planck \cite{Ade:2015xua,Aghanim:2018eyx} at a $2-3\sigma$ 
level. This tension, if not due to systematics of the dynamical probes or CMB 
data, could be interpreted as an indication of a weaker gravitational growth 
of perturbations than the growth indicated by GR in the context of a \lcdm 
model 
with the  \plcdm parameter values, which are shown in the following Table 
\ref{PKtab:plcdm18}.
\begin{table}[ht!]
\centering
\begin{tabular}{cc}
 \hline
  Parameter & \plcdm \cite{Aghanim:2018eyx}\\
    \hline
$\Omega_b h^2$ & $0.02237 \pm 0.00015$ \\
$\Omega_c h^2$ & $0.1200 \pm 0.0012$  \\
$n_s$ & $ 0.9649 \pm 0.0042$\\
$H_0$ & $67.36 \pm 0.54$ \\
$\Omega_{m0}$ & $0.3153 \pm 0.0073$ \\
$w$ & $-1$  \\
$\sigma_8$ & $ 0.8111 \pm 0.0060$\\
\hline
\end{tabular}
\caption{\plcdm  parameters values from Ref. \cite{Aghanim:2018eyx}, based on  
TT,TE,EE, lowE and lensing likelihoods.}
\label{PKtab:plcdm18}
\end{table}

In addition to a possible evolution of the effective Newton constant discussed 
below,  non-gravitational mechanisms can also reduce the $\sigma_8$ tension 
(see, e.g., Ref. \cite{Ishak:2018his} for a  recent review). Such effects 
include the 
following:
\begin{itemize}
\item
Interacting dark energy models, which modify the equation for the evolution of 
linear matter fluctuations in a given $H(z)$ cosmological background 
\cite{Pourtsidou:2016ico,Barros:2018efl,Camera:2019vbp}.
\item
Dynamical dark energy models 
\cite{Melia:2016djn,Lambiase:2018ows,Ooba:2018dzf,Yang:2018qmz,Joudaki:2016kym,
Barros:2018efl} and running vacuum models  
\cite{Gomez-Valent:2017idt,Gomez-Valent:2018nib},  which modify the 
cosmological 
background  $H(z)$ to a form different from \lcdmnospace.
\item
Effects of massive neutrinos \cite{Joudaki:2016kym,DiazRivero:2019ukx}.   
Neutrinos are relativistic at early times (contribute to radiation) while at 
late times they become non-relativistic but with significant velocities (hot 
dark matter), while they constitute a non-negligible fraction of the dark 
matter 
of the Universe. The conversion of radiation
to hot dark matter plays a role in the Hubble expansion.  At the same time the  
residual streaming velocities are still large enough
at late times to slow down the growth of structure.
Thus, neutrinos affect both background expansion and the growth of cosmological 
perturbations in such a way as to slow down the growth as required by the RSD 
data. Their effects on easing the $\sigma_8$ tension coming from WL data has 
been questioned by the recent analysis of Ref. \cite{DiValentino:2018gcu} (see 
also Fig. \ref{fig:S8const})
\end{itemize}
Besides these categories, alternative parameters beyond the standard ones, such 
as a running  scalar spectral index, a modified matter expansion rate or a bulk 
viscosity coefficient may have the potential to ease the $\sigma_8$ tension 
\cite{Wang:2019acf}.

In addition to these non-gravitational effects that can slow down growth at low 
redshifts, modified  gravity theories can also contribute in the same direction 
in a manner that is more generic and fundamental. The feature required from 
this 
class of theories is a reduced effective Newton's constant $\Geff$ at low 
redshifts. It turns out that this behaviour cannot be achieved in a \lcdm 
background for most scalar-tensor and $f(R)$ theories 
\cite{Tsujikawa:2007gd,Gannouji:2018ncm,Nesseris:2017vor}. However, it is 
possible in other less generic modified gravity theories, including 
telleparallel 
theories of gravity \cite{DAgostino:2018ngy,Gonzalez-Espinoza:2018gyl}, 
Horndeski theories \cite{Kennedy:2018gtx,Linder:2018jil,Gannouji:2020ylf}, or 
theories beyond 
Horndeski  \cite{DAmico:2016ntq}.

Clearly, a reduced (compared to GR)  evolving effective Newton's constant would 
have important signatures  on low $z$ cosmological observations. In particular:
\begin{itemize}
\item
It would affect \cite{Nesseris:2017vor,Kazantzidis:2018rnb} the  low $l$ CMB  
\label{CMBrefs8}
power spectrum through the Integrated Sachs-Wolfe (ISW) effect 
\cite{Pogosian:2005ez,Ho:2008bz}.
\item
It would affect \cite{Nesseris:2017vor,Kazantzidis:2018rnb} the growth  rate of 
cosmological fluctuations as detected through the RSD 
\cite{Macaulay:2013swa,Johnson:2015aaa,Basilakos:2016nyg,Nesseris:2017vor,
Kazantzidis:2018rnb}, WL 
\cite{Hildebrandt:2016iqg,Joudaki:2017zdt,Troxel:2017xyo,Kohlinger:2017sxk,
Abbott:2017wau,Abbott:2018xao} and CC data 
\cite{Rozo:2009jj,Rapetti:2008rm,Ade:2015fva,Bocquet:2014lmj,Ruiz:2014hma}.
\item
It would induce an evolution of the SnIa absolute magnitude, which depends on 
the 
 magnitude of Newton's constant 
\cite{Amendola:1999vu,Gaztanaga:2001fh,Nesseris:2006jc,Wright:2017rsu,
Sapone:2020wwz}. 
Notice, 
however, that the value of the effective Newton's constant here should be 
obtained from a strong gravity calculation in the context of a modified gravity 
theory and thus is not   generally identical to the $\Geff$ derived for the 
growth of cosmological perturbation that involves a perturbative calculation.
\end{itemize}
For a viable modified gravity mechanism there should be consistency with 
respect 
to  the type and magnitude of Newton's constant evolution favoured by the above 
cosmological observations, keeping in mind the strong gravity effects involved 
in 
the SnIa physics. It will be seen in what follows that indeed all of the above 
probes mildly favour a reduced value of Newton's constant at low $z$. However, 
the favoured magnitude and statistical significance of such a reduction varies  
among the above observational probes.

A Newton's constant \label{Geffref1} \Geffz evolving with redshift,  may be 
parametrised in the 
context of a wide  range of parametrisations, including theoretically motivated 
\cite{Bertschinger:2008zb,DiValentino:2015bja,Ade:2015rim,Baker:2014zva,
Li:2018tfg} and model-independent \cite{Nesseris:2017vor,Kazantzidis:2018rnb}  
\label{madelindepeefs6}
forms, and leads to a modification of the linear growth of cosmological 
perturbations. \label{Scalarperfrref6}

This modified growth equation is obtained by considering the perturbed 
Friedmann–\\Lema\^itre-Robertson–Walker  (FLRW) metric in the Newtonian gauge, 
which is given by \cite{Bardeen:1980kt,Ma:1995ey,EspositoFarese:2000ij}
\be
ds^2 = -(1+2\Psi)dt^2 + a^2(1-2\Phi) d{\vec{x}}^2\ , \label{PKeq:pertfrw}
\ee
where $a$ is the scale factor that is connected to the redshift $z$ through 
$a=1/(1+z)$, and $\Psi$,$\Phi$ correspond to the Bardeen potentials in 
the 
Newtonian gauge \cite{Ma:1995ey}. Einstein's equations in  Fourier space at 
linear order take the form  
\cite{Huterer:2013xky,Pogosian:2010tj,Perenon:2019dpc}
\ba
 k^2 \Psi &=& -4\pi G_N \mu (a, k) a^2 \rho_m \Delta \ ,
\label{PKeq:poissonmg1} \\
 \Phi &=& \eta(a, k) \Psi \ ,
\label{PKeq:poissonmg2} \\
 k^2(\Phi +\Psi) &=& -8\pi G_N\, \Sigma(a, k)\, a^2 \rho_m \Delta \ ,
\label{PKeq:poissonmg3}
\ea
where $\mu\equiv \Geff/G_{\textrm{N}}$ ($G_N$ is Newton's constant as measured 
by local experiments  and Solar System observations), $\Delta$ is the comoving 
density contrast defined as $\Delta \equiv \delta+3 a H u/k$. $\delta \equiv 
\frac{\delta \rho_m}{\rho_m}$ is the linear matter growth  factor, $u$ is the 
irrotational component potential of the peculiar velocity, $\rho_m$ is the  
matter 
density of the background, $\eta$ is the gravitational slip and $\Sigma\equiv 
G_{\textrm{L}}/G_{\textrm{N}}$ is the lensing normalized Newton constant.

In GR, $\Geff$, which is connected with the growth of matter  perturbations and 
$G_L$, which is related with the lensing of light through the Weyl potential 
$\frac{\Phi_+}{2}=\Phi+\Psi$, coincide with $G_N$. The Weyl potential can be 
connected with lensing, since   the cosmic convergence of null geodesics with 
respect to unperturbed geodesics is given by \cite{Tereno:2010dt}
\be
\kappa(r,\theta)=\frac{1}{4}\int_0^{r} dr'\,\left(\frac{r-r'}{r}\right)r'
\nabla^2\Phi_+(\theta,r'),
\ee
where $r$ and $\theta$ are the comoving coordinates of the source.
The parameters $\mu$, $\eta$ and $\Sigma$ are connected as
\be
\Sigma(a, k)=\frac{\mu(a, k)\, \left[1+\eta (a, k) \right]}{2} \ ,
\ee
and they are key parameters in detecting deviations from GR where their value 
coincides with unity.

In what follows we focus on the parameter $\mu$. This parameter is associated 
with the linear matter growth   factor through the growth equation 
\cite{Perenon:2019dpc}
\be
\ddot{\delta} +2H\dot{\delta}= \frac{3}{2}\,H^2\,\Omega_m \,\mu \, \delta \ ,
\label{deltatevol}
\ee
where the dot denotes differentiation with respect to cosmic time $t$. Eq. 
(\ref{deltatevol}) is derived  using the conservation of the matter energy 
momentum tensor and the modified Poisson Eq. \eqref{PKeq:poissonmg1}, assuming 
scales much smaller than the Hubble scale.  For most modified gravity models 
the 
scale dependence of $\mu$ is very weak in scales much smaller than the Hubble 
scale. In addition, most growth data do not report scale dependence but only 
redshift dependence.  Thus, we only parametrise the dependence of  $\mu$ on the 
the scale factor, i.e. $\mu(a,k)=\mu(a)$, and  we obtain the growth equation in 
redshift space as
\be
\delta'' +\left(\frac{H'}{H}-\frac{1}{1+z}\right)\, \delta'  
=\frac{3}{2}\frac{\Omega_m }{\left(1+z\right)^2}  \,\mu\, \delta \ , 
\label{PKeq:oddD}
\ee
where in Eq. \eqref{PKeq:oddD}, the prime denotes differentiation with respect 
to the redshift $z$.   The equation for the growth rate $f(z)\equiv 
\frac{dln\delta}{dlna}$ may also be obtained from Eq. \eqref{PKeq:oddD} as
\be
\left(1+z\right) f'-f^2+\left[ (1+z)\frac{H'}{H}-2 \right] f=-\frac{3}{2}\, 
\Omega_m \,\mu\ , \label{PKeq:oddf}
\ee
Fixing the background $H(z)$ and considering a specific parametrisation for 
$\mu$, Eq. (\ref{PKeq:oddf}) can be solved (either numerically or analytically) 
with initial conditions deep in the matter era, where $\delta\sim a$ (assuming 
GR 
is restored at early times). Combining this solution with the rms density 
fluctuations on scales of $8Mpc$, $\sigma_8$, which evolves as $\sigma_8(z)= 
\sigma_8 (z=0) \frac{\delta(z)}{\delta(z=0)}$, we obtain  theoretical 
prediction 
for the product ${\rm{\it f\sigma}}_8$ given $\sigma_8(z=0)\equiv \sigma_8$, 
$H(z)$ and $\mu(z)$. In particular,
\be
{\rm{\it f\sigma}}_8(a)\equiv f(a)\cdot 
\sigma(a)=\frac{\sigma_8}{\delta(1)}~a~\delta'(a) \label{PK:eqfs8}
\ee
is reported by many surveys since 2006, leading to collections of data which 
can 
be used to simultaneously constrain   $H(z)$ and $\mu(z)$.

On the other hand, $H(z)$ is usually parametrised as $wCDM$ i.e.
\be
H^2(z)=H_0^2 \left[\Omega_{m0}(1+z)^3 +(1-\Omega_{m0})(1+z)^{3(1+w)}\right] ,
\label{PKeq:wcdm}
\ee
which reduces to \lcdm\, for an  equation of state parameter $w=-1$, the 
effective Newton's constant parameter $\mu$ does not have a commonly accepted 
parametrisation. Some authors motivated from the predictions of scalar-tensor 
theories use a  scale dependent parametrisation for $\mu$ and $\eta$ as 
\cite{Bertschinger:2008zb,Li:2018tfg}
\ba
\mu(a,k) &=& \frac{1+\beta_1 \lambda_1^2 k^2 a^s}{1+\lambda_1^2k^2a^s}, \nn \\
\eta(a,k) &=& \frac{1+\beta_2 \lambda_2^2 k^2 a^s}{1+\lambda_2^2k^2a^s}.
\label{PKeq:muetabzmod}
\ea
In the special case of $f(R)$ theories the parameters that appear in Eq.  
\eqref{PKeq:muetabzmod} are
\ba
\beta_1=4/3; \  \ \ \beta_2=1/2; \ \ \ \lambda_2^2/\lambda_1^2=4/3.
\ea
Clearly for $\beta_1>1$ (as is the case for $f(R)$ theories  \label{fRref7}
\cite{Hu:2007nk,Starobinsky:2007hu} and  in most scalar-tensor theories) we 
have 
$\mu>1$ at low $z$, and thus gravity is stronger than in GR at low $z$ in these 
classes of theories 
\cite{Hu:2007nk,Starobinsky:2007hu,Tsujikawa:2015mga,Gannouji:2018ncm,
Polarski:2016ieb}.

Another parametrisation that has been studied in the literature is the 
parametrisation of  no-slip gravity \cite{Linder:2018jil}, a subclass of 
Horndeski theories which remains viable after the binary star collision 
GW170817 
\cite{TheLIGOScientific:2017qsa}. In this case $\mu$ takes the form 
\cite{Linder:2018jil}
\be
\mu=\frac{2}{2+b+b \, \tanh \left[\frac{\tau}{2} \, 
\log_{10}(\frac{a}{a_t})\right]},  \label{PK:eqnoslipmu}
\ee
where $b, \tau$ and $a_t$ correspond to parameters that describe  the 
amplitude, 
the  rapidity and the scale factor at the time when $\mu$ shifts from unity in 
the early Universe to $\mu=1+b$.

An alternative scale-dependent class of parametrisations for $\mu$ and $\eta$ 
is 
of the form \cite{Ade:2015rim,DiValentino:2015bja}
\ba
\mu(a,k) &=& 1 + f_1(a) \frac{1 + c_1 (\lambda H/k)^2}{1+(\lambda H/k)^2}; \\
\eta(a,k) &=& 1 + f_2(a) \frac{1 + c_2 (\lambda H/k)^2}{1+(\lambda H/k)^2}.
\ea
For sub-Hubble scales this parametrisation becomes scale-independent and has 
been expressed as \cite{DiValentino:2015bja}
\ba
\mu(a,k) &=& 1 + E_{\rm{11}} \Omega_{\rm{DE}}(a)\,; \\
\eta(a,k) &=& 1 + E_{\rm{22}} \Omega_{\rm{DE}}(a)\,.
\ea
where $\Omega_{\rm{DE}}(a)$ is the density parameter of the dark energy.
For $E_{11}<0$, gravity is weaker compared to GR at low $z$, and indeed the 
best 
fit value  obtained in Ref. \cite{DiValentino:2015bja} when dynamical probes 
are 
taken into account is negative ($E_{11}=-0.21^{+0.19}_{-0.45}$ when CMB and WL 
data are taken into account).

A model and scale-independent parametrisation 
\cite{Nesseris:2017vor,Kazantzidis:2018rnb} for $\mu$, which reduces to the GR 
value at low and high $z$, while respecting the constraints from Solar Systems  
\label{solarsystemref9}
tests and from the nucleosynthesis \cite{Gannouji:2006jm,Nesseris:2006hp} is
\be
\mu =1+g_a(1-a)^n - g_a(1-a)^{2n}= 1+g_a \left(\frac{z}{1+z} \right)^n - g_a 
\left(\frac{z}{1+z} \right)^{2n}\ , \label{PKeq:geffansatz}
\ee
where $g_a$ and $n$ integer with $n\geq 2$ are parameters to be fit from data. 
A 
distinguishing feature of this parametrisation is that it naturally and 
generically respects Solar System and nucleosynthesis constraints 
($\frac{d\mu}{dz}\vert_{z=0}=0, \mu(z =0)=1$, $\mu(z\rightarrow \infty)=1$) 
\cite{Muller:2005sr,Nesseris:2006hp,Gannouji:2006jm,Pitjeva:2013chs}.

An alternative approach for the parametrisation of deviations from GR is based 
directly on the growth rate $f$ of density fluctuations. The growth rate $f$ is 
usually parametrised using the ``growth index'' $\gamma$ as
\be
f(z)=\frac{dln\delta}{dlna} \approx \Omega_m(z)^\gamma,
\ee
where $\gamma$ in most dark energy models based on GR is $\gamma \approx 0.55$. 
For many modified gravity theories this quantity is not constant and is 
parameterised instead as a function of the redshift $z$ (see, e.g., Ref. 
\cite{Gannouji:2008wt}, and for updated observational constraints of this 
parameter Refs. 
\cite{Shafieloo:2018gin,Gannouji:2018col,Gannouji:2018ncm,Basilakos:2019hlb}). 
In particular, recent observations indicate that $\gamma >0.55$ (weaker growth 
rate) in contrast to the usual theoretical prediction of $\gamma <0.55$  that 
is 
supported by many modified gravity  models such as $f(R)$ theories 
\cite{Yin:2018mvu} and indicates stronger gravity at low $z$. In  what follows 
we focus on the parametrisation \eqref{PKeq:geffansatz}.

The $\mu$ parametrisation \eqref{PKeq:geffansatz} has been extensively studied 
in Refs.  \cite{Nesseris:2017vor,Kazantzidis:2018rnb, 
Perivolaropoulos:2019vkb}, 
where it was shown that in the context of a wide range of different RSD 
datasets, a negative value of the parameter $g_a$ is favoured in the context of 
{\plcdm background expansion rate $H(z)$ ($g_a=-0.68\pm 0.18$), indicating 
weaker 
gravity than the GR prediction at low $z$.

This trend for weaker gravity at low $z$ is also supported by WL data 
\cite{Hildebrandt:2016iqg,Kohlinger:2017sxk,Joudaki:2017zdt,Abbott:2017wau,
DiValentino:2018gcu}, even though in these references this trend was expressed 
as 
a trend for lower values of $\sigma_8$ and \omom (or equivalently $S_8\equiv  
\sigma_8 \sqrt{\Omega_{m0}/3}$) compared to the \plcdm best fit, since $\mu$ 
was 
fixed to unity. This tension level which can not be released even by the 
inclusion of massive neutrinos, is demonstrated in Fig.  \ref{fig:S8const}. In 
Fig. \ref{fig:S8const} $A_{lens}$ is an effective parameter that rescales the 
lensing amplitude in the CMB spectra. The extension of \lcdm involving the 
parameter $A_{lens}$ is degenerate with a modified gravity extension and can 
clearly decrease the $\sigma_8$ tension implied by the WL data, as shown in 
Fig. 
\ref{fig:S8const} \cite{DiValentino:2018gcu}. In contrast, the  introduction of 
massive sterile neutrinos appears to have little effect on the tension level.
\begin{figure}[ht!]
\centering
\includegraphics[width = 0.99\textwidth]{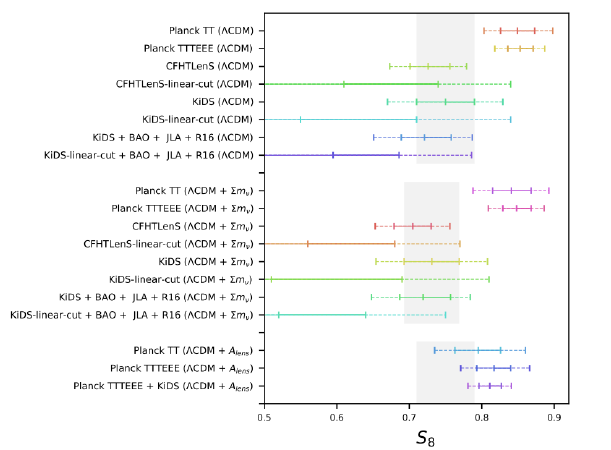}
\caption{{\it{
$1-2 \sigma$ constraints for $S_8=\sigma_8 
\sqrt{\frac{\Omega_{m0}}{0.3}}$  for various combinations of datasets and 
models 
superimposed with the Kilo Degree Survey (KiDS) \cite{Joudaki:2016kym} survey 
bounds (grey regions) for each cosmological model (adopted from Ref. 
\cite{DiValentino:2018gcu}). In particular, the CFHTLenS linear cut model 
corresponds to the conservative cut of the cosmic shear data of the 
Canada-France-Hawaii Telescope Lensing (CFHTLens) survey 
\cite{Heymans:2012gg,Erben:2012zw} in order to reduce the non-linear scale 
contribution \cite{Ade:2015rim}, the JLA acronym corresponds to the Supernovae 
Data from Ref. \cite{Betoule:2014frx}, the R16 stands for the $H_0$ measurement 
given in Ref. \cite{Riess:2016jrr}, the BAO data correspond to the data used in 
Ref. \cite{Ade:2015xua}, while the KiDS linear cut model describes the 
conservative cut of Ref. \cite{Joudaki:2017zdt} and $\sum m_\nu$ stands for the 
inclusion of massive neutrinos. Notice that only the introduction of the 
$A_{lens}$ parameter, which is degenerate with the evolution of $\Geff$ 
\cite{DiValentino:2015bja}, can lead to a reduction of the tension between 
Planck 
and dynamical probes.}}}
\label{fig:S8const}
\end{figure}

As discussed above, an evolving $\mu$ can also affect the $H_0$ tension 
problem.  \label{H0tensionjref3}
Indeed,  local measurements of $H_0$ are heavily based on SnIa as distance 
indicators and on the assumption that after proper calibration the SnIa 
absolute 
magnitude $M$ may be assumed to be constant.  The peak luminosity of SnIa, is 
related to the gravitational constant as $L \propto G^{-3/2}$ 
\cite{Gaztanaga:2001fh}, which leads to an absolute magnitude $M$ that is 
associated with $\mu$ through \cite{Amendola:1999vu,Gaztanaga:2001fh}
\be
M-M_0=\frac{15}{4} \, \log_{10} \left(\mu \right),
\label{PKeq:mgeffconn}
\ee
where $M_0$ is a reference asymptotic value of the absolute magnitude. A more 
detailed  and accurate approach for determing the dependence of $M$ of SnIa on 
the Newton's constant has been implemented in Ref. \cite{Wright:2017rsu} 
through 
a semi-analytical method of light curve fitting that uses the standardised 
intrinsic luminosity L instead of  the peak luminosity of individual events to 
find the dependence of $M$ on the value of $G$. In this model the sign of the 
power index $-\frac{3}{2}$ appearing above, is indicated to be positive 
instead. 
This possibility will be discussed in a following publication 
\cite{Kazantzidisinprog}. Usually, $M$ is considered to be a constant nuisance 
parameter and is marginalised. However, since dynamical probes favour a $\mu$ 
smaller than the GR value, similar trends (perhaps not of the same magnitude 
due 
to the strong gravitational fields involved) are expected for the absolute 
magnitude $M$. In what follows we present a short preliminary analysis 
attempting to address this issue and identify possible trends and constraints 
in 
the absolute $M$ of the SnIa.

In the context of the above discussion, the following questions arise:
\begin{itemize}
\item What is the current level of the ${\rm{\it f\sigma}}_8$ tension and what 
is the implied evolution of $\mu$ in the context of \lcdmnospace?
\item Are there hints of a similar evolution of $\mu$ in the Pantheon SnIa 
dataset?
\item What is the allowed evolution of $\mu$ from the low $l$ CMB data?
\end{itemize}
These questions will be addressed in what follows.

The structure of this brief review is the following: In Sec. 
\ref{sec:fstensionmodgrav} we review the ${\rm{\it f\sigma}}_8$ tension and the 
implications of a dynamical $\mu(z)$ for modified gravity theories. In Sec. 
\ref{sec:Pantheon} a tomographic analysis of the SnIa absolute magnitude of the 
Pantheon dataset is performed and the constraints on possible evolution at low 
$z$ are specified. Finally, in Sec. \ref{sec:ISW} the constraints on an 
evolving 
$\mu$ from the low $l$ angular CMB spectrum and the ISW effect are presented in 
the context of a \lcdm background, while in Sec. \ref{sec:Conclusions} we 
outline and discuss our results.

\section{The \texorpdfstring{$f\sigma_8$}{fs8} Tension and Modified Gravity.}
\label{sec:fstensionmodgrav}

\subsection{Observational Evidence}

The solution of Eq. \eqref{PKeq:oddD} with initial conditions deep in the 
matter 
era, a $wCDM$ background \eqref{PKeq:wcdm} and an evolving parameter $\mu(z)$ 
of 
the form \eqref{PKeq:geffansatz} with $n=2$ respects both nucleosynthesis 
constraints and Solar System constraints. The theoretical prediction for 
${\rm{\it f\sigma}}_8(z)$ obtained from such a solution using also Eq. 
\eqref{PK:eqfs8} depends on the parameters $\Omega_{m0}$, $w$ and $g_a$ and is 
shown in Fig. \ref{fig:fs8z}, along with a large compilation of corresponding 
datapoints \cite{Kazantzidis:2018rnb} (the different colours correspond to 
early 
or more recent times of publication). Clearly, the parameter values 
($\Omega_{m0}, 
w,\sigma_8, g_a)=(0.31,-1,0.83,0)$ corresponding to \plcdm  lead to larger 
growth (${\rm{\it f\sigma}}_8(z)$) than most data would imply, especially at 
redshifts $z<1$ (red line).
\begin{figure}[ht!]
\centering
\includegraphics[width = 
0.75\textwidth]{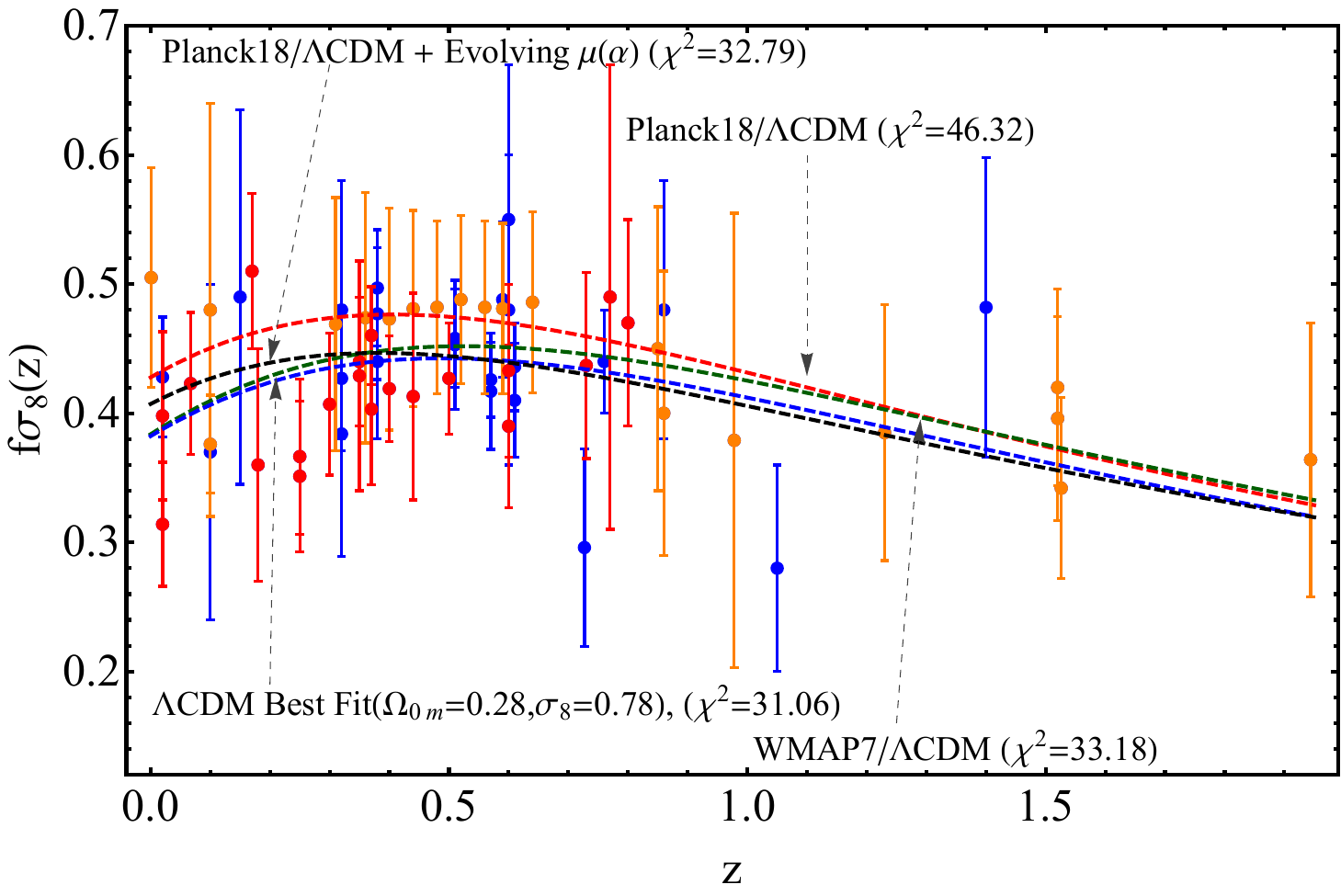}
\caption{{\it{Evolution of ${\rm{\it f\sigma}}_8$ as a function of redshift. 
The red 
dashed line corresponds to the \plcdm model ($\Omega_{m0}=0.315 \pm 0.007, 
\sigma_8= 0.811 \pm 0.006$), the green one to the \wlcdm ($\Omega_{m0}=0.266 
\pm 
0.025, \sigma_8= 0.801 \pm 0.030$), the black one to an evolving $\mu$ with a 
\plcdm background ($\Omega_{m0}=0.315 \pm 0.007, \sigma_8= 0.811 \pm 0.006$, 
$g_a =-0.681 \pm 0.177$), while the blue one describes the best fit \lcdm 
coming 
from the 63 compilation of Ref. \cite{Kazantzidis:2018rnb} ($\Omega_{m0}=0.279 
\pm 0.028, \sigma_8= 0.775 \pm 0.018$). The orange points correspond to the 20 
latest datapoints, while the red ones correspond to the 20 earliest of this 
compilation. 
The blue points account for the rest of the growth data.}}}
\label{fig:fs8z}
\end{figure}

The fit to the data may be improved either by 
modifying the background expansion rate $H(z)$ (\eg ~lowering $\Omega_{m0}$) 
and/or by lowering the strength of gravity at low $z$. Fixing the background 
$H(z)$ to \plcdm and allowing $g_a$ in Eq. \eqref{PKeq:geffansatz} to vary we 
obtain \cite{Nesseris:2017vor,Kazantzidis:2018rnb} a best fit value of 
$g_a=-0.68\pm 0.18$ for $n=2$ which is approximately $3.7\sigma$ away from the 
GR value $g_a=0$.

The trend for weaker gravity at low redshifts is also evident in Fig.  
\ref{fig:Geffaplot}, which shows the best fit form of $\mu(a)$ as a function of 
the scale factor $a$ for the best fit values of $g_a$ coming from the robust 
RSD 
data compilation of Ref. \cite{Nesseris:2017vor} for different values of $n$. 
The required drop of $\mu(a)$ becomes stronger and localised to low $z$ as $n$ 
increases. As discussed in Section \ref{sec:ISW}, however, such a large drop is 
not consistent with the low $l$ CMB angular power spectrum and the ISW 
effect.\label{CMBrefs9}
\begin{figure}[!ht]
\centering
\includegraphics[width = 0.75\textwidth]{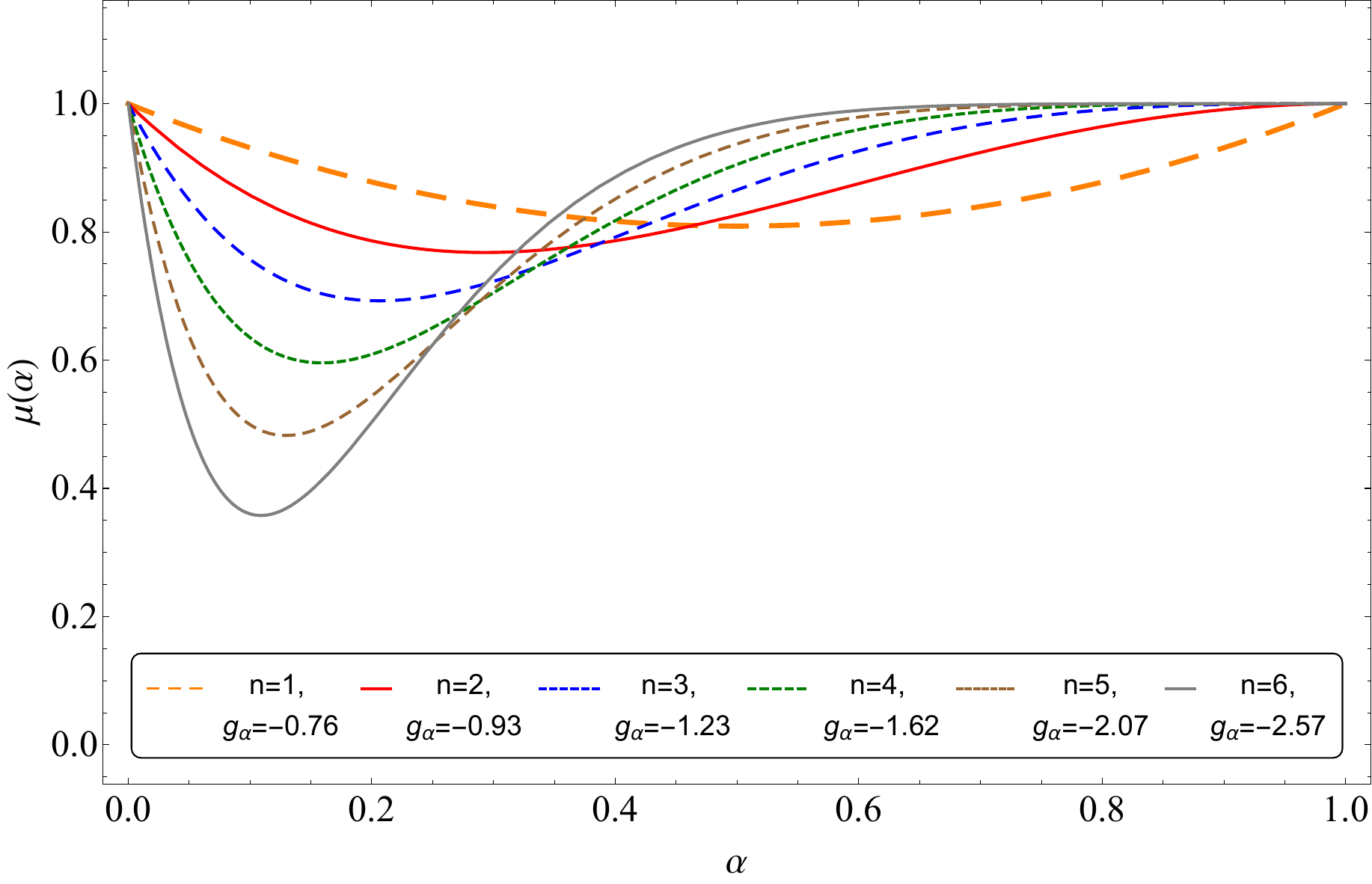}
\caption{{\it{Evolution of $\mu$ as a function of the scale factor $a$ 
considering  
the best fit values for $g_a$ and various values of $n$ using the robust 
collection of Ref. \cite{Nesseris:2017vor}}.}}
\label{fig:Geffaplot}
\end{figure}

An interesting feature of the theoretical model predictions for ${\rm{\it 
f\sigma}}_8(z)$ shown in Fig. \ref{fig:fs8z} is the degeneracy among these 
predictions for $z>1$. This degeneracy has been investigated in some detail in 
\cite{Kazantzidis:2018jtb} for ${\rm{\it f\sigma}}_8(z)$ and for other 
cosmological observables. It was found that there are blind redshift spots 
where 
observables are degenerate with respect to specific cosmological parameters. 
For 
${\rm{\it f\sigma}}_8(z)$ with respect to the parameter $g_a$ there is a blind 
spot at $z\simeq 2.5$ and its constraining power is significantly reduced for 
$z>1$. Thus ${\rm{\it f\sigma}}_8(z)$ datapoints with $z<1$ can constrain 
$\mu(z)$ (or equivalently $g_a$) much more efficiently than points at higher 
redshifts. This is demonstrated in Fig. \ref{fig:deltafs8},  which shows the 
difference between the growth rate in the context of an evolving ${\rm{\it 
f\sigma}}_8(z)$ from the \plcdm ${\rm{\it f\sigma}}_8(z)$ 
\cite{Kazantzidis:2018jtb}  for various values of $g_a$. This difference is 
defined as
\begin{figure}[ht!]
\centering
\includegraphics[width = 0.5\textwidth]{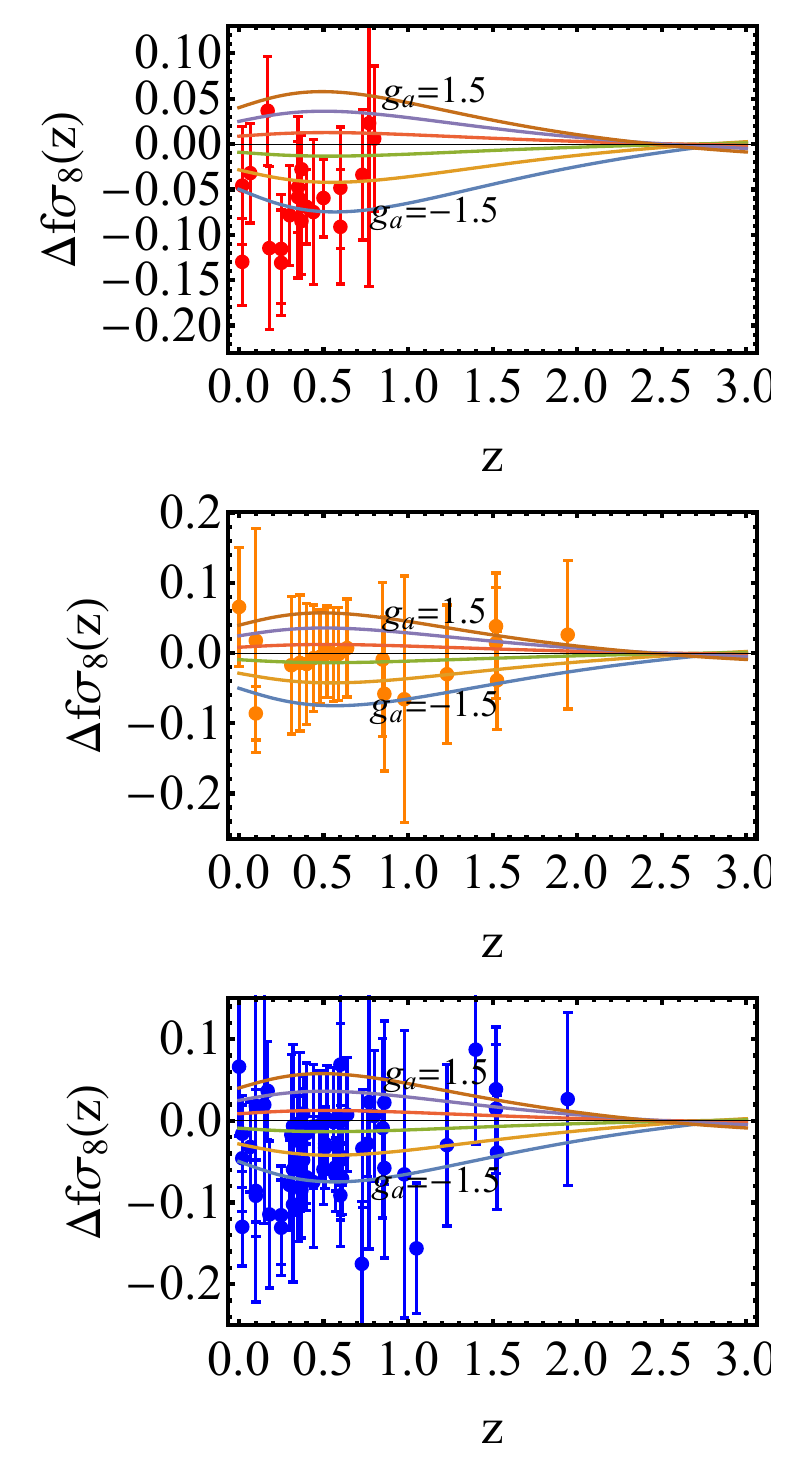}
\caption{{\it{Evolution of $\Delta {\rm{\it f\sigma}}_8$ as a function of the 
redshift $z$ for different values of $g_a$. These curves are superimposed with 
the 20 earliest datapoints (upper panel), the 20 latest (middle panel) and the 
full compilation (lower panel) of Ref. \cite{Kazantzidis:2018rnb}. Notice that 
early lower $z$ datapoints are much more efficient in detecting hints of 
modified gravity (a non-zero value of $g_a$).}}}
\label{fig:deltafs8}
\end{figure}
\be
\Delta {\rm{\it f\sigma}}_8= {\rm{\it f\sigma}}_8 
(z,\Omega_{m0}^{Planck18},-1,g_a) -{\rm{\it f\sigma}}_8 
(z,\Omega_{m0}^{Planck18},-1,0).
\ee
Clearly, early published datapoints that tend to have lower redshifts (right 
panel) have more constraining power than more recently published datapoints 
(middle panel), which have higher $z$ and larger errorbars. The tension level 
comes mainly from early datapoints, which appear to favour $\Delta {\rm{\it 
f\sigma}}_8<0$, i.e. weaker growth. \footnote{The RSD datapoints  of Fig. 
\ref{fig:deltafs8} include a $1-3\%$ ``fiducial cosmology'' Alcock-Paczynski 
correction 
\cite{Alcock:1979mp,Macaulay:2013swa,Kazantzidis:2018rnb,Nesseris:2017vor}, 
i.e. 
they have been multiplied by a factor $\frac{H(z) 
d_A(z)}{H_{fid}(z)d_{A_{fid}}(z)}$ where the subscript $_{fid}$ indicates the  
fiducial cosmology used in each survey to convert angles and redshift to 
distances for evaluating the correlation function and $H(z), d_A(z)$ correspond 
to the Hubble parameter and the angular diameter distance of the true 
cosmology.}

The trend for weaker growth of matter perturbation than the growth favoured by 
\plcdm has been pointed out  in a wide range of studies in the context of 
different dynamical probe data. One of the first analyses that pointed out the 
weak growth tension was that of Ref. \cite{Macaulay:2013swa}, where it was 
pointed out that RSD measurements are consistently lower than the values 
expected from Planck in the context of \lcdmnospace cosmology. It was also 
pointed out 
that other dynamical probes like the Sunyaev-Zeldovich (SZ) cluster counts 
\cite{Ade:2013lmv} also indicate weaker growth ($\sigma_8=0.77\pm 0.02$, 
$\Omega_{m0}=0.29\pm 0.02$). Similar trends were found earlier, using the 
measurement of the galaxy cluster   mass function in the redshift range 
$z\in [0,0.9]$ \cite{Vikhlinin:2008ym}, where lower values of $\Omega_{m0}$ and 
$\sigma_8$ were favoured. Later studies confirmed this trend by pointing out 
that 
best fit cosmological parameters like the matter density $\Omega_{m0}$ and the 
dark energy equation of state $w$ differ at a level of $2-3\sigma$ between 
geometric probes (SnIa, BAO and CMB peak locations) and dynamical probes (RSD 
data, CC and WL) \cite{Ruiz:2014hma,Bernal:2015zom}. The dynamical probes of 
growth pointed consistently towards lower values of $\Omega_{m0}$ and thus 
weaker growth. It was also realised that in particular, WL data indicated 
consistently a $2-3\sigma$ tension with the Planck parameter values of 
$\Omega_{m0}-\sigma_8$ \cite{Ade:2015xua,Joudaki:2017zdt,DiValentino:2018gcu} 
(for updated constraints see also Fig. \ref{fig:S8const} adopted from Ref. 
\cite{DiValentino:2018gcu}). For example, the Kilo-Degree Survey (KiDs-450) 
\cite{Joudaki:2017zdt,Joudaki:2016kym} finds $S_8\equiv \sigma_8 
\sqrt{\Omega_{m0}/0.3}=0.74\pm 0.035$, which is smaller at a $2.6\sigma$ 
tension 
compared to the corresponding Planck best fit value  $S_8=0.832\pm 0.013$ 
\cite{Aghanim:2018eyx}. More recent WL cosmic shear data from the Dark Energy 
Survey \cite{Troxel:2017xyo,Abbott:2017wau} indicate $S_8=0.792\pm 0.024$, 
i.e., a 
weaker tension with geometric probes and Planck (about $1-2\sigma$), albeit in 
the same direction of weaker growth and lower  $\Omega_{m0}-\sigma_8$ (DES 
indicates that $\Omega_{m0}=0.264^{+0.032}_{-0.019}$ \cite{Abbott:2017wau} to 
be 
compared with Planck best fit $\Omega_{m0}=0.315\pm0.007$  
\cite{Aghanim:2018eyx}). Reduced value of $\sigma_8$ ($\sigma_8=0.77\pm 0.02$) 
is also indicated by high $l$ measurements ($l>2000$) of the E-mode angular 
auto-power spectrum (EE) and the temperature-E-mode cross-power spectrum (TE)  
taken with the SPTpol instrument \cite{Henning:2017nuy}.

The tension level between geometric and dynamical probes has recently been  
quantified by using specific statistics designed to probe the tension in a more 
efficient and quantitative manner 
\cite{Raveri:2015maa,Lin:2017ikq,Sagredo:2018rvc}.  These studies have verified 
the statistical significance of the tension between geometric and dynamical 
probes and demonstrated that even though the dynamical probes (RSD, WL and CC) 
are consistent with each other, pointing towards weaker growth than GR, they 
are 
in discordance with the geometric probes in the context of GR.

\subsection{Theoretical Implications}

The most generic approach to the ``weak growth'' tension is the modified 
gravity 
approach. If this tension is in fact due to a modification of GR on 
cosmological 
scales the following question arises: ``What observationally viable modified 
gravity models can reproduce a weaker gravity than that predicted by GR at low 
redshifts?'' A naive response to this question would indicate that any viable 
modified gravity model can lead to weaker gravity than GR at late times with 
proper choice of its parameters. However, it may be shown that this is not the 
case. Recent studies have addressed this question for $f(R)$ theories, for 
minimal scalar tensor theories  
\cite{Gannouji:2018ncm,Perivolaropoulos:2019vkb}, 
for Horndeski theories  \cite{Arjona:2019rfn,Linder:2018jil} and beyond 
Horndeski  Gleyzes-Langlois-Piazza-Vernizzi (GLPV) theories 
\cite{Tsujikawa:2015mga}. \label{beyondHorndeskiref4}

We will consider  $f(R)$ models with an action of the form
\be
S=\int d^4x \sqrt{-g} \, \frac{f(R)}{2}+ S_m,\label{PKeq:mogaction}
\ee
where from now on we set $8\pi G_{\textrm{N}}=1$. The predicted $\mu(z,k)$ is 
given as \cite{Tsujikawa:2007gd}
\be
\mu(z,k)=\left(\frac{d f}{dR} \right)^{-1}
\left[ \frac{1+4 \left(\frac{ d^2 f}{d R^2}/ \frac{ d f}{d R}\right) \cdot k^2 
\,
(1+z)^2}{1+3 \left(\frac{ d^2 f}{d R^2}/ \frac{ d f}{d R}\right) \cdot k^2 \, 
(1+z)^2} \right],
\label{PKeq:gefffr}
\ee
where in this case $\mu$ depends on both the redshift $z$ and the scale $k$. In 
addition, the stability conditions
\ba
\frac{ d^2 f}{d R^2}&>&0 \nn \\
\frac{ d f}{d R}&>&0
\ea
should be satisfied \cite{Starobinsky:2007hu}. Also in viable $f(R)$  models 
$\frac{d f}{d R} \simeq 1$ at early times deep in the matter era (high $R$) 
\cite{Amendola:2006we}. Thus, since $\frac{ d^2 f}{d R^2}>0$ we must have 
$\frac{ d f}{d R}<1$ at late times (low $R$). It follows that both factors of 
Eq. \eqref{PKeq:gefffr} are larger than unity and we have generically in $f(R)$ 
theories that $\mu(z)\geq 1$. This is a generic result independent of the 
background $H(z)$, indicating that $f(R)$ theories are unable to resolve the 
weak 
growth tension because they predict stronger gravity than GR.

A similar result is true for minimal scalar-tensor theories, provided that the 
expansion background is close to \lcdmnospace.
The minimal scalar-tensor action has the form 
\cite{Boisseau:2000pr}
\be
S= \int d^4 x \sqrt{-g} \left[ \frac{1}{2}F(\phi)R - \frac{1}{2} g^{\mu\nu} 
\partial_\mu \phi \partial_\nu \phi - U(\phi) \right] + S_m.
\label{PKeq:actionscalten}
\ee
The dynamical equations obtained by variation of this action in the context of 
a 
flat  FRW metric are of the form
\cite{EspositoFarese:2000ij,Boisseau:2000pr}
\begin{eqnarray}
3F H^2 &=&  \rho_m +{\frac{1}{2}} \dot\phi^2 - 3 H \dot F + U \label{PKeq:fe1}\\
-2F \dot H  &=& (\rho_m+p_m) + \dot \phi^2 +\ddot F - H \dot F, \label{PKeq:fe2}
\end{eqnarray}
where the dot represents differentiation with respect to cosmic time $t$. After 
rewriting the equations  of motion in terms of the redshift, defining the 
rescaled square Hubble parameter as $q(z)=\frac{H^2(z)}{H_0^2}$  and 
eliminating 
the scalar field potential $U(\phi)$, we obtain a differential equation that 
associates the coupling function $F(\phi)$ and the scalar field $\phi$ as
\be F^{\prime\prime}(z)+ \left[\frac{q^\prime(z)}{2q(z)}-\frac{2}{1+z}\right] 
F^{\prime}(z)   - \frac{1}{(1+z)}\frac{q^\prime(z)}{q(z)} F(z)+3 
\frac{1+z}{q(z)} \Omega_{m0}= -\phi^{\prime}(z)^2 \label{PKeq:fe1a},
\ee
where the prime stands for differentiation with respect to redshift $z$.
In scalar tensor theories $\mu$ is expressed as 
\cite{EspositoFarese:2000ij,Nesseris:2006jc}
\be
\mu(z)= \frac{1}{F(z)}\frac{F(z)+2F_{,\phi}^2}
{F(z)+\frac{3}{2}F_{,\phi}^2}.
\label{PK:eqgeffsclatens}
\ee
Using Eq. \eqref{PK:eqgeffsclatens} in the differential equation  
\eqref{PKeq:fe1a} and expanding around $z=0$, while using the Solar System 
constraint $\mu'(z=0)=0$ \cite{Gannouji:2006jm,Nesseris:2006hp}, we find in the 
context of a $wCDM$ background  \cite{Gannouji:2018ncm}
\be
\mu''(0)=9(1+w)(-1+\Omega_{m0})+\frac{9(1+w)^2 (-1+\Omega_{m0})^2}{\phi'(0)^2} 
+2\phi'(0)^2.   \label{PKeq:geffdp0sct}
\ee
For a \lcdm background, Eq. \eqref{PKeq:geffdp0sct} leads to the low $z$ 
expansion
\be
\mu(z) \approx  \mu(0)+ \frac{1}{2} \mu''(0)z^2=1+\phi^{\prime}(0)^2 \, 
z^2+\ldots, \label{eq:geffform}
\ee
which implies that $\mu(z)$ can only increase with redshift around $z=0$ in the 
context of a \lcdm backround.  In fact this result ($\mu''(0)>0$) is also 
applicable for $w<-1$, as   can be seen from Eq. \eqref{PKeq:geffdp0sct}, 
while 
for $w>-1$ it is possible to have $\mu''(0)<0$, as shown in Fig. 
\ref{fig:Geffdoubprime}.
\begin{figure}[ht!]
\centering
\includegraphics[width = 
0.68\textwidth]{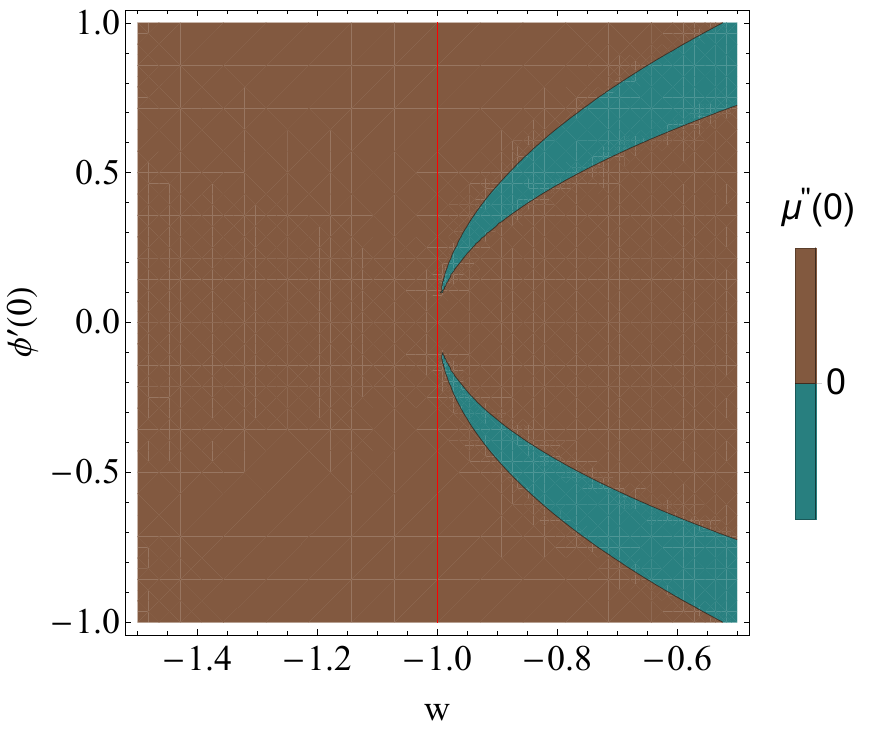}
\caption{{\it{The second derivative of $\mu(0)$ in the parametric space of 
$\phi^{\prime}(0)$ and $w$.  The brown region describes the parameter values 
for 
$\mu''(0)>0$, while the blue region describes the parameter values for 
$\mu''(0)<0$ (from Ref. \cite{Gannouji:2018ncm}).}}}
\label{fig:Geffdoubprime}
\end{figure}
Thus, the increasing nature of $\mu(z)$ in scalar 
tensor theories that respect 
the Solar System constraints has been demonstrated analytically in the context 
of a \lcdm background around $z=0$.

  This result is also demonstrated numerically  by using Eqs. 
\eqref{PKeq:fe1a}, \eqref{PK:eqgeffsclatens} in the context of the best fit 
parametrisation \eqref{PKeq:geffansatz} obtained from the RSD growth data 
\cite{Nesseris:2017vor} ($g_a<0$). Fig. \ref{fig:phizplot} shows the 
corresponding evolution of $\phi'(z)^2$, demonstrating that, as expected, for a 
decreasing $\mu(z)<1$ we obtain $\phi'(z)^2<0$ (ghost instabilities) at least 
close to $z=0$ when the Solar System constraints are respected (this does not 
include the $n=1$ case).  In the case of more general scalar-tensor theories 
(Horndeski and beyond Horndeski) it has been shown that weaker gravity may be 
possible, provided specific constraints among the terms of the Lagrangian are 
applicable \cite{Tsujikawa:2015mga,Linder:2018jil}.
\begin{figure}[ht!]
\centering
\includegraphics[width = 0.72\textwidth]{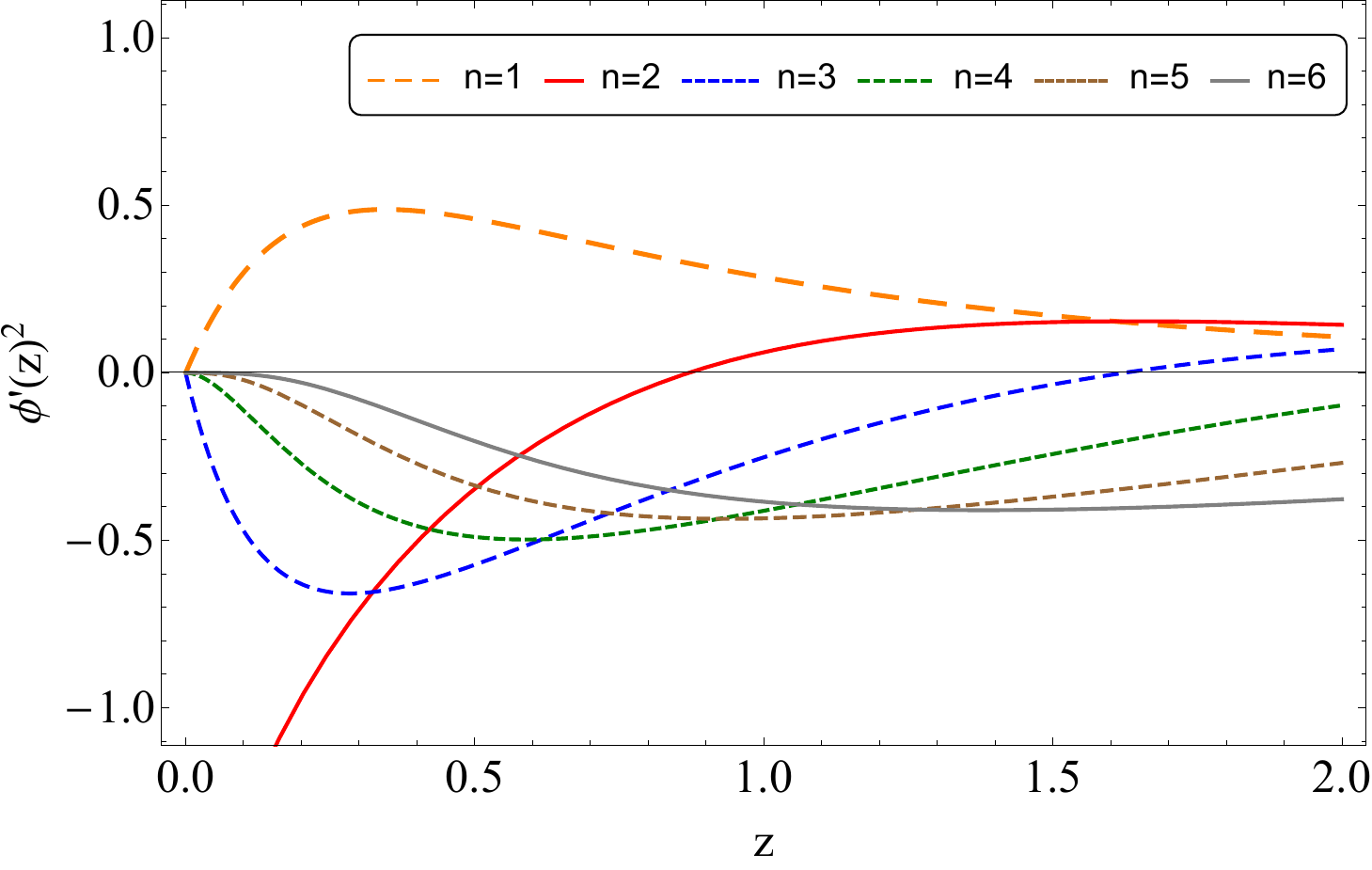}
\caption{{\it{Evolution of the scalar field $\phi$ as a function of redshift 
$z$ 
corresponding to the best fit values of $g_a$ and various values $n$ using the 
robust compilation of Ref. \cite{Nesseris:2017vor}. Notice that the $n=1$ case 
shows no ghost  instabilities ($\phi'(z\simeq 0)>0$) but it does not satisfy 
the 
Solar System constraint $\mu'(z=0)=0$) and thus
 Eq. (\ref{eq:geffform}) is not applicable for $n=1$.}}}
\label{fig:phizplot}
\end{figure}

We therefore conclude that from the theoretical point of view it is highly 
challenging  to construct a viable theoretical model that allows for weaker 
gravity than GR at low redshifts while at the same time it respects solar 
system 
and other observational constraints with an $H(z)$ background close to 
\lcdmnospace. This challenge, however, may prove a useful discriminating tool 
among modified gravity models if the weak growth tension persists and gets 
verified by future cosmological data.

The issue of weak growth tension is expected to be  clarified within the next 
decade 
due to a wide range of  upcoming surveys. The surveys include Euclid 
\cite{Laureijs:2011gra,Amendola:2016saw} (aiming at mapping the geometry of the 
Universe when dark energy leads to its accelerated expansion), Square Kilometer 
Array (SKA) \cite{Jarvis:2015tqa,Bacon:2015dqe}  (aiming at analysing 
radiosignals from various galactic sources), Large Synoptic Survey Telescope 
(LSST) \cite{Marshall:2017wph} (aiming at mapping and cataloguing galaxies, in 
order to study their impact on the distortion of spacetime), Cosmic Origins 
Explorer (COrE) (aiming at mapping the polarisation of the CMB) 
\cite{Bouchet:2015arn},  Dark Energy Spectroscopic Instrument (DESI) 
\cite{Aghamousa:2016zmz,Aghamousa:2016sne} (aiming at studying the effects of 
dark energy and obtaining the optical spectra of galaxies and quasars)  and 
Wide 
Field Infrared Survey Telescope  (WFIRST) 
\cite{Spergel:2015sza,Hounsell:2017ejq} (aiming at answering key questions in 
cosmology, probing BAO, WL and Supernovae data simultaneously). These surveys 
are expected to provide new and  more detailed measurements of the dark energy 
probes 
BAO, SnIa, RSD, WL and CC extending to both dynamical and geometrical probes. 
They are expected to either confirm or eliminate the weak growth tension. In 
the 
first case, they will also provide a concrete discriminator among the modified 
gravity models and non-gravitational models that constitute candidate 
extensions 
of the standard \lcdm model and are motivated by the weak growth tension.

\section[Evolving \texorpdfstring{$G_{eff}$}{Geff} and the Pantheon SNeIa 
Dataset]{Evolving  \texorpdfstring{$G_{eff}$}{Geff} and the Pantheon SNeIa  
Dataset}
\label{sec:Pantheon}

If the effective Newton's constant $\mu=\Geff/G_{\textrm{N}}$ is indeed   
\label{Geffref4}
evolving 
with redshift  on cosmological timescales it is expected to lead to an 
evolution 
of the absolute luminosity and absolute magnitude of SnIa. In this section we 
present preliminary work searching for such evolution of the  SnIa absolute 
magnitude with redshift. We use the Pantheon SnIa dataset 
\cite{Scolnic:2017caz}, which is the latest compilation of SnIa. It consists of 
1048 data points with redshifts spanning the region $z \in 
\left[0.01,2.3\right]$.  
This dataset is a combined set of the PS1 SnIa dataset \cite{Scolnic:2013efb}, 
which consists of 279 SnIa with redshifts spanned in the region $z \in 
\left[0.03,0.68\right]$ along with
probes of low redshifts ($z \in \left[0.01,0.1\right]$), including the  
CfA1-CfA4 \cite{Riess:1998dv,Jha:2005jg,Hicken:2009df,Hicken:2012zr} and CSP 
surveys \cite{Contreras:2009nt,Stritzinger:2011qd}, as well as high redshifts 
($z >0.1$), probed by SDSS  \cite{Sako:2014qmj,Kessler:2009ys}, SNLS 
\cite{Sullivan:2011kv,Conley:2011ku} and HST surveys 
\cite{Riess:2006fw,Suzuki:2011hu}.

The measured apparent magnitude $m$ for SnIa data is connected to cosmological 
parameters through the relation
\be
m_{th}(z)=M+5 \, \log_{10} \left[\frac{d_L(z)}{Mpc} \right]+25 ,
\label{PKeq:appmagn}
\ee
where $d_L(z)$ is the  luminosity distances and $M$ is the absolute magnitude. 
The luminosity  distance for a flat FLRW metric, is given by
\be
d_L=c(1+z) \int_0^{z} \frac{dz'}{H(z')}.
\ee

The Pantheon dataset provides the apparent magnitude $m_{obs}(z_i)$ after 
corrections over  the stretch, colour and possible biases from simulations 
\cite{Scolnic:2017caz}. Following the usual method of maximum likelihood  
\cite{Arjona:2018jhh}  we can obtain the best fit parameters, minimising the 
quantity
\be
\chi^2 (M,\Omega_{m0},w,h)=V^i_{Panth.} C_{ij}^{-1} V^j_{Panth.},
\ee
where $V^i_{Panth.}\equiv m_{obs}(z_i)-m_{th}(z)$, $C_{ij}$ provided in 
\cite{Scolnic:2017caz}, is the  covariance matrix and $h$ is the dimensionless 
parameter of the Hubble constant, which is defined as $h \equiv H_0/100 \, 
(km/s)/Mpc$.

Usually, the absolute magnitude $M$ is considered a nuisance parameter and is 
marginali-\\sed along  with $h$, due to a clear degeneracy between the two 
parameters. However, in the context of modified gravity with an evolving 
Newton's constant the absolute magnitude is expected to evolve with redshift in 
accordance with Eq. \eqref{PKeq:mgeffconn} and may contain useful information 
on 
fundamental physics. In an effort to identify such evolution we minimise 
$\chi^2$ with respect to the parameter $M$ with fixed background corresponding 
to the best fit \lcdm $H(z)$ as obtained from the full Pantheon dataset with 
$M$ 
marginalization (we fix $w=-1$ and $\Omega_{m0}=0.28$ \cite{Scolnic:2013efb} 
and 
set $h=1$ for simplicity). In this context, we identify the best fit value and 
$1\sigma$ error of $M$ for various subsets of the full Pantheon dataset.
In Fig. \ref{fig:erroplotMcut} we show the best fit absolute magnitude $M$  
using subsamples of the Pantheon  dataset in the redshift range $z \in 
\left[0.01, z_{max}\right]$. The $1 \sigma$ range for the $M$ parameter for 
various cutoffs $z_{max}$ is also shown.
\begin{figure}[ht!]
\centering
\includegraphics[width = 
0.72\textwidth]{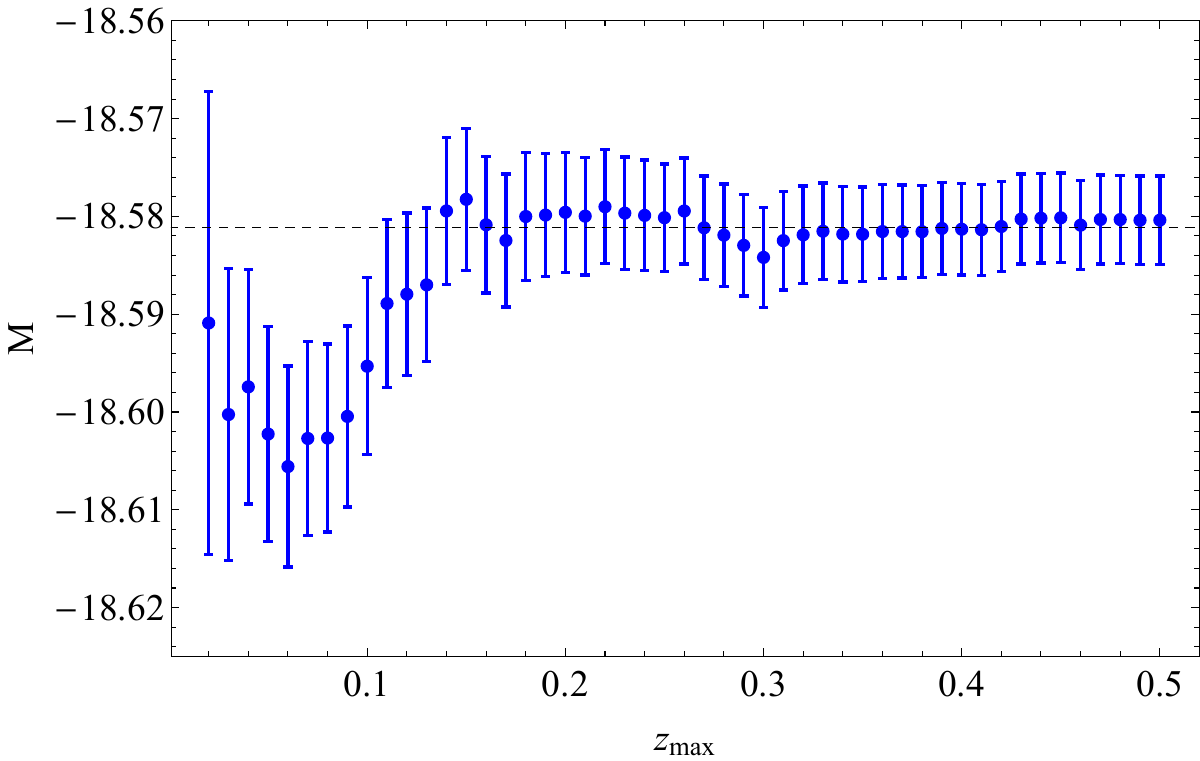}
\caption{{\it{Evolution of the absolute magnitude $M$ as a function of the 
cutoff 
$z_{max.}$. We have set $h=1$  and thus the value of $M$ is shifted compared to 
its usual value of $M=-19.3$.}}}
\label{fig:erroplotMcut}
\end{figure}

It is clear from Fig. \ref{fig:erroplotMcut} that low resdhift data in the 
redshift range $z \in \left[0.01, 0.1\right]$ seem to favour a value $M$ 
smaller 
than its best fit asymptotic value based on the full dataset ($z_{max}=2.3$) at 
a level of about $2\sigma$. At redshifts $z>0.2$, $M$ approaches its asymptotic 
value (dashed line). Our results are consistent with the analysis of Ref. 
\cite{Colgain:2019pck}, where the best fit parameters of $\Omega_{m0}$ and 
$H_0$ 
were investigated as a function of the redshift cutoff $z_{max}$. In agreement 
with our results it was found that the low  $z$ Pantheon data appear to have 
interesting features, which may indicate the presence of either systematics or 
new physics.

In Fig. \ref{fig:mu} (left panel) we show the 100-point moving best fit value 
of 
$M$ along with its  $1\sigma$ errors. To construct this plot we rank the 
Pantheon datapoints from lowest to highest redshift. We start  with the first 
100 datapoints (lowest redshift points 1 to 100) and use them to obtain the 
best 
fit value of $M$ (assuming the fixed \lcdm background) with its $1\sigma$ 
error. 
The corresponding $z$ coordinate of this point is the mean redshift of the 
first 
100 points. The $i^{th}$ point is obtained by repeating the above procedure for 
the datapoints from $i$ to $i+100$.
\begin{figure}[ht!]
\centering
\includegraphics[width = 
0.99\textwidth]{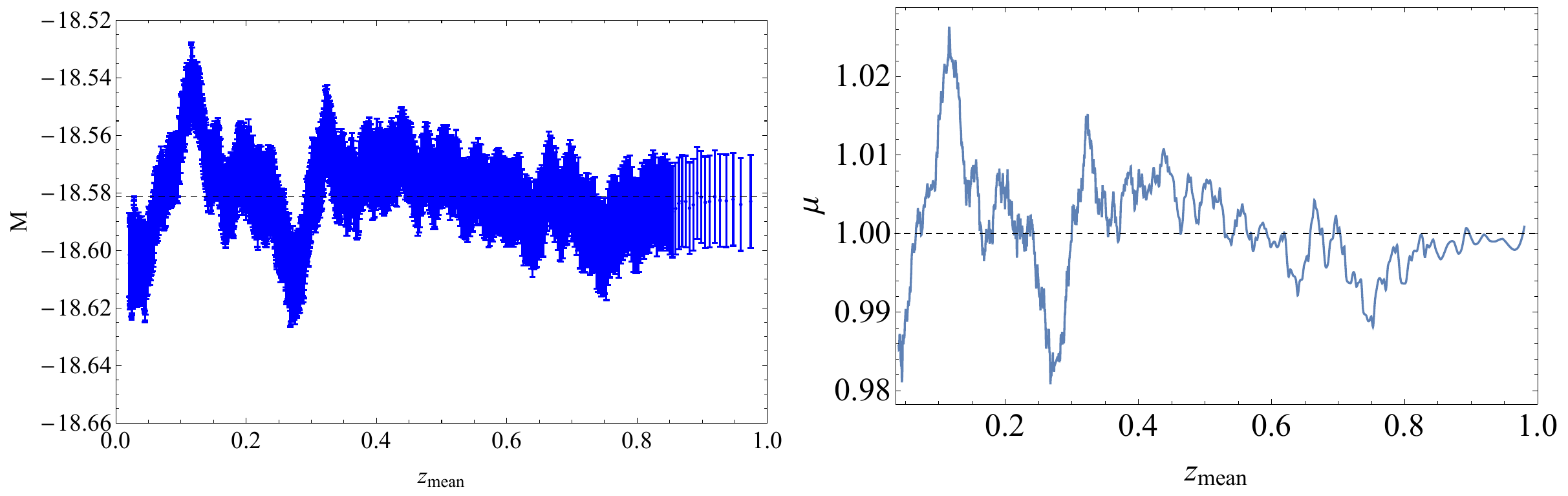}
\caption{{\it{ Left Panel: Variation of the absolute magnitude $M$ as a 
function of 
$z_{mean}$ for 100 point  Pantheon subsamples.   Right Panel: The 
corresponding 
variation of $\mu$ as a function of $z_{mean}$ for 100 datapoint Pantheon 
subsamples.}}}
\label{fig:mu}
\end{figure}

Using the best fit value of $M$ obtained from each 100-point subsample, we can 
calculate $\mu$ using Eq. \eqref{PKeq:mgeffconn} (right panel of Fig. 
\ref{fig:mu}) setting $M_0$ equal to the best fit value of $M$ obtained from 
the 
full Pantheon dataset. Clearly, an oscillating effect is evident for the 
absolute magnitude M at low $z$, that is eased at high redshifts. The same 
oscillating effect was also observed in Refs. 
\cite{Kazantzidis:2020xta,Sapone:2020wwz,Kazantzidis:2020tko}, where a 
binning method was used instead.

The absolute magnitude $M$ is degenerate with $h$. Therefore, the value of $h$ 
obtained under the assumption of a constant M would not be the same as the 
value 
of $h$ that would be obtained if M was allowed to evolve. In particular, using 
the ``Hubble constant free'' luminosity distance, that is defined as 
\cite{Nesseris:2005ur}
\be
D_L(z)=\frac{H_0 \, d_L(z)}{c},
\ee
we can rewrite  Eq. \eqref{PKeq:appmagn} as \cite{Nesseris:2005ur}
\be
m_{th}(z)=M+5 \, \log_{10} \left(D_L (z)\right)+5 \, \log_{10} 
\left(\frac{c/H_0}{1 Mpc}  \right)+25. \label{PKeq:appmagndL}
\ee
In terms of $h$ and taking into account the possible evolution of 
$\mu(z)$\footnote{The $\mu$ here is the evolving normalized  Newton's constant 
and should not be confused with the distance modulus.}, Eq. 
\eqref{PKeq:appmagndL} takes the form
\be
m_{th}(z)=M_0+\frac{15}{4} \, \log_{10} \left(\mu (z) \right)+5  \, \log_{10} 
\left(D_L (z)\right)-5 \, \log_{10} \left(h\right)+42.38  .
\label{PKeq:appmagndLh}
\ee
Using Eq. \eqref{PKeq:appmagndLh} it is easy to show that a change  of $\mu$ by 
a small amount $\Delta \mu$ around $\mu=1$ is equivalent to a small change of 
$h$ by
\be
\Delta h= -\frac{3}{4}\, h\, \Delta\mu.
\label{deltah}
\ee
Thus, a decrease of $\mu$ at low $z$ ($\Delta\mu<0$) is equivalent to an 
increase of $h$ by $\Delta h$ compared to the true value of $h$. The value of 
$\Delta \mu \simeq -2\times 10^{-2}$ indicated in Fig. \ref{fig:mu} could be 
interpreted as   a shift of $h$ by about $1.5\%$ if $\mu$ was assumed fixed to 
1. This artificial increase of $h$ is in the right direction but does not 
appear 
to be enough to explain a tension of about $8\%$ between the value indicated by 
the CMB and the value indicated by the SnIa sample. We stress, however, that 
the 
above analysis is heuristic and a more detailed analysis is required to include 
the possible effects of a varying $\mu$ in the derivation of $H_0$ from SnIa. 
In 
particular, the effects on Cepheid period-luminosity  relation, used in the 
determination of $H_0$ have not been taken into account, the effects of strong 
gravity in the interior of the progenitor stars have been ignored and the 
background cosmology has been assumed fixed to \lcdmnospace. These effects 
should be taken into account in a more complete and detailed analysis.

\section{Constraints on Evolving \texorpdfstring{$G_{eff}$}{Geff} from Low 
\texorpdfstring{$l$}{l} CMB  Spectrum and the ISW Effect.}\label{CMBrefs10}
\label{sec:ISW} \label{ISWref4}

As stated in the previous sections an evolving Newton's constant $\mu(z)$ would 
help resolve the weak growth and the $H_0$ tensions. However, such an evolution 
would also affect \cite{Giannantonio:2009gi} other dynamical probes, and in 
particular the low $l$ (large scale) CMB angular power spectrum through the 
Integrated-Sachs Wolfe (ISW) effect created as the CMB photons travel through  
time varying gravitational potential that would be modified by the evolving 
$\mu(z)$. Any such  modification is constrained by the Planck data. The 
questions that we address in this section are the following:
\begin{itemize}
\item
What are the constraints imposed by the Planck CMB TT power spectrum data on 
the 
 parameter $g_a$ of the parametrisation \eqref{PKeq:geffansatz}, assuming a 
fixed 
slip parameter to its GR value $\eta=1$?
\item
Are these constraints consistent with the value of $g_a$ required to resolve 
the 
weak growth tension?
\end{itemize}
In order to address these questions  we use the 2019 version 
\cite{Zucca:2019xhg} of MGCAMB \cite{Hojjati:2011ix,Zhao:2008bn}, which  is a 
modified version of the CAMB code \cite{Lewis:1999bs}, that it is designed to 
produce the CMB spectrum in the context of modified gravity theories with a 
given background model $H(z)$ and a given scale dependent evolution of $\mu$ 
and 
$\eta$. We fix $H(z)$ to Planck15/$\Lambda$CDM, since the Planck18 likelihood 
chains that are implemented in COSMOMC and MGCOSMOMC are not yet publicly 
available, $\eta=1$ and for $\mu(z)$ we use the parametrisation  
\eqref{PKeq:geffansatz}. The values of the parameters for the 
Planck15/$\Lambda$CDM model are shown in Table  \ref{PKtab:plcdm15}.
\label{multipolesref3}
 
\begin{table}[ht!]
\centering
\begin{tabular}{cc}
 \hline
 \rule{0pt}{3ex}
  Parameter & Planck15/$\Lambda$CDM \cite{Ade:2015xua}\\
    \hline
    \rule{0pt}{3ex}
$\Omega_b h^2$ & $0.02225\pm0.00016$ \\
$\Omega_c h^2$ & $0.1198\pm0.0015$  \\
$n_s$ & $0.9645\pm0.0049$\\
$H_0$ & $67.27\pm0.66$ \\
$\Omega_{m0}$ & $0.3156\pm0.0091$ \\
$w$ & $-1$  \\
$\sigma_8$ & $0.831\pm0.013$ \\
\hline
\end{tabular}
\caption{Planck15/$\Lambda$CDM  parameters values from Ref. \cite{Ade:2015xua} 
based on TT,TE,EE and lowP likelihoods. Notice that $\sigma_8$ is larger for 
the 
$2015$ data release, which implies a stronger $\sigma_8$ tension that the 
Planck18/$\Lambda$CDM best fit model.}
\label{PKtab:plcdm15}
\end{table} 
 The predicted form of the CMB angular power spectrum for various 
values of $g_a$ is shown in  Fig. \ref{fig:mgcamb} along with the corresponding 
Planck datapoints.
\begin{figure}[ht!]
\centering
\includegraphics[width = 
0.83\textwidth]{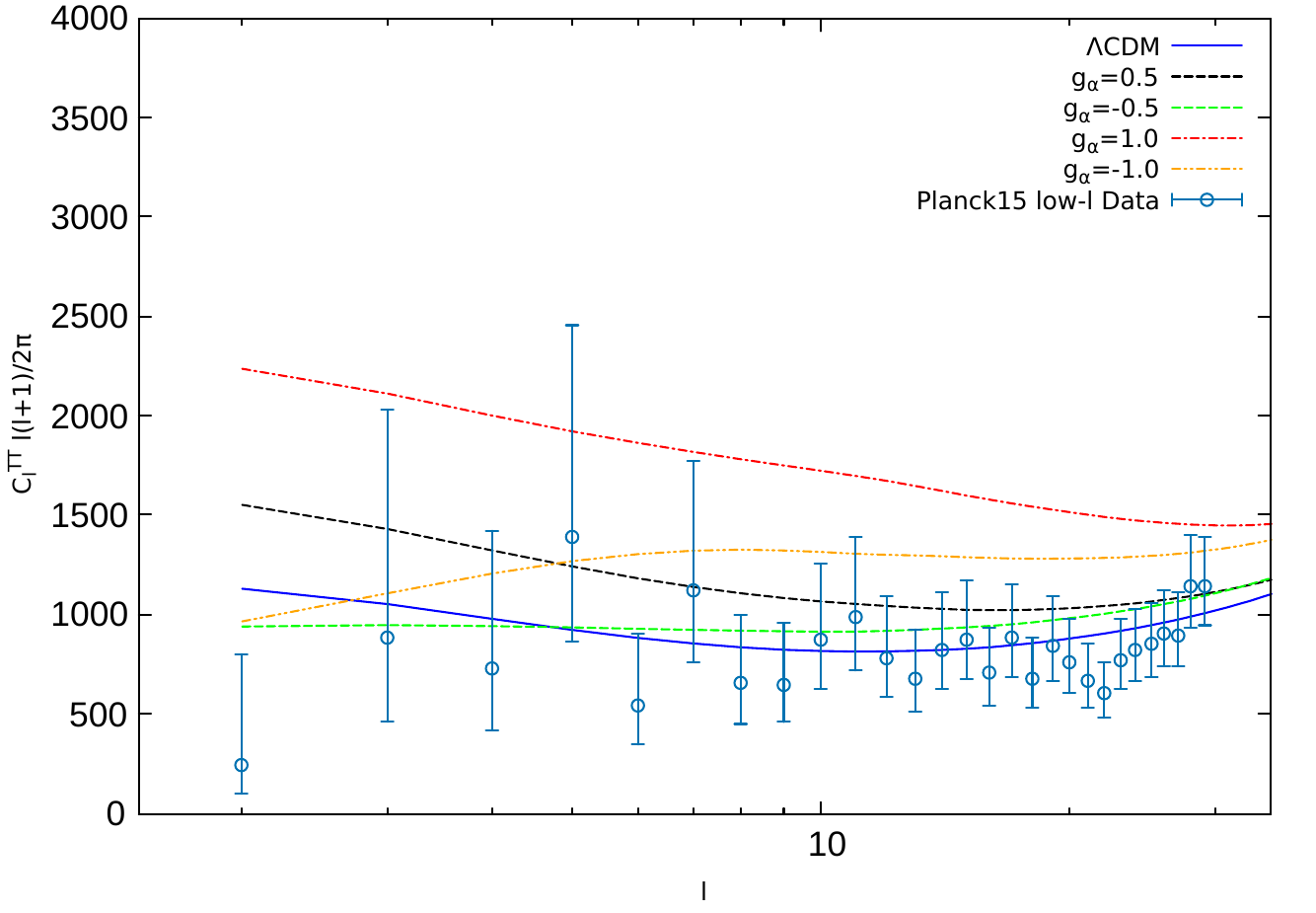}
\caption{{\it{The theoretically predicted form of the CMB power spectrum for a 
Planck15/$\Lambda$CDM background  in the context of a varying $\mu$ cosmology 
described by  Eq. \eqref{PKeq:geffansatz} for various values of $g_a$ (obtained 
using MGCAMB).}}}
\label{fig:mgcamb}
\end{figure}

As we can see, the low-l Planck data do not allow significant variations in the 
parameter $g_a$ and imply strong constraints on it. These constraints can be 
made precise using MGCOSMOMC \cite{Hojjati:2011ix,Zhao:2008bn}, the 2019 
modified version \cite{Zucca:2019xhg} of the COSMOMC code \cite{Lewis:2002ah}. 
Allowing variation of the parameters $\left(\Omega_{m0}, \sigma_8, g_a\right)$ 
while fixing the rest to their Planck15/$\Lambda$CDM values, we obtain the 
parameter contour constraints shown in Fig. \ref{fig:mgcosmomccontplc}.
\begin{figure}[ht!]
\centering
\hspace{-1cm}
\includegraphics[width = 
1.04\textwidth]{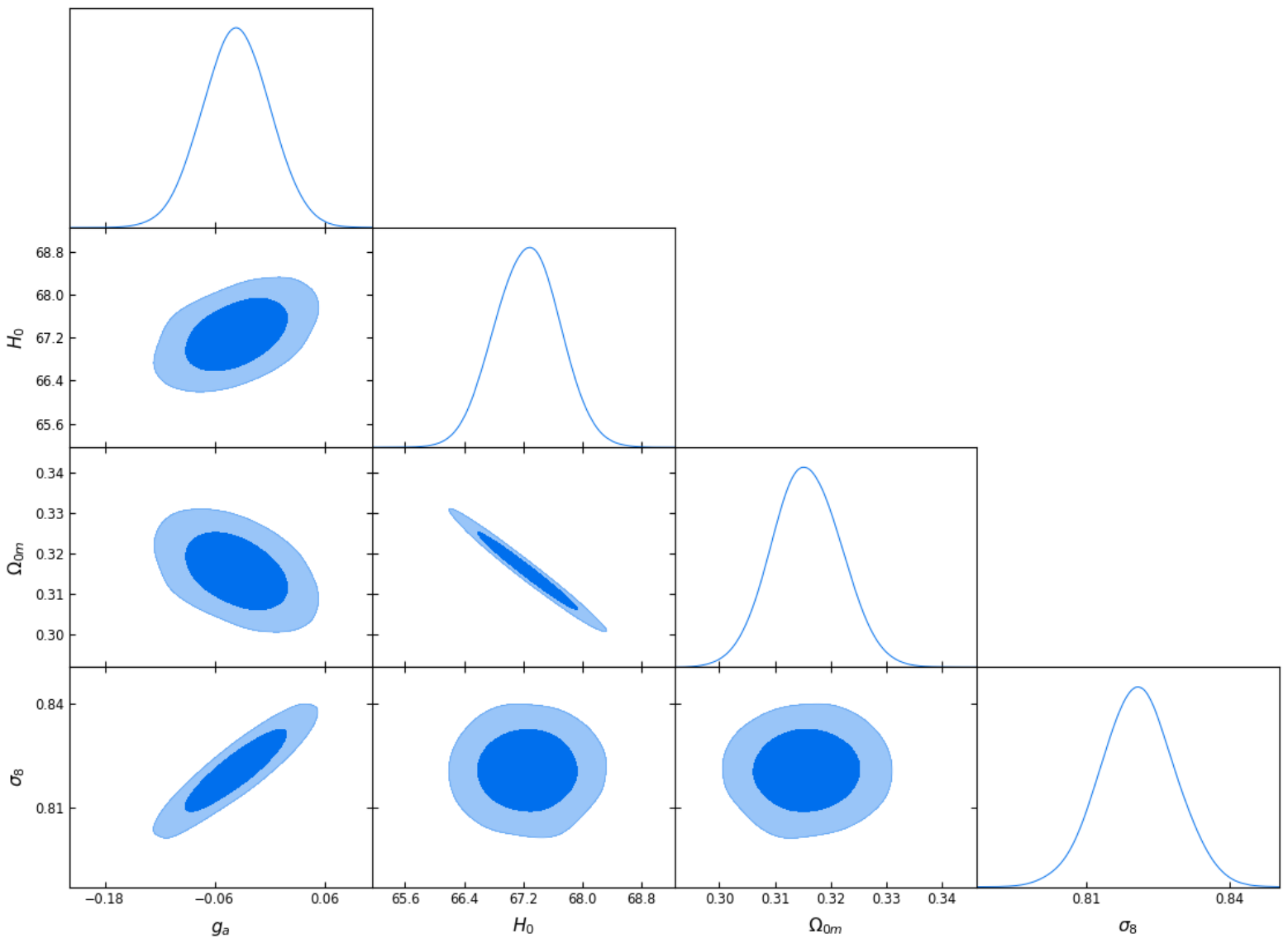}
\caption{{\it{The $1\sigma-2\sigma$ contour ranges of cosmological parameters 
in the context of the parametrisation  \eqref{PKeq:geffansatz}, using the 
Planck15/$\Lambda$CDM data and setting $n=2$.}}}
\label{fig:mgcosmomccontplc}
\end{figure}

  Clearly, even though negative values of $g_a$ are mildly favoured and 
are consistent  with the small $\mu$ variation implied by the Pantheon SnIa 
data 
of the previous section, this parameter is constrained to be larger than $-0.1$ 
at a $3\sigma$ level. This range is barely overlapping with the $2\sigma$ range 
of $g_a$ indicated by the compilation of the RSD growth data shown in Fig. 
\ref{fig:mtrsdcontour}. Thus, as pointed out also in previous studies 
\cite{Nesseris:2017vor}, the low $l$ CMB spectrum strongly  constrains   
the 
evolution of $\mu$ in the context of the parametrisation 
\eqref{PKeq:geffansatz} 
and implies that additional parameters and/or systematic effects are required 
for the resolution of the weak growth tension (e.g., the extension of the 
\lcdm 
$H(z)$ to $wCDM$ or the introduction of a sterile massive neutrino).
\begin{figure}[ht!]
\centering
\includegraphics[width = 
0.99\textwidth]{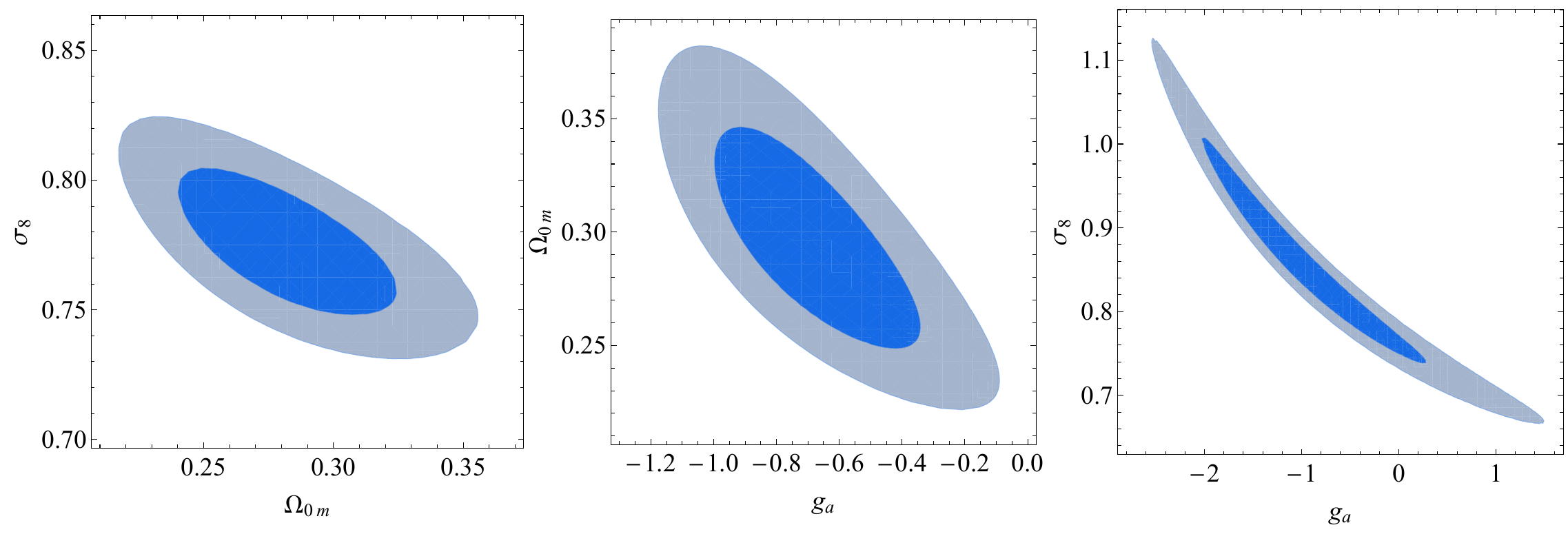}
\caption{
{\it{
The $1\sigma-2\sigma$ parameter constraints in the context of an 
evolving $\mu$ described by  the parametrisation  \eqref{PKeq:geffansatz} with 
$n=2$. The full RSD data compilation of Ref. \cite{Kazantzidis:2018rnb} was 
used. The third parameter in each plot was fixed to the corresponding 
Planck15/$\Lambda$CDM. Notice the strong indication for weaker gravity at low 
$z$ whose magnitude is marginally consistent with  the corresponding indication 
from 
CMB data (Fig. \ref{fig:mgcosmomccontplc}).}}}
\label{fig:mtrsdcontour}
\end{figure}

\section{Conclusions}
\label{sec:Conclusions}

Assuming a Planck/$\Lambda$CDM background expansion $H(z)$ and fixing the slip 
parameter $\eta$ to unity,  we have investigated the constraints on a possible 
evolution of Newton's constant expressed through the parameter $\mu$ using 
three 
observational probes: large RSD data compilations, the Pantheon SnIa distance 
indicators and the TT CMB angular power spectrum from the Planck mission.  We 
have shown that all three probes mildly favour a Newton's constant that is 
weaker 
at low $z$ compared to GR. For RSD data this trend is at the $2-3\sigma$ level 
at $z<0.3$, for SnIa it is at about $2\sigma$ at $z<0.1$ and for the CMB it is 
at less than $1\sigma$. In the case of RSD and CMB data we have assumed a 
specific parametrisation that respects the Solar System and nucleosynthesis 
constraints while reducing to GR at $z=0$ and at high $z$. The magnitude of 
suggested and allowed variation of $\mu$ is much smaller for the SnIa and CMB 
data ($1-2\%$) compared to the corresponding magnitude suggested by the RSD 
data 
(about $50\%$). This inconsistency suggests that a  variation of Newton's 
constant in the context of a modified gravity scenario for the parametrisation 
and the background considered may not by itself be able to explain the weak 
growth tension indicated by dynamical observational probes.

The simultaneous mild indication for weaker gravity at low $z$ by  independent 
probes suggests the more  careful investigation of the scenario of an evolving 
Newton's constant in the context of different $\mu$ and $\eta$ 
parametrisations, 
different $H(z)$ backgrounds and further dynamical observational probes. Such 
probes may include dynamical probes such as updated RSD data, WL and CC, as well 
as geometrical probes including CMB spectrum peaks, BAO and updated SnIa 
datasets.

The difficulty of viable modified gravity theories ($f(R)$ and scalar-tensor, 
Horndeski and beyond Hornseski)  to provide a weaker gravity at low redshifts is 
an interesting point that may be used as a powerful discriminator among modified 
gravity theories.





\chapter[Testing Gravity with Standard Sirens: Challenges and Opportunities]
{Testing Gravity with Standard \\ Sirens: Challenges and Opportunities}
\label{sec:Ezquiaga}

{\em Jose Mar\'ia Ezquiaga}




\section{Gravitational Wave Propagation Beyond General Relativity}
\label{gravitationalwavrefs3}

Gravitational wave (GW) astronomy presents promising opportunities to test 
gravity 
at new  scales and regimes. Theories beyond General Relativity (GR) generically 
modify the emission, propagation and detection of GWs with respect to Einstein's 
gravity \cite{Ezquiaga:2018btd}. These modifications can be imprinted either in 
the phase, amplitude or polarisation of the GW.
Among the possible probes, we will focus on tests of the propagation of GWs 
because  they are clean (compared to effects in the emission) and precise 
(since 
small modification accumulate over large travel distances).
For the interested reader in modifications of the strong gravity regime close 
to 
the merger  of a compact binary, we refer to the recent roadmap 
\cite{Barack:2018yly}.

The propagation of GWs in gravity theories beyond GR depends on the number of 
additional fields and  the background in which they propagate. We will restrict 
to cosmological environments respecting the symmetries of 
Friedmann-Robertson-Walker (FRW) metrics. Conveniently, at leading order in 
perturbations over FRW backgrounds, the tensor modes coupled to matter $h_{ij}$ 
only interact with other tensor perturbations $t_{ij}$. Assuming first that 
there are only the transverse-traceless polarisations $h_{ij}$, we can 
parametrise the modified cosmological GW evolution by
 \begin{equation}
 \label{eq:ModProp_JME}
 h''_{ij}+(2+\nu)\mathcal{H} h'_{ij}+(c_g^2k^2+m_g^2a^2)h_{ij}=0\,.
 \end{equation}
GR corresponds to the limit in which $\nu=m_g=0$ and $c_g=c$, so that GWs 
propagate at the  speed of light and they are only affected by the Hubble 
friction (we are timing in conformal time $\eta$).
However, the additional friction $\nu$, the anomalous speed $c_g$ and the 
effective mass $m_g$  generically alter the GR prediction 
$h^{_{\mathrm{GR}}}_{ij}$ by
 \begin{equation} \label{eq:schematic_mod_wave-form_JME}
 h_{ij}\sim h_{ij}^{_{\mathrm{GR}}}\,\underbrace{e^{-\frac{1}{2}\int   
\nu\mathcal{H}d\eta}}_\text{Affects 
amplitude}\,\underbrace{e^{ik\int(c_g^2-1+a^2m_g^2/k^2)^{1/2}d\eta}}_\text{
Affects phase}\,.
 \end{equation}
Mainly, the additional friction will modify the amplitude, which affects the 
inferred  GW luminosity distance $d^\text{gw}_{L}$, while the anomalous speed 
and the effective mass change the phase.
In analogy with optics, these effects could be interpreted as arising from a  
dia-gravitational medium for GWs \cite{Cembranos:2018lcs}.
Note that we are assuming that there is not a modification in the emission in 
order to take the GR signal as an initial condition. For a compact binary 
coalescence, the wave-form $h^{_{\mathrm{GR}}}$ depends at leading order on the 
chirp mass $\mathcal{M}_c$, the frequency $f$, the redshift $z$, the cosmic 
expansion history $H(z)$, the polarisation $+,\,\times$ and the inclination 
angle $\iota$ by~\cite{maggiore2008gravitational}\footnote{The signal at the 
detector $h(t)$ also depends  on the orientation (defined by two angles 
$\theta$ 
and $\phi$) when tacking into account the antenna pattern functions 
$F_{+,\times}$, i.e. 
$h(t)=h_+(t)F_+(\theta,\phi)+h_\times(t)F_\times(\theta,\phi)$.}
\begin{equation} \label{eq:amplitude_polarisations_JME}
h_+^{_{\mathrm{GR}}}=\vert 
h^{_{\mathrm{GR}}}\vert\,\frac{1+\cos^2\iota}{2}\,\cos\Phi\quad  \text{and} 
\quad h_\times^{_{\mathrm{GR}}}=\vert h^{_{\mathrm{GR}}}\vert\, 
\cos\iota\,\sin\Phi\,,
\end{equation}
where $\Phi$ is the phase that depends on the time to coalescence and the 
chirp 
mass.  The chirp amplitude $\vert h^{_{\mathrm{GR}}}\vert$ is given by
\begin{equation} \label{eq:chirp_amplitude_JME}
\vert h^{_{\mathrm{GR}}}\vert=\frac{4}{d_L^{_{\mathrm{GR}}}} 
\left(\frac{G\,\mathcal{M}_z}{c^2}\right)^{5/3}\left(\frac{\pi\,f}{c}\right)^{
2/3}\,,
\end{equation}
where $\mathcal{M}_z$ is the redshifted chirp mass, 
$\mathcal{M}_z=(1+z)\mathcal{M}_c$,  and the GW luminosity distance in GR is 
equal to the standard electromagnetic (EM) one, which is determined by the 
Hubble parameter,
\begin{equation} \label{eq:dl_GR_JME}
d_{L}^{_{\mathrm{GR}}}=d^\text{em}_{L}=(1+z)\int_0^z\frac{c}{H(z)}dz\,.
\end{equation}
Therefore, within GR, a measurement of $d^\text{gw}_{L}$ directly  contains    
information of the cosmic expansion history.

Whenever there are additional tensor modes $t_{ij}$ the phenomenology enriches 
because now there could be mixings with the tensor modes coupled to matter 
$h_{ij}$. In analogy with neutrino physics, this would induce GW oscillations 
along the propagation. For instance, in the case in which there are 
interactions 
in the friction and in the mass terms, the propagation equations can be 
parametrised by \cite{Jimenez:2019lrk}
\begin{eqnarray}
\label{eq:generalequation_JME}
\left[ \frac{d^2}{d\eta^2} + 2\begin{pmatrix} \mathcal{H} & -\alpha \\  \alpha 
& 
\mathcal{H}+2\Delta\nu  \end{pmatrix}\frac{d}{d\eta} +c^2 k^2   + 
m_g^2a^2\begin{pmatrix} 1 & -1 \\ -\tan^{-2}\theta_g &  
\tan^{-2}\theta_g\end{pmatrix}\right] \begin{pmatrix} h_{ij} \\ t_{ij} 
\end{pmatrix} =0\,,  
\end{eqnarray}
which is similar to \eqref{eq:ModProp_JME} but promoting $\nu$ and $m_g$ to 
matrices. Nonetheless, there  could  also  be  mixings in the velocities or 
  the $+$ and $\times$ polarisations could be affected differently,  which we 
are not 
covering here (see \cite{Jimenez:2019lrk}). It is important to note that the 
additional tensor modes could arise from extra dynamical metric fields such as 
in bigravity \cite{Narikawa:2014fua,Max:2017flc}, but also from   effective 
configurations of vector fields with an internal symmetry 
\cite{Caldwell:2016sut,BeltranJimenez:2018ymu}.

In broad terms, modifications of the cosmological propagation of GWs can induce 
frequency and redshift dependent  deviations w.r.t. GR. Frequency-dependent 
phenomena are, 
for instance, the modified dispersion relation of an effective mass 
\cite{Will:1997bb} or the modulation of the wave-form when there is a mass 
mixing \cite{Narikawa:2014fua,Max:2017flc}. Conversely, redshift-dependent 
phenomena  are 
the time delay produced by an anomalous speed or the modified luminosity 
distance. Frequency-dependent modifications can be constrained with GWs alone, 
while redshift-dependent ones require an independent determination of $z$ to 
break the degeneracy with the theory parameters. In this contribution we will 
concentrate on redshift-dependent modifications. We will first present how GWs 
can become standard sirens in Section \ref{sec:standard_sirens_JME}. Then, we 
will discuss how to constrain gravity theories with the GW speed, the 
luminosity 
distance and GW oscillations in sections \ref{sec:gw_speed_JME}, 
\ref{sec:gw_dl_JME} and \ref{sec:gw_oscillations_JME} respectively. We will 
conclude with the future prospects in Section \ref{sec:future_JME}.

\section{Standard Sirens}
\label{sec:standard_sirens_JME}
\label{earlyDEefs4}

A GW becomes a standard siren when it probes the cosmological evolution of the  
luminosity distance \cite{Schutz:1986gp,Holz:2005df}. From the amplitude of the 
GW of a compact binary (see 
\ref{eq:amplitude_polarisations_JME}-\ref{eq:chirp_amplitude_JME}), it is 
clear 
that $d^\text{gw}_{L}$ is degenerate with the inclination angle. Moreover, 
since 
in the detector frame   only   redshifted masses  are measured, the 
cosmological 
evolution of $d^\text{gw}_{L}$ is degenerate with the redshift of the source, 
which is unknown a priori. The distance-inclination degeneracy can in principle 
be broken if the $+$ and $\times$ polarisations can be distinguished. This 
requires at least a three detector network and a good sky localisation.
Regarding the redshift identification, there are two main methods:
\begin{itemize}
\item \emph{EM counterpart:} the first method entails an EM counterpart of the 
GW.  The redshift is then obtained from the EM signal itself or from the host 
galaxy. Binary neutron stars (BNS) produce a short gamma-ray burst (sGRB) after 
the merger. sGRB have a beaming angle $\theta_j$ typically expected to be 
$\theta_j\leq30^\circ$. This implies that, given the random orientation of the 
sources, both GW and EM signals will only be detected in a small fraction of 
events. Observing a bright afterglow or kilonovae \cite{Metzger:2016pju} could 
increase the chances of detecting a counterpart.
Of course, the key parameter for an efficient search is the localisation in the 
sky provided  by the GW signal itself. This improves with higher signal-to-noise
ratios (SNR) and with a network of ground-based detectors.
BNS will be the primary source of standard sirens for LIGO-Virgo  
\label{eLIGOfref1}
\cite{Dalal:2006qt},  although black-hole neutron star mergers may be relevant 
too \cite{Vitale:2018wlg}. Super massive black-holes (SMBHs) are thought to be 
good standard sirens for LISA \cite{Holz:2005df}.

\item \emph{Statistical method:}
alternatively, we could associate every GW event with all the galaxies within 
the error  box in the localisation and infer the cosmology 
\cite{Schutz:1986gp,DelPozzo:2011yh}. As more events accumulate, the true 
cosmology will statistically prevail. This method applies to all types of 
sources, including binary black-holes (BBH) which are the most common events 
and 
can be observed at further distances (see left panel of Fig. 
\ref{fig:distance_loc_JME}). BBH standard sirens are also known as dark sirens. 
For very loud events, there might be only few galaxies in the localisation 
box~\cite{Chen:2016tys}. On the con side, this method relies on a complete 
galaxy catalogue.
\end{itemize}
When the binary involves neutron stars, \label{neutronstarsref7} there are also 
other possibilities to 
infer the  redshift; see 
\cite{Messenger:2011gi,Taylor:2011fs,Messenger:2013fya}. However, they entail a 
good knowledge of the population of neutron stars and their equation of state.

GW170817 has become the first GW event detected together with an EM counterpart 
\cite{TheLIGOScientific:2017qsa}.  The redshift, $z=0.008^{+0.002}_{-0.003}$, 
was obtained identifying the host galaxy NGC4993 \cite{GBM:2017lvd}. On the 
other hand, from the amplitude of the GW signal the luminosity distance was 
obtained, $d_{_\text{L}}=40^{+8}_{-14}$Mpc. Assuming GR holds 
\eqref{eq:dl_GR_JME}, the first standard siren measurement of the Hubble 
constant $H_0$ was performed \cite{Abbott:2017xzu}. The error was 14$\%$, 
mainly 
caused by the uncertainty in the determination of the GW amplitude. Since the 
event had a large SNR with a precise localisation (see right panel of Fig. 
\ref{fig:distance_loc_JME}), the statistical method could be applied to obtain 
$H_0$, although the error is significantly larger \cite{Fishbach:2018gjp}. The 
statistical method has also been applied  to  a dark siren: GW170814, a three 
detector BBH event with good sky localisation \cite{Soares-Santos:2019irc}.
\begin{figure}[ht!]
\centering
\includegraphics[width=0.49\textwidth]{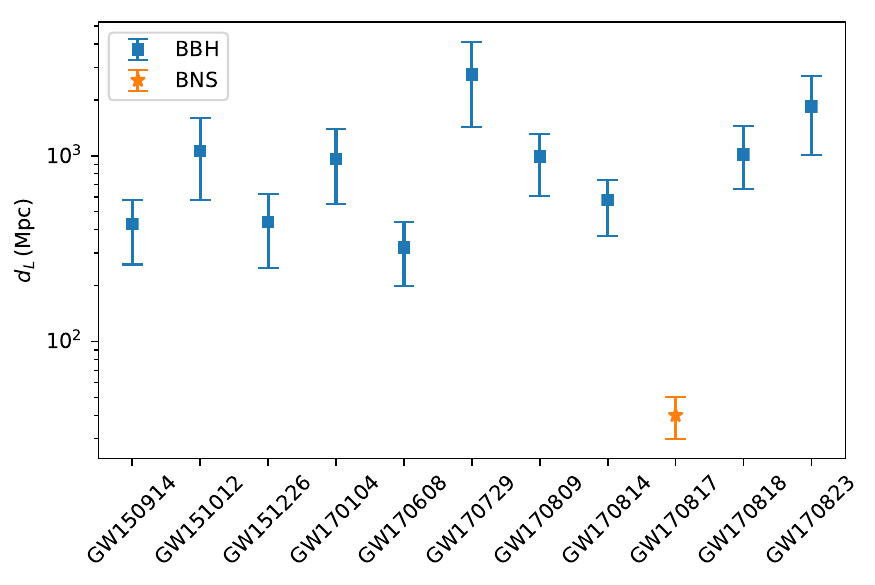}
\includegraphics[width=0.49\textwidth]{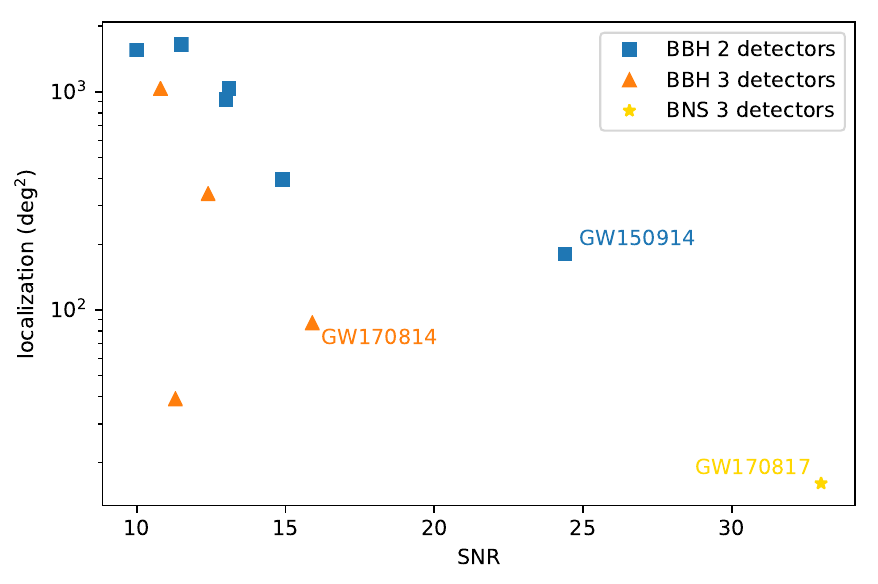}
\caption{{\it{On the left, luminosity distance $d_L$ of the GW events reported 
by  the LIGO-Virgo collaboration in runs O1-2 \cite{LIGOScientific:2018mvr}. On 
the right, $90\%$ credible localisation for the 11 events as a function of 
signal-to-noise ratio (SNR). We distinguish those events detected only with the 
two LIGO detectors and those that were found also by Virgo. For an updated list 
of all the candidate events in O3 not included in the figures see 
\textcolor{cyan}{\href{https://gracedb.ligo.org/latest/}{
https://gracedb.ligo.org/latest/}}.}}}
 \label{fig:distance_loc_JME}
\end{figure}

\section{The Speed of GWs}
\label{sec:gw_speed_JME}
\label{anomGEspeedef1}

The speed of GWs is determined by the second-order action for the tensor 
perturbations, which depends on the underlying gravity theory and the 
background fields. The key quantity is the \emph{effective metric} 
$\mathcal{G}^{\mu\nu}$ for the GWs, shaping the leading derivative interactions
\begin{equation} \label{eq:effective_metric_JME}
\mathcal{L}\propto\,h_{\mu\nu}\,\mathcal{G}^{\alpha\beta}
\nabla_\alpha\nabla_\beta\,h^{\mu\nu}=h_{\mu\nu}\left(\mathcal{C}\Box+\mathcal{D
}^{\alpha\beta}\nabla_\alpha\nabla_\beta\right) h^{\mu\nu}\,.
\end{equation}
GR predicts that the effective metric for the GWs is equal to the background 
metric $g^{\mu\nu}$. Accordingly, the local speed of GWs is equal to the speed 
of light, $c_g=c$. Other gravity theories have different predictions. The 
effective metric $\mathcal{G}^{\mu\nu}$ can be decomposed in a part 
proportional to $g^{\mu\nu}$ and another that is not (see $\mathcal{C}$ and 
$\mathcal{D}^{\alpha\beta}$ terms respectively 
in~\eqref{eq:effective_metric_JME}).
Whenever the two metrics are not conformally related, the \emph{GW-cone} and 
the light-cone will not match inducing an anomalous propagation speed $c_g\neq 
c$ and a time delay between both signals.

In order to have a non-luminal GW speed, there must be terms in the action 
affecting the second derivative interactions of the metric and not vanishing in 
the given background. For instance, in scalar-tensor gravity, two conditions 
have to be fulfilled \cite{Bettoni:2016mij}:
\begin{itemize}
\item[\emph{i)}] There is a non-trivial scalar field configuration   
\label{loclinref8}
spontaneously breaking Lorentz invariance. In order to explain DE, one typically 
demands $\dot{\phi}\sim H_0$.
\item[\emph{ii)}] There is a derivative coupling to the curvature. This 
highlights the presence of a modified gravity coupling leading to 
$\mathcal{D}^{\alpha\beta}\sim\partial^\alpha\phi\,\partial^\beta\phi$.
\end{itemize}
Whenever these two conditions are fulfilled, the GW and the EM signals will 
travel at different speeds. For example, $1\%$ differences in the speed for 
sources at $100$Mpc induce time delays of tens of millions of  years, 
frustrating 
any multi-messenger detection.

\subsection{Constraints After GW170817}
\label{sec:constraints_speed_JME}

GW170817 was followed by a sGRB only $\Delta t=1.74\pm0.05$s after  
\cite{Monitor:2017mdv}. Since the BNS was at $\mathcal{O}(40\text{Mpc})$, this 
event led to the impressive constraint on the speed of GWs
\begin{equation}
-3\cdot 10^{-15}\leq c_g/c-1\leq 7\cdot 10^{-16}\,.
\end{equation}
This result has profound implications for many gravity theories and dark energy 
models  (see \cite{Ezquiaga:2018btd} for a recent review).
In particular, a large sector of the scalar-tensor dark energy models is now 
ruled out  
\cite{Ezquiaga:2017ekz,Creminelli:2017sry,Baker:2017hug,Sakstein:2017xjx}. 
These 
include cosmologically viable models with screening and self-accelerating like 
the covariant Galileon \cite{Nicolis:2008in,Deffayet:2009wt}, or proposals to 
self-tune the cosmological constant like the Fab-Four \cite{Charmousis:2011bf}.
For vector-tensor theories the situation is very similar. In order to describe 
DE and to  pass the GW test, some couplings of the theory have to be eliminated 
\cite{Ezquiaga:2017ekz,Baker:2017hug,BeltranJimenez:2018ymu}.
The implications of GW170817 extend to other gravity theories such as 
doubly-coupled  bigravity \cite{Akrami:2018yjz}, $f(T)$ gravity 
\cite{Cai:2018rzd}, Ho\v{r}ava gravity \cite{Gumrukcuoglu:2017ijh} or 
Born-Infeld models \cite{Jana:2017ost}.

Nonetheless, it is important to remark possible caveats and ways to avoid these 
constraints:
\begin{itemize}
\item \emph{Constraints apply to dark energy models:} constraints after 
GW170817 
generically  apply only to gravity theories in which the additional fields have 
a relevant role in cosmology. For instance, we could take as an example the 
case 
of Galileons. There, the tensor speed speed excess is proportional to 
\cite{Ezquiaga:2017ekz}
\begin{equation}
(c_g^2-1)\propto\left(\frac{\dot{\phi}}{H_0}\right)^4\,.
\end{equation}
Accordingly, for models in which the Galileon triggers the present expansion,  
$\dot{\phi}\sim H_0$, there are $\mathcal{O}(1)$ deviations in $c_g$. However, 
if we resign from this goal, Galileons could be in agreement with GW170817, 
simply choosing $\dot{\phi}<10^{-4}H_0$. Definitely, this sector of the theory 
is 
less interesting a priori but serves to exemplify that, in most cases, GW170817 
constrains DE models rather than gravity theories. Another example considered 
recently is scalar Gauss-Bonnet gravity \cite{Franchini:2019npi}, where it was 
shown that if the scalar does not have a dominant energy density, it passes the 
constraints on the speed.

\item \emph{Cosmological tuning of $c_g(z=0)=c$ is not viable:} although 
tuning  by hand the parameters of the theory to pass GW170817 is not appealing, 
it is tempting to devise a dynamical mechanism that leads to this tunning. One 
could therefore think that a cosmological model with such mechanism could be 
viable. However, to avoid the constraint on the speed of GWs, one has to fix 
$c_g=c$ on arbitrary backgrounds \cite{Ezquiaga:2017ekz}. This is because when 
the GW travels from the source to the observer, it will cross backgrounds 
deviating from FRW, e.g. when they cross the Milky Way or due to the 
large-scale 
structure. Therefore, delays between the GW and the EM radiation will be 
accumulated again. Given the strength of the GW170817 constraint, any small 
deviation from the cosmological background would kill the tunning mechanism. 
This reasoning has been applied lately to an interesting sector of Horndeski 
gravity in which the scalar EoM dynamically cancel the anomalous speed 
\cite{Copeland:2018yuh}. Nevertheless, large-scale inhomogeneities are 
sufficient to make the mechanism fail.

\item \emph{Constraints assume EFT validity:} when computing the speed  of GWs 
or any other GW observable from a dark energy model, we are assuming that the 
effective field theory is valid. In other words, we are assuming that 
higher-order operators do not modify the action.
However, the frequency of GW170817 was of the same order of the typical strong 
coupling  scale of the EFT of DE \label{eftrefs2}
\begin{equation}
\Lambda_\text{strong}\sim (M_\text{pl}H_0^2)^{1/3}\sim260\,\text{Hz}\,.
\end{equation}
Taking the cut-off of the theory around the strong coupling scale 
$M_\text{cutoff}\sim\Lambda_\text{strong}$,  higher dimensional operators could 
modify the dispersion relation. Nevertheless, one would not expect them to 
conspire to completely erase the anomalous speed at the level of 
$\mathcal{O}(10^{-15})$ \cite{Creminelli:2017sry}.
On the contrary, when the cut-off scale is parametrically smaller, 
$M_\text{cutoff}\ll\Lambda_\text{strong}$,  the situation could change 
\cite{deRham:2018red}.
Lorentz invariance in the ultra-violet (UV) imposes luminal GW propagation. 
Therefore, higher dimensional operators  would tend to cancel any anomalous 
speed beyond the cut-off scale, which with these assumptions might already 
happen 
in the LIGO band. The computation of the speed of GWs beyond $M_\text{cutoff}$ 
requires, however, knowledge of the UV completion.
In any case, the possibility that $c_g(k_{_\text{LIGO}})=c$ could be tested 
detecting GWs at different frequencies,  for example with LISA. 
\end{itemize}

\section{GW Luminosity Distance}
\label{sec:gw_dl_JME}
 \label{luminositydisdef1}

Another powerful observable to constrain modifications in the GW propagation is 
 
the luminosity distance $d_L^{\text{gw}}$, which can be obtained from the 
inverse of the amplitude $\vert h_{+,\times}\vert\propto 1/d^\text{gw}_{L}$.
GR predicts that GWs are only sensitive to the Hubble friction along the 
propagation  so that the GW luminosity distance is equal to the EM one, cf. 
\eqref{eq:dl_GR_JME}.
However, in other theories of gravity, the cosmic medium could be, for 
instance, 
more absorptive.  This would dim the received signal, which would be 
interpreted 
as the source being further apart. In other words, the GW luminosity distance 
$d^\text{gw}_{L}$ would be larger than the EM luminosity distance 
$d^\text{em}_{L}$.

Following the propagation equation (\ref{eq:ModProp_JME}), in the case in which 
$c_g=c$,\footnote{For  the general formula of luminosity distance ratio 
$d^\text{gw}_{L}/d^\text{em}_{L}$ when $c_g\neq c$ see Section 2.2.3 of Ref. 
\cite{Belgacem:2019pkk}.} the ratio of the GW and EM luminosity distance is 
given by
\begin{equation} \label{eq:dl_cg_c_JME}
\frac{d_L^{\text{gw}}(z)}{d_L^{\text{em}}(z)}=\text{exp}\left[\frac{1}{2}
\int_0^z   \frac{\nu(z')}{1+z'}dz'\right]\,.
\end{equation}
For instance, in scalar-tensor gravity, the additional friction is equal to the 
effective  Planck mass run rate $\alpha_{_\text{M}}$
\begin{equation}
\nu=\alpha_{_\text{M}}=\frac{d\ln M_*^2}{d\ln a}\,,
\end{equation}
where $M_*$ is the local, effective Planck mass, i.e., the normalisation of the 
kinetic  term of the tensor perturbations. Then, we arrive  at
\begin{equation}
\frac{d_L^{\text{gw}}(z)}{d_L^{\text{em}}(z)}=\frac{M_*(0)}{M_*(z)}\,,
\end{equation}
where $M_*(0)$ and $M_*(z)$ are the effective Planck masses at the time of 
observation  and emission respectively.
Instead of fixing the theory, one could parametrise the friction term. One 
possibility  is to assume that $\nu(z)$ scales with the DE 
\cite{Lagos:2019kds}, 
i.e.
\begin{equation}
\nu(z)=c_{_M}\,\frac{\Omega_{_{DE}}(z)}{\Omega_{_{DE}}(0)}\,,
\end{equation}
where $c_{_M}$ is constant. This is a reasonable parametrisation for  modified 
gravity theories trying to explain the late-time cosmic acceleration. Assuming 
that the background cosmology is $\Lambda$CDM, we obtain 
\begin{equation}
\frac{d^\text{gw}_{L}}{d^\text{em}_{L}}=\exp\left\{\frac{1}{2}\frac{c_{_M}}{
\Omega_{_{DE}}(0)}\log\left\{\frac{1+z}{[\Omega_{_{m}}(0)(1+z)^3 + 
\Omega_{_{DE}}(0)]^{1/3}}\right\}\right\}\,.
\end{equation}
For illustration, in Fig.~\ref{fig:dl_JME} we depict how the ratio 
$d_L^{\text{gw}}(z)/d_L^{\text{em}}(z)$ varies  for different values of 
$c_{_M}$. Positive values of $c_{_M}$ make $d^\text{gw}_{L}$ to be larger than 
$d^\text{em}_{L}$ and vice versa. We compare the effects at low redshift (left 
panel) with high redshift (right panel) expected for ground-base and space-base 
interferometers respectively. Clearly, low redshift signals are less effective 
at constraining a modification in the propagation of GWs. In the case of 
scalar-tensor gravity, present cosmological surveys place  constraints of order 
one on 
$c_{_M}$, while future ones could reach $\mathcal{O}(0.1)$ 
\cite{Alonso:2016suf}.
\begin{figure}[ht!]
\centering
\includegraphics[width=0.63\textwidth]
{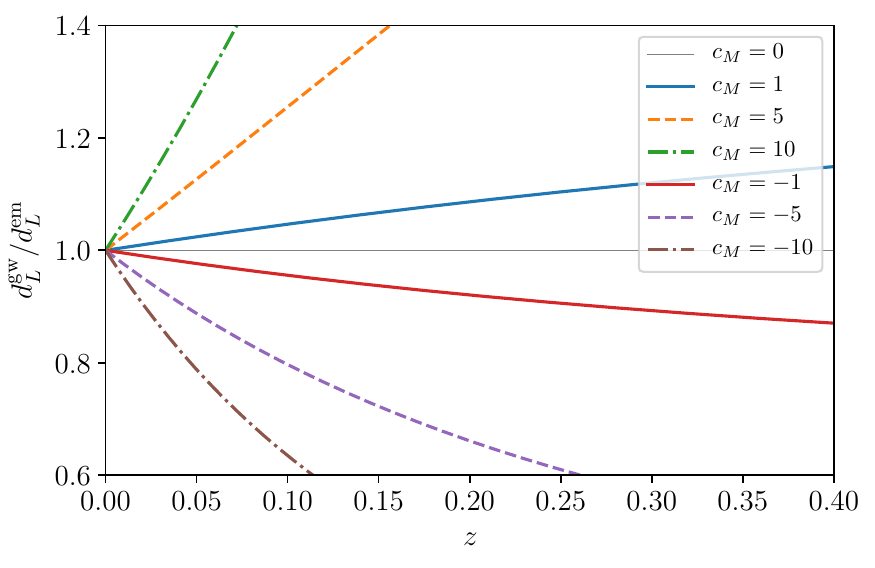}\\
\includegraphics[width=0.63\textwidth]
{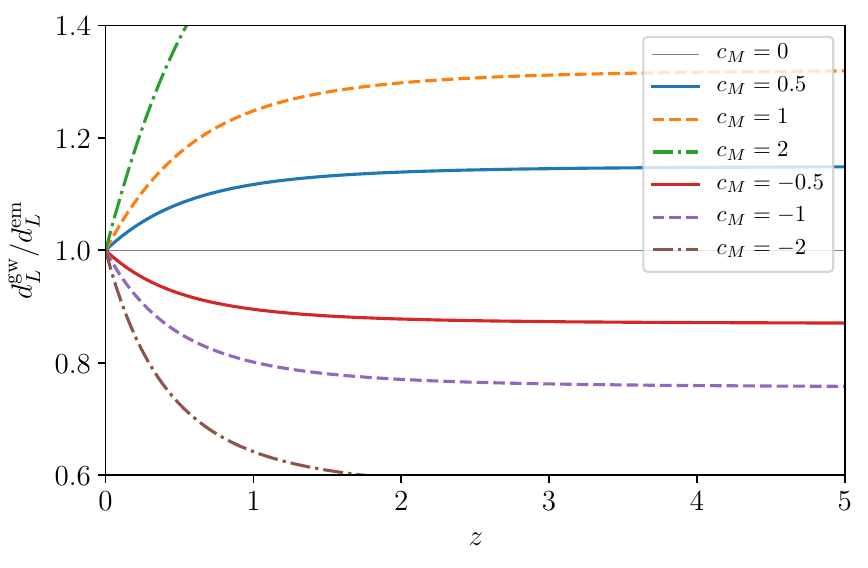}
\caption{{\it{Ratio between the GW and the EM luminosity distances  due to an 
additional friction term in the propagation $\nu$ scaling with the DE, 
$\nu(z)=c_{_M}\Omega_{_{DE}}(z)/\Omega_{_{DE}}(0)$. We compare low- (upper) and 
high-redshift (lower) signals expected for ground- and space-base GW detectors 
respectively.}}}
 \label{fig:dl_JME}
\end{figure}

An alternative possibility is to parametrise the ratio of luminosity distances,
which is more directly linked to observations. A convenient choice is 
\cite{Belgacem:2018lbp}
\begin{equation} \label{eq:dLratio_parametrisation_JME}
\frac{d_L^{\text{gw}}(z)}{d_L^{\text{em}}(z)}=\Xi_0+\frac{1-\Xi_0}{(1+z)^n}\,,
\end{equation}
which resembles the usual $(w_0,w_a)$ parametrisation of the equation of state 
of DE.  This parametrisation interpolates between 1 at $z=0$ and a constant 
value $\Xi_0$ at high redshift.
From Fig. \ref{fig:dl_JME} one can already see that this is the expected 
behaviour when  the modification follows the DE.
Ref. \cite{Belgacem:2019pkk} studied in detail how this parametrisation adjusts 
to  several modified gravity theories, including degenerate higher order 
scalar-tensor theories and non-local infrared modifications of gravity. It was 
found a good fit in most of the cases, specially at high redshift.

Present constraints on a modified GW luminosity distance are still very loose  
\cite{Arai:2017hxj,Belgacem:2018lbp}. This is because at the moment there is 
only one standard siren, GW170817. This is expected to change significantly in 
the coming years as more detections accumulate in the next observing runs of 
advanced LIGO-Virgo. Looking further to the future, using thousands of BNS 
standard sirens, the  Einstein Telescope could reach a $1\%$ measurement of 
$\Xi_0$ 
\cite{Belgacem:2018lbp}.
In the same manner, using SMBHs at  $z\sim2-5$ as standard sirens, LISA could 
constrain $\Xi_0$ to the $1-5\%$ level \cite{Belgacem:2019pkk}.

\section{GW Oscillations}
\label{sec:gw_oscillations_JME}
\label{GWoscillationsdef1}

If GWs mix with other tensor modes $t_{ij}$ along the propagation,  there will 
be modifications of the GW luminosity distance. This is because GW detectors 
are 
only sensitive to $h_{ij}$. Therefore, if $h_{ij}$ converts partially into 
$t_{ij}$, the amplitude detected will be lower and the inferred 
$d^\text{gw}_{L}$ larger. If the conversion is complete, no GW would be 
detected, making $d^\text{gw}_{L}$ effectively to diverge. A series of 
consecutive mixings imprints a characteristic oscillatory pattern in the GW 
luminosity distance that can be probed with standard sirens.
\begin{figure}[ht!]
\centering
\includegraphics[width=.63\textwidth]{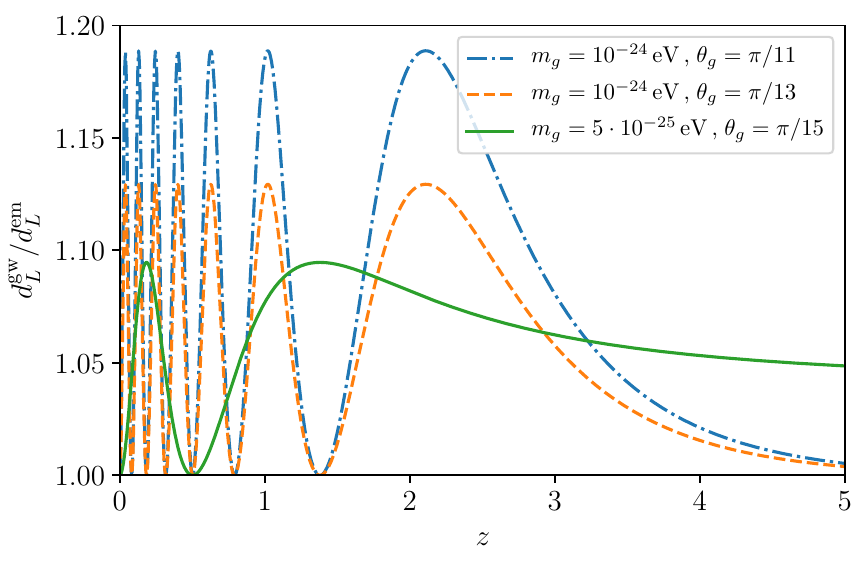}\\
\includegraphics[width=.63\textwidth]
{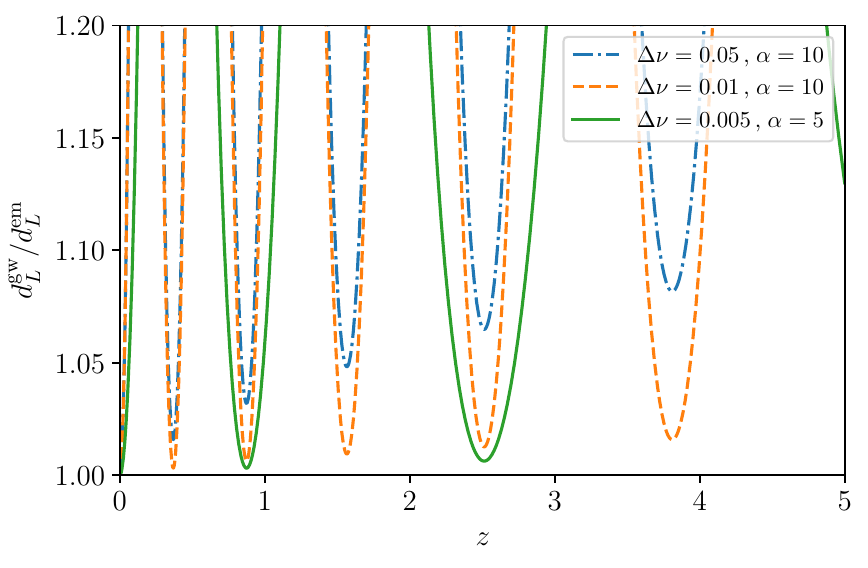}
 \caption{
 {\it{
 Modified GW luminosity distance as a function of redshift for 
theories 
with a  mass mixing (upper) or a friction mixing (lower). The ratio 
$d^\text{gw}_{L}/d^\text{em}_{L}$ is plotted for different values of the 
effective mass $m_g$ and mixing angle $\theta_g$; and damping factor $\Delta 
\nu$ and friction mixing $\alpha$. Figures reproduced from 
\cite{Jimenez:2019lrk}.}}}
 \label{fig:dL_oscillations_JME}
\end{figure}

We can proceed as in the previous section and compute the ratio of the GW and 
EM 
luminosity distances.  Considering only the mass mixing described in 
\eqref{eq:generalequation_JME}, which is characteristic of bigravity theory, it 
would be given by \cite{Belgacem:2019pkk,Jimenez:2019lrk}
\begin{equation} \label{eq:dLratio_mass_JME}
\frac{d^\text{gw}_{L}}{d^\text{em}_{L}}=\frac{1}{\cos^2\theta_g} 
\left\{1+\tan^4\theta_g+2\tan^2\theta_g\,\cos\left[\frac{m_g^2}{2k}\int_{0}
^z\frac{dz}{(1+z)^2H(z)}\right]\right\}^{-1/2}\,.
\end{equation}
The mixing angle $\theta_g$ controls the amount of conversion while the  
effective mass $m_g$ determines the frequency of oscillation. The mixing angles 
range from 0 to $\pi/2$, having the maximum mixing at $\theta_g=\pi/4$ when a 
complete oscillation of $h_{ij}$ into $t_{ij}$ occurs.
Note that the frequency of oscillation also depends on the frequency of the GW. 
For example, to have an $\mathcal{O}(1)$ frequency of oscillation at low 
redshift, $m_g^2/2k$ should be of the  order of $H_0$ to compensate the Hubble 
parameter 
in the denominator. Since $H_0\sim10^{-33}\text{eV}\sim10^{-18}\text{Hz}$, this 
means that present ground-based interferometer, $f_{\text{LIGO}}\sim100$Hz, can 
 
test $m_g\sim 10^{-23}$eV through the oscillations of $d^\text{gw}_{L}$. 

In the \label{LISAref3}
same manner, the future space-based detector LISA, $f_{\text{LISA}}\sim10$mHz, 
will be sensitive to $m_g\sim 10^{-25}$eV.
In the upper panel of Fig. \ref{fig:dL_oscillations_JME} we present how 
the 
GW luminosity  distance varies with $\theta_g$ and $m_g$. As the mixing angle 
approaches to $\pi/4$, the amplitude of the oscillation increases. Similarly, 
the frequency of oscillation increases with $m_g$.
The plot corresponds to the expected redshift range, $z\sim2-6$, and 
sensitivity, $\Delta d_L/d_L\sim10\%$, of LISA standard sirens. A recent 
analysis studied the capability of LISA to detect GW oscillations in 
$d^\text{gw}_{L}$  \cite{Belgacem:2019pkk}, showing that LISA could probe 
masses 
of $m_g\gtrsim 2\cdot10^{-25}$eV with mixing angles 
$0.05\pi\lesssim\theta_g\lesssim0.45\pi$.

Moving now to the friction mixing in \eqref{eq:generalequation_JME},  the ratio 
of the GW and EM luminosity distance reads \cite{Jimenez:2019lrk}
\begin{equation}
\frac{d^\text{gw}_{L}}{d^\text{em}_{L}}= 
(1+z)^{\Delta\nu}\left\{\cos\left[\omega_\nu\,\log(1+z)\right]+\frac{\Delta\nu}{
\omega_\nu}\sin\left[\omega_\nu\,\log(1+z)\right]\right\}^{-1}\,.
\end{equation}
This expression contains a first global friction term induced by $\Delta\nu$, 
the  difference between the friction term of $t_{ij}$ and $h_{ij}$. The 
oscillation rate is determined by $\omega_\nu=\sqrt{\alpha^2-\Delta \nu^2}$, 
where $\alpha$ is the non-diagonal term producing the mixing. In the limit of 
no mixing, $\alpha\rightarrow0$, one recovers 
$d^\text{gw}_{L}/d^\text{em}_{L}\rightarrow1$. Moreover, there is always a 
complete conversion of $h_{ij}$ into $t_{ij}$, implying that at certain 
redshifts there will be no GW events. This in turn translates into periodic 
divergences of $d^\text{gw}_{L}/d^\text{em}_{L}$ that makes this models easier 
to probe.
This can be seen in the lower panel of Fig. \ref{fig:dL_oscillations_JME}, 
where 
we depict $d^\text{gw}_{L}/d^\text{em}_{L}$ for different values of 
$\Delta\nu$ and $\alpha$. The global friction term $(1+z)^{\Delta \nu}$ makes 
the minimum values of $d^\text{gw}_{L}/d^\text{em}_{L}$ to increase away from 1 
as $\Delta \nu$ increases. On the other hand, the mixing parameter $\alpha$ 
controls the frequency of oscillation. Future GW detectors could also probe 
this 
type of GW oscillations, which occur, for instance, in theories with 
cosmological 
gauge fields  \cite{Caldwell:2016sut,BeltranJimenez:2018ymu}.

\section{Future Prospects}
\label{sec:future_JME}

In the last four years we have witnessed a vibrant beginning of GW astronomy:
from the excitement of the first detection, to the routine of the following 
events and back to  the euphoria of a multi-messenger historic achievement.
BBH mergers have taught us not only  about a population of heavy BHs 
\cite{LIGOScientific:2018mvr} but also that their signals have served to 
strongly 
constrain the mass of the graviton \cite{LIGOScientific:2019fpa}. More 
notoriously, GW170817, a BNS event, brought the first standard siren 
measurement 
of the Hubble constant (see Section \ref{sec:standard_sirens_JME}) and an 
impressive bound on the speed of GWs, $\vert c_g/c-1\vert \leq 10^{-15}$, that 
swept out many DE contenders (see Section~\ref{sec:gw_speed_JME}).

The following years are no less exciting, with observational  improvements 
pointing in three main directions: number of detections, sky localisation and 
distance range (see Fig.~\ref{fig:MMtimeline_JME}). All of them will boost the 
constraining power of standard siren tests of gravity and cosmology.
Moreover, any deviation of the ratio of luminosity distances  
$d^\text{gw}_{L}/d^\text{em}_{L}$ from being 1 would be a smoking gun of 
physics 
beyond GR and $\Lambda$CDM.
However, these observational opportunities will also pose theoretical 
challenges, as  we will be crossing both discovery and precision frontiers.
In what comes next, we overview these challenges and opportunities, focusing on 
 
those related to unveil the nature of gravity and dark energy.
\begin{figure}[ht!]
\includegraphics[width=\textwidth]{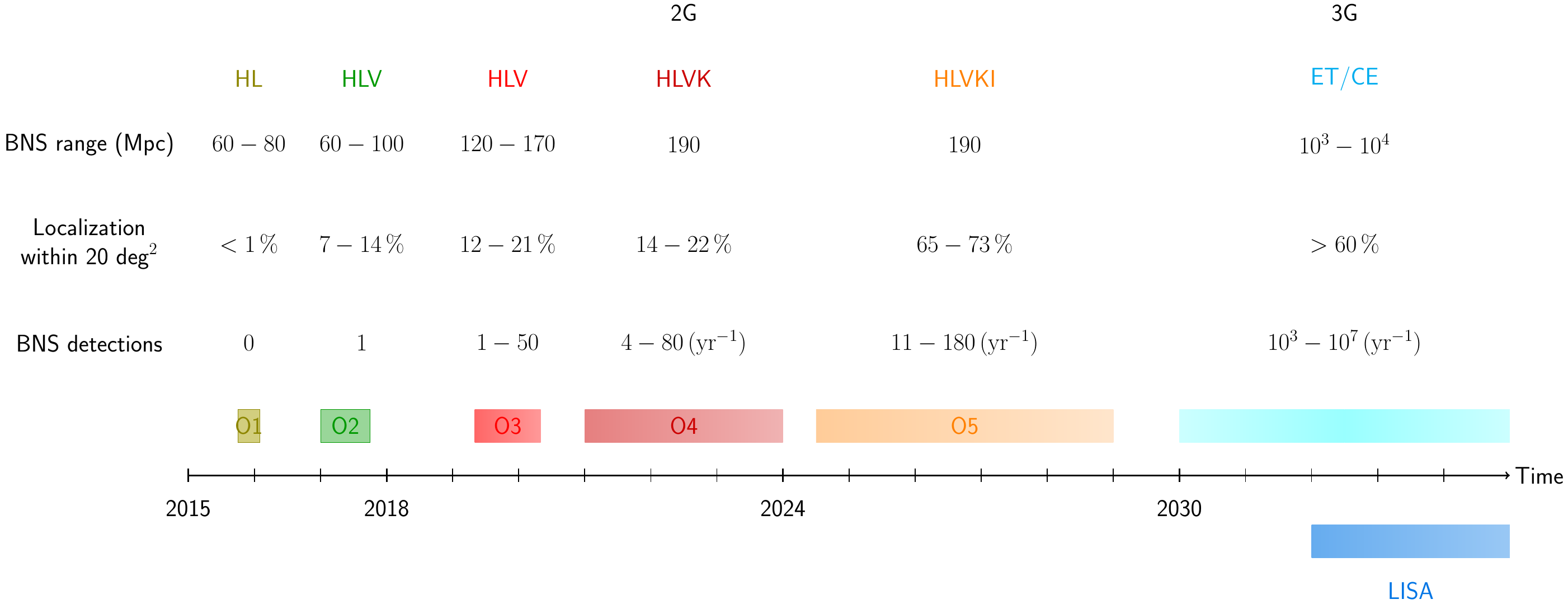}
\vspace{-15pt}
\caption{
{\it{
Schematic multi-messenger GW astronomy timeline. The binary neutron 
star \label{neutronstarsref8} (BNS) rate, the localisation area in the sky, and 
the number of BNS 
detections are given for past and future observation runs. Second generation 
(2G) ground-based detectors organise in five runs (O1-O5) with a  different 
number 
of detectors online (from 2 to 5) \cite{Aasi:2013wya}. The nomenclature used is 
H=Hanford, L=Livingston, V=Virgo, K=KAGRA, I=IndIGO. Third generation (3G) 
detectors projected are Einstein Telescope (ET) \cite{Sathyaprakash:2012jk} or 
Cosmic Explorer (CE) \cite{Evans:2016mbw}. The localisation in 3G depends on 
the 
network of detectors, which is still uncertain \cite{Mills:2017urp}. For 
reference, we include the timeline space-based detector LISA 
\cite{AmaroSeoane:2012km}. The reader should note that these numbers correspond 
to present expectations. Figure updated from Ref. \cite{Ezquiaga:2018btd}.} }}
\label{fig:MMtimeline_JME}
\end{figure}
\subsection{Theoretical Challenges}
\label{sec:theory_challenges_JME}
\label{gphenomenologicalprfs5}
In this chapter we have focused on the cosmological propagation of GWs. 
However,  
modifications in the propagation are not restricted to cosmological 
backgrounds. 
For instance, when a GW crosses a screened region, the spatially dependent 
background might affect the GW signal.
From a phenomenological perspective there are two possible GW observables 
beyond 
 what we have considered here: additional polarisations and frequency dependent 
modifications.

\emph{Additional polarisations:}
on FRW backgrounds GWs can only mix with other tensor perturbations. 
Nevertheless,  in more general spacetimes there could be an interplay between 
the scalar, vector and tensor polarisations.
In particular, non-GR polarisations could be excited along the propagation even 
if  they were not originally produced. If those polarisations couple to matter, 
they could be detected with a network of ground-based detectors.
This would be a clear indication of a modification of gravity.
In addition, one could also probe if the two tensor polarisations have the same 
properties,  constraining parity-violating gravity \cite{Nishizawa:2018srh}.

\emph{Frequency-dependent modifications:}
the propagation of GWs depends on space and time, but may also depend on the 
frequency  of the wave.
The simplest example is having a graviton mass, which modifies the dispersion 
relation.
The advantage of this type of modification is that they can be probed with GWs 
alone and on  single events, due to the chirping of the GW signal.
Moreover, GW detectors are very sensitive to frequency-dependent modifications.
These effects might be caused by the breaking of the EFT at the frequencies of 
GW detectors  \cite{deRham:2018red}, or by the decay of GWs into DE 
\cite{Creminelli:2018xsv}.
It would be interesting to search further for these phenomena in the landscape 
of modified  gravity.

Finally, a theoretical challenge across the different probes would be to 
understand possible  degeneracies with astrophysics. Along the same lines, the 
separation between modifications in the emission and propagation might not be 
possible with the improved precision of future detectors.

----------

\subsection{Observational Opportunities}
\label{sec:observ_opportunities_JME}

The future of gravitational wave astronomy is promising. With a larger  
catalogue 
of events, one could perform statistically significant studies at the same time 
of precision measurements from rare events with high SNR. Moreover, new 
frequencies will be opened, namely the mHz from space-based detectors, 
certainly 
leading to discoveries as well as enabling new techniques such us multi-band GW 
astronomy.
Here we highlight how some of these observational opportunities could 
contribute 
to  the quest of the dark side of the Universe.

\emph{2G detectors:}
second generation (2G) detectors are  still on their way to their design 
sensitivity (see Fig. \ref{fig:MMtimeline_JME}). Thus, merger rates are 
expected 
to increase in future observational runs. In fact, we are now witnessing one of 
such steps forward in O3 compared to the previous runs O1-2.
Collecting statistics has implications for the study of the cosmic expansion,  
enlarging the number of multi-messenger events to be used as standard sirens or 
the number of BBHs to be used as dark sirens. This will improve the measurement 
of the Hubble constant (see Section \ref{sec:standard_sirens_JME}) as well as 
the constraints on the propagation of GWs (see, for example, the upper panel of 
Fig. \ref{fig:dl_JME}).
Moreover, all of the planned 2G detectors are not yet online. KAGRA is expected 
to  join by the end of O3 and IndIGO just in the last run (again, see Fig. 
\ref{fig:MMtimeline_JME}). Having a large network of detectors is essential to 
test the number of polarisations, a fundamental property of each gravity 
theory. 
This will open new opportunities to test, for instance, chiral GWs.

\emph{3G detectors:}
third generation (3G) detectors will improve their  predecessors in multiple 
ways \cite{Kalogera:2019sui}: observing more distant binaries, in a wider 
frequency range and with higher precision.
All of these improvements will be relevant to test gravity.
Detecting mergers at high-redshift will serve to determine the evolution of the 
GW  luminosity distance.
Enlarging the frequency band will be interesting at both ends. At low 
frequencies, reaching down to 1Hz,  3G detectors will be sensitive to higher 
masses, possibly above $100M_\odot$. If some events have small mass ratios, 
precession effects could be observed.
At high frequencies, reaching up to 5kHz, the merger phase of BNS could be much 
better constrained,  including tidal effects.
As the sensitivity will improve significantly, 3G detectors could hear many 
orders of magnitude more events.  This will ensure precision measurements of 
$H_0$ and the propagation of GWs beyond 2G capabilities. Moreover, it is 
expected that some signals will be strongly lensed \cite{Piorkowska:2013eww}. 
This could be a perfect arena to test modifications of~gravity.

\emph{Space-based detectors:}
a million kilometer interferometers in space could hear GWs in the mHz band. 
This 
is the goal of LISA,  which will inaugurate a completely new channel to detect 
very massive BBHs, galactic binaries and possibly cosmological backgrounds.
If identified with an EM counterpart, SMBH binaries would become standard 
sirens 
at cosmological scales,  $z\sim2-5$. LISA standard sirens could bound generic 
modifications of the GW luminosity distance (see the  lower panel of Fig. 
\ref{fig:dl_JME}) and GW oscillations (see Fig. \ref{fig:dL_oscillations_JME}).
Extreme mass ratio inspirals and galactic binaries are also good candidates to  
test gravity \cite{Gair:2012nm}.

\emph{Multi-band GW astronomy:}
the reach of GW astronomy can be enhanced if GW events are heard at  different 
frequencies.
Heavy BBHs like GW150914 could be detected months in advance by LISA 
\cite{Sesana:2016ljz}.
This will again open new opportunities. For instance, with multi-band events 
one 
could combine  the information that is better measured in space, such as 
eccentricities, with the inference on the mass and spin obtained at the merger. 
This could improve the source characterisation.
Similarly, anticipating the detection of a signal from space will improve its 
localisation on Earth,  making   the search for an EM counterpart  more 
effective. 
At the same time, multi-band events with precise localisation are perfect 
candidates to probe additional polarisations.

\emph{Synergies with cosmological surveys:}
GW observatories will not be alone in  probing the dark universe.  Cosmological 
surveys such as LSST, Euclid or WFIRST will have a fundamental role (see recent 
reviews of their science cases in \cite{Ishak:2019aay,Bechtol:2019acd}, 
\cite{Amendola:2016saw} and \cite{Dore:2019pld} respectively). With respect to 
the theoretical challenges, they will constrain, among others, the DE equation 
of 
state and possible deviations of GR in the growth of structure 
\cite{Amendola:2016saw,Ishak:2019aay,Dore:2019pld}. These complementary 
observations could be useful in breaking degeneracies.
From the observational side, a key task of future surveys will be to  deliver 
complete and uniform galaxy catalogues. This is essential in order to properly 
identify the redshift of a GW signal with the statistical method (see Section 
\ref{sec:standard_sirens_JME}).
We expect that the most challenging questions about DE will only be addressed  
with a powerful synergy between GW astronomy and other cosmological probes.


%






\chapter[Testing the Dark Universe with Cosmic Shear]{Testing the Dark Universe 
with \\ Cosmic Shear}
\label{sec:Pettorino}

{\em Valeria Pettorino,
Alessio Spurio 
Mancini}\\





Although a cosmological  constant framework is still in agreement with current 
data, several other cosmological models in which gravity is modified are also 
still viable. There are several approaches that one can adopt in order to 
distinguish $\Lambda$CDM from modified gravity models. One can try to: a) use or 
combine different probes, b) get more data, c) improve the analysis to extract 
more information from the available data. Below we focus on weak lensing, its 
different approaches and the impact of statistics we use on constraining or 
distinguishing cosmological models.
 \label{stronglensefs4}
 
Weak lensing describes, in particular, small  distortions in the observed image 
of galaxy shapes, induced by the presence of massive structures along the line 
of sight. Weak lensing can be typically described in terms of shear and 
convergence fields, quantifying anisotropic and isotropic distortions 
respectively. Convergence can be derived from shear, up to a constant, and both 
depend on the angular position $\theta$ on the sky. Given a convergence map 
$\kappa(\theta)$ for a particular realisation of a model, one can also compute  
the aperture mass map \cite{1996MNRAS.283..837S, 1998MNRAS.296..873S} by 
applying a filter (see, for example, \cite{2018A&A...619A..38P} for a review of 
different filters adopted in literature).

In \ref{sec:weaklensing} we will recall different  weak lensing methodologies; 
in \ref{sec:forecasts} we will describe how well we can use current and future 
probes (in particular including cross-correlations or combining with galaxy 
clustering) \label{galaxyclurefs3}to test modified gravity models; in 
\ref{sec:higherorder} we recall  \label{hoghstatisticsef1}
how higher order statistics, and in particular peak counts, can help in breaking 
degeneracies between parameters; finally in \ref{sec:machinelearning} we 
illustrate recent results using machine learning techniques to improve the 
discrimination efficiency between $\Lambda$CDM and alternative theories in which 
gravity is modified with respect to General Relativity.


\section{2D, Tomographic and 3D Weak 
Lensing}\label{sec:weaklensing}\label{weaklensefs4}
Here we provide a mathematical description of cosmic shear in a general modified 
gravity context, similar to the one presented in \cite{2018MNRAS.480.3725S}.~We 
focus on two different formalisms commonly used to study the evolution in 
redshift of the lensing effect, so-called `tomography' and `3D cosmic shear'.~We 
assume spatial flatness throughout, and consider scalar linear perturbations on  
a Friedmann-Robertson-Walker metric, such that the line element in Newtonian 
gauge can be written as
\begin{align}\label{eq:lineelementMG_alessio}
{d}s^2 = -\left( 1 + 2 \Psi  \right)  {d}t^2  + 
a^2(t) \left( 1 - 2 \Phi   \right) {d} \textbf{x}^2,
\end{align}
with the scale factor $a(t)$ and the Bardeen potentials $\Psi$ and $\Phi$. 
In 
General Relativity   $\Psi=\Phi$ in absence of anisotropic stress, but this 
is 
in general not true in a modified gravity theory.~Poisson's equation links one 
of the Bardeen potentials to the overdensity field $\delta(k, \chi)$,
\begin{align}\label{eq:poisson_alessio}
\Psi (k, \chi) = - \frac{3}{2} \, \frac{\Omega_m}{(k  \chi_H)^2} \, 
\frac{\delta(k, \chi)}{a(\chi)} \, \mu(k, a(\chi)),
\end{align}
with the Hubble radius $\chi_H \equiv 1/H_0$ and the function $\mu(k, a(\chi))$ 
parameterising variations from General Relativity, its value being 1 in 
standard 
gravity. We also define
\begin{align}\label{eq:eta_alessio}
\eta(k, a(\chi)) \equiv \frac{\Psi(k, a(\chi))}{\Phi(k, a(\chi))},
\end{align}
as the ratio of the Bardeen potentials, again identically equal to 1  in 
General 
Relativity in absence of anisotropic stress. Other choices (such as $\Sigma$, 
defined in terms of the lensing potential $\Psi +\Phi$), of such two 
functions 
of time and scale are also possible, and may be more or less convenient; see 
\cite{Amendola:2007rr} or \cite{Ade2016} for a review.
\label{shearref1}

A quantitative description of cosmic shear starts with the definition of the 
lensing potential
\begin{align}\label{eq:lensingpotential_alessio}
\Psi(\chi, \hat{\mathbf{n}}) = \int_0^\chi \, {d}\chi' \,  
\frac{\chi-\chi'}{\chi \chi'} \,  [\Psi(\chi, \hat{\mathbf{n}})+\Phi(\chi, 
\hat{\mathbf{n}})],
\end{align}
as a weighted projection of the sum of the Bardeen potentials along the line of 
sight. In Eq.~\ref{eq:lensingpotential_alessio}, $\chi$ is the comoving 
distance 
and the normalised vector $\hat{\textbf{n}}$ selects a direction in the sky. 
From its definition in Eq.~\ref{eq:lensingpotential_alessio} we notice that the 
lensing potential is sensitive to the growth of perturbations of the 
gravitational potentials, as well as to the geometry of the Universe through 
the 
weighting factor $\frac{\chi - \chi'}{\chi \chi'}$. We will assume that the 
integration in Eq.~\ref{eq:lensingpotential_alessio} is carried out along the 
unperturbed light path, following the Born approximation. The lensing 
observables, i.e., convergence and shear, are derived from the lensing 
potential 
through linear relations, so that these three fields share the same statistical 
properties.~Hence, cosmic shear is sensitive to structure growth and the 
geometry of the Universe.

The sensitivity of cosmic shear to the growth of structure is particularly 
important in studies of cosmic acceleration, as different dark energy and 
modified gravity models are endowed with different predictions for structure 
growth. As a consequence, it is crucial to include redshift information in a 
cosmic shear analysis, so that the effect of dark energy on structure growth 
can 
be studied in its evolution with redshift.~A two-dimensional analysis (like the 
one 
carried out in \cite{2013MNRAS.430.2200K}, for example) can achieve this goal 
only to a 
limited extent, as it projects quantities along the line of sight; this implies 
loss of redshift information, due to the mixing of spatial scales and to the 
reduced sensitivity to those parameters that, entering the model in a nonlinear 
way, may produce different effects on the lensing signal at different redshifts 
\cite{2012MNRAS.423.3445S}.

To overcome the limitations of a purely two-dimensional analysis, a formalism 
was 
first introduced in \cite{1999ApJ...522L..21H}, which assigns galaxies to 
different redshift bins according to their estimated (photometric) redshift, 
and 
calculates correlations of the lensing signal through redshift bins.~This 
approach is commonly known as \emph{tomography} and is the most common 
methodology to analyse a cosmic shear survey (as used, e.g.,   in 
\cite{2013MNRAS.431.1547B}). The integration along the line of sight that 
characterises a two-dimensional analysis is here reduced to the width of the 
redshift bin; the correlation among different redshift bins provides 
information 
on the evolution in redshift of the lensing signal.~Defining the matter power 
spectrum $P_\delta(k)$ as
\begin{align}\label{eq:matterpowerspectrum_alessio}
\left\langle \delta(\mathbf{k}, z) \delta(\mathbf{k}', z) \right\rangle = (2 
\pi)^3 P_\delta(k, z
) \delta^D (\mathbf{k} - \mathbf{k}'),
\end{align}
and making use of the Limber approximation  \cite{1953ApJ...117..134L, 
1992ApJ...388..272K, LoVerde:2008re}, one can write down the flat-sky 
tomographic convergence power spectrum between tomographic bins $i$ and $j$ as
\begin{align}\label{eq:cij_alessio}
C_{ij}^{\kappa} (\ell) = \int \, \frac{{d}\chi}{\chi^2} \, W_{i}(\ell/\chi, 
\chi) \, W_{j}(\ell/\chi, \chi) \, P_{\delta}(\ell/\chi, \chi),
\end{align}
where the lensing efficiency function $W_i(\ell/\chi, \chi)$ is defined as
\begin{eqnarray}
\label{eq:lensingefficiency_alessio}
W_i(\ell/\chi,\chi) =
\frac{3\Omega_m}{4\chi_H^2} \, \int_\chi^\infty \, 
{d} \chi^\prime \, \frac{{d} z}{{d}\chi^\prime} \, 
\frac{n_i(z(\chi^\prime))}{a(\chi')} \left(\frac{\chi-\chi'}{\chi\chi'} \right) 
\left[1+\frac{1}{\eta(\ell/\chi,\chi^\prime)}\right] \, \mu 
(\ell/\chi,\chi^\prime),
\end{eqnarray}
with $n_i(z(\chi))$ the distribution of sources in the $i$-th bin,  normalized 
to one, $\int \, {d} \chi \, n_i(z(\chi)) = 1$. Clearly, this approach 
still remains an approximation to a purely 3-dimensional treatment of the 
cosmic 
shear field, as it is still characterised by an averaging in redshift, which 
produces loss of information.

An alternative formalism, commonly known as \emph{3D cosmic shear},  makes use 
of a spherical Fourier-Bessel decomposition of the cosmic shear field, to 
include all of the redshift information in the analysis. First introduced in 
\cite{2003MNRAS.343.1327H} and subsequently refined in 
\cite{2005PhRvD..72b3516C, 2006MNRAS.373..105H, 2011MNRAS.413.2923K}, this 
method has so far been applied to real data only in \cite{2014MNRAS.442.1326K}. 
A code comparison between available codes and numerical challenges have been 
discussed in \cite{2018PhRvD..98j3507S}. 3D cosmic shear is based on a 
decomposition of the cosmic shear field in a suitable basis of functions, given 
by a combination of spin-2 spherical harmonics $_2 Y_{\ell m} 
(\hat{\mathbf{n}})$ for the angular components, and spherical Bessel functions 
for the radial coordinate $j_\ell(k \chi)$; together, these functions 
constitute 
the spherical Fourier-Bessel basis. The shear tensor $\gamma (\chi, 
\mathbf{\hat{n}})$ is defined as the second $\slashed{\partial}$ derivative of 
the lensing potential $\Psi$
\begin{align}\label{eq:edth-derivative_alessio}
\gamma (\chi, \hat{\mathbf{n}}) = \frac{1}{2} \slashed{\partial} 
\slashed{\partial}  \Psi (\chi, \hat{\mathbf{n}}).
\end{align}
The shear $\gamma$ can be expanded in the spherical Fourier-Bessel basis as
\begin{align}\label{eq:sphericalFourier-Bessel_alessio}
\gamma (\chi, \hat{\mathbf{n}}) = \sqrt{\frac{2}{\pi}} \sum_{\ell m} \int k^2 
\, 
 {d} k \, \gamma_{\ell m}(k) \, _2 Y_{\ell m} (\hat{\mathbf{n}}) \, 
j_\ell(k \chi),
\end{align}
where the coefficients $\gamma_{\ell m}(k)$ are given by
\begin{align}\label{eq:coefficients_gamma_alessio}
\gamma_{\ell m}(k) = \sqrt{\frac{2}{\pi}} \int \chi^2 {d}\chi \int 
{d}\Omega \,  \gamma(\chi, \hat{\mathbf{n}}) \, j_\ell(k \chi) \, _2 
Y_{\ell m}^*(\hat{\mathbf{n}}).
\end{align}
The covariance of shear modes can be related to the matter power spectrum  
\cite{2016MNRAS.459.1586Z, 2018MNRAS.480.3725S, 2018PhRvD..98j3507S},
\begin{align}
\left\langle\bar{\gamma}_{lm}(k)\bar{\gamma}_{\ell'm'}^*(k')\right\rangle =  
\frac{9 \Omega_m^2}{16 \pi^4 \chi_H^4}\frac{(\ell+2)!}{(\ell-2)!} \int 
\frac{{d}\tilde{k}}{\tilde{k}^2} \, G_\ell(k, \tilde{k}) \, G_\ell(k', 
\tilde{k}) \, \delta_{\ell \ell'}^K \, \delta_{m m'}^K.\nonumber
\end{align}
where
\begin{align}
G_\ell(k,k') &= \int {d}z \, n_z(z) \, F_\ell(z,k) \, 
U_\ell(z,k'),\label{eq:G_alessio}\\
F_\ell(z,k) &= \int {d}z_p \, p(z_p|z) \, j_\ell[k 
\chi^0(z_p)],\label{eq:F_alessio}\\
U_\ell(z,k) &= \frac{1}{2}\int_0^{\chi(z)} \frac{{d} \chi'}{a(\chi')} 
\left(\frac{\chi-\chi'}{\chi \chi'}\right) j_\ell(k \chi')  \, P_\delta^{1/2} 
\left( k, z 
\left( \chi \right) \right) \mu(k, a(\chi)) \left[1 + \frac{1}{\eta(k, 
a(\chi'))}\right] \label{eq:U_alessio} .
\end{align}
The estimates $\bar{\gamma}$ of the pure cosmic shear field $\gamma$ keep into 
account observational effects such  as the redshift distribution $n_z(z)$ of 
the 
lensed galaxies and the conditional probability $p(z_p|z)$ of estimating the 
redshift $z_p$ given the true redshift $z$.
More recently, 3D cosmic shear was used in \cite{2018MNRAS.480.3725S} to 
forecast modified gravity predictions, with a  quantitative comparison with a 
tomographic analysis, whose results we recall below.

\section{Current Data and Forecasts on Horndeski Gravity}\label{sec:forecasts}
\label{cforecastsefs2}
\label{Scalartensoref4}\label{Horndeskiref3}

The Horndeski Lagrangian \cite{1974IJTP...10..363H} is the most general 
scalar-tensor theory of gravity  with a scalar degree of freedom in addition to 
the metric, that respects the following conditions: it is four-dimensional, 
Lorentz-invariant, local and has equations of motion with derivatives not 
higher 
than second order. The latter condition guarantees that the theory is safe from 
Ostrogradski instabilities \cite{Woodard:2006nt}. We will consider only 
universal coupling between the metric and the matter fields (collectively 
described by $\Phi_m$ and contained in the matter Lagrangian $\mathcal{L}_m$), 
which are therefore uncoupled to the scalar field.~The Horndeski action can be 
written as follows:
\begin{align}\label{eq:Horndeski_action_alessio}
S [ g_{\mu \nu}, \Psi ] &= \int {d}^4 x \sqrt{- g} \left[ \sum_{i=2}^5 
\frac{1}{8 \mathrm{\pi} G_N}  \mathcal{L}_i[g_{\mu \nu}, \Psi] + 
\mathcal{L}_m[g_{\mu \nu}, \Phi_M] \right],\\
\mathcal{L}_2 &= G_2 (\Psi, X), \nonumber \\
\mathcal{L}_3 &= -G_3(\Psi, X) \Box \Psi, \nonumber \\
\mathcal{L}_4 &=  G_4 (\Psi, X) R + G_{4X}(\Psi, X) \left[ (\Box 
\Psi)^2  - 
\Psi_{;\mu\nu} \Psi^{;\mu\nu} \right],\nonumber \\
\mathcal{L}_5 &= G_5 (\Psi, X) G_{\mu\nu} \Psi^{;\mu\nu} \nonumber \\
&- \frac{1}{6}G_{5X} (\Psi, X) \left[ (\Box \Psi)^3 + 2 
\Psi_{;\mu}{}^{\nu} 
\Psi_{;\nu}{}^{\alpha} \Psi_{;\alpha}{}^{\mu} -  3 \Psi_{;\mu\nu} 
\Psi^{;\mu\nu} 
\Box \Psi \right]. \nonumber
\end{align}
The subscripts $\Psi, X$ denote partial derivatives, e.g. $G_{iX} = 
\frac{\partial G_i}{\partial X}$.~The choice of the arbitrary functions 
$G_i(\Psi, X)$ of the scalar field $\Psi$ and its kinetic term $X = 
-\frac{1}{2} 
\partial_\mu \Psi \, \partial^{\mu} \Psi$ determines the specific gravity 
model 
considered within this class. Several known models of dark energy and modified 
gravity are contained  within this class, such as quintessence, $f(R)$ and 
Galileon models.


The evolution of linear perturbations in Horndeski gravity can be fully 
described by  four functions of (conformal) time $\tau$ only \cite{Gleyzes2013, 
Bellini:2014fua}:
\begin{itemize}\label{eq:alpha_alessio}
    \item[i)] $\alpha_K$ is the \textit{kineticity} function, representing the 
kinetic  energy of the scalar perturbations arising directly from the action;
    \item[ii)] $\alpha_B$ is the \textit{braiding} function, which describes 
mixing of  the scalar field with the metric kinetic term;
    \item[iii)] $\alpha_M$ is the \textit{Planck mass run rate}, describing the 
rate of  evolution of the effective Planck mass;
    \item[iv)] $\alpha_T$ is the \textit{tensor speed excess}, describing 
deviations of  the propagation speed of gravitational waves from the speed of 
light. This function has recently been constrained to be very close to 0, its 
General Relativity value, by the detection of the binary neutron star 
\label{neutronstarsref9} merger 
GW170817 and the associated gamma ray burst GRB170817A 
\cite{2017PhRvL.119p1101A, GBM:2017lvd}.
\end{itemize}
Constraints on these functions can be obtained from large-scale structure  
\label{LSSefs4}
observations by  choosing a time parametrization, such as the one that traces 
the evolution of the dark energy component $\Omega_{\rm DE}(\tau)$:
\begin{align}\label{eq:alphaparam_alessio}
\alpha_i (\tau) = \hat{\alpha}_i \Omega_{\rm DE}(\tau)\, \quad i = K, B, M, T
\end{align}
and getting constraints on the proportionality coefficients $\hat{\alpha}_i$. 
All of these  functions are identically vanishing in General Relativity, so that 
any detection of a value different from 0 would be a clear signal of deviations 
from Einstein's gravity. This is the idea developed in 
\cite{2018MNRAS.480.3725S} and \cite{2019arXiv190103686S}, using cosmic shear as 
the cosmological probe (alone and in cross-correlation with other observables) 
to constrain Horndeski gravity.

In \cite{2018MNRAS.480.3725S}, the authors present a Fisher matrix forecast for 
the Euclid survey,  with the goal of quantitatively predicting its constraining 
power on Horndeski parameters as introduced above 
Eq.~\ref{eq:alpha_alessio}.
The parameterization chosen for the evolution of the $\alpha$ functions is the 
one described by  Eq.~\ref{eq:alphaparam_alessio}. The authors fix the values 
of 
$\alpha_K$ and $\alpha_T$ to 0, the former being largely uncorrelated with the 
other three functions and unconstrained by large-scale structure probes, the 
latter being strongly constrained by gravitational wave experiments. Moreover, 
they 
present a forecast comparing tomography and 3d cosmic shear, presenting 
expressions for both formalisms in a general modified gravity setting 
(similarly 
to the description provided in Sec.~\ref{sec:weaklensing}). They
simultaneously place constraints on a set of cosmological parameters describing 
the evolution of the background (assumed to be well modelled by a $\Lambda$CDM 
model), as well as on the Horndeski parameters $\alpha_M$ and $\alpha_B$, which 
act at the perturbation level. They find that a 3D analysis can constrain 
Horndeski theories better than a tomographic one, with a reduction of the 
errors 
of the order of 20$\%$ on the Horndeski parameters. 
\begin{figure}[ht!]	
\includegraphics[width=\textwidth]
{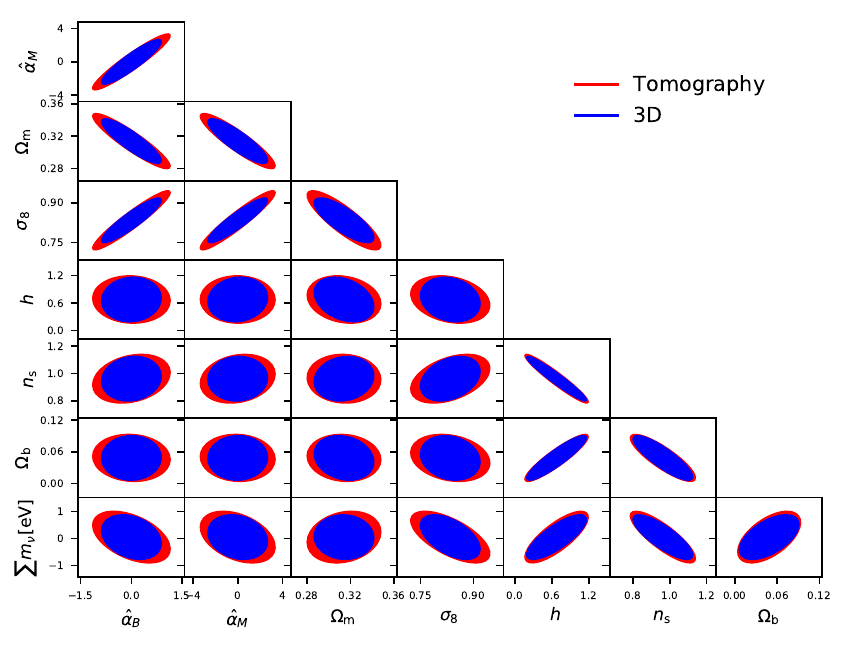}
    \caption{{\it{ 1-$\sigma$ Fisher forecast 
contours for  Euclid-like survey, obtained with tomography (\textit{red}) and 
3D cosmic shear (\textit{blue}). The parameters constrained are a set of 
standard cosmological parameters describing the evolution of the background and 
the Horndeski $\hat{\alpha}_B$ and $\hat{\alpha}_M$ parameters acting on the 
perturbations.~As discussed in \cite{2018MNRAS.480.3725S}, a 3D analysis 
tightens constraints on all standard and Horndeski parameters of about $20\%$ 
with respect to a tomographic analysis. The figure is taken from 
\cite{2018MNRAS.480.3725S}.}}}
    \label{fig:SpurioManciniEtal2018_fig6}
\end{figure}

Despite performing a 
conservative cut in angular and radial scales and only using a linear matter 
power spectrum for the calculation of the covariance of the cosmic shear modes, 
3D cosmic shear performs better than tomography in constraining both standard 
and Horndeski parameters (as shown in Fig.~\ref{fig:SpurioManciniEtal2018_fig6}, 
taken from \cite{2018MNRAS.480.3725S}).~The two methods show similar 
degeneracies, despite being completely independent in their implementation and 
based on two different formalisms.~To illustrate the importance of non-linear 
corrections, the authors produce constraints with 3D cosmic shear and a 
prescription for the non-linear matter power spectrum based on 
\cite{2015MNRAS.454.1958M}; the resulting increase in sensitivity from the 
non-linear corrections calls for the development of nonlinear prescriptions for 
general dark energy models in view of applications to future datasets.

\label{crpssprobesref1}
In \cite{2019arXiv190103686S}, the authors present a cross-correlation analysis  
of cosmic shear, galaxy-galaxy lensing and galaxy clustering tomographic power 
spectra from $\sim$450 deg$^2$ of cosmic shear data from the Kilo Degree Survey 
(KiDS) and two overlapping spectroscopic samples from the GAlaxy and Mass 
Assembly (GAMA) survey.~The goal of this analysis is to provide the first 
constraints on Honrdeski parameters achieved from currently available cosmic 
shear data (alone and in cross-correlation with the other two probes).~The 
methodology followed to model the power spectra extends to a Horndeski gravity 
setting the analysis performed in \cite{2018MNRAS.476.4662V}, carried out in 
$\Lambda$CDM on the same power spectra dataset.~The authors adopt the same 
parameterization for the Horndeski $\alpha_B, \alpha_M$ functions chosen in 
\cite{2018MNRAS.480.3725S} (and given by Eq.~\ref{eq:alphaparam_alessio}), 
finding values for $\hat{\alpha}_B$ and $\hat{\alpha}_M$ compatible with 
$\Lambda$CDM. Interestingly, the values found for $S_8 \equiv \sigma_8 
\sqrt{\Omega_m/0.3}$ (a combination of the parameters $\Omega_m$ and $\sigma_8$ 
particularly well probed by lensing) are in better agreement with the Planck CMB 
\label{CMBrefs11} \label{sigmaref6}
values when the analysis is carried out in Horndeski gravity, rather than in 
$\Lambda$CDM; the tension in the $\Omega_m - \sigma_8$ plane between large-scale 
structure and CMB measurements is largely reduced in Horndeski gravity (see 
Fig.~\ref{fig:SpurioManciniEtAl2019_fig5}).
\begin{figure}[ht!]
	\centering
	\includegraphics[scale=0.53]{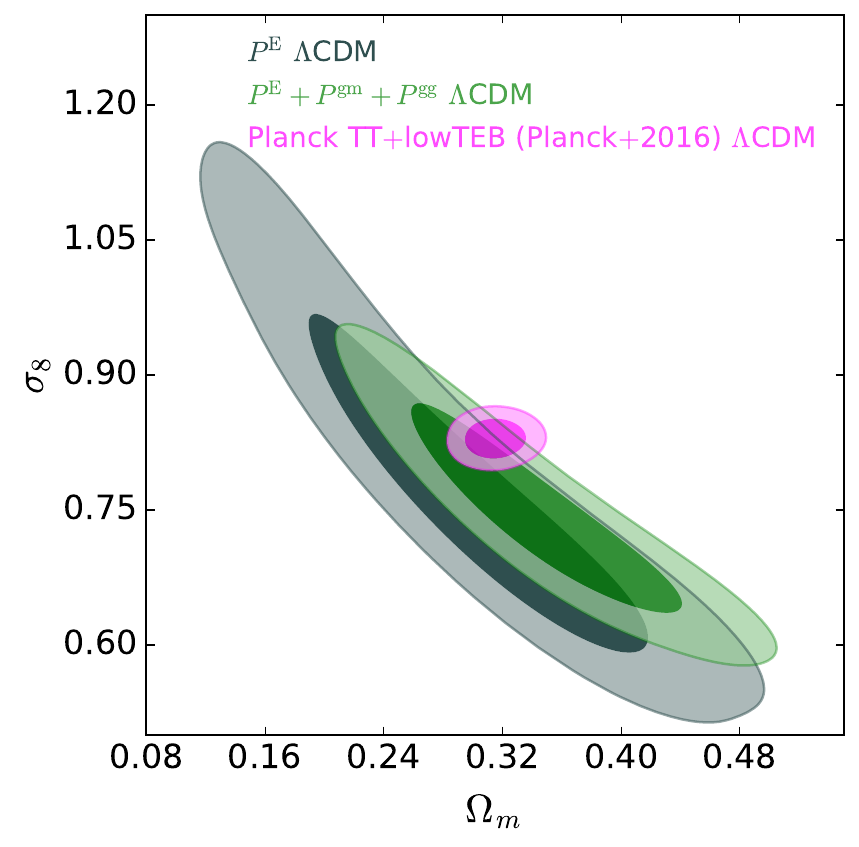}~~~
    \includegraphics[scale=0.53] 
{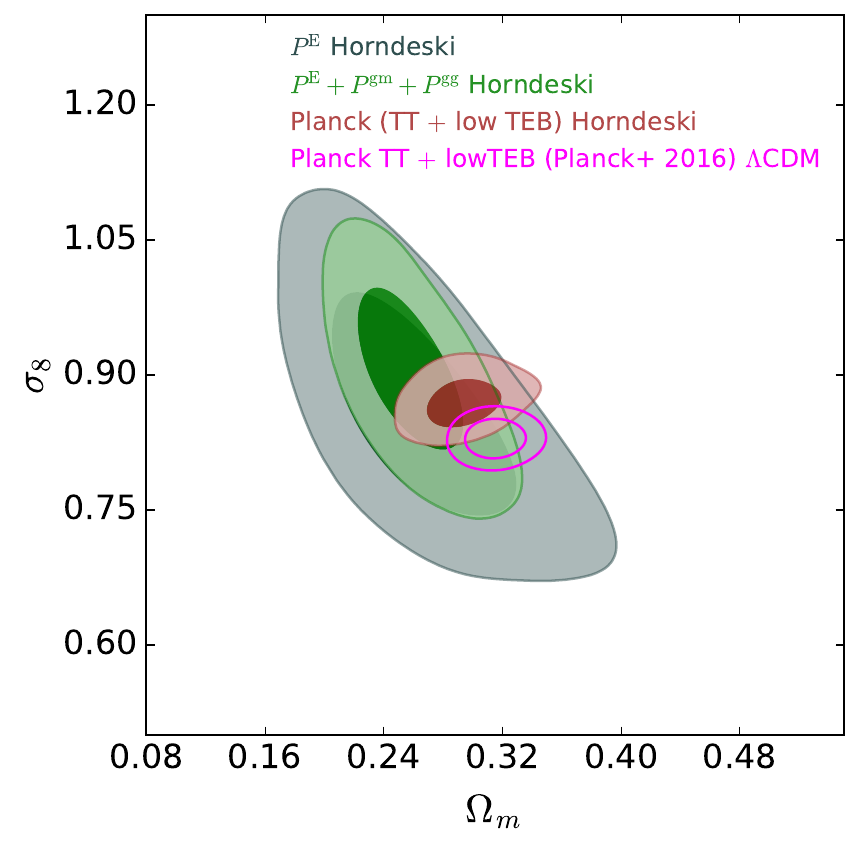}
	\caption{{\it{ 68\% and 95\%  contours 
 on the cosmological parameters $\Omega_{\mathrm{m}}$ and $\sigma_8$. The 
\textit{grey} contours are obtained considering $\sim$450 deg$^2$ of cosmic 
shear data from the KiDS survey; the $\textit{green}$ contours are obtained from 
a joint analysis of cosmic shear - galaxy-galaxy lensing and galaxy clustering 
from the same KiDS samples and two overlapping spectroscopic samples from the 
GAMA survey. In the \textit{left} panel, large-scale structure and CMB probes 
are analysed assuming a $\Lambda$CDM model (the \textit{Planck} contours in 
\textit{magenta} are the same as in \cite{2016A&A...594A..13P}). In the 
\textit{right} panel, the large-scale structure constraints are obtained 
assuming Horndeski gravity; in \textit{brown} we plot the \textit{Planck} 
contours assuming Horndeski gravity, whereas in \textit{magenta} the 
$\Lambda$CDM contours of \cite{2016A&A...594A..13P} (the same as in the left 
panel) are reproduced for comparison.}}}
\label{fig:SpurioManciniEtAl2019_fig5}
\end{figure}

\section{Higher-order Statistics and Lensing Peak Counts}  
\label{sec:higherorder}\label{weaklensefs5}

Using different statistics, beyond the second-order Gaussian power spectrum, 
can 
 help to capture non Gaussian content and better discriminate among different 
cosmological models. An analysis of a variety of different statistics in weak 
lensing observables has been extensively presented in 
\cite{2018A&A...619A..38P}.
In particular, it is relevant to ask the following questions: if a non-standard  
gravity cosmology is mimicking a cosmological constant, can we distinguish the 
two scenarios using weak lensing? Which statistic best discriminates them?
Massive neutrinos are degenerate with the strength of a fifth force  
\label{Fifthref4}
gravitational interaction: higher values of the neutrino mass suppress the 
growth of structure, and can therefore compensate higher values of the strength 
of the fifth force interaction, which would enhance the growth. For example, an 
$f(R)$ model with amplitude $f_{R0} \sim 10^{-5}$ and massive neutrinos of 
$m_\nu \sim 1.5$eV can mimic the  matter power spectrum of a cosmological 
constant model with a neutrino mass of $0.06 $ eV (as currently typically fixed 
in $\Lambda$CDM).
Authors in \cite{2018A&A...619A..38P} then used hydro simulations for 
$\Lambda$CDM  and different $f(R)$ cosmologies (of the type Hu-Sawicki), built 
on purpose to be degenerate in their matter power spectra. They then compared 
different statistics in weak lensing observables, including variance, skewness, 
kurtosis and peak counts, i.e. the number count of lensing peaks in their 
aperture mass maps. \label{peaksef1}
\begin{figure}[ht!]
\centering
	\includegraphics[width=243pt]{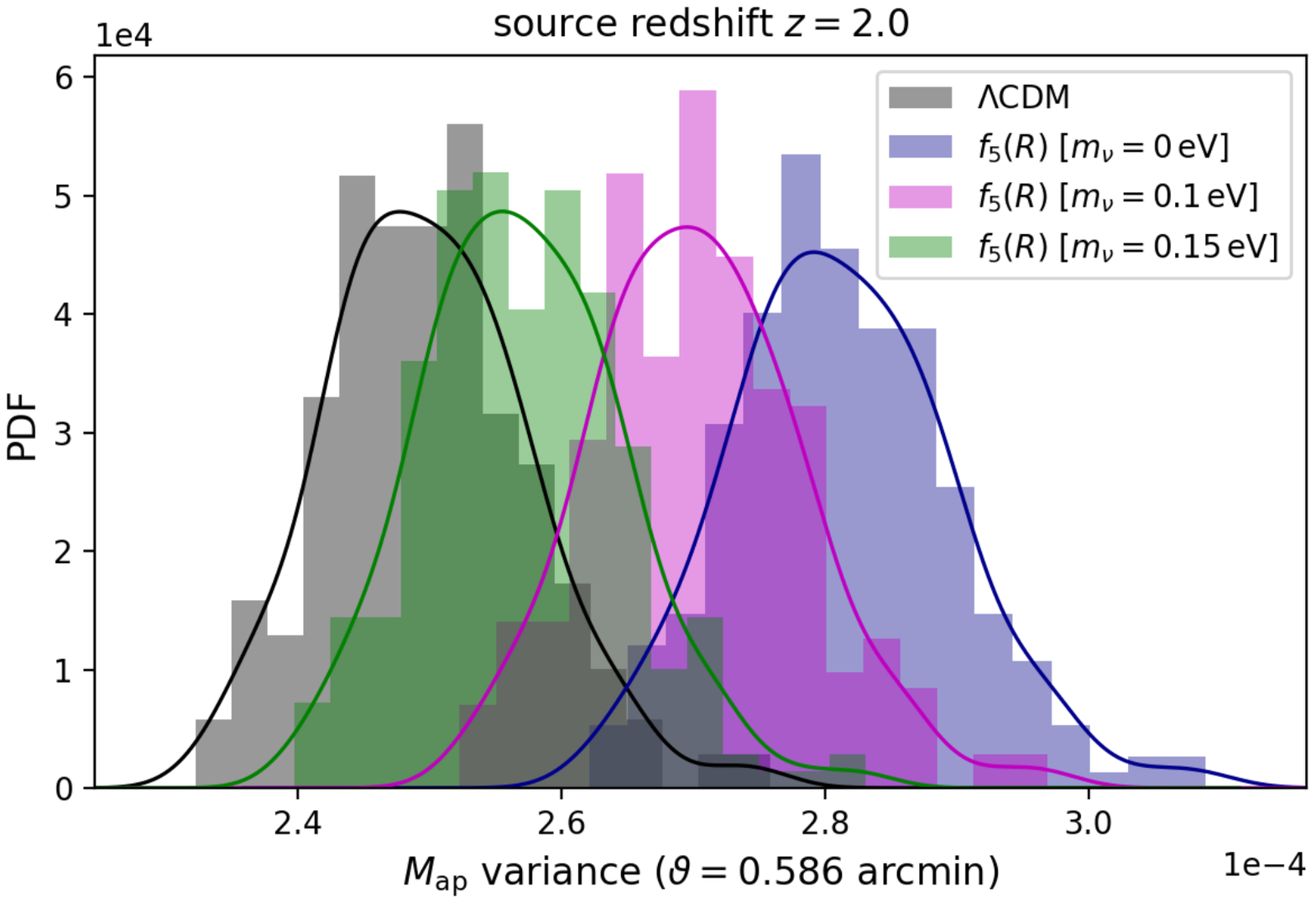}\\
~~~\\
    \includegraphics[width=243pt]{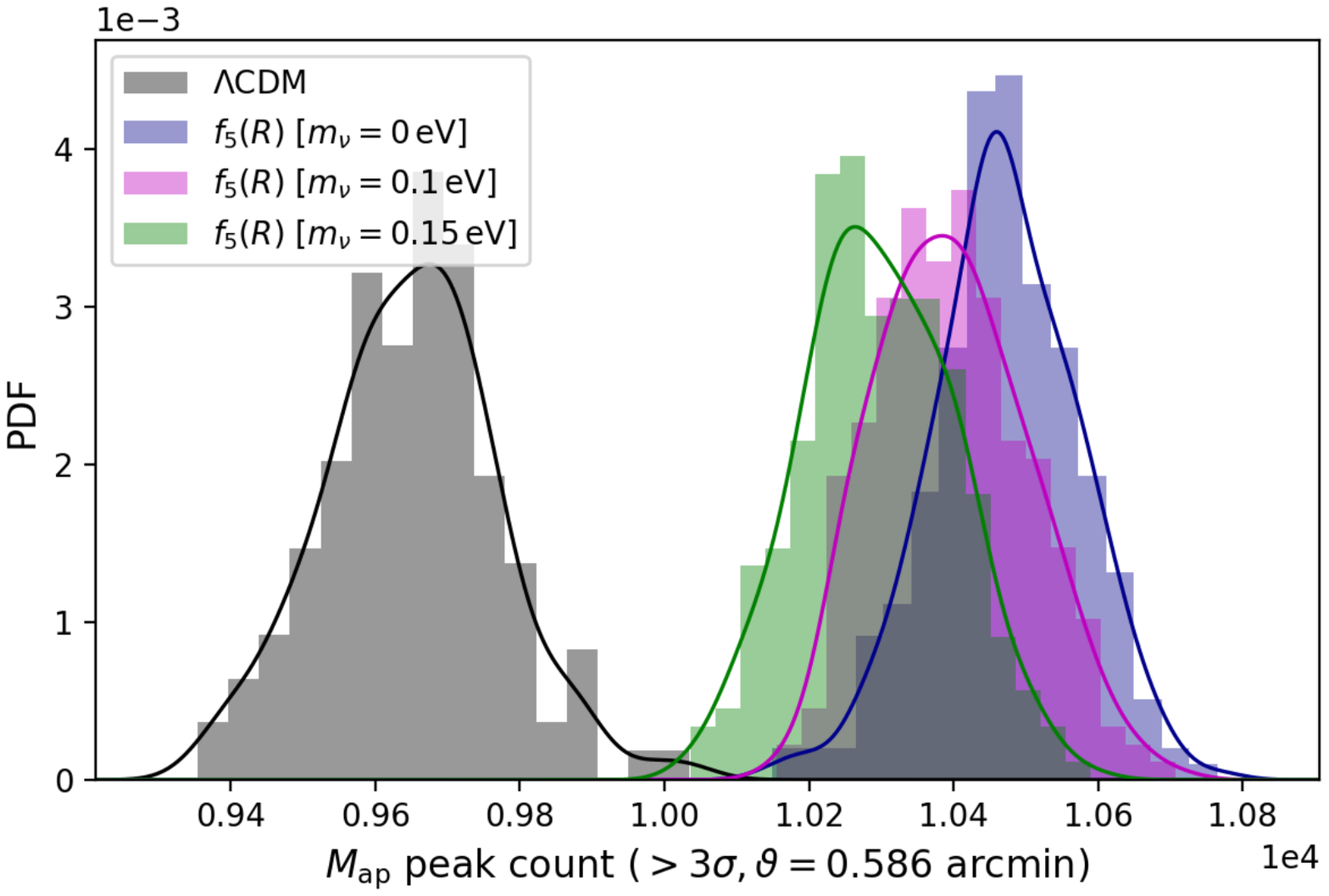}
    \caption{
    {\it{
   Histograms of  
aperture mass statistics for $\Lambda$CDM and $f_5(R)$ models (i.e. Hu-Sawicki 
models with amplitude $f_{R0} = 10^{-5}$) and different values of the neutrino 
mass $m_\nu$. Each histogram, with area normalised to one, comprises 256 
samples 
of the statistic computed at a filtering scale of $\vartheta=0.586'$ and for 
sources at redshift $z_s=2.0$. Solid lines represent the result of smoothing 
the 
distribution by KDE (cf. Sect.~5.3 in \cite{2018A&A...619A..38P}).
  Considering the most degenerate case with $\Lambda$CDM, $f_5(R)$ with 
$m_\nu=0.15~\mathrm{eV}$, second- and higher-order moments of $M_\mathrm{ap}$ 
do 
not appear able to distinguish the models. Peak counts, on the other hand, 
shown 
here for a $3\sigma$ threshold, cleanly separate the two distributions. It is 
interesting to note that peak counts separate all $f_5(R)$ cases from 
$\Lambda$CDM by  approximately the same amount, independent of $m_\nu$.   
The figure is taken from  \cite{2018A&A...619A..38P}.
}}}
    \label{fig:PeelEtal2018_fig9}
\end{figure}

Results show that peak counts best capture non-Gaussian information and 
represent  the statistic that has a higher chance to discriminate between $f(R)$ 
and $\Lambda$CDM models, with a discrimination efficiency that depends on 
redshift and angular scale of observation. Figure \ref{fig:PeelEtal2018_fig9} 
from \cite{2018A&A...619A..38P} nicely shows this effect for a specific 
filtering scale.

Peak counts are therefore a promising tool for future weak lensing surveys.  In 
addition, as shown in forecasts presented    in 
\cite{2019PhRvD..99f3527L}, 
combining peak
counts with lensing power spectrum can improve the constraints on the sum of  
neutrino masses, on the relative matter density $\Omega_m$, and on the 
primordial amplitude $A_s$ by factors $39\%$, $32\%$, and $60\%$ respectively, 
as compared to constraints derived from the power spectrum alone 
\cite{Ajani:2020dvu}. More recently, in  \cite{Ajani:2021pgp} the authors 
proposed a new statistics that joins peaks and voids, and avoids the problem of 
defining what is a peak or what is a void.

\section{Machine Learning and the Dark Universe} \label{sec:machinelearning}
\label{machineref1}

Machine learning has recently seen an increase in applications in all fields,  
including cosmology for which new opportunities and challenges have been 
recently summarised in \cite{Ntampaka:2019udw}.
Convolutional Neural Networks (CNN) have been used in particular on weak lensing 
observables,  trained on convergence maps, to discriminate models along the 
$\Omega_m, \sigma_8$ degeneracy \cite{2017arXiv170705167S, 2018PhRvD..97j3515G, 
2018PhRvD..98l3518F}. In \cite{2019NatAs...3...93R} the authors also showed that 
the network can exploit information related to the steepness of local peaks, 
rather than to their amplitude. More recently, it was shown in 
\cite{2019PhRvD.100b3508P} that CNN can break the degeneracy discussed above 
between neutrino masses and the dark universe, significantly outperforming all 
statistics, including peak counts.

We briefly recall here the main result developed in \cite{2019PhRvD.100b3508P}, 
as this  directly compares with what discussed in \ref{sec:higherorder} and the 
python code used in the analysis has been made publicly available: specifically, 
authors apply CNN to discriminating between $\Lambda$CDM cosmologies and $f(R)$ 
(Hu-Sawicki) models with massive neutrinos. The authors start from one 
simulation 
per model: this is possible for a classification problem, for which simulations 
are done on purpose for models which are degenerate at the level of the power 
spectrum. Convergence maps are then obtained with random reorientation in the 
same simulation run; furthermore, a compressed representation of the input is 
used, which reduces the dimentionality of the data and speeds up the training. 
As known, the network learning procedure consists in updating the parameters 
(weights) in the cost function via gradient descent and back-propagation in 
order to match the desired output. This learning (training) process is done on 
75$\%$ of the available input data (for which labels are known) and tested on 
the remaining 25$\%$ of input data. Validation accuracy (i.e. the ratio of 
correct predictions to the total number of test observations) has been shown to 
go from 92$\%$ (for a noiseless case) down to 48$\%$ for a pessimistic noise 
level. 

The results in \cite{2019PhRvD.100b3508P} show that the CNN is able to 
discriminate $\Lambda$CDM from $f(R)$ gravity better than other statistics, 
including peak counts, for all choices of noise levels. For example, for an 
intermediate/optimistic noise level ($\sigma = 0.35$ standard deviation in 
Gaussian random noise), $\Lambda$CDM can be discriminated with 79$\%$ accuracy 
(against $30 \%$ maximum accuracy for peak count statistic, for the same 
redshift). Including all four source redshifts available $\{0.5, 1, 1.5, 2\}$ 
further increases CNN accuracy to $87\%$
for the same noise level. With respect to peak count statistic, CNN also seems  
to be more efficient in discriminating among different neutrino masses, within 
$f(R)$ scenarios. Different types of machine learning techniques were also 
tested on the same simulations in \cite{2019MNRAS.487..104M}, finding that CNN 
is the one that best performs, among the ones tested.

While the results are promising for classification problems, this proof of 
concept  
opens the path to new challenges. First, one may want to also address a 
regression problem, i.e. infer cosmological parameters from real data: in this 
case, one can expect many more simulations to be needed, and a different 
architecture to serve for regression. Second, one may expect weak lensing 
systematics to also play a role when dealing with real data, and it is not clear 
at this stage if machine learning will be robust to these systematics. This has 
to be investigated in the future.

\chapter[Galaxy Clusters and Modified Gravity]{Galaxy Clusters and Modified \\ 
Gravity}
\label{sec:Saltas}
\label{galaxyclurefs4}

{\em Ippocratis D.~Saltas and Lorenzo Pizzuti}

 



\section[What Makes Galaxy Clusters Interesting for Testing Gravity?]{What Makes 
Galaxy Clusters Interesting for Testing \\Gravity?}

Put simply, a galaxy cluster is a self-gravitating system  built out of dark 
matter, hot gas and baryonic tracers in the form of galaxies.
What makes galaxy clusters attractive as tests of gravity is that they lie on 
the borderline  between astrophysical and cosmological scales, allowing to test 
different aspects and predictions of gravity theories.

Traditionally, the combination of kinematical/dynamical measurements  in a 
cluster with lensing observations allows to reconstruct the underlying mass 
distribution. In General Relativity (GR) the two are expected to be the same, 
however, this is no longer true within general families of theories beyond GR 
which predict that pressureless matter and light respond differently to gravity, 
in turn implying different predictions for the cluster's inferred 
dynamical/kinematical and lensing mass profiles. This idea has formed the basis 
for the construction of consistency checks studied predominantly within 
scalar--tensor theories such as $f(R)$, Brans--Dicke and (Beyond Horndeski) 
theories.
 
Disentangling genuine gravitational effects from the complicated astrophysical  
processes at cluster scales is, however, a subtle task and requires an adequate 
knowledge of the underlying systematics. In addition, since the bulk of the 
cluster's mass comes from dark matter, an adequate modelling of the dark matter 
density distribution is necessary, which proves challenging without knowledge of 
the actual physics of the dark sector. Therefore, the underlying assumptions for 
constraints on gravity in this regard have to be challenged and testes before 
conclusive statements are to be made.

Our goal here is to briefly review progress on the methods and constraints on 
theories  beyond GR with the physics of galaxy clusters, along with a discussion 
on the associated hurdles from astrophysical and observational systematics.

\section[Consistency Conditions Based on the Mass Profiles of Galaxy Clusters] 
{Consistency Conditions Based on the Mass \\ Profiles of Galaxy Clusters}

\subsection{Generalities}

It is instructive to start with a recap of some useful concepts from 
cosmological perturbation theory.  In galaxy clusters, matter collapses in a 
non-linear fashion, however, gravity remains linear implying that, the 
gravitational wells are sufficiently small for linear perturbation theory to 
hold. In this regard, the Poisson equation relates the Newtonian potential 
$\Psi$ to the matter density as $\nabla^2 \Psi = 8 \pi G_N \rho_{m}$, while 
the gravitational slip parameter relates $\Psi$ to the relativistic potential 
$\Phi$ through the gravitational slip parameter $\eta \equiv \Psi/\Phi$. The 
weak-lensing potential satisfies a Poisson-like equation as $\nabla^{2}(\Phi + 
\Psi) = 8 \pi G_N \rho_{lens}$, where $\rho_{lens}$ is the matter density 
inferred 
through lensing probes -- we will get back to this later.

Hydrostatic equilibrium in a galaxy cluster is achieved through the balance 
between gravity  and the gas pressure as
\begin{equation}
\frac{d P_{\text{tot.}}}{dr} = \rho_{\text{gas}}\frac{d \Psi}{dr}, 
\label{ippo-hydro}
\end{equation}
with the total pressure given by the sum of thermal and non-thermal pressure 
$P_{\text{tot.}}  = P_{\text{therm.}} +P _{\text{non-therm.}}$. The thermal 
pressure comes predominantly from the hot gas, as  $P_{\text{therm.}} = 
n_{\text{gas}} k T_{\text{gas}}$. Writing $d \Psi/dr = G_N M(r)/r^2$, equation 
(\ref{ippo-hydro}) then provides a definition of the (non-)thermal mass profile.
 The bulk of the pressureless matter sourcing the potential $\Psi$ comes from 
dark matter. The most popular density profile to model it is the 
Navarro-Frenk-White (NFW), which is a 2-parameter profile depending on a 
characteristic density and radius as
\begin{align}
\rho_{\text{NFW}} = \frac{\rho_{s}}{(r/r_s)(1 + r/r_s)^2}, \; \; \; \;  \rho_s 
\equiv \frac{1}{3} \cdot \frac{\rho_{c}(z) \Delta_{\text{vir}} \cdot 
c^3}{\log(1+c) - c(1+c)^{-1}},
\end{align}
with $\rho_{c} \equiv 3H^{2}(z)/(8 \pi G_N)$ corresponds to the critical 
density 
of the Universe. The concentration parameter, $c \equiv r_{\text{vir}}/r_{s}$, 
relates the viral radius with the  scale $r_s$, with the former defining the 
region of the cluster that encloses a mean over density equal to $\rho_{c} 
\Delta_{\text{vir}}$. The popularity of the NFW profile relies  in that it 
provides a good fit to haloes within N-body simulations, and for large range of 
masses both in Newtonian and modified gravity. In particular, Ref. 
\cite{Schmidt:2008tn} showed that the NFW profile provides equally good fits to 
N-body simulations in $f(R)$ gravity as the standard Newtonian case, which was 
later confirmed by Ref. \cite{Lombriser:2012nn} considering the particular case 
of the Hu-Sawicki $f(R)$ model. Since here we will be mostly interested in 
constraints on the theory space of modified gravity, for more details on actual 
the hallo modelling beyond GR we refer to \cite{Schmidt:2008tn, Li:2011uw,  
Lombriser:2012nn, Lombriser:2013wta, Lombriser:2013eza}.


\subsection{Probes Based on Mass Profiles from Galaxy Kinematics and Lensing}
\label{massprofef1}
\label{weaklensefs7}

Under the assumption of spherical symmetry, the Poisson equations for the 
potentials $\Psi$ and $\Phi + \Psi$ allow us  to derive the following 
expressions
\begin{align}
& \Psi(r) = G_N \int_{r_0}^{r} \frac{ds}{s^2} M_{dyn}(s), \label{ippo-Psi} \\
& \Phi(r) = G_N \int^r \frac{ds}{s^2} \left( 2M_{lens} - M_{dyn} \right). 
\label{ippo-Phi}
\end{align}
Equation (\ref{ippo-Psi}) serves as a definition of the dynamical mass of the 
cluster, i.e the mass inferred from dynamical  probes, while it is easy to see 
that for the lensing mass $M_{lens} = \frac{r^2}{2G} \frac{d}{dr} (\Phi + 
\Psi)$. Assuming that member galaxies are collision less tracers of the 
underlying gravitational field $\Psi$, their velocity dispersion field 
satisfyies the Jeans equation
\begin{equation}\label{ippo-jeans}
\frac{\partial (\nu\sigma_r^2)}{\partial 
t}+2\beta(r)\frac{\nu\sigma^2_r}{r}=-\nu(r)\frac{\partial \Psi}{\partial r} \, ,
\end{equation}
where $\nu(r)$ is the number density of tracers, $\sigma^2_r$ the velocity 
dispersion along the radial direction and $\beta \equiv 
1-(\sigma_{\theta}^2+\sigma^2_{\phi})/2\sigma^2_r$ the velocity anisotropy 
profile. The latter, accounts for the neglected velocity component in  
along-the-light-of-sight observations, and it introduces an important source of 
uncertainty which has to be accounted for, as we will discuss in Section 
\ref{sec:syst}. In principle, the choice of $\nu(r)$ is also model dependent 
and 
its effect on any constraints needs to be investigated.

The combination of lensing and dynamics of a galaxy cluster can provide a 
powerful test based on the gravitational slip parameter $\eta$.  Combining 
(\ref{ippo-Psi}) and (\ref{ippo-Phi}) we may derive an expression for the 
gravitational slip in terms of the dynamical and lensing mass profiles as
\begin{equation}
\eta(r) =\frac{\int^r 
\frac{ds}{s^2}\left[2M_{\text{lens}}(s)-M_{\text{dyn}}(s)\right]}{\int^r 
\frac{ds}{s^2}M_{\text{dyn}}(s)} \, .  \label{ippo-eta}
\end{equation}
Therefore, the existence of gravitational slip can be viewed as a consistency 
condition between the dynamical and lensing mass of the cluster --   In GR, and 
in the presence of perfect fluid matter, it is $M_{dyn} = M_{lens}$ and $\eta = 
1$, but this is not true anymore as soon as gravity is modified. In this view, 
the existence of new gravitational degrees of freedom will manifest itself as a 
tension in the mass profile inferred from kinematics and lensing.

The above idea formed the basis of Ref. \cite{Pizzuti:2016ouw} which combined 
kinematical and lensing observations of the relaxed cluster  MACS J1206.2-0847 
(hereafter MACS 1206), at redshift $z=0.44$ (at redshift $z = 0.44$) from the 
CLASH\footnote{http://www.stsci.edu/~postman/CLASH/} and 
CLASH-VLT\footnote{https://kyle.na.astro.it/CLASH-VLT/Public/index.html} 
surveys 
to reconstruct the slip parameter $\eta$ as a function of the distance from the 
cluster's center. In particular, it considered line-of-sight velocity 
measurements and projected positions for $592$ member galaxies to perform a 
phase-space analysis using the code \emph{MAMPOSSt} of Ref. \cite{Mamon01}, 
which solves the Jeans equation (\ref{ippo-jeans}) to provide a maximum 
likelihood fit to the mass profile parameters. Assuming an NFW profile, a model 
for $\beta(r)$, and a Newtonian form for $\Psi$, combination of dynamics and 
lensing led to the constraint
\begin{align}
\eta(r = 1.96 \text{Mpc}) = 1.00^{+0.31}_{-0.28} \; (\text{statistical}) \pm  
0.35 \; (\text{systematic}).
\end{align}
The assumption of the NFW profile for the total matter distribution was 
challenged by repeating the analysis  with an Hernquist and Burkert profile, 
where it was found that NFW provided the highest likelihood fit to the 
kinematic 
data.
The same concept was followed in Ref. \cite{Pizzuti:2019wte}, which forecasted 
the ability of galaxy clusters to constraint $\eta$ using the procedure 
outlined 
above. In particular, dynamical mass profiles were re-constructed from a set of 
$15$ spherical mock clusters in equilibrium solving (\ref{ippo-jeans}), 
followed 
by a maximum likelihood analysis for the NFW parameters $(r_s, r_{200})$. 
Lensing information was simulated from the based on the observations of MACS 
1206. Results showed that $\eta$ can be constrained at the $\sim 9 \%$ level 
($2 
\sigma$) when assumed to be scale-independent, and at $21 \%$ when 
scale-dependence is accounted for.

Ref. \cite{Pizzuti:2017diz} implemented a similar combination of kinematics and
lensing for MACS1206 with real data, but introducing the effect of the 
fifth-force in the gravitational potential within $f(R)$ gravity and a 
simplified approach for screening. Under the assumption of an NFW profile and 
the form of the anisotropy profile, it quoted the upper bound on the fifth 
force's Compton wavelength as
\begin{align}
\lambda_{f(R)} \leq 1.61\; (\text{statistical}) + 0.30 \; (\text{systematic}) \; 
\; \text{Mpc}.
\end{align}
Notice that, $\lambda_{f(R)}$ is related to the mass of the scalar field as 
$\lambda_{f(R)} \sim 1/ m^{2}_{f(R)} \sim f_{RR}$.

We notice that, constraints on gravity from statistics of a sample of galaxy 
clusters should be in principle weighted over an appropriate mass function. In 
fact, the abundance of clusters in modified gravity have provided tight 
constraints on the allowed theory space of $f(R)$ scalar-tensor theories, since 
the fifth force modifies the collapse of matter at large scales, leading to an 
enhancement of the mass function. This has been the topic of investigation in 
Refs \cite{Schmidt2009, Rapetti2010, Lombriser:2010mp, Rapetti2011, 
Cataneo2015, 
Ferraro2011, Cataneo:2016iav}. In particular, Ref. \cite{Lombriser:2010mp} 
combined geometrical probes with CMB and cluster abundance data to quote an 
upper bound on the parameter $B$ related to the scalaron's Compton wavelength 
\footnote{$B \equiv \frac{f_{RR}}{1+f_R} R' \frac{H}{H'}$, with  $' \equiv 
d/dlna$ where $a(t)$ is the  scale factor and $R$ the Ricci scalar.} 
for the so--called designer $f(R)$ model as $B(z = 0) < 1.1 \cdot 10^{-3}$ ($95 
\% \; \text{C.L}$). An updated analysis using a similar combination of 
observables and galaxy clusters up to $z \sim 0.5$ derived the tighter 
constraint $B(z= 0) < 1.78 \cdot 10^{-4}$ ($95.4 \% \; \text{C.L}$) 
\cite{Cataneo2015}. It should be noticed that, the designer model does not 
assume a fixed, a priori functional form for $f(R)$, but rather fixes it 
implicitly by requiring that the predicted background evolution of the Universe 
matches with observations.

\subsection{Probes Based on Thermal and Lensing Mass Profiles} \label{Bransref6}

 An alternative route to test scalar-tensor theories with a  conformal coupling 
between the scalar field and curvature, such as Brans--Dicke or $f(R)$ 
theories, 
can be followed through the construction of tests based on the cluster's 
thermal 
and lensing mass. In particular, the coupling between the scalar field and 
matter is expected to have a direct impact on the cluster's inferred thermal 
mass, but not on the lensing one, since photons travel on null geodesics. This 
is the main idea followed in \cite{Terukina2012}, which considered the 
phenomenological implications of the coupling between the chameleon scalar 
field 
with the baryonic and dark matter component in the cluster. 

The chameleon field   \label{chameleonkiref6}
is sourced by the scalar potential and matter density according to
\begin{align}
\nabla^2 \phi = \frac{\partial V(\phi)}{\partial \phi} +  \beta\sqrt{8\pi 
G_N} 
\rho e^{\beta\sqrt{8\pi G_N} \phi},
\end{align} \label{Fifthref5}
with $\beta$ a dimensionless coupling strength, and $ \beta = 1/\sqrt{6}$ for 
the case of $f(R)$ gravity.  Sufficiently deep within the cluster, $\nabla^2 
\phi \approx 0$, and the scalar field acquires a minimum, $\phi_{0}$. In this 
region, the fifth-force is screened. Towards the outskirts of the cluster, a 
sizeable field gradient builds up, leading to a fifth-force effect with 
$F_{\phi} = - \beta\sqrt{8\pi G_N} \frac{d\phi}{dr}$. The contribution of the 
fifth-force to the r.h.s of the hydrostatic equilibrium (\ref{ippo-hydro}) will 
in turn affect the gas and temperature profiles of the cluster.
Typically, one assumes that outside the cluster the chameleon scalar acquires 
its ambient  cosmological value, $\phi_{cosm.}$, which is the free, model 
parameter to be constrained. In the language of $f(R)$ gravity, \label{fRref8}
\begin{align}
\frac{\partial f}{\partial R} \equiv f_{R} = - \sqrt{\frac{16\pi G_N}{3}} 
 \phi . \label{ippo-phi-f_R}
\end{align}
Ref. \cite{Terukina2012} started with a generalised NFW profile for dark 
matter, $\rho = \rho_{s}/\left[(r/r_s)  (1 + r/r_s)\right]^b$, and a polytropic 
one for the gas. In the presence of the chameleon field, the gas distribution 
becomes steeper at the outskirts of the cluster (compared to GR), where the 
fifth force is operative. From an observational viewpoint,  this in turn causes 
a decrease in the predicted X-ray surface brightness of the cluster at large 
radii. Using the X-ray temperature profile observations from the Hydra A 
cluster, Ref. \cite{Terukina2012} was able to derive the bound $\phi_{cosm.} < 
10^{-4}/\sqrt{8\pi G_N}$ (at redshift $z = 0$) assuming $\beta = 1$. Notice 
that, 
this result is insensitive to the details of the potential $V(\phi)$.

The case of $f(R)$ gravity ($\beta = 1/\sqrt{6}$) was studied with a more 
thorough analysis in \cite{Terukina:2013eqa}  adopting a conceptually similar 
strategy. The work reconstructed the 3-dimensional X-ray temperature and 
surface brightness profiles, as well as the expected Sunyaev-Zel’dovich (SZ) 
effect in the presence of the chameleon force under sufficiently general 
assumptions for the modeling of the gas temperature profile and pressure. With 
the aid of an MCMC analysis, and the observations of the Coma cluster, the 
best-fit values for the gas/dark-matter and modified gravity parameter space 
were inferred, leading to the $2\sigma$ constraints $\phi_{cosm.} \lesssim 7 
\cdot 10^{-5} /\sqrt{8\pi G_N}$. Under (\ref{ippo-phi-f_R}) this translates to
\begin{align}
f_{R_0} \lesssim 6 \cdot 10^{-5}.
\end{align}

\label{screeningref3}
In a similar context, Ref. \cite{Wilcox:2015kna} extended the analysis for 
multiple clusters,  analysing  $58$ stacked cluster profiles observed within 
$0.1 < z < 1.2$, using X-ray and lensing data from the XMM Cluster Survey and 
the Canada France Hawaii Telescope Lensing Survey respectively. It is important 
that, the clusters were found to be living in unscreened environments, since 
otherwise the fifth force would be environmentally screened. The analysis 
concluded that, $f_{R_0} \lesssim 6 \cdot 10^{-5}$ at $2\sigma$. The 
methodology of the latter works was further verified in \cite{Wilcox:2016guw}, 
through the study of hydrodynamical simulations in $f(R)$ gravity. Most notably, 
it confirmed the validity of the NFW profile when modelling weak-lensing mass 
profiles in $f(R)$ gravity and the spherical symmetry of the stacked cluster 
profiles.

Galaxy clusters have been also employed to probe a broader part of the theory 
space  of scalar-tensor theories, beyond a conformal coupling. In particular, 
the so--called Beyond Hortndeski theories have been shown to exhibit an 
intriguing breaking of the Vainshtein mechanism within compact matter sources 
such as stars or galaxy clusters. A fundamental difference with 
conformally-coupled theories is that, here, lensing is directly affected by the 
fifth force. In particular, for the Beyond Horndeski theories exhibiting a  
\label{beyondHorndeskiref5}
breaking of the Vainshtein mechanism, the two scalar gravitational potentials 
{\it within} a compact object \label{compactobrefs5} can be shown to be,
\begin{align}
& \frac{d \Psi(r)}{dr} =  - \frac{G_N M(r)}{r^2} +  \frac{Y_{1} G_N}{4} \cdot 
\frac{d^{2} M(r)}{dr^2}, \label{ippo-BH-Phi} \\
&\frac{d \Phi(r)}{dr} =  - \frac{G_N M(r)}{r^2} +   \frac{5 Y_{2} G_N}{4 r} 
\cdot \frac{d M(r)}{dr},  \label{ippo-BH-Psi}
\end{align}
where $Y_1$ and $Y_2$ are dimensionless couplings. For $Y_1 = 0 = Y_2$ the 
standard  expressions are recovered. The gas in the cluster responds to $\Psi$, 
hence will be sensitive to $Y_1$, whereas the lensing potential will probe both 
$Y_1, Y_2$. Ref. \cite{Sakstein:2016ggl} used the same data and methodology of 
\cite{Wilcox:2015kna} for the modelling of the X-ray and lensing profiles to 
produce stacked profiles of $58$ clusters. A simultaneous fit of the X-ray and 
lensing to the data with an MCMC analysis, lead to the following constraints for 
the fifth-force couplings,
\begin{align}
Y_{1} = -0.11^{+0.93}_{-0.67}  \; \; \; \text{and} \; \; \; Y_{2} = -0.22^{+1.22}_{-1.19}.
\end{align}

In the context of Beyond Horndeski theories it has been also investigated 
whether the  fifth force associated with the breaking of Vainshtein mechanism 
could mimic the effect of dark matter in galaxy clusters. In particular, Ref. 
\cite{Salzano:2016udu} considered a sample of galaxy clusters from the CLASH 
survey, and reconstructed their gas density from the observed X-ray profiles. It 
was then shown that for the particular sample of clusters, and assuming a 
$\Lambda CDM$ background cosmology, the model can provide a good fit to the 
lensing of observations without the introduction of dark matter component. In a 
similar follow up analysis by the same authors, and assuming that $Y_1 = Y_2$, 
an upper bound at $2\sigma$ was derived as \cite{Salzano:2017qac}
\begin{align}
Y_{1} \leq 0.16.
\end{align}

\section{A Brief Discussion on Systematics}\label{sec:syst}

The previously discussed constraints rely on various simplifications regarding 
the  modeling of the cluster, e.g the assumption of relaxation. Departures from 
these assumptions in realistic observations introduce systematics which need to 
be accounted for, if constraints need to be consistent and robust. Here, we will 
briefly discuss the impact of such systematics.

The Jean's equation is a powerful method to infer the local potential $\Phi$ in 
the cluster, however, applications of the method are limited by observational 
constraints, such as the fact that only the velocity dispersion along the line 
of sight $\sigma^2_r$ and the projected number density profile of galaxies can 
be measured directly. Tangential velocities are generally small and direct 
measurements of the velocity anisotropy are complicated. To infer $\beta(r)$ one 
can proceeds parametrically, assuming a model for the anisotropy and determining 
the parameters of the profile along with the mass profile with a Maximum 
Likelihood fit to the data (see e.g. Ref. \cite{Biviano01}). Nevertheless, some 
non-parametric techniques based on inverting the Jeans' equation can be found in 
literature (e.g., Refs. \cite{Binney1982,Host2}), but their application 
generally 
requires additional information and assumptions. From both observations of 
galaxy clusters (e.g., Ref. \cite{Host2009}) and analyses of halos in 
cosmological simulations (e.g., Refs.  \cite{Hansen2006,mamon10}) it has been 
found that generally orbits tend to be isotropic in the center (i.e. $\beta=0$) 
while the anisotropy grows with radius.
\begin{figure}[ht!]
\centering
\vspace{-0.9cm}
\includegraphics[width=1\textwidth]{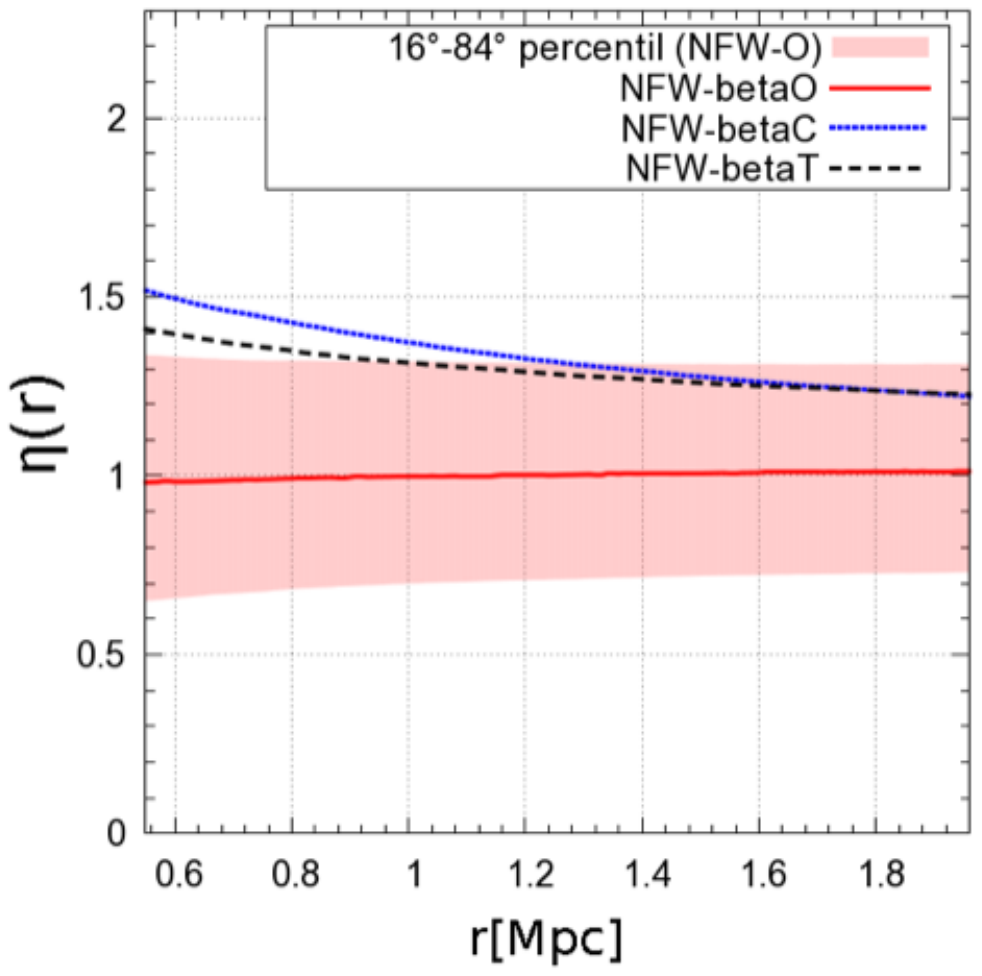}
\vspace{-1.4cm}
\caption{ {\it{From Figure 3 of Ref. \cite{Pizzuti16}:  
constraints on $\eta(r)$ obtained for the CLASH galaxy cluster MACS 1206 
combining lensing and dynamics mass profile determinations. The mass profile is 
parametrized as a NFW model; different lines correspond to different ansatz in 
the  velocity anisotropy parametrisation. The red shaded area indicates the 
region between the 16th and 84th percentile for the reference model, ``O'' 
profile.}}}
\label{fig:etabeta}
\end{figure}

In Fig. \ref{fig:etabeta} (right panel of Figure 3 in Ref. \cite{Pizzuti16}) 
we  show the constraints on the gravitational slip, obtained by combining 
lensing and dynamics mass profiles of the cluster MACS 1206 by Ref. 
\cite{Pizzuti16}, changing the parametrisation of the velocity anisotropy 
profile in the dynamical analysis. All the three models used in the fit - the 
constant anisotropy ``C'', the Tiret model ``T'' of Ref. \cite{Tiret01} and the 
Opposite model ``O'' of Ref. \cite{Biviano01} - produce bounds on $\eta$ in 
agreement within the 68\% C.L. given the current uncertainties. However, with 
the expected precision achievable from future surveys (see below), these effects 
would become a relevant source of systematics.

A natural question arising when reconstructing mass profiles through a Jeans 
analysis is the dependence  of the results on the number of tracers considered, 
since realistic observations typically come with a restricted number of 
spectroscopic velocity measurements. In addition, understanding of this may aid 
the optimisation of future cluster surveys. This question was investigated 
\cite{Pizzuti:2019wte} in the context of forecasting constraints on $\eta$ with 
future surveys, through a combination of simulated kinematical and lensing 
information. It was found that, keeping the NFW parameters fixed and assuming a 
scale--independent $\eta$, the effect on the forecasted errors is moderate when 
varying the tracers number between $100 \leq N_{\text{galaxies}} \leq 500$, 
while it becomes practically negligible for $N_{\text{galaxies}} > 500$.

Departures from dynamical relaxation and spherical symmetry produce a bias in 
the estimation of the mass  profile, introducing systematics in the constraints 
of modified gravity parameters. In particular, this has been shown in Ref. 
\cite{Pizzuti19b}, which performed a detailed study on cosmological N-body 
simulations showing a strong correlation between the constraints in modified 
gravity models and the effect of the aforementioned departures; the analysis 
further identified two observational criteria, linked to the cluster's dynamical 
properties, to be used in the selection of those clusters  suitable for the 
application of this kind of methods. Moreover the assumption of dynamical 
equilibrium limits the validity of the Jeans' equation out to the virial radius, 
which at $z=0$ corresponds to the radius $r_{200}$. It is possible to employ 
other techniques which doesn't rely on the dynamical state of the cluster and 
thus they can be used to reconstruct the mass profile in more external regions 
$r>r_{200}$ (e.g., the Caustic method of Ref. \cite{Diaferio01}); nevertheless, 
these methods suffer different kinds of systematics, and the application can be 
more or less appropriate with respect to the Jeans' analysis depending on the 
case studied. 

It is worth to notice that since both galaxies and the hot X-ray emitting gas of 
the Intra Cluster Medium (ICM) perceive the same gravitational potential, the 
two methods for reconstructing the mass profile are sensitive in the same way to 
modification of gravity. However, diffuse gas and galaxies dynamics suffer 
different systematics; for instance non-thermal pressure, e.g. associated with 
unthermalized gas motions, leads to a biased estimate of the cluster mass from 
X-ray analyses compared to other methods (see, e.g., Ref. \cite{Ettori2013} and 
references therein, Ref. \cite{Biffi16}), especially in the cluster outskirts 
where the contribution of non-thermal pressure becomes large (e.g., Ref. 
\cite{Martizzi2016}).

Moreover, analyses based on the dynamics of cluster member galaxies allow to  
constrain the gravitational potential out to the virial radius $r_{200}$ (or 
beyond, as discussed above) while X-ray observations are generally limited to 
$r_{500}$; in addition, gas clumping produces biased X-ray measurements in the 
outskirt of galaxy clusters (see, e.g., Ref. \cite{Nagai11}).  Clearly, 
combined 
X-ray and Jeans' analyses to infer the gravitational potential in the central 
region of relaxed clusters could in principle help in tightening the constraints 
on the inner shape of the mass profile and to break possible degeneracies 
between the dynamical parameters and additional degrees of freedom in 
non-standard theories of gravity.

Finally, as regards tests based on the slip parameter $\eta$, it is worth  
pointing out that $\eta = 1$ in GR only if relativistic corrections to the 
gravitational potentials $\Psi$ and $\Phi$ can be neglected. In fact, future 
constraints of $\eta$ based on galaxy cluster observations are expected to bring 
the statistical uncertainties down to few percents. At this level of precision, 
tiny departures form $\Psi=\Phi$  sourced by non-linear effects in GR, and not 
by a modification of gravity, are no more negligible and could constitute a 
severe limitation of the measurements of $\eta$. The contribution of these 
relativistic terms should be taken into account as systematic effect from future 
analyses.

\section{Future Outlook}

In the next years, new generation surveys will provide a significant amount  of 
imaging and spectroscopic data covering a large portion of the sky, within a 
broad redshift range. Both ground-based (e.g., 
LSST
and space-borne telescopes (e.g., 
Euclid
are expected to observe several 
billions galaxies in different bands, aiming at probing the nature of dark 
energy and gravity at large scales. In particular, the forecasting analysis of 
Ref. \cite{Martinelli:2010wn} showed that Euclid's weak lensing measurements 
will be able to tighten the constraints on MG parameters placed by the Planck 
satellite mission by two orders of magnitude.

As for mass determinations of 
galaxy clusters, the combination of data from the aforementioned surveys with 
spectroscopic observations coming from next generation high-multiplexing 
spectrographs on 8m-class telescopes will provide joint dynamics and lensing 
mass reconstruction of thousands clusters. The signal-to-noise ratio will be 
much lower than what has been already achieved by current surveys such as the 
CLASH and CLASH-VLT. Therefore, a good understanding of systematic effects, as 
discussed in the previous section, is required in order to take full advantage 
of galaxy cluster analyses for testing gravity on cosmological scales.

\chapter[Probing Screening Modified Gravity with Non-linear Structure Formation]
{Probing Screening Modified \\ Gravity with Non-linear \\ Structure Formation}
\label{sec:Mota}
\label{screeningref4}

{\em David F. Mota}\\






Extended Theories of Gravity have considered a new paradigm to cure 
shortcomings of General Relativity  at infrared and ultraviolet scales. They are 
an approach that, by preserving the undoubtedly positive results of Einstein's 
theory, is aimed to address problems recently emerged in astrophysics, 
cosmology 
and high energy physics \cite{Clifton:2011jh}; in particular,  problems like 
dark energy and dark matter.
Several very good reviews have been written about the topic. Bellow I summarise 
some of them:

In Capozziello $\&$ De Laurentis \cite{Capozziello:2011et} principles of such 
modifications are presented, focusing on specific classes of theories like 
f(R)-gravity and scalar-tensor gravity in the metric and Palatini approaches. 
The special role of torsion is also discussed. The conceptual features of these 
theories are fully explored and attention is payed to the issues of dynamical 
and conformal equivalence between them considering also the initial value 
problem. A number of viability criteria are presented considering the 
post-Newtonian and the post-Minkowskian limits. In particular, the authors 
discuss the problems of neutrino oscillations and gravitational waves in 
Extended Gravity.

In Cai et al. \cite{Cai:2015emx} the role of torsion in gravity has been 
extensively  investigated along the main direction of bringing gravity closer 
to its gauge formulation and incorporating spin in a geometric description.  
From teleparallel, to Einstein$-$Cartan, and metric-affine gauge theories, 
resulting in extending torsional gravity in the paradigm of $f(\mathbb{T})$ 
gravity, where 
$f(\mathbb{T})$ is an arbitrary function of the torsion scalar. Based on this 
theory, the 
authors review the corresponding cosmological and astrophysical applications. 
In particular, they study cosmological solutions arising from $f(\mathbb{T})$ 
gravity, both 
at the background and perturbation levels, in different eras along the cosmic 
expansion. The $f(\mathbb{T})$ gravity construction can provide a theoretical 
interpretation of the late-time universe acceleration, alternative to a 
cosmological constant, and it can easily accommodate with the regular thermal 
expanding history including the radiation and cold dark matter dominated 
phases. Furthermore, if one traces back to very early times, for a certain class 
of $f(\mathbb{T})$ models, a sufficiently long period of inflation can be 
achieved and 
hence can be investigated by cosmic microwave background observations? Or, 
alternatively, the Big Bang singularity can be avoided at even earlier moments 
due to the appearance of non-singular bounces. Various observational 
constraints, especially the bounds coming from the large-scale structure data 
\label{LSSefs5} in 
the case of $f(\mathbb{T})$ cosmology, as well as the behavior of gravitational 
waves, are 
described in detail. Moreover, the spherically symmetric and black hole 
solutions of the theory are reviewed. Additionally, the auhtors discuss various 
extensions of the $f(\mathbb{T})$ paradigm. Finally, we consider the relation 
with other 
modified gravitational theories, such as those based on curvature, like f(R) 
gravity, trying to illuminate the subject of which formulation, or combination 
of formulations, might be more suitable for quantization ventures and 
cosmological applications.

While the above two review articles investigate modified gravity models as 
alternatives to dark energy, in Capozziello $\&$ De Laurentis 
\cite{Capozziello:2012ie}, the authors review an alternative view to the dark 
matter puzzle that is represented by Extended Theories of Gravity. The approach 
consists in addressing issues like dark components from the point of view of 
gravitational field instead of requiring new material ingredients that, up to 
now, have not been detected at fundamental level. In this review paper, by 
extending the Hilbert-Einstein action of gravitational field to more general 
actions (e.g. f(R) gravity), it is shown that several gravitating structures 
like stars, spiral galaxies, elliptical galaxies and clusters of galaxies can be 
self-consistently described without asking for dark matter. It is also shown 
that standard General Relativity tests and Equivalence Principle constraints can 
be evaded at Solar System scales.

Gravity theories beyond General Relativity may possibly explain several 
cosmological puzzles, specifically the present accelerated expansion of the 
Universe 
\cite{Clifton:2011jh,Koivisto:2008xf,Koivisto:2012za,Akrami:2013ffa,
Thorsrud:2012mu,Barrow:2002zh,DeFelice:2009rw,Li:2008fa}. Modified Gravity must, 
however, comply with strong requirements: One is that the model must have 
similar cosmological predictions to those of  $\Lambda$CDM for the background 
evolution and the linear large scale structures \cite{Ade:2015rim}. Another 
condition is that the modifications to General Relativity are suppressed at 
small scales \cite{Will:2014kxa}. This requirement is assured through the 
so-called screening mechanisms \cite{Brax:2012bsa}.

Since different modified gravity theories can be degenerate with regard to both 
the background cosmology and the growth rate of linear perturbations, it is 
crucial to identify new probes that can be used to break these degeneracies.  In 
this Chapter we study the effects of a class of screened modified gravity 
models 
in the nonlinear regime of structure formation. The aim is to predict possible 
smoking guns of modified gravity and of screening mechanisms at cluster of 
galaxy scales.

\section{Theoretical Models}
\label{sect:theo}

Scalar-tensor theories are an extension of General Relativity that add a scalar 
field $\varphi$ to the standard Einstein-Hilbert Lagrangian.
A general Lagrangian for the scalar field is
\be
{\cal L} = -\frac{1}{2} \partial_{\mu} \varphi \partial_{\nu} \varphi 
-V(\varphi)
+ \beta(\varphi) T^{\mu}_{\mu},
\ee
where  $V(\varphi)$ is the self interacting potential, $\beta(\varphi)$ is a 
coupling function and $T^{\mu}_{\mu}$ is the trace of the matter energy-momentum 
tensor.  The scalar field gives rise to an additional {\it fifth force}, which 
can be quantified by $\gamma\equiv |{\bf F}_{\rm Fifth}|/|{\bf F}_{\rm N}|$, 
where ${\bf F}_{\rm N}$ is the Newtonian force. Experiments constrain $\gamma 
\ll 1$ in the Solar System.  A screening mechanism with the aim to hide this 
field from local gravity experiments can be realised in two different ways:

{\it{Density dependent mass:}}
If the mass of the field $m^2({\varphi})$ is large in dense environments, then 
the fifth force mediated by the scalar field is suppressed on scales above its 
Compton wavelength. On the other hand, in low density environments, the mass can 
be light and the scalar field mediates a long range fifth force. This is the 
so-called {\it Chameleon} screening \cite{Khoury:2003aq};

{\it{Density dependent coupling:}}
If the coupling to matter $\beta({\varphi})$ is small in the region of high 
density, the strength of the fifth force ${\bf  F}_{\rm Fifth}$ is weak and the 
modifications to gravity are suppressed. On the other hand, in low density 
environments, the size of the fifth force can be of the same order as standard 
gravity. This idea is the so-called Symmetron mechanism 
\cite{Hinterbichler:2010es}.

\subsection{Chameleon-$f(R)$ Gravity}
\label{chameleonkiref1}
\label{fRref9}

The Lagrangian for this theory is
\be
S=\int {d}^4 x \sqrt{-g} \frac{1}{16\pi G_N}\Big(R + f(R) \Big)+ S_m(g_{\mu 
\nu}, 
\psi_i).
\ee
In the quasi-static and weak-field limits the  equations of motion become
\be
\label{eq:poisson-fr}\nabla^2\Phi = \frac{16\pi G_N}{3}a^2\delta\rho_m + 
\frac{1}{6}a^2\delta R, \qquad
\nabla^2 f_R = -\frac{a^2}{3}\left[\delta R + 8\pi G_N\delta\rho_m\right],
\ee
where $\delta\rho_m = \rho_m - \bar{\rho}_m$ and $\delta R = R - \bar{R}$ are 
the density and Ricci scalar perturbations (overbars denote background 
quantities), and $f_R = {d}f(R)/{d}R$.  In this formulation, $f_R$ plays the 
role of the scalar degree of freedom $\varphi$ that determines the fifth force.

We choose the Hu-Sawicki model \cite{Hu:2007nk} as a working example, which is 
given by
\be
\label{eq:hs}f(R) = -m^2\frac{c_1(R/m^2)^n}{c_2(R/m^2)^n + 1},
\ee
where $m^2 = H_0^2\Omega_m$ is a mass scale and $c_1$, $c_2$ and $n$ are model 
parameters. One recovers a $\Lambda$CDM expansion history by setting $c_1/c_2 = 
6\Omega_\Lambda/\Omega_m$.
In this paper we consider $n = 1$, and we consider models with 
$|\bar{f}_{R0}|=10^{-4}$, $|\bar{f}_{R0}|=10^{-5}$ and  
$|\bar{f}_{R0}|=10^{-6}$.

Notice that the modified Poisson equation, equation~(\ref{eq:poisson-fr}), can 
also be written as
\be\label{eq:poissonmod-fr}
\nabla^2\Phi = \nabla^2\Phi_N - \frac{1}{2}\nabla^2f_R .
\ee
This makes explicit that in $f(R)$ models the total gravitational force is 
governed by a modified gravitational potential $\Phi = \Phi_N - \frac{1}{2}f_R$.
It is the nonlinearity of $f(R)$  in equation (\ref{eq:hs}) that gives rise to 
the Chameleon screening, and the screening of the fifth force is determined by 
the depth of the gravitational potential $\Phi_N$.

\subsection{Symmetron}
 \label{symmetronref1}
 
The Symmetron model \cite{Hinterbichler:2010es} action is given by
\be\label{eq:action_symm}
S = \int {d}x^4 \sqrt{-g}\left[\frac{R}{16\pi G_N} - 
\frac{1}{2}(\partial\varphi)^2 -
V(\varphi)\right]
+ S_m(\tilde{g}_{\mu\nu},\psi).
\ee
The matter fields $\psi$ couple to the Jordan frame \label{Jordanrref5} metric 
$\tilde{g}_{\mu\nu}$, 
which relates to the Einstein frame metric $g_{\mu\nu}$ as
$\tilde{g}_{\mu\nu} = A^2(\varphi)g_{\mu\nu}.$
The coupling function $A(\varphi)$ is
\be\label{eq:coupling_function}
A(\varphi) = 1 + \frac{1}{2}\left(\frac{\varphi}{M}\right)^2,
\ee
where $M$ is a mass scale. The total force felt by matter is given by
\be
{\bf F} = \nabla\left(\Phi_N + \frac{1}{2}\frac{\varphi^2}{M^2}\right) = 
\nabla\Phi_N + \frac{\varphi\nabla\varphi}{M^2}.
\ee
The potential is taken to be
\bq\label{potential}
V(\varphi) = V_0 -\frac{1}{2}\mu^2\varphi^2 + \frac{1}{4}\lambda\varphi^4,
\eq
where  the value of $V_0$ is determined by the condition that the model gives 
rise to the observed accelerated expansion of the Universe \cite{Davis:2011pj}. 
The field equation for $\varphi$  reads
\be\label{eq:eom_symm}
\square\varphi = V_{\rm eff,\varphi},
\ee
where for nonrelativistic matter the effective potential is given by
\be\label{eq:veff_symm}
V_{\rm eff}(\varphi) =  V_0 + 
\frac{1}{2}\left(\frac{\rho_m}{M^2}-\mu^2\right)\varphi^2 + 
\frac{1}{4}\lambda\varphi^4.
\ee
In the quasi-static limit, equation (\ref{eq:eom_symm}) becomes
\be
\nabla^2\chi = \frac{a^2}{2\lambda_0^2}\left( \frac{\rho_m}{\rho_{\rm SSB}} - 1 
+ \chi^2\right)\chi,
\ee
where $\chi = {\varphi}/{\varphi_0}$.

Screening in the Symmetron model is very similar to the Chameleon-$f(R)$ case in 
the sense that the condition for screening is determined by the local 
gravitational potential. The important difference is that the coupling 
$\beta(\varphi) = \frac{\beta_0 \varphi}{\varphi_0}$, which is constant in 
$f(R)$ gravity, now depends on the local field value. In high density regions, 
$\rho_m  > \rho_{\rm SSB}$, the field falls into the minima $\varphi = 0$, and 
since the coupling is proportional to $\varphi$, the fifth force vanishes.

We define three physical  parameters $L$, $\beta$ and $z_{\rm SSB}$ which are 
the  range of the field, the coupling strength to matter and the redshift of 
symmetry breaking:
\be
L = \frac{\lambda_{0}}{\text{Mpc}/h} = \frac{3000 H_0}{\sqrt{2}\mu},\qquad
\beta = \frac{\phi_0 M_{\rm pl}}{M^2} = \frac{\mu M_{\rm 
pl}}{\sqrt{\lambda}M^2},\qquad
(1+z_{\rm SSB})^3 = \frac{\mu^2 M^2}{\rho_{m 0}}.
\ee
We simulate four symmetron models: symm$\_A$ ($z_{ssb}=1,\beta=1, L=1$), 
symm$\_B$ ($z_{ssb}=2,\beta=1, L=1$), symm$\_C$ ($z_{ssb}=1,\beta=2, L=1$), and 
symm$\_D$ ($z_{ssb}=3,\beta=1, L=1$).

\section{Efficiency of Screening Mechanisms}

The common feature to all the screening mechanisms proposed in the literature is 
that they are built, and their efficiency tested, assuming the so called 
quasi-static approximation for the field equations.  For instance,  in 
scalar-tensor theories, a scalar degree of freedom is introduced into the 
standard Einstein-Hilbert action. This field follows the Klein-Gordon equation 
of motion, which determines both its time and spatial variations. When 
constructing screening mechanisms to hide the scalar field within the accurately 
tested regimes, the quasi-static approximation is invariably applied to the 
equations of motion for the field. This simplifies the calculations by implying 
that the scalar field is at rest in the minimum of the local effective 
potential 
at all points in space and time. This reduces the equation of motion to a 
Poisson-like equation, which is readily solved to find the approximated scalar 
field value at any point.

Notice, however, that the full equation of motion for the scalar field is,  in 
fact, a second order differential equation in time, more similar to a wave 
equation. Therefore, ignoring the time evolution of the field, via the 
quasi-static approximation, is to shortfall effects that are only possible to 
realize when considering the full equation of motion 
\cite{Hagala:2016fks,Hagala:2015paa}.

\subsection{Solar System Constraints}
\label{solarsystemref10}

In order to test how screening mechanisms work in the Solar System, the 
community generally  chooses a static, spherically symmetric matter 
distribution 
to mimic the Galaxy. We follow this approach and choose the Navarro-Frenk-White 
(NFW) density profile with the characteristics to represent the Milky Way 
Galaxy, specifically with a virial radius of $r_\mathrm{vir} = 
137\,\mathrm{kpc}/h$ and concentration $c = 28$, resulting in a halo mass of 
$1.0\times10^{12} M_\odot$ and a circular velocity of 220~km/s at 8~kpc. The 
reason for the high value of the concentration is simply that we are modeling 
not only dark matter, but the total matter of the Milky Way, which is more 
concentrated than the pure dark matter halo.
We also did the calculations with an Einasto profile with identical virial mass,
and found that the results presented  are not very sensitive to the choice of 
distribution.
Because of limitations of spherical symmetry, we did not model a galactic disc.

One of the most precisely measured gravity parameters to probe deviations from 
general relativity is the parametrised post-Newtonian (PPN) parameter $\gamma$. 
It can be expressed as the ratio of the metric perturbations in the Jordan 
frame, $\Psi_J$ and $\Phi_J$.  We find the expression for $\gamma - 1$ to be
\begin{equation}
\gamma-1=-\frac{\phi^{2}}{M^{2}}\frac{2}{\frac{\phi^{2}}{M^{2}}-2\Psi_{E}-2\Psi_
{E}\frac{\phi^{2}}{M^{2}}}.
\end{equation}
In general relativity, $\gamma  = 1$ exactly. The strongest constraint to date, 
measured by the Cassini spacecraft \cite{Bertotti:2003rm}, is
$\gamma - 1 = \left( 2.1 \pm 2.3 \right ) \times 10^{-5}.$

The screening mechanism of the symmetron model works by modifying the effective 
potential  such that the field value is pushed towards zero in high density 
regions -- like the inner regions of the Galaxy. This results in $\gamma - 1 
\rightarrow 0$, such that the deviations from general relativity in the 
proximity of the Solar System are small.  The same occurs for the fifth force 
$F_{\phi}$ associated to the scalar field.

We calculate the $\gamma$ parameter arising from the smoothed matter 
distribution of the Milky Way. Note that, by using this method, we find an 
upper 
bound on the actual value of $|\gamma - 1|$ in the inner Solar System. This is 
because we do not include the presence of massive bodies like the Sun, which 
will increase the screening to some degree.
Nevertheless, most of the screening is believed to come from the matter 
distribution of the Galaxy because, in the symmetron model, the Solar System 
cannot screen itself in vacuum, and therefore, the theory depends on a working 
screening from the Galaxy.

\subsection{Simulations}

Since the equation of motion is a hyperbolic partial differential equation, it 
can be solved as an initial value and boundary condition problem. The initial 
condition at $t=0$ is chosen to be the static solution of the nonlinear 
Klein-Gordon equation of motion. With a constant boundary condition, this would 
imply that the field will stay at rest forever. The boundary condition at the 
edge of the simulation at $r_\mathrm{max}$ is chosen to emulate incoming 
sinusoidal waves in the scalar field, specifically $\chi(r_\mathrm{max}, t) = 
\chi_0 (r_\mathrm{max}) + A\sin\left( \omega t \right)$. Possible sources of 
such waves will be discussed later.

We set up a radial grid, divided into linearly spaced steps $\Delta r$ up to 
$r_\mathrm{max} = 4\,\mathrm{Mpc}/h$. On each of the grid points we specify the 
matter density according to the NFW halo.  Starting from the initial value and 
with the inclusion of incoming waves, we evolve Eq. \eqref{eq:eom_symm} forward 
in time steps of size $\Delta t$, using the leapfrog algorithm for time 
integration in each grid point. Tests of this technique applied to the 
symmetron 
are presented in \cite{Llinares:2013qbh,Llinares:2013jua}.
We are only interested in events that happen during the last few megayears of 
cosmic time, meaning that we take the approximations $z\approx0$ and 
$a\approx1$ 
in all computations.  Spatial derivatives are found using a finite difference 
method in spherical coordinates, assuming all derivatives in the tangential 
directions vanish.
The code outputs the evolution of the scalar field and, more importantly, the 
value of $\left | \gamma - 1 \right |$ at 8 kpc from the center---corresponding 
to the position of the Solar System in the Milky Way. Finally, we confirm that 
the values used for technical parameters of our solver give a 
stable solution by running convergence tests. These are performed by increasing 
the resolution in factors of two (both temporal and spatial resolution 
separately) until the resulting scalar field at some later time 
$t_\mathrm{max}$ 
does not change significantly with resolution.

\subsection{Results}

We are interested in investigating  how the PPN parameter $\gamma$ 
changes when a wave enters the inner 100 kpc of the Milky Way.    The 
modifications to gravity are initially screened very 
well in the regions around this position, with $|\gamma - 1| < 10^{-8}$. 
However, after the wave has arrived  the 
scalar 
field is perturbed enough to breach the Solar System constraints, $|\gamma - 1| 
> 2\times 10^{-5}$. In other words, the screening mechanism breaks down under 
these circumstances.

When measuring $\gamma$ arising from a single sinusoidal wave with low 
frequency,
there is a possibility that the local wave is between two extrema at the time 
of measurement.
This could render this kind of detection difficult for several thousand years.
Nevertheless, given that various astrophysical events---such as 
supernovae---can 
generate waves,
the probability that one of the wavefronts would bring us away from the minima 
at the present time is not negligible.

 Hence, it is possible to conclude that  higher 
frequencies and amplitudes for the incoming scalar waves give larger deviations 
from the general relativity result (i.e., $\gamma = 1$).  The limit where 
amplitude and frequency go to zero is equivalent to the quasi-static limit, 
where 
no waves are produced and their energy is zero.  As one goes into the high 
frequency and amplitude regime, the waves carry more energy, and therefore, the 
PPN parameter $\gamma$ starts deviating significantly from the quasi-static 
limit.  Note that, since in the symmetron model, the fifth force is 
$F_{\phi}\propto \nabla\phi^2/M^2$, these values can be immediately 
extrapolated 
to the impact of the waves on this quantity.

The  dependence of the $\gamma$ PPN parameter on the wave amplitude is 
straightforward to understand: When a wave propagates through the screened 
regions of the halo, a larger amplitude wave will lead to larger displacements 
of the field  from the screening value $\phi \approx 0$. Therefore,  $|\gamma - 
1| \propto \phi^2$ will increase accordingly.

The frequency dependence of the $\gamma$ parameter is a consequence of the 
following:  The effective potential of the symmetron grows steeper and narrower 
in high density areas. In other words, the mass of the field increases towards 
the center of the halo. Therefore,  it becomes more difficult to perturb the 
field away from the minimum, and a higher wave energy is needed to displace it. 
Specifically, if the energy of the external waves is small compared to the mass 
of the field, the field will not be perturbed and the $\gamma$ parameter 
will not be affected.

The results obtained imply that if waves with sufficient amplitude or frequency 
can somehow be  generated in a given model for modified gravity, they 
will have to be taken into account when constraining the model 
parameters. Cosmic tsunamis, resulting from extreme events, could even 
completely ruin the screening mechanisms in modified gravity by increasing the 
deviations from general relativity by several orders of magnitude compared to 
the quasi-static case.  A subject that that must be discussed now is the 
generation of such waves.
Extreme events on small scales, such as collision of neutron stars, 
\label{neutronstarsref10} stellar, or 
super-massive  black holes are obvious examples.  Generation of waves by 
pulsating stars are another possibility.

In the specific case of the symmetron model, it is possible to obtain waves 
from 
events that occur  on cosmological scales.  First, the symmetron model 
undergoes 
a phase transition when the density falls below a specific threshold.  This 
transition first occurs in voids when the expansion factor is close to 
$a_\mathrm{SSB}$ \cite{Llinares:2013qbh, Llinares:2013jua}.  When this happens, 
the scalar field receives a kick, which produces waves traveling from the 
center 
of the voids towards the dark matter halos.  By doing postprocessing of 
simulations presented in \cite{Hagala:2015paa}, we find that, in a symmetron 
model with slightly different parameters, the amplitude of cosmological waves 
is 
typically smaller that 0.1 and the associated frequencies are of the order of 
1/Myr.  Note that these values depend on the model parameters and, hence, must 
be taken only as indicative. Scalar waves can also be created through the 
collapse of topological defects, which are known to exist in any model in which 
such phase transition occurs.

\section{Distribution of Fifth Force in Dark Matter Haloes}
 \label{Fifthref7}
\label{Darkmatterharef3}

 To understand the main features of nonlinear structure formation within 
screened modified gravity,  it is important to investigate the magnitude of the 
fifth force inside the dark matter haloes 
\cite{Gronke:2013mea,Gronke:2016lfd,Gronke:2015ama,Gronke:2014gaa}.  For that, 
we run N-body simulations for each model.
The simulations were run with the code \texttt{Isis} \cite{Llinares:2013jza}.   
The background cosmology for all the models is $(\Omega_{m0}, \Omega_{\Lambda 
0}, H_0)=(0.267,\,0.733,\,71.9)$.   The data sets are taken at $z=0$, and  all 
simulations were run with the same initial conditions.

The halos were identified using the spherical overdensity halo finder 
\texttt{AHF}  \cite{Knollmann:2009pb}. For the analysis we 
used only halos consisting of at least $100$ particles which limits the 
smallest 
halo we can probe to $M\sim 3\times 10^{12}M_{\odot}h^{-1}$. In the high mass 
end the simulation-box limits the maximum halo-masses we can study and the 
largest halos in our simulations has mass $M \sim 2-3\times 10^{15} 
M_{\odot}h^{-1}$. We have checked that the mass-function of our simulation 
agrees to $\sim 10\%-20\%$ to simulations with larger box-size and also to the 
fit to the mass-function in the range $M\in [ 10^{13},8\times 10^{14}] 
M_{\odot}h^{-1}$. The total number of halos in this mass-range in our 
simulations, which is what we used for the upcoming analysis, is $\sim 8000$.
\label{Fifthref6}
In the $f(R)$ case, the maximum 
of 
the fifth force profile moves towards larger radii when increasing mass, and 
decreasing  $|f_{R0}|$.  This is a  consequence of the screening that is 
activated in the centre of the haloes, and  when decreasing $|f_{R0}|$.  For 
the 
symmetron model, an increase of $\beta$ or $z_{SSB}$ leads to a stronger fifth 
force. A greater $\beta$ value increases the $|\vek{\ddot  x_{\text{Fifth}}}|$ 
values by a constant factor, while 
altering $z_{SSB}$ changes the shape of the fifth force profile in general.
   It is clear that the fifth force is screened in large massive haloes 
(since they have a higher density) while the force is active in the low mass 
range (where the density is lower).

The dependence of the symmetron fifth force on the redshift of  symmetry 
breaking $z_{SSB}$ is thus clear. The higher the symmetry breaking, 
the higher the masses that are unscreened and affected by the fifth force.  The 
low mass end of the distribution is  insensitive to changes in $z_{SSB}$, since 
in there the density is low and the force is unscreened.  Differences in the 
models are also dependent of the strength of the coupling $\beta$.

\section{The Matter and the Velocity Power Spectra}
\label{sec:results}

Due to the presence of the fifth force, it is expected in modified theories of 
gravity,  that the acceleration felt by particles is in general higher than in 
General Relativity \cite{Gronke:2014gaa}. Therefore, a promising observable and 
probe of modified gravity is the measure of velocity distributions in galaxy 
clusters.

The global statistical properties of the velocity field are described by  the 
velocity divergence power spectrum.  Normalising the divergence of the velocity 
field $\vec{\nabla}\cdot \mathbf{v}$ with the Hubble parameter gives the 
dimensionless expansion scalar
$\theta = \frac{1}{H}\vec{\nabla}\cdot \mathbf{v}.  $

We compute the power-spectrum of $\theta$ from our simulations.
To characterize this quantity, we study
the relative particle velocity, which is simply defined as
$v_{\rm rel} = \sqrt{(\mathbf{v} - \mathbf{v}_{\mathcal H})^2}$,
where $\mathbf{v}$ and $\mathbf{v}_{\mathcal H}$ are the particle velocity and 
its halo velocity, respectively. For the latter, we use the core velocity of 
the 
halo. 
For our $f(R)$ simulations we find that the difference with respect to 
$\Lambda$CDM in  the velocity divergence spectrum can be roughly two times as 
large as the difference in the matter power spectrum.
For the Symmetron the difference can be much larger. For the symm$\_$C model 
(which is the model  with the largest value of the coupling strength $\beta$), 
we see that $(\Delta P/P)_{m}\approx 10\%$ at $k=1h/Mpc$ while $(\Delta 
P/P)_{\theta}\approx 200\%$. The \symmC model has a fifth force in unscreened 
regions  four times that of the other Symmetron models and this is likely why 
we 
get this extreme signal.

The reason why generally we have $(\Delta P/P)_{\theta}\gtrsim(\Delta P/P)_{m}$ 
is that the  velocity divergence field is not mass-weighted in any way. Hence, 
low-density regions (voids) will contribute a large part of the signal 
in the velocity divergence power-spectrum (since voids contribute a large part 
of the volume in the Universe), which is not the case for the matter 
power-spectrum. Now the fifth-force is generally not screened in low-density 
regions, so consequently velocities are boosted to significantly higher 
relative 
values (when compared with $\Lambda$CDM) in voids opposed to in clusters.
This indicates that low-density regions like cosmic voids, as we would
expect, is the place where the strongest signals of modified gravity
can be found.

\section{The Dynamical and Lensing Masses}
\label{sec2}\label{weaklensefs6}

A general prediction of Modified Gravity theories with screening  mechanisms is 
violations of equivalence principle.  \label{equivprinref10} The equivalence 
principle can be tested by 
observing differences between the gravitational mass and the inertial mass of 
objects, in
cosmology through measurements of galaxy clusters masses. Astronomers  infer 
the 
masses of galaxy clusters using lensing observations and  dynamical methods. 
The 
former is a probe of the gravitational mass, while the later is a measure of 
the 
inertial mass. Differences between these two masses would be a smoking gun of 
modified gravity and screening mechanisms.  In order to make predictions within 
the chamleon-f(R) and the symmetron gravity, we first need to compute the 
Bardeen potentials.

In the Jordan-frame we have
\begin{equation}
ds^2 = a^2(\eta)\left[-d\eta^2(1+2\Phi) + (1-2\Psi)d\vec{x}^2\right],
\end{equation}
where $\Phi  \simeq \Phi_N+ \delta A(\phi)$, $\Psi  \simeq \Phi_N -  \delta 
A(\phi)$,
with
\begin{align}
 \delta A(\phi) \equiv A(\phi) - 1 = \frac{1}{2}\left(\frac{\phi}{M}\right)^2.
\end{align}

 The fifth-force potential is given by the difference in the above two 
potentials
\begin{align}
\Phi_- = \frac{\Phi - \Psi}{2} =  \delta A(\phi).
\end{align}
Lensing on the other hand is affected by the lensing potential
\begin{align}
\Phi_+ = \frac{\Phi + \Psi}{2} = \Phi_N,
\end{align}
which satisfies the Poisson equation
\begin{equation}
\nabla^2\Phi_+ = 4\pi G_N a^2 \delta\rho_m.
\end{equation}
This is the same equation as in GR since it is conformally invariant and 
therefore photons are not affected by the fifth-force.  The lensing mass is 
defined as
\begin{equation}
M_L = \frac{1}{4\pi G_N a^2}\int \nabla^2\Phi_+ dV,
\end{equation}
which is the gravitational mass of the halo. It is determined from the 
simulations by counting the number of particles within a given radius. For 
spherical symmetry,  using the Stokes theorem,  we have
\begin{equation}
M_L(r) \propto r^2\frac{d\Phi_+}{dr}.
\end{equation}
The dynamical mass $M_D(r)$ of a halo is defined as the mass contained within a 
radius $r$ as inferred from the gravitational potential $\Phi$, i.e.
\begin{equation}
M_D(r) = \frac{1}{4\pi G_N a^2}\int \nabla^2\Phi dV,
\end{equation}
where the integration is over the volume of the body out to radius $r$. For 
spherical symmetry, and using Stokes theorem, we find
\begin{equation}\label{dyn_mass}
M_D(r) \propto \int r^2 \frac{d\Phi(r)}{dr} = r^2\left(\frac{d\Phi_N}{dr} + 
\frac{\phi}{M^2}\frac{d\phi}{dr}\right).
\end{equation}
The terms in the brackets are recognised as the sum of the gravitational force 
and the fifth-force.  Observationally, $M_D$ can be determined from 
measurements of velocity dispersion of galaxies in clusters.

In GR the lensing mass is the same as the dynamical mass, but they can be 
significantly different in modified gravity. We follow \cite{Winther:2011qb} 
and define the relative difference
\begin{equation}
\Delta_M(r) = \frac{M_D}{M_L}-1 = \frac{d\Phi_-/dr}{d\Phi_+/dr}.
\end{equation}
This allows us to quantify the difference between the two masses in the 
simulations. In GR we have $\Delta_M \equiv 0$ while in the symmetron model 
$\Delta_M$ will vary depending on the mass of the halo and its environment. The 
theoretical maximum is achieved for small objects in a low-density environment 
where the screening is negligible and reads
\begin{equation}
\Delta_M^{\rm Max}(r) = 2\beta^2.
\end{equation}
For low-mass halos we obtain a significant 
dispersion of $\Delta_M$ from 0 to the maximum value obtained in low-density 
environments for the same mass ranges. This is because low-mass halos cannot 
efficiently screen themselves and must rely on the environment to get the 
screening.  As expected, there is a clear trend that the small halos which are 
efficiently screened generally reside in high-density environments, while those 
which are less screened lie in low-density environments.
Massive halos on the other hand can screen themselves efficiently and the 
environment only plays a small role in their total screening.
Additionally,  there is a large difference between large halos 
in dense environments and small halos in low density environments. The 
$r$-dependence of $\Delta_M(r)$ is shown to be rather weak in high density 
environments since the value of the scalar field inside the halo is mainly 
determined by the environment, while in low-density environments the value of 
the scalar field mainly depends on the mass of the halo, which leads to a 
stronger $r$-dependence. Finally, note also that  the deviation from GR is 
stronger for 
higher symmetry-breaking redshift $z_{\rm SSB}$, and for larger values of the 
coupling $\beta$, which implies a stronger fifth-force and therefore a stronger 
effect.
 
\section[Thermal Versus Lensing Mass Measurements]{Thermal Versus Lensing Mass 
Measurements}
\label{sec:therm-mass-meas}

In order to compare the lensing and thermal mass of the clusters we took 
measurements from \cite{Zhang:2010hma} and \cite{Mahdavi:2012zy}. These two 
data-sets provide both thermal and lensing mass measurements and uncertainties 
for a total of $58$ clusters in the mass range $M/M_{\odot} h^{-1}\in[5\times 
10^{13},\,3\times 10^{15}]$ so there was no need to combine the mass estimates 
in a similar fashion as in the previous section. We divided the data for the 
thermal mass measurements by the data for the lensing mass measurements while 
properly propagating the error. As we're interested in a systematic deviation, 
we binned the data in six (lensing) mass bins which we stratified so that 
roughly the same number of halos are in each bin.

An important point to bear in mind when working with thermal mass estimates is 
the fact that the measured quantity in this case is the temperature of the 
intracluster gas. The conversion to a mass assumes hydrostatic-equilibrium 
\cite{Zhang:2010hma,Mahdavi:2012zy}. However, it has been 
shown that in reality the pressure of the intracluster medium will have a 
significant non-thermal component generated by random gas motions and 
turbulence.
This means the inferred thermal mass given a temperature $T$ will be slightly 
lower than the true mass of the cluster.

While empirical models exist in order to quantify the magnitude of this 
deviation (where the non-thermal component yield variations to the mass from 
10\% to 30\% ) we want to stress that these were calibrated against pure 
$\Lambda$CDM simulations, and thus their results cannot be taken into account 
when dealing with modified gravity.
One has to consider instead that if gravity is truly enhanced, the temperature 
of the intra-cluster medium will be hotter and, thus, the inferred thermal mass 
will be greater (as shown in Sec.~\ref{sec:results}).
This means the effect of any non-thermal physics (such as cosmic rays) is 
degenerate with modified gravity and, consequently, at the present time thermal 
measurements cannot be used to constrain modified gravity 
\cite{Hammami:2015ela,Hammami:2016npf,Hammami:2015iwa}.

Let is now compare 
the $M_{\rm therm}/M_{\rm lens}$ results of the Symmetron D model and our 
$\Lambda$CDM simulation to hypothetical measurements where we modelled the 
contribution of non-thermal pressure as
\begin{equation}
P_{\rm non-thermal} = P_{\rm total} \tilde g(M_{\rm 200}),
\label{eq:Pnontherm1}
\end{equation}
which resembles the functional forms fitted to $\Lambda$CDM simulations (see 
also subsection \ref{sec:appendix}). Thus, our proposed non-thermal 
contribution 
is not unreasonable. Modelling a non-thermal contribution as given by 
Eq.~\eqref{eq:Pnontherm1} while keeping the total pressure $P_{\rm total} = 
P_{\rm therm} + P_{\rm non-thermal}$ (and, thus, the halo structure) constant 
is equivalent to rescaling the temperature as $T\rightarrow T (1 - \tilde g)$ 
since naturally $P_{\rm therm}\propto T$.
In the case of a non-thermal 
contribution the $M_{\rm therm}$ measurement (which is carried out the same way 
as done by observations, i.e., assuming no non-thermal contribution) matches 
the lensing mass reasonably well \textit{in the case of modified gravity}. We 
achieved this by choosing the functional form of $\tilde g$ in 
Eq.~\eqref{eq:Pnontherm1} as
\begin{equation}
\tilde g(M) = \frac{1}{1 + \tilde a M_{13}^{\tilde\alpha}}\;,
\end{equation}
with $M_{13}\equiv M/(10^{13}M_\odot h^{-1})$ and $(\tilde a,\,\tilde 
\alpha)=(3/4,\,2/3)$.
This serves as an example of how unknown non-thermal physics can cancel out any 
signal originating from modified gravity -- which is a severe problem when 
trying to place constraints on the modifications of gravity using thermal 
measurements.

This problem will be alleviated once the contribution of non-thermal effects 
can be directly quantified using observational data (e.g., by measuring 
directly the intra-cluster turbulence).
In the sequel of this subsection, we assume this has been done and is has been 
shown the contribution of the non-thermal components is negligible. We do this 
in order to show which constraints on modified gravity can be placed 
hypothetically using thermal mass estimates.

\subsection{Including the Non-thermal Pressure Component}
\label{sec:appendix}

It is known that the pressure of the intracluster medium will have a 
significant  non-thermal component generated by random gas motions and 
turbulence, so that the total pressure $P_{\rm Tot}$ of a cluster is
\begin{align}
P_{\rm Tot}(<r) = P_{\rm thermal}(r)+P_{\rm non-thermal}(r).
\end{align}
This results in the mass estimates \cite{Will:2014kxa} consist of a thermal and 
non-thermal component as well
\begin{align}
M(<r) = M_{\rm thermal}(r)+M_{\rm non-thermal}(r)
\end{align}
where
\begin{align}
 M_{\rm thermal}(r) &=  -\frac{r^2}{G_N\rho_{\rm gas}(r)}\frac{dP_{\rm 
thermal}(r)}{dr}\\
 M_{\rm non-thermal}(r) &= -\frac{r^2}{G_N\rho_{\rm gas}(r)}\frac{dP_{\rm 
non-thermal}(r)}{dr}.
\end{align}

\noindent By using $P_{\rm thermal}=kn_{\rm gas}T_{\rm gas}$, where $\rho_{\rm 
gas}=\mu m_p n_{\rm gas}$, we find that
\begin{align}
 \frac{dP_{\rm thermal}}{dr} =  \frac{kT_{\rm gas}(r)}{\mu 
m_p}\left[\frac{d\rho_{\rm gas}(r)}{dr} + \frac{\rho_{\rm gas}(r)}{T_{\rm 
gas}(r)}\frac{dT_{\rm gas}(r)}{dr}\right],
\end{align}
and hence
\begin{align}
M_{\rm therm} &= -\frac{k_Br^2T_{\rm thermal}(r)}{\mu m_p 
G_N}\left(\frac{d\ln\rho_{\rm thermal}}{dr} + \frac{d\ln T_{\rm 
thermal}}{dr}\right),
\end{align}
as show earlier in the paper.

Often, the non-thermal pressure is expressed as a fraction of the total pressure
\begin{align}
P_{\rm non-thermal}(r) = g(r)P_{\rm total}(r) = \frac{g(r)}{1-g(r)}P_{\rm 
thermal},
\end{align}
with the derivative
\begin{align}
\frac{d P_{\rm non-thermal}}{dr} &= \frac{1}{1-g(r)}\left[g(r)\frac{dP_{\rm 
thermal}}{dr} + \frac{dg(r)}{dr}P_{\rm thermal}  + 
\frac{g(r)}{1-g(r)}\frac{dg(r)}{dr}P_{\rm thermal}\right].
\end{align}

A fit to the $g$-function has been found from  $\Lambda$CDM simulations to be
\begin{align}
 g(r) = \alpha_{\rm nt}(1+z)^{\beta_{\rm 
nt}}\left(\frac{r}{r_{500}}\right)^{n_{\rm 
nt}}\left(\frac{M_{200}}{3\times10^{14}M_{\odot}}\right)^{n_m},
\end{align}
where the free variables have the $\Lambda$CDM best-fit values $\alpha_{\rm 
nt}=0.18$, $\beta_{\rm nt}=0.5$, $n_{\rm nt}=0.8$, and $n_M=0.2$. The 
derivative of the $g$-factor is
\begin{align}
 \frac{dg(r)}{dr} = \frac{n_{\rm nt}}{r}g(r).
\end{align}

Using the best fit we redo the analysis from before, now including the 
non-thermal pressure contribution. The results now differs: with 
the non-thermal 
pressure component 
having introduced a strong mass dependence. However, we want to stress that 
this is just one particular example as the current expression of the 
non-thermal pressure contribution is derived from standard gravity simulations 
is strongly model dependent. Thus, we cannot simply use the expression as is 
for the modified gravity models.

In spite of this complication, we want to note that \textit{in principle} it is 
possible to use the ratio between the thermal and lensing mass to constrain 
screened modified gravity theories, and also -- when including the kinetic mass 
-- to rule out certain combinations of non-universal coupling. All this, 
however, requires the contribution of the non-thermal pressure to be `under 
control', i.e., the magnitude of the intra-cluster turbulence are at least 
limited by observations.

\section{Modelling Void Abundance in Modified Gravity}

Due to the fact that screening mechanisms are much less efficient in low 
density regions, then voids are clearly one of the most promising astrophysical 
objects to observe signatures of a strong and active fifth force, and, 
therefore, to probe possible deviations from General Relativity and evidence 
for 
Modified Gravity. In this section we review in detail how one can use voids 
statistics to perform such work.

\subsection{Linear Power Spectrum}
\label{Scalarperfrref7}


	The spherical evolution model is usually the first step to investigate
	the abundance of virialized objects tracing the Universe structure,
	such as halos, and likewise it is a promising tool for voids 
\cite{Voivodic:2016kog}. It also offers a
	starting point to study the collapse of non-spherical structures
	and  the parameters required to quantify the
	abundance of these objects within extended models.

	The large scale structure of the Universe is well characterized by the
	evolution of dark matter, which interacts only gravitationally and can be 
approximated
	by a pressureless perfect fluid. The line element for a perturbed 
Friedmann-Lema\^itre-Robertson-Walker (FLRW)
	metric in the Newtonian gauge is given by
\begin{equation}
ds ^{2} = -a^{2}(1 + 2\Psi)d\tau ^{2} + a^{2}(1 - 2 \Phi)dl^{2}\,,
\label{Pertubed_metric}
\end{equation}	
where $a$ is the scale factor, $\tau$ is the conformal time related to the
physical time $t$ by $ad\tau = dt$, $dl^2$ is the line element for the spatial 
metric in
a homogeneous and isotropic Universe and $\Psi$ and $\Phi$ are the
gravitational potentials.
	
	For a large class of modified gravity models, the perturbed fluid equations
	in Fourier space are given by
\begin{eqnarray}
\dot{\delta}  &=& -(1+\delta)\theta \,,\\
\dot{\theta} +2H\theta + \frac{1}{3} \theta ^{2}  &=&  k^{2} \Phi\,, \\
-k^{2} \Phi &=& 4 \pi G_N \mu (k,a) \bar{\rho}_{m}\delta\,,
\label{eq:perturbations}
\end{eqnarray}
where $\delta = (\rho_{m} - \bar{\rho}_{m})/\bar{\rho}_{m}$ is the 
matter density contrast, $\theta$
 is the velocity divergence,
  $H = \dot{a}/a$ is the Hubble parameter and dots denote derivatives with 
respect to physical time $t$.

The first is the continuity equation, the second the Euler equation and the 
last is the modified Poisson equation, where modified gravity effects are 
incorporated within
the function $\mu (a,k)$. In general this function depends on scale factor $a$
as well as physical scale or wave number $k$ in Fourier space.

	Combining these equations we obtain an evolution equation for spherical 
perturbations in modified gravity given by
\begin{equation}
\delta '' + \left( \frac{3}{a} + \frac{E'}{E} \right) \delta ' - 
\frac{4}{3}\frac{(\delta ')^{2}}{1+\delta}  = \frac{3}{2} \frac{\Omega 
_{m}}{a^{5} E^{2}}\mu(k,a) \delta (1+\delta)\,,
\label{delta_eq}
\end{equation}
where primes denote derivatives with respect to the scale factor $a$,
$E(a) = H(a)/H_{0}$, $H(a)$ is the Hubble parameter at $a$,
$H_0$ is the Hubble constant and $\Omega_m$ is the present matter density 
relative
to critical. Clearly the growth of perturbations is scale-dependent -- a
general feature of modified theories of gravity.
	
	The linearized version of Eq.~\eqref{delta_eq}  is given by
\begin{equation}
\delta '' + \left( \frac{3}{a} + \frac{E'}{E} \right) \delta ' = \frac{3}{2} 
\frac{\Omega _{m}}{a^{5} E^{2}}\mu(k,a) \delta\,,
\label{delta_lin}
\end{equation}
and can be used to determine linear quantities, such as the linear power 
spectrum.
Notice that this matter linear equation is valid more generally and does not not 
require spherical perturbations.

	The function $\mu(k,a)$ above is given by
\begin{equation}
\mu (k,a) = \frac{(1+2\beta ^{2})k^{2} + m^{2}a^{2}}{k^{2} + m^{2}a^{2}}\,,
\label{mu}
\end{equation}
where $\beta$ is the coupling between matter and the fifth force and $m$ is the 
mass of the scalar field propagating the extra force.

	It is important to stress that the parameterisation in Eq.~\eqref{mu}
	does not fully account for modified gravity perturbative effects, 
containing 
only effects of the background
	and linear perturbations for extra fields related to modified gravity.
	This is enough for the linearized Eq.~\eqref{delta_lin}, but is
	only an approximation in Eq.~\eqref{delta_eq}. For instance the 
parameterisation in Eq.~\eqref{mu}
	does not contain effects from the screening mechanisms, which would turn
	$\mu$ into a function not only of scale $k$, but of e.g. the
	local density or gravitational potential.

	We start by defining the
	linear density contrast field $\delta (R)$
	smoothed on a scale $R$ around  $\textbf{x}=0$ \footnote{The choice  
$\textbf{x}=0$ is irrelevant
	because of translational invariance in a homogeneous Universe, and is
	used for simplificity here,
	as we are interested in the behaviour of $\delta$ as a function of
	scale $R$.}
\begin{equation}
\delta (R) = \int 
\frac{d^{3}k}{(2\pi)^{3}}\tilde{\delta}(\textbf{k})\tilde{W}(k,R) \,,
\label{window}
\end{equation}
where tildes denote quantities in Fourier space and $W(\textbf{x},R)$ is the
window function that smooths the original field $\delta(\textbf{x})$ on scale 
$R$.

	The variance $S(R)=\sigma^2(R)$ of the linear density field can be written 
as
\begin{equation}
S(R) = \langle |\delta (R)|^2 \rangle = \int \frac{dk}{2\pi ^{2}} k^{2} P(k) 
|\tilde{W}(k,R)|^{2}\,,
\label{sigma}
\end{equation}	
where $P(k)$ is the linear power spectrum defined via
\begin{equation}
\langle \tilde{\delta}(\textbf{k})\tilde{\delta}(\textbf{k}')\rangle = 
(2\pi)^{3} \delta _{D}(\textbf{k} - \textbf{k}')P(k)\,,
\end{equation}
and $\delta _{D}(\textbf{k} - \textbf{k}')$ is a Dirac delta function.
	Clearly the linear power spectrum will play a key role in describing the 
effects of modified gravity on void properties.
	For GR computations, we use  \texttt{CAMB} to compute the linear
	power spectrum.
	For modified gravity, we may use \texttt{MGCAMB} , a modified
	version of \texttt{CAMB} which
	generates the linear  spectrum for a number of alternative models, such as 
the
	Hu \& Sawicki $f(R)$ model \cite{Hu:2007nk} and others.
	However it does not compute the linear spectrum
	for instance for the symmetron model. Therefore we also construct the 
linear 
power
	spectrum independently for an arbitrary gravity theory parametrized by
	Eqs.~\eqref{delta_lin} and ~\eqref{mu}.
	
	Our independent estimation of the spectrum is accomplished by evolving 
Eqs.~\eqref{delta_lin}
	and \eqref{mu} with
	parameters from specific gravity theories (e.g.
	for $f(R)$ and for
	symmetron models) for a set of initial conditions at matter domination.
	Since at sufficiently high redshifts
	viable gravity models reduce to GR, we take initial conditions given by 
\texttt{CAMB} at
	high redshifts ($z \approx 100$), when gravity is not yet modified and the 
Universe is
	deep into matter domination. We also compute
	initial conditions for $\dot{\delta}$ numerically by using the $\Lambda$CDM 
power spectrum at
	two closeby redshifts, e.g. at $z=99$ and $z=100$.
	
Comparing the results of using this procedure     with the results 
  for
	the Hu \& Sawicki model we can see that solving
	Eq.~\eqref{delta_lin} for the power spectrum produces results nearly 
identical to the full solution from
	\texttt{MGCAMB} on all scales of interest. The percent level differences 
may 
be traced
	to the fact that the
	simplified equation solved does not contain information about photons
	and baryons, but only dark matter.
	For our purposes, this procedure can
	be used to compute
	the linear power spectrum for other modified gravity models that reduce to
	GR at high redshifts, such as the symmetron model. Additionally, the 
relative 
difference of $\sigma(R)=S(R)^{1/2}$ for the $f(R)$ model
	with respect to GR can be significant
	on the scales of interest ($1$ Mpc/$h < R < 20$ Mpc/$h$). Therefore we 
expect a
	similar impact on void properties derived from $\sigma$ and the linear 
power 
spectrum.

\subsection{Spherical Collapse}
\label{citationalcollapss2}

	Because of the void-in-cloud effect, namely that voids inside 
halos are eventually swallowed and disappear,
	the linearly extrapolated density contrast $\delta _{c}$ for the formation 
of halos is important in describing the properties of voids
	as both are clearly connected.
	Within theoretical calculations of the void abundance using the excursion 
set formalism, $\delta_c$ corresponds to another
	absorbing barrier, whose equivalent is not present for halo abundance. 
Therefore calculating $\delta_c$ in the gravity theory
	of interest gives us
	important hints into the properties of both halos and voids.
	
	The computation of $\delta_c$ is done similarly to that of the GR case, but 
using
	Eqs.~\eqref{delta_eq} and
	\eqref{delta_lin} with the appropriate modified gravity parameterisation 
$\mu (k,a)$ (GR is recovered with $\mu (k,a)=1$).
		 We start with appropriate initial conditions \footnote{These initial 
conditions are actually determined by a shooting method, evolving the 
nonlinear Eq.~\eqref{delta_eq}
	for multiple initial values and checking when collapse happens ($\delta 
\rightarrow \infty$)
	at $a=a_{c}$.}
	 for $\delta $ and
	$\dot{\delta} $ and
	evolve the the linear Eq.~\eqref{delta_lin} until $a_c $.
	The value of $\delta$ obtained is $\delta _{c}$, the density contrast 
linearly extrapolated for halo formation
	at $a=a_c$. In this work, since we only study simulation outputs at $z=0$, 
we take
	$a_{c}=1$ in all calculations.
	The only modification introduced by a nontrivial parameterisation $\mu 
(k,a)$ is that the collapse parameters will depend on the scale $k$ of the
	halo. As mentioned previously, the parameterisation of Eq.~\eqref{mu} only 
takes into account
	the evolution of the scalar field in the
	background, and does not account for the
	dependence of the collapse parameters on screening effects.
	Even though our calculation is approximated, it
	does approach the correct limits at sufficiently large and small scales.

	For a Universe with only cold dark matter (CDM) under GR, the collapse 
equations can be solved analytically
	yielding $\delta_c=1.686$. For a $\Lambda$CDM Universe, still within GR, 
$\delta_c$ changes to a
	slightly lower value, whereas for stronger gravity it becomes slightly 
larger. In particular, the value of $\delta _{c}$ starts at
	its $\Lambda$CDM value $\delta_c=1.675$ on scales larger than
	the Compton scale ($k/a \ll m$; weak field limit where $\mu\approx 1$) and 
approaches
	the totally modified value $\delta_c=1.693$ on smaller scales ($k/a \gg m$; 
strong field
	limit where $\mu\approx 1+2\beta^2=4/3$) where the modification to the 
strength of
	gravitational force is maximal. These values were computed at the background 
cosmology described in
	\S~\ref{sec:void_sim}.
	Note that $\delta _{c}$ reaches its strong field limit faster for
	larger values of $|f_{R0}|$ (value of the extra scalar field today), as 
expected.
	In the approximation of Eq.~\eqref{delta_eq}, $\delta _{c}$ varies with $k$ 
less than in the full collapse,
	indicating that the no-screening approximation may not be sufficient. As a 
full exact calculation is beyond
	the scope of this work and given that $\delta_c$ does not change 
appreciably, in our abundance models we will fix
	$\delta_c$ to its $\Lambda$CDM value and encapsulate modified gravity 
effects  on the linear
	power spectrum and on
	other model parameters.

\subsubsection{Spherical Expansion}

	We now compute $\delta _{v}$, the analog of $\delta _{c}$ for voids, i.e. 
the density contrast linearly extrapolated to today
	for the formation of a void.
	We follow a procedure similar to spherical collapse, but in this case the 
initial values for
	$\delta _{i}$ are negative. We also set a
	criterium in the nonlinear
	field $\delta$ for the formation of a void to be  $\delta _{sc} = -0.8$
	or equivalently
	$\Delta_{sc}=1+\delta_{sc}=0.2$ .
	This quantity is somewhat the analogue for voids of the virial overdensity 
$\Delta_{vir}\approx 180$ for halo formation in
	an  Einstein-de-Sitter (EdS) Universe.
	Despite the value of $\Delta_{vir}$ being only strictly appropriate for an 
EdS Universe, halos are often defined
	with this overdensity or other arbitrary values that may be more 
appropriate 
for specific observations.
	Similarly, $\delta_{sc}=-0.8$ is only strictly appropriate for 
shell-crossing in an EdS Universe. Here we will employ
	$\delta_{sc}=-0.8$, but we should keep in mind that this is an arbitrary 
definition of
	our spherical voids.
	When we fix this criterium for void formation we also fix the factor by 
which the
	void radius $R$ expands with respect to its linear theory radius $R_L$. 
This 
factor is given by
	$R/R_L=(1+\delta _{sc})^{-1/3} = 1.717$,
	and comes about from mass conservation throughout the expansion.
        Differently from halos, voids are not virialized structures and 
continue 
to expand faster than the background. 	
	Again environmental dependences are not incorporated in our computations as 
these values will depend
	only on scale factor $a$ and the scale $k$ or size of the void.
Note that      the behaviour of $\delta 
_{v}$ as a function of $k$,   is very similar to that
	of $\delta _{c}$.
	This is important when modelling the absorbing barriers used for 	
evaluating the void abundance distribution function. Again the values of 
$\delta 
_{v}$ vary with $k$ less than in the full calculation.

	The spherical collapse and expansion calculations can be performed 
similarly 
for the symmetron model, with the
	appropriate change in the expression for the mass and coupling of the 
scalar 
field. For $f(R)$ gravity the change in parameters
	does not seem to be relevant and we fix these parameters to their 
$\Lambda$CDM values. In order to treat
	both gravity models in the same way, we do the same for the symmetron 
model. 
Therefore we do not show explicit
	calculations of $\delta_c$ and $\delta_v$ for symmetron.

\subsection{Void Abundance Function} \label{sec:void_dist}

	We now compute the void abundance distribution function as a function of 
void size
	using an extended Excursion Set formalism, which
	consists in solving the Fokker-Planck equation with appropriate boundary 
conditions. This procedure is valid
	when the barrier (boundary conditions) is linear in $S$ and the random walk 
motion is Markovian.
	
	Differently from the halo description, for voids it is necessary to use two 
boundary conditions, because of the
	void-in-cloud effect. In this case we use two Markovian stochastic barriers 
with linear dependence in the
	density variance $S$, which is a simple generalization from the 
conventional 
problem with a constant barrier.
	The barriers can be described statistically as
\begin{eqnarray}
\langle B_{c}(S) \rangle &=& \delta _{c} + \beta _{c} S\,, \nonumber \\
\langle B_{c}(S)B_{c}(S')\rangle &=& D_{c} \min(S,S')\,, \nonumber \\
\langle B_{v}(S)\rangle &=& \delta _{v} + \beta _{v} S\,, \nonumber \\
\langle B_{v}(S)B_{v}(S')\rangle &=& D_{v} \min(S,S')\,,
\label{barries}
\end{eqnarray}
where $B_{c}(S)$ is the barrier associated with halos and $B_{v}(S)$ the 
barrier 
associated with voids. Notice that
the two barriers are uncorrelated, i.e. $\langle B_{c}(S)B_{v}(S') \rangle=0$.
Here $\beta_c$ describes the linear relation between the mean barrier and the 
variance $S$,
$\delta_{c,v}$ is the mean barrier as $S\rightarrow 0$ ($R\rightarrow \infty$), 
and $D_{c,v}$
describes the barrier diffusion coefficient.

	As we consider different scales $R$, the smoothed density field $\delta(R)$ 
performs a random walk with respect
	to a \emph{time coordinate} $S$, and we have 
\begin{eqnarray}
\langle \delta (S)\rangle &=& 0\,, \nonumber \\
\langle \delta (S) \delta (S')\rangle &=& \min(S,S')\,.
\label{delta}
\end{eqnarray}
We mention  that this occurs when the 
window function in Eq.~\eqref{window} $S$ is sharp in $k$-space.
	For a window that is sharp in real space the motion of $\delta$ is not 
Markovian and the second equation in \eqref{delta} is not true.
	In that case a more sophisticated method is necessary, and the solution 
presented above
	represents the zero-order approximation for the full solution.

	The field $\delta$ satisfies a Langevin equation with white noise and 
therefore the probability density $\Pi(\delta,S)$ to find the
	value $\delta$ at variance $S$ is a solution of the Fokker-Planck equation
\begin{equation}
\frac{\partial \Pi}{\partial S} = \frac{1}{2}\frac{\partial ^{2} \Pi}{\partial 
\delta ^{2}}\,,
\label{FP1}
\end{equation}
with boundary conditions
\begin{equation}
\Pi(\delta = B_{c}(S),S) = 0 \quad \mbox{and} \quad \Pi(\delta = B_{v}(S),S) = 
0\,,
\label{boundary1}
\end{equation}
and  initial condition
\begin{equation}
\Pi(\delta ,S=0) = \delta _{D}(\delta)\,,
\label{initial1}
\end{equation}
where $\delta_{D}$ is a Dirac delta function and notice that $S\rightarrow 0$
corresponds to void radius $R\rightarrow \infty$.
In order to solve this problem, it is convenient to introduce the
variable
\begin{equation}
Y(S) = B_{v}(S) - \delta (S) \,.
\end{equation}
Making the {\it simplifying} assumption that $\beta \equiv \beta _{c} = \beta 
_{v}$ \footnote{Notice that $\beta$ here should not be
confused with the coupling between matter and the extra scalar in 
Eq.~\eqref{mu}}
and using the fact that all variances can be added in quadrature, the 
Fokker-Planck Eq.~\eqref{FP1} becomes
\begin{equation}
\frac{\partial \Pi}{\partial S} =  -\beta \frac{\partial \Pi}{\partial Y} + 
\frac{1+D}{2}\frac{\partial ^{2} \Pi}{\partial Y ^{2}},
\label{FP2}
\end{equation}
where $D=D_{v}+D_{c}$.

	 We define $\delta _{T} = |\delta _{v}| + \delta _{c}$ and notice that 
$\delta (S) = B_{v} (S)$ implies $Y(S) = 0$, $\delta (S) = B_{c}(S)$
	 implies  $Y(S) = -\delta _{T}$ (only occurs because we set $\beta _{c} = 
\beta _{v}$) and $\delta (0) = 0$ implies $Y(0) = \delta _{v} $.
	 Therefore, the boundary conditions become
\begin{equation}
\Pi (Y =0,S) = 0 \quad \mbox{and} \quad \Pi (Y =-\delta _{T},S) = 0\,,
\label{boundary2}
\end{equation}
and the initial conditions
\begin{equation}
\Pi (Y,0) = \delta _{D}(Y-\delta _{v})\,.
\end{equation}

	Rescaling the variable $Y \rightarrow \tilde{Y} = Y/\sqrt{1+D}$ and 
factoring the solution in the form $\Pi(\tilde{Y},S) = U(\tilde{Y},S)\exp 
[c(\tilde{Y} - cS/2 - \tilde{Y}_{0})]$ where $c=\beta /\sqrt{1+D}$ and 
$\tilde{Y}_{0} = \delta _{v}/\sqrt{1 + D}$. The function $U(\tilde{Y},S)$ obeys 
a Fokker-Planck equation like Eq.~\eqref{FP1}, for which the solution is known. 
Putting it all together the probability distribution function becomes
\begin{equation}
\Pi (Y,S) = \exp \left[  \frac{\beta}{1\!+\!D}\left(Y - \frac{\beta S}{2} - 
\delta 
_{v}\right) \right]\cdot  \sum _{n=1} ^{\infty} \frac{2}{\delta _{T}} \sin 
\left( 
\frac{n \pi \delta _{v}}{\delta _{T}} \right) \sin \left( \frac{n\pi}{\delta 
_{T}} Y \right) \exp \left[ - \frac{n^{2} \pi ^{2} (1\!+\!D)}{2 \delta _{T} 
^{2}} S 
\right]\,. 
\end{equation}

	The ratio of walkers that cross the barrier $B_{v}(S)$ is then given by
\begin{equation}
\mathcal{F}(S) = \frac{\partial}{\partial S} \int _{\infty} ^{0} dY \Pi (Y,S) = 
\frac{1+D}{2} \left. \frac{\partial \Pi}{\partial Y} \right|_{Y=0}\,,
\end{equation}
where we used the modified Fokker-Planck equation Eq.~\eqref{FP2} and the first 
boundary
condition from Eq.~\eqref{boundary2}.
The void abundance function, defined as $f(S) = 2S\mathcal{F}(S)$, for this 
model is then given by
\begin{equation}
f(S) = 2 (1+D) \exp \left[ -\frac{\beta ^{2} S}{2(1+D)} + \frac{\beta \delta 
_{v}}{(1+D)} \right] \cdot
\sum _{n=1} ^{\infty} \frac{n\pi }{\delta _{T}^{2}} 
S \sin \left( \frac{n \pi \delta _{v}}{\delta _{T}} \right) \exp \left[- 
\frac{n^{2} \pi ^{2} (1+D)}{2 \delta _{T} ^{2}} S \right]  .
\label{my}
\end{equation}

	There are four important limiting cases to consider:
	
\begin{itemize}

\item $D=\beta =0$: This is the simplest case of two static barriers.  It is 
given by
\begin{eqnarray}
f_{D=\beta =0}(S) = 2 \sum _{n=1} ^{\infty} \frac{n\pi }{\delta _{T}^{2}} S 
\sin \left( \frac{n \pi \delta _{v}}{\delta _{T}} \right) 
\times  \exp \left(- \frac{n^{2} \pi ^{2} }{2 \delta _{T} ^{2}} S \right)\,.
\label{2SB}
\end{eqnarray}
This is one of the functional forms tested in this work and the only case with 
no free parameters. We refer to this case as that of
two static barriers (2SB).

\item $D=0$ and $\beta \ne 0$: This case considers that the barriers depend 
linearly on $S$ but are not difusive. In this case the expression is given
by
\begin{eqnarray}
f_{D=0}(S) = 2 e^{-\frac{\beta ^{2} S}{2}} e^{\beta \delta _{v}} \sum _{n=1} 
^{\infty} \frac{n\pi }{\delta _{T}^{2}} S \sin \left( \frac{n \pi \delta 
_{v}}{\delta _{T}} \right) 
\times  \exp \left(- \frac{n^{2} \pi ^{2} }{2 \delta _{T} ^{2}} S \right).
\end{eqnarray}
	Note that these authors define the barrier with a negative slope,
	therefore our $\beta$ is equal to their $-\beta$, but $\delta _{v} <0$ in 
our case;

\item $\beta = 0$ and $D \ne 0$: Here we have a barrier that does not depend on 
$S$ but which is diffusive. In this case we have
\begin{eqnarray}
f_{\beta = 0}(S) = 2 (1+D) \sum _{n=1} ^{\infty} \frac{n\pi }{\delta _{T}^{2}} 
S \sin \left( \frac{n \pi \delta _{v}}{\delta _{T}} \right) \cdot
\exp \left[- 
\frac{n^{2} \pi ^{2} (1+D)}{2 \delta _{T} ^{2}} S \right].
\end{eqnarray}

\item {\it Large void radius}: As discussed in \cite{Sheth:1999su,Sheth:2003py} 
and \cite{Jennings:2013nsa}, for large radii $R$ the void-in-cloud effect is not
important  as we do not expected to find big voids inside halos. In others 
words, when $S \rightarrow 0  ( R\rightarrow \infty )$ the abundance becomes
equal to that of a one-barrier problem. Even though we do not attempt to 
properly consider the limit
of Eq.~\eqref{my} when
$S\rightarrow 0$, this expression can be directly compared to the function of 
the problem with one
linear diffusive barrier (1LDB), given by
\begin{equation}
f_{1{\rm LDB}}(S) = \frac{|\delta _{v}|}{\sqrt{S(1+D_v)}}\sqrt{\frac{2}{\pi}} 
\exp \left[- \frac{(|\delta _{v}| + \beta_v S)^{2}}{2S(1+D_v)} \right].
\label{One_barier}
\end{equation}

\end{itemize}

We can now  compare the void abundance from multiple cases by 
taking their ratio
	with respect to the abundance of the 2SB model. The abundance of the model 
with
	$D\neq 0$ is
	substantially higher than 2SB, whereas that of the model with $\beta \neq 
0$ 
is 	significantly lower.
	The cases with two linear diffusive barriers (2LDB) Eq.~\eqref{my}
	and one linear diffusive barrier (1LDB) Eq.~\eqref{One_barier} are the main 
models
	considered in this work. The void abundance of the 1LDB and 2LDB models
	are nearly identical for $R>4$ Mpc/$h$, when the same values of $\beta$ and 
$D$ are used.

	Given the ratio of walkers that cross the barrier $B_{v}(S)$ with a radius 
given by $S(R)$, the number density
	of voids with radius between $R_L$ and $R_L+dR_L$ in linear theory is given 
by
\begin{equation}
\frac{dn_{L}}{d\ln R_{L}} = \left. \frac{f(\sigma)}{V(R_{L})} \frac{d \ln 
\sigma 
^{-1}}{d \ln R_{L}} \right|_{R_{L}(R)},
\end{equation}
where the subscript $L$ denotes linear theory quantities, $V(R_L)$ is the 
volume 
of the spherical void of linear radius $R_{L}$ and recall $S=\sigma ^{2}$.

	Whereas for halos the number density in linear theory is equal to the final 
nonlinear number density, for voids this is not the case.
	In fact, Jennings et al. \cite{Jennings:2013nsa} shows that such criterium 
produces nonphysical void abundances, in which the volume fraction of the 
Universe occupied by
	voids becomes larger than unity. Instead, to ensure that the void volume 
fraction
	is physical (less than unity) the authors of \cite{Jennings:2013nsa} impose 
that the volume density is the conserved quantity when going from the
	linear-theory calculation to the nonlinear abundance. Therefore, when a 
void 
expands from $R_{L} \rightarrow R$ it combines
	with its neighbours to conserve volume and not number. This assumption is 
quantified  by the equation
\begin{equation}
V(R)dn = \left. V(R_{L})dn_{L} \right|_{R_{L}(R)} \,,
\end{equation}
which implies
\begin{equation}
\frac{dn}{d\ln R} = \left. \frac{f(\sigma)}{V(R)} \frac{d\ln \sigma ^{-1}}{d\ln 
R_{L}}\frac{d \ln R_{L}}{d\ln R}\right|_{R_{L}(R)} \,,
\label{void_abundance}
\end{equation}
where recall in our case $R=(1+\delta _{sc})^{-1/3}R_L=1.717R_{L}$ is the 
expansion factor for voids. Therefore
we have trivially $d\ln R_L/d\ln R =1$ above.

	The expression in Eq.~\eqref{void_abundance} -- referred as the Vdn model 
-- 
along with the function in Eq.~\eqref{my} provide the theoretical prediction 
for 
the void
	abundance distribution in terms of void radius, which will be compared to 
the abundance of spherical voids found in N-body simulations
	of GR and modified gravity.

\subsection{Voids from Simulations}\label{sec:void_sim}
\label{voidsaref1}

	We use the N-body simulations that were run with the Isis code 
\cite{Llinares:2013jza} for $\Lambda$CDM, $f(R)$
	Hu-Sawicki and symmetron cosmological models. For the $f(R)$ case we fixed 
$n=1$ and considered $|f_{R0}|=10^{-4}$,
	$10^{-5}$ and $10^{-6}$. For symmetron, we fix $\beta _{0} = 1$ and $L=1$ 
and used simulations SymmA, SymmB, SymmD,
	which have $z_{SSB} = 1, 2, 3$ respectively. Each simulation has $512^{3}$ 
particles in a box of size
	$256$ Mpc/$h$, and cosmological parameters $(\Omega _{b}, \Omega _{dm}, 
\Omega _{\Lambda}, \Omega _{\nu}, h, T_{CMB}, n_{s}, \sigma _{8}) = (0.045, 
0.222, 0.733, 0.0, 0.72, 2.726 {\rm K}, 1.0, 0.8)$.
	These represent the baryon density relative to critical, dark matter 
density, effective cosmological constant density, neutrino density,
	Hubble constant, CMB temperature, scalar spectrum index and spectrum 
normalization.
	The normalization is actually fixed at high redshifts, so that
	$\sigma _{8}=0.8$ is derived for the $\Lambda$CDM simulation, but is larger 
for  \label{sigmaref7}
	the modified gravity simulations. In terms of spatial resolution, seven 
levels of refinement were employed on top of a uniform grid with 512 nodes per 
dimension.  This gives an effective resolution of of 32,678 nodes per 
dimension, 
which corresponds to 7.8 kpc/$h$.  The particle mass is $9.26\times 10^9 
M_{\odot}/h$.

We ran the {\tt ZOBOV} void-finder algorithm -- based on Voronoi tessellation --
	 on the simulation outputs
	 at $z=0$ in order to find underdense regions and define voids, and 
compared 
our findings
	 to the Vdn model of Eq.~\eqref{void_abundance} \cite{Jennings:2013nsa} 
with 
the various multiplicity
	 functions $f(\sigma)$ proposed above (2SB, 1LDB and 2LDB models).

	 First, we used {\tt ZOBOV} to determine the position of the density minima 
locations within the
	  simulations
	  and rank them
	  by signal-to-noise S/N significance. Next, we started from the minimum 
density point of
	  highest significance and grew a sphere around this point, adding one
	   particle at a time in each step, until the overdensity $\Delta=1+\delta$ 
enclosed within the sphere
	   was $0.2$ times the mean background density of the simulation at $z=0$. 
Therefore we
	   defined {\it spherical} voids, which are more closely related to our
	   theoretical predictions based on spherical expansion.
	
	We also considered growing voids around the {\it center-of-volume} from the 
central Voronoi
	zones. The center-of-volume is defined similarly to the center-of-mass, but 
each particle position
	is weighted by the volume of the Voronoi cell enclosing the particle,
	instead of the particle mass. Using the center-of-volume produces results 
very similar to the
	previous prescription, so we only present results for the centers fixed at 
the density minima.
	
	We can now  compare the void abundance inferred from 
simulations for the
	three $f(R)$ and the three symmetron
	theories relative to the $\Lambda$CDM model. Since the differential 
abundance as a function
	of void radius is denoted by $dn/d\ln R$, we
	denote the relative difference between the $f(R)$ and $\Lambda$CDM 
abundances by
	$dn_{f(R)}/dn_{\Lambda{\rm CDM}} - 1$ and show
	the results in terms of percent differences.   
	In the $f(R)$ simulation this relative difference is around $100\%$ at 
radii 
$R>10$ Mpc$/h$
	(for the $|f_{R0}|=10^{-4}$ case). In the symmetron simulation, the 
difference is around
	$40\%$ (for the $z_{SSB} = 3$ case), for radii $R\sim8$ Mpc$/h$. This 
indicates that void
	abundance is a potentially powerful tool for constraining modified gravity
	parameters.

\subsection{Results}\label{sec:res}

\subsubsection{Fitting $\beta$ and $D$ from Simulations}

	In order to use the theoretical expression in Eq.~\eqref{my} to predict the 
void abundance
	we need values for the parameters $\beta$ and $D$. The usual interpretation 
of $\beta$ is
	that it encodes, at the linear level, the fact that the true barrier in 
real 
cases is not constant.
	In other words, the contrast density for the void (or halo) formation 
depends on its size/scale. 	
	This can occur because halos/voids are not perfectly spherical and/or 
because the expansion
	(or collapse) intrinsically depends on scale (Birkhoff's theorem is 
generally not valid in
	modified gravity).
	The scale dependency induced by modified gravity can be calculated using 
our 
model
	for spherical collapse (expansion), described in sections II.C and II.D, by 
fitting a linear relationship between $\delta_{c}$ ($\delta _{v}$) or average 
barrier
	$\langle B_c\rangle$ ($\langle B_v \rangle$) as a function of the variance
	$S(R)$. Here we use $k = 2\pi /R$ to convert wave number to scale $R$.	
 
Focusing on the average barriers $\langle B_c 
\rangle$, $\langle B_v \rangle$
        as functions of variance $S$ for multiple gravity theories, and 
empirical fits for the parameters
        $\delta_c, \delta_v, \beta_c, \beta_v$ from Eqs.~\eqref{barries}, we 
find that the corresponding  fits 	indicate that the barriers depend 
weakly on scale in the range 
of interest. The values of
	$\delta_c, \delta_v$ are nearly constant and those of
	$\beta_c, \beta_v$ are of order $10^{-3}$ while the corresponding values for 
halos in
	$\Lambda$CDM are of order $10^{-1}$.
	Even though voids are quite spherical, the small values of $\beta$ indicate 
that 	the main contribution to
	$\beta$ may come from more general aspects of nonspherical evolution. The
	small fitted values of $\beta$ can also be due to errors induced by the
	approximations in the nonlinear equation
	Eq.~\eqref{delta_eq}, which does not capture screening effects of modified 
gravity.
	
Given these issues, and as it is beyond the scope of this work to consider 
more general collapse models or study the exact modified gravity equations, 	
we will instead keep the values of $\delta_c$ and	$\delta_v$ fixed to their 
$\Lambda$CDM values and treat $\beta$ as a free parameter to be fitted from the 
abundance of voids detected in the simulations.
	
	Likewise, the usual interpretation of $D$ is that it encodes stochastic effects of possible
	problems in our void (halo) finder \cite{Maggiore:2009rw}, such as an intrinsic incompleteness or impurity
	of the void sample, or other peculiarities of the finder, which may even differ from one algorithm to another.
	Therefore $D$ is also taken as a free parameter in our abundance models.
	
Focusing on  the abundance of voids $dn/d\ln R$ 
as measured from simulations, as well as three theoretical
	models, namely the 2SB\cite{Jennings:2013nsa}, 1LDB Eq.~\eqref{One_barier} and 2LDB  Eq.~\eqref{my}
	models, we can see that linear-diffusive-barrier models (1LDB and 2LDB) 
work best in all gravities, relative to the static
	barriers model (2SB).  In fact, these two models describe the void abundance distribution within $10\%$ precision for
	$R\lesssim 10$ Mpc/$h$.	
	We mention that the model with two linear diffusive barriers (2LDB) better 
describes the abundance
	of small voids  ($R\lesssim 3 $ Mpc/$h$), due to the void-in-cloud effect, more relevant for
	small voids \cite{Sheth:1999su,Sheth:2003py}.

	As both parameters $\beta$ and $D$ have an explicit dependence on the modified gravity strength,
	next we fit
	a relationship between the abundance parameters $\beta$ and $D$ and the gravity
	parameters $\log _{10} |f_{R0}|$  and $z_{SSB}$.
	In these fits we set the value $\log _{10} |f_{R0}| = -8$ to represent the case of
	$\Lambda$CDM cosmology, as this is indeed nearly identical to $\Lambda$CDM for purposes
	of large-scale structure observables, \label{LSSefs6} i.e. $\log _{10} 
|f_{R0}| = -8 \simeq - \infty $.
	
        As we expect $\beta$ and $D$ to depend monotonically on the modified gravity
        parameters, we fit for them using simple
        two-parameter functions. For $\beta$ case we use a straight line, and for $D$ a
        second order polynomial with maximum fixed by the $\Lambda$CDM value.
     Additionally, our values of $\beta$ and $D$ as a function of gravity 
parameters
fluctuate considerably around the best fit. This occurs at least partially 
because we have used only one simulation for each gravity model, and we expect 
this oscillation to be reduced with a larger number of simulations. At present, 
the use of the fits is likely more robust than the use of exact values obtained 
for each parameter/case.  	
	
\subsubsection{Constraining Modified Gravity}

	Given the discussion on $\beta$ and $D$  of the last subsection, we 
now check for the power
	of constraining modified gravity from the void distribution function in each of the three void
	abundance models considered, namely 2SB, 1LDB and 2LDB.
	We take the abundance of voids actually found in simulations (described in the \S IV) to
	represent a hypothetical real
	measurement of voids and compare it to the model predictions, evaluating the
	posterior for $\log_{10}|f_{R0}|$ and $z_{SSB}$,
	thus assessing the constraining power
	of each abundance model in each gravity theory.
	Obviously the constraints obtained in this comparison are optimistic -- since we are taking
	as real data the same simulations used to fit for the abundance model parameters -- but
	they provide us with idealized constraints
	similar in spirit to a Fisher analysis around a fiducial model.
One can show that the 2SB model predicts values 
for the $f(R)$ parameter
	($\log _{10} |f_{R0}|$), which are incorrect by more than 3$\sigma$ for all 
cases. In fact, this
	model predicts incorrect values even for general relativity. This is not 
surprising given the bad
	$\chi^2$ fits. Therefore we find this model to be highly 
inappropriate to describe
	the abundance of dark matter voids, and focus on
	models with linear diffusive barriers.

Both the 1LDB and 2LDB models predict correct values for the gravity parameters 
within 1$\sigma$ in most cases.
We find that the 1LDB model presents results similar to 2LDB, despite being a 
simpler model and providing a worse fit to the data (larger reduced $\chi 
^{2}$). For $\Lambda$CDM both posteriors go to
$\log _{10} |f_{R0}| = 10^{-8}$, which represents the GR case by assumption.
This shows that within the $f(R)$ framework, we can also constrain GR with 
reasonable precision
from void abundance, using
one of these two abundance models with diffusive barriers (1LDB, 2LDB). 
Finally, for the symmetron model,   the 
parameter $z_{SSB}$
is also well constrained, similarly to $f_{R0}$ in $f(R)$.
Again the 2SB model has the worst result in all cases, and the 1LDB and 2LDB 
models produce
similar results.

\subsubsection{Voids in Galaxy Samples}

	In real observations it is much harder to have direct access to the the dark matter density field.
	Instead we observe the galaxy field, a biased tracer of the dark matter. Therefore it is
	important to investigate the abundance of voids defined by galaxies and the possibility of
	constraining cosmology and modified gravity in this case.
	
	We introduce galaxies in the original dark matter simulations using the
	Halo Occupation Distribution (HOD) model from \cite{Zheng:2007zg}. In \cite{Nadathur:2015qua,Nadathur:2015lha,Nadathur:2014uua} the authors
	investigated
	similar void properties but did not considered spherical voids, using instead the
	direct outputs of the \texttt{VIDE} \cite{Sutter:2014haa} void finder.
	In our implementation, first we find the dark matter halos in the simulations using the
	overdensities outputted by
	\texttt{ZOBOV}. We grow a sphere around each of the densest particles until its enclosed
	density is $200$ times the mean density of the simulation. This process is the
	reverse analog of the spherical void finder described in \S~IV, the only difference being
	the criterium used to sort the list of potential halo centers. Here we sort them using the value of
	the point density, not a S/N significance, as the latter is not provided by
	\texttt{ZOBOV} in the case of halos.
	
	We populate these halos with galaxies using the HOD model of \cite{Zheng:2007zg}.
	This model consist of a mean occupation function of central galaxies given by
\begin{equation}
\langle N_{cen}(M)\rangle = \frac{1}{2} \left[ 1 + {\rm erf}\left( \frac{\log M - \log M_{min}}{\sigma _{\log M}}\right) \right],
\end{equation}
with a nearest-integer distribution. The satellite galaxies follow a Poisson distribution with mean given by
\begin{equation}
\langle N_{sat}(M)\rangle = \langle N_{cen}(M)\rangle \left( \frac{M-M_{0}}{M_{1}'}\right) ^{\alpha}.
\end{equation}
Central galaxies are put in the center of halo, and the satellite galaxies 
are distributed following a
Navarro Frenk and White profile.

	We use parameter values representing the sample {\it Main 1} of  
\cite{Nadathur:2015qua,Nadathur:2015lha,Nadathur:2014uua}, namely:
	$(\log M_{min}$, $\sigma _{\log M}, \log M_{0}, \log M_{1}', \alpha) = 
(12.14, 0.17, 11.62, 13.43, 1.15)$.
	These parameters give a mock galaxy catalogue with galaxy bias $b_g = 1.3$ and mean
	galaxy density $\bar{n}_g = 5.55 \times 10^{-3} (h/$Mpc$)^{3}$ in $\Lambda$CDM.
	
	We then find voids in this galaxy catalogue using the same algorithm applied to the dark matter
	catalogue (described in \S~IV). We use the same criterium that a void is a spherical,
	non-overlapping structure with overdensity equal to $0.2$ times the background
	galaxy density. However, as the galaxies are a biased
	tracer of the dark matter field, if we find galaxy voids with $0.2$ times the mean density, we are
	really finding regions which are denser in the dark matter field. In fact, if
	$\delta _{g}  = b_{g} \delta$ is the galaxy overdensity, with galaxy bias $b_{g}$ and $\delta$ is the dark matter overdensity
	we have
\begin{equation}
\Delta = 1+\delta=1+\frac{\delta _{g}}{b_{g}}.
\end{equation}

	Therefore, if we find voids with $\delta _{g} = -0.8$ and $b_{g} = 1.3$ we have $\Delta = 0.38$,
	i.e. the galaxy voids enclose a region of density $0.38$ times the mean density of the dark matter field.
	Therefore it is this value that must be used in the previous theoretical 
predictions.
	Using this value, the relation between linear and nonlinear radii is $R = 1.37 R_{L}$,
	 and the density parameter for the spherical void formation -- calculated using the spherical
	 expansion equations (\S~II.D) -- is $\delta _{v} = -1.33$. We insert these new values
	 into the theoretical predictions and compare to the measured galaxy void abundance.
	 We mention that  both original models, 2SB and 2LDB (blue 
curves), with $R=1.71R_{L}$ and
	 $\delta _{v} = -2.788$, provide incorrect predictions for the abundance of galaxy voids.
	 However when corrected for the galaxy bias (red curves), these models are in good agreement
	 with the data. We also see that the 2LDB provides a slightly better fit, which is not
	 significant given the error bars.

	The main problem of our galaxy catalogues is the low number density of objects.
	 Larger box sizes (or a galaxy population intrinsically denser)
	 might help decrease the error bars sufficiently in order to constrain modified
	 gravity parameters. Focusing on the  relative difference between the 
abundance for the
	 three modified gravity models and GR as inferred from our simulations, we 
see that it is not possible to constrain the gravity model using the abundance 
of	 galaxy voids, as extracted from mock galaxy catalogues of the size 
considered here,
	 due to limited statistics.
	 Further investigations using larger or multiple boxes, or else considering a
	 galaxy population with larger intrinsic number density should decrease Poisson errors
	 significantly, allowing for a better investigation of void abundance in the large
	 data sets expected for current and upcoming surveys, such as the SDSS-IV, DES, DESI,
	 Euclid and LSST.

\section{Conclusions and Perspectives}\label{perspectives}

In this chapter we briefly review gravity theories beyond General  Relativity 
which may possibly explain several cosmological puzzles, specifically the 
present accelerated expansion of the Universe \cite{Clifton:2011jh}. Modified 
Gravity must comply with strong requirements: One is that the model must have 
similar cosmological predictions to those of  $\Lambda$CDM for the background 
evolution and the linear large scale structures \cite{Ade:2015rim}. Another 
condition is that the modifications to General Relativity are suppressed at 
small scales \cite{Will:2014kxa}. This requirement is assured through the 
so-called screening mechanisms \cite{Brax:2012bsa}. Since different modified 
gravity theories can be degenerate with regard to both the background cosmology 
and the growth rate of linear perturbations, it is crucial to identify new 
probes that can be used to break these degeneracies.  In this paper we study the 
effects of a class of screened modified gravity models in the nonlinear regime 
of structure formation. The aim is to predict possible smoking guns of modified 
gravity and of screening mechanisms at cluster of galaxy scales.

The rise of a fifth force in modified gravity theories leads to a stronger 
clustering of matter.  Therefore, the matter power spectra has in general a 
higher amplitude than in GR. In screened modified gravity theories, the range of 
the fifth force at cosmological scales is  around Mpc  (in order to avoid the 
strong local gravity constraints); therefore, the linear power spectra is 
similar to $\Lambda$CDM, and the differences occur in the small scales of the 
nonlinear regime.

Overall, we find that halo velocity profiles are an excellent direct tracer of 
the fifth force: large deviations in the relative velocity of particles are 
found. Moreover, we find that the velocity field in modified gravity simulations 
is more affected by the presence of the fifth force than the density field for 
$f(R)$-gravity. For the Symmetron model we found this to be even more apparent. 
A particular  striking example of this is the Symm$\_C$ model, with boosts of up 
to $(\Delta P/P)_{\theta}\gtrsim 3$, whereas $(\Delta P/P)_{m}\sim 0.1$.

In order to find smoking guns of screening mechanisms, we  studied the 
environment dependence of  the masses of dark matter halos   
\label{Darkmatterharef2}in the symmetron 
modified gravity scenario. The potential governing the dynamics of the matter 
fields $(\Phi_- + \Phi_+)$ can differ significantly from the lensing potential 
$\Phi_+$ in this model, which leads to a clear difference between the mass of 
the halo as obtained from dynamical measurements and that obtained from 
gravitational lensing. Such an effect found in the symmetron model can be 
significantly stronger than in $f(R)$ gravity. This signature, which is unique 
to modified gravity, can in practice be measured by combining dynamical (e.g., 
velocity dispersion) and lensing mass measurements of clusters of galaxies or 
even single galaxies. We find that the environmental dependence is strongest for 
small halos as very large halos are sufficiently massive to be able to screen 
themselves.

This discovered feature of environmental dependence also allows us, in 
principle, to  distinguish between different modified gravity scenarios such as 
$f(R)$, more general chameleons, and the symmetron. In $f(R)$ the maximum 
fraction of the fifth-force to the Newtonian force in halos are around $30\%$ 
while in chameleon/symmetron scenarios this fraction can be either smaller or 
larger, depending on the value of the coupling strength $\beta$.

Although the theoretical nature of screening mechanisms is
different, we find common features in both the matter and the velocity
properties. In particular, our findings suggest that one can classify screening  
mechanisms into three general
categories: 1) the fully screened regime where GR is recovered, 2) an unscreened 
regime where the strength of the fifth force is large, and, 3) a partially 
screened regime where screening occurs in the inner part of a halo, but the 
fifth force is active at larger radii.  Any observable sensitive to this regimes 
and environments can be a future probe of screening mechanisms and modified 
gravity theories beyond General Relativity.

While the above results concerned modified gravity models which could be an 
alternative to a dark  energy fluid, there are similar astrophysical probes 
which could also indicate that an extended theory of gravity could be an 
alternative to a dark matter particle.

For instance, in Salzano et al. \cite{Mota:2012zw}, motivated by chameleon, 
symmetron and $f(R)$ gravity models,  it was studied a phenomenological scenario 
where the scalar field has both a mass (i.e. interaction length) and a coupling 
constant to the ordinary matter which scale with the local properties of the 
considered astrophysical system. The authors analysed the feasibility of this 
scenario using the modified gravitational potential obtained in its context and 
applied it to the galactic and hot gas/stellar dynamics in galaxy clusters and 
elliptical/spiral galaxies respectively. 

The main results are: 1. The velocity 
dispersion of elliptical galaxies can be fitted remarkably well by the suggested 
scalar field, with model significance similar to a classical Navarro-Frenk-White 
dark halo profile; 2. The analysis of the stellar dynamics and the gas 
equilibrium in elliptical galaxies has shown that the scalar field can couple 
with ordinary matter with different strengths (different coupling constants) 
producing and/or depending on the different clustering state of matter 
components; 3. Elliptical and spiral galaxies, combined with clusters of 
galaxies, show evident correlations among theory parameters which suggest the 
general validity of our results at all scales and a way toward a possible 
unification of the theory for all types of gravitational systems we considered. 

All these results demonstrate that the proposed scalar field scenario can work 
fairly well as an alternative to dark matter. Moreover, Salzano et al. 
\cite{Salzano:2017qac, Salzano:2016udu} investigate an extension of covariant 
Galileon models in the so-called ``beyond Horndeski'' scenario, 
\label{beyondHorndeskiref6} where a breaking 
of the Vainshtein mechanism is possible inside large astrophysical objects, thus 
having possibly detectable observational signatures. It is found that for those 
clusters which are very close to be relaxed, and thus less perturbed by possible 
astrophysical local processes, the Galileon model gives a quite good fit to both 
x-ray and  lensing observations.  When these models are applied  to a sample of 
clusters of galaxies observed under the CLASH survey, using both new data from 
gravitational lensing events and archival data from X-ray intra-cluster hot gas 
observations, one finds that the assumption of having only gas and no Dark 
Matter at all in the clusters is able to match observations. In particular, the 
authors find that, the general relativity limit is excluded at 2$\sigma$ 
confidence 
level, thus making the this type of Galileon model clearly statistically 
different and competitive with respect to general relativity.





\newpage

\phantomsection
\addcontentsline{toc}{part}{\bf Conclusions}
\begin{center}
{\Huge \bf Conclusions}
\end{center}
\begin{center}
Editors:   Emmanuel N. Saridakis,   Paulo Vargas Moniz
\end{center}

\chapter[The End of the Beginning]{The End of the Beginning}

{\em Emmanuel N. Saridakis}\\
 \label{paradishiftref1}

Once  upon a time there was a concrete, holistic, self-consistent 
and perfect
cosmological paradigm in agreement with all observations, a model so 
successful that became the most long-lived scientific system in history. This 
model was 
introduced by Aristotle (incorporating 
previous considerations) around 340 BC  \cite{AristotleHeavens} and took its 
final form after 
the works of Autolycus, 
Eudoxus, Callippus, Apollonius, Conon, Hipparchus, Sosigenes and other Greek 
astronomers, 
and in particular of Claudios Ptolemy \cite{PtolemyAlmagest}.
\label{paradigmref1} \label{Aripstotelftref1}

\begin{figure}[ht]
\begin{center}
\mbox{\includegraphics[width=.75\textwidth]
{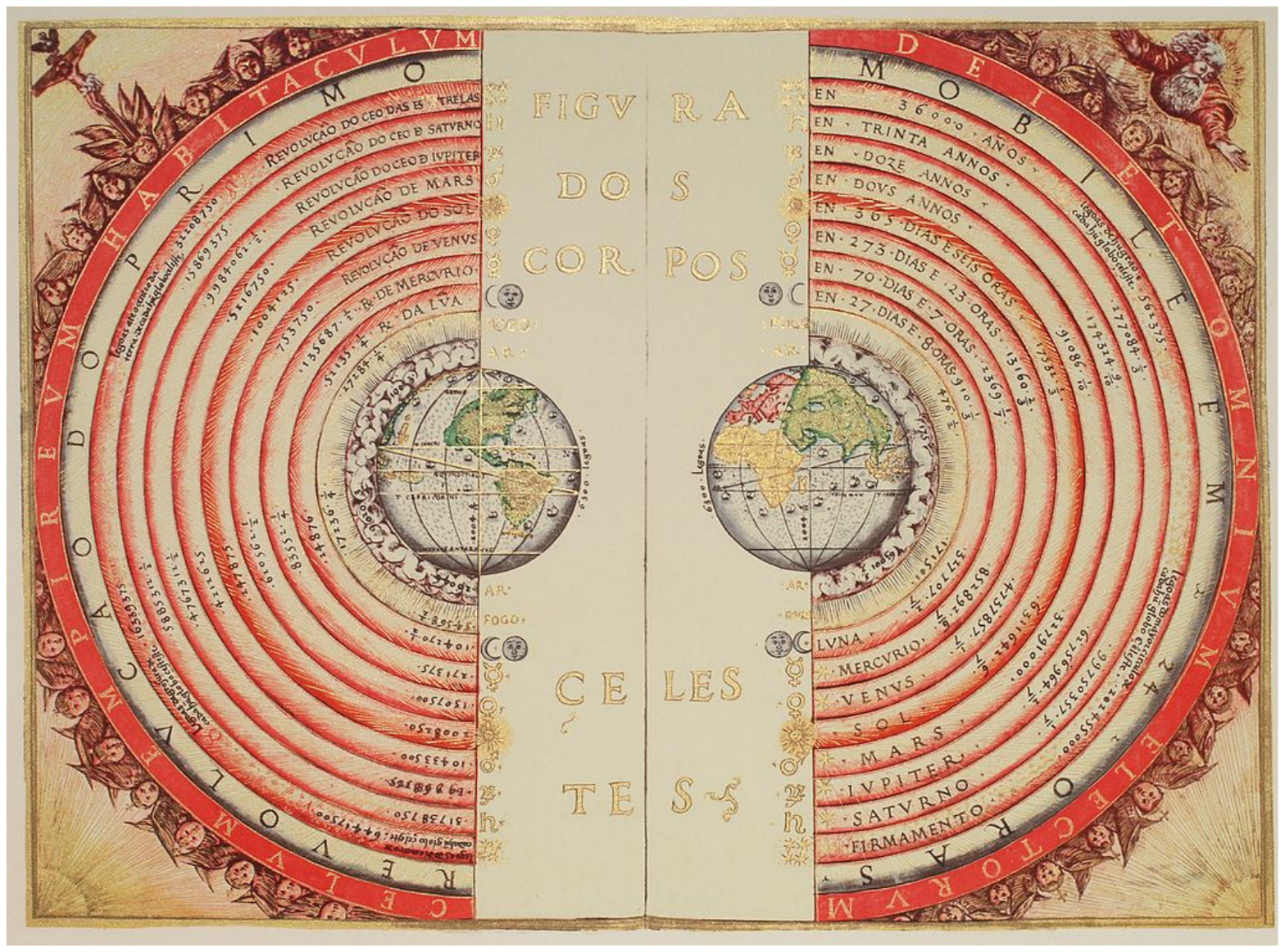}}
\end{center}
\caption{{\it{An illustration of the Aristotelian-Ptolemaic geocentric system 
of 1568  by 
Portuguese 
cosmographer and cartographer Bartolomeu Velho (Biblioth\'eque 
Nationale, Paris).}}}
\label{saridfig}
\end{figure}

In the Aristotelian-Ptolemaic cosmological and physical 
paradigm  the Earth is spherical and exists in the center of the universe 
(geocentrism), it 
does not revolve 
around anything else neither it rotates around its own axis. It 
is composed of four elements and their combinations:  Earth, Water, Fire and 
Air. It is surrounded by 
ten concentric spheres made of a perfectly transparent substance known as 
``quintessence'' (the ``fifth'' element). These spheres revolve around the 
earth, carrying the other celestial bodies. The first    is the sphere of the 
Moon (Lunae), the second of the Mercury (Mercurii), the third of the Venus 
(Veneris), the fourth is the Sun (Solis), the fifth of the Mars (Martis), the 
sixth of the  Jupiter (Iovis), the  seventh of the Saturn (Saturni), and 
spheres 
eight, nine and ten hold the ``fixed stars''. The fixed stars were given their 
name since
they do not move relative to each other, unlike the planets, which move among
and independently of the fixed stars (the Greek word planet means exactly 
that: wanderer). Even the subtle and puzzling observation of the ``retrograde 
motion'' was incorporated in the model by Ptolemy through the epicycles 
modification. Finally, the Universe is spherical, finite and eternal.  A 
Renaissance illustration of the  Aristotelian-Ptolemaic 
cosmological paradigm can be seen in Fig. \ref{saridfig}.

The Aristotelian-Ptolemaic system was remarkably plausible and powerful as a 
scientific theory, intuitively explaining all observations and being 
consistent with the main philosophical considerations. Hence, despite the 
existence of alternative models, and in particular of the Earth-rotational 
model 
of Heraclides Ponticos (4rth century BC), and of the heliocentric model of 
Aristarchos of Samos (3rd century BC) and Seleucus of Seleucia (2nd century 
BC), the Aristotelian-Ptolemaic paradigm remained the absolute cosmological 
model for more than one thousand five hundred years.

It was only after the 11th century AD, where various Arab and Persian scholars 
such as 
Alhazen, Al-Zarkali, Fakhr al-Din al-Razi, Al-Tusi and Ibn al-Shatir, first 
through philosophical considerations and thinking and then incorporating 
observations performed in Maragha observatory, started putting into doubt the 
details of Aristotelian-Ptolemaic 
model such as the epicyles and the Earth's non-rotation 
\cite{HistoryArabicAstronomy},
although never actually disputing geocentrism \cite{HuffModernScience}.
 The actual ``paradigm shift'' to heliocentrism was initialized by Copernicus 
in 1543 \cite{Revolutionibus}
 (the influence of the Islamic, and in particular of Maragha, school on 
Copernicus 
remains speculative \cite{EncyclopaediaIslamic},
however it has been verified that Copernicus had read in detail pieces of 
Aristarchos of Samos \cite{ThomasHeath1913}\footnote{Just before Copernicus 
sent 
his manuscript to be published in 1543, afraid of reactions he removed 
two pages from the submitted manuscript that acknowledged his indebtedness to 
the ancient Greek astronomers. However, he kept the two pages in his own 
personal copy, 
which was not discovered until 300 years later, towards the end of the 19th 
century \cite{RosenAristarchus1978}.}).

We stress here that  the Aristotelian-Ptolemaic
 system was still the standard cosmological 
paradigm for many years after Copernicus, as at the time the Copernican system, 
which still used circular orbits, did not offer better predictions than 
the geocentric system, and it posed problems for both natural 
philosophy and Christian scripture.
It was only after many decades, where detailed observations made by Tycho 
Brahe, Johannes Kepler, and Galileo Galilei
postulated elliptical orbits and established heliocentrism 
\cite{KeplerAstronomia} (although still not 
completely accepted, as Giordano Bruno's fate revealed).

Finally, the theoretical justification for the above ``paradigm shift'' was 
offered one century later by Isaac Newton, since for the first time he offered 
a physical theory as the underlying foundation of the cosmological model, 
namely the theory of gravitation \cite{Newtonprincipia}. In order to achieve 
this, Newton had first to 
use the detailed  results of gravitation experiments made by Galileo, which 
allowed him to dispute the Aristotelian ideas on gravity (bodies do not fall 
towards Earth in order to come back to their ``initial state'' but due to the 
attraction between masses), as well as the detailed astronomical observations 
of Brahe and Kepler which allowed him to conclude that it is the same force 
that makes the celestial bodies to move and the bodies on Earth to fall. As he 
himself wrote:
``If I have seen further it is by standing on the shoulders of Giants'' 
\cite{NewtonLetterHooke}.
 
 Newtonian gravity and the Newtonian-Keplerian astronomical model was very 
successful in explaining observations. Nevertheless, theoretical investigation 
and progress in the fields of mechanics and especially in the novel field of 
electromagnetism \cite{Lorentzoriginal},
as well as the technological advance that allowed for radical increase in 
experimental capabilities (e.g. the Michelson-Morley experiment 
\cite{Michelson:1887zz})
and in observational accuracy (e.g. observation of the 
precession  of Mercury's perihelion \cite{1880AnPar..15...23T}),
started putting it into doubt. Despite the fact that some of the findings 
could be explained without a changing in the foundations of the Newtonian system
\footnote{Urbain Le Verrier was able to describe the discrepancies with 
Uranus's orbit by predicting through purely theoretical calculations  the 
existence and current position of a new planet, namely Neptune, which was 
indeed discovered by Johann Gottfried Galle in 1846  at exactly the predicted 
coordinates. This astonishing validation of celestial Newtonian mechanics led  
Verrier in 1859 to propose that the  precession of Mercury's orbit was a result 
of another extra planet between Mercury and Sun, namely the Vulcan, or even of 
a 
series of smaller ``corpuscules'' \cite{2005ASTPS...9.....H}.}
the complete and self-consistent incorporation of both theoretical research and 
all experimental and observational data  led Einstein 
to construct a new gravitational theory. The theory of General Relativity, had 
different foundations, physical interpretation and mathematical structure from 
Newtonian gravity, nevertheless at the level of predictions it was served as 
its modification, exhibiting the latter as a particular limit and offering 
corrections that were larger at scales that had just then started to become 
  accessible  by technological advance.

The shift in the description of gravity soon started having effects in the 
description of astronomy, since from Newton times it   was established that 
astronomical behavior was the result of gravitational interactions. 
Additionally, many people, amongst them Einstein \cite{1917SPAW.......142E}, 
realized that an 
understanding of gravity offered the way to understand not only astronomy (the 
``laws of stars'') but also quantitatively describe the behavior of the 
universe 
as a whole  and thus bringing cosmology (``laws of cosmos'')
 from the sphere of philosophy, to which it belonged for 
thousands of years, to the sphere of natural sciences.

 The following decades were characterized by a significant advance in the 
quality and quantity of observations, leading to corrections and improvements 
in the new cosmological paradigm. The discovery of extra galaxies beyond the 
Milky Way,
that are moreover moving away from each other \cite{Hubble:1929ig},
established the framework of an expanding and cooling universe originating from 
a primordial super-dense and super-hot state.
Although this ``Big Bang'' theory offered verified quantitative predictions 
(e.g. the abundance of primordial elements and the cosmic microwave background 
radiation)
and was able to describe all observations,
theoretical investigation revealed that it might have some theoretical 
``problems'' (or at least issues whose explanation was not ``natural''), such 
as the horizon, the flatness and the magnetic monopole ones. Hence,   
after 1980 the  phase of inflation 
\cite{Kazanas:1980tx,Guth:1980zm,Linde:1981mu}
was established as a necessary ingredient of the cosmological paradigm at its 
early stages
\footnote{We will not enter here into the discussion of whether inflation is an 
over-predictable, non-falsifiable theory, and whether it 
completely satisfies the criteria of being a successful physical theory. The 
reader could see \cite{Ijjas:2014nta} and references therein.}.
Finally, the last   decades the cosmological paradigm underwent through another 
modification, in order to incorporate the ``indirect observation'' of the dark 
matter sector, as well as the direct observation of accelerated expansion. 
Hence,  the concordance model, the paradigm of $\Lambda$CDM  cosmology,
is now well established. Namely, a universe filled with the Standard 
Model particles plus the dark matter sector, which is governed by the 
gravitational theory of General Relativity with the addition of the 
cosmological 
constant, and which has originated from an inflationary Big Bang.

As one can see, exhaustive and diligent observations in various areas, 
in a dialectical relation with pure theoretical thinking and investigation, led 
to a series of paradigm shifts, each time offering a better and deeper 
description of Nature. As a result, the  concrete, holistic, self-consistent 
and 
perfect Aristotelian-Ptolemaic cosmological and physical paradigm, in agreement 
with all observations and theoretical considerations for more than one and a 
half thousand years, gave its place to the  concrete, holistic, self-consistent 
and perfect cosmological and physical  paradigm of $\Lambda$CDM cosmology and 
General Relativity,    in agreement 
with all observations and theoretical considerations for many decades.

A question thus arises naturally: Since accumulated observational, 
experimental and theoretical research and results are able (after a period of
challenge of validity and dispute) to transmute viable and established 
theories, scenarios and paradigms into non-viable ones, which are then forced 
to 
give their place to new paradigms that are able to share the successes of their 
ancestors without sharing their failures, what will be the future of 
$\Lambda$CDM cosmology and 
General Relativity? Are these the final theories that will remain valid 
perpetually? They are not the final theories but for the moment there is no
theoretical or
observational need for their shift? They are not the final theories and we have 
already entered 
the phase of their challenge of validity and dispute? Has the paradigm shift 
already happened?
  
General Relativity and  $\Lambda$CDM cosmology are remarkably  powerful 
and successful in describing Nature. From the experimental point of view they 
have been completely verified in every experiment made. From the observational 
point of view they are to a large extent (if not completely) in agreement with 
observations. From the theoretical point of view, they are on an acceptable 
level, though presenting various disadvantages. Nevertheless, intense 
theoretical investigation over the last 50 years, as well as the radical 
improvement 
and advance in astronomical, astrophysical and cosmological observations over 
the 
last 25 years, has raised various questions.

Some of the theoretical questions are 
the following:
i) General Relativity is non-renormalizable and hence it cannot get quantized 
with standard approaches. Definitely, there is not any fundamental principle 
that gravity should be quantum, however it is almost a consensus in 
theoretical physics community that the fact that the other three fundamental 
interactions are, offers a strong indication that gravity should   have a 
quantum nature too.\footnote{Note that the AdS/CFT correspondence 
\cite{Maldacena:1997re} revealed the 
possibility that the quantum behavior could be of effective nature arising from 
classical foundations. Although in this specific framework this did not turn 
out to be the case (at the ``gravity side'' gravity still needs to be quantum) 
it offered us the idea that ``quantum'' may be more complex than current 
knowledge suggests.}
 ii) The cosmological constant problem, namely what is the cosmological 
constant, what is its microscopic nature, and more importantly why is
its observed value is many orders of magnitude smaller than the result of any 
field theoretical calculation.
iii) What is the microscopic nature of dark matter.
iv) What is the exact mechanism/scenario that leads to the successful 
inflationary realization.
 
Some of the observational issues (unless these   are wrong due 
to unknown systematics) may include: \label{H0tensionjref4}
i) The $H_0$ tension: The direct measurements of the present value of the 
Hubble 
parameter give a value that is
larger ($\sim5\sigma$) than the one estimated by the Planck Probe through the 
CMB spectrum in the context of $\Lambda$CDM cosmology.
ii) The $\sigma_8$ tension:   possible deviation was noticed between 
measurements of CMB and LSS surveys. iii)  The cuspy halo problem,
the dwarf galaxy problem,  and other 
clustering-related problems that seem to puzzle   standard collisionless dark 
matter.
  \label{sigmaref8}

Even if the above observational issues, as well as others of less significance, 
  will be proven to be wrong and $\Lambda$CDM cosmology with General Relativity 
will remain in perfect agreement with observations, still the theoretical 
issues 
are quite strong and seem to suggest that some kind of modification is still 
needed. In principle, the community  follows two main roads to alleviate them. 
The first is to remain in the framework of General Relativity and modify the 
content of the universe, namely to consider extra fields and sectors, such as 
the inflaton, quintessence etc, or other  particles
beyond the Standard Model of particles physics. The second direction is to 
extend/modify the gravitational theory itself, namely construct
gravitational theories which possess General Relativity as a particular limit, 
but which in general provide the extra degree(s) of freedom that could 
incorporate the aforementioned issues. Definitely, one may follow 
combinations of both
main directions. 

As one may see, the first direction, even if it successfully solved all 
observational and cosmology-related theoretical issues it cannot offer a 
solution to the theoretical issues related to General Relativity itself, since 
it leaves the gravitational theory unchanged. On the other hand, the ultimate 
goal of the second direction is to provide a complete and coherent solution to 
all open issues, namely both to cosmological as well as to purely theoretical 
ones.

If the modifications remain in the first direction, strictly speaking  it 
will not constitute a paradigm shift, but rather an improvement of the existing 
paradigm (in the same lines that the discovery of extra particles in the 
framework of Standard Model, e.g. the top quark or the tau neutrino, is 
considered as its modification/completion and not as a paradigm shift). On the 
other hand, in the case of a novel gravitational theory, and depending on the 
amount of deviation from General Relativity in terms of  foundations, physical 
interpretation and mathematical structure, a paradigm shift might seem secured. 
The reader must have already noticed the interesting similarity with the 
explanation of the perihelion shift of Mercury, where the paradigm shift of the 
new 
gravitation theory of General Relativity was proved to be the case instead of 
the existence of an extra planet, unseen till then and thus ``dark'', in the 
framework of Newtonian gravity.

In the previous chapters of this Review we presented in detail the state of the 
art of gravitational modifications and extensions, as well as their 
cosmological and astrophysical implications, a topic that was the main field of 
interest of CANTATA Collaboration. In particular, we tried to provide   
the classification and definition of theoretical and phenomenological aspects 
of gravitational interaction that cannot be enclosed in the concordance model 
but might be considered as signs of alternative theories of gravity, to present 
 the confrontation of the theoretical predictions with observations at both the 
background and the perturbation levels, and to examine   how extended and 
modified theories of gravity could emerge from quantum field theory.
We hope that the reader has obtained a clear picture of 
the motivations, the mathematical structure, the resulting physical behavior, 
and their advantages and disadvantages.

As we saw, the theories of gravitational extension and modification that are 
efficient in incorporating observations on an advanced level, and that attract 
the interest of the cosmological community, do not seem to alter the 
foundations 
of General Relativity. Gravity is still classical, it is related to the 
geometry 
of spacetime (even if this geometry is different or richer than the Riemannian 
one), and its Lagrangian is constructed by various invariants and fields. On 
the other hand, gravitational theories of radical difference from General 
Relativity, with novel foundations, mathematical structure, and physical 
interpretation, like string theory \cite{Aharony:1999ti}, loop quantum gravity 
\cite{Rovelli:1997yv}, 
and supergravity \cite{Nilles:1983ge},
  which attract  the interest of the 
theoretical-physics community, although efficient in solving the 
theoretical problems of   General Relativity, for the moment seem to 
be far away from providing better explanations and predictions for the 
high-accuracy 
near-future observations and experiments  comparing to $\Lambda$CDM cosmology 
and 
General Relativity. Elaborating these progresses, and, in addition, knowing 
  the 
structure of the previous scientific revolutions, we may deduce that the 
construction of the   gravitational theory that could replace General 
Relativity will be a complex procedure that would require the superposition of 
argumentation and ideas from geometry, quantum field theory, statistical 
physics, complexity theory, known unknowns, as well as   unknown unknowns.

So what can we say about the previously raised question? What is the future of
$\Lambda$CDM cosmology and General Relativity? Clearly, there has not been any 
paradigm shift, and they remain the widely accepted lore system for the 
description of Nature. Additionally, still we have not entered into the phase 
of their challenge of validity and dispute, since their predictions are in 
general the best amongst various alternative theories and scenarios. 
Nevertheless, pieces of observational evidences, as well as 
diligent theoretical investigation, seem to have opened the door towards such a 
phase. The fact that we still  lack robust and successful alternative paradigms 
is definitely an inhibitory and suspending factor, however the relevant 
discussion attracts a considerable part of the cosmological community.

If we were to use historical appositions with the Aristotelian-Ptolemaic
paradigm, Newton  has not appeared. The detailed observations of Brahe and  
Kepler and the experiments of Galileo have not appeared either. Copernicus  
ideas and proposition might or might not have appeared, it is too early to 
conclude on that. However, the argumentation and early observations 
of the Arab and Islamic school have definitely appeared. 

From these it is revealed that we are not in the end. We are not in the 
beginning of the end either. It is better to say that we are in the end of the 
beginning. The investigation of Nature as a whole, and not only of specific 
parts of it, in order to reveal the truth, namely the 
physical laws that govern objective reality, is a hard and arduous procedure. 
A procedure that is not a solitary or instant play but it demands the 
collective 
work of scientific community in many generations. Nature is out there, it is 
knowable, and the 
physical laws, the first principles and the 
highest causes, are waiting to be discovered. It is up to us to accept the 
challenge.\\

\begin{displayquote}

  {\it{
 The investigation of the truth is in one way hard, in another easy. An 
indication of this is found in the fact that no one is able to attain the truth 
adequately, while, on the other hand, no one fails entirely, but everyone says 
something true about the nature of things. And while individually they 
contribute little or nothing to the truth, by the union of all a considerable 
amount is amassed. Therefore, since the truth seems to be like the proverbial 
door, which no one can fail to hit, in this respect it is easy, but the fact 
that we can have a whole truth and not the particular part we aim at, shows the 
difficulty of it. Perhaps, too, as difficulties are of two kinds, the cause of 
the present difficulty is not in the facts but in us 
\cite{AristotleMetaphysicsb}.

 There is a science which investigates being as being and the attributes which 
belong to this in virtue of its own nature. This is not the same as any of 
the so-called special sciences; for none of these treats universally of being 
as being. They cut off a part of being and investigate the attribute of this 
part. Now since we are seeking the first principles and the highest causes, 
clearly there must be some thing to which these belong in virtue of its own 
nature. If then those who sought the elements of existing things were seeking 
these same principles, it is necessary that the elements must be elements of 
being not by accident but just because it 
is being. Therefore it is of being as being that we also must grasp the first 
causes 
\cite{AristotleMetaphysicsg}. \\}}
 
 \hspace{9.5cm} Aristotle,  in {\it{Metaphysics}}.
  
\end{displayquote}

\newpage
\section*{Acknowledgements}

This publication is supported by COST (European Cooperation in
Science and Technology)  and is based upon work from CANTATA COST action 
CA15117, EU Framework Programme Horizon 2020.

Y.A. is supported by 
LabEx ENS-ICFP: ANR-10-LABX-0010/ANR-10-IDEX-0001-02 PSL*.

J.B.J. would like to thank David Figueruelo, for his help in compiling and 
unifying parts of the manuscript.

J.L.B-S, B.K. and J.K. gratefully acknowledge support by the DFG Research 
Training Group 1620 {\sl Models of Gravity} and the COST Action CA16104. JLBS 
would like to acknowledge support from the DFG project BL 1553, and the FCT 
projects PTDC/FISOUT/28407/2017 and PTDC/FIS-AST/3041/2020.

M.B.L.'s work is supported by the Basque Foundation of Science Ikerbasque. She 
also would like to acknowledge the partial support from the Basque government 
Grant No. IT956-16 (Spain) and from the project FIS2017-85076-P 
(MINECO/AEI/FEDER, UE).

G.C. thanks M.\ Maggiore for useful feedback; his work is supported by the I+D 
grant FIS2017-86497-C2-2-P of the Spanish Ministry of Science, Innovation and 
Universities. 

J.A.R.C.  acknowledges financial support from the MINECO (Spain) Projects No. 
FIS2016-78859-P (AEI/ FEDER) 
and No. PID2019-107394GB-I00.

A.d.l.C.-D. acknowledges financial support from NRF Grants No.120390, Reference: 
BSFP190-416431035, No.120396, Reference: CSRP190405427545, and No 101775, 
Reference: SFH1507-27131568; Project No. FPA2014-53375-C2-1-P from the Spanish 
Ministry of Economy and Science, MICINN Project No. PID2019-108655GB-I00, 
Project No. FIS2016-78859-P from the European Regional Development Fund and 
Spanish Research Agency (AEI).

A.-C. D. is supported by a PhD contract of the program \emph{FPU 2015} 
(SpanishMinistry of Economy and Competitiveness) with reference FPU15/05406, and 
acknowledges further support from the Spanish project No. FIS2017-84440-C2-1-P 
(MINECO/FEDER,EU), i-LINK1215 (CSIC), and the projects SEJI/2017/042 and 
PROMETEO/2020/079(Generalitat Valenciana).

 J.M.E. is grateful to M. Zumalac\'arregui, J. Beltr\'an, L. Heisenberg and 
J. Garc\'ia-Bellido for fruitful discussions and collaborations on topics 
covered in this contribution; he has been supported by the Spanish FPU Grant 
No. FPU14/01618, the Research Project FPA2015-68048-03-3P (MINECO-FEDER) and 
the Centro de Excelencia Severo Ochoa Program SEV- 2016-0597, by NASA through 
the NASA Hubble Fellowship grant HST-HF2-51435.001-A awarded by the Space 
Telescope Science Institute, which is operated by the Association of 
Universities for Research in Astronomy, Inc., for NASA, under contract 
NAS5-26555, and by the Kavli Institute for Cosmological Physics 
through an endowment from the Kavli Foun- dation and its founder Fred Kavli.
 
   N.F. and  S.P. thank Louis Perenon for useful discussions;  the research 
of NF is supported by Funda\c{c}\~{a}o para a  
Ci\^{e}ncia e a Tecnologia (FCT) through national funds
   (UID/FIS/04434/2013), by FEDER through COMPETE2020  
(POCI-01-0145-FEDER-007672) and by FCT project ``DarkRipple -- Spacetime
   ripples in the dark gravitational Universe" with ref.~number 
PTDC/FIS-OUT/29048/2017, and S.P. acknowledges support from the NWO and the 
Dutch Ministry of Education, Culture and Science (OCW), and also from the D-ITP
   consortium, a program of the NWO that is funded by the OCW.
 
L.\'{A}.G. is grateful to Cec\'{\i}lia Nagy for generating the figure, to
Roberto Casadio for discussions o and to David Hobill
for collaboration in the subject over many years; he was supported by the
Hungarian National Research Development and Innovation Office (NKFIH) in the
form of the Grant No. 123996.
 
L.K. and L.P. thank Levon Pogosian, Alex Zucca, Savvas Nesseris and George 
Alestas for their help with the MGCAMB and MGCOSMOMC packages; all the runs 
were performed in the Hydra Cluster of the Institute of Theoretical Physics 
(IFT) in Madrid, their research is co-financed by Greece and the European Union 
(European Social Fund- ESF) through the Operational Programme ``Human Resources 
Development, Education and Lifelong Learning" in the context of the project 
``Strengthening Human Resources Research Potential via Doctorate Research" 
(MIS-5000432), implemented by the State Scholarships Foundation (IKY).

J.P.M. and F.S.N.L. acknowledge funding from the  Funda\c c\~{a}o para a 
Ci\^encia e a Tecnologia (FCT, Portugal) research projects No. 
UID/FIS/04434/2020, No. PTDC/FIS-OUT/29048/2017 and No. CERN/FIS-PAR/0037/2019; 
F.S.N.L. additionally acknowledges support from the FCT Scientific Employment 
Stimulus 
contract with reference CEECIND/04057/2017.

 P.M.M. acknowledges financial support from the project FIS2016-78859- P 
(AEI/FEDER, UE).

M.M. has received the support of a fellowship from ``la Caixa'' Foundation (ID 
100010434), with fellowship code LCF/BQ/PI19/11690015. 
 
D.F.M. thanks M. Gronke, R. Hagala, A. Hammami, C. Llineares, M. Lima, R. 
Voivodic, 
and H. Winther whose collaborations resulted in the articles which this review 
is based on; his work was partially supported by the Research Council of 
Norway and the simulations where performed in NOTUR. 
 
G.J.O. is supported by the 
Spanish project FIS2017-84440-C2-1-P (MINECO/FEDER, EU), the project 
H2020-MSCA-RISE-2017 Grant FunFiCO-777740, the project PROMETEO/2020/079 
(Generalitat Valenciana), and the Edital 006/2018 PRONEX (FAPESQ-PB/CNPQ, 
Brazil, Grant 0015/2019). 

 V.P. and A.S.  thank Jean-Luc Starck for useful comments.
 
 D.R.G. is funded by the \emph{Atracci\'on de Talento Investigador} programme of 
the Comunidad de Madrid (Spain) No. 2018-T1/TIC-10431, 
and acknowledges further support from the Ministerior de Ciencia, Innovaci\'on y 
Universidades (Spain) project No. 
PID2019-108485GB-I00/AEI/10.13039/501100011033, the FCT projects No. 
PTDC/FIS-PAR/31938/2017 and PTDC/FIS-OUT/29048/2017, the Spanish project
No. FIS2017-84440-C2-1-P (MINECO/FEDER, EU), the project PROMETEO/2020/079 
(Generalitat Valenciana), and the Edital 006/2018 PRONEX
(FAPESQ-PB/CNPQ, Brazil) Grant No. 0015/2019.

E.N.S is supported in part by the USTC Fellowship for international
professors.

I.S. is partially supported by a 2019 ``Research and Education" grant from 
Fondazione CRT. The OAVdA is managed by the Fondazione Cl\'ement 
Fillietroz-ONLUS, which is supported by the Regional Government of the Aosta 
Valley, the Town Municipality of Nus and the ``Unit\'e des Communes 
vald\^otaines Mont-\'Emilius''. I.D.S. is funded by European Structural and 
Investment Funds and the Czech Ministry of Education, Youth and Sports (Project 
CoGraDS — CZ$.02.1.01/0.0/0.0/15_003/0000437$).

 A. W. is supported by the EU through the European Regional Development Fund CoE 
program TK133 ``The Dark Side of the Universe''.

\newpage
\section*{Authors' Institutions}

\noindent {$ ^1${National Observatory of Athens, Lofos Nymfon, 11852 Athens, 
Greece}

\noindent {$ ^2${CAS Key Laboratory for Researches in Galaxies and Cosmology,
Department of Astronomy, University of Science and Technology of China, Hefei,
Anhui 230026, P.R. China}
 
\noindent {$ ^3$ School of Astronomy, School of Physical Sciences,
University of Science and Technology of China, Hefei 230026, P.R. China}

\noindent {$ ^4${Department of Theoretical Physics, University of 
the Basque Country UPV/EHU, P.O. Box 644, 48080 Bilbao, Spain}}

\noindent {$ ^{5}${Institute of Physics, University of Szczecin, Wielkopolska 
15, 70-451 Szczecin, Poland}}

\noindent {$ ^6 ${Departamento de F\'{i}sica, Centro de Matem\'atica e 
Aplica\c{c}\~oes
(CMA-UBI), Universidade da Beira Interior, 6200 Covilh\~a, Portugal}}

\noindent {$ ^7 ${DAMTP,  Centre for Mathematical Sciences, University of 
Cambridge CB3 0WA, United 
Kingdom   }}

\noindent {$ ^8${
Dipartimento di Fisica "E. Pancini", Universit\'a di Napoli "Federico II",
80126 Napoli, Italy}}

\noindent {$ ^9${INFN, Sezione di Napoli, Complesso Universitario di Monte S. 
Angelo, Via 
Cintia Edificio 6, 80126 Napoli, Italy}}

\noindent {$ ^{10}${Laboratory for Theoretical Cosmology,
Tomsk State University of Control Systems and Radioelectronics (TUSUR),
634050 Tomsk, Russia}}

\noindent {$ ^{11}${ Departamento de F{\'i}sica Fundamental and IUFFyM, 
Universidad de 
Salamanca, E-37008 Salamanca, Spain}}

\noindent {$ ^{12}${Istituto Nazionale di Fisica Nucleare (INFN), Sezione di 
Torino, Via P. Giuria 
1, I-10125 Torino, Italy}}

\noindent {$ ^{13}${Depto. de F\'isica Te\'orica and IFIC, Centro Mixto 
Universidad de 
Valencia-CSIC, \\ Burjassot-46100, Valencia, Spain}}

\noindent {$ ^{14}${Departamento de F\'isica, Universidade Federal da 
Para\'iba, 58051-900 Jo\~ao Pessoa, Para\'iba, Brazil}}

\noindent {$ ^{15}${Laboratoire de Physique de l'\'Ecole Normale Sup\'erieure, 
ENS, Universit\'e PSL, CNRS, Sorbonne Universit\'e, Universit\'e de Paris, 
F-75005 Paris, France}}

\noindent {$ ^{16}${Observatoire de Paris, Universit\'e PSL, Sorbonne 
Universit\'e, LERMA, 75014 Paris, France}}

 
\noindent {$ ^{17}${Laboratory of Theoretical Physics, Institute of Physics, 
University  of Tartu, W. Ostwaldi 1, 50411 Tartu, Estonia}}

\noindent{$ ^{18}$ {  Department of Mathematics, University College London, 
Gower Street, London, WC1E 6BT, United Kingdom}

\noindent {$ ^{19}${ Departamento de  F\'{\i}sica Te\'orica and IPARCOS,
Universidad Complutense de Madrid, E-28040 Madrid, Spain}}


\noindent{$ ^{20}$ {D\'{e}partment de Physique Th\'{e}orique and Center for 
Astroparticle Physics, Universit\'{e} de Gen\`{e}ve, 24 quai Ernest-Ansermet, 
CH-1211 Gen\`{e}ve 4, Switzerland}}

\noindent{$ ^{21}$
{IKERBASQUE, Basque Foundation for Science, Bilbao 48011, Spain}}

\noindent {$ ^{22}${ Institut de Physique Th{\'e}orique, Universit{\'e} 
Paris-Saclay, CEA, CNRS, F-91191 Gif/Yvette Cedex, France}}

\noindent {$ ^{23}${ Instituto de Estructura de la Materia, CSIC, Serrano 121, 
28006 Madrid, Spain}}

\noindent {$ ^{24}${ Dipartimento di Fisica e Astronomia, Universit\`a di 
Bologna, via Irnerio~46, 40126 Bologna, Italy}}

\noindent {$ ^{25}${ I.N.F.N., Sezione di Bologna, I.S.~FLAG viale 
B.~Pichat~6/2, 40127 Bologna, Italy}}


\noindent {$ ^{26}${ Cosmology and Gravity Group, Department of Mathematics and 
Applied Mathematics, University of Cape Town, Rondebosch 7701, Cape Town, South 
Africa}}



\noindent {$ ^{27}${Jodrell Bank Center for Astrophysics, School of Physics 
and Astronomy,  University of  Manchester, Oxford Road, Manchester, M13 9PL, 
United Kingdom}}

\noindent {$ ^{28}${ Center for Gravitation and Cosmology, College of Physical 
Science and Technology, Yangzhou University, Yangzhou 225009, China}}

\noindent {$ ^{29}${ School of Physics and Astronomy, University of Nottingham, 
Nottingham, NG7 2RD, United Kingdom}}

\noindent {$ ^{30}${
Department of Physics and Astronomy, University of Hawai'i,
Watanabe Hall, 2505 Correa Road, Honolulu, HI, 96822, USA}

\noindent {$ ^{31}${NASA Einstein Fellow, Kavli Institute for Cosmological 
Physics and Enrico Fermi Institute, The University of Chicago, Chicago, IL 
60637, USA}}

\noindent {$ ^{32}${ Instituto de Astrofísica e Ci\^encias do  Espa\c{c}o, 
Faculdade 
de Ci\^encias da Universidade de Lisboa, Edif\'icio C8, Campo Grande, 
P-1749-016, Lisbon, Portugal}}


\noindent {$ ^{33}${ Universit\`{a} degli Studi di Bergamo, 
Dipartimento di Ingegneria e Scienze Applicate, Viale Marconi, 5 24044 Dalmine 
(Bergamo)  Italy}}

\noindent {$ ^{34}${ Istituto Nazionale di Fisica Nucleare (INFN), Sezione di  
Milano, Milan, Italy}}

\noindent {$ ^{35}${ 
 Institute of Physics, University of Szeged, D\'om t\'er 9, H-6720 Szeged, 
Hungary}}
 
\noindent {$ ^{36}${ Department of Physics and Astronomy, 
Bishop's University, Sherbrooke, Qu\'ebec, Canada J1M~1Z7
}}

\noindent {$ ^{37}${Institute for Theoretical Physics, ETH Zurich, 
Wolfgang-Pauli-Strasse 27, 8093, Zurich, Switzerland}}


\noindent {$ ^{38}${
  Department of Physics,
Aristotle University of Thessaloniki, 54124 Thessaloniki, Greece}}

\noindent {$ ^{39}${
Department of Physics, University of Ioannina, GR-45110, Ioannina, Greece}}

\noindent{$ ^{40}$ 
{Institut f\"ur Physik, Universit\"at Oldenburg, Postfach 2503,
D-26111 Oldenburg, Germany}}



\noindent {$ ^{41}${National Institute of Chemical Physics and Biophysics, 
R\"avala pst. 10, 10143 Tallinn, Estonia}}

\noindent {$ ^{42}${Helsinki Institute of Physics, P.O. Box 64, FIN-00014 
Helsinki, Finland}}

\noindent {$ ^{43}${Department of Physical Sciences, Helsinki University, P.O. 
Box 64, FIN-00014 Helsinki, Finland}}


 \noindent {$ ^{44}${ Institute Lorentz, Leiden University, PO Box 9506, Leiden 
2300 RA, The 
Netherlands}}

  \noindent {$ ^{45}$
{Instituto de F\'isica T\'eorica UAM-CSIC, Campus de Cantoblanco, E-28049 
Madrid, Spain}}


\noindent {$ ^{46}${ 
 Institute of Theoretical Astrophysics, University of Oslo P.O. Box 1029 
Blindern, N-0315 Oslo, Norway}}



\noindent {$ ^{47}${ AIM, CEA, CNRS, Universit{\'e} Paris-Saclay, 
Universit{\'e} Paris Diderot, Sorbonne Paris Cit{\'e}, F-91191 
Gif-sur-Yvette, France}}        


\noindent {$ ^{48}${
{Osservatorio Astronomico della Regione Autonoma Valle d’Aosta, Loc. Lignan 39, 
I-11020, Nus, Italy}}


\noindent {$ ^{49}${ Institute of Space Sciences and Astronomy, University of 
Malta, Msida, MSD 2080, Malta}}

\noindent {$ ^{50}${Department of Physics, University of Malta, Msida, MSD 
2080, Malta}}

\noindent {$ ^{51}${ 
 Theoretical Particle Physics and Cosmology Group, Department of Physics, King’s 
College London, University of London, Strand, London, WC2R 2LS, UK}}

\noindent {$ ^{52}${ 
CEICO, Institute of Physics of the Czech Academy of Sciences, Na Slovance 2, 
182 21 Praha 8, Czechia}}

\noindent {$ ^{53}${Mullard Space Science Laboratory, University College 
London, 
Holmbury St. Mary, Dorking, Surrey, RH5 6NT, UK}}

  \noindent {$ ^{54}${
  Faculty of Mathematics and Computer Science, Transilvania University, Iuliu 
Maniu Str. 50, 500091 Brasov, Romania}
}


\newpage
 
\phantomsection
\addcontentsline{toc}{part}{\bf Bibliography}

\bibliographystyle{JHEP}
\bibliography{CANTATAwhitepaper}

\providecommand{\href}[2]{#2}\begingroup\raggedright\begin{thebibliography}{1000}

\bibitem{DiValentino:2019qzk}
E.~Di~Valentino, A.~Melchiorri, and J.~Silk, {\it {Planck evidence for a closed
  Universe and a possible crisis for cosmology}},  {\em Nature Astron.} {\bf 4}
  (2019), no.~2 196--203, [\href{http://arxiv.org/abs/1911.02087}{{\tt
  arXiv:1911.02087}}].

\bibitem{Vagnozzi:2020dfn}
S.~Vagnozzi, A.~Loeb, and M.~Moresco, {\it {Eppurè piatto? The cosmic
  chronometer take on spatial curvature and cosmic concordance}},
  \href{http://arxiv.org/abs/2011.11645}{{\tt arXiv:2011.11645}}.

\bibitem{Jarv:2018bgs}
L.~Järv, M.~Rünkla, M.~Saal, and O.~Vilson, {\it {Nonmetricity formulation of
  general relativity and its scalar-tensor extension}},  {\em Phys. Rev. D}
  {\bf 97} (2018), no.~12 124025, [\href{http://arxiv.org/abs/1802.00492}{{\tt
  arXiv:1802.00492}}].

\bibitem{BeltranJimenez:2019tjy}
J.~B. Jiménez, L.~Heisenberg, and T.~S. Koivisto, {\it {The Geometrical
  Trinity of Gravity}},  {\em Universe} {\bf 5} (2019), no.~7 173,
  [\href{http://arxiv.org/abs/1903.06830}{{\tt arXiv:1903.06830}}].

\bibitem{Olmo:2012yv}
G.~J. Olmo, {\em {Introduction to Palatini theories of gravity and nonsingular
  cosmologies}}, pp.~157--184.
\newblock INTECH, 2012.
\newblock \href{http://arxiv.org/abs/1212.6393}{{\tt arXiv:1212.6393}}.

\bibitem{Hehl:1994ue}
F.~W. Hehl, J.~McCrea, E.~W. Mielke, and Y.~Ne'eman, {\it {Metric affine gauge
  theory of gravity: Field equations, Noether identities, world spinors, and
  breaking of dilation invariance}},  {\em Phys. Rept.} {\bf 258} (1995)
  1--171, [\href{http://arxiv.org/abs/gr-qc/9402012}{{\tt gr-qc/9402012}}].

\bibitem{Khoury:2010xi}
J.~Khoury, {\it {Theories of Dark Energy with Screening Mechanisms}},
  \href{http://arxiv.org/abs/1011.5909}{{\tt arXiv:1011.5909}}.

\bibitem{Belgacem:2017ihm}
E.~Belgacem, Y.~Dirian, S.~Foffa, and M.~Maggiore, {\it {Gravitational-wave
  luminosity distance in modified gravity theories}},  {\em Phys. Rev. D} {\bf
  97} (2018), no.~10 104066, [\href{http://arxiv.org/abs/1712.08108}{{\tt
  arXiv:1712.08108}}].

\bibitem{Gray:2019ksv}
R.~Gray et~al., {\it {Cosmological inference using gravitational wave standard
  sirens: A mock data analysis}},  {\em Phys. Rev. D} {\bf 101} (2020), no.~12
  122001, [\href{http://arxiv.org/abs/1908.06050}{{\tt arXiv:1908.06050}}].

\bibitem{Chan:2019ukj}
M.~H. Chan, {\it {A universal constant for dark matter-baryon interplay}},
  {\em Sci. Rep.} {\bf 9} (2019), no.~1 3570,
  [\href{http://arxiv.org/abs/1902.03786}{{\tt arXiv:1902.03786}}].

\bibitem{Bonvin:2020cxp}
C.~Bonvin, F.~O. Franco, and P.~Fleury, {\it {A null test of the equivalence
  principle using relativistic effects in galaxy surveys}},  {\em JCAP} {\bf
  08} (2020) 004, [\href{http://arxiv.org/abs/2004.06457}{{\tt
  arXiv:2004.06457}}].

\bibitem{Einstein:1917ce}
A.~Einstein, {\it {Cosmological Considerations in the General Theory of
  Relativity}},  {\em Sitzungsber. Preuss. Akad. Wiss. Berlin (Math. Phys.)}
  {\bf 1917} (1917) 142--152.

\bibitem{Rindler:2006km}
W.~Rindler, {\em {Relativity: Special, general, and cosmological}}.
\newblock 2006.

\bibitem{Einstein:1915ca}
A.~Einstein, {\it {The Field Equations of Gravitation}},  {\em Sitzungsber.
  Preuss. Akad. Wiss. Berlin (Math. Phys.)} {\bf 1915} (1915) 844--847.

\bibitem{Cai:2015emx}
Y.-F. Cai, S.~Capozziello, M.~De~Laurentis, and E.~N. Saridakis, {\it {f(T)
  teleparallel gravity and cosmology}},  {\em Rept. Prog. Phys.} {\bf 79}
  (2016), no.~10 106901, [\href{http://arxiv.org/abs/1511.07586}{{\tt
  arXiv:1511.07586}}].

\bibitem{dInverno:1992gxs}
R.~d'Inverno, {\em {Introducing Einstein's relativity}}.
\newblock 1992.

\bibitem{Einstein:1915by}
A.~Einstein, {\it {On the General Theory of Relativity}},  {\em Sitzungsber.
  Preuss. Akad. Wiss. Berlin (Math. Phys.)} {\bf 1915} (1915) 778--786.
  [Addendum: Sitzungsber. Preuss. Akad. Wiss. Berlin (Math.
  Phys.)1915,799(1915)].

\bibitem{Barrow:1991fj}
J.~D. Barrow, A.~B. Henriques, M.~T. V.~T. Lago, and M.~S. Longair, {\it {The
  Physical universe: The interface between cosmology, astrophysics and particle
  physics. Proceedings, 12th Autumn School of Physics, Lisbon, Portugal,
  October 1-5, 1990}},  {\em Lect. Notes Phys.} {\bf 383} (1991) pp.1--312.

\bibitem{1916AbhKP1916..189S}
K.~{Schwarzschild}, {\it {On the Gravitational Field of a Mass Point According
  to Einstein's Theory}},  {\em Abh. Konigl. Preuss. Akad. Wissenschaften Jahre
  1906,92, Berlin,1907} {\bf 1916} (Jan, 1916) 189--196.

\bibitem{Misner:1964je}
C.~W. Misner and D.~H. Sharp, {\it {Relativistic equations for adiabatic,
  spherically symmetric gravitational collapse}},  {\em Phys. Rev.} {\bf 136}
  (1964) B571--B576.

\bibitem{Brans:1961sx}
C.~Brans and R.~H. Dicke, {\it {Mach's principle and a relativistic theory of
  gravitation}},  {\em Phys. Rev.} {\bf 124} (1961) 925--935. [,142(1961)].

\bibitem{Will:1993ns}
C.~M. Will, {\em {Theory and experiment in gravitational physics}}.
\newblock 1993.

\bibitem{Will:2014kxa}
C.~M. Will, {\it {The Confrontation between General Relativity and
  Experiment}},  {\em Living Rev. Rel.} {\bf 17} (2014) 4,
  [\href{http://arxiv.org/abs/1403.7377}{{\tt arXiv:1403.7377}}].

\bibitem{Abbott:2016blz}
{\bf LIGO Scientific, Virgo} Collaboration, B.~P. Abbott et~al., {\it
  {Observation of Gravitational Waves from a Binary Black Hole Merger}},  {\em
  Phys. Rev. Lett.} {\bf 116} (2016), no.~6 061102,
  [\href{http://arxiv.org/abs/1602.03837}{{\tt arXiv:1602.03837}}].

\bibitem{Akiyama:2019eap}
{\bf Event Horizon Telescope} Collaboration, K.~Akiyama et~al., {\it {First M87
  Event Horizon Telescope Results. VI. The Shadow and Mass of the Central Black
  Hole}},  {\em Astrophys. J.} {\bf 875} (2019), no.~1 L6,
  [\href{http://arxiv.org/abs/1906.11243}{{\tt arXiv:1906.11243}}].

\bibitem{Friedman:1922kd}
A.~Friedman, {\it {On the Curvature of space}},  {\em Z. Phys.} {\bf 10} (1922)
  377--386. [Gen. Rel. Grav.31,1991(1999)].

\bibitem{Lemaitre:1931zz}
G.~Lemaitre, {\it {The Expanding Universe}},  {\em Mon. Not. Roy. Astron. Soc.}
  {\bf 91} (1931) 490--501.

\bibitem{Lemaitre:1931zzb}
G.~Lemaitre, {\it {Republication of: The beginning of the world from the point
  of view of quantum theory}},  {\em Nature} {\bf 127} (1931) 706. [Gen. Rel.
  Grav.43,2929(2011)].

\bibitem{Hubble:1929ig}
E.~Hubble, {\it {A relation between distance and radial velocity among
  extra-galactic nebulae}},  {\em Proc. Nat. Acad. Sci.} {\bf 15} (1929)
  168--173.

\bibitem{Lemaitre:1933gd}
G.~Lemaitre, {\it {The expanding universe}},  {\em Gen. Rel. Grav.} {\bf 29}
  (1997) 641--680. [Annales Soc. Sci. Bruxelles A53,51(1933)].

\bibitem{deSitter:1916zza}
W.~de~Sitter, {\it {Einstein's theory of gravitation and its astronomical
  consequences, First Paper}},  {\em Mon. Not. Roy. Astron. Soc.} {\bf 76}
  (1916) 699--728.

\bibitem{deSitter:1916zz}
W.~de~Sitter, {\it {Einstein's theory of gravitation and its astronomical
  consequences, Second Paper}},  {\em Mon. Not. Roy. Astron. Soc.} {\bf 77}
  (1916) 155--184.

\bibitem{deSitter:1917zz}
W.~de~Sitter, {\it {Einstein's theory of gravitation and its astronomical
  consequences, Third Paper}},  {\em Mon. Not. Roy. Astron. Soc.} {\bf 78}
  (1917) 3--28.

\bibitem{Bernstein:1988bw}
J.~Bernstein, {\em {KINETIC THEORY IN THE EXPANDING UNIVERSE}}.
\newblock Cambridge Monographs on Mathematical Physics. Cambridge University
  Press, Cambridge, U.K., 1988.

\bibitem{Kolb:1990vq}
E.~W. Kolb and M.~S. Turner, {\it {The Early Universe}},  {\em Front. Phys.}
  {\bf 69} (1990) 1--547.

\bibitem{Ellis:1973jva}
G.~F.~R. Ellis, {\it {Relativistic cosmology}},  {\em Cargese Lect. Phys.} {\bf
  6} (1973) 1--60.

\bibitem{Ellis:1998ct}
G.~F.~R. Ellis and H.~van Elst, {\it Cosmological models: {Cargese} lectures
  1998},  {\em NATO Sci. Ser. C} {\bf 541} (1999) 1--116,
  [\href{http://arxiv.org/abs/gr-qc/9812046}{{\tt gr-qc/9812046}}].

\bibitem{Gamow:1948pob}
G.~Gamow, {\it {The Evolution of the Universe}},  {\em Nature} {\bf 162}
  (1948), no.~4122 680--682.

\bibitem{Perlmutter:1998np}
{\bf Supernova Cosmology Project} Collaboration, S.~Perlmutter et~al., {\it
  {Measurements of $\Omega$ and $\Lambda$ from 42 high redshift supernovae}},
  {\em Astrophys. J.} {\bf 517} (1999) 565--586,
  [\href{http://arxiv.org/abs/astro-ph/9812133}{{\tt astro-ph/9812133}}].

\bibitem{Riess:1998cb}
{\bf Supernova Search Team} Collaboration, A.~G. Riess et~al., {\it
  {Observational evidence from supernovae for an accelerating universe and a
  cosmological constant}},  {\em Astron. J.} {\bf 116} (1998) 1009--1038,
  [\href{http://arxiv.org/abs/astro-ph/9805201}{{\tt astro-ph/9805201}}].

\bibitem{Copeland:2006wr}
E.~J. Copeland, M.~Sami, and S.~Tsujikawa, {\it {Dynamics of dark energy}},
  {\em Int. J. Mod. Phys.} {\bf D15} (2006) 1753--1936,
  [\href{http://arxiv.org/abs/hep-th/0603057}{{\tt hep-th/0603057}}].

\bibitem{Clifton:2011jh}
T.~Clifton, P.~G. Ferreira, A.~Padilla, and C.~Skordis, {\it {Modified Gravity
  and Cosmology}},  {\em Phys. Rept.} {\bf 513} (2012) 1--189,
  [\href{http://arxiv.org/abs/1106.2476}{{\tt arXiv:1106.2476}}].

\bibitem{Dicke:1900mn}
R.~Dicke and P.~Peebles, {\it {The big bang cosmology: Enigmas and nostrums}},
  1979.

\bibitem{Guth:1980zm}
A.~H. Guth, {\it {The Inflationary Universe: A Possible Solution to the Horizon
  and Flatness Problems}},  {\em Adv. Ser. Astrophys. Cosmol.} {\bf 3} (1987)
  139--148.

\bibitem{Starobinsky:1980te}
A.~A. Starobinsky, {\it {A New Type of Isotropic Cosmological Models Without
  Singularity}},  {\em Adv. Ser. Astrophys. Cosmol.} {\bf 3} (1987) 130--133.

\bibitem{Brandenberger:1999sw}
R.~H. Brandenberger, {\it {Inflationary cosmology: Progress and problems}},  in
  {\em {IPM School on Cosmology 1999: Large Scale Structure Formation}}, 1,
  1999.
\newblock \href{http://arxiv.org/abs/hep-ph/9910410}{{\tt hep-ph/9910410}}.

\bibitem{Bertschinger:1998tv}
E.~Bertschinger, {\it {Simulations of structure formation in the universe}},
  {\em Ann. Rev. Astron. Astrophys.} {\bf 36} (1998) 599--654.

\bibitem{Alpher:1948ve}
R.~A. Alpher, H.~Bethe, and G.~Gamow, {\it {The origin of chemical elements}},
  {\em Phys. Rev.} {\bf 73} (1948) 803--804.

\bibitem{Penzias:1965wn}
A.~A. Penzias and R.~W. Wilson, {\it {A Measurement of excess antenna
  temperature at 4080-Mc/s}},  {\em Astrophys. J.} {\bf 142} (1965) 419--421.

\bibitem{Dicke:1965zz}
R.~H. Dicke, P.~J.~E. Peebles, P.~G. Roll, and D.~T. Wilkinson, {\it {Cosmic
  Black-Body Radiation}},  {\em Astrophys. J.} {\bf 142} (1965) 414--419.

\bibitem{Ade:2013sjv}
{\bf Planck} Collaboration, P.~A.~R. Ade et~al., {\it {Planck 2013 results. I.
  Overview of products and scientific results}},  {\em Astron. Astrophys.} {\bf
  571} (2014) A1, [\href{http://arxiv.org/abs/1303.5062}{{\tt
  arXiv:1303.5062}}].

\bibitem{Percival:2009xn}
{\bf SDSS} Collaboration, W.~J. Percival et~al., {\it {Baryon Acoustic
  Oscillations in the Sloan Digital Sky Survey Data Release 7 Galaxy Sample}},
  {\em Mon. Not. Roy. Astron. Soc.} {\bf 401} (2010) 2148--2168,
  [\href{http://arxiv.org/abs/0907.1660}{{\tt arXiv:0907.1660}}].

\bibitem{Bartelmann:1999yn}
M.~Bartelmann and P.~Schneider, {\it {Weak gravitational lensing}},  {\em Phys.
  Rept.} {\bf 340} (2001) 291--472,
  [\href{http://arxiv.org/abs/astro-ph/9912508}{{\tt astro-ph/9912508}}].

\bibitem{Ferreira:2019xrr}
P.~G. Ferreira, {\it {Cosmological Tests of Gravity}},  {\em Ann. Rev. Astron.
  Astrophys.} {\bf 57} (2019) 335--374,
  [\href{http://arxiv.org/abs/1902.10503}{{\tt arXiv:1902.10503}}].

\bibitem{Aubourg:2014yra}
\'E.~Aubourg et~al., {\it {Cosmological implications of baryon acoustic
  oscillation measurements}},  {\em Phys. Rev.} {\bf D92} (2015), no.~12
  123516, [\href{http://arxiv.org/abs/1411.1074}{{\tt arXiv:1411.1074}}].

\bibitem{Mukhanov:1990me}
V.~F. Mukhanov, H.~A. Feldman, and R.~H. Brandenberger, {\it {Theory of
  cosmological perturbations. Part 1. Classical perturbations. Part 2. Quantum
  theory of perturbations. Part 3. Extensions}},  {\em Phys. Rept.} {\bf 215}
  (1992) 203--333.

\bibitem{Efstathiou:1989yr}
G.~Efstathiou, {\it {Cosmological perturbations}},  in {\em {36th Scottish
  Universities Summer School in Physics, a NATO ASI: Physics of the Early
  Universe}}, pp.~361--463, 1989.

\bibitem{Debono:2016vkp}
I.~Debono and G.~F. Smoot, {\it {General Relativity and Cosmology: Unsolved
  Questions and Future Directions}},  {\em Universe} {\bf 2} (2016), no.~4 23,
  [\href{http://arxiv.org/abs/1609.09781}{{\tt arXiv:1609.09781}}].

\bibitem{Goenner:2004se}
H.~Goenner, {\it {On the history of unified field theories}},  {\em Living Rev.
  Rel.} {\bf 7} (2004) 2.

\bibitem{Green:1987sp}
M.~B. Green, J.~Schwarz, and E.~Witten, {\em {SUPERSTRING THEORY. VOL. 1:
  INTRODUCTION}}.
\newblock Cambridge Monographs on Mathematical Physics. 7, 1988.

\bibitem{Green:1987mn}
M.~B. Green, J.~Schwarz, and E.~Witten, {\em {SUPERSTRING THEORY. VOL. 2: LOOP
  AMPLITUDES, ANOMALIES AND PHENOMENOLOGY}}.
\newblock 7, 1988.

\bibitem{Olmo:2004hj}
G.~J. Olmo and W.~Komp, {\it {Nonlinear gravity theories in the metric and
  Palatini formalisms}},  \href{http://arxiv.org/abs/gr-qc/0403092}{{\tt
  gr-qc/0403092}}.

\bibitem{Koivisto:2019jra}
T.~Koivisto, M.~Hohmann, and L.~Marzola, {\it {An Axiomatic Purification of
  Gravity}},  \href{http://arxiv.org/abs/1909.10415}{{\tt arXiv:1909.10415}}.

\bibitem{Lovelock:1971yv}
D.~Lovelock, {\it {The Einstein tensor and its generalizations}},  {\em J.
  Math. Phys.} {\bf 12} (1971) 498--501.

\bibitem{Avelino:2016lpj}
P.~Avelino et~al., {\it {Unveiling the Dynamics of the Universe}},  {\em
  Symmetry} {\bf 8} (2016), no.~8 70,
  [\href{http://arxiv.org/abs/1607.02979}{{\tt arXiv:1607.02979}}].

\bibitem{Kaluza:1984ws}
T.~Kaluza, {\it {Zum Unitätsproblem der Physik}},  {\em Sitzungsber. Preuss.
  Akad. Wiss. Berlin (Math. Phys.)} {\bf 1921} (1921) 966--972,
  [\href{http://arxiv.org/abs/1803.08616}{{\tt arXiv:1803.08616}}]. [Int. J.
  Mod. Phys.D27, no.14, 1870001(2018)].

\bibitem{Klein:1926tv}
O.~Klein, {\it {Quantum Theory and Five-Dimensional Theory of Relativity. (In
  German and English)}},  {\em Z. Phys.} {\bf 37} (1926) 895--906.

\bibitem{Goenner2004}
H.~F.~M. Goenner, {\it On the history of unified field theories},  {\em Living
  Reviews in Relativity} {\bf 7} (Feb, 2004) 2.

\bibitem{Goenner2014}
H.~F.~M. Goenner, {\it On the history of unified field theories. part ii. (ca.
  1930--ca. 1965)},  {\em Living Reviews in Relativity} {\bf 17} (Jun, 2014) 5.

\bibitem{Hehl:1976kj}
F.~W. Hehl, P.~Von Der~Heyde, G.~D. Kerlick, and J.~M. Nester, {\it {General
  Relativity with Spin and Torsion: Foundations and Prospects}},  {\em Rev.
  Mod. Phys.} {\bf 48} (1976) 393--416.

\bibitem{Bohmer:2016ome}
C.~G. B{\"o}hmer, {\em {Introduction to General Relativity and Cosmology}}.
\newblock Essential Textbooks in Physics Vol.~2. World Scientific (Europe),
  2016.

\bibitem{Blagojevic:2013xpa}
M.~Blagojević and F.~W. Hehl, eds., {\em {Gauge Theories of Gravitation}}.
\newblock World Scientific, Singapore, 2013.

\bibitem{Obukhov:2006gea}
Y.~N. Obukhov, {\it {Poincare gauge gravity: Selected topics}},  {\em Int. J.
  Geom. Meth. Mod. Phys.} {\bf 3} (2006) 95--138,
  [\href{http://arxiv.org/abs/gr-qc/0601090}{{\tt gr-qc/0601090}}].

\bibitem{Aldrovandi:2013wha}
R.~Aldrovandi and J.~G. Pereira, {\em {Teleparallel Gravity}}, vol.~173.
\newblock Springer, Dordrecht, 2013.

\bibitem{Ashtekar:1986yd}
A.~Ashtekar, {\it {New Variables for Classical and Quantum Gravity}},  {\em
  Phys. Rev. Lett.} {\bf 57} (1986) 2244--2247.

\bibitem{Ashtekar:1987gu}
A.~Ashtekar, {\it {New Hamiltonian Formulation of General Relativity}},  {\em
  Phys. Rev.} {\bf D36} (1987) 1587--1602.

\bibitem{Barbero:1994ap}
J.~F. Barbero~G., {\it {Real Ashtekar variables for Lorentzian signature space
  times}},  {\em Phys. Rev.} {\bf D51} (1995) 5507--5510,
  [\href{http://arxiv.org/abs/gr-qc/9410014}{{\tt gr-qc/9410014}}].

\bibitem{Holst:1995pc}
S.~Holst, {\it {Barbero's Hamiltonian derived from a generalized
  Hilbert-Palatini action}},  {\em Phys. Rev.} {\bf D53} (1996) 5966--5969,
  [\href{http://arxiv.org/abs/gr-qc/9511026}{{\tt gr-qc/9511026}}].

\bibitem{Immirzi:1996di}
G.~Immirzi, {\it {Real and complex connections for canonical gravity}},  {\em
  Class. Quant. Grav.} {\bf 14} (1997) L177--L181,
  [\href{http://arxiv.org/abs/gr-qc/9612030}{{\tt gr-qc/9612030}}].

\bibitem{thiemann2007}
T.~Thiemann, {\em Modern Canonical Quantum General Relativity}.
\newblock Cambridge Monographs on Mathematical Physics. Cambridge University
  Press, 2007.

\bibitem{Maluf:2013gaa}
J.~W. Maluf, {\it {The teleparallel equivalent of general relativity}},  {\em
  Annalen Phys.} {\bf 525} (2013) 339--357,
  [\href{http://arxiv.org/abs/1303.3897}{{\tt arXiv:1303.3897}}].

\bibitem{Obukhov:2002tm}
{\relax Yu}.~N. Obukhov and J.~G. Pereira, {\it {Metric affine approach to
  teleparallel gravity}},  {\em Phys. Rev.} {\bf D67} (2003) 044016,
  [\href{http://arxiv.org/abs/gr-qc/0212080}{{\tt gr-qc/0212080}}].

\bibitem{Maluf:2003fs}
J.~W. Maluf, {\it {Dirac spinor fields in the teleparallel gravity: Comment on
  `Metric affine approach to teleparallel gravity'}},  {\em Phys. Rev.} {\bf
  D67} (2003) 108501, [\href{http://arxiv.org/abs/gr-qc/0304005}{{\tt
  gr-qc/0304005}}].

\bibitem{Mielke:2004gg}
E.~W. Mielke, {\it {Consistent coupling to Dirac fields in teleparallelism:
  Comment on `Metric-affine approach to teleparallel gravity'}},  {\em Phys.
  Rev.} {\bf D69} (2004) 128501.

\bibitem{Obukhov:2004hv}
{\relax Yu}.~N. Obukhov and J.~G. Pereira, {\it {Lessons of spin and torsion:
  Reply to `Consistent coupling to Dirac fields in teleparallelism'}},  {\em
  Phys. Rev.} {\bf D69} (2004) 128502,
  [\href{http://arxiv.org/abs/gr-qc/0406015}{{\tt gr-qc/0406015}}].

\bibitem{Leclerc:2004uu}
M.~Leclerc, {\it {On the teleparallel limit of Poincare gauge theory}},  {\em
  Phys. Rev.} {\bf D71} (2005) 027503,
  [\href{http://arxiv.org/abs/gr-qc/0411119}{{\tt gr-qc/0411119}}].

\bibitem{Barrow:1983rx}
J.~D. Barrow and A.~C. Ottewill, {\it {The Stability of General Relativistic
  Cosmological Theory}},  {\em J. Phys.} {\bf A16} (1983) 2757.

\bibitem{Capozziello:2002rd}
S.~Capozziello, {\it {Curvature quintessence}},  {\em Int. J. Mod. Phys.} {\bf
  D11} (2002) 483--492, [\href{http://arxiv.org/abs/gr-qc/0201033}{{\tt
  gr-qc/0201033}}].

\bibitem{Capozziello:2003tk}
S.~Capozziello, S.~Carloni, and A.~Troisi, {\it {Quintessence without scalar
  fields}},  {\em Recent Res. Dev. Astron. Astrophys.} {\bf 1} (2003) 625,
  [\href{http://arxiv.org/abs/astro-ph/0303041}{{\tt astro-ph/0303041}}].

\bibitem{Sotiriou:2008rp}
T.~P. Sotiriou and V.~Faraoni, {\it {f(R) Theories Of Gravity}},  {\em Rev.
  Mod. Phys.} {\bf 82} (2010) 451--497,
  [\href{http://arxiv.org/abs/0805.1726}{{\tt arXiv:0805.1726}}].

\bibitem{DeFelice:2010aj}
A.~De~Felice and S.~Tsujikawa, {\it {f(R) theories}},  {\em Living Rev. Rel.}
  {\bf 13} (2010) 3, [\href{http://arxiv.org/abs/1002.4928}{{\tt
  arXiv:1002.4928}}].

\bibitem{Nojiri:2010wj}
S.~Nojiri and S.~D. Odintsov, {\it {Unified cosmic history in modified gravity:
  from F(R) theory to Lorentz non-invariant models}},  {\em Phys. Rept.} {\bf
  505} (2011) 59--144, [\href{http://arxiv.org/abs/1011.0544}{{\tt
  arXiv:1011.0544}}].

\bibitem{Harko:2018ayt}
T.~Harko and F.~S.~N. Lobo, {\em {Extensions of f(R) Gravity}}.
\newblock Cambridge University Press, 2018.

\bibitem{Ferraro:2006jd}
R.~Ferraro and F.~Fiorini, {\it {Modified teleparallel gravity: Inflation
  without inflaton}},  {\em Phys. Rev.} {\bf D75} (2007) 084031,
  [\href{http://arxiv.org/abs/gr-qc/0610067}{{\tt gr-qc/0610067}}].

\bibitem{Capozziello:2011et}
S.~Capozziello and M.~De~Laurentis, {\it {Extended Theories of Gravity}},  {\em
  Phys. Rept.} {\bf 509} (2011) 167--321,
  [\href{http://arxiv.org/abs/1108.6266}{{\tt arXiv:1108.6266}}].

\bibitem{Bahamonde:2015zma}
S.~Bahamonde, C.~G. B{\"o}hmer, and M.~Wright, {\it {Modified teleparallel
  theories of gravity}},  {\em Phys. Rev.} {\bf D92} (2015), no.~10 104042,
  [\href{http://arxiv.org/abs/1508.05120}{{\tt arXiv:1508.05120}}].

\bibitem{Krssak:2018ywd}
M.~Krssak, R.~J. van~den Hoogen, J.~G. Pereira, C.~G. B{\"o}hmer, and A.~A.
  Coley, {\it {Teleparallel theories of gravity: illuminating a fully invariant
  approach}},  {\em Class. Quant. Grav.} {\bf 36} (2019), no.~18 183001,
  [\href{http://arxiv.org/abs/1810.12932}{{\tt arXiv:1810.12932}}].

\bibitem{Ferraro:2018tpu}
R.~Ferraro and M.~J. Guzmán, {\it {Hamiltonian formalism for f(T) gravity}},
  {\em Phys. Rev.} {\bf D97} (2018), no.~10 104028,
  [\href{http://arxiv.org/abs/1802.02130}{{\tt arXiv:1802.02130}}].

\bibitem{Ferraro:2018axk}
R.~Ferraro and M.~J. Guzmán, {\it {Quest for the extra degree of freedom in
  $f(T)$ gravity}},  {\em Phys. Rev.} {\bf D98} (2018), no.~12 124037,
  [\href{http://arxiv.org/abs/1810.07171}{{\tt arXiv:1810.07171}}].

\bibitem{Blagojevic:2020dyq}
M.~Blagojevi\'c and J.~M. Nester, {\it {Local symmetries and physical degrees
  of freedom in $f(T)$ gravity: a Dirac Hamiltonian constraint analysis}},
  {\em Phys. Rev. D} {\bf 102} (2020), no.~6 064025,
  [\href{http://arxiv.org/abs/2006.15303}{{\tt arXiv:2006.15303}}].

\bibitem{Linden:1972up}
T.~L.~J. Linden, {\it {A scalar field theory of gravitation}},  {\em Int. J.
  Theor. Phys.} {\bf 5} (1972) 359--368.

\bibitem{Faraoni:2004pi}
V.~Faraoni, {\em {Cosmology in scalar tensor gravity}}, vol.~139.
\newblock Springer, 2004.

\bibitem{Stelle:1976gc}
K.~S. Stelle, {\it {Renormalization of Higher Derivative Quantum Gravity}},
  {\em Phys. Rev.} {\bf D16} (1977) 953--969.

\bibitem{Utiyama:1962sn}
R.~Utiyama and B.~S. DeWitt, {\it {Renormalization of a classical gravitational
  field interacting with quantized matter fields}},  {\em J. Math. Phys.} {\bf
  3} (1962) 608--618.

\bibitem{Birrell:1982ix}
N.~D. Birrell and P.~C.~W. Davies, {\em {Quantum Fields in Curved Space}}.
\newblock Cambridge Monographs on Mathematical Physics. Cambridge Univ. Press,
  Cambridge, UK, 1984.

\bibitem{Buchbinder:1992rb}
I.~L. Buchbinder, S.~D. Odintsov, and I.~L. Shapiro, {\em {Effective action in
  quantum gravity}}.
\newblock Bristol, UK: IOP (1992) 413 p, 1992.

\bibitem{Vilkovisky:1992pb}
G.~A. Vilkovisky, {\it {Effective action in quantum gravity}},  {\em Class.
  Quant. Grav.} {\bf 9} (1992) 895--903.

\bibitem{Brandenberger:1992sy}
R.~H. Brandenberger, {\it {A Nonsingular universe}},  in {\em {International
  School of Astrophysics, 'D. Chalonge': 2nd Course: Current Topics in
  Astrofundamental Physics Erice, Italy, September 6-13, 1992}}, pp.~102--112,
  1992.
\newblock \href{http://arxiv.org/abs/gr-qc/9210014}{{\tt gr-qc/9210014}}.

\bibitem{Resco:2016upv}
M.~Aparicio~Resco, {\'A}.~de~la Cruz-Dombriz, F.~J. Llanes~Estrada, and
  V.~Zapatero~Castrillo, {\it {On neutron stars in $f(R)$ theories: Small
  radii, large masses and large energy emitted in a merger}},  {\em Phys. Dark
  Univ.} {\bf 13} (2016) 147--161, [\href{http://arxiv.org/abs/1602.03880}{{\tt
  arXiv:1602.03880}}].

\bibitem{Eisenstein:2005su}
{\bf SDSS} Collaboration, D.~J. Eisenstein et~al., {\it {Detection of the
  Baryon Acoustic Peak in the Large-Scale Correlation Function of SDSS Luminous
  Red Galaxies}},  {\em Astrophys. J.} {\bf 633} (2005) 560--574,
  [\href{http://arxiv.org/abs/astro-ph/0501171}{{\tt astro-ph/0501171}}].

\bibitem{Riess:2004nr}
{\bf Supernova Search Team} Collaboration, A.~G. Riess et~al., {\it {Type Ia
  supernova discoveries at z > 1 from the Hubble Space Telescope: Evidence for
  past deceleration and constraints on dark energy evolution}},  {\em
  Astrophys. J.} {\bf 607} (2004) 665--687,
  [\href{http://arxiv.org/abs/astro-ph/0402512}{{\tt astro-ph/0402512}}].

\bibitem{Spergel:2006hy}
{\bf WMAP} Collaboration, D.~N. Spergel et~al., {\it {Wilkinson Microwave
  Anisotropy Probe (WMAP) three year results: implications for cosmology}},
  {\em Astrophys. J. Suppl.} {\bf 170} (2007) 377,
  [\href{http://arxiv.org/abs/astro-ph/0603449}{{\tt astro-ph/0603449}}].

\bibitem{Martin:2012bt}
J.~Martin, {\it {Everything You Always Wanted To Know About The Cosmological
  Constant Problem (But Were Afraid To Ask)}},  {\em Comptes Rendus Physique}
  {\bf 13} (2012) 566--665, [\href{http://arxiv.org/abs/1205.3365}{{\tt
  arXiv:1205.3365}}].

\bibitem{Capozziello:2010zz}
V.~Faraoni and S.~Capozziello, {\em {Beyond Einstein Gravity}}, vol.~170.
\newblock Springer, Dordrecht, 2011.

\bibitem{Cognola:2006eg}
G.~Cognola, E.~Elizalde, S.~Nojiri, S.~D. Odintsov, and S.~Zerbini, {\it {Dark
  energy in modified Gauss-Bonnet gravity: Late-time acceleration and the
  hierarchy problem}},  {\em Phys. Rev.} {\bf D73} (2006) 084007,
  [\href{http://arxiv.org/abs/hep-th/0601008}{{\tt hep-th/0601008}}].

\bibitem{Nojiri:2005jg}
S.~Nojiri and S.~D. Odintsov, {\it {Modified Gauss-Bonnet theory as
  gravitational alternative for dark energy}},  {\em Phys. Lett.} {\bf B631}
  (2005) 1--6, [\href{http://arxiv.org/abs/hep-th/0508049}{{\tt
  hep-th/0508049}}].

\bibitem{delaCruzDombriz:2011wn}
A.~de~la Cruz-Dombriz and D.~Saez-Gomez, {\it {On the stability of the
  cosmological solutions in $f(R,G)$ gravity}},  {\em Class. Quant. Grav.} {\bf
  29} (2012) 245014, [\href{http://arxiv.org/abs/1112.4481}{{\tt
  arXiv:1112.4481}}].

\bibitem{Brans:1962zz}
C.~H. Brans, {\it {Mach's Principle and a Relativistic Theory of Gravitation.
  II}},  {\em Phys. Rev.} {\bf 125} (1962) 2194--2201.

\bibitem{GarciaBellido:1993wn}
J.~Garcia-Bellido, A.~D. Linde, and D.~A. Linde, {\it {Fluctuations of the
  gravitational constant in the inflationary Brans-Dicke cosmology}},  {\em
  Phys. Rev.} {\bf D50} (1994) 730--750,
  [\href{http://arxiv.org/abs/astro-ph/9312039}{{\tt astro-ph/9312039}}].

\bibitem{Cembranos:2009ds}
J.~A.~R. Cembranos, K.~A. Olive, M.~Peloso, and J.-P. Uzan, {\it {Quantum
  Corrections to the Cosmological Evolution of Conformally Coupled Fields}},
  {\em JCAP} {\bf 0907} (2009) 025, [\href{http://arxiv.org/abs/0905.1989}{{\tt
  arXiv:0905.1989}}].

\bibitem{Ford:1989me}
L.~H. Ford, {\it {INFLATION DRIVEN BY A VECTOR FIELD}},  {\em Phys. Rev.} {\bf
  D40} (1989) 967.

\bibitem{Jimenez:2008au}
J.~Beltran~Jimenez and A.~L. Maroto, {\it {A cosmic vector for dark energy}},
  {\em Phys. Rev.} {\bf D78} (2008) 063005,
  [\href{http://arxiv.org/abs/0801.1486}{{\tt arXiv:0801.1486}}].

\bibitem{Koivisto:2008xf}
T.~Koivisto and D.~F. Mota, {\it {Vector Field Models of Inflation and Dark
  Energy}},  {\em JCAP} {\bf 0808} (2008) 021,
  [\href{http://arxiv.org/abs/0805.4229}{{\tt arXiv:0805.4229}}].

\bibitem{Alcaraz:2002iu}
J.~Alcaraz, J.~A.~R. Cembranos, A.~Dobado, and A.~L. Maroto, {\it {Limits on
  the brane fluctuations mass and on the brane tension scale from electron
  positron colliders}},  {\em Phys. Rev.} {\bf D67} (2003) 075010,
  [\href{http://arxiv.org/abs/hep-ph/0212269}{{\tt hep-ph/0212269}}].

\bibitem{Dvali:2000hr}
G.~R. Dvali, G.~Gabadadze, and M.~Porrati, {\it {4-D gravity on a brane in 5-D
  Minkowski space}},  {\em Phys. Lett.} {\bf B485} (2000) 208--214,
  [\href{http://arxiv.org/abs/hep-th/0005016}{{\tt hep-th/0005016}}].

\bibitem{Blaschke:2004wa}
D.~Blaschke and M.~P. Dabrowski, {\it {Conformal relativity versus Brans-Dicke
  and superstring theories}},  {\em Entropy} {\bf 14} (2012) 1978--1996,
  [\href{http://arxiv.org/abs/hep-th/0407078}{{\tt hep-th/0407078}}].

\bibitem{Khoury:2003rn}
J.~Khoury and A.~Weltman, {\it {Chameleon cosmology}},  {\em Phys. Rev.} {\bf
  D69} (2004) 044026, [\href{http://arxiv.org/abs/astro-ph/0309411}{{\tt
  astro-ph/0309411}}].

\bibitem{Perivolaropoulos:2009ak}
L.~Perivolaropoulos, {\it {PPN Parameter gamma and Solar System Constraints of
  Massive Brans-Dicke Theories}},  {\em Phys. Rev.} {\bf D81} (2010) 047501,
  [\href{http://arxiv.org/abs/0911.3401}{{\tt arXiv:0911.3401}}].

\bibitem{Hohmann:2013rba}
M.~Hohmann, L.~Jarv, P.~Kuusk, and E.~Randla, {\it {Post-Newtonian parameters
  $\gamma$ and $\beta$ of scalar-tensor gravity with a general potential}},
  {\em Phys. Rev.} {\bf D88} (2013), no.~8 084054,
  [\href{http://arxiv.org/abs/1309.0031}{{\tt arXiv:1309.0031}}]. [Erratum:
  Phys. Rev.D89,no.6,069901(2014)].

\bibitem{Capozziello:2005mj}
S.~Capozziello, S.~Nojiri, and S.~D. Odintsov, {\it {Dark energy: The Equation
  of state description versus scalar-tensor or modified gravity}},  {\em Phys.
  Lett.} {\bf B634} (2006) 93--100,
  [\href{http://arxiv.org/abs/hep-th/0512118}{{\tt hep-th/0512118}}].

\bibitem{Sotiriou:2006hs}
T.~P. Sotiriou, {\it {f(R) gravity and scalar-tensor theory}},  {\em Class.
  Quant. Grav.} {\bf 23} (2006) 5117--5128,
  [\href{http://arxiv.org/abs/gr-qc/0604028}{{\tt gr-qc/0604028}}].

\bibitem{Sotiriou:2006sf}
T.~P. Sotiriou, {\it {Curvature scalar instability in f(R) gravity}},  {\em
  Phys. Lett.} {\bf B645} (2007) 389--392,
  [\href{http://arxiv.org/abs/gr-qc/0611107}{{\tt gr-qc/0611107}}].

\bibitem{Carroll:2003wy}
S.~M. Carroll, V.~Duvvuri, M.~Trodden, and M.~S. Turner, {\it {Is cosmic speed
  - up due to new gravitational physics?}},  {\em Phys.Rev.} {\bf D70} (2004)
  043528, [\href{http://arxiv.org/abs/astro-ph/0306438}{{\tt
  astro-ph/0306438}}].

\bibitem{Bertotti:2003rm}
B.~Bertotti, L.~Iess, and P.~Tortora, {\it {A test of general relativity using
  radio links with the Cassini spacecraft}},  {\em Nature} {\bf 425} (2003)
  374--376.

\bibitem{Amendola:2006kh}
L.~Amendola, D.~Polarski, and S.~Tsujikawa, {\it {Are f(R) dark energy models
  cosmologically viable ?}},  {\em Phys. Rev. Lett.} {\bf 98} (2007) 131302,
  [\href{http://arxiv.org/abs/astro-ph/0603703}{{\tt astro-ph/0603703}}].

\bibitem{Amendola:2006eh}
L.~Amendola, D.~Polarski, and S.~Tsujikawa, {\it {Power-laws f(R) theories are
  cosmologically unacceptable}},  {\em Int. J. Mod. Phys.} {\bf D16} (2007)
  1555--1561, [\href{http://arxiv.org/abs/astro-ph/0605384}{{\tt
  astro-ph/0605384}}].

\bibitem{Amendola:2006we}
L.~Amendola, R.~Gannouji, D.~Polarski, and S.~Tsujikawa, {\it {Conditions for
  the cosmological viability of f(R) dark energy models}},  {\em Phys. Rev.}
  {\bf D75} (2007) 083504, [\href{http://arxiv.org/abs/gr-qc/0612180}{{\tt
  gr-qc/0612180}}].

\bibitem{Hu:2007nk}
W.~Hu and I.~Sawicki, {\it {Models of f(R) Cosmic Acceleration that Evade
  Solar-System Tests}},  {\em Phys. Rev.} {\bf D76} (2007) 064004,
  [\href{http://arxiv.org/abs/0705.1158}{{\tt arXiv:0705.1158}}].

\bibitem{Nojiri:2006gh}
S.~Nojiri and S.~D. Odintsov, {\it {Modified f(R) gravity consistent with
  realistic cosmology: From matter dominated epoch to dark energy universe}},
  {\em Phys. Rev.} {\bf D74} (2006) 086005,
  [\href{http://arxiv.org/abs/hep-th/0608008}{{\tt hep-th/0608008}}].

\bibitem{Nojiri:2006be}
S.~Nojiri and S.~D. Odintsov, {\it {Modified gravity and its reconstruction
  from the universe expansion history}},  {\em J. Phys. Conf. Ser.} {\bf 66}
  (2007) 012005, [\href{http://arxiv.org/abs/hep-th/0611071}{{\tt
  hep-th/0611071}}].

\bibitem{Evans:2007ch}
J.~D. Evans, L.~M.~H. Hall, and P.~Caillol, {\it {Standard Cosmological
  Evolution in a Wide Range of f(R) Models}},  {\em Phys. Rev.} {\bf D77}
  (2008) 083514, [\href{http://arxiv.org/abs/0711.3695}{{\tt
  arXiv:0711.3695}}].

\bibitem{delaCruzDombriz:2006fj}
A.~de~la Cruz-Dombriz and A.~Dobado, {\it {A f(R) gravity without cosmological
  constant}},  {\em Phys. Rev.} {\bf D74} (2006) 087501,
  [\href{http://arxiv.org/abs/gr-qc/0607118}{{\tt gr-qc/0607118}}].

\bibitem{Dunsby:2010wg}
P.~K.~S. Dunsby, E.~Elizalde, R.~Goswami, S.~Odintsov, and D.~S. Gomez, {\it
  {On the LCDM Universe in f(R) gravity}},  {\em Phys. Rev.} {\bf D82} (2010)
  023519, [\href{http://arxiv.org/abs/1005.2205}{{\tt arXiv:1005.2205}}].

\bibitem{DeWitt:1965jb}
B.~S. DeWitt, {\it {Dynamical theory of groups and fields}},  {\em Conf. Proc.}
  {\bf C630701} (1964) 585--820. [Les Houches Lect. Notes13,585(1964)].

\bibitem{Olmo:2011uz}
G.~J. Olmo, {\it {Palatini Approach to Modified Gravity: f(R) Theories and
  Beyond}},  {\em Int. J. Mod. Phys.} {\bf D20} (2011) 413--462,
  [\href{http://arxiv.org/abs/1101.3864}{{\tt arXiv:1101.3864}}].

\bibitem{Sotiriou:2006qn}
T.~P. Sotiriou and S.~Liberati, {\it {Metric-affine f(R) theories of gravity}},
   {\em Annals Phys.} {\bf 322} (2007) 935--966,
  [\href{http://arxiv.org/abs/gr-qc/0604006}{{\tt gr-qc/0604006}}].

\bibitem{Chiba:2003ir}
T.~Chiba, {\it {1/R gravity and scalar - tensor gravity}},  {\em Phys. Lett.}
  {\bf B575} (2003) 1--3, [\href{http://arxiv.org/abs/astro-ph/0307338}{{\tt
  astro-ph/0307338}}].

\bibitem{OHanlon:1972sdp}
J.~O'~Hanlon, {\it {Mach's principle and a new gauge freedom in Brans-Dicke
  theory}},  {\em J. Phys.} {\bf A5} (1972) 803--811.

\bibitem{Teyssandier:1983zz}
P.~Teyssandier and P.~Tourrenc, {\it {The Cauchy problem for the R+R**2
  theories of gravity without torsion}},  {\em J. Math. Phys.} {\bf 24} (1983)
  2793.

\bibitem{Wands:1993uu}
D.~Wands, {\it {Extended gravity theories and the Einstein-Hilbert action}},
  {\em Class. Quant. Grav.} {\bf 11} (1994) 269--280,
  [\href{http://arxiv.org/abs/gr-qc/9307034}{{\tt gr-qc/9307034}}].

\bibitem{Lazkoz:2018aqk}
R.~Lazkoz, M.~Ortiz-Ba{\~n}os, and V.~Salzano, {\it {$f(R)$ gravity
  modifications: from the action to the data}},  {\em Eur. Phys. J.} {\bf C78}
  (2018), no.~3 213, [\href{http://arxiv.org/abs/1803.05638}{{\tt
  arXiv:1803.05638}}].

\bibitem{Basilakos:2017rgc}
S.~Basilakos and S.~Nesseris, {\it {Conjoined constraints on modified gravity
  from the expansion history and cosmic growth}},  {\em Phys. Rev.} {\bf D96}
  (2017), no.~6 063517, [\href{http://arxiv.org/abs/1705.08797}{{\tt
  arXiv:1705.08797}}].

\bibitem{Jaime:2012nj}
L.~Jaime, M.~Salgado, and L.~Patino, {\it {Cosmology in $\mathcal{f}$(R)
  exponential gravity}},  {\em Springer Proc. Phys.} {\bf 157} (2014) 363--371,
  [\href{http://arxiv.org/abs/1211.0015}{{\tt arXiv:1211.0015}}].

\bibitem{Miranda:2009rs}
V.~Miranda, S.~E. Joras, I.~Waga, and M.~Quartin, {\it {Viable Singularity-Free
  f(R) Gravity Without a Cosmological Constant}},  {\em Phys. Rev. Lett.} {\bf
  102} (2009) 221101, [\href{http://arxiv.org/abs/0905.1941}{{\tt
  arXiv:0905.1941}}].

\bibitem{Chiba:2006jp}
T.~Chiba, T.~L. Smith, and A.~L. Erickcek, {\it {Solar System constraints to
  general f(R) gravity}},  {\em Phys. Rev.} {\bf D75} (2007) 124014,
  [\href{http://arxiv.org/abs/astro-ph/0611867}{{\tt astro-ph/0611867}}].

\bibitem{Khoury:2003aq}
J.~Khoury and A.~Weltman, {\it {Chameleon fields: Awaiting surprises for tests
  of gravity in space}},  {\em Phys. Rev. Lett.} {\bf 93} (2004) 171104,
  [\href{http://arxiv.org/abs/astro-ph/0309300}{{\tt astro-ph/0309300}}].

\bibitem{Sokolowski:2008kf}
L.~M. Sokolowski, {\it {Stability of a metric f(R) gravity theory implies the
  Newtonian limit}},  {\em Acta Phys. Polon.} {\bf B39} (2008) 2879--2901,
  [\href{http://arxiv.org/abs/0810.2554}{{\tt arXiv:0810.2554}}].

\bibitem{Appleby:2009uf}
S.~A. Appleby, R.~A. Battye, and A.~A. Starobinsky, {\it {Curing singularities
  in cosmological evolution of F(R) gravity}},  {\em JCAP} {\bf 1006} (2010)
  005, [\href{http://arxiv.org/abs/0909.1737}{{\tt arXiv:0909.1737}}].

\bibitem{Cline:2003gs}
J.~M. Cline, S.~Jeon, and G.~D. Moore, {\it {The Phantom menaced: Constraints
  on low-energy effective ghosts}},  {\em Phys. Rev.} {\bf D70} (2004) 043543,
  [\href{http://arxiv.org/abs/hep-ph/0311312}{{\tt hep-ph/0311312}}].

\bibitem{Nariai:1973eg}
H.~Nariai, {\it {Gravitational instability of regular model-universes in a
  modified theory of general relativity}},  {\em Prog. Theor. Phys.} {\bf 49}
  (1973) 165--180.

\bibitem{Gurovich:1979xg}
V.~T. Gurovich and A.~A. Starobinsky, {\it {QUANTUM EFFECTS AND REGULAR
  COSMOLOGICAL MODELS}},  {\em Sov. Phys. JETP} {\bf 50} (1979) 844--852. [Zh.
  Eksp. Teor. Fiz.77,1683(1979)].

\bibitem{Dolgov:2003px}
A.~D. Dolgov and M.~Kawasaki, {\it {Can modified gravity explain accelerated
  cosmic expansion?}},  {\em Phys. Lett.} {\bf B573} (2003) 1--4,
  [\href{http://arxiv.org/abs/astro-ph/0307285}{{\tt astro-ph/0307285}}].

\bibitem{Faraoni:2006sy}
V.~Faraoni, {\it {Matter instability in modified gravity}},  {\em Phys. Rev.}
  {\bf D74} (2006) 104017, [\href{http://arxiv.org/abs/astro-ph/0610734}{{\tt
  astro-ph/0610734}}].

\bibitem{Faraoni:2007yn}
V.~Faraoni, {\it {de Sitter space and the equivalence between f(R) and
  scalar-tensor gravity}},  {\em Phys. Rev.} {\bf D75} (2007) 067302,
  [\href{http://arxiv.org/abs/gr-qc/0703044}{{\tt gr-qc/0703044}}].

\bibitem{Faraoni:2008mf}
V.~Faraoni, {\it {f(R) gravity: Successes and challenges}},  in {\em {18th
  SIGRAV Conference Cosenza, Italy, September 22-25, 2008}}, 2008.
\newblock \href{http://arxiv.org/abs/0810.2602}{{\tt arXiv:0810.2602}}.

\bibitem{Frolov:2008uf}
A.~V. Frolov, {\it {A Singularity Problem with f(R) Dark Energy}},  {\em Phys.
  Rev. Lett.} {\bf 101} (2008) 061103,
  [\href{http://arxiv.org/abs/0803.2500}{{\tt arXiv:0803.2500}}].

\bibitem{Appleby:2007vb}
S.~A. Appleby and R.~A. Battye, {\it {Do consistent $F(R)$ models mimic General
  Relativity plus $\Lambda$?}},  {\em Phys. Lett.} {\bf B654} (2007) 7--12,
  [\href{http://arxiv.org/abs/0705.3199}{{\tt arXiv:0705.3199}}].

\bibitem{Starobinsky:2007hu}
A.~A. Starobinsky, {\it {Disappearing cosmological constant in f(R) gravity}},
  {\em JETP Lett.} {\bf 86} (2007) 157--163,
  [\href{http://arxiv.org/abs/0706.2041}{{\tt arXiv:0706.2041}}].

\bibitem{Tsujikawa:2007xu}
S.~Tsujikawa, {\it {Observational signatures of $f(R)$ dark energy models that
  satisfy cosmological and local gravity constraints}},  {\em Phys. Rev.} {\bf
  D77} (2008) 023507, [\href{http://arxiv.org/abs/0709.1391}{{\tt
  arXiv:0709.1391}}].

\bibitem{Jaime:2012gc}
L.~G. Jaime, L.~Patino, and M.~Salgado, {\it {f(R) Cosmology revisited}},
  \href{http://arxiv.org/abs/1206.1642}{{\tt arXiv:1206.1642}}.

\bibitem{Ezquiaga:2017ekz}
J.~M. Ezquiaga and M.~Zumalac{\'a}rregui, {\it {Dark Energy After GW170817:
  Dead Ends and the Road Ahead}},  {\em Phys. Rev. Lett.} {\bf 119} (2017),
  no.~25 251304, [\href{http://arxiv.org/abs/1710.05901}{{\tt
  arXiv:1710.05901}}].

\bibitem{Creminelli:2017sry}
P.~Creminelli and F.~Vernizzi, {\it {Dark Energy after GW170817 and
  GRB170817A}},  {\em Phys. Rev. Lett.} {\bf 119} (2017), no.~25 251302,
  [\href{http://arxiv.org/abs/1710.05877}{{\tt arXiv:1710.05877}}].

\bibitem{Zhang:2005vt}
P.~Zhang, {\it {Testing $f(R)$ gravity against the large scale structure of the
  universe.}},  {\em Phys. Rev.} {\bf D73} (2006) 123504,
  [\href{http://arxiv.org/abs/astro-ph/0511218}{{\tt astro-ph/0511218}}].

\bibitem{Boisseau:2000pr}
B.~Boisseau, G.~Esposito-Farese, D.~Polarski, and A.~A. Starobinsky, {\it
  {Reconstruction of a scalar tensor theory of gravity in an accelerating
  universe}},  {\em Phys. Rev. Lett.} {\bf 85} (2000) 2236,
  [\href{http://arxiv.org/abs/gr-qc/0001066}{{\tt gr-qc/0001066}}].

\bibitem{EspositoFarese:2000ij}
G.~Esposito-Farese and D.~Polarski, {\it {Scalar tensor gravity in an
  accelerating universe}},  {\em Phys. Rev.} {\bf D63} (2001) 063504,
  [\href{http://arxiv.org/abs/gr-qc/0009034}{{\tt gr-qc/0009034}}].

\bibitem{Tsujikawa:2007gd}
S.~Tsujikawa, {\it {Matter density perturbations and effective gravitational
  constant in modified gravity models of dark energy}},  {\em Phys. Rev.} {\bf
  D76} (2007) 023514, [\href{http://arxiv.org/abs/0705.1032}{{\tt
  arXiv:0705.1032}}].

\bibitem{Bean:2006up}
R.~Bean, D.~Bernat, L.~Pogosian, A.~Silvestri, and M.~Trodden, {\it {Dynamics
  of Linear Perturbations in f(R) Gravity}},  {\em Phys. Rev.} {\bf D75} (2007)
  064020, [\href{http://arxiv.org/abs/astro-ph/0611321}{{\tt
  astro-ph/0611321}}].

\bibitem{delaCruzDombriz:2008cp}
A.~de~la Cruz-Dombriz, A.~Dobado, and A.~L. Maroto, {\it {On the evolution of
  density perturbations in f(R) theories of gravity}},  {\em Phys. Rev.} {\bf
  D77} (2008) 123515, [\href{http://arxiv.org/abs/0802.2999}{{\tt
  arXiv:0802.2999}}].

\bibitem{Salgado:2005hx}
M.~Salgado, {\it {The Cauchy problem of scalar tensor theories of gravity}},
  {\em Class. Quant. Grav.} {\bf 23} (2006) 4719--4742,
  [\href{http://arxiv.org/abs/gr-qc/0509001}{{\tt gr-qc/0509001}}].

\bibitem{Ehlers:1966ad}
J.~Ehlers, P.~Geren, and R.~K. Sachs, {\it {Isotropic solutions of the
  Einstein-Liouville equations}},  {\em J. Math. Phys.} {\bf 9} (1968)
  1344--1349.

\bibitem{Stoeger:1994qs}
S.~J. Stoeger, William~R., R.~Maartens, and G.~F.~R. Ellis, {\it {Proving
  almost homogeneity of the universe: An Almost Ehlers-Geren-Sachs theorem}},
  {\em Astrophys. J.} {\bf 443} (1995) 1.

\bibitem{Clarkson:2001qc}
C.~A. Clarkson, A.~A. Coley, and E.~S.~D. O'Neill, {\it {The Cosmic microwave
  background and scalar tensor theories of gravity}},  {\em Phys. Rev.} {\bf
  D64} (2001) 063510, [\href{http://arxiv.org/abs/gr-qc/0105026}{{\tt
  gr-qc/0105026}}].

\bibitem{Maartens:1994pb}
R.~Maartens and D.~R. Taylor, {\it {Fluid dynamics in higher order gravity}},
  {\em Gen. Rel. Grav.} {\bf 26} (1994) 599--613.

\bibitem{Ehlers:1993gf}
J.~Ehlers, {\it {Contributions to the relativistic mechanics of continuous
  media}},  {\em Gen. Rel. Grav.} {\bf 25} (1993) 1225--1266. [Abh. Akad. Wiss.
  Lit. Mainz. Nat. Kl.11,793(1961)].

\bibitem{Maartens:1996hb}
R.~Maartens, {\it {Linearization instability of gravity waves?}},  {\em Phys.
  Rev.} {\bf D55} (1997) 463--467,
  [\href{http://arxiv.org/abs/astro-ph/9609198}{{\tt astro-ph/9609198}}].

\bibitem{Carloni:2007br}
S.~Carloni, A.~Troisi, and P.~K.~S. Dunsby, {\it {Some remarks on the dynamical
  systems approach to fourth order gravity}},  {\em Gen. Rel. Grav.} {\bf 41}
  (2009) 1757--1776, [\href{http://arxiv.org/abs/0706.0452}{{\tt
  arXiv:0706.0452}}].

\bibitem{Carloni:2007yv}
S.~Carloni, P.~K.~S. Dunsby, and A.~Troisi, {\it {The Evolution of density
  perturbations in f(R) gravity}},  {\em Phys. Rev.} {\bf D77} (2008) 024024,
  [\href{http://arxiv.org/abs/0707.0106}{{\tt arXiv:0707.0106}}].

\bibitem{Abebe:2013zua}
A.~Abebe, A.~de~la Cruz-Dombriz, and P.~K.~S. Dunsby, {\it {Large Scale
  Structure Constraints for a Class of f(R) Theories of Gravity}},  {\em Phys.
  Rev.} {\bf D88} (2013) 044050, [\href{http://arxiv.org/abs/1304.3462}{{\tt
  arXiv:1304.3462}}].

\bibitem{Abebe:2011ry}
A.~Abebe, M.~Abdelwahab, A.~de~la Cruz-Dombriz, and P.~K.~S. Dunsby, {\it
  {Covariant gauge-invariant perturbations in multifluid f(R) gravity}},  {\em
  Class. Quant. Grav.} {\bf 29} (2012) 135011,
  [\href{http://arxiv.org/abs/1110.1191}{{\tt arXiv:1110.1191}}].

\bibitem{Kodama:1985bj}
H.~Kodama and M.~Sasaki, {\it {Cosmological Perturbation Theory}},  {\em
  Prog.Theor.Phys.Suppl.} {\bf 78} (1984) 1--166.

\bibitem{delaCruz-Dombriz:2013gfa}
A.~de~la Cruz-Dombriz, P.~K.~S. Dunsby, V.~C. Busti, and S.~Kandhai, {\it {On
  tidal forces in f(R) theories of gravity}},  {\em Phys. Rev.} {\bf D89}
  (2014), no.~6 064029, [\href{http://arxiv.org/abs/1312.2022}{{\tt
  arXiv:1312.2022}}].

\bibitem{delaCruz-Dombriz:2015tye}
{\'A}.~de~la Cruz-Dombriz, P.~K.~S. Dunsby, S.~Kandhai, and
  D.~S{\'a}ez-G{\'o}mez, {\it {Theoretical and observational constraints of
  viable f(R) theories of gravity}},  {\em Phys. Rev.} {\bf D93} (2016), no.~8
  084016, [\href{http://arxiv.org/abs/1511.00102}{{\tt arXiv:1511.00102}}].

\bibitem{Albareti:2012va}
F.~D. Albareti, J.~A.~R. Cembranos, A.~de~la Cruz-Dombriz, and A.~Dobado, {\it
  {On the non-attractive character of gravity in f(R) theories}},  {\em JCAP}
  {\bf 1307} (2013) 009, [\href{http://arxiv.org/abs/1212.4781}{{\tt
  arXiv:1212.4781}}].

\bibitem{Albareti:2012se}
F.~D. Albareti, J.~A.~R. Cembranos, and A.~de~la Cruz-Dombriz, {\it {Focusing
  of geodesic congruences in an accelerated expanding Universe}},  {\em JCAP}
  {\bf 1212} (2012) 020, [\href{http://arxiv.org/abs/1208.4201}{{\tt
  arXiv:1208.4201}}].

\bibitem{ArmendarizPicon:2000ah}
C.~Armendariz-Picon, V.~F. Mukhanov, and P.~J. Steinhardt, {\it {Essentials of
  k essence}},  {\em Phys. Rev.} {\bf D63} (2001) 103510,
  [\href{http://arxiv.org/abs/astro-ph/0006373}{{\tt astro-ph/0006373}}].

\bibitem{Deffayet:2010qz}
C.~Deffayet, O.~Pujolas, I.~Sawicki, and A.~Vikman, {\it {Imperfect Dark Energy
  from Kinetic Gravity Braiding}},  {\em JCAP} {\bf 1010} (2010) 026,
  [\href{http://arxiv.org/abs/1008.0048}{{\tt arXiv:1008.0048}}].

\bibitem{Woodard:2006nt}
R.~P. Woodard, {\it {Avoiding dark energy with 1/r modifications of gravity}},
  {\em Lect.Notes Phys.} {\bf 720} (2007) 403--433,
  [\href{http://arxiv.org/abs/astro-ph/0601672}{{\tt astro-ph/0601672}}].

\bibitem{Nicolis:2008in}
A.~Nicolis, R.~Rattazzi, and E.~Trincherini, {\it {The Galileon as a local
  modification of gravity}},  {\em Phys. Rev.} {\bf D79} (2009) 064036,
  [\href{http://arxiv.org/abs/0811.2197}{{\tt arXiv:0811.2197}}].

\bibitem{Deffayet:2009wt}
C.~Deffayet, G.~Esposito-Farese, and A.~Vikman, {\it {Covariant Galileon}},
  {\em Phys. Rev.} {\bf D79} (2009) 084003,
  [\href{http://arxiv.org/abs/0901.1314}{{\tt arXiv:0901.1314}}].

\bibitem{Deffayet:2011gz}
C.~Deffayet, X.~Gao, D.~Steer, and G.~Zahariade, {\it {From k-essence to
  generalised Galileons}},  {\em Phys.Rev.} {\bf D84} (2011) 064039,
  [\href{http://arxiv.org/abs/1103.3260}{{\tt arXiv:1103.3260}}].

\bibitem{Kobayashi:2011nu}
T.~Kobayashi, M.~Yamaguchi, and J.~Yokoyama, {\it {Generalized G-inflation:
  Inflation with the most general second-order field equations}},  {\em Prog.
  Theor. Phys.} {\bf 126} (2011) 511--529,
  [\href{http://arxiv.org/abs/1105.5723}{{\tt arXiv:1105.5723}}].

\bibitem{Horndeski:1974wa}
G.~W. Horndeski, {\it {Second-order scalar-tensor field equations in a
  four-dimensional space}},  {\em Int.J.Theor.Phys.} {\bf 10} (1974) 363--384.

\bibitem{Zumalacarregui:2013pma}
M.~Zumalac{\'a}rregui and J.~Garc{\'\i}a-Bellido, {\it {Transforming gravity:
  from derivative couplings to matter to second-order scalar-tensor theories
  beyond the Horndeski Lagrangian}},  {\em Phys.Rev.} {\bf D89} (2014) 064046,
  [\href{http://arxiv.org/abs/1308.4685}{{\tt arXiv:1308.4685}}].

\bibitem{Gleyzes:2014dya}
J.~Gleyzes, D.~Langlois, F.~Piazza, and F.~Vernizzi, {\it {Healthy theories
  beyond Horndeski}},  \href{http://arxiv.org/abs/1404.6495}{{\tt
  arXiv:1404.6495}}.

\bibitem{Gleyzes:2014qga}
J.~Gleyzes, D.~Langlois, F.~Piazza, and F.~Vernizzi, {\it {Exploring
  gravitational theories beyond Horndeski}},
  \href{http://arxiv.org/abs/1408.1952}{{\tt arXiv:1408.1952}}.

\bibitem{Charmousis:2011ea}
C.~Charmousis, E.~J. Copeland, A.~Padilla, and P.~M. Saffin, {\it {Self-tuning
  and the derivation of a class of scalar-tensor theories}},  {\em Phys. Rev.}
  {\bf D85} (2012) 104040, [\href{http://arxiv.org/abs/1112.4866}{{\tt
  arXiv:1112.4866}}].

\bibitem{Martin-Moruno:2015bda}
P.~Mart\'in-Moruno, N.~J. Nunes, and F.~S.~N. Lobo, {\it {Horndeski theories
  self-tuning to a de Sitter vacuum}},  {\em Phys. Rev.} {\bf D91} (2015),
  no.~8 084029, [\href{http://arxiv.org/abs/1502.03236}{{\tt
  arXiv:1502.03236}}].

\bibitem{Germani:2017pwt}
C.~Germani and P.~Mart{\'\i}n-Moruno, {\it {Tracking our Universe to de Sitter
  by a Horndeski scalar}},  {\em Phys. Dark Univ.} {\bf 18} (2017) 1--5,
  [\href{http://arxiv.org/abs/1707.03741}{{\tt arXiv:1707.03741}}].

\bibitem{Silva:2009km}
F.~P. Silva and K.~Koyama, {\it {Self-Accelerating Universe in Galileon
  Cosmology}},  {\em Phys. Rev.} {\bf D80} (2009) 121301,
  [\href{http://arxiv.org/abs/0909.4538}{{\tt arXiv:0909.4538}}].

\bibitem{DeFelice:2010pv}
A.~De~Felice and S.~Tsujikawa, {\it {Cosmology of a covariant Galileon field}},
   {\em Phys. Rev. Lett.} {\bf 105} (2010) 111301,
  [\href{http://arxiv.org/abs/1007.2700}{{\tt arXiv:1007.2700}}].

\bibitem{DeFelice:2010nf}
A.~De~Felice and S.~Tsujikawa, {\it {Generalized Galileon cosmology}},  {\em
  Phys. Rev.} {\bf D84} (2011) 124029,
  [\href{http://arxiv.org/abs/1008.4236}{{\tt arXiv:1008.4236}}].

\bibitem{Appleby:2011aa}
S.~Appleby and E.~V. Linder, {\it {The Paths of Gravity in Galileon
  Cosmology}},  {\em JCAP} {\bf 1203} (2012) 043,
  [\href{http://arxiv.org/abs/1112.1981}{{\tt arXiv:1112.1981}}].

\bibitem{Martin-Moruno:2015lha}
P.~Mart{\'\i}n-Moruno, N.~J. Nunes, and F.~S.~N. Lobo, {\it {Attracted to de
  Sitter: cosmology of the linear Horndeski models}},  {\em JCAP} {\bf 1505}
  (2015), no.~05 033, [\href{http://arxiv.org/abs/1502.05878}{{\tt
  arXiv:1502.05878}}].

\bibitem{Martin-Moruno:2015kaa}
P.~Mart\'in-Moruno and N.~J. Nunes, {\it {Attracted to de Sitter II: cosmology
  of the shift-symmetric Horndeski models}},  {\em JCAP} {\bf 1509} (2015),
  no.~09 056, [\href{http://arxiv.org/abs/1506.02497}{{\tt arXiv:1506.02497}}].

\bibitem{Easson:2011zy}
D.~A. Easson, I.~Sawicki, and A.~Vikman, {\it {G-Bounce}},  {\em JCAP} {\bf
  1111} (2011) 021, [\href{http://arxiv.org/abs/1109.1047}{{\tt
  arXiv:1109.1047}}].

\bibitem{Qiu:2015nha}
T.~Qiu and Y.-T. Wang, {\it {G-Bounce Inflation: Towards Nonsingular Inflation
  Cosmology with Galileon Field}},  {\em JHEP} {\bf 04} (2015) 130,
  [\href{http://arxiv.org/abs/1501.03568}{{\tt arXiv:1501.03568}}].

\bibitem{Bellini:2014fua}
E.~Bellini and I.~Sawicki, {\it {Maximal freedom at minimum cost: linear
  large-scale structure in general modifications of gravity}},  {\em JCAP} {\bf
  1407} (2014) 050, [\href{http://arxiv.org/abs/1404.3713}{{\tt
  arXiv:1404.3713}}].

\bibitem{Zumalacarregui:2016pph}
M.~Zumalac\'arregui, E.~Bellini, I.~Sawicki, J.~Lesgourgues, and P.~G.
  Ferreira, {\it {hi$\_$class: Horndeski in the Cosmic Linear Anisotropy
  Solving System}},  {\em JCAP} {\bf 1708} (2017), no.~08 019,
  [\href{http://arxiv.org/abs/1605.06102}{{\tt arXiv:1605.06102}}].

\bibitem{Gubitosi:2012hu}
G.~Gubitosi, F.~Piazza, and F.~Vernizzi, {\it {The Effective Field Theory of
  Dark Energy}},  {\em JCAP} {\bf 1302} (2013) 032,
  [\href{http://arxiv.org/abs/1210.0201}{{\tt arXiv:1210.0201}}].
  [JCAP1302,032(2013)].

\bibitem{Hu:2013twa}
B.~Hu, M.~Raveri, N.~Frusciante, and A.~Silvestri, {\it {Effective Field Theory
  of Cosmic Acceleration: an implementation in CAMB}},  {\em Phys. Rev.} {\bf
  D89} (2014), no.~10 103530, [\href{http://arxiv.org/abs/1312.5742}{{\tt
  arXiv:1312.5742}}].

\bibitem{Baker:2017hug}
T.~Baker, E.~Bellini, P.~G. Ferreira, M.~Lagos, J.~Noller, and I.~Sawicki, {\it
  {Strong constraints on cosmological gravity from GW170817 and GRB 170817A}},
  {\em Phys. Rev. Lett.} {\bf 119} (2017), no.~25 251301,
  [\href{http://arxiv.org/abs/1710.06394}{{\tt arXiv:1710.06394}}].

\bibitem{deRham:2018red}
C.~de~Rham and S.~Melville, {\it {Gravitational Rainbows: LIGO and Dark Energy
  at its Cutoff}},  {\em Phys. Rev. Lett.} {\bf 121} (2018), no.~22 221101,
  [\href{http://arxiv.org/abs/1806.09417}{{\tt arXiv:1806.09417}}].

\bibitem{Jimenez:2015bwa}
J.~Beltr\'a~Jim\'enez, F.~Piazza, and H.~Velten, {\it {Evading the Vainshtein
  Mechanism with Anomalous Gravitational Wave Speed: Constraints on Modified
  Gravity from Binary Pulsars}},  {\em Phys. Rev. Lett.} {\bf 116} (2016),
  no.~6 061101, [\href{http://arxiv.org/abs/1507.05047}{{\tt
  arXiv:1507.05047}}].

\bibitem{Damour:1992kf}
T.~Damour and K.~Nordtvedt, {\it {General relativity as a cosmological
  attractor of tensor scalar theories}},  {\em Phys. Rev. Lett.} {\bf 70}
  (1993) 2217--2219.

\bibitem{Joyce:2014kja}
A.~Joyce, B.~Jain, J.~Khoury, and M.~Trodden, {\it {Beyond the Cosmological
  Standard Model}},  {\em Phys. Rept.} {\bf 568} (2015) 1--98,
  [\href{http://arxiv.org/abs/1407.0059}{{\tt arXiv:1407.0059}}].

\bibitem{Heisenberg:2018vsk}
L.~Heisenberg, {\it {A systematic approach to generalisations of General
  Relativity and their cosmological implications}},  {\em Phys. Rept.} {\bf
  796} (2019) 1--113, [\href{http://arxiv.org/abs/1807.01725}{{\tt
  arXiv:1807.01725}}].

\bibitem{Sotiriou:2010wn}
T.~P. Sotiriou, {\it {Horava-Lifshitz gravity: a status report}},  {\em J.
  Phys. Conf. Ser.} {\bf 283} (2011) 012034,
  [\href{http://arxiv.org/abs/1010.3218}{{\tt arXiv:1010.3218}}].

\bibitem{Sotiriou:2009bx}
T.~P. Sotiriou, M.~Visser, and S.~Weinfurtner, {\it {Quantum gravity without
  Lorentz invariance}},  {\em JHEP} {\bf 10} (2009) 033,
  [\href{http://arxiv.org/abs/0905.2798}{{\tt arXiv:0905.2798}}].

\bibitem{Li:2010cg}
B.~Li, T.~P. Sotiriou, and J.~D. Barrow, {\it {$f(T)$ gravity and local Lorentz
  invariance}},  {\em Phys. Rev.} {\bf D83} (2011) 064035,
  [\href{http://arxiv.org/abs/1010.1041}{{\tt arXiv:1010.1041}}].

\bibitem{BeltranJimenez:2017tkd}
J.~B. Jim{\'e}nez, L.~Heisenberg, and T.~Koivisto, {\it {Coincident General
  Relativity}},  {\em Phys. Rev.} {\bf D98} (2018), no.~4 044048,
  [\href{http://arxiv.org/abs/1710.03116}{{\tt arXiv:1710.03116}}].

\bibitem{Jimenez:2019ovq}
J.~B. Jim{\'e}nez, L.~Heisenberg, T.~S. Koivisto, and S.~Pekar, {\it {Cosmology
  in $f(Q)$ geometry}},  \href{http://arxiv.org/abs/1906.10027}{{\tt
  arXiv:1906.10027}}.

\bibitem{Heisenberg:2014rta}
L.~Heisenberg, {\it {Generalization of the Proca Action}},  {\em JCAP} {\bf
  1405} (2014) 015, [\href{http://arxiv.org/abs/1402.7026}{{\tt
  arXiv:1402.7026}}].

\bibitem{Jimenez:2016isa}
J.~Beltran~Jimenez and L.~Heisenberg, {\it {Derivative self-interactions for a
  massive vector field}},  {\em Phys. Lett.} {\bf B757} (2016) 405--411,
  [\href{http://arxiv.org/abs/1602.03410}{{\tt arXiv:1602.03410}}].

\bibitem{deRham:2010kj}
C.~de~Rham, G.~Gabadadze, and A.~J. Tolley, {\it {Resummation of Massive
  Gravity}},  {\em Phys.Rev.Lett.} {\bf 106} (2011) 231101,
  [\href{http://arxiv.org/abs/1011.1232}{{\tt arXiv:1011.1232}}].

\bibitem{Hassan:2011zd}
S.~Hassan and R.~A. Rosen, {\it {Bimetric Gravity from Ghost-free Massive
  Gravity}},  {\em JHEP} {\bf 1202} (2012) 126,
  [\href{http://arxiv.org/abs/1109.3515}{{\tt arXiv:1109.3515}}].

\bibitem{Hassan:2011vm}
S.~Hassan and R.~A. Rosen, {\it {On Non-Linear Actions for Massive Gravity}},
  {\em JHEP} {\bf 1107} (2011) 009, [\href{http://arxiv.org/abs/1103.6055}{{\tt
  arXiv:1103.6055}}].

\bibitem{deRham:2010ik}
C.~de~Rham and G.~Gabadadze, {\it {Generalization of the Fierz-Pauli Action}},
  {\em Phys.Rev.} {\bf D82} (2010) 044020,
  [\href{http://arxiv.org/abs/1007.0443}{{\tt arXiv:1007.0443}}].

\bibitem{deRham:2012ew}
C.~de~Rham, G.~Gabadadze, L.~Heisenberg, and D.~Pirtskhalava, {\it
  {Nonrenormalization and naturalness in a class of scalar-tensor theories}},
  {\em Phys.Rev.} {\bf D87} (2013), no.~8 085017,
  [\href{http://arxiv.org/abs/1212.4128}{{\tt arXiv:1212.4128}}].

\bibitem{Buchbinder:2012wb}
I.~L. Buchbinder, D.~D. Pereira, and I.~L. Shapiro, {\it {One-loop divergences
  in massive gravity theory}},  {\em Phys. Lett.} {\bf B712} (2012) 104--108,
  [\href{http://arxiv.org/abs/1201.3145}{{\tt arXiv:1201.3145}}].

\bibitem{deRham:2013qqa}
C.~de~Rham, L.~Heisenberg, and R.~H. Ribeiro, {\it {Quantum Corrections in
  Massive Gravity}},  {\em Phys.Rev.} {\bf D88} (2013) 084058,
  [\href{http://arxiv.org/abs/1307.7169}{{\tt arXiv:1307.7169}}].

\bibitem{Heisenberg:2014rka}
L.~Heisenberg, {\it {Quantum corrections in massive bigravity and new effective
  composite metrics}},  {\em Class. Quant. Grav.} {\bf 32} (2015), no.~10
  105011, [\href{http://arxiv.org/abs/1410.4239}{{\tt arXiv:1410.4239}}].

\bibitem{vanDam:1970vg}
H.~van Dam and M.~Veltman, {\it {Massive and massless Yang-Mills and
  gravitational fields}},  {\em Nucl.Phys.} {\bf B22} (1970) 397--411.

\bibitem{Zakharov:1970cc}
V.~Zakharov, {\it {Linearized gravitation theory and the graviton mass}},  {\em
  JETP Lett.} {\bf 12} (1970) 312.

\bibitem{Babichev:2009jt}
E.~Babichev, C.~Deffayet, and R.~Ziour, {\it {Recovering General Relativity
  from massive gravity}},  {\em Phys. Rev. Lett.} {\bf 103} (2009) 201102,
  [\href{http://arxiv.org/abs/0907.4103}{{\tt arXiv:0907.4103}}].

\bibitem{Babichev:2010jd}
E.~Babichev, C.~Deffayet, and R.~Ziour, {\it {The Recovery of General
  Relativity in massive gravity via the Vainshtein mechanism}},  {\em Phys.
  Rev.} {\bf D82} (2010) 104008, [\href{http://arxiv.org/abs/1007.4506}{{\tt
  arXiv:1007.4506}}].

\bibitem{deRham:2010tw}
C.~de~Rham, G.~Gabadadze, L.~Heisenberg, and D.~Pirtskhalava, {\it {Cosmic
  Acceleration and the Helicity-0 Graviton}},  {\em Phys. Rev.} {\bf D83}
  (2011) 103516, [\href{http://arxiv.org/abs/1010.1780}{{\tt
  arXiv:1010.1780}}].

\bibitem{DAmico:2011eto}
G.~D'Amico, C.~de~Rham, S.~Dubovsky, G.~Gabadadze, D.~Pirtskhalava, and A.~J.
  Tolley, {\it {Massive Cosmologies}},  {\em Phys. Rev.} {\bf D84} (2011)
  124046, [\href{http://arxiv.org/abs/1108.5231}{{\tt arXiv:1108.5231}}].

\bibitem{Gumrukcuoglu:2011ew}
A.~E. G{\"u}mr{\"u}k{\c c}{\"u}o{\u g}lu, C.~Lin, and S.~Mukohyama, {\it {Open
  FRW universes and self-acceleration from nonlinear massive gravity}},  {\em
  JCAP} {\bf 1111} (2011) 030, [\href{http://arxiv.org/abs/1109.3845}{{\tt
  arXiv:1109.3845}}].

\bibitem{Gumrukcuoglu:2011zh}
A.~E. G{\"u}mr{\"u}k{\c c}{\"u}o{\u g}lu, C.~Lin, and S.~Mukohyama, {\it
  {Cosmological perturbations of self-accelerating universe in nonlinear
  massive gravity}},  {\em JCAP} {\bf 1203} (2012) 006,
  [\href{http://arxiv.org/abs/1111.4107}{{\tt arXiv:1111.4107}}].

\bibitem{Comelli:2011zm}
D.~Comelli, M.~Crisostomi, F.~Nesti, and L.~Pilo, {\it {FRW Cosmology in Ghost
  Free Massive Gravity}},  {\em JHEP} {\bf 1203} (2012) 067,
  [\href{http://arxiv.org/abs/1111.1983}{{\tt arXiv:1111.1983}}].

\bibitem{vonStrauss:2011mq}
M.~von Strauss, A.~Schmidt-May, J.~Enander, E.~M{\"o}rtsell, and S.~Hassan,
  {\it {Cosmological Solutions in Bimetric Gravity and their Observational
  Tests}},  {\em JCAP} {\bf 1203} (2012) 042,
  [\href{http://arxiv.org/abs/1111.1655}{{\tt arXiv:1111.1655}}].

\bibitem{Akrami:2012vf}
Y.~Akrami, T.~S. Koivisto, and M.~Sandstad, {\it {Accelerated expansion from
  ghost-free bigravity: a statistical analysis with improved generality}},
  {\em JHEP} {\bf 1303} (2013) 099, [\href{http://arxiv.org/abs/1209.0457}{{\tt
  arXiv:1209.0457}}].

\bibitem{Akrami:2013ffa}
Y.~Akrami, T.~S. Koivisto, D.~F. Mota, and M.~Sandstad, {\it {Bimetric gravity
  doubly coupled to matter: theory and cosmological implications}},  {\em JCAP}
  {\bf 1310} (2013) 046, [\href{http://arxiv.org/abs/1306.0004}{{\tt
  arXiv:1306.0004}}].

\bibitem{Solomon:2014dua}
A.~R. Solomon, Y.~Akrami, and T.~S. Koivisto, {\it {Linear growth of structure
  in massive bigravity}},  {\em JCAP} {\bf 1410} (2014) 066,
  [\href{http://arxiv.org/abs/1404.4061}{{\tt arXiv:1404.4061}}].

\bibitem{Enander:2015vja}
J.~Enander, Y.~Akrami, E.~M{\"o}rtsell, M.~Renneby, and A.~R. Solomon, {\it
  {Integrated Sachs-Wolfe effect in massive bigravity}},  {\em Phys. Rev.} {\bf
  D91} (2015) 084046, [\href{http://arxiv.org/abs/1501.02140}{{\tt
  arXiv:1501.02140}}].

\bibitem{Konnig:2014xva}
F.~K{\"o}nnig, Y.~Akrami, L.~Amendola, M.~Motta, and A.~R. Solomon, {\it
  {Stable and unstable cosmological models in bimetric massive gravity}},  {\em
  Phys.Rev.} {\bf D90} (2014) 124014,
  [\href{http://arxiv.org/abs/1407.4331}{{\tt arXiv:1407.4331}}].

\bibitem{DeFelice:2014nja}
A.~De~Felice, A.~E. G{\"u}mr{\"u}k{\c c}{\"u}o{\u g}lu, S.~Mukohyama,
  N.~Tanahashi, and T.~Tanaka, {\it {Viable cosmology in bimetric theory}},
  {\em JCAP} {\bf 1406} (2014) 037, [\href{http://arxiv.org/abs/1404.0008}{{\tt
  arXiv:1404.0008}}].

\bibitem{Cusin:2014psa}
G.~Cusin, R.~Durrer, P.~Guarato, and M.~Motta, {\it {Gravitational waves in
  bigravity cosmology}},  {\em JCAP} {\bf 1505} (2015), no.~05 030,
  [\href{http://arxiv.org/abs/1412.5979}{{\tt arXiv:1412.5979}}].

\bibitem{Comelli:2012db}
D.~Comelli, M.~Crisostomi, and L.~Pilo, {\it {Perturbations in Massive Gravity
  Cosmology}},  {\em JHEP} {\bf 1206} (2012) 085,
  [\href{http://arxiv.org/abs/1202.1986}{{\tt arXiv:1202.1986}}].

\bibitem{Comelli:2014bqa}
D.~Comelli, M.~Crisostomi, and L.~Pilo, {\it {FRW Cosmological Perturbations in
  Massive Bigravity}},  {\em Phys.Rev.} {\bf D90} (2014), no.~8 084003,
  [\href{http://arxiv.org/abs/1403.5679}{{\tt arXiv:1403.5679}}].

\bibitem{DeFelice:2013nba}
A.~De~Felice, T.~Nakamura, and T.~Tanaka, {\it {Possible existence of viable
  models of bi-gravity with detectable graviton oscillations by gravitational
  wave detectors}},  {\em PTEP} {\bf 2014} (2014) 043E01,
  [\href{http://arxiv.org/abs/1304.3920}{{\tt arXiv:1304.3920}}].

\bibitem{Akrami:2015qga}
Y.~Akrami, S.~F. Hassan, F.~K{\"o}nnig, A.~Schmidt-May, and A.~R. Solomon, {\it
  {Bimetric gravity is cosmologically viable}},  {\em Phys. Lett.} {\bf B748}
  (2015) 37--44, [\href{http://arxiv.org/abs/1503.07521}{{\tt
  arXiv:1503.07521}}].

\bibitem{deRham:2014naa}
C.~de~Rham, L.~Heisenberg, and R.~H. Ribeiro, {\it {On couplings to matter in
  massive (bi-)gravity}},  {\em Class.Quant.Grav.} {\bf 32} (2015) 035022,
  [\href{http://arxiv.org/abs/1408.1678}{{\tt arXiv:1408.1678}}].

\bibitem{Enander:2014xga}
J.~Enander, A.~R. Solomon, Y.~Akrami, and E.~M{\"o}rtsell, {\it {Cosmic
  expansion histories in massive bigravity with symmetric matter coupling}},
  {\em JCAP} {\bf 1501} (2015) 006, [\href{http://arxiv.org/abs/1409.2860}{{\tt
  arXiv:1409.2860}}].

\bibitem{Solomon:2014iwa}
A.~R. Solomon, J.~Enander, Y.~Akrami, T.~S. Koivisto, F.~K{\"o}nnig, and
  E.~M{\"o}rtsell, {\it {Cosmological viability of massive gravity with
  generalized matter coupling}},  \href{http://arxiv.org/abs/1409.8300}{{\tt
  arXiv:1409.8300}}.

\bibitem{Akrami:2018yjz}
Y.~Akrami, P.~Brax, A.-C. Davis, and V.~Vardanyan, {\it {Neutron star merger
  GW170817 strongly constrains doubly coupled bigravity}},  {\em Phys. Rev. D}
  {\bf 97} (2018), no.~12 124010, [\href{http://arxiv.org/abs/1803.09726}{{\tt
  arXiv:1803.09726}}].

\bibitem{Maxwell:1865zz}
J.~C. Maxwell, {\it {A dynamical theory of the electromagnetic field}},  {\em
  Phil. Trans. Roy. Soc. Lond.} {\bf 155} (1865) 459--512.

\bibitem{Cembranos:2003PhD}
J.~A.~R. Cembranos, {\it {PhD thesis: Lagrangianos efectivos en teorias con
  dimensiones extras}}, .

\bibitem{Bailin:1987jd}
D.~Bailin and A.~Love, {\it {KALUZA-KLEIN THEORIES}},  {\em Rept. Prog. Phys.}
  {\bf 50} (1987) 1087--1170.

\bibitem{Han:1998sg}
T.~Han, J.~D. Lykken, and R.-J. Zhang, {\it {On Kaluza-Klein states from large
  extra dimensions}},  {\em Phys. Rev.} {\bf D59} (1999) 105006,
  [\href{http://arxiv.org/abs/hep-ph/9811350}{{\tt hep-ph/9811350}}].

\bibitem{Antoniadis:1998ig}
I.~Antoniadis, N.~Arkani-Hamed, S.~Dimopoulos, and G.~R. Dvali, {\it {New
  dimensions at a millimeter to a Fermi and superstrings at a TeV}},  {\em
  Phys. Lett.} {\bf B436} (1998) 257--263,
  [\href{http://arxiv.org/abs/hep-ph/9804398}{{\tt hep-ph/9804398}}].

\bibitem{Andrianov:2003hx}
A.~A. Andrianov, V.~A. Andrianov, P.~Giacconi, and R.~Soldati, {\it {Domain
  wall generation by fermion selfinteraction and light particles}},  {\em JHEP}
  {\bf 07} (2003) 063, [\href{http://arxiv.org/abs/hep-ph/0305271}{{\tt
  hep-ph/0305271}}].

\bibitem{ArkaniHamed:1998nn}
N.~Arkani-Hamed, S.~Dimopoulos, and G.~R. Dvali, {\it {Phenomenology,
  astrophysics and cosmology of theories with submillimeter dimensions and TeV
  scale quantum gravity}},  {\em Phys. Rev.} {\bf D59} (1999) 086004,
  [\href{http://arxiv.org/abs/hep-ph/9807344}{{\tt hep-ph/9807344}}].

\bibitem{Giudice:1998ck}
G.~F. Giudice, R.~Rattazzi, and J.~D. Wells, {\it {Quantum gravity and extra
  dimensions at high-energy colliders}},  {\em Nucl. Phys.} {\bf B544} (1999)
  3--38, [\href{http://arxiv.org/abs/hep-ph/9811291}{{\tt hep-ph/9811291}}].

\bibitem{Hewett:1998sn}
J.~L. Hewett, {\it {Indirect collider signals for extra dimensions}},  {\em
  Phys. Rev. Lett.} {\bf 82} (1999) 4765--4768,
  [\href{http://arxiv.org/abs/hep-ph/9811356}{{\tt hep-ph/9811356}}].

\bibitem{Giudice:2003tu}
G.~F. Giudice and A.~Strumia, {\it {Constraints on extra dimensional theories
  from virtual graviton exchange}},  {\em Nucl. Phys.} {\bf B663} (2003)
  377--393, [\href{http://arxiv.org/abs/hep-ph/0301232}{{\tt hep-ph/0301232}}].

\bibitem{Randall:1999ee}
L.~Randall and R.~Sundrum, {\it {A Large mass hierarchy from a small extra
  dimension}},  {\em Phys. Rev. Lett.} {\bf 83} (1999) 3370--3373,
  [\href{http://arxiv.org/abs/hep-ph/9905221}{{\tt hep-ph/9905221}}].

\bibitem{Randall:1999vf}
L.~Randall and R.~Sundrum, {\it {An Alternative to compactification}},  {\em
  Phys. Rev. Lett.} {\bf 83} (1999) 4690--4693,
  [\href{http://arxiv.org/abs/hep-th/9906064}{{\tt hep-th/9906064}}].

\bibitem{Davoudiasl:1999jd}
H.~Davoudiasl, J.~L. Hewett, and T.~G. Rizzo, {\it {Phenomenology of the
  Randall-Sundrum Gauge Hierarchy Model}},  {\em Phys. Rev. Lett.} {\bf 84}
  (2000) 2080, [\href{http://arxiv.org/abs/hep-ph/9909255}{{\tt
  hep-ph/9909255}}].

\bibitem{Allanach:2000nr}
B.~C. Allanach, K.~Odagiri, M.~A. Parker, and B.~R. Webber, {\it {Searching for
  narrow graviton resonances with the ATLAS detector at the Large Hadron
  Collider}},  {\em JHEP} {\bf 09} (2000) 019,
  [\href{http://arxiv.org/abs/hep-ph/0006114}{{\tt hep-ph/0006114}}].

\bibitem{Davoudiasl:2000wi}
H.~Davoudiasl, J.~L. Hewett, and T.~G. Rizzo, {\it {Experimental probes of
  localized gravity: On and off the wall}},  {\em Phys. Rev.} {\bf D63} (2001)
  075004, [\href{http://arxiv.org/abs/hep-ph/0006041}{{\tt hep-ph/0006041}}].

\bibitem{Dobado:2000gr}
A.~Dobado and A.~L. Maroto, {\it {The Dynamics of the Goldstone bosons on the
  brane}},  {\em Nucl. Phys.} {\bf B592} (2001) 203--218,
  [\href{http://arxiv.org/abs/hep-ph/0007100}{{\tt hep-ph/0007100}}].

\bibitem{Cembranos:2001rp}
J.~A.~R. Cembranos, A.~Dobado, and A.~L. Maroto, {\it {Brane skyrmions and
  wrapped states}},  {\em Phys. Rev.} {\bf D65} (2002) 026005,
  [\href{http://arxiv.org/abs/hep-ph/0106322}{{\tt hep-ph/0106322}}].

\bibitem{Cembranos:2004eb}
J.~A.~R. Cembranos, A.~Dobado, and A.~L. Maroto, {\it {Dark geometry}},  {\em
  Int. J. Mod. Phys.} {\bf D13} (2004) 2275--2280,
  [\href{http://arxiv.org/abs/hep-ph/0405165}{{\tt hep-ph/0405165}}].

\bibitem{Cembranos:2016jun}
J.~A.~R. Cembranos and A.~L. Maroto, {\it {Disformal scalars as dark matter
  candidates: Branon phenomenology}},  {\em Int. J. Mod. Phys.} {\bf 31}
  (2016), no.~14n15 1630015, [\href{http://arxiv.org/abs/1602.07270}{{\tt
  arXiv:1602.07270}}].

\bibitem{Brax:2014vva}
P.~Brax and C.~Burrage, {\it {Constraining Disformally Coupled Scalar Fields}},
   {\em Phys. Rev.} {\bf D90} (2014), no.~10 104009,
  [\href{http://arxiv.org/abs/1407.1861}{{\tt arXiv:1407.1861}}].

\bibitem{Achard:2004uu}
{\bf L3} Collaboration, P.~Achard et~al., {\it {Search for branons at LEP}},
  {\em Phys. Lett.} {\bf B597} (2004) 145--154,
  [\href{http://arxiv.org/abs/hep-ex/0407017}{{\tt hep-ex/0407017}}].

\bibitem{Cembranos:2004jp}
J.~A.~R. Cembranos, A.~Dobado, and A.~L. Maroto, {\it {Branon search in
  hadronic colliders}},  {\em Phys. Rev.} {\bf D70} (2004) 096001,
  [\href{http://arxiv.org/abs/hep-ph/0405286}{{\tt hep-ph/0405286}}].

\bibitem{Cembranos:2011cm}
J.~A.~R. Cembranos, J.~L. Diaz-Cruz, and L.~Prado, {\it {Impact of DM direct
  searches and the LHC analyses on branon phenomenology}},  {\em Phys. Rev.}
  {\bf D84} (2011) 083522, [\href{http://arxiv.org/abs/1110.0542}{{\tt
  arXiv:1110.0542}}].

\bibitem{Landsberg:2015pka}
G.~Landsberg, {\it {Searches for Extra Spatial Dimensions with the CMS Detector
  at the LHC}},  {\em Mod. Phys. Lett.} {\bf A30} (2015), no.~15 1540017,
  [\href{http://arxiv.org/abs/1506.00024}{{\tt arXiv:1506.00024}}].

\bibitem{Khachatryan:2014rwa}
{\bf CMS} Collaboration, V.~Khachatryan et~al., {\it {Search for new phenomena
  in monophoton final states in proton-proton collisions at $\sqrt s =$ 8
  TeV}},  {\em Phys. Lett.} {\bf B755} (2016) 102--124,
  [\href{http://arxiv.org/abs/1410.8812}{{\tt arXiv:1410.8812}}].

\bibitem{Cembranos:2005sr}
J.~A.~R. Cembranos, A.~Dobado, and A.~L. Maroto, {\it {Dark matter clues in the
  muon anomalous magnetic moment}},  {\em Phys. Rev.} {\bf D73} (2006) 057303,
  [\href{http://arxiv.org/abs/hep-ph/0507066}{{\tt hep-ph/0507066}}].

\bibitem{Cembranos:2005jc}
J.~A.~R. Cembranos, A.~Dobado, and A.~L. Maroto, {\it {Branon radiative
  corrections to collider physics and precision observables}},  {\em Phys.
  Rev.} {\bf D73} (2006) 035008,
  [\href{http://arxiv.org/abs/hep-ph/0510399}{{\tt hep-ph/0510399}}].

\bibitem{Cembranos:2003mr}
J.~A.~R. Cembranos, A.~Dobado, and A.~L. Maroto, {\it {Brane world dark
  matter}},  {\em Phys. Rev. Lett.} {\bf 90} (2003) 241301,
  [\href{http://arxiv.org/abs/hep-ph/0302041}{{\tt hep-ph/0302041}}].

\bibitem{Cembranos:2003fu}
J.~A.~R. Cembranos, A.~Dobado, and A.~L. Maroto, {\it {Cosmological and
  astrophysical limits on brane fluctuations}},  {\em Phys. Rev.} {\bf D68}
  (2003) 103505, [\href{http://arxiv.org/abs/hep-ph/0307062}{{\tt
  hep-ph/0307062}}].

\bibitem{Maroto:2003gm}
A.~L. Maroto, {\it {The Nature of branon dark matter}},  {\em Phys. Rev.} {\bf
  D69} (2004) 043509, [\href{http://arxiv.org/abs/hep-ph/0310272}{{\tt
  hep-ph/0310272}}].

\bibitem{Preskill:1982cy}
J.~Preskill, M.~B. Wise, and F.~Wilczek, {\it {Cosmology of the Invisible
  Axion}},  {\em Phys. Lett.} {\bf 120B} (1983) 127--132.

\bibitem{Abbott:1982af}
L.~F. Abbott and P.~Sikivie, {\it {A Cosmological Bound on the Invisible
  Axion}},  {\em Phys. Lett.} {\bf 120B} (1983) 133--136.

\bibitem{Dine:1982ah}
M.~Dine and W.~Fischler, {\it {The Not So Harmless Axion}},  {\em Phys. Lett.}
  {\bf 120B} (1983) 137--141.

\bibitem{Frieman:1991qv}
J.~A. Frieman and A.~H. Jaffe, {\it {Cosmological constraints on
  pseudoNambu-Goldstone bosons}},  {\em Phys. Rev.} {\bf D45} (1992)
  2674--2684.

\bibitem{Hu:2000ke}
W.~Hu, R.~Barkana, and A.~Gruzinov, {\it {Cold and fuzzy dark matter}},  {\em
  Phys. Rev. Lett.} {\bf 85} (2000) 1158--1161,
  [\href{http://arxiv.org/abs/astro-ph/0003365}{{\tt astro-ph/0003365}}].

\bibitem{Cembranos:2012kk}
J.~A.~R. Cembranos, C.~Hallabrin, A.~L. Maroto, and S.~J.~N. Jareno, {\it
  {Isotropy theorem for cosmological vector fields}},  {\em Phys. Rev.} {\bf
  D86} (2012) 021301, [\href{http://arxiv.org/abs/1203.6221}{{\tt
  arXiv:1203.6221}}].

\bibitem{Cembranos:2016ugq}
J.~A.~R. Cembranos, A.~L. Maroto, and S.~J. N{\'u}{\~n}ez~Jare{\~n}o, {\it
  {Perturbations of ultralight vector field dark matter}},  {\em JHEP} {\bf 02}
  (2017) 064, [\href{http://arxiv.org/abs/1611.03793}{{\tt arXiv:1611.03793}}].

\bibitem{Cembranos:2015oya}
J.~A.~R. Cembranos, A.~L. Maroto, and S.~J. N{\'u}{\~n}ez~Jare{\~n}o, {\it
  {Cosmological perturbations in coherent oscillating scalar field models}},
  {\em JHEP} {\bf 03} (2016) 013, [\href{http://arxiv.org/abs/1509.08819}{{\tt
  arXiv:1509.08819}}].

\bibitem{Cembranos:2018ulm}
J.~A.~R. Cembranos, A.~L. Maroto, S.~J. N{\'u}{\~n}ez~Jare{\~n}o, and
  H.~Villarrubia-Rojo, {\it {Constraints on anharmonic corrections of Fuzzy
  Dark Matter}},  {\em JHEP} {\bf 08} (2018) 073,
  [\href{http://arxiv.org/abs/1805.08112}{{\tt arXiv:1805.08112}}].

\bibitem{Brax:2019fzb}
P.~Brax, J.~A.~R. Cembranos, and P.~Valageas, {\it {Impact of kinetic and
  potential self-interactions on scalar dark matter}},  {\em Phys. Rev.} {\bf
  D100} (2019), no.~2 023526, [\href{http://arxiv.org/abs/1906.00730}{{\tt
  arXiv:1906.00730}}].

\bibitem{Brax:2019npi}
P.~Brax, P.~Valageas, and J.~A.~R. Cembranos, {\it {Fate of scalar dark matter
  solitons around supermassive galactic black holes}},
  \href{http://arxiv.org/abs/1909.02614}{{\tt arXiv:1909.02614}}.

\bibitem{Cembranos:2008bw}
J.~A.~R. Cembranos and L.~E. Strigari, {\it {Diffuse MeV Gamma-rays and
  Galactic 511 keV Line from Decaying WIMP Dark Matter}},  {\em Phys. Rev.}
  {\bf D77} (2008) 123519, [\href{http://arxiv.org/abs/0801.0630}{{\tt
  arXiv:0801.0630}}].

\bibitem{Cembranos:2011hi}
J.~A.~R. Cembranos, A.~de~la Cruz-Dombriz, V.~Gammaldi, and A.~L. Maroto, {\it
  {Detection of branon dark matter with gamma ray telescopes}},  {\em Phys.
  Rev.} {\bf D85} (2012) 043505, [\href{http://arxiv.org/abs/1111.4448}{{\tt
  arXiv:1111.4448}}].

\bibitem{Cembranos:2010dm}
J.~A.~R. Cembranos, A.~de~la Cruz-Dombriz, A.~Dobado, R.~A. Lineros, and A.~L.
  Maroto, {\it {Photon spectra from WIMP annihilation}},  {\em Phys. Rev.} {\bf
  D83} (2011) 083507, [\href{http://arxiv.org/abs/1009.4936}{{\tt
  arXiv:1009.4936}}].

\bibitem{Cembranos:2012nj}
J.~A.~R. Cembranos, V.~Gammaldi, and A.~L. Maroto, {\it {Possible dark matter
  origin of the gamma ray emission from the galactic center observed by HESS}},
   {\em Phys. Rev.} {\bf D86} (2012) 103506,
  [\href{http://arxiv.org/abs/1204.0655}{{\tt arXiv:1204.0655}}].

\bibitem{Cembranos:2013fya}
J.~A.~R. Cembranos, V.~Gammaldi, and A.~L. Maroto, {\it {Spectral Study of the
  HESS J1745-290 Gamma-Ray Source as Dark Matter Signal}},  {\em JCAP} {\bf
  1304} (2013) 051, [\href{http://arxiv.org/abs/1302.6871}{{\tt
  arXiv:1302.6871}}].

\bibitem{Cembranos:2014yqa}
J.~A.~R. Cembranos, V.~Gammaldi, and A.~L. Maroto, {\it {Neutrino fluxes from
  Dark Matter in the HESS J1745-290 source at the Galactic Center}},  {\em
  Phys. Rev.} {\bf D90} (2014), no.~4 043004,
  [\href{http://arxiv.org/abs/1403.6018}{{\tt arXiv:1403.6018}}].

\bibitem{Cembranos:2014wza}
J.~A.~R. Cembranos, V.~Gammaldi, and A.~L. Maroto, {\it {Antiproton signatures
  from astrophysical and dark matter sources at the galactic center}},  {\em
  JCAP} {\bf 1503} (2015), no.~03 041,
  [\href{http://arxiv.org/abs/1410.6689}{{\tt arXiv:1410.6689}}].

\bibitem{Bull:2018lat}
A.~Weltman et~al., {\it {Fundamental Physics with the Square Kilometre Array}},
   \href{http://arxiv.org/abs/1810.02680}{{\tt arXiv:1810.02680}}.

\bibitem{Cembranos:2019noa}
J.~A.~R. Cembranos, {\'A}.~De~La Cruz-Dombriz, V.~Gammaldi, and
  M.~M{\'e}ndez-Isla, {\it {SKA-Phase 1 sensitivity for synchrotron radio
  emission from multi-TeV Dark Matter candidates}},
  \href{http://arxiv.org/abs/1905.11154}{{\tt arXiv:1905.11154}}.

\bibitem{Cembranos:2019amc}
J.~A.~R. Cembranos, A.~de~la Cruz-Dombriz, P.~K.~S. Dunsby, and M.~Mendez-Isla,
  {\it {Analysis of branon dark matter and extra-dimensional models with
  AMS-02}},  {\em Phys. Lett.} {\bf B790} (2019) 345--353,
  [\href{http://arxiv.org/abs/1709.09819}{{\tt arXiv:1709.09819}}].

\bibitem{Rizzo:2001sd}
T.~G. Rizzo, {\it {Probes of universal extra dimensions at colliders}},  {\em
  Phys. Rev.} {\bf D64} (2001) 095010,
  [\href{http://arxiv.org/abs/hep-ph/0106336}{{\tt hep-ph/0106336}}].

\bibitem{Cheng:2002ej}
H.-C. Cheng, J.~L. Feng, and K.~T. Matchev, {\it {Kaluza-Klein dark matter}},
  {\em Phys. Rev. Lett.} {\bf 89} (2002) 211301,
  [\href{http://arxiv.org/abs/hep-ph/0207125}{{\tt hep-ph/0207125}}].

\bibitem{Servant:2002aq}
G.~Servant and T.~M.~P. Tait, {\it {Is the lightest Kaluza-Klein particle a
  viable dark matter candidate?}},  {\em Nucl. Phys.} {\bf B650} (2003)
  391--419, [\href{http://arxiv.org/abs/hep-ph/0206071}{{\tt hep-ph/0206071}}].

\bibitem{Kong:2005hn}
K.~Kong and K.~T. Matchev, {\it {Precise calculation of the relic density of
  Kaluza-Klein dark matter in universal extra dimensions}},  {\em JHEP} {\bf
  01} (2006) 038, [\href{http://arxiv.org/abs/hep-ph/0509119}{{\tt
  hep-ph/0509119}}].

\bibitem{Cembranos:2006gt}
J.~A.~R. Cembranos, J.~L. Feng, and L.~E. Strigari, {\it {Exotic Collider
  Signals from the Complete Phase Diagram of Minimal Universal Extra
  Dimensions}},  {\em Phys. Rev.} {\bf D75} (2007) 036004,
  [\href{http://arxiv.org/abs/hep-ph/0612157}{{\tt hep-ph/0612157}}].

\bibitem{Cembranos:2007fj}
J.~A.~R. Cembranos, J.~L. Feng, and L.~E. Strigari, {\it {Resolving Cosmic
  Gamma Ray Anomalies with Dark Matter Decaying Now}},  {\em Phys. Rev. Lett.}
  {\bf 99} (2007) 191301, [\href{http://arxiv.org/abs/0704.1658}{{\tt
  arXiv:0704.1658}}].

\bibitem{Masip:1999mk}
M.~Masip and A.~Pomarol, {\it {Effects of SM Kaluza-Klein excitations on
  electroweak observables}},  {\em Phys. Rev.} {\bf D60} (1999) 096005,
  [\href{http://arxiv.org/abs/hep-ph/9902467}{{\tt hep-ph/9902467}}].

\bibitem{Rizzo:1999br}
T.~G. Rizzo and J.~D. Wells, {\it {Electroweak precision measurements and
  collider probes of the standard model with large extra dimensions}},  {\em
  Phys. Rev.} {\bf D61} (2000) 016007,
  [\href{http://arxiv.org/abs/hep-ph/9906234}{{\tt hep-ph/9906234}}].

\bibitem{delAguila:2003bh}
F.~del Aguila, M.~Perez-Victoria, and J.~Santiago, {\it {Bulk fields with
  general brane kinetic terms}},  {\em JHEP} {\bf 02} (2003) 051,
  [\href{http://arxiv.org/abs/hep-th/0302023}{{\tt hep-th/0302023}}].

\bibitem{Gorbunov:2005zk}
D.~Gorbunov, K.~Koyama, and S.~Sibiryakov, {\it {More on ghosts in DGP model}},
   {\em Phys. Rev.} {\bf D73} (2006) 044016,
  [\href{http://arxiv.org/abs/hep-th/0512097}{{\tt hep-th/0512097}}].

\bibitem{Fang:2008kc}
W.~Fang, S.~Wang, W.~Hu, Z.~Haiman, L.~Hui, and M.~May, {\it {Challenges to the
  DGP Model from Horizon-Scale Growth and Geometry}},  {\em Phys. Rev.} {\bf
  D78} (2008) 103509, [\href{http://arxiv.org/abs/0808.2208}{{\tt
  arXiv:0808.2208}}].

\bibitem{Calcagni:2017sdq}
G.~Calcagni, {\em {Classical and Quantum Cosmology}}.
\newblock Graduate Texts in Physics. Springer, 2017.

\bibitem{Pais:1950za}
A.~Pais and G.~E. Uhlenbeck, {\it On field theories with non-localized action},
   {\em Phys. Rev.} {\bf 79} (Jul, 1950) 145--165.

\bibitem{Calcagni:2018lyd}
G.~Calcagni, L.~Modesto, and G.~Nardelli, {\it {Initial conditions and degrees
  of freedom of non-local gravity}},  {\em JHEP} {\bf 05} (2018) 087,
  [\href{http://arxiv.org/abs/1803.00561}{{\tt arXiv:1803.00561}}]. [Erratum:
  JHEP05,095(2019)].

\bibitem{Calcagni:2018gke}
G.~Calcagni, L.~Modesto, and G.~Nardelli, {\it {Non-perturbative spectrum of
  non-local gravity}},  {\em Phys. Lett.} {\bf B795} (2019) 391--397,
  [\href{http://arxiv.org/abs/1803.07848}{{\tt arXiv:1803.07848}}].

\bibitem{Wataghin:1934ann}
G.~Wataghin, {\it {Bemerkung {\"u}ber die Selbstenergie der Elektronen}},  {\em
  Z. Phys.} {\bf 88} (1934), no.~1-2 92--98.

\bibitem{Yuk49}
H.~Yukawa, {\it On the radius of the elementary particle},  {\em Phys. Rev.}
  {\bf 76} (Jul, 1949) 300--301.

\bibitem{Yukawa:1950eq}
H.~Yukawa, {\it Quantum theory of non-local fields. part i. free fields},  {\em
  Phys. Rev.} {\bf 77} (Jan, 1950) 219--226.

\bibitem{Pau33}
W.~Pauli, {\em Die allgemeinen Prinzipien der Wellenmechanik}.
\newblock Handb. Phys. XXIV/1 83,
  https://www.springer.com/gp/book/9783642525650, 1933.

\bibitem{Efimov:1967dpd}
G.~V. Efimov, {\it {Analytic properties of Euclidean amplitudes}},  {\em Sov.
  J. Nucl. Phys.} {\bf 4} (1967), no.~2 309--315. [Yad. Fiz.4,no.2,432(1966)].

\bibitem{Efimov:1967pjn}
G.~V. Efimov, {\it {Non-local quantum theory of the scalar field}},  {\em
  Commun. Math. Phys.} {\bf 5} (1967), no.~1 42--56.

\bibitem{Alebastrov:1973vw}
V.~A. Alebastrov and G.~V. Efimov, {\it {A proof of the unitarity of S-matrix
  in a nonlocal quantum field theory}},  {\em Commun. Math. Phys.} {\bf 31}
  (1973), no.~1 1--24.

\bibitem{Alebastrov:1973np}
V.~A. Alebastrov and G.~V. Efimov, {\it {Causality in quantum field theory with
  nonlocal interaction}},  {\em Commun. Math. Phys.} {\bf 38} (1974), no.~1
  11--28.

\bibitem{Efimov:1974cmh}
G.~V. Efimov, {\it {Quantization of non-local field theory}},  {\em Int. J.
  Theor. Phys.} {\bf 10} (1974), no.~1 19--37.

\bibitem{Efimov:1977bpa}
G.~V. Efimov, {\em {Nonlocal Interactions of Quantized Fields [in Russian]}}.
\newblock Nauka, Moscow, 1977.

\bibitem{Krasnikov:1987yj}
N.~V. Krasnikov, {\it {NONLOCAL GAUGE THEORIES}},  {\em Theor. Math. Phys.}
  {\bf 73} (1987) 1184--1190. [Teor. Mat. Fiz.73,235(1987)].

\bibitem{Kuzmin:1989sp}
{\relax Yu}.~V. Kuzmin, {\it {THE CONVERGENT NONLOCAL GRAVITATION. (IN
  RUSSIAN)}},  {\em Sov. J. Nucl. Phys.} {\bf 50} (1989) 1011--1014. [Yad.
  Fiz.50,1630(1989)].

\bibitem{Tseytlin:1995uq}
A.~A. Tseytlin, {\it {On singularities of spherically symmetric backgrounds in
  string theory}},  {\em Phys. Lett.} {\bf B363} (1995) 223--229,
  [\href{http://arxiv.org/abs/hep-th/9509050}{{\tt hep-th/9509050}}].

\bibitem{Tomboulis:1997gg}
E.~T. Tomboulis, {\it {Superrenormalizable gauge and gravitational theories}},
  \href{http://arxiv.org/abs/hep-th/9702146}{{\tt hep-th/9702146}}.

\bibitem{Siegel:2003vt}
W.~Siegel, {\it {Stringy gravity at short distances}},
  \href{http://arxiv.org/abs/hep-th/0309093}{{\tt hep-th/0309093}}.

\bibitem{Biswas:2005qr}
T.~Biswas, A.~Mazumdar, and W.~Siegel, {\it {Bouncing universes in
  string-inspired gravity}},  {\em JCAP} {\bf 0603} (2006) 009,
  [\href{http://arxiv.org/abs/hep-th/0508194}{{\tt hep-th/0508194}}].

\bibitem{Khoury:2006fg}
J.~Khoury, {\it {Fading gravity and self-inflation}},  {\em Phys. Rev.} {\bf
  D76} (2007) 123513, [\href{http://arxiv.org/abs/hep-th/0612052}{{\tt
  hep-th/0612052}}].

\bibitem{Calcagni:2010ab}
G.~Calcagni and G.~Nardelli, {\it {Non-local gravity and the diffusion
  equation}},  {\em Phys. Rev.} {\bf D82} (2010) 123518,
  [\href{http://arxiv.org/abs/1004.5144}{{\tt arXiv:1004.5144}}].

\bibitem{Biswas:2010zk}
T.~Biswas, T.~Koivisto, and A.~Mazumdar, {\it {Towards a resolution of the
  cosmological singularity in non-local higher derivative theories of
  gravity}},  {\em JCAP} {\bf 1011} (2010) 008,
  [\href{http://arxiv.org/abs/1005.0590}{{\tt arXiv:1005.0590}}].

\bibitem{Moffat:2010bh}
J.~W. Moffat, {\it {Ultraviolet Complete Quantum Gravity}},  {\em Eur. Phys. J.
  Plus} {\bf 126} (2011) 43, [\href{http://arxiv.org/abs/1008.2482}{{\tt
  arXiv:1008.2482}}].

\bibitem{Modesto:2010uh}
L.~Modesto, J.~W. Moffat, and P.~Nicolini, {\it {Black holes in an ultraviolet
  complete quantum gravity}},  {\em Phys. Lett.} {\bf B695} (2011) 397--400,
  [\href{http://arxiv.org/abs/1010.0680}{{\tt arXiv:1010.0680}}].

\bibitem{Modesto:2011kw}
L.~Modesto, {\it {Super-renormalizable Quantum Gravity}},  {\em Phys. Rev.}
  {\bf D86} (2012) 044005, [\href{http://arxiv.org/abs/1107.2403}{{\tt
  arXiv:1107.2403}}].

\bibitem{Biswas:2011ar}
T.~Biswas, E.~Gerwick, T.~Koivisto, and A.~Mazumdar, {\it {Towards singularity
  and ghost free theories of gravity}},  {\em Phys. Rev. Lett.} {\bf 108}
  (2012) 031101, [\href{http://arxiv.org/abs/1110.5249}{{\tt
  arXiv:1110.5249}}].

\bibitem{Alexander:2012aw}
S.~Alexander, A.~Marciano, and L.~Modesto, {\it {The Hidden Quantum Groups
  Symmetry of Super-renormalizable Gravity}},  {\em Phys. Rev.} {\bf D85}
  (2012) 124030, [\href{http://arxiv.org/abs/1202.1824}{{\tt
  arXiv:1202.1824}}].

\bibitem{Modesto:2012ys}
L.~Modesto, {\it {Super-renormalizable Multidimensional Quantum Gravity}},
  {\em Astron. Rev.} {\bf 8} (2013), no.~2 4--33,
  [\href{http://arxiv.org/abs/1202.3151}{{\tt arXiv:1202.3151}}].

\bibitem{Biswas:2012bp}
T.~Biswas, A.~S. Koshelev, A.~Mazumdar, and S.~{\relax Yu}. Vernov, {\it
  {Stable bounce and inflation in non-local higher derivative cosmology}},
  {\em JCAP} {\bf 1208} (2012) 024, [\href{http://arxiv.org/abs/1206.6374}{{\tt
  arXiv:1206.6374}}].

\bibitem{Briscese:2012ys}
F.~Briscese, A.~Marcian{\`o}, L.~Modesto, and E.~N. Saridakis, {\it {Inflation
  in (Super-)renormalizable Gravity}},  {\em Phys. Rev.} {\bf D87} (2013),
  no.~8 083507, [\href{http://arxiv.org/abs/1212.3611}{{\tt arXiv:1212.3611}}].

\bibitem{Modesto:2013ioa}
L.~Modesto, {\it {Super-renormalizable Gravity}},  in {\em {Proceedings, 13th
  Marcel Grossmann Meeting on Recent Developments in Theoretical and
  Experimental General Relativity, Astrophysics, and Relativistic Field
  Theories (MG13): Stockholm, Sweden, July 1-7, 2012}}, pp.~1128--1130, 2015.
\newblock \href{http://arxiv.org/abs/1302.6348}{{\tt arXiv:1302.6348}}.

\bibitem{Calcagni:2013vra}
G.~Calcagni, L.~Modesto, and P.~Nicolini, {\it {Super-accelerating bouncing
  cosmology in asymptotically-free non-local gravity}},  {\em Eur. Phys. J.}
  {\bf C74} (2014), no.~8 2999, [\href{http://arxiv.org/abs/1306.5332}{{\tt
  arXiv:1306.5332}}].

\bibitem{Modesto:2013jea}
L.~Modesto and S.~Tsujikawa, {\it {Non-local massive gravity}},  {\em Phys.
  Lett.} {\bf B727} (2013) 48--56, [\href{http://arxiv.org/abs/1307.6968}{{\tt
  arXiv:1307.6968}}].

\bibitem{Briscese:2013lna}
F.~Briscese, L.~Modesto, and S.~Tsujikawa, {\it {Super-renormalizable or finite
  completion of the Starobinsky theory}},  {\em Phys. Rev.} {\bf D89} (2014),
  no.~2 024029, [\href{http://arxiv.org/abs/1308.1413}{{\tt arXiv:1308.1413}}].

\bibitem{Biswas:2013cha}
T.~Biswas, A.~Conroy, A.~S. Koshelev, and A.~Mazumdar, {\it {Generalized
  ghost-free quadratic curvature gravity}},  {\em Class. Quant. Grav.} {\bf 31}
  (2014) 015022, [\href{http://arxiv.org/abs/1308.2319}{{\tt
  arXiv:1308.2319}}]. [Erratum: Class. Quant. Grav.31,159501(2014)].

\bibitem{Modesto:2014xta}
L.~Modesto, {\it {Multidimensional finite quantum gravity}},
  \href{http://arxiv.org/abs/1402.6795}{{\tt arXiv:1402.6795}}.

\bibitem{Calcagni:2014vxa}
G.~Calcagni, L.~Modesto, and P.~Nicolini, {\it {Super-accelerating bouncing
  cosmology in asymptotically-free non-local gravity}},  {\em Eur. Phys. J.}
  {\bf C74} (2014), no.~8 2999, [\href{http://arxiv.org/abs/1306.5332}{{\tt
  arXiv:1306.5332}}].

\bibitem{Biswas:2014yia}
T.~Biswas and N.~Okada, {\it {Towards LHC physics with nonlocal Standard
  Model}},  {\em Nucl. Phys.} {\bf B898} (2015) 113--131,
  [\href{http://arxiv.org/abs/1407.3331}{{\tt arXiv:1407.3331}}].

\bibitem{Modesto:2014lga}
L.~Modesto and L.~Rachwal, {\it {Super-renormalizable and finite gravitational
  theories}},  {\em Nucl. Phys.} {\bf B889} (2014) 228--248,
  [\href{http://arxiv.org/abs/1407.8036}{{\tt arXiv:1407.8036}}].

\bibitem{Conroy:2014dja}
A.~Conroy, A.~S. Koshelev, and A.~Mazumdar, {\it {Geodesic completeness and
  homogeneity condition for cosmic inflation}},  {\em Phys. Rev.} {\bf D90}
  (2014), no.~12 123525, [\href{http://arxiv.org/abs/1408.6205}{{\tt
  arXiv:1408.6205}}].

\bibitem{Talaganis:2014ida}
S.~Talaganis, T.~Biswas, and A.~Mazumdar, {\it {Towards understanding the
  ultraviolet behavior of quantum loops in infinite-derivative theories of
  gravity}},  {\em Class. Quant. Grav.} {\bf 32} (2015), no.~21 215017,
  [\href{http://arxiv.org/abs/1412.3467}{{\tt arXiv:1412.3467}}].

\bibitem{Modesto:2015lna}
L.~Modesto and L.~Rachwal, {\it {Universally finite gravitational and gauge
  theories}},  {\em Nucl. Phys.} {\bf B900} (2015) 147--169,
  [\href{http://arxiv.org/abs/1503.00261}{{\tt arXiv:1503.00261}}].

\bibitem{Dona:2015tra}
P.~Don{\`a}, S.~Giaccari, L.~Modesto, L.~Rachwal, and Y.~Zhu, {\it {Scattering
  amplitudes in super-renormalizable gravity}},  {\em JHEP} {\bf 08} (2015)
  038, [\href{http://arxiv.org/abs/1506.04589}{{\tt arXiv:1506.04589}}].

\bibitem{Modesto:2015foa}
L.~Modesto, M.~Piva, and L.~Rachwal, {\it {Finite quantum gauge theories}},
  {\em Phys. Rev.} {\bf D94} (2016), no.~2 025021,
  [\href{http://arxiv.org/abs/1506.06227}{{\tt arXiv:1506.06227}}].

\bibitem{Li:2015bqa}
Y.-D. Li, L.~Modesto, and L.~Rachwal, {\it {Exact solutions and spacetime
  singularities in nonlocal gravity}},  {\em JHEP} {\bf 12} (2015) 173,
  [\href{http://arxiv.org/abs/1506.08619}{{\tt arXiv:1506.08619}}].

\bibitem{Tomboulis:2015esa}
E.~T. Tomboulis, {\it {Renormalization and unitarity in higher derivative and
  nonlocal gravity theories}},  {\em Mod. Phys. Lett.} {\bf A30} (2015),
  no.~03n04 1540005.

\bibitem{Talaganis:2016ovm}
S.~Talaganis and A.~Mazumdar, {\it {High-Energy Scatterings in
  Infinite-Derivative Field Theory and Ghost-Free Gravity}},  {\em Class.
  Quant. Grav.} {\bf 33} (2016), no.~14 145005,
  [\href{http://arxiv.org/abs/1603.03440}{{\tt arXiv:1603.03440}}].

\bibitem{Edholm:2016hbt}
J.~Edholm, A.~S. Koshelev, and A.~Mazumdar, {\it {Behavior of the Newtonian
  potential for ghost-free gravity and singularity-free gravity}},  {\em Phys.
  Rev.} {\bf D94} (2016), no.~10 104033,
  [\href{http://arxiv.org/abs/1604.01989}{{\tt arXiv:1604.01989}}].

\bibitem{Giaccari:2016kzy}
S.~Giaccari and L.~Modesto, {\it {Nonlocal supergravity}},  {\em Phys. Rev.}
  {\bf D96} (2017), no.~6 066021, [\href{http://arxiv.org/abs/1605.03906}{{\tt
  arXiv:1605.03906}}].

\bibitem{Modesto:2016max}
L.~Modesto and L.~Rachwal, {\it {Finite Conformal Quantum Gravity and
  Nonsingular Spacetimes}},  \href{http://arxiv.org/abs/1605.04173}{{\tt
  arXiv:1605.04173}}.

\bibitem{Biswas:2016egy}
T.~Biswas, A.~S. Koshelev, and A.~Mazumdar, {\it {Consistent higher derivative
  gravitational theories with stable de Sitter and anti--de Sitter
  backgrounds}},  {\em Phys. Rev.} {\bf D95} (2017), no.~4 043533,
  [\href{http://arxiv.org/abs/1606.01250}{{\tt arXiv:1606.01250}}].

\bibitem{Koshelev:2017ebj}
A.~S. Koshelev, K.~Sravan~Kumar, L.~Modesto, and L.~Rachwal, {\it {Finite
  quantum gravity in dS and AdS spacetimes}},  {\em Phys. Rev.} {\bf D98}
  (2018), no.~4 046007, [\href{http://arxiv.org/abs/1710.07759}{{\tt
  arXiv:1710.07759}}].

\bibitem{Calcagni:2017sov}
G.~Calcagni and L.~Modesto, {\it {Stability of Schwarzschild singularity in
  non-local gravity}},  {\em Phys. Lett.} {\bf B773} (2017) 596--600,
  [\href{http://arxiv.org/abs/1707.01119}{{\tt arXiv:1707.01119}}].

\bibitem{Cornell:2017irh}
A.~S. Cornell, G.~Harmsen, G.~Lambiase, and A.~Mazumdar, {\it {Rotating metric
  in nonsingular infinite derivative theories of gravity}},  {\em Phys. Rev.}
  {\bf D97} (2018), no.~10 104006, [\href{http://arxiv.org/abs/1710.02162}{{\tt
  arXiv:1710.02162}}].

\bibitem{Edholm:2018wjh}
J.~Edholm, {\it {Revealing infinite derivative gravity's true potential: The
  weak-field limit around de Sitter backgrounds}},  {\em Phys. Rev.} {\bf D97}
  (2018), no.~6 064011, [\href{http://arxiv.org/abs/1801.00834}{{\tt
  arXiv:1801.00834}}].

\bibitem{Buoninfante:2018xiw}
L.~Buoninfante, A.~S. Koshelev, G.~Lambiase, and A.~Mazumdar, {\it {Classical
  properties of non-local, ghost- and singularity-free gravity}},  {\em JCAP}
  {\bf 1809} (2018), no.~09 034, [\href{http://arxiv.org/abs/1802.00399}{{\tt
  arXiv:1802.00399}}].

\bibitem{Koshelev:2018hpt}
A.~S. Koshelev, J.~Marto, and A.~Mazumdar, {\it {Schwarzschild
  $1/r$-singularity is not permissible in ghost free quadratic curvature
  infinite derivative gravity}},  {\em Phys. Rev.} {\bf D98} (2018), no.~6
  064023, [\href{http://arxiv.org/abs/1803.00309}{{\tt arXiv:1803.00309}}].

\bibitem{Koshelev:2018rau}
A.~S. Koshelev, J.~Marto, and A.~Mazumdar, {\it {Towards resolution of
  anisotropic cosmological singularity in infinite derivative gravity}},  {\em
  JCAP} {\bf 1902} (2019) 020, [\href{http://arxiv.org/abs/1803.07072}{{\tt
  arXiv:1803.07072}}].

\bibitem{Calcagni:2018pro}
G.~Calcagni, L.~Modesto, and Y.~S. Myung, {\it {Black-hole stability in
  non-local gravity}},  {\em Phys. Lett.} {\bf B783} (2018) 19--23,
  [\href{http://arxiv.org/abs/1803.08388}{{\tt arXiv:1803.08388}}].

\bibitem{Giaccari:2018nzr}
S.~Giaccari and L.~Modesto, {\it {Causality in Nonlocal Gravity}},
  \href{http://arxiv.org/abs/1803.08748}{{\tt arXiv:1803.08748}}.

\bibitem{Briscese:2018oyx}
F.~Briscese and L.~Modesto, {\it {Cutkosky rules and perturbative unitarity in
  Euclidean nonlocal quantum field theories}},  {\em Phys. Rev.} {\bf D99}
  (2019), no.~10 104043, [\href{http://arxiv.org/abs/1803.08827}{{\tt
  arXiv:1803.08827}}].

\bibitem{Buoninfante:2018rlq}
L.~Buoninfante, A.~S. Koshelev, G.~Lambiase, J.~Marto, and A.~Mazumdar, {\it
  {Conformally-flat, non-singular static metric in infinite derivative
  gravity}},  {\em JCAP} {\bf 1806} (2018), no.~06 014,
  [\href{http://arxiv.org/abs/1804.08195}{{\tt arXiv:1804.08195}}].

\bibitem{Buoninfante:2018stt}
L.~Buoninfante, G.~Harmsen, S.~Maheshwari, and A.~Mazumdar, {\it {Nonsingular
  metric for an electrically charged point-source in ghost-free infinite
  derivative gravity}},  {\em Phys. Rev.} {\bf D98} (2018), no.~8 084009,
  [\href{http://arxiv.org/abs/1804.09624}{{\tt arXiv:1804.09624}}].

\bibitem{Buoninfante:2018mre}
L.~Buoninfante, G.~Lambiase, and A.~Mazumdar, {\it {Ghost-free infinite
  derivative quantum field theory}},  {\em Nucl. Phys.} {\bf B944} (2019)
  114646, [\href{http://arxiv.org/abs/1805.03559}{{\tt arXiv:1805.03559}}].

\bibitem{Briscese:2018bny}
F.~Briscese and L.~Modesto, {\it {Nonlinear stability of Minkowski spacetime in
  Nonlocal Gravity}},  {\em JCAP} {\bf 1907} (2019), no.~07 009,
  [\href{http://arxiv.org/abs/1811.05117}{{\tt arXiv:1811.05117}}].

\bibitem{Buoninfante:2018lnh}
L.~Buoninfante, G.~Lambiase, and M.~Yamaguchi, {\it {Nonlocal generalization of
  Galilean theories and gravity}},  {\em Phys. Rev.} {\bf D100} (2019), no.~2
  026019, [\href{http://arxiv.org/abs/1812.10105}{{\tt arXiv:1812.10105}}].

\bibitem{Briscese:2019rii}
F.~Briscese, G.~Calcagni, and L.~Modesto, {\it {Nonlinear stability in nonlocal
  gravity}},  {\em Phys. Rev.} {\bf D99} (2019), no.~8 084041,
  [\href{http://arxiv.org/abs/1901.03267}{{\tt arXiv:1901.03267}}].

\bibitem{Stelle:1977ry}
K.~S. Stelle, {\it {Classical Gravity with Higher Derivatives}},  {\em Gen.
  Rel. Grav.} {\bf 9} (1978) 353--371.

\bibitem{Asorey:1996hz}
M.~Asorey, J.~L. Lopez, and I.~L. Shapiro, {\it {Some remarks on high
  derivative quantum gravity}},  {\em Int. J. Mod. Phys.} {\bf A12} (1997)
  5711--5734, [\href{http://arxiv.org/abs/hep-th/9610006}{{\tt
  hep-th/9610006}}].

\bibitem{Salles:2014rua}
F.~d.~O. Salles and I.~L. Shapiro, {\it {Do we have unitary and
  (super)renormalizable quantum gravity below the Planck scale?}},  {\em Phys.
  Rev.} {\bf D89} (2014), no.~8 084054,
  [\href{http://arxiv.org/abs/1401.4583}{{\tt arXiv:1401.4583}}]. [Erratum:
  Phys. Rev.D90,no.12,129903(2014)].

\bibitem{Monitor:2017mdv}
{\bf LIGO Scientific, Virgo, Fermi-GBM, INTEGRAL} Collaboration, B.~P. Abbott
  et~al., {\it {Gravitational Waves and Gamma-rays from a Binary Neutron Star
  Merger: GW170817 and GRB 170817A}},  {\em Astrophys. J.} {\bf 848} (2017),
  no.~2 L13, [\href{http://arxiv.org/abs/1710.05834}{{\tt arXiv:1710.05834}}].

\bibitem{Akrami:2018vks}
{\bf Planck} Collaboration, Y.~Akrami et~al., {\it {Planck 2018 results. I.
  Overview and the cosmological legacy of Planck}},
  \href{http://arxiv.org/abs/1807.06205}{{\tt arXiv:1807.06205}}.

\bibitem{Belgacem:2019pkk}
{\bf LISA Cosmology Working Group} Collaboration, E.~Belgacem et~al., {\it
  {Testing modified gravity at cosmological distances with LISA standard
  sirens}},  {\em JCAP} {\bf 1907} (2019), no.~07 024,
  [\href{http://arxiv.org/abs/1906.01593}{{\tt arXiv:1906.01593}}].

\bibitem{Calcagni:2019kzo}
G.~Calcagni, S.~Kuroyanagi, S.~Marsat, M.~Sakellariadou, N.~Tamanini, and
  G.~Tasinato, {\it {Gravitational-wave luminosity distance in quantum
  gravity}},  {\em Phys. Lett.} {\bf B798} (2019) 135000,
  [\href{http://arxiv.org/abs/1904.00384}{{\tt arXiv:1904.00384}}].

\bibitem{Calcagni:2019ngc}
G.~Calcagni, S.~Kuroyanagi, S.~Marsat, M.~Sakellariadou, N.~Tamanini, and
  G.~Tasinato, {\it {Quantum gravity and gravitational-wave astronomy}},  {\em
  JCAP} {\bf 1910} (2019), no.~10 012,
  [\href{http://arxiv.org/abs/1907.02489}{{\tt arXiv:1907.02489}}].

\bibitem{Calcagni:2020tvw}
G.~Calcagni and S.~Kuroyanagi, {\it {Stochastic gravitational-wave background
  in quantum gravity}},  \href{http://arxiv.org/abs/2012.00170}{{\tt
  arXiv:2012.00170}}.

\bibitem{Maggiore:1900zz}
M.~Maggiore, {\em {Gravitational Waves. Vol. 1: Theory and Experiments}}.
\newblock Oxford Master Series in Physics. Oxford University Press, 2007.

\bibitem{Kostelecky:1990mi}
V.~A. Kostelecky and S.~Samuel, {\it {Collective Physics in the Closed Bosonic
  String}},  {\em Phys. Rev.} {\bf D42} (1990) 1289--1292.

\bibitem{Dalal:2006qt}
N.~Dalal, D.~E. Holz, S.~A. Hughes, and B.~Jain, {\it {Short grb and binary
  black hole standard sirens as a probe of dark energy}},  {\em Phys. Rev.}
  {\bf D74} (2006) 063006, [\href{http://arxiv.org/abs/astro-ph/0601275}{{\tt
  astro-ph/0601275}}].

\bibitem{Nissanke:2009kt}
S.~Nissanke, D.~E. Holz, S.~A. Hughes, N.~Dalal, and J.~L. Sievers, {\it
  {Exploring short gamma-ray bursts as gravitational-wave standard sirens}},
  {\em Astrophys. J.} {\bf 725} (2010) 496--514,
  [\href{http://arxiv.org/abs/0904.1017}{{\tt arXiv:0904.1017}}].

\bibitem{Camera:2013xfa}
S.~Camera and A.~Nishizawa, {\it {Beyond Concordance Cosmology with
  Magnification of Gravitational-Wave Standard Sirens}},  {\em Phys. Rev.
  Lett.} {\bf 110} (2013), no.~15 151103,
  [\href{http://arxiv.org/abs/1303.5446}{{\tt arXiv:1303.5446}}].

\bibitem{Tamanini:2016zlh}
N.~Tamanini, C.~Caprini, E.~Barausse, A.~Sesana, A.~Klein, and A.~Petiteau,
  {\it {Science with the space-based interferometer eLISA. III: Probing the
  expansion of the Universe using gravitational wave standard sirens}},  {\em
  JCAP} {\bf 1604} (2016), no.~04 002,
  [\href{http://arxiv.org/abs/1601.07112}{{\tt arXiv:1601.07112}}].

\bibitem{Frolov:1979tu}
V.~P. Frolov and G.~A. Vilkovisky, {\it {QUANTUM GRAVITY REMOVES CLASSICAL
  SINGULARITIES AND SHORTENS THE LIFE OF BLACK HOLES}},  in {\em {The Second
  Marcel Grossmann Meeting on the Recent Developments of General Relativity (In
  Honor of Albert Einstein) Trieste, Italy, July 5-11, 1979}}, p.~0455, 1979.

\bibitem{Barvinsky:1985an}
A.~O. Barvinsky and G.~A. Vilkovisky, {\it {The Generalized Schwinger-Dewitt
  Technique in Gauge Theories and Quantum Gravity}},  {\em Phys. Rept.} {\bf
  119} (1985) 1--74.

\bibitem{Shapiro:2008sf}
I.~L. Shapiro, {\it {Effective Action of Vacuum: Semiclassical Approach}},
  {\em Class. Quant. Grav.} {\bf 25} (2008) 103001,
  [\href{http://arxiv.org/abs/0801.0216}{{\tt arXiv:0801.0216}}].

\bibitem{Barvinsky:1987uw}
A.~O. Barvinsky and G.~A. Vilkovisky, {\it {Beyond the Schwinger-Dewitt
  Technique: Converting Loops Into Trees and In-In Currents}},  {\em Nucl.
  Phys.} {\bf B282} (1987) 163--188.

\bibitem{Barvinsky:1990up}
A.~O. Barvinsky and G.~A. Vilkovisky, {\it {Covariant perturbation theory. 2:
  Second order in the curvature. General algorithms}},  {\em Nucl. Phys.} {\bf
  B333} (1990) 471--511.

\bibitem{Barvinsky:1994hw}
A.~O. Barvinsky, {\relax Yu}.~V. Gusev, G.~A. Vilkovisky, and V.~V. Zhytnikov,
  {\it {The Basis of nonlocal curvature invariants in quantum gravity theory.
  (Third order.)}},  {\em J. Math. Phys.} {\bf 35} (1994) 3525--3542,
  [\href{http://arxiv.org/abs/gr-qc/9404061}{{\tt gr-qc/9404061}}].

\bibitem{Barvinsky:1994ic}
A.~O. Barvinsky, {\relax Yu}.~V. Gusev, G.~A. Vilkovisky, and V.~V. Zhytnikov,
  {\it {Asymptotic behaviors of the heat kernel in covariant perturbation
  theory}},  {\em J. Math. Phys.} {\bf 35} (1994) 3543--3559,
  [\href{http://arxiv.org/abs/gr-qc/9404063}{{\tt gr-qc/9404063}}].

\bibitem{Gorbar:2002pw}
E.~V. Gorbar and I.~L. Shapiro, {\it {Renormalization group and decoupling in
  curved space}},  {\em JHEP} {\bf 02} (2003) 021,
  [\href{http://arxiv.org/abs/hep-ph/0210388}{{\tt hep-ph/0210388}}].

\bibitem{Gorbar:2003yt}
E.~V. Gorbar and I.~L. Shapiro, {\it {Renormalization group and decoupling in
  curved space. 2. The Standard model and beyond}},  {\em JHEP} {\bf 06} (2003)
  004, [\href{http://arxiv.org/abs/hep-ph/0303124}{{\tt hep-ph/0303124}}].

\bibitem{Maggiore:2016fbn}
M.~Maggiore, {\it {Perturbative loop corrections and nonlocal gravity}},  {\em
  Phys. Rev.} {\bf D93} (2016), no.~6 063008,
  [\href{http://arxiv.org/abs/1603.01515}{{\tt arXiv:1603.01515}}].

\bibitem{Belgacem:2017cqo}
E.~Belgacem, Y.~Dirian, S.~Foffa, and M.~Maggiore, {\it {Nonlocal gravity.
  Conceptual aspects and cosmological predictions}},  {\em JCAP} {\bf 1803}
  (2018), no.~03 002, [\href{http://arxiv.org/abs/1712.07066}{{\tt
  arXiv:1712.07066}}].

\bibitem{Barvinsky:2003kg}
A.~O. Barvinsky, {\it {Nonlocal action for long distance modifications of
  gravity theory}},  {\em Phys. Lett.} {\bf B572} (2003) 109--116,
  [\href{http://arxiv.org/abs/hep-th/0304229}{{\tt hep-th/0304229}}].

\bibitem{Foffa:2013sma}
S.~Foffa, M.~Maggiore, and E.~Mitsou, {\it {Apparent ghosts and spurious
  degrees of freedom in non-local theories}},  {\em Phys. Lett.} {\bf B733}
  (2014) 76--83, [\href{http://arxiv.org/abs/1311.3421}{{\tt
  arXiv:1311.3421}}].

\bibitem{Zhang:2016ykx}
Y.-l. Zhang, K.~Koyama, M.~Sasaki, and G.-B. Zhao, {\it {Acausality in Nonlocal
  Gravity Theory}},  {\em JHEP} {\bf 03} (2016) 039,
  [\href{http://arxiv.org/abs/1601.03808}{{\tt arXiv:1601.03808}}].

\bibitem{Deser:2007jk}
S.~Deser and R.~P. Woodard, {\it {Nonlocal Cosmology}},  {\em Phys. Rev. Lett.}
  {\bf 99} (2007) 111301, [\href{http://arxiv.org/abs/0706.2151}{{\tt
  arXiv:0706.2151}}].

\bibitem{Maggiore:2013mea}
M.~Maggiore, {\it {Phantom dark energy from nonlocal infrared modifications of
  general relativity}},  {\em Phys. Rev.} {\bf D89} (2014), no.~4 043008,
  [\href{http://arxiv.org/abs/1307.3898}{{\tt arXiv:1307.3898}}].

\bibitem{Maggiore:2014sia}
M.~Maggiore and M.~Mancarella, {\it {Non-local gravity and dark energy}},  {\em
  Phys.Rev.} {\bf D90} (2014) 023005,
  [\href{http://arxiv.org/abs/1402.0448}{{\tt arXiv:1402.0448}}].

\bibitem{Nojiri:2007uq}
S.~Nojiri and S.~D. Odintsov, {\it {Modified non-local-F(R) gravity as the key
  for the inflation and dark energy}},  {\em Phys. Lett.} {\bf B659} (2008)
  821--826, [\href{http://arxiv.org/abs/0708.0924}{{\tt arXiv:0708.0924}}].

\bibitem{Koivisto:2008xfa}
T.~Koivisto, {\it {Dynamics of Nonlocal Cosmology}},  {\em Phys. Rev.} {\bf
  D77} (2008) 123513, [\href{http://arxiv.org/abs/0803.3399}{{\tt
  arXiv:0803.3399}}].

\bibitem{Koivisto:2008dh}
T.~S. Koivisto, {\it {Newtonian limit of nonlocal cosmology}},  {\em Phys.
  Rev.} {\bf D78} (2008) 123505, [\href{http://arxiv.org/abs/0807.3778}{{\tt
  arXiv:0807.3778}}].

\bibitem{Koshelev:2008ie}
N.~A. Koshelev, {\it {Comments on scalar-tensor representation of nonlocally
  corrected gravity}},  {\em Grav. Cosmol.} {\bf 15} (2009) 220--223,
  [\href{http://arxiv.org/abs/0809.4927}{{\tt arXiv:0809.4927}}].

\bibitem{Deffayet:2009ca}
C.~Deffayet and R.~P. Woodard, {\it {Reconstructing the Distortion Function for
  Nonlocal Cosmology}},  {\em JCAP} {\bf 0908} (2009) 023,
  [\href{http://arxiv.org/abs/0904.0961}{{\tt arXiv:0904.0961}}].

\bibitem{Elizalde:2012ja}
E.~Elizalde, E.~O. Pozdeeva, and S.~{\relax Yu}. Vernov, {\it {Reconstruction
  Procedure in Nonlocal Models}},  {\em Class. Quant. Grav.} {\bf 30} (2013)
  035002, [\href{http://arxiv.org/abs/1209.5957}{{\tt arXiv:1209.5957}}].

\bibitem{Elizalde:2013dlt}
E.~Elizalde, E.~O. Pozdeeva, S.~{\relax Yu}. Vernov, and Y.-l. Zhang, {\it
  {Cosmological Solutions of a Nonlocal Model with a Perfect Fluid}},  {\em
  JCAP} {\bf 1307} (2013) 034, [\href{http://arxiv.org/abs/1302.4330}{{\tt
  arXiv:1302.4330}}].

\bibitem{Deser:2013uya}
S.~Deser and R.~P. Woodard, {\it {Observational Viability and Stability of
  Nonlocal Cosmology}},  {\em JCAP} {\bf 1311} (2013) 036,
  [\href{http://arxiv.org/abs/1307.6639}{{\tt arXiv:1307.6639}}].

\bibitem{Dodelson:2013sma}
S.~Dodelson and S.~Park, {\it {Nonlocal Gravity and Structure in the
  Universe}},  {\em Phys. Rev.} {\bf D90} (2014) 043535,
  [\href{http://arxiv.org/abs/1310.4329}{{\tt arXiv:1310.4329}}]. [Erratum:
  Phys. Rev.D98,no.2,029904(2018)].

\bibitem{Park:2016jym}
S.~Park and A.~Shafieloo, {\it {Growth of perturbations in nonlocal gravity
  with non-$\Lambda$CDM background}},  {\em Phys. Rev.} {\bf D95} (2017), no.~6
  064061, [\href{http://arxiv.org/abs/1608.02541}{{\tt arXiv:1608.02541}}].

\bibitem{Nersisyan:2017mgj}
H.~Nersisyan, A.~F. Cid, and L.~Amendola, {\it {Structure formation in the
  Deser-Woodard nonlocal gravity model: a reappraisal}},  {\em JCAP} {\bf 1704}
  (2017), no.~04 046, [\href{http://arxiv.org/abs/1701.00434}{{\tt
  arXiv:1701.00434}}].

\bibitem{Park:2017zls}
S.~Park, {\it {Revival of the Deser-Woodard nonlocal gravity model: Comparison
  of the original nonlocal form and a localized formulation}},  {\em Phys.
  Rev.} {\bf D97} (2018), no.~4 044006,
  [\href{http://arxiv.org/abs/1711.08759}{{\tt arXiv:1711.08759}}].

\bibitem{Belgacem:2018lbp}
E.~Belgacem, Y.~Dirian, S.~Foffa, and M.~Maggiore, {\it {Modified
  gravitational-wave propagation and standard sirens}},  {\em Phys. Rev.} {\bf
  D98} (2018), no.~2 023510, [\href{http://arxiv.org/abs/1805.08731}{{\tt
  arXiv:1805.08731}}].

\bibitem{Park:2019btx}
S.~Park and R.~P. Woodard, {\it {Exciting the scalar ghost mode through time
  evolution}},  {\em Phys. Rev.} {\bf D99} (2019), no.~2 024014,
  [\href{http://arxiv.org/abs/1809.06841}{{\tt arXiv:1809.06841}}].

\bibitem{Belgacem:2018wtb}
E.~Belgacem, A.~Finke, A.~Frassino, and M.~Maggiore, {\it {Testing nonlocal
  gravity with Lunar Laser Ranging}},  {\em JCAP} {\bf 1902} (2019) 035,
  [\href{http://arxiv.org/abs/1812.11181}{{\tt arXiv:1812.11181}}].

\bibitem{Amendola:2019fhc}
L.~Amendola, Y.~Dirian, H.~Nersisyan, and S.~Park, {\it {Observational
  Constraints in Nonlocal Gravity: the Deser-Woodard Case}},  {\em JCAP} {\bf
  1903} (2019), no.~03 045, [\href{http://arxiv.org/abs/1901.07832}{{\tt
  arXiv:1901.07832}}].

\bibitem{Chen:2019wlu}
C.-Y. Chen, P.~Chen, and S.~Park, {\it {Primordial bouncing cosmology in the
  Deser-Woodard nonlocal gravity}},  {\em Phys. Lett.} {\bf B796} (2019)
  112--116, [\href{http://arxiv.org/abs/1905.04557}{{\tt arXiv:1905.04557}}].

\bibitem{Wetterich:1997bz}
C.~Wetterich, {\it {Effective nonlocal Euclidean gravity}},  {\em Gen. Rel.
  Grav.} {\bf 30} (1998) 159--172,
  [\href{http://arxiv.org/abs/gr-qc/9704052}{{\tt gr-qc/9704052}}].

\bibitem{Foffa:2013vma}
S.~Foffa, M.~Maggiore, and E.~Mitsou, {\it {Cosmological dynamics and dark
  energy from nonlocal infrared modifications of gravity}},  {\em Int. J. Mod.
  Phys.} {\bf A29} (2014) 1450116, [\href{http://arxiv.org/abs/1311.3435}{{\tt
  arXiv:1311.3435}}].

\bibitem{Nesseris:2014mea}
S.~Nesseris and S.~Tsujikawa, {\it {Cosmological perturbations and
  observational constraints on nonlocal massive gravity}},  {\em Phys. Rev.}
  {\bf D90} (2014), no.~2 024070, [\href{http://arxiv.org/abs/1402.4613}{{\tt
  arXiv:1402.4613}}].

\bibitem{Dirian:2014bma}
Y.~Dirian, S.~Foffa, M.~Kunz, M.~Maggiore, and V.~Pettorino, {\it {Non-local
  gravity and comparison with observational datasets}},  {\em JCAP} {\bf 1504}
  (2015), no.~04 044, [\href{http://arxiv.org/abs/1411.7692}{{\tt
  arXiv:1411.7692}}].

\bibitem{Dirian:2016puz}
Y.~Dirian, S.~Foffa, M.~Kunz, M.~Maggiore, and V.~Pettorino, {\it {Non-local
  gravity and comparison with observational datasets. II. Updated results and
  Bayesian model comparison with $\Lambda$CDM}},  {\em JCAP} {\bf 1605} (2016),
  no.~05 068, [\href{http://arxiv.org/abs/1602.03558}{{\tt arXiv:1602.03558}}].

\bibitem{Belgacem:2019lwx}
E.~Belgacem, Y.~Dirian, A.~Finke, S.~Foffa, and M.~Maggiore, {\it {Nonlocal
  gravity and gravitational-wave observations}},  {\em JCAP} {\bf 1911} (2019),
  no.~11 022, [\href{http://arxiv.org/abs/1907.02047}{{\tt arXiv:1907.02047}}].

\bibitem{Jaccard:2013gla}
M.~Jaccard, M.~Maggiore, and E.~Mitsou, {\it {Nonlocal theory of massive
  gravity}},  {\em Phys. Rev.} {\bf D88} (2013), no.~4 044033,
  [\href{http://arxiv.org/abs/1305.3034}{{\tt arXiv:1305.3034}}].

\bibitem{Dirian:2014ara}
Y.~Dirian, S.~Foffa, N.~Khosravi, M.~Kunz, and M.~Maggiore, {\it {Cosmological
  perturbations and structure formation in nonlocal infrared modifications of
  general relativity}},  {\em JCAP} {\bf 1406} (2014) 033,
  [\href{http://arxiv.org/abs/1403.6068}{{\tt arXiv:1403.6068}}].

\bibitem{Barreira:2014kra}
A.~Barreira, B.~Li, W.~A. Hellwing, C.~M. Baugh, and S.~Pascoli, {\it
  {Nonlinear structure formation in Nonlocal Gravity}},  {\em JCAP} {\bf 1409}
  (2014), no.~09 031, [\href{http://arxiv.org/abs/1408.1084}{{\tt
  arXiv:1408.1084}}].

\bibitem{Nersisyan:2016hjh}
H.~Nersisyan, Y.~Akrami, L.~Amendola, T.~S. Koivisto, and J.~Rubio, {\it
  {Dynamical analysis of $R\dfrac{1}{\Box^{2}}R$ cosmology: Impact of initial
  conditions and constraints from supernovae}},  {\em Phys. Rev. D} {\bf 94}
  (2016), no.~4 043531, [\href{http://arxiv.org/abs/1606.04349}{{\tt
  arXiv:1606.04349}}].

\bibitem{Ferreira:2013tqn}
P.~G. Ferreira and A.~L. Maroto, {\it {A few cosmological implications of
  tensor nonlocalities}},  {\em Phys. Rev.} {\bf D88} (2013), no.~12 123502,
  [\href{http://arxiv.org/abs/1310.1238}{{\tt arXiv:1310.1238}}].

\bibitem{Nersisyan:2016jta}
H.~Nersisyan, Y.~Akrami, L.~Amendola, T.~S. Koivisto, J.~Rubio, and A.~R.
  Solomon, {\it {Instabilities in tensorial nonlocal gravity}},  {\em Phys.
  Rev. D} {\bf 95} (2017), no.~4 043539,
  [\href{http://arxiv.org/abs/1610.01799}{{\tt arXiv:1610.01799}}].

\bibitem{Barvinsky:2005db}
A.~O. Barvinsky, {\it {On covariant long-distance modifications of Einstein
  theory and strong coupling problem}},  {\em Phys. Rev.} {\bf D71} (2005)
  084007, [\href{http://arxiv.org/abs/hep-th/0501093}{{\tt hep-th/0501093}}].

\bibitem{Barvinsky:2011hd}
A.~O. Barvinsky, {\it {Dark energy and dark matter from nonlocal ghost-free
  gravity theory}},  {\em Phys. Lett.} {\bf B710} (2012) 12--16,
  [\href{http://arxiv.org/abs/1107.1463}{{\tt arXiv:1107.1463}}].

\bibitem{Barvinsky:2011rk}
A.~O. Barvinsky, {\it {Serendipitous discoveries in nonlocal gravity theory}},
  {\em Phys. Rev.} {\bf D85} (2012) 104018,
  [\href{http://arxiv.org/abs/1112.4340}{{\tt arXiv:1112.4340}}].

\bibitem{Cusin:2015rex}
G.~Cusin, S.~Foffa, M.~Maggiore, and M.~Mancarella, {\it {Nonlocal gravity with
  a Weyl-square term}},  {\em Phys. Rev.} {\bf D93} (2016), no.~4 043006,
  [\href{http://arxiv.org/abs/1512.06373}{{\tt arXiv:1512.06373}}].

\bibitem{Cusin:2016nzi}
G.~Cusin, S.~Foffa, M.~Maggiore, and M.~Mancarella, {\it {Conformal symmetry
  and nonlinear extensions of nonlocal gravity}},  {\em Phys. Rev.} {\bf D93}
  (2016), no.~8 083008, [\href{http://arxiv.org/abs/1602.01078}{{\tt
  arXiv:1602.01078}}].

\bibitem{Deser:2019lmm}
S.~Deser and R.~P. Woodard, {\it {Nonlocal Cosmology II --- Cosmic acceleration
  without fine tuning or dark energy}},  {\em JCAP} {\bf 1906} (2019), no.~06
  034, [\href{http://arxiv.org/abs/1902.08075}{{\tt arXiv:1902.08075}}].

\bibitem{Ding:2019rlp}
J.-C. Ding and J.-B. Deng, {\it {Structure formation in the new Deser-Woodard
  nonlocal gravity model}},  {\em JCAP} {\bf 1912} (2019), no.~12 054,
  [\href{http://arxiv.org/abs/1908.11223}{{\tt arXiv:1908.11223}}].

\bibitem{Calcagni:2021ljs}
G.~Calcagni, {\it {Scalar and gravity quantum field theories with fractional
  operators}},  \href{http://arxiv.org/abs/2102.03363}{{\tt arXiv:2102.03363}}.

\bibitem{eisenhart1997riemannian}
L.~P. Eisenhart, {\em Riemannian geometry}.
\newblock Princeton university press, 1997.

\bibitem{eisenhart2012non}
L.~P. Eisenhart, {\em Non-riemannian geometry}.
\newblock Courier Corporation, 2012.

\bibitem{Nester:1998mp}
J.~M. Nester and H.-J. Yo, {\it {Symmetric teleparallel general relativity}},
  {\em Chin. J. Phys.} {\bf 37} (1999) 113,
  [\href{http://arxiv.org/abs/gr-qc/9809049}{{\tt gr-qc/9809049}}].

\bibitem{BeltranJimenez:2018vdo}
J.~Beltrán~Jiménez, L.~Heisenberg, and T.~S. Koivisto, {\it {Teleparallel
  Palatini theories}},  {\em JCAP} {\bf 1808} (2018) 039,
  [\href{http://arxiv.org/abs/1803.10185}{{\tt arXiv:1803.10185}}].

\bibitem{Jimenez:2019ghw}
J.~B. Jiménez, L.~Heisenberg, D.~Iosifidis, A.~Jiménez-Cano, and T.~S.
  Koivisto, {\it {General Teleparallel Quadratic Gravity}},  {\em Phys. Lett.}
  {\bf B805} (2020) 135422, [\href{http://arxiv.org/abs/1909.09045}{{\tt
  arXiv:1909.09045}}].

\bibitem{Percacci:2020bzf}
R.~Percacci, {\it {Towards Metric-Affine Quantum Gravity}},  2020.
\newblock \href{http://arxiv.org/abs/2003.09486}{{\tt arXiv:2003.09486}}.

\bibitem{Hehl:1999sb}
F.~W. Hehl and A.~Macias, {\it {Metric affine gauge theory of gravity. 2. Exact
  solutions}},  {\em Int. J. Mod. Phys.} {\bf D8} (1999) 399--416,
  [\href{http://arxiv.org/abs/gr-qc/9902076}{{\tt gr-qc/9902076}}].

\bibitem{Iosifidis:2019jgi}
D.~Iosifidis, {\em {Metric-Affine Gravity and Cosmology/Aspects of Torsion and
  non-Metricity in Gravity Theories}}.
\newblock PhD thesis, Thessaloniki U., 2019.
\newblock \href{http://arxiv.org/abs/1902.09643}{{\tt arXiv:1902.09643}}.

\bibitem{Iosifidis:2018jwu}
D.~Iosifidis, {\it {Exactly Solvable Connections in Metric-Affine Gravity}},
  {\em Class. Quant. Grav.} {\bf 36} (2019), no.~8 085001,
  [\href{http://arxiv.org/abs/1812.04031}{{\tt arXiv:1812.04031}}].

\bibitem{Obukhov:1996ka}
{\relax Yu}.~N. Obukhov, E.~J. Vlachynsky, W.~Esser, and F.~W. Hehl, {\it
  {Effective Einstein theory from metric affine gravity models via irreducible
  decompositions}},  {\em Phys. Rev.} {\bf D56} (1997) 7769--7778.

\bibitem{Vitagliano:2010sr}
V.~Vitagliano, T.~P. Sotiriou, and S.~Liberati, {\it {The dynamics of
  metric-affine gravity}},  {\em Annals Phys.} {\bf 326} (2011) 1259--1273,
  [\href{http://arxiv.org/abs/1008.0171}{{\tt arXiv:1008.0171}}]. [Erratum:
  Annals Phys.329,186(2013)].

\bibitem{schouten2013ricci}
J.~A. Schouten, {\em Ricci-calculus: an introduction to tensor analysis and its
  geometrical applications}, vol.~10.
\newblock Springer Science \& Business Media, 2013.

\bibitem{hehl1976hypermomentum}
F.~W. Hehl, G.~D. Kerlick, and P.~von~der Heyde, {\it On hypermomentum in
  general relativity i. the notion of hypermomentum},  {\em Zeitschrift fuer
  Naturforschung A} {\bf 31} (1976), no.~2 111--114.

\bibitem{Obukhov:2014nja}
Y.~N. Obukhov and D.~Puetzfeld, {\it {Conservation laws in gravity: A unified
  framework}},  {\em Phys. Rev.} {\bf D90} (2014), no.~2 024004,
  [\href{http://arxiv.org/abs/1405.4003}{{\tt arXiv:1405.4003}}].

\bibitem{Obukhov:2013ona}
Y.~N. Obukhov and D.~Puetzfeld, {\it {Conservation laws in gravitational
  theories with general nonminimal coupling}},  {\em Phys. Rev.} {\bf D87}
  (2013), no.~8 081502, [\href{http://arxiv.org/abs/1303.6050}{{\tt
  arXiv:1303.6050}}].

\bibitem{Babourova:2004xx}
O.~V. Babourova and B.~N. Frolov, {\it {Perfect hypermomentum fluid:
  Variational theory and equations of motion}},  {\em Int. J. Mod. Phys.} {\bf
  A13} (1998) 5391--5407, [\href{http://arxiv.org/abs/gr-qc/0405124}{{\tt
  gr-qc/0405124}}].

\bibitem{Iosifidis:2020gth}
D.~Iosifidis, {\it {Cosmological Hyperfluids, Torsion and Non-metricity}},
  \href{http://arxiv.org/abs/2003.07384}{{\tt arXiv:2003.07384}}.

\bibitem{Puetzfeld:2001ur}
D.~Puetzfeld, {\it {A Cosmological model in Weyl-Cartan space-time. 1. Field
  equations and solutions}},  {\em Class. Quant. Grav.} {\bf 19} (2002)
  3263--3280, [\href{http://arxiv.org/abs/gr-qc/0111014}{{\tt gr-qc/0111014}}].

\bibitem{Puetzfeld:2004yg}
D.~Puetzfeld, {\it {Status of non-Riemannian cosmology}},  {\em New Astron.
  Rev.} {\bf 49} (2005) 59--64, [\href{http://arxiv.org/abs/gr-qc/0404119}{{\tt
  gr-qc/0404119}}].

\bibitem{Shimada:2018lnm}
K.~Shimada, K.~Aoki, and K.-i. Maeda, {\it {Metric-affine Gravity and
  Inflation}},  {\em Phys. Rev.} {\bf D99} (2019), no.~10 104020,
  [\href{http://arxiv.org/abs/1812.03420}{{\tt arXiv:1812.03420}}].

\bibitem{Obukhov:1996mg}
Y.~N. Obukhov, {\it {On a model of an unconstrained hyperfluid}},  {\em Phys.
  Lett.} {\bf A210} (1996) 163--167,
  [\href{http://arxiv.org/abs/gr-qc/0008014}{{\tt gr-qc/0008014}}].

\bibitem{1979PhLA...75...27T}
M.~{Tsamparlis}, {\it {Cosmological principle and torsion}},  {\em Physics
  Letters A} {\bf 75} (Dec., 1979) 27--28.

\bibitem{Minkevich:1998cv}
A.~V. Minkevich and A.~S. Garkun, {\it {Isotropic cosmology in metric - affine
  gauge theory of gravity}},  \href{http://arxiv.org/abs/gr-qc/9805007}{{\tt
  gr-qc/9805007}}.

\bibitem{einstein2015meaning}
A.~Einstein, {\em The Meaning of Relativity}.
\newblock Routledge Classics Series. Routledge, 2015.

\bibitem{Adak:2005cd}
M.~Adak, M.~Kalay, and O.~Sert, {\it {Lagrange formulation of the symmetric
  teleparallel gravity}},  {\em Int. J. Mod. Phys.} {\bf D15} (2006) 619--634,
  [\href{http://arxiv.org/abs/gr-qc/0505025}{{\tt gr-qc/0505025}}].

\bibitem{1916SPAW......1111E}
A.~{Einstein}, {\it {HAMILTONsches Prinzip und allgemeine
  Relativit{\"a}tstheorie}},  {\em Sitzungsberichte der K{\"o}niglich
  Preu{\ss}ischen Akademie der Wissenschaften (Berlin), Seite 1111-1116.}
  (1916).

\bibitem{Noether1918}
E.~Noether, {\it Invariante variationsprobleme},  {\em Nachrichten von der
  Gesellschaft der Wissenschaften zu G{\"o}ttingen, Mathematisch-Physikalische
  Klasse} {\bf 1918} (1918) 235--257.

\bibitem{Jimenez:2019yyx}
J.~B. Jim{\'e}nez, L.~Heisenberg, and T.~S. Koivisto, {\it {The canonical frame
  of purified gravity}},  \href{http://arxiv.org/abs/1903.12072}{{\tt
  arXiv:1903.12072}}.

\bibitem{Ruegg:2003ps}
H.~Ruegg and M.~Ruiz-Altaba, {\it {The Stueckelberg field}},  {\em Int. J. Mod.
  Phys.} {\bf A19} (2004) 3265--3348,
  [\href{http://arxiv.org/abs/hep-th/0304245}{{\tt hep-th/0304245}}].

\bibitem{Isham:1971dv}
C.~J. Isham, A.~Salam, and J.~A. Strathdee, {\it {Nonlinear realizations of
  space-time symmetries. Scalar and tensor gravity}},  {\em Annals Phys.} {\bf
  62} (1971) 98--119.

\bibitem{Percacci:1990wy}
R.~Percacci, {\it {The Higgs phenomenon in quantum gravity}},  {\em Nucl.
  Phys.} {\bf B353} (1991) 271--290,
  [\href{http://arxiv.org/abs/0712.3545}{{\tt arXiv:0712.3545}}].

\bibitem{Percacci:2009ij}
R.~Percacci, {\it {Gravity from a Particle Physicists' perspective}},  {\em
  PoS} {\bf ISFTG} (2009) 011, [\href{http://arxiv.org/abs/0910.5167}{{\tt
  arXiv:0910.5167}}].

\bibitem{Pagani:2015ema}
C.~Pagani and R.~Percacci, {\it {Quantum gravity with torsion and
  non-metricity}},  {\em Class. Quant. Grav.} {\bf 32} (2015), no.~19 195019,
  [\href{http://arxiv.org/abs/1506.02882}{{\tt arXiv:1506.02882}}].

\bibitem{Tresguerres:2000qn}
R.~Tresguerres and E.~W. Mielke, {\it {Gravitational Goldstone fields from
  affine gauge theory}},  {\em Phys. Rev.} {\bf D62} (2000) 044004,
  [\href{http://arxiv.org/abs/gr-qc/0007072}{{\tt gr-qc/0007072}}].

\bibitem{Leclerc:2005qc}
M.~Leclerc, {\it {The Higgs sector of gravitational gauge theories}},  {\em
  Annals Phys.} {\bf 321} (2006) 708--743,
  [\href{http://arxiv.org/abs/gr-qc/0502005}{{\tt gr-qc/0502005}}].

\bibitem{Tiemblo:2005js}
A.~Tiemblo and R.~Tresguerres, {\it {Gauge theories of gravity: The Nonlinear
  framework}},  {\em Recent Res. Devel. Phys.} {\bf 5} (2004) 1255,
  [\href{http://arxiv.org/abs/gr-qc/0510089}{{\tt gr-qc/0510089}}].

\bibitem{Ali:2007hu}
S.~A. Ali and S.~Capozziello, {\it {Nonlinear realization of the local
  conform-affine symmetry group for gravity in the composite fiber bundle
  formalism}},  {\em Int. J. Geom. Meth. Mod. Phys.} {\bf 4} (2007) 1041--1074,
  [\href{http://arxiv.org/abs/0705.4609}{{\tt arXiv:0705.4609}}].

\bibitem{Westman:2014yca}
H.~F. Westman and T.~G. Zlosnik, {\it {An introduction to the physics of Cartan
  gravity}},  {\em Annals Phys.} {\bf 361} (2015) 330--376,
  [\href{http://arxiv.org/abs/1411.1679}{{\tt arXiv:1411.1679}}].

\bibitem{Zlosnik:2018qvg}
T.~Zlosnik, F.~Urban, L.~Marzola, and T.~Koivisto, {\it {Spacetime and dark
  matter from spontaneous breaking of Lorentz symmetry}},  {\em Class. Quant.
  Grav.} {\bf 35} (2018), no.~23 235003,
  [\href{http://arxiv.org/abs/1807.01100}{{\tt arXiv:1807.01100}}].

\bibitem{Koivisto:2019ejt}
T.~Koivisto, M.~Hohmann, and T.~Zlosnik, {\it {The General Linear Cartan
  Khronon}},  {\em Universe} {\bf 5} (2019), no.~6 153,
  [\href{http://arxiv.org/abs/1905.02967}{{\tt arXiv:1905.02967}}].

\bibitem{Cheung:2017pzi}
C.~Cheung, {\it {TASI Lectures on Scattering Amplitudes}},  in {\em
  {Proceedings, Theoretical Advanced Study Institute in Elementary Particle
  Physics : Anticipating the Next Discoveries in Particle Physics (TASI 2016):
  Boulder, CO, USA, June 6-July 1, 2016}}, pp.~571--623, 2018.
\newblock \href{http://arxiv.org/abs/1708.03872}{{\tt arXiv:1708.03872}}.

\bibitem{Koivisto:2018aip}
T.~Koivisto, {\it {An integrable geometrical foundation of gravity}},  {\em
  Int. J. Geom. Meth. Mod. Phys.} {\bf 15} (2018) 1840006,
  [\href{http://arxiv.org/abs/1802.00650}{{\tt arXiv:1802.00650}}].

\bibitem{Janssen:2018exh}
B.~Janssen and A.~Jim{\'e}nez-Cano, {\it {Projective symmetries and induced
  electromagnetism in metric-affine gravity}},  {\em Phys. Lett.} {\bf B786}
  (2018) 462--465, [\href{http://arxiv.org/abs/1807.10168}{{\tt
  arXiv:1807.10168}}].

\bibitem{Adak:2002pq}
M.~Adak, T.~Dereli, and L.~H. Ryder, {\it {Dirac equation in space-times with
  nonmetricity and torsion}},  {\em Int. J. Mod. Phys.} {\bf D12} (2003)
  145--156, [\href{http://arxiv.org/abs/gr-qc/0208042}{{\tt gr-qc/0208042}}].

\bibitem{Adak:2003qg}
M.~Adak, T.~Dereli, and L.~H. Ryder, {\it {Possible effects of spacetime
  non-metricity on neutrino oscillations}},  {\em Phys. Rev.} {\bf D69} (2004)
  123002, [\href{http://arxiv.org/abs/gr-qc/0303080}{{\tt gr-qc/0303080}}].

\bibitem{Adak:2008gd}
M.~Adak, {\"O}.~Sert, M.~Kalay, and M.~Sari, {\it {Symmetric Teleparallel
  Gravity: Some exact solutions and spinor couplings}},  {\em Int. J. Mod.
  Phys.} {\bf A28} (2013) 1350167, [\href{http://arxiv.org/abs/0810.2388}{{\tt
  arXiv:0810.2388}}].

\bibitem{Formiga:2012ns}
J.~B. Formiga and C.~Romero, {\it {Dirac equation in non-Riemannian
  geometries}},  {\em Int. J. Geom. Meth. Mod. Phys.} {\bf 10} (2013) 1320012,
  [\href{http://arxiv.org/abs/1210.1615}{{\tt arXiv:1210.1615}}].

\bibitem{Koivisto:2006ie}
T.~Koivisto, {\it {The matter power spectrum in f(r) gravity}},  {\em Phys.
  Rev.} {\bf D73} (2006) 083517,
  [\href{http://arxiv.org/abs/astro-ph/0602031}{{\tt astro-ph/0602031}}].

\bibitem{Koivisto:2007sq}
T.~Koivisto, {\it {Viable Palatini-f(R) cosmologies with generalized dark
  matter}},  {\em Phys. Rev. D} {\bf 76} (2007) 043527,
  [\href{http://arxiv.org/abs/0706.0974}{{\tt arXiv:0706.0974}}].

\bibitem{Harko:2011nh}
T.~Harko, T.~S. Koivisto, F.~S.~N. Lobo, and G.~J. Olmo, {\it {Metric-Palatini
  gravity unifying local constraints and late-time cosmic acceleration}},  {\em
  Phys. Rev.} {\bf D85} (2012) 084016,
  [\href{http://arxiv.org/abs/1110.1049}{{\tt arXiv:1110.1049}}].

\bibitem{Leanizbarrutia:2017xyd}
I.~Leanizbarrutia, F.~S.~N. Lobo, and D.~Saez-Gomez, {\it {Crossing SNe Ia and
  BAO observational constraints with local ones in hybrid metric-Palatini
  gravity}},  {\em Phys. Rev.} {\bf D95} (2017), no.~8 084046,
  [\href{http://arxiv.org/abs/1701.08980}{{\tt arXiv:1701.08980}}].

\bibitem{VargasdosSantos:2017ggl}
M.~Vargas~dos Santos, J.~S. Alcaniz, D.~F. Mota, and S.~Capozziello, {\it
  {Screening mechanisms in hybrid metric-Palatini gravity}},  {\em Phys. Rev.}
  {\bf D97} (2018), no.~10 104010, [\href{http://arxiv.org/abs/1712.03831}{{\tt
  arXiv:1712.03831}}].

\bibitem{Danila:2018xya}
B.~D{\v a}nil{\v a}, T.~Harko, F.~S.~N. Lobo, and M.~K. Mak, {\it {Spherically
  symmetric static vacuum solutions in hybrid metric-Palatini gravity}},  {\em
  Phys. Rev.} {\bf D99} (2019), no.~6 064028,
  [\href{http://arxiv.org/abs/1811.02742}{{\tt arXiv:1811.02742}}].

\bibitem{Wojnar:2018hal}
A.~Wojnar, {\it {Polytropic stars in Palatini gravity}},  {\em Eur. Phys. J.}
  {\bf C79} (2019), no.~1 51, [\href{http://arxiv.org/abs/1808.04188}{{\tt
  arXiv:1808.04188}}].

\bibitem{Bronnikov:2019ugl}
K.~A. Bronnikov, {\it {Spherically symmetric black holes and wormholes in
  hybrid metric-Palatini gravity}},
  \href{http://arxiv.org/abs/1908.02012}{{\tt arXiv:1908.02012}}.

\bibitem{Rosa:2019ejh}
J.~L. Rosa, S.~Carloni, and J.~P.~S. Lemos, {\it {The cosmological phase space
  of generalized hybrid metric-Palatini theories of gravity}},
  \href{http://arxiv.org/abs/1908.07778}{{\tt arXiv:1908.07778}}.

\bibitem{Amendola:2010bk}
L.~Amendola, K.~Enqvist, and T.~Koivisto, {\it {Unifying Einstein and Palatini
  gravities}},  {\em Phys.Rev.} {\bf D83} (2011) 044016,
  [\href{http://arxiv.org/abs/1010.4776}{{\tt arXiv:1010.4776}}].

\bibitem{Afonso:2018bpv}
V.~I. Afonso, G.~J. Olmo, and D.~Rubiera-Garcia, {\it {Mapping Ricci-based
  theories of gravity into general relativity}},  {\em Phys. Rev.} {\bf D97}
  (2018), no.~2 021503, [\href{http://arxiv.org/abs/1801.10406}{{\tt
  arXiv:1801.10406}}].

\bibitem{BeltranJimenez:2019acz}
J.~Beltr{\'a}n~Jim{\'e}nez and A.~Delhom, {\it {Ghosts in metric-affine higher
  order curvature gravity}},  \href{http://arxiv.org/abs/1901.08988}{{\tt
  arXiv:1901.08988}}.

\bibitem{BeltranJimenez:2017doy}
J.~Beltran~Jimenez, L.~Heisenberg, G.~J. Olmo, and D.~Rubiera-Garcia, {\it
  {Born--Infeld inspired modifications of gravity}},  {\em Phys. Rept.} {\bf
  727} (2018) 1--129, [\href{http://arxiv.org/abs/1704.03351}{{\tt
  arXiv:1704.03351}}].

\bibitem{Lu:2019hra}
J.~Lu, X.~Zhao, and G.~Chee, {\it {Cosmology in symmetric teleparallel gravity
  and its dynamical system}},  {\em Eur. Phys. J.} {\bf C79} (2019), no.~6 530,
  [\href{http://arxiv.org/abs/1906.08920}{{\tt arXiv:1906.08920}}].

\bibitem{Lazkoz:2019sjl}
R.~Lazkoz, F.~S.~N. Lobo, M.~Ortiz-Ba\~{n}o, and V.~Salzano, {\it
  {Observational constraints of $f(Q)$ gravity}},
  \href{http://arxiv.org/abs/1907.13219}{{\tt arXiv:1907.13219}}.

\bibitem{Harko:2018gxr}
T.~Harko, T.~S. Koivisto, F.~S.~N. Lobo, G.~J. Olmo, and D.~Rubiera-Garcia,
  {\it {Coupling matter in modified $Q$ gravity}},  {\em Phys. Rev.} {\bf D98}
  (2018), no.~8 084043, [\href{http://arxiv.org/abs/1806.10437}{{\tt
  arXiv:1806.10437}}].

\bibitem{Lobo:2019xwp}
T.~Harko, T.~S. Koivisto, G.~J. Olmo, F.~S.~N. Lobo, and R.-G. Diego, {\it
  {Novel couplings between nonmetricity and matter}},  in {\em {15th Marcel
  Grossmann Meeting on Recent Developments in Theoretical and Experimental
  General Relativity, Astrophysics, and Relativistic Field Theories (MG15)
  Rome, Italy, July 1-7, 2018}}, 2019.
\newblock \href{http://arxiv.org/abs/1901.00805}{{\tt arXiv:1901.00805}}.

\bibitem{Xu:2019sbp}
Y.~Xu, G.~Li, T.~Harko, and S.-D. Liang, {\it {$f(Q,T)$ gravity}},  {\em Eur.
  Phys. J.} {\bf C79} (2019), no.~8 708,
  [\href{http://arxiv.org/abs/1908.04760}{{\tt arXiv:1908.04760}}].

\bibitem{Milgrom:2019rtd}
M.~Milgrom, {\it {Noncovariance at low accelerations as a route to MOND}},
  \href{http://arxiv.org/abs/1908.01691}{{\tt arXiv:1908.01691}}.

\bibitem{DAmbrosio:2020nev}
F.~D'Ambrosio, M.~Garg, and L.~Heisenberg, {\it {Non-linear extension of
  non-metricity scalar for MOND}},  \href{http://arxiv.org/abs/2004.00888}{{\tt
  arXiv:2004.00888}}.

\bibitem{Dialektopoulos:2019mtr}
K.~F. Dialektopoulos, T.~S. Koivisto, and S.~Capozziello, {\it {Noether
  symmetries in Symmetric Teleparallel Cosmology}},  {\em Eur. Phys. J.} {\bf
  C79} (2019), no.~7 606, [\href{http://arxiv.org/abs/1905.09019}{{\tt
  arXiv:1905.09019}}].

\bibitem{Hohmann:2018wxu}
M.~Hohmann, C.~Pfeifer, J.~L. Said, and U.~Ualikhanova, {\it {Propagation of
  gravitational waves in symmetric teleparallel gravity theories}},  {\em Phys.
  Rev.} {\bf D99} (2019), no.~2 024009,
  [\href{http://arxiv.org/abs/1808.02894}{{\tt arXiv:1808.02894}}].

\bibitem{Soudi:2018dhv}
I.~Soudi, G.~Farrugia, V.~Gakis, J.~Levi~Said, and E.~N. Saridakis, {\it
  {Polarization of gravitational waves in symmetric teleparallel theories of
  gravity and their modifications}},  {\em Phys. Rev. D} {\bf 100} (2019),
  no.~4 044008, [\href{http://arxiv.org/abs/1810.08220}{{\tt
  arXiv:1810.08220}}].

\bibitem{Conroy:2017yln}
A.~Conroy and T.~Koivisto, {\it {The spectrum of symmetric teleparallel
  gravity}},  {\em Eur. Phys. J.} {\bf C78} (2018), no.~11 923,
  [\href{http://arxiv.org/abs/1710.05708}{{\tt arXiv:1710.05708}}].

\bibitem{Iosifidis:2018zwo}
D.~Iosifidis and T.~Koivisto, {\it {Scale transformations in metric-affine
  geometry}},  \href{http://arxiv.org/abs/1810.12276}{{\tt arXiv:1810.12276}}.

\bibitem{Gakis:2019rdd}
V.~Gakis, M.~Kr\v{s}\v{s}\'ak, J.~Levi~Said, and E.~N. Saridakis, {\it
  {Conformal gravity and transformations in the symmetric teleparallel
  framework}},  {\em Phys. Rev. D} {\bf 101} (2020), no.~6 064024,
  [\href{http://arxiv.org/abs/1908.05741}{{\tt arXiv:1908.05741}}].

\bibitem{Jimenez:2019tkx}
J.~B. Jim{\'e}nez and K.~F. Dialektopoulos, {\it {Non-Linear Obstructions for
  Consistent New General Relativity}},
  \href{http://arxiv.org/abs/1907.10038}{{\tt arXiv:1907.10038}}.

\bibitem{Cheng:1988zg}
W.-H. Cheng, D.-C. Chern, and J.~M. Nester, {\it {Canonical Analysis of the One
  Parameter Teleparallel Theory}},  {\em Phys. Rev.} {\bf D38} (1988)
  2656--2658.

\bibitem{Koivisto:2018loq}
T.~Koivisto and G.~Tsimperis, {\it {The spectrum of teleparallel gravity}},
  \href{http://arxiv.org/abs/1810.11847}{{\tt arXiv:1810.11847}}.

\bibitem{Afonso:2017bxr}
V.~I. Afonso, C.~Bejarano, J.~Beltran~Jimenez, G.~J. Olmo, and E.~Orazi, {\it
  {The trivial role of torsion in projective invariant theories of gravity with
  non-minimally coupled matter fields}},  {\em Class. Quant. Grav.} {\bf 34}
  (2017), no.~23 235003, [\href{http://arxiv.org/abs/1705.03806}{{\tt
  arXiv:1705.03806}}].

\bibitem{Jimenez:2020iok}
J.~B. Jim\'enez, D.~de~Andr\'es, and A.~Delhom, {\it {Anisotropic deformations
  in a class of projectively-invariant metric-affine theories of gravity}},
  {\em Class. Quant. Grav.} {\bf 37} (2020), no.~22 225013,
  [\href{http://arxiv.org/abs/2006.07406}{{\tt arXiv:2006.07406}}].

\bibitem{Afonso:2018mxn}
V.~I. Afonso, G.~J. Olmo, E.~Orazi, and D.~Rubiera-Garcia, {\it {Mapping
  nonlinear gravity into General Relativity with nonlinear electrodynamics}},
  {\em Eur. Phys. J.} {\bf C78} (2018), no.~10 866,
  [\href{http://arxiv.org/abs/1807.06385}{{\tt arXiv:1807.06385}}].

\bibitem{Afonso:2018hyj}
V.~I. Afonso, G.~J. Olmo, E.~Orazi, and D.~Rubiera-Garcia, {\it {Correspondence
  between modified gravity and general relativity with scalar fields}},  {\em
  Phys. Rev.} {\bf D99} (2019), no.~4 044040,
  [\href{http://arxiv.org/abs/1810.04239}{{\tt arXiv:1810.04239}}].

\bibitem{Delhom:2019zrb}
A.~Delhom, G.~J. Olmo, and E.~Orazi, {\it {Ricci-Based Gravity theories and
  their impact on Maxwell and nonlinear electromagnetic models}},  {\em JHEP}
  {\bf 11} (2019) 149, [\href{http://arxiv.org/abs/1907.04183}{{\tt
  arXiv:1907.04183}}].

\bibitem{Olmo:2008nf}
G.~J. Olmo and P.~Singh, {\it {Effective Action for Loop Quantum Cosmology a la
  Palatini}},  {\em JCAP} {\bf 0901} (2009) 030,
  [\href{http://arxiv.org/abs/0806.2783}{{\tt arXiv:0806.2783}}].

\bibitem{Barragan:2009sq}
C.~Barragan, G.~J. Olmo, and H.~Sanchis-Alepuz, {\it {Bouncing Cosmologies in
  Palatini f(R) Gravity}},  {\em Phys. Rev.} {\bf D80} (2009) 024016,
  [\href{http://arxiv.org/abs/0907.0318}{{\tt arXiv:0907.0318}}].

\bibitem{Bombacigno:2018tyw}
F.~Bombacigno and G.~Montani, {\it {Big bounce cosmology for Palatini $R^2$
  gravity with a Nieh--Yan term}},  {\em Eur. Phys. J.} {\bf C79} (2019), no.~5
  405, [\href{http://arxiv.org/abs/1809.07563}{{\tt arXiv:1809.07563}}].

\bibitem{Koivisto:2010jj}
T.~S. Koivisto, {\it {Bouncing Palatini cosmologies and their perturbations}},
  {\em Phys. Rev.} {\bf D82} (2010) 044022,
  [\href{http://arxiv.org/abs/1004.4298}{{\tt arXiv:1004.4298}}].

\bibitem{Banados:2010ix}
M.~Banados and P.~G. Ferreira, {\it {Eddington's theory of gravity and its
  progeny}},  {\em Phys. Rev. Lett.} {\bf 105} (2010) 011101,
  [\href{http://arxiv.org/abs/1006.1769}{{\tt arXiv:1006.1769}}]. [Erratum:
  Phys. Rev. Lett.113,no.11,119901(2014)].

\bibitem{EscamillaRivera:2012vz}
C.~Escamilla-Rivera, M.~Banados, and P.~G. Ferreira, {\it {A tensor instability
  in the Eddington inspired Born-Infeld Theory of Gravity}},  {\em Phys. Rev.}
  {\bf D85} (2012) 087302, [\href{http://arxiv.org/abs/1204.1691}{{\tt
  arXiv:1204.1691}}].

\bibitem{Avelino:2012ue}
P.~P. Avelino and R.~Z. Ferreira, {\it {Bouncing Eddington-inspired Born-Infeld
  cosmologies: an alternative to Inflation ?}},  {\em Phys. Rev.} {\bf D86}
  (2012) 041501, [\href{http://arxiv.org/abs/1205.6676}{{\tt
  arXiv:1205.6676}}].

\bibitem{Yang:2013hsa}
K.~Yang, X.-L. Du, and Y.-X. Liu, {\it {Linear perturbations in
  Eddington-inspired Born-Infeld gravity}},  {\em Phys. Rev.} {\bf D88} (2013)
  124037, [\href{http://arxiv.org/abs/1307.2969}{{\tt arXiv:1307.2969}}].

\bibitem{Cho:2012vg}
I.~Cho, H.-C. Kim, and T.~Moon, {\it {Universe Driven by Perfect Fluid in
  Eddington-inspired Born-Infeld Gravity}},  {\em Phys. Rev.} {\bf D86} (2012)
  084018, [\href{http://arxiv.org/abs/1208.2146}{{\tt arXiv:1208.2146}}].

\bibitem{Odintsov:2014yaa}
S.~D. Odintsov, G.~J. Olmo, and D.~Rubiera-Garcia, {\it {Born-Infeld gravity
  and its functional extensions}},  {\em Phys. Rev.} {\bf D90} (2014) 044003,
  [\href{http://arxiv.org/abs/1406.1205}{{\tt arXiv:1406.1205}}].

\bibitem{Jimenez:2014fla}
J.~Beltran~Jimenez, L.~Heisenberg, and G.~J. Olmo, {\it {Infrared lessons for
  ultraviolet gravity: the case of massive gravity and Born-Infeld}},  {\em
  JCAP} {\bf 1411} (2014) 004, [\href{http://arxiv.org/abs/1409.0233}{{\tt
  arXiv:1409.0233}}].

\bibitem{BeltranJimenez:2017uwv}
J.~Beltran~Jimenez, L.~Heisenberg, G.~J. Olmo, and D.~Rubiera-Garcia, {\it {On
  gravitational waves in Born-Infeld inspired non-singular cosmologies}},  {\em
  JCAP} {\bf 1710} (2017), no.~10 029,
  [\href{http://arxiv.org/abs/1707.08953}{{\tt arXiv:1707.08953}}]. [Erratum:
  JCAP1808,no.08,E01(2018)].

\bibitem{Makarenko:2014lxa}
A.~N. Makarenko, S.~Odintsov, and G.~J. Olmo, {\it {Born-Infeld-$f(R)$
  gravity}},  {\em Phys. Rev.} {\bf D90} (2014) 024066,
  [\href{http://arxiv.org/abs/1403.7409}{{\tt arXiv:1403.7409}}].

\bibitem{Makarenko:2014cca}
A.~N. Makarenko, S.~D. Odintsov, G.~J. Olmo, and D.~Rubiera-Garcia, {\it
  {Early-time cosmic dynamics in $f(R)$ and $f(|\hat\Omega|)$ extensions of
  Born-Infeld gravity}},  {\em TSPU Bulletin} {\bf 12} (2014) 158--163,
  [\href{http://arxiv.org/abs/1411.6193}{{\tt arXiv:1411.6193}}].

\bibitem{Li:2017ttl}
S.-L. Li and H.~Wei, {\it {Stability of the Einstein static universe in
  Eddington-inspired Born-Infeld theory}},  {\em Phys. Rev.} {\bf D96} (2017),
  no.~2 023531, [\href{http://arxiv.org/abs/1705.06819}{{\tt
  arXiv:1705.06819}}].

\bibitem{Bouhmadi-Lopez:2014jfa}
M.~Bouhmadi-L{\'o}pez, C.-Y. Chen, and P.~Chen, {\it {Eddington--Born--Infeld
  cosmology: a cosmographic approach, a tale of doomsdays and the fate of bound
  structures}},  {\em Eur. Phys. J.} {\bf C75} (2015) 90,
  [\href{http://arxiv.org/abs/1406.6157}{{\tt arXiv:1406.6157}}].

\bibitem{Bouhmadi-Lopez:2018tel}
M.~Bouhmadi-L{\'o}pez, C.-Y. Chen, and P.~Chen, {\it {On the Consistency of the
  Wheeler-DeWitt Equation in the Quantized Eddington-inspired Born-Infeld
  Gravity}},  {\em JCAP} {\bf 1812} (2018), no.~12 032,
  [\href{http://arxiv.org/abs/1810.10918}{{\tt arXiv:1810.10918}}].

\bibitem{Albarran:2018mpg}
I.~Albarran, M.~Bouhmadi-L{\'o}pez, C.-Y. Chen, and P.~Chen, {\it {Quantum
  cosmology of Eddington-Born--Infeld gravity fed by a scalar field: The big
  rip case}},  {\em Phys. Dark Univ.} {\bf 23} (2019) 100255,
  [\href{http://arxiv.org/abs/1811.05041}{{\tt arXiv:1811.05041}}].

\bibitem{Stachowski:2016zio}
A.~Stachowski, M.~Szydlowski, and A.~Borowiec, {\it {Starobinsky cosmological
  model in Palatini formalism}},  {\em Eur. Phys. J.} {\bf C77} (2017), no.~6
  406, [\href{http://arxiv.org/abs/1608.03196}{{\tt arXiv:1608.03196}}].

\bibitem{Meng:2004yf}
X.-H. Meng and P.~Wang, {\it {R**2 corrections to the cosmological dynamics of
  inflation in the Palatini formulation}},  {\em Class. Quant. Grav.} {\bf 21}
  (2004) 2029--2036, [\href{http://arxiv.org/abs/gr-qc/0402011}{{\tt
  gr-qc/0402011}}].

\bibitem{Antoniadis:2018ywb}
I.~Antoniadis, A.~Karam, A.~Lykkas, and K.~Tamvakis, {\it {Palatini inflation
  in models with an $R^2$ term}},  {\em JCAP} {\bf 1811} (2018), no.~11 028,
  [\href{http://arxiv.org/abs/1810.10418}{{\tt arXiv:1810.10418}}].

\bibitem{Fu:2017iqg}
C.~Fu, P.~Wu, and H.~Yu, {\it {Inflationary dynamics and preheating of the
  nonminimally coupled inflaton field in the metric and Palatini formalisms}},
  {\em Phys. Rev.} {\bf D96} (2017), no.~10 103542,
  [\href{http://arxiv.org/abs/1801.04089}{{\tt arXiv:1801.04089}}].

\bibitem{Almeida:2018oid}
J.~P.~B. Almeida, N.~Bernal, J.~Rubio, and T.~Tenkanen, {\it {Hidden Inflaton
  Dark Matter}},  {\em JCAP} {\bf 1903} (2019) 012,
  [\href{http://arxiv.org/abs/1811.09640}{{\tt arXiv:1811.09640}}].

\bibitem{Racioppi:2017spw}
A.~Racioppi, {\it {Coleman-Weinberg linear inflation: metric vs. Palatini
  formulation}},  {\em JCAP} {\bf 1712} (2017), no.~12 041,
  [\href{http://arxiv.org/abs/1710.04853}{{\tt arXiv:1710.04853}}].

\bibitem{Enckell:2018hmo}
V.-M. Enckell, K.~Enqvist, S.~Rasanen, and L.-P. Wahlman, {\it {Inflation with
  $R^2$ term in the Palatini formalism}},  {\em JCAP} {\bf 1902} (2019) 022,
  [\href{http://arxiv.org/abs/1810.05536}{{\tt arXiv:1810.05536}}].

\bibitem{Antoniadis:2018yfq}
I.~Antoniadis, A.~Karam, A.~Lykkas, T.~Pappas, and K.~Tamvakis, {\it {Rescuing
  Quartic and Natural Inflation in the Palatini Formalism}},  {\em JCAP} {\bf
  1903} (2019) 005, [\href{http://arxiv.org/abs/1812.00847}{{\tt
  arXiv:1812.00847}}].

\bibitem{Rubio:2019ypq}
J.~Rubio and E.~S. Tomberg, {\it {Preheating in Palatini Higgs inflation}},
  {\em JCAP} {\bf 1904} (2019), no.~04 021,
  [\href{http://arxiv.org/abs/1902.10148}{{\tt arXiv:1902.10148}}].

\bibitem{Jinno:2018jei}
R.~Jinno, K.~Kaneta, K.-y. Oda, and S.~C. Park, {\it {Hillclimbing inflation in
  metric and Palatini formulations}},  {\em Phys. Lett.} {\bf B791} (2019)
  396--402, [\href{http://arxiv.org/abs/1812.11077}{{\tt arXiv:1812.11077}}].

\bibitem{Fay:2007gg}
S.~Fay, R.~Tavakol, and S.~Tsujikawa, {\it {f(R) gravity theories in Palatini
  formalism: Cosmological dynamics and observational constraints}},  {\em Phys.
  Rev.} {\bf D75} (2007) 063509,
  [\href{http://arxiv.org/abs/astro-ph/0701479}{{\tt astro-ph/0701479}}].

\bibitem{Cho:2013pea}
I.~Cho, H.-C. Kim, and T.~Moon, {\it {Precursor of Inflation}},  {\em Phys.
  Rev. Lett.} {\bf 111} (2013) 071301,
  [\href{http://arxiv.org/abs/1305.2020}{{\tt arXiv:1305.2020}}].

\bibitem{Cho:2014jta}
I.~Cho and N.~K. Singh, {\it {Tensor-to-scalar ratio in Eddington-inspired
  Born--Infeld inflation}},  {\em Eur. Phys. J.} {\bf C74} (2014), no.~11 3155,
  [\href{http://arxiv.org/abs/1408.2652}{{\tt arXiv:1408.2652}}].

\bibitem{Cho:2014xaa}
I.~Cho and N.~K. Singh, {\it {Scalar perturbation produced at the
  pre-inflationary stage in Eddington-inspired Born--Infeld gravity}},  {\em
  Eur. Phys. J.} {\bf C75} (2015), no.~6 240,
  [\href{http://arxiv.org/abs/1412.6344}{{\tt arXiv:1412.6344}}].

\bibitem{Cho:2014ija}
I.~Cho and H.-C. Kim, {\it {Inflationary tensor perturbation in
  Eddington-inspired born-infeld gravity}},  {\em Phys. Rev.} {\bf D90} (2014),
  no.~2 024063, [\href{http://arxiv.org/abs/1404.6081}{{\tt arXiv:1404.6081}}].

\bibitem{Kim:2013noa}
H.-C. Kim, {\it {Origin of the universe: A hint from Eddington-inspired
  Born-Infeld gravity}},  {\em J. Korean Phys. Soc.} {\bf 65} (2014), no.~6
  840--845, [\href{http://arxiv.org/abs/1312.0703}{{\tt arXiv:1312.0703}}].

\bibitem{Cho:2015yza}
I.~Cho and N.~K. Singh, {\it {Primordial Power Spectra of EiBI Inflation in
  Strong Gravity Limit}},  {\em Phys. Rev.} {\bf D92} (2015), no.~2 024038,
  [\href{http://arxiv.org/abs/1506.02213}{{\tt arXiv:1506.02213}}].

\bibitem{Jana:2016uvq}
S.~Jana and S.~Kar, {\it {Born-Infeld cosmology with scalar Born-Infeld
  matter}},  {\em Phys. Rev.} {\bf D94} (2016), no.~6 064016,
  [\href{http://arxiv.org/abs/1605.00820}{{\tt arXiv:1605.00820}}].

\bibitem{Jimenez:2015jqa}
J.~Beltran~Jimenez, L.~Heisenberg, G.~J. Olmo, and C.~Ringeval, {\it {Cascading
  dust inflation in Born-Infeld gravity}},  {\em JCAP} {\bf 1511} (2015) 046,
  [\href{http://arxiv.org/abs/1509.01188}{{\tt arXiv:1509.01188}}].

\bibitem{Banik:2018nyz}
D.~K. Banik, S.~K. Banik, and K.~Bhuyan, {\it {Dynamical system approach to
  Born-Infeld $f(R)$ gravity in Palatini formalism}},  {\em Phys. Rev.} {\bf
  D97} (2018), no.~12 124041.

\bibitem{Allemandi:2004wn}
G.~Allemandi, A.~Borowiec, and M.~Francaviglia, {\it {Accelerated cosmological
  models in Ricci squared gravity}},  {\em Phys. Rev.} {\bf D70} (2004) 103503,
  [\href{http://arxiv.org/abs/hep-th/0407090}{{\tt hep-th/0407090}}].

\bibitem{Vollick:2003aw}
D.~N. Vollick, {\it {1/R Curvature corrections as the source of the
  cosmological acceleration}},  {\em Phys. Rev.} {\bf D68} (2003) 063510,
  [\href{http://arxiv.org/abs/astro-ph/0306630}{{\tt astro-ph/0306630}}].

\bibitem{Flanagan:2003rb}
E.~E. Flanagan, {\it {Palatini form of 1/R gravity}},  {\em Phys. Rev. Lett.}
  {\bf 92} (2004) 071101, [\href{http://arxiv.org/abs/astro-ph/0308111}{{\tt
  astro-ph/0308111}}].

\bibitem{Vollick:2004ws}
D.~N. Vollick, {\it {On the Dirac field in the Palatini form of 1/R gravity}},
  {\em Phys. Rev.} {\bf D71} (2005) 044020,
  [\href{http://arxiv.org/abs/gr-qc/0409068}{{\tt gr-qc/0409068}}].

\bibitem{Meng:2003uv}
X.~Meng and P.~Wang, {\it {Cosmological evolution in 1/r-gravity theory}},
  {\em Class. Quant. Grav.} {\bf 21} (2004) 951--960,
  [\href{http://arxiv.org/abs/astro-ph/0308031}{{\tt astro-ph/0308031}}].

\bibitem{Nojiri:2003wx}
S.~Nojiri and S.~D. Odintsov, {\it {The Minimal curvature of the universe in
  modified gravity and conformal anomaly resolution of the instabilities}},
  {\em Mod. Phys. Lett.} {\bf A19} (2004) 627--638,
  [\href{http://arxiv.org/abs/hep-th/0310045}{{\tt hep-th/0310045}}].

\bibitem{Olmo:2008ye}
G.~J. Olmo, {\it {Hydrogen atom in Palatini theories of gravity}},  {\em Phys.
  Rev.} {\bf D77} (2008) 084021, [\href{http://arxiv.org/abs/0802.4038}{{\tt
  arXiv:0802.4038}}].

\bibitem{Li:2006vi}
B.~Li and M.~C. Chu, {\it {CMB and Matter Power Spectra of Early f(R) Cosmology
  in Palatini Formalism}},  {\em Phys. Rev.} {\bf D74} (2006) 104010,
  [\href{http://arxiv.org/abs/astro-ph/0610486}{{\tt astro-ph/0610486}}].

\bibitem{Li:2006ag}
B.~Li, K.~C. Chan, and M.~C. Chu, {\it {Constraints on f(R) Cosmology in the
  Palatini Formalism}},  {\em Phys. Rev.} {\bf D76} (2007) 024002,
  [\href{http://arxiv.org/abs/astro-ph/0610794}{{\tt astro-ph/0610794}}].

\bibitem{Amarzguioui:2005zq}
M.~Amarzguioui, O.~Elgaroy, D.~F. Mota, and T.~Multamaki, {\it {Cosmological
  constraints on f(r) gravity theories within the palatini approach}},  {\em
  Astron. Astrophys.} {\bf 454} (2006) 707--714,
  [\href{http://arxiv.org/abs/astro-ph/0510519}{{\tt astro-ph/0510519}}].

\bibitem{Lee:2008ek}
S.~Lee, {\it {Palatini f(R) Cosmology}},  {\em Mod. Phys. Lett.} {\bf A23}
  (2008) 1388--1396, [\href{http://arxiv.org/abs/0801.4606}{{\tt
  arXiv:0801.4606}}].

\bibitem{Pires:2010fv}
N.~Pires, J.~Santos, and J.~S. Alcaniz, {\it {Cosmographic constraints on a
  class of Palatini f(R) gravity}},  {\em Phys. Rev.} {\bf D82} (2010) 067302,
  [\href{http://arxiv.org/abs/1006.0264}{{\tt arXiv:1006.0264}}].

\bibitem{Movahed:2007cs}
M.~S. Movahed, S.~Baghram, and S.~Rahvar, {\it {Consistency of
  $f(R)=\sqrt{R^{2}-R_{0}^2}$ Gravity with the Cosmological Observations in
  Palatini Formalism}},  {\em Phys. Rev.} {\bf D76} (2007) 044008,
  [\href{http://arxiv.org/abs/0705.0889}{{\tt arXiv:0705.0889}}].

\bibitem{Baghram:2009fr}
S.~Baghram, M.~S. Movahed, and S.~Rahvar, {\it {Observational tests of a two
  parameter power-law class modified gravity in Palatini formalism}},  {\em
  Phys. Rev.} {\bf D80} (2009) 064003,
  [\href{http://arxiv.org/abs/0904.4390}{{\tt arXiv:0904.4390}}].

\bibitem{Astier:2005qq}
{\bf SNLS} Collaboration, P.~Astier et~al., {\it {The Supernova Legacy Survey:
  Measurement of $\Omega_M$, $\Omega_\Lambda$ and ${\cal w}$ from the first
  year data set}},  {\em Astron. Astrophys.} {\bf 447} (2006) 31--48,
  [\href{http://arxiv.org/abs/astro-ph/0510447}{{\tt astro-ph/0510447}}].

\bibitem{Amanullah:2010vv}
R.~Amanullah et~al., {\it {Spectra and Light Curves of Six Type Ia Supernovae
  at 0.511 < z < 1.12 and the Union2 Compilation}},  {\em Astrophys. J.} {\bf
  716} (2010) 712--738, [\href{http://arxiv.org/abs/1004.1711}{{\tt
  arXiv:1004.1711}}].

\bibitem{Pinto:2018rfg}
P.~Pinto, L.~Del~Vecchio, L.~Fatibene, and M.~Ferraris, {\it {Extended
  cosmology in Palatini f(R)-theories}},  {\em JCAP} {\bf 1811} (2018), no.~11
  044, [\href{http://arxiv.org/abs/1807.00397}{{\tt arXiv:1807.00397}}].

\bibitem{DelVecchio:2018abv}
L.~Del~Vecchio, L.~Fatibene, S.~Capozziello, M.~Ferraris, P.~Pinto, and
  S.~Camera, {\it {Hubble drift in Palatini $f(\mathcal{R})$-theories}},  {\em
  Eur. Phys. J. Plus} {\bf 134} (2019), no.~1 5,
  [\href{http://arxiv.org/abs/1810.10754}{{\tt arXiv:1810.10754}}].

\bibitem{Li:2007xw}
B.~Li, J.~D. Barrow, and D.~F. Mota, {\it {The Cosmology of
  Ricci-Tensor-Squared gravity in the Palatini variational approach}},  {\em
  Phys. Rev.} {\bf D76} (2007) 104047,
  [\href{http://arxiv.org/abs/0707.2664}{{\tt arXiv:0707.2664}}].

\bibitem{Capozziello:2018aba}
S.~Capozziello, R.~D'Agostino, and O.~Luongo, {\it {Kinematic model-independent
  reconstruction of Palatini $f(R)$ cosmology}},  {\em Gen. Rel. Grav.} {\bf
  51} (2019), no.~1 2, [\href{http://arxiv.org/abs/1806.06385}{{\tt
  arXiv:1806.06385}}].

\bibitem{Capozziello:2019cav}
S.~Capozziello, R.~D'Agostino, and O.~Luongo, {\it {Extended Gravity
  Cosmography}},  \href{http://arxiv.org/abs/1904.01427}{{\tt
  arXiv:1904.01427}}.

\bibitem{TeppaPannia:2018ale}
F.~A. Teppa~Pannia, S.~E. Perez~Bergliaffa, and N.~Manske, {\it {Cosmography
  and the redshift drift in Palatini $f({\cal R})$ theories}},  {\em Eur. Phys.
  J.} {\bf C79} (2019), no.~3 267, [\href{http://arxiv.org/abs/1811.08176}{{\tt
  arXiv:1811.08176}}].

\bibitem{Du:2014jka}
X.-L. Du, K.~Yang, X.-H. Meng, and Y.-X. Liu, {\it {Large Scale Structure
  Formation in Eddington-inspired Born-Infeld Gravity}},  {\em Phys. Rev.} {\bf
  D90} (2014) 044054, [\href{http://arxiv.org/abs/1403.0083}{{\tt
  arXiv:1403.0083}}].

\bibitem{Avelino:2012ge}
P.~P. Avelino, {\it {Eddington-inspired Born-Infeld gravity: astrophysical and
  cosmological constraints}},  {\em Phys. Rev.} {\bf D85} (2012) 104053,
  [\href{http://arxiv.org/abs/1201.2544}{{\tt arXiv:1201.2544}}].

\bibitem{Avelino:2012qe}
P.~P. Avelino, {\it {Eddington-inspired Born-Infeld gravity: nuclear physics
  constraints and the validity of the continuous fluid approximation}},  {\em
  JCAP} {\bf 1211} (2012) 022, [\href{http://arxiv.org/abs/1207.4730}{{\tt
  arXiv:1207.4730}}].

\bibitem{Avelino:2019esh}
P.~P. Avelino, {\it {Probing gravity at sub-femtometer scales through the
  pressure distribution inside the proton}},  {\em Phys. Lett.} {\bf B795}
  (2019) 627--631, [\href{http://arxiv.org/abs/1902.01318}{{\tt
  arXiv:1902.01318}}].

\bibitem{Latorre:2017uve}
A.~D.~I. Latorre, G.~J. Olmo, and M.~Ronco, {\it {Observable traces of
  non-metricity: new constraints on metric-affine gravity}},  {\em Phys. Lett.}
  {\bf B780} (2018) 294--299, [\href{http://arxiv.org/abs/1709.04249}{{\tt
  arXiv:1709.04249}}].

\bibitem{Delhom:2019wir}
A.~Delhom, V.~Miralles, and A.~Pe{\~n}uelas, {\it {Effective interactions in
  Ricci-Based Gravity models below the non-metricity scale}},
  \href{http://arxiv.org/abs/1907.05615}{{\tt arXiv:1907.05615}}.

\bibitem{Banados:2008rm}
M.~Banados, {\it {Eddington-Born-Infeld action for dark matter and dark
  energy}},  {\em Phys. Rev.} {\bf D77} (2008) 123534,
  [\href{http://arxiv.org/abs/0801.4103}{{\tt arXiv:0801.4103}}].

\bibitem{Skordis:2009zza}
C.~Skordis, {\it {Eddington-Born-Infeld theory and the dark sector}},  {\em
  Nucl. Phys. Proc. Suppl.} {\bf 194} (2009) 338--343.

\bibitem{Banados:2008fj}
M.~Banados, P.~G. Ferreira, and C.~Skordis, {\it {Eddington-Born-Infeld gravity
  and the large scale structure of the Universe}},  {\em Phys. Rev.} {\bf D79}
  (2009) 063511, [\href{http://arxiv.org/abs/0811.1272}{{\tt
  arXiv:0811.1272}}].

\bibitem{DeFelice:2012hq}
A.~De~Felice, B.~Gumjudpai, and S.~Jhingan, {\it {Cosmological constraints for
  an Eddington-Born-Infeld field}},  {\em Phys. Rev.} {\bf D86} (2012) 043525,
  [\href{http://arxiv.org/abs/1205.1168}{{\tt arXiv:1205.1168}}].

\bibitem{Rodrigues:2008kv}
D.~C. Rodrigues, {\it {Evolution of Anisotropies in Eddington-Born-Infeld
  Cosmology}},  {\em Phys. Rev.} {\bf D78} (2008) 063013,
  [\href{http://arxiv.org/abs/0806.3613}{{\tt arXiv:0806.3613}}].

\bibitem{Wu:2018idg}
J.~Wu, G.~Li, T.~Harko, and S.-D. Liang, {\it {Palatini formulation of $f(R,T)$
  gravity theory, and its cosmological implications}},  {\em Eur. Phys. J.}
  {\bf C78} (2018), no.~5 430, [\href{http://arxiv.org/abs/1805.07419}{{\tt
  arXiv:1805.07419}}].

\bibitem{Allemandi:2005qs}
G.~Allemandi, A.~Borowiec, M.~Francaviglia, and S.~D. Odintsov, {\it {Dark
  energy dominance and cosmic acceleration in first order formalism}},  {\em
  Phys. Rev.} {\bf D72} (2005) 063505,
  [\href{http://arxiv.org/abs/gr-qc/0504057}{{\tt gr-qc/0504057}}].

\bibitem{Koivisto:2005yc}
T.~Koivisto and H.~Kurki-Suonio, {\it {Cosmological perturbations in the
  palatini formulation of modified gravity}},  {\em Class. Quant. Grav.} {\bf
  23} (2006) 2355--2369, [\href{http://arxiv.org/abs/astro-ph/0509422}{{\tt
  astro-ph/0509422}}].

\bibitem{Capozziello:2009mq}
S.~Capozziello and S.~Vignolo, {\it {Metric-affine f(R)-gravity with torsion:
  An Overview}},  {\em Annalen Phys.} {\bf 19} (2010) 238--248,
  [\href{http://arxiv.org/abs/0910.5230}{{\tt arXiv:0910.5230}}].

\bibitem{Capozziello:2007tj}
S.~Capozziello, R.~Cianci, C.~Stornaiolo, and S.~Vignolo, {\it {f(R) gravity
  with torsion: The Metric-affine approach}},  {\em Class. Quant. Grav.} {\bf
  24} (2007) 6417--6430, [\href{http://arxiv.org/abs/0708.3038}{{\tt
  arXiv:0708.3038}}].

\bibitem{Galtsov:2018xuc}
D.~Gal'tsov and S.~Zhidkova, {\it {Ghost-free Palatini derivative
  scalar--tensor theory: Desingularization and the speed test}},  {\em Phys.
  Lett.} {\bf B790} (2019) 453--457,
  [\href{http://arxiv.org/abs/1808.00492}{{\tt arXiv:1808.00492}}].

\bibitem{AndresBletranDelhom}
A.~D. D.~Andres, J. Beltran~Jimenez, {\it {Anisotropic deformations in
  projectively invariant metric-affine theories of gravity}},  {\em {To
  Appear}} (2020).

\bibitem{Olmo:2006zu}
G.~J. Olmo, {\it {Violation of the Equivalence Principle in Modified Theories
  of Gravity}},  {\em Phys. Rev. Lett.} {\bf 98} (2007) 061101,
  [\href{http://arxiv.org/abs/gr-qc/0612002}{{\tt gr-qc/0612002}}].

\bibitem{Capozziello:2013uya}
S.~Capozziello, T.~Harko, F.~S.~N. Lobo, and G.~J. Olmo, {\it {Hybrid modified
  gravity unifying local tests, galactic dynamics and late-time cosmic
  acceleration}},  {\em Int. J. Mod. Phys.} {\bf D22} (2013) 1342006,
  [\href{http://arxiv.org/abs/1305.3756}{{\tt arXiv:1305.3756}}].

\bibitem{Capozziello:2012ny}
S.~Capozziello, T.~Harko, T.~S. Koivisto, F.~S.~N. Lobo, and G.~J. Olmo, {\it
  {Cosmology of hybrid metric-Palatini f(X)-gravity}},  {\em JCAP} {\bf 1304}
  (2013) 011, [\href{http://arxiv.org/abs/1209.2895}{{\tt arXiv:1209.2895}}].

\bibitem{Boehmer:2013oxa}
C.~G. B{\"o}hmer, F.~S.~N. Lobo, and N.~Tamanini, {\it {Einstein static
  Universe in hybrid metric-Palatini gravity}},  {\em Phys. Rev.} {\bf D88}
  (2013), no.~10 104019, [\href{http://arxiv.org/abs/1305.0025}{{\tt
  arXiv:1305.0025}}]. [Phys. Rev.D88,104019(2013)].

\bibitem{Borowiec:2014wva}
A.~Borowiec, S.~Capozziello, M.~De~Laurentis, F.~S.~N. Lobo, A.~Paliathanasis,
  M.~Paolella, and A.~Wojnar, {\it {Invariant solutions and Noether symmetries
  in Hybrid Gravity}},  {\em Phys. Rev.} {\bf D91} (2015), no.~2 023517,
  [\href{http://arxiv.org/abs/1407.4313}{{\tt arXiv:1407.4313}}].

\bibitem{Carloni:2015bua}
S.~Carloni, T.~Koivisto, and F.~S.~N. Lobo, {\it {Dynamical system analysis of
  hybrid metric-Palatini cosmologies}},  {\em Phys. Rev.} {\bf D92} (2015),
  no.~6 064035, [\href{http://arxiv.org/abs/1507.04306}{{\tt
  arXiv:1507.04306}}].

\bibitem{Olmo:2005zr}
G.~J. Olmo, {\it {The Gravity Lagrangian according to solar system
  experiments}},  {\em Phys. Rev. Lett.} {\bf 95} (2005) 261102,
  [\href{http://arxiv.org/abs/gr-qc/0505101}{{\tt gr-qc/0505101}}].

\bibitem{Olmo:2005hc}
G.~J. Olmo, {\it {Post-Newtonian constraints on f(R) cosmologies in metric and
  Palatini formalism}},  {\em Phys. Rev.} {\bf D72} (2005) 083505,
  [\href{http://arxiv.org/abs/gr-qc/0505135}{{\tt gr-qc/0505135}}].

\bibitem{Koivisto:2009jn}
T.~S. Koivisto, {\it {Cosmology of modified (but second order) gravity}},  {\em
  AIP Conf. Proc.} {\bf 1206} (2010) 79--96,
  [\href{http://arxiv.org/abs/0910.4097}{{\tt arXiv:0910.4097}}].

\bibitem{fRgravity6}
S.~Capozziello and M.~De~Laurentis, {\it {Extended Theories of Gravity}},  {\em
  Phys. Rept.} {\bf 509} (2011) 167--321,
  [\href{http://arxiv.org/abs/1108.6266}{{\tt arXiv:1108.6266}}].

\bibitem{Lima:2014aza}
N.~A. Lima, {\it {Dynamics of Linear Perturbations in the hybrid
  metric-Palatini gravity}},  {\em Phys. Rev.} {\bf D89} (2014), no.~8 083527,
  [\href{http://arxiv.org/abs/1402.4458}{{\tt arXiv:1402.4458}}].

\bibitem{Lima:2015nma}
N.~A. Lima and V.~S.~Barreto, {\it {Constraints on Hybrid Metric-palatini
  Gravity from Background Evolution}},  {\em Astrophys. J.} {\bf 818} (2016),
  no.~2 186, [\href{http://arxiv.org/abs/1501.05786}{{\tt arXiv:1501.05786}}].

\bibitem{Ma:1995ey}
C.-P. Ma and E.~Bertschinger, {\it {Cosmological perturbation theory in the
  synchronous and conformal Newtonian gauges}},  {\em Astrophys. J.} {\bf 455}
  (1995) 7--25, [\href{http://arxiv.org/abs/astro-ph/9506072}{{\tt
  astro-ph/9506072}}].

\bibitem{coupled}
T.~S. Koivisto, E.~N. Saridakis, and N.~Tamanini, {\it {Scalar-Fluid theories:
  cosmological perturbations and large-scale structure}},  {\em JCAP} {\bf
  1509} (2015) 047, [\href{http://arxiv.org/abs/1505.07556}{{\tt
  arXiv:1505.07556}}].

\bibitem{Capozziello:2012qt}
S.~Capozziello, T.~Harko, T.~S. Koivisto, F.~S.~N. Lobo, and G.~J. Olmo, {\it
  {The virial theorem and the dark matter problem in hybrid metric-Palatini
  gravity}},  {\em JCAP} {\bf 1307} (2013) 024,
  [\href{http://arxiv.org/abs/1212.5817}{{\tt arXiv:1212.5817}}].

\bibitem{Capozziello:2013yha}
S.~Capozziello, T.~Harko, T.~S. Koivisto, F.~S.~N. Lobo, and G.~J. Olmo, {\it
  {Galactic rotation curves in hybrid metric-Palatini gravity}},  {\em
  Astropart. Phys.} {\bf 50-52} (2013) 65--75,
  [\href{http://arxiv.org/abs/1307.0752}{{\tt arXiv:1307.0752}}].

\bibitem{Lobo:2008sg}
F.~S.~N. Lobo, {\it {The Dark side of gravity: Modified theories of gravity}},
  \href{http://arxiv.org/abs/0807.1640}{{\tt arXiv:0807.1640}}.

\bibitem{Einstein1928}
A.~Einstein, {\it {Riemann-Geometrie mit Aufrechterhaltung des Begriffes des
  Fernparallelismus}},  {\em Sitzber. Preuss. Akad. Wiss.} {\bf 17} (1928)
  217--221.

\bibitem{Krssak:2015oua}
M.~Kr{\v s}{\v s}{\'a}k and E.~N. Saridakis, {\it {The covariant formulation of
  f(T) gravity}},  {\em Class. Quant. Grav.} {\bf 33} (2016), no.~11 115009,
  [\href{http://arxiv.org/abs/1510.08432}{{\tt arXiv:1510.08432}}].

\bibitem{Golovnev:2017dox}
A.~Golovnev, T.~Koivisto, and M.~Sandstad, {\it {On the covariance of
  teleparallel gravity theories}},  {\em Class. Quant. Grav.} {\bf 34} (2017),
  no.~14 145013, [\href{http://arxiv.org/abs/1701.06271}{{\tt
  arXiv:1701.06271}}].

\bibitem{Blixt:2018znp}
D.~Blixt, M.~Hohmann, and C.~Pfeifer, {\it {Hamiltonian and primary constraints
  of new general relativity}},  {\em Phys. Rev.} {\bf D99} (2019), no.~8
  084025, [\href{http://arxiv.org/abs/1811.11137}{{\tt arXiv:1811.11137}}].

\bibitem{Blixt:2019mkt}
D.~Blixt, M.~Hohmann, and C.~Pfeifer, {\it {On the gauge fixing in the
  Hamiltonian analysis of general teleparallel theories}},  {\em Universe} {\bf
  5} (2019), no.~6 143, [\href{http://arxiv.org/abs/1905.01048}{{\tt
  arXiv:1905.01048}}].

\bibitem{Hayashi:1967se}
K.~Hayashi and T.~Nakano, {\it {Extended translation invariance and associated
  gauge fields}},  {\em Prog. Theor. Phys.} {\bf 38} (1967) 491--507.
  [,354(1967)].

\bibitem{Cho:1975dh}
Y.~M. Cho, {\it {Einstein Lagrangian as the Translational Yang-Mills
  Lagrangian}},  {\em Phys. Rev.} {\bf D14} (1976) 2521. [,393(1975)].

\bibitem{Hayashi:1979qx}
K.~Hayashi and T.~Shirafuji, {\it {New General Relativity}},  {\em Phys. Rev.}
  {\bf D19} (1979) 3524--3553. [,409(1979)].

\bibitem{Baez:2012bn}
J.~C. Baez and D.~K. Wise, {\it {Teleparallel Gravity as a Higher Gauge
  Theory}},  {\em Commun. Math. Phys.} {\bf 333} (2015), no.~1 153--186,
  [\href{http://arxiv.org/abs/1204.4339}{{\tt arXiv:1204.4339}}].

\bibitem{Fontanini:2018krt}
M.~Fontanini, E.~Huguet, and M.~Le~Delliou, {\it {Teleparallel gravity
  equivalent of general relativity as a gauge theory: Translation or Cartan
  connection?}},  {\em Phys. Rev.} {\bf D99} (2019), no.~6 064006,
  [\href{http://arxiv.org/abs/1811.03810}{{\tt arXiv:1811.03810}}].

\bibitem{Pereira:2019woq}
J.~G. Pereira and Y.~N. Obukhov, {\it {Gauge Structure of Teleparallel
  Gravity}},  {\em Universe} {\bf 5} (2019), no.~6 139,
  [\href{http://arxiv.org/abs/1906.06287}{{\tt arXiv:1906.06287}}].

\bibitem{Blagojevic:2002du}
M.~Blagojevic, {\em Gravitation and Gauge Symmetries}.
\newblock Series in High Energy Physics, Cosmology and Gravitation. CRC Press,
  2001.

\bibitem{Sotiriou:2010mv}
T.~P. Sotiriou, B.~Li, and J.~D. Barrow, {\it {Generalizations of teleparallel
  gravity and local Lorentz symmetry}},  {\em Phys. Rev.} {\bf D83} (2011)
  104030, [\href{http://arxiv.org/abs/1012.4039}{{\tt arXiv:1012.4039}}].

\bibitem{Ferraro:2011us}
R.~Ferraro and F.~Fiorini, {\it {Non trivial frames for f(T) theories of
  gravity and beyond}},  {\em Phys. Lett.} {\bf B702} (2011) 75--80,
  [\href{http://arxiv.org/abs/1103.0824}{{\tt arXiv:1103.0824}}].

\bibitem{Tamanini:2012hg}
N.~Tamanini and C.~G. Boehmer, {\it {Good and bad tetrads in f(T) gravity}},
  {\em Phys. Rev.} {\bf D86} (2012) 044009,
  [\href{http://arxiv.org/abs/1204.4593}{{\tt arXiv:1204.4593}}].

\bibitem{Tamanini:2013xya}
N.~Tamanini and C.~G. B{\"o}hmer, {\it {Definition of Good Tetrads for f(T)
  Gravity}},  in {\em {Proceedings, 13th Marcel Grossmann Meeting on Recent
  Developments in Theoretical and Experimental General Relativity,
  Astrophysics, and Relativistic Field Theories (MG13): Stockholm, Sweden, July
  1-7, 2012}}, pp.~1282--1284, 2015.
\newblock \href{http://arxiv.org/abs/1304.0672}{{\tt arXiv:1304.0672}}.

\bibitem{Kopczynski_1982}
W.~Kopczynski, {\it Problems with metric-teleparallel theories of gravitation},
   {\em Journal of Physics A: Mathematical and General} {\bf 15} (feb, 1982)
  493--506.

\bibitem{Nester_1988}
J.~M. Nester, {\it Is there really a problem with the teleparallel theory?},
  {\em Classical and Quantum Gravity} {\bf 5} (jul, 1988) 1003--1010.

\bibitem{Chen:1998ad}
H.~Chen, J.~M. Nester, and H.-J. Yo, {\it {Acausal PGT modes and the nonlinear
  constraint effect}},  {\em Acta Phys. Polon.} {\bf B29} (1998) 961--970.

\bibitem{Ong:2013qja}
Y.~C. Ong, K.~Izumi, J.~M. Nester, and P.~Chen, {\it {Problems with Propagation
  and Time Evolution in f(T) Gravity}},  {\em Phys. Rev.} {\bf D88} (2013)
  024019, [\href{http://arxiv.org/abs/1303.0993}{{\tt arXiv:1303.0993}}].

\bibitem{Ferraro:2014owa}
R.~Ferraro and F.~Fiorini, {\it {Remnant group of local Lorentz transformations
  in $\mathcal{f}(T)$ theories}},  {\em Phys. Rev.} {\bf D91} (2015), no.~6
  064019, [\href{http://arxiv.org/abs/1412.3424}{{\tt arXiv:1412.3424}}].

\bibitem{Chen:2014qtl}
P.~Chen, K.~Izumi, J.~M. Nester, and Y.~C. Ong, {\it {Remnant Symmetry,
  Propagation and Evolution in $f$(T) Gravity}},  {\em Phys. Rev.} {\bf D91}
  (2015), no.~6 064003, [\href{http://arxiv.org/abs/1412.8383}{{\tt
  arXiv:1412.8383}}].

\bibitem{Bejarano:2019fii}
C.~Bejarano, R.~Ferraro, F.~Fiorini, and M.~J. Guzm{\'a}n, {\it {Reflections on
  the covariance of modified teleparallel theories of gravity}},  {\em
  Universe} {\bf 5} (2019) 158, [\href{http://arxiv.org/abs/1905.09913}{{\tt
  arXiv:1905.09913}}].

\bibitem{Ferraro:2020tqk}
R.~Ferraro and M.~J. Guzmán, {\it {Pseudoinvariance and the extra degree of
  freedom in $f(T)$ gravity}},  {\em Phys. Rev. D} {\bf 101} (2020), no.~8
  084017, [\href{http://arxiv.org/abs/2001.08137}{{\tt arXiv:2001.08137}}].

\bibitem{Hohmann:2018rwf}
M.~Hohmann, L.~J{\"a}rv, and U.~Ualikhanova, {\it {Covariant formulation of
  scalar-torsion gravity}},  {\em Phys. Rev.} {\bf D97} (2018), no.~10 104011,
  [\href{http://arxiv.org/abs/1801.05786}{{\tt arXiv:1801.05786}}].

\bibitem{Formiga:2013dja}
J.~B. Formiga, {\it {Comment on ``Metric-affine approach to teleparallel
  gravity''}},  {\em Phys. Rev.} {\bf D88} (2013), no.~6 068501,
  [\href{http://arxiv.org/abs/1306.4964}{{\tt arXiv:1306.4964}}].

\bibitem{BeltranJimenez:2020sih}
J.~Beltran~Jimenez, L.~Heisenberg, and T.~Koivisto, {\it {The coupling of
  matter and spacetime geometry}},  \href{http://arxiv.org/abs/2004.04606}{{\tt
  arXiv:2004.04606}}.

\bibitem{Obukhov:2006ge}
Y.~N. Obukhov and G.~F. Rubilar, {\it {Invariant conserved currents in gravity
  theories with local Lorentz and diffeomorphism symmetry}},  {\em Phys. Rev.}
  {\bf D74} (2006) 064002, [\href{http://arxiv.org/abs/gr-qc/0608064}{{\tt
  gr-qc/0608064}}].

\bibitem{Krssak:2015lba}
M.~Kr{\v s}{\v s}{\'a}k, {\it {Holographic Renormalization in Teleparallel
  Gravity}},  {\em Eur. Phys. J.} {\bf C77} (2017), no.~1 44,
  [\href{http://arxiv.org/abs/1510.06676}{{\tt arXiv:1510.06676}}].

\bibitem{Maluf:2018coz}
J.~W. Maluf, S.~C. Ulhoa, and J.~F. da~Rocha-Neto, {\it {Difficulties of
  Teleparallel Theories of Gravity with Local Lorentz Symmetry}},
  \href{http://arxiv.org/abs/1811.06876}{{\tt arXiv:1811.06876}}.

\bibitem{Nojiri:2006ri}
S.~Nojiri and S.~D. Odintsov, {\it {Introduction to modified gravity and
  gravitational alternative for dark energy}},  {\em eConf} {\bf C0602061}
  (2006) 06, [\href{http://arxiv.org/abs/hep-th/0601213}{{\tt
  hep-th/0601213}}]. [Int. J. Geom. Meth. Mod. Phys.4,115(2007)].

\bibitem{Myung:2016zdl}
Y.~S. Myung, {\it {Propagating Degrees of Freedom in $f(R)$ Gravity}},  {\em
  Adv. High Energy Phys.} {\bf 2016} (2016) 3901734,
  [\href{http://arxiv.org/abs/1608.01764}{{\tt arXiv:1608.01764}}].

\bibitem{Berry:2011pb}
C.~P.~L. Berry and J.~R. Gair, {\it {Linearized f(R) Gravity: Gravitational
  Radiation and Solar System Tests}},  {\em Phys. Rev.} {\bf D83} (2011)
  104022, [\href{http://arxiv.org/abs/1104.0819}{{\tt arXiv:1104.0819}}].
  [Erratum: Phys. Rev.D85,089906(2012)].

\bibitem{Capozziello:2008rq}
S.~Capozziello, C.~Corda, and M.~F. De~Laurentis, {\it {Massive gravitational
  waves from f(R) theories of gravity: Potential detection with LISA}},  {\em
  Phys. Lett.} {\bf B669} (2008) 255--259,
  [\href{http://arxiv.org/abs/0812.2272}{{\tt arXiv:0812.2272}}].

\bibitem{Bengochea:2008gz}
G.~R. Bengochea and R.~Ferraro, {\it {Dark torsion as the cosmic speed-up}},
  {\em Phys. Rev.} {\bf D79} (2009) 124019,
  [\href{http://arxiv.org/abs/0812.1205}{{\tt arXiv:0812.1205}}].

\bibitem{Ferraro:2008ey}
R.~Ferraro and F.~Fiorini, {\it {On Born-Infeld Gravity in Weitzenbock
  spacetime}},  {\em Phys. Rev.} {\bf D78} (2008) 124019,
  [\href{http://arxiv.org/abs/0812.1981}{{\tt arXiv:0812.1981}}].

\bibitem{Nassur:2016yhc}
S.~B. Nassur, C.~Ainamon, M.~J.~S. Houndjo, and J.~Tossa, {\it {Unimodular f(T)
  gravity}},  {\em Eur. Phys. J. Plus} {\bf 131} (2016), no.~12 420,
  [\href{http://arxiv.org/abs/1602.03172}{{\tt arXiv:1602.03172}}].

\bibitem{Krssak:2017nlv}
M.~Krssak, {\it {Variational Problem and Bigravity Nature of Modified
  Teleparallel Theories}},  \href{http://arxiv.org/abs/1705.01072}{{\tt
  arXiv:1705.01072}}.

\bibitem{Hohmann:2017duq}
M.~Hohmann, L.~J{\"a}rv, M.~Kr{\v s}{\v s}{\'a}k, and C.~Pfeifer, {\it
  {Teleparallel theories of gravity as analogue of nonlinear electrodynamics}},
   {\em Phys. Rev.} {\bf D97} (2018), no.~10 104042,
  [\href{http://arxiv.org/abs/1711.09930}{{\tt arXiv:1711.09930}}].

\bibitem{Hohmann:2019nat}
M.~Hohmann, L.~Järv, M.~Kr\v~s\v sák, and C.~Pfeifer, {\it {Modified
  teleparallel theories of gravity in symmetric spacetimes}},  {\em Phys. Rev.
  D} {\bf 100} (2019), no.~8 084002,
  [\href{http://arxiv.org/abs/1901.05472}{{\tt arXiv:1901.05472}}].

\bibitem{MuellerHoissen:1983vc}
F.~Mueller-Hoissen and J.~Nitsch, {\it {TELEPARALLELISM - A VIABLE THEORY OF
  GRAVITY?}},  {\em Phys. Rev.} {\bf D28} (1983) 718--728.

\bibitem{1982JPhA...15..493K}
W.~{Kopczynski}, {\it {Problems with metric-teleparallel theories of
  gravitation}},  {\em Journal of Physics A Mathematical General} {\bf 15}
  (Feb., 1982) 493--506.

\bibitem{1988CQGra...5.1003N}
J.~M. {Nester}, {\it {Is there really a problem with the teleparallel
  theory?}},  {\em Classical and Quantum Gravity} {\bf 5} (July, 1988)
  1003--1010.

\bibitem{Kawai:1989qt}
T.~Kawai and N.~Toma, {\it {Singularities of 'Schwarzschild Like' Space-time in
  New General Relativity}},  {\em Prog. Theor. Phys.} {\bf 83} (1990) 1.

\bibitem{Fukui:1981si}
M.~Fukui and K.~Hayashi, {\it {Axially Symmetric Solutions of New General
  Relativity}},  {\em Prog. Theor. Phys.} {\bf 66} (1981) 1500.

\bibitem{Fukui:1984rj}
M.~Fukui and J.~Masukawa, {\it {WEAK FIELD APPROXIMATION OF NEW GENERAL
  RELATIVITY}},  {\em Prog. Theor. Phys.} {\bf 73} (1985) 973.

\bibitem{Hohmann:2018jso}
M.~Hohmann, M.~Kr{\v s}{\v s}{\'a}k, C.~Pfeifer, and U.~Ualikhanova, {\it
  {Propagation of gravitational waves in teleparallel gravity theories}},  {\em
  Phys. Rev.} {\bf D98} (2018), no.~12 124004,
  [\href{http://arxiv.org/abs/1807.04580}{{\tt arXiv:1807.04580}}].

\bibitem{Okolow:2011np}
A.~Okolow and J.~Swiezewski, {\it {Hamiltonian formulation of a simple theory
  of the teleparallel geometry}},  {\em Class. Quant. Grav.} {\bf 29} (2012)
  045008, [\href{http://arxiv.org/abs/1111.5490}{{\tt arXiv:1111.5490}}].

\bibitem{Nester:2017wau}
Y.~C. Ong and J.~M. Nester, {\it {Counting Components in the Lagrange
  Multiplier Formulation of Teleparallel Theories}},  {\em Eur. Phys. J.} {\bf
  C78} (2018), no.~7 568, [\href{http://arxiv.org/abs/1709.00068}{{\tt
  arXiv:1709.00068}}].

\bibitem{ortin2004gravity}
T.~Ort{\'\i}n, {\em Gravity and Strings}.
\newblock Cambridge Monographs on Mathematical Physics. Cambridge University
  Press, 2004.

\bibitem{Maluf:2011kf}
J.~W. Maluf and F.~F. Faria, {\it {Conformally invariant teleparallel theories
  of gravity}},  {\em Phys. Rev.} {\bf D85} (2012) 027502,
  [\href{http://arxiv.org/abs/1110.3095}{{\tt arXiv:1110.3095}}].

\bibitem{Bahamonde:2017wwk}
S.~Bahamonde, C.~G. B{\"o}hmer, and M.~Kr{\v s}{\v s}{\'a}k, {\it {New classes
  of modified teleparallel gravity models}},  {\em Phys. Lett.} {\bf B775}
  (2017) 37--43, [\href{http://arxiv.org/abs/1706.04920}{{\tt
  arXiv:1706.04920}}].

\bibitem{Bamba:2013jqa}
K.~Bamba, S.~D. Odintsov, and D.~S{\'a}ez-G{\'o}mez, {\it {Conformal symmetry
  and accelerating cosmology in teleparallel gravity}},  {\em Phys. Rev.} {\bf
  D88} (2013) 084042, [\href{http://arxiv.org/abs/1308.5789}{{\tt
  arXiv:1308.5789}}].

\bibitem{Yang:2010ji}
R.-J. Yang, {\it {Conformal transformation in $f(T)$ theories}},  {\em EPL}
  {\bf 93} (2011), no.~6 60001, [\href{http://arxiv.org/abs/1010.1376}{{\tt
  arXiv:1010.1376}}].

\bibitem{Bahamonde:2016cul}
S.~Bahamonde, M.~Zubair, and G.~Abbas, {\it {Thermodynamics and cosmological
  reconstruction in $f(T,B)$ gravity}},  {\em Phys. Dark Univ.} {\bf 19} (2018)
  78--90, [\href{http://arxiv.org/abs/1609.08373}{{\tt arXiv:1609.08373}}].

\bibitem{Bahamonde:2016grb}
S.~Bahamonde and S.~Capozziello, {\it {Noether Symmetry Approach in $f(T,B)$
  teleparallel cosmology}},  {\em Eur. Phys. J.} {\bf C77} (2017), no.~2 107,
  [\href{http://arxiv.org/abs/1612.01299}{{\tt arXiv:1612.01299}}].

\bibitem{Otalora:2016dxe}
G.~Otalora and E.~N. Saridakis, {\it {Modified teleparallel gravity with
  higher-derivative torsion terms}},  {\em Phys. Rev.} {\bf D94} (2016), no.~8
  084021, [\href{http://arxiv.org/abs/1605.04599}{{\tt arXiv:1605.04599}}].

\bibitem{Capozziello:2016eaz}
S.~Capozziello, M.~De~Laurentis, and K.~F. Dialektopoulos, {\it {Noether
  symmetries in Gauss--Bonnet-teleparallel cosmology}},  {\em Eur. Phys. J.}
  {\bf C76} (2016), no.~11 629, [\href{http://arxiv.org/abs/1609.09289}{{\tt
  arXiv:1609.09289}}].

\bibitem{Bahamonde:2016kba}
S.~Bahamonde and C.~G. B{\"o}hmer, {\it {Modified teleparallel theories of
  gravity: Gauss--Bonnet and trace extensions}},  {\em Eur. Phys. J.} {\bf C76}
  (2016), no.~10 578, [\href{http://arxiv.org/abs/1606.05557}{{\tt
  arXiv:1606.05557}}].

\bibitem{Kofinas:2014owa}
G.~Kofinas and E.~N. Saridakis, {\it {Teleparallel equivalent of Gauss-Bonnet
  gravity and its modifications}},  {\em Phys. Rev.} {\bf D90} (2014) 084044,
  [\href{http://arxiv.org/abs/1404.2249}{{\tt arXiv:1404.2249}}].

\bibitem{Kofinas:2014aka}
G.~Kofinas, G.~Leon, and E.~N. Saridakis, {\it Dynamical behavior in $f(t,t_g)$
  cosmology},  {\em Classical and Quantum Gravity} {\bf 31} (2014), no.~17
  175011, [\href{http://arxiv.org/abs/1404.7100}{{\tt arXiv:1404.7100}}].

\bibitem{Harko:2014aja}
T.~Harko, F.~S.~N. Lobo, G.~Otalora, and E.~N. Saridakis, {\it
  {$f(T,\mathcal{T})$ gravity and cosmology}},  {\em JCAP} {\bf 1412} (2014)
  021, [\href{http://arxiv.org/abs/1405.0519}{{\tt arXiv:1405.0519}}].

\bibitem{Harko:2011kv}
T.~Harko, F.~S.~N. Lobo, S.~Nojiri, and S.~D. Odintsov, {\it {$f(R,T)$
  gravity}},  {\em Phys. Rev.} {\bf D84} (2011) 024020,
  [\href{http://arxiv.org/abs/1104.2669}{{\tt arXiv:1104.2669}}].

\bibitem{Harko:2014sja}
T.~Harko, F.~S.~N. Lobo, G.~Otalora, and E.~N. Saridakis, {\it {Nonminimal
  torsion-matter coupling extension of f(T) gravity}},  {\em Phys. Rev.} {\bf
  D89} (2014) 124036, [\href{http://arxiv.org/abs/1404.6212}{{\tt
  arXiv:1404.6212}}].

\bibitem{Bahamonde:2017ifa}
S.~Bahamonde, {\it {Generalised nonminimally gravity-matter coupled theory}},
  {\em Eur. Phys. J.} {\bf C78} (2018), no.~4 326,
  [\href{http://arxiv.org/abs/1709.05319}{{\tt arXiv:1709.05319}}].

\bibitem{Donoghue:1994dn}
J.~F. Donoghue, {\it {General relativity as an effective field theory: The
  leading quantum corrections}},  {\em Phys. Rev.} {\bf D50} (1994) 3874--3888,
  [\href{http://arxiv.org/abs/gr-qc/9405057}{{\tt gr-qc/9405057}}].

\bibitem{Giddings:2006sj}
S.~B. Giddings, {\it {Black hole information, unitarity, and nonlocality}},
  {\em Phys. Rev.} {\bf D74} (2006) 106005,
  [\href{http://arxiv.org/abs/hep-th/0605196}{{\tt hep-th/0605196}}].

\bibitem{Bahamonde:2017bps}
S.~Bahamonde, S.~Capozziello, M.~Faizal, and R.~C. Nunes, {\it {Nonlocal
  Teleparallel Cosmology}},  {\em Eur. Phys. J.} {\bf C77} (2017), no.~9 628,
  [\href{http://arxiv.org/abs/1709.02692}{{\tt arXiv:1709.02692}}].

\bibitem{Bahamonde:2017sdo}
S.~Bahamonde, S.~Capozziello, and K.~F. Dialektopoulos, {\it {Constraining
  Generalized Non-local Cosmology from Noether Symmetries}},  {\em Eur. Phys.
  J.} {\bf C77} (2017), no.~11 722,
  [\href{http://arxiv.org/abs/1708.06310}{{\tt arXiv:1708.06310}}].

\bibitem{Perrotta:1999am}
F.~Perrotta, C.~Baccigalupi, and S.~Matarrese, {\it {Extended quintessence}},
  {\em Phys. Rev.} {\bf D61} (1999) 023507,
  [\href{http://arxiv.org/abs/astro-ph/9906066}{{\tt astro-ph/9906066}}].

\bibitem{Bahamonde:2019shr}
S.~Bahamonde, K.~F. Dialektopoulos, and J.~L. Said, {\it {Can Horndeski Theory
  be recast using Teleparallel Gravity?}},
  \href{http://arxiv.org/abs/1904.10791}{{\tt arXiv:1904.10791}}.

\bibitem{Bahamonde:2019ipm}
S.~Bahamonde, K.~F. Dialektopoulos, V.~Gakis, and J.~L. Said, {\it {Reviving
  Horndeski Theory using Teleparallel Gravity after GW170817}},
  \href{http://arxiv.org/abs/1907.10057}{{\tt arXiv:1907.10057}}.

\bibitem{Hohmann:2019gmt}
M.~Hohmann, {\it {Disformal Transformations in Scalar-Torsion Gravity}},  {\em
  Universe} {\bf 5} (2019), no.~7 167,
  [\href{http://arxiv.org/abs/1905.00451}{{\tt arXiv:1905.00451}}].

\bibitem{Kreisch:2017uet}
C.~D. Kreisch and E.~Komatsu, {\it {Cosmological Constraints on Horndeski
  Gravity in Light of GW170817}},  {\em JCAP} {\bf 1812} (2018), no.~12 030,
  [\href{http://arxiv.org/abs/1712.02710}{{\tt arXiv:1712.02710}}].

\bibitem{Gong:2017kim}
Y.~Gong, E.~Papantonopoulos, and Z.~Yi, {\it {Constraints on scalar--tensor
  theory of gravity by the recent observational results on gravitational
  waves}},  {\em Eur. Phys. J.} {\bf C78} (2018), no.~9 738,
  [\href{http://arxiv.org/abs/1711.04102}{{\tt arXiv:1711.04102}}].

\bibitem{TheLIGOScientific:2017qsa}
{\bf LIGO Scientific, Virgo} Collaboration, B.~P. Abbott et~al., {\it
  {GW170817: Observation of Gravitational Waves from a Binary Neutron Star
  Inspiral}},  {\em Phys. Rev. Lett.} {\bf 119} (2017), no.~16 161101,
  [\href{http://arxiv.org/abs/1710.05832}{{\tt arXiv:1710.05832}}].

\bibitem{Geng:2011aj}
C.-Q. Geng, C.-C. Lee, E.~N. Saridakis, and Y.-P. Wu, {\it {``Teleparallel''
  dark energy}},  {\em Phys. Lett.} {\bf B704} (2011) 384--387,
  [\href{http://arxiv.org/abs/1109.1092}{{\tt arXiv:1109.1092}}].

\bibitem{Kofinas:2015hla}
G.~Kofinas, E.~Papantonopoulos, and E.~N. Saridakis, {\it {Self-Gravitating
  Spherically Symmetric Solutions in Scalar-Torsion Theories}},  {\em Phys.
  Rev.} {\bf D91} (2015), no.~10 104034,
  [\href{http://arxiv.org/abs/1501.00365}{{\tt arXiv:1501.00365}}].

\bibitem{Geng:2011ka}
C.-Q. Geng, C.-C. Lee, and E.~N. Saridakis, {\it {Observational Constraints on
  Teleparallel Dark Energy}},  {\em JCAP} {\bf 1201} (2012) 002,
  [\href{http://arxiv.org/abs/1110.0913}{{\tt arXiv:1110.0913}}].

\bibitem{Zubair:2016uhx}
M.~Zubair, S.~Bahamonde, and M.~Jamil, {\it {Generalized Second Law of
  Thermodynamic in Modified Teleparallel Theory}},  {\em Eur. Phys. J.} {\bf
  C77} (2017), no.~7 472, [\href{http://arxiv.org/abs/1604.02996}{{\tt
  arXiv:1604.02996}}].

\bibitem{Bahamonde:2018miw}
S.~Bahamonde, M.~Marciu, and P.~Rudra, {\it {Generalised teleparallel quintom
  dark energy non-minimally coupled with the scalar torsion and a boundary
  term}},  {\em JCAP} {\bf 1804} (2018), no.~04 056,
  [\href{http://arxiv.org/abs/1802.09155}{{\tt arXiv:1802.09155}}].

\bibitem{Jarv:2015odu}
L.~Jarv and A.~Toporensky, {\it {General relativity as an attractor for
  scalar-torsion cosmology}},  {\em Phys. Rev.} {\bf D93} (2016), no.~2 024051,
  [\href{http://arxiv.org/abs/1511.03933}{{\tt arXiv:1511.03933}}].

\bibitem{Kofinas:2015zaa}
G.~Kofinas, {\it {Hyperscaling violating black holes in scalar-torsion
  theories}},  {\em Phys. Rev.} {\bf D92} (2015), no.~8 084022,
  [\href{http://arxiv.org/abs/1507.07434}{{\tt arXiv:1507.07434}}].

\bibitem{Horvat:2014xwa}
D.~Horvat, S.~Iliji{\'c}, A.~Kirin, and Z.~Naran{\v c}i{\'c}, {\it
  {Nonminimally coupled scalar field in teleparallel gravity: boson stars}},
  {\em Class. Quant. Grav.} {\bf 32} (2015), no.~3 035023,
  [\href{http://arxiv.org/abs/1407.2067}{{\tt arXiv:1407.2067}}].

\bibitem{Jamil:2012vb}
M.~Jamil, D.~Momeni, and R.~Myrzakulov, {\it {Stability of a non-minimally
  conformally coupled scalar field in F(T) cosmology}},  {\em Eur. Phys. J.}
  {\bf C72} (2012) 2075, [\href{http://arxiv.org/abs/1208.0025}{{\tt
  arXiv:1208.0025}}].

\bibitem{Wu:2011kh}
Y.-P. Wu and C.-Q. Geng, {\it {Primordial Fluctuations within
  Teleparallelism}},  {\em Phys. Rev.} {\bf D86} (2012) 104058,
  [\href{http://arxiv.org/abs/1110.3099}{{\tt arXiv:1110.3099}}].

\bibitem{Wei:2011yr}
H.~Wei, {\it {Dynamics of Teleparallel Dark Energy}},  {\em Phys. Lett.} {\bf
  B712} (2012) 430--436, [\href{http://arxiv.org/abs/1109.6107}{{\tt
  arXiv:1109.6107}}].

\bibitem{Bahamonde:2015hza}
S.~Bahamonde and M.~Wright, {\it {Teleparallel quintessence with a nonminimal
  coupling to a boundary term}},  {\em Phys. Rev.} {\bf D92} (2015), no.~8
  084034, [\href{http://arxiv.org/abs/1508.06580}{{\tt arXiv:1508.06580}}].
  [Erratum: Phys. Rev.D93,no.10,109901(2016)].

\bibitem{Hohmann:2018vle}
M.~Hohmann, {\it {Scalar-torsion theories of gravity I: general formalism and
  conformal transformations}},  {\em Phys. Rev.} {\bf D98} (2018), no.~6
  064002, [\href{http://arxiv.org/abs/1801.06528}{{\tt arXiv:1801.06528}}].

\bibitem{Hohmann:2018dqh}
M.~Hohmann and C.~Pfeifer, {\it {Scalar-torsion theories of gravity II: $L(T,
  X, Y, \phi)$ theory}},  {\em Phys. Rev.} {\bf D98} (2018), no.~6 064003,
  [\href{http://arxiv.org/abs/1801.06536}{{\tt arXiv:1801.06536}}].

\bibitem{Hohmann:2018ijr}
M.~Hohmann, {\it {Scalar-torsion theories of gravity III: analogue of
  scalar-tensor gravity and conformal invariants}},  {\em Phys. Rev.} {\bf D98}
  (2018), no.~6 064004, [\href{http://arxiv.org/abs/1801.06531}{{\tt
  arXiv:1801.06531}}].

\bibitem{Hayward:1981bk}
J.~Hayward, {\it {SCALAR TETRAD THEORIES OF GRAVITY}},  {\em Gen. Rel. Grav.}
  {\bf 13} (1981) 43--55.

\bibitem{Chen:2014qsa}
Z.-C. Chen, Y.~Wu, and H.~Wei, {\it {Post-Newtonian Approximation of
  Teleparallel Gravity Coupled with a Scalar Field}},  {\em Nucl. Phys.} {\bf
  B894} (2015) 422--438, [\href{http://arxiv.org/abs/1410.7715}{{\tt
  arXiv:1410.7715}}].

\bibitem{Li:2013oef}
J.-T. Li, Y.-P. Wu, and C.-Q. Geng, {\it {Parametrized post-Newtonian limit of
  the teleparallel dark energy model}},  {\em Phys. Rev.} {\bf D89} (2014),
  no.~4 044040, [\href{http://arxiv.org/abs/1312.4332}{{\tt arXiv:1312.4332}}].

\bibitem{Ualikhanova:2019ygl}
U.~Ualikhanova and M.~Hohmann, {\it {Parameterized post-Newtonian limit of
  general teleparallel gravity theories}},
  \href{http://arxiv.org/abs/1907.08178}{{\tt arXiv:1907.08178}}.

\bibitem{Iorio:2015rla}
L.~Iorio, N.~Radicella, and M.~L. Ruggiero, {\it {Constraining f(T) gravity in
  the Solar System}},  {\em JCAP} {\bf 1508} (2015), no.~08 021,
  [\href{http://arxiv.org/abs/1505.06996}{{\tt arXiv:1505.06996}}].

\bibitem{Iorio:2012cm}
L.~Iorio and E.~N. Saridakis, {\it {Solar system constraints on f(T) gravity}},
   {\em Mon. Not. Roy. Astron. Soc.} {\bf 427} (2012) 1555,
  [\href{http://arxiv.org/abs/1203.5781}{{\tt arXiv:1203.5781}}].

\bibitem{Farrugia:2016xcw}
G.~Farrugia, J.~L. Said, and M.~L. Ruggiero, {\it {Solar System tests in f(T)
  gravity}},  {\em Phys. Rev.} {\bf D93} (2016), no.~10 104034,
  [\href{http://arxiv.org/abs/1605.07614}{{\tt arXiv:1605.07614}}].

\bibitem{Sadjadi:2016kwj}
H.~Mohseni~Sadjadi, {\it {Parameterized post-Newtonian approximation in a
  teleparallel model of dark energy with a boundary term}},  {\em Eur. Phys.
  J.} {\bf C77} (2017), no.~3 191, [\href{http://arxiv.org/abs/1606.04362}{{\tt
  arXiv:1606.04362}}].

\bibitem{Emtsova:2019qsl}
E.~D. Emtsova and M.~Hohmann, {\it {Post-Newtonian limit of scalar-torsion
  theories of gravity as analogue to scalar-curvature theories}},  {\em Phys.
  Rev. D} {\bf 101} (2020), no.~2 024017,
  [\href{http://arxiv.org/abs/1909.09355}{{\tt arXiv:1909.09355}}].

\bibitem{Flathmann:2019khc}
K.~Flathmann and M.~Hohmann, {\it {Post-Newtonian Limit of Generalized
  Scalar-Torsion Theories of Gravity}},  {\em Phys. Rev. D} {\bf 101} (2020),
  no.~2 024005, [\href{http://arxiv.org/abs/1910.01023}{{\tt
  arXiv:1910.01023}}].

\bibitem{Bahamonde:2020cfv}
S.~Bahamonde, K.~F. Dialektopoulos, M.~Hohmann, and J.~Levi~Said, {\it
  {Post-Newtonian limit of Teleparallel Horndeski gravity}},
  \href{http://arxiv.org/abs/2003.11554}{{\tt arXiv:2003.11554}}.

\bibitem{Farrugia:2020fcu}
G.~Farrugia, J.~Levi~Said, and A.~Finch, {\it {Gravitoelectromagnetism, Solar
  System Test and Weak-Field Solutions in $f(T,B)$ Gravity with Observational
  Constraints}},  {\em Universe} {\bf 6} (2020), no.~2 34,
  [\href{http://arxiv.org/abs/2002.08183}{{\tt arXiv:2002.08183}}].

\bibitem{Capozziello:2019msc}
S.~Capozziello, M.~Capriolo, and L.~Caso, {\it {Weak Field Limit and
  Gravitational Waves in $f(T,B)$ Teleparallel Gravity}},  {\em Eur. Phys. J.
  C} {\bf 80} (2020), no.~2 156, [\href{http://arxiv.org/abs/1912.12469}{{\tt
  arXiv:1912.12469}}].

\bibitem{Pourbagher:2019zhq}
A.~Pourbagher and A.~Amani, {\it {Thermodynamics and stability of $f(T,B)$
  gravity with viscous fluid by observational constraints}},  {\em Astrophys.
  Space Sci.} {\bf 364} (2019), no.~8 140,
  [\href{http://arxiv.org/abs/1908.11595}{{\tt arXiv:1908.11595}}].

\bibitem{Cai:2018rzd}
Y.-F. Cai, C.~Li, E.~N. Saridakis, and L.~Xue, {\it {$f(T)$ gravity after
  GW170817 and GRB170817A}},  \href{http://arxiv.org/abs/1801.05827}{{\tt
  arXiv:1801.05827}}.

\bibitem{Farrugia:2018gyz}
G.~Farrugia, J.~L. Said, V.~Gakis, and E.~N. Saridakis, {\it {Gravitational
  Waves in Modified Teleparallel Theories}},  {\em Phys. Rev.} {\bf D97}
  (2018), no.~12 124064, [\href{http://arxiv.org/abs/1804.07365}{{\tt
  arXiv:1804.07365}}].

\bibitem{Bamba:2013ooa}
K.~Bamba, S.~Capozziello, M.~De~Laurentis, S.~Nojiri, and
  D.~S{\'a}ez-G{\'o}mez, {\it {No further gravitational wave modes in $F(T)$
  gravity}},  {\em Phys. Lett.} {\bf B727} (2013) 194--198,
  [\href{http://arxiv.org/abs/1309.2698}{{\tt arXiv:1309.2698}}].

\bibitem{Finch:2018gkh}
A.~Finch and J.~L. Said, {\it {Galactic Rotation Dynamics in f(T) gravity}},
  {\em Eur. Phys. J.} {\bf C78} (2018), no.~7 560,
  [\href{http://arxiv.org/abs/1806.09677}{{\tt arXiv:1806.09677}}].

\bibitem{Aghanim:2018eyx}
{\bf Planck} Collaboration, N.~Aghanim et~al., {\it {Planck 2018 results. VI.
  Cosmological parameters}},  \href{http://arxiv.org/abs/1807.06209}{{\tt
  arXiv:1807.06209}}.

\bibitem{Riess:2019cxk}
A.~G. Riess, S.~Casertano, W.~Yuan, L.~M. Macri, and D.~Scolnic, {\it {Large
  Magellanic Cloud Cepheid Standards Provide a 1\% Foundation for the
  Determination of the Hubble Constant and Stronger Evidence for Physics beyond
  $\Lambda$CDM}},  {\em Astrophys. J.} {\bf 876} (2019), no.~1 85,
  [\href{http://arxiv.org/abs/1903.07603}{{\tt arXiv:1903.07603}}].

\bibitem{Wong:2019kwg}
K.~C. Wong et~al., {\it {H0LiCOW XIII. A 2.4\% measurement of $H_{0}$ from
  lensed quasars: $5.3\sigma$ tension between early and late-Universe probes}},
   \href{http://arxiv.org/abs/1907.04869}{{\tt arXiv:1907.04869}}.

\bibitem{Zheng:2010am}
R.~Zheng and Q.-G. Huang, {\it {Growth factor in $f(T)$ gravity}},  {\em JCAP}
  {\bf 1103} (2011) 002, [\href{http://arxiv.org/abs/1010.3512}{{\tt
  arXiv:1010.3512}}].

\bibitem{Linder:2010py}
E.~V. Linder, {\it {Einstein's Other Gravity and the Acceleration of the
  Universe}},  {\em Phys. Rev.} {\bf D81} (2010) 127301,
  [\href{http://arxiv.org/abs/1005.3039}{{\tt arXiv:1005.3039}}]. [Erratum:
  Phys. Rev.D82,109902(2010)].

\bibitem{Farrugia:2016qqe}
G.~Farrugia and J.~L. Said, {\it {Stability of the flat FLRW metric in $f(T)$
  gravity}},  {\em Phys. Rev.} {\bf D94} (2016), no.~12 124054,
  [\href{http://arxiv.org/abs/1701.00134}{{\tt arXiv:1701.00134}}].

\bibitem{de2018towards}
{\'A}.~de~la Cruz~Dombriz, {\it Towards new constraints in extended theories of
  gravity: Cosmography and gravitational-wave signals from neutron stars},
  {\em Galaxies} {\bf 6} (2018), no.~1 28.

\bibitem{Aviles:2013nga}
A.~Aviles, A.~Bravetti, S.~Capozziello, and O.~Luongo, {\it {Cosmographic
  reconstruction of $f(\mathcal{T})$ cosmology}},  {\em Phys. Rev.} {\bf D87}
  (2013), no.~6 064025, [\href{http://arxiv.org/abs/1302.4871}{{\tt
  arXiv:1302.4871}}].

\bibitem{Capozziello:2011hj}
S.~Capozziello, V.~F. Cardone, H.~Farajollahi, and A.~Ravanpak, {\it
  {Cosmography in f(T)-gravity}},  {\em Phys. Rev.} {\bf D84} (2011) 043527,
  [\href{http://arxiv.org/abs/1108.2789}{{\tt arXiv:1108.2789}}].

\bibitem{Suzuki:2011hu}
N.~Suzuki, D.~Rubin, C.~Lidman, G.~Aldering, R.~Amanullah, et~al., {\it {The
  Hubble Space Telescope Cluster Supernova Survey: V. Improving the Dark Energy
  Constraints Above $z>1$ and Building an Early-Type-Hosted Supernova Sample}},
   {\em Astrophys.J.} {\bf 746} (2012) 85,
  [\href{http://arxiv.org/abs/1105.3470}{{\tt arXiv:1105.3470}}].

\bibitem{Riess:2009pu}
A.~G. Riess et~al., {\it {Cepheid Calibrations of Modern Type Ia
  Supernovae:Implications for the Hubble Constant}},  {\em Astrophys. J.
  Suppl.} {\bf 183} (2009) 109--141,
  [\href{http://arxiv.org/abs/0905.0697}{{\tt arXiv:0905.0697}}].

\bibitem{Myrzakulov2011}
R.~Myrzakulov, {\it Accelerating universe from f(t) gravity},  {\em The
  European Physical Journal C} {\bf 71} (Sep, 2011) 1752.

\bibitem{Chakrabarti:2019bed}
S.~Chakrabarti, J.~L. Said, and K.~Bamba, {\it {On Reconstruction of Extended
  Teleparallel Gravity from the Cosmological Jerk Parameter}},  {\em Eur. Phys.
  J. C} {\bf 79} (2019), no.~6 454,
  [\href{http://arxiv.org/abs/1905.09711}{{\tt arXiv:1905.09711}}].

\bibitem{ElHanafy:2019zhr}
W.~El~Hanafy and G.~Nashed, {\it {Phenomenological Reconstruction of $f(T)$
  Teleparallel Gravity}},  {\em Phys. Rev. D} {\bf 100} (2019), no.~8 083535,
  [\href{http://arxiv.org/abs/1910.04160}{{\tt arXiv:1910.04160}}].

\bibitem{ElHanafy:2020pek}
W.~El~Hanafy and E.~N. Saridakis, {\it {$f(T)$ cosmology: From Pseudo-Bang to
  Pseudo-Rip}},  \href{http://arxiv.org/abs/2011.15070}{{\tt
  arXiv:2011.15070}}.

\bibitem{Chen:2010va}
S.-H. Chen, J.~B. Dent, S.~Dutta, and E.~N. Saridakis, {\it {Cosmological
  perturbations in f(T) gravity}},  {\em Phys. Rev.} {\bf D83} (2011) 023508,
  [\href{http://arxiv.org/abs/1008.1250}{{\tt arXiv:1008.1250}}].

\bibitem{Izumi:2012qj}
K.~Izumi and Y.~C. Ong, {\it {Cosmological Perturbation in f(T) Gravity
  Revisited}},  {\em JCAP} {\bf 1306} (2013) 029,
  [\href{http://arxiv.org/abs/1212.5774}{{\tt arXiv:1212.5774}}].

\bibitem{Farrugia:2016pjh}
G.~Farrugia and J.~L. Said, {\it {Growth factor in $f(T,\mathcal{T})$
  gravity}},  {\em Phys. Rev.} {\bf D94} (2016), no.~12 124004,
  [\href{http://arxiv.org/abs/1612.00974}{{\tt arXiv:1612.00974}}].

\bibitem{Golovnev:2018wbh}
A.~Golovnev and T.~Koivisto, {\it {Cosmological perturbations in modified
  teleparallel gravity models}},  {\em JCAP} {\bf 1811} (2018), no.~11 012,
  [\href{http://arxiv.org/abs/1808.05565}{{\tt arXiv:1808.05565}}].

\bibitem{Wu:2012hs}
Y.-P. Wu and C.-Q. Geng, {\it {Matter Density Perturbations in Modified
  Teleparallel Theories}},  {\em JHEP} {\bf 11} (2012) 142,
  [\href{http://arxiv.org/abs/1211.1778}{{\tt arXiv:1211.1778}}].

\bibitem{Nunes:2016qyp}
R.~C. Nunes, S.~Pan, and E.~N. Saridakis, {\it {New observational constraints
  on f(T) gravity from cosmic chronometers}},  {\em JCAP} {\bf 1608} (2016),
  no.~08 011, [\href{http://arxiv.org/abs/1606.04359}{{\tt arXiv:1606.04359}}].

\bibitem{Nunes:2018xbm}
R.~C. Nunes, {\it {Structure formation in $f(T)$ gravity and a solution for
  $H_0$ tension}},  \href{http://arxiv.org/abs/1802.02281}{{\tt
  arXiv:1802.02281}}.

\bibitem{Nunes:2018evm}
R.~C. Nunes, S.~Pan, and E.~N. Saridakis, {\it {New observational constraints
  on $f(T)$ gravity through gravitational-wave astronomy}},  {\em Phys. Rev.}
  {\bf D98} (2018), no.~10 104055, [\href{http://arxiv.org/abs/1810.03942}{{\tt
  arXiv:1810.03942}}].

\bibitem{Nunes:2019bjq}
R.~C. Nunes, M.~E.~S. Alves, and J.~C.~N. de~Araujo, {\it {Forecast constraints
  on $f(T)$ gravity with gravitational waves from compact binary
  coalescences}},  \href{http://arxiv.org/abs/1905.03237}{{\tt
  arXiv:1905.03237}}.

\bibitem{dodelson2003modern}
S.~Dodelson, {\em Modern Cosmology}.
\newblock Academic Press, 2003.

\bibitem{Alam:2016hwk}
{\bf BOSS} Collaboration, S.~Alam et~al., {\it {The clustering of galaxies in
  the completed SDSS-III Baryon Oscillation Spectroscopic Survey: cosmological
  analysis of the DR12 galaxy sample}},  {\em Mon. Not. Roy. Astron. Soc.} {\bf
  470} (2017), no.~3 2617--2652, [\href{http://arxiv.org/abs/1607.03155}{{\tt
  arXiv:1607.03155}}].

\bibitem{Wang:2020zfv}
D.~Wang and D.~Mota, {\it {Can $f(T)$ gravity resolve the $H_0$ tension?}},
  \href{http://arxiv.org/abs/2003.10095}{{\tt arXiv:2003.10095}}.

\bibitem{Nesseris:2013jea}
S.~Nesseris, S.~Basilakos, E.~N. Saridakis, and L.~Perivolaropoulos, {\it
  {Viable $f(T)$ models are practically indistinguishable from $\Lambda$CDM}},
  {\em Phys. Rev.} {\bf D88} (2013) 103010,
  [\href{http://arxiv.org/abs/1308.6142}{{\tt arXiv:1308.6142}}].

\bibitem{El-Zant:2018bsc}
A.~El-Zant, W.~El~Hanafy, and S.~Elgammal, {\it {$H_0$ Tension and the Phantom
  Regime: A Case Study in Terms of an Infrared $f(T)$ Gravity}},  {\em
  Astrophys. J.} {\bf 871} (2019), no.~2 210,
  [\href{http://arxiv.org/abs/1809.09390}{{\tt arXiv:1809.09390}}].

\bibitem{2018ApJ...855..136R}
A.~G. {Riess}, S.~{Casertano}, W.~{Yuan}, L.~{Macri}, J.~{Anderson}, J.~W.
  {MacKenty}, J.~B. {Bowers}, K.~I. {Clubb}, A.~V. {Filippenko}, D.~O. {Jones},
  and B.~E. {Tucker}, {\it {New Parallaxes of Galactic Cepheids from Spatially
  Scanning the Hubble Space Telescope: Implications for the Hubble Constant}},
  {\em The Astrophysical Journal} {\bf 855} (Mar, 2018) 136,
  [\href{http://arxiv.org/abs/1801.01120}{{\tt arXiv:1801.01120}}].

\bibitem{Anagnostopoulos:2019miu}
F.~K. Anagnostopoulos, S.~Basilakos, and E.~N. Saridakis, {\it {Bayesian
  analysis of $f(T)$ gravity using $f\sigma_8$ data}},
  \href{http://arxiv.org/abs/1907.07533}{{\tt arXiv:1907.07533}}.

\bibitem{Sagredo:2018ahx}
B.~Sagredo, S.~Nesseris, and D.~Sapone, {\it {Internal Robustness of Growth
  Rate data}},  {\em Phys. Rev.} {\bf D98} (2018), no.~8 083543,
  [\href{http://arxiv.org/abs/1806.10822}{{\tt arXiv:1806.10822}}].

\bibitem{Yu:2017iju}
H.~Yu, B.~Ratra, and F.-Y. Wang, {\it {Hubble Parameter and Baryon Acoustic
  Oscillation Measurement Constraints on the Hubble Constant, the Deviation
  from the Spatially Flat LCDM Model, the Deceleration--Acceleration Transition
  Redshift, and Spatial Curvature}},  {\em Astrophys. J.} {\bf 856} (2018),
  no.~1 3, [\href{http://arxiv.org/abs/1711.03437}{{\tt arXiv:1711.03437}}].

\bibitem{Scolnic:2017caz}
D.~M. Scolnic et~al., {\it {The Complete Light-curve Sample of
  Spectroscopically Confirmed SNe Ia from Pan-STARRS1 and Cosmological
  Constraints from the Combined Pantheon Sample}},  {\em Astrophys. J.} {\bf
  859} (2018), no.~2 101, [\href{http://arxiv.org/abs/1710.00845}{{\tt
  arXiv:1710.00845}}].

\bibitem{1100705}
H.~{Akaike}, {\it A new look at the statistical model identification},  {\em
  IEEE Transactions on Automatic Control} {\bf 19} (December, 1974) 716--723.

\bibitem{schwarz1978}
G.~Schwarz, {\it Estimating the dimension of a model},  {\em Ann. Statist.}
  {\bf 6} (03, 1978) 461--464.

\bibitem{burnham2004multimodel}
K.~P. Burnham and D.~R. Anderson, {\it Multimodel inference: understanding aic
  and bic in model selection},  {\em Sociological methods \& research} {\bf 33}
  (2004), no.~2 261--304.

\bibitem{Cai:2019bdh}
Y.-F. Cai, M.~Khurshudyan, and E.~N. Saridakis, {\it {Model-independent
  reconstruction of $f(T)$ gravity from Gaussian Processes}},  {\em Astrophys.
  J.} {\bf 888} (2020) 62, [\href{http://arxiv.org/abs/1907.10813}{{\tt
  arXiv:1907.10813}}].

\bibitem{Yan:2019gbw}
S.-F. Yan, P.~Zhang, J.-W. Chen, X.-Z. Zhang, Y.-F. Cai, and E.~N. Saridakis,
  {\it {Interpreting cosmological tensions from the effective field theory of
  torsional gravity}},  {\em Phys. Rev. D} {\bf 101} (2020), no.~12 121301,
  [\href{http://arxiv.org/abs/1909.06388}{{\tt arXiv:1909.06388}}].

\bibitem{Briffa:2020qli}
R.~Briffa, S.~Capozziello, J.~Levi~Said, J.~Mifsud, and E.~N. Saridakis, {\it
  {Constraining Teleparallel Gravity through Gaussian Processes}},
  \href{http://arxiv.org/abs/2009.14582}{{\tt arXiv:2009.14582}}.

\bibitem{Escamilla-Rivera:2019ulu}
C.~Escamilla-Rivera and J.~Levi~Said, {\it {Cosmological viable models in
  f(T,B) gravity as solutions to the $H_0$ tension}},
  \href{http://arxiv.org/abs/1909.10328}{{\tt arXiv:1909.10328}}.

\bibitem{Bamba:2014zra}
K.~Bamba, S.~Nojiri, and S.~D. Odintsov, {\it {Trace-anomaly driven inflation
  in $f(T)$ gravity and in minimal massive bigravity}},  {\em Phys. Lett.} {\bf
  B731} (2014) 257--264, [\href{http://arxiv.org/abs/1401.7378}{{\tt
  arXiv:1401.7378}}].

\bibitem{Nashed:2014lva}
G.~G.~L. Nashed and W.~El~Hanafy, {\it {A Built-in Inflation in the
  $f(T)$-Cosmology}},  {\em Eur. Phys. J.} {\bf C74} (2014) 3099,
  [\href{http://arxiv.org/abs/1403.0913}{{\tt arXiv:1403.0913}}].

\bibitem{Rezazadeh:2015dza}
K.~Rezazadeh, A.~Abdolmaleki, and K.~Karami, {\it {Power-law and intermediate
  inflationary models in f(T)-gravity}},  {\em JHEP} {\bf 01} (2016) 131,
  [\href{http://arxiv.org/abs/1509.08769}{{\tt arXiv:1509.08769}}].

\bibitem{Bamba:2016gbu}
K.~Bamba, G.~G.~L. Nashed, W.~El~Hanafy, and S.~K. Ibraheem, {\it {Bounce
  inflation in $f(T)$ Cosmology: A unified inflaton-quintessence field}},  {\em
  Phys. Rev.} {\bf D94} (2016), no.~8 083513,
  [\href{http://arxiv.org/abs/1604.07604}{{\tt arXiv:1604.07604}}].

\bibitem{Rezazadeh:2017edd}
K.~Rezazadeh, A.~Abdolmaleki, and K.~Karami, {\it {Logamediate Inflation in
  f(T) Teleparallel Gravity}},  {\em Astrophys. J.} {\bf 836} (2017), no.~2
  228, [\href{http://arxiv.org/abs/1702.07877}{{\tt arXiv:1702.07877}}].

\bibitem{Keskin:2018gev}
A.~I. Keskin, {\it {Viable super inflation scenario from F(T) modified
  teleparallel gravity}},  {\em Eur. Phys. J.} {\bf C78} (2018), no.~9 705.

\bibitem{Bamba:2012mi}
K.~Bamba, C.-Q. Geng, and L.-W. Luo, {\it {Generation of large-scale magnetic
  fields from inflation in teleparallelism}},  {\em JCAP} {\bf 1210} (2012)
  058, [\href{http://arxiv.org/abs/1208.0665}{{\tt arXiv:1208.0665}}].

\bibitem{Keskin:2017yzy}
A.~I. Keskin, {\it {Super inflation mechanism and dark energy in $F(T,T_{G})$
  gravity}},  {\em Astrophys. Space Sci.} {\bf 362} (2017), no.~3 50.

\bibitem{Bamba:2016wjm}
K.~Bamba, S.~D. Odintsov, and E.~N. Saridakis, {\it {Inflationary cosmology in
  unimodular $F(T)$ gravity}},  {\em Mod. Phys. Lett.} {\bf A32} (2017), no.~21
  1750114, [\href{http://arxiv.org/abs/1605.02461}{{\tt arXiv:1605.02461}}].

\bibitem{Jamil:2013nca}
M.~Jamil, D.~Momeni, and R.~Myrzakulov, {\it {Warm Intermediate Inflation in
  $F(T)$ Gravity}},  {\em Int. J. Theor. Phys.} {\bf 54} (2015), no.~4
  1098--1112, [\href{http://arxiv.org/abs/1309.3269}{{\tt arXiv:1309.3269}}].

\bibitem{Goodarzi:2018feh}
P.~Goodarzi and H.~Mohseni~Sadjadi, {\it {Reheating in a modified teleparallel
  model of inflation}},  {\em Eur. Phys. J.} {\bf C79} (2019), no.~3 193,
  [\href{http://arxiv.org/abs/1808.01225}{{\tt arXiv:1808.01225}}].

\bibitem{Abedi:2017ijd}
H.~Abedi, M.~Wright, and A.~M. Abbassi, {\it {Nonminimal coupling in
  anisotropic teleparallel inflation}},  {\em Phys. Rev.} {\bf D95} (2017),
  no.~6 064020.

\bibitem{Awad:2017ign}
A.~Awad, W.~El~Hanafy, G.~G.~L. Nashed, S.~D. Odintsov, and V.~K. Oikonomou,
  {\it {Constant-roll Inflation in $f(T)$ Teleparallel Gravity}},
  \href{http://arxiv.org/abs/1710.00682}{{\tt arXiv:1710.00682}}.

\bibitem{Gonzalez-Espinoza:2019ajd}
M.~Gonzalez-Espinoza, G.~Otalora, N.~Videla, and J.~Saavedra, {\it {Slow-roll
  inflation in generalized scalar-torsion gravity}},
  \href{http://arxiv.org/abs/1904.08068}{{\tt arXiv:1904.08068}}.

\bibitem{Raatikainen:2019qey}
S.~Raatikainen and S.~Rasanen, {\it {Higgs inflation and teleparallel
  gravity}},  {\em JCAP} {\bf 12} (2019), no.~12 021,
  [\href{http://arxiv.org/abs/1910.03488}{{\tt arXiv:1910.03488}}].

\bibitem{Akbarieh:2018oie}
A.~Rezaei~Akbarieh and Y.~Izadi, {\it {Tachyon Inflation in Teleparallel
  Gravity}},  {\em Eur. Phys. J.} {\bf C79} (2019), no.~4 366,
  [\href{http://arxiv.org/abs/1812.06649}{{\tt arXiv:1812.06649}}].

\bibitem{Bahamonde:2019gjk}
S.~Bahamonde, M.~Marciu, and J.~L. Said, {\it {Generalized Tachyonic
  Teleparallel cosmology}},  {\em Eur. Phys. J.} {\bf C79} (2019), no.~4 324,
  [\href{http://arxiv.org/abs/1901.04973}{{\tt arXiv:1901.04973}}].

\bibitem{Bahamonde:2017ize}
S.~Bahamonde, C.~G. B{\"o}hmer, S.~Carloni, E.~J. Copeland, W.~Fang, and
  N.~Tamanini, {\it {Dynamical systems applied to cosmology: dark energy and
  modified gravity}},  {\em Phys. Rept.} {\bf 775-777} (2018) 1--122,
  [\href{http://arxiv.org/abs/1712.03107}{{\tt arXiv:1712.03107}}].

\bibitem{arrowsmith}
D.~K. Arrowsmith and C.~M. Place, {\em An Introduction to Dynamical Systems}.
\newblock Cambridge University Press, 1990.

\bibitem{Coley:2003mj}
A.~A. Coley, {\em {Dynamical systems and cosmology}}.
\newblock Kluwer Academic Publishers, Dordrecht Boston London, 2003.

\bibitem{WainwrightEllis}
J.~Wainwright and G.~F.~R. Ellis, {\em Dynamical systems in cosmology}.
\newblock Cambridge University Press, 1997.

\bibitem{Wu:2010xk}
P.~Wu and H.~W. Yu, {\it {The dynamical behavior of $f(T)$ theory}},  {\em
  Phys.Lett.} {\bf B692} (2010) 176--179,
  [\href{http://arxiv.org/abs/1007.2348}{{\tt arXiv:1007.2348}}].

\bibitem{Zhang:2011qp}
Y.~Zhang, H.~Li, Y.~Gong, and Z.-H. Zhu, {\it {Notes on $f(T)$ Theories}},
  {\em JCAP} {\bf 1107} (2011) 015, [\href{http://arxiv.org/abs/1103.0719}{{\tt
  arXiv:1103.0719}}].

\bibitem{Jamil:2012nma}
M.~Jamil, D.~Momeni, and R.~Myrzakulov, {\it {Attractor Solutions in $f(T)$
  Cosmology}},  {\em Eur. Phys. J.} {\bf C72} (2012) 1959,
  [\href{http://arxiv.org/abs/1202.4926}{{\tt arXiv:1202.4926}}].

\bibitem{Jamil:2012yz}
M.~Jamil, K.~Yesmakhanova, D.~Momeni, and R.~Myrzakulov, {\it {Phase space
  analysis of interacting dark energy in f(T) cosmology}},  {\em Central Eur.
  J. Phys.} {\bf 10} (2012) 1065--1071,
  [\href{http://arxiv.org/abs/1207.2735}{{\tt arXiv:1207.2735}}].

\bibitem{Biswas:2015cva}
S.~K. Biswas and S.~Chakraborty, {\it {Interacting Dark Energy in $f(T)$
  cosmology : A Dynamical System analysis}},  {\em Int. J. Mod. Phys.} {\bf
  D24} (2015), no.~07 1550046, [\href{http://arxiv.org/abs/1504.02431}{{\tt
  arXiv:1504.02431}}].

\bibitem{Feng:2014fsa}
C.-J. Feng, X.-Z. Li, and L.-Y. Liu, {\it {Bifurcation and global dynamical
  behavior of the $f(T)$ theory}},  {\em Mod.Phys.Lett.} {\bf A29} (2014),
  no.~7 1450033, [\href{http://arxiv.org/abs/1403.4328}{{\tt
  arXiv:1403.4328}}].

\bibitem{Mirza:2017vrk}
B.~Mirza and F.~Oboudiat, {\it {Constraining f(T) gravity by dynamical system
  analysis}},  {\em JCAP} {\bf 1711} (2017), no.~11 011,
  [\href{http://arxiv.org/abs/1704.02593}{{\tt arXiv:1704.02593}}].

\bibitem{Hohmann:2017jao}
M.~Hohmann, L.~Jarv, and U.~Ualikhanova, {\it {Dynamical systems approach and
  generic properties of $f(T)$ cosmology}},  {\em Phys. Rev.} {\bf D96} (2017),
  no.~4 043508, [\href{http://arxiv.org/abs/1706.02376}{{\tt
  arXiv:1706.02376}}].

\bibitem{Awad:2017yod}
A.~Awad, W.~El~Hanafy, G.~G.~L. Nashed, and E.~N. Saridakis, {\it {Phase
  Portraits of general $f(T)$ Cosmology}},  {\em JCAP} {\bf 1802} (2018),
  no.~02 052, [\href{http://arxiv.org/abs/1710.10194}{{\tt arXiv:1710.10194}}].

\bibitem{Karpathopoulos:2017arc}
L.~Karpathopoulos, S.~Basilakos, G.~Leon, A.~Paliathanasis, and M.~Tsamparlis,
  {\it {Cartan symmetries and global dynamical systems analysis in a
  higher-order modified teleparallel theory}},  {\em Gen. Rel. Grav.} {\bf 50}
  (2018), no.~7 79, [\href{http://arxiv.org/abs/1709.02197}{{\tt
  arXiv:1709.02197}}].

\bibitem{Paliathanasis:2017flf}
A.~Paliathanasis, {\it {de Sitter and Scaling solutions in a higher-order
  modified teleparallel theory}},  {\em JCAP} {\bf 1708} (2017), no.~08 027,
  [\href{http://arxiv.org/abs/1706.02662}{{\tt arXiv:1706.02662}}].

\bibitem{Paliathanasis:2017efk}
A.~Paliathanasis, {\it {Cosmological Evolution and Exact Solutions in a
  Fourth-order Theory of Gravity}},  {\em Phys. Rev.} {\bf D95} (2017), no.~6
  064062, [\href{http://arxiv.org/abs/1701.04360}{{\tt arXiv:1701.04360}}].

\bibitem{Gonzalez:2019tky}
P.~A. Gonz{\'a}lez, S.~Reyes, and Y.~V{\'a}squez, {\it {Teleparallel Equivalent
  of Lovelock Gravity, Generalizations and Cosmological Applications}},
  \href{http://arxiv.org/abs/1905.07633}{{\tt arXiv:1905.07633}}.

\bibitem{Bahamonde:2020vfj}
S.~Bahamonde, M.~Marciu, S.~D. Odintsov, and P.~Rudra, {\it {String-inspired
  Teleparallel Cosmology}},  \href{http://arxiv.org/abs/2003.13434}{{\tt
  arXiv:2003.13434}}.

\bibitem{Bamba:2017ufh}
K.~Bamba, D.~Momeni, and M.~A. Ajmi, {\it {Phase Space description of Nonlocal
  Teleparallel Gravity}},  {\em Eur. Phys. J.} {\bf C78} (2018), no.~9 771,
  [\href{http://arxiv.org/abs/1711.10475}{{\tt arXiv:1711.10475}}].

\bibitem{Bohmer:2019qfi}
C.~G. Böhmer, F.~Fiorini, P.~González, and Y.~Vásquez, {\it {$D=11$
  cosmologies with teleparallel structure}},  {\em Phys. Rev. D} {\bf 100}
  (2019), no.~8 084007, [\href{http://arxiv.org/abs/1908.03680}{{\tt
  arXiv:1908.03680}}].

\bibitem{Otalora:2013tba}
G.~Otalora, {\it {Scaling attractors in interacting teleparallel dark energy}},
   {\em JCAP} {\bf 1307} (2013) 044,
  [\href{http://arxiv.org/abs/1305.0474}{{\tt arXiv:1305.0474}}].

\bibitem{Xu:2012jf}
C.~Xu, E.~N. Saridakis, and G.~Leon, {\it {Phase-Space analysis of Teleparallel
  Dark Energy}},  {\em JCAP} {\bf 1207} (2012) 005,
  [\href{http://arxiv.org/abs/1202.3781}{{\tt arXiv:1202.3781}}].

\bibitem{Skugoreva:2014ena}
M.~A. Skugoreva, E.~N. Saridakis, and A.~V. Toporensky, {\it {Dynamical
  features of scalar-torsion theories}},  {\em Phys. Rev.} {\bf D91} (2015)
  044023, [\href{http://arxiv.org/abs/1412.1502}{{\tt arXiv:1412.1502}}].

\bibitem{Skugoreva:2014gka}
M.~A. Skugoreva, A.~V. Toporensky, and S.~Y. Vernov, {\it Global stability
  analysis for cosmological models with nonminimally coupled scalar fields},
  {\em Phys. Rev. D} {\bf 90} (Sep, 2014) 064044,
  [\href{http://arxiv.org/abs/1404.6226}{{\tt arXiv:1404.6226}}].

\bibitem{Sadjadi:2015fca}
H.~Mohseni~Sadjadi, {\it {Onset of acceleration in a universe initially filled
  by dark and baryonic matters in a nonminimally coupled teleparallel model}},
  {\em Phys. Rev.} {\bf D92} (2015), no.~12 123538,
  [\href{http://arxiv.org/abs/1510.02085}{{\tt arXiv:1510.02085}}].

\bibitem{Marciu:2017sji}
M.~Marciu, {\it {Dynamical properties of scaling solutions in teleparallel dark
  energy cosmologies with nonminimal coupling}},  {\em Int. J. Mod. Phys.} {\bf
  D26} (2017), no.~09 1750103.

\bibitem{Otalora:2013dsa}
G.~Otalora, {\it {Cosmological dynamics of tachyonic teleparallel dark
  energy}},  {\em Phys.Rev.} {\bf D88} (2013) 063505,
  [\href{http://arxiv.org/abs/1305.5896}{{\tt arXiv:1305.5896}}].

\bibitem{Banijamali:2016ozr}
A.~Banijamali and E.~Ghasemi, {\it {Dynamical Characteristics of a
  Non-canonical Scalar-Torsion Model of Dark Energy}},  {\em Int. J. Theor.
  Phys.} {\bf 55} (2016), no.~8 3752--3760.

\bibitem{Fazlpour:2014qaa}
B.~Fazlpour and A.~Banijamali, {\it {Dynamics of Generalized Tachyon Field in
  Teleparallel Gravity}},  {\em Adv. High Energy Phys.} {\bf 2015} (2015)
  283273, [\href{http://arxiv.org/abs/1408.0203}{{\tt arXiv:1408.0203}}].

\bibitem{Fazlpour:2014qla}
B.~Fazlpour and A.~Banijamali, {\it {Non-minimally Coupled Tachyon Field in
  Teleparallel Gravity}},  {\em JCAP} {\bf 1504} (2015), no.~04 030,
  [\href{http://arxiv.org/abs/1410.4446}{{\tt arXiv:1410.4446}}].

\bibitem{Noether:1918zz}
E.~Noether, {\it {Invariant Variation Problems}},  {\em Gott. Nachr.} {\bf
  1918} (1918) 235--257, [\href{http://arxiv.org/abs/physics/0503066}{{\tt
  physics/0503066}}]. [Transp. Theory Statist. Phys.1,186(1971)].

\bibitem{Dialektopoulos:2018qoe}
K.~F. Dialektopoulos and S.~Capozziello, {\it {Noether Symmetries as a
  geometric criterion to select theories of gravity}},  {\em Int. J. Geom.
  Meth. Mod. Phys.} {\bf 15} (2018), no.~supp01 1840007,
  [\href{http://arxiv.org/abs/1808.03484}{{\tt arXiv:1808.03484}}].

\bibitem{Paliathanasis:2015mxa}
A.~Paliathanasis, {\em {Symmetries of Differential equations and Applications
  in Relativistic Physics}}.
\newblock PhD thesis, Athens U., 2014.
\newblock \href{http://arxiv.org/abs/1501.05129}{{\tt arXiv:1501.05129}}.

\bibitem{Wei:2011aa}
H.~Wei, X.-J. Guo, and L.-F. Wang, {\it {Noether Symmetry in $f(T)$ Theory}},
  {\em Phys. Lett.} {\bf B707} (2012) 298--304,
  [\href{http://arxiv.org/abs/1112.2270}{{\tt arXiv:1112.2270}}].

\bibitem{Atazadeh:2011aa}
K.~Atazadeh and F.~Darabi, {\it {$f(T)$ cosmology via Noether symmetry}},  {\em
  Eur. Phys. J.} {\bf C72} (2012) 2016,
  [\href{http://arxiv.org/abs/1112.2824}{{\tt arXiv:1112.2824}}].

\bibitem{Sadjadi:2012xa}
H.~Mohseni~Sadjadi, {\it {Generalized Noether symmetry in f(T) gravity}},  {\em
  Phys. Lett.} {\bf B718} (2012) 270--275,
  [\href{http://arxiv.org/abs/1210.0937}{{\tt arXiv:1210.0937}}].

\bibitem{Basilakos:2013rua}
S.~Basilakos, S.~Capozziello, M.~De~Laurentis, A.~Paliathanasis, and
  M.~Tsamparlis, {\it {Noether symmetries and analytical solutions in
  f(T)-cosmology: A complete study}},  {\em Phys. Rev.} {\bf D88} (2013)
  103526, [\href{http://arxiv.org/abs/1311.2173}{{\tt arXiv:1311.2173}}].

\bibitem{Jamil:2012fs}
M.~Jamil, D.~Momeni, and R.~Myrzakulov, {\it {Noether symmetry of F(T)
  cosmology with quintessence and phantom scalar fields}},  {\em Eur. Phys. J.}
  {\bf C72} (2012) 2137, [\href{http://arxiv.org/abs/1210.0001}{{\tt
  arXiv:1210.0001}}].

\bibitem{Paliathanasis:2014iva}
A.~Paliathanasis, S.~Basilakos, E.~N. Saridakis, S.~Capozziello, K.~Atazadeh,
  F.~Darabi, and M.~Tsamparlis, {\it {New Schwarzschild-like solutions in f(T)
  gravity through Noether symmetries}},  {\em Phys. Rev.} {\bf D89} (2014)
  104042, [\href{http://arxiv.org/abs/1402.5935}{{\tt arXiv:1402.5935}}].

\bibitem{Bahamonde:2019jkf}
S.~Bahamonde and U.~Camci, {\it {Exact Spherically Symmetric Solutions in
  Modified Teleparallel gravity}},  {\em Symmetry} {\bf 11} (2019), no.~12
  1462, [\href{http://arxiv.org/abs/1911.03965}{{\tt arXiv:1911.03965}}].

\bibitem{Bahamonde:2018ibz}
S.~Bahamonde, U.~Camci, and S.~Capozziello, {\it {Noether symmetries and
  boundary terms in extended Teleparallel gravity cosmology}},  {\em Class.
  Quant. Grav.} {\bf 36} (2019), no.~6 065013,
  [\href{http://arxiv.org/abs/1807.02891}{{\tt arXiv:1807.02891}}].

\bibitem{Sharif:2014fla}
M.~Sharif and I.~Shafique, {\it {Noether symmetries in a modified scalar-tensor
  gravity}},  {\em Phys. Rev.} {\bf D90} (2014), no.~8 084033.

\bibitem{Kucukakca:2013mya}
Y.~Kucukakca, {\it {Scalar tensor teleparallel dark gravity via Noether
  symmetry}},  {\em Eur. Phys. J.} {\bf C73} (2013), no.~2 2327,
  [\href{http://arxiv.org/abs/1404.7315}{{\tt arXiv:1404.7315}}].

\bibitem{Gecim:2017hmn}
G.~Gecim and Y.~Kucukakca, {\it {Scalar--tensor teleparallel gravity with
  boundary term by Noether symmetries}},  {\em Int. J. Geom. Meth. Mod. Phys.}
  {\bf 15} (2018), no.~09 1850151, [\href{http://arxiv.org/abs/1708.07430}{{\tt
  arXiv:1708.07430}}].

\bibitem{Bahamonde:2016jqq}
S.~Bahamonde, U.~Camci, S.~Capozziello, and M.~Jamil, {\it {Scalar-Tensor
  Teleparallel Wormholes by Noether Symmetries}},  {\em Phys. Rev.} {\bf D94}
  (2016), no.~8 084042, [\href{http://arxiv.org/abs/1608.03918}{{\tt
  arXiv:1608.03918}}].

\bibitem{Motavalli:2018ien}
H.~Motavalli and A.~Rezaei~Akbarieh, {\it {Teleparallel gravity with scalar and
  vector fields}},  {\em Astrophys. Space Sci.} {\bf 363} (2018), no.~10 200.

\bibitem{Tajahmad:2016bjs}
B.~Tajahmad, {\it {Noether Symmetries of a Modified Model in Teleparallel
  Gravity and a New Approach for Exact Solutions}},  {\em Eur. Phys. J.} {\bf
  C77} (2017), no.~4 211, [\href{http://arxiv.org/abs/1610.08099}{{\tt
  arXiv:1610.08099}}].

\bibitem{Kucukakca:2014vja}
Y.~Kucukakca, {\it {Teleparallel dark energy model with a fermionic field via
  Noether symmetry}},  {\em Eur. Phys. J.} {\bf C74} (2014), no.~10 3086,
  [\href{http://arxiv.org/abs/1407.1188}{{\tt arXiv:1407.1188}}].

\bibitem{Tajahmad:2017ywa}
B.~Tajahmad, {\it {Studying the intervention of an unusual term in $f(T)$
  gravity via the Noether symmetry approach}},  {\em Eur. Phys. J.} {\bf C77}
  (2017), no.~8 510, [\href{http://arxiv.org/abs/1701.01620}{{\tt
  arXiv:1701.01620}}].

\bibitem{Brandenberger:2016vhg}
R.~Brandenberger and P.~Peter, {\it {Bouncing Cosmologies: Progress and
  Problems}},  {\em Found. Phys.} {\bf 47} (2017), no.~6 797--850,
  [\href{http://arxiv.org/abs/1603.05834}{{\tt arXiv:1603.05834}}].

\bibitem{Cai:2011tc}
Y.-F. Cai, S.-H. Chen, J.~B. Dent, S.~Dutta, and E.~N. Saridakis, {\it {Matter
  Bounce Cosmology with the f(T) Gravity}},  {\em Class. Quant. Grav.} {\bf 28}
  (2011) 215011, [\href{http://arxiv.org/abs/1104.4349}{{\tt
  arXiv:1104.4349}}].

\bibitem{Astashenok:2013kka}
A.~V. Astashenok, {\it {Effective dark energy models and dark energy models
  with bounce in frames of $F(T)$ gravity}},  {\em Astrophys. Space Sci.} {\bf
  351} (2014) 377--383, [\href{http://arxiv.org/abs/1308.0581}{{\tt
  arXiv:1308.0581}}].

\bibitem{Odintsov:2015uca}
S.~D. Odintsov, V.~K. Oikonomou, and E.~N. Saridakis, {\it {Superbounce and
  Loop Quantum Ekpyrotic Cosmologies from Modified Gravity: $F(R)$, $F(G)$ and
  $F(T)$ Theories}},  {\em Annals Phys.} {\bf 363} (2015) 141--163,
  [\href{http://arxiv.org/abs/1501.06591}{{\tt arXiv:1501.06591}}].

\bibitem{Qiu:2018nle}
T.~Qiu, K.~Tian, and S.~Bu, {\it {Perturbations of bounce inflation scenario
  from $f(T)$ modified gravity revisited}},  {\em Eur. Phys. J.} {\bf C79}
  (2019), no.~3 261, [\href{http://arxiv.org/abs/1810.04436}{{\tt
  arXiv:1810.04436}}].

\bibitem{Haro:2014wha}
J.~Haro and J.~Amoros, {\it {Viability of the matter bounce scenario in $F(T)$
  gravity and Loop Quantum Cosmology for general potentials}},  {\em JCAP} {\bf
  1412} (2014), no.~12 031, [\href{http://arxiv.org/abs/1406.0369}{{\tt
  arXiv:1406.0369}}].

\bibitem{Amoros:2013nxa}
J.~Amor{\'o}s, J.~de~Haro, and S.~D. Odintsov, {\it {Bouncing loop quantum
  cosmology from $F(T)$ gravity}},  {\em Phys. Rev.} {\bf D87} (2013) 104037,
  [\href{http://arxiv.org/abs/1305.2344}{{\tt arXiv:1305.2344}}].

\bibitem{deHaro:2017yll}
J.~De~Haro and J.~Amor{\'o}s, {\it {Bouncing cosmologies via modified gravity
  in the ADM formalism: Application to Loop Quantum Cosmology}},  {\em Phys.
  Rev.} {\bf D97} (2018), no.~6 064014,
  [\href{http://arxiv.org/abs/1712.08399}{{\tt arXiv:1712.08399}}].

\bibitem{delaCruz-Dombriz:2018nvt}
A.~de~la Cruz-Dombriz, G.~Farrugia, J.~L. Said, and
  D.~S{\'a}ez-Chill{\'o}n~G{\'o}mez, {\it {Cosmological bouncing solutions in
  extended teleparallel gravity theories}},  {\em Phys. Rev.} {\bf D97} (2018),
  no.~10 104040, [\href{http://arxiv.org/abs/1801.10085}{{\tt
  arXiv:1801.10085}}].

\bibitem{delaCruz-Dombriz:2017lvj}
A.~de~la Cruz-Dombriz, G.~Farrugia, J.~L. Said, and D.~Saez-Gomez, {\it
  {Cosmological reconstructed solutions in extended teleparallel gravity
  theories with a teleparallel Gauss--Bonnet term}},  {\em Class. Quant. Grav.}
  {\bf 34} (2017), no.~23 235011, [\href{http://arxiv.org/abs/1705.03867}{{\tt
  arXiv:1705.03867}}].

\bibitem{Fiorini:2013kba}
F.~Fiorini, {\it {Nonsingular Promises from Born-Infeld Gravity}},  {\em Phys.
  Rev. Lett.} {\bf 111} (2013) 041104,
  [\href{http://arxiv.org/abs/1306.4392}{{\tt arXiv:1306.4392}}].

\bibitem{Bouhmadi-Lopez:2014tna}
M.~Bouhmadi-Lopez, C.-Y. Chen, and P.~Chen, {\it {Cosmological singularities in
  Born-Infeld determinantal gravity}},  {\em Phys. Rev.} {\bf D90} (2014)
  123518, [\href{http://arxiv.org/abs/1407.5114}{{\tt arXiv:1407.5114}}].

\bibitem{Boehmer:2019uxv}
C.~G. Boehmer and F.~Fiorini, {\it {The regular black hole in four dimensional
  Born-Infeld gravity}},  {\em Class. Quant. Grav.} {\bf 36} (2019) 12,
  [\href{http://arxiv.org/abs/1901.02965}{{\tt arXiv:1901.02965}}].

\bibitem{2011SHPMP..42..264K}
E.~{Knox}, {\it {Newton-Cartan theory and teleparallel gravity: The force of a
  formulation}},  {\em Studies in the History and Philosophy of Modern Physics}
  {\bf 42} (2011) 264--275.

\bibitem{Hammad:2019oyb}
F.~Hammad, D.~Dijamco, A.~Torres-Rivas, and D.~Bérubé, {\it {Noether charge
  and black hole entropy in teleparallel gravity}},  {\em Phys. Rev. D} {\bf
  100} (2019), no.~12 124040, [\href{http://arxiv.org/abs/1912.08811}{{\tt
  arXiv:1912.08811}}].

\bibitem{Ferraro:2012wp}
R.~Ferraro, {\it {f(R) and f(T) theories of modified gravity}},  {\em AIP Conf.
  Proc.} {\bf 1471} (2012) 103--110,
  [\href{http://arxiv.org/abs/1204.6273}{{\tt arXiv:1204.6273}}].

\bibitem{Nunes:2016plz}
R.~C. Nunes, A.~Bonilla, S.~Pan, and E.~N. Saridakis, {\it {Observational
  Constraints on $f(T)$ gravity from varying fundamental constants}},  {\em
  Eur. Phys. J.} {\bf C77} (2017), no.~4 230,
  [\href{http://arxiv.org/abs/1608.01960}{{\tt arXiv:1608.01960}}].

\bibitem{Oikonomou:2016jjh}
V.~K. Oikonomou and E.~N. Saridakis, {\it {$f(T)$ gravitational baryogenesis}},
   {\em Phys. Rev.} {\bf D94} (2016), no.~12 124005,
  [\href{http://arxiv.org/abs/1607.08561}{{\tt arXiv:1607.08561}}].

\bibitem{Capozziello:2017bxm}
S.~Capozziello, G.~Lambiase, and E.~N. Saridakis, {\it {Constraining f(T)
  teleparallel gravity by Big Bang Nucleosynthesis}},  {\em Eur. Phys. J.} {\bf
  C77} (2017), no.~9 576, [\href{http://arxiv.org/abs/1702.07952}{{\tt
  arXiv:1702.07952}}].

\bibitem{Basilakos:2018arq}
S.~Basilakos, S.~Nesseris, F.~Anagnostopoulos, and E.~Saridakis, {\it {Updated
  constraints on $f(T)$ models using direct and indirect measurements of the
  Hubble parameter}},  {\em JCAP} {\bf 08} (2018) 008,
  [\href{http://arxiv.org/abs/1803.09278}{{\tt arXiv:1803.09278}}].

\bibitem{Li:2018ixg}
C.~Li, Y.~Cai, Y.-F. Cai, and E.~N. Saridakis, {\it {The effective field theory
  approach of teleparallel gravity, $f(T)$ gravity and beyond}},  {\em JCAP}
  {\bf 10} (2018) 001, [\href{http://arxiv.org/abs/1803.09818}{{\tt
  arXiv:1803.09818}}].

\bibitem{Abedi:2017jqx}
H.~Abedi and S.~Capozziello, {\it {Gravitational waves in modified teleparallel
  theories of gravity}},  \href{http://arxiv.org/abs/1712.05933}{{\tt
  arXiv:1712.05933}}.

\bibitem{Chen:2019ftv}
Z.~Chen, W.~Luo, Y.-F. Cai, and E.~N. Saridakis, {\it {New test on general
  relativity and $f(T)$ torsional gravity from galaxy-galaxy weak lensing
  surveys}},  {\em Phys. Rev. D} {\bf 102} (2020), no.~10 104044,
  [\href{http://arxiv.org/abs/1907.12225}{{\tt arXiv:1907.12225}}].

\bibitem{2017ApJ...836..152L}
F.~{Lelli}, S.~S. {McGaugh}, J.~M. {Schombert}, and M.~S. {Pawlowski}, {\it
  {One Law to Rule Them All: The Radial Acceleration Relation of Galaxies}},
  {\em The Astrophysical Journal} {\bf 836} (Feb, 2017) 152,
  [\href{http://arxiv.org/abs/1610.08981}{{\tt arXiv:1610.08981}}].

\bibitem{Dupuis:2019unm}
M.~Dupuis, F.~Girelli, A.~Osumanu, and W.~Wieland, {\it {First-order
  formulation of teleparallel gravity and dual loop gravity}},  {\em Class.
  Quant. Grav.} {\bf 37} (2020), no.~8 085023,
  [\href{http://arxiv.org/abs/1906.02801}{{\tt arXiv:1906.02801}}].

\bibitem{Falls:2018ylp}
K.~G. Falls, D.~F. Litim, and J.~Schr{\"o}der, {\it {Aspects of asymptotic
  safety for quantum gravity}},  {\em Phys. Rev.} {\bf D99} (2019), no.~12
  126015, [\href{http://arxiv.org/abs/1810.08550}{{\tt arXiv:1810.08550}}].

\bibitem{Niedermaier:2006wt}
M.~Niedermaier and M.~Reuter, {\it {The Asymptotic Safety Scenario in Quantum
  Gravity}},  {\em Living Rev. Rel.} {\bf 9} (2006) 5--173.

\bibitem{Hohmann:2018rpp}
M.~Hohmann, C.~Pfeifer, and N.~Voicu, {\it {Finsler gravity action from
  variational completion}},  {\em Phys. Rev.} {\bf D100} (2019), no.~6 064035,
  [\href{http://arxiv.org/abs/1812.11161}{{\tt arXiv:1812.11161}}].

\bibitem{Pfeifer:2019wus}
C.~Pfeifer, {\it {Finsler spacetime geometry in Physics}},
  \href{http://arxiv.org/abs/1903.10185}{{\tt arXiv:1903.10185}}.

\bibitem{Hohmann:2019sni}
M.~Hohmann, C.~Pfeifer, and N.~Voicu, {\it {Relativistic kinetic gases as
  direct sources of gravity}},  \href{http://arxiv.org/abs/1910.14044}{{\tt
  arXiv:1910.14044}}.

\bibitem{Raetzel:2010je}
D.~Raetzel, S.~Rivera, and F.~P. Schuller, {\it {Geometry of physical
  dispersion relations}},  {\em Phys. Rev.} {\bf D83} (2011) 044047,
  [\href{http://arxiv.org/abs/1010.1369}{{\tt arXiv:1010.1369}}].

\bibitem{Amelino-Camelia:2014rga}
G.~Amelino-Camelia, L.~Barcaroli, G.~Gubitosi, S.~Liberati, and N.~Loret, {\it
  {Realization of doubly special relativistic symmetries in Finsler
  geometries}},  {\em Phys. Rev.} {\bf D90} (2014), no.~12 125030,
  [\href{http://arxiv.org/abs/1407.8143}{{\tt arXiv:1407.8143}}].

\bibitem{Lobo:2020qoa}
I.~P. Lobo and C.~Pfeifer, {\it {Reaching the Planck scale with muon lifetime
  measurements}},  \href{http://arxiv.org/abs/2011.10069}{{\tt
  arXiv:2011.10069}}.

\bibitem{Gurlebeck:2018nme}
N.~G{\"u}rlebeck and C.~Pfeifer, {\it {Observers' measurements in premetric
  electrodynamics: Time and radar length}},  {\em Phys. Rev.} {\bf D97} (2018),
  no.~8 084043, [\href{http://arxiv.org/abs/1801.07724}{{\tt
  arXiv:1801.07724}}].

\bibitem{Gibbons:2011ib}
G.~W. Gibbons and C.~M. Warnick, {\it {The Geometry of sound rays in a wind}},
  {\em Contemp. Phys.} {\bf 52} (2011) 197--209,
  [\href{http://arxiv.org/abs/1102.2409}{{\tt arXiv:1102.2409}}].

\bibitem{Yajima2009}
T.~Yajima and H.~Nagahama, {\it Finsler geometry of seismic ray path in
  anisotropic media},  {\em Proceedings of the Royal Society A: Mathematical,
  Physical and Engineering Sciences} {\bf 465} (2009), no.~2106 1763--1777.

\bibitem{MARKVORSEN2016208}
S.~Markvorsen, {\it A finsler geodesic spray paradigm for wildfire spread
  modelling},  {\em Nonlinear Analysis: Real World Applications} {\bf 28}
  (2016).

\bibitem{Barcaroli:2017gvg}
L.~Barcaroli, L.~K. Brunkhorst, G.~Gubitosi, N.~Loret, and C.~Pfeifer, {\it
  {Curved spacetimes with local $\kappa$-Poincar{\'e} dispersion relation}},
  {\em Phys. Rev.} {\bf D96} (2017), no.~8 084010,
  [\href{http://arxiv.org/abs/1703.02058}{{\tt arXiv:1703.02058}}].

\bibitem{Kostelecky:2011qz}
A.~Kostelecky, {\it {Riemann-Finsler geometry and Lorentz-violating
  kinematics}},  {\em Phys. Lett.} {\bf B701} (2011) 137--143,
  [\href{http://arxiv.org/abs/1104.5488}{{\tt arXiv:1104.5488}}].

\bibitem{Cohen:2006ky}
A.~G. Cohen and S.~L. Glashow, {\it {Very special relativity}},  {\em Phys.
  Rev. Lett.} {\bf 97} (2006) 021601,
  [\href{http://arxiv.org/abs/hep-ph/0601236}{{\tt hep-ph/0601236}}].

\bibitem{Bogoslovsky1977}
G.~Bogoslovsky, {\it A special-relativistic theory of the locally anisotropic
  space-time},  {\em Il Nuovo Cimento B Series 11} {\bf 40} (1977) 99.

\bibitem{Gibbons:2007iu}
G.~W. Gibbons, J.~Gomis, and C.~N. Pope, {\it {General very special relativity
  is Finsler geometry}},  {\em Phys. Rev.} {\bf D76} (2007) 081701,
  [\href{http://arxiv.org/abs/0707.2174}{{\tt arXiv:0707.2174}}].

\bibitem{Fuster:2018djw}
A.~Fuster, C.~Pabst, and C.~Pfeifer, {\it {Berwald spacetimes and very special
  relativity}},  {\em Phys. Rev.} {\bf D98} (2018), no.~8 084062,
  [\href{http://arxiv.org/abs/1804.09727}{{\tt arXiv:1804.09727}}].

\bibitem{Randers}
G.~Randers, {\it On an asymmetrical metric in the four-space of general
  relativity},  {\em Phys. Rev.} {\bf 59} (1941) 195--199.

\bibitem{Javaloyes:2018lex}
M.~A. Javaloyes and M.~S{\'a}nchez, {\it {On the definition and examples of
  cones and Finsler spacetimes}},  \href{http://arxiv.org/abs/1805.06978}{{\tt
  arXiv:1805.06978}}.

\bibitem{Punzi:2007di}
R.~Punzi, M.~N. Wohlfarth, and F.~P. Schuller, {\it {Propagation of light in
  area metric backgrounds}},  {\em Class.Quant.Grav.} {\bf 26} (2009) 035024,
  [\href{http://arxiv.org/abs/0711.3771}{{\tt arXiv:0711.3771}}].

\bibitem{Ehlers2012}
J.~Ehlers, F.~Pirani, and A.~Schild, {\it Republication of: The geometry of
  free fall and light propagation},  {\em General Relativity and Gravitation}
  {\bf 44} (06, 2012).

\bibitem{Tavakol2009}
R.~Tavakol, {\it Geometry of spacetime and finsler geometry},  {\em
  International Journal of Modern Physics A - IJMPA} {\bf 24} (04, 2009)
  1678--1685.

\bibitem{Bernal:2020bul}
A.~Bernal, M.~\'A.~Javaloyes and M.~S\'anchez, {\it {Foundations of Finsler
  spacetimes from the Observers' Viewpoint}},  {\em Universe} {\bf 6} (2020),
  no.~4 55, [\href{http://arxiv.org/abs/2003.00455}{{\tt arXiv:2003.00455}}].

\bibitem{Clowe:2006eq}
D.~Clowe, M.~Bradac, A.~H. Gonzalez, M.~Markevitch, S.~W. Randall, C.~Jones,
  and D.~Zaritsky, {\it {A direct empirical proof of the existence of dark
  matter}},  {\em Astrophys. J.} {\bf 648} (2006) L109--L113,
  [\href{http://arxiv.org/abs/astro-ph/0608407}{{\tt astro-ph/0608407}}].

\bibitem{Corbelli:1999af}
E.~Corbelli and P.~Salucci, {\it {The Extended Rotation Curve and the Dark
  Matter Halo of M33}},  {\em Mon. Not. Roy. Astron. Soc.} {\bf 311} (2000)
  441--447, [\href{http://arxiv.org/abs/astro-ph/9909252}{{\tt
  astro-ph/9909252}}].

\bibitem{Peebles:2002gy}
P.~J.~E. Peebles and B.~Ratra, {\it {The Cosmological constant and dark
  energy}},  {\em Rev. Mod. Phys.} {\bf 75} (2003) 559--606,
  [\href{http://arxiv.org/abs/astro-ph/0207347}{{\tt astro-ph/0207347}}].
  [,592(2002)].

\bibitem{Beem}
J.~K. Beem, {\it Indefinite finsler spaces and timelike spaces},  {\em Canadian
  Journal of Mathematics} {\bf 22} (1970), no.~5 1035--1039.

\bibitem{Rutz}
S.~Rutz, {\it A finsler generalisation of einstein's vacuum field equations},
  {\em General Relativity and Gravitation} {\bf 25} (01, 1993) 1139--1158.

\bibitem{Pfeifer:2011tk}
C.~Pfeifer and M.~N.~R. Wohlfarth, {\it {Causal structure and electrodynamics
  on Finsler spacetimes}},  {\em Phys. Rev.} {\bf D84} (2011) 044039,
  [\href{http://arxiv.org/abs/1104.1079}{{\tt arXiv:1104.1079}}].

\bibitem{Pfeifer:2011xi}
C.~Pfeifer and M.~N.~R. Wohlfarth, {\it {Finsler geometric extension of
  Einstein gravity}},  {\em Phys. Rev.} {\bf D85} (2012) 064009,
  [\href{http://arxiv.org/abs/1112.5641}{{\tt arXiv:1112.5641}}].

\bibitem{Basilakos:2013hua}
S.~Basilakos, A.~P. Kouretsis, E.~N. Saridakis, and P.~Stavrinos, {\it
  {Resembling dark energy and modified gravity with Finsler-Randers
  cosmology}},  {\em Phys. Rev.} {\bf D88} (2013) 123510,
  [\href{http://arxiv.org/abs/1311.5915}{{\tt arXiv:1311.5915}}].

\bibitem{Pfeifer:2014yua}
C.~Pfeifer, {\it {Radar orthogonality and radar length in Finsler and metric
  spacetime geometry}},  {\em Phys. Rev.} {\bf D90} (2014), no.~6 064052,
  [\href{http://arxiv.org/abs/1408.5306}{{\tt arXiv:1408.5306}}].

\bibitem{Hohmann:2020mgs}
M.~Hohmann, C.~Pfeifer, and N.~Voicu, {\it {Cosmological Finsler Spacetimes}},
  {\em Universe} {\bf 6} (2020), no.~5 65,
  [\href{http://arxiv.org/abs/2003.02299}{{\tt arXiv:2003.02299}}].

\bibitem{Hasse:2019zqi}
W.~Hasse and V.~Perlick, {\it {Redshift in Finsler spacetimes}},  {\em Phys.
  Rev.} {\bf D100} (2019), no.~2 024033,
  [\href{http://arxiv.org/abs/1904.08521}{{\tt arXiv:1904.08521}}].

\bibitem{Hohmann:2016pyt}
M.~Hohmann and C.~Pfeifer, {\it {Geodesics and the magnitude-redshift relation
  on cosmologically symmetric Finsler spacetimes}},  {\em Phys. Rev.} {\bf D95}
  (2017), no.~10 104021, [\href{http://arxiv.org/abs/1612.08187}{{\tt
  arXiv:1612.08187}}].

\bibitem{Minas:2019urp}
G.~Minas, E.~N. Saridakis, P.~C. Stavrinos, and A.~Triantafyllopoulos, {\it
  {Bounce cosmology in generalized modified gravities}},  {\em Universe} {\bf
  5} (2019) 74, [\href{http://arxiv.org/abs/1902.06558}{{\tt
  arXiv:1902.06558}}].

\bibitem{Ikeda:2019ckp}
S.~Ikeda, E.~N. Saridakis, P.~C. Stavrinos, and A.~Triantafyllopoulos, {\it
  {Cosmology of Lorentz fiber-bundle induced scalar-tensor theories}},  {\em
  Phys. Rev. D} {\bf 100} (2019), no.~12 124035,
  [\href{http://arxiv.org/abs/1907.10950}{{\tt arXiv:1907.10950}}].

\bibitem{Riemann}
B.~Riemann, {\it \"uber die hypothesen, welche der geometrie zu grunde liegen},
   {\em Abhandlungen der K\"oniglichen Gesellschaft der Wissenschaften zu
  G\"ottingen} {\bf 13} (1867).

\bibitem{Finsler}
P.~Finsler, {\em \"Uber Kurven und Fl\"achen in allgemeinen R\"aumen}.
\newblock PhD thesis, Georg-August Universit\"at zu G\"ottingen, 1918.

\bibitem{Bao}
D.~Bao, S.-S. Chern, and Z.~Shen, {\em An Introduction to Finsler-Riemann
  Geometry}.
\newblock Springer, New York, 2000.

\bibitem{Krupka-book}
D.~Krupka, {\em Introduction to Global Variational Geometry}.
\newblock Springer, 2015.

\bibitem{Hohmann:2020yia}
M.~Hohmann, C.~Pfeifer, and N.~Voicu, {\it {The kinetic gas universe}},  {\em
  Eur. Phys. J. C} {\bf 80} (2020), no.~9 809,
  [\href{http://arxiv.org/abs/2005.13561}{{\tt arXiv:2005.13561}}].

\bibitem{Ehlers2011}
J.~Ehlers, {\em General-Relativistc Kinetic Theory Of Gases}, pp.~301--388.
\newblock Springer Berlin Heidelberg, Berlin, Heidelberg, 2011.

\bibitem{Rezzolla2013}
L.~Rezzolla and O.~Zanotti, {\em Relativistic Hydrodynamics}.
\newblock Oxford University Press, Oxford, 2013.

\bibitem{Voicu-Krupka}
N.~Voicu and D.~Krupka, {\it Canonical variational completion of differential
  equations},  {\em Journal of Mathematical Physics} {\bf 56} (2015), no.~4
  043507, [\href{http://arxiv.org/abs/https://doi.org/10.1063/1.4918789}{{\tt
  https://doi.org/10.1063/1.4918789}}].

\bibitem{DeWitt:1967yk}
B.~S. DeWitt, {\it {Quantum Theory of Gravity. 1. The Canonical Theory}},  {\em
  Phys. Rev.} {\bf 160} (1967) 1113--1148.

\bibitem{Garattini:2004zu}
R.~Garattini, {\it {Casimir energy and the cosmological constant}},  {\em TSPU
  Bulletin} {\bf 44N7} (2004) 72--80,
  [\href{http://arxiv.org/abs/gr-qc/0409016}{{\tt gr-qc/0409016}}].

\bibitem{AmelinoCamelia:2002tc}
G.~Amelino-Camelia, F.~D'Andrea, and G.~Mandanici, {\it {Group velocity in
  noncommutative space-time}},  {\em JCAP} {\bf 0309} (2003) 006,
  [\href{http://arxiv.org/abs/hep-th/0211022}{{\tt hep-th/0211022}}].

\bibitem{Arnowitt:1962hi}
R.~L. Arnowitt, S.~Deser, and C.~W. Misner, {\it {The Dynamics of general
  relativity}},  {\em Gen. Rel. Grav.} {\bf 40} (2008) 1997--2027,
  [\href{http://arxiv.org/abs/gr-qc/0405109}{{\tt gr-qc/0405109}}].

\bibitem{Vilenkin:1987kf}
A.~Vilenkin, {\it {Quantum Cosmology and the Initial State of the Universe}},
  {\em Phys. Rev.} {\bf D37} (1988) 888.

\bibitem{Capozziello:2007gm}
S.~Capozziello and R.~Garattini, {\it {The Cosmological constant as an
  eigenvalue of f(R)-gravity Hamiltonian constraint}},  {\em Class. Quant.
  Grav.} {\bf 24} (2007) 1627--1646,
  [\href{http://arxiv.org/abs/gr-qc/0702075}{{\tt gr-qc/0702075}}].

\bibitem{Garattini:2015aca}
R.~Garattini and M.~Faizal, {\it {Cosmological constant from a deformation of
  the Wheeler--DeWitt equation}},  {\em Nucl. Phys.} {\bf B905} (2016)
  313--326, [\href{http://arxiv.org/abs/1510.04423}{{\tt arXiv:1510.04423}}].

\bibitem{Garattini:2012ca}
R.~Garattini and M.~Sakellariadou, {\it {Does gravity's rainbow induce
  inflation without an inflaton?}},  {\em Phys. Rev.} {\bf D90} (2014), no.~4
  043521, [\href{http://arxiv.org/abs/1212.4987}{{\tt arXiv:1212.4987}}].

\bibitem{Garattini:2011kp}
R.~Garattini and G.~Mandanici, {\it {Modified Dispersion Relations lead to a
  finite Zero Point Gravitational Energy}},  {\em Phys. Rev.} {\bf D83} (2011)
  084021, [\href{http://arxiv.org/abs/1102.3803}{{\tt arXiv:1102.3803}}].

\bibitem{Magueijo:2002xx}
J.~Magueijo and L.~Smolin, {\it {Gravity's rainbow}},  {\em Class. Quant.
  Grav.} {\bf 21} (2004) 1725--1736,
  [\href{http://arxiv.org/abs/gr-qc/0305055}{{\tt gr-qc/0305055}}].

\bibitem{Ling:2006az}
Y.~Ling, {\it {Rainbow universe}},  {\em JCAP} {\bf 0708} (2007) 017,
  [\href{http://arxiv.org/abs/gr-qc/0609129}{{\tt gr-qc/0609129}}].

\bibitem{Ling:2005bp}
Y.~Ling, X.~Li, and H.-b. Zhang, {\it {Thermodynamics of modified black holes
  from gravity's rainbow}},  {\em Mod. Phys. Lett.} {\bf A22} (2007)
  2749--2756, [\href{http://arxiv.org/abs/gr-qc/0512084}{{\tt gr-qc/0512084}}].

\bibitem{Garattini:2010dn}
R.~Garattini and P.~Nicolini, {\it {A Noncommutative approach to the
  cosmological constant problem}},  {\em Phys. Rev.} {\bf D83} (2011) 064021,
  [\href{http://arxiv.org/abs/1006.5418}{{\tt arXiv:1006.5418}}].

\bibitem{Horava:2008ih}
P.~Horava, {\it {Membranes at Quantum Criticality}},  {\em JHEP} {\bf 03}
  (2009) 020, [\href{http://arxiv.org/abs/0812.4287}{{\tt arXiv:0812.4287}}].

\bibitem{Garattini:2014rwa}
R.~Garattini and E.~N. Saridakis, {\it {Gravity's Rainbow: a bridge towards
  Ho{\v r}ava--Lifshitz gravity}},  {\em Eur. Phys. J.} {\bf C75} (2015), no.~7
  343, [\href{http://arxiv.org/abs/1411.7257}{{\tt arXiv:1411.7257}}].

\bibitem{Horava:2009uw}
P.~Horava, {\it {Quantum Gravity at a Lifshitz Point}},  {\em Phys. Rev.} {\bf
  D79} (2009) 084008, [\href{http://arxiv.org/abs/0901.3775}{{\tt
  arXiv:0901.3775}}].

\bibitem{Garattini:2005ky}
R.~Garattini, {\it {Casimir energy, the cosmological constant and massive
  gravitons}},  {\em J. Phys. Conf. Ser.} {\bf 33} (2006) 215--220,
  [\href{http://arxiv.org/abs/gr-qc/0510062}{{\tt gr-qc/0510062}}].

\bibitem{Garattini:2012ec}
R.~Garattini, {\it {Distorting General Relativity: Gravity's Rainbow and f(R)
  theories at work}},  {\em JCAP} {\bf 1306} (2013) 017,
  [\href{http://arxiv.org/abs/1210.7760}{{\tt arXiv:1210.7760}}].

\bibitem{Olmo:2011sw}
G.~J. Olmo, {\it {Palatini Actions and Quantum Gravity Phenomenology}},  {\em
  JCAP} {\bf 1110} (2011) 018, [\href{http://arxiv.org/abs/1101.2841}{{\tt
  arXiv:1101.2841}}].

\bibitem{kiefer2007quantum}
C.~Kiefer, {\em Quantum Gravity}.
\newblock International Series of Monographs on Physics. OUP Oxford, 2007.

\bibitem{Vilenkin:1985md}
A.~Vilenkin, {\it {Classical and Quantum Cosmology of the Starobinsky
  Inflationary Model}},  {\em Phys. Rev.} {\bf D32} (1985) 2511.

\bibitem{Hawking:1984ph}
S.~W. Hawking and J.~C. Luttrell, {\it {Higher Derivatives in Quantum
  Cosmology. 1. The Isotropic Case}},  {\em Nucl. Phys.} {\bf B247} (1984) 250.
  [Adv. Ser. Astrophys. Cosmol.3,256(1987)].

\bibitem{Horowitz:1984wv}
G.~T. Horowitz, {\it {Quantum Cosmology With a Positive Definite Action}},
  {\em Phys. Rev.} {\bf D31} (1985) 1169. [Adv. Ser. Astrophys.
  Cosmol.3,292(1987)].

\bibitem{Alonso-Serrano:2018zpi}
A.~Alonso-Serrano, M.~Bouhmadi-López, and P.~Martín-Moruno, {\it {$f(R)$
  quantum cosmology: avoiding the Big Rip}},  {\em Phys. Rev.} {\bf D98}
  (2018), no.~10 104004, [\href{http://arxiv.org/abs/1802.03290}{{\tt
  arXiv:1802.03290}}].

\bibitem{Vasilev:2019iuh}
T.~Borislavov~Vasilev, M.~Bouhmadi-López, and P.~Martín-Moruno, {\it
  {Classical and quantum fate of the little sibling of the big rip in $f(R)$
  cosmology}},  {\em Phys. Rev.} {\bf D100} (2019), no.~8 084016,
  [\href{http://arxiv.org/abs/1907.13081}{{\tt arXiv:1907.13081}}].

\bibitem{Bouhmadi-Lopez:2019zvz}
M.~Bouhmadi-López, C.~Kiefer, and P.~Martín-Moruno, {\it {Phantom
  singularities and their quantum fate: general relativity and beyond—a
  CANTATA COST action topic}},  {\em Gen. Rel. Grav.} {\bf 51} (2019), no.~10
  135, [\href{http://arxiv.org/abs/1904.01836}{{\tt arXiv:1904.01836}}].

\bibitem{Deser:1998rj}
S.~Deser and G.~W. Gibbons, {\it {Born-Infeld-Einstein actions?}},  {\em Class.
  Quant. Grav.} {\bf 15} (1998) L35--L39,
  [\href{http://arxiv.org/abs/hep-th/9803049}{{\tt hep-th/9803049}}].

\bibitem{Bouhmadi-Lopez:2016dcf}
M.~Bouhmadi-López and C.-Y. Chen, {\it {Towards the Quantization of
  Eddington-inspired-Born-Infeld Theory}},  {\em JCAP} {\bf 1611} (2016) 023,
  [\href{http://arxiv.org/abs/1609.00700}{{\tt arXiv:1609.00700}}].

\bibitem{Arroja:2016ffm}
F.~Arroja, C.-Y. Chen, P.~Chen, and D.-h. Yeom, {\it {Singular Instantons in
  Eddington-inspired-Born-Infeld Gravity}},  {\em JCAP} {\bf 1703} (2017) 044,
  [\href{http://arxiv.org/abs/1612.00674}{{\tt arXiv:1612.00674}}].

\bibitem{Albarran:2017swy}
I.~Albarran, M.~Bouhmadi-López, C.-Y. Chen, and P.~Chen, {\it {Doomsdays in a
  modified theory of gravity: A classical and a quantum approach}},  {\em Phys.
  Lett.} {\bf B772} (2017) 814--818,
  [\href{http://arxiv.org/abs/1703.09263}{{\tt arXiv:1703.09263}}].

\bibitem{Bouhmadi-Lopez:2018sto}
M.~Bouhmadi-López, C.-Y. Chen, P.~Chen, and D.-h. Yeom, {\it {Regular
  Instantons in the Eddington-inspired-Born-Infeld Gravity: Lorentzian
  Wormholes from Bubble Nucleations}},  {\em JCAP} {\bf 1810} (2018) 056,
  [\href{http://arxiv.org/abs/1809.06579}{{\tt arXiv:1809.06579}}].

\bibitem{Brax:2017idh}
P.~Brax, {\it {What makes the Universe accelerate? A review on what dark energy
  could be and how to test it}},  {\em Rept. Prog. Phys.} {\bf 81} (2018),
  no.~1 016902.

\bibitem{Kapner:2006si}
D.~J. Kapner, T.~S. Cook, E.~G. Adelberger, J.~H. Gundlach, B.~R. Heckel, C.~D.
  Hoyle, and H.~E. Swanson, {\it {Tests of the gravitational inverse-square law
  below the dark-energy length scale}},  {\em Phys. Rev. Lett.} {\bf 98} (2007)
  021101, [\href{http://arxiv.org/abs/hep-ph/0611184}{{\tt hep-ph/0611184}}].

\bibitem{Lamoreaux:1996wh}
S.~K. Lamoreaux, {\it {Demonstration of the Casimir force in the 0.6 to 6
  micrometers range}},  {\em Phys. Rev. Lett.} {\bf 78} (1997) 5--8. [Erratum:
  Phys. Rev. Lett.81,5475(1998)].

\bibitem{Hamilton:2015zga}
P.~Hamilton, M.~Jaffe, P.~Haslinger, Q.~Simmons, H.~Müller, and J.~Khoury,
  {\it {Atom-interferometry constraints on dark energy}},  {\em Science} {\bf
  349} (2015) 849--851, [\href{http://arxiv.org/abs/1502.03888}{{\tt
  arXiv:1502.03888}}].

\bibitem{Sabulsky:2018jma}
D.~O. Sabulsky, I.~Dutta, E.~A. Hinds, B.~Elder, C.~Burrage, and E.~J.
  Copeland, {\it {Experiment to detect dark energy forces using atom
  interferometry}},  {\em Phys. Rev. Lett.} {\bf 123} (2019), no.~6 061102,
  [\href{http://arxiv.org/abs/1812.08244}{{\tt arXiv:1812.08244}}].

\bibitem{Lemmel:2015kwa}
H.~Lemmel, P.~Brax, A.~N. Ivanov, T.~Jenke, G.~Pignol, M.~Pitschmann,
  T.~Potocar, M.~Wellenzohn, M.~Zawisky, and H.~Abele, {\it {Neutron
  Interferometry constrains dark energy chameleon fields}},  {\em Phys. Lett.}
  {\bf B743} (2015) 310--314, [\href{http://arxiv.org/abs/1502.06023}{{\tt
  arXiv:1502.06023}}].

\bibitem{Nesvizhevsky:2003ww}
V.~V. Nesvizhevsky et~al., {\it {Measurement of quantum states of neutrons in
  the earth's gravitational field}},  {\em Phys. Rev.} {\bf D67} (2003) 102002,
  [\href{http://arxiv.org/abs/hep-ph/0306198}{{\tt hep-ph/0306198}}].

\bibitem{Brax:2010gp}
P.~Brax and C.~Burrage, {\it {Atomic Precision Tests and Light Scalar
  Couplings}},  {\em Phys.Rev.} {\bf D83} (2011) 035020,
  [\href{http://arxiv.org/abs/1010.5108}{{\tt arXiv:1010.5108}}].

\bibitem{Wong:2017jer}
L.~K. Wong and A.-C. Davis, {\it {One-electron atoms in screened modified
  gravity}},  {\em Phys. Rev.} {\bf D95} (2017), no.~10 104010,
  [\href{http://arxiv.org/abs/1703.05659}{{\tt arXiv:1703.05659}}].

\bibitem{Brax:2018zfb}
P.~Brax, A.-C. Davis, B.~Elder, and L.~K. Wong, {\it {Constraining screened
  fifth forces with the electron magnetic moment}},  {\em Phys. Rev.} {\bf D97}
  (2018), no.~8 084050, [\href{http://arxiv.org/abs/1802.05545}{{\tt
  arXiv:1802.05545}}].

\bibitem{Brax:2018iyo}
P.~Brax, C.~Burrage, and A.-C. Davis, {\it {Laboratory constraints}},  {\em
  Int. J. Mod. Phys.} {\bf D27} (2018), no.~15 1848009.

\bibitem{Brax:2004qh}
P.~Brax, C.~van~de Bruck, A.-C. Davis, J.~Khoury, and A.~Weltman, {\it
  {Detecting dark energy in orbit: The cosmological chameleon}},  {\em Phys.
  Rev.} {\bf D70} (2004) 123518,
  [\href{http://arxiv.org/abs/astro-ph/0408415}{{\tt astro-ph/0408415}}].

\bibitem{Elder:2016yxm}
B.~Elder, J.~Khoury, P.~Haslinger, M.~Jaffe, H.~Müller, and P.~Hamilton, {\it
  {Chameleon Dark Energy and Atom Interferometry}},  {\em Phys. Rev.} {\bf D94}
  (2016), no.~4 044051, [\href{http://arxiv.org/abs/1603.06587}{{\tt
  arXiv:1603.06587}}].

\bibitem{Schlogel:2015uea}
S.~Schlögel, S.~Clesse, and A.~Füzfa, {\it {Probing Modified Gravity with
  Atom-Interferometry: a Numerical Approach}},  {\em Phys. Rev.} {\bf D93}
  (2016), no.~10 104036, [\href{http://arxiv.org/abs/1507.03081}{{\tt
  arXiv:1507.03081}}].

\bibitem{Burrage:2016bwy}
C.~Burrage and J.~Sakstein, {\it {A Compendium of Chameleon Constraints}},
  {\em JCAP} {\bf 1611} (2016), no.~11 045,
  [\href{http://arxiv.org/abs/1609.01192}{{\tt arXiv:1609.01192}}].

\bibitem{ChuAI}
A.~Peters, K.~Y. Chung, and S.~Chu, {\it Measurement of gravitational
  acceleration by dropping atoms},  {\em Nature} {\bf 400} (1999), no.~6747
  849--852.

\bibitem{Burrage:2014oza}
C.~Burrage, E.~J. Copeland, and E.~Hinds, {\it {Probing Dark Energy with Atom
  Interferometry}},  \href{http://arxiv.org/abs/1408.1409}{{\tt
  arXiv:1408.1409}}.

\bibitem{Burrage:2015lya}
C.~Burrage and E.~J. Copeland, {\it {Using Atom Interferometry to Detect Dark
  Energy}},  {\em Contemp. Phys.} {\bf 57} (2016), no.~2 164--176,
  [\href{http://arxiv.org/abs/1507.07493}{{\tt arXiv:1507.07493}}].

\bibitem{Burrage:2016rkv}
C.~Burrage, A.~Kuribayashi-Coleman, J.~Stevenson, and B.~Thrussell, {\it
  {Constraining symmetron fields with atom interferometry}},  {\em JCAP} {\bf
  1612} (2016) 041, [\href{http://arxiv.org/abs/1609.09275}{{\tt
  arXiv:1609.09275}}].

\bibitem{Jaffe:2016fsh}
M.~Jaffe, P.~Haslinger, V.~Xu, P.~Hamilton, A.~Upadhye, B.~Elder, J.~Khoury,
  and H.~Müller, {\it {Testing sub-gravitational forces on atoms from a
  miniature, in-vacuum source mass}},  {\em Nature Phys.} {\bf 13} (2017) 938,
  [\href{http://arxiv.org/abs/1612.05171}{{\tt arXiv:1612.05171}}].

\bibitem{Adelberger:2003zx}
E.~G. Adelberger, B.~R. Heckel, and A.~E. Nelson, {\it {Tests of the
  gravitational inverse square law}},  {\em Ann. Rev. Nucl. Part. Sci.} {\bf
  53} (2003) 77--121, [\href{http://arxiv.org/abs/hep-ph/0307284}{{\tt
  hep-ph/0307284}}].

\bibitem{Upadhye:2012qu}
A.~Upadhye, {\it {Dark energy fifth forces in torsion pendulum experiments}},
  {\em Phys. Rev.} {\bf D86} (2012) 102003,
  [\href{http://arxiv.org/abs/1209.0211}{{\tt arXiv:1209.0211}}].

\bibitem{Upadhye:2012rc}
A.~Upadhye, {\it {Symmetron dark energy in laboratory experiments}},  {\em
  Phys. Rev. Lett.} {\bf 110} (2013), no.~3 031301,
  [\href{http://arxiv.org/abs/1210.7804}{{\tt arXiv:1210.7804}}].

\bibitem{Lamoreaux:2005zza}
S.~K. Lamoreaux and W.~T. Buttler, {\it {Thermal noise limitations to force
  measurements with torsion pendulums: Applications to the measurement of the
  Casimir force and its thermal correction}},  {\em Phys. Rev.} {\bf E71}
  (2005) 036109, [\href{http://arxiv.org/abs/quant-ph/0408027}{{\tt
  quant-ph/0408027}}].

\bibitem{Brax:2007vm}
P.~Brax, C.~van~de Bruck, A.-C. Davis, D.~F. Mota, and D.~J. Shaw, {\it
  {Detecting chameleons through Casimir force measurements}},  {\em Phys.Rev.}
  {\bf D76} (2007) 124034, [\href{http://arxiv.org/abs/0709.2075}{{\tt
  arXiv:0709.2075}}].

\bibitem{Brax:2010xx}
P.~Brax, C.~van~de Bruck, A.~C. Davis, D.~J. Shaw, and D.~Iannuzzi, {\it
  {Tuning the Mass of Chameleon Fields in Casimir Force Experiments}},  {\em
  Phys. Rev. Lett.} {\bf 104} (2010) 241101,
  [\href{http://arxiv.org/abs/1003.1605}{{\tt arXiv:1003.1605}}].

\bibitem{Chen:2014oda}
Y.~J. Chen, W.~K. Tham, D.~E. Krause, D.~Lopez, E.~Fischbach, and R.~S. Decca,
  {\it {Stronger Limits on Hypothetical Yukawa Interactions in the 30–8000 nm
  Range}},  {\em Phys. Rev. Lett.} {\bf 116} (2016), no.~22 221102,
  [\href{http://arxiv.org/abs/1410.7267}{{\tt arXiv:1410.7267}}].

\bibitem{Brax:2013cfa}
P.~Brax, G.~Pignol, and D.~Roulier, {\it {Probing Strongly Coupled Chameleons
  with Slow Neutrons}},  {\em Phys. Rev.} {\bf D88} (2013) 083004,
  [\href{http://arxiv.org/abs/1306.6536}{{\tt arXiv:1306.6536}}].

\bibitem{Hinterbichler:2010es}
K.~Hinterbichler and J.~Khoury, {\it {Symmetron Fields: Screening Long-Range
  Forces Through Local Symmetry Restoration}},  {\em Phys. Rev. Lett.} {\bf
  104} (2010) 231301, [\href{http://arxiv.org/abs/1001.4525}{{\tt
  arXiv:1001.4525}}].

\bibitem{Elder:2019yyp}
B.~Elder, V.~Vardanyan, Y.~Akrami, P.~Brax, A.-C. Davis, and R.~S. Decca, {\it
  {Classical symmetron force in Casimir experiments}},  {\em Phys. Rev. D} {\bf
  101} (2020), no.~6 064065, [\href{http://arxiv.org/abs/1912.10015}{{\tt
  arXiv:1912.10015}}].

\bibitem{Najafi:2018apy}
S.~Najafi, M.~T. Mirtorabi, Z.~Ansari, and D.~F. Mota, {\it {Red Giant
  evolution in Modified Gravity}},  {\em JCAP} {\bf 1902} (2019) 011,
  [\href{http://arxiv.org/abs/1802.04001}{{\tt arXiv:1802.04001}}].

\bibitem{Brax:2017wcj}
P.~Brax, A.-C. Davis, and R.~Jha, {\it {Neutron Stars in Screened Modified
  Gravity: Chameleon vs Dilaton}},  {\em Phys. Rev.} {\bf D95} (2017), no.~8
  083514, [\href{http://arxiv.org/abs/1702.02983}{{\tt arXiv:1702.02983}}].

\bibitem{Ho:2018byw}
A.~Ho, M.~Gronke, B.~Falck, and D.~F. Mota, {\it {Probing modified gravity in
  cosmic filaments}},  {\em Astron. Astrophys.} {\bf 619} (2018) A122,
  [\href{http://arxiv.org/abs/1807.07287}{{\tt arXiv:1807.07287}}].

\bibitem{Brax:2018zbd}
P.~Brax and P.~Valageas, {\it {Lyman-$\alpha$ power spectrum as a probe of
  modified gravity}},  {\em JCAP} {\bf 1901} (2019) 049,
  [\href{http://arxiv.org/abs/1810.06661}{{\tt arXiv:1810.06661}}].

\bibitem{Aviles:2018qot}
A.~Aviles, J.~L. Cervantes-Cota, and D.~F. Mota, {\it {Screenings in Modified
  Gravity: a perturbative approach}},  {\em Astron. Astrophys.} {\bf 622}
  (2019) A62, [\href{http://arxiv.org/abs/1810.02652}{{\tt arXiv:1810.02652}}].

\bibitem{Brax:2017hna}
P.~Brax and M.~Pitschmann, {\it {Exact solutions to nonlinear symmetron theory:
  One- and two-mirror systems}},  {\em Phys. Rev.} {\bf D97} (2018), no.~6
  064015, [\href{http://arxiv.org/abs/1712.09852}{{\tt arXiv:1712.09852}}].

\bibitem{Llinares:2018mzl}
C.~Llinares and P.~Brax, {\it {Detecting Coupled Domain Walls in Laboratory
  Experiments}},  {\em Phys. Rev. Lett.} {\bf 122} (2019), no.~9 091102,
  [\href{http://arxiv.org/abs/1807.06870}{{\tt arXiv:1807.06870}}].

\bibitem{Brax:2018grq}
P.~Brax and S.~Fichet, {\it {Quantum Chameleons}},  {\em Phys. Rev.} {\bf D99}
  (2019), no.~10 104049, [\href{http://arxiv.org/abs/1809.10166}{{\tt
  arXiv:1809.10166}}].

\bibitem{Salzano:2017qac}
V.~Salzano, D.~F. Mota, S.~Capozziello, and M.~Donahue, {\it {Breaking the
  Vainshtein screening in clusters of galaxies}},  {\em Phys. Rev.} {\bf D95}
  (2017), no.~4 044038, [\href{http://arxiv.org/abs/1701.03517}{{\tt
  arXiv:1701.03517}}].

\bibitem{Brax:2019koq}
P.~Brax, S.~Fichet, and P.~Tanedo, {\it {The Warped Dark Sector}},
  \href{http://arxiv.org/abs/1906.02199}{{\tt arXiv:1906.02199}}.

\bibitem{Jimenez:2020dpn}
J.~B. Jim\'enez and A.~Delhom, {\it {Instabilities in metric-affine theories of
  gravity with higher order curvature terms}},  {\em Eur. Phys. J. C} {\bf 80}
  (2020), no.~6 585, [\href{http://arxiv.org/abs/2004.11357}{{\tt
  arXiv:2004.11357}}].

\bibitem{Vollick:2003ic}
D.~N. Vollick, {\it {On the viability of the Palatini form of 1/R gravity}},
  {\em Class. Quant. Grav.} {\bf 21} (2004) 3813--3816,
  [\href{http://arxiv.org/abs/gr-qc/0312041}{{\tt gr-qc/0312041}}].

\bibitem{Thorne:1970wv}
K.~S. Thorne and C.~M. Will, {\it {Theoretical Frameworks for Testing
  Relativistic Gravity. I. Foundations}},  {\em Astrophys. J.} {\bf 163} (1971)
  595--610.

\bibitem{will_2018}
C.~M. Will, {\em {Theory and Experiment in Gravitational Physics}}.
\newblock {Cambridge University Press}, 2~ed., 2018.

\bibitem{Allemandi:2005tg}
G.~Allemandi, M.~Francaviglia, M.~L. Ruggiero, and A.~Tartaglia, {\it
  {Post-Newtonian parameters from alternative theories of gravity}},  {\em Gen.
  Rel. Grav.} {\bf 37} (2005) 1891--1904,
  [\href{http://arxiv.org/abs/gr-qc/0506123}{{\tt gr-qc/0506123}}].

\bibitem{Sotiriou:2005xe}
T.~P. Sotiriou, {\it {The Nearly Newtonian regime in non-linear theories of
  gravity}},  {\em Gen. Rel. Grav.} {\bf 38} (2006) 1407--1417,
  [\href{http://arxiv.org/abs/gr-qc/0507027}{{\tt gr-qc/0507027}}].

\bibitem{Pani:2012qb}
P.~Pani, T.~Delsate, and V.~Cardoso, {\it {Eddington-inspired Born-Infeld
  gravity. Phenomenology of non-linear gravity-matter coupling}},  {\em Phys.
  Rev.} {\bf D85} (2012) 084020, [\href{http://arxiv.org/abs/1201.2814}{{\tt
  arXiv:1201.2814}}].

\bibitem{Boos:2016cey}
J.~Boos and F.~W. Hehl, {\it {Gravity-induced four-fermion contact interaction
  implies gravitational intermediate W and Z type gauge bosons}},  {\em Int. J.
  Theor. Phys.} {\bf 56} (2017), no.~3 751--756,
  [\href{http://arxiv.org/abs/1606.09273}{{\tt arXiv:1606.09273}}].

\bibitem{Cartan1}
E.~Cartan {\em C. R. Acad. Sci. (Paris)} {\bf 174} (1922) 593.

\bibitem{Cartan2}
E.~Cartan {\em Ann.Ec. Norm. Sup.} {\bf 40} (1923) 325.

\bibitem{Cartan3}
E.~Cartan {\em Ann.Ec. Norm. Sup.} {\bf 41} (1924) 1.

\bibitem{Cartan4}
E.~Cartan {\em Ann.Ec. Norm. Sup.} {\bf 42} (1925) 17.

\bibitem{Utiyama:1956sy}
R.~Utiyama, {\it {Invariant theoretical interpretation of interaction}},  {\em
  Phys. Rev.} {\bf 101} (1956) 1597--1607. [,157(1956)].

\bibitem{Kibble:1961ba}
T.~W.~B. Kibble, {\it {Lorentz invariance and the gravitational field}},  {\em
  J. Math. Phys.} {\bf 2} (1961) 212--221. [,168(1961)].

\bibitem{Sciama_1962}
D.~W. {Sciama}, {\em {On the analogy between charge and spin in general
  relativity}}.
\newblock 1962.

\bibitem{Sciama:1964wt}
D.~W. Sciama, {\it {The Physical structure of general relativity}},  {\em Rev.
  Mod. Phys.} {\bf 36} (1964) 463--469. [Erratum: Rev. Mod.
  Phys.36,1103(1964)].

\bibitem{Poplawski:2010kb}
N.~J. Poplawski, {\it {Cosmology with torsion: An alternative to cosmic
  inflation}},  {\em Phys. Lett.} {\bf B694} (2010) 181--185,
  [\href{http://arxiv.org/abs/1007.0587}{{\tt arXiv:1007.0587}}]. [Erratum:
  Phys. Lett.B701,672(2011)].

\bibitem{Shapiro:2001rz}
I.~L. Shapiro, {\it {Physical aspects of the space-time torsion}},  {\em Phys.
  Rept.} {\bf 357} (2002) 113, [\href{http://arxiv.org/abs/hep-th/0103093}{{\tt
  hep-th/0103093}}].

\bibitem{Carroll:1994dq}
S.~M. Carroll and G.~B. Field, {\it {Consequences of propagating torsion in
  connection dynamic theories of gravity}},  {\em Phys. Rev.} {\bf D50} (1994)
  3867--3873, [\href{http://arxiv.org/abs/gr-qc/9403058}{{\tt gr-qc/9403058}}].

\bibitem{Belyaev:1997zv}
A.~S. Belyaev and I.~L. Shapiro, {\it {The Action for the (propagating) torsion
  and the limits on the torsion parameters from present experimental data}},
  {\em Phys. Lett.} {\bf B425} (1998) 246--254,
  [\href{http://arxiv.org/abs/hep-ph/9712503}{{\tt hep-ph/9712503}}].

\bibitem{HEHL1971225}
F.~Hehl, {\it How does one measure torsion of space-time?},  {\em Physics
  Letters A} {\bf 36} (1971), no.~3 225 -- 226.

\bibitem{Rumpf:1979vh}
H.~Rumpf, {\it {QUASICLASSICAL LIMIT OF THE DIRAC EQUATION AND THE EQUIVALENCE
  PRINCIPLE IN THE RIEMANN-CARTAN GEOMETRY.}},  {\em NATO Sci. Ser. B} {\bf 58}
  (1980) 93--104.

\bibitem{Audretsch:1981xn}
J.~Audretsch, {\it {Dirac Electron in Space-times With Torsion: Spinor
  Propagation, Spin Precession, and Nongeodesic Orbits}},  {\em Phys. Rev.}
  {\bf D24} (1981) 1470--1477.

\bibitem{Nomura:1991yx}
K.~Nomura, T.~Shirafuji, and K.~Hayashi, {\it {Spinning test particles in
  space-time with torsion}},  {\em Prog. Theor. Phys.} {\bf 86} (1991)
  1239--1258.

\bibitem{Obukhov:2014fta}
Y.~N. Obukhov, A.~J. Silenko, and O.~V. Teryaev, {\it {Spin-torsion coupling
  and gravitational moments of Dirac fermions: theory and experimental
  bounds}},  {\em Phys. Rev.} {\bf D90} (2014), no.~12 124068,
  [\href{http://arxiv.org/abs/1410.6197}{{\tt arXiv:1410.6197}}].

\bibitem{Mashhoon:2003ax}
B.~Mashhoon, {\it {Gravitoelectromagnetism: A Brief review}},
  \href{http://arxiv.org/abs/gr-qc/0311030}{{\tt gr-qc/0311030}}.

\bibitem{Ni:2009fg}
W.-T. Ni, {\it {Searches for the role of spin and polarization in gravity}},
  {\em Rept. Prog. Phys.} {\bf 73} (2010) 056901,
  [\href{http://arxiv.org/abs/0912.5057}{{\tt arXiv:0912.5057}}].

\bibitem{Ni:2015poa}
W.-T. Ni, {\it {Searches for the role of spin and polarization in gravity: a
  five-year update}},  {\em Int. J. Mod. Phys. Conf. Ser.} {\bf 40} (2016)
  1660010, [\href{http://arxiv.org/abs/1501.07696}{{\tt arXiv:1501.07696}}].

\bibitem{AudretschLammerzahl_1983}
J.~{Audretsch} and C.~{Lammerzahl}, {\it {Neutron interference: general theory
  of the influence of gravity, inertia and space-time torsion}},  {\em Journal
  of Physics A Mathematical General} {\bf 16} (Aug., 1983) 2457--2477.

\bibitem{Lammerzahl:1997wk}
C.~Lammerzahl, {\it {Constraints on space-time torsion from Hughes-Drever
  experiments}},  {\em Phys. Lett.} {\bf A228} (1997) 223,
  [\href{http://arxiv.org/abs/gr-qc/9704047}{{\tt gr-qc/9704047}}].

\bibitem{Mohanty:1998vs}
S.~Mohanty and U.~Sarkar, {\it {Constraints on background torsion field from K
  physics}},  {\em Phys. Lett.} {\bf B433} (1998) 424--428,
  [\href{http://arxiv.org/abs/hep-ph/9804259}{{\tt hep-ph/9804259}}].

\bibitem{Singh_1997}
P.~Singh and L.~H. Ryder, {\it Einstein - cartan - dirac theory in the
  low-energy limit},  {\em Classical and Quantum Gravity} {\bf 14} (dec, 1997)
  3513--3525.

\bibitem{Kostelecky:2007kx}
V.~A. Kostelecky, N.~Russell, and J.~Tasson, {\it {New Constraints on Torsion
  from Lorentz Violation}},  {\em Phys. Rev. Lett.} {\bf 100} (2008) 111102,
  [\href{http://arxiv.org/abs/0712.4393}{{\tt arXiv:0712.4393}}].

\bibitem{Foster:2016uui}
J.~Foster, V.~A. Kosteleck{\'y}, and R.~Xu, {\it {Constraints on Nonmetricity
  from Bounds on Lorentz Violation}},  {\em Phys. Rev.} {\bf D95} (2017), no.~8
  084033, [\href{http://arxiv.org/abs/1612.08744}{{\tt arXiv:1612.08744}}].

\bibitem{Delhom:2019gxg}
A.~Delhom, J.~R. Nascimento, G.~J. Olmo, A.~Y. Petrov, and P.~J. Porf\'\i{}rio,
  {\it {Quantum corrections in weak metric-affine bumblebee gravity}},
  \href{http://arxiv.org/abs/1911.11605}{{\tt arXiv:1911.11605}}.

\bibitem{Delhom:2020gfv}
A.~Delhom, J.~R. Nascimento, G.~J. Olmo, A.~Y. Petrov, and P.~J. Porf\'\i{}rio,
  {\it {Metric-affine bumblebee gravity: quantum aspects}},
  \href{http://arxiv.org/abs/2010.06391}{{\tt arXiv:2010.06391}}.

\bibitem{Duan:2015zmf}
X.-C. Duan, M.-K. Zhou, X.-B. Deng, H.-B. Yao, C.-G. Shao, J.~Luo, and Z.-K.
  Hu, {\it {Test of the universality of free fall with atoms in different spin
  Orientations}},  {\em Phys. Rev. Lett.} {\bf 117} (2016) 023001,
  [\href{http://arxiv.org/abs/1503.00433}{{\tt arXiv:1503.00433}}].

\bibitem{deAlmeida:2008dk}
F.~M.~L. de~Almeida, Jr., A.~A. Nepomuceno, and M.~A.~B. do~Vale, {\it {Torsion
  Discovery Potential and Its Discrimination at CERN LHC}},  {\em Phys. Rev.}
  {\bf D79} (2009) 014029, [\href{http://arxiv.org/abs/0811.0291}{{\tt
  arXiv:0811.0291}}].

\bibitem{Belyaev:2007fn}
A.~S. Belyaev, I.~L. Shapiro, and M.~A.~B. do~Vale, {\it {Torsion phenomenology
  at the LHC}},  {\em Phys. Rev.} {\bf D75} (2007) 034014,
  [\href{http://arxiv.org/abs/hep-ph/0701002}{{\tt hep-ph/0701002}}].

\bibitem{Lehnert:2013jsa}
R.~Lehnert, W.~M. Snow, and H.~Yan, {\it {A First Experimental Limit on
  In-matter Torsion from Neutron Spin Rotation in Liquid $^4He$}},  {\em Phys.
  Lett.} {\bf B730} (2014) 353--356,
  [\href{http://arxiv.org/abs/1311.0467}{{\tt arXiv:1311.0467}}]. [Erratum:
  Phys. Lett.B744,415(2015)].

\bibitem{Lattimer:2004pg}
J.~M. Lattimer and M.~Prakash, {\it {The physics of neutron stars}},  {\em
  Science} {\bf 304} (2004) 536--542,
  [\href{http://arxiv.org/abs/astro-ph/0405262}{{\tt astro-ph/0405262}}].

\bibitem{Olmo:2019flu}
G.~J. Olmo, D.~Rubiera-Garcia, and A.~Wojnar, {\it {Stellar structure models in
  modified theories of gravity: lessons and challenges}},
  \href{http://arxiv.org/abs/1912.05202}{{\tt arXiv:1912.05202}}.

\bibitem{Erickcek:2006vf}
A.~L. Erickcek, T.~L. Smith, and M.~Kamionkowski, {\it {Solar System tests do
  rule out 1/R gravity}},  {\em Phys. Rev.} {\bf D74} (2006) 121501,
  [\href{http://arxiv.org/abs/astro-ph/0610483}{{\tt astro-ph/0610483}}].

\bibitem{Kainulainen:2007bt}
K.~Kainulainen, J.~Piilonen, V.~Reijonen, and D.~Sunhede, {\it {Spherically
  symmetric spacetimes in f(R) gravity theories}},  {\em Phys. Rev.} {\bf D76}
  (2007) 024020, [\href{http://arxiv.org/abs/0704.2729}{{\tt
  arXiv:0704.2729}}].

\bibitem{Faulkner:2006ub}
T.~Faulkner, M.~Tegmark, E.~F. Bunn, and Y.~Mao, {\it {Constraining f(R)
  Gravity as a Scalar Tensor Theory}},  {\em Phys. Rev.} {\bf D76} (2007)
  063505, [\href{http://arxiv.org/abs/astro-ph/0612569}{{\tt
  astro-ph/0612569}}].

\bibitem{Multamaki:2006ym}
T.~Multamaki and I.~Vilja, {\it {Static spherically symmetric perfect fluid
  solutions in f(R) theories of gravity}},  {\em Phys. Rev.} {\bf D76} (2007)
  064021, [\href{http://arxiv.org/abs/astro-ph/0612775}{{\tt
  astro-ph/0612775}}].

\bibitem{Henttunen:2007bz}
K.~Henttunen, T.~Multamaki, and I.~Vilja, {\it {Stellar configurations in f(R)
  theories of gravity}},  {\em Phys. Rev.} {\bf D77} (2008) 024040,
  [\href{http://arxiv.org/abs/0705.2683}{{\tt arXiv:0705.2683}}].

\bibitem{Pani:2011mg}
P.~Pani, V.~Cardoso, and T.~Delsate, {\it {Compact stars in Eddington inspired
  gravity}},  {\em Phys. Rev. Lett.} {\bf 107} (2011) 031101,
  [\href{http://arxiv.org/abs/1106.3569}{{\tt arXiv:1106.3569}}].

\bibitem{Barausse:2007pn}
E.~Barausse, T.~P. Sotiriou, and J.~C. Miller, {\it {A No-go theorem for
  polytropic spheres in Palatini f(R) gravity}},  {\em Class. Quant. Grav.}
  {\bf 25} (2008) 062001, [\href{http://arxiv.org/abs/gr-qc/0703132}{{\tt
  gr-qc/0703132}}].

\bibitem{Barausse:2007ys}
E.~Barausse, T.~P. Sotiriou, and J.~C. Miller, {\it {Curvature singularities,
  tidal forces and the viability of Palatini f(R) gravity}},  {\em Class.
  Quant. Grav.} {\bf 25} (2008) 105008,
  [\href{http://arxiv.org/abs/0712.1141}{{\tt arXiv:0712.1141}}].

\bibitem{Olmo:2008pv}
G.~J. Olmo, {\it {Re-examination of Polytropic Spheres in Palatini f(R)
  Gravity}},  {\em Phys. Rev.} {\bf D78} (2008) 104026,
  [\href{http://arxiv.org/abs/0810.3593}{{\tt arXiv:0810.3593}}].

\bibitem{Lattimer:2006xb}
J.~M. Lattimer and M.~Prakash, {\it {Neutron Star Observations: Prognosis for
  Equation of State Constraints}},  {\em Phys. Rept.} {\bf 442} (2007)
  109--165, [\href{http://arxiv.org/abs/astro-ph/0612440}{{\tt
  astro-ph/0612440}}].

\bibitem{Llanes-Estrada:2019wmz}
F.~J. Llanes-Estrada and E.~Lope-Oter, {\it {Hadron matter in neutron stars in
  view of gravitational wave observations}},  {\em Prog. Part. Nucl. Phys.}
  {\bf 109} (2019) 103715, [\href{http://arxiv.org/abs/1907.12760}{{\tt
  arXiv:1907.12760}}].

\bibitem{Shao:2019gjj}
L.~Shao, {\it {Degeneracy in Studying the Supranuclear Equation of State and
  Modified Gravity with Neutron Stars}},  {\em AIP Conf. Proc.} {\bf 2127}
  (2019), no.~1 020016, [\href{http://arxiv.org/abs/1901.07546}{{\tt
  arXiv:1901.07546}}].

\bibitem{Antoniadis:2013pzd}
J.~Antoniadis et~al., {\it {A Massive Pulsar in a Compact Relativistic
  Binary}},  {\em Science} {\bf 340} (2013) 6131,
  [\href{http://arxiv.org/abs/1304.6875}{{\tt arXiv:1304.6875}}].

\bibitem{Crawford:2006xb}
F.~Crawford, M.~S.~E. Roberts, J.~W.~T. Hessels, S.~M. Ransom, M.~Livingstone,
  C.~R. Tam, and V.~M. Kaspi, {\it {A Survey of 56 Mid-latitude EGRET Error
  Boxes for Radio Pulsars}},  {\em Astrophys. J.} {\bf 652} (2006) 1499--1507,
  [\href{http://arxiv.org/abs/astro-ph/0608225}{{\tt astro-ph/0608225}}].

\bibitem{Linares:2018ppq}
M.~Linares, T.~Shahbaz, and J.~Casares, {\it {Peering into the dark side:
  Magnesium lines establish a massive neutron star in PSR J2215+5135}},  {\em
  Astrophys. J.} {\bf 859} (2018), no.~1 54,
  [\href{http://arxiv.org/abs/1805.08799}{{\tt arXiv:1805.08799}}].

\bibitem{Cromartie:2019kug}
{\bf NANOGrav} Collaboration, H.~Cromartie et~al., {\it {Relativistic Shapiro
  delay measurements of an extremely massive millisecond pulsar}},  {\em Nature
  Astron.} {\bf 4} (2019), no.~1 72--76,
  [\href{http://arxiv.org/abs/1904.06759}{{\tt arXiv:1904.06759}}].

\bibitem{Cooney:2009rr}
A.~Cooney, S.~DeDeo, and D.~Psaltis, {\it {Neutron Stars in f(R) Gravity with
  Perturbative Constraints}},  {\em Phys. Rev.} {\bf D82} (2010) 064033,
  [\href{http://arxiv.org/abs/0910.5480}{{\tt arXiv:0910.5480}}].

\bibitem{Orellana:2013gn}
M.~Orellana, F.~Garcia, F.~A. Teppa~Pannia, and G.~E. Romero, {\it {Structure
  of neutron stars in $R$-squared gravity}},  {\em Gen. Rel. Grav.} {\bf 45}
  (2013) 771--783, [\href{http://arxiv.org/abs/1301.5189}{{\tt
  arXiv:1301.5189}}].

\bibitem{Yazadjiev:2014cza}
S.~S. Yazadjiev, D.~D. Doneva, K.~D. Kokkotas, and K.~V. Staykov, {\it
  {Non-perturbative and self-consistent models of neutron stars in R-squared
  gravity}},  {\em JCAP} {\bf 1406} (2014) 003,
  [\href{http://arxiv.org/abs/1402.4469}{{\tt arXiv:1402.4469}}].

\bibitem{Astashenok:2017dpo}
A.~V. Astashenok, S.~D. Odintsov, and A.~de~la Cruz-Dombriz, {\it {The
  realistic models of relativistic stars in $f(R) = R + \alpha R^2$ gravity}},
  {\em Class. Quant. Grav.} {\bf 34} (2017), no.~20 205008,
  [\href{http://arxiv.org/abs/1704.08311}{{\tt arXiv:1704.08311}}].

\bibitem{Astashenok:2018iav}
A.~V. Astashenok, A.~S. Baigashov, and S.~A. Lapin, {\it {Neutron stars in
  frames of $R^2$-gravity and gravitational waves}},  {\em Int. J. Geom. Meth.
  Mod. Phys.} {\bf 16} (2018), no.~01 1950004,
  [\href{http://arxiv.org/abs/1812.10439}{{\tt arXiv:1812.10439}}].

\bibitem{Folomeev:2018ioy}
V.~Folomeev, {\it {Anisotropic neutron stars in $R^2$ gravity}},  {\em Phys.
  Rev.} {\bf D97} (2018), no.~12 124009,
  [\href{http://arxiv.org/abs/1802.01801}{{\tt arXiv:1802.01801}}].

\bibitem{Capozziello:2015yza}
S.~Capozziello, M.~De~Laurentis, R.~Farinelli, and S.~D. Odintsov, {\it
  {Mass-radius relation for neutron stars in f(R) gravity}},  {\em Phys. Rev.}
  {\bf D93} (2016), no.~2 023501, [\href{http://arxiv.org/abs/1509.04163}{{\tt
  arXiv:1509.04163}}].

\bibitem{Cognola:2007zu}
G.~Cognola, E.~Elizalde, S.~Nojiri, S.~D. Odintsov, L.~Sebastiani, and
  S.~Zerbini, {\it {A Class of viable modified f(R) gravities describing
  inflation and the onset of accelerated expansion}},  {\em Phys. Rev.} {\bf
  D77} (2008) 046009, [\href{http://arxiv.org/abs/0712.4017}{{\tt
  arXiv:0712.4017}}].

\bibitem{Astashenok:2013vza}
A.~V. Astashenok, S.~Capozziello, and S.~D. Odintsov, {\it {Further stable
  neutron star models from f(R) gravity}},  {\em JCAP} {\bf 1312} (2013) 040,
  [\href{http://arxiv.org/abs/1309.1978}{{\tt arXiv:1309.1978}}].

\bibitem{Alavirad:2013paa}
H.~Alavirad and J.~M. Weller, {\it {Modified gravity with logarithmic curvature
  corrections and the structure of relativistic stars}},  {\em Phys. Rev.} {\bf
  D88} (2013), no.~12 124034, [\href{http://arxiv.org/abs/1307.7977}{{\tt
  arXiv:1307.7977}}].

\bibitem{DeLaurentis:2018odx}
M.~De~Laurentis, {\it {Noether's stars in $f(\cal {R})$ gravity}},  {\em Phys.
  Lett.} {\bf B780} (2018) 205--210,
  [\href{http://arxiv.org/abs/1802.09073}{{\tt arXiv:1802.09073}}].

\bibitem{Kase:2019dqc}
R.~Kase and S.~Tsujikawa, {\it {Neutron stars in $f(R)$ gravity and
  scalar-tensor theories}},  {\em JCAP} {\bf 1909} (2019), no.~09 054,
  [\href{http://arxiv.org/abs/1906.08954}{{\tt arXiv:1906.08954}}].

\bibitem{Astashenok:2014pua}
A.~V. Astashenok, S.~Capozziello, and S.~D. Odintsov, {\it {Maximal neutron
  star mass and the resolution of the hyperon puzzle in modified gravity}},
  {\em Phys. Rev.} {\bf D89} (2014), no.~10 103509,
  [\href{http://arxiv.org/abs/1401.4546}{{\tt arXiv:1401.4546}}].

\bibitem{Horbatsch:2010hj}
M.~W. Horbatsch and C.~P. Burgess, {\it {Semi-Analytic Stellar Structure in
  Scalar-Tensor Gravity}},  {\em JCAP} {\bf 1108} (2011) 027,
  [\href{http://arxiv.org/abs/1006.4411}{{\tt arXiv:1006.4411}}].

\bibitem{Cisterna:2015yla}
A.~Cisterna, T.~Delsate, and M.~Rinaldi, {\it {Neutron stars in general second
  order scalar-tensor theory: The case of nonminimal derivative coupling}},
  {\em Phys. Rev.} {\bf D92} (2015), no.~4 044050,
  [\href{http://arxiv.org/abs/1504.05189}{{\tt arXiv:1504.05189}}].

\bibitem{Wojnar:2016bzk}
A.~Wojnar and H.~Velten, {\it {Equilibrium and stability of relativistic stars
  in extended theories of gravity}},  {\em Eur. Phys. J. C} {\bf 76} (2016),
  no.~12 697, [\href{http://arxiv.org/abs/1604.04257}{{\tt arXiv:1604.04257}}].

\bibitem{Sotani:2017pfj}
H.~Sotani and K.~D. Kokkotas, {\it {Maximum mass limit of neutron stars in
  scalar-tensor gravity}},  {\em Phys. Rev.} {\bf D95} (2017), no.~4 044032,
  [\href{http://arxiv.org/abs/1702.00874}{{\tt arXiv:1702.00874}}].

\bibitem{Silva:2014ora}
H.~O. Silva, H.~Sotani, E.~Berti, and M.~Horbatsch, {\it {Torsional
  oscillations of neutron stars in scalar-tensor theory of gravity}},  {\em
  Phys. Rev.} {\bf D90} (2014), no.~12 124044,
  [\href{http://arxiv.org/abs/1410.2511}{{\tt arXiv:1410.2511}}].

\bibitem{Horbatsch:2015bua}
M.~Horbatsch, H.~O. Silva, D.~Gerosa, P.~Pani, E.~Berti, L.~Gualtieri, and
  U.~Sperhake, {\it {Tensor-multi-scalar theories: relativistic stars and 3 + 1
  decomposition}},  {\em Class. Quant. Grav.} {\bf 32} (2015), no.~20 204001,
  [\href{http://arxiv.org/abs/1505.07462}{{\tt arXiv:1505.07462}}].

\bibitem{Novak:1997hw}
J.~Novak, {\it {Spherical neutron star collapse in tensor - scalar theory of
  gravity}},  {\em Phys. Rev.} {\bf D57} (1998) 4789--4801,
  [\href{http://arxiv.org/abs/gr-qc/9707041}{{\tt gr-qc/9707041}}].

\bibitem{Crisostomi:2017pjs}
M.~Crisostomi and K.~Koyama, {\it {Self-accelerating universe in scalar-tensor
  theories after GW170817}},  {\em Phys. Rev.} {\bf D97} (2018), no.~8 084004,
  [\href{http://arxiv.org/abs/1712.06556}{{\tt arXiv:1712.06556}}].

\bibitem{Babichev:2016jom}
E.~Babichev, K.~Koyama, D.~Langlois, R.~Saito, and J.~Sakstein, {\it
  {Relativistic Stars in Beyond Horndeski Theories}},  {\em Class. Quant.
  Grav.} {\bf 33} (2016), no.~23 235014,
  [\href{http://arxiv.org/abs/1606.06627}{{\tt arXiv:1606.06627}}].

\bibitem{Chagoya:2018lmv}
J.~Chagoya and G.~Tasinato, {\it {Compact objects in scalar-tensor theories
  after GW170817}},  {\em JCAP} {\bf 1808} (2018), no.~08 006,
  [\href{http://arxiv.org/abs/1803.07476}{{\tt arXiv:1803.07476}}].

\bibitem{Kase:2017egk}
R.~Kase, M.~Minamitsuji, and S.~Tsujikawa, {\it {Relativistic stars in
  vector-tensor theories}},  {\em Phys. Rev.} {\bf D97} (2018), no.~8 084009,
  [\href{http://arxiv.org/abs/1711.08713}{{\tt arXiv:1711.08713}}].

\bibitem{Pani:2011xm}
P.~Pani, E.~Berti, V.~Cardoso, and J.~Read, {\it {Compact stars in alternative
  theories of gravity. Einstein-Dilaton-Gauss-Bonnet gravity}},  {\em Phys.
  Rev.} {\bf D84} (2011) 104035, [\href{http://arxiv.org/abs/1109.0928}{{\tt
  arXiv:1109.0928}}].

\bibitem{Panotopoulos:2019zxv}
G.~Panotopoulos and \'A.~Rinc\'on, {\it {Relativistic strange quark stars in
  Lovelock gravity}},  {\em Eur. Phys. J. Plus} {\bf 134} (2019), no.~9 472,
  [\href{http://arxiv.org/abs/1907.03545}{{\tt arXiv:1907.03545}}].

\bibitem{Silva:2017uqg}
H.~O. Silva, J.~Sakstein, L.~Gualtieri, T.~P. Sotiriou, and E.~Berti, {\it
  {Spontaneous scalarization of black holes and compact stars from a
  Gauss-Bonnet coupling}},  {\em Phys. Rev. Lett.} {\bf 120} (2018), no.~13
  131104, [\href{http://arxiv.org/abs/1711.02080}{{\tt arXiv:1711.02080}}].

\bibitem{Doneva:2017duq}
D.~D. Doneva and S.~S. Yazadjiev, {\it {Neutron star solutions with curvature
  induced scalarization in the extended Gauss-Bonnet scalar-tensor theories}},
  {\em JCAP} {\bf 1804} (2018), no.~04 011,
  [\href{http://arxiv.org/abs/1712.03715}{{\tt arXiv:1712.03715}}].

\bibitem{Blazquez-Salcedo:2015ets}
J.~L. Blazquez-Salcedo, L.~M. Gonzalez-Romero, J.~Kunz, S.~Mojica, and
  F.~Navarro-Lerida, {\it {Axial quasinormal modes of
  Einstein-Gauss-Bonnet-dilaton neutron stars}},  {\em Phys. Rev.} {\bf D93}
  (2016), no.~2 024052, [\href{http://arxiv.org/abs/1511.03960}{{\tt
  arXiv:1511.03960}}].

\bibitem{Katsuragawa:2015lbl}
T.~Katsuragawa, S.~Nojiri, S.~D. Odintsov, and M.~Yamazaki, {\it {Relativistic
  stars in de Rham-Gabadadze-Tolley massive gravity}},  {\em Phys. Rev.} {\bf
  D93} (2016) 124013, [\href{http://arxiv.org/abs/1512.00660}{{\tt
  arXiv:1512.00660}}].

\bibitem{Kareeso:2018xum}
P.~Kareeso, P.~Burikham, and T.~Harko, {\it {Mass-radius ratio bounds for
  compact objects in Lorentz-violating dRGT massive gravity theory}},  {\em
  Eur. Phys. J.} {\bf C78} (2018), no.~11 941,
  [\href{http://arxiv.org/abs/1802.01017}{{\tt arXiv:1802.01017}}].

\bibitem{Hendi:2017ibm}
S.~H. Hendi, G.~H. Bordbar, B.~Eslam~Panah, and S.~Panahiyan, {\it {Neutron
  stars structure in the context of massive gravity}},  {\em JCAP} {\bf 1707}
  (2017) 004, [\href{http://arxiv.org/abs/1701.01039}{{\tt arXiv:1701.01039}}].

\bibitem{Enander:2015kda}
J.~Enander and E.~Mortsell, {\it {On stars, galaxies and black holes in massive
  bigravity}},  {\em JCAP} {\bf 1511} (2015), no.~11 023,
  [\href{http://arxiv.org/abs/1507.00912}{{\tt arXiv:1507.00912}}].

\bibitem{Aoki:2016eov}
K.~Aoki, K.-i. Maeda, and M.~Tanabe, {\it {Relativistic stars in bigravity
  theory}},  {\em Phys. Rev.} {\bf D93} (2016), no.~6 064054,
  [\href{http://arxiv.org/abs/1602.02227}{{\tt arXiv:1602.02227}}].

\bibitem{Hendi:2015vta}
S.~H. Hendi, G.~H. Bordbar, B.~E. Panah, and S.~Panahiyan, {\it {Modified TOV
  in gravity's rainbow: properties of neutron stars and dynamical stability
  conditions}},  {\em JCAP} {\bf 1609} (2016), no.~09 013,
  [\href{http://arxiv.org/abs/1509.05145}{{\tt arXiv:1509.05145}}].

\bibitem{Ilijic:2018ulf}
S.~Ilijic and M.~Sossich, {\it {Compact stars in $f(T)$ extended theory of
  gravity}},  {\em Phys. Rev.} {\bf D98} (2018), no.~6 064047,
  [\href{http://arxiv.org/abs/1807.03068}{{\tt arXiv:1807.03068}}].

\bibitem{DeBenedictis:2016aze}
A.~DeBenedictis and S.~Ilijic, {\it {Spherically symmetric vacuum in covariant
  $F(T) = T + \frac{\alpha}{2}T^{2} + \mathcal{O}(T^{\gamma})$ gravity
  theory}},  {\em Phys. Rev.} {\bf D94} (2016), no.~12 124025,
  [\href{http://arxiv.org/abs/1609.07465}{{\tt arXiv:1609.07465}}].

\bibitem{Deb:2018voz}
D.~Deb, S.~Ghosh, S.~K. Maurya, M.~Khlopov, and S.~Ray, {\it {Anisotropic
  compact stars in $f(T)$ gravity under Karmarkar condition}},
  \href{http://arxiv.org/abs/1811.11797}{{\tt arXiv:1811.11797}}.

\bibitem{Astashenok:2015qzw}
A.~V. Astashenok and S.~D. Odintsov, {\it {From neutron stars to quark stars in
  mimetic gravity}},  {\em Phys. Rev.} {\bf D94} (2016), no.~6 063008,
  [\href{http://arxiv.org/abs/1512.07279}{{\tt arXiv:1512.07279}}].

\bibitem{Fabris:2019qvy}
J.~C. Fabris, H.~Velten, and A.~Wojnar, {\it {Existence of static
  spherically-symmetric objects in action-dependent Lagrangian theories}},
  {\em Phys. Rev. D} {\bf 99} (2019), no.~12 124031,
  [\href{http://arxiv.org/abs/1903.12193}{{\tt arXiv:1903.12193}}].

\bibitem{Xu:2019gua}
R.~Xu, J.~Zhao, and L.~Shao, {\it {Neutron Star Structure in the Minimal
  Gravitational Standard-Model Extension and the Implication to Continuous
  Gravitational Waves}},  \href{http://arxiv.org/abs/1909.10372}{{\tt
  arXiv:1909.10372}}.

\bibitem{Kim:2018dbs}
K.~Kim, J.~J. Oh, C.~Park, and E.~J. Son, {\it {Neutron Star Structure in
  Ho\v{r}ava-Lifshitz Gravity}},  \href{http://arxiv.org/abs/1810.07497}{{\tt
  arXiv:1810.07497}}.

\bibitem{Eling:2007xh}
C.~Eling, T.~Jacobson, and M.~Coleman~Miller, {\it {Neutron stars in
  Einstein-aether theory}},  {\em Phys. Rev.} {\bf D76} (2007) 042003,
  [\href{http://arxiv.org/abs/0705.1565}{{\tt arXiv:0705.1565}}]. [Erratum:
  Phys. Rev.D80,129906(2009)].

\bibitem{Barausse:2019yuk}
E.~Barausse, {\it {Neutron star sensitivities in Horava gravity after
  GW170817}},  {\em Phys. Rev.} {\bf D100} (2019), no.~8 084053,
  [\href{http://arxiv.org/abs/1907.05958}{{\tt arXiv:1907.05958}}].

\bibitem{Moraes:2015uxq}
P.~H. R.~S. Moraes, J.~D.~V. Arbañil, and M.~Malheiro, {\it {Stellar
  equilibrium configurations of compact stars in $f(R,T)$ gravity}},  {\em
  JCAP} {\bf 1606} (2016) 005, [\href{http://arxiv.org/abs/1511.06282}{{\tt
  arXiv:1511.06282}}].

\bibitem{Das:2016mxq}
A.~Das, F.~Rahaman, B.~K. Guha, and S.~Ray, {\it {Compact stars in
  $f(R,\mathcal {T})$ gravity}},  {\em Eur. Phys. J.} {\bf C76} (2016), no.~12
  654, [\href{http://arxiv.org/abs/1608.00566}{{\tt arXiv:1608.00566}}].

\bibitem{Deb:2018gzt}
D.~Deb, S.~V. Ketov, M.~Khlopov, and S.~Ray, {\it {Study on charged strange
  stars in $f(R, T)$ gravity}},  {\em JCAP} {\bf 1910} (2019), no.~10 070,
  [\href{http://arxiv.org/abs/1812.11736}{{\tt arXiv:1812.11736}}].

\bibitem{Maurya:2019hds}
S.~K. Maurya and F.~Tello-Ortiz, {\it {Charged anisotropic compact star in
  $f(R,\mathcal{T})$ gravity: A minimal geometric deformation gravitational
  decoupling approach}},  \href{http://arxiv.org/abs/1905.13519}{{\tt
  arXiv:1905.13519}}.

\bibitem{Maurya:2019sfm}
S.~K. Maurya, A.~Errehymy, D.~Deb, F.~Tello-Ortiz, and M.~Daoud, {\it {Study of
  anisotropic strange stars in $f(R,T)$ gravity: An embedding approach under
  the simplest linear functional of the matter-geometry coupling}},  {\em Phys.
  Rev.} {\bf D100} (2019), no.~4 044014,
  [\href{http://arxiv.org/abs/1907.10149}{{\tt arXiv:1907.10149}}].

\bibitem{Carvalho:2019gzs}
G.~A. Carvalho, S.~I.~d. Santos, P.~H. R.~S. Moraes, and M.~Malheiro, {\it
  {Strange stars in energy-momentum-conserved $f(R,T)$ gravity}},
  \href{http://arxiv.org/abs/1911.02484}{{\tt arXiv:1911.02484}}.

\bibitem{Oliveira:2015lka}
A.~M. Oliveira, H.~E.~S. Velten, J.~C. Fabris, and L.~Casarini, {\it {Neutron
  Stars in Rastall Gravity}},  {\em Phys. Rev.} {\bf D92} (2015), no.~4 044020,
  [\href{http://arxiv.org/abs/1506.00567}{{\tt arXiv:1506.00567}}].

\bibitem{Hansraj:2018reh}
S.~Hansraj and A.~Banerjee, {\it {Equilibrium stellar configurations in Rastall
  theory and linear equation of state}},
  \href{http://arxiv.org/abs/1807.00812}{{\tt arXiv:1807.00812}}.

\bibitem{Kainulainen:2006wz}
K.~Kainulainen, V.~Reijonen, and D.~Sunhede, {\it {The Interior spacetimes of
  stars in Palatini f(R) gravity}},  {\em Phys. Rev.} {\bf D76} (2007) 043503,
  [\href{http://arxiv.org/abs/gr-qc/0611132}{{\tt gr-qc/0611132}}].

\bibitem{Pannia:2016qbj}
F.~A. Teppa~Pannia, F.~García, S.~E. Perez~Bergliaffa, M.~Orellana, and G.~E.
  Romero, {\it {Structure of Compact Stars in R-squared Palatini Gravity}},
  {\em Gen. Rel. Grav.} {\bf 49} (2017), no.~2 25,
  [\href{http://arxiv.org/abs/1607.03508}{{\tt arXiv:1607.03508}}].

\bibitem{Panotopoulos:2017jdc}
G.~Panotopoulos, {\it {Strange stars in $f(R)$ theories of gravity in the
  Palatini formalism}},  {\em Gen. Rel. Grav.} {\bf 49} (2017), no.~5 69,
  [\href{http://arxiv.org/abs/1704.04961}{{\tt arXiv:1704.04961}}].

\bibitem{Wojnar:2017tmy}
A.~Wojnar, {\it {On stability of a neutron star system in Palatini gravity}},
  {\em Eur. Phys. J. C} {\bf 78} (2018), no.~5 421,
  [\href{http://arxiv.org/abs/1712.01943}{{\tt arXiv:1712.01943}}].

\bibitem{Bull:2015stt}
P.~Bull et~al., {\it {Beyond $\Lambda$CDM: Problems, solutions, and the road
  ahead}},  {\em Phys. Dark Univ.} {\bf 12} (2016) 56--99,
  [\href{http://arxiv.org/abs/1512.05356}{{\tt arXiv:1512.05356}}].

\bibitem{Berti:2015itd}
E.~Berti et~al., {\it {Testing General Relativity with Present and Future
  Astrophysical Observations}},  {\em Class. Quant. Grav.} {\bf 32} (2015)
  243001, [\href{http://arxiv.org/abs/1501.07274}{{\tt arXiv:1501.07274}}].

\bibitem{Stergioulas:2003yp}
N.~Stergioulas, {\it {Rotating Stars in Relativity}},  {\em Living Rev. Rel.}
  {\bf 6} (2003) 3, [\href{http://arxiv.org/abs/gr-qc/0302034}{{\tt
  gr-qc/0302034}}].

\bibitem{Staykov:2014mwa}
K.~V. Staykov, D.~D. Doneva, S.~S. Yazadjiev, and K.~D. Kokkotas, {\it {Slowly
  rotating neutron and strange stars in $R^2$ gravity}},  {\em JCAP} {\bf 1410}
  (2014), no.~10 006, [\href{http://arxiv.org/abs/1407.2180}{{\tt
  arXiv:1407.2180}}].

\bibitem{Staykov:2015kwa}
K.~V. Staykov, D.~D. Doneva, and S.~S. Yazadjiev, {\it {Orbital and epicyclic
  frequencies around neutron and strange stars in $R^2$ gravity}},  {\em Eur.
  Phys. J.} {\bf C75} (2015), no.~12 607,
  [\href{http://arxiv.org/abs/1508.07790}{{\tt arXiv:1508.07790}}].

\bibitem{Staykov:2015cfa}
K.~V. Staykov, D.~D. Doneva, S.~S. Yazadjiev, and K.~D. Kokkotas, {\it
  {Gravitational wave asteroseismology of neutron and strange stars in R$^2$
  gravity}},  {\em Phys. Rev.} {\bf D92} (2015), no.~4 043009,
  [\href{http://arxiv.org/abs/1503.04711}{{\tt arXiv:1503.04711}}].

\bibitem{Silva:2018yxz}
H.~O. Silva and N.~Yunes, {\it {Neutron star pulse profiles in scalar-tensor
  theories of gravity}},  {\em Phys. Rev.} {\bf D99} (2019), no.~4 044034,
  [\href{http://arxiv.org/abs/1808.04391}{{\tt arXiv:1808.04391}}].

\bibitem{Staykov:2015mma}
K.~Staykov, K.~Y. Ekşi, S.~S. Yazadjiev, M.~M. Türkoğlu, and A.~S.
  Arapoğlu, {\it {Moment of inertia of neutron star crust in alternative and
  modified theories of gravity}},  {\em Phys. Rev.} {\bf D94} (2016), no.~2
  024056, [\href{http://arxiv.org/abs/1507.05878}{{\tt arXiv:1507.05878}}].

\bibitem{Popchev:2018fwu}
D.~Popchev, K.~V. Staykov, D.~D. Doneva, and S.~S. Yazadjiev, {\it {Moment of
  inertia–mass universal relations for neutron stars in scalar-tensor theory
  with self-interacting massive scalar field}},  {\em Eur. Phys. J.} {\bf C79}
  (2019), no.~2 178, [\href{http://arxiv.org/abs/1812.00347}{{\tt
  arXiv:1812.00347}}].

\bibitem{Staykov:2018hhc}
K.~V. Staykov, D.~Popchev, D.~D. Doneva, and S.~S. Yazadjiev, {\it {Static and
  slowly rotating neutron stars in scalar–tensor theory with self-interacting
  massive scalar field}},  {\em Eur. Phys. J.} {\bf C78} (2018), no.~7 586,
  [\href{http://arxiv.org/abs/1805.07818}{{\tt arXiv:1805.07818}}].

\bibitem{Silva:2016smx}
H.~O. Silva, A.~Maselli, M.~Minamitsuji, and E.~Berti, {\it {Compact objects in
  Horndeski gravity}},  {\em Int. J. Mod. Phys.} {\bf D25} (2016), no.~09
  1641006, [\href{http://arxiv.org/abs/1602.05997}{{\tt arXiv:1602.05997}}].

\bibitem{Cisterna:2016vdx}
A.~Cisterna, T.~Delsate, L.~Ducobu, and M.~Rinaldi, {\it {Slowly rotating
  neutron stars in the nonminimal derivative coupling sector of Horndeski
  gravity}},  {\em Phys. Rev.} {\bf D93} (2016), no.~8 084046,
  [\href{http://arxiv.org/abs/1602.06939}{{\tt arXiv:1602.06939}}].

\bibitem{Maselli:2016gxk}
A.~Maselli, H.~O. Silva, M.~Minamitsuji, and E.~Berti, {\it {Neutron stars in
  Horndeski gravity}},  {\em Phys. Rev.} {\bf D93} (2016), no.~12 124056,
  [\href{http://arxiv.org/abs/1603.04876}{{\tt arXiv:1603.04876}}].

\bibitem{Sullivan:2017kwo}
A.~Sullivan and N.~Yunes, {\it {Slowly-Rotating Neutron Stars in Massive
  Bigravity}},  {\em Class. Quant. Grav.} {\bf 35} (2018), no.~4 045003,
  [\href{http://arxiv.org/abs/1709.03311}{{\tt arXiv:1709.03311}}].

\bibitem{Silva:2014fca}
H.~O. Silva, C.~F.~B. Macedo, E.~Berti, and L.~C.~B. Crispino, {\it {Slowly
  rotating anisotropic neutron stars in general relativity and scalar–tensor
  theory}},  {\em Class. Quant. Grav.} {\bf 32} (2015) 145008,
  [\href{http://arxiv.org/abs/1411.6286}{{\tt arXiv:1411.6286}}].

\bibitem{Hartle:1967he}
J.~B. Hartle, {\it {Slowly rotating relativistic stars. 1. Equations of
  structure}},  {\em Astrophys. J.} {\bf 150} (1967) 1005--1029.

\bibitem{Doneva:2013qva}
D.~D. Doneva, S.~S. Yazadjiev, N.~Stergioulas, and K.~D. Kokkotas, {\it
  {Rapidly rotating neutron stars in scalar-tensor theories of gravity}},  {\em
  Phys. Rev.} {\bf D88} (2013), no.~8 084060,
  [\href{http://arxiv.org/abs/1309.0605}{{\tt arXiv:1309.0605}}].

\bibitem{Yazadjiev:2015zia}
S.~S. Yazadjiev, D.~D. Doneva, and K.~D. Kokkotas, {\it {Rapidly rotating
  neutron stars in R-squared gravity}},  {\em Phys. Rev.} {\bf D91} (2015),
  no.~8 084018, [\href{http://arxiv.org/abs/1501.04591}{{\tt
  arXiv:1501.04591}}].

\bibitem{Yazadjiev:2017vpg}
S.~S. Yazadjiev, D.~D. Doneva, and K.~D. Kokkotas, {\it {Oscillation modes of
  rapidly rotating neutron stars in scalar-tensor theories of gravity}},  {\em
  Phys. Rev.} {\bf D96} (2017), no.~6 064002,
  [\href{http://arxiv.org/abs/1705.06984}{{\tt arXiv:1705.06984}}].

\bibitem{Kleihaus:2016dui}
B.~Kleihaus, J.~Kunz, S.~Mojica, and M.~Zagermann, {\it {Rapidly Rotating
  Neutron Stars in Dilatonic Einstein-Gauss-Bonnet Theory}},  {\em Phys. Rev.}
  {\bf D93} (2016), no.~6 064077, [\href{http://arxiv.org/abs/1601.05583}{{\tt
  arXiv:1601.05583}}].

\bibitem{Doneva:2018ouu}
D.~D. Doneva, S.~S. Yazadjiev, N.~Stergioulas, and K.~D. Kokkotas, {\it
  {Differentially rotating neutron stars in scalar-tensor theories of
  gravity}},  {\em Phys. Rev.} {\bf D98} (2018), no.~10 104039,
  [\href{http://arxiv.org/abs/1807.05449}{{\tt arXiv:1807.05449}}].

\bibitem{Pani:2014jra}
P.~Pani and E.~Berti, {\it {Slowly rotating neutron stars in scalar-tensor
  theories}},  {\em Phys. Rev.} {\bf D90} (2014), no.~2 024025,
  [\href{http://arxiv.org/abs/1405.4547}{{\tt arXiv:1405.4547}}].

\bibitem{Barausse:2012da}
E.~Barausse, C.~Palenzuela, M.~Ponce, and L.~Lehner, {\it {Neutron-star mergers
  in scalar-tensor theories of gravity}},  {\em Phys. Rev.} {\bf D87} (2013)
  081506, [\href{http://arxiv.org/abs/1212.5053}{{\tt arXiv:1212.5053}}].

\bibitem{Yagi:2013mbt}
K.~Yagi, L.~C. Stein, N.~Yunes, and T.~Tanaka, {\it {Isolated and Binary
  Neutron Stars in Dynamical Chern-Simons Gravity}},  {\em Phys. Rev.} {\bf
  D87} (2013) 084058, [\href{http://arxiv.org/abs/1302.1918}{{\tt
  arXiv:1302.1918}}]. [Erratum: Phys. Rev.D93,no.8,089909(2016)].

\bibitem{Okounkova:2019dfo}
M.~Okounkova, L.~C. Stein, M.~A. Scheel, and S.~A. Teukolsky, {\it {Numerical
  binary black hole collisions in dynamical Chern-Simons gravity}},  {\em Phys.
  Rev.} {\bf D100} (2019), no.~10 104026,
  [\href{http://arxiv.org/abs/1906.08789}{{\tt arXiv:1906.08789}}].

\bibitem{Carson:2019fxr}
Z.~Carson, B.~C. Seymour, and K.~Yagi, {\it {Future Prospects for Probing
  Scalar-Tensor Theories with Gravitational Waves from Mixed Binaries}},
  \href{http://arxiv.org/abs/1907.03897}{{\tt arXiv:1907.03897}}.

\bibitem{Zhang:2019iim}
C.~Zhang, X.~Zhao, A.~Wang, B.~Wang, K.~Yagi, N.~Yunes, W.~Zhao, and T.~Zhu,
  {\it {Gravitational waves from the quasi-circular inspiral of compact
  binaries in Einstein-aether theory}},
  \href{http://arxiv.org/abs/1911.10278}{{\tt arXiv:1911.10278}}.

\bibitem{Zhao:2019kif}
X.~Zhao et~al., {\it {Gravitational waveforms and radiation powers of the
  triple system PSR J0337+1715 in modified theories of gravity}},  {\em Phys.
  Rev.} {\bf D100} (2019), no.~8 083012,
  [\href{http://arxiv.org/abs/1903.09865}{{\tt arXiv:1903.09865}}].

\bibitem{Capozziello:2011nr}
S.~Capozziello, M.~De~Laurentis, S.~D. Odintsov, and A.~Stabile, {\it
  {Hydrostatic equilibrium and stellar structure in f(R)-gravity}},  {\em Phys.
  Rev.} {\bf D83} (2011) 064004, [\href{http://arxiv.org/abs/1101.0219}{{\tt
  arXiv:1101.0219}}].

\bibitem{Farinelli:2013pza}
R.~Farinelli, M.~De~Laurentis, S.~Capozziello, and S.~D. Odintsov, {\it
  {Numerical solutions of the modified Lane–Emden equation in
  $f(R)$-gravity}},  {\em Mon. Not. Roy. Astron. Soc.} {\bf 440} (2014)
  2909--2915, [\href{http://arxiv.org/abs/1311.2744}{{\tt arXiv:1311.2744}}].

\bibitem{Andre:2017usv}
R.~André and G.~M. Kremer, {\it {Stellar structure model in hydrostatic
  equilibrium in the context of $f(\mathcal{R})$-gravity}},  {\em Res. Astron.
  Astrophys.} {\bf 17} (2017), no.~12 122,
  [\href{http://arxiv.org/abs/1707.07675}{{\tt arXiv:1707.07675}}].

\bibitem{Koyama:2015oma}
K.~Koyama and J.~Sakstein, {\it {Astrophysical Probes of the Vainshtein
  Mechanism: Stars and Galaxies}},  {\em Phys. Rev.} {\bf D91} (2015) 124066,
  [\href{http://arxiv.org/abs/1502.06872}{{\tt arXiv:1502.06872}}].

\bibitem{Saito:2015fza}
R.~Saito, D.~Yamauchi, S.~Mizuno, J.~Gleyzes, and D.~Langlois, {\it {Modified
  gravity inside astrophysical bodies}},  {\em JCAP} {\bf 1506} (2015) 008,
  [\href{http://arxiv.org/abs/1503.01448}{{\tt arXiv:1503.01448}}].

\bibitem{Wibisono:2017dkt}
C.~Wibisono and A.~Sulaksono, {\it {Information-entropic method for studying
  the stability bound of nonrelativistic polytropic stars within modified
  gravity theories}},  {\em Int. J. Mod. Phys.} {\bf D27} (2018), no.~05
  1850051, [\href{http://arxiv.org/abs/1712.07587}{{\tt arXiv:1712.07587}}].

\bibitem{Sergyeyev:2019aul}
A.~Sergyeyev and A.~Wojnar, {\it {The Palatini star: exact solutions of the
  modified Lane-Emden equation}},  \href{http://arxiv.org/abs/1901.10448}{{\tt
  arXiv:1901.10448}}.

\bibitem{Chandrasekhar:1935zz}
S.~Chandrasekhar, {\it {The highly collapsed configurations of a stellar mass
  (Second paper)}},  {\em Mon. Not. Roy. Astron. Soc.} {\bf 95} (1935)
  207--225.

\bibitem{Jain:2015edg}
R.~K. Jain, C.~Kouvaris, and N.~G. Nielsen, {\it {White Dwarf Critical Tests
  for Modified Gravity}},  {\em Phys. Rev. Lett.} {\bf 116} (2016), no.~15
  151103, [\href{http://arxiv.org/abs/1512.05946}{{\tt arXiv:1512.05946}}].

\bibitem{Chowdhury:2018qrf}
S.~Chowdhury and T.~Sarkar, {\it {Small Anisotropy in Stellar Objects in
  Modified Theories of Gravity}},  \href{http://arxiv.org/abs/1811.07685}{{\tt
  arXiv:1811.07685}}.

\bibitem{Carvalho:2017pgk}
G.~A. Carvalho, R.~V. Lobato, P.~H. R.~S. Moraes, J.~D.~V. Arbañil, R.~M.
  Marinho, E.~Otoniel, and M.~Malheiro, {\it {Stellar equilibrium
  configurations of white dwarfs in the f(R, T) gravity}},  {\em Eur. Phys. J.}
  {\bf C77} (2017), no.~12 871, [\href{http://arxiv.org/abs/1706.03596}{{\tt
  arXiv:1706.03596}}].

\bibitem{Crisostomi:2019yfo}
M.~Crisostomi, M.~Lewandowski, and F.~Vernizzi, {\it {Vainshtein regime in
  scalar-tensor gravity: Constraints on degenerate higher-order scalar-tensor
  theories}},  {\em Phys. Rev.} {\bf D100} (2019), no.~2 024025,
  [\href{http://arxiv.org/abs/1903.11591}{{\tt arXiv:1903.11591}}].

\bibitem{Kalita:2019yaj}
S.~Kalita, B.~Mukhopadhyay, and T.~R. Govindarajan, {\it {Violation of
  Chandrasekhar mass-limit in noncommutative geometry: A strong possible
  explanation for the super-Chandrasekhar limiting mass white dwarfs}},
  \href{http://arxiv.org/abs/1912.00900}{{\tt arXiv:1912.00900}}.

\bibitem{Banerjee:2017uwz}
S.~Banerjee, S.~Shankar, and T.~P. Singh, {\it {Constraints on Modified Gravity
  Models from White Dwarfs}},  {\em JCAP} {\bf 1710} (2017), no.~10 004,
  [\href{http://arxiv.org/abs/1705.01048}{{\tt arXiv:1705.01048}}].

\bibitem{Olmo:2019qsj}
G.~J. Olmo, D.~Rubiera-Garcia, and A.~Wojnar, {\it {Minimum main sequence mass
  in quadratic Palatini $f(R)$ gravity}},  {\em Phys. Rev.} {\bf D100} (2019),
  no.~4 044020, [\href{http://arxiv.org/abs/1906.04629}{{\tt
  arXiv:1906.04629}}].

\bibitem{Kumarnum}
S.~Jumar, {\it {The Structure of Stars of Very Low Mass}},  {\em Astrophysical
  Journal} {\bf 137} (1963) 1121.

\bibitem{Burrows:1992fg}
A.~Burrows and J.~Liebert, {\it {The Science of brown dwarfs}},  {\em Rev. Mod.
  Phys.} {\bf 65} (1993) 301--336.

\bibitem{Sakstein:2015zoa}
J.~Sakstein, {\it {Hydrogen Burning in Low Mass Stars Constrains Scalar-Tensor
  Theories of Gravity}},  {\em Phys. Rev. Lett.} {\bf 115} (2015) 201101,
  [\href{http://arxiv.org/abs/1510.05964}{{\tt arXiv:1510.05964}}].

\bibitem{Segransan:2000jq}
D.~Segransan, X.~Delfosse, T.~Forveille, J.~L. Beuzit, S.~Udry, C.~Perrier, and
  M.~Mayor, {\it {Accurate masses of very low mass stars. 3. 16 New or improved
  masses}},  {\em Astron. Astrophys.} {\bf 364} (2000) 665,
  [\href{http://arxiv.org/abs/astro-ph/0010585}{{\tt astro-ph/0010585}}].

\bibitem{Brown:2018dum}
{\bf Gaia} Collaboration, A.~G.~A. Brown et~al., {\it {Gaia Data Release 2}},
  {\em Astron. Astrophys.} {\bf 616} (2018) A1,
  [\href{http://arxiv.org/abs/1804.09365}{{\tt arXiv:1804.09365}}].

\bibitem{Saumon:1995bn}
D.~Saumon, W.~B. Hubbard, A.~Burrows, T.~Guillot, J.~I. Lunine, and
  G.~Chabrier, {\it {A Theory of extrasolar giant planets}},  {\em Astrophys.
  J.} {\bf 460} (1996) 993--1018,
  [\href{http://arxiv.org/abs/astro-ph/9510046}{{\tt astro-ph/9510046}}].

\bibitem{aneta4}
A.~Wojnar, {\it {Stability of polytropic stars in Palatini gravity}},  {\em In
  preparation.} (2020).

\bibitem{chang2011stellar}
P.~Chang and L.~Hui, {\it Stellar structure and tests of modified gravity},
  {\em The Astrophysical Journal} {\bf 732} (2011), no.~1 25.

\bibitem{davis2012modified}
A.-C. Davis, E.~A. Lim, J.~Sakstein, and D.~J. Shaw, {\it Modified gravity
  makes galaxies brighter},  {\em Physical Review D} {\bf 85} (2012), no.~12
  123006.

\bibitem{Chowdhury:2020wfr}
S.~Chowdhury and T.~Sarkar, {\it {Modified gravity in the interior of
  population II stars}},  \href{http://arxiv.org/abs/2008.12264}{{\tt
  arXiv:2008.12264}}.

\bibitem{Wojnar:2020txr}
A.~Wojnar, {\it {Early evolutionary tracks of low-mass stellar objects in
  modified gravity}},  {\em Phys. Rev. D} {\bf 102} (2020) 124045,
  [\href{http://arxiv.org/abs/2007.13451}{{\tt arXiv:2007.13451}}].

\bibitem{Wojnar:2020frr}
A.~Wojnar, {\it {Lithium abundance is a gravitational model dependent
  quantity}},  \href{http://arxiv.org/abs/2009.10983}{{\tt arXiv:2009.10983}}.

\bibitem{Masuda_2019}
K.~Masuda, H.~Kawahara, D.~W. Latham, A.~Bieryla, M.~Kunitomo, M.~MacLeod, and
  W.~Aoki, {\it Self-lensing discovery of a 0.2 m $\odot$ white dwarf in an
  unusually wide orbit around a sun-like star},  {\em The Astrophysical
  Journal} {\bf 881} (aug, 2019) L3.

\bibitem{Laughlin_1997}
G.~Laughlin, P.~Bodenheimer, and F.~C. Adams, {\it The end of the main
  sequence},  {\em The Astrophysical Journal} {\bf 482} (jun, 1997) 420--432.

\bibitem{Abbott:2016nmj}
{\bf LIGO Scientific, Virgo} Collaboration, B.~P. Abbott et~al., {\it
  {GW151226: Observation of Gravitational Waves from a 22-Solar-Mass Binary
  Black Hole Coalescence}},  {\em Phys. Rev. Lett.} {\bf 116} (2016), no.~24
  241103, [\href{http://arxiv.org/abs/1606.04855}{{\tt arXiv:1606.04855}}].

\bibitem{Abbott:2017oio}
{\bf LIGO Scientific, Virgo} Collaboration, B.~P. Abbott et~al., {\it
  {GW170814: A Three-Detector Observation of Gravitational Waves from a Binary
  Black Hole Coalescence}},  {\em Phys. Rev. Lett.} {\bf 119} (2017), no.~14
  141101, [\href{http://arxiv.org/abs/1709.09660}{{\tt arXiv:1709.09660}}].

\bibitem{Abbott:2017vtc}
{\bf LIGO Scientific, VIRGO} Collaboration, B.~P. Abbott et~al., {\it
  {GW170104: Observation of a 50-Solar-Mass Binary Black Hole Coalescence at
  Redshift 0.2}},  {\em Phys. Rev. Lett.} {\bf 118} (2017), no.~22 221101,
  [\href{http://arxiv.org/abs/1706.01812}{{\tt arXiv:1706.01812}}]. [Erratum:
  Phys. Rev. Lett.121,no.12,129901(2018)].

\bibitem{LIGOScientific:2018mvr}
{\bf LIGO Scientific, Virgo} Collaboration, B.~P. Abbott et~al., {\it {GWTC-1:
  A Gravitational-Wave Transient Catalog of Compact Binary Mergers Observed by
  LIGO and Virgo during the First and Second Observing Runs}},
  \href{http://arxiv.org/abs/1811.12907}{{\tt arXiv:1811.12907}}.

\bibitem{Abbott:2017xzg}
{\bf LIGO Scientific, Virgo} Collaboration, B.~P. Abbott et~al., {\it
  {GW170817: Implications for the Stochastic Gravitational-Wave Background from
  Compact Binary Coalescences}},  {\em Phys. Rev. Lett.} {\bf 120} (2018),
  no.~9 091101, [\href{http://arxiv.org/abs/1710.05837}{{\tt
  arXiv:1710.05837}}].

\bibitem{Coulter:2017wya}
D.~A. Coulter et~al., {\it {Swope Supernova Survey 2017a (SSS17a), the Optical
  Counterpart to a Gravitational Wave Source}},  {\em Science} (2017)
  [\href{http://arxiv.org/abs/1710.05452}{{\tt arXiv:1710.05452}}].
  [Science358,1556(2017)].

\bibitem{GBM:2017lvd}
{\bf LIGO Scientific, Virgo, Fermi GBM, INTEGRAL, IceCube, AstroSat Cadmium
  Zinc Telluride Imager Team, IPN, Insight-Hxmt, ANTARES, Swift, AGILE Team,
  1M2H Team, Dark Energy Camera GW-EM, DES, DLT40, GRAWITA, Fermi-LAT, ATCA,
  ASKAP, Las Cumbres Observatory Group, OzGrav, DWF (Deeper Wider Faster
  Program), AST3, CAASTRO, VINROUGE, MASTER, J-GEM, GROWTH, JAGWAR,
  CaltechNRAO, TTU-NRAO, NuSTAR, Pan-STARRS, MAXI Team, TZAC Consortium, KU,
  Nordic Optical Telescope, ePESSTO, GROND, Texas Tech University, SALT Group,
  TOROS, BOOTES, MWA, CALET, IKI-GW Follow-up, H.E.S.S., LOFAR, LWA, HAWC,
  Pierre Auger, ALMA, Euro VLBI Team, Pi of Sky, Chandra Team at McGill
  University, DFN, ATLAS Telescopes, High Time Resolution Universe Survey,
  RIMAS, RATIR, SKA South Africa/MeerKAT} Collaboration, B.~P. Abbott et~al.,
  {\it {Multi-messenger Observations of a Binary Neutron Star Merger}},  {\em
  Astrophys. J.} {\bf 848} (2017), no.~2 L12,
  [\href{http://arxiv.org/abs/1710.05833}{{\tt arXiv:1710.05833}}].

\bibitem{Abbott:2018wiz}
{\bf LIGO Scientific, Virgo} Collaboration, B.~P. Abbott et~al., {\it
  {Properties of the binary neutron star merger GW170817}},  {\em Phys. Rev.}
  {\bf X9} (2019), no.~1 011001, [\href{http://arxiv.org/abs/1805.11579}{{\tt
  arXiv:1805.11579}}].

\bibitem{Akiyama:2019cqa}
{\bf Event Horizon Telescope} Collaboration, K.~Akiyama et~al., {\it {First M87
  Event Horizon Telescope Results. I. The Shadow of the Supermassive Black
  Hole}},  {\em Astrophys. J.} {\bf 875} (2019), no.~1 L1,
  [\href{http://arxiv.org/abs/1906.11238}{{\tt arXiv:1906.11238}}].

\bibitem{Akiyama:2019fyp}
{\bf Event Horizon Telescope} Collaboration, K.~Akiyama et~al., {\it {First M87
  Event Horizon Telescope Results. V. Physical Origin of the Asymmetric Ring}},
   {\em Astrophys. J.} {\bf 875} (2019), no.~1 L5,
  [\href{http://arxiv.org/abs/1906.11242}{{\tt arXiv:1906.11242}}].

\bibitem{Lattimer:2012nd}
J.~M. Lattimer, {\it {The nuclear equation of state and neutron star masses}},
  {\em Ann. Rev. Nucl. Part. Sci.} {\bf 62} (2012) 485--515,
  [\href{http://arxiv.org/abs/1305.3510}{{\tt arXiv:1305.3510}}].

\bibitem{Ozel:2016oaf}
F.~Ozel and P.~Freire, {\it {Masses, Radii, and the Equation of State of
  Neutron Stars}},  {\em Ann. Rev. Astron. Astrophys.} {\bf 54} (2016)
  401--440, [\href{http://arxiv.org/abs/1603.02698}{{\tt arXiv:1603.02698}}].

\bibitem{Baym:2017whm}
G.~Baym, T.~Hatsuda, T.~Kojo, P.~D. Powell, Y.~Song, and T.~Takatsuka, {\it
  {From hadrons to quarks in neutron stars: a review}},  {\em Rept. Prog.
  Phys.} {\bf 81} (2018), no.~5 056902,
  [\href{http://arxiv.org/abs/1707.04966}{{\tt arXiv:1707.04966}}].

\bibitem{Yagi:2016bkt}
K.~Yagi and N.~Yunes, {\it {Approximate Universal Relations for Neutron Stars
  and Quark Stars}},  {\em Phys. Rept.} {\bf 681} (2017) 1--72,
  [\href{http://arxiv.org/abs/1608.02582}{{\tt arXiv:1608.02582}}].

\bibitem{Doneva:2017jop}
D.~D. Doneva and G.~Pappas, {\it {Universal Relations and Alternative Gravity
  Theories}},  {\em Astrophys. Space Sci. Libr.} {\bf 457} (2018) 737--806,
  [\href{http://arxiv.org/abs/1709.08046}{{\tt arXiv:1709.08046}}].

\bibitem{Will:2005va}
C.~M. Will, {\it {The Confrontation between general relativity and
  experiment}},  {\em Living Rev. Rel.} {\bf 9} (2006) 3,
  [\href{http://arxiv.org/abs/gr-qc/0510072}{{\tt gr-qc/0510072}}].

\bibitem{Demorest:2010bx}
P.~Demorest, T.~Pennucci, S.~Ransom, M.~Roberts, and J.~Hessels, {\it {Shapiro
  Delay Measurement of A Two Solar Mass Neutron Star}},  {\em Nature} {\bf 467}
  (2010) 1081--1083, [\href{http://arxiv.org/abs/1010.5788}{{\tt
  arXiv:1010.5788}}].

\bibitem{BlazquezSalcedo:2012pd}
J.~L. Blazquez-Salcedo, L.~M. Gonzalez-Romero, and F.~Navarro-Lerida, {\it
  {Phenomenological relations for axial quasinormal modes of neutron stars with
  realistic equations of state}},  {\em Phys. Rev.} {\bf D87} (2013), no.~10
  104042, [\href{http://arxiv.org/abs/1207.4651}{{\tt arXiv:1207.4651}}].

\bibitem{Blazquez-Salcedo:2013jka}
J.~L. Blazquez-Salcedo, L.~M. Gonzalez-Romero, and F.~Navarro-Lerida, {\it
  {Polar quasi-normal modes of neutron stars with equations of state satisfying
  the $2 M_{\odot}$ constraint}},  {\em Phys. Rev.} {\bf D89} (2014), no.~4
  044006, [\href{http://arxiv.org/abs/1307.1063}{{\tt arXiv:1307.1063}}].

\bibitem{Motahar:2017blm}
Z.~Altaha~Motahar, J.~L. Blazquez-Salcedo, B.~Kleihaus, and J.~Kunz, {\it
  {Scalarization of neutron stars with realistic equations of state}},  {\em
  Phys. Rev.} {\bf D96} (2017), no.~6 064046,
  [\href{http://arxiv.org/abs/1707.05280}{{\tt arXiv:1707.05280}}].

\bibitem{AltahaMotahar:2019ekm}
Z.~Altaha~Motahar, J.~L. Blazquez-Salcedo, D.~D. Doneva, J.~Kunz, and S.~S.
  Yazadjiev, {\it {Axial quasinormal modes of scalarized neutron stars with
  massive self-interacting scalar field}},  {\em Phys. Rev.} {\bf D99} (2019),
  no.~10 104006, [\href{http://arxiv.org/abs/1902.01277}{{\tt
  arXiv:1902.01277}}].

\bibitem{Yagi:2013bca}
K.~Yagi and N.~Yunes, {\it {I-Love-Q}},  {\em Science} {\bf 341} (2013)
  365--368, [\href{http://arxiv.org/abs/1302.4499}{{\tt arXiv:1302.4499}}].

\bibitem{Yagi:2013awa}
K.~Yagi and N.~Yunes, {\it {I-Love-Q Relations in Neutron Stars and their
  Applications to Astrophysics, Gravitational Waves and Fundamental Physics}},
  {\em Phys. Rev.} {\bf D88} (2013), no.~2 023009,
  [\href{http://arxiv.org/abs/1303.1528}{{\tt arXiv:1303.1528}}].

\bibitem{Yagi:2014qua}
K.~Yagi, L.~C. Stein, G.~Pappas, N.~Yunes, and T.~A. Apostolatos, {\it {Why
  I-Love-Q: Explaining why universality emerges in compact objects}},  {\em
  Phys. Rev.} {\bf D90} (2014), no.~6 063010,
  [\href{http://arxiv.org/abs/1406.7587}{{\tt arXiv:1406.7587}}].

\bibitem{Andersson:1997rn}
N.~Andersson and K.~D. Kokkotas, {\it {Towards gravitational wave
  asteroseismology}},  {\em Mon. Not. Roy. Astron. Soc.} {\bf 299} (1998)
  1059--1068, [\href{http://arxiv.org/abs/gr-qc/9711088}{{\tt gr-qc/9711088}}].

\bibitem{Kokkotas:1999bd}
K.~D. Kokkotas and B.~G. Schmidt, {\it {Quasinormal modes of stars and black
  holes}},  {\em Living Rev. Rel.} {\bf 2} (1999) 2,
  [\href{http://arxiv.org/abs/gr-qc/9909058}{{\tt gr-qc/9909058}}].

\bibitem{Geroch:1970cd}
R.~P. Geroch, {\it {Multipole moments. II. Curved space}},  {\em J. Math.
  Phys.} {\bf 11} (1970) 2580--2588.

\bibitem{Hansen:1974zz}
R.~O. Hansen, {\it {Multipole moments of stationary space-times}},  {\em J.
  Math. Phys.} {\bf 15} (1974) 46--52.

\bibitem{Hoenselaers:1992bm}
C.~Hoenselaers and Z.~Perjes, {\it {Remarks on the Robinson-Trautman
  solutions}},  {\em Class. Quant. Grav.} {\bf 10} (1993) 375--384.

\bibitem{Sotiriou:2004ud}
T.~P. Sotiriou and T.~A. Apostolatos, {\it {Corrected multipole moments of
  axisymmetric electrovacuum spacetimes}},  {\em Class. Quant. Grav.} {\bf 21}
  (2004) 5727--5733, [\href{http://arxiv.org/abs/gr-qc/0407064}{{\tt
  gr-qc/0407064}}].

\bibitem{Thorne:1980ru}
K.~S. Thorne, {\it {Multipole Expansions of Gravitational Radiation}},  {\em
  Rev. Mod. Phys.} {\bf 52} (1980) 299--339.

\bibitem{Flanagan:2007ix}
E.~E. Flanagan and T.~Hinderer, {\it {Constraining neutron star tidal Love
  numbers with gravitational wave detectors}},  {\em Phys. Rev.} {\bf D77}
  (2008) 021502, [\href{http://arxiv.org/abs/0709.1915}{{\tt
  arXiv:0709.1915}}].

\bibitem{Hinderer:2007mb}
T.~Hinderer, {\it {Tidal Love numbers of neutron stars}},  {\em Astrophys. J.}
  {\bf 677} (2008) 1216--1220, [\href{http://arxiv.org/abs/0711.2420}{{\tt
  arXiv:0711.2420}}].

\bibitem{Damour:2009vw}
T.~Damour and A.~Nagar, {\it {Relativistic tidal properties of neutron stars}},
   {\em Phys. Rev.} {\bf D80} (2009) 084035,
  [\href{http://arxiv.org/abs/0906.0096}{{\tt arXiv:0906.0096}}].

\bibitem{Stein:2013ofa}
L.~C. Stein, K.~Yagi, and N.~Yunes, {\it {Three-Hair Relations for Rotating
  Stars: Nonrelativistic Limit}},  {\em Astrophys. J.} {\bf 788} (2014) 15,
  [\href{http://arxiv.org/abs/1312.4532}{{\tt arXiv:1312.4532}}].

\bibitem{Doneva:2013rha}
D.~D. Doneva, S.~S. Yazadjiev, N.~Stergioulas, and K.~D. Kokkotas, {\it
  {Breakdown of I-Love-Q universality in rapidly rotating relativistic stars}},
   {\em Astrophys. J.} {\bf 781} (2013) L6,
  [\href{http://arxiv.org/abs/1310.7436}{{\tt arXiv:1310.7436}}].

\bibitem{Pappas:2013naa}
G.~Pappas and T.~A. Apostolatos, {\it {Effectively universal behavior of
  rotating neutron stars in general relativity makes them even simpler than
  their Newtonian counterparts}},  {\em Phys. Rev. Lett.} {\bf 112} (2014)
  121101, [\href{http://arxiv.org/abs/1311.5508}{{\tt arXiv:1311.5508}}].

\bibitem{Chakrabarti:2013tca}
S.~Chakrabarti, T.~Delsate, N.~Gurlebeck, and J.~Steinhoff, {\it {I-Q relation
  for rapidly rotating neutron stars}},  {\em Phys. Rev. Lett.} {\bf 112}
  (2014) 201102, [\href{http://arxiv.org/abs/1311.6509}{{\tt
  arXiv:1311.6509}}].

\bibitem{Yagi:2014bxa}
K.~Yagi, K.~Kyutoku, G.~Pappas, N.~Yunes, and T.~A. Apostolatos, {\it
  {Effective No-Hair Relations for Neutron Stars and Quark Stars: Relativistic
  Results}},  {\em Phys. Rev.} {\bf D89} (2014), no.~12 124013,
  [\href{http://arxiv.org/abs/1403.6243}{{\tt arXiv:1403.6243}}].

\bibitem{Kokkotas:2003mh}
K.~D. Kokkotas and B.~F. Schutz, {\it {W-modes: A New family of normal modes of
  pulsating relativistic stars}},  {\em Mon. Not. Roy. Astron. Soc.} {\bf 255}
  (1992) 119.

\bibitem{Andersson:1996pn}
N.~Andersson and K.~D. Kokkotas, {\it {Gravitational waves and pulsating stars:
  What can we learn from future observations?}},  {\em Phys. Rev. Lett.} {\bf
  77} (1996) 4134--4137, [\href{http://arxiv.org/abs/gr-qc/9610035}{{\tt
  gr-qc/9610035}}].

\bibitem{Kokkotas:1999mn}
K.~D. Kokkotas, T.~A. Apostolatos, and N.~Andersson, {\it {The Inverse problem
  for pulsating neutron stars: A 'Fingerprint analysis' for the supranuclear
  equation of state}},  {\em Mon. Not. Roy. Astron. Soc.} {\bf 320} (2001)
  307--315, [\href{http://arxiv.org/abs/gr-qc/9901072}{{\tt gr-qc/9901072}}].

\bibitem{Benhar:1998au}
O.~Benhar, E.~Berti, and V.~Ferrari, {\it {The Imprint of the equation of state
  on the axial w modes of oscillating neutron stars}},  {\em Mon. Not. Roy.
  Astron. Soc.} {\bf 310} (1999) 797--803,
  [\href{http://arxiv.org/abs/gr-qc/9901037}{{\tt gr-qc/9901037}}]. [ICTP Lect.
  Notes Ser.3,35(2001)].

\bibitem{Benhar:2004xg}
O.~Benhar, V.~Ferrari, and L.~Gualtieri, {\it {Gravitational wave
  asteroseismology revisited}},  {\em Phys. Rev.} {\bf D70} (2004) 124015,
  [\href{http://arxiv.org/abs/astro-ph/0407529}{{\tt astro-ph/0407529}}].

\bibitem{Tsui:2004qd}
L.~K. Tsui and P.~T. Leung, {\it {Universality in quasi-normal modes of neutron
  stars}},  {\em Mon. Not. Roy. Astron. Soc.} {\bf 357} (2005) 1029--1037,
  [\href{http://arxiv.org/abs/gr-qc/0412024}{{\tt gr-qc/0412024}}].

\bibitem{Lau:2009bu}
H.~K. Lau, P.~T. Leung, and L.~M. Lin, {\it {Inferring physical parameters of
  compact stars from their f-mode gravitational wave signals}},  {\em
  Astrophys. J.} {\bf 714} (2010) 1234--1238,
  [\href{http://arxiv.org/abs/0911.0131}{{\tt arXiv:0911.0131}}].

\bibitem{Chirenti:2015dda}
C.~Chirenti, G.~H. de~Souza, and W.~Kastaun, {\it {Fundamental oscillation
  modes of neutron stars: validity of universal relations}},  {\em Phys. Rev.}
  {\bf D91} (2015), no.~4 044034, [\href{http://arxiv.org/abs/1501.02970}{{\tt
  arXiv:1501.02970}}].

\bibitem{Gaertig:2010kc}
E.~Gaertig and K.~D. Kokkotas, {\it {Gravitational wave asteroseismology with
  fast rotating neutron stars}},  {\em Phys. Rev.} {\bf D83} (2011) 064031,
  [\href{http://arxiv.org/abs/1005.5228}{{\tt arXiv:1005.5228}}].

\bibitem{Kruger:2019zuz}
C.~Kr\"uger and K.~Kokkotas, {\it {Fast Rotating Relativistic Stars: Spectra
  and Stability without Approximation}},  {\em Phys. Rev. Lett.} {\bf 125}
  (2020), no.~11 111106, [\href{http://arxiv.org/abs/1910.08370}{{\tt
  arXiv:1910.08370}}].

\bibitem{Damour:1993hw}
T.~Damour and G.~Esposito-Farese, {\it {Nonperturbative strong field effects in
  tensor - scalar theories of gravitation}},  {\em Phys. Rev. Lett.} {\bf 70}
  (1993) 2220--2223.

\bibitem{Damour:1992we}
T.~Damour and G.~Esposito-Farese, {\it {Tensor multiscalar theories of
  gravitation}},  {\em Class. Quant. Grav.} {\bf 9} (1992) 2093--2176.

\bibitem{Fujii:2003pa}
Y.~Fujii and K.~Maeda, {\em {The scalar-tensor theory of gravitation}}.
\newblock Cambridge Monographs on Mathematical Physics. Cambridge University
  Press, 2007.

\bibitem{Damour:1996ke}
T.~Damour and G.~Esposito-Farese, {\it {Tensor - scalar gravity and binary
  pulsar experiments}},  {\em Phys. Rev.} {\bf D54} (1996) 1474--1491,
  [\href{http://arxiv.org/abs/gr-qc/9602056}{{\tt gr-qc/9602056}}].

\bibitem{Harada:1998ge}
T.~Harada, {\it {Neutron stars in scalar tensor theories of gravity and
  catastrophe theory}},  {\em Phys. Rev.} {\bf D57} (1998) 4802--4811,
  [\href{http://arxiv.org/abs/gr-qc/9801049}{{\tt gr-qc/9801049}}].

\bibitem{Harada:1997mr}
T.~Harada, {\it {Stability analysis of spherically symmetric star in scalar -
  tensor theories of gravity}},  {\em Prog. Theor. Phys.} {\bf 98} (1997)
  359--379, [\href{http://arxiv.org/abs/gr-qc/9706014}{{\tt gr-qc/9706014}}].

\bibitem{Salgado:1998sg}
M.~Salgado, D.~Sudarsky, and U.~Nucamendi, {\it {On spontaneous
  scalarization}},  {\em Phys. Rev.} {\bf D58} (1998) 124003,
  [\href{http://arxiv.org/abs/gr-qc/9806070}{{\tt gr-qc/9806070}}].

\bibitem{Sotani:2012eb}
H.~Sotani, {\it {Slowly Rotating Relativistic Stars in Scalar-Tensor Gravity}},
   {\em Phys. Rev.} {\bf D86} (2012) 124036,
  [\href{http://arxiv.org/abs/1211.6986}{{\tt arXiv:1211.6986}}].

\bibitem{Doneva:2014faa}
D.~D. Doneva, S.~S. Yazadjiev, K.~V. Staykov, and K.~D. Kokkotas, {\it
  {Universal I-Q relations for rapidly rotating neutron and strange stars in
  scalar-tensor theories}},  {\em Phys. Rev.} {\bf D90} (2014), no.~10 104021,
  [\href{http://arxiv.org/abs/1408.1641}{{\tt arXiv:1408.1641}}].

\bibitem{Staykov:2016mbt}
K.~V. Staykov, D.~D. Doneva, and S.~S. Yazadjiev, {\it
  {Moment-of-inertia–compactness universal relations in scalar-tensor
  theories and $\mathcal{R}^2$ gravity}},  {\em Phys. Rev.} {\bf D93} (2016),
  no.~8 084010, [\href{http://arxiv.org/abs/1602.00504}{{\tt
  arXiv:1602.00504}}].

\bibitem{Yazadjiev:2016pcb}
S.~S. Yazadjiev, D.~D. Doneva, and D.~Popchev, {\it {Slowly rotating neutron
  stars in scalar-tensor theories with a massive scalar field}},  {\em Phys.
  Rev.} {\bf D93} (2016), no.~8 084038,
  [\href{http://arxiv.org/abs/1602.04766}{{\tt arXiv:1602.04766}}].

\bibitem{Doneva:2016xmf}
D.~D. Doneva and S.~S. Yazadjiev, {\it {Rapidly rotating neutron stars with a
  massive scalar field—structure and universal relations}},  {\em JCAP} {\bf
  1611} (2016), no.~11 019, [\href{http://arxiv.org/abs/1607.03299}{{\tt
  arXiv:1607.03299}}].

\bibitem{Freire:2012mg}
P.~C.~C. Freire, N.~Wex, G.~Esposito-Farese, J.~P.~W. Verbiest, M.~Bailes,
  B.~A. Jacoby, M.~Kramer, I.~H. Stairs, J.~Antoniadis, and G.~H. Janssen, {\it
  {The relativistic pulsar-white dwarf binary PSR J1738+0333 II. The most
  stringent test of scalar-tensor gravity}},  {\em Mon. Not. Roy. Astron. Soc.}
  {\bf 423} (2012) 3328, [\href{http://arxiv.org/abs/1205.1450}{{\tt
  arXiv:1205.1450}}].

\bibitem{Archibald:2018oxs}
A.~M. Archibald, N.~V. Gusinskaia, J.~W.~T. Hessels, A.~T. Deller, D.~L.
  Kaplan, D.~R. Lorimer, R.~S. Lynch, S.~M. Ransom, and I.~H. Stairs, {\it
  {Universality of free fall from the orbital motion of a pulsar in a stellar
  triple system}},  {\em Nature} {\bf 559} (2018), no.~7712 73--76,
  [\href{http://arxiv.org/abs/1807.02059}{{\tt arXiv:1807.02059}}].

\bibitem{Pappas:2018csu}
G.~Pappas, D.~D. Doneva, T.~P. Sotiriou, S.~S. Yazadjiev, and K.~D. Kokkotas,
  {\it {Multipole moments and universal relations for scalarized neutron
  stars}},  {\em Phys. Rev.} {\bf D99} (2019), no.~10 104014,
  [\href{http://arxiv.org/abs/1812.01117}{{\tt arXiv:1812.01117}}].

\bibitem{Danchev:2020zwn}
V.~I. Danchev and D.~D. Doneva, {\it {Constraining the equation of state in
  modified gravity via universal relations}},  {\em Phys. Rev. D} {\bf 103}
  (2021), no.~2 024049, [\href{http://arxiv.org/abs/2010.07392}{{\tt
  arXiv:2010.07392}}].

\bibitem{Sotani:2004rq}
H.~Sotani and K.~D. Kokkotas, {\it {Probing strong-field scalar-tensor gravity
  with gravitational wave asteroseismology}},  {\em Phys. Rev.} {\bf D70}
  (2004) 084026, [\href{http://arxiv.org/abs/gr-qc/0409066}{{\tt
  gr-qc/0409066}}].

\bibitem{Sotani:2005qx}
H.~Sotani and K.~D. Kokkotas, {\it {Stellar oscillations in scalar-tensor
  theory of gravity}},  {\em Phys. Rev.} {\bf D71} (2005) 124038,
  [\href{http://arxiv.org/abs/gr-qc/0506060}{{\tt gr-qc/0506060}}].

\bibitem{AltahaMotahar:2018djk}
Z.~Altaha~Motahar, J.~L. Blazquez-Salcedo, B.~Kleihaus, and J.~Kunz, {\it
  {Axial quasinormal modes of scalarized neutron stars with realistic equations
  of state}},  {\em Phys. Rev.} {\bf D98} (2018), no.~4 044032,
  [\href{http://arxiv.org/abs/1807.02598}{{\tt arXiv:1807.02598}}].

\bibitem{Mendes:2018qwo}
R.~F.~P. Mendes and N.~Ortiz, {\it {New class of quasinormal modes of neutron
  stars in scalar-tensor gravity}},  {\em Phys. Rev. Lett.} {\bf 120} (2018),
  no.~20 201104, [\href{http://arxiv.org/abs/1802.07847}{{\tt
  arXiv:1802.07847}}].

\bibitem{Doneva:2015hsa}
D.~D. Doneva, S.~S. Yazadjiev, and K.~D. Kokkotas, {\it {The I-Q relations for
  rapidly rotating neutron stars in $f(R)$ gravity}},  {\em Phys. Rev.} {\bf
  D92} (2015), no.~6 064015, [\href{http://arxiv.org/abs/1507.00378}{{\tt
  arXiv:1507.00378}}].

\bibitem{Yazadjiev:2018xxk}
S.~S. Yazadjiev, D.~D. Doneva, and K.~D. Kokkotas, {\it {Tidal Love numbers of
  neutron stars in $f(R)$ gravity}},  {\em Eur. Phys. J.} {\bf C78} (2018),
  no.~10 818, [\href{http://arxiv.org/abs/1803.09534}{{\tt arXiv:1803.09534}}].

\bibitem{Blazquez-Salcedo:2018qyy}
J.~L. Blazquez-Salcedo, D.~D. Doneva, J.~Kunz, K.~V. Staykov, and S.~S.
  Yazadjiev, {\it {Axial quasinormal modes of neutron stars in $R^2$ gravity}},
   {\em Phys. Rev.} {\bf D98} (2018), no.~10 104047,
  [\href{http://arxiv.org/abs/1804.04060}{{\tt arXiv:1804.04060}}].

\bibitem{Blazquez-Salcedo:2018pxo}
J.~L. Blazquez-Salcedo, Z.~Altaha~Motahar, D.~D. Doneva, F.~S. Khoo, J.~Kunz,
  S.~Mojica, K.~V. Staykov, and S.~S. Yazadjiev, {\it {Quasinormal modes of
  compact objects in alternative theories of gravity}},  {\em Eur. Phys. J.
  Plus} {\bf 134} (2019), no.~1 46,
  [\href{http://arxiv.org/abs/1810.09432}{{\tt arXiv:1810.09432}}].

\bibitem{Blazquez-Salcedo:2020ibb}
J.~L. Bl\'azquez-Salcedo, F.~Scen~Khoo, and J.~Kunz, {\it {Ultra-long-lived
  quasi-normal modes of neutron stars in massive scalar-tensor gravity}},  {\em
  EPL} {\bf 130} (2020), no.~5 50002,
  [\href{http://arxiv.org/abs/2001.09117}{{\tt arXiv:2001.09117}}].

\bibitem{Kobayashi:2019hrl}
T.~Kobayashi, {\it {Horndeski theory and beyond: a review}},
  \href{http://arxiv.org/abs/1901.07183}{{\tt arXiv:1901.07183}}.

\bibitem{Charmousis:2011bf}
C.~Charmousis, E.~J. Copeland, A.~Padilla, and P.~M. Saffin, {\it {General
  second order scalar-tensor theory, self tuning, and the Fab Four}},  {\em
  Phys. Rev. Lett.} {\bf 108} (2012) 051101,
  [\href{http://arxiv.org/abs/1106.2000}{{\tt arXiv:1106.2000}}].

\bibitem{Blazquez-Salcedo:2018tyn}
J.~L. Blazquez-Salcedo and K.~Eickhoff, {\it {Axial quasinormal modes of static
  neutron stars in the nonminimal derivative coupling sector of Horndeski
  gravity: spectrum and universal relations for realistic equations of state}},
   {\em Phys. Rev.} {\bf D97} (2018), no.~10 104002,
  [\href{http://arxiv.org/abs/1803.01655}{{\tt arXiv:1803.01655}}].

\bibitem{Gross:1986mw}
D.~J. Gross and J.~H. Sloan, {\it {The Quartic Effective Action for the
  Heterotic String}},  {\em Nucl. Phys.} {\bf B291} (1987) 41--89.

\bibitem{Metsaev:1987zx}
R.~R. Metsaev and A.~A. Tseytlin, {\it {Order alpha-prime (Two Loop)
  Equivalence of the String Equations of Motion and the Sigma Model Weyl
  Invariance Conditions: Dependence on the Dilaton and the Antisymmetric
  Tensor}},  {\em Nucl. Phys.} {\bf B293} (1987) 385--419.

\bibitem{Kleihaus:2014lba}
B.~Kleihaus, J.~Kunz, and S.~Mojica, {\it {Quadrupole Moments of Rapidly
  Rotating Compact Objects in Dilatonic Einstein-Gauss-Bonnet Theory}},  {\em
  Phys. Rev.} {\bf D90} (2014), no.~6 061501,
  [\href{http://arxiv.org/abs/1407.6884}{{\tt arXiv:1407.6884}}].

\bibitem{Doneva:2017bvd}
D.~D. Doneva and S.~S. Yazadjiev, {\it {New Gauss-Bonnet Black Holes with
  Curvature-Induced Scalarization in Extended Scalar-Tensor Theories}},  {\em
  Phys. Rev. Lett.} {\bf 120} (2018), no.~13 131103,
  [\href{http://arxiv.org/abs/1711.01187}{{\tt arXiv:1711.01187}}].

\bibitem{Antoniou:2017acq}
G.~Antoniou, A.~Bakopoulos, and P.~Kanti, {\it {Evasion of No-Hair Theorems and
  Novel Black-Hole Solutions in Gauss-Bonnet Theories}},  {\em Phys. Rev.
  Lett.} {\bf 120} (2018), no.~13 131102,
  [\href{http://arxiv.org/abs/1711.03390}{{\tt arXiv:1711.03390}}].

\bibitem{Alexander:2009tp}
S.~Alexander and N.~Yunes, {\it {Chern-Simons Modified General Relativity}},
  {\em Phys. Rept.} {\bf 480} (2009) 1--55,
  [\href{http://arxiv.org/abs/0907.2562}{{\tt arXiv:0907.2562}}].

\bibitem{Yunes:2009ch}
N.~Yunes, D.~Psaltis, F.~Ozel, and A.~Loeb, {\it {Constraining Parity Violation
  in Gravity with Measurements of Neutron-Star Moments of Inertia}},  {\em
  Phys. Rev.} {\bf D81} (2010) 064020,
  [\href{http://arxiv.org/abs/0912.2736}{{\tt arXiv:0912.2736}}].

\bibitem{AliHaimoud:2011fw}
Y.~Ali-Haimoud and Y.~Chen, {\it {Slowly-rotating stars and black holes in
  dynamical Chern-Simons gravity}},  {\em Phys. Rev.} {\bf D84} (2011) 124033,
  [\href{http://arxiv.org/abs/1110.5329}{{\tt arXiv:1110.5329}}].

\bibitem{Yagi:2011xp}
K.~Yagi, L.~C. Stein, N.~Yunes, and T.~Tanaka, {\it {Post-Newtonian,
  Quasi-Circular Binary Inspirals in Quadratic Modified Gravity}},  {\em Phys.
  Rev.} {\bf D85} (2012) 064022, [\href{http://arxiv.org/abs/1110.5950}{{\tt
  arXiv:1110.5950}}]. [Erratum: Phys. Rev.D93,no.2,029902(2016)].

\bibitem{Gupta:2017vsl}
T.~Gupta, B.~Majumder, K.~Yagi, and N.~Yunes, {\it {I-Love-Q Relations for
  Neutron Stars in dynamical Chern Simons Gravity}},  {\em Class. Quant. Grav.}
  {\bf 35} (2018), no.~2 025009, [\href{http://arxiv.org/abs/1710.07862}{{\tt
  arXiv:1710.07862}}].

\bibitem{Chrusciel:2012jk}
P.~T. Chrusciel, J.~Lopes~Costa, and M.~Heusler, {\it {Stationary Black Holes:
  Uniqueness and Beyond}},  {\em Living Rev. Rel.} {\bf 15} (2012) 7,
  [\href{http://arxiv.org/abs/1205.6112}{{\tt arXiv:1205.6112}}].

\bibitem{Cardoso:2016ryw}
V.~Cardoso and L.~Gualtieri, {\it {Testing the black hole ‘no-hair’
  hypothesis}},  {\em Class. Quant. Grav.} {\bf 33} (2016), no.~17 174001,
  [\href{http://arxiv.org/abs/1607.03133}{{\tt arXiv:1607.03133}}].

\bibitem{Nollert:1999ji}
H.-P. Nollert, {\it {TOPICAL REVIEW: Quasinormal modes: the characteristic
  `sound' of black holes and neutron stars}},  {\em Class. Quant. Grav.} {\bf
  16} (1999) R159--R216.

\bibitem{Rezzolla:2003ua}
L.~Rezzolla, {\it {Gravitational waves from perturbed black holes and
  relativistic stars}},  {\em ICTP Lect. Notes Ser.} {\bf 14} (2003) 255--316,
  [\href{http://arxiv.org/abs/gr-qc/0302025}{{\tt gr-qc/0302025}}].

\bibitem{Berti:2009kk}
E.~Berti, V.~Cardoso, and A.~O. Starinets, {\it {Quasinormal modes of black
  holes and black branes}},  {\em Class. Quant. Grav.} {\bf 26} (2009) 163001,
  [\href{http://arxiv.org/abs/0905.2975}{{\tt arXiv:0905.2975}}].

\bibitem{Konoplya:2011qq}
R.~A. Konoplya and A.~Zhidenko, {\it {Quasinormal modes of black holes: From
  astrophysics to string theory}},  {\em Rev. Mod. Phys.} {\bf 83} (2011)
  793--836, [\href{http://arxiv.org/abs/1102.4014}{{\tt arXiv:1102.4014}}].

\bibitem{Whiting:1988vc}
B.~F. Whiting, {\it {Mode Stability of the Kerr Black Hole}},  {\em J. Math.
  Phys.} {\bf 30} (1989) 1301.

\bibitem{Nollert:1993zz}
H.-P. Nollert, {\it {Quasinormal modes of Schwarzschild black holes: The
  determination of quasinormal frequencies with very large imaginary parts}},
  {\em Phys. Rev.} {\bf D47} (1993) 5253--5258.

\bibitem{Ferrari:1984zz}
V.~Ferrari and B.~Mashhoon, {\it {New approach to the quasinormal modes of a
  black hole}},  {\em Phys. Rev.} {\bf D30} (1984) 295--304.

\bibitem{Yang:2012he}
H.~Yang, D.~A. Nichols, F.~Zhang, A.~Zimmerman, Z.~Zhang, and Y.~Chen, {\it
  {Quasinormal-mode spectrum of Kerr black holes and its geometric
  interpretation}},  {\em Phys. Rev.} {\bf D86} (2012) 104006,
  [\href{http://arxiv.org/abs/1207.4253}{{\tt arXiv:1207.4253}}].

\bibitem{Synge:1966okc}
J.~L. Synge, {\it {The Escape of Photons from Gravitationally Intense Stars}},
  {\em Mon. Not. Roy. Astron. Soc.} {\bf 131} (1966), no.~3 463--466.

\bibitem{Bardeen:1973}
J.~Bardeen, {\it Timelike and null geodesics in the {K}err metric},  in {\em
  Black Holes} (C.~DeWitt and B.~DeWitt, eds.), p.~215.
\newblock Gordon and Breach, New York, 1973.

\bibitem{Perlick:2004tq}
V.~Perlick, {\it {Gravitational lensing from a spacetime perspective}},  {\em
  Living Rev. Rel.} {\bf 7} (2004) 9.

\bibitem{Grenzebach:2014fha}
A.~Grenzebach, V.~Perlick, and C.~Lammerzahl, {\it {Photon Regions and Shadows
  of Kerr-Newman-NUT Black Holes with a Cosmological Constant}},  {\em Phys.
  Rev.} {\bf D89} (2014), no.~12 124004,
  [\href{http://arxiv.org/abs/1403.5234}{{\tt arXiv:1403.5234}}].

\bibitem{Volkov:1998cc}
M.~S. Volkov and D.~V. Gal'tsov, {\it {Gravitating nonAbelian solitons and
  black holes with Yang-Mills fields}},  {\em Phys. Rept.} {\bf 319} (1999)
  1--83, [\href{http://arxiv.org/abs/hep-th/9810070}{{\tt hep-th/9810070}}].

\bibitem{Kleihaus:2016rgf}
B.~Kleihaus, J.~Kunz, and F.~Navarro-Lerida, {\it {Rotating black holes with
  non-Abelian hair}},  {\em Class. Quant. Grav.} {\bf 33} (2016), no.~23
  234002, [\href{http://arxiv.org/abs/1609.07357}{{\tt arXiv:1609.07357}}].

\bibitem{Herdeiro:2014goa}
C.~A.~R. Herdeiro and E.~Radu, {\it {Kerr black holes with scalar hair}},  {\em
  Phys. Rev. Lett.} {\bf 112} (2014) 221101,
  [\href{http://arxiv.org/abs/1403.2757}{{\tt arXiv:1403.2757}}].

\bibitem{Herdeiro:2015waa}
C.~A.~R. Herdeiro and E.~Radu, {\it {Asymptotically flat black holes with
  scalar hair: a review}},  {\em Int. J. Mod. Phys.} {\bf D24} (2015), no.~09
  1542014, [\href{http://arxiv.org/abs/1504.08209}{{\tt arXiv:1504.08209}}].

\bibitem{Brito:2015pxa}
R.~Brito, V.~Cardoso, C.~A.~R. Herdeiro, and E.~Radu, {\it {Proca stars:
  Gravitating Bose–Einstein condensates of massive spin 1 particles}},  {\em
  Phys. Lett.} {\bf B752} (2016) 291--295,
  [\href{http://arxiv.org/abs/1508.05395}{{\tt arXiv:1508.05395}}].

\bibitem{Herdeiro:2014jaa}
C.~Herdeiro and E.~Radu, {\it {Ergosurfaces for Kerr black holes with scalar
  hair}},  {\em Phys. Rev.} {\bf D89} (2014), no.~12 124018,
  [\href{http://arxiv.org/abs/1406.1225}{{\tt arXiv:1406.1225}}].

\bibitem{Herdeiro:2015gia}
C.~Herdeiro and E.~Radu, {\it {Construction and physical properties of Kerr
  black holes with scalar hair}},  {\em Class. Quant. Grav.} {\bf 32} (2015),
  no.~14 144001, [\href{http://arxiv.org/abs/1501.04319}{{\tt
  arXiv:1501.04319}}].

\bibitem{Kleihaus:2015iea}
B.~Kleihaus, J.~Kunz, and S.~Yazadjiev, {\it {Scalarized Hairy Black Holes}},
  {\em Phys. Lett.} {\bf B744} (2015) 406--412,
  [\href{http://arxiv.org/abs/1503.01672}{{\tt arXiv:1503.01672}}].

\bibitem{Cunha:2015yba}
P.~V.~P. Cunha, C.~A.~R. Herdeiro, E.~Radu, and H.~F. Runarsson, {\it {Shadows
  of Kerr black holes with scalar hair}},  {\em Phys. Rev. Lett.} {\bf 115}
  (2015), no.~21 211102, [\href{http://arxiv.org/abs/1509.00021}{{\tt
  arXiv:1509.00021}}].

\bibitem{Shen:2016acv}
T.~Shen, M.~Zhou, C.~Bambi, C.~A.~R. Herdeiro, and E.~Radu, {\it {Iron
  K$\alpha$ line of Proca stars}},  {\em JCAP} {\bf 1708} (2017) 014,
  [\href{http://arxiv.org/abs/1701.00192}{{\tt arXiv:1701.00192}}].

\bibitem{Babichev:2017guv}
E.~Babichev, C.~Charmousis, and A.~Leh{\'e}bel, {\it {Asymptotically flat black
  holes in Horndeski theory and beyond}},  {\em JCAP} {\bf 1704} (2017), no.~04
  027, [\href{http://arxiv.org/abs/1702.01938}{{\tt arXiv:1702.01938}}].

\bibitem{Babichev:2013cya}
E.~Babichev and C.~Charmousis, {\it {Dressing a black hole with a
  time-dependent Galileon}},  {\em JHEP} {\bf 08} (2014) 106,
  [\href{http://arxiv.org/abs/1312.3204}{{\tt arXiv:1312.3204}}].

\bibitem{Kobayashi:2014eva}
T.~Kobayashi and N.~Tanahashi, {\it {Exact black hole solutions in shift
  symmetric scalar–tensor theories}},  {\em PTEP} {\bf 2014} (2014) 073E02,
  [\href{http://arxiv.org/abs/1403.4364}{{\tt arXiv:1403.4364}}].

\bibitem{Tattersall:2017erk}
O.~J. Tattersall, P.~G. Ferreira, and M.~Lagos, {\it {General theories of
  linear gravitational perturbations to a Schwarzschild Black Hole}},  {\em
  Phys. Rev.} {\bf D97} (2018), no.~4 044021,
  [\href{http://arxiv.org/abs/1711.01992}{{\tt arXiv:1711.01992}}].

\bibitem{Tattersall:2018nve}
O.~J. Tattersall and P.~G. Ferreira, {\it {Quasinormal modes of black holes in
  Horndeski gravity}},  {\em Phys. Rev.} {\bf D97} (2018), no.~10 104047,
  [\href{http://arxiv.org/abs/1804.08950}{{\tt arXiv:1804.08950}}].

\bibitem{Tattersall:2019pvx}
O.~J. Tattersall and P.~G. Ferreira, {\it {Forecasts for Low Spin Black Hole
  Spectroscopy in Horndeski Gravity}},  {\em Phys. Rev.} {\bf D99} (2019),
  no.~10 104082, [\href{http://arxiv.org/abs/1904.05112}{{\tt
  arXiv:1904.05112}}].

\bibitem{Tattersall:2019nmh}
O.~J. Tattersall, {\it {Quasi-Normal Modes of Hairy Scalar Tensor Black Holes:
  Odd Parity}},  {\em Class. Quant. Grav.} {\bf 37} (2020), no.~11 115007,
  [\href{http://arxiv.org/abs/1911.07593}{{\tt arXiv:1911.07593}}].

\bibitem{Mignemi:1992nt}
S.~Mignemi and N.~R. Stewart, {\it {Charged black holes in effective string
  theory}},  {\em Phys. Rev.} {\bf D47} (1993) 5259--5269,
  [\href{http://arxiv.org/abs/hep-th/9212146}{{\tt hep-th/9212146}}].

\bibitem{Mignemi:1993ce}
S.~Mignemi, {\it {Dyonic black holes in effective string theory}},  {\em Phys.
  Rev.} {\bf D51} (1995) 934--937,
  [\href{http://arxiv.org/abs/hep-th/9303102}{{\tt hep-th/9303102}}].

\bibitem{Kanti:1995vq}
P.~Kanti, N.~E. Mavromatos, J.~Rizos, K.~Tamvakis, and E.~Winstanley, {\it
  {Dilatonic black holes in higher curvature string gravity}},  {\em Phys.
  Rev.} {\bf D54} (1996) 5049--5058,
  [\href{http://arxiv.org/abs/hep-th/9511071}{{\tt hep-th/9511071}}].

\bibitem{Torii:1996yi}
T.~Torii, H.~Yajima, and K.-i. Maeda, {\it {Dilatonic black holes with
  Gauss-Bonnet term}},  {\em Phys. Rev.} {\bf D55} (1997) 739--753,
  [\href{http://arxiv.org/abs/gr-qc/9606034}{{\tt gr-qc/9606034}}].

\bibitem{Guo:2008hf}
Z.-K. Guo, N.~Ohta, and T.~Torii, {\it {Black Holes in the Dilatonic
  Einstein-Gauss-Bonnet Theory in Various Dimensions. I. Asymptotically Flat
  Black Holes}},  {\em Prog. Theor. Phys.} {\bf 120} (2008) 581--607,
  [\href{http://arxiv.org/abs/0806.2481}{{\tt arXiv:0806.2481}}].

\bibitem{Pani:2009wy}
P.~Pani and V.~Cardoso, {\it {Are black holes in alternative theories serious
  astrophysical candidates? The Case for Einstein-Dilaton-Gauss-Bonnet black
  holes}},  {\em Phys. Rev.} {\bf D79} (2009) 084031,
  [\href{http://arxiv.org/abs/0902.1569}{{\tt arXiv:0902.1569}}].

\bibitem{Pani:2011gy}
P.~Pani, C.~F.~B. Macedo, L.~C.~B. Crispino, and V.~Cardoso, {\it {Slowly
  rotating black holes in alternative theories of gravity}},  {\em Phys. Rev.}
  {\bf D84} (2011) 087501, [\href{http://arxiv.org/abs/1109.3996}{{\tt
  arXiv:1109.3996}}].

\bibitem{Ayzenberg:2014aka}
D.~Ayzenberg and N.~Yunes, {\it {Slowly-Rotating Black Holes in
  Einstein-Dilaton-Gauss-Bonnet Gravity: Quadratic Order in Spin Solutions}},
  {\em Phys. Rev.} {\bf D90} (2014) 044066,
  [\href{http://arxiv.org/abs/1405.2133}{{\tt arXiv:1405.2133}}]. [Erratum:
  Phys. Rev.D91,no.6,069905(2015)].

\bibitem{Maselli:2015tta}
A.~Maselli, P.~Pani, L.~Gualtieri, and V.~Ferrari, {\it {Rotating black holes
  in Einstein-Dilaton-Gauss-Bonnet gravity with finite coupling}},  {\em Phys.
  Rev.} {\bf D92} (2015), no.~8 083014,
  [\href{http://arxiv.org/abs/1507.00680}{{\tt arXiv:1507.00680}}].

\bibitem{Kleihaus:2011tg}
B.~Kleihaus, J.~Kunz, and E.~Radu, {\it {Rotating Black Holes in Dilatonic
  Einstein-Gauss-Bonnet Theory}},  {\em Phys. Rev. Lett.} {\bf 106} (2011)
  151104, [\href{http://arxiv.org/abs/1101.2868}{{\tt arXiv:1101.2868}}].

\bibitem{Kleihaus:2015aje}
B.~Kleihaus, J.~Kunz, S.~Mojica, and E.~Radu, {\it {Spinning black holes in
  Einstein–Gauss-Bonnet–dilaton theory: Nonperturbative solutions}},  {\em
  Phys. Rev.} {\bf D93} (2016), no.~4 044047,
  [\href{http://arxiv.org/abs/1511.05513}{{\tt arXiv:1511.05513}}].

\bibitem{Chen:2018jed}
B.~Chen and L.~C. Stein, {\it {Deformation of extremal black holes from stringy
  interactions}},  {\em Phys. Rev.} {\bf D97} (2018), no.~8 084012,
  [\href{http://arxiv.org/abs/1802.02159}{{\tt arXiv:1802.02159}}].

\bibitem{Cunha:2016wzk}
P.~V.~P. Cunha, C.~A.~R. Herdeiro, B.~Kleihaus, J.~Kunz, and E.~Radu, {\it
  {Shadows of Einstein–dilaton–Gauss–Bonnet black holes}},  {\em Phys.
  Lett.} {\bf B768} (2017) 373--379,
  [\href{http://arxiv.org/abs/1701.00079}{{\tt arXiv:1701.00079}}].

\bibitem{Wald:1993nt}
R.~M. Wald, {\it {Black hole entropy is the Noether charge}},  {\em Phys. Rev.}
  {\bf D48} (1993), no.~8 R3427--R3431,
  [\href{http://arxiv.org/abs/gr-qc/9307038}{{\tt gr-qc/9307038}}].

\bibitem{Kanti:1997br}
P.~Kanti, N.~E. Mavromatos, J.~Rizos, K.~Tamvakis, and E.~Winstanley, {\it
  {Dilatonic black holes in higher curvature string gravity. 2: Linear
  stability}},  {\em Phys. Rev.} {\bf D57} (1998) 6255--6264,
  [\href{http://arxiv.org/abs/hep-th/9703192}{{\tt hep-th/9703192}}].

\bibitem{Ayzenberg:2013wua}
D.~Ayzenberg, K.~Yagi, and N.~Yunes, {\it {Linear Stability Analysis of
  Dynamical Quadratic Gravity}},  {\em Phys. Rev.} {\bf D89} (2014), no.~4
  044023, [\href{http://arxiv.org/abs/1310.6392}{{\tt arXiv:1310.6392}}].

\bibitem{Blazquez-Salcedo:2016enn}
J.~L. Blazquez-Salcedo, C.~F.~B. Macedo, V.~Cardoso, V.~Ferrari, L.~Gualtieri,
  F.~S. Khoo, J.~Kunz, and P.~Pani, {\it {Perturbed black holes in
  Einstein-dilaton-Gauss-Bonnet gravity: Stability, ringdown, and
  gravitational-wave emission}},  {\em Phys. Rev.} {\bf D94} (2016), no.~10
  104024, [\href{http://arxiv.org/abs/1609.01286}{{\tt arXiv:1609.01286}}].

\bibitem{Blazquez-Salcedo:2017txk}
J.~L. Blazquez-Salcedo, F.~S. Khoo, and J.~Kunz, {\it {Quasinormal modes of
  Einstein-Gauss-Bonnet-dilaton black holes}},  {\em Phys. Rev.} {\bf D96}
  (2017), no.~6 064008, [\href{http://arxiv.org/abs/1706.03262}{{\tt
  arXiv:1706.03262}}].

\bibitem{Konoplya:2019hml}
R.~A. Konoplya, A.~F. Zinhailo, and Z.~Stuchlík, {\it {Quasinormal modes,
  scattering, and Hawking radiation in the vicinity of an
  Einstein-dilaton-Gauss-Bonnet black hole}},  {\em Phys. Rev.} {\bf D99}
  (2019), no.~12 124042, [\href{http://arxiv.org/abs/1903.03483}{{\tt
  arXiv:1903.03483}}].

\bibitem{Zinhailo:2019rwd}
A.~F. Zinhailo, {\it {Quasinormal modes of Dirac field in the
  Einstein–Dilaton–Gauss–Bonnet and Einstein–Weyl gravities}},  {\em
  Eur. Phys. J.} {\bf C79} (2019), no.~11 912,
  [\href{http://arxiv.org/abs/1909.12664}{{\tt arXiv:1909.12664}}].

\bibitem{Blazquez-Salcedo:2018jnn}
J.~L. Blazquez-Salcedo, D.~D. Doneva, J.~Kunz, and S.~S. Yazadjiev, {\it
  {Radial perturbations of the scalarized Einstein-Gauss-Bonnet black holes}},
  {\em Phys. Rev.} {\bf D98} (2018), no.~8 084011,
  [\href{http://arxiv.org/abs/1805.05755}{{\tt arXiv:1805.05755}}].

\bibitem{Antoniou:2017hxj}
G.~Antoniou, A.~Bakopoulos, and P.~Kanti, {\it {Black-Hole Solutions with
  Scalar Hair in Einstein-Scalar-Gauss-Bonnet Theories}},  {\em Phys. Rev.}
  {\bf D97} (2018), no.~8 084037, [\href{http://arxiv.org/abs/1711.07431}{{\tt
  arXiv:1711.07431}}].

\bibitem{Doneva:2018rou}
D.~D. Doneva, S.~Kiorpelidi, P.~G. Nedkova, E.~Papantonopoulos, and S.~S.
  Yazadjiev, {\it {Charged Gauss-Bonnet black holes with curvature induced
  scalarization in the extended scalar-tensor theories}},  {\em Phys. Rev.}
  {\bf D98} (2018), no.~10 104056, [\href{http://arxiv.org/abs/1809.00844}{{\tt
  arXiv:1809.00844}}].

\bibitem{Silva:2018qhn}
H.~O. Silva, C.~F.~B. Macedo, T.~P. Sotiriou, L.~Gualtieri, J.~Sakstein, and
  E.~Berti, {\it {Stability of scalarized black hole solutions in
  scalar-Gauss-Bonnet gravity}},  {\em Phys. Rev.} {\bf D99} (2019), no.~6
  064011, [\href{http://arxiv.org/abs/1812.05590}{{\tt arXiv:1812.05590}}].

\bibitem{Cunha:2019dwb}
P.~V.~P. Cunha, C.~A.~R. Herdeiro, and E.~Radu, {\it {Spontaneously Scalarized
  Kerr Black Holes in Extended Scalar-Tensor–Gauss-Bonnet Gravity}},  {\em
  Phys. Rev. Lett.} {\bf 123} (2019), no.~1 011101,
  [\href{http://arxiv.org/abs/1904.09997}{{\tt arXiv:1904.09997}}].

\bibitem{Macedo:2019sem}
C.~F.~B. Macedo, J.~Sakstein, E.~Berti, L.~Gualtieri, H.~O. Silva, and T.~P.
  Sotiriou, {\it {Self-interactions and Spontaneous Black Hole Scalarization}},
   {\em Phys. Rev.} {\bf D99} (2019), no.~10 104041,
  [\href{http://arxiv.org/abs/1903.06784}{{\tt arXiv:1903.06784}}].

\bibitem{Collodel:2019kkx}
L.~G. Collodel, B.~Kleihaus, J.~Kunz, and E.~Berti, {\it {Spinning and excited
  black holes in Einstein-scalar-Gauss\textendash{}Bonnet theory}},  {\em
  Class. Quant. Grav.} {\bf 37} (2020), no.~7 075018,
  [\href{http://arxiv.org/abs/1912.05382}{{\tt arXiv:1912.05382}}].

\bibitem{Macedo:2020tbm}
C.~F. Macedo, {\it {Scalar modes, spontaneous scalarization and circular
  null-geodesics of black holes in scalar-Gauss\textendash{}Bonnet gravity}},
  {\em Int. J. Mod. Phys. D} {\bf 29} (2020), no.~11 2041006,
  [\href{http://arxiv.org/abs/2002.12719}{{\tt arXiv:2002.12719}}].

\bibitem{Blazquez-Salcedo:2020rhf}
J.~L. Bl\'azquez-Salcedo, D.~D. Doneva, S.~Kahlen, J.~Kunz, P.~Nedkova, and
  S.~S. Yazadjiev, {\it {Axial perturbations of the scalarized
  Einstein-Gauss-Bonnet black holes}},  {\em Phys. Rev. D} {\bf 101} (2020),
  no.~10 104006, [\href{http://arxiv.org/abs/2003.02862}{{\tt
  arXiv:2003.02862}}].

\bibitem{Blazquez-Salcedo:2020caw}
J.~L. Bl\'azquez-Salcedo, D.~D. Doneva, S.~Kahlen, J.~Kunz, P.~Nedkova, and
  S.~S. Yazadjiev, {\it {Polar quasinormal modes of the scalarized
  Einstein-Gauss-Bonnet black holes}},  {\em Phys. Rev. D} {\bf 102} (2020),
  no.~2 024086, [\href{http://arxiv.org/abs/2006.06006}{{\tt
  arXiv:2006.06006}}].

\bibitem{Dima:2020yac}
A.~Dima, E.~Barausse, N.~Franchini, and T.~P. Sotiriou, {\it {Spin-induced
  black hole spontaneous scalarization}},  {\em Phys. Rev. Lett.} {\bf 125}
  (2020), no.~23 231101, [\href{http://arxiv.org/abs/2006.03095}{{\tt
  arXiv:2006.03095}}].

\bibitem{Hod:2020jjy}
S.~Hod, {\it {Onset of spontaneous scalarization in spinning Gauss-Bonnet black
  holes}},  {\em Phys. Rev. D} {\bf 102} (2020), no.~8 084060,
  [\href{http://arxiv.org/abs/2006.09399}{{\tt arXiv:2006.09399}}].

\bibitem{Doneva:2020nbb}
D.~D. Doneva, L.~G. Collodel, C.~J. Kr\"uger, and S.~S. Yazadjiev, {\it {Black
  hole scalarization induced by the spin: 2+1 time evolution}},  {\em Phys.
  Rev. D} {\bf 102} (2020), no.~10 104027,
  [\href{http://arxiv.org/abs/2008.07391}{{\tt arXiv:2008.07391}}].

\bibitem{Herdeiro:2020wei}
C.~A.~R. Herdeiro, E.~Radu, H.~O. Silva, T.~P. Sotiriou, and N.~Yunes, {\it
  {Spin-induced scalarized black holes}},  {\em Phys. Rev. Lett.} {\bf 126}
  (2021), no.~1 011103, [\href{http://arxiv.org/abs/2009.03904}{{\tt
  arXiv:2009.03904}}].

\bibitem{Berti:2020kgk}
E.~Berti, L.~G. Collodel, B.~Kleihaus, and J.~Kunz, {\it {Spin-induced
  black-hole scalarization in Einstein-scalar-Gauss-Bonnet theory}},  {\em
  Phys. Rev. Lett.} {\bf 126} (2021), no.~1 011104,
  [\href{http://arxiv.org/abs/2009.03905}{{\tt arXiv:2009.03905}}].

\bibitem{Cardoso:2009pk}
V.~Cardoso and L.~Gualtieri, {\it {Perturbations of Schwarzschild black holes
  in Dynamical Chern-Simons modified gravity}},  {\em Phys. Rev.} {\bf D80}
  (2009) 064008, [\href{http://arxiv.org/abs/0907.5008}{{\tt
  arXiv:0907.5008}}]. [Erratum: Phys. Rev.D81,089903(2010)].

\bibitem{Molina:2010fb}
C.~Molina, P.~Pani, V.~Cardoso, and L.~Gualtieri, {\it {Gravitational signature
  of Schwarzschild black holes in dynamical Chern-Simons gravity}},  {\em Phys.
  Rev.} {\bf D81} (2010) 124021, [\href{http://arxiv.org/abs/1004.4007}{{\tt
  arXiv:1004.4007}}].

\bibitem{Kimura:2018nxk}
M.~Kimura, {\it {Stability analysis of Schwarzschild black holes in dynamical
  Chern-Simons gravity}},  {\em Phys. Rev.} {\bf D98} (2018), no.~2 024048,
  [\href{http://arxiv.org/abs/1807.05029}{{\tt arXiv:1807.05029}}].

\bibitem{Yunes:2009hc}
N.~Yunes and F.~Pretorius, {\it {Dynamical Chern-Simons Modified Gravity. I.
  Spinning Black Holes in the Slow-Rotation Approximation}},  {\em Phys. Rev.}
  {\bf D79} (2009) 084043, [\href{http://arxiv.org/abs/0902.4669}{{\tt
  arXiv:0902.4669}}].

\bibitem{Konno:2009kg}
K.~Konno, T.~Matsuyama, and S.~Tanda, {\it {Rotating black hole in extended
  Chern-Simons modified gravity}},  {\em Prog. Theor. Phys.} {\bf 122} (2009)
  561--568, [\href{http://arxiv.org/abs/0902.4767}{{\tt arXiv:0902.4767}}].

\bibitem{Yagi:2012ya}
K.~Yagi, N.~Yunes, and T.~Tanaka, {\it {Slowly Rotating Black Holes in
  Dynamical Chern-Simons Gravity: Deformation Quadratic in the Spin}},  {\em
  Phys. Rev.} {\bf D86} (2012) 044037,
  [\href{http://arxiv.org/abs/1206.6130}{{\tt arXiv:1206.6130}}]. [Erratum:
  Phys. Rev.D89,049902(2014)].

\bibitem{Delsate:2018ome}
T.~Delsate, C.~Herdeiro, and E.~Radu, {\it {Non-perturbative spinning black
  holes in dynamical Chern–Simons gravity}},  {\em Phys. Lett.} {\bf B787}
  (2018) 8--15, [\href{http://arxiv.org/abs/1806.06700}{{\tt
  arXiv:1806.06700}}].

\bibitem{Shibata:2013pra}
M.~Shibata, K.~Taniguchi, H.~Okawa, and A.~Buonanno, {\it {Coalescence of
  binary neutron stars in a scalar-tensor theory of gravity}},  {\em Phys.
  Rev.} {\bf D89} (2014), no.~8 084005,
  [\href{http://arxiv.org/abs/1310.0627}{{\tt arXiv:1310.0627}}].

\bibitem{Taniguchi:2014fqa}
K.~Taniguchi, M.~Shibata, and A.~Buonanno, {\it {Quasiequilibrium sequences of
  binary neutron stars undergoing dynamical scalarization}},  {\em Phys. Rev.}
  {\bf D91} (2015), no.~2 024033, [\href{http://arxiv.org/abs/1410.0738}{{\tt
  arXiv:1410.0738}}].

\bibitem{Sennett:2016rwa}
N.~Sennett and A.~Buonanno, {\it {Modeling dynamical scalarization with a
  resummed post-Newtonian expansion}},  {\em Phys. Rev.} {\bf D93} (2016),
  no.~12 124004, [\href{http://arxiv.org/abs/1603.03300}{{\tt
  arXiv:1603.03300}}].

\bibitem{Ponce:2014hha}
M.~Ponce, C.~Palenzuela, E.~Barausse, and L.~Lehner, {\it {Electromagnetic
  outflows in a class of scalar-tensor theories: Binary neutron star
  coalescence}},  {\em Phys. Rev.} {\bf D91} (2015), no.~8 084038,
  [\href{http://arxiv.org/abs/1410.0638}{{\tt arXiv:1410.0638}}].

\bibitem{Sennett:2017lcx}
N.~Sennett, L.~Shao, and J.~Steinhoff, {\it {Effective action model of
  dynamically scalarizing binary neutron stars}},  {\em Phys. Rev.} {\bf D96}
  (2017), no.~8 084019, [\href{http://arxiv.org/abs/1708.08285}{{\tt
  arXiv:1708.08285}}].

\bibitem{Sagunski:2017nzb}
L.~Sagunski, J.~Zhang, M.~C. Johnson, L.~Lehner, M.~Sakellariadou, S.~L.
  Liebling, C.~Palenzuela, and D.~Neilsen, {\it {Neutron star mergers as a
  probe of modifications of general relativity with finite-range scalar
  forces}},  {\em Phys. Rev.} {\bf D97} (2018), no.~6 064016,
  [\href{http://arxiv.org/abs/1709.06634}{{\tt arXiv:1709.06634}}].

\bibitem{Berti:2018cxi}
E.~Berti, K.~Yagi, and N.~Yunes, {\it {Extreme Gravity Tests with Gravitational
  Waves from Compact Binary Coalescences: (I) Inspiral-Merger}},  {\em Gen.
  Rel. Grav.} {\bf 50} (2018), no.~4 46,
  [\href{http://arxiv.org/abs/1801.03208}{{\tt arXiv:1801.03208}}].

\bibitem{Berti:2018vdi}
E.~Berti, K.~Yagi, H.~Yang, and N.~Yunes, {\it {Extreme Gravity Tests with
  Gravitational Waves from Compact Binary Coalescences: (II) Ringdown}},  {\em
  Gen. Rel. Grav.} {\bf 50} (2018), no.~5 49,
  [\href{http://arxiv.org/abs/1801.03587}{{\tt arXiv:1801.03587}}].

\bibitem{Vivanco:2019qnt}
F.~Hernandez~Vivanco, R.~Smith, E.~Thrane, P.~D. Lasky, C.~Talbot, and
  V.~Raymond, {\it {Measuring the neutron star equation of state with
  gravitational waves: The first forty binary neutron star merger
  observations}},  {\em Phys. Rev.} {\bf D100} (2019), no.~10 103009,
  [\href{http://arxiv.org/abs/1909.02698}{{\tt arXiv:1909.02698}}].

\bibitem{Sakstein:2017xjx}
J.~Sakstein and B.~Jain, {\it {Implications of the Neutron Star Merger GW170817
  for Cosmological Scalar-Tensor Theories}},  {\em Phys. Rev. Lett.} {\bf 119}
  (2017), no.~25 251303, [\href{http://arxiv.org/abs/1710.05893}{{\tt
  arXiv:1710.05893}}].

\bibitem{Kobayashi:2012kh}
T.~Kobayashi, H.~Motohashi, and T.~Suyama, {\it {Black hole perturbation in the
  most general scalar-tensor theory with second-order field equations I: the
  odd-parity sector}},  {\em Phys. Rev. D} {\bf 85} (2012) 084025,
  [\href{http://arxiv.org/abs/1202.4893}{{\tt arXiv:1202.4893}}]. [Erratum:
  Phys.Rev.D 96, 109903 (2017)].

\bibitem{Kobayashi:2014wsa}
T.~Kobayashi, H.~Motohashi, and T.~Suyama, {\it {Black hole perturbation in the
  most general scalar-tensor theory with second-order field equations II: the
  even-parity sector}},  {\em Phys. Rev. D} {\bf 89} (2014), no.~8 084042,
  [\href{http://arxiv.org/abs/1402.6740}{{\tt arXiv:1402.6740}}].

\bibitem{Antoniou:2019awm}
G.~Antoniou, A.~Bakopoulos, P.~Kanti, B.~Kleihaus, and J.~Kunz, {\it {Novel
  Einstein\textendash{}scalar-Gauss-Bonnet wormholes without exotic matter}},
  {\em Phys. Rev. D} {\bf 101} (2020), no.~2 024033,
  [\href{http://arxiv.org/abs/1904.13091}{{\tt arXiv:1904.13091}}].

\bibitem{Eddington:1922}
A.~S. Eddington, {\em {The mathematical theory of relativity}}.
\newblock Cambridge University Press, Cambridge, 1922.

\bibitem{Robertson:1962}
H.~P. Robertson, {\it {Relativity and Cosmology}},  in {\em Space Age
  Astronomy} (A.~J. Deutsch and W.~B. Klemperer, eds.), p.~228, 1962.

\bibitem{Schiff:1967}
L.~I. Schiff, {\it {Comparison of Theory and Observation in General
  Relativity}},  in {\em Relativity Theory and Astrophysics.~Vol.1: Relativity
  and Cosmology} (J.~Ehlers, ed.), p.~105, 1967.

\bibitem{Baierlein:1967zz}
R.~Baierlein, {\it {Testing General Relativity with Laser Ranging to the
  Moon}},  {\em Phys. Rev.} {\bf 162} (1967) 1275--1288.

\bibitem{Nordtvedt:1968qr}
K.~Nordtvedt, {\it {Equivalence Principle for Massive Bodies. 1.
  Phenomenology}},  {\em Phys. Rev.} {\bf 169} (1968) 1014--1016.

\bibitem{Nordtvedt:1968qs}
K.~Nordtvedt, {\it {Equivalence Principle for Massive Bodies. 2. Theory}},
  {\em Phys. Rev.} {\bf 169} (1968) 1017--1025.

\bibitem{Will:1971zzb}
C.~M. Will, {\it {Theoretical Frameworks for Testing Relativistic Gravity. 2.
  Parametrized Post-Newtonian Hydrodynamics, and the Nordtvedt Effect}},  {\em
  Astrophys. J.} {\bf 163} (1971) 611--627.

\bibitem{Will:1971wt}
C.~M. Will, {\it {Theoretical Frameworks for Testing Relativistic Gravity. 3.
  Conservation Laws, Lorentz Invariance and Values of the PPN Parameters}},
  {\em Astrophys. J.} {\bf 169} (1971) 125--140.

\bibitem{Will:1972zz}
C.~M. Will and K.~Nordtvedt, Jr., {\it {Conservation Laws and Preferred Frames
  in Relativistic Gravity. I. Preferred-Frame Theories and an Extended PPN
  Formalism}},  {\em Astrophys. J.} {\bf 177} (1972) 757.

\bibitem{Nordtvedt:1972zz}
K.~J. Nordtvedt and C.~M. Will, {\it {Conservation Laws and Preferred Frames in
  Relativistic Gravity. II. Experimental Evidence to Rule Out Preferred-Frame
  Theories of Gravity}},  {\em Astrophys. J.} {\bf 177} (1972) 775--792.

\bibitem{Will:1973zz}
C.~M. Will, {\it {Relativistic Gravity tn the Solar System. III. Experimental
  Disproof of a Class of Linear Theories of Gravitation}},  {\em Astrophys. J.}
  {\bf 185} (1973) 31--42.

\bibitem{Fomalont:2009zg}
E.~Fomalont, S.~Kopeikin, G.~Lanyi, and J.~Benson, {\it {Progress in
  Measurements of the Gravitational Bending of Radio Waves Using the VLBA}},
  {\em Astrophys. J.} {\bf 699} (2009) 1395--1402,
  [\href{http://arxiv.org/abs/0904.3992}{{\tt arXiv:0904.3992}}].

\bibitem{Verma:2013ata}
A.~Verma, A.~Fienga, J.~Laskar, H.~Manche, and M.~Gastineau, {\it {Use of
  MESSENGER radioscience data to improve planetary ephemeris and to test
  general relativity}},  {\em Astron. Astrophys.} {\bf 561} (2014) A115,
  [\href{http://arxiv.org/abs/1306.5569}{{\tt arXiv:1306.5569}}].

\bibitem{Viswanathan:2017inp}
V.~{Viswanathan}, A.~{Fienga}, M.~{Gastineau}, and J.~{Laskar}, {\it {INPOP17a
  planetary ephemerides}},  {\em Notes Scientifiques et Techniques de
  l'Institut de Mecanique Celeste} {\bf 108} (Aug, 2017).

\bibitem{Viswanathan:2017vob}
V.~Viswanathan, A.~Fienga, O.~Minazzoli, L.~Bernus, J.~Laskar, and
  M.~Gastineau, {\it {The new lunar ephemeris INPOP17a and its application to
  fundamental physics}},  {\em Mon. Not. Roy. Astron. Soc.} {\bf 476} (2018),
  no.~2 1877--1888, [\href{http://arxiv.org/abs/1710.09167}{{\tt
  arXiv:1710.09167}}].

\bibitem{Kreuzer:1968zz}
L.~B. Kreuzer, {\it {Experimental Measurement of the Equivalence of Active and
  Passive Gravitational Mass}},  {\em Phys. Rev.} {\bf 169} (1968) 1007--1012.

\bibitem{Will:1976zza}
C.~M. Will, {\it {Active mass in relativistic gravity - Theoretical
  interpretation of the Kreuzer experiment}},  {\em Astrophys. J.} {\bf 204}
  (1976) 224--234.

\bibitem{Will:2018bme}
C.~M. Will, {\em {Theory and Experiment in Gravitational Physics}}.
\newblock Cambridge University Press, 2018.

\bibitem{Lin:2012bs}
K.~Lin, S.~Mukohyama, and A.~Wang, {\it {Solar system tests and interpretation
  of gauge field and Newtonian prepotential in general covariant
  Ho\v{r}ava-Lifshitz gravity}},  {\em Phys. Rev.} {\bf D86} (2012) 104024,
  [\href{http://arxiv.org/abs/1206.1338}{{\tt arXiv:1206.1338}}].

\bibitem{Lin:2012ea}
K.~Lin and A.~Wang, {\it {Static post-Newtonian limits in nonprojectable
  Hořava-Lifshitz gravity with an extra U(1) symmetry}},  {\em Phys. Rev.}
  {\bf D87} (2013), no.~8 084041, [\href{http://arxiv.org/abs/1212.6794}{{\tt
  arXiv:1212.6794}}].

\bibitem{Lin:2013tua}
K.~Lin, S.~Mukohyama, A.~Wang, and T.~Zhu, {\it {Post-Newtonian approximations
  in the Hořava-Lifshitz gravity with extra U(1) symmetry}},  {\em Phys. Rev.}
  {\bf D89} (2014), no.~8 084022, [\href{http://arxiv.org/abs/1310.6666}{{\tt
  arXiv:1310.6666}}].

\bibitem{Zaglauer:1990yh}
H.~W. Zaglauer, {\em {Phenomenological aspects of scalar fields in
  astrophysics, cosmology and particle physics}}.
\newblock PhD thesis, Washington U., St. Louis, 1990.

\bibitem{Helbig:1991pk}
T.~Helbig, {\it {Gravitational effects of light scalar particles}},  {\em
  Astrophys. J.} {\bf 382} (1991) 223--232.

\bibitem{Gladchenko:1990nw}
M.~S. Gladchenko, V.~N. Ponomarev, and V.~V. Zhytnikov, {\it {PPN metric and
  PPN torsion in the quadratic Poincare gauge theory of gravity}},  {\em Phys.
  Lett.} {\bf B241} (1990) 67--69.

\bibitem{Alexander:2007zg}
S.~Alexander and N.~Yunes, {\it {A New PPN parameter to test Chern-Simons
  gravity}},  {\em Phys. Rev. Lett.} {\bf 99} (2007) 241101,
  [\href{http://arxiv.org/abs/hep-th/0703265}{{\tt hep-th/0703265}}].

\bibitem{Alexander:2007vt}
S.~Alexander and N.~Yunes, {\it {Parametrized post-Newtonian expansion of
  Chern-Simons gravity}},  {\em Phys. Rev.} {\bf D75} (2007) 124022,
  [\href{http://arxiv.org/abs/0704.0299}{{\tt arXiv:0704.0299}}].

\bibitem{Vainshtein:1972sx}
A.~Vainshtein, {\it {To the problem of nonvanishing gravitation mass}},  {\em
  Phys.Lett.} {\bf B39} (1972) 393--394.

\bibitem{Babichev:2013usa}
E.~Babichev and C.~Deffayet, {\it {An introduction to the Vainshtein
  mechanism}},  {\em Class.Quant.Grav.} {\bf 30} (2013) 184001,
  [\href{http://arxiv.org/abs/1304.7240}{{\tt arXiv:1304.7240}}].

\bibitem{Avilez-Lopez:2015dja}
A.~Avilez-Lopez, A.~Padilla, P.~M. Saffin, and C.~Skordis, {\it {The
  Parametrized Post-Newtonian-Vainshteinian Formalism}},  {\em JCAP} {\bf 1506}
  (2015), no.~06 044, [\href{http://arxiv.org/abs/1501.01985}{{\tt
  arXiv:1501.01985}}].

\bibitem{Hees:2011mu}
A.~Hees and A.~Fuzfa, {\it {Combined cosmological and solar system constraints
  on chameleon mechanism}},  {\em Phys. Rev.} {\bf D85} (2012) 103005,
  [\href{http://arxiv.org/abs/1111.4784}{{\tt arXiv:1111.4784}}].

\bibitem{Scharer:2014kya}
A.~Schärer, R.~Angélil, R.~Bondarescu, P.~Jetzer, and A.~Lundgren, {\it
  {Testing scalar-tensor theories and parametrized post-Newtonian parameters in
  Earth orbit}},  {\em Phys. Rev.} {\bf D90} (2014), no.~12 123005,
  [\href{http://arxiv.org/abs/1410.7914}{{\tt arXiv:1410.7914}}].

\bibitem{Burrage:2017qrf}
C.~Burrage and J.~Sakstein, {\it {Tests of Chameleon Gravity}},  {\em Living
  Rev. Rel.} {\bf 21} (2018), no.~1 1,
  [\href{http://arxiv.org/abs/1709.09071}{{\tt arXiv:1709.09071}}].

\bibitem{McManus:2017itv}
R.~McManus, L.~Lombriser, and J.~Peñarrubia, {\it {Parameterised
  Post-Newtonian Expansion in Screened Regions}},  {\em JCAP} {\bf 1712}
  (2017), no.~12 031, [\href{http://arxiv.org/abs/1705.05324}{{\tt
  arXiv:1705.05324}}].

\bibitem{Sanghai:2016tbi}
V.~A.~A. Sanghai and T.~Clifton, {\it {Parameterized Post-Newtonian
  Cosmology}},  {\em Class. Quant. Grav.} {\bf 34} (2017), no.~6 065003,
  [\href{http://arxiv.org/abs/1610.08039}{{\tt arXiv:1610.08039}}].

\bibitem{Rosen:1974ua}
N.~Rosen, {\it {A theory of gravitation}},  {\em Annals Phys.} {\bf 84} (1974)
  455--473.

\bibitem{Lee:1975kc}
D.~L. Lee, C.~M. Caves, W.-T. Ni, and C.~M. Will, {\it {Theoretical Frameworks
  for Testing Relativistic Gravity. 5. PostNewtonian Limit of Rosen's Theory}},
   {\em Astrophys. J.} {\bf 206} (1976) 555--558.

\bibitem{Hinterbichler:2011tt}
K.~Hinterbichler, {\it {Theoretical Aspects of Massive Gravity}},  {\em
  Rev.Mod.Phys.} {\bf 84} (2012) 671--710,
  [\href{http://arxiv.org/abs/1105.3735}{{\tt arXiv:1105.3735}}].

\bibitem{deRham:2014zqa}
C.~de~Rham, {\it {Massive Gravity}},  {\em Living Rev.Rel.} {\bf 17} (2014) 7,
  [\href{http://arxiv.org/abs/1401.4173}{{\tt arXiv:1401.4173}}].

\bibitem{Clifton:2010hz}
T.~Clifton, M.~Banados, and C.~Skordis, {\it {The Parameterised Post-Newtonian
  Limit of Bimetric Theories of Gravity}},  {\em Class. Quant. Grav.} {\bf 27}
  (2010) 235020, [\href{http://arxiv.org/abs/1006.5619}{{\tt
  arXiv:1006.5619}}].

\bibitem{Hohmann:2010ni}
M.~Hohmann and M.~N.~R. Wohlfarth, {\it {Multimetric extension of the PPN
  formalism: experimental consistency of repulsive gravity}},  {\em Phys. Rev.}
  {\bf D82} (2010) 084028, [\href{http://arxiv.org/abs/1007.4945}{{\tt
  arXiv:1007.4945}}].

\bibitem{Hohmann:2013oca}
M.~Hohmann, {\it {Parameterized post-Newtonian formalism for multimetric
  gravity}},  {\em Class. Quant. Grav.} {\bf 31} (2014) 135003,
  [\href{http://arxiv.org/abs/1309.7787}{{\tt arXiv:1309.7787}}].

\bibitem{Smalley:1980em}
L.~L. Smalley, {\it {Postnewtonian approximation of the Poincare gauge theory
  of gravitation}},  {\em Phys. Rev.} {\bf D21} (1980) 328--331.

\bibitem{Nitsch:1979qn}
J.~Nitsch and F.~W. Hehl, {\it {Translational Gauge Theory of Gravity:
  Postnewtonian Approximation and Spin Precession}},  {\em Phys. Lett.} {\bf
  90B} (1980) 98--102.

\bibitem{Bardeen:1980kt}
J.~M. Bardeen, {\it {Gauge Invariant Cosmological Perturbations}},  {\em
  Phys.Rev.} {\bf D22} (1980) 1882--1905.

\bibitem{Malik:2008im}
K.~A. Malik and D.~Wands, {\it {Cosmological perturbations}},  {\em Phys.
  Rept.} {\bf 475} (2009) 1--51, [\href{http://arxiv.org/abs/0809.4944}{{\tt
  arXiv:0809.4944}}].

\bibitem{Nakamura:2004rm}
K.~Nakamura, {\it {Second-order gauge invariant cosmological perturbation
  theory: Einstein equations in terms of gauge invariant variables}},  {\em
  Prog. Theor. Phys.} {\bf 117} (2007) 17--74,
  [\href{http://arxiv.org/abs/gr-qc/0605108}{{\tt gr-qc/0605108}}].

\bibitem{Nakamura:2006rk}
K.~Nakamura, {\it {Gauge-invariant formulation of the second-order cosmological
  perturbations}},  {\em Phys. Rev.} {\bf D74} (2006) 101301,
  [\href{http://arxiv.org/abs/gr-qc/0605107}{{\tt gr-qc/0605107}}].

\bibitem{Hohmann:2019qgo}
M.~Hohmann, {\it {Gauge-invariant approach to the parametrized post-Newtonian
  formalism}},  {\em Phys. Rev. D} {\bf 101} (2020), no.~2 024061,
  [\href{http://arxiv.org/abs/1910.09245}{{\tt arXiv:1910.09245}}].

\bibitem{Flanagan:2004bz}
E.~E. Flanagan, {\it {The Conformal frame freedom in theories of gravitation}},
   {\em Class. Quant. Grav.} {\bf 21} (2004) 3817,
  [\href{http://arxiv.org/abs/gr-qc/0403063}{{\tt gr-qc/0403063}}].

\bibitem{Nordtvedt:1970uv}
K.~Nordtvedt, Jr., {\it {Post-Newtonian metric for a general class of scalar
  tensor gravitational theories and observational consequences}},  {\em
  Astrophys. J.} {\bf 161} (1970) 1059--1067.

\bibitem{Minazzoli:2010pr}
O.~Minazzoli and B.~Chauvineau, {\it {Scalar-tensor propagation of light in the
  inner solar system at the millimetric level}},  {\em Class. Quant. Grav.}
  {\bf 28} (2011) 085010, [\href{http://arxiv.org/abs/1007.3942}{{\tt
  arXiv:1007.3942}}].

\bibitem{Deng:2012rx}
X.-M. Deng and Y.~Xie, {\it {Two-post-Newtonian Light Propagation in the
  Scalar-Tensor Theory: an $N$-Point-Masses Case}},  {\em Phys. Rev.} {\bf D86}
  (2012) 044007, [\href{http://arxiv.org/abs/1207.3138}{{\tt
  arXiv:1207.3138}}].

\bibitem{Xie:2007gq}
Y.~Xie, W.-T. Ni, P.~Dong, and T.-Y. Huang, {\it {Second post-Newtonian
  approximation of scalar-tensor theory of gravity}},  {\em Adv. Space Res.}
  {\bf 43} (2009) 171--180, [\href{http://arxiv.org/abs/0704.2991}{{\tt
  arXiv:0704.2991}}].

\bibitem{Jarv:2014hma}
L.~Järv, P.~Kuusk, M.~Saal, and O.~Vilson, {\it {Invariant quantities in the
  scalar-tensor theories of gravitation}},  {\em Phys. Rev.} {\bf D91} (2015),
  no.~2 024041, [\href{http://arxiv.org/abs/1411.1947}{{\tt arXiv:1411.1947}}].

\bibitem{Hohmann:2017qje}
M.~Hohmann and A.~Schärer, {\it {Post-Newtonian parameters $\gamma$ and
  $\beta$ of scalar-tensor gravity for a homogeneous gravitating sphere}},
  {\em Phys. Rev.} {\bf D96} (2017), no.~10 104026,
  [\href{http://arxiv.org/abs/1708.07851}{{\tt arXiv:1708.07851}}].

\bibitem{Moffat:2010ek}
J.~W. Moffat and V.~T. Toth, {\it {Modified Jordan-Brans-Dicke theory with
  scalar current and the Eddington-Robertson $\gamma$-parameter}},  {\em Int.
  J. Mod. Phys.} {\bf D21} (2012) 1250084,
  [\href{http://arxiv.org/abs/1001.1564}{{\tt arXiv:1001.1564}}].

\bibitem{Saaidi:2011zza}
K.~Saaidi, A.~Mohammadi, and H.~Sheikhahmadi, {\it {$\gamma$ Parameter and
  Solar System constraint in Chameleon Brans Dick theory}},  {\em Phys. Rev.}
  {\bf D83} (2011) 104019, [\href{http://arxiv.org/abs/1201.0271}{{\tt
  arXiv:1201.0271}}].

\bibitem{Minazzoli:2012ym}
O.~Minazzoli, {\it {$\gamma$ parameter and Solar System constraint in
  Scalar-Tensor theory with a power law potential and universal scalar/matter
  coupling}},  \href{http://arxiv.org/abs/1208.2372}{{\tt arXiv:1208.2372}}.

\bibitem{Devi:2011zz}
N.~C. Devi, S.~Panda, and A.~A. Sen, {\it {Solar System Constraints on Scalar
  Tensor Theories with Non-Standard Action}},  {\em Phys. Rev.} {\bf D84}
  (2011) 063521, [\href{http://arxiv.org/abs/1104.0152}{{\tt
  arXiv:1104.0152}}].

\bibitem{Roshan:2011kz}
M.~Roshan and F.~Shojai, {\it {Notes on the post-Newtonian limit of massive
  Brans-Dicke theory}},  {\em Class. Quant. Grav.} {\bf 28} (2011) 145012,
  [\href{http://arxiv.org/abs/1106.1264}{{\tt arXiv:1106.1264}}].

\bibitem{Minazzoli:2013ara}
O.~Minazzoli and A.~Hees, {\it {Intrinsic Solar System decoupling of a
  scalar-tensor theory with a universal coupling between the scalar field and
  the matter Lagrangian}},  {\em Phys. Rev.} {\bf D88} (2013), no.~4 041504,
  [\href{http://arxiv.org/abs/1308.2770}{{\tt arXiv:1308.2770}}].

\bibitem{Capozziello:2005bu}
S.~Capozziello and A.~Troisi, {\it {PPN-limit of fourth order gravity inspired
  by scalar-tensor gravity}},  {\em Phys. Rev.} {\bf D72} (2005) 044022,
  [\href{http://arxiv.org/abs/astro-ph/0507545}{{\tt astro-ph/0507545}}].

\bibitem{Capozziello:2006jj}
S.~Capozziello, A.~Stabile, and A.~Troisi, {\it {Fourth-order gravity and
  experimental constraints on Eddington parameters}},  {\em Mod. Phys. Lett.}
  {\bf A21} (2006) 2291--2301, [\href{http://arxiv.org/abs/gr-qc/0603071}{{\tt
  gr-qc/0603071}}].

\bibitem{Capozziello:2007ms}
S.~Capozziello, A.~Stabile, and A.~Troisi, {\it {The Newtonian Limit of f(R)
  gravity}},  {\em Phys. Rev.} {\bf D76} (2007) 104019,
  [\href{http://arxiv.org/abs/0708.0723}{{\tt arXiv:0708.0723}}].

\bibitem{Capozziello:2010wt}
S.~Capozziello, A.~Stabile, and A.~Troisi, {\it {Comparing scalar-tensor
  gravity and f(R)-gravity in the Newtonian limit}},  {\em Phys. Lett.} {\bf
  B686} (2010) 79--83, [\href{http://arxiv.org/abs/1002.1364}{{\tt
  arXiv:1002.1364}}].

\bibitem{Clifton:2008jq}
T.~Clifton, {\it {The Parameterised Post-Newtonian Limit of Fourth-Order
  Theories of Gravity}},  {\em Phys. Rev.} {\bf D77} (2008) 024041,
  [\href{http://arxiv.org/abs/0801.0983}{{\tt arXiv:0801.0983}}].

\bibitem{Capone:2009xk}
M.~Capone and M.~L. Ruggiero, {\it {Jumping from Metric f(R) to Scalar-Tensor
  Theories and the relations between their post-Newtonian Parameters}},  {\em
  Class. Quant. Grav.} {\bf 27} (2010) 125006,
  [\href{http://arxiv.org/abs/0910.0434}{{\tt arXiv:0910.0434}}].

\bibitem{Capozziello:2010gu}
S.~Capozziello and A.~Stabile, {\it {The Weak Field Limit of Fourth Order
  Gravity}},  in {\em {In *Frignanni, Vincent R. (ed.): Classical and Quantum
  Gravity: Theory and Applications* Chapter 2}}, 2010.
\newblock \href{http://arxiv.org/abs/1009.3441}{{\tt arXiv:1009.3441}}.

\bibitem{Deng:2016moh}
X.-M. Deng and Y.~Xie, {\it {Solar System tests of a scalar-tensor gravity with
  a general potential: Insensitivity of light deflection and Cassini
  tracking}},  {\em Phys. Rev.} {\bf D93} (2016), no.~4 044013.

\bibitem{Zhang:2016njn}
X.~Zhang, W.~Zhao, H.~Huang, and Y.~Cai, {\it {Post-Newtonian parameters and
  cosmological constant of screened modified gravity}},  {\em Phys. Rev.} {\bf
  D93} (2016), no.~12 124003, [\href{http://arxiv.org/abs/1603.09450}{{\tt
  arXiv:1603.09450}}].

\bibitem{Berkin:1993bt}
A.~L. Berkin and R.~W. Hellings, {\it {Multiple field scalar - tensor theories
  of gravity and cosmology}},  {\em Phys. Rev.} {\bf D49} (1994) 6442--6449,
  [\href{http://arxiv.org/abs/gr-qc/9401033}{{\tt gr-qc/9401033}}].

\bibitem{Randla:2014saa}
E.~Randla, {\it {PPN parameters for multiscalar-tensor gravity without a
  potential}},  {\em J. Phys. Conf. Ser.} {\bf 532} (2014) 012024.

\bibitem{Kuusk:2015dda}
P.~Kuusk, L.~Jarv, and O.~Vilson, {\it {Invariant quantities in the
  multiscalar-tensor theories of gravitation}},  {\em Int. J. Mod. Phys.} {\bf
  A31} (2016), no.~02n03 1641003, [\href{http://arxiv.org/abs/1509.02903}{{\tt
  arXiv:1509.02903}}].

\bibitem{Hohmann:2016yfd}
M.~Hohmann, L.~Jarv, P.~Kuusk, E.~Randla, and O.~Vilson, {\it {Post-Newtonian
  parameter $\gamma$ for multiscalar-tensor gravity with a general potential}},
   {\em Phys. Rev.} {\bf D94} (2016), no.~12 124015,
  [\href{http://arxiv.org/abs/1607.02356}{{\tt arXiv:1607.02356}}].

\bibitem{Koivisto:2011tp}
T.~S. Koivisto, {\it {The post-Newtonian limit in C-theories of gravitation}},
  {\em Phys. Rev.} {\bf D84} (2011) 121502,
  [\href{http://arxiv.org/abs/1109.4585}{{\tt arXiv:1109.4585}}].

\bibitem{Conroy:2014eja}
A.~Conroy, T.~Koivisto, A.~Mazumdar, and A.~Teimouri, {\it {Generalized
  quadratic curvature, non-local infrared modifications of gravity and
  Newtonian potentials}},  {\em Class. Quant. Grav.} {\bf 32} (2015), no.~1
  015024, [\href{http://arxiv.org/abs/1406.4998}{{\tt arXiv:1406.4998}}].

\bibitem{Hohmann:2015kra}
M.~Hohmann, {\it {Parametrized post-Newtonian limit of Horndeski’s gravity
  theory}},  {\em Phys. Rev.} {\bf D92} (2015), no.~6 064019,
  [\href{http://arxiv.org/abs/1506.04253}{{\tt arXiv:1506.04253}}].

\bibitem{Hou:2017cjy}
S.~Hou and Y.~Gong, {\it {Constraints on Horndeski Theory Using the
  Observations of Nordtvedt Effect, Shapiro Time Delay and Binary Pulsars}},
  {\em Eur. Phys. J.} {\bf C78} (2018), no.~3 247,
  [\href{http://arxiv.org/abs/1711.05034}{{\tt arXiv:1711.05034}}].

\bibitem{Kase:2013uja}
R.~Kase and S.~Tsujikawa, {\it {Screening the fifth force in the Horndeski's
  most general scalar-tensor theories}},  {\em JCAP} {\bf 1308} (2013) 054,
  [\href{http://arxiv.org/abs/1306.6401}{{\tt arXiv:1306.6401}}].

\bibitem{Hinterbichler:2012cn}
K.~Hinterbichler and R.~A. Rosen, {\it {Interacting Spin-2 Fields}},  {\em
  JHEP} {\bf 1207} (2012) 047, [\href{http://arxiv.org/abs/1203.5783}{{\tt
  arXiv:1203.5783}}].

\bibitem{Schmidt-May:2015vnx}
A.~Schmidt-May and M.~von Strauss, {\it {Recent developments in bimetric
  theory}},  {\em J. Phys.} {\bf A49} (2016), no.~18 183001,
  [\href{http://arxiv.org/abs/1512.00021}{{\tt arXiv:1512.00021}}].

\bibitem{Hohmann:2017uxe}
M.~Hohmann, {\it {Post-Newtonian parameter $\gamma$ and the deflection of light
  in ghost-free massive bimetric gravity}},  {\em Phys. Rev.} {\bf D95} (2017),
  no.~12 124049, [\href{http://arxiv.org/abs/1701.07700}{{\tt
  arXiv:1701.07700}}].

\bibitem{Einstein:1928}
A.~Einstein, {\it {Riemann-Geometrie mit Aufrechterhaltung des Begriffes des
  Fernparallelismus}},  {\em Sitzber. Preuss. Akad. Wiss.} {\bf 17} (1928)
  217--221.

\bibitem{TheLIGOScientific:2016htt}
{\bf LIGO Scientific, Virgo} Collaboration, B.~P. Abbott et~al., {\it
  {Astrophysical Implications of the Binary Black-Hole Merger GW150914}},  {\em
  Astrophys. J.} {\bf 818} (2016), no.~2 L22,
  [\href{http://arxiv.org/abs/1602.03846}{{\tt arXiv:1602.03846}}].

\bibitem{Caldwell:2019vru}
R.~Caldwell et~al., {\it {Astro2020 Science White Paper: Cosmology with a
  Space-Based Gravitational Wave Observatory}},
  \href{http://arxiv.org/abs/1903.04657}{{\tt arXiv:1903.04657}}.

\bibitem{Sathyaprakash:2019nnu}
B.~S. Sathyaprakash et~al., {\it {Cosmology and the Early Universe}},
  \href{http://arxiv.org/abs/1903.09260}{{\tt arXiv:1903.09260}}.

\bibitem{Auclair:2019wcv}
P.~Auclair et~al., {\it {Probing the gravitational wave background from cosmic
  strings with LISA}},  \href{http://arxiv.org/abs/1909.00819}{{\tt
  arXiv:1909.00819}}.

\bibitem{Ringeval:2017eww}
C.~Ringeval and T.~Suyama, {\it {Stochastic gravitational waves from cosmic
  string loops in scaling}},  {\em JCAP} {\bf 1712} (2017), no.~12 027,
  [\href{http://arxiv.org/abs/1709.03845}{{\tt arXiv:1709.03845}}].

\bibitem{Abbott:2017mem}
{\bf LIGO Scientific, Virgo} Collaboration, B.~P. Abbott et~al., {\it
  {Constraints on cosmic strings using data from the first Advanced LIGO
  observing run}},  {\em Phys. Rev.} {\bf D97} (2018), no.~10 102002,
  [\href{http://arxiv.org/abs/1712.01168}{{\tt arXiv:1712.01168}}].

\bibitem{Jenkins:2018lvb}
A.~C. Jenkins and M.~Sakellariadou, {\it {Anisotropies in the stochastic
  gravitational-wave background: Formalism and the cosmic string case}},  {\em
  Phys. Rev.} {\bf D98} (2018), no.~6 063509,
  [\href{http://arxiv.org/abs/1802.06046}{{\tt arXiv:1802.06046}}].

\bibitem{Brdar:2018num}
V.~Brdar, A.~J. Helmboldt, and J.~Kubo, {\it {Gravitational Waves from
  First-Order Phase Transitions: LIGO as a Window to Unexplored Seesaw
  Scales}},  {\em JCAP} {\bf 1902} (2019) 021,
  [\href{http://arxiv.org/abs/1810.12306}{{\tt arXiv:1810.12306}}].

\bibitem{Caprini:2019egz}
C.~Caprini et~al., {\it {Detecting gravitational waves from cosmological phase
  transitions with LISA: an update}},
  \href{http://arxiv.org/abs/1910.13125}{{\tt arXiv:1910.13125}}.

\bibitem{TheLIGOScientific:2016src}
{\bf LIGO Scientific, Virgo} Collaboration, B.~P. Abbott et~al., {\it {Tests of
  general relativity with GW150914}},  {\em Phys. Rev. Lett.} {\bf 116} (2016),
  no.~22 221101, [\href{http://arxiv.org/abs/1602.03841}{{\tt
  arXiv:1602.03841}}]. [Erratum: Phys. Rev. Lett.121,no.12,129902(2018)].

\bibitem{Callister:2017ocg}
T.~Callister, A.~S. Biscoveanu, N.~Christensen, M.~Isi, A.~Matas, O.~Minazzoli,
  T.~Regimbau, M.~Sakellariadou, J.~Tasson, and E.~Thrane, {\it
  {Polarization-based Tests of Gravity with the Stochastic Gravitational-Wave
  Background}},  {\em Phys. Rev.} {\bf X7} (2017), no.~4 041058,
  [\href{http://arxiv.org/abs/1704.08373}{{\tt arXiv:1704.08373}}].

\bibitem{Abbott:2018lct}
{\bf LIGO Scientific, Virgo} Collaboration, B.~P. Abbott et~al., {\it {Tests of
  General Relativity with GW170817}},  {\em Phys. Rev. Lett.} {\bf 123} (2019),
  no.~1 011102, [\href{http://arxiv.org/abs/1811.00364}{{\tt
  arXiv:1811.00364}}].

\bibitem{LIGOScientific:2019fpa}
{\bf LIGO Scientific, Virgo} Collaboration, B.~P. Abbott et~al., {\it {Tests of
  General Relativity with the Binary Black Hole Signals from the LIGO-Virgo
  Catalog GWTC-1}},  {\em Phys. Rev.} {\bf D100} (2019), no.~10 104036,
  [\href{http://arxiv.org/abs/1903.04467}{{\tt arXiv:1903.04467}}].

\bibitem{Berti:2019xgr}
E.~Berti et~al., {\it {Tests of General Relativity and Fundamental Physics with
  Space-based Gravitational Wave Detectors}},
  \href{http://arxiv.org/abs/1903.02781}{{\tt arXiv:1903.02781}}.

\bibitem{LIGOScientific:2019vic}
{\bf LIGO Scientific, Virgo} Collaboration, B.~P. Abbott et~al., {\it {Search
  for the isotropic stochastic background using data from Advanced LIGO’s
  second observing run}},  {\em Phys. Rev.} {\bf D100} (2019), no.~6 061101,
  [\href{http://arxiv.org/abs/1903.02886}{{\tt arXiv:1903.02886}}].

\bibitem{Colladay:1998fq}
D.~Colladay and V.~A. Kostelecky, {\it {Lorentz violating extension of the
  standard model}},  {\em Phys. Rev.} {\bf D58} (1998) 116002,
  [\href{http://arxiv.org/abs/hep-ph/9809521}{{\tt hep-ph/9809521}}].

\bibitem{Minazzoli:2019ugi}
O.~Minazzoli, N.~K. Johnson-Mcdaniel, and M.~Sakellariadou, {\it {Shortcomings
  of Shapiro delay-based tests of the equivalence principle on cosmological
  scales}},  {\em Phys. Rev.} {\bf D100} (2019), no.~10 104047,
  [\href{http://arxiv.org/abs/1907.12453}{{\tt arXiv:1907.12453}}].

\bibitem{Soldner1804}
J.~G.~V. Soldner, {\it {On the deflection of a light ray from its rectilinear
  motion, by the attraction of a celestial body at which it nearly passes by}},
   {\em Phys. Rev. Lett.} {\bf 3} (1959) 439--441.

\bibitem{PoundRebka1959}
R.~V. Pound and G.~A. Rebka, {\it {Gravitational Red-Shift in Nuclear
  Resonance}},  {\em Berliner Astronomisches Jahrbuch}.

\bibitem{TPChengRel2010}
T.-P. Cheng, {\em {Relativity, Gravitation and Cosmology, A Basic
  Introduction}}.
\newblock Oxford University Press, 2010.

\bibitem{Dyson:1920cwa}
F.~W. Dyson, A.~S. Eddington, and C.~Davidson, {\it {A Determination of the
  Deflection of Light by the Sun's Gravitational Field, from Observations Made
  at the Total Eclipse of May 29, 1919}},  {\em Phil. Trans. Roy. Soc. Lond.}
  {\bf A220} (1920) 291--333.

\bibitem{Dadhich:2000am}
N.~Dadhich, R.~Maartens, P.~Papadopoulos, and V.~Rezania, {\it {Black holes on
  the brane}},  {\em Phys. Lett.} {\bf B487} (2000) 1--6,
  [\href{http://arxiv.org/abs/hep-th/0003061}{{\tt hep-th/0003061}}].

\bibitem{Shiromizu:1999wj}
T.~Shiromizu, K.-i. Maeda, and M.~Sasaki, {\it {The Einstein equation on the
  3-brane world}},  {\em Phys. Rev.} {\bf D62} (2000) 024012,
  [\href{http://arxiv.org/abs/gr-qc/9910076}{{\tt gr-qc/9910076}}].

\bibitem{Gergely:2003pn}
L.~A. Gergely, {\it {Generalized Friedmann branes}},  {\em Phys. Rev.} {\bf
  D68} (2003) 124011, [\href{http://arxiv.org/abs/gr-qc/0308072}{{\tt
  gr-qc/0308072}}].

\bibitem{Gergely:2008fw}
L.~A. Gergely, {\it {Friedmann branes with variable tension}},  {\em Phys.
  Rev.} {\bf D78} (2008) 084006, [\href{http://arxiv.org/abs/0806.3857}{{\tt
  arXiv:0806.3857}}].

\bibitem{Rindler:2007zz}
W.~Rindler and M.~Ishak, {\it {Contribution of the cosmological constant to the
  relativistic bending of light revisited}},  {\em Phys. Rev.} {\bf D76} (2007)
  043006, [\href{http://arxiv.org/abs/0709.2948}{{\tt arXiv:0709.2948}}].

\bibitem{Kar:2003gh}
S.~Kar and M.~Sinha, {\it {Bending of light and gravitational signals in
  certain on-brane and bulk geometries}},  {\em Gen. Rel. Grav.} {\bf 35}
  (2003) 1775--1784.

\bibitem{Majumdar:2004mz}
A.~S. Majumdar and N.~Mukherjee, {\it {Gravitational lensing in the weak field
  limit by a braneworld black hole}},  {\em Mod. Phys. Lett.} {\bf A20} (2005)
  2487--2496, [\href{http://arxiv.org/abs/astro-ph/0403405}{{\tt
  astro-ph/0403405}}].

\bibitem{Gergely:2006qt}
L.~A. Gergely and B.~Darazs, {\it {Weak gravitational lensing in
  brane-worlds}},  {\em Publ. Astron. Dep. Eotvos Univ.} {\bf 17} (2006)
  213--219, [\href{http://arxiv.org/abs/astro-ph/0602427}{{\tt
  astro-ph/0602427}}].

\bibitem{Briet:2008mz}
J.~Briet and D.~Hobill, {\it {Determining the Dimensionality of Spacetime by
  Gravitational Lensing}},  \href{http://arxiv.org/abs/0801.3859}{{\tt
  arXiv:0801.3859}}.

\bibitem{Boehmer:2008zh}
C.~G. Boehmer, T.~Harko, and F.~S.~N. Lobo, {\it {Solar system tests of brane
  world models}},  {\em Class. Quant. Grav.} {\bf 25} (2008) 045015,
  [\href{http://arxiv.org/abs/0801.1375}{{\tt arXiv:0801.1375}}].

\bibitem{Gergely:2009xg}
L.~A. Gergely, Z.~Keresztes, and M.~Dwornik, {\it {Second-order light
  deflection by tidal charged black holes}},  {\em Class. Quant. Grav.} {\bf
  26} (2009) 145002, [\href{http://arxiv.org/abs/0903.1558}{{\tt
  arXiv:0903.1558}}].

\bibitem{Frittelli:1998hr}
S.~Frittelli and E.~T. Newman, {\it {An Exact universal gravitational lensing
  equation}},  {\em Phys. Rev.} {\bf D59} (1999) 124001,
  [\href{http://arxiv.org/abs/gr-qc/9810017}{{\tt gr-qc/9810017}}].

\bibitem{Horvath:2010xq}
Z.~Horvath, L.~A. Gergely, and D.~Hobill, {\it {Image formation in weak
  gravitational lensing by tidal charged black holes}},  {\em Class. Quant.
  Grav.} {\bf 27} (2010) 235006, [\href{http://arxiv.org/abs/1005.2286}{{\tt
  arXiv:1005.2286}}].

\bibitem{Virbhadra:2008ws}
K.~S. Virbhadra, {\it {Relativistic images of Schwarzschild black hole
  lensing}},  {\em Phys. Rev.} {\bf D79} (2009) 083004,
  [\href{http://arxiv.org/abs/0810.2109}{{\tt arXiv:0810.2109}}].

\bibitem{Virbhadra:1999nm}
K.~S. Virbhadra and G.~F.~R. Ellis, {\it {Schwarzschild black hole lensing}},
  {\em Phys. Rev.} {\bf D62} (2000) 084003,
  [\href{http://arxiv.org/abs/astro-ph/9904193}{{\tt astro-ph/9904193}}].

\bibitem{Dabrowski:1998ac}
M.~P. Dabrowski and F.~E. Schunck, {\it {Boson stars as gravitational lenses}},
   {\em Astrophys. J.} {\bf 535} (2000) 316--324,
  [\href{http://arxiv.org/abs/astro-ph/9807039}{{\tt astro-ph/9807039}}].

\bibitem{Capozziello:2006dp}
S.~Capozziello, V.~F. Cardone, and A.~Troisi, {\it {Gravitational lensing in
  fourth order gravity}},  {\em Phys. Rev.} {\bf D73} (2006) 104019,
  [\href{http://arxiv.org/abs/astro-ph/0604435}{{\tt astro-ph/0604435}}].

\bibitem{Cramer:1994qj}
J.~G. Cramer, R.~L. Forward, M.~S. Morris, M.~Visser, G.~Benford, and G.~A.
  Landis, {\it {Natural wormholes as gravitational lenses}},  {\em Phys. Rev.}
  {\bf D51} (1995) 3117--3120,
  [\href{http://arxiv.org/abs/astro-ph/9409051}{{\tt astro-ph/9409051}}].

\bibitem{Sereno:2003nd}
M.~Sereno, {\it {Weak field limit of Reissner-Nordstrom black hole lensing}},
  {\em Phys. Rev.} {\bf D69} (2004) 023002,
  [\href{http://arxiv.org/abs/gr-qc/0310063}{{\tt gr-qc/0310063}}].

\bibitem{Schneider1984AA}
P.~Schneider, {\it {The amplification caused by gravitational bending of
  light}},  {\em Astr. Ap.} {\bf 140} (1984) 119.

\bibitem{BlandfordNarayan1986}
R.~Blandford and R.~Narayan, {\it {Fermat's Principle, Caustics, and the
  Classification of Gravitational Lens Images}},  {\em Astroph. J} {\bf 310}
  (1986) 568.

\bibitem{Bozza:2001xd}
V.~Bozza, S.~Capozziello, G.~Iovane, and G.~Scarpetta, {\it {Strong field limit
  of black hole gravitational lensing}},  {\em Gen. Rel. Grav.} {\bf 33} (2001)
  1535--1548, [\href{http://arxiv.org/abs/gr-qc/0102068}{{\tt gr-qc/0102068}}].

\bibitem{Bozza:2002zj}
V.~Bozza, {\it {Gravitational lensing in the strong field limit}},  {\em Phys.
  Rev.} {\bf D66} (2002) 103001,
  [\href{http://arxiv.org/abs/gr-qc/0208075}{{\tt gr-qc/0208075}}].

\bibitem{BinNun:2010se}
A.~Y. Bin-Nun, {\it {Lensing By Sgr A* as a Probe of Modified Gravity}},  {\em
  Phys. Rev.} {\bf D82} (2010) 064009,
  [\href{http://arxiv.org/abs/1004.0379}{{\tt arXiv:1004.0379}}].

\bibitem{Horvath:2012ru}
Z.~Horvath and L.~A. Gergely, {\it {Black hole tidal charge constrained by
  strong gravitational lensing}},  {\em Astron. Nachr.} {\bf 334} (2013)
  1047--1050, [\href{http://arxiv.org/abs/1203.6576}{{\tt arXiv:1203.6576}}].

\bibitem{Gillessen:2010ei}
S.~Gillessen et~al., {\it {GRAVITY: a four-telescope beam combiner instrument
  for the VLTI}},  {\em Proc. SPIE Int. Soc. Opt. Eng.} {\bf 7734} (2010)
  77340Y, [\href{http://arxiv.org/abs/1007.1612}{{\tt arXiv:1007.1612}}].

\bibitem{Garriga:1999yh}
J.~Garriga and T.~Tanaka, {\it {Gravity in the brane world}},  {\em Phys. Rev.
  Lett.} {\bf 84} (2000) 2778--2781,
  [\href{http://arxiv.org/abs/hep-th/9911055}{{\tt hep-th/9911055}}].

\bibitem{BinNun:2009jr}
A.~Y. Bin-Nun, {\it {Relativistic Images in Randall-Sundrum II Braneworld
  Lensing}},  {\em Phys. Rev.} {\bf D81} (2010) 123011,
  [\href{http://arxiv.org/abs/0912.2081}{{\tt arXiv:0912.2081}}].

\bibitem{Guedens:2002km}
R.~Guedens, D.~Clancy, and A.~R. Liddle, {\it {Primordial black holes in
  braneworld cosmologies: Formation, cosmological evolution and evaporation}},
  {\em Phys. Rev.} {\bf D66} (2002) 043513,
  [\href{http://arxiv.org/abs/astro-ph/0205149}{{\tt astro-ph/0205149}}].

\bibitem{Eiroa:2004gh}
E.~F. Eiroa, {\it {A Braneworld black hole gravitational lens: Strong field
  limit analysis}},  {\em Phys. Rev.} {\bf D71} (2005) 083010,
  [\href{http://arxiv.org/abs/gr-qc/0410128}{{\tt gr-qc/0410128}}].

\bibitem{Whisker:2004gq}
R.~Whisker, {\it {Strong gravitational lensing by braneworld black holes}},
  {\em Phys. Rev.} {\bf D71} (2005) 064004,
  [\href{http://arxiv.org/abs/astro-ph/0411786}{{\tt astro-ph/0411786}}].

\bibitem{Casadio:2001jg}
R.~Casadio, A.~Fabbri, and L.~Mazzacurati, {\it {New black holes in the brane
  world?}},  {\em Phys. Rev.} {\bf D65} (2002) 084040,
  [\href{http://arxiv.org/abs/gr-qc/0111072}{{\tt gr-qc/0111072}}].

\bibitem{Majumdar:2005ba}
A.~S. Majumdar and N.~Mukherjee, {\it {Braneworld black holes in cosmology and
  astrophysics}},  {\em Int. J. Mod. Phys.} {\bf D14} (2005) 1095,
  [\href{http://arxiv.org/abs/astro-ph/0503473}{{\tt astro-ph/0503473}}].

\bibitem{Myers:1986un}
R.~C. Myers and M.~J. Perry, {\it {Black Holes in Higher Dimensional
  Space-Times}},  {\em Annals Phys.} {\bf 172} (1986) 304.

\bibitem{Pal:2007ap}
S.~Pal and S.~Kar, {\it {Gravitational lensing in braneworld gravity: Formalism
  and applications}},  {\em Class. Quant. Grav.} {\bf 25} (2008) 045003,
  [\href{http://arxiv.org/abs/0707.0223}{{\tt arXiv:0707.0223}}].

\bibitem{Visser:2011mf}
M.~Visser, {\it {Status of Horava gravity: A personal perspective}},  {\em J.
  Phys. Conf. Ser.} {\bf 314} (2011) 012002,
  [\href{http://arxiv.org/abs/1103.5587}{{\tt arXiv:1103.5587}}].

\bibitem{Kehagias:2009is}
A.~Kehagias and K.~Sfetsos, {\it {The Black hole and FRW geometries of
  non-relativistic gravity}},  {\em Phys. Lett.} {\bf B678} (2009) 123--126,
  [\href{http://arxiv.org/abs/0905.0477}{{\tt arXiv:0905.0477}}].

\bibitem{Horvath:2011xr}
Z.~Horvath, L.~A. Gergely, Z.~Keresztes, T.~Harko, and F.~S.~N. Lobo, {\it
  {Constraining Ho\v{r}ava-Lifshitz gravity by weak and strong gravitational
  lensing}},  {\em Phys. Rev.} {\bf D84} (2011) 083006,
  [\href{http://arxiv.org/abs/1105.0765}{{\tt arXiv:1105.0765}}].

\bibitem{Trippe:2009hf}
S.~Trippe, R.~Davies, F.~Eisenhauer, N.~M.~F. Schreiber, T.~K. Fritz, and
  R.~Genzel, {\it {High Precision Astrometry with MICADO at the European
  Extremely Large Telescope}},  {\em Mon. Not. Roy. Astron. Soc.} {\bf 402}
  (2010) 1126, [\href{http://arxiv.org/abs/0910.5114}{{\tt arXiv:0910.5114}}].

\bibitem{Sotiriou:2011dz}
T.~P. Sotiriou and V.~Faraoni, {\it {Black holes in scalar-tensor gravity}},
  {\em Phys. Rev. Lett.} {\bf 108} (2012) 081103,
  [\href{http://arxiv.org/abs/1109.6324}{{\tt arXiv:1109.6324}}].

\bibitem{Capozziello:2006ph}
S.~Capozziello, V.~F. Cardone, and A.~Troisi, {\it {Low surface brightness
  galaxies rotation curves in the low energy limit of r**n gravity: no need for
  dark matter?}},  {\em Mon. Not. Roy. Astron. Soc.} {\bf 375} (2007)
  1423--1440, [\href{http://arxiv.org/abs/astro-ph/0603522}{{\tt
  astro-ph/0603522}}].

\bibitem{Horvath:2012kf}
Z.~Horvath, L.~A. Gergely, D.~Hobill, S.~Capozziello, and M.~De~Laurentis, {\it
  {Weak gravitational lensing by compact objects in fourth order gravity}},
  {\em Phys. Rev.} {\bf D88} (2013), no.~6 063009,
  [\href{http://arxiv.org/abs/1207.1823}{{\tt arXiv:1207.1823}}].

\bibitem{Janis:1968zz}
A.~I. Janis, E.~T. Newman, and J.~Winicour, {\it {Reality of the Schwarzschild
  Singularity}},  {\em Phys. Rev. Lett.} {\bf 20} (1968) 878--880.

\bibitem{GautreauNeuvoC1969}
R.~Gautreau, {\it {Gautreau, Coupled weyl gravitational and zero-rest-mass
  scalar fields}},  {\em Nuovo Cimento} {\bf B62} (1969) 360--370.

\bibitem{Wyman:1981bd}
M.~Wyman, {\it {Static Spherically Symmetric Scalar Fields in General
  Relativity}},  {\em Phys. Rev.} {\bf D24} (1981) 839--841.

\bibitem{Virbhadra:1997ie}
K.~S. Virbhadra, {\it {Janis-Newman-Winicour and Wyman solutions are the
  same}},  {\em Int. J. Mod. Phys.} {\bf A12} (1997) 4831--4836,
  [\href{http://arxiv.org/abs/gr-qc/9701021}{{\tt gr-qc/9701021}}].

\bibitem{Virbhadra:1998dy}
K.~S. Virbhadra, D.~Narasimha, and S.~M. Chitre, {\it {Role of the scalar field
  in gravitational lensing}},  {\em Astron. Astrophys.} {\bf 337} (1998) 1--8,
  [\href{http://arxiv.org/abs/astro-ph/9801174}{{\tt astro-ph/9801174}}].

\bibitem{Younas:2015sva}
A.~Younas, S.~Hussain, M.~Jamil, and S.~Bahamonde, {\it {Strong Gravitational
  Lensing by Kiselev Black Hole}},  {\em Phys. Rev.} {\bf D92} (2015), no.~8
  084042, [\href{http://arxiv.org/abs/1502.01676}{{\tt arXiv:1502.01676}}].

\bibitem{Kiselev:2002dx}
V.~V. Kiselev, {\it {Quintessence and black holes}},  {\em Class. Quant. Grav.}
  {\bf 20} (2003) 1187--1198, [\href{http://arxiv.org/abs/gr-qc/0210040}{{\tt
  gr-qc/0210040}}].

\bibitem{Mandelbaum:2017dvy}
R.~Mandelbaum et~al., {\it {The first-year shear catalog of the Subaru Hyper
  Suprime-Cam SSP Survey}},  \href{http://arxiv.org/abs/1705.06745}{{\tt
  arXiv:1705.06745}}.

\bibitem{deJong:2012zb}
{\bf Astro-WISE, KiDS} Collaboration, J.~T.~A. de~Jong, G.~A.~V. Kleijn, K.~H.
  Kuijken, E.~A. Valentijn, and KiDS, {\it {The Kilo-Degree Survey}},  {\em
  Exper. Astron.} {\bf 35} (2013) 25--44,
  [\href{http://arxiv.org/abs/1206.1254}{{\tt arXiv:1206.1254}}].

\bibitem{Zhang:2007nk}
P.~Zhang, M.~Liguori, R.~Bean, and S.~Dodelson, {\it {Probing Gravity at
  Cosmological Scales by Measurements which Test the Relationship between
  Gravitational Lensing and Matter Overdensity}},  {\em Phys. Rev. Lett.} {\bf
  99} (2007) 141302, [\href{http://arxiv.org/abs/0704.1932}{{\tt
  arXiv:0704.1932}}].

\bibitem{Mak:2004hv}
M.~K. Mak and T.~Harko, {\it {Can the galactic rotation curves be explained in
  brane world models?}},  {\em Phys. Rev.} {\bf D70} (2004) 024010,
  [\href{http://arxiv.org/abs/gr-qc/0404104}{{\tt gr-qc/0404104}}].

\bibitem{Harko:2005fq}
T.~Harko and M.~K. Mak, {\it {Conformally symmetric vacuum solutions of the
  gravitational field equations in the brane-world models}},  {\em Annals
  Phys.} {\bf 319} (2005) 471--492,
  [\href{http://arxiv.org/abs/gr-qc/0503072}{{\tt gr-qc/0503072}}].

\bibitem{Harko:2005yk}
T.~Harko and K.~S. Cheng, {\it {Galactic metric, dark radiation, dark pressure
  and gravitational lensing in brane world models}},  {\em Astrophys. J.} {\bf
  636} (2005) 8--20, [\href{http://arxiv.org/abs/astro-ph/0509576}{{\tt
  astro-ph/0509576}}].

\bibitem{Boehmer:2007xh}
C.~G. Boehmer and T.~Harko, {\it {Galactic dark matter as a bulk effect on the
  brane}},  {\em Class. Quant. Grav.} {\bf 24} (2007) 3191--3210,
  [\href{http://arxiv.org/abs/0705.2496}{{\tt arXiv:0705.2496}}].

\bibitem{Gergely:2011df}
L.~A. Gergely, T.~Harko, M.~Dwornik, G.~Kupi, and Z.~Keresztes, {\it {Galactic
  rotation curves in brane world models}},  {\em Mon. Not. Roy. Astron. Soc.}
  {\bf 415} (2011) 3275--3290, [\href{http://arxiv.org/abs/1105.0159}{{\tt
  arXiv:1105.0159}}].

\bibitem{Wong:2012sd}
K.~C. Wong, T.~Harko, K.~S. Cheng, and L.~A. Gergely, {\it {Weyl fluid dark
  matter model tested on the galactic scale by weak gravitational lensing}},
  {\em Phys. Rev.} {\bf D86} (2012) 044038,
  [\href{http://arxiv.org/abs/1207.3167}{{\tt arXiv:1207.3167}}].

\bibitem{Harte:2019tid}
A.~I. Harte, {\it {Gravitational lensing beyond geometric optics: II. Metric
  independence}},  {\em Gen. Rel. Grav.} {\bf 51} (2019), no.~12 160,
  [\href{http://arxiv.org/abs/1906.10708}{{\tt arXiv:1906.10708}}].

\bibitem{Hawking:1969sw}
S.~W. Hawking and R.~Penrose, {\it {The Singularities of gravitational collapse
  and cosmology}},  {\em Proc. Roy. Soc. Lond.} {\bf A314} (1970) 529--548.

\bibitem{Hossenfelder:2012jw}
S.~Hossenfelder, {\it {Minimal Length Scale Scenarios for Quantum Gravity}},
  {\em Living Rev. Rel.} {\bf 16} (2013) 2,
  [\href{http://arxiv.org/abs/1203.6191}{{\tt arXiv:1203.6191}}].

\bibitem{Casadio:2013tma}
R.~Casadio, {\it {Localised particles and fuzzy horizons: A tool for probing
  Quantum Black Holes}},  \href{http://arxiv.org/abs/1305.3195}{{\tt
  arXiv:1305.3195}}.

\bibitem{Casadio:2013aua}
R.~Casadio and F.~Scardigli, {\it {Horizon wave-function for single localized
  particles: GUP and quantum black hole decay}},  {\em Eur. Phys. J.} {\bf C74}
  (2014), no.~1 2685, [\href{http://arxiv.org/abs/1306.5298}{{\tt
  arXiv:1306.5298}}].

\bibitem{Casadio:2015qaq}
R.~Casadio, A.~Giugno, and O.~Micu, {\it {Horizon quantum mechanics: A
  hitchhiker’s guide to quantum black holes}},  {\em Int. J. Mod. Phys.} {\bf
  D25} (2016), no.~02 1630006, [\href{http://arxiv.org/abs/1512.04071}{{\tt
  arXiv:1512.04071}}].

\bibitem{Casadio:2016fev}
R.~Casadio, A.~Giugno, and A.~Giusti, {\it {Global and Local Horizon Quantum
  Mechanics}},  {\em Gen. Rel. Grav.} {\bf 49} (2017), no.~2 32,
  [\href{http://arxiv.org/abs/1605.06617}{{\tt arXiv:1605.06617}}].

\bibitem{Casadio:2013uga}
R.~Casadio, O.~Micu, and F.~Scardigli, {\it {Quantum hoop conjecture: Black
  hole formation by particle collisions}},  {\em Phys. Lett.} {\bf B732} (2014)
  105--109, [\href{http://arxiv.org/abs/1311.5698}{{\tt arXiv:1311.5698}}].

\bibitem{Casadio:2015sda}
R.~Casadio, O.~Micu, and D.~Stojkovic, {\it {Horizon Wave-Function and the
  Quantum Cosmic Censorship}},  {\em Phys. Lett.} {\bf B747} (2015) 68--72,
  [\href{http://arxiv.org/abs/1503.02858}{{\tt arXiv:1503.02858}}].

\bibitem{Casadio:2015rwa}
R.~Casadio, O.~Micu, and D.~Stojkovic, {\it {Inner horizon of the quantum
  Reissner-Nordstr{\"o}m black holes}},  {\em JHEP} {\bf 05} (2015) 096,
  [\href{http://arxiv.org/abs/1503.01888}{{\tt arXiv:1503.01888}}].

\bibitem{Casadio:2017nfg}
R.~Casadio, A.~Giugno, A.~Giusti, and O.~Micu, {\it {Horizon Quantum Mechanics
  of Rotating Black Holes}},  {\em Eur. Phys. J.} {\bf C77} (2017), no.~5 322,
  [\href{http://arxiv.org/abs/1701.05778}{{\tt arXiv:1701.05778}}].

\bibitem{Giugno:2017xtl}
A.~Giugno, A.~Giusti, and A.~Helou, {\it {Horizon quantum fuzziness for
  non-singular black holes}},  {\em Eur. Phys. J.} {\bf C78} (2018), no.~3 208,
  [\href{http://arxiv.org/abs/1711.06209}{{\tt arXiv:1711.06209}}].

\bibitem{Casadio:2018vae}
R.~Casadio, A.~Giugno, A.~Giusti, and M.~Lenzi, {\it {Quantum Formation of
  Primordial Black holes}},  {\em Gen. Rel. Grav.} {\bf 51} (2019), no.~8 103,
  [\href{http://arxiv.org/abs/1810.05185}{{\tt arXiv:1810.05185}}].

\bibitem{Casadio:2018rlc}
R.~Casadio and O.~Micu, {\it {Horizon Quantum Mechanics of collapsing shells}},
   {\em Eur. Phys. J.} {\bf C78} (2018), no.~10 852,
  [\href{http://arxiv.org/abs/1806.05944}{{\tt arXiv:1806.05944}}].

\bibitem{Dvali:2011aa}
G.~Dvali and C.~Gomez, {\it {Black Hole's Quantum N-Portrait}},  {\em Fortsch.
  Phys.} {\bf 61} (2013) 742--767, [\href{http://arxiv.org/abs/1112.3359}{{\tt
  arXiv:1112.3359}}].

\bibitem{Dvali:2012rt}
G.~Dvali and C.~Gomez, {\it {Black Hole's 1/N Hair}},  {\em Phys. Lett.} {\bf
  B719} (2013) 419--423, [\href{http://arxiv.org/abs/1203.6575}{{\tt
  arXiv:1203.6575}}].

\bibitem{Dvali:2012en}
G.~Dvali and C.~Gomez, {\it {Black Holes as Critical Point of Quantum Phase
  Transition}},  {\em Eur. Phys. J.} {\bf C74} (2014) 2752,
  [\href{http://arxiv.org/abs/1207.4059}{{\tt arXiv:1207.4059}}].

\bibitem{Casadio:2015bna}
R.~Casadio, A.~Giugno, and A.~Orlandi, {\it {Thermal corpuscular black holes}},
   {\em Phys. Rev.} {\bf D91} (2015), no.~12 124069,
  [\href{http://arxiv.org/abs/1504.05356}{{\tt arXiv:1504.05356}}].

\bibitem{Casadio:2015lis}
R.~Casadio, A.~Giugno, O.~Micu, and A.~Orlandi, {\it {Thermal BEC black
  holes}},  {\em Entropy} {\bf 17} (2015) 6893--6924,
  [\href{http://arxiv.org/abs/1511.01279}{{\tt arXiv:1511.01279}}].

\bibitem{Casadio:2018ukt}
R.~Casadio, A.~Giusti, and J.~Mureika, {\it {Lower dimensional corpuscular
  gravity and the end of black hole evaporation}},  {\em Mod. Phys. Lett.} {\bf
  A34} (2019), no.~22 1950174, [\href{http://arxiv.org/abs/1805.10444}{{\tt
  arXiv:1805.10444}}].

\bibitem{Giusti:2019wdx}
A.~Giusti, {\it {On the corpuscular theory of gravity}},  {\em Int. J. Geom.
  Meth. Mod. Phys.} {\bf 16} (2019), no.~03 1930001.

\bibitem{Casadio:2014vja}
R.~Casadio, A.~Giugno, O.~Micu, and A.~Orlandi, {\it {Black holes as
  self-sustained quantum states, and Hawking radiation}},  {\em Phys. Rev.}
  {\bf D90} (2014), no.~8 084040, [\href{http://arxiv.org/abs/1405.4192}{{\tt
  arXiv:1405.4192}}].

\bibitem{Dvali:2010jz}
G.~Dvali, G.~F. Giudice, C.~Gomez, and A.~Kehagias, {\it {UV-Completion by
  Classicalization}},  {\em JHEP} {\bf 08} (2011) 108,
  [\href{http://arxiv.org/abs/1010.1415}{{\tt arXiv:1010.1415}}].

\bibitem{Casadio:2019tfz}
R.~Casadio and A.~Giusti, {\it {The role of collapsed matter in the decay of
  black holes}},  {\em Phys. Lett.} {\bf B797} (2019) 134915,
  [\href{http://arxiv.org/abs/1904.12663}{{\tt arXiv:1904.12663}}].

\bibitem{Hawking:1974sw}
S.~W. Hawking, {\it {Particle Creation by Black Holes}},  {\em Commun. Math.
  Phys.} {\bf 43} (1975) 199--220. [,167(1975)].

\bibitem{Muck:2016stv}
W.~M{\"u}ck, {\it {Hawking radiation is corpuscular}},  {\em Eur. Phys. J.}
  {\bf C76} (2016), no.~7 374, [\href{http://arxiv.org/abs/1606.01790}{{\tt
  arXiv:1606.01790}}].

\bibitem{Casadio:2016zpl}
R.~Casadio, A.~Giugno, and A.~Giusti, {\it {Matter and gravitons in the
  gravitational collapse}},  {\em Phys. Lett.} {\bf B763} (2016) 337--340,
  [\href{http://arxiv.org/abs/1606.04744}{{\tt arXiv:1606.04744}}].

\bibitem{Mueck:2013mha}
W.~M{\"u}ck, {\it {On the number of soft quanta in classical field
  configurations}},  {\em Can. J. Phys.} {\bf 92} (2014), no.~9 973--975,
  [\href{http://arxiv.org/abs/1306.6245}{{\tt arXiv:1306.6245}}].

\bibitem{Casadio:2017cdv}
R.~Casadio, A.~Giugno, A.~Giusti, and M.~Lenzi, {\it {Quantum corpuscular
  corrections to the Newtonian potential}},  {\em Phys. Rev.} {\bf D96} (2017),
  no.~4 044010, [\href{http://arxiv.org/abs/1702.05918}{{\tt
  arXiv:1702.05918}}].

\bibitem{Casadio:2018qeh}
R.~Casadio, M.~Lenzi, and O.~Micu, {\it {Bootstrapping Newtonian gravity}},
  {\em Phys. Rev.} {\bf D98} (2018), no.~10 104016,
  [\href{http://arxiv.org/abs/1806.07639}{{\tt arXiv:1806.07639}}].

\bibitem{Casadio:2019cux}
R.~Casadio, M.~Lenzi, and O.~Micu, {\it {Bootstrapped Newtonian stars and black
  holes}},  {\em Eur. Phys. J.} {\bf C79} (2019), no.~11 894,
  [\href{http://arxiv.org/abs/1904.06752}{{\tt arXiv:1904.06752}}].

\bibitem{Dvali:2013eja}
G.~Dvali and C.~Gomez, {\it {Quantum Compositeness of Gravity: Black Holes, AdS
  and Inflation}},  {\em JCAP} {\bf 1401} (2014) 023,
  [\href{http://arxiv.org/abs/1312.4795}{{\tt arXiv:1312.4795}}].

\bibitem{Casadio:2017twg}
R.~Casadio, A.~Giugno, and A.~Giusti, {\it {Corpuscular slow-roll inflation}},
  {\em Phys. Rev.} {\bf D97} (2018), no.~2 024041,
  [\href{http://arxiv.org/abs/1708.09736}{{\tt arXiv:1708.09736}}].

\bibitem{Starobinsky:1979ty}
A.~A. Starobinsky, {\it {Spectrum of relict gravitational radiation and the
  early state of the universe}},  {\em JETP Lett.} {\bf 30} (1979) 682--685.
  [Pisma Zh. Eksp. Teor. Fiz.30,719(1979); ,767(1979)].

\bibitem{Giugno:2018zty}
A.~Giugno and A.~Giusti, {\it {Domestic Corpuscular Inflaton}},  {\em Int. J.
  Geom. Meth. Mod. Phys.} {\bf 16} (2019), no.~07 1950108,
  [\href{http://arxiv.org/abs/1806.11168}{{\tt arXiv:1806.11168}}].

\bibitem{Dvali:2018fqu}
G.~Dvali and C.~Gomez, {\it {On Exclusion of Positive Cosmological Constant}},
  {\em Fortsch. Phys.} {\bf 67} (2019), no.~1-2 1800092,
  [\href{http://arxiv.org/abs/1806.10877}{{\tt arXiv:1806.10877}}].

\bibitem{Casadio:2019rrc}
R.~Casadio, A.~Giugno, A.~Giusti, and V.~Faraoni, {\it {Is de Sitter space
  always excluded in semiclassical f(R) gravity?}},  {\em JCAP} {\bf 1906}
  (2019), no.~06 005, [\href{http://arxiv.org/abs/1903.07685}{{\tt
  arXiv:1903.07685}}].

\bibitem{Cadoni:2017evg}
M.~Cadoni, R.~Casadio, A.~Giusti, W.~M{\"u}ck, and M.~Tuveri, {\it {Effective
  Fluid Description of the Dark Universe}},  {\em Phys. Lett.} {\bf B776}
  (2018) 242--248, [\href{http://arxiv.org/abs/1707.09945}{{\tt
  arXiv:1707.09945}}].

\bibitem{Cadoni:2018dnd}
M.~Cadoni, R.~Casadio, A.~Giusti, and M.~Tuveri, {\it {Emergence of a Dark
  Force in Corpuscular Gravity}},  {\em Phys. Rev.} {\bf D97} (2018), no.~4
  044047, [\href{http://arxiv.org/abs/1801.10374}{{\tt arXiv:1801.10374}}].

\bibitem{Tuveri:2019zor}
M.~Tuveri and M.~Cadoni, {\it {Galactic dynamics and long-range quantum
  gravity}},  {\em Phys. Rev.} {\bf D100} (2019), no.~2 024029,
  [\href{http://arxiv.org/abs/1904.11835}{{\tt arXiv:1904.11835}}].

\bibitem{Caldwell:2009ix}
R.~R. Caldwell and M.~Kamionkowski, {\it {The Physics of Cosmic Acceleration}},
   {\em Ann. Rev. Nucl. Part. Sci.} {\bf 59} (2009) 397--429,
  [\href{http://arxiv.org/abs/0903.0866}{{\tt arXiv:0903.0866}}].

\bibitem{Weinberg:2012es}
D.~H. Weinberg, M.~J. Mortonson, D.~J. Eisenstein, C.~Hirata, A.~G. Riess, and
  E.~Rozo, {\it {Observational Probes of Cosmic Acceleration}},  {\em Phys.
  Rept.} {\bf 530} (2013) 87--255, [\href{http://arxiv.org/abs/1201.2434}{{\tt
  arXiv:1201.2434}}].

\bibitem{2010deto.book.....A}
L.~{Amendola} and S.~{Tsujikawa}, {\em {Dark Energy: Theory and Observations}}.
\newblock {Cambridge University Press}, 2010.

\bibitem{Koyama:2015vza}
K.~Koyama, {\it {Cosmological Tests of Modified Gravity}},  {\em Rept. Prog.
  Phys.} {\bf 79} (2016), no.~4 046902,
  [\href{http://arxiv.org/abs/1504.04623}{{\tt arXiv:1504.04623}}].

\bibitem{Ishak:2018his}
M.~Ishak, {\it {Testing General Relativity in Cosmology}},  {\em Living Rev.
  Rel.} {\bf 22} (2019), no.~1 1, [\href{http://arxiv.org/abs/1806.10122}{{\tt
  arXiv:1806.10122}}].

\bibitem{Weinberg:1988cp}
S.~Weinberg, {\it {The Cosmological Constant Problem}},  {\em Rev.Mod.Phys.}
  {\bf 61} (1989) 1--23.

\bibitem{Goldstein:2017mmi}
A.~Goldstein et~al., {\it {An Ordinary Short Gamma-Ray Burst with Extraordinary
  Implications: Fermi-GBM Detection of GRB 170817A}},  {\em Astrophys. J.} {\bf
  848} (2017), no.~2 L14, [\href{http://arxiv.org/abs/1710.05446}{{\tt
  arXiv:1710.05446}}].

\bibitem{Nojiri:2017hai}
S.~Nojiri and S.~D. Odintsov, {\it {Cosmological Bound from the Neutron Star
  Merger GW170817 in Modified Gravity}},
  \href{http://arxiv.org/abs/1711.00492}{{\tt arXiv:1711.00492}}.

\bibitem{Boran:2017rdn}
S.~Boran, S.~Desai, E.~O. Kahya, and R.~P. Woodard, {\it {GW170817 Falsifies
  Dark Matter Emulators}},  {\em Phys. Rev.} {\bf D97} (2018), no.~4 041501,
  [\href{http://arxiv.org/abs/1710.06168}{{\tt arXiv:1710.06168}}].

\bibitem{Amendola:2017orw}
L.~Amendola, M.~Kunz, I.~D. Saltas, and I.~Sawicki, {\it {Fate of Large-Scale
  Structure in Modified Gravity After GW170817 and GRB170817A}},  {\em Phys.
  Rev. Lett.} {\bf 120} (2018), no.~13 131101,
  [\href{http://arxiv.org/abs/1711.04825}{{\tt arXiv:1711.04825}}].

\bibitem{Crisostomi:2017lbg}
M.~Crisostomi and K.~Koyama, {\it {Vainshtein mechanism after GW170817}},  {\em
  Phys. Rev.} {\bf D97} (2018), no.~2 021301,
  [\href{http://arxiv.org/abs/1711.06661}{{\tt arXiv:1711.06661}}].

\bibitem{Langlois:2017dyl}
D.~Langlois, R.~Saito, D.~Yamauchi, and K.~Noui, {\it {Scalar-tensor theories
  and modified gravity in the wake of GW170817}},  {\em Phys. Rev.} {\bf D97}
  (2018), no.~6 061501, [\href{http://arxiv.org/abs/1711.07403}{{\tt
  arXiv:1711.07403}}].

\bibitem{Gumrukcuoglu:2017ijh}
A.~E. Gumrukcuoglu, M.~Saravani, and T.~P. Sotiriou, {\it {Ho\v{r}ava Gravity
  after GW170817}},  {\em Phys. Rev.} {\bf D97} (2017), no.~2 024032,
  [\href{http://arxiv.org/abs/1711.08845}{{\tt arXiv:1711.08845}}].

\bibitem{Heisenberg:2017qka}
L.~Heisenberg and S.~Tsujikawa, {\it {Dark energy survivals in massive gravity
  after GW170817: SO(3) invariant}},  {\em JCAP} {\bf 1801} (2017), no.~01 044,
  [\href{http://arxiv.org/abs/1711.09430}{{\tt arXiv:1711.09430}}].

\bibitem{Dima:2017pwp}
A.~Dima and F.~Vernizzi, {\it {Vainshtein Screening in Scalar-Tensor Theories
  before and after GW170817: Constraints on Theories beyond Horndeski}},  {\em
  Phys. Rev.} {\bf D97} (2018), no.~10 101302,
  [\href{http://arxiv.org/abs/1712.04731}{{\tt arXiv:1712.04731}}].

\bibitem{Peirone:2017ywi}
S.~Peirone, K.~Koyama, L.~Pogosian, M.~Raveri, and A.~Silvestri, {\it
  {Large-scale structure phenomenology of viable Horndeski theories}},  {\em
  Phys. Rev.} {\bf D97} (2018), no.~4 043519,
  [\href{http://arxiv.org/abs/1712.00444}{{\tt arXiv:1712.00444}}].

\bibitem{Linder:2018jil}
E.~V. Linder, {\it {No Slip Gravity}},  {\em JCAP} {\bf 1803} (2018), no.~03
  005, [\href{http://arxiv.org/abs/1801.01503}{{\tt arXiv:1801.01503}}].

\bibitem{Kase:2018iwp}
R.~Kase and S.~Tsujikawa, {\it {Dark energy scenario consistent with GW170817
  in theories beyond Horndeski gravity}},  {\em Phys. Rev.} {\bf D97} (2018),
  no.~10 103501, [\href{http://arxiv.org/abs/1802.02728}{{\tt
  arXiv:1802.02728}}].

\bibitem{Battye:2018ssx}
R.~A. Battye, F.~Pace, and D.~Trinh, {\it {Gravitational wave constraints on
  dark sector models}},  {\em Phys. Rev.} {\bf D98} (2018), no.~2 023504,
  [\href{http://arxiv.org/abs/1802.09447}{{\tt arXiv:1802.09447}}].

\bibitem{Lombriser:2015sxa}
L.~Lombriser and A.~Taylor, {\it {Breaking a Dark Degeneracy with Gravitational
  Waves}},  {\em JCAP} {\bf 1603} (2016), no.~03 031,
  [\href{http://arxiv.org/abs/1509.08458}{{\tt arXiv:1509.08458}}].

\bibitem{Brax:2015dma}
P.~Brax, C.~Burrage, and A.-C. Davis, {\it {The Speed of Galileon Gravity}},
  {\em JCAP} {\bf 1603} (2016), no.~03 004,
  [\href{http://arxiv.org/abs/1510.03701}{{\tt arXiv:1510.03701}}].

\bibitem{Lombriser:2016yzn}
L.~Lombriser and N.~A. Lima, {\it {Challenges to Self-Acceleration in Modified
  Gravity from Gravitational Waves and Large-Scale Structure}},  {\em Phys.
  Lett.} {\bf B765} (2017) 382--385,
  [\href{http://arxiv.org/abs/1602.07670}{{\tt arXiv:1602.07670}}].

\bibitem{Pogosian:2016pwr}
L.~Pogosian and A.~Silvestri, {\it {What can cosmology tell us about gravity?
  Constraining Horndeski gravity with $\Sigma$ and $\mu$}},  {\em Phys. Rev.}
  {\bf D94} (2016), no.~10 104014, [\href{http://arxiv.org/abs/1606.05339}{{\tt
  arXiv:1606.05339}}].

\bibitem{Bettoni:2016mij}
D.~Bettoni, J.~M. Ezquiaga, K.~Hinterbichler, and M.~Zumalac·rregui, {\it
  {Speed of Gravitational Waves and the Fate of Scalar-Tensor Gravity}},  {\em
  Phys. Rev.} {\bf D95} (2017), no.~8 084029,
  [\href{http://arxiv.org/abs/1608.01982}{{\tt arXiv:1608.01982}}].

\bibitem{Creminelli:2008wc}
P.~Creminelli, G.~D'Amico, J.~Norena, and F.~Vernizzi, {\it {The Effective
  Theory of Quintessence: the w<-1 Side Unveiled}},  {\em JCAP} {\bf 0902}
  (2009) 018, [\href{http://arxiv.org/abs/0811.0827}{{\tt arXiv:0811.0827}}].

\bibitem{Bloomfield:2012ff}
J.~K. Bloomfield, {\'E}.~{\'E}. Flanagan, M.~Park, and S.~Watson, {\it {Dark
  energy or modified gravity? An effective field theory approach}},  {\em JCAP}
  {\bf 1308} (2013) 010, [\href{http://arxiv.org/abs/1211.7054}{{\tt
  arXiv:1211.7054}}].

\bibitem{Gleyzes:2013ooa}
J.~Gleyzes, D.~Langlois, F.~Piazza, and F.~Vernizzi, {\it {Essential Building
  Blocks of Dark Energy}},  {\em JCAP} {\bf 1308} (2013) 025,
  [\href{http://arxiv.org/abs/1304.4840}{{\tt arXiv:1304.4840}}].

\bibitem{Hu:2007pj}
W.~Hu and I.~Sawicki, {\it {A Parameterized Post-Friedmann Framework for
  Modified Gravity}},  {\em Phys. Rev.} {\bf D76} (2007) 104043,
  [\href{http://arxiv.org/abs/0708.1190}{{\tt arXiv:0708.1190}}].

\bibitem{Bertschinger:2008zb}
E.~Bertschinger and P.~Zukin, {\it {Distinguishing Modified Gravity from Dark
  Energy}},  {\em Phys. Rev.} {\bf D78} (2008) 024015,
  [\href{http://arxiv.org/abs/0801.2431}{{\tt arXiv:0801.2431}}].

\bibitem{Amin:2007wi}
M.~A. Amin, R.~V. Wagoner, and R.~D. Blandford, {\it {A sub-horizon framework
  for probing the relationship between the cosmological matter distribution and
  metric perturbations}},  {\em Mon. Not. Roy. Astron. Soc.} {\bf 390} (2008)
  131--142, [\href{http://arxiv.org/abs/0708.1793}{{\tt arXiv:0708.1793}}].

\bibitem{Baker:2014zva}
T.~Baker, P.~G. Ferreira, C.~D. Leonard, and M.~Motta, {\it {New Gravitational
  Scales in Cosmological Surveys}},  {\em Phys.Rev.} {\bf D90} (2014), no.~12
  124030, [\href{http://arxiv.org/abs/1409.8284}{{\tt arXiv:1409.8284}}].

\bibitem{DeFelice:2011hq}
A.~De~Felice, T.~Kobayashi, and S.~Tsujikawa, {\it {Effective gravitational
  couplings for cosmological perturbations in the most general scalar-tensor
  theories with second-order field equations}},  {\em Phys.Lett.} {\bf B706}
  (2011) 123--133, [\href{http://arxiv.org/abs/1108.4242}{{\tt
  arXiv:1108.4242}}].

\bibitem{Casas:2017eob}
S.~Casas, M.~Kunz, M.~Martinelli, and V.~Pettorino, {\it {Linear and non-linear
  Modified Gravity forecasts with future surveys}},  {\em Phys. Dark Univ.}
  {\bf 18} (2017) 73--104, [\href{http://arxiv.org/abs/1703.01271}{{\tt
  arXiv:1703.01271}}].

\bibitem{Baker:2015bva}
T.~Baker and P.~Bull, {\it {Observational signatures of modified gravity on
  ultra-large scales}},  {\em Astrophys. J.} {\bf 811} (2015) 116,
  [\href{http://arxiv.org/abs/1506.00641}{{\tt arXiv:1506.00641}}].

\bibitem{dalal:inprep}
M.~Martinelli, R.~Dalal, F.~Majidi, Y.~Akrami, S.~Camera, and E.~Sellentin,
  {\it {Ultra-large-scale approximations and galaxy clustering: debiasing
  constraints on cosmological parameters}},  {\em To appear} (2021).

\bibitem{Okamoto:2003zw}
T.~Okamoto and W.~Hu, {\it {CMB lensing reconstruction on the full sky}},  {\em
  Phys. Rev.} {\bf D67} (2003) 083002,
  [\href{http://arxiv.org/abs/astro-ph/0301031}{{\tt astro-ph/0301031}}].

\bibitem{Blanchard:2019oqi}
{\bf Euclid} Collaboration, A.~Blanchard et~al., {\it {Euclid preparation: VII.
  Forecast validation for Euclid cosmological probes}},
  \href{http://arxiv.org/abs/1910.09273}{{\tt arXiv:1910.09273}}.

\bibitem{Desjacques:2016bnm}
V.~Desjacques, D.~Jeong, and F.~Schmidt, {\it {Large-Scale Galaxy Bias}},  {\em
  Phys. Rept.} {\bf 733} (2018) 1--193,
  [\href{http://arxiv.org/abs/1611.09787}{{\tt arXiv:1611.09787}}].

\bibitem{Kaiser:1987qv}
N.~Kaiser, {\it {Clustering in real space and in redshift space}},  {\em Mon.
  Not. Roy. Astron. Soc.} {\bf 227} (1987) 1--27.

\bibitem{Ballinger:1996cd}
W.~E. Ballinger, J.~A. Peacock, and A.~F. Heavens, {\it {Measuring the
  cosmological constant with redshift surveys}},  {\em Mon. Not. Roy. Astron.
  Soc.} {\bf 282} (1996) 877--888,
  [\href{http://arxiv.org/abs/astro-ph/9605017}{{\tt astro-ph/9605017}}].

\bibitem{Kaiser:1991qi}
N.~Kaiser, {\it {Weak gravitational lensing of distant galaxies}},  {\em
  Astrophys. J.} {\bf 388} (1992) 272.

\bibitem{LoVerde:2008re}
M.~LoVerde and N.~Afshordi, {\it {Extended Limber Approximation}},  {\em Phys.
  Rev.} {\bf D78} (2008) 123506, [\href{http://arxiv.org/abs/0809.5112}{{\tt
  arXiv:0809.5112}}].

\bibitem{Giannantonio:2011ya}
T.~Giannantonio, C.~Porciani, J.~Carron, A.~Amara, and A.~Pillepich, {\it
  {Constraining primordial non-Gaussianity with future galaxy surveys}},  {\em
  Mon. Not. Roy. Astron. Soc.} {\bf 422} (2012) 2854--2877,
  [\href{http://arxiv.org/abs/1109.0958}{{\tt arXiv:1109.0958}}].

\bibitem{Kitching:2016zkn}
T.~D. Kitching, J.~Alsing, A.~F. Heavens, R.~Jimenez, J.~D. McEwen, and
  L.~Verde, {\it {The Limits of Cosmic Shear}},  {\em Mon. Not. Roy. Astron.
  Soc.} {\bf 469} (2017), no.~3 2737--2749,
  [\href{http://arxiv.org/abs/1611.04954}{{\tt arXiv:1611.04954}}].

\bibitem{Kilbinger:2017lvu}
M.~Kilbinger et~al., {\it {Precision calculations of the cosmic shear power
  spectrum projection}},  {\em Mon. Not. Roy. Astron. Soc.} {\bf 472} (2017),
  no.~2 2126--2141, [\href{http://arxiv.org/abs/1702.05301}{{\tt
  arXiv:1702.05301}}].

\bibitem{Lemos:2017arq}
P.~Lemos, A.~Challinor, and G.~Efstathiou, {\it {The effect of Limber and
  flat-sky approximations on galaxy weak lensing}},  {\em JCAP} {\bf 05} (2017)
  014, [\href{http://arxiv.org/abs/1704.01054}{{\tt arXiv:1704.01054}}].

\bibitem{Taylor:2018qda}
P.~L. Taylor, T.~D. Kitching, J.~D. McEwen, and T.~Tram, {\it {Testing the
  Cosmic Shear Spatially-Flat Universe Approximation with Generalized Lensing
  and Shear Spectra}},  {\em Phys. Rev. D} {\bf 98} (2018), no.~2 023522,
  [\href{http://arxiv.org/abs/1804.03668}{{\tt arXiv:1804.03668}}].

\bibitem{SpurioMancini:2019rxy}
A.~Spurio~Mancini, F.~Köhlinger, B.~Joachimi, V.~Pettorino, B.~M. Schäfer,
  R.~Reischke, S.~Brieden, M.~Archidiacono, and J.~Lesgourgues, {\it
  {KiDS+GAMA: Constraints on Horndeski gravity from combined large-scale
  structure probes}},  \href{http://arxiv.org/abs/1901.03686}{{\tt
  arXiv:1901.03686}}.

\bibitem{Lewis:1999bs}
A.~Lewis, A.~Challinor, and A.~Lasenby, {\it {Efficient computation of CMB
  anisotropies in closed FRW models}},  {\em Astrophys. J.} {\bf 538} (2000)
  473--476, [\href{http://arxiv.org/abs/astro-ph/9911177}{{\tt
  astro-ph/9911177}}].

\bibitem{Blas:2011rf}
D.~Blas, J.~Lesgourgues, and T.~Tram, {\it {The Cosmic Linear Anisotropy
  Solving System (CLASS) II: Approximation schemes}},  {\em JCAP} {\bf 1107}
  (2011) 034, [\href{http://arxiv.org/abs/1104.2933}{{\tt arXiv:1104.2933}}].

\bibitem{Zhao:2008bn}
G.-B. Zhao, L.~Pogosian, A.~Silvestri, and J.~Zylberberg, {\it {Searching for
  modified growth patterns with tomographic surveys}},  {\em Phys. Rev.} {\bf
  D79} (2009) 083513, [\href{http://arxiv.org/abs/0809.3791}{{\tt
  arXiv:0809.3791}}].

\bibitem{Hojjati:2011ix}
A.~Hojjati, L.~Pogosian, and G.-B. Zhao, {\it {Testing gravity with CAMB and
  CosmoMC}},  {\em JCAP} {\bf 1108} (2011) 005,
  [\href{http://arxiv.org/abs/1106.4543}{{\tt arXiv:1106.4543}}].

\bibitem{Zucca:2019xhg}
A.~Zucca, L.~Pogosian, A.~Silvestri, and G.-B. Zhao, {\it {MGCAMB with massive
  neutrinos and dynamical dark energy}},  {\em JCAP} {\bf 2019} (2020), no.~05
  001, [\href{http://arxiv.org/abs/1901.05956}{{\tt arXiv:1901.05956}}].

\bibitem{Ade:2015rim}
{\bf Planck} Collaboration, P.~A.~R. Ade et~al., {\it {Planck 2015 results.
  XIV. Dark energy and modified gravity}},  {\em Astron. Astrophys.} {\bf 594}
  (2016) A14, [\href{http://arxiv.org/abs/1502.01590}{{\tt arXiv:1502.01590}}].

\bibitem{Lewis:2002ah}
A.~Lewis and S.~Bridle, {\it {Cosmological parameters from CMB and other data:
  A Monte Carlo approach}},  {\em Phys. Rev.} {\bf D66} (2002) 103511,
  [\href{http://arxiv.org/abs/astro-ph/0205436}{{\tt astro-ph/0205436}}].

\bibitem{Zuntz:2014csq}
J.~Zuntz, M.~Paterno, E.~Jennings, D.~Rudd, A.~Manzotti, S.~Dodelson,
  S.~Bridle, S.~Sehrish, and J.~Kowalkowski, {\it {CosmoSIS: Modular
  Cosmological Parameter Estimation}},  {\em Astron. Comput.} {\bf 12} (2015)
  45--59, [\href{http://arxiv.org/abs/1409.3409}{{\tt arXiv:1409.3409}}].

\bibitem{Audren:2012wb}
B.~Audren, J.~Lesgourgues, K.~Benabed, and S.~Prunet, {\it {Conservative
  Constraints on Early Cosmology: an illustration of the Monte Python
  cosmological parameter inference code}},  {\em JCAP} {\bf 1302} (2013) 001,
  [\href{http://arxiv.org/abs/1210.7183}{{\tt arXiv:1210.7183}}].

\bibitem{Chisari:2018prw}
N.~E. Chisari, M.~L.~A. Richardson, J.~Devriendt, Y.~Dubois, A.~Schneider,
  M.~C. Brun, Amandine~Le, R.~S. Beckmann, S.~Peirani, A.~Slyz, and C.~Pichon,
  {\it {The impact of baryons on the matter power spectrum from the Horizon-AGN
  cosmological hydrodynamical simulation}},  {\em Mon. Not. Roy. Astron. Soc.}
  {\bf 480} (2018), no.~3 3962--3977,
  [\href{http://arxiv.org/abs/1801.08559}{{\tt arXiv:1801.08559}}].

\bibitem{Alsing:2019xrx}
J.~Alsing, T.~Charnock, S.~Feeney, and B.~Wandelt, {\it {Fast likelihood-free
  cosmology with neural density estimators and active learning}},  {\em Mon.
  Not. Roy. Astron. Soc.} {\bf 488} (2019), no.~3 4440--4458,
  [\href{http://arxiv.org/abs/1903.00007}{{\tt arXiv:1903.00007}}].

\bibitem{Ntampaka:2019udw}
M.~Ntampaka et~al., {\it {The Role of Machine Learning in the Next Decade of
  Cosmology}},  {\em BAAS} {\bf 51} (May, 2019) 14,
  [\href{http://arxiv.org/abs/1902.10159}{{\tt arXiv:1902.10159}}].

\bibitem{He:2018ggn}
S.~He, Y.~Li, Y.~Feng, S.~Ho, S.~Ravanbakhsh, W.~Chen, and B.~Póczos, {\it
  {Learning to Predict the Cosmological Structure Formation}},  {\em Proc. Nat.
  Acad. Sci.} {\bf 116} (2019), no.~28 13825--13832,
  [\href{http://arxiv.org/abs/1811.06533}{{\tt arXiv:1811.06533}}].

\bibitem{Chartier:2020pmu}
N.~Chartier, B.~Wandelt, Y.~Akrami, and F.~Villaescusa-Navarro, {\it {CARPool:
  fast, accurate computation of large-scale structure statistics by pairing
  costly and cheap cosmological simulations}},
  \href{http://arxiv.org/abs/2009.08970}{{\tt arXiv:2009.08970}}.

\bibitem{Jain:1993jh}
B.~Jain and E.~Bertschinger, {\it {Second order power spectrum and nonlinear
  evolution at high redshift}},  {\em Astrophys. J.} {\bf 431} (1994) 495,
  [\href{http://arxiv.org/abs/astro-ph/9311070}{{\tt astro-ph/9311070}}].

\bibitem{1986ApJ...311....6G}
M.~H. {Goroff}, B.~{Grinstein}, S.~J. {Rey}, and M.~B. {Wise}, {\it {Coupling
  of modes of cosmological mass density fluctuations}},  {\em \apj} {\bf 311}
  (Dec., 1986) 6--14.

\bibitem{Bouchet:1994xp}
F.~Bouchet, S.~Colombi, E.~Hivon, and R.~Juszkiewicz, {\it {Perturbative
  Lagrangian approach to gravitational instability}},  {\em Astron. Astrophys.}
  {\bf 296} (1995) 575, [\href{http://arxiv.org/abs/astro-ph/9406013}{{\tt
  astro-ph/9406013}}].

\bibitem{Matsubara:2007wj}
T.~Matsubara, {\it {Resumming Cosmological Perturbations via the Lagrangian
  Picture: One-loop Results in Real Space and in Redshift Space}},  {\em Phys.
  Rev. D} {\bf 77} (2008) 063530, [\href{http://arxiv.org/abs/0711.2521}{{\tt
  arXiv:0711.2521}}].

\bibitem{Crocce:2005xy}
M.~Crocce and R.~Scoccimarro, {\it {Renormalized cosmological perturbation
  theory}},  {\em Phys. Rev. D} {\bf 73} (2006) 063519,
  [\href{http://arxiv.org/abs/astro-ph/0509418}{{\tt astro-ph/0509418}}].

\bibitem{Carrasco:2012cv}
J.~J.~M. Carrasco, M.~P. Hertzberg, and L.~Senatore, {\it {The Effective Field
  Theory of Cosmological Large Scale Structures}},  {\em JHEP} {\bf 09} (Sept.,
  2012) 082, [\href{http://arxiv.org/abs/1206.2926}{{\tt arXiv:1206.2926}}].

\bibitem{Vlah:2015sea}
Z.~Vlah, M.~White, and A.~Aviles, {\it {A Lagrangian effective field theory}},
  {\em JCAP} {\bf 09} (2015) 014, [\href{http://arxiv.org/abs/1506.05264}{{\tt
  arXiv:1506.05264}}].

\bibitem{2016arXiv161009321P}
A.~{Perko}, L.~{Senatore}, E.~{Jennings}, and R.~H. {Wechsler}, {\it {Biased
  Tracers in Redshift Space in the EFT of Large-Scale Structure}},  {\em arXiv
  e-prints} (Oct., 2016) arXiv:1610.09321,
  [\href{http://arxiv.org/abs/1610.09321}{{\tt arXiv:1610.09321}}].

\bibitem{Cusin:2017mzw}
G.~Cusin, M.~Lewandowski, and F.~Vernizzi, {\it {Nonlinear Effective Theory of
  Dark Energy}},  {\em JCAP} {\bf 1804} (2018), no.~04 061,
  [\href{http://arxiv.org/abs/1712.02782}{{\tt arXiv:1712.02782}}].

\bibitem{Cusin:2017wjg}
G.~Cusin, M.~Lewandowski, and F.~Vernizzi, {\it {Dark Energy and Modified
  Gravity in the Effective Field Theory of Large-Scale Structure}},  {\em JCAP}
  {\bf 1804} (2018), no.~04 005, [\href{http://arxiv.org/abs/1712.02783}{{\tt
  arXiv:1712.02783}}].

\bibitem{Bartelmann:2019unp}
M.~Bartelmann, E.~Kozlikin, R.~Lilow, C.~Littek, F.~Fabis, I.~Kostyuk,
  C.~Viermann, L.~Heisenberg, S.~Konrad, and D.~Geiss, {\it {Cosmic Structure
  Formation with Kinetic Field Theory}},
  \href{http://arxiv.org/abs/1905.01179}{{\tt arXiv:1905.01179}}.

\bibitem{Bernardeau:2001qr}
F.~Bernardeau, S.~Colombi, E.~Gaztanaga, and R.~Scoccimarro, {\it {Large scale
  structure of the universe and cosmological perturbation theory}},  {\em Phys.
  Rept.} {\bf 367} (2002) 1--248,
  [\href{http://arxiv.org/abs/astro-ph/0112551}{{\tt astro-ph/0112551}}].

\bibitem{Beutler:2011hx}
F.~Beutler, C.~Blake, M.~Colless, D.~H. Jones, L.~Staveley-Smith, et~al., {\it
  {The 6dF Galaxy Survey: Baryon Acoustic Oscillations and the Local Hubble
  Constant}},  {\em Mon.Not.Roy.Astron.Soc.} {\bf 416} (2011) 3017--3032,
  [\href{http://arxiv.org/abs/1106.3366}{{\tt arXiv:1106.3366}}].

\bibitem{Ross:2014qpa}
A.~J. Ross, L.~Samushia, C.~Howlett, W.~J. Percival, A.~Burden, and M.~Manera,
  {\it {The clustering of the SDSS DR7 main Galaxy sample – I. A 4 per cent
  distance measure at $z = 0.15$}},  {\em Mon. Not. Roy. Astron. Soc.} {\bf
  449} (2015), no.~1 835--847, [\href{http://arxiv.org/abs/1409.3242}{{\tt
  arXiv:1409.3242}}].

\bibitem{Abbott:2017wau}
{\bf DES} Collaboration, T.~M.~C. Abbott et~al., {\it {Dark Energy Survey year
  1 results: Cosmological constraints from galaxy clustering and weak
  lensing}},  {\em Phys. Rev.} {\bf D98} (2018), no.~4 043526,
  [\href{http://arxiv.org/abs/1708.01530}{{\tt arXiv:1708.01530}}].

\bibitem{Joudaki:2016kym}
S.~Joudaki et~al., {\it {KiDS-450: Testing extensions to the standard
  cosmological model}},  {\em Mon. Not. Roy. Astron. Soc.} {\bf 471} (2017),
  no.~2 1259--1279, [\href{http://arxiv.org/abs/1610.04606}{{\tt
  arXiv:1610.04606}}].

\bibitem{Abbott:2018xao}
{\bf DES} Collaboration, T.~M.~C. Abbott et~al., {\it {Dark Energy Survey Year
  1 Results: Constraints on Extended Cosmological Models from Galaxy Clustering
  and Weak Lensing}},  {\em Phys. Rev.} {\bf D99} (2019), no.~12 123505,
  [\href{http://arxiv.org/abs/1810.02499}{{\tt arXiv:1810.02499}}].

\bibitem{Asgari:2020wuj}
{\bf KiDS} Collaboration, M.~Asgari et~al., {\it {KiDS-1000 Cosmology: Cosmic
  shear constraints and comparison between two point statistics}},  7, 2020.
\newblock \href{http://arxiv.org/abs/2007.15633}{{\tt arXiv:2007.15633}}.

\bibitem{Verde:2019ivm}
L.~Verde, T.~Treu, and A.~G. Riess, {\it {Tensions between the Early and the
  Late Universe}},  in {\em {Nature Astronomy 2019}}, 2019.
\newblock \href{http://arxiv.org/abs/1907.10625}{{\tt arXiv:1907.10625}}.

\bibitem{Espejo:2018hxa}
J.~Espejo, S.~Peirone, M.~Raveri, K.~Koyama, L.~Pogosian, and A.~Silvestri,
  {\it {Phenomenology of Large Scale Structure in scalar-tensor theories: joint
  prior covariance of $w_{\textrm{DE}}$, $\Sigma$ and $\mu$ in Horndeski}},
  {\em Phys. Rev.} {\bf D99} (2018), no.~2 023512,
  [\href{http://arxiv.org/abs/1809.01121}{{\tt arXiv:1809.01121}}].

\bibitem{Aghamousa:2016zmz}
{\bf DESI} Collaboration, A.~Aghamousa et~al., {\it {The DESI Experiment Part
  I: Science,Targeting, and Survey Design}},
  \href{http://arxiv.org/abs/1611.00036}{{\tt arXiv:1611.00036}}.

\bibitem{Aghamousa:2016sne}
{\bf DESI} Collaboration, A.~Aghamousa et~al., {\it {The DESI Experiment Part
  II: Instrument Design}},  \href{http://arxiv.org/abs/1611.00037}{{\tt
  arXiv:1611.00037}}.

\bibitem{Ivezic:2008fe}
{\bf LSST} Collaboration, Z.~Ivezic et~al., {\it {LSST: from Science Drivers to
  Reference Design and Anticipated Data Products}},  {\em Astrophys. J.} {\bf
  873} (2019), no.~2 111, [\href{http://arxiv.org/abs/0805.2366}{{\tt
  arXiv:0805.2366}}].

\bibitem{Abell:2009aa}
{\bf LSST Science, LSST Project} Collaboration, P.~A. Abell et~al., {\it {LSST
  Science Book, Version 2.0}},  \href{http://arxiv.org/abs/0912.0201}{{\tt
  arXiv:0912.0201}}.

\bibitem{Mandelbaum:2018ouv}
{\bf LSST Dark Energy Science} Collaboration, D.~Alonso et~al., {\it {The LSST
  Dark Energy Science Collaboration (DESC) Science Requirements Document}},
  \href{http://arxiv.org/abs/1809.01669}{{\tt arXiv:1809.01669}}.

\bibitem{Bull:2014rha}
P.~Bull, P.~G. Ferreira, P.~Patel, and M.~G. Santos, {\it {Late-time cosmology
  with 21cm intensity mapping experiments}},
  \href{http://arxiv.org/abs/1405.1452}{{\tt arXiv:1405.1452}}.

\bibitem{Jarvis:2015tqa}
M.~J. Jarvis, D.~Bacon, C.~Blake, M.~L. Brown, S.~N. Lindsay, A.~Raccanelli,
  M.~Santos, and D.~Schwarz, {\it {Cosmology with SKA Radio Continuum
  Surveys}},  \href{http://arxiv.org/abs/1501.03825}{{\tt arXiv:1501.03825}}.

\bibitem{Bacon:2015dqe}
D.~Bacon et~al., {\it {Synergy between the Large Synoptic Survey Telescope and
  the Square Kilometre Array}},  {\em PoS} {\bf AASKA14} (2015) 145,
  [\href{http://arxiv.org/abs/1501.03977}{{\tt arXiv:1501.03977}}].

\bibitem{Kitching:2015fra}
T.~D. Kitching, D.~Bacon, M.~L. Brown, P.~Bull, J.~D. McEwen, M.~Oguri,
  R.~Scaramella, K.~Takahashi, K.~Wu, and D.~Yamauchi, {\it {Euclid \& SKA
  Synergies}},  \href{http://arxiv.org/abs/1501.03978}{{\tt arXiv:1501.03978}}.

\bibitem{Yahya:2014yva}
S.~Yahya, P.~Bull, M.~G. Santos, M.~Silva, R.~Maartens, P.~Okouma, and
  B.~Bassett, {\it {Cosmological performance of SKA HI galaxy surveys}},  {\em
  Mon. Not. Roy. Astron. Soc.} {\bf 450} (2015), no.~3 2251--2260,
  [\href{http://arxiv.org/abs/1412.4700}{{\tt arXiv:1412.4700}}].

\bibitem{Santos:2015gra}
M.~G. Santos et~al., {\it {Cosmology with a SKA HI intensity mapping survey}},
  \href{http://arxiv.org/abs/1501.03989}{{\tt arXiv:1501.03989}}.

\bibitem{Bacon:2018dui}
{\bf SKA} Collaboration, D.~J. Bacon et~al., {\it {Cosmology with Phase 1 of
  the Square Kilometre Array: Red Book 2018: Technical specifications and
  performance forecasts}},  {\em Submitted to: Publ. Astron. Soc. Austral.}
  (2018) [\href{http://arxiv.org/abs/1811.02743}{{\tt arXiv:1811.02743}}].

\bibitem{Aihara:2017paw}
H.~Aihara et~al., {\it {The Hyper Suprime-Cam SSP Survey: Overview and Survey
  Design}},  {\em Publ. Astron. Soc. Jap.} {\bf 70} (2018) S4,
  [\href{http://arxiv.org/abs/1704.05858}{{\tt arXiv:1704.05858}}].

\bibitem{Tamura:2016wsg}
N.~Tamura et~al., {\it {Prime Focus Spectrograph (PFS) for the Subaru
  Telescope: Overview, recent progress, and future perspectives}},  {\em Proc.
  SPIE Int. Soc. Opt. Eng.} {\bf 9908} (2016) 99081M,
  [\href{http://arxiv.org/abs/1608.01075}{{\tt arXiv:1608.01075}}].

\bibitem{Laureijs:2011gra}
{\bf EUCLID Collaboration} Collaboration, R.~Laureijs et~al., {\it {Euclid
  Definition Study Report}},  \href{http://arxiv.org/abs/1110.3193}{{\tt
  arXiv:1110.3193}}.

\bibitem{Amendola:2016saw}
L.~Amendola et~al., {\it {Cosmology and fundamental physics with the Euclid
  satellite}},  {\em Living Rev. Rel.} {\bf 21} (2018), no.~1 2,
  [\href{http://arxiv.org/abs/1606.00180}{{\tt arXiv:1606.00180}}].

\bibitem{Spergel:2015sza}
D.~Spergel et~al., {\it {Wide-Field InfrarRed Survey Telescope-Astrophysics
  Focused Telescope Assets WFIRST-AFTA 2015 Report}},
  \href{http://arxiv.org/abs/1503.03757}{{\tt arXiv:1503.03757}}.

\bibitem{Hounsell:2017ejq}
R.~Hounsell et~al., {\it {Simulations of the WFIRST Supernova Survey and
  Forecasts of Cosmological Constraints}},  {\em Astrophys. J.} {\bf 867}
  (2017), no.~1 23, [\href{http://arxiv.org/abs/1702.01747}{{\tt
  arXiv:1702.01747}}].

\bibitem{Dore:2014cca}
O.~Dor\'e et~al., {\it {Cosmology with the SPHEREX All-Sky Spectral Survey}},
  \href{http://arxiv.org/abs/1412.4872}{{\tt arXiv:1412.4872}}.

\bibitem{Dore:2018kgp}
O.~Dor\'e et~al., {\it {Science Impacts of the SPHEREx All-Sky Optical to
  Near-Infrared Spectral Survey II: Report of a Community Workshop on the
  Scientific Synergies Between the SPHEREx Survey and Other Astronomy
  Observatories}},  \href{http://arxiv.org/abs/1805.05489}{{\tt
  arXiv:1805.05489}}.

\bibitem{Abazajian:2016yjj}
{\bf CMB-S4} Collaboration, K.~N. Abazajian et~al., {\it {CMB-S4 Science Book,
  First Edition}},  \href{http://arxiv.org/abs/1610.02743}{{\tt
  arXiv:1610.02743}}.

\bibitem{Abazajian:2020dmr}
{\bf CMB-S4} Collaboration, K.~Abazajian et~al., {\it {CMB-S4: Forecasting
  Constraints on Primordial Gravitational Waves}},
  \href{http://arxiv.org/abs/2008.12619}{{\tt arXiv:2008.12619}}.

\bibitem{Ade:2018sbj}
{\bf Simons Observatory} Collaboration, P.~Ade et~al., {\it {The Simons
  Observatory: Science goals and forecasts}},  {\em JCAP} {\bf 02} (2019) 056,
  [\href{http://arxiv.org/abs/1808.07445}{{\tt arXiv:1808.07445}}].

\bibitem{Sehgal:2019ewc}
N.~Sehgal et~al., {\it {CMB-HD: An Ultra-Deep, High-Resolution Millimeter-Wave
  Survey Over Half the Sky}},  \href{http://arxiv.org/abs/1906.10134}{{\tt
  arXiv:1906.10134}}.

\bibitem{Sehgal:2020yja}
N.~Sehgal et~al., {\it {CMB-HD: Astro2020 RFI Response}},
  \href{http://arxiv.org/abs/2002.12714}{{\tt arXiv:2002.12714}}.

\bibitem{BOSSWeb}
``{BOSS webpage: http://www.sdss3.org/surveys/boss.php}.''

\bibitem{WigglezWeb}
``{WiggleZ webpage: http://wigglez.swin.edu.au/site/}.''

\bibitem{Parkinson_2012}
D.~Parkinson, S.~Riemer-S¿rensen, C.~Blake, G.~B. Poole, T.~M. Davis,
  S.~Brough, M.~Colless, C.~Contreras, W.~Couch, S.~Croom, and et~al., {\it The
  wigglez dark energy survey: Final data release and cosmological results},
  {\em Physical Review D} {\bf 86} (Nov, 2012).

\bibitem{Yoo:2009au}
J.~Yoo, A.~L. Fitzpatrick, and M.~Zaldarriaga, {\it {A New Perspective on
  Galaxy Clustering as a Cosmological Probe: General Relativistic Effects}},
  {\em Phys. Rev.} {\bf D80} (2009) 083514,
  [\href{http://arxiv.org/abs/0907.0707}{{\tt arXiv:0907.0707}}].

\bibitem{Bonvin:2011bg}
C.~Bonvin and R.~Durrer, {\it {What galaxy surveys really measure}},  {\em
  Phys. Rev.} {\bf D84} (2011) 063505,
  [\href{http://arxiv.org/abs/1105.5280}{{\tt arXiv:1105.5280}}].

\bibitem{Challinor:2011bk}
A.~Challinor and A.~Lewis, {\it {The linear power spectrum of observed source
  number counts}},  {\em Phys. Rev.} {\bf D84} (2011) 043516,
  [\href{http://arxiv.org/abs/1105.5292}{{\tt arXiv:1105.5292}}].

\bibitem{Bonvin:2018ckp}
C.~Bonvin and P.~Fleury, {\it {Testing the equivalence principle on
  cosmological scales}},  {\em JCAP} {\bf 1805} (2018), no.~05 061,
  [\href{http://arxiv.org/abs/1803.02771}{{\tt arXiv:1803.02771}}].

\bibitem{Scranton:2005ci}
{\bf SDSS} Collaboration, R.~Scranton et~al., {\it {Detection of cosmic
  magnification with the Sloan Digital Sky Survey}},  {\em Astrophys. J.} {\bf
  633} (2005) 589--602, [\href{http://arxiv.org/abs/astro-ph/0504510}{{\tt
  astro-ph/0504510}}].

\bibitem{Duncan:2013haa}
C.~Duncan, B.~Joachimi, A.~Heavens, C.~Heymans, and H.~Hildebrandt, {\it {On
  the complementarity of galaxy clustering with cosmic shear and flux
  magnification}},  {\em Mon. Not. Roy. Astron. Soc.} {\bf 437} (2014), no.~3
  2471--2487, [\href{http://arxiv.org/abs/1306.6870}{{\tt arXiv:1306.6870}}].

\bibitem{Garcia-Fernandez:2016oud}
{\bf DES} Collaboration, M.~Garcia-Fernandez et~al., {\it {Weak lensing
  magnification in the Dark Energy Survey Science Verification Data}},  {\em
  Mon. Not. Roy. Astron. Soc.} {\bf 476} (2018), no.~1 1071--1085,
  [\href{http://arxiv.org/abs/1611.10326}{{\tt arXiv:1611.10326}}].

\bibitem{Asorey:2012rd}
J.~Asorey, M.~Crocce, E.~Gaztanaga, and A.~Lewis, {\it {Recovering 3D
  clustering information with angular correlations}},  {\em Mon. Not. Roy.
  Astron. Soc.} {\bf 427} (2012) 1891,
  [\href{http://arxiv.org/abs/1207.6487}{{\tt arXiv:1207.6487}}].

\bibitem{Xu_2013}
X.~Xu, A.~J. Cuesta, N.~Padmanabhan, D.~J. Eisenstein, and C.~K. McBride, {\it
  Measuring da and h at z=0.35 from the sdss dr7 lrgs using baryon acoustic
  oscillations},  {\em Monthly Notices of the Royal Astronomical Society} {\bf
  431} (Mar, 2013) 2834--2860.

\bibitem{Bonvin:2013ogt}
C.~Bonvin, L.~Hui, and E.~Gazta\~naga, {\it {Asymmetric galaxy correlation
  functions}},  {\em Phys. Rev.} {\bf D89} (2014), no.~8 083535,
  [\href{http://arxiv.org/abs/1309.1321}{{\tt arXiv:1309.1321}}].

\bibitem{Yamamoto:2005dz}
K.~Yamamoto, M.~Nakamichi, A.~Kamino, B.~A. Bassett, and H.~Nishioka, {\it {A
  Measurement of the quadrupole power spectrum in the clustering of the 2dF QSO
  Survey}},  {\em Publ. Astron. Soc. Jap.} {\bf 58} (2006) 93--102,
  [\href{http://arxiv.org/abs/astro-ph/0505115}{{\tt astro-ph/0505115}}].

\bibitem{Tansella:2017rpi}
V.~Tansella, C.~Bonvin, R.~Durrer, B.~Ghosh, and E.~Sellentin, {\it {The
  full-sky relativistic correlation function and power spectrum of galaxy
  number counts: I. Theoretical aspects}},
  \href{http://arxiv.org/abs/1708.00492}{{\tt arXiv:1708.00492}}.

\bibitem{Reimberg:2015jma}
P.~H.~F. Reimberg, F.~Bernardeau, and C.~Pitrou, {\it {Redshift-space
  distortions with wide angular separations}},  {\em JCAP} {\bf 1601} (2016),
  no.~01 048, [\href{http://arxiv.org/abs/1506.06596}{{\tt arXiv:1506.06596}}].

\bibitem{Gaztanaga:2015jrs}
E.~Gazta\~naga, C.~Bonvin, and L.~Hui, {\it {Measurement of the dipole in the
  cross-correlation function of galaxies}},  {\em JCAP} {\bf 1701} (2017),
  no.~01 032, [\href{http://arxiv.org/abs/1512.03918}{{\tt arXiv:1512.03918}}].

\bibitem{Kodwani:2019onz}
D.~Kodwani and H.~Desmond, {\it {Screened fifth forces in parity-breaking
  correlation functions}},  {\em Phys. Rev.} {\bf D100} (2019), no.~6 064030,
  [\href{http://arxiv.org/abs/1904.12310}{{\tt arXiv:1904.12310}}].

\bibitem{Kehagias:2013rpa}
A.~Kehagias, J.~Norena, H.~Perrier, and A.~Riotto, {\it {Consequences of
  Symmetries and Consistency Relations in the Large-Scale Structure of the
  Universe for Non-local bias and Modified Gravity}},  {\em Nucl. Phys.} {\bf
  B883} (2014) 83--106, [\href{http://arxiv.org/abs/1311.0786}{{\tt
  arXiv:1311.0786}}].

\bibitem{Creminelli:2013nua}
P.~Creminelli, J.~Gleyzes, L.~Hui, M.~Simonovi?, and F.~Vernizzi, {\it
  {Single-Field Consistency Relations of Large Scale Structure. Part III: Test
  of the Equivalence Principle}},  {\em JCAP} {\bf 1406} (2014) 009,
  [\href{http://arxiv.org/abs/1312.6074}{{\tt arXiv:1312.6074}}].

\bibitem{Hall:2016bmm}
A.~Hall and C.~Bonvin, {\it {Measuring cosmic velocities with 21 cm intensity
  mapping and galaxy redshift survey cross-correlation dipoles}},  {\em Phys.
  Rev.} {\bf D95} (2017), no.~4 043530,
  [\href{http://arxiv.org/abs/1609.09252}{{\tt arXiv:1609.09252}}].

\bibitem{SKAWeb}
``{SKA webpage: https://www.skatelescope.org}.''

\bibitem{Bull:2015lja}
P.~Bull, {\it {Extending cosmological tests of General Relativity with the
  Square Kilometre Array}},  {\em Astrophys. J.} {\bf 817} (2016), no.~1 26,
  [\href{http://arxiv.org/abs/1509.07562}{{\tt arXiv:1509.07562}}].

\bibitem{Gleyzes:2014rba}
J.~Gleyzes, D.~Langlois, and F.~Vernizzi, {\it {A unifying description of dark
  energy}},  {\em Int. J. Mod. Phys.} {\bf D23} (2015), no.~13 1443010,
  [\href{http://arxiv.org/abs/1411.3712}{{\tt arXiv:1411.3712}}].

\bibitem{Piazza:2013pua}
F.~Piazza, H.~Steigerwald, and C.~Marinoni, {\it {Phenomenology of dark energy:
  exploring the space of theories with future redshift surveys}},  {\em JCAP}
  {\bf 1405} (2014) 043, [\href{http://arxiv.org/abs/1312.6111}{{\tt
  arXiv:1312.6111}}].

\bibitem{Hu:2014oga}
B.~Hu, M.~Raveri, N.~Frusciante, and A.~Silvestri, {\it {EFTCAMB/EFTCosmoMC:
  Numerical Notes v2.0}},  \href{http://arxiv.org/abs/1405.3590}{{\tt
  arXiv:1405.3590}}.

\bibitem{Tsujikawa:2014mba}
S.~Tsujikawa, {\it {The effective field theory of inflation/dark energy and the
  Horndeski theory}},  {\em Lect. Notes Phys.} {\bf 892} (2015) 97--136,
  [\href{http://arxiv.org/abs/1404.2684}{{\tt arXiv:1404.2684}}].

\bibitem{Kase:2014cwa}
R.~Kase and S.~Tsujikawa, {\it {Effective field theory approach to modified
  gravity including Horndeski theory and Horava Lifshitz gravity}},  {\em Int.
  J. Mod. Phys.} {\bf D23} (2015), no.~13 1443008,
  [\href{http://arxiv.org/abs/1409.1984}{{\tt arXiv:1409.1984}}].

\bibitem{Frusciante:2015maa}
N.~Frusciante, M.~Raveri, D.~Vernieri, B.~Hu, and A.~Silvestri, {\it {Horava
  Gravity in the Effective Field Theory formalism: From cosmology to
  observational constraints}},  {\em Phys. Dark Univ.} {\bf 13} (2016) 7--24,
  [\href{http://arxiv.org/abs/1508.01787}{{\tt arXiv:1508.01787}}].

\bibitem{Frusciante:2016xoj}
N.~Frusciante, G.~Papadomanolakis, and A.~Silvestri, {\it {An Extended action
  for the effective field theory of dark energy: a stability analysis and a
  complete guide to the mapping at the basis of EFTCAMB}},  {\em JCAP} {\bf
  1607} (2016), no.~07 018, [\href{http://arxiv.org/abs/1601.04064}{{\tt
  arXiv:1601.04064}}].

\bibitem{Gleyzes:2015pma}
J.~Gleyzes, D.~Langlois, M.~Mancarella, and F.~Vernizzi, {\it {Effective Theory
  of Interacting Dark Energy}},  {\em JCAP} {\bf 1508} (2015), no.~08 054,
  [\href{http://arxiv.org/abs/1504.05481}{{\tt arXiv:1504.05481}}].

\bibitem{Tsujikawa:2015upa}
S.~Tsujikawa, {\it {Cosmological disformal transformations to the Einstein
  frame and gravitational couplings with matter perturbations}},  {\em Phys.
  Rev.} {\bf D92} (2015), no.~6 064047,
  [\href{http://arxiv.org/abs/1506.08561}{{\tt arXiv:1506.08561}}].

\bibitem{DAmico:2016ntq}
G.~D'Amico, Z.~Huang, M.~Mancarella, and F.~Vernizzi, {\it {Weakening Gravity
  on Redshift-Survey Scales with Kinetic Matter Mixing}},  {\em JCAP} {\bf
  1702} (2017) 014, [\href{http://arxiv.org/abs/1609.01272}{{\tt
  arXiv:1609.01272}}].

\bibitem{Lagos:2016wyv}
M.~Lagos, T.~Baker, P.~G. Ferreira, and J.~Noller, {\it {A general theory of
  linear cosmological perturbations: scalar-tensor and vector-tensor
  theories}},  {\em JCAP} {\bf 1608} (2016), no.~08 007,
  [\href{http://arxiv.org/abs/1604.01396}{{\tt arXiv:1604.01396}}].

\bibitem{Lagos:2017hdr}
M.~Lagos, E.~Bellini, J.~Noller, P.~G. Ferreira, and T.~Baker, {\it {A general
  theory of linear cosmological perturbations: stability conditions, the
  quasistatic limit and dynamics}},  {\em JCAP} {\bf 1803} (2018), no.~03 021,
  [\href{http://arxiv.org/abs/1711.09893}{{\tt arXiv:1711.09893}}].

\bibitem{Bellini:2015wfa}
E.~Bellini, R.~Jimenez, and L.~Verde, {\it {Signatures of Horndeski gravity on
  the Dark Matter Bispectrum}},  {\em JCAP} {\bf 1505} (2015), no.~05 057,
  [\href{http://arxiv.org/abs/1504.04341}{{\tt arXiv:1504.04341}}].

\bibitem{Frusciante:2017nfr}
N.~Frusciante and G.~Papadomanolakis, {\it {Tackling non-linearities with the
  effective field theory of dark energy and modified gravity}},
  \href{http://arxiv.org/abs/1706.02719}{{\tt arXiv:1706.02719}}.

\bibitem{Yamauchi:2017ibz}
D.~Yamauchi, S.~Yokoyama, and H.~Tashiro, {\it {Constraining modified theories
  of gravity with the galaxy bispectrum}},  {\em Phys. Rev.} {\bf D96} (2017),
  no.~12 123516, [\href{http://arxiv.org/abs/1709.03243}{{\tt
  arXiv:1709.03243}}].

\bibitem{Kennedy:2019nie}
J.~Kennedy, L.~Lombriser, and A.~Taylor, {\it {Screening and degenerate kinetic
  self-acceleration from the nonlinear freedom of reconstructed Horndeski
  theories}},  \href{http://arxiv.org/abs/1902.09853}{{\tt arXiv:1902.09853}}.

\bibitem{Bloomfield:2013efa}
J.~Bloomfield, {\it {A Simplified Approach to General Scalar-Tensor Theories}},
   {\em JCAP} {\bf 1312} (2013) 044,
  [\href{http://arxiv.org/abs/1304.6712}{{\tt arXiv:1304.6712}}].

\bibitem{Deffayet:2009mn}
C.~Deffayet, S.~Deser, and G.~Esposito-Farese, {\it {Generalized Galileons: All
  scalar models whose curved background extensions maintain second-order field
  equations and stress-tensors}},  {\em Phys. Rev.} {\bf D80} (2009) 064015,
  [\href{http://arxiv.org/abs/0906.1967}{{\tt arXiv:0906.1967}}].

\bibitem{DeFelice:2016ucp}
A.~De~Felice, N.~Frusciante, and G.~Papadomanolakis, {\it {On the stability
  conditions for theories of modified gravity in the presence of matter
  fields}},  {\em JCAP} {\bf 1703} (2017), no.~03 027,
  [\href{http://arxiv.org/abs/1609.03599}{{\tt arXiv:1609.03599}}].

\bibitem{DeFelice:2017mwa}
A.~De~Felice, N.~Frusciante, and G.~Papadomanolakis, {\it {A de Sitter limit
  analysis for dark energy and modified gravity models}},
  \href{http://arxiv.org/abs/1705.01960}{{\tt arXiv:1705.01960}}.

\bibitem{Perenon:2015sla}
L.~Perenon, F.~Piazza, C.~Marinoni, and L.~Hui, {\it {Phenomenology of dark
  energy: general features of large-scale perturbations}},  {\em JCAP} {\bf
  1511} (2015), no.~11 029, [\href{http://arxiv.org/abs/1506.03047}{{\tt
  arXiv:1506.03047}}].

\bibitem{Perenon:2016blf}
L.~Perenon, C.~Marinoni, and F.~Piazza, {\it {Diagnostic of Horndeski
  Theories}},  {\em JCAP} {\bf 1701} (2017), no.~01 035,
  [\href{http://arxiv.org/abs/1609.09197}{{\tt arXiv:1609.09197}}].

\bibitem{Lombriser:2015cla}
L.~Lombriser and A.~Taylor, {\it {Semi-dynamical perturbations of unified dark
  energy}},  {\em JCAP} {\bf 1511} (2015), no.~11 040,
  [\href{http://arxiv.org/abs/1505.05915}{{\tt arXiv:1505.05915}}].

\bibitem{Frusciante:2018vht}
N.~Frusciante, G.~Papadomanolakis, S.~Peirone, and A.~Silvestri, {\it {The role
  of the tachyonic instability in Horndeski gravity}},
  \href{http://arxiv.org/abs/1810.03461}{{\tt arXiv:1810.03461}}.

\bibitem{Kennedy:2017sof}
J.~Kennedy, L.~Lombriser, and A.~Taylor, {\it {Reconstructing Horndeski models
  from the effective field theory of dark energy}},  {\em Phys. Rev.} {\bf D96}
  (2017), no.~8 084051, [\href{http://arxiv.org/abs/1705.09290}{{\tt
  arXiv:1705.09290}}].

\bibitem{Kennedy:2018gtx}
J.~Kennedy, L.~Lombriser, and A.~Taylor, {\it {Reconstructing Horndeski
  theories from phenomenological modified gravity and dark energy models on
  cosmological scales}},  {\em Phys. Rev.} {\bf D98} (2018), no.~4 044051,
  [\href{http://arxiv.org/abs/1804.04582}{{\tt arXiv:1804.04582}}].

\bibitem{Lombriser:2018olq}
L.~Lombriser, C.~Dalang, J.~Kennedy, and A.~Taylor, {\it {Inherently stable
  effective field theory for dark energy and modified gravity}},
  \href{http://arxiv.org/abs/1810.05225}{{\tt arXiv:1810.05225}}.

\bibitem{Tsujikawa:2015mga}
S.~Tsujikawa, {\it {Possibility of realizing weak gravity in redshift space
  distortion measurements}},  {\em Phys. Rev.} {\bf D92} (2015), no.~4 044029,
  [\href{http://arxiv.org/abs/1505.02459}{{\tt arXiv:1505.02459}}].

\bibitem{Traykova:2019oyx}
D.~Traykova, E.~Bellini, and P.~G. Ferreira, {\it {The phenomenology of beyond
  Horndeski gravity}},  \href{http://arxiv.org/abs/1902.10687}{{\tt
  arXiv:1902.10687}}.

\bibitem{Linder:2019bqp}
E.~V. Linder, {\it {No Run Gravity}},
  \href{http://arxiv.org/abs/1903.02010}{{\tt arXiv:1903.02010}}.

\bibitem{Amendola:2014wma}
L.~Amendola, G.~Ballesteros, and V.~Pettorino, {\it {Effects of modified
  gravity on B-mode polarization}},  {\em Phys. Rev.} {\bf D90} (2014) 043009,
  [\href{http://arxiv.org/abs/1405.7004}{{\tt arXiv:1405.7004}}].

\bibitem{Salvatelli:2016mgy}
V.~Salvatelli, F.~Piazza, and C.~Marinoni, {\it {Constraints on modified
  gravity from Planck 2015: when the health of your theory makes the
  difference}},  {\em JCAP} {\bf 1609} (2016), no.~09 027,
  [\href{http://arxiv.org/abs/1602.08283}{{\tt arXiv:1602.08283}}].

\bibitem{Renk:2016olm}
J.~Renk, M.~Zumalacarregui, and F.~Montanari, {\it {Gravity at the horizon: on
  relativistic effects, CMB-LSS correlations and ultra-large scales in
  Horndeski's theory}},  {\em JCAP} {\bf 1607} (2016), no.~07 040,
  [\href{http://arxiv.org/abs/1604.03487}{{\tt arXiv:1604.03487}}].

\bibitem{Brush:2018dhg}
M.~Brush, E.~V. Linder, and M.~ZumalacÃ¡rregui, {\it {No Slip CMB}},
  \href{http://arxiv.org/abs/1810.12337}{{\tt arXiv:1810.12337}}.

\bibitem{Garcia-Garcia:2018hlc}
C.~García-García, E.~V. Linder, P.~Ruíz-Lapuente, and M.~Zumalacárregui,
  {\it {Dark energy from $\alpha$-attractors: phenomenology and observational
  constraints}},  {\em JCAP} {\bf 1808} (2018) 022,
  [\href{http://arxiv.org/abs/1803.00661}{{\tt arXiv:1803.00661}}].

\bibitem{Frusciante:2018jzw}
N.~Frusciante, S.~Peirone, S.~Casas, and N.~A. Lima, {\it {The road ahead of
  Horndeski: cosmology of surviving scalar-tensor theories}},
  \href{http://arxiv.org/abs/1810.10521}{{\tt arXiv:1810.10521}}.

\bibitem{Hirano:2018uar}
S.~Hirano, T.~Kobayashi, H.~Tashiro, and S.~Yokoyama, {\it {Matter bispectrum
  beyond Horndeski theories}},  {\em Phys. Rev.} {\bf D97} (2018), no.~10
  103517, [\href{http://arxiv.org/abs/1801.07885}{{\tt arXiv:1801.07885}}].

\bibitem{Duniya:2019mpr}
D.~Duniya, T.~Moloi, C.~Clarkson, J.~Larena, R.~Maartens, B.~Mongwane, and
  A.~Weltman, {\it {Probing beyond-Horndeski gravity on ultra-large scales}},
  \href{http://arxiv.org/abs/1902.09919}{{\tt arXiv:1902.09919}}.

\bibitem{Pace:2019uow}
F.~Pace, R.~A. Battye, B.~Bolliet, and D.~Trinh, {\it {Dark sector evolution in
  Horndeski models}},  \href{http://arxiv.org/abs/1905.06795}{{\tt
  arXiv:1905.06795}}.

\bibitem{Peebles:1970ag}
P.~J.~E. Peebles and J.~T. Yu, {\it {Primeval adiabatic perturbation in an
  expanding universe}},  {\em Astrophys. J.} {\bf 162} (1970) 815--836.

\bibitem{Wilson:1981yi}
M.~L. Wilson and J.~Silk, {\it {On the Anisotropy of the cosomological
  background matter and radiation distribution. 1. The Radiation anisotropy in
  a spatially flat universe}},  {\em Astrophys. J.} {\bf 243} (1981) 14--25.

\bibitem{Seljak:1996is}
U.~Seljak and M.~Zaldarriaga, {\it {A Line of sight integration approach to
  cosmic microwave background anisotropies}},  {\em Astrophys. J.} {\bf 469}
  (1996) 437--444, [\href{http://arxiv.org/abs/astro-ph/9603033}{{\tt
  astro-ph/9603033}}].

\bibitem{Seljak:2003th}
U.~Seljak, N.~Sugiyama, M.~J. White, and M.~Zaldarriaga, {\it {A Comparison of
  cosmological Boltzmann codes: Are we ready for high precision cosmology?}},
  {\em Phys. Rev.} {\bf D68} (2003) 083507,
  [\href{http://arxiv.org/abs/astro-ph/0306052}{{\tt astro-ph/0306052}}].

\bibitem{Kaplinghat:2002mh}
M.~Kaplinghat, L.~Knox, and C.~Skordis, {\it {Rapid calculation of theoretical
  cmb angular power spectra}},  {\em Astrophys. J.} {\bf 578} (2002) 665,
  [\href{http://arxiv.org/abs/astro-ph/0203413}{{\tt astro-ph/0203413}}].

\bibitem{Doran:2003sy}
M.~Doran, {\it {CMBEASY: an object oriented code for the cosmic microwave
  background}},  {\em JCAP} {\bf 0510} (2005) 011,
  [\href{http://arxiv.org/abs/astro-ph/0302138}{{\tt astro-ph/0302138}}].

\bibitem{Raveri:2014cka}
M.~Raveri, B.~Hu, N.~Frusciante, and A.~Silvestri, {\it {Effective Field Theory
  of Cosmic Acceleration: constraining dark energy with CMB data}},  {\em Phys.
  Rev.} {\bf D90} (2014), no.~4 043513,
  [\href{http://arxiv.org/abs/1405.1022}{{\tt arXiv:1405.1022}}].

\bibitem{Hu:2016zrh}
B.~Hu, M.~Raveri, M.~Rizzato, and A.~Silvestri, {\it {Testing Hu \& Sawicki
  f(R) gravity with the effective field theory approach}},  {\em Mon. Not. Roy.
  Astron. Soc.} {\bf 459} (2016), no.~4 3880--3889,
  [\href{http://arxiv.org/abs/1601.07536}{{\tt arXiv:1601.07536}}].

\bibitem{Peirone:2017vcq}
S.~Peirone, N.~Frusciante, B.~Hu, M.~Raveri, and A.~Silvestri, {\it {Do current
  cosmological observations rule out all Covariant Galileons?}},  {\em Phys.
  Rev.} {\bf D97} (2018), no.~6 063518,
  [\href{http://arxiv.org/abs/1711.04760}{{\tt arXiv:1711.04760}}].

\bibitem{Benevento:2018xcu}
G.~Benevento, M.~Raveri, A.~Lazanu, N.~Bartolo, M.~Liguori, P.~Brax, and
  P.~Valageas, {\it {K-mouflage Imprints on Cosmological Observables and Data
  Constraints}},  \href{http://arxiv.org/abs/1809.09958}{{\tt
  arXiv:1809.09958}}.

\bibitem{Peirone:2019aua}
S.~Peirone, G.~Benevento, N.~Frusciante, and S.~Tsujikawa, {\it {Cosmological
  data favor Galileon ghost condensate over $\Lambda$CDM}},
  \href{http://arxiv.org/abs/1905.05166}{{\tt arXiv:1905.05166}}.

\bibitem{Peirone:2019yjs}
S.~Peirone, G.~Benevento, N.~Frusciante, and S.~Tsujikawa, {\it {First
  cosmological constraints and phenomenology of a beyond-Horndeski model}},
  \href{http://arxiv.org/abs/1905.11364}{{\tt arXiv:1905.11364}}.

\bibitem{Frusciante:2019puu}
N.~Frusciante, S.~Peirone, L.~Atayde, and A.~De~Felice, {\it {Phenomenology of
  the generalized cubic covariant Galileon model and cosmological bounds}},
  {\em Phys. Rev. D} {\bf 101} (2020), no.~6 064001,
  [\href{http://arxiv.org/abs/1912.07586}{{\tt arXiv:1912.07586}}].

\bibitem{Hu:2014sea}
B.~Hu, M.~Raveri, A.~Silvestri, and N.~Frusciante, {\it {Exploring massive
  neutrinos in dark cosmologies with $\scriptsize{EFTCAMB}$/ EFTCosmoMC}},
  {\em Phys. Rev.} {\bf D91} (2015), no.~6 063524,
  [\href{http://arxiv.org/abs/1410.5807}{{\tt arXiv:1410.5807}}].

\bibitem{Hu:2015rva}
B.~Hu and M.~Raveri, {\it {Can modified gravity models reconcile the tension
  between the CMB anisotropy and lensing maps in Planck-like observations?}},
  {\em Phys. Rev.} {\bf D91} (2015), no.~12 123515,
  [\href{http://arxiv.org/abs/1502.06599}{{\tt arXiv:1502.06599}}].

\bibitem{Raveri:2017qvt}
M.~Raveri, P.~Bull, A.~Silvestri, and L.~Pogosian, {\it {Priors on the
  effective Dark Energy equation of state in scalar-tensor theories}},  {\em
  Phys. Rev.} {\bf D96} (2017), no.~8 083509,
  [\href{http://arxiv.org/abs/1703.05297}{{\tt arXiv:1703.05297}}].

\bibitem{Peirone:2017lgi}
S.~Peirone, M.~Martinelli, M.~Raveri, and A.~Silvestri, {\it {Impact of
  theoretical priors in cosmological analyses: the case of single field
  quintessence}},  {\em Phys. Rev.} {\bf D96} (2017), no.~6 063524,
  [\href{http://arxiv.org/abs/1702.06526}{{\tt arXiv:1702.06526}}].

\bibitem{Brinckmann:2018cvx}
T.~Brinckmann and J.~Lesgourgues, {\it {MontePython 3: boosted MCMC sampler and
  other features}},  \href{http://arxiv.org/abs/1804.07261}{{\tt
  arXiv:1804.07261}}.

\bibitem{Bellini:2015xja}
E.~Bellini, A.~J. Cuesta, R.~Jimenez, and L.~Verde, {\it {Constraints on
  deviations from $\Lambda$CDM within Horndeski gravity}},  {\em JCAP} {\bf
  1602} (2016), no.~02 053, [\href{http://arxiv.org/abs/1509.07816}{{\tt
  arXiv:1509.07816}}]. [Erratum: JCAP1606,no.06,E01(2016)].

\bibitem{Alonso:2016suf}
D.~Alonso, E.~Bellini, P.~G. Ferreira, and M.~Zumalac{\'a}rregui, {\it
  {Observational future of cosmological scalar-tensor theories}},  {\em Phys.
  Rev.} {\bf D95} (2017), no.~6 063502,
  [\href{http://arxiv.org/abs/1610.09290}{{\tt arXiv:1610.09290}}].

\bibitem{Renk:2017rzu}
J.~Renk, M.~Zumalac{\'a}rregui, F.~Montanari, and A.~Barreira, {\it {Galileon
  gravity in light of ISW, CMB, BAO and H$_0$ data}},  {\em JCAP} {\bf 1710}
  (2017), no.~10 020, [\href{http://arxiv.org/abs/1707.02263}{{\tt
  arXiv:1707.02263}}].

\bibitem{Huang:2012mt}
Z.~Huang, {\it {A Cosmology Forecast Toolkit -- CosmoLib}},  {\em JCAP} {\bf
  1206} (2012) 012, [\href{http://arxiv.org/abs/1201.5961}{{\tt
  arXiv:1201.5961}}].

\bibitem{Hu:1997hp}
W.~Hu and M.~J. White, {\it {CMB anisotropies: Total angular momentum method}},
   {\em Phys. Rev.} {\bf D56} (1997) 596--615,
  [\href{http://arxiv.org/abs/astro-ph/9702170}{{\tt astro-ph/9702170}}].

\bibitem{Battye:2013aaa}
R.~A. Battye and J.~A. Pearson, {\it {Parametrizing dark sector perturbations
  via equations of state}},  {\em Phys. Rev.} {\bf D88} (2013), no.~6 061301,
  [\href{http://arxiv.org/abs/1306.1175}{{\tt arXiv:1306.1175}}].

\bibitem{Battye:2015hza}
R.~A. Battye, B.~Bolliet, and J.~A. Pearson, {\it {$f(R)$ gravity as a dark
  energy fluid}},  {\em Phys. Rev.} {\bf D93} (2016), no.~4 044026,
  [\href{http://arxiv.org/abs/1508.04569}{{\tt arXiv:1508.04569}}].

\bibitem{Bellini:2017avd}
E.~Bellini et~al., {\it {Comparison of Einstein-Boltzmann solvers for testing
  general relativity}},  {\em Phys. Rev.} {\bf D97} (2018), no.~2 023520,
  [\href{http://arxiv.org/abs/1709.09135}{{\tt arXiv:1709.09135}}].

\bibitem{Sawicki:2015zya}
I.~Sawicki and E.~Bellini, {\it {Limits of quasistatic approximation in
  modified-gravity cosmologies}},  {\em Phys. Rev.} {\bf D92} (2015), no.~8
  084061, [\href{http://arxiv.org/abs/1503.06831}{{\tt arXiv:1503.06831}}].

\bibitem{Sachs:1967er}
R.~K. Sachs and A.~M. Wolfe, {\it {Perturbations of a cosmological model and
  angular variations of the microwave background}},  {\em Astrophys. J.} {\bf
  147} (1967) 73--90. [Gen. Rel. Grav.39,1929(2007)].

\bibitem{Kofman:1985fp}
L.~Kofman and A.~A. Starobinsky, {\it {Effect of the cosmological constant on
  large scale anisotropies in the microwave backbround}},  {\em Sov. Astron.
  Lett.} {\bf 11} (1985) 271--274. [Pisma Astron. Zh.11,643(1985)].

\bibitem{Hu:1996vq}
W.~Hu and M.~J. White, {\it {Acoustic signatures in the cosmic microwave
  background}},  {\em Astrophys. J.} {\bf 471} (1996) 30--51,
  [\href{http://arxiv.org/abs/astro-ph/9602019}{{\tt astro-ph/9602019}}].

\bibitem{Acquaviva:2005xz}
V.~Acquaviva and C.~Baccigalupi, {\it {Dark energy records in lensed cosmic
  microwave background}},  {\em Phys. Rev.} {\bf D74} (2006) 103510,
  [\href{http://arxiv.org/abs/astro-ph/0507644}{{\tt astro-ph/0507644}}].

\bibitem{Amendola:2011ie}
L.~Amendola, V.~Pettorino, C.~Quercellini, and A.~Vollmer, {\it {Testing
  coupled dark energy with next-generation large-scale observations}},  {\em
  Phys. Rev.} {\bf D85} (2012) 103008,
  [\href{http://arxiv.org/abs/1111.1404}{{\tt arXiv:1111.1404}}].

\bibitem{Nunes:2018zot}
R.~C. Nunes, M.~E.~S. Alves, and J.~C.~N. de~Araujo, {\it {Primordial
  gravitational waves in Horndeski gravity}},
  \href{http://arxiv.org/abs/1811.12760}{{\tt arXiv:1811.12760}}.

\bibitem{Abbott:2017xzu}
{\bf LIGO Scientific, Virgo, 1M2H, Dark Energy Camera GW-E, DES, DLT40, Las
  Cumbres Observatory, VINROUGE, MASTER} Collaboration, B.~P. Abbott et~al.,
  {\it {A gravitational-wave standard siren measurement of the Hubble
  constant}},  {\em Nature} {\bf 551} (2017), no.~7678 85--88,
  [\href{http://arxiv.org/abs/1710.05835}{{\tt arXiv:1710.05835}}].

\bibitem{Nissanke:2013fka}
S.~Nissanke, D.~E. Holz, N.~Dalal, S.~A. Hughes, J.~L. Sievers, and C.~M.
  Hirata, {\it {Determining the Hubble constant from gravitational wave
  observations of merging compact binaries}},
  \href{http://arxiv.org/abs/1307.2638}{{\tt arXiv:1307.2638}}.

\bibitem{Lagos:2019kds}
M.~Lagos, M.~Fishbach, P.~Landry, and D.~E. Holz, {\it {Standard sirens with a
  running Planck mass}},  {\em Phys. Rev.} {\bf D99} (2019), no.~8 083504,
  [\href{http://arxiv.org/abs/1901.03321}{{\tt arXiv:1901.03321}}].

\bibitem{Ezquiaga:2018btd}
J.~M. Ezquiaga and M.~Zumalacrregui, {\it {Dark Energy in light of
  Multi-Messenger Gravitational-Wave astronomy}},  {\em Front. Astron. Space
  Sci.} {\bf 5} (2018) 44, [\href{http://arxiv.org/abs/1807.09241}{{\tt
  arXiv:1807.09241}}].

\bibitem{Perenon:2019dpc}
L.~Perenon, J.~Bel, R.~Maartens, and A.~de~la Cruz-Dombriz, {\it {Optimising
  growth of structure constraints on modified gravity}},
  \href{http://arxiv.org/abs/1901.11063}{{\tt arXiv:1901.11063}}.

\bibitem{Frusciante:2019xia}
N.~Frusciante and L.~Perenon, {\it {Effective Field Theory of Dark Energy: a
  Review}},  \href{http://arxiv.org/abs/1907.03150}{{\tt arXiv:1907.03150}}.

\bibitem{Chevallier:2000qy}
M.~Chevallier and D.~Polarski, {\it {Accelerating universes with scaling dark
  matter}},  {\em Int.J.Mod.Phys.} {\bf D10} (2001) 213--224,
  [\href{http://arxiv.org/abs/gr-qc/0009008}{{\tt gr-qc/0009008}}].

\bibitem{Linder:2002et}
E.~V. Linder, {\it {Exploring the expansion history of the universe}},  {\em
  Phys.Rev.Lett.} {\bf 90} (2003) 091301,
  [\href{http://arxiv.org/abs/astro-ph/0208512}{{\tt astro-ph/0208512}}].

\bibitem{Riess:2011yx}
A.~G. Riess, L.~Macri, S.~Casertano, H.~Lampeitl, H.~C. Ferguson, A.~V.
  Filippenko, S.~W. Jha, W.~Li, and R.~Chornock, {\it {A 3\% Solution:
  Determination of the Hubble Constant with the Hubble Space Telescope and Wide
  Field Camera 3}},  {\em Astrophys. J.} {\bf 730} (2011) 119,
  [\href{http://arxiv.org/abs/1103.2976}{{\tt arXiv:1103.2976}}]. [Erratum:
  Astrophys. J.732,129(2011)].

\bibitem{Huang:2015srv}
Z.~Huang, {\it {Observational effects of a running Planck mass}},  {\em Phys.
  Rev.} {\bf D93} (2016), no.~4 043538,
  [\href{http://arxiv.org/abs/1511.02808}{{\tt arXiv:1511.02808}}].

\bibitem{Brando:2019xbv}
G.~Brando, F.~T. Falciano, E.~V. Linder, and H.~E.~S. Velten, {\it {Modified
  Gravity Away from a $\Lambda$CDM Background}},
  \href{http://arxiv.org/abs/1904.12903}{{\tt arXiv:1904.12903}}.

\bibitem{Noller:2018wyv}
J.~Noller and A.~Nicola, {\it {Cosmological parameter constraints for Horndeski
  scalar-tensor gravity}},  \href{http://arxiv.org/abs/1811.12928}{{\tt
  arXiv:1811.12928}}.

\bibitem{Reischke:2018ooh}
R.~Reischke, A.~Spurio~Mancini, B.~M. Schfer, and P.~M. Merkel, {\it
  {Investigating scalar-tensor-gravity with statistics of the cosmic
  large-scale structure}},  \href{http://arxiv.org/abs/1804.02441}{{\tt
  arXiv:1804.02441}}.

\bibitem{Pettorino:2014bka}
V.~Pettorino and L.~Amendola, {\it {Friction in Gravitational Waves: a test for
  early-time modified gravity}},  {\em Phys. Lett.} {\bf B742} (2015) 353--357,
  [\href{http://arxiv.org/abs/1408.2224}{{\tt arXiv:1408.2224}}].

\bibitem{Raveri:2014eea}
M.~Raveri, C.~Baccigalupi, A.~Silvestri, and S.-Y. Zhou, {\it {Measuring the
  speed of cosmological gravitational waves}},  {\em Phys. Rev.} {\bf D91}
  (2015), no.~6 061501, [\href{http://arxiv.org/abs/1405.7974}{{\tt
  arXiv:1405.7974}}].

\bibitem{Perenon:2019qmd}
L.~Perenon and H.~Velten, {\it {The effective field theory of dark energy
  diagnostic of linear Horndeski theories after GW170817 and GRB170817A}},
  {\em Universe} {\bf 5} (2019), no.~6 138,
  [\href{http://arxiv.org/abs/1903.08088}{{\tt arXiv:1903.08088}}].

\bibitem{Creminelli:2018xsv}
P.~Creminelli, M.~Lewandowski, G.~Tambalo, and F.~Vernizzi, {\it {Gravitational
  Wave Decay into Dark Energy}},  {\em JCAP} {\bf 1812} (2018), no.~12 025,
  [\href{http://arxiv.org/abs/1809.03484}{{\tt arXiv:1809.03484}}].

\bibitem{Hui:2009kc}
L.~Hui, A.~Nicolis, and C.~Stubbs, {\it {Equivalence Principle Implications of
  Modified Gravity Models}},  {\em Phys. Rev.} {\bf D80} (2009) 104002,
  [\href{http://arxiv.org/abs/0905.2966}{{\tt arXiv:0905.2966}}].

\bibitem{Gleyzes:2015rua}
J.~Gleyzes, D.~Langlois, M.~Mancarella, and F.~Vernizzi, {\it {Effective Theory
  of Dark Energy at Redshift Survey Scales}},  {\em JCAP} {\bf 1602} (2016),
  no.~02 056, [\href{http://arxiv.org/abs/1509.02191}{{\tt arXiv:1509.02191}}].

\bibitem{Mancini:2018qtb}
A.~S. Mancini, R.~Reischke, V.~Pettorino, B.~M. SchÃ¡efer, and
  M.~ZumalacÃ¡rregui, {\it {Testing (modified) gravity with 3D and
  tomographic cosmic shear}},  {\em Mon. Not. Roy. Astron. Soc.} {\bf 480}
  (2018) 3725, [\href{http://arxiv.org/abs/1801.04251}{{\tt
  arXiv:1801.04251}}].

\bibitem{Hildebrandt:2016iqg}
H.~Hildebrandt et~al., {\it {KiDS-450: Cosmological parameter constraints from
  tomographic weak gravitational lensing}},  {\em Mon. Not. Roy. Astron. Soc.}
  {\bf 465} (2017) 1454, [\href{http://arxiv.org/abs/1606.05338}{{\tt
  arXiv:1606.05338}}].

\bibitem{deJong:2015wca}
J.~T.~A. de~Jong et~al., {\it {The first and second data releases of the
  Kilo-Degree Survey}},  {\em Astron. Astrophys.} {\bf 582} (2015) A62,
  [\href{http://arxiv.org/abs/1507.00742}{{\tt arXiv:1507.00742}}].

\bibitem{Kuijken:2015vca}
K.~Kuijken et~al., {\it {Gravitational Lensing Analysis of the Kilo Degree
  Survey}},  {\em Mon. Not. Roy. Astron. Soc.} {\bf 454} (2015), no.~4
  3500--3532, [\href{http://arxiv.org/abs/1507.00738}{{\tt arXiv:1507.00738}}].

\bibitem{Conti:2016gav}
I.~Fenech~Conti, R.~Herbonnet, H.~Hoekstra, J.~Merten, L.~Miller, and M.~Viola,
  {\it {Calibration of weak-lensing shear in the Kilo-Degree Survey}},  {\em
  Mon. Not. Roy. Astron. Soc.} {\bf 467} (2017), no.~2 1627--1651,
  [\href{http://arxiv.org/abs/1606.05337}{{\tt arXiv:1606.05337}}].

\bibitem{Linder:2005in}
E.~V. Linder, {\it {Cosmic growth history and expansion history}},  {\em Phys.
  Rev.} {\bf D72} (2005) 043529,
  [\href{http://arxiv.org/abs/astro-ph/0507263}{{\tt astro-ph/0507263}}].

\bibitem{Gleyzes:2017kpi}
J.~Gleyzes, {\it {Parametrizing modified gravity for cosmological surveys}},
  {\em Phys. Rev.} {\bf D96} (2017), no.~6 063516,
  [\href{http://arxiv.org/abs/1705.04714}{{\tt arXiv:1705.04714}}].

\bibitem{Linder:2015rcz}
E.~V. Linder, G.~SengÃ¶r, and S.~Watson, {\it {Is the Effective Field Theory
  of Dark Energy Effective?}},  {\em JCAP} {\bf 1605} (2016), no.~05 053,
  [\href{http://arxiv.org/abs/1512.06180}{{\tt arXiv:1512.06180}}].

\bibitem{Linder:2016wqw}
E.~V. Linder, {\it {Challenges in connecting modified gravity theory and
  observations}},  {\em Phys. Rev.} {\bf D95} (2017), no.~2 023518,
  [\href{http://arxiv.org/abs/1607.03113}{{\tt arXiv:1607.03113}}].

\bibitem{Raveri:2019mxg}
M.~Raveri, {\it {Reconstructing Gravity on Cosmological Scales}},
  \href{http://arxiv.org/abs/1902.01366}{{\tt arXiv:1902.01366}}.

\bibitem{Kase:2018aps}
R.~Kase and S.~Tsujikawa, {\it {Dark energy in Horndeski theories after
  GW170817: A review}},  \href{http://arxiv.org/abs/1809.08735}{{\tt
  arXiv:1809.08735}}.

\bibitem{Audley:2017drz}
{\bf LISA} Collaboration, H.~Audley et~al., {\it {Laser Interferometer Space
  Antenna}},  \href{http://arxiv.org/abs/1702.00786}{{\tt arXiv:1702.00786}}.

\bibitem{Creminelli:2019kjy}
P.~Creminelli, G.~Tambalo, F.~Vernizzi, and V.~Yingcharoenrat, {\it
  {Dark-Energy Instabilities induced by Gravitational Waves}},
  \href{http://arxiv.org/abs/1910.14035}{{\tt arXiv:1910.14035}}.

\bibitem{Uzan2011}
J.-P. Uzan, {\it Varying constants, gravitation and cosmology},  {\em Living
  Reviews in Relativity} {\bf 14} (Mar, 2011) 2.

\bibitem{Will2014}
C.~M. Will, {\it {The Confrontation between General Relativity and
  Experiment}},  {\em Living Rev. Rel.} {\bf 17} (Jun, 2014) 4,
  [\href{http://arxiv.org/abs/1403.7377}{{\tt arXiv:1403.7377}}].

\bibitem{Koyama:2013paa}
K.~Koyama, G.~Niz, and G.~Tasinato, {\it {Effective theory for the Vainshtein
  mechanism from the Horndeski action}},  {\em Phys. Rev.} {\bf D88} (2013)
  021502, [\href{http://arxiv.org/abs/1305.0279}{{\tt arXiv:1305.0279}}].

\bibitem{Kimura:2011dc}
R.~Kimura, T.~Kobayashi, and K.~Yamamoto, {\it {Vainshtein screening in a
  cosmological background in the most general second-order scalar-tensor
  theory}},  {\em Phys. Rev.} {\bf D85} (2012) 024023,
  [\href{http://arxiv.org/abs/1111.6749}{{\tt arXiv:1111.6749}}].

\bibitem{Kobayashi:2014ida}
T.~Kobayashi, Y.~Watanabe, and D.~Yamauchi, {\it {Breaking of Vainshtein
  screening in scalar-tensor theories beyond Horndeski}},  {\em Phys. Rev.}
  {\bf D91} (2015), no.~6 064013, [\href{http://arxiv.org/abs/1411.4130}{{\tt
  arXiv:1411.4130}}].

\bibitem{Sakstein:2016ggl}
J.~Sakstein, H.~Wilcox, D.~Bacon, K.~Koyama, and R.~C. Nichol, {\it {Testing
  Gravity Using Galaxy Clusters: New Constraints on Beyond Horndeski
  Theories}},  {\em JCAP} {\bf 1607} (2016), no.~07 019,
  [\href{http://arxiv.org/abs/1603.06368}{{\tt arXiv:1603.06368}}].

\bibitem{Sakstein:2016oel}
J.~Sakstein, E.~Babichev, K.~Koyama, D.~Langlois, and R.~Saito, {\it {Towards
  Strong Field Tests of Beyond Horndeski Gravity Theories}},  {\em Phys. Rev.}
  {\bf D95} (2017), no.~6 064013, [\href{http://arxiv.org/abs/1612.04263}{{\tt
  arXiv:1612.04263}}].

\bibitem{Romer:1999qt}
A.~K. Romer, P.~T.~P. Viana, A.~R. Liddle, and R.~G. Mann, {\it {A
  serendipitous galaxy cluster survey with xmm: expected catalogue properties
  and scientific applications}},
  \href{http://arxiv.org/abs/astro-ph/9911499}{{\tt astro-ph/9911499}}.

\bibitem{Heymans:2012gg}
C.~Heymans et~al., {\it {CFHTLenS: The Canada-France-Hawaii Telescope Lensing
  Survey}},  {\em Mon. Not. Roy. Astron. Soc.} {\bf 427} (2012) 146,
  [\href{http://arxiv.org/abs/1210.0032}{{\tt arXiv:1210.0032}}].

\bibitem{Riess:2016jrr}
A.~G. Riess et~al., {\it {A 2.4\% Determination of the Local Value of the
  Hubble Constant}},  {\em Astrophys. J.} {\bf 826} (2016), no.~1 56,
  [\href{http://arxiv.org/abs/1604.01424}{{\tt arXiv:1604.01424}}].

\bibitem{Riess:2018uxu}
A.~G. Riess et~al., {\it {New Parallaxes of Galactic Cepheids from Spatially
  Scanning the Hubble Space Telescope: Implications for the Hubble Constant}},
  {\em Astrophys. J.} {\bf 855} (2018), no.~2 136,
  [\href{http://arxiv.org/abs/1801.01120}{{\tt arXiv:1801.01120}}].

\bibitem{Troxel:2017xyo}
{\bf DES} Collaboration, M.~A. Troxel et~al., {\it {Dark Energy Survey Year 1
  results: Cosmological constraints from cosmic shear}},  {\em Phys. Rev.} {\bf
  D98} (2018), no.~4 043528, [\href{http://arxiv.org/abs/1708.01538}{{\tt
  arXiv:1708.01538}}].

\bibitem{Krause:2017ekm}
{\bf DES} Collaboration, E.~Krause et~al., {\it {Dark Energy Survey Year 1
  Results: Multi-Probe Methodology and Simulated Likelihood Analyses}},  {\em
  Submitted to: Phys. Rev. D} (2017)
  [\href{http://arxiv.org/abs/1706.09359}{{\tt arXiv:1706.09359}}].

\bibitem{Cooke:2017cwo}
R.~J. Cooke, M.~Pettini, and C.~C. Steidel, {\it {One Percent Determination of
  the Primordial Deuterium Abundance}},  {\em Astrophys. J.} {\bf 855} (2018),
  no.~2 102, [\href{http://arxiv.org/abs/1710.11129}{{\tt arXiv:1710.11129}}].

\bibitem{Birrer:2018vtm}
S.~Birrer et~al., {\it {H0LiCOW - IX. Cosmographic analysis of the doubly
  imaged quasar SDSS 1206+4332 and a new measurement of the Hubble constant}},
  {\em Mon. Not. Roy. Astron. Soc.} {\bf 484} (2019) 4726,
  [\href{http://arxiv.org/abs/1809.01274}{{\tt arXiv:1809.01274}}].

\bibitem{Cardona:2016ems}
W.~Cardona, M.~Kunz, and V.~Pettorino, {\it {Determining $H_0$ with Bayesian
  hyper-parameters}},  {\em JCAP} {\bf 1703} (2017), no.~03 056,
  [\href{http://arxiv.org/abs/1611.06088}{{\tt arXiv:1611.06088}}].

\bibitem{Camarena:2019moy}
D.~Camarena and V.~Marra, {\it {Local determination of the Hubble constant and
  the deceleration parameter}},  \href{http://arxiv.org/abs/1906.11814}{{\tt
  arXiv:1906.11814}}.

\bibitem{Freedman:2019jwv}
W.~L. Freedman et~al., {\it {The Carnegie-Chicago Hubble Program. VIII. An
  Independent Determination of the Hubble Constant Based on the Tip of the Red
  Giant Branch}},  \href{http://arxiv.org/abs/1907.05922}{{\tt
  arXiv:1907.05922}}.

\bibitem{Yuan:2019npk}
W.~Yuan, A.~G. Riess, L.~M. Macri, S.~Casertano, and D.~Scolnic, {\it
  {Consistent Calibration of the Tip of the Red Giant Branch in the Large
  Magellanic Cloud on the Hubble Space Telescope Photometric System and a
  Re-determination of the Hubble Constant}},  {\em Astrophys. J.} {\bf 886}
  (2019) 61, [\href{http://arxiv.org/abs/1908.00993}{{\tt arXiv:1908.00993}}].

\bibitem{Freedman:2020dne}
W.~L. Freedman, B.~F. Madore, T.~Hoyt, I.~S. Jang, R.~Beaton, M.~G. Lee,
  A.~Monson, J.~Neeley, and J.~Rich, {\it {Calibration of the Tip of the Red
  Giant Branch (TRGB)}},  \href{http://arxiv.org/abs/2002.01550}{{\tt
  arXiv:2002.01550}}.

\bibitem{Huang:2019yhh}
C.~D. Huang, A.~G. Riess, W.~Yuan, L.~M. Macri, N.~L. Zakamska, S.~Casertano,
  P.~A. Whitelock, S.~L. Hoffmann, A.~V. Filippenko, and D.~Scolnic, {\it
  {Hubble Space Telescope Observations of Mira Variables in the Type Ia
  Supernova Host NGC 1559: An Alternative Candle to Measure the Hubble
  Constant}},  \href{http://arxiv.org/abs/1908.10883}{{\tt arXiv:1908.10883}}.

\bibitem{Shajib:2019toy}
{\bf DES} Collaboration, A.~J. Shajib et~al., {\it {STRIDES: A 3.9 Per Cent
  Measurement of the Hubble Constant from the Strong Lens System DES
  J0408-5354}},  \href{http://arxiv.org/abs/1910.06306}{{\tt
  arXiv:1910.06306}}.

\bibitem{Riess:2020sih}
A.~G. Riess, {\it {The Expansion of the Universe is Faster than Expected}},
  {\em Nature Rev. Phys.} {\bf 2} (2019), no.~1 10--12,
  [\href{http://arxiv.org/abs/2001.03624}{{\tt arXiv:2001.03624}}].

\bibitem{DiValentino:2015ola}
E.~Di~Valentino, A.~Melchiorri, and J.~Silk, {\it {Beyond six parameters:
  extending $\Lambda$CDM}},  {\em Phys. Rev.} {\bf D92} (2015), no.~12 121302,
  [\href{http://arxiv.org/abs/1507.06646}{{\tt arXiv:1507.06646}}].

\bibitem{DiValentino:2016hlg}
E.~Di~Valentino, A.~Melchiorri, and J.~Silk, {\it {Reconciling Planck with the
  local value of $H_0$ in extended parameter space}},  {\em Phys. Lett.} {\bf
  B761} (2016) 242--246, [\href{http://arxiv.org/abs/1606.00634}{{\tt
  arXiv:1606.00634}}].

\bibitem{Bernal:2016gxb}
J.~L. Bernal, L.~Verde, and A.~G. Riess, {\it {The trouble with $H_0$}},  {\em
  JCAP} {\bf 1610} (2016), no.~10 019,
  [\href{http://arxiv.org/abs/1607.05617}{{\tt arXiv:1607.05617}}].

\bibitem{Kumar:2016zpg}
S.~Kumar and R.~C. Nunes, {\it {Probing the interaction between dark matter and
  dark energy in the presence of massive neutrinos}},  {\em Phys. Rev.} {\bf
  D94} (2016), no.~12 123511, [\href{http://arxiv.org/abs/1608.02454}{{\tt
  arXiv:1608.02454}}].

\bibitem{Kumar:2017dnp}
S.~Kumar and R.~C. Nunes, {\it {Echo of interactions in the dark sector}},
  {\em Phys. Rev.} {\bf D96} (2017), no.~10 103511,
  [\href{http://arxiv.org/abs/1702.02143}{{\tt arXiv:1702.02143}}].

\bibitem{DiValentino:2017iww}
E.~Di~Valentino, A.~Melchiorri, and O.~Mena, {\it {Can interacting dark energy
  solve the $H_0$ tension?}},  {\em Phys. Rev.} {\bf D96} (2017), no.~4 043503,
  [\href{http://arxiv.org/abs/1704.08342}{{\tt arXiv:1704.08342}}].

\bibitem{DiValentino:2017oaw}
E.~Di~Valentino, C.~Bøehm, E.~Hivon, and F.~R. Bouchet, {\it {Reducing the
  $H_0$ and $\sigma_8$ tensions with Dark Matter-neutrino interactions}},  {\em
  Phys. Rev.} {\bf D97} (2018), no.~4 043513,
  [\href{http://arxiv.org/abs/1710.02559}{{\tt arXiv:1710.02559}}].

\bibitem{DiValentino:2017rcr}
E.~Di~Valentino, E.~V. Linder, and A.~Melchiorri, {\it {Vacuum phase transition
  solves the $H_0$ tension}},  {\em Phys. Rev.} {\bf D97} (2018), no.~4 043528,
  [\href{http://arxiv.org/abs/1710.02153}{{\tt arXiv:1710.02153}}].

\bibitem{Binder:2017lkj}
T.~Binder, M.~Gustafsson, A.~Kamada, S.~M.~R. Sandner, and M.~Wiesner, {\it
  {Reannihilation of self-interacting dark matter}},  {\em Phys. Rev.} {\bf
  D97} (2018), no.~12 123004, [\href{http://arxiv.org/abs/1712.01246}{{\tt
  arXiv:1712.01246}}].

\bibitem{Khosravi:2017hfi}
N.~Khosravi, S.~Baghram, N.~Afshordi, and N.~Altamirano, {\it {$H_0$ tension as
  a hint for a transition in gravitational theory}},  {\em Phys. Rev.} {\bf
  D99} (2019), no.~10 103526, [\href{http://arxiv.org/abs/1710.09366}{{\tt
  arXiv:1710.09366}}].

\bibitem{DiValentino:2017zyq}
E.~Di~Valentino, A.~Melchiorri, E.~V. Linder, and J.~Silk, {\it {Constraining
  Dark Energy Dynamics in Extended Parameter Space}},  {\em Phys. Rev.} {\bf
  D96} (2017), no.~2 023523, [\href{http://arxiv.org/abs/1704.00762}{{\tt
  arXiv:1704.00762}}].

\bibitem{DiValentino:2017gzb}
E.~Di~Valentino, {\it {Crack in the cosmological paradigm}},  {\em Nat.
  Astron.} {\bf 1} (2017), no.~9 569--570,
  [\href{http://arxiv.org/abs/1709.04046}{{\tt arXiv:1709.04046}}].

\bibitem{Sola:2017znb}
J.~Solà, A.~Gómez-Valent, and J.~de~Cruz~Pérez, {\it {The $H_0$ tension in
  light of vacuum dynamics in the Universe}},  {\em Phys. Lett.} {\bf B774}
  (2017) 317--324, [\href{http://arxiv.org/abs/1705.06723}{{\tt
  arXiv:1705.06723}}].

\bibitem{Yang:2018euj}
W.~Yang, S.~Pan, E.~Di~Valentino, R.~C. Nunes, S.~Vagnozzi, and D.~F. Mota,
  {\it {Tale of stable interacting dark energy, observational signatures, and
  the $H_0$ tension}},  {\em JCAP} {\bf 1809} (2018), no.~09 019,
  [\href{http://arxiv.org/abs/1805.08252}{{\tt arXiv:1805.08252}}].

\bibitem{Colgain:2018wgk}
E.~Ó~Colgáin, M.~H. P.~M. van Putten, and H.~Yavartanoo, {\it {de Sitter
  Swampland, $H_0$ tension \& observation}},  {\em Phys. Lett.} {\bf B793}
  (2019) 126--129, [\href{http://arxiv.org/abs/1807.07451}{{\tt
  arXiv:1807.07451}}].

\bibitem{DEramo:2018vss}
F.~D'Eramo, R.~Z. Ferreira, A.~Notari, and J.~L. Bernal, {\it {Hot Axions and
  the $H_0$ tension}},  {\em JCAP} {\bf 1811} (2018), no.~11 014,
  [\href{http://arxiv.org/abs/1808.07430}{{\tt arXiv:1808.07430}}].

\bibitem{Yang:2018uae}
W.~Yang, A.~Mukherjee, E.~Di~Valentino, and S.~Pan, {\it {Interacting dark
  energy with time varying equation of state and the $H_0$ tension}},  {\em
  Phys. Rev.} {\bf D98} (2018), no.~12 123527,
  [\href{http://arxiv.org/abs/1809.06883}{{\tt arXiv:1809.06883}}].

\bibitem{Guo:2018ans}
R.-Y. Guo, J.-F. Zhang, and X.~Zhang, {\it {Can the $H_0$ tension be resolved
  in extensions to $\Lambda$CDM cosmology?}},  {\em JCAP} {\bf 1902} (2019)
  054, [\href{http://arxiv.org/abs/1809.02340}{{\tt arXiv:1809.02340}}].

\bibitem{Lin:2018nxe}
M.-X. Lin, M.~Raveri, and W.~Hu, {\it {Phenomenology of Modified Gravity at
  Recombination}},  {\em Phys. Rev.} {\bf D99} (2019), no.~4 043514,
  [\href{http://arxiv.org/abs/1810.02333}{{\tt arXiv:1810.02333}}].

\bibitem{Yang:2018qmz}
W.~Yang, S.~Pan, E.~Di~Valentino, E.~N. Saridakis, and S.~Chakraborty, {\it
  {Observational constraints on one-parameter dynamical dark-energy
  parametrizations and the $H_0$ tension}},  {\em Phys. Rev.} {\bf D99} (2019),
  no.~4 043543, [\href{http://arxiv.org/abs/1810.05141}{{\tt
  arXiv:1810.05141}}].

\bibitem{Poulin:2018cxd}
V.~Poulin, T.~L. Smith, T.~Karwal, and M.~Kamionkowski, {\it {Early Dark Energy
  Can Resolve The Hubble Tension}},  {\em Phys. Rev. Lett.} {\bf 122} (2019),
  no.~22 221301, [\href{http://arxiv.org/abs/1811.04083}{{\tt
  arXiv:1811.04083}}].

\bibitem{Banihashemi:2018oxo}
A.~Banihashemi, N.~Khosravi, and A.~H. Shirazi, {\it {Ups and Downs in Dark
  Energy: phase transition in dark sector as a proposal to lessen cosmological
  tensions}},  \href{http://arxiv.org/abs/1808.02472}{{\tt arXiv:1808.02472}}.

\bibitem{Banihashemi:2018has}
A.~Banihashemi, N.~Khosravi, and A.~H. Shirazi, {\it {Ginzburg-Landau Theory of
  Dark Energy: A Framework to Study Both Temporal and Spatial Cosmological
  Tensions Simultaneously}},  {\em Phys. Rev.} {\bf D99} (2019), no.~8 083509,
  [\href{http://arxiv.org/abs/1810.11007}{{\tt arXiv:1810.11007}}].

\bibitem{Mortsell:2018mfj}
E.~Mörtsell and S.~Dhawan, {\it {Does the Hubble constant tension call for new
  physics?}},  {\em JCAP} {\bf 1809} (2018), no.~09 025,
  [\href{http://arxiv.org/abs/1801.07260}{{\tt arXiv:1801.07260}}].

\bibitem{Kreisch:2019yzn}
C.~D. Kreisch, F.-Y. Cyr-Racine, and O.~Doré, {\it {The Neutrino Puzzle:
  Anomalies, Interactions, and Cosmological Tensions}},
  \href{http://arxiv.org/abs/1902.00534}{{\tt arXiv:1902.00534}}.

\bibitem{Martinelli:2019dau}
M.~Martinelli, N.~B. Hogg, S.~Peirone, M.~Bruni, and D.~Wands, {\it
  {Constraints on the interacting vacuum–geodesic CDM scenario}},  {\em Mon.
  Not. Roy. Astron. Soc.} {\bf 488} (2019), no.~3 3423--3438,
  [\href{http://arxiv.org/abs/1902.10694}{{\tt arXiv:1902.10694}}].

\bibitem{Vattis:2019efj}
K.~Vattis, S.~M. Koushiappas, and A.~Loeb, {\it {Late universe decaying dark
  matter can relieve the $H_0$ tension}},
  \href{http://arxiv.org/abs/1903.06220}{{\tt arXiv:1903.06220}}.

\bibitem{Kumar:2019wfs}
S.~Kumar, R.~C. Nunes, and S.~K. Yadav, {\it {Dark sector interaction: a remedy
  of the tensions between CMB and LSS data}},
  \href{http://arxiv.org/abs/1903.04865}{{\tt arXiv:1903.04865}}.

\bibitem{Agrawal:2019lmo}
P.~Agrawal, F.-Y. Cyr-Racine, D.~Pinner, and L.~Randall, {\it {Rock 'n' Roll
  Solutions to the Hubble Tension}},
  \href{http://arxiv.org/abs/1904.01016}{{\tt arXiv:1904.01016}}.

\bibitem{Yang:2019jwn}
W.~Yang, S.~Pan, A.~Paliathanasis, S.~Ghosh, and Y.~Wu, {\it {Observational
  constraints of a new unified dark fluid and the $H_0$ tension}},
  \href{http://arxiv.org/abs/1904.10436}{{\tt arXiv:1904.10436}}.

\bibitem{Yang:2019qza}
W.~Yang, S.~Pan, E.~Di~Valentino, A.~Paliathanasis, and J.~Lu, {\it
  {Challenging bulk viscous unified scenarios with cosmological observations}},
   {\em Phys. Rev.} {\bf D100} (2019), no.~10 103518,
  [\href{http://arxiv.org/abs/1906.04162}{{\tt arXiv:1906.04162}}].

\bibitem{Yang:2019uzo}
W.~Yang, O.~Mena, S.~Pan, and E.~Di~Valentino, {\it {Dark sectors with
  dynamical coupling}},  {\em Phys. Rev.} {\bf D100} (2019), no.~8 083509,
  [\href{http://arxiv.org/abs/1906.11697}{{\tt arXiv:1906.11697}}].

\bibitem{DiValentino:2019exe}
E.~Di~Valentino, R.~Z. Ferreira, L.~Visinelli, and U.~Danielsson, {\it {Late
  time transitions in the quintessence field and the $H_0$ tension}},  {\em
  Phys. Dark Univ.} {\bf 26} (2019) 100385,
  [\href{http://arxiv.org/abs/1906.11255}{{\tt arXiv:1906.11255}}].

\bibitem{Desmond:2019ygn}
H.~Desmond, B.~Jain, and J.~Sakstein, {\it {Local resolution of the Hubble
  tension: The impact of screened fifth forces on the cosmic distance ladder}},
   {\em Phys. Rev.} {\bf D100} (2019), no.~4 043537,
  [\href{http://arxiv.org/abs/1907.03778}{{\tt arXiv:1907.03778}}].

\bibitem{Yang:2019nhz}
W.~Yang, S.~Pan, S.~Vagnozzi, E.~Di~Valentino, D.~F. Mota, and S.~Capozziello,
  {\it {Dawn of the dark: unified dark sectors and the EDGES Cosmic Dawn 21-cm
  signal}},  {\em JCAP} {\bf 1911} (2019) 044,
  [\href{http://arxiv.org/abs/1907.05344}{{\tt arXiv:1907.05344}}].

\bibitem{Pan:2019gop}
S.~Pan, W.~Yang, E.~Di~Valentino, E.~N. Saridakis, and S.~Chakraborty, {\it
  {Interacting scenarios with dynamical dark energy: Observational constraints
  and alleviation of the $H_0$ tension}},  {\em Phys. Rev.} {\bf D100} (2019),
  no.~10 103520, [\href{http://arxiv.org/abs/1907.07540}{{\tt
  arXiv:1907.07540}}].

\bibitem{Visinelli:2019qqu}
L.~Visinelli, S.~Vagnozzi, and U.~Danielsson, {\it {Revisiting a negative
  cosmological constant from low-redshift data}},  {\em Symmetry} {\bf 11}
  (2019), no.~8 1035, [\href{http://arxiv.org/abs/1907.07953}{{\tt
  arXiv:1907.07953}}].

\bibitem{Martinelli:2019krf}
M.~Martinelli and I.~Tutusaus, {\it {CMB tensions with low-redshift $H_0$ and
  $S_8$ measurements: impact of a redshift-dependent type-Ia supernovae
  intrinsic luminosity}},  {\em Symmetry} {\bf 11} (2019), no.~8 986,
  [\href{http://arxiv.org/abs/1906.09189}{{\tt arXiv:1906.09189}}].

\bibitem{Pan:2019hac}
S.~Pan, W.~Yang, E.~Di~Valentino, A.~Shafieloo, and S.~Chakraborty, {\it
  {Reconciling $H_0$ tension in a six parameter space?}},
  \href{http://arxiv.org/abs/1907.12551}{{\tt arXiv:1907.12551}}.

\bibitem{DiValentino:2019dzu}
E.~Di~Valentino, A.~Melchiorri, and J.~Silk, {\it {Cosmological constraints in
  extended parameter space from the Planck 2018 Legacy release}},
  \href{http://arxiv.org/abs/1908.01391}{{\tt arXiv:1908.01391}}.

\bibitem{DiValentino:2019ffd}
E.~Di~Valentino, A.~Melchiorri, O.~Mena, and S.~Vagnozzi, {\it {Interacting
  dark energy after the latest Planck, DES, and $H_0$ measurements: an
  excellent solution to the $H_0$ and cosmic shear tensions}},
  \href{http://arxiv.org/abs/1908.04281}{{\tt arXiv:1908.04281}}.

\bibitem{DiValentino:2019jae}
E.~Di~Valentino, A.~Melchiorri, O.~Mena, and S.~Vagnozzi, {\it {Non-minimal
  dark sector physics and cosmological tensions}},
  \href{http://arxiv.org/abs/1910.09853}{{\tt arXiv:1910.09853}}.

\bibitem{Schoneberg:2019wmt}
N.~Schöneberg, J.~Lesgourgues, and D.~C. Hooper, {\it {The BAO+BBN take on the
  Hubble tension}},  {\em JCAP} {\bf 1910} (2019), no.~10 029,
  [\href{http://arxiv.org/abs/1907.11594}{{\tt arXiv:1907.11594}}].

\bibitem{Shafieloo:2016bpk}
A.~Shafieloo, D.~K. Hazra, V.~Sahni, and A.~A. Starobinsky, {\it {Metastable
  Dark Energy with Radioactive-like Decay}},  {\em Mon. Not. Roy. Astron. Soc.}
  {\bf 473} (2018), no.~2 2760--2770,
  [\href{http://arxiv.org/abs/1610.05192}{{\tt arXiv:1610.05192}}].

\bibitem{Li:2019san}
X.~Li, A.~Shafieloo, V.~Sahni, and A.~A. Starobinsky, {\it {Revisiting
  Metastable Dark Energy and Tensions in the Estimation of Cosmological
  Parameters}},  \href{http://arxiv.org/abs/1904.03790}{{\tt
  arXiv:1904.03790}}.

\bibitem{Cuceu:2019for}
A.~Cuceu, J.~Farr, P.~Lemos, and A.~Font-Ribera, {\it {Baryon Acoustic
  Oscillations and the Hubble Constant: Past, Present and Future}},  {\em JCAP}
  {\bf 1910} (2019), no.~10 044, [\href{http://arxiv.org/abs/1906.11628}{{\tt
  arXiv:1906.11628}}].

\bibitem{Colgain:2019joh}
E.~. Colgáin and H.~Yavartanoo, {\it {Testing the Swampland: $H_0$ tension}},
  {\em Phys. Lett.} {\bf B797} (2019) 134907,
  [\href{http://arxiv.org/abs/1905.02555}{{\tt arXiv:1905.02555}}].

\bibitem{Pan:2019jqh}
S.~Pan, W.~Yang, C.~Singha, and E.~N. Saridakis, {\it {Observational
  constraints on sign-changeable interaction models and alleviation of the
  $H_0$ tension}},  {\em Phys. Rev.} {\bf D100} (2019), no.~8 083539,
  [\href{http://arxiv.org/abs/1903.10969}{{\tt arXiv:1903.10969}}].

\bibitem{Berghaus:2019cls}
K.~V. Berghaus and T.~Karwal, {\it {Thermal Friction as a Solution to the
  Hubble Tension}},  \href{http://arxiv.org/abs/1911.06281}{{\tt
  arXiv:1911.06281}}.

\bibitem{Knox:2019rjx}
L.~Knox and M.~Millea, {\it {Hubble constant hunter\textquoteright{}s guide}},
  {\em Phys. Rev. D} {\bf 101} (2020), no.~4 043533,
  [\href{http://arxiv.org/abs/1908.03663}{{\tt arXiv:1908.03663}}].

\bibitem{Pandey:2019plg}
K.~L. Pandey, T.~Karwal, and S.~Das, {\it {Alleviating the $H_0$ and $\sigma_8$
  anomalies with a decaying dark matter model}},
  \href{http://arxiv.org/abs/1902.10636}{{\tt arXiv:1902.10636}}.

\bibitem{Vagnozzi:2019ezj}
S.~Vagnozzi, {\it {New physics in light of the $H_0$ tension: an alternative
  view}},  \href{http://arxiv.org/abs/1907.07569}{{\tt arXiv:1907.07569}}.

\bibitem{Adhikari:2019fvb}
S.~Adhikari and D.~Huterer, {\it {Super-CMB fluctuations can resolve the Hubble
  tension}},  \href{http://arxiv.org/abs/1905.02278}{{\tt arXiv:1905.02278}}.

\bibitem{Lancaster:2017ksf}
L.~Lancaster, F.-Y. Cyr-Racine, L.~Knox, and Z.~Pan, {\it {A tale of two modes:
  Neutrino free-streaming in the early universe}},  {\em JCAP} {\bf 1707}
  (2017), no.~07 033, [\href{http://arxiv.org/abs/1704.06657}{{\tt
  arXiv:1704.06657}}].

\bibitem{Niedermann:2019olb}
F.~Niedermann and M.~S. Sloth, {\it {New Early Dark Energy}},
  \href{http://arxiv.org/abs/1910.10739}{{\tt arXiv:1910.10739}}.

\bibitem{Hart:2019dxi}
L.~Hart and J.~Chluba, {\it {Updated fundamental constant constraints from
  Planck 2018 data and possible relations to the Hubble tension}},
  \href{http://arxiv.org/abs/1912.03986}{{\tt arXiv:1912.03986}}.

\bibitem{Yadav:2019jio}
S.~K. Yadav, {\it {Constraints on Dark Matter-Photon Coupling in the Presence
  of Time-Varying Dark Energy}},  {\em Mod. Phys. Lett.} {\bf A33} (2019)
  1950358, [\href{http://arxiv.org/abs/1907.05886}{{\tt arXiv:1907.05886}}].

\bibitem{Kasai:2019yqn}
M.~Kasai and T.~Futamase, {\it {A possible solution to the Hubble constant
  discrepancy -- Cosmology where the local volume expansion is driven by the
  domain average density}},  {\em PTEP} {\bf 2019} (2019), no.~7 073E01,
  [\href{http://arxiv.org/abs/1904.09689}{{\tt arXiv:1904.09689}}].

\bibitem{Amirhashchi:2020qep}
H.~Amirhashchi and A.~K. Yadav, {\it {Interacting Dark Sectors in Anisotropic
  Universe: Observational Constraints and $H_{0}$ Tension}},
  \href{http://arxiv.org/abs/2001.03775}{{\tt arXiv:2001.03775}}.

\bibitem{Perez:2020cwa}
A.~Perez, D.~Sudarsky, and E.~Wilson-Ewing, {\it {Resolving the $H_0$ tension
  with diffusion}},  \href{http://arxiv.org/abs/2001.07536}{{\tt
  arXiv:2001.07536}}.

\bibitem{Pan:2020bur}
S.~Pan, W.~Yang, and A.~Paliathanasis, {\it {Non-linear interacting
  cosmological models after Planck 2018 legacy release and the $H_0$ tension}},
   {\em Mon. Not. Roy. Astron. Soc.} {\bf 493} (2020), no.~3 3114--3131,
  [\href{http://arxiv.org/abs/2002.03408}{{\tt arXiv:2002.03408}}].

\bibitem{DAgostino:2020dhv}
R.~D'Agostino and R.~C. Nunes, {\it {Measurements of $H_0$ in modified gravity
  theories}},  \href{http://arxiv.org/abs/2002.06381}{{\tt arXiv:2002.06381}}.

\bibitem{Benevento:2020fev}
G.~Benevento, W.~Hu, and M.~Raveri, {\it {Can Late Dark Energy Transitions
  Raise the Hubble constant?}},  \href{http://arxiv.org/abs/2002.11707}{{\tt
  arXiv:2002.11707}}.

\bibitem{DiValentino:2020zio}
E.~Di~Valentino et~al., {\it {Cosmology Intertwined II: The Hubble Constant
  Tension}},  \href{http://arxiv.org/abs/2008.11284}{{\tt arXiv:2008.11284}}.

\bibitem{Tram:2016rcw}
T.~Tram, R.~Vallance, and V.~Vennin, {\it {Inflation Model Selection meets Dark
  Radiation}},  {\em JCAP} {\bf 01} (2017) 046,
  [\href{http://arxiv.org/abs/1606.09199}{{\tt arXiv:1606.09199}}].

\bibitem{DiValentino:2016ziq}
E.~Di~Valentino and L.~Mersini-Houghton, {\it {Testing Predictions of the
  Quantum Landscape Multiverse 2: The Exponential Inflationary Potential}},
  {\em JCAP} {\bf 03} (2017) 020, [\href{http://arxiv.org/abs/1612.08334}{{\tt
  arXiv:1612.08334}}].

\bibitem{Gomez-Valent:2020mqn}
A.~G\'omez-Valent, V.~Pettorino, and L.~Amendola, {\it {Update on coupled dark
  energy and the $H_0$ tension}},  {\em Phys. Rev. D} {\bf 101} (2020), no.~12
  123513, [\href{http://arxiv.org/abs/2004.00610}{{\tt arXiv:2004.00610}}].

\bibitem{Lucca:2020zjb}
M.~Lucca and D.~C. Hooper, {\it {Tensions in the dark: shedding light on Dark
  Matter-Dark Energy interactions}},
  \href{http://arxiv.org/abs/2002.06127}{{\tt arXiv:2002.06127}}.

\bibitem{vandeBruck:2017idm}
C.~Van De~Bruck and J.~Mifsud, {\it {Searching for dark matter - dark energy
  interactions: going beyond the conformal case}},  {\em Phys. Rev. D} {\bf 97}
  (2018), no.~2 023506, [\href{http://arxiv.org/abs/1709.04882}{{\tt
  arXiv:1709.04882}}].

\bibitem{DiValentino:2020naf}
E.~Di~Valentino, A.~Mukherjee, and A.~A. Sen, {\it {Dark Energy with Phantom
  Crossing and the $H_0$ tension}},
  \href{http://arxiv.org/abs/2005.12587}{{\tt arXiv:2005.12587}}.

\bibitem{Keeley:2019esp}
R.~E. Keeley, S.~Joudaki, M.~Kaplinghat, and D.~Kirkby, {\it {Implications of a
  transition in the dark energy equation of state for the $H_0$ and $\sigma_8$
  tensions}},  {\em JCAP} {\bf 12} (2019) 035,
  [\href{http://arxiv.org/abs/1905.10198}{{\tt arXiv:1905.10198}}].

\bibitem{Yang:2018prh}
W.~Yang, S.~Pan, E.~Di~Valentino, and E.~N. Saridakis, {\it {Observational
  constraints on dynamical dark energy with pivoting redshift}},  {\em
  Universe} {\bf 5} (2019), no.~11 219,
  [\href{http://arxiv.org/abs/1811.06932}{{\tt arXiv:1811.06932}}].

\bibitem{Karwal:2016vyq}
T.~Karwal and M.~Kamionkowski, {\it {Dark energy at early times, the Hubble
  parameter, and the string axiverse}},  {\em Phys. Rev. D} {\bf 94} (2016),
  no.~10 103523, [\href{http://arxiv.org/abs/1608.01309}{{\tt
  arXiv:1608.01309}}].

\bibitem{Smith:2019ihp}
T.~L. Smith, V.~Poulin, and M.~A. Amin, {\it {Oscillating scalar fields and the
  Hubble tension: a resolution with novel signatures}},  {\em Phys. Rev. D}
  {\bf 101} (2020), no.~6 063523, [\href{http://arxiv.org/abs/1908.06995}{{\tt
  arXiv:1908.06995}}].

\bibitem{Lucca:2020fgp}
M.~Lucca, {\it {The role of CMB spectral distortions in the Hubble tension: a
  proof of principle}},  \href{http://arxiv.org/abs/2008.01115}{{\tt
  arXiv:2008.01115}}.

\bibitem{Lin:2019qug}
M.-X. Lin, G.~Benevento, W.~Hu, and M.~Raveri, {\it {Acoustic Dark Energy:
  Potential Conversion of the Hubble Tension}},  {\em Phys. Rev. D} {\bf 100}
  (2019), no.~6 063542, [\href{http://arxiv.org/abs/1905.12618}{{\tt
  arXiv:1905.12618}}].

\bibitem{Kumar:2018yhh}
S.~Kumar, R.~C. Nunes, and S.~K. Yadav, {\it {Cosmological bounds on dark
  matter-photon coupling}},  {\em Phys. Rev. D} {\bf 98} (2018), no.~4 043521,
  [\href{http://arxiv.org/abs/1803.10229}{{\tt arXiv:1803.10229}}].

\bibitem{Yang:2020zuk}
W.~Yang, E.~Di~Valentino, S.~Pan, S.~Basilakos, and A.~Paliathanasis, {\it
  {Metastable dark energy models in light of Planck 2018: Alleviating the $H_0$
  tension}},  \href{http://arxiv.org/abs/2001.04307}{{\tt arXiv:2001.04307}}.

\bibitem{Pan:2020zza}
S.~Pan, G.~S. Sharov, and W.~Yang, {\it {Field theoretic interpretations of
  interacting dark energy scenarios and recent observations}},  {\em Phys. Rev.
  D} {\bf 101} (2020), no.~10 103533,
  [\href{http://arxiv.org/abs/2001.03120}{{\tt arXiv:2001.03120}}].

\bibitem{Wu:2020nxz}
W.~K. Wu, P.~Motloch, W.~Hu, and M.~Raveri, {\it {Hubble constant tension
  between CMB lensing and BAO measurements}},
  \href{http://arxiv.org/abs/2004.10207}{{\tt arXiv:2004.10207}}.

\bibitem{Blinov:2020hmc}
N.~Blinov and G.~Marques-Tavares, {\it {Interacting radiation after Planck and
  its implications for the Hubble Tension}},
  \href{http://arxiv.org/abs/2003.08387}{{\tt arXiv:2003.08387}}.

\bibitem{Alestas:2020mvb}
G.~Alestas, L.~Kazantzidis, and L.~Perivolaropoulos, {\it {$H_0$ Tension,
  Phantom Dark Energy and Cosmological Parameter Degeneracies}},  {\em Phys.
  Rev. D} {\bf 101} (2020), no.~12 123516,
  [\href{http://arxiv.org/abs/2004.08363}{{\tt arXiv:2004.08363}}].

\bibitem{Clark:2020miy}
S.~J. Clark, K.~Vattis, and S.~M. Koushiappas, {\it {CMB constraints on
  late-universe decaying dark matter as a solution to the $H_0$ tension}},
  \href{http://arxiv.org/abs/2006.03678}{{\tt arXiv:2006.03678}}.

\bibitem{Keeley:2020rmo}
R.~E. Keeley, A.~Shafieloo, D.~K. Hazra, and T.~Souradeep, {\it {Inflation
  Wars: A New Hope}},  \href{http://arxiv.org/abs/2006.12710}{{\tt
  arXiv:2006.12710}}.

\bibitem{Hazra:2018opk}
D.~K. Hazra, A.~Shafieloo, and T.~Souradeep, {\it {Parameter discordance in
  Planck CMB and low-redshift measurements: projection in the primordial power
  spectrum}},  {\em JCAP} {\bf 04} (2019) 036,
  [\href{http://arxiv.org/abs/1810.08101}{{\tt arXiv:1810.08101}}].

\bibitem{Niedermann:2020dwg}
F.~Niedermann and M.~S. Sloth, {\it {Resolving the Hubble Tension with New
  Early Dark Energy}},  \href{http://arxiv.org/abs/2006.06686}{{\tt
  arXiv:2006.06686}}.

\bibitem{Archidiacono:2020yey}
M.~Archidiacono, S.~Gariazzo, C.~Giunti, S.~Hannestad, and T.~Tram, {\it
  {Sterile neutrino self-interactions: $H_0$ tension and short-baseline
  anomalies}},  \href{http://arxiv.org/abs/2006.12885}{{\tt arXiv:2006.12885}}.

\bibitem{DiValentino:2020kha}
E.~Di~Valentino, E.~V. Linder, and A.~Melchiorri, {\it {$H_0$ Ex Machina:
  Vacuum Metamorphosis and Beyond $H_0$}},
  \href{http://arxiv.org/abs/2006.16291}{{\tt arXiv:2006.16291}}.

\bibitem{Capozziello:2020nyq}
S.~Capozziello, M.~Benetti, and A.~D. Spallicci, {\it {Addressing the
  cosmological $H_0$ tension by the Heisenberg uncertainty}},
  \href{http://arxiv.org/abs/2007.00462}{{\tt arXiv:2007.00462}}.

\bibitem{Anchordoqui:2019yzc}
L.~A. Anchordoqui and S.~E. Perez~Bergliaffa, {\it {Hot thermal universe
  endowed with massive dark vector fields and the Hubble tension}},  {\em Phys.
  Rev. D} {\bf 100} (2019), no.~12 123525,
  [\href{http://arxiv.org/abs/1910.05860}{{\tt arXiv:1910.05860}}].

\bibitem{Ivanov:2020mfr}
M.~M. Ivanov, Y.~Ali-Haïmoud, and J.~Lesgourgues, {\it {H0 tension or T0
  tension?}},  \href{http://arxiv.org/abs/2005.10656}{{\tt arXiv:2005.10656}}.

\bibitem{Gonzalez:2020fdy}
M.~Gonzalez, M.~P. Hertzberg, and F.~Rompineve, {\it {Ultralight Scalar Decay
  and the Hubble Tension}},  \href{http://arxiv.org/abs/2006.13959}{{\tt
  arXiv:2006.13959}}.

\bibitem{Hryczuk:2020jhi}
A.~Hryczuk and K.~Jodlowski, {\it {Self-interacting dark matter from late
  decays and the $H_0$ tension}},  \href{http://arxiv.org/abs/2006.16139}{{\tt
  arXiv:2006.16139}}.

\bibitem{Carneiro:2018xwq}
S.~Carneiro, P.~C. de~Holanda, C.~Pigozzo, and F.~Sobreira, {\it {Is the $H_0$
  tension suggesting a fourth neutrino generation?}},  {\em Phys. Rev.} {\bf
  D100} (2019), no.~2 023505, [\href{http://arxiv.org/abs/1812.06064}{{\tt
  arXiv:1812.06064}}].

\bibitem{Paul:2018njm}
A.~Paul, A.~Ghoshal, A.~Chatterjee, and S.~Pal, {\it {Inflation, (P)reheating
  and Neutrino Anomalies: Production of Sterile Neutrinos with Secret
  Interactions}},  {\em Eur. Phys. J.} {\bf C79} (2019), no.~10 818,
  [\href{http://arxiv.org/abs/1808.09706}{{\tt arXiv:1808.09706}}].

\bibitem{Gelmini:2019deq}
G.~B. Gelmini, A.~Kusenko, and V.~Takhistov, {\it {Hints of Sterile Neutrinos
  in Recent Measurements of the Hubble Parameter}},
  \href{http://arxiv.org/abs/1906.10136}{{\tt arXiv:1906.10136}}.

\bibitem{Anchordoqui:2020znj}
L.~A. Anchordoqui, {\it {Hubble Hullabaloo and String Cosmology}},  5, 2020.
\newblock \href{http://arxiv.org/abs/2005.01217}{{\tt arXiv:2005.01217}}.

\bibitem{Sakstein:2019fmf}
J.~Sakstein and M.~Trodden, {\it {Early Dark Energy from Massive Neutrinos as a
  Natural Resolution of the Hubble Tension}},  {\em Phys. Rev. Lett.} {\bf 124}
  (2020), no.~16 161301, [\href{http://arxiv.org/abs/1911.11760}{{\tt
  arXiv:1911.11760}}].

\bibitem{Das:2020wfe}
A.~Gogoi, P.~Chanda, and S.~Das, {\it {Dark matter nugget and new early dark
  energy from interacting neutrino: A promising solution to Hubble anomaly}},
  \href{http://arxiv.org/abs/2005.11889}{{\tt arXiv:2005.11889}}.

\bibitem{Ye:2020btb}
G.~Ye and Y.-S. Piao, {\it {Is the Hubble tension a hint of AdS around
  recombination?}},  \href{http://arxiv.org/abs/2001.02451}{{\tt
  arXiv:2001.02451}}.

\bibitem{Hart:2017ndk}
L.~Hart and J.~Chluba, {\it {New constraints on time-dependent variations of
  fundamental constants using Planck data}},  {\em Mon. Not. Roy. Astron. Soc.}
  {\bf 474} (2018), no.~2 1850--1861,
  [\href{http://arxiv.org/abs/1705.03925}{{\tt arXiv:1705.03925}}].

\bibitem{Chiang:2018xpn}
C.-T. Chiang and A.~z. Slosar, {\it {Inferences of $H_0$ in presence of a
  non-standard recombination}},  \href{http://arxiv.org/abs/1811.03624}{{\tt
  arXiv:1811.03624}}.

\bibitem{Jedamzik:2020krr}
K.~Jedamzik and L.~Pogosian, {\it {Relieving the Hubble tension with primordial
  magnetic fields}},  \href{http://arxiv.org/abs/2004.09487}{{\tt
  arXiv:2004.09487}}.

\bibitem{Yang:2020myd}
W.~Yang, E.~Di~Valentino, S.~Pan, and O.~Mena, {\it {A complete model of
  Phenomenologically Emergent Dark Energy}},
  \href{http://arxiv.org/abs/2007.02927}{{\tt arXiv:2007.02927}}.

\bibitem{Chudaykin:2020igl}
A.~Chudaykin, D.~Gorbunov, and N.~Nedelko, {\it {Exploring Early Dark Energy
  solution to the Hubble tension with Planck and SPTPol data}},
  \href{http://arxiv.org/abs/2011.04682}{{\tt arXiv:2011.04682}}.

\bibitem{Sekiguchi:2020teg}
T.~Sekiguchi and T.~Takahashi, {\it {Early recombination as a solution to the
  $H_0$ tension}},  \href{http://arxiv.org/abs/2007.03381}{{\tt
  arXiv:2007.03381}}.

\bibitem{Bose:2020cjb}
B.~Bose and L.~Lombriser, {\it {Easing cosmic tensions with an open and hotter
  universe}},  \href{http://arxiv.org/abs/2006.16149}{{\tt arXiv:2006.16149}}.

\bibitem{Agrawal:2019dlm}
P.~Agrawal, G.~Obied, and C.~Vafa, {\it {$H_0$ Tension, Swampland Conjectures
  and the Epoch of Fading Dark Matter}},
  \href{http://arxiv.org/abs/1906.08261}{{\tt arXiv:1906.08261}}.

\bibitem{Anchordoqui:2019amx}
L.~A. Anchordoqui, I.~Antoniadis, D.~Lüst, J.~F. Soriano, and T.~R. Taylor,
  {\it {$H_0$ tension and the String Swampland}},  {\em Phys. Rev. D} {\bf 101}
  (2020) 083532, [\href{http://arxiv.org/abs/1912.00242}{{\tt
  arXiv:1912.00242}}].

\bibitem{Braglia:2020iik}
M.~Braglia, M.~Ballardini, W.~T. Emond, F.~Finelli, A.~E. Gumrukcuoglu,
  K.~Koyama, and D.~Paoletti, {\it {A larger value for $H_0$ by an evolving
  gravitational constant}},  \href{http://arxiv.org/abs/2004.11161}{{\tt
  arXiv:2004.11161}}.

\bibitem{Ballardini:2020iws}
M.~Ballardini, M.~Braglia, F.~Finelli, D.~Paoletti, A.~A. Starobinsky, and
  C.~Umiltà, {\it {Scalar-tensor theories of gravity, neutrino physics, and
  the $H_0$ tension}},  \href{http://arxiv.org/abs/2004.14349}{{\tt
  arXiv:2004.14349}}.

\bibitem{Li:2019yem}
X.~Li and A.~Shafieloo, {\it {A Simple Phenomenological Emergent Dark Energy
  Model can Resolve the Hubble Tension}},  {\em Astrophys. J. Lett.} {\bf 883}
  (2019), no.~1 L3, [\href{http://arxiv.org/abs/1906.08275}{{\tt
  arXiv:1906.08275}}].

\bibitem{Rezaei:2020mrj}
M.~Rezaei, T.~Naderi, M.~Malekjani, and A.~Mehrabi, {\it {A Bayesian comparison
  between $\Lambda$CDM and phenomenologically emergent dark energy models}},
  {\em Eur. Phys. J. C} {\bf 80} (2020), no.~5 374,
  [\href{http://arxiv.org/abs/2004.08168}{{\tt arXiv:2004.08168}}].

\bibitem{Li:2020ybr}
X.~Li and A.~Shafieloo, {\it {Generalised Emergent Dark Energy Model:
  Confronting $\Lambda$ and PEDE}},
  \href{http://arxiv.org/abs/2001.05103}{{\tt arXiv:2001.05103}}.

\bibitem{Choi:2019jck}
G.~Choi, M.~Suzuki, and T.~T. Yanagida, {\it {Quintessence Axion Dark Energy
  and a Solution to the Hubble Tension}},  {\em Phys. Lett. B} {\bf 805} (2020)
  135408, [\href{http://arxiv.org/abs/1910.00459}{{\tt arXiv:1910.00459}}].

\bibitem{Choi:2020tqp}
G.~Choi, M.~Suzuki, and T.~T. Yanagida, {\it {Degenerate Sub-keV Fermion Dark
  Matter from a Solution to the Hubble Tension}},  {\em Phys. Rev. D} {\bf 101}
  (2020), no.~7 075031, [\href{http://arxiv.org/abs/2002.00036}{{\tt
  arXiv:2002.00036}}].

\bibitem{Berezhiani:2015yta}
Z.~Berezhiani, A.~Dolgov, and I.~Tkachev, {\it {Reconciling Planck results with
  low redshift astronomical measurements}},  {\em Phys. Rev. D} {\bf 92}
  (2015), no.~6 061303, [\href{http://arxiv.org/abs/1505.03644}{{\tt
  arXiv:1505.03644}}].

\bibitem{Anchordoqui:2015lqa}
L.~A. Anchordoqui, V.~Barger, H.~Goldberg, X.~Huang, D.~Marfatia, L.~H.~M.
  da~Silva, and T.~J. Weiler, {\it {IceCube neutrinos, decaying dark matter,
  and the Hubble constant}},  {\em Phys. Rev. D} {\bf 92} (2015), no.~6 061301,
  [\href{http://arxiv.org/abs/1506.08788}{{\tt arXiv:1506.08788}}]. [Erratum:
  Phys.Rev.D 94, 069901 (2016)].

\bibitem{Desai:2019pvs}
A.~Desai, K.~R. Dienes, and B.~Thomas, {\it {Constraining Dark-Matter Ensembles
  with Supernova Data}},  {\em Phys. Rev. D} {\bf 101} (2020), no.~3 035031,
  [\href{http://arxiv.org/abs/1909.07981}{{\tt arXiv:1909.07981}}].

\bibitem{Alcaniz:2019kah}
J.~Alcaniz, N.~Bernal, A.~Masiero, and F.~S. Queiroz, {\it {Light Dark Matter:
  A Common Solution to the Lithium and ${H_0}$ Problems}},
  \href{http://arxiv.org/abs/1912.05563}{{\tt arXiv:1912.05563}}.

\bibitem{Chudaykin:2016yfk}
A.~Chudaykin, D.~Gorbunov, and I.~Tkachev, {\it {Dark matter component decaying
  after recombination: Lensing constraints with Planck data}},  {\em Phys. Rev.
  D} {\bf 94} (2016) 023528, [\href{http://arxiv.org/abs/1602.08121}{{\tt
  arXiv:1602.08121}}].

\bibitem{Chudaykin:2017ptd}
A.~Chudaykin, D.~Gorbunov, and I.~Tkachev, {\it {Dark matter component decaying
  after recombination: Sensitivity to baryon acoustic oscillation and redshift
  space distortion probes}},  {\em Phys. Rev. D} {\bf 97} (2018), no.~8 083508,
  [\href{http://arxiv.org/abs/1711.06738}{{\tt arXiv:1711.06738}}].

\bibitem{Hill:2020osr}
J.~C. Hill, E.~McDonough, M.~W. Toomey, and S.~Alexander, {\it {Early Dark
  Energy Does Not Restore Cosmological Concordance}},
  \href{http://arxiv.org/abs/2003.07355}{{\tt arXiv:2003.07355}}.

\bibitem{Ivanov:2020ril}
M.~M. Ivanov, E.~McDonough, J.~C. Hill, M.~Simonovi\'c, M.~W. Toomey,
  S.~Alexander, and M.~Zaldarriaga, {\it {Constraining Early Dark Energy with
  Large-Scale Structure}},  \href{http://arxiv.org/abs/2006.11235}{{\tt
  arXiv:2006.11235}}.

\bibitem{Rezaei:2020lfy}
M.~Rezaei, S.~P. Ojaghi, and M.~Malekjani, {\it {Cosmography approach to dark
  energy cosmologies: new constrains using the Hubble diagrams of supernovae,
  quasars and gamma-ray bursts}},  \href{http://arxiv.org/abs/2008.03092}{{\tt
  arXiv:2008.03092}}.

\bibitem{Wang:2020dsc}
D.~Wang, {\it {Can $f(R)$ gravity relieve $H_0$ and $\sigma_8$ tensions?}},
  \href{http://arxiv.org/abs/2008.03966}{{\tt arXiv:2008.03966}}.

\bibitem{Leonhardt:2020qam}
U.~Leonhardt and D.~Berechya, {\it {Observed Hubble constant is consistent with
  physics of the quantum vacuum}},  \href{http://arxiv.org/abs/2008.04789}{{\tt
  arXiv:2008.04789}}.

\bibitem{Ballesteros:2020sik}
G.~Ballesteros, A.~Notari, and F.~Rompineve, {\it {The $H_0$ tension: $\Delta
  G_N$ vs. $\Delta N_{\rm eff}$}},  \href{http://arxiv.org/abs/2004.05049}{{\tt
  arXiv:2004.05049}}.

\bibitem{Blinov:2019gcj}
N.~Blinov, K.~J. Kelly, G.~Z. Krnjaic, and S.~D. McDermott, {\it {Constraining
  the Self-Interacting Neutrino Interpretation of the Hubble Tension}},  {\em
  Phys. Rev. Lett.} {\bf 123} (2019), no.~19 191102,
  [\href{http://arxiv.org/abs/1905.02727}{{\tt arXiv:1905.02727}}].

\bibitem{Hernandez-Almada:2020uyr}
A.~Hernández-Almada, G.~Leon, J.~Magaña, M.~A. García-Aspeitia, and
  V.~Motta, {\it {Generalized Emergent Dark Energy: observational Hubble data
  constraints and stability analysis}},
  \href{http://arxiv.org/abs/2002.12881}{{\tt arXiv:2002.12881}}.

\bibitem{Feeney:2017sgx}
S.~M. Feeney, D.~J. Mortlock, and N.~Dalmasso, {\it {Clarifying the Hubble
  constant tension with a Bayesian hierarchical model of the local distance
  ladder}},  {\em Mon. Not. Roy. Astron. Soc.} {\bf 476} (2018), no.~3
  3861--3882, [\href{http://arxiv.org/abs/1707.00007}{{\tt arXiv:1707.00007}}].

\bibitem{Banerjee:2020xcn}
A.~Banerjee, H.~Cai, L.~Heisenberg, E.~O. Colg\'ain, M.~Sheikh-Jabbari, and
  T.~Yang, {\it {Hubble Sinks In The Low-Redshift Swampland}},
  \href{http://arxiv.org/abs/2006.00244}{{\tt arXiv:2006.00244}}.

\bibitem{Adler:2019fnp}
S.~L. Adler, {\it {Implications of a frame dependent dark energy for the
  spacetime metric, cosmography, and effective Hubble constant}},  {\em Phys.
  Rev. D} {\bf 100} (2019), no.~12 123503,
  [\href{http://arxiv.org/abs/1905.08228}{{\tt arXiv:1905.08228}}].

\bibitem{Gu:2020ozv}
Y.~Gu, M.~Khlopov, L.~Wu, J.~M. Yang, and B.~Zhu, {\it {Light gravitino dark
  matter for Hubble tension and LHC}},
  \href{http://arxiv.org/abs/2006.09906}{{\tt arXiv:2006.09906}}.

\bibitem{Akarsu:2019pwn}
{\"O}.~Akarsu, S.~Kumar, S.~Sharma, and L.~Tedesco, {\it {Constraints on a
  Bianchi type I spacetime extension of the standard $\Lambda$CDM model}},
  {\em Phys. Rev. D} {\bf 100} (2019), no.~2 023532,
  [\href{http://arxiv.org/abs/1905.06949}{{\tt arXiv:1905.06949}}].

\bibitem{Nygaard:2020sow}
A.~Nygaard, T.~Tram, and S.~Hannestad, {\it {Updated constraints on decaying
  cold dark matter}},  \href{http://arxiv.org/abs/2011.01632}{{\tt
  arXiv:2011.01632}}.

\bibitem{Benaoum:2020qsi}
H.~Benaoum, W.~Yang, S.~Pan, and E.~Di~Valentino, {\it {Modified Emergent Dark
  Energy and its Astronomical Constraints}},
  \href{http://arxiv.org/abs/2008.09098}{{\tt arXiv:2008.09098}}.

\bibitem{Krishnan:2020vaf}
C.~Krishnan, E.~O. Colgain, M.~Sheikh-Jabbari, and T.~Yang, {\it {Running
  Hubble Tension and a H0 Diagnostic}},
  \href{http://arxiv.org/abs/2011.02858}{{\tt arXiv:2011.02858}}.

\bibitem{DiValentino:2020vnx}
E.~Di~Valentino, {\it {A (brave) combined analysis of the $H_0$ late time
  direct measurements and the impact on the Dark Energy sector}},
  \href{http://arxiv.org/abs/2011.00246}{{\tt arXiv:2011.00246}}.

\bibitem{Kitazawa:2020qdx}
N.~Kitazawa, {\it {Polarizations of CMB and the Hubble tension}},
  \href{http://arxiv.org/abs/2010.12164}{{\tt arXiv:2010.12164}}.

\bibitem{Anchordoqui:2020djl}
L.~A. Anchordoqui, {\it {Decaying dark matter, the $H_0$ tension, and the
  lithium problem}},  \href{http://arxiv.org/abs/2010.09715}{{\tt
  arXiv:2010.09715}}.

\bibitem{Yao:2020hkw}
Y.-H. Yao and X.-H. Meng, {\it {A new coupled three-form dark energy model and
  implications for the H0 tension}},  {\em Phys. Dark Univ.} {\bf 30} (2020)
  100729.

\bibitem{Artymowski:2020zwy}
M.~Artymowski, I.~Ben-Dayan, and U.~Kumar, {\it {Emergent dark energy from
  unparticles}},  \href{http://arxiv.org/abs/2010.02998}{{\tt
  arXiv:2010.02998}}.

\bibitem{Jedamzik:2020zmd}
K.~Jedamzik, L.~Pogosian, and G.-B. Zhao, {\it {Why reducing the cosmic sound
  horizon can not fully resolve the Hubble tension}},
  \href{http://arxiv.org/abs/2010.04158}{{\tt arXiv:2010.04158}}.

\bibitem{Smith:2020rxx}
T.~L. Smith, V.~Poulin, J.~L. Bernal, K.~K. Boddy, M.~Kamionkowski, and
  R.~Murgia, {\it {Early dark energy is not excluded by current large-scale
  structure data}},  \href{http://arxiv.org/abs/2009.10740}{{\tt
  arXiv:2009.10740}}.

\bibitem{Murgia:2020ryi}
R.~Murgia, G.~F. Abell\'an, and V.~Poulin, {\it {The early dark energy
  resolution to the Hubble tension in light of weak lensing surveys and lensing
  anomalies}},  \href{http://arxiv.org/abs/2009.10733}{{\tt arXiv:2009.10733}}.

\bibitem{LinaresCedeno:2020uxx}
F.~X. Linares Cede\~no and U.~Nucamendi, {\it {Revisiting cosmological
  diffusion models in Unimodular Gravity and the $H_0$ tension}},
  \href{http://arxiv.org/abs/2009.10268}{{\tt arXiv:2009.10268}}.

\bibitem{Lin:2020jcb}
M.-X. Lin, W.~Hu, and M.~Raveri, {\it {Testing $H_0$ in Acoustic Dark Energy
  with Planck and ACT Polarization}},
  \href{http://arxiv.org/abs/2009.08974}{{\tt arXiv:2009.08974}}.

\bibitem{DeFelice:2020cpt}
A.~De~Felice, S.~Mukohyama, and M.~C. Pookkillath, {\it {Addressing $H_0$
  tension by means of VCDM}},  \href{http://arxiv.org/abs/2009.08718}{{\tt
  arXiv:2009.08718}}.

\bibitem{Alvarez:2020xmk}
P.~D. Alvarez, B.~Koch, C.~Laporte, and A.~Rincon, {\it {Can scale--dependent
  cosmology alleviate the $H_0$ tension?}},
  \href{http://arxiv.org/abs/2009.02311}{{\tt arXiv:2009.02311}}.

\bibitem{Arias-Aragon:2020qip}
F.~Arias-Aragon, E.~Fernandez-Martinez, M.~Gonzalez-Lopez, and L.~Merlo, {\it
  {Neutrino Masses and Hubble Tension via a Majoron in MFV}},
  \href{http://arxiv.org/abs/2009.01848}{{\tt arXiv:2009.01848}}.

\bibitem{Kameli:2020kao}
H.~Kameli and S.~Baghram, {\it {Merger history of dark matter halos in the
  light of $H_0$ tension}},  \href{http://arxiv.org/abs/2008.13175}{{\tt
  arXiv:2008.13175}}.

\bibitem{Ortiz:2020noa}
C.~Ortiz, {\it {Surface Tension: Accelerated Expansion, Coincidence Problem \&
  Hubble Tension}},  \href{http://arxiv.org/abs/2011.02317}{{\tt
  arXiv:2011.02317}}.

\bibitem{Mandal:2020buf}
S.~Mandal, D.~Wang, and P.~Sahoo, {\it {Cosmography in $f(Q)$ gravity}},
  \href{http://arxiv.org/abs/2011.00420}{{\tt arXiv:2011.00420}}.

\bibitem{Hashim:2020sez}
M.~Hashim, W.~El~Hanafy, A.~Golovnev, and A.~El-Zant, {\it {Toward a
  concordance teleparallel Cosmology I: Background Dynamics}},
  \href{http://arxiv.org/abs/2010.14964}{{\tt arXiv:2010.14964}}.

\bibitem{Braglia:2020bym}
M.~Braglia, W.~T. Emond, F.~Finelli, A.~E. Gumrukcuoglu, and K.~Koyama, {\it
  {Unified framework for early dark energy from $\alpha$-attractors}},  {\em
  Phys. Rev. D} {\bf 102} (2020), no.~8 083513,
  [\href{http://arxiv.org/abs/2005.14053}{{\tt arXiv:2005.14053}}].

\bibitem{Chudaykin:2020acu}
A.~Chudaykin, D.~Gorbunov, and N.~Nedelko, {\it {Combined analysis of Planck
  and SPTPol data favors the early dark energy models}},
  \href{http://arxiv.org/abs/2004.13046}{{\tt arXiv:2004.13046}}.

\bibitem{Odintsov:2020qzd}
S.~D. Odintsov, D.~S.-C. G\'omez, and G.~S. Sharov, {\it {Analyzing the $H_0$
  tension in $F(R)$ gravity models}},
  \href{http://arxiv.org/abs/2011.03957}{{\tt arXiv:2011.03957}}.

\bibitem{Yao:2020pji}
Y.~Yao and X.~Meng, {\it {Relieve the $H_0$ tension with a new coupled
  generalized three-form dark energy model}},
  \href{http://arxiv.org/abs/2011.09160}{{\tt arXiv:2011.09160}}.

\bibitem{Cruz:2020cje}
N.~J. Cruz and C.~Escamilla-Rivera, {\it {Late-time and Big Bang
  nucleosynthesis constraints for generic modify gravity surveys}},
  \href{http://arxiv.org/abs/2011.09623}{{\tt arXiv:2011.09623}}.

\bibitem{daSilva:2020mvk}
W.~da~Silva and R.~Silva, {\it {Growth of matter perturbations in the extended
  viscous dark energy models}},  \href{http://arxiv.org/abs/2011.09516}{{\tt
  arXiv:2011.09516}}.

\bibitem{daSilva:2020bdc}
W.~da~Silva and R.~Silva, {\it {Cosmological Perturbations in the Tsallis
  Holographic Dark Energy Scenarios}},
  \href{http://arxiv.org/abs/2011.09520}{{\tt arXiv:2011.09520}}.

\bibitem{Dhawan:2020xmp}
S.~Dhawan, D.~Brout, D.~Scolnic, A.~Goobar, A.~Riess, and V.~Miranda, {\it
  {Cosmological Model Insensitivity of Local $H_0$ from the Cepheid Distance
  Ladder}},  {\em Astrophys. J.} {\bf 894} (2020), no.~1 54,
  [\href{http://arxiv.org/abs/2001.09260}{{\tt arXiv:2001.09260}}].

\bibitem{Hinshaw:2012aka}
{\bf WMAP} Collaboration, G.~Hinshaw et~al., {\it {Nine-Year Wilkinson
  Microwave Anisotropy Probe (WMAP) Observations: Cosmological Parameter
  Results}},  {\em Astrophys. J. Suppl.} {\bf 208} (2013) 19,
  [\href{http://arxiv.org/abs/1212.5226}{{\tt arXiv:1212.5226}}].

\bibitem{Henning:2017nuy}
{\bf SPT} Collaboration, J.~Henning et~al., {\it {Measurements of the
  Temperature and E-Mode Polarization of the CMB from 500 Square Degrees of
  SPTpol Data}},  {\em Astrophys. J.} {\bf 852} (2018), no.~2 97,
  [\href{http://arxiv.org/abs/1707.09353}{{\tt arXiv:1707.09353}}].

\bibitem{Aiola:2020azj}
{\bf ACT} Collaboration, S.~Aiola et~al., {\it {The Atacama Cosmology
  Telescope: DR4 Maps and Cosmological Parameters}},
  \href{http://arxiv.org/abs/2007.07288}{{\tt arXiv:2007.07288}}.

\bibitem{Ivanov:2019pdj}
M.~M. Ivanov, M.~Simonovi\'c, and M.~Zaldarriaga, {\it {Cosmological Parameters
  from the BOSS Galaxy Power Spectrum}},  {\em JCAP} {\bf 05} (2020) 042,
  [\href{http://arxiv.org/abs/1909.05277}{{\tt arXiv:1909.05277}}].

\bibitem{DAmico:2019fhj}
G.~D'Amico, J.~Gleyzes, N.~Kokron, D.~Markovic, L.~Senatore, P.~Zhang,
  F.~Beutler, and H.~Gil-Marín, {\it {The Cosmological Analysis of the
  SDSS/BOSS data from the Effective Field Theory of Large-Scale Structure}},
  {\em JCAP} {\bf 05} (2020) 005, [\href{http://arxiv.org/abs/1909.05271}{{\tt
  arXiv:1909.05271}}].

\bibitem{Alam:2020sor}
{\bf eBOSS} Collaboration, S.~Alam et~al., {\it {The Completed SDSS-IV extended
  Baryon Oscillation Spectroscopic Survey: Cosmological Implications from two
  Decades of Spectroscopic Surveys at the Apache Point observatory}},
  \href{http://arxiv.org/abs/2007.08991}{{\tt arXiv:2007.08991}}.

\bibitem{Zhang:2018air}
X.~Zhang and Q.-G. Huang, {\it {Constraints on $H_0$ from WMAP and BAO
  Measurements}},  {\em Commun. Theor. Phys.} {\bf 71} (2019), no.~7 826--830,
  [\href{http://arxiv.org/abs/1812.01877}{{\tt arXiv:1812.01877}}].

\bibitem{Pogosian:2020ded}
L.~Pogosian, G.-B. Zhao, and K.~Jedamzik, {\it {Recombination-independent
  determination of the sound horizon and the Hubble constant from BAO}},
  \href{http://arxiv.org/abs/2009.08455}{{\tt arXiv:2009.08455}}.

\bibitem{Philcox:2020xbv}
O.~H. Philcox, B.~D. Sherwin, G.~S. Farren, and E.~J. Baxter, {\it {Determining
  the Hubble Constant without the Sound Horizon: Measurements from Galaxy
  Surveys}},  \href{http://arxiv.org/abs/2008.08084}{{\tt arXiv:2008.08084}}.

\bibitem{Reid:2019tiq}
M.~Reid, D.~Pesce, and A.~Riess, {\it {An Improved Distance to NGC 4258 and its
  Implications for the Hubble Constant}},  {\em Astrophys. J. Lett.} {\bf 886}
  (2019), no.~2 L27, [\href{http://arxiv.org/abs/1908.05625}{{\tt
  arXiv:1908.05625}}].

\bibitem{Bonvin:2016crt}
V.~Bonvin et~al., {\it {H0LiCOW \textendash{} V. New COSMOGRAIL time delays of
  HE 0435\ensuremath{-}1223: $H_0$ to 3.8 per cent precision from strong
  lensing in a flat \ensuremath{\Lambda}CDM model}},  {\em Mon. Not. Roy.
  Astron. Soc.} {\bf 465} (2017), no.~4 4914--4930,
  [\href{http://arxiv.org/abs/1607.01790}{{\tt arXiv:1607.01790}}].

\bibitem{Birrer:2020tax}
S.~Birrer et~al., {\it {TDCOSMO IV: Hierarchical time-delay cosmography --
  joint inference of the Hubble constant and galaxy density profiles}},
  \href{http://arxiv.org/abs/2007.02941}{{\tt arXiv:2007.02941}}.

\bibitem{Birrer:2020jyr}
S.~Birrer and T.~Treu, {\it {TDCOSMO V: strategies for precise and accurate
  measurements of the Hubble constant with strong lensing}},
  \href{http://arxiv.org/abs/2008.06157}{{\tt arXiv:2008.06157}}.

\bibitem{Dhawan:2017ywl}
S.~Dhawan, S.~W. Jha, and B.~Leibundgut, {\it {Measuring the Hubble constant
  with Type Ia supernovae as near-infrared standard candles}},  {\em Astron.
  Astrophys.} {\bf 609} (2018) A72,
  [\href{http://arxiv.org/abs/1707.00715}{{\tt arXiv:1707.00715}}].

\bibitem{Freedman_2012}
W.~L. Freedman, B.~F. Madore, V.~Scowcroft, C.~Burns, A.~Monson, S.~E. Persson,
  M.~Seibert, and J.~Rigby, {\it Carnegie hubble program: A mid-infrared
  calibration of the hubble constant},  {\em The Astrophysical Journal} {\bf
  758} (Sep, 2012) 24.

\bibitem{Kim:2020gai}
Y.~J. Kim, J.~Kang, M.~G. Lee, and I.~S. Jang, {\it {Determination of the Local
  Hubble Constant from Virgo Infall Using TRGB Distances}},
  \href{http://arxiv.org/abs/2010.01364}{{\tt arXiv:2010.01364}}.

\bibitem{Gomez-Valent:2019lny}
A.~Gómez-Valent and L.~Amendola, {\it {$H_0$ from cosmic chronometers and Type
  Ia supernovae, with Gaussian processes and the weighted polynomial regression
  method}},  in {\em {15th Marcel Grossmann Meeting on Recent Developments in
  Theoretical and Experimental General Relativity, Astrophysics, and
  Relativistic Field Theories}}, 5, 2019.
\newblock \href{http://arxiv.org/abs/1905.04052}{{\tt arXiv:1905.04052}}.

\bibitem{Haridasu:2018gqm}
B.~S. Haridasu, V.~V. Lukovi\'c, M.~Moresco, and N.~Vittorio, {\it {An improved
  model-independent assessment of the late-time cosmic expansion}},  {\em JCAP}
  {\bf 10} (2018) 015, [\href{http://arxiv.org/abs/1805.03595}{{\tt
  arXiv:1805.03595}}].

\bibitem{Dutta:2019pio}
K.~Dutta, A.~Roy, Ruchika, A.~A. Sen, and M.~Sheikh-Jabbari, {\it {Cosmology
  with low-redshift observations: No signal for new physics}},  {\em Phys. Rev.
  D} {\bf 100} (2019), no.~10 103501,
  [\href{http://arxiv.org/abs/1908.07267}{{\tt arXiv:1908.07267}}].

\bibitem{Nunes:2020uex}
R.~C. Nunes and A.~Bernui, {\it {$\theta_{\rm BAO}$ estimates and the $H_0$
  tension}},  \href{http://arxiv.org/abs/2008.03259}{{\tt arXiv:2008.03259}}.

\bibitem{Liao:2019qoc}
K.~Liao, A.~Shafieloo, R.~E. Keeley, and E.~V. Linder, {\it {A
  model-independent determination of the Hubble constant from lensed quasars
  and supernovae using Gaussian process regression}},  {\em Astrophys. J.
  Lett.} {\bf 886} (2019), no.~1 L23,
  [\href{http://arxiv.org/abs/1908.04967}{{\tt arXiv:1908.04967}}].

\bibitem{Liao:2020zko}
K.~Liao, A.~Shafieloo, R.~E. Keeley, and E.~V. Linder, {\it {Determining $H_0$
  Model-Independently and Consistency Tests}},
  \href{http://arxiv.org/abs/2002.10605}{{\tt arXiv:2002.10605}}.

\bibitem{Kourkchi:2020iyz}
E.~Kourkchi, R.~B. Tully, G.~S. Anand, H.~M. Courtois, A.~Dupuy, J.~D. Neill,
  L.~Rizzi, and M.~Seibert, {\it {Cosmicflows-4: The Calibration of Optical and
  Infrared Tully--Fisher Relations}},  {\em Astrophys. J.} {\bf 896} (2020),
  no.~1 3, [\href{http://arxiv.org/abs/2004.14499}{{\tt arXiv:2004.14499}}].

\bibitem{Schombert:2020pxm}
J.~Schombert, S.~McGaugh, and F.~Lelli, {\it {Using The Baryonic Tully-Fisher
  Relation to Measure $H_o$}},  {\em Astron. J.} {\bf 160} (2020), no.~2 71,
  [\href{http://arxiv.org/abs/2006.08615}{{\tt arXiv:2006.08615}}].

\bibitem{deJaeger:2020zpb}
T.~de~Jaeger, B.~Stahl, W.~Zheng, A.~Filippenko, A.~Riess, and L.~Galbany, {\it
  {A measurement of the Hubble constant from Type II supernovae}},
  \href{http://arxiv.org/abs/2006.03412}{{\tt arXiv:2006.03412}}.

\bibitem{Fernandez-Arenas:2017isq}
D.~Fernández~Arenas, E.~Terlevich, R.~Terlevich, J.~Melnick, R.~Chávez,
  F.~Bresolin, E.~Telles, M.~Plionis, and S.~Basilakos, {\it {An independent
  determination of the local Hubble constant}},  {\em Mon. Not. Roy. Astron.
  Soc.} {\bf 474} (2018), no.~1 1250--1276,
  [\href{http://arxiv.org/abs/1710.05951}{{\tt arXiv:1710.05951}}].

\bibitem{Pesce:2020xfe}
D.~Pesce et~al., {\it {The Megamaser Cosmology Project. XIII. Combined Hubble
  constant constraints}},  {\em Astrophys. J. Lett.} {\bf 891} (2020), no.~1
  L1, [\href{http://arxiv.org/abs/2001.09213}{{\tt arXiv:2001.09213}}].

\bibitem{Qi:2020rmm}
J.-Z. Qi, J.-W. Zhao, S.~Cao, M.~Biesiada, and Y.~Liu, {\it {Measurements of
  the Hubble constant and cosmic curvature with quasars: ultra-compact radio
  structure and strong gravitational lensing}},
  \href{http://arxiv.org/abs/2011.00713}{{\tt arXiv:2011.00713}}.

\bibitem{Baxter:2020qlr}
E.~J. Baxter and B.~D. Sherwin, {\it {Determining the Hubble Constant without
  the Sound Horizon Scale: Measurements from CMB Lensing}},
  \href{http://arxiv.org/abs/2007.04007}{{\tt arXiv:2007.04007}}.

\bibitem{Freedman:2000cf}
{\bf HST} Collaboration, W.~Freedman et~al., {\it {Final results from the
  Hubble Space Telescope key project to measure the Hubble constant}},  {\em
  Astrophys. J.} {\bf 553} (2001) 47--72,
  [\href{http://arxiv.org/abs/astro-ph/0012376}{{\tt astro-ph/0012376}}].

\bibitem{Gayathri:2020fbl}
V.~Gayathri, J.~Healy, J.~Lange, B.~O'Brien, M.~Szczepanczyk, I.~Bartos,
  M.~Campanelli, S.~Klimenko, C.~Lousto, and R.~O'Shaughnessy, {\it {Hubble
  Constant Measurement with GW190521 as an Eccentric Black Hole Merger}},
  \href{http://arxiv.org/abs/2009.14247}{{\tt arXiv:2009.14247}}.

\bibitem{Mukherjee:2020kki}
S.~Mukherjee, A.~Ghosh, M.~J. Graham, C.~Karathanasis, M.~M. Kasliwal,
  I.~Maga\~na Hernandez, S.~M. Nissanke, A.~Silvestri, and B.~D. Wandelt, {\it
  {First measurement of the Hubble parameter from bright binary black hole
  GW190521}},  \href{http://arxiv.org/abs/2009.14199}{{\tt arXiv:2009.14199}}.

\bibitem{Ashton:2020kyr}
G.~Ashton, K.~Ackley, I.~M.~n. Hernandez, and B.~Piotrzkowski, {\it {Current
  observations are insufficient to confidently associate the binary black hole
  merger GW190521 with AGN J124942.3+344929}},
  \href{http://arxiv.org/abs/2009.12346}{{\tt arXiv:2009.12346}}.

\bibitem{Bonilla:2020wbn}
A.~Bonilla, S.~Kumar, and R.~C. Nunes, {\it {Measurements of $H_0$ and
  reconstruction of the dark energy properties from a model-independent joint
  analysis}},  \href{http://arxiv.org/abs/2011.07140}{{\tt arXiv:2011.07140}}.

\bibitem{Harvey:2020lwf}
D.~Harvey, {\it {A 4\% measurement of $H_0$ using the cumulative distribution
  of strong-lensing time delays in doubly-imaged quasars}},
  \href{http://arxiv.org/abs/2011.09488}{{\tt arXiv:2011.09488}}.

\bibitem{Renzi:2020fnx}
F.~Renzi and A.~Silvestri, {\it {A look at the Hubble speed from first
  principles}},  \href{http://arxiv.org/abs/2011.10559}{{\tt
  arXiv:2011.10559}}.

\bibitem{Wang:2020hqq}
D.~Wang, {\it {Assessing the potential of cluster edges as a standard ruler}},
  \href{http://arxiv.org/abs/2011.11924}{{\tt arXiv:2011.11924}}.

\bibitem{Mangano:2005cc}
G.~Mangano, G.~Miele, S.~Pastor, T.~Pinto, O.~Pisanti, and P.~D. Serpico, {\it
  {Relic neutrino decoupling including flavor oscillations}},  {\em Nucl.
  Phys.} {\bf B729} (2005) 221--234,
  [\href{http://arxiv.org/abs/hep-ph/0506164}{{\tt hep-ph/0506164}}].

\bibitem{deSalas:2016ztq}
P.~F. de~Salas and S.~Pastor, {\it {Relic neutrino decoupling with flavour
  oscillations revisited}},  {\em JCAP} {\bf 1607} (2016), no.~07 051,
  [\href{http://arxiv.org/abs/1606.06986}{{\tt arXiv:1606.06986}}].

\bibitem{Archidiacono:2013fha}
M.~Archidiacono, E.~Giusarma, S.~Hannestad, and O.~Mena, {\it {Cosmic dark
  radiation and neutrinos}},  {\em Adv. High Energy Phys.} {\bf 2013} (2013)
  191047, [\href{http://arxiv.org/abs/1307.0637}{{\tt arXiv:1307.0637}}].

\bibitem{DiValentino:2015sam}
E.~Di~Valentino, E.~Giusarma, O.~Mena, A.~Melchiorri, and J.~Silk, {\it
  {Cosmological limits on neutrino unknowns versus low redshift priors}},  {\em
  Phys. Rev.} {\bf D93} (2016), no.~8 083527,
  [\href{http://arxiv.org/abs/1511.00975}{{\tt arXiv:1511.00975}}].

\bibitem{Green:2019glg}
D.~Green et~al., {\it {Messengers from the Early Universe: Cosmic Neutrinos and
  Other Light Relics}},  {\em Bull. Am. Astron. Soc.} {\bf 51} (2019), no.~7
  159, [\href{http://arxiv.org/abs/1903.04763}{{\tt arXiv:1903.04763}}].

\bibitem{Ferreira:2018vjj}
R.~Z. Ferreira and A.~Notari, {\it {Observable Windows for the QCD Axion
  Through the Number of Relativistic Species}},  {\em Phys. Rev. Lett.} {\bf
  120} (2018), no.~19 191301, [\href{http://arxiv.org/abs/1801.06090}{{\tt
  arXiv:1801.06090}}].

\bibitem{DiValentino:2015wba}
E.~Di~Valentino, E.~Giusarma, M.~Lattanzi, O.~Mena, A.~Melchiorri, and J.~Silk,
  {\it {Cosmological Axion and neutrino mass constraints from Planck 2015
  temperature and polarization data}},  {\em Phys. Lett.} {\bf B752} (2016)
  182--185, [\href{http://arxiv.org/abs/1507.08665}{{\tt arXiv:1507.08665}}].

\bibitem{Parker:2000pr}
L.~Parker and A.~Raval, {\it {New quantum aspects of a vacuum dominated
  universe}},  {\em Phys. Rev.} {\bf D62} (2000) 083503,
  [\href{http://arxiv.org/abs/gr-qc/0003103}{{\tt gr-qc/0003103}}]. [Erratum:
  Phys. Rev.D67,029903(2003)].

\bibitem{Parker:2003as}
L.~Parker and D.~A.~T. Vanzella, {\it {Acceleration of the universe, vacuum
  metamorphosis, and the large time asymptotic form of the heat kernel}},  {\em
  Phys. Rev.} {\bf D69} (2004) 104009,
  [\href{http://arxiv.org/abs/gr-qc/0312108}{{\tt gr-qc/0312108}}].

\bibitem{Caldwell:2005xb}
R.~R. Caldwell, W.~Komp, L.~Parker, and D.~A.~T. Vanzella, {\it {A Sudden
  gravitational transition}},  {\em Phys. Rev.} {\bf D73} (2006) 023513,
  [\href{http://arxiv.org/abs/astro-ph/0507622}{{\tt astro-ph/0507622}}].

\bibitem{Calabrese:2008rt}
E.~Calabrese, A.~Slosar, A.~Melchiorri, G.~F. Smoot, and O.~Zahn, {\it {Cosmic
  Microwave Weak lensing data as a test for the dark universe}},  {\em Phys.
  Rev.} {\bf D77} (2008) 123531, [\href{http://arxiv.org/abs/0803.2309}{{\tt
  arXiv:0803.2309}}].

\bibitem{Kamionkowski:2014zda}
M.~Kamionkowski, J.~Pradler, and D.~G.~E. Walker, {\it {Dark energy from the
  string axiverse}},  {\em Phys. Rev. Lett.} {\bf 113} (2014), no.~25 251302,
  [\href{http://arxiv.org/abs/1409.0549}{{\tt arXiv:1409.0549}}].

\bibitem{Aghanim:2015xee}
{\bf Planck} Collaboration, N.~Aghanim et~al., {\it {Planck 2015 results. XI.
  CMB power spectra, likelihoods, and robustness of parameters}},  {\em Astron.
  Astrophys.} {\bf 594} (2016) A11,
  [\href{http://arxiv.org/abs/1507.02704}{{\tt arXiv:1507.02704}}].

\bibitem{Alexander:2019rsc}
S.~Alexander and E.~McDonough, {\it {Axion-Dilaton Destabilization and the
  Hubble Tension}},  {\em Phys. Lett.} {\bf B797} (2019) 134830,
  [\href{http://arxiv.org/abs/1904.08912}{{\tt arXiv:1904.08912}}].

\bibitem{Pettorino:2013ia}
V.~Pettorino, L.~Amendola, and C.~Wetterich, {\it {How early is early dark
  energy?}},  {\em Phys. Rev.} {\bf D87} (2013) 083009,
  [\href{http://arxiv.org/abs/1301.5279}{{\tt arXiv:1301.5279}}].

\bibitem{Heinesen:2019phg}
A.~Heinesen, C.~Blake, and D.~L. Wiltshire, {\it {Quantifying the accuracy of
  the Alcock-Paczynski scaling of baryon acoustic oscillation measurements}},
  {\em JCAP} {\bf 2001} (2020), no.~01 038,
  [\href{http://arxiv.org/abs/1908.11508}{{\tt arXiv:1908.11508}}].

\bibitem{Arendse:2019hev}
N.~Arendse et~al., {\it {Cosmic dissonance: new physics or systematics behind a
  short sound horizon?}},  {\em Astron. Astrophys.} {\bf 639} (2020) A57,
  [\href{http://arxiv.org/abs/1909.07986}{{\tt arXiv:1909.07986}}].

\bibitem{DiValentino:2018jbh}
E.~Di~Valentino, D.~E. Holz, A.~Melchiorri, and F.~Renzi, {\it {The
  cosmological impact of future constraints on $H_0$ from gravitational-wave
  standard sirens}},  \href{http://arxiv.org/abs/1806.07463}{{\tt
  arXiv:1806.07463}}.

\bibitem{Palmese:2019ehe}
A.~Palmese et~al., {\it {Gravitational Wave Cosmology and Astrophysics with
  Large Spectroscopic Galaxy Surveys}},
  \href{http://arxiv.org/abs/1903.04730}{{\tt arXiv:1903.04730}}.

\bibitem{DiValentino:2017clw}
E.~Di~Valentino and A.~Melchiorri, {\it {First cosmological constraints
  combining Planck with the recent gravitational-wave standard siren
  measurement of the Hubble constant}},  {\em Phys. Rev.} {\bf D97} (2018),
  no.~4 041301, [\href{http://arxiv.org/abs/1710.06370}{{\tt
  arXiv:1710.06370}}].

\bibitem{Chen:2017rfc}
H.-Y. Chen, M.~Fishbach, and D.~E. Holz, {\it {A two per cent Hubble constant
  measurement from standard sirens within five years}},  {\em Nature} {\bf 562}
  (2018), no.~7728 545--547, [\href{http://arxiv.org/abs/1712.06531}{{\tt
  arXiv:1712.06531}}].

\bibitem{Carroll:2000fy}
S.~M. Carroll, {\it {The Cosmological constant}},  {\em Living Rev. Rel.} {\bf
  4} (2001) 1, [\href{http://arxiv.org/abs/astro-ph/0004075}{{\tt
  astro-ph/0004075}}].

\bibitem{Betoule:2014frx}
{\bf SDSS} Collaboration, M.~Betoule et~al., {\it {Improved cosmological
  constraints from a joint analysis of the SDSS-II and SNLS supernova
  samples}},  {\em Astron. Astrophys.} {\bf 568} (2014) A22,
  [\href{http://arxiv.org/abs/1401.4064}{{\tt arXiv:1401.4064}}].

\bibitem{Ade:2015xua}
{\bf Planck} Collaboration, P.~A.~R. Ade et~al., {\it {Planck 2015 results.
  XIII. Cosmological parameters}},  {\em Astron. Astrophys.} {\bf 594} (2016)
  A13, [\href{http://arxiv.org/abs/1502.01589}{{\tt arXiv:1502.01589}}].

\bibitem{Rozo:2009jj}
{\bf DSDD} Collaboration, E.~Rozo et~al., {\it {Cosmological Constraints from
  the SDSS maxBCG Cluster Catalog}},  {\em Astrophys. J.} {\bf 708} (2010)
  645--660, [\href{http://arxiv.org/abs/0902.3702}{{\tt arXiv:0902.3702}}].

\bibitem{Rapetti:2008rm}
D.~Rapetti, S.~W. Allen, A.~Mantz, and H.~Ebeling, {\it {Constraints on
  modified gravity from the observed X-ray luminosity function of galaxy
  clusters}},  {\em Mon. Not. Roy. Astron. Soc.} {\bf 400} (2009) 699,
  [\href{http://arxiv.org/abs/0812.2259}{{\tt arXiv:0812.2259}}].

\bibitem{Ade:2015fva}
{\bf Planck} Collaboration, P.~A.~R. Ade et~al., {\it {Planck 2015 results.
  XXIV. Cosmology from Sunyaev-Zeldovich cluster counts}},  {\em Astron.
  Astrophys.} {\bf 594} (2016) A24,
  [\href{http://arxiv.org/abs/1502.01597}{{\tt arXiv:1502.01597}}].

\bibitem{Bocquet:2014lmj}
{\bf SPT} Collaboration, S.~Bocquet et~al., {\it {Mass Calibration and
  Cosmological Analysis of the SPT-SZ Galaxy Cluster Sample Using Velocity
  Dispersion $\sigma_v$ and X-ray $Y_\textrm{X}$ Measurements}},  {\em
  Astrophys. J.} {\bf 799} (2015), no.~2 214,
  [\href{http://arxiv.org/abs/1407.2942}{{\tt arXiv:1407.2942}}].

\bibitem{Ruiz:2014hma}
E.~J. Ruiz and D.~Huterer, {\it {Testing the dark energy consistency with
  geometry and growth}},  {\em Phys. Rev.} {\bf D91} (2015) 063009,
  [\href{http://arxiv.org/abs/1410.5832}{{\tt arXiv:1410.5832}}].

\bibitem{Joudaki:2017zdt}
S.~Joudaki et~al., {\it {KiDS-450 + 2dFLenS: Cosmological parameter constraints
  from weak gravitational lensing tomography and overlapping redshift-space
  galaxy clustering}},  {\em Mon. Not. Roy. Astron. Soc.} {\bf 474} (2018),
  no.~4 4894--4924, [\href{http://arxiv.org/abs/1707.06627}{{\tt
  arXiv:1707.06627}}].

\bibitem{Kohlinger:2017sxk}
F.~Köhlinger et~al., {\it {KiDS-450: The tomographic weak lensing power
  spectrum and constraints on cosmological parameters}},  {\em Mon. Not. Roy.
  Astron. Soc.} {\bf 471} (2017), no.~4 4412--4435,
  [\href{http://arxiv.org/abs/1706.02892}{{\tt arXiv:1706.02892}}].

\bibitem{Macaulay:2013swa}
E.~Macaulay, I.~K. Wehus, and H.~K. Eriksen, {\it {Lower Growth Rate from
  Recent Redshift Space Distortion Measurements than Expected from Planck}},
  {\em Phys.Rev.Lett.} {\bf 111} (2013), no.~16 161301,
  [\href{http://arxiv.org/abs/1303.6583}{{\tt arXiv:1303.6583}}].

\bibitem{Johnson:2015aaa}
A.~Johnson, C.~Blake, J.~Dossett, J.~Koda, D.~Parkinson, and S.~Joudaki, {\it
  {Searching for Modified Gravity: Scale and Redshift Dependent Constraints
  from Galaxy Peculiar Velocities}},  {\em Mon. Not. Roy. Astron. Soc.} {\bf
  458} (2016), no.~3 2725--2744, [\href{http://arxiv.org/abs/1504.06885}{{\tt
  arXiv:1504.06885}}].

\bibitem{Basilakos:2016nyg}
S.~Basilakos and S.~Nesseris, {\it {Testing Einstein’s gravity and dark
  energy with growth of matter perturbations: Indications for new physics?}},
  {\em Phys. Rev.} {\bf D94} (2016), no.~12 123525,
  [\href{http://arxiv.org/abs/1610.00160}{{\tt arXiv:1610.00160}}].

\bibitem{Nesseris:2017vor}
S.~Nesseris, G.~Pantazis, and L.~Perivolaropoulos, {\it {Tension and
  constraints on modified gravity parametrizations of $G_{\textrm{eff}}(z)$
  from growth rate and Planck data}},  {\em Phys. Rev.} {\bf D96} (2017), no.~2
  023542, [\href{http://arxiv.org/abs/1703.10538}{{\tt arXiv:1703.10538}}].

\bibitem{Kazantzidis:2018rnb}
L.~Kazantzidis and L.~Perivolaropoulos, {\it {Evolution of the $f\sigma_8$
  tension with the Planck15/$\Lambda$CDM determination and implications for
  modified gravity theories}},  {\em Phys. Rev.} {\bf D97} (2018), no.~10
  103503, [\href{http://arxiv.org/abs/1803.01337}{{\tt arXiv:1803.01337}}].

\bibitem{Nesseris:2006er}
S.~Nesseris and L.~Perivolaropoulos, {\it {Crossing the Phantom Divide:
  Theoretical Implications and Observational Status}},  {\em JCAP} {\bf 0701}
  (2007) 018, [\href{http://arxiv.org/abs/astro-ph/0610092}{{\tt
  astro-ph/0610092}}].

\bibitem{Basilakos:2013nfa}
S.~Basilakos, S.~Nesseris, and L.~Perivolaropoulos, {\it {Observational
  constraints on viable f(R) parametrizations with geometrical and dynamical
  probes}},  {\em Phys. Rev.} {\bf D87} (2013), no.~12 123529,
  [\href{http://arxiv.org/abs/1302.6051}{{\tt arXiv:1302.6051}}].

\bibitem{Tammann:2008xf}
G.~A. Tammann, A.~Sandage, and B.~Reindl, {\it {The expansion field: The value
  of $H_0$}},  {\em Astron. Astrophys. Rev.} {\bf 15} (2008) 289--331,
  [\href{http://arxiv.org/abs/0806.3018}{{\tt arXiv:0806.3018}}].

\bibitem{Margalef-Bentabol:2012kwa}
B.~Margalef-Bentabol, J.~Margalef-Bentabol, and J.~Cepa, {\it {Evolution of the
  Cosmological Horizons in a Concordance Universe}},  {\em JCAP} {\bf 1212}
  (2012) 035, [\href{http://arxiv.org/abs/1302.1609}{{\tt arXiv:1302.1609}}].

\bibitem{Zarrouk:2018vwy}
P.~Zarrouk et~al., {\it {The clustering of the SDSS-IV extended Baryon
  Oscillation Spectroscopic Survey DR14 quasar sample: measurement of the
  growth rate of structure from the anisotropic correlation function between
  redshift 0.8 and 2.2}},  {\em Mon. Not. Roy. Astron. Soc.} {\bf 477} (2018),
  no.~2 1639--1663, [\href{http://arxiv.org/abs/1801.03062}{{\tt
  arXiv:1801.03062}}].

\bibitem{Bautista:2017zgn}
J.~E. Bautista et~al., {\it {Measurement of baryon acoustic oscillation
  correlations at $z=2.3$ with SDSS DR12 Ly$\alpha$-Forests}},  {\em Astron.
  Astrophys.} {\bf 603} (2017) A12,
  [\href{http://arxiv.org/abs/1702.00176}{{\tt arXiv:1702.00176}}].

\bibitem{Huterer:2017buf}
D.~Huterer and D.~L. Shafer, {\it {Dark energy two decades after: Observables,
  probes, consistency tests}},  {\em Rept. Prog. Phys.} {\bf 81} (2018), no.~1
  016901, [\href{http://arxiv.org/abs/1709.01091}{{\tt arXiv:1709.01091}}].

\bibitem{Efstathiou:2013via}
G.~Efstathiou, {\it {H0 Revisited}},  {\em Mon. Not. Roy. Astron. Soc.} {\bf
  440} (2014), no.~2 1138--1152, [\href{http://arxiv.org/abs/1311.3461}{{\tt
  arXiv:1311.3461}}].

\bibitem{Zhang:2017aqn}
B.~R. Zhang, M.~J. Childress, T.~M. Davis, N.~V. Karpenka, C.~Lidman, B.~P.
  Schmidt, and M.~Smith, {\it {A blinded determination of $H_0$ from
  low-redshift Type Ia supernovae, calibrated by Cepheid variables}},  {\em
  Mon. Not. Roy. Astron. Soc.} {\bf 471} (2017), no.~2 2254--2285,
  [\href{http://arxiv.org/abs/1706.07573}{{\tt arXiv:1706.07573}}].

\bibitem{Suyu2012}
S.~S.~H. et~al., {\it {Two accurate time-delay distances from strong lensing:
  Implications for cosmology}},  {\em Astrophys. J.} {\bf 766} (2013) 70,
  [\href{http://arxiv.org/abs/1208.3311}{{\tt arXiv:1208.3311}}].

\bibitem{Sorce_2012}
J.~G. Sorce, R.~B. Tully, and H.~M. Courtois, {\it {THE} {MID}-{INFRARED}
  {TULLY}-{FISHER} {RELATION}: {CALIBRATION} {OF} {THE} {TYPE} ia {SUPERNOVA}
  {SCALE} {ANDH}0},  {\em The Astrophysical Journal} {\bf 758} (sep, 2012) L12.

\bibitem{Tammann:2012ut}
G.~A. Tammann and B.~Reindl, {\it {The luminosity of supernovae of type Ia from
  TRGB distances and the value of $H_0$}},  {\em Astron. Astrophys.} {\bf 549}
  (2013) A136, [\href{http://arxiv.org/abs/1208.5054}{{\tt arXiv:1208.5054}}].

\bibitem{Wu:2017fpr}
H.-Y. Wu and D.~Huterer, {\it {Sample variance in the local measurements of the
  Hubble constant}},  {\em Mon. Not. Roy. Astron. Soc.} {\bf 471} (2017), no.~4
  4946--4955, [\href{http://arxiv.org/abs/1706.09723}{{\tt arXiv:1706.09723}}].

\bibitem{Kazantzidis:2020tko}
L.~Kazantzidis and L.~Perivolaropoulos, {\it {Hints of a Local Matter
  Underdensity or Modified Gravity in the Low $z$ Pantheon data}},  {\em Phys.
  Rev. D} {\bf 102} (2020), no.~2 023520,
  [\href{http://arxiv.org/abs/2004.02155}{{\tt arXiv:2004.02155}}].

\bibitem{Zhao:2017urm}
M.-M. Zhao, D.-Z. He, J.-F. Zhang, and X.~Zhang, {\it {Search for sterile
  neutrinos in holographic dark energy cosmology: Reconciling Planck
  observation with the local measurement of the Hubble constant}},  {\em Phys.
  Rev.} {\bf D96} (2017), no.~4 043520,
  [\href{http://arxiv.org/abs/1703.08456}{{\tt arXiv:1703.08456}}].

\bibitem{Lukovic:2018ljo}
V.~V. Luković, B.~S. Haridasu, and N.~Vittorio, {\it {Cosmological constraints
  from low-redshift data}},  {\em Found. Phys.} {\bf 48} (2018), no.~10
  1446--1485, [\href{http://arxiv.org/abs/1801.05765}{{\tt arXiv:1801.05765}}].

\bibitem{Skara:2019usd}
F.~Skara and L.~Perivolaropoulos, {\it {Tension of the $E_G$ statistic and
  redshift space distortion data with the Planck - $\Lambda CDM$ model and
  implications for weakening gravity}},  {\em Phys. Rev. D} {\bf 101} (2020),
  no.~6 063521, [\href{http://arxiv.org/abs/1911.10609}{{\tt
  arXiv:1911.10609}}].

\bibitem{Pourtsidou:2016ico}
A.~Pourtsidou and T.~Tram, {\it {Reconciling CMB and structure growth
  measurements with dark energy interactions}},  {\em Phys. Rev.} {\bf D94}
  (2016), no.~4 043518, [\href{http://arxiv.org/abs/1604.04222}{{\tt
  arXiv:1604.04222}}].

\bibitem{Barros:2018efl}
B.~J. Barros, L.~Amendola, T.~Barreiro, and N.~J. Nunes, {\it {Coupled
  quintessence with a $\Lambda$CDM background: removing the $\sigma_8$
  tension}},  {\em JCAP} {\bf 1901} (2019), no.~01 007,
  [\href{http://arxiv.org/abs/1802.09216}{{\tt arXiv:1802.09216}}].

\bibitem{Camera:2019vbp}
S.~Camera, M.~Martinelli, and D.~Bertacca, {\it {Does quartessence ease cosmic
  tensions?}},  {\em Phys. Dark Univ.} {\bf 23} (2019) 100247,
  [\href{http://arxiv.org/abs/1704.06277}{{\tt arXiv:1704.06277}}].

\bibitem{Melia:2016djn}
F.~Melia, {\it {The Linear Growth of Structure in the $R_h=ct$ Universe}},
  {\em Mon. Not. Roy. Astron. Soc.} {\bf 464} (2017), no.~2 1966--1976,
  [\href{http://arxiv.org/abs/1609.08576}{{\tt arXiv:1609.08576}}].

\bibitem{Lambiase:2018ows}
G.~Lambiase, S.~Mohanty, A.~Narang, and P.~Parashari, {\it {Testing dark energy
  models in the light of $\sigma _8$ tension}},  {\em Eur. Phys. J.} {\bf C79}
  (2019), no.~2 141, [\href{http://arxiv.org/abs/1804.07154}{{\tt
  arXiv:1804.07154}}].

\bibitem{Ooba:2018dzf}
J.~Ooba, B.~Ratra, and N.~Sugiyama, {\it {Planck 2015 constraints on
  spatially-flat dynamical dark energy models}},
  \href{http://arxiv.org/abs/1802.05571}{{\tt arXiv:1802.05571}}.

\bibitem{Gomez-Valent:2017idt}
A.~Gomez-Valent and J.~Sola, {\it {Relaxing the $\sigma_8$-tension through
  running vacuum in the Universe}},  {\em EPL} {\bf 120} (2017), no.~3 39001,
  [\href{http://arxiv.org/abs/1711.00692}{{\tt arXiv:1711.00692}}].

\bibitem{Gomez-Valent:2018nib}
A.~Gómez-Valent and J.~Solà~Peracaula, {\it {Density perturbations for
  running vacuum: a successful approach to structure formation and to the
  $\sigma_8$-tension}},  {\em Mon. Not. Roy. Astron. Soc.} {\bf 478} (2018),
  no.~1 126--145, [\href{http://arxiv.org/abs/1801.08501}{{\tt
  arXiv:1801.08501}}].

\bibitem{DiazRivero:2019ukx}
A.~Diaz~Rivero, V.~Miranda, and C.~Dvorkin, {\it {Observable Predictions for
  Massive-Neutrino Cosmologies with Model-Independent Dark Energy}},
  \href{http://arxiv.org/abs/1903.03125}{{\tt arXiv:1903.03125}}.

\bibitem{DiValentino:2018gcu}
E.~Di~Valentino and S.~Bridle, {\it {Exploring the Tension between Current
  Cosmic Microwave Background and Cosmic Shear Data}},  {\em Symmetry} {\bf 10}
  (2018), no.~11 585.

\bibitem{Wang:2019acf}
D.~Wang, {\it {Dark Energy Survey Year 1: Exploring New Physics Beyond the
  Standard Cosmology}},  \href{http://arxiv.org/abs/1904.00657}{{\tt
  arXiv:1904.00657}}.

\bibitem{Gannouji:2018ncm}
R.~Gannouji, L.~Kazantzidis, L.~Perivolaropoulos, and D.~Polarski, {\it
  {Consistency of modified gravity with a decreasing $G_{\rm eff}(z)$ in a
  $\Lambda$CDM background}},  {\em Phys. Rev.} {\bf D98} (2018), no.~10 104044,
  [\href{http://arxiv.org/abs/1809.07034}{{\tt arXiv:1809.07034}}].

\bibitem{DAgostino:2018ngy}
R.~D'Agostino and O.~Luongo, {\it {Growth of matter perturbations in nonminimal
  teleparallel dark energy}},  {\em Phys. Rev.} {\bf D98} (2018), no.~12
  124013, [\href{http://arxiv.org/abs/1807.10167}{{\tt arXiv:1807.10167}}].

\bibitem{Gonzalez-Espinoza:2018gyl}
M.~Gonzalez-Espinoza, G.~Otalora, J.~Saavedra, and N.~Videla, {\it {Growth of
  matter overdensities in non-minimal torsion-matter coupling theories}},  {\em
  Eur. Phys. J.} {\bf C78} (2018), no.~10 799,
  [\href{http://arxiv.org/abs/1808.01941}{{\tt arXiv:1808.01941}}].

\bibitem{Gannouji:2020ylf}
R.~Gannouji, L.~Perivolaropoulos, D.~Polarski, and F.~Skara, {\it {Weak gravity
  on a $\Lambda$CDM background}},  \href{http://arxiv.org/abs/2011.01517}{{\tt
  arXiv:2011.01517}}.

\bibitem{Pogosian:2005ez}
L.~Pogosian, P.~S. Corasaniti, C.~Stephan-Otto, R.~Crittenden, and R.~Nichol,
  {\it {Tracking dark energy with the ISW effect: Short and long-term
  predictions}},  {\em Phys. Rev.} {\bf D72} (2005) 103519,
  [\href{http://arxiv.org/abs/astro-ph/0506396}{{\tt astro-ph/0506396}}].

\bibitem{Ho:2008bz}
S.~Ho, C.~Hirata, N.~Padmanabhan, U.~Seljak, and N.~Bahcall, {\it {Correlation
  of CMB with large-scale structure: I. ISW Tomography and Cosmological
  Implications}},  {\em Phys. Rev.} {\bf D78} (2008) 043519,
  [\href{http://arxiv.org/abs/0801.0642}{{\tt arXiv:0801.0642}}].

\bibitem{Amendola:1999vu}
L.~Amendola, P.~S. Corasaniti, and F.~Occhionero, {\it {Time variability of the
  gravitational constant and type Ia supernovae}},
  \href{http://arxiv.org/abs/astro-ph/9907222}{{\tt astro-ph/9907222}}.

\bibitem{Gaztanaga:2001fh}
E.~Gaztanaga, E.~Garcia-Berro, J.~Isern, E.~Bravo, and I.~Dominguez, {\it
  {Bounds on the possible evolution of the gravitational constant from
  cosmological type Ia supernovae}},  {\em Phys. Rev.} {\bf D65} (2002) 023506,
  [\href{http://arxiv.org/abs/astro-ph/0109299}{{\tt astro-ph/0109299}}].

\bibitem{Nesseris:2006jc}
S.~Nesseris and L.~Perivolaropoulos, {\it {Evolving newton's constant, extended
  gravity theories and snia data analysis}},  {\em Phys. Rev.} {\bf D73} (2006)
  103511, [\href{http://arxiv.org/abs/astro-ph/0602053}{{\tt
  astro-ph/0602053}}].

\bibitem{Wright:2017rsu}
B.~S. Wright and B.~Li, {\it {Type Ia supernovae, standardizable candles, and
  gravity}},  {\em Phys. Rev.} {\bf D97} (2018), no.~8 083505,
  [\href{http://arxiv.org/abs/1710.07018}{{\tt arXiv:1710.07018}}].

\bibitem{Sapone:2020wwz}
D.~Sapone, S.~Nesseris, and C.~A. Bengaly, {\it {Is there any measurable
  redshift dependence on the SN Ia absolute magnitude?}},
  \href{http://arxiv.org/abs/2006.05461}{{\tt arXiv:2006.05461}}.

\bibitem{DiValentino:2015bja}
E.~Di~Valentino, A.~Melchiorri, and J.~Silk, {\it {Cosmological hints of
  modified gravity?}},  {\em Phys. Rev.} {\bf D93} (2016), no.~2 023513,
  [\href{http://arxiv.org/abs/1509.07501}{{\tt arXiv:1509.07501}}].

\bibitem{Li:2018tfg}
J.~Li and G.-B. Zhao, {\it {Cosmological Tests of Gravity with the Latest
  Observations}},  {\em Astrophys. J.} {\bf 871} (2019), no.~2 196,
  [\href{http://arxiv.org/abs/1806.05022}{{\tt arXiv:1806.05022}}].

\bibitem{Huterer:2013xky}
D.~Huterer et~al., {\it {Growth of Cosmic Structure: Probing Dark Energy Beyond
  Expansion}},  {\em Astropart. Phys.} {\bf 63} (2015) 23--41,
  [\href{http://arxiv.org/abs/1309.5385}{{\tt arXiv:1309.5385}}].

\bibitem{Pogosian:2010tj}
L.~Pogosian, A.~Silvestri, K.~Koyama, and G.-B. Zhao, {\it {How to optimally
  parametrize deviations from General Relativity in the evolution of
  cosmological perturbations?}},  {\em Phys. Rev.} {\bf D81} (2010) 104023,
  [\href{http://arxiv.org/abs/1002.2382}{{\tt arXiv:1002.2382}}].

\bibitem{Tereno:2010dt}
I.~Tereno, E.~Semboloni, and T.~Schrabback, {\it {COSMOS weak-lensing
  constraints on modified gravity}},  {\em Astron. Astrophys.} {\bf 530} (2011)
  A68, [\href{http://arxiv.org/abs/1012.5854}{{\tt arXiv:1012.5854}}].

\bibitem{Polarski:2016ieb}
D.~Polarski, A.~A. Starobinsky, and H.~Giacomini, {\it {When is the growth
  index constant?}},  {\em JCAP} {\bf 1612} (2016), no.~12 037,
  [\href{http://arxiv.org/abs/1610.00363}{{\tt arXiv:1610.00363}}].

\bibitem{Gannouji:2006jm}
R.~Gannouji, D.~Polarski, A.~Ranquet, and A.~A. Starobinsky, {\it
  {Scalar-Tensor Models of Normal and Phantom Dark Energy}},  {\em JCAP} {\bf
  0609} (2006) 016, [\href{http://arxiv.org/abs/astro-ph/0606287}{{\tt
  astro-ph/0606287}}].

\bibitem{Nesseris:2006hp}
S.~Nesseris and L.~Perivolaropoulos, {\it {The Limits of Extended
  Quintessence}},  {\em Phys. Rev.} {\bf D75} (2007) 023517,
  [\href{http://arxiv.org/abs/astro-ph/0611238}{{\tt astro-ph/0611238}}].

\bibitem{Muller:2005sr}
J.~Muller, J.~G. Williams, and S.~G. Turyshev, {\it {Lunar laser ranging
  contributions to relativity and geodesy}},  {\em Astrophys. Space Sci. Libr.}
  {\bf 349} (2008) 457--472, [\href{http://arxiv.org/abs/gr-qc/0509114}{{\tt
  gr-qc/0509114}}].

\bibitem{Pitjeva:2013chs}
E.~V. Pitjeva and N.~P. Pitjev, {\it {Relativistic effects and dark matter in
  the Solar system from observations of planets and spacecraft}},  {\em Mon.
  Not. Roy. Astron. Soc.} {\bf 432} (2013) 3431,
  [\href{http://arxiv.org/abs/1306.3043}{{\tt arXiv:1306.3043}}].

\bibitem{Gannouji:2008wt}
R.~Gannouji, B.~Moraes, and D.~Polarski, {\it {The growth of matter
  perturbations in f(R) models}},  {\em JCAP} {\bf 0902} (2009) 034,
  [\href{http://arxiv.org/abs/0809.3374}{{\tt arXiv:0809.3374}}].

\bibitem{Shafieloo:2018gin}
A.~Shafieloo, B.~L'Huillier, and A.~A. Starobinsky, {\it {Falsifying
  $\Lambda$CDM: Model-independent tests of the concordance model with eBOSS
  DR14Q and Pantheon}},  {\em Phys. Rev.} {\bf D98} (2018), no.~8 083526,
  [\href{http://arxiv.org/abs/1804.04320}{{\tt arXiv:1804.04320}}].

\bibitem{Gannouji:2018col}
R.~Gannouji and D.~Polarski, {\it {Consistency of the expansion of the Universe
  with density perturbations}},  {\em Phys. Rev.} {\bf D98} (2018), no.~8
  083533, [\href{http://arxiv.org/abs/1805.08230}{{\tt arXiv:1805.08230}}].

\bibitem{Basilakos:2019hlb}
S.~Basilakos and F.~K. Anagnostopoulos, {\it {Growth index of matter
  perturbations in the light of Dark Energy Survey}},
  \href{http://arxiv.org/abs/1903.10758}{{\tt arXiv:1903.10758}}.

\bibitem{Yin:2018mvu}
Z.-Y. Yin and H.~Wei, {\it {Non-parametric Reconstruction of Growth Index via
  Gaussian Processes}},  {\em Sci. China Phys. Mech. Astron.} {\bf 62} (2019),
  no.~9 999811, [\href{http://arxiv.org/abs/1808.00377}{{\tt
  arXiv:1808.00377}}].

\bibitem{Perivolaropoulos:2019vkb}
L.~Perivolaropoulos and L.~Kazantzidis, {\it {Hints of modified gravity in
  cosmos and in the lab?}},  {\em Int. J. Mod. Phys.} {\bf D28} (2019), no.~05
  1942001, [\href{http://arxiv.org/abs/1904.09462}{{\tt arXiv:1904.09462}}].

\bibitem{Erben:2012zw}
T.~Erben et~al., {\it {CFHTLenS: The Canada-France-Hawaii Telescope Lensing
  Survey - Imaging Data and Catalogue Products}},  {\em Mon. Not. Roy. Astron.
  Soc.} {\bf 433} (2013) 2545, [\href{http://arxiv.org/abs/1210.8156}{{\tt
  arXiv:1210.8156}}].

\bibitem{Kazantzidisinprog}
L.~Kazantzidis and L.~Perivolaropoulos, ``Work in progress.''

\bibitem{Kazantzidis:2018jtb}
L.~Kazantzidis, L.~Perivolaropoulos, and F.~Skara, {\it {Constraining power of
  cosmological observables: blind redshift spots and optimal ranges}},  {\em
  Phys. Rev.} {\bf D99} (2019), no.~6 063537,
  [\href{http://arxiv.org/abs/1812.05356}{{\tt arXiv:1812.05356}}].

\bibitem{Alcock:1979mp}
C.~Alcock and B.~Paczynski, {\it {An evolution free test for non-zero
  cosmological constant}},  {\em Nature} {\bf 281} (1979) 358--359.

\bibitem{Ade:2013lmv}
{\bf Planck} Collaboration, P.~A.~R. Ade et~al., {\it {Planck 2013 results. XX.
  Cosmology from Sunyaev–Zeldovich cluster counts}},  {\em Astron.
  Astrophys.} {\bf 571} (2014) A20, [\href{http://arxiv.org/abs/1303.5080}{{\tt
  arXiv:1303.5080}}].

\bibitem{Vikhlinin:2008ym}
A.~Vikhlinin et~al., {\it {Chandra Cluster Cosmology Project III: Cosmological
  Parameter Constraints}},  {\em Astrophys. J.} {\bf 692} (2009) 1060--1074,
  [\href{http://arxiv.org/abs/0812.2720}{{\tt arXiv:0812.2720}}].

\bibitem{Bernal:2015zom}
J.~L. Bernal, L.~Verde, and A.~J. Cuesta, {\it {Parameter splitting in dark
  energy: is dark energy the same in the background and in the cosmic
  structures?}},  {\em JCAP} {\bf 1602} (2016), no.~02 059,
  [\href{http://arxiv.org/abs/1511.03049}{{\tt arXiv:1511.03049}}].

\bibitem{Raveri:2015maa}
M.~Raveri, {\it {Are cosmological data sets consistent with each other within
  the $\Lambda$ cold dark matter model?}},  {\em Phys. Rev.} {\bf D93} (2016),
  no.~4 043522, [\href{http://arxiv.org/abs/1510.00688}{{\tt
  arXiv:1510.00688}}].

\bibitem{Lin:2017ikq}
W.~Lin and M.~Ishak, {\it {Cosmological discordances: A new measure,
  marginalization effects, and application to geometry versus growth current
  data sets}},  {\em Phys. Rev.} {\bf D96} (2017), no.~2 023532,
  [\href{http://arxiv.org/abs/1705.05303}{{\tt arXiv:1705.05303}}].

\bibitem{Sagredo:2018rvc}
B.~Sagredo, J.~S. Lafaurie, and D.~Sapone, {\it {Comparing Dark Energy models
  with Hubble versus Growth Rate data}},
  \href{http://arxiv.org/abs/1808.05660}{{\tt arXiv:1808.05660}}.

\bibitem{Arjona:2019rfn}
R.~Arjona, W.~Cardona, and S.~Nesseris, {\it {Designing Horndeski and the
  effective fluid approach}},  \href{http://arxiv.org/abs/1904.06294}{{\tt
  arXiv:1904.06294}}.

\bibitem{Marshall:2017wph}
{\bf LSST} Collaboration, P.~Marshall et~al., {\it {Science-Driven Optimization
  of the LSST Observing Strategy}},
  \href{http://arxiv.org/abs/1708.04058}{{\tt arXiv:1708.04058}}.

\bibitem{Bouchet:2015arn}
F.~R. Bouchet et~al., {\it {COrE: Cosmic Origins Explorer - A White Paper}}, .

\bibitem{Scolnic:2013efb}
D.~Scolnic et~al., {\it {Systematic Uncertainties Associated with the
  Cosmological Analysis of the First Pan-STARRS1 Type Ia Supernova Sample}},
  {\em Astrophys. J.} {\bf 795} (2014), no.~1 45,
  [\href{http://arxiv.org/abs/1310.3824}{{\tt arXiv:1310.3824}}].

\bibitem{Riess:1998dv}
A.~G. Riess et~al., {\it {BV RI light curves for 22 type Ia supernovae}},  {\em
  Astron. J.} {\bf 117} (1999) 707--724,
  [\href{http://arxiv.org/abs/astro-ph/9810291}{{\tt astro-ph/9810291}}].

\bibitem{Jha:2005jg}
S.~Jha et~al., {\it {Ubvri light curves of 44 type ia supernovae}},  {\em
  Astron. J.} {\bf 131} (2006) 527--554,
  [\href{http://arxiv.org/abs/astro-ph/0509234}{{\tt astro-ph/0509234}}].

\bibitem{Hicken:2009df}
M.~Hicken, P.~Challis, S.~Jha, R.~P. Kirsher, T.~Matheson, M.~Modjaz, A.~Rest,
  and W.~M. Wood-Vasey, {\it {CfA3: 185 Type Ia Supernova Light Curves from the
  CfA}},  {\em Astrophys. J.} {\bf 700} (2009) 331--357,
  [\href{http://arxiv.org/abs/0901.4787}{{\tt arXiv:0901.4787}}].

\bibitem{Hicken:2012zr}
M.~Hicken et~al., {\it {CfA4: Light Curves for 94 Type Ia Supernovae}},  {\em
  Astrophys. J. Suppl.} {\bf 200} (2012) 12,
  [\href{http://arxiv.org/abs/1205.4493}{{\tt arXiv:1205.4493}}].

\bibitem{Contreras:2009nt}
C.~Contreras et~al., {\it {The Carnegie Supernova Project: First Photometry
  Data Release of Low-Redshift Type Ia Supernovae}},  {\em Astron. J.} {\bf
  139} (2010) 519--539, [\href{http://arxiv.org/abs/0910.3330}{{\tt
  arXiv:0910.3330}}].

\bibitem{Stritzinger:2011qd}
M.~D. Stritzinger et~al., {\it {The Carnegie Supernova Project: Second
  Photometry Data Release of Low-Redshift Type Ia Supernovae}},  {\em Astron.
  J.} {\bf 142} (2011) 156, [\href{http://arxiv.org/abs/1108.3108}{{\tt
  arXiv:1108.3108}}].

\bibitem{Sako:2014qmj}
{\bf SDSS} Collaboration, M.~Sako et~al., {\it {The Data Release of the Sloan
  Digital Sky Survey-II Supernova Survey}},  {\em Publ. Astron. Soc. Pac.} {\bf
  130} (2018), no.~988 064002, [\href{http://arxiv.org/abs/1401.3317}{{\tt
  arXiv:1401.3317}}].

\bibitem{Kessler:2009ys}
R.~Kessler et~al., {\it {First-year Sloan Digital Sky Survey-II (SDSS-II)
  Supernova Results: Hubble Diagram and Cosmological Parameters}},  {\em
  Astrophys. J. Suppl.} {\bf 185} (2009) 32--84,
  [\href{http://arxiv.org/abs/0908.4274}{{\tt arXiv:0908.4274}}].

\bibitem{Sullivan:2011kv}
{\bf SNLS} Collaboration, M.~Sullivan et~al., {\it {SNLS3: Constraints on Dark
  Energy Combining the Supernova Legacy Survey Three Year Data with Other
  Probes}},  {\em Astrophys. J.} {\bf 737} (2011) 102,
  [\href{http://arxiv.org/abs/1104.1444}{{\tt arXiv:1104.1444}}].

\bibitem{Conley:2011ku}
{\bf SNLS} Collaboration, A.~Conley et~al., {\it {Supernova Constraints and
  Systematic Uncertainties from the First 3 Years of the Supernova Legacy
  Survey}},  {\em Astrophys. J. Suppl.} {\bf 192} (2011) 1,
  [\href{http://arxiv.org/abs/1104.1443}{{\tt arXiv:1104.1443}}].

\bibitem{Riess:2006fw}
A.~G. Riess et~al., {\it {New Hubble Space Telescope Discoveries of Type Ia
  Supernovae at z>=1: Narrowing Constraints on the Early Behavior of Dark
  Energy}},  {\em Astrophys. J.} {\bf 659} (2007) 98--121,
  [\href{http://arxiv.org/abs/astro-ph/0611572}{{\tt astro-ph/0611572}}].

\bibitem{Arjona:2018jhh}
R.~Arjona, W.~Cardona, and S.~Nesseris, {\it {Unraveling the effective fluid
  approach for $f(R)$ models in the subhorizon approximation}},  {\em Phys.
  Rev.} {\bf D99} (2019), no.~4 043516,
  [\href{http://arxiv.org/abs/1811.02469}{{\tt arXiv:1811.02469}}].

\bibitem{Colgain:2019pck}
E.~\'O.~Colg\'ain, {\it {A hint of matter underdensity at low z?}},  {\em JCAP}
  {\bf 1909} (2019) 006, [\href{http://arxiv.org/abs/1903.11743}{{\tt
  arXiv:1903.11743}}].

\bibitem{Kazantzidis:2020xta}
L.~Kazantzidis, H.~Koo, S.~Nesseris, L.~Perivolaropoulos, and A.~Shafieloo,
  {\it {Hints for possible low redshift oscillation around the best fit
  $\Lambda$CDM model in the expansion history of the universe}},
  \href{http://arxiv.org/abs/2010.03491}{{\tt arXiv:2010.03491}}.

\bibitem{Nesseris:2005ur}
S.~Nesseris and L.~Perivolaropoulos, {\it {Comparison of the legacy and gold
  snia dataset constraints on dark energy models}},  {\em Phys. Rev.} {\bf D72}
  (2005) 123519, [\href{http://arxiv.org/abs/astro-ph/0511040}{{\tt
  astro-ph/0511040}}].

\bibitem{Giannantonio:2009gi}
T.~Giannantonio, M.~Martinelli, A.~Silvestri, and A.~Melchiorri, {\it {New
  constraints on parametrised modified gravity from correlations of the CMB
  with large scale structure}},  {\em JCAP} {\bf 1004} (2010) 030,
  [\href{http://arxiv.org/abs/0909.2045}{{\tt arXiv:0909.2045}}].

\bibitem{Barack:2018yly}
L.~Barack et~al., {\it {Black holes, gravitational waves and fundamental
  physics: a roadmap}},  \href{http://arxiv.org/abs/1806.05195}{{\tt
  arXiv:1806.05195}}.

\bibitem{Cembranos:2018lcs}
J.~A.~R. Cembranos, M.~Coma~D{\'\i}az, and P.~Mart{\'\i}n-Moruno, {\it
  {Modified gravity as a diagravitational medium}},  {\em Phys. Lett.} {\bf
  B788} (2019) 336--340, [\href{http://arxiv.org/abs/1805.09629}{{\tt
  arXiv:1805.09629}}].

\bibitem{maggiore2008gravitational}
M.~Maggiore, {\em Gravitational Waves: Volume 1: Theory and Experiments},
  vol.~1.
\newblock Oxford university press, 2008.

\bibitem{Jimenez:2019lrk}
J.~B. Jiménez, J.~M. Ezquiaga, and L.~Heisenberg, {\it {Probing cosmological
  fields with gravitational wave oscillations}},
  \href{http://arxiv.org/abs/1912.06104}{{\tt arXiv:1912.06104}}.

\bibitem{Narikawa:2014fua}
T.~Narikawa, K.~Ueno, H.~Tagoshi, T.~Tanaka, N.~Kanda, and T.~Nakamura, {\it
  {Detectability of bigravity with graviton oscillations using gravitational
  wave observations}},  {\em Phys. Rev.} {\bf D91} (2015) 062007,
  [\href{http://arxiv.org/abs/1412.8074}{{\tt arXiv:1412.8074}}].

\bibitem{Max:2017flc}
K.~Max, M.~Platscher, and J.~Smirnov, {\it {Gravitational Wave Oscillations in
  Bigravity}},  {\em Phys. Rev. Lett.} {\bf 119} (2017), no.~11 111101,
  [\href{http://arxiv.org/abs/1703.07785}{{\tt arXiv:1703.07785}}].

\bibitem{Caldwell:2016sut}
R.~R. Caldwell, C.~Devulder, and N.~A. Maksimova, {\it {Gravitational
  wave--Gauge field oscillations}},  {\em Phys. Rev.} {\bf D94} (2016), no.~6
  063005, [\href{http://arxiv.org/abs/1604.08939}{{\tt arXiv:1604.08939}}].

\bibitem{BeltranJimenez:2018ymu}
J.~Beltrn~Jimnez and L.~Heisenberg, {\it {Non-trivial gravitational waves and
  structure formation phenomenology from dark energy}},  {\em JCAP} {\bf 1809}
  (2018), no.~09 035, [\href{http://arxiv.org/abs/1806.01753}{{\tt
  arXiv:1806.01753}}].

\bibitem{Will:1997bb}
C.~M. Will, {\it {Bounding the mass of the graviton using gravitational wave
  observations of inspiralling compact binaries}},  {\em Phys. Rev.} {\bf D57}
  (1998) 2061--2068, [\href{http://arxiv.org/abs/gr-qc/9709011}{{\tt
  gr-qc/9709011}}].

\bibitem{Schutz:1986gp}
B.~F. Schutz, {\it {Determining the Hubble Constant from Gravitational Wave
  Observations}},  {\em Nature} {\bf 323} (1986) 310--311.

\bibitem{Holz:2005df}
D.~E. Holz and S.~A. Hughes, {\it {Using gravitational-wave standard sirens}},
  {\em Astrophys. J.} {\bf 629} (2005) 15--22,
  [\href{http://arxiv.org/abs/astro-ph/0504616}{{\tt astro-ph/0504616}}].

\bibitem{Metzger:2016pju}
B.~D. Metzger, {\it {Kilonovae}},  {\em Living Rev. Rel.} {\bf 20} (2017) 3,
  [\href{http://arxiv.org/abs/1610.09381}{{\tt arXiv:1610.09381}}].

\bibitem{Vitale:2018wlg}
S.~Vitale and H.-Y. Chen, {\it {Measuring the Hubble constant with neutron star
  black hole mergers}},  {\em Phys. Rev. Lett.} {\bf 121} (2018), no.~2 021303,
  [\href{http://arxiv.org/abs/1804.07337}{{\tt arXiv:1804.07337}}].

\bibitem{DelPozzo:2011yh}
W.~Del~Pozzo, {\it {Inference of the cosmological parameters from gravitational
  waves: application to second generation interferometers}},  {\em Phys. Rev.}
  {\bf D86} (2012) 043011, [\href{http://arxiv.org/abs/1108.1317}{{\tt
  arXiv:1108.1317}}].

\bibitem{Chen:2016tys}
H.-Y. Chen and D.~E. Holz, {\it {Finding the One: Identifying the Host Galaxies
  of Gravitational-Wave Sources}},  \href{http://arxiv.org/abs/1612.01471}{{\tt
  arXiv:1612.01471}}.

\bibitem{Messenger:2011gi}
C.~Messenger and J.~Read, {\it {Measuring a cosmological distance-redshift
  relationship using only gravitational wave observations of binary neutron
  star coalescences}},  {\em Phys. Rev. Lett.} {\bf 108} (2012) 091101,
  [\href{http://arxiv.org/abs/1107.5725}{{\tt arXiv:1107.5725}}].

\bibitem{Taylor:2011fs}
S.~R. Taylor, J.~R. Gair, and I.~Mandel, {\it {Hubble without the Hubble:
  Cosmology using advanced gravitational-wave detectors alone}},  {\em Phys.
  Rev.} {\bf D85} (2012) 023535, [\href{http://arxiv.org/abs/1108.5161}{{\tt
  arXiv:1108.5161}}].

\bibitem{Messenger:2013fya}
C.~Messenger, K.~Takami, S.~Gossan, L.~Rezzolla, and B.~S. Sathyaprakash, {\it
  {Source Redshifts from Gravitational-Wave Observations of Binary Neutron Star
  Mergers}},  {\em Phys. Rev.} {\bf X4} (2014), no.~4 041004,
  [\href{http://arxiv.org/abs/1312.1862}{{\tt arXiv:1312.1862}}].

\bibitem{Fishbach:2018gjp}
{\bf Virgo, members of the LIGO Scientific} Collaboration, M.~Fishbach,
  R.~Gray, I.~M. Hernandez, H.~Qi, and A.~Sur, {\it {A standard siren
  measurement of the Hubble constant from GW170817 without the electromagnetic
  counterpart}},  \href{http://arxiv.org/abs/1807.05667}{{\tt
  arXiv:1807.05667}}.

\bibitem{Soares-Santos:2019irc}
{\bf DES, LIGO Scientific, Virgo} Collaboration, M.~Soares-Santos et~al., {\it
  {First Measurement of the Hubble Constant from a Dark Standard Siren using
  the Dark Energy Survey Galaxies and the LIGO/Virgo BinaryBlack-hole Merger
  GW170814}},  {\em Astrophys. J.} {\bf 876} (2019), no.~1 L7,
  [\href{http://arxiv.org/abs/1901.01540}{{\tt arXiv:1901.01540}}].

\bibitem{Jana:2017ost}
S.~Jana, G.~K. Chakravarty, and S.~Mohanty, {\it {Constraints on Born-Infeld
  gravity from the speed of gravitational waves after GW170817 and GRB
  170817A}},  {\em Phys. Rev.} {\bf D97} (2018), no.~8 084011,
  [\href{http://arxiv.org/abs/1711.04137}{{\tt arXiv:1711.04137}}].

\bibitem{Franchini:2019npi}
N.~Franchini and T.~P. Sotiriou, {\it {Cosmology with subdominant Horndeski
  scalar field}},  \href{http://arxiv.org/abs/1903.05427}{{\tt
  arXiv:1903.05427}}.

\bibitem{Copeland:2018yuh}
E.~J. Copeland, M.~Kopp, A.~Padilla, P.~M. Saffin, and C.~Skordis, {\it {Dark
  energy after GW170817, revisited}},  {\em Phys. Rev. Lett.} {\bf 122} (2018),
  no.~6 061301, [\href{http://arxiv.org/abs/1810.08239}{{\tt
  arXiv:1810.08239}}].

\bibitem{Arai:2017hxj}
S.~Arai and A.~Nishizawa, {\it {Generalized framework for testing gravity with
  gravitational-wave propagation. II. Constraints on Horndeski theory}},  {\em
  Phys. Rev.} {\bf D97} (2018), no.~10 104038,
  [\href{http://arxiv.org/abs/1711.03776}{{\tt arXiv:1711.03776}}].

\bibitem{Aasi:2013wya}
{\bf VIRGO, KAGRA, LIGO Scientific} Collaboration, B.~P. Abbott et~al., {\it
  {Prospects for Observing and Localizing Gravitational-Wave Transients with
  Advanced LIGO, Advanced Virgo and KAGRA}},  {\em Living Rev. Rel.} {\bf 21}
  (2018) 3, [\href{http://arxiv.org/abs/1304.0670}{{\tt arXiv:1304.0670}}].
  [Living Rev. Rel.19,1(2016)].

\bibitem{Sathyaprakash:2012jk}
B.~Sathyaprakash et~al., {\it {Scientific Objectives of Einstein Telescope}},
  {\em Class. Quant. Grav.} {\bf 29} (2012) 124013,
  [\href{http://arxiv.org/abs/1206.0331}{{\tt arXiv:1206.0331}}]. [Erratum:
  Class. Quant. Grav.30,079501(2013)].

\bibitem{Evans:2016mbw}
{\bf LIGO Scientific} Collaboration, B.~P. Abbott et~al., {\it {Exploring the
  Sensitivity of Next Generation Gravitational Wave Detectors}},  {\em Class.
  Quant. Grav.} {\bf 34} (2017), no.~4 044001,
  [\href{http://arxiv.org/abs/1607.08697}{{\tt arXiv:1607.08697}}].

\bibitem{Mills:2017urp}
C.~Mills, V.~Tiwari, and S.~Fairhurst, {\it {Localization of binary neutron
  star mergers with second and third generation gravitational-wave detectors}},
   {\em Phys. Rev.} {\bf D97} (2018), no.~10 104064,
  [\href{http://arxiv.org/abs/1708.00806}{{\tt arXiv:1708.00806}}].

\bibitem{AmaroSeoane:2012km}
P.~Amaro-Seoane et~al., {\it {eLISA/NGO: Astrophysics and cosmology in the
  gravitational-wave millihertz regime}},  {\em GW Notes} {\bf 6} (2013)
  4--110, [\href{http://arxiv.org/abs/1201.3621}{{\tt arXiv:1201.3621}}].

\bibitem{Nishizawa:2018srh}
A.~Nishizawa and T.~Kobayashi, {\it {Parity-violating gravity and GW170817}},
  {\em Phys. Rev.} {\bf D98} (2018), no.~12 124018,
  [\href{http://arxiv.org/abs/1809.00815}{{\tt arXiv:1809.00815}}].

\bibitem{Kalogera:2019sui}
V.~Kalogera et~al., {\it {Deeper, Wider, Sharper: Next-Generation Ground-Based
  Gravitational-Wave Observations of Binary Black Holes}},
  \href{http://arxiv.org/abs/1903.09220}{{\tt arXiv:1903.09220}}.

\bibitem{Piorkowska:2013eww}
A.~Pirkowska, M.~Biesiada, and Z.-H. Zhu, {\it {Strong gravitational lensing of
  gravitational waves in Einstein Telescope}},  {\em JCAP} {\bf 1310} (2013)
  022, [\href{http://arxiv.org/abs/1309.5731}{{\tt arXiv:1309.5731}}].

\bibitem{Gair:2012nm}
J.~R. Gair, M.~Vallisneri, S.~L. Larson, and J.~G. Baker, {\it {Testing General
  Relativity with Low-Frequency, Space-Based Gravitational-Wave Detectors}},
  {\em Living Rev. Rel.} {\bf 16} (2013) 7,
  [\href{http://arxiv.org/abs/1212.5575}{{\tt arXiv:1212.5575}}].

\bibitem{Sesana:2016ljz}
A.~Sesana, {\it {The promise of multi-band gravitational wave astronomy}},
  {\em Phys. Rev. Lett.} {\bf 116} (2016), no.~23 231102,
  [\href{http://arxiv.org/abs/1602.06951}{{\tt arXiv:1602.06951}}].

\bibitem{Ishak:2019aay}
M.~Ishak et~al., {\it {Modified Gravity and Dark Energy models Beyond $w(z)$CDM
  Testable by LSST}},  \href{http://arxiv.org/abs/1905.09687}{{\tt
  arXiv:1905.09687}}.

\bibitem{Bechtol:2019acd}
K.~Bechtol et~al., {\it {Dark Matter Science in the Era of LSST}},
  \href{http://arxiv.org/abs/1903.04425}{{\tt arXiv:1903.04425}}.

\bibitem{Dore:2019pld}
O.~Dor et~al., {\it {WFIRST: The Essential Cosmology Space Observatory for the
  Coming Decade}},  \href{http://arxiv.org/abs/1904.01174}{{\tt
  arXiv:1904.01174}}.

\bibitem{1996MNRAS.283..837S}
P.~{Schneider}, {\it {Detection of (dark) matter concentrations via weak
  gravitational lensing}},  {\em \mnras} {\bf 283} (Dec., 1996) 837--853,
  [\href{http://arxiv.org/abs/astro-ph/9601039}{{\tt astro-ph/9601039}}].

\bibitem{1998MNRAS.296..873S}
P.~{Schneider}, L.~{van Waerbeke}, B.~{Jain}, and G.~{Kruse}, {\it {A new
  measure for cosmic shear}},  {\em \mnras} {\bf 296} (Jun, 1998) 873--892,
  [\href{http://arxiv.org/abs/astro-ph/9708143}{{\tt astro-ph/9708143}}].

\bibitem{2018A&A...619A..38P}
A.~{Peel}, V.~{Pettorino}, C.~{Giocoli}, J.-L. {Starck}, and M.~{Baldi}, {\it
  {Breaking degeneracies in modified gravity with higher (than 2nd) order
  weak-lensing statistics}},  {\em Astron. Astrophys.} {\bf 619} (Nov, 2018)
  A38, [\href{http://arxiv.org/abs/1805.05146}{{\tt arXiv:1805.05146}}].

\bibitem{2018MNRAS.480.3725S}
A.~{Spurio Mancini}, R.~{Reischke}, V.~{Pettorino}, B.~M. {Sch{\"a}fer}, and
  M.~{Zumalac{\'a}rregui}, {\it {Testing (modified) gravity with 3D and
  tomographic cosmic shear}},  {\em \mnras} {\bf 480} (Nov, 2018) 3725--3738,
  [\href{http://arxiv.org/abs/1801.04251}{{\tt arXiv:1801.04251}}].

\bibitem{Amendola:2007rr}
L.~Amendola, M.~Kunz, and D.~Sapone, {\it {Measuring the dark side (with weak
  lensing)}},  {\em JCAP} {\bf 0804} (Apr, 2008) 013,
  [\href{http://arxiv.org/abs/0704.2421}{{\tt arXiv:0704.2421}}].

\bibitem{Ade2016}
{\bf Planck} Collaboration, P.~A.~R. Ade et~al., {\it {Planck 2015 results.
  XIV. Dark energy and modified gravity}},  {\em Astron. Astrophys.} {\bf 594}
  (Sep, 2016) A14, [\href{http://arxiv.org/abs/1502.01590}{{\tt
  arXiv:1502.01590}}].

\bibitem{2013MNRAS.430.2200K}
M.~{Kilbinger}, L.~{Fu}, C.~{Heymans}, F.~{Simpson}, J.~{Benjamin}, T.~{Erben},
  J.~{Harnois-D{\'e}raps}, H.~{Hoekstra}, H.~{Hildebrand t}, T.~D. {Kitching},
  Y.~{Mellier}, L.~{Miller}, L.~{Van Waerbeke}, K.~{Benabed}, C.~{Bonnett},
  J.~{Coupon}, M.~J. {Hudson}, K.~{Kuijken}, B.~{Rowe}, T.~{Schrabback},
  E.~{Semboloni}, S.~{Vafaei}, and M.~{Velander}, {\it {CFHTLenS: combined
  probe cosmological model comparison using 2D weak gravitational lensing}},
  {\em \mnras} {\bf 430} (Apr, 2013) 2200--2220,
  [\href{http://arxiv.org/abs/1212.3338}{{\tt arXiv:1212.3338}}].

\bibitem{2012MNRAS.423.3445S}
B.~M. {Sch{\"a}fer} and L.~{Heisenberg}, {\it {Weak lensing tomography with
  orthogonal polynomials}},  {\em \mnras} {\bf 423} (Jul, 2012) 3445--3457,
  [\href{http://arxiv.org/abs/1107.2213}{{\tt arXiv:1107.2213}}].

\bibitem{1999ApJ...522L..21H}
W.~{Hu}, {\it {Power Spectrum Tomography with Weak Lensing}},  {\em The
  Astrophysical Journal Letter} {\bf 522} (Sep, 1999) L21--L24,
  [\href{http://arxiv.org/abs/astro-ph/9904153}{{\tt astro-ph/9904153}}].

\bibitem{2013MNRAS.431.1547B}
J.~{Benjamin}, L.~{Van Waerbeke}, C.~{Heymans}, M.~{Kilbinger}, T.~{Erben},
  H.~{Hildebrandt}, H.~{Hoekstra}, T.~D. {Kitching}, Y.~{Mellier}, L.~{Miller},
  B.~{Rowe}, T.~{Schrabback}, F.~{Simpson}, J.~{Coupon}, L.~{Fu},
  J.~{Harnois-D{\'e}raps}, M.~J. {Hudson}, K.~{Kuijken}, E.~{Semboloni},
  S.~{Vafaei}, and M.~{Velander}, {\it {CFHTLenS tomographic weak lensing:
  quantifying accurate redshift distributions}},  {\em \mnras} {\bf 431} (May,
  2013) 1547--1564, [\href{http://arxiv.org/abs/1212.3327}{{\tt
  arXiv:1212.3327}}].

\bibitem{1953ApJ...117..134L}
D.~N. {Limber}, {\it {The Analysis of Counts of the Extragalactic Nebulae in
  Terms of a Fluctuating Density Field.}},  {\em The Astrophysical Journal}
  {\bf 117} (Jan., 1953) 134.

\bibitem{1992ApJ...388..272K}
N.~{Kaiser}, {\it {Weak gravitational lensing of distant galaxies}},  {\em The
  Astrophysical Journal} {\bf 388} (Apr., 1992) 272--286.

\bibitem{2003MNRAS.343.1327H}
A.~{Heavens}, {\it {3D weak lensing}},  {\em \mnras} {\bf 343} (Aug, 2003)
  1327--1334, [\href{http://arxiv.org/abs/astro-ph/0304151}{{\tt
  astro-ph/0304151}}].

\bibitem{2005PhRvD..72b3516C}
P.~G. {Castro}, A.~F. {Heavens}, and T.~D. {Kitching}, {\it {Weak lensing
  analysis in three dimensions}},  {\em Physical Review D} {\bf 72} (Jul, 2005)
  023516, [\href{http://arxiv.org/abs/astro-ph/0503479}{{\tt
  astro-ph/0503479}}].

\bibitem{2006MNRAS.373..105H}
A.~F. {Heavens}, T.~D. {Kitching}, and A.~N. {Taylor}, {\it {Measuring dark
  energy properties with 3D cosmic shear}},  {\em \mnras} {\bf 373} (Nov, 2006)
  105--120, [\href{http://arxiv.org/abs/astro-ph/0606568}{{\tt
  astro-ph/0606568}}].

\bibitem{2011MNRAS.413.2923K}
T.~D. {Kitching}, A.~F. {Heavens}, and L.~{Miller}, {\it {3D photometric cosmic
  shear}},  {\em \mnras} {\bf 413} (Jun, 2011) 2923--2934,
  [\href{http://arxiv.org/abs/1007.2953}{{\tt arXiv:1007.2953}}].

\bibitem{2014MNRAS.442.1326K}
T.~D. {Kitching}, A.~F. {Heavens}, J.~{Alsing}, T.~{Erben}, C.~{Heymans},
  H.~{Hildebrandt}, H.~{Hoekstra}, A.~{Jaffe}, A.~{Kiessling}, Y.~{Mellier},
  L.~{Miller}, L.~{van Waerbeke}, J.~{Benjamin}, J.~{Coupon}, L.~{Fu}, M.~J.
  {Hudson}, M.~{Kilbinger}, K.~{Kuijken}, B.~T.~P. {Rowe}, T.~{Schrabback},
  E.~{Semboloni}, and M.~{Veland er}, {\it {3D cosmic shear: cosmology from
  CFHTLenS}},  {\em \mnras} {\bf 442} (Aug, 2014) 1326--1349,
  [\href{http://arxiv.org/abs/1401.6842}{{\tt arXiv:1401.6842}}].

\bibitem{2018PhRvD..98j3507S}
A.~{Spurio Mancini}, P.~L. {Taylor}, R.~{Reischke}, T.~{Kitching},
  V.~{Pettorino}, B.~M. {Sch{\"a}fer}, B.~{Zieser}, and P.~M. {Merkel}, {\it
  {3D cosmic shear: Numerical challenges, 3D lensing random fields generation,
  and Minkowski functionals for cosmological inference}},  {\em Physical Review
  D} {\bf 98} (Nov, 2018) 103507, [\href{http://arxiv.org/abs/1807.11461}{{\tt
  arXiv:1807.11461}}].

\bibitem{2016MNRAS.459.1586Z}
B.~{Zieser} and P.~M. {Merkel}, {\it {The cross-correlation between 3D cosmic
  shear and the integrated Sachs-Wolfe effect}},  {\em \mnras} {\bf 459} (Jun,
  2016) 1586--1595, [\href{http://arxiv.org/abs/1603.06406}{{\tt
  arXiv:1603.06406}}].

\bibitem{1974IJTP...10..363H}
G.~W. {Horndeski}, {\it {Second-Order Scalar-Tensor Field Equations in a
  Four-Dimensional Space}},  {\em International Journal of Theoretical Physics}
  {\bf 10} (Sept., 1974) 363--384.

\bibitem{Gleyzes2013}
J.~Gleyzes, D.~Langlois, F.~Piazza, and F.~Vernizzi, {\it {Essential Building
  Blocks of Dark Energy}},  {\em JCAP} {\bf 1308} (Aug, 2013) 025,
  [\href{http://arxiv.org/abs/1304.4840}{{\tt arXiv:1304.4840}}].

\bibitem{2017PhRvL.119p1101A}
{\bf LIGO Scientific, Virgo} Collaboration, B.~P. {Abbott}, R.~{Abbott}, T.~D.
  {Abbott}, F.~{Acernese}, K.~{Ackley}, C.~{Adams}, T.~{Adams}, P.~{Addesso},
  R.~X. {Adhikari}, V.~B. {Adya}, and et~al., {\it {GW170817: Observation of
  Gravitational Waves from a Binary Neutron Star Inspiral}},  {\em Physical
  Review Letters} {\bf 119} (Oct, 2017) 161101,
  [\href{http://arxiv.org/abs/1710.05832}{{\tt arXiv:1710.05832}}].

\bibitem{2019arXiv190103686S}
A.~{Spurio Mancini}, F.~{K{\"o}hlinger}, B.~{Joachimi}, V.~{Pettorino}, B.~M.
  {Sch{\"a}fer}, R.~{Reischke}, S.~{Brieden}, M.~{Archidiacono}, and
  J.~{Lesgourgues}, {\it {KiDS+GAMA: Constraints on Horndeski gravity from
  combined large-scale structure probes}},  {\em arXiv e-prints} (Jan, 2019)
  arXiv:1901.03686, [\href{http://arxiv.org/abs/1901.03686}{{\tt
  arXiv:1901.03686}}].

\bibitem{2015MNRAS.454.1958M}
A.~J. {Mead}, J.~A. {Peacock}, C.~{Heymans}, S.~{Joudaki}, and A.~F. {Heavens},
  {\it {An accurate halo model for fitting non-linear cosmological power
  spectra and baryonic feedback models}},  {\em \mnras} {\bf 454} (Dec, 2015)
  1958--1975, [\href{http://arxiv.org/abs/1505.07833}{{\tt arXiv:1505.07833}}].

\bibitem{2018MNRAS.476.4662V}
E.~{van Uitert}, B.~{Joachimi}, S.~{Joudaki}, A.~{Amon}, C.~{Heymans},
  F.~{K{\"o}hlinger}, M.~{Asgari}, C.~{Blake}, A.~{Choi}, T.~{Erben}, D.~J.
  {Farrow}, J.~{Harnois-D{\'e}raps}, H.~{Hildebrandt}, H.~{Hoekstra}, T.~D.
  {Kitching}, D.~{Klaes}, K.~{Kuijken}, J.~{Merten}, L.~{Miller},
  R.~{Nakajima}, P.~{Schneider}, E.~{Valentijn}, and M.~{Viola}, {\it
  {KiDS+GAMA: cosmology constraints from a joint analysis of cosmic shear,
  galaxy-galaxy lensing, and angular clustering}},  {\em \mnras} {\bf 476}
  (Jun, 2018) 4662--4689, [\href{http://arxiv.org/abs/1706.05004}{{\tt
  arXiv:1706.05004}}].

\bibitem{2016A&A...594A..13P}
{Planck Collaboration}, P.~A.~R. {Ade}, N.~{Aghanim}, M.~{Arnaud},
  M.~{Ashdown}, J.~{Aumont}, C.~{Baccigalupi}, A.~J. {Banday}, R.~B.
  {Barreiro}, J.~G. {Bartlett}, and et~al., {\it {Planck 2015 results. XIII.
  Cosmological parameters}},  {\em \aap} {\bf 594} (Sep, 2016) A13,
  [\href{http://arxiv.org/abs/1502.01589}{{\tt arXiv:1502.01589}}].

\bibitem{2019PhRvD..99f3527L}
Z.~{Li}, J.~{Liu}, J.~M.~Z. {Matilla}, and W.~R. {Coulton}, {\it {Constraining
  neutrino mass with tomographic weak lensing peak counts}},  {\em Physical
  Review D} {\bf 99} (Mar, 2019) 063527,
  [\href{http://arxiv.org/abs/1810.01781}{{\tt arXiv:1810.01781}}].

\bibitem{Ajani:2020dvu}
V.~Ajani, A.~Peel, V.~Pettorino, J.-L. Starck, Z.~Li, and J.~Liu, {\it
  {Constraining neutrino masses with weak-lensing multiscale peak counts}},
  {\em Phys. Rev. D} {\bf 102} (2020), no.~10 103531,
  [\href{http://arxiv.org/abs/2001.10993}{{\tt arXiv:2001.10993}}].

\bibitem{Ajani:2021pgp}
V.~Ajani, J.-L. Starck, and V.~Pettorino, {\it {Starlet l1-norm for weak
  lensing cosmology}},  {\em Astron. Astrophys.} {\bf 645} (2021) L11,
  [\href{http://arxiv.org/abs/2101.01542}{{\tt arXiv:2101.01542}}].

\bibitem{2017arXiv170705167S}
J.~{Schmelzle}, A.~{Lucchi}, T.~{Kacprzak}, A.~{Amara}, R.~{Sgier},
  A.~{R{\'e}fr{\'e}gier}, and T.~{Hofmann}, {\it {Cosmological model
  discrimination with Deep Learning}},  {\em arXiv e-prints} (Jul, 2017)
  arXiv:1707.05167, [\href{http://arxiv.org/abs/1707.05167}{{\tt
  arXiv:1707.05167}}].

\bibitem{2018PhRvD..97j3515G}
A.~{Gupta}, J.~M.~Z. {Matilla}, D.~{Hsu}, and Z.~{Haiman}, {\it {Non-Gaussian
  information from weak lensing data via deep learning}},  {\em Physical Review
  D} {\bf 97} (May, 2018) 103515, [\href{http://arxiv.org/abs/1802.01212}{{\tt
  arXiv:1802.01212}}].

\bibitem{2018PhRvD..98l3518F}
J.~{Fluri}, T.~{Kacprzak}, A.~{Refregier}, A.~{Amara}, A.~{Lucchi}, and
  T.~{Hofmann}, {\it {Cosmological constraints from noisy convergence maps
  through deep learning}},  {\em Physical Review D} {\bf 98} (Dec, 2018)
  123518, [\href{http://arxiv.org/abs/1807.08732}{{\tt arXiv:1807.08732}}].

\bibitem{2019NatAs...3...93R}
D.~{Ribli}, B.~{\'A}. {Pataki}, and I.~{Csabai}, {\it {An improved cosmological
  parameter inference scheme motivated by deep learning}},  {\em Nature
  Astronomy} {\bf 3} (Jan, 2019) 93--98,
  [\href{http://arxiv.org/abs/1806.05995}{{\tt arXiv:1806.05995}}].

\bibitem{2019PhRvD.100b3508P}
A.~{Peel}, F.~{Lalande}, J.-L. {Starck}, V.~{Pettorino}, J.~{Merten},
  C.~{Giocoli}, M.~{Meneghetti}, and M.~{Baldi}, {\it {Distinguishing standard
  and modified gravity cosmologies with machine learning}},  {\em \prd} {\bf
  100} (Jul, 2019) 023508, [\href{http://arxiv.org/abs/1810.11030}{{\tt
  arXiv:1810.11030}}].

\bibitem{2019MNRAS.487..104M}
J.~{Merten}, C.~{Giocoli}, M.~{Baldi}, M.~{Meneghetti}, A.~{Peel},
  F.~{Lalande}, J.-L. {Starck}, and V.~{Pettorino}, {\it {On the dissection of
  degenerate cosmologies with machine learning}},  {\em \mnras} {\bf 487} (Jul,
  2019) 104--122, [\href{http://arxiv.org/abs/1810.11027}{{\tt
  arXiv:1810.11027}}].

\bibitem{Schmidt:2008tn}
F.~Schmidt, M.~V. Lima, H.~Oyaizu, and W.~Hu, {\it {Non-linear Evolution of
  f(R) Cosmologies III: Halo Statistics}},  {\em Phys. Rev.} {\bf D79} (2009)
  083518, [\href{http://arxiv.org/abs/0812.0545}{{\tt arXiv:0812.0545}}].

\bibitem{Lombriser:2012nn}
L.~Lombriser, K.~Koyama, G.-B. Zhao, and B.~Li, {\it {Chameleon f(R) gravity in
  the virialized cluster}},  {\em Phys. Rev.} {\bf D85} (2012) 124054,
  [\href{http://arxiv.org/abs/1203.5125}{{\tt arXiv:1203.5125}}].

\bibitem{Li:2011uw}
Y.~Li and W.~Hu, {\it {Chameleon Halo Modeling in f(R) Gravity}},  {\em Phys.
  Rev.} {\bf D84} (2011) 084033, [\href{http://arxiv.org/abs/1107.5120}{{\tt
  arXiv:1107.5120}}].

\bibitem{Lombriser:2013wta}
L.~Lombriser, B.~Li, K.~Koyama, and G.-B. Zhao, {\it {Modeling halo mass
  functions in chameleon f(R) gravity}},  {\em Phys. Rev.} {\bf D87} (2013),
  no.~12 123511, [\href{http://arxiv.org/abs/1304.6395}{{\tt
  arXiv:1304.6395}}].

\bibitem{Lombriser:2013eza}
L.~Lombriser, K.~Koyama, and B.~Li, {\it {Halo modelling in chameleon
  theories}},  {\em JCAP} {\bf 1403} (2014) 021,
  [\href{http://arxiv.org/abs/1312.1292}{{\tt arXiv:1312.1292}}].

\bibitem{Pizzuti:2016ouw}
L.~Pizzuti et~al., {\it {CLASH-VLT: Testing the Nature of Gravity with Galaxy
  Cluster Mass Profiles}},  {\em JCAP} {\bf 1604} (2016), no.~04 023,
  [\href{http://arxiv.org/abs/1602.03385}{{\tt arXiv:1602.03385}}].

\bibitem{Mamon01}
G.~A. {Mamon}, A.~{Biviano}, and G.~{Bou{\'e}}, {\it {MAMPOSSt: Modelling
  Anisotropy and Mass Profiles of Observed Spherical Systems - I. Gaussian 3D
  velocities}},  {\em Mon. Not. Roy. Astron. Soc.} {\bf 429} (Mar., 2013)
  3079--3098, [\href{http://arxiv.org/abs/1212.1455}{{\tt arXiv:1212.1455}}].

\bibitem{Pizzuti:2019wte}
L.~Pizzuti, I.~D. Saltas, S.~Casas, L.~Amendola, and A.~Biviano, {\it {Future
  constraints on the gravitational slip with the mass profiles of galaxy
  clusters}},  {\em Mon. Not. Roy. Astron. Soc.} {\bf 486} (2019), no.~1
  596--607, [\href{http://arxiv.org/abs/1901.01961}{{\tt arXiv:1901.01961}}].

\bibitem{Pizzuti:2017diz}
L.~Pizzuti et~al., {\it {CLASH-VLT: constraints on $f(R)$ gravity models with
  galaxy clusters using lensing and kinematic analyses}},  {\em JCAP} {\bf
  1707} (2017), no.~07 023, [\href{http://arxiv.org/abs/1705.05179}{{\tt
  arXiv:1705.05179}}].

\bibitem{Schmidt2009}
F.~Schmidt, A.~Vikhlinin, and W.~Hu, {\it Cluster constraints on $f(r)$
  gravity},  {\em Phys. Rev. D} {\bf 80} (Oct, 2009) 083505.

\bibitem{Rapetti2010}
D.~Rapetti, S.~W. Allen, A.~Mantz, and H.~Ebeling, {\it {The observed growth of
  massive galaxy clusters – III. Testing general relativity on cosmological
  scales}},  {\em Monthly Notices of the Royal Astronomical Society} {\bf 406}
  (08, 2010) 1796--1804.

\bibitem{Lombriser:2010mp}
L.~Lombriser, A.~Slosar, U.~Seljak, and W.~Hu, {\it {Constraints on f(R)
  gravity from probing the large-scale structure}},  {\em Phys. Rev.} {\bf D85}
  (2012) 124038, [\href{http://arxiv.org/abs/1003.3009}{{\tt
  arXiv:1003.3009}}].

\bibitem{Rapetti2011}
D.~{Rapetti}, S.~W. {Allen}, A.~{Mantz}, and H.~{Ebeling}, {\it {Testing
  General Relativity on Cosmic Scales with the Observed Abundance of Massive
  Clusters}},  {\em Progress of Theoretical Physics Supplement} {\bf 190}
  (2011) 179--187.

\bibitem{Cataneo2015}
M.~Cataneo, D.~Rapetti, F.~Schmidt, A.~Mantz, S.~Allen, D.~Applegate, P.~Kelly,
  A.~{Von Der Linden}, and R.~Morris, {\it New constraints on f (r) gravity
  from clusters of galaxies},  {\em Physical Review D - Particles, Fields,
  Gravitation and Cosmology} {\bf 92} (8, 2015).

\bibitem{Ferraro2011}
S.~Ferraro, F.~Schmidt, and W.~Hu, {\it Cluster abundance in $f(r)$ gravity
  models},  {\em Phys. Rev. D} {\bf 83} (Mar, 2011) 063503.

\bibitem{Cataneo:2016iav}
M.~Cataneo, D.~Rapetti, L.~Lombriser, and B.~Li, {\it {Cluster abundance in
  chameleon $f(R)$ gravity I: toward an accurate halo mass function
  prediction}},  {\em JCAP} {\bf 1612} (2016), no.~12 024,
  [\href{http://arxiv.org/abs/1607.08788}{{\tt arXiv:1607.08788}}].

\bibitem{Terukina2012}
A.~Terukina and K.~Yamamoto, {\it Gas density profile in dark matter halo in
  chameleon cosmology},  {\em Phys. Rev. D} {\bf 86} (Nov, 2012) 103503.

\bibitem{Terukina:2013eqa}
A.~Terukina, L.~Lombriser, K.~Yamamoto, D.~Bacon, K.~Koyama, and R.~C. Nichol,
  {\it {Testing chameleon gravity with the Coma cluster}},  {\em JCAP} {\bf
  1404} (2014) 013, [\href{http://arxiv.org/abs/1312.5083}{{\tt
  arXiv:1312.5083}}].

\bibitem{Wilcox:2015kna}
H.~Wilcox et~al., {\it {The XMM Cluster Survey: Testing chameleon gravity using
  the profiles of clusters}},  {\em Mon. Not. Roy. Astron. Soc.} {\bf 452}
  (2015), no.~2 1171--1183, [\href{http://arxiv.org/abs/1504.03937}{{\tt
  arXiv:1504.03937}}].

\bibitem{Wilcox:2016guw}
H.~Wilcox, R.~C. Nichol, G.-B. Zhao, D.~Bacon, K.~Koyama, and A.~K. Romer, {\it
  {Simulation tests of galaxy cluster constraints on chameleon gravity}},  {\em
  Mon. Not. Roy. Astron. Soc.} {\bf 462} (2016), no.~1 715--725,
  [\href{http://arxiv.org/abs/1603.05911}{{\tt arXiv:1603.05911}}].

\bibitem{Salzano:2016udu}
V.~Salzano, D.~F. Mota, M.~P. Dabrowski, and S.~Capozziello, {\it {No need for
  dark matter in galaxy clusters within Galileon theory}},  {\em JCAP} {\bf
  1610} (2016), no.~10 033, [\href{http://arxiv.org/abs/1607.02606}{{\tt
  arXiv:1607.02606}}].

\bibitem{Biviano01}
A.~{Biviano}, P.~{Rosati}, I.~{Balestra}, A.~{Mercurio}, M.~{Girardi},
  M.~{Nonino}, C.~{Grillo}, M.~{Scodeggio}, D.~{Lemze}, D.~{Kelson}, and
  {others}, {\it {CLASH-VLT: The mass, velocity-anisotropy, and
  pseudo-phase-space density profiles of the z = 0.44 galaxy cluster MACS
  J1206.2-0847}},  {\em Astron. Astrophys.} {\bf 558} (Oct., 2013) A1,
  [\href{http://arxiv.org/abs/1307.5867}{{\tt arXiv:1307.5867}}].

\bibitem{Binney1982}
J.~{Binney} and G.~A. {Mamon}, {\it {M/L and velocity anisotropy from
  observations of spherical galaxies, or must M87 have a massive black hole}},
  {\em Mon. Not. Roy. Astron. Soc.} {\bf 200} (July, 1982) 361--375.

\bibitem{Host2}
O.~{Host}, S.~H. {Hansen}, R.~{Piffaretti}, A.~{Morandi}, S.~{Ettori}, S.~T.
  {Kay}, and R.~{Valdarnini}, {\it {Measurement of the Dark Matter Velocity
  Anisotropy in Galaxy Clusters}},  {\em Astrophys. J.} {\bf 690} (Jan., 2009)
  358--366, [\href{http://arxiv.org/abs/0808.2049}{{\tt arXiv:0808.2049}}].

\bibitem{Host2009}
O.~{Host}, {\it {Measurement of the dark matter velocity anisotropy profile in
  galaxy clusters}},  {\em Nuclear Physics B Proceedings Supplements} {\bf 194}
  (Oct., 2009) 111--115, [\href{http://arxiv.org/abs/0810.3676}{{\tt
  arXiv:0810.3676}}].

\bibitem{Hansen2006}
S.~H. {Hansen} and B.~{Moore}, {\it {A universal density slope Velocity
  anisotropy relation for relaxed structures}},  {\em New Astron.} {\bf 11}
  (Mar., 2006) 333--338, [\href{http://arxiv.org/abs/astro-ph/0411473}{{\tt
  astro-ph/0411473}}].

\bibitem{mamon10}
G.~A. {Mamon}, A.~{Biviano}, and G.~{Murante}, {\it {The universal distribution
  of halo interlopers in projected phase space. Bias in galaxy cluster
  concentration and velocity anisotropy?}},  {\em Astron. Astrophys.} {\bf 520}
  (Sept., 2010) A30, [\href{http://arxiv.org/abs/1003.0033}{{\tt
  arXiv:1003.0033}}].

\bibitem{Pizzuti16}
L.~{Pizzuti}, B.~{Sartoris}, S.~{Borgani}, L.~{Amendola}, K.~{Umetsu},
  A.~{Biviano}, M.~{Girardi}, P.~{Rosati}, I.~{Balestra}, G.~B. {Caminha},
  B.~{Frye}, A.~{Koekemoer}, C.~{Grillo}, M.~{Lombardi}, A.~{Mercurio}, and
  M.~{Nonino}, {\it {CLASH-VLT: testing the nature of gravity with galaxy
  cluster mass profiles}},  {\em \jcap} {\bf 2016} (Apr, 2016) 023,
  [\href{http://arxiv.org/abs/1602.03385}{{\tt arXiv:1602.03385}}].

\bibitem{Tiret01}
O.~{Tiret}, F.~{Combes}, G.~W. {Angus}, B.~{Famaey}, and H.~S. {Zhao}, {\it
  {Velocity dispersion around ellipticals in MOND}},  {\em \aap} {\bf 476}
  (Dec., 2007) L1--L4, [\href{http://arxiv.org/abs/0710.4070}{{\tt
  arXiv:0710.4070}}].

\bibitem{Pizzuti19b}
L.~{Pizzuti}, B.~{Sartoris}, S.~{Borgani}, and A.~{Biviano}, {\it {Calibration
  of systematics in constraining modified gravity models with galaxy cluster
  mass profiles}},  {\em arXiv e-prints} (Dec, 2019) arXiv:1912.09096,
  [\href{http://arxiv.org/abs/1912.09096}{{\tt arXiv:1912.09096}}].

\bibitem{Diaferio01}
A.~{Diaferio} and M.~J. {Geller}, {\it {Infall Regions of Galaxy Clusters}},
  {\em Astrophys. J.} {\bf 481} (May, 1997) 633--643,
  [\href{http://arxiv.org/abs/astro-ph/9701034}{{\tt astro-ph/9701034}}].

\bibitem{Ettori2013}
S.~{Ettori}, A.~{Donnarumma}, E.~{Pointecouteau}, T.~H. {Reiprich},
  S.~{Giodini}, L.~{Lovisari}, and R.~W. {Schmidt}, {\it {Mass Profiles of
  Galaxy Clusters from X-ray Analysis}},  {\em Space Sci. Rev.} {\bf 177}
  (Aug., 2013) 119--154, [\href{http://arxiv.org/abs/1303.3530}{{\tt
  arXiv:1303.3530}}].

\bibitem{Biffi16}
V.~{Biffi}, S.~{Borgani}, G.~{Murante}, E.~{Rasia}, S.~{Planelles}, G.~L.
  {Granato}, C.~{Ragone-Figueroa}, A.~M. {Beck}, M.~{Gaspari}, and K.~{Dolag},
  {\it {On the Nature of Hydrostatic Equilibrium in Galaxy Clusters}},  {\em
  Astrophys. J.} {\bf 827} (Aug., 2016) 112,
  [\href{http://arxiv.org/abs/1606.02293}{{\tt arXiv:1606.02293}}].

\bibitem{Martizzi2016}
D.~{Martizzi} and H.~{Agrusa}, {\it {Mass modeling of galaxy clusters:
  quantifying hydrostatic bias and contribution from non-thermal pressure}},
  {\em ArXiv e-prints} (Aug., 2016)
  [\href{http://arxiv.org/abs/1608.04388}{{\tt arXiv:1608.04388}}].

\bibitem{Nagai11}
D.~{Nagai} and E.~T. {Lau}, {\it {Gas Clumping in the Outskirts of
  {$\Lambda$}CDM Clusters}},  {\em Astrophys. J.} {\bf 731} (Apr., 2011) L10,
  [\href{http://arxiv.org/abs/1103.0280}{{\tt arXiv:1103.0280}}].

\bibitem{Martinelli:2010wn}
M.~Martinelli, E.~Calabrese, F.~De~Bernardis, A.~Melchiorri, L.~Pagano, and
  R.~Scaramella, {\it {Constraining Modified Gravity with Euclid}},  {\em Phys.
  Rev.} {\bf D83} (2011) 023012, [\href{http://arxiv.org/abs/1010.5755}{{\tt
  arXiv:1010.5755}}].

\bibitem{Capozziello:2012ie}
S.~Capozziello and M.~De~Laurentis, {\it {The dark matter problem from f(R)
  gravity viewpoint}},  {\em Annalen Phys.} {\bf 524} (2012) 545--578.

\bibitem{Koivisto:2012za}
T.~S. Koivisto, D.~F. Mota, and M.~Zumalac{\'a}rregui, {\it {Screening
  Modifications of Gravity through Disformally Coupled Fields}},  {\em
  Phys.Rev.Lett.} {\bf 109} (2012) 241102,
  [\href{http://arxiv.org/abs/1205.3167}{{\tt arXiv:1205.3167}}].

\bibitem{Thorsrud:2012mu}
M.~Thorsrud, D.~F. Mota, and S.~Hervik, {\it {Cosmology of a Scalar Field
  Coupled to Matter and an Isotropy-Violating Maxwell Field}},  {\em JHEP} {\bf
  10} (2012) 066, [\href{http://arxiv.org/abs/1205.6261}{{\tt
  arXiv:1205.6261}}].

\bibitem{Barrow:2002zh}
J.~D. Barrow and D.~F. Mota, {\it {Gauge invariant perturbations of varying
  alpha cosmologies}},  {\em Class. Quant. Grav.} {\bf 20} (2003) 2045--2062,
  [\href{http://arxiv.org/abs/gr-qc/0212032}{{\tt gr-qc/0212032}}].

\bibitem{DeFelice:2009rw}
A.~De~Felice, D.~F. Mota, and S.~Tsujikawa, {\it {Matter instabilities in
  general Gauss-Bonnet gravity}},  {\em Phys. Rev.} {\bf D81} (2010) 023532,
  [\href{http://arxiv.org/abs/0911.1811}{{\tt arXiv:0911.1811}}].

\bibitem{Li:2008fa}
B.~Li, D.~F. Mota, and D.~J. Shaw, {\it {Microscopic and Macroscopic Behaviors
  of Palatini Modified Gravity Theories}},  {\em Phys. Rev.} {\bf D78} (2008)
  064018, [\href{http://arxiv.org/abs/0805.3428}{{\tt arXiv:0805.3428}}].

\bibitem{Brax:2012bsa}
P.~Brax, {\it {Screened modified gravity}},  {\em Acta Phys. Polon.} {\bf B43}
  (2012) 2307--2329, [\href{http://arxiv.org/abs/1211.5237}{{\tt
  arXiv:1211.5237}}].

\bibitem{Davis:2011pj}
A.-C. Davis, B.~Li, D.~F. Mota, and H.~A. Winther, {\it {Structure Formation in
  the Symmetron Model}},  {\em Astrophys. J.} {\bf 748} (2012) 61,
  [\href{http://arxiv.org/abs/1108.3081}{{\tt arXiv:1108.3081}}].

\bibitem{Hagala:2016fks}
R.~Hagala, C.~Llinares, and D.~F. Mota, {\it {Cosmic Tsunamis in Modified
  Gravity: Scalar waves disrupting screening mechanisms}},
  \href{http://arxiv.org/abs/1607.02600}{{\tt arXiv:1607.02600}}.

\bibitem{Hagala:2015paa}
R.~Hagala, C.~Llinares, and D.~F. Mota, {\it {Cosmological simulations with
  disformally coupled symmetron fields}},  {\em Astron. Astrophys.} {\bf 585}
  (2016) A37, [\href{http://arxiv.org/abs/1504.07142}{{\tt arXiv:1504.07142}}].

\bibitem{Llinares:2013qbh}
C.~Llinares and D.~Mota, {\it {Releasing scalar fields: cosmological
  simulations of scalar-tensor theories for gravity beyond the static
  approximation}},  {\em Phys.Rev.Lett.} {\bf 110} (2013), no.~16 161101,
  [\href{http://arxiv.org/abs/1302.1774}{{\tt arXiv:1302.1774}}].

\bibitem{Llinares:2013jua}
C.~Llinares and D.~F. Mota, {\it {Cosmological simulations of screened modified
  gravity out of the static approximation: effects on matter distribution}},
  {\em Phys.Rev.} {\bf D89} (2014) 084023,
  [\href{http://arxiv.org/abs/1312.6016}{{\tt arXiv:1312.6016}}].

\bibitem{Gronke:2013mea}
M.~B. Gronke, C.~Llinares, and D.~F. Mota, {\it {Gravitational redshift
  profiles in the $f(R)$ and symmetron models}},  {\em Astron. Astrophys.} {\bf
  562} (2014) A9, [\href{http://arxiv.org/abs/1307.6994}{{\tt
  arXiv:1307.6994}}].

\bibitem{Gronke:2016lfd}
M.~Gronke, A.~Hammami, D.~F. Mota, and H.~A. Winther, {\it {Estimates of
  cluster masses in screened modified gravity}},  {\em Astron. Astrophys.} {\bf
  595} (2016) A78, [\href{http://arxiv.org/abs/1609.02937}{{\tt
  arXiv:1609.02937}}].

\bibitem{Gronke:2015ama}
M.~Gronke, D.~F. Mota, and H.~A. Winther, {\it {Universal predictions of
  screened modified gravity on cluster scales}},  {\em Astron. Astrophys.} {\bf
  583} (2015) A123, [\href{http://arxiv.org/abs/1505.07129}{{\tt
  arXiv:1505.07129}}].

\bibitem{Gronke:2014gaa}
M.~Gronke, C.~Llinares, D.~F. Mota, and H.~A. Winther, {\it {Halo velocity
  profiles in screened modified gravity theories}},  {\em Mon. Not. Roy.
  Astron. Soc.} {\bf 449} (2015), no.~3 2837--2844,
  [\href{http://arxiv.org/abs/1412.0066}{{\tt arXiv:1412.0066}}].

\bibitem{Llinares:2013jza}
C.~Llinares, D.~F. Mota, and H.~A. Winther, {\it {ISIS: a new N-body
  cosmological code with scalar fields based on RAMSES. Code presentation and
  application to the shapes of clusters}},  {\em Astron. Astrophys.} {\bf 562}
  (2014) A78, [\href{http://arxiv.org/abs/1307.6748}{{\tt arXiv:1307.6748}}].

\bibitem{Knollmann:2009pb}
S.~R. Knollmann and A.~Knebe, {\it {Ahf: Amiga's Halo Finder}},  {\em
  Astrophys. J. Suppl.} {\bf 182} (2009) 608--624,
  [\href{http://arxiv.org/abs/0904.3662}{{\tt arXiv:0904.3662}}].

\bibitem{Winther:2011qb}
H.~A. Winther, D.~F. Mota, and B.~Li, {\it {Environment Dependence of Dark
  Matter Halos in Symmetron Modified Gravity}},  {\em Astrophys. J.} {\bf 756}
  (2012) 166, [\href{http://arxiv.org/abs/1110.6438}{{\tt arXiv:1110.6438}}].

\bibitem{Zhang:2010hma}
Y.-Y. Zhang et~al., {\it {LoCuSS: A Comparison of Cluster Mass Measurements
  from XMM-Newton and Subaru - Testing Deviation from Hydrostatic Equilibrium
  and Non-Thermal Pressure Support}},  {\em Astrophys. J.} {\bf 711} (2010)
  1033--1043, [\href{http://arxiv.org/abs/1001.0780}{{\tt arXiv:1001.0780}}].

\bibitem{Mahdavi:2012zy}
A.~Mahdavi, H.~Hoekstra, A.~Babul, C.~Bildfell, T.~Jeltema, and J.~P. Henry,
  {\it {Joint Analysis of Cluster Observations: II. Chandra/XMM-Newton X-ray
  and Weak Lensing Scaling Relations for a Sample of 50 Rich Clusters of
  Galaxies}},  {\em Astrophys. J.} {\bf 767} (2013), no.~2 116,
  [\href{http://arxiv.org/abs/1210.3689}{{\tt arXiv:1210.3689}}].

\bibitem{Hammami:2015ela}
A.~Hammami and D.~F. Mota, {\it {Cosmological simulations with hydrodynamics of
  screened scalar-tensor gravity with non-universal coupling}},  {\em Astron.
  Astrophys.} {\bf 584} (2015) A57,
  [\href{http://arxiv.org/abs/1505.06803}{{\tt arXiv:1505.06803}}].

\bibitem{Hammami:2016npf}
A.~Hammami and D.~F. Mota, {\it {Probing modified gravity via the
  mass-temperature relation of galaxy clusters}},  {\em Astron. Astrophys.}
  {\bf 598} (2017) A132, [\href{http://arxiv.org/abs/1603.08662}{{\tt
  arXiv:1603.08662}}].

\bibitem{Hammami:2015iwa}
A.~Hammami, C.~Llinares, D.~F. Mota, and H.~A. Winther, {\it {Hydrodynamic
  Effects in the Symmetron and $f(R)$-gravity Models}},  {\em Mon. Not. Roy.
  Astron. Soc.} {\bf 449} (2015), no.~4 3635--3644,
  [\href{http://arxiv.org/abs/1503.02004}{{\tt arXiv:1503.02004}}].

\bibitem{Voivodic:2016kog}
R.~Voivodic, M.~Lima, C.~Llinares, and D.~F. Mota, {\it {Modelling Void
  Abundance in Modified Gravity}},  {\em Phys. Rev.} {\bf D95} (2017), no.~2
  024018, [\href{http://arxiv.org/abs/1609.02544}{{\tt arXiv:1609.02544}}].

\bibitem{Sheth:1999su}
R.~K. Sheth, H.~J. Mo, and G.~Tormen, {\it {Ellipsoidal collapse and an
  improved model for the number and spatial distribution of dark matter
  haloes}},  {\em Mon. Not. Roy. Astron. Soc.} {\bf 323} (2001) 1,
  [\href{http://arxiv.org/abs/astro-ph/9907024}{{\tt astro-ph/9907024}}].

\bibitem{Sheth:2003py}
R.~K. Sheth and R.~van~de Weygaert, {\it {A Hierarchy of voids: Much ado about
  nothing}},  {\em Mon. Not. Roy. Astron. Soc.} {\bf 350} (2004) 517,
  [\href{http://arxiv.org/abs/astro-ph/0311260}{{\tt astro-ph/0311260}}].

\bibitem{Jennings:2013nsa}
E.~Jennings, Y.~Li, and W.~Hu, {\it {The abundance of voids and the excursion
  set formalism}},  \href{http://arxiv.org/abs/1304.6087}{{\tt
  arXiv:1304.6087}}. [Mon. Not. Roy. Astron. Soc.434,2167(2013)].

\bibitem{Maggiore:2009rw}
M.~Maggiore and A.~Riotto, {\it {The Halo mass function from excursion set
  theory. II. The diffusing barrier}},  {\em Astrophys. J.} {\bf 717} (2010)
  515--525, [\href{http://arxiv.org/abs/0903.1250}{{\tt arXiv:0903.1250}}].

\bibitem{Zheng:2007zg}
Z.~Zheng, A.~L. Coil, and I.~Zehavi, {\it {Galaxy Evolution from Halo
  Occupation Distribution Modeling of DEEP2 and SDSS Galaxy Clustering}},  {\em
  Astrophys. J.} {\bf 667} (2007) 760--779,
  [\href{http://arxiv.org/abs/astro-ph/0703457}{{\tt astro-ph/0703457}}].

\bibitem{Nadathur:2015qua}
S.~Nadathur and S.~Hotchkiss, {\it {The nature of voids – I. Watershed void
  finders and their connection with theoretical models}},  {\em Mon. Not. Roy.
  Astron. Soc.} {\bf 454} (2015), no.~2 2228--2241,
  [\href{http://arxiv.org/abs/1504.06510}{{\tt arXiv:1504.06510}}].

\bibitem{Nadathur:2015lha}
S.~Nadathur and S.~Hotchkiss, {\it {The nature of voids – II. Tracing
  underdensities with biased galaxies}},  {\em Mon. Not. Roy. Astron. Soc.}
  {\bf 454} (2015), no.~1 889--901,
  [\href{http://arxiv.org/abs/1507.00197}{{\tt arXiv:1507.00197}}].

\bibitem{Nadathur:2014uua}
S.~Nadathur, S.~Hotchkiss, J.~M. Diego, I.~T. Iliev, S.~Gottlöber, W.~A.
  Watson, and G.~Yepes, {\it {Universal void density profiles from simulation
  and SDSS}},  {\em IAU Symp.} {\bf 308} (2014) 542--545,
  [\href{http://arxiv.org/abs/1412.8372}{{\tt arXiv:1412.8372}}].

\bibitem{Sutter:2014haa}
P.~M. Sutter, G.~Lavaux, N.~Hamaus, A.~Pisani, B.~D. Wandelt, M.~S. Warren,
  F.~Villaescusa-Navarro, P.~Zivick, Q.~Mao, and B.~B. Thompson, {\it {VIDE:
  The Void IDentification and Examination toolkit}},  {\em Astron. Comput.}
  {\bf 9} (2015) 1--9, [\href{http://arxiv.org/abs/1406.1191}{{\tt
  arXiv:1406.1191}}].

\bibitem{Mota:2012zw}
V.~Salzano, S.~Capozziello, N.~R. Napolitano, and D.~F. Mota, {\it {Unifying
  static analysis of gravitational structures with a scale-dependent scalar
  field gravity as an alternative to dark matter}},  {\em Astron. Astrophys.}
  {\bf 561} (2014) A131, [\href{http://arxiv.org/abs/1211.1019}{{\tt
  arXiv:1211.1019}}].

\bibitem{AristotleHeavens}
Aristotle, {\em {On the Heavens}}.
\newblock Volume VI, Aristotle in twenty-three volumes, Loeb classical library,
  translated by W. K. C. Guthrie. Harvard University Press, Cambridge, MA,
  1939.

\bibitem{PtolemyAlmagest}
{\em {Ptolemy's Almagest}}.
\newblock translated and annotated by Gerald J. Toomer. Princeton University
  Press, New Jersey, 1984.

\bibitem{HistoryArabicAstronomy}
G.~Saliba, {\em {History of Arabic Astronomy: Planetary Theories During the
  Golden Age of Islam}}.
\newblock New York University Press, New York, 1994.

\bibitem{HuffModernScience}
T.~E. Huff, {\em {The Rise of Early Modern Science: Islam, China and the
  West}}.
\newblock Cambridge University Press, Cambridge, 2003.

\bibitem{Revolutionibus}
N.~Copernicus, {\em {De Revolutionibus Orbium Coelestium}}.
\newblock Latin and English Edition, translated by Edward Rosen. Octavo
  Publishing, New York, 1999.

\bibitem{EncyclopaediaIslamic}
{N. K. Singh, M. Zaki Kirmani}, {\em {Encyclopaedia of Islamic Science and
  Scientists}}.
\newblock {Global Vision Publishing House}, New Delhi, 2005.

\bibitem{ThomasHeath1913}
{Thomas Heath}, {\em {Aristarchus of Samos: The Ancient Copernicus}}.
\newblock {Clarendon Press}, Oxford, 1913.

\bibitem{RosenAristarchus1978}
E.~Rosen, {\it {Aristarchus of Samos and Copernicus}},  {\em The Bulletin of
  the American Society of Papyrologists} {\bf 15} (1978), no.~1/2, 85--93.

\bibitem{KeplerAstronomia}
J.~Kepler, {\em {Astronomia Nova}}.
\newblock translated by W. H. Donahue. Green Lion Press, Michigan, 2015.

\bibitem{Newtonprincipia}
I.~Newton, {\em {Philosophiae Naturalis Principia Mathematica}}.
\newblock The third edition (1726), edited by Alexandre Koyr\'e and I. Bernard
  Cohen. Cambridge University Press, Cambridge, 1972.

\bibitem{NewtonLetterHooke}
{Isaac Newton}, {\em {Letter from Sir Isaac Newton to Robert Hooke, February 5,
  1675}}.
\newblock Historical Society of Pennsylvania, retrieved 7 June 2018.

\bibitem{Lorentzoriginal}
H.~A. Lorentz, {\it {La Th\'eorie Électromagn\'e tique de Maxwell et Son
  Application Aux Corps Mouvants}},  {\em Archives n\'eerlandaises des Sciences
  exactes et naturelles} (1892) 451.

\bibitem{Michelson:1887zz}
A.~A. Michelson and E.~W. Morley, {\it {On the Relative Motion of the Earth and
  the Luminiferous Ether}},  {\em Am. J. Sci.} {\bf 34} (1887) 333--345.

\bibitem{1880AnPar..15...23T}
F.~{Tisserand}, {\it {Les travaux de Le Verrier}},  {\em Annales de
  l'Observatoire de Paris} {\bf 15} (Jan., 1880) 23--43.

\bibitem{2005ASTPS...9.....H}
J.-P. {Hsu} and D.~{Fine}, {\em {100 Years Of Gravity And Accelerated Frames}}.
\newblock Advanced Series on Theoretical Physical Science. 2005.

\bibitem{1917SPAW.......142E}
A.~{Einstein}, {\it {Kosmologische Betrachtungen zur allgemeinen
  Relativit{\"a}tstheorie}},  {\em Sitzungsberichte der K{\"o}niglich
  Preu{\ss}ischen Akademie der Wissenschaften (Berlin} (Jan, 1917) 142--152.

\bibitem{Kazanas:1980tx}
D.~Kazanas, {\it {Dynamics of the Universe and Spontaneous Symmetry Breaking}},
   {\em Astrophys. J.} {\bf 241} (1980) L59--L63.

\bibitem{Linde:1981mu}
A.~D. Linde, {\it {A New Inflationary Universe Scenario: A Possible Solution of
  the Horizon, Flatness, Homogeneity, Isotropy and Primordial Monopole
  Problems}},  {\em Phys. Lett.} {\bf 108B} (1982) 389--393.

\bibitem{Ijjas:2014nta}
A.~Ijjas, P.~J. Steinhardt, and A.~Loeb, {\it {Inflationary schism}},  {\em
  Phys. Lett.} {\bf B736} (2014) 142--146,
  [\href{http://arxiv.org/abs/1402.6980}{{\tt arXiv:1402.6980}}].

\bibitem{Maldacena:1997re}
J.~M. Maldacena, {\it {The Large N limit of superconformal field theories and
  supergravity}},  {\em Int. J. Theor. Phys.} {\bf 38} (1999) 1113--1133,
  [\href{http://arxiv.org/abs/hep-th/9711200}{{\tt hep-th/9711200}}]. [Adv.
  Theor. Math. Phys. 2, 231 (1998)].

\bibitem{Aharony:1999ti}
O.~Aharony, S.~S. Gubser, J.~M. Maldacena, H.~Ooguri, and Y.~Oz, {\it {Large N
  field theories, string theory and gravity}},  {\em Phys. Rept.} {\bf 323}
  (2000) 183--386, [\href{http://arxiv.org/abs/hep-th/9905111}{{\tt
  hep-th/9905111}}].

\bibitem{Rovelli:1997yv}
C.~Rovelli, {\it {Loop quantum gravity}},  {\em Living Rev. Rel.} {\bf 1}
  (1998) 1, [\href{http://arxiv.org/abs/gr-qc/9710008}{{\tt gr-qc/9710008}}].

\bibitem{Nilles:1983ge}
H.~P. Nilles, {\it {Supersymmetry, Supergravity and Particle Physics}},  {\em
  Phys. Rept.} {\bf 110} (1984) 1--162.

\bibitem{AristotleMetaphysicsb}
Aristotle, {\em {Metaphysics}}.
\newblock Volume I, Book B, translated by W.D. Ross. Oxford University Press,
  Oxford, 1924.

\bibitem{AristotleMetaphysicsg}
Aristotle, {\em {Metaphysics}}.
\newblock Volume I, Book $\Gamma$, translated by W.D. Ross. Oxford University
  Press, Oxford, 1924.

\end{thebibliography}\endgroup
 
\newpage
\begin{theindex}

\addcontentsline{toc}{chapter}{Index}

\thispagestyle{empty}
   
   
     
 
 \item  1+3 formalism \pageref{1p3formaref1}
 \item 
1-particle distribution function  \pageref{Finslsdistributionref1}

   \item   $\alpha$-basis   \pageref{alphabasisrefs}
   \item  affine connection  \pageref{metaffconref1}, \pageref{metaffconref2}, 
 \pageref{metaffconref3},  \pageref{metaffconref4}

 \item   Aristotelian-Ptolemaic cosmology   \pageref{Aripstotelftref1}

 \item   atom interferometry  \pageref{atominteref1}

 \item autoparallel curves  \pageref{autoparallelsref1}
\item   averaged conservation laws    \pageref{Finslconservationnref1}

 \item beyond Horndeski theories 
 \pageref{beyondHorndeskiref2},  \pageref{beyondHorndeskiref1}, 
  \pageref{beyondHorndeskiref4}, 
 \pageref{beyondHorndeskiref5},  \pageref{beyondHorndeskiref6}
 
\item Bianchi identities  \pageref{Bianchiref1}

\item bigravity \pageref{bigravityrefs1},  \pageref{bigravityrefs2}

\item black hole 
 \pageref{BHref1}, \pageref{BHref2}, \pageref{BHref3}, \pageref{BHref4},
 \pageref{BHref5},  \pageref{BHref6},   \pageref{BHref7},   \pageref{BHref8}, 
 \pageref{BHref9},  \pageref{BHref10},  \pageref{BHref11},  \pageref{BHref12}

\item black hole hair  \pageref{hairtheref1}
 
    \item   Bounce cosmology   \pageref{Bouncerefs1},  \pageref{Bouncerefs2}, 
 \pageref{Bouncerefs3}
 

   \item braneworld cosmology \pageref{braneworldsrefs1}

\item  Brans-Dicke theory  \pageref{Bransref1},  \pageref{Bransref2},  
 \pageref{Bransref3},  \pageref{Bransref4},   \pageref{Bransref5},   
 \pageref{Bransref6}
   
   \item Casimir effect  \pageref{Casimirref1}

 \item  chameleon mechanism/models   \pageref{chameleonkiref2},  
 \pageref{chameleonkiref1},   \pageref{chameleonkiref3}, 
 \pageref{chameleonkiref4},   \pageref{chameleonkiref5},  
\pageref{chameleonkiref6}

\item  classicalizing gravity  \pageref{classicalizingprfs1}

 \item cosmic microwave background (CMB)  \pageref{CMBrefs1},  
 \pageref{CMBrefs2},  \pageref{CMBrefs3},   \pageref{CMBrefs4},  
\pageref{CMBrefs5},
 \pageref{CMBrefs6},  \pageref{CMBrefs7},   \pageref{CMBrefs8}, 
 \pageref{CMBrefs9},  \pageref{CMBrefs10},   \pageref{CMBrefs11}
  \item   cosmic shear  \pageref{shearref1},   \pageref{shearref2}

 \item cosmological data  analysis  \pageref{cosmoldatanalysefs1}
 \item  cosmological paradigm    \pageref{paradigmref1}
 
   \item cosmological surveys   \pageref{cosmosurveysefs1},  
 \pageref{cosmosurveysefs2}

 \item  compact objects    \pageref{compactobrefs1}, \pageref{compactobrefs2},  
 \pageref{compactobrefs3},  \pageref{compactobrefs4},  \pageref{compactobrefs5}
 \item  corpuscular gravity   \pageref{corpuscularrfs1}

\item cross-correlations of cosmological probes  \pageref{crpssprobesref1}, 
 \pageref{crpssprobesref2}

\item dark matter halo   \pageref{Darkmatterharef1}, 
 \pageref{Darkmatterharef2},   \pageref{Darkmatterharef3}
\item   deflection of light  \pageref{deflectionlirefs1}, 
 \pageref{deflectionlirefs2}
    \item  dipole moment tests   \pageref{dipolemomref1}
  
\item dwarf stars  \pageref{dwarfsref1}
 
 \item   dynamical system approach  \pageref{Dynamicalref1},  
 \pageref{Dynamicalref2}
 
\item  early dark energy  \pageref{earlyDEefs001}

 \item    Eddington-inspired Born-Infeld gravity \pageref{EddingtonBIref1}, 
 \pageref{EddingtonBIref2}

\item    effective field theory of dark energy    \pageref{eftrefs1},   
 \pageref{eftrefs2}
 \item effective Newton's constant   \pageref{Geffref2},  \pageref{Geffref1}, 
 \pageref{Geffref3},  \pageref{Geffref4}

\item  effective theories, \pageref{effectivetherrefs1}

  \item Einstein frame  \pageref{Einsteinfrref1}, \pageref{Einsteinfrref4},  
 \pageref{Einsteinfrref5},  \pageref{Einsteinfrref6},  \pageref{Einsteinfrref7}, 
 
 \pageref{Einsteinfrref8},  \pageref{Einsteinfrref9}
 \item   Einstein-Boltzmann codes for dark energy and modified gravity    
  \pageref{EinBoltcodesref1}
  \item  Einstein-Hilbert action  \pageref{EHactionref1}, \pageref{EHactionref2}
  
\item  energy-momentum tensor \pageref{Enmomtenref1},  \pageref{Enmomtenref2}, 
 \pageref{Enmomtenref3},  \pageref{Enmomtenref4}
  

  \item  E\"{o}t-Wash experiment  \pageref{Washref1}
    \item equivalence principle \pageref{equivprinref1}, 
 \pageref{equivprinref2}, 
 \pageref{equivprinref3},  \pageref{equivprinref4},  \pageref{equivprinref5},  
 \pageref{equivprinref6},  \pageref{equivprinref7},   \pageref{equivprinref9}, 
 \pageref{equivprinref10}
\subitem test of  \pageref{equivprinref8}

\item   extra dimensions \pageref{extradimrefs1}
 \item Fifth Force  \pageref{Fifthref1},   \pageref{Fifthref2},  
 \pageref{Fifthref3},  \pageref{Fifthref4},  \pageref{Fifthref5}, 
 \pageref{Fifthref6},  \pageref{Fifthref7}
 
 \item  Finsler gravity  \pageref{Finslergrref1}
\item  Finsler Ricci scalar  \pageref{Finslerriccigrref1}
\item  Finsler spacetimes  \pageref{Finslspacetibeccigrref1}
     \item   forecast constraints    \pageref{cforecastsefs1},  
 \pageref{cforecastsefs2}
 \item      $ f(\mathbb{Q})$ gravity    \pageref{fQref1}
 \item $f(R)$ gravity  \pageref{fRref1}, \pageref{fRref2},  \pageref{fRref3},  
 \pageref{fRref4},  \pageref{fRref5},  \pageref{fRref6},  \pageref{fRref7}, 
 \pageref{fRref8},  \pageref{fRref9}

   \item Friedmann-Lema\^{\i}tre-Robertson-Walker (FLRW) 
metrics  \pageref{FRWref1}

\item $f(\mathbb{T})$ gravity  \pageref{fTgraref1},  \pageref{fTgraref2}, 
 \pageref{fTgraref3}

\item galaxy clustering   \pageref{galaxyclurefs1},   \pageref{galaxyclurefs2}, 
 \pageref{galaxyclurefs3},  \pageref{galaxyclurefs4}
\item  gauge theory of gravity  \pageref{gaugetheref1},  \pageref{gaugetheref2}

\item  general relativity  \pageref{GRref}, \pageref{GRref2}
\item  
 generalized Galileons  \pageref{Galileonsref1}

 \item geodesic deviation
   \subitem in $f(R)$ gravity  \pageref{geodesicdevref1}
      \subitem in    $f(\mathbb{T})$ gravity   \pageref{geodesicdevfTref1}
    \subitem in   Finsler gravity   \pageref{geodeFinslerfTref1}
     \subitem in symmetric Teleparallel Equivalent of General Relativity  
 \pageref{geodesymmtegTref1}


\item geodesics  \pageref{geodesicsref1},  \pageref{geodesicsref2}

\item geometrical optics approximation   \pageref{geometricaloprfs1}
 
  \item  geometrical trinity   \pageref{trinityref1}

    \item   Gleyzes-Langlois-Piazza-Vernizzi (GLPV) theories  \pageref{GLPV1}
      \item  gravitational collapse  \pageref{citationalcollapss1}, 
 \pageref{citationalcollapss2}
  
 \item   gravitational wave anomalous speed  \pageref{anomGEspeedef1}

\item  gravitational wave  oscillations  \pageref{GWoscillationsdef1}
 \item   gravitational waves \pageref{gravitationalwavrefs1}, 
 \pageref{gravitationalwavrefs2},  \pageref{gravitationalwavrefs3}, 
 \pageref{gravitationalwavrefs4},  \pageref{gravitationalwavrefs5},  
 \pageref{gravitationalwavrefs6},   \pageref{gravitationalwavrefs7}, 
 \pageref{gravitationalwavrefs8},  \pageref{gravitationalwavrefs9}
 \item gravity’s rainbow    \pageref{rainbowref1}
    \item growth factor  \pageref{growthref2}
  \subitem in teleparallel gravity  \pageref{growthref1}
 

    \item  $ H_0$ tension  \pageref{H0tensionjref1},  \pageref{H0tensionjref2}, 
 
 \pageref{H0tensionjref3},  \pageref{H0tensionjref4}

\item helicity  \pageref{helicityref1}
\item   higher order statistics   \pageref{hoghstatisticsef1}
 \item Ho\v{r}ava-Lifshitz  \pageref{Lifshitzref1},  \pageref{Lifshitzref2}, 
 \pageref{Lifshitzref3}

\item  
 Horndeski theory  \pageref{Horndeskiref1},  \pageref{Horndeskiref2}, 
 \pageref{Horndeskiref3},  \pageref{Horndeskiref4}
\item   Hubble Constant \pageref{Hubbleefs1},  \pageref{Hubbleefs2}, 
 \pageref{Hubbleefs3}

\item   hybrid metric-Palatini gravity  \pageref{metricPalatress1}
 
 \item  hyperfluids  \pageref{hyperfluidsref1}
 
\item  image separation  \pageref{limagesepfs1}

  \item   inflation \pageref{inflationrefs1},  \pageref{inflationrefs2}  
 \item  integrated Sachs-Wolfe effect (ISW)  \pageref{ISWref1},  
 \pageref{ISWref2},  \pageref{ISWref3},  \pageref{ISWref4}
\item   interacting dark energy  \pageref{IDErefs1}

        \item Jordan frame   \pageref{Jordanrref1},  \pageref{Jordanrref2},  
 \pageref{Jordanrref3}, \pageref{Jordanrref4}, \pageref{Jordanrref5}, 
, \pageref{Jordanrref6}
\item kinetic gas    \pageref{Finslskineticrref1}

\item  laboratory constraints   \pageref{laboratoryref1}, 
\pageref{laboratoryref2}
  \item Lane-Emden equation  \pageref{maEmdenusref1}
 
\item  large-scale structure  \pageref{LSSefs1},  \pageref{LSSefs2},  
 \pageref{LSSefs3},  \pageref{LSSefs4},   \pageref{LSSefs5},  \pageref{LSSefs6}
  \item  late-time solutions  \pageref{latetimerefss1}
 
 \item  lens equation  \pageref{lensrefs1}
 \item    lensing, strong  \pageref{stronglensefs1},  \pageref{stronglensefs2}, 
 \pageref{stronglensefs3},  \pageref{stronglensefs4}
\item    lensing, weak  \pageref{weaklensefs1},  \pageref{weaklensefs2}, 
 \pageref{weaklensefs3},  \pageref{weaklensefs4},  \pageref{weaklensefs5}, 
 \pageref{weaklensefs7},  \pageref{weaklensefs6}
   
  \item limiting mass  \pageref{masslimitingusref1}

\item    LIGO/Virgo   \pageref{eLIGOfref1},   \pageref{eLIGOfref2}, 
 \pageref{eLIGOfref3},  \pageref{eLIGOfref4},   \pageref{eLIGOfref5}
\item  LISA  \pageref{LISAref1},  \pageref{LISAref2},  \pageref{LISAref3}

\item   local Lorentz invariance/violation  \pageref{loclinref1}, 
 \pageref{loclinref2},  \pageref{loclinref3},   \pageref{loclinref4},  
 \pageref{loclinref5},   \pageref{loclinref6},   \pageref{loclinref7}, 
 \pageref{loclinref8}  
  \item  Lovelock’s theorem  \pageref{Lovelocktheref1}
  
\item   machine learning  \pageref{machineref1}

\item  magnification ratio  \pageref{magnificationratrefs1}
    \item   mass profiles  \pageref{massprofef1}

    \item mass-radius relations  \pageref{massradiausref1}

 \item massive gravity  \pageref{massivegravrefs1}
 
 \item  metric-affine gravity  \pageref{metaffgrref1},  \pageref{metaffgrref2}, 
 \pageref{metaffgrref3},   \pageref{metaffgrref4}
  \item matter couplings  \pageref{mattercouplingsref1},  
 \pageref{mattercouplingsref2}

 \item     model-independent constraints  \pageref{madelindepeefs1},  
 \pageref{madelindepeefs2},  \pageref{madelindepeefs3}, 
 \pageref{madelindepeefs4},  \pageref{madelindepeefs5},  
\pageref{madelindepeefs6}

\item  modified luminosity distance  \pageref{luminositydisdef1}

  \item multipoles   \pageref{multipolesref2},  \pageref{multipolesref1},
 \pageref{multipolesref3}

\item neutron stars  \pageref{neutronstarsref1},  \pageref{neutronstarsref2}, 
 \pageref{neutronstarsref3},  \pageref{neutronstarsref4},
  \pageref{neutronstarsref5},  \pageref{neutronstarsref6}, 
 \pageref{neutronstarsref7},  \pageref{neutronstarsref8},  
 \pageref{neutronstarsref9},  \pageref{neutronstarsref10}
\item no hair theorem    \pageref{nohairtheref2},   

   \item  Noether symmetry approach  \pageref{NoetherTref1}
 \item non-local gravity    \pageref{nonloccurnref1}
    \subitem in torsional theories  \pageref{nonloctorref1}
 \item non-metricity \pageref{nonmetricityref1}, \pageref{nonmetricityref2}, 
 \pageref{nonmetricityref3},  \pageref{nonmetricityref4}

 \item non-Riemannian geometry 
 \pageref{nonRiemannianref1}, \pageref{nonRiemannianref2}, 
 \pageref{nonRiemannianref3},  \pageref{nonRiemannianref4}

 \item  number counts   \pageref{numbercountsefs1}
  
 \item observational tensions  \pageref{tensionsrefs1},  
\pageref{tensionsrefs2},   \pageref{tensionsrefs3}
 \item   observer space  \pageref{Finslsobserverrref1}

\item  Palatini formalism  \pageref{Palatiniformref1}, 
 \pageref{Palatiniformref2},  \pageref{Palatiniformref3}, 
 \pageref{Palatiniformref4},   \pageref{Palatiniformref5}

\item   paradigm shift   \pageref{paradishiftref1}
\item  parametrized post-Newtonian formalism   \pageref{PPNref1}
 \item parity-violating terms   \pageref{parityef1},  \pageref{parityef2}, 
 \pageref{parityef3}
 
\item peak counts   \pageref{peaksef1}
\item  phenomenological constraints   \pageref{gphenomenologicalprfs1},  
 \pageref{gphenomenologicalprfs2},   \pageref{gphenomenologicalprfs3},
 \pageref{gphenomenologicalprfs4},  \pageref{gphenomenologicalprfs5}

 \item   quantum  effects   \pageref{quanteffectsfs1},  
 \pageref{quanteffectsfs2},   \pageref{quanteffectsfs3}
 
 \item    quantum  gravity   \pageref{quantumgrefs1},  \pageref{quantumgrefs2}, 
 \pageref{quantumgrefs3}

  \item quantum metric gravity   \pageref{quantumetricf1}
 
\item  quantum Palatini gravity   \pageref{quantumPalef1}
 \item quasi-normal mode  \pageref{tuasnormiref1}
\item   relativistic effects  \pageref{relativistiffectcefs1}
  \item Riemann tensor \pageref{Riemanntenref}, \pageref{Riemanntenref2}, 
 \pageref{Riemanntenref3}

\item   $\sigma_8$ tension  \pageref{sigmaref1},  \pageref{sigmaref2}, 
 \pageref{sigmaref3},  \pageref{sigmaref4},   \pageref{sigmaref5},  
 \pageref{sigmaref6},  \pageref{sigmaref7},  \pageref{sigmaref8}

\item  scalar perturbations 
 \pageref{Scalarperfrref1}, \pageref{Scalarperfrref2}, 
\pageref{Scalarperfrref3}, 
 \pageref{Scalarperfrref4},  \pageref{Scalarperfrref5}, 
 \pageref{Scalarperfrref6},  \pageref{Scalarperfrref7}

\item   scalar-tensor theories   
 \pageref{Scalartensoref1}, \pageref{Scalartensoref2}, 
\pageref{Scalartensoref3}, 
 \pageref{Scalartensoref4},  \pageref{Scalartensoref5}

  \item scalarization   \pageref{scalarizationiref1}

 \item Schwarzschild metric \pageref{Schwarzschild1}
 \item  screening mechanisms   \pageref{screeningref1},  
\pageref{screeningref2}, 
 \pageref{screeningref3},  \pageref{screeningref4}
   
  \item  semiclassical gravity  \pageref{csemiclassicalrfs1}
 \item  
shift-symmetric theories  \pageref{siftsymref1}

\item     singularities
\subitem black hole  \pageref{singularitiesrefs2}
\subitem cosmological   \pageref{singularitiesrefs1}

\item   small-scale effects  \pageref{smallscaleeffetsfs1}
 
 \item  solar system constraints   \pageref{solarsystemref1},  
 \pageref{solarsystemref2},  \pageref{solarsystemref3}, 
 \pageref{solarsystemref4},  \pageref{solarsystemref5}, 
 \pageref{solarsystemref6},   \pageref{solarsystemref7}, 
 \pageref{solarsystemref8},  \pageref{solarsystemref9},  
\pageref{solarsystemref10}
      
\item  standard sirens  \pageref{earlyDEefs1},   \pageref{earlyDEefs2},   
 \pageref{earlyDEefs3},  \pageref{earlyDEefs4}
\item  stellar structure \pageref{stellarstrref1}
 \item Sturm-Liouville problem  \pageref{Liouvilleref1}
 \item  symmetron models   \pageref{symmetronref1}
 
 \item  tangent bundle \pageref{Finslcobundlelref1}
  \item  teleparallel cosmology   \pageref{telecosmoref1}
   \item   teleparallel covariance  \pageref{telecovarref1}
 \item teleparallel dark energy \pageref{teleDEref1}
 \item Teleparallel Equivalent of General Relativity (TEGR)  \pageref{TEGRref1}
\item teleparallel Horndeski gravity \pageref{teleHornref1}
\item  teleparallel gravity   \pageref{telegraref1},  \pageref{telegraref2}
\item  teleparallel gravity with  higher-order derivatives
     \pageref{higherordtorref1}

 \item   tensor perturbations \pageref{tensorpertref1}, 
 \pageref{tensorpertref2},   \pageref{tensorpertref3}
\item  tests of general relativity     \pageref{testsGRefs1},  
 \pageref{testsGRefs2}
 \item three hair relation  \pageref{threehairef1}
\item Tolman-Oppenheimer-Volkoff (TOV)  equations  \pageref{TOVtrref1}

  \item torsion  \pageref{torsionref1}, \pageref{torsionref4}, 
 \pageref{torsionref5},  \pageref{torsionref7},  \pageref{torsionref8}

\item  torsion 
gravity  \pageref{torsionref2}, \pageref{torsionref3}, \pageref{torsionref6}

 \item  universal relation  \pageref{cuniversalefs1}

 \item voids  \pageref{voidsaref1}

 \item  Wheeler-DeWitt equation  \pageref{DeWittref1},   \pageref{DeWittref2},  
 \pageref{DeWittref3}

      \item zero point energy  \pageref{zeropointref1}

\end{theindex}

\end{document}